\ifx\noparts\undefined
\else
\fi

\documentclass[reqno,11pt]{amsbook}
\usepackage{amsmidx}
\usepackage{graphicx}
\usepackage{amscd}
\usepackage{slashed}
\usepackage{amssymb}
\usepackage{esint}
\usepackage{enumitem}
\usepackage[off]{auto-pst-pdf} 
\usepackage{pst-grad} 
\usepackage{pst-plot} 
\usepackage[mathscr]{eucal}

\numberwithin{equation}{section}
\allowdisplaybreaks[1]

\title[The Continuum Limit of Causal Fermion Systems]{The Continuum Limit of Causal Fermion Systems}
\author[F.\ Finster]{\huge{Felix Finster}}
\date{\today}

\newtheorem{Def}{Definition}[section]
\newtheorem{Thm}[Def]{Theorem}
\newtheorem{Prp}[Def]{Proposition}
\newtheorem{Lemma}[Def]{Lemma}
\newtheorem{Remark}[Def]{Remark}
\newtheorem{Exercise}{Exercise}[chapter]
\newtheorem{Corollary}[Def]{Corollary}
\newtheorem{Example}[Def]{Example}
\newtheorem{Assumption}[Def]{Assumption}

\newcommand{\beq}{\begin{equation}}
\newcommand{\eeq}{\end{equation}}
\newcommand{\Proof}{\begin{proof}}
\newcommand{\QED}{\end{proof} \noindent}
\newcommand{\QEDrem}{\ \hspace*{1em} \hfill $\Diamond$}
\newcommand{\la}{\langle}
\newcommand{\ra}{\rangle}
\newcommand{\bra}{\mathopen{<}}
\newcommand{\ket}{\mathclose{>}}
\newcommand{\Sl}{\mbox{$\prec \!\!$ \nolinebreak}}
\newcommand{\Sr}{\mbox{\nolinebreak $\succ$}}

\newcommand{\C}{\mathbb{C}}
\newcommand{\R}{\mathbb{R}}
\newcommand{\1}{\mbox{\rm 1 \hspace{-1.05 em} 1}}
\newcommand{\Z}{\mathbb{Z}}
\newcommand{\N}{\mathbb{N}}
\newcommand{\G}{\mathscr{G}}
\newcommand{\Pdd}{\slashed{\partial}}
\newcommand{\iotaslsh}{\mbox{ $\!\!\iota$ \hspace{-1.02 em} $/$}}
\newcommand{\reg}{\text{\rm{reg}}}
\newcommand{\np}{n_{\mathrm{p}}}
\newcommand{\na}{n_{\mathrm{a}}}
\newcommand{\Dir}{{\mathcal{D}}}
\newcommand{\K}{{\mathscr{K}}}
\newcommand{\D}{\mathscr{D}}
\newcommand{\even}{\text{\rm{\tiny{even}}}}
\newcommand{\sproj}{\check{\pi}}

\newcommand{\A}{\mathscr{A}}
\newcommand{\mix}{\text{\rm{\tiny{mix}}}}
\newcommand{\free}{\text{\rm{\tiny{free}}}}
\newcommand{\Cl}{{\mathscr{C}}\ell}
\renewcommand{\H}{\mathscr{H}}
\newcommand{\F}{{\mathscr{F}}}
\newcommand{\Lin}{\text{\rm{L}}}
\newcommand{\PP}{\text{\rm{PP}}}
\newcommand{\itemD}{\item[{\raisebox{0.125em}{\tiny $\blacktriangleright$}}]}

\newcommand{\as}{{\mathfrak{a}}}
\newcommand{\bs}{{\mathfrak{b}}}
\DeclareMathOperator{\re}{Re}
\DeclareMathOperator{\im}{Im}
\DeclareMathOperator{\Tr}{Tr}
\DeclareMathOperator{\tr}{tr}
\DeclareMathOperator{\Pexp}{Pexp}
\DeclareMathOperator{\Pe}{Pe}
\DeclareMathOperator{\Symm}{Symm}
\DeclareMathOperator{\pr}{\text{\rm{pr}}}

\renewcommand{\O}{{\mathscr{O}}}
\renewcommand{\L}{{\mathcal{L}}}
\newcommand{\LDirac}{{\mathcal{L}}_{\text{\tiny{\rm{Dirac}}}}}
\newcommand{\LYM}{{\mathcal{L}}_{\text{\tiny{\rm{YM}}}}}
\newcommand{\LEH}{{\mathcal{L}}_{\text{\tiny{\rm{EH}}}}}
\newcommand{\Sact}{{\mathcal{S}}}
\newcommand{\T}{{\mathcal{T}}}
\newcommand\B{{\mathscr{B}}}
\newcommand{\U}{\text{\rm{U}}}
\newcommand{\SU}{\text{\rm{SU}}}
\renewcommand{\u}{\mathfrak{u}}
\newcommand{\su}{\mathfrak{su}}
\newcommand{\h}{{\mathfrak{h}}}
\newcommand{\g}{{\mathfrak{g}}}

\newcommand{\nf}{{\mathfrak{n}}}
\newcommand{\LR}{{L\!/\!R}}
\newcommand{\I}{{\mathbb{I}}}
\newcommand{\UMNS}{U_\text{\rm{\tiny{MNS}}}}
\newcommand{\UCKM}{U_\text{\rm{\tiny{CKM}}}}
\newcommand{\NORM}{\|}
\newcommand{\back}{\!}
\DeclareMathOperator{\norm}{| \hspace*{-0.1em}| \hspace*{-0.1em}|}
\DeclareMathOperator{\supp}{supp}
\newcommand{\pseudo}{\Gamma}
\newcommand{\scrM}{\myscr M}
\newcommand{\scrN}{\myscr N}
\newcommand{\hscrM}{\,\,\hat{\!\!\scrM}}

\newcommand{\f}{{\mathfrak{f}}}
\newcommand{\scrU}{{\mathscr{U}}}
\newcommand{\scrA}{{\mathscr{A}}}
\newcommand{\Sig}{\mathscr{S}}
\newcommand{\sea}{\text{\rm{sea}}}
\newcommand{\res}{\text{\rm{res}}}
\newcommand{\lec}{\text{\rm{le}}}
\newcommand{\hec}{\text{\rm{he}}}

\DeclareFontFamily{OT1}{rsfso}{}
\DeclareFontShape{OT1}{rsfso}{m}{n}{ <-7> rsfso5 <7-10> rsfso7 <10-> rsfso10}{}
\DeclareMathAlphabet{\myscr}{OT1}{rsfso}{m}{n}

\newcommand{\subsectionn}[1]{\subsection{#1}
\addcontentsline{toc}{section}{\tocsection {}{\thesubsection}
{\hspace*{0.5em} #1}}}

\newcommand{\nsubsubsection}[1]{\subsubsection*{\\[-0.5em] \noindent {\bf{#1}}}
\addcontentsline{toc}{section}{
{\hspace*{4.5em} #1}}}

\newcommand{\nindex}{\index{notation}}
\newcommand{\sindex}{\index{subject}}
\makeindex{notation}
\makeindex{subject}

\begin{document}

\frontmatter
\title{{{\vspace*{2.0cm}
\Huge{The Continuum Limit of \\ Causal Fermion Systems} \\[.7em]
{\Large{\bf{From Planck Scale Structures to Macroscopic Physics}}\\[12cm]}}}}
\maketitle


\noindent
Felix Finster \\
Fakult\"at f\"ur Mathematik \\
Universit\"at Regensburg \\
Regensburg \\
Germany \\[0.2em]
{\tt{finster@ur.de}} \\[16cm]

\setcounter{page}{4}

\mainmatter

\frontmatter
\setcounter{page}{5}

\chapter*{Preface to the First Online Edition}
In the two years since this book appeared, I got a lot of feedback 
which led to a number of small corrections and improvements.
Although all changes are minor, I decided to replace the book on the arXiv,
simply in order to improve readability.
I also took the opportunity to correct a few typos.
Apart from these small changes, the present online version coincides precisely
with the Springer book. In particular, all equation numbers are the same.

I am grateful to all the readers who gave me feedback.
More specifically, I would like to thank Jos{\'e} Isidro, Maximilian Jokel and Andreas Platzer
for helpful comments.

\hspace*{1cm}

\hfill Felix Finster, Berlin, October 2018

\chapter*{Preface}
This book is devoted to explaining how the causal action principle gives rise to
the interactions of the standard model plus gravity on the level of second-quantized fermionic
fields coupled to classical bosonic fields.
\sindex{quantum field!fermionic}%
\sindex{fermionic field!quantized}%
\sindex{classical field!bosonic}%
\sindex{bosonic field!classical}%
\sindex{gauge field!classical}%
It is the result of an endeavor which I was occupied with for many years.
Publishing the methods and results as a book gives me the opportunity to present the material in
a coherent and comprehensible way.

The four chapters of this book evolved differently.
Chapters~\ref{introduction} and~\ref{tools} are based on the notes
of my lecture ``The fermionic projector and causal variational principles''
given at the University of Regensburg in the summer semester 2014.
The intention of this lecture was to introduce the basic concepts. Most of the material in these two chapters
has been published previously, as is made clear in the text by references to the corresponding
research articles. We also included exercises in order to facilitate the self-study.
Chapters~\ref{sector}--\ref{quark}, however, are extended versions of
three consecutive research papers written in the years 2007-2014 (arXiv:0908.1542 [math-ph],
arXiv:1211.3351 [math-ph], arXiv:1409.2568 [math-ph]). Thus the results
of these chapters are new and have not been published elsewhere.
Similarly, the appendix is formed of the appendices of the above-mentioned papers
and also contains results of original research.

The fact that Chapters~\ref{sector}--\ref{quark} originated from separate research papers
is still visible in their style. In particular, each chapter has its own short introduction,
where the notation is fixed and some important formulas are stated.
Although this leads to some redundancy and a few repetitions, I decided to leave these
introductions unchanged, because they might help the reader to revisit the prerequisites
of each chapter.

We remark that, having the explicit analysis of the continuum limit in mind, the focus
of this book is on the computational side. This entails that more theoretical questions
like the existence and uniqueness of solutions of Cauchy problems
or the non-perturbative methods for constructing the fermionic projector are omitted.
To the reader interested in mathematical concepts 
from functional analysis and partial differential equations, we can recommend the
book ``An Introduction to the Fermionic Projector and Causal Fermion Systems''~\cite{intro}.
The intention is that the book~\cite{intro} explains the physical ideas in a non-technical way
and introduces the mathematical background from a conceptual point of view.
It also includes the non-perturbative construction of the fermionic projector in the presence of an
external potential and introduces spinors in curved space-time.
The present book, on the other hand, focuses on getting a rigorous connection between
causal fermion systems and physical systems in Minkowski space.
Here we also introduce the mathematical tools and give all the technical and computational details
needed for the analysis of the continuum limit. With this different perspective, the
two books should complement each other and when combined should give a
mathematically and physically convincing introduction to causal fermion systems
and to the analysis of the causal action principle in the continuum limit.

We point out that the connection to quantum field theory (in particular to second-quantized
bosonic fields) is not covered in this book. The reader interested in this direction
is referred to~\cite{qft} and~\cite{qftlimit}.
\sindex{quantum field!bosonic}%
\sindex{bosonic field!quantized}%
\sindex{gauge field!quantized}%

I would like to thank the participants of the spring school ``Causal fermion systems''
hold in Regensburg in March 2016 for their interest and feedback.
Moreover, I am grateful to David Cherney, Andreas Grotz, Christian Hainzl, Johannes Kleiner, Simone Murro, 
Joel Smoller and Alexander Strohmaier for helpful discussions and valuable comments on the manuscript.
Special thanks goes to Johannes Kleiner for suggesting many of the exercises.
I would also like to thank the Max Planck Institute for Mathematics in the Sciences in Leipzig
and the Center of Mathematical Sciences and Applications at
Harvard University for hospitality while I was working on the manuscript.
I am grateful to the Deutsche Forschungsgemeinschaft (DFG) for financial support.

\hspace*{1cm}

\hfill Felix Finster, Regensburg, May 2016

\tableofcontents

\mainmatter
\setcounter{chapter}{0}

\chapter{Causal Fermion Systems -- An Overview}  \label{introduction}
Causal fermion systems were introduced in~\cite{rrev} as a reformulation and generalization
of the setting used in the fermionic projector approach~\cite{PFP}.
In the meantime, the theory of causal fermion systems has evolved to an approach to fundamental physics.
It gives quantum mechanics, general relativity and quantum field theory as limiting cases and is therefore a
candidate for a unified physical theory. In this chapter,
we introduce the mathematical framework
and give an overview of the different limiting cases.
The presentation is self-contained and includes references to
the corresponding research papers.
The aim is not only to convey the underlying physical picture,
but also to lay the mathematical foundations in a conceptually convincing way.
This includes technical issues like specifying the topologies on the different spaces of
functions and operators, giving a mathematical definition of an ultraviolet regularization,
or specifying the maps which identify the objects of the
causal fermion system with corresponding objects in Minkowski space.
Also, we use a basis-independent notation whenever possible.
The reader interested in a non-technical introduction is referred to~\cite{dice2014}.

\section{The Abstract Framework} \label{secframe}

\subsectionn{Basic Definitions} \label{secbasicdef}
For conceptual clarity, we begin with the general definitions.
\begin{Def} \label{defparticle} (causal fermion system) {\em{ \hspace*{1em}
Given a separable complex Hilbert space~$\H$ with scalar product~$\la .|. \ra_\H$
and a parameter~$n \in \N$ (the {\em{``spin dimension''}}), we let~$\F \subset \Lin(\H)$ be the set of all
self-adjoint operators on~$\H$ of finite rank, which (counting multiplicities) have
at most~$n$ positive and at most~$n$ negative eigenvalues. On~$\F$ we are given
a positive measure~$\rho$ (defined on a $\sigma$-algebra of subsets of~$\F$), the so-called
{\em{universal measure}}. We refer to~$(\H, \F, \rho)$ as a {\em{causal fermion system}}.
}}
\end{Def} \noindent
\sindex{causal fermion system}%
\sindex{spin dimension}%
\sindex{universal measure}%
\sindex{measure!universal}%
\nindex{a00@$(\H, \la . \vert . \ra_\H)$ -- complex Hilbert space}%
\nindex{a02@$\Lin(\H)$ -- bounded linear operators on~$\H$}%
\nindex{a04@$\F \subset \Lin(\H)$ -- set of linear operators on~$\H$}%
\nindex{a06@$\rho$ -- universal measure, measure on~$\F$}%
We remark that the separability of the Hilbert space (i.e.\ the assumption that~$\H$ admits an at most
countable Hilbert space basis) is not essential and could be left out.
We included the separability assumption because it seems to cover all cases of physical interest
and is useful if one wants to work with basis representations.
A simple example of a causal fermion system is given in Exercise~\ref{exm4}.
\sindex{Hilbert space!separable}%

As will be explained in detail in this book, a causal fermion system describes a space-time together
with all structures and objects therein (like the causal and metric structures, spinors
and interacting quantum fields).
In order to single out the physically admissible
causal fermion systems, one must formulate physical equations. To this end, we impose that
the universal measure should be a minimizer of the causal action principle,
which we now introduce. For any~$x, y \in \F$, the product~$x y$ is an operator
of rank at most~$2n$. We denote its non-trivial eigenvalues counting algebraic multiplicities
by~$\lambda^{xy}_1, \ldots, \lambda^{xy}_{2n} \in \C$
(more specifically,
denoting the rank of~$xy$ by~$k \leq 2n$, we choose~$\lambda^{xy}_1, \ldots, \lambda^{xy}_{k}$ as all
the non-zero eigenvalues and set~$\lambda^{xy}_{k+1}, \ldots, \lambda^{xy}_{2n}=0$).
\nindex{a08@$\lambda^{xy}_1, \ldots, \lambda^{xy}_{2n}$ -- non-trivial eigenvalues of~$xy$}%
We introduce the {\em{spectral weight}}~$| \,.\, |$
\nindex{aa1@$\vert \,.\, \vert$ -- spectral weight}%
\sindex{spectral weight}%
of an operator as the sum of the absolute values
of its eigenvalues. In particular, the spectral weights of the operator
products~$xy$ and~$(xy)^2$ are defined by
\[ |xy| = \sum_{i=1}^{2n} \big| \lambda^{xy}_i \big|
\qquad \text{and} \qquad \big| (xy)^2 \big| = \sum_{i=1}^{2n} \big| \lambda^{xy}_i \big|^2 \:. \]
We introduce the Lagrangian and the causal action by
\begin{align}
\text{\em{Lagrangian:}} && \L(x,y) &= \big| (xy)^2 \big| - \frac{1}{2n}\: |xy|^2 \label{Lagrange} \\
\text{\em{causal action:}} && \Sact(\rho) &= \iint_{\F \times \F} \L(x,y)\: d\rho(x)\, d\rho(y) \:. \label{Sdef}
\end{align}
\nindex{aa2@$\L(x,y)$ -- Lagrangian}%
\nindex{aa4@$\Sact(\rho)$ -- causal action}%
\sindex{Lagrangian}%
\sindex{causal action}%
\sindex{action!causal}%
The {\em{causal action principle}} is to minimize~$\Sact$ by varying the universal measure
\sindex{measure!universal|see{universal measure}}%
\sindex{universal measure!variation of}%
\sindex{variation of universal measure}%
under the following constraints:
\sindex{causal action principle}%
\sindex{causal variational principle}%
\sindex{variational principle!causal}%
\begin{align}
\text{\em{volume constraint:}} && \rho(\F) = \text{const} \quad\;\; & \label{volconstraint} \\
\text{\em{trace constraint:}} && \int_\F \tr(x)\: d\rho(x) = \text{const}& \label{trconstraint} \\
\text{\em{boundedness constraint:}} && \T(\rho) := \iint_{\F \times \F} |xy|^2\: d\rho(x)\, d\rho(y) &\leq C \:, \label{Tdef}
\end{align}
where~$C$ is a given parameter (and~$\tr$ denotes the trace of a linear operator on~$\H$).
\sindex{constraint!volume}%
\sindex{constraint!trace}%
\sindex{constraint!boundedness}%
\nindex{aa6@$\tr$ -- trace on~$\H$}%
\nindex{aa8@$\T(\rho)$ -- functional in boundedness constraint}%

In order to make the causal action principle mathematically well-defined, one needs
to specify the class of measures in which to vary~$\rho$. To this end,
on~$\F$ we consider the topology induced by the operator norm
\beq \label{supnorm}
\|A\| := \sup \big\{ \|A u \|_\H \text{ with } \| u \|_\H = 1 \big\} \:.
\eeq
\nindex{ab0@$\NORM . \NORM$ -- $\sup$-norm on~$\Lin(\H)$}%
In this topology, the Lagrangian as well as the integrands in~\eqref{trconstraint}
and~\eqref{Tdef} are continuous.
The $\sigma$-algebra generated by the open sets of~$\F$ consists of the so-called Borel sets.
A {\em{regular Borel measure}}
\sindex{measure!regular Borel}%
is a measure on the Borel sets with the property that
it is continuous under approximations by compact sets from inside and by open sets from outside
(for basics see for example~\cite[\S52]{halmosmt}).
The right prescription is to vary~$\rho$ within the class of regular Borel measures on~$\F$.
In the so-called {\em{finite-dimensional setting}}
\sindex{causal action principle!finite-dimensional setting}%
when~$\H$ is finite-dimensional and the total volume~$\rho(\F)$
is finite, the existence of minimizers is proven in~\cite{discrete, continuum}, and the properties of
minimizing measures are analyzed in~\cite{support, lagrange}.

The causal action principle is ill-posed if
the total volume~$\rho(\F)$ is finite and the Hilbert space~$\H$ is infinite-dimensional
(see Exercises~\ref{exm3} and~\ref{exm2}).
But the causal action principle does make mathematical sense in the so-called
{\em{infinite-dimensional setting}}
\sindex{causal action principle!infinite-dimensional setting}%
when~$\H$ is infinite-dimensional and the total volume~$\rho(\F)$
is infinite. In this case, the volume constraint~\eqref{volconstraint}
is implemented by demanding that all variations~$(\rho(\tau))_{\tau \in (-\varepsilon, \varepsilon)}$
should for all~$\tau, \tau' \in (-\varepsilon, \varepsilon)$ satisfy the conditions
\beq \label{totvol}
\big| \rho(\tau) - \rho(\tau') \big|(\F) < \infty \qquad \text{and} \qquad
\big( \rho(\tau) - \rho(\tau') \big) (\F) = 0
\eeq
(where~$|.|$ denotes the total variation of a measure; see~\cite[\S28]{halmosmt}).
The existence theory in the infinite-dimensional setting has not yet been developed.
But it is known that the Euler-Lagrange equations corresponding to the
causal action principle still have a mathematical meaning (as will be explained in~\S\ref{secvary}
below).
This will make it possible to analyze the causal action principle without restrictions on the
dimension of~$\H$ nor on the total volume.
One way of getting along without an existence theory in the infinite-dimensional
setting is to take the point of view that on a fundamental physical level,
the total volume is finite and
the Hilbert space~$\H$ is finite-dimensional, whereas the
infinite-dimensional setting merely is a mathematical idealization needed in
order to describe systems in infinite volume involving an infinite number of quantum particles.

We finally explain the significance of the constraints. Generally speaking,
the constraints~\eqref{volconstraint}--\eqref{Tdef} are needed to avoid trivial minimizers and in order for the variational principle to be well-posed.
More specifically, if we dropped the constraint of fixed total volume~\eqref{volconstraint}, the measure~$\rho=0$
would be a trivial minimizer. Without the boundedness constraint~\eqref{Tdef},
the loss of compactness discussed in~\cite[Section~2.2]{continuum} implies that no minimizers exist
(see Exercises~\ref{exm3} and~\ref{exm1}).
If, on the other hand, we dropped the trace constraint~\eqref{trconstraint},
a trivial minimizer could be constructed as follows: We let~$x$ be the operator with the matrix representation
\beq \label{xtrivial}
x = \text{diag} \big( \underbrace{1, \ldots, 1}_{\text{$n$ times}}, 
\underbrace{-1, \ldots, -1}_{\text{$n$ times}}, 0, 0, \ldots \big)
\eeq
and choose~$\rho$ as a multiple of the Dirac measure supported at~$x$.
Then~$\T > 0$ but~$\Sact=0$.

\subsectionn{Space-Time and Causal Structure} \label{seccausal}
A causal fermion system~$(\H, \F, \rho)$ encodes a large amount of information.
In order to recover this information, one can for example form
products of linear operators in~$\F$, compute the eigenvalues of such operator products
and integrate expressions involving these eigenvalues with respect to the universal measure.
However, it is not obvious what all this information means. In order to clarify the situation, we now
introduce additional mathematical objects.
\sindex{inherent structures}%
These objects are {\em{inherent}} in the sense that we only use information already encoded
in the causal fermion system.

We define {\em{space-time}}, denoted by~$M$, as the support of the universal measure\footnote{The
{\em{support}} of a measure is defined as the complement of the largest open set of measure zero, i.e.
\[ \supp \rho := \F \setminus \bigcup \big\{ \text{$\Omega \subset \F$ \,\big|\,
$\Omega$ is open and $\rho(\Omega)=0$} \big\} \:. \]
It is by definition a closed set. This definition is illustrated in Exercise~\ref{exm41}.},
\sindex{space-time}%
\nindex{ab2@$\supp$ -- support of measure}%
\sindex{measure!support of}%
\nindex{ab4@$M := \text{supp} \rho$ -- space-time}%
\[ M := \text{supp}\, \rho \subset \F \:. \]
Thus the space-time points are symmetric linear operators on~$\H$.
On~$M$ we consider the topology induced by~$\F$ (generated by the $\sup$-norm~\eqref{supnorm}
on~$\Lin(\H)$). Moreover, the universal measure~$\rho|_M$ restricted to~$M$ can be regarded as a volume
measure on space-time. This makes space-time to a {\em{topological measure space}}.
\sindex{topological measure space}%
Furthermore, one has the following notion of causality:

\begin{Def} (causal structure) \label{def2}
\sindex{causality}%
\sindex{causal structure}%
{\em{ For any~$x, y \in \F$, the product~$x y$ is an operator
of rank at most~$2n$. We denote its non-trivial eigenvalues (counting algebraic multiplicities)
by~$\lambda^{xy}_1, \ldots, \lambda^{xy}_{2n}$.
\nindex{a08@$\lambda^{xy}_1, \ldots, \lambda^{xy}_{2n}$ -- non-trivial eigenvalues of~$xy$}%
The points~$x$ and~$y$ are
called {\em{spacelike}} separated if all the~$\lambda^{xy}_j$ have the same absolute value.
They are said to be {\em{timelike}} separated if the~$\lambda^{xy}_j$ are all real and do not all 
have the same absolute value.
In all other cases (i.e.\ if the~$\lambda^{xy}_j$ are not all real and do not all 
have the same absolute value),
the points~$x$ and~$y$ are said to be {\em{lightlike}} separated. }}
\end{Def} \noindent
\sindex{timelike}%
\sindex{spacelike}%
\sindex{lightlike}%
Restricting the causal structure of~$\F$ to~$M$, we get causal relations in space-time.
To avoid confusion, we remark that in earlier papers (see~\cite{lqg}, \cite{rrev})
a slightly different definition of the causal structure was used.
But the modified definition used here seems preferable.

The Lagrangian~\eqref{Lagrange} is compatible with the above notion of causality in the
following sense (the correspondence to the causal structure of Minkowski space and the notion
of lightlike separation will be explained in Section~\ref{seccorcs} below).
Suppose that two points~$x, y \in \F$ are spacelike separated.
Then the eigenvalues~$\lambda^{xy}_i$ all have the same absolute value.
Rewriting~\eqref{Lagrange} as
\beq \label{Ldiff}
\L = \sum_{i=1}^{2n} |\lambda^{xy}_i|^2 - \frac{1}{2n} \sum_{i,j=1}^{2n}
 |\lambda^{xy}_i|\: |\lambda^{xy}_j|
 = \frac{1}{4n} \sum_{i,j=1}^{2n} \Big( \big|\lambda^{xy}_i \big| - \big|\lambda^{xy}_j \big| \Big)^2 \:,
\eeq
one concludes that the Lagrangian vanishes. Thus pairs of points with spacelike
separation do not enter the action. This can be seen in analogy to the usual notion of causality where
points with spacelike separation cannot influence each other\footnote{For clarity, we point
out that our notion of causality does allow for nonlocal correlations and entanglement
between regions with space-like separation. This will become clear in~\S\ref{secwave}
and~\S\ref{secQFT}.}.
This analogy is the reason for the notion ``causal'' in ``causal fermion system''
and ``causal action principle.''

The above notion of causality is {\em{symmetric}} in~$x$ and~$y$, as we now explain.
Since the trace is invariant under cyclic permutations, we know that
\beq \label{trsymm}
\tr \big( (xy)^p \big) = \tr \big( x \,(yx)^{p-1}\, y \big) = \tr \big( (yx)^{p-1} \,yx \big)
= \tr \big( (yx)^p \big)
\eeq
(where $\tr$ again denotes the trace of a linear operator on~$\H$).
Since all our operators have finite rank, there is a finite-dimensional subspace~$I$
of~$\H$ such that~$xy$ maps~$I$ to itself and vanishes on the orthogonal complement of~$I$.
Then the non-trivial eigenvalues of the operator product~$xy$ are given
as the zeros of the characteristic polynomial of the restriction~$xy|_I : I \rightarrow I$.
The coefficients of this characteristic polynomial (like the trace, the determinant, etc.)
are symmetric polynomials in the eigenvalues and can therefore be expressed in terms of traces of
powers of~$xy$. 
As a consequence, the identity~\eqref{trsymm} implies that the operators~$xy$
and~$yx$ have the same characteristic polynomial and are thus
isospectral. This shows that our notions of causality are indeed symmetric in the
sense that~$x$ and~$y$ are spacelike separated if and only if~$y$ and~$x$ are
 (and similarly for timelike and lightlike separation). One also sees that the Lagrangian~$\L(x,y)$ is symmetric
in its two arguments.

A causal fermion system also distinguishes a {\em{direction of time}}.
\sindex{time direction}%
To this end, we let~$\pi_x$ be the orthogonal projection in~$\H$ on the subspace~$x(\H) \subset \H$
and introduce the functional
\beq \label{Cform}
{\mathscr{C}} \::\: M \times M \rightarrow \R\:,\qquad
{\mathscr{C}}(x, y) := i \tr \big( y\,x \,\pi_y\, \pi_x - x\,y\,\pi_x \,\pi_y \big)
\eeq
\nindex{ab8@${\mathscr{C}}(x, y)$ -- distinguishes time direction}%
(this functional was first stated in~\cite[Section~8.5]{topology}, motivated by constructions
in~\cite[Section~3.5]{lqg}).
Obviously, this functional is anti-symmetric in its two arguments.
This makes it possible to introduce the notions
\beq \label{tdir}
\left\{ \begin{array}{cl} \text{$y$ lies in the {\em{future}} of~$x$} &\quad \text{if~${\mathscr{C}}(x, y)>0$} \\[0.2em]
\text{$y$ lies in the {\em{past}} of~$x$} &\quad \text{if~${\mathscr{C}}(x, y)<0$}\:. \end{array} \right.
\eeq
By distinguishing a direction of time, we get a structure similar to
a causal set (see for example~\cite{sorkin}). But in contrast to a causal set, our notion of
``lies in the future of'' is not necessarily transitive.
This corresponds to our physical conception that the transitivity of the causal relations
could be violated both on the cosmological scale (there might be closed timelike curves)
and on the microscopic scale (there seems no compelling reason why the causal
relations should be transitive down to the Planck scale).
This is the reason why we consider other structures (namely the universal measure and the causal
action principle) as being more fundamental. In our setting, causality merely is a
derived structure encoded in the causal fermion system.

In Exercise~\ref{exm5}, the causal structure is studied in the example of Exercise~\ref{exm4}.

\subsectionn{The Kernel of the Fermionic Projector} \label{secker}
The causal action principle depends crucially on the eigenvalues of the operator
product~$xy$ with~$x,y \in \F$. For computing these eigenvalues, it is convenient not to
consider this operator product on the (possibly infinite-dimensional) Hilbert space~$\H$, 
but instead to restrict attention to a finite-dimensional subspace of~$\H$, chosen such that
the operator product vanishes on the orthogonal complement of this subspace.
This construction leads us to the spin spaces and to the kernel of the fermionic projector,
which we now introduce.
For every~$x \in \F$ we define the {\em{spin space}}~$S_x$ by~$S_x = x(\H)$; it is a subspace of~$\H$ of dimension at most~$2n$. 
\sindex{spin space}%
\nindex{ac0@$S_x := x(\H)$ -- spin space}%
\nindex{ac2@$(S_xM, \Sl . \vert . \Sr_x)$ -- spin space}%
For any~$x, y \in M$ we define the
{\em{kernel of the fermionic operator}}~$P(x,y)$ by
\beq \label{Pxydef}
P(x,y) = \pi_x \,y|_{S_y} \::\: S_y \rightarrow S_x
\eeq
(where~$\pi_x$ is again the orthogonal projection on the subspace~$x(\H) \subset \H$).
\sindex{fermionic projector!kernel of}%
\sindex{fermionic operator!kernel of}%
\nindex{ac4@$\pi_x$ -- orthogonal projection on spin space}%
\nindex{ac6@$P(x,y)$ -- kernel of fermionic projector}%
Taking the trace of~\eqref{Pxydef} in the case~$x=y$, one
finds that~$\tr(x) = \Tr_{S_x}(P(x,x))$,
\nindex{ac8@$\Tr_{S_x}$ -- trace on spin space}%
making it possible to express
the integrand of the trace constraint~\eqref{trconstraint} in terms of the kernel of the fermionic operator.
In order to also express the eigenvalues of the operator~$xy$, we introduce the
{\em{closed chain}}~$A_{xy}$ as the product
\sindex{closed chain}%
\nindex{ad0@$A_{xy}$ -- closed chain}%
\beq \label{Axydef}
A_{xy} = P(x,y)\, P(y,x) \::\: S_x \rightarrow S_x\:.
\eeq
Computing powers of the closed chain, one obtains
\[ A_{xy} = (\pi_x y)(\pi_y x)|_{S_x} = \pi_x\, yx|_{S_x} \:,\qquad
(A_{xy})^p = \pi_x\, (yx)^p|_{S_x} \:. \]
Taking the trace, one sees in particular that~$\Tr_{S_x}(A_{xy}^p) = \tr \big((yx)^p \big)$.
Repeating the arguments after~\eqref{trsymm}, one concludes that
the eigenvalues of the closed chain coincide with the non-trivial
eigenvalues~$\lambda^{xy}_1, \ldots, \lambda^{xy}_{2n}$ of the operator~$xy$ in Definition~\ref{def2}.
\nindex{a08@$\lambda^{xy}_1, \ldots, \lambda^{xy}_{2n}$ -- non-trivial eigenvalues of~$xy$}%
Therefore, the kernel of the fermionic operator encodes the causal structure of~$M$.
The main advantage of working with the kernel of the fermionic operator is that the closed chain~\eqref{Axydef}
is a linear operator on a vector space of dimension at most~$2n$, making it possible
to compute the~$\lambda^{xy}_1, \ldots, \lambda^{xy}_{2n}$ as the eigenvalues of 
a finite matrix. 

Next, it is very convenient to arrange that the kernel of the fermionic operator is symmetric in the sense that
\beq \label{Pxysymm}
P(x,y)^* = P(y,x) \:.
\eeq
To this end, one chooses on the spin space~$S_x$ the {\em{spin scalar product}} $\Sl .|. \Sr_x$ by
\beq \label{ssp}
\Sl u | v \Sr_x = -\la u | x v \ra_\H \qquad \text{(for all $u,v \in S_x$)}\:.
\eeq
\sindex{spin scalar product}%
\nindex{ad4@$\Sl . \vert . \Sr_x$ -- spin scalar product}%
\nindex{ac2@$(S_xM, \Sl . \vert . \Sr_x)$ -- spin space}%
Due to the factor~$x$ on the right, this definition really makes the kernel of the fermionic operator symmetric,
as is verified by the computation
\begin{align*}
\Sl u \,|\, P(x,y) \,v \Sr_x &= - \la u \,|\, x\, P(x,y) \,v \ra_\H = - \la u \,|\, x y \,v \ra_\H \\
&= -\la \pi_y \,x\, u \,|\, y \,v \ra_\H = \Sl P(y,x)\, u \,|\,  v \Sr_y
\end{align*}
(where~$u \in S_x$ and~$v \in S_y$).
The spin space~$(S_x, \Sl .|. \Sr_x)$ is an {\em{indefinite}} inner product of signature~$(p,q)$ with~$p,q \leq n$
(for textbooks on indefinite inner product spaces see~\cite{bognar, GLR}).
In this way, indefinite inner product spaces arise naturally when analyzing the
mathematical structure of the causal action principle.

The kernel of the fermionic operator as defined by~\eqref{Pxydef} is also referred to as
the kernel of the {\em{fermionic projector}},
\sindex{fermionic operator!kernel of}%
\sindex{fermionic projector!kernel of}%
provided that suitable normalization conditions are
satisfied. Different normalization conditions have been proposed and analyzed
(see the discussion in~\cite[Section~2.2]{norm}). More recently, it was observed in~\cite{noether} that
one of these normalization conditions is automatically satisfied if the universal measure
is a minimizer of the causal action principle (see~\S\ref{secnoether} below).
With this in mind, we no longer need to be so careful about the normalization.
For notational simplicity, we always refer to~$P(x,y)$ as the kernel of the fermionic projector.

\subsectionn{Wave Functions and Spinors} \label{secwave}
For clarity, we sometimes denote the spin space~$S_x$ at a space-time point~$x \in M$
by~$S_xM$.
\nindex{ac2@$(S_xM, \Sl . \vert . \Sr_x)$ -- spin space}%
A {\em{wave function}}~$\psi$
\sindex{wave function}%
is defined as a function
which to every~$x \in M$ associates a vector of the corresponding spin space,
\beq \label{psirep}
\psi \::\: M \rightarrow \H \qquad \text{with} \qquad \psi(x) \in S_xM \quad \text{for all~$x \in M$}\:. 
\eeq
We now want to define what we mean by {\em{continuity}} of a wave function.
For the notion of continuity, we need to compare the wave function at different space-time points,
being vectors~$\psi(x) \in S_xM$ and~$\psi(y) \in S_yM$ in different spin spaces.
Using that both spin spaces~$S_xM$ and~$S_yM$ are subspaces of the same
Hilbert space~$\H$, an obvious idea is to simply work with the Hilbert space norm
$\|\psi(x) - \psi(y)\|_\H$. However, in view of the factor~$x$ in the spin scalar product~\eqref{ssp},
it is preferable to insert a corresponding power of the operator~$x$.
Namely, the natural norm on the spin space~$(S_x, \Sl .|. \Sr_x)$ is given by
\[ \big| \psi(x) \big|_x^2 := \big\la \psi(x) \,\big|\, |x|\, \psi(x) \big\ra_\H = \Big\| \sqrt{|x|} \,\psi(x) \Big\|_\H^2 \]
(where~$|x|$ is the absolute value of the symmetric operator~$x$ on~$\H$, and~$\sqrt{|x|}$
is the square root thereof).
\nindex{ae0@$\vert \,.\, \vert$ -- absolute value of symmetric operator on~$\H$}%
This leads us to defining that
the wave function~$\psi$ is {\em{continuous}} at~$x$ if
\sindex{wave function!continuous}
for every~$\varepsilon>0$ there is~$\delta>0$ such that
\[ \big\| \sqrt{|y|} \,\psi(y) -  \sqrt{|x|}\, \psi(x) \big\|_\H < \varepsilon
\qquad \text{for all~$y \in M$ with~$\|y-x\| \leq \delta$} \:. \]
Likewise, $\psi$ is said to be continuous on~$M$ if it is continuous at every~$x \in M$.
We denote the set of continuous wave functions by~$C^0(M, SM)$.
\nindex{ae2@$C^0(M, SM)$ -- continuous wave functions}%
Clearly, the space of continuous wave functions is a complex vector space with pointwise operations, i.e.\
$(\alpha \psi + \beta \phi)(x) := \alpha \psi(x) + \beta \phi(x)$ with~$\alpha, \beta \in \C$.

It is an important observation that every vector~$u \in \H$ of the Hilbert space gives rise to a unique
wave function. To obtain this wave function, denoted by~$\psi^u$, we simply project the vector~$u$
to the corresponding spin spaces,
\beq \label{psiudef}
\psi^u \::\: M \rightarrow \H\:,\qquad \psi^u(x) = \pi_x u \in S_xM \:.
\eeq
We refer to~$\psi^u$ as the {\em{physical wave function}} of~$u \in \H$.
\sindex{wave function!physical}
\nindex{ae4@$\psi^u$ -- physical wave function}
The estimate
\beq \label{ineqstar}
\begin{split}
\Big\| &\sqrt{|y|} \,\psi^u(y) -  \sqrt{|x|}\, \psi^u(x) \Big\|_\H = \Big\| \sqrt{|y|} \,u -  \sqrt{|x|}\,u \Big\|_\H \\
&\leq \Big\| \sqrt{|y|} - \sqrt{|x|} \Big\| \, \|u\|_\H \overset{(\star)}{\leq}
\|y-x\|^\frac{1}{4} \:\|y+x\|^\frac{1}{4} \: \|u\|_\H
\end{split}
\eeq
shows that~$\psi^u$ is indeed continuous
(for the inequality~$(\star)$ see Exercise~\ref{ex0}).
The physical picture is that the physical wave functions~$\psi^u$ are those wave functions
which are realized in the physical system.
Using a common physical notion, one could say that the vectors in~$\H$ correspond
to the ``occupied states'' of the system, and that an occupied state~$u \in \H$ is represented
in space-time by the corresponding physical wave function~$\psi^u$.
The shortcoming of this notion is that an ``occupied state'' is defined only
for free quantum fields, whereas the
physical wave functions are defined also in the interacting theory.
For this reason, we prefer not use the notion of ``occupied states.''

For a convenient notation, we also introduce the {\em{wave evaluation operator}}~$\Psi$
as an operator which to every Hilbert space vector associates the corresponding physical wave function,
\beq \label{weo}
\Psi \::\: \H \rightarrow C^0(M, SM)\:, \qquad u \mapsto \psi^u \:.
\eeq
\sindex{wave evaluation operator}
\nindex{ae6@$\Psi$ -- wave evaluation operator}
Evaluating at a fixed space-time point gives the mapping
\[ \Psi(x) \::\: \H \rightarrow S_xM\:, \qquad u \mapsto \psi^u(x) \:. \]
The kernel of the fermionic projector can be expressed in terms of the wave evaluation operator:
\begin{Lemma} \label{lemmaPxyrep} For any~$x, y \in M$,
\begin{align}
x &= - \Psi(x)^* \,\Psi(x) \label{Fid} \\
P(x,y) &= -\Psi(x)\, \Psi(y)^*\:. \label{Pid}
\end{align}
\nindex{ac6@$P(x,y)$ -- kernel of fermionic projector}%
\end{Lemma}
\Proof For any~$v \in S_xM$ and~$u \in \H$,
\[ \Sl v \,|\, \Psi(x)\, u \Sr_x = \Sl v \,|\, \pi_x\, u \Sr_x
\overset{\eqref{ssp}}{=} -\la v \,|\, x\, u \ra_\H = \la (-x)\, v \,|\, u \ra_\H \]
and thus
\[ \Psi(x)^* = -x|_{S_xM} \::\: S_xM \rightarrow \H \:. \]
Hence
\[ \Psi(x)^* \,\Psi(x) \,u = \Psi(x)^* \,\psi^u_x = -x \,\psi^u_x \overset{\eqref{psiudef}}{=} -x \,\pi_x u = -x u \:, \]
proving~\eqref{Fid}. Similarly, the relation~\eqref{Pid} follows from the
computation
\[ \Psi(x)\, \Psi(y)^* = -\pi_x\, y|_{S_y} = -P(x,y) \:. \]
This completes the proof.
\QED

The structure of the wave functions~\eqref{psirep} taking values in the spin spaces
is reminiscent of sections of a vector bundle. The only difference is that our setting
is more general in that the base space~$M$ does not need to be a manifold, and the
fibres~$S_xM$ do not need to depend smoothly on the base point~$x$.
However, comparing to the setting of spinors in Minkowski space or on a Lorentzian
manifold, one important structure is missing: we have no Dirac matrices
and no notion of Clifford multiplication. The following definition is a step towards
introducing these additional structures.

\begin{Def} (Clifford subspace) \label{defcliffsubspace} {\em{
We denote the space of symmetric linear operators on~$(S_x, \Sl .|. \Sr_x)$
by~$\Symm(S_x) \subset \Lin(S_x)$.
\nindex{af0@$\Symm(S_x)$ -- symmetric linear operators on~$S_x$}%
A subspace~$K \subset \Symm(S_x)$ is called
a {\em{Clifford subspace}} of signature~$(r,s)$ at the point~$x$ (with~$r,s \in \N_0$)
\sindex{Clifford subspace}%
if the following conditions hold:
\begin{itemize}
\item[(i)] For any~$u, v \in K$, the anti-commutator~$\{ u,v \} \equiv u v + v u$ is a multiple
of the identity on~$S_x$.
\item[(ii)] The bilinear form~$\la .,. \ra$ on~$K$ defined by
\nindex{af2@$\la .,. \ra$ -- inner product on Clifford subspace}%
\beq \label{anticomm}
\frac{1}{2} \left\{ u,v \right\} = \la u,v \ra \, \1 \qquad {\text{for all~$u,v \in K$}}
\eeq
is non-degenerate and has signature~$(r,s)$.
\end{itemize} }}
\end{Def}
In view of the anti-commutation relations~\eqref{anticomm}, a Clifford subspace can be
regarded as a generalization of the space spanned by the usual Dirac matrices.
However, the above definition has two shortcomings:
First, there are many different Clifford subspaces, so that there is no unique
notion of Clifford multiplication. Second, we are missing the structure of tangent vectors
as well as a mapping which would associate a tangent vector to an element of
the Clifford subspace.

These shortcomings can be overcome by using either geometric or measure-theoretic
methods. In the geometric approach, one gets along with the non-uniqueness of the Clifford subspaces
by working with suitable equivalence classes. Using geometric information
encoded in the causal fermion system, one can then construct mappings between
the equivalence classes at different space-time points. This method will be
outlined in~\S\ref{secgeom}. In the measure-theoretic approach, on the other hand, one
uses the local form of the universal measure with the aim of 
constructing a unique Clifford subspace at every space-time point.
This will be outlined in~\S\ref{sectopology}.
Before entering these geometric and measure-theoretic constructions, we
introduce additional structures on the space of wave functions.

\subsectionn{The Fermionic Projector on the Krein Space} \label{secKrein}
The space of wave functions can be endowed with an inner product and a topology.
The inner product is defined by
\beq \label{Sprod}
\bra \psi | \phi \ket = \int_M \Sl \psi(x) | \phi(x) \Sr_x \: d\rho(x) \:.
\eeq
\nindex{af4@$\bra . \vert . \ket$ -- inner product on Krein space}%
In order to ensure that the last integral converges, we also introduce the scalar product~$\la\!\la .|. \ra\!\ra$
by
\beq \label{ndef}
\la\!\la \psi | \phi \ra\!\ra = \int_M \la \psi(x) |\, |x|\, \phi(x) \ra_\H \:d\rho(x)
\eeq
\nindex{af6@$\la\back\la . \vert . \ra\back\ra$ -- scalar product on Krein space}%
(where~$|x|$ is again the absolute value of the symmetric operator~$x$ on~$\H$).
\nindex{ae0@$\vert \,.\, \vert$ -- absolute value of symmetric operator on~$\H$}%
The {\em{one-particle space}}~$(\K, \bra .|. \ket)$ is defined as the space of wave functions for which
the corresponding norm~$\norm . \norm$ is finite, with the topology induced by this norm, and endowed with
the inner product~$\bra .|. \ket$. Such an indefinite inner product space with a topology induced
by an additional scalar product is referred to as a {\em{Krein space}} (see for example~\cite{bognar, langer}).
\nindex{ag0@$(\K, \bra . \vert . \ket)$ -- Krein space}%
\sindex{Krein space}%

When working with the one-particle Krein space, one must keep in mind that
the physical wave function~$\psi^u$ of a vector~$u \in \H$ does not need to be a vector in~$\K$
because the corresponding integral in~\eqref{Sprod} may diverge. Similarly, the
scalar product~$\la\!\la \psi^u | \psi^u \ra\!\ra$ may be infinite.
One could impose conditions on the causal fermion system which ensure that the integrals
in~\eqref{Sprod} and~\eqref{ndef} are finite for all physical wave functions.
Then the mapping~$u \mapsto \psi^u$ would give rise to an embedding~$\H \hookrightarrow \K$
of the Hilbert space~$\H$ into the one-particle Krein space. However, such conditions seem too
restrictive and are not really needed. Therefore, here we shall not impose any conditions on the causal
fermion systems but simply keep in mind that the physical wave functions are in general no
Krein vectors.

Despite this shortcoming, the Krein space is useful because the kernel of the fermionic projector
gives rise to an operator on~$\K$. Namely, choosing a suitable dense domain of
definition\footnote{For example, one may choose~$\D(P)$ as the set of all vectors~$\psi \in \K$
satisfying the conditions
\[ \phi := \int_M x\, \psi(x)\, d\rho(x) \:\in \: \H \qquad \text{and} \qquad \norm \phi \norm < \infty\:. \]
}~$\D(P)$, we can regard~$P(x,y)$ as the integral kernel of a corresponding operator~$P$,
\beq \label{Pdef}
P \::\: \D(P) \subset \K \rightarrow \K \:,\qquad (P \psi)(x) =
\int_M P(x,y)\, \psi(y)\, d\rho(y)\:,
\eeq
referred to as the {\em{fermionic projector}}.
\nindex{ag1@$P$ -- fermionic projector}%
\sindex{fermionic projector}%
The fermionic projector has the following two useful
properties:
\begin{itemize}[leftmargin=2em]
\itemD $P$ is {\em{symmetric}} in the sense that~$\bra P \psi | \phi \ket = \bra \psi | P \phi \ket$
for all~$\psi, \phi \in \D(P)$: \\[0.2em] \label{ABdef}
The symmetry of the kernel of the fermionic projector~\eqref{Pxysymm} implies that
\[ \Sl P(x,y) \,\psi(y)\,|\, \psi(x) \Sr_x = \Sl \psi(y)\,|\, P(y,x) \,\psi(x) \Sr_y \:. \]
Integrating over~$x$ and~$y$ and
applying~\eqref{Pdef} and~\eqref{Sprod} gives the result.
\itemD $(-P)$ is {\em{positive}} in the sense that~$\bra \psi | (-P) \psi \ket \geq 0$~for
all $\psi \in \D(P)$: \\[0.2em]
This follows immediately from the calculation
\begin{align*}
\bra \psi | (-P) \psi \ket &= - \iint_{M \times M} \Sl \psi(x) \,|\, P(x,y)\, \psi(y) \Sr_x\:
d\rho(x)\, d\rho(y) \\
&= \iint_{M \times M} \la \psi(x) \,|\, x \, \pi_x \,y\, \psi(y) \ra_\H \:
d\rho(x)\, d\rho(y) = \la \phi | \phi \ra_\H \geq 0 \:,
\end{align*}
where we again used~\eqref{Sprod} and~\eqref{Pxydef} and set
\[ \phi = \int_M x\, \psi(x)\: d\rho(x)\:. \]
\end{itemize}
In Exercise~\ref{ex01} the wave functions and the Krein structure are
studied in the example of Exercise~\ref{exm4}.

\subsectionn{Geometric Structures} \label{secgeom}
A causal fermion system also encodes geometric information on space-time.
More specifically, in the paper~\cite{lqg} notions of connection and curvature are introduced
and analyzed. We now outline a few constructions from this paper.
Recall that the kernel of the fermionic projector~\eqref{Pxydef} is a mapping from
one spin space to another, thereby inducing relations between different space-time points.
The idea is to use these relations for the construction of a spin connection~$D_{x,y}$, being a unitary
mapping between the corresponding spin spaces,
\sindex{spin connection}%
\nindex{ag2@$D_{x,y}$ -- spin connection}%
\[ D_{x,y} \::\: S_y \rightarrow S_x \]
(we consistently use the notation that
the subscript~$_{xy}$ denotes an object at the point~$x$, whereas
the additional comma $_{x,y}$ denotes an operator which maps an object at~$y$
to an object at~$x$).
The simplest method for constructing the spin connection would be to
form a polar decomposition, $P(x,y) = A_{xy}^{-\frac{1}{2}}\,U$, and to
introduce the spin connection as the unitary part, $D_{x,y} = U$.
However, this method is too naive, because we want the spin connection to be compatible
with a corresponding metric connection~$\nabla_{x,y}$ which should map Clifford subspaces
at~$x$ and~$y$ (see Definition~\ref{defcliffsubspace} above) isometrically to each other.
A complication is that, as discussed at the end of~\S\ref{secwave},
the Clifford subspaces at~$x$ and~$y$ are not unique.
The method to bypass these problems is to work with several Clifford subspaces
and to use so-called splice maps, as we now briefly explain.

First, it is useful to restrict the freedom in choosing the Clifford subspaces
with the following construction.
Recall that for any~$x \in M$, the operator~$(-x)$ on~$\H$ has at most $n$~positive
and at most~$n$ negative eigenvalues. We denote its positive and negative spectral subspaces
by~$S_x^+$ and~$S_x^-$, respectively. In view of~\eqref{ssp}, these subspaces are also orthogonal
with respect to the spin scalar product,
\[ S_x = S_x^+ \oplus S_x^- \:. \]
We introduce the {\em{Euclidean sign operator}}~$s_x$ as a symmetric operator on~$S_x$
whose eigenspaces corresponding to the eigenvalues~$\pm 1$ are the spaces~$S_x^+$
and~$S_x^-$, respectively.
\sindex{sign operator!Euclidean}%
\nindex{ag4@$s_x$ -- Euclidean sign operator}%
Since~$s_x^2=\1$, the span of the Euclidean sign operator is a one-dimensional Clifford subspace of
signature~$(1,0$). The idea is to extend~$s_x$ to obtain higher-dimensional Clifford subspaces.
We thus define a {\em{Clifford extension}} as a Clifford subspace which contains~$s_x$.
\sindex{Clifford extension}%
By restricting attention to Clifford extensions, we have reduced the freedom in choosing
Clifford subspaces. However, still there is not a unique Clifford extension, even for fixed
dimension and signature. But one can define the {\em{tangent space}}~$T_x$
as an equivalence class of Clifford extensions; for details see~\cite[Section~3.1]{lqg}.
\sindex{tangent space}%
\nindex{ag6@$T_x$ -- tangent space}%
The bilinear form~$\la .,. \ra$ in~\eqref{anticomm} induces a Lorentzian metric on the
tangent space.
\nindex{af2@$\la .,. \ra$ -- inner product on Clifford subspace}%

Next, for our constructions to work, we need to
assume that the points~$x$ and~$y$ are both regular and are properly timelike
separated, defined as follows:
\begin{Def} \label{defregular} {\em{
A space-time point~$x \in M$ is said to be {\em{regular}} if~$x$ has the maximal possible rank,
i.e.~$\dim x(\H) = 2n$. Otherwise, the space-time point is called {\em{singular}}.}}
\end{Def} \noindent
\sindex{space-time point!regular}%
\sindex{space-time point!singular}%
In most situations of physical interest (like Dirac sea configurations to be
discussed in Sections~\ref{secmink} and~\ref{seclimit} below),
all space-time points are regular. Singular points, on the other hand,
should be regarded as exceptional points or ``singularities'' of space-time.

\begin{Def} \label{defproptl} {\em{
The space-time points~$x,y \in M$ are {\em{properly timelike}}
\sindex{timelike!properly}%
separated if the closed chain~$A_{xy}$, \eqref{Axydef}, has a strictly positive spectrum and if all
eigenspaces are definite subspaces of~$(S_x, \Sl .|. \Sr_x)$. }}
\end{Def} \noindent
By a definite subspace of~$S_x$ we mean a subspace on which
the inner product~$\Sl .|. \Sr_x$ is either positive or negative definite.

The two following observations explain why the last definition makes sense:
\begin{itemize}[leftmargin=2em]
\itemD Properly timelike separation implies timelike separation (see Definition~\ref{def2}): \\[0.2em]
Before entering the proof, we give a simple counter example which shows why the
assumption of definite eigenspaces in Definition~\ref{defproptl} is necessary for the implication
to hold. Namely, if the point~$x$ is regular and~$A_{xy}$ is the identity, then
the eigenvalues~$\lambda_1,\ldots, \lambda_{2n}$ are all strictly positive, but they are all equal.

If~$I \subset S_x$ is a definite invariant subspace of~$A_{xy}$,
then the restriction~$A_{xy}|_I$ is a symmetric operator on the Hilbert space~$(I, \pm \Sl .|. \Sr_{I \times I})$,
which is diagonalizable with real eigenvalues. Moreover, the orthogonal complement~$I^\perp$ of~$I \subset S_x$
is again invariant. If~$I^\perp$ is non-trivial, the restriction~$A_{xy}|_{I^\perp}$ has at least one eigenspace. Therefore, the assumption in Definition~\ref{defproptl} that all eigenspaces are definite makes it possible to proceed
inductively to conclude that the operator~$A_{xy}$ is diagonalizable and has real eigenvalues.

If~$x$ and~$y$ are properly timelike separated, then its eigenvalues are
by definition all real and positive. Thus it remains to show that they are not all the same.
If conversely they were all the same, i.e.~$\lambda_1 = \cdots = \lambda_{2n} = \lambda>0$,
then~$S_x$ would necessarily have the maximal dimension~$2n$.
Moreover, the fact that~$A_{xy}$ is diagonalizable implies that~$A_{xy}$ would be a multiple
of the identity on~$S_x$. Therefore, the spin space~$(S_x, \Sl .|. \Sr_x)$ would have to be definite, in contradiction to
the fact that it has signature~$(n,n)$.
\itemD The notion is symmetric in~$x$ and~$y$: \\[0.2em]
Suppose that~$A_{xy} u = \lambda u$ with~$u \in S_x$ and~$\lambda>0$.
Then the vector~$w := P(y,x) \,u \in S_y$ is an eigenvector of~$A_{yx}$ again to the
eigenvalue~$\lambda$,
\begin{align*}
A_{yx} \,w &= P(y,x) P(x,y) \,P(y,x) \, u \\
&= P(y,x) \,A_{xy} \,u = \lambda\, P(y,x)\, u = \lambda w \:.
\end{align*}
Moreover, the calculation
\begin{align*}
\lambda \,\Sl u | u \Sr &= \Sl u | A_{xy} u \Sr = \Sl u \,|\, P(x,y)\, P(y,x) \,u \Sr \\
&= \Sl P(y,x) u \,|\, P(y,x) u \Sr = \Sl w | w \Sr
\end{align*}
shows that~$w$ is a definite vector if and only if~$u$ is.
We conclude that~$A_{yx}$ has positive eigenvalues and definite eigenspaces
if and only if~$A_{xy}$ has these properties.
\end{itemize}

So far, the construction of the spin connection has been worked out only in the case
of spin dimension~$n=2$. Then for two regular and properly timelike separated points~$x,y \in M$,
the spin space~$S_x$ can be decomposed uniquely
into an orthogonal direct sum~$S_x = I^+ \oplus I^-$ of a two-dimensional
positive definite subspace~$I^+$ and a two-dimensional negative definite subspace~$I^-$ of~$A_{xy}$.
We define the {\em{directional sign operator}}~$v_{xy}$ of~$A_{xy}$ as the
\sindex{sign operator!directional}%
unique operator on~$S_xM$ such that
the eigenspaces corresponding to the eigenvalues~$\pm 1$
are the subspaces~$I^\pm$.

Having the Euclidean sign operator~$s_x$ and the directional sign operator~$v_{xy}$
to our disposal, under generic assumptions one can distinguish two Clifford subspaces at
the point~$x$: a Clifford subspace~$K_{xy}$ containing~$v_{xy}$ and a Clifford extension~$K_x^{(y)}$
(for details see~\cite[Lemma~3.12]{lqg}). Similarly, at the point~$y$ we have a
distinguished Clifford subspace~$K_{yx}$ (which contains~$v_{yx}$)
and a distinguished Clifford extension~$K_y^{(x)}$.
For the construction of the {\em{spin connection}}~$D_{x,y} : S_y \rightarrow S_x$ one works
\sindex{spin connection}%
\nindex{ag2@$D_{x,y}$ -- spin connection}%
with the Clifford subspaces~$K_{xy}$ and~$K_{yx}$ and demands that these are mapped
to each other. More precisely, the spin connection is uniquely characterized by the following
properties (see~\cite[Theorem~3.20]{lqg}):
\begin{itemize}
\item[(i)] $D_{x,y}$ is of the form
\[ D_{x,y} = e^{i \varphi_{xy}\, v_{xy}}\: A_{xy}^{-\frac{1}{2}}\: P(x,y) \quad \text{with} \quad
\varphi_{xy} \in \Big( -\frac{3\pi}{4}, -\frac{\pi}{2} \Big) \cup \Big( \frac{\pi}{2}, \frac{3\pi}{4} \Big) \:. \]
\item[(ii)] The spin connection maps the Clifford subspaces~$K_{xy}$ and~$K_{yx}$ to each other, i.e.\
\[ D_{y,x} \,K_{xy}\, D_{x,y} = K_{yx} \:. \]
\end{itemize}
The spin connection has the properties
\[ D_{y,x} = (D_{x,y})^{-1} = (D_{x,y})^* \qquad \text{and} \qquad
A_{xy} = D_{x,y}\, A_{yx}\, D_{y,x} \:. \]
All the assumptions needed for the construction of the spin connection are combined
in the notion that~$x$ and~$y$ must be {\em{spin-connectable}}
\sindex{spin-connectable}%
(see~\cite[Definition~3.17]{lqg}). We remark that in the limiting case of a Lorentzian manifold,
the points~$x$ and~$y$ are spin-connectable if they are timelike separated
and sufficiently close to each other (see~\cite[Section~5]{lqg}).

By composing the spin connection along a discrete ``path'' of space-time points,
one obtains a ``parallel transport'' of spinors.
When doing so, it is important to keep track of the different Clifford subspaces and to carefully transform
them to each other. In order to illustrate in an example how this works,
suppose that we want to compose the spin connection~$D_{y,z}$
with~$D_{z,x}$. As mentioned above, the spin connection~$D_{z,x}$ at the point~$z$
is constructed using the Clifford subspace~$K_{zx}$. The spin connection~$D_{y,z}$, however,
takes at the same space-time point~$z$ the Clifford subspace~$K_{zy}$ as reference.
This entails that before applying~$D_{y,z}$ we must transform
from the Clifford subspace~$K_{zx}$ to the Clifford subspace~$K_{zy}$.
This is accomplished by the {\em{splice map}}~$U_z^{(y|x)}$,
\sindex{splice map}%
being a uniquely defined unitary transformation of~$S_x$ with the property that
\[ K_{zy} = U_z^{(y|x)} \,K_{zx}\, \big( U_z^{(y|x)} \big)^* \:. \]
The splice map must be sandwiched between the spin connections in combinations like
\[ D_{y,z}\, U_z^{(y|x)} \,D_{z,x} \:. \]

In order to construct a corresponding metric connection~$\nabla_{x,y}$,
\nindex{ah2@$\nabla_{x,y}$ -- metric connection}%
one uses a similar procedure to
relate the Clifford subspaces to corresponding Clifford extensions. More precisely, one
first unitarily transforms the Clifford extension~$K_y^{(x)}$ to the Clifford subspace~$K_{yx}$.
Unitarily transforming with the spin connection~$D_{xy}$ gives the Clifford subspace~$K_{xy}$.
Finally, one unitarily transforms to the Clifford extension~$K_x^{(y)}$.
Since the Clifford extensions at the beginning and end are
representatives of the corresponding tangent spaces, we thus obtain an isometry
\[ \nabla_{x,y} \::\: T_y \rightarrow T_x \]
between the tangent spaces (for details see~\cite[Section~3.4]{lqg}).

In this setting, {\em{curvature}} is defined as usual as the holonomy of the connection.
Thus the curvature of the spin connection is given by
\[ \mathfrak{R}(x,y,z) = U_x^{(z|y)} \:D_{x,y}\: U_y^{(x|z)} \:D_{y,z}\: U_z^{(y|x)}
\:D_{z,x} \::\: S_x \rightarrow S_x \:, \]
\nindex{ah4@$\mathfrak{R}$ -- curvature of causal fermion system}%
\sindex{curvature!of causal fermion system}%
and similarly for the metric connection.
In~\cite[Sections~4 and~5]{lqg} it is proven that the above notions in fact
reduce to the spinorial Levi-Civita connection and the Riemannian curvature on a globally hyperbolic Lorentzian manifold if the causal fermion system is constructed by regularizing solutions of the Dirac equation
(similar as will explained in the next section for the Minkowski vacuum) and removing the regularization
in a suitable way. These results show that the notions of connection and curvature
defined above indeed generalize the corresponding notions in Lorentzian spin geometry.

\subsectionn{Topological Structures} \label{sectopology}
From a mathematical perspective, causal fermion systems provide a framework for
non-smooth geometries or generalized ``quantum geometries.''
In this context, it is of interest how the topological notions on a differentiable manifold
or a spin manifold generalize to causal fermion systems.
Such topological questions are analyzed in~\cite{topology}, as we now briefly summarize.

By definition, space-time~$M$ is a topological space (see~\S\ref{seccausal}).
Attaching to every space-time point~$x \in M$ the corresponding spin space~$S_x$
gives the structure of a {\em{sheaf}}, making it possible to describe the topology by sheaf cohomology.
\sindex{sheaf}%
If one assumes in addition that all space-time points are regular (see Definition~\ref{defregular}), then
all spin spaces are isomorphic, giving rise to a {\em{topological vector bundle}}.
\sindex{topological vector bundle}%

In order to get the connection to spinor bundles, one needs the additional structure of
Clifford multiplication. As explained in~\S\ref{secwave}, the notion of a Clifford subspace
(see Definition~\ref{defcliffsubspace}) makes it possible to define Clifford structures at
every space-time point, but the definition is not unique and does not give the
connection to tangent vectors of the base space.
In~\S\ref{secgeom} these shortcomings where bypassed by working with suitable equivalence
classes of Clifford subspaces.
From the topological point of view, the basic question is whether one can choose
a representative of this equivalence class at each space-time point in such a way that
the representative depends continuously on the base point.
This leads to the notion of a {\em{Clifford section}}~$\Cl$, being a continuous
\nindex{ah6@$\Cl$ -- Clifford section}%
\sindex{Clifford section}%
mapping which to every space-time point~$x \in M$ associates a corresponding Clifford
subspace~$\Cl_x$ (for details see~\cite[Section~4.1]{topology}).
Choosing a Clifford section leads to the structure of a so-called {\em{topological spinor bundle}}.
\sindex{topological spinor bundle}%
\sindex{Clifford subspace}%
An advantage of working with topological spinor bundles is
that no notion of differentiability is required.

If~$M$ has a differentiable structure, one would like to associate a tangent vector~$u \in T_xM$
to a corresponding element of the Clifford subspace~$\Cl_x$.
This leads to the notion of a {\em{spin structure}}~$\gamma$ on a topological spinor bundle,
being a continuous mapping which to every~$x \in M$ associates a mapping~$\gamma_x :
T_xM \rightarrow \Cl_x$.
\sindex{spin structure}%
The topological obstructions for the existence of a spin structure on a topological spinor bundle
generalize the spin condition on a spin manifold (for details see~\cite[Sections~4.2 and~4.5]{topology}).

A useful analytic tool for the construction of Clifford sections are so-called
{\em{tangent cone measures}} (see~\cite[Section~6]{topology}).
\sindex{tangent cone measure}%
These measures make it possible to analyze the local structure of space-time in a neighborhood of a
point~$x \in M$ (again without any differentiability assumptions).
The tangent cone measures can be used to distinguish a specific Clifford subspace~$\Cl_x$
and to relate~$\Cl_x$ to neighboring space-time points.

We close with two remarks. First, all the above constructions generalize to the
{\em{Riemannian setting}} if the definition of causal fermion systems is extended to
so-called {\em{topological fermion systems}} (see~\cite[Definition~2.1]{topology}).
\sindex{topological fermion system}%
We thus obtain a mathematical framework to describe {\em{spinors on singular spaces}}
(see~\cite[Sections~8 and~9]{topology} for many examples).
Second, one can introduce nontrivial topological notions even for discrete space-times
by constructing neighborhoods of~$M$ in~$\F$ (using the metric structure of~$\F$
induced by the norm on the Banach space~$\Lin(\H)$) and by studying the topology of these
neighborhoods (see~\cite[Section~9.4]{topology}).

\section{Correspondence to Minkowski Space} \label{secmink}
\sindex{correspondence to Minkowski space|(}%
In order to put the abstract framework into a simple and concrete context, we
now explain how to describe Dirac spinors in Minkowski space by a causal fermion system.

\subsectionn{Concepts Behind the Construction of Causal Fermion Systems}
\label{secuvintro}
We let~$(\scrM, \la .,. \ra)$ be Minkowski space 
(with the signature convention~$(+ - - -)$) and~$d\mu$ the standard volume measure
(thus~$d\mu = d^4x$ in a reference frame~$x= (x^0, \ldots, x^3)$).
\sindex{Minkowski space}%
\nindex{ah8@$(\scrM, \la .,. \ra)$ -- Minkowski space}%
\nindex{ai0@$d\mu = d^4x$ -- volume measure in Minkowski space}%
We denote the spinor space at a point~$x \in \scrM$ by~$S_x\scrM$,
\sindex{spinor space}%
\nindex{ai2@$(S_x\scrM, \Sl . \vert . \Sr_x)$ -- spinor space}%
so that a Dirac wave function~$\psi$
\sindex{Dirac wave function}%
\sindex{wave function!Dirac}%
takes values in
\[ \psi(x) \in S_x\scrM \simeq \C^4 \:. \]
The spinor space at~$x$ is endowed with an indefinite inner product of signature~$(2,2)$,
which as in physics textbooks we denote by~$\overline{\psi} \phi$
(where~$\overline{\psi} = \psi^\dagger \gamma^0$ is the usual adjoint spinor).
\sindex{adjoint spinor}%
\nindex{ai4@$\overline{\psi} \phi = \Sl \psi \vert \phi \Sr$ -- inner product on spinors}%
Clearly, in Minkowski space one has a trivial parallel transport of spinors, making it possible
to identify the spinor spaces at different space-time points. Thus the
space-time index~$S_x\scrM$ of the spinor space is added only for notational clarity.

We now consider solutions of the Dirac equation of mass~$m$,
\beq \label{Dirfree}
(i \gamma^j \partial_j - m) \psi = 0 \:.
\eeq
\sindex{Dirac equation!vacuum}%
\nindex{ai6@$\gamma^j$ -- Dirac matrices}%
For a solution~$\psi$, the function~$(\overline{\psi} \gamma^0 \psi)(t,\vec{x})$ has the interpretation as
the probability density of the Dirac particle at time~$t$ to be at the position~$\vec{x}$.
The spatial integral of this probability density is time independent
(for more details in the presence of an external potential see Exercise~\ref{ex2.3}).
Considering the bilinear form corresponding to this probability integral gives the
scalar product
\beq \label{sprodMin}
( \psi | \phi) := 2 \pi \int_{\R^3} (\overline{\psi} \gamma^0 \phi)(t, \vec{x})\: d^3x \:.
\eeq
\nindex{ai8@$(. \vert .)$ -- scalar product on Dirac wave functions}%
This scalar product is indeed independent of time and does not depend on
the choice of the reference frame.
In order to ensure that the integral in~\eqref{sprodMin} is well-defined and finite,
we first consider solutions which at time~$t$ are smooth and have compact support.
Taking the completion, the solution space becomes a separable Hilbert space.
We note that the factor~$2 \pi$ in~\eqref{sprodMin} is not quite standard,
but our convention has the advantage that many formulas become simpler.

Next, we choose~$\H$ as a closed subspace of this Hilbert space
\nindex{a00@$(\H, \la . \vert . \ra_\H)$ -- complex Hilbert space}%
with the induced scalar product~$\la .|. \ra_\H := (.|.)|_{\H \times \H}$.
Clearly, $\H$ is again a separable Hilbert space.
\sindex{Hilbert space!separable}%
In order to describe the vacuum (i.e.\ the physical system where no particles and anti-particles are present),
one chooses~$\H$ as the subspace spanned by all the negative-energy solutions (the ``Dirac sea vacuum'').
\sindex{Dirac sea configuration}%
\sindex{Dirac sea}%
To describe particles or anti-particles, one includes
positive-energy solutions or leaves out negative-energy solutions, respectively.
But any other closed subspace of the solution space may be chosen as well.
We remark for clarity that in this section, we only consider the vacuum
Dirac equation~\eqref{Dirfree}, so that
the Dirac particles do not interact (interacting systems will be discussed in Section~\ref{seclimit} below).

In order to get into the framework of causal fermion systems, to every space-time point~$x \in \scrM$
we want to associate a linear operator~$F(x) \in \F$. Once this has been accomplished, the resulting mapping
\beq \label{Fmap}
F \::\: \scrM \rightarrow \F
\eeq
\nindex{aj2@$F(x)$ -- local correlation operator}%
\sindex{local correlation operator}%
can be used to introduce a measure~$\rho$ on~$\F$. Namely, we say that a
subset~$\Omega \subset \F$
is measurable if and only if its pre-image~$F^{-1}(\Omega)$ is a measurable subset of~$\scrM$.
Moreover, we define the measure of~$\Omega$ as the space-time volume of the pre-image,
$\rho(\Omega) := \mu ( F^{-1}(\Omega) )$. This construction is commonly used in mathematical
analysis and is referred to as the {\em{push-forward measure}}, denoted by
\[ \rho= F_* \mu \]
\sindex{measure!push-forward}%
\nindex{aj4@$F_* \mu$ -- push-forward measure}%
(see for example~\cite[Section~3.6]{bogachev} or Exercise~\ref{exm41}~(b)).
Then~$(\H, \F, \rho)$ will be a causal fermion system.

The basic idea for constructing~$F(x)$ is to represent the inner product on the spinors 
at the space-time point~$x$ in terms of the Hilbert space scalar product, i.e.
\beq \label{Fdefnaive}
\la \psi | F(x) \phi \ra_\H =  -\overline{\psi(x)} \phi(x)  \qquad \text{for all~$\psi, \phi \in \H$}\:.
\eeq
The operator~$F(x)$ gives information on the densities and correlations of the
Dirac wave functions at the space-time point~$x$. It is referred to as the {\em{local correlation operator}}
at~$x$.
\nindex{aj6@$F(x)$ -- local correlation operator}%
\sindex{local correlation operator}%
Since the spinor space at~$x$ is four-dimensional, it follows that the operator~$F(x)$ has
rank at most four. Moreover, the fact that the spin scalar product has signature~$(2,2)$
implies that~$F(x)$ has at most two positive
and at most two negative eigenvalues. Therefore, the local correlation operator~$F(x)$ is indeed
an element of~$\F$ if we choose the spin dimension~$n=2$.
However, the equation~\eqref{Fdefnaive} is problematic
because Dirac solutions~$\psi, \phi \in \H$ are in general not continuous, so that the
pointwise evaluation on the right side of~\eqref{Fdefnaive} makes no mathematical sense.
This is the reason why we need to introduce an {\em{ultraviolet regularization}} (UV regularization).
\sindex{regularization!ultraviolet (UV)}%
Before entering the analysis, we first outline our method and
explain the physical picture in a few remarks. The mathematical construction will be
given afterwards in~\S\ref{secuvreg}.

In order to put our constructions in the general physical context, we first
note that UV regularizations are frequently used in relativistic quantum field theory
as a technical tool to remove divergences.
A common view is that the appearance of such divergences
indicates that the physical theory is incomplete and should be replaced for very small distances
by another, more fundamental theory.
The renormalization program is a method to get along with standard quantum field theory
by finding a way of dealing with the divergences.
The first step is the UV regularization,
which is usually a set of prescriptions which make divergent integrals finite.
The next step of the renormalization program is to show that the 
UV regularization can be taken out if other parameters of the theory (like masses
and coupling constants) are suitably rescaled.
Conceptually, in the renormalization program the UV regularization merely is a technical tool.
All predictions of theory should be independent of how the regularization is carried out.

In the context of causal fermion systems, however, the physical picture behind the UV regularization
is quite different. Namely, in our setting the {\em{regularized}} objects are to be considered as the
fundamental physical objects. Therefore, the regularization has a physical significance. It
should describe the microscopic structure of physical space-time.

Before explaining this physical picture in more detail, we need to introduce a microscopic length
scale~$\varepsilon>0$ on which the UV regularization should come into play.
\nindex{aj8@$\varepsilon$ -- regularization length}%
\sindex{regularization length}%
Regularization lengths are often associated to the Planck length
$\ell_P \approx1.6 \cdot 10^{-35} \:{\mbox{m}}$.
\sindex{Planck length}%
\nindex{aj9@$\ell_P$ -- Planck length}%
The analysis of the gravitational field
in this book suggests that~$\varepsilon$ should be chosen even much smaller than the Planck length
(see Section~\ref{l:secgrav} and~\S\ref{q:secgrav}).
Even without entering a detailed discussion
of the length scales, it is clear that~$\varepsilon$ will be by many orders of magnitude smaller than most other
physical length scales of the system.
Therefore, it is a sensible method to analyze the causal action principle in the asymptotics
when~$\varepsilon$ is very small. In order to make such an asymptotics mathematically precise,
it is necessary to consider the {\em{regularization length}}~$\varepsilon$ as a
{\em{variable parameter}} taking values in an
interval~$(0, \varepsilon_{\max})$. Only for such a variable parameter,
it will be possible later in this book to 
analyze the asymptotics as~$\varepsilon \searrow 0$.

For any~$\varepsilon \in (0, \varepsilon_{\max})$, similar to~\eqref{Fmap}
we shall construct a mapping~$F^\varepsilon : \scrM \rightarrow \F$
by suitably inserting an UV regularization into~\eqref{Fdefnaive}.
Then we construct the corresponding universal measure
as the push-forward by~$F^\varepsilon$, i.e.
\beq \label{pushforward}
\rho^\varepsilon := F^\varepsilon_* \mu \:.
\eeq
This will give rise to a causal fermion system~$(\H, \F, \rho^\varepsilon)$.
We will also explain how to identify the objects in
Minkowski space with corresponding objects of the causal fermion system: \\

\begin{center}
\begin{tabular}{|c|c|}
\hline & \\[-0.8em]
$\quad$ {\bf{Minkowski space}} $\quad$ & {\bf{causal fermion system}} \\[0.2em]
\hline & \\[-0.8em]
space-time point~$x \in \scrM$ & space-time point~$x \in M^\varepsilon:=\supp \rho^\varepsilon$ \\[0.2em]
topology of~$\scrM$ & topology of~$M^\varepsilon$ \\[0.2em]
spinor space $S_x\scrM$ & spin space $S_xM^\varepsilon$ \\[0.2em]
causal structure of Minkowski space & causal structure of Definition~\ref{def2} \\[0.2em]
\hline
\end{tabular}
\end{center}
\hspace*{1em}
\nindex{ac2@$(S_xM, \Sl . \vert . \Sr_x)$ -- spin space}%
\nindex{ai2@$(S_x\scrM, \Sl . \vert . \Sr_x)$ -- spinor space}%

\noindent
With these identifications made, the structures of Minkowski space will no longer be needed.
They are encoded in the causal fermion system, and we may describe the physical space-time
exclusively by the causal fermion system. We consider the objects with UV regularization
as described by the causal fermion system as the fundamental physical objects.

In the following remarks we elaborate on the physical picture behind the UV regularization
and explain why our setting is sufficiently general to describe the physical situation we have in mind.
\begin{Remark} {\bf{(method of variable regularization)}} \label{remmvr} {\em{
\sindex{regularization!method of variable}%
As just explained, the only reason for considering a family of causal fermion systems
is to give the asymptotics~$\varepsilon \searrow 0$ a precise mathematical meaning.
But from the physical point of view, a specific regularization for a specific value of~$\varepsilon$
should be distinguished by the fact that the corresponding causal fermion system~$(\H, \F, \rho^\varepsilon)$
describes our physical space-time. We again point out that this concept is different from standard quantum field
theory, where the regularization merely is a technical tool used in order to remove divergences.
In our setting, the regularization has a physical significance. The {\em{regularized}} objects are to be
considered as the {\em{fundamental}} physical objects, and the regularization is a method to describe
the microscopic structure of physical space-time.

This concept immediately raises the question how the ``physical regularization'' should look like.
Generally speaking, the regularized space-time should look like Minkowski space down to distances
of the scale~$\varepsilon$. For distances smaller than~$\varepsilon$, the structure of
space-time may be completely different. The simplest method of regularizing is to ``smear out''
or ``mollify'' all wave functions on the scale~$\varepsilon$
(this corresponds to Example~\ref{exmollify} below). But it is also conceivable that
space-time has a non-trivial microstructure on the scale~$\varepsilon$, which cannot be
guessed or extrapolated from the structures of Minkowski space.
Since experiments on the length scale~$\varepsilon$ seem out of reach, it is completely
unknown what the microscopic structure of space-time is.
Nevertheless, we can hope that we can get along without knowing this micro-structure,
because the detailed form of this micro-structure might have no influence
on the effective physical equations which are valid on the energy scales accessible to experiments.
More precisely, the picture is that the general structure of the effective physical equations should be independent
of the micro-structure of space-time. Values of mass ratios or coupling constants, however, may well
depend on the micro-structure (a typical example is the gravitational constant, which is
closely tied to the Planck length, which in turn is related to~$\varepsilon$
as explained in~Section~\ref{l:secgrav} below).
In more general terms, the unknown micro-structure
of space-time should enter the effective physical equations only by a finite (hopefully
small) number of free parameters, which can then be taken as empirical free parameters
of the effective macroscopic theory.

Clearly, the above picture must be questioned and supported by mathematical results.
To this end, one needs to analyze in detail how the effective macroscopic theory
depends on the regularization. For this reason, it is not sufficient to consider a specific family of regularizations.
Instead, one must analyze a whole class of regularizations which is so large that it covers all relevant
regularization effects. This strategy is referred to as the
{\em{method of variable regularization}} (for a longer explanation see~\cite[Section~4.1]{PFP}).
It is the reason why in Definition~\ref{defreg} below we shall only state properties of the regularization,
but we do not specify how precisely it should look like.
}} \QEDrem
\end{Remark}

\begin{Remark} {\bf{(sequences of finite-dimensional regularizations)}} \label{remdiscrete}
{\em{ The critical reader may wonder why we consider a family of regularizations~$(\H, \F, \rho^\varepsilon)$
parametrized by a continuous parameter~$(0, \varepsilon_{\max})$.
Would it not be more suitable to consider instead a sequence of
causal fermion systems~$(\H_\ell, \F_\ell, \rho_\ell)$ which asymptotically as~$\ell \rightarrow \infty$
describes Minkowski space?
A related question is why we constructed the measure~$\rho$ as the push-forward of the
Lebesgue measure~\eqref{pushforward}. Would it not be better to work with more general
measures such as to allow for the possibility of discrete micro-structures?
The answer to these questions is that it is no loss of generality and a simply a matter
of convenience to work with the family~$(\H, \F, \rho^\varepsilon)$ with~$\varepsilon \in (0, \varepsilon_{\max})$,
as we now explain.

We first point out that we do not demand our family~$(\H, \F, \rho^\varepsilon)$
to be in any sense ``continuous'' in the parameter~$\varepsilon$. Therefore, one can
also describe a sequence~$(\H, \F, \rho_\ell)$ simply by choosing the family~$\rho^\varepsilon$ to be
piecewise constant, for example
\[ \rho^\varepsilon = \rho_\ell \qquad \text{if} \qquad \frac{1}{\ell} \leq \varepsilon < \frac{1}{\ell+1}\:. \]
Similarly, it is no loss of generality to take~$\rho$ as the push-forward measure of the Lebesgue measure
because~$F^\varepsilon(x)$ need not depend continuously on~$x \in M$.
For example, one can arrange a discrete space-time like a space-time lattice by
choosing~$F^\varepsilon$ as a mapping which is piecewise constant on little cubes of
Minkowski space. Clearly, this mapping is not continuous, but it is continuous almost everywhere. Moreover, 
its image is a discrete set, corresponding to a discrete micro-structure of space-time.
For the method for representing a general measure~$\rho$ as the push-forward of
for example the Lebesgue measure we refer the interested reader
to the proof of~\cite[Lemma~1.4]{continuum}.

The remaining question is why we may keep the Hilbert space~$\H$ fixed.
In particular, we noted in~\S\ref{secbasicdef} that the existence of minimizers
of the causal action principle has been proven only if~$\H$ is finite-dimensional.
Therefore, should one not consider a filtration~$\H_1 \subset \H_2 \subset \cdots \subset \H$
of~$\H$ by finite-dimensional subspaces? Indeed, from the conceptual point of view,
this would be the correct way to proceed.
Nevertheless, the following consideration explains why we can just as well replace all
the Hilbert spaces~$\H_\ell$ by the larger space~$\H$:
For a given causal fermion system~$(\H_\ell, \F_\ell, \rho_\ell)$
with~$\H_\ell \subset \H$, by extending all operators by zero to the orthogonal complement
of~$\H_\ell$, one obtains the so-called {\em{extended causal fermion system}}~$(\H, \F, \rho_\ell)$.
The fact that the causal fermion system was extended can still be seen by forming the
so-called {\em{effective Hilbert space}} as
\sindex{Hilbert space!effective}%
\[ \H^\text{eff} = \overline{ \text{span} \{ x(\H) \:|\: x \in \supp \rho \} }\:. \]
Namely, for an extended causal fermion system, the effective Hilbert space still is a subset
of the original Hilbert space, $\H^\text{eff} \subset \H_\ell$.
Moreover, the support of the extended causal fermion system is still contained
in~$\F_\ell \subset \Lin(\H_\ell)$. Therefore, we do not lose any
information by extending a causal fermion system. Conversely, when analyzing a causal fermion system,
it seems preferable to always make the Hilbert space as small as possible by taking~$\H^\text{eff}$
as the underlying Hilbert space.

The delicate point about extending causal fermion systems is that the causal action principle does
depend sensitively on the dimension of the underlying Hilbert space~$\H$.
More specifically, the infimum of the action is known to be strictly decreasing in the dimension of~$\H$
(see the estimates in~\cite[Lemma~5.1]{discrete}, which apply similarly in the 
more general setting of~\cite{continuum}).
Therefore, a minimizer~$\rho$ of the causal action principle will no longer be a minimizer
if the causal fermion system is extended.
However, the first order {\em{Euler-Lagrange equations}} (for details see~\S\ref{secvary} below)
are still satisfied for the extended causal fermion system,
and this is all we need for the analysis in this book.
Therefore, for convenience we fix the Hilbert space~$\H$ and consider
a family of causal fermion systems~$(\H, \F, \rho^\varepsilon)$ thereon.
In order for the causal action principle to be well-defined and for~$\rho^\varepsilon$ to be
a minimizer, one should replace~$\H$ by the corresponding effective Hilbert space~$\H^\text{eff}$,
which may depend on~$\varepsilon$ and should be arranged to be finite-dimensional.
For the analysis of the Euler-Lagrange equations, however, the restriction to~$\H^\text{eff}$ is unnecessary,
and it is preferable to work with the extended Hilbert space~$\H$.
}} \QEDrem
\end{Remark}

We finally remark that the hurried reader who wants to skip the following constructions
may read instead the introductory section~\cite[Section~1.1]{rrev}
where formal considerations without UV regularization are given.
Moreover, a more explicit analysis of four-dimensional Minkowski space with a particularly
convenient regularization is presented in~\cite[Section~4]{lqg}. For a somewhat simpler analysis of
two-dimensional Minkowski space we refer to~\cite[Section~8.2]{topology}.

\subsectionn{Introducing an Ultraviolet Regularization} \label{secuvreg}
We now enter the construction of the UV regularization.
\sindex{regularization!ultraviolet (UV)}%
We denote the continuous Dirac wave functions (i.e.\ the continuous sections of the spinor bundle,
not necessarily solutions of the Dirac equation) by~$C^0(\scrM, S\scrM)$.
\nindex{ak4@$C^0(K, S\scrM)$ -- continuous Dirac wave functions on~$K \subset \scrM$}%
Similarly, the smooth wave functions with compact support in
a subset~$K \subset \scrM$ are denoted by~$C^\infty_0(K, S\scrM)$.
\nindex{ak6@$C^\infty_0(K, S\scrM)$ -- smooth and compactly supported
Dirac wave functions on~$K \subset \scrM$}%
For the $C^k$-norms we use the notation
\[ |\eta|_{C^k(K)} = \sum_{|\alpha| \leq k}\: \sup_{x \in K} |\partial^\alpha \eta(x)| \qquad
\text{for~$\eta \in C^\infty_0(K, S\scrM)$}\:, \]
where the~$\alpha$ are multi-indices.
\nindex{ak8@$\vert \,.\, \vert_{C^k(K)}$ -- $C^k$-norm on Dirac wave functions}%
Here~$|.|$ is any pointwise norm on the spinor spaces (we again
identify all spinor spaces via the trivial parallel transport). Since any two such norms
can be estimated from above and below by a constant,
the $C^k$-norms corresponding to different choices of the norms~$|.|$ are also equivalent.
For example, one can choose~$|\psi|^2 := \overline{\psi} \gamma^0 \psi$ similar to the
integrand in the scalar product~\eqref{sprodMin}. But clearly, other choices are possible just as well.

The UV regularization is performed most conveniently with so-called regularization operators,
which we now  define.
\begin{Def} \label{defreg} Consider a family of linear operators~$({\mathfrak{R}}_\varepsilon)$
with~$0 < \varepsilon < \varepsilon_{\max}$
which map~$\H$ to the continuous wave functions,
\[ {\mathfrak{R}}_\varepsilon \::\: \H \rightarrow C^0(\scrM, S\scrM) \:. \]
The family is called a family of {\bf{regularization operators}} if the following conditions hold:
\sindex{regularization operator}%
\nindex{al0@${\mathfrak{R}}_\varepsilon$ -- regularization operator}%
\begin{itemize}[leftmargin=2em]
\item[(i)] The image of every regularization operator is pointwise bounded,
meaning that for every~$\varepsilon \in (0, \varepsilon_{\max})$ and all~$x \in \scrM$
there is a constant~$c>0$ such that for all~$u \in \H$,
\beq \label{reges0}
\big| \big({\mathfrak{R}}_\varepsilon u \big)(x) \big| \leq c \:\|u\|_\H\ \:.
\eeq
\item[(ii)] The image of every regularization operator is equicontinuous almost everywhere
in the sense that
for every~$\varepsilon \in (0, \varepsilon_{\max})$, almost all~$x \in \scrM$ and every~$\delta>0$,
there is an open neighborhood~$U \subset \scrM$ of~$x$ such that for all~$u \in \H$ and all~$y \in U$,
\beq \label{reges1}
\big| \big({\mathfrak{R}}_\varepsilon u \big)(x) - \big({\mathfrak{R}}_\varepsilon u \big)(y) \big| \leq \delta \:\|u\|_\H\ \:.
\eeq
\item[(iii)] In the limit~$\varepsilon \searrow 0$, the family converges weakly to the identity,
meaning that for every compact subset~$K \subset \scrM$ and every~$\delta>0$
there is a constant~$\varepsilon_0>0$, such that
for all~$\varepsilon \in (0, \varepsilon_0)$, $u \in \H$ and~$\eta \in C^\infty_0(K, S\scrM)$,
\beq \label{reges2}
\Big| \int_\scrM \overline{\eta(x)} \big( {\mathfrak{R}}_\varepsilon(u) - u \big)(x)\: d^4x \Big| \leq \delta \:
\|u\|_\H\, |\eta|_{C^1(K)} \:.
\eeq
\end{itemize}
\end{Def} \noindent
We point out that we do not demand that the regularized wave function~${\mathfrak{R}}_\varepsilon \psi$
is again a solution of the Dirac equation. This could be imposed (as is done in~\cite[Section~4]{finite}),
but doing so seems too restrictive for the physical applications.
We also note that ``almost all'' in~(ii) refers to the standard volume measure~$d\mu$ on~$\scrM$.

For the mathematically interested reader we remark that the above properties~(i) and~(ii)
are very similar to the assumptions in the Arzel{\`a}-Ascoli theorem
(see for example~\cite[Section~VII.5]{dieudonne1} or~\cite[Theorem~7.25]{rudinprinciples}).
In fact, if we replaced ``almost all'' in~(ii) by ``all'', one could apply the Arzel{\`a}-Ascoli theorem
and restate the properties~(i) and~(ii) equivalently by
saying that taking the image~${\mathfrak{R}}_\varepsilon(B_1(0))$ of the unit ball in~$\H$
and restricting the resulting family of functions to any compact set~$K \subset \scrM$,
one obtains a relatively compact subset of~$C^0(K, S\scrM)$. It is remarkable that
the properties~(i) and~(ii) come up naturally as conditions for a sensible UV regularization,
although we shall never use compactness arguments in our proofs.
Weakening ``all'' by ``almost all'' in~(ii) makes it possible to describe discrete
space-times like space-time lattices, as was mentioned in Remark~\ref{remdiscrete} above.

Simple examples of regularization operators are obtained by mollifying the wave functions
on the scale~$\varepsilon$:
\begin{Example} {\bf{(regularization by mollification)}} \label{exmollify}
\sindex{regularization!by mollification}%
{\em{Let~$h \in C^\infty_0(\scrM, \R)$ be a non-negative test function with
\[ \int_\scrM h(x)\: d^4x=1 \:. \]
We define the operators~${\mathfrak{R}}_\varepsilon$ for~$\varepsilon>0$ as the
convolution operators
\[ ({\mathfrak{R}}_\varepsilon u)(x) := \frac{1}{\varepsilon^4}
\int_\scrM h\Big(\frac{x-y}{\varepsilon}\Big)\: u(y)\: d^4y \:. \]

Let us prove that the family~$({\mathfrak{R}}_\varepsilon)_{0<\varepsilon<1}$ is a family of regularization operators. First, 
\[ \big| \big( {\mathfrak{R}}_\varepsilon u \big)(x) \big|
\leq \frac{|h|_{C^0}}{\varepsilon^4}\: \int_K |u(y)|\: d^4y \leq \frac{|h|_{C^0}}{\varepsilon^4}\:\sqrt{\mu(K)}\:
\Big( \int_K |u(y)|^2\: d^4y \Big)^\frac{1}{2}\:, \]
where in the last step we used the Schwarz inequality.
We now rewrite the obtained space-time integral of~$|u|^2$ with the help of Fubini's theorem
as a bounded time integral and a spatial integral. In view of~\eqref{sprodMin},
the spatial integral can be estimated by the Hilbert space norm. We thus obtain
\beq \label{intes}
\int_K |u(y)|^2 \: d^4y \leq C \int_K \big(\overline{u} \gamma^0 u \big)(y)\: d^4y \leq
C \int_{t_0}^{t_1} \|u\|_\H^2 = C \,(t_1-t_0)\: \|u\|_\H^2 \:,
\eeq
where~$t_0$ and~$t_1$ are chosen such that~$K$ is contained in the time strip~$t_0 < t < t_1$. We conclude that
\[ \big| \big( {\mathfrak{R}}_\varepsilon u \big) \big|
\leq \frac{|h|_{C^0}}{\varepsilon^4}\:\sqrt{\mu(K)\:C\, (t_1-t_0)}\: \|u\|_\H^2 \:, \]
proving~\eqref{reges0}.

In order to derive the inequality~\eqref{reges1}, we begin with the estimate
\[ \big| \big( {\mathfrak{R}}_\varepsilon u \big)(x) - \big( {\mathfrak{R}}_\varepsilon u \big)(y) \big|
\leq \frac{1}{\varepsilon^4}\: \sup_{z \in \scrM} \Big|
h\Big(\frac{x-z}{\varepsilon}\Big) - h\Big(\frac{y-z}{\varepsilon}\Big) \Big|
\int_K |u(y)|\: d^4y \:. \]
Again applying~\eqref{intes} and using that~$h$ is uniformly continuous, one obtains~\eqref{reges1}.

It remains to prove~\eqref{reges2}. We first write the integral on the left as
\beq \label{vorhol}
\int_\scrM \overline{\eta(x)} \big( {\mathfrak{R}}_\varepsilon(u) - u \big)(x)\: d^4x
= \int_\scrM \overline{ \big( \eta_\varepsilon(y) - \eta(y) \big) }\: u(y)\: d^4y \:,
\eeq
where we set
\[ \eta_\varepsilon(y) = \frac{1}{\varepsilon^4} \int_\scrM \eta(x) \:h\Big(\frac{x-y}{\varepsilon}\Big)\: d^4x\:. \]
Now we use the standard estimate for convolutions
\begin{align*}
| & \eta_\varepsilon(y) - \eta(y)| = \frac{1}{\varepsilon^4}
\bigg| \int_\scrM \big( \eta(x)-\eta(y) \big)\:h\Big(\frac{x-y}{\varepsilon}\Big)\: d^4x \bigg| \\
&= \bigg| \int_\scrM \Big( \eta(y+\varepsilon z)-\eta(y) \Big)\:h(z)\: d^4z \bigg|
\leq |\eta|_{C^1(K)} \int_\scrM |\varepsilon z|\: \:h(z)\: d^4z
\end{align*}
(where in the last step we used the mean value theorem).
This gives rise to the estimate
\[ |\eta_\varepsilon - \eta|_{C^0(K)} \leq c\, \varepsilon\, |\eta|_{C^1(K)} \:, \]
where~$c$ may depend on~$K$ and the choice of~$h$, but is independent of~$\eta$.
This makes it possible to estimate~\eqref{vorhol} by
\[ \Big| \int_\scrM \overline{\eta(x)} \big( {\mathfrak{R}}_\varepsilon(u) - u \big)(x)\: d^4x \Big| \\
\leq \varepsilon\, |\eta|_{C^1(K)} \int_K |u(y)|_y \: d^4y \:. \]
Again applying~\eqref{intes}, we conclude that
\[ \Big| \int_\scrM \overline{\eta(x)} \big( {\mathfrak{R}}_\varepsilon(u) - u \big)(x)\: d^4x \Big|
\leq \delta\, |\eta|_{C^1(K)}\: \sqrt{\mu(K)} \: \sqrt{C\, (t_1-t_0)}\; \|u\|_\H \:, \]
proving~\eqref{reges2}.
}} \QEDrem
\end{Example}

Given a family of regularization operators, we can construct
causal fermion systems as follows. We fix~$\varepsilon \in (0, \varepsilon_{\max})$.
For any~$x \in \scrM$, we consider the bilinear form
\beq \label{bxdef}
b_x \::\: \H \times \H \rightarrow \C\:,\quad
b_x(u, v) = - \overline{({\mathfrak{R}}_\varepsilon \,u)(x)} ({\mathfrak{R}}_\varepsilon \,v)(x) \:.
\eeq
This bilinear form is well-defined and bounded because~${\mathfrak{R}}_\varepsilon$ 
is defined pointwise and because evaluation at~$x$ gives a linear operator of finite rank.
Thus for any~$v \in \H$, the anti-linear form~$b_x(.,v) : \H \rightarrow \C$
is continuous. By the Fr{\'e}chet-Riesz theorem (see for example~\cite[Section~6.3]{lax}),
there is a unique vector~$w \in \H$
such that~$b_x(u,v) = \la u | w \ra_\H$ for all~$u \in \H$.
The mapping~$v \mapsto w$ is linear and bounded. We thus obtain a bounded linear
operator~$F^\varepsilon(x)$ on~$\H$ such that
\[ b_x(u, v) = \la u \,|\, F^\varepsilon(x)\, v \ra_\H \qquad \text{for all~$u,v \in \H$}\:, \]
referred to as the {\em{local correlation operator}}.
\sindex{local correlation operator}%
\nindex{al2@$F^\varepsilon(x)$ -- local correlation operator with UV regularization}%
Taking into account that the inner product on the Dirac spinors at~$x$ has signature~$(2,2)$,
the local correlation operator~$F^\varepsilon(x)$ is a symmetric operator on~$\H$
of rank at most four, which has at most two positive and at most two negative eigenvalues.
Finally, we introduce the {\em{universal measure}}~$\rho^\varepsilon=
F^\varepsilon_* \mu$ as the push-forward
of the volume measure on~$\scrM$ under the mapping~$F^\varepsilon$.
\sindex{universal measure}%
\nindex{al4@$\rho^\varepsilon$ -- universal measure with UV regularization}%
In this way, for every~$\varepsilon \in (0, \varepsilon_0)$ we obtain 
a causal fermion system~$(\H, \F, \rho^\varepsilon)$ of spin dimension~$n=2$.

\subsectionn{Correspondence of Space-Time} \label{seccorst}
We now explain the connection between points of Minkowski space and points
of space-time~$M^\varepsilon := \supp \rho^\varepsilon$ of the corresponding
causal fermion system~$(\H, \F, \rho^\varepsilon)$. We begin with a general
characterization of~$M^\varepsilon$.
\nindex{al6@$M^\varepsilon$ -- regularized space-time}%
\sindex{correspondence to Minkowski space!$M^\varepsilon \leftrightarrow \scrM$}%

\begin{Prp} \label{prpscrMM}
For any~$\varepsilon \in (0, \varepsilon_{\max})$, there is a subset~$E \subset \scrM$
of $\mu$-measure zero such that the mapping~$F^\varepsilon|_{\scrM \setminus E} \::\: \scrM \setminus E \rightarrow
 \F$ is continuous. Moreover, the support of the universal
measure~$M^\varepsilon:= \supp \rho^\varepsilon$ is given by
\beq \label{suppprop}
M^\varepsilon = \overline{F^\varepsilon(\scrM \setminus E)}^{\Lin(\H)} \:.
\eeq
\end{Prp}
\Proof In order to show continuity, we need to estimate the sup-norm~$\|F^\varepsilon(x)-F^\varepsilon(y)\|$.
We first write the expectation value of the corresponding operator by
\begin{align*}
\la u &\,|\, \big(F^\varepsilon(x)-F^\varepsilon(y) \big) \,v \ra_\H
= - \overline{({\mathfrak{R}}_\varepsilon \,u)(x)} ({\mathfrak{R}}_\varepsilon \,v)(x)
+ \overline{({\mathfrak{R}}_\varepsilon \,u)(y)} ({\mathfrak{R}}_\varepsilon \,v)(y) \\
&= - \overline{({\mathfrak{R}}_\varepsilon \,u)(x)} \big( ({\mathfrak{R}}_\varepsilon \,v)(x) - ({\mathfrak{R}}_\varepsilon \,v)(y) \big) 
- \overline{ \big( ({\mathfrak{R}}_\varepsilon \,u)(x) - ({\mathfrak{R}}_\varepsilon \,u)(y) \big)} 
({\mathfrak{R}}_\varepsilon \,v)(y) \:,
\end{align*}
giving rise to the estimate
\begin{align*}
\big| \la u &\,|\, \big(F^\varepsilon(x)-F^\varepsilon(y) \big) v \ra_\H \big| \\
&\leq | ({\mathfrak{R}}_\varepsilon \,u)(x)|\: \big| ({\mathfrak{R}}_\varepsilon \,v)(x)
- {\mathfrak{R}}_\varepsilon \,v)(y) \big|
+ \big| ({\mathfrak{R}}_\varepsilon \,u)(x) - ({\mathfrak{R}}_\varepsilon \,u)(y) \big|\:
|({\mathfrak{R}}_\varepsilon \,v)(y)|\:.
\end{align*}

We now estimate the resulting spinor norms with the help of properties~(i) and~(ii)
of Definition~\ref{defreg}. First, we denote the exceptional set of $\mu$-measure zero where~\eqref{reges1}
does not hold by~$E \subset \scrM$. Combining~\eqref{reges0} and~\eqref{reges1},
one immediately sees that every point~$x \in \scrM \setminus E$ has a neighborhood~$U$
such that the boundedness property~\eqref{reges0} holds uniformly
on~$U$ (i.e.\ $|({\mathfrak{R}}_\varepsilon u)(y)| \leq c \,\|u\|_\H$ for all~$y \in U$).
We thus obtain the estimate
\begin{align*}
\big| \la u &\,|\, \big(F^\varepsilon(x)-F^\varepsilon(y) \big) v \ra_\H \big| \leq
2 c\, \delta \: \|u\|_\H\: \|v\|_\H \:,
\end{align*}
valid for all~$y \in U$ and~$u,v \in \H$.
Hence the sup-norm is bounded by~$\|F^\varepsilon(x)-F^\varepsilon(y)\| \leq 2 c \delta$,
showing that~$F^\varepsilon$ is continuous on~$\scrM \setminus E$.

It remains to prove~\eqref{suppprop}.
Since~$\mu(E)=0$, the set~$E$ can be disregarded when forming the push-forward measure.
Therefore, taking into account that the support of a measure is by definition a closed set,
it suffices to show that for every~$x \in \scrM \setminus E$,
the operator~$p:=F^\varepsilon(x)$ lies in the support of~$\rho^\varepsilon$.
Let~$U \subset \F$ be an open neighborhood of~$p$.
Then the continuity of~$F^\varepsilon$ at~$x$ implies that the
preimage~$(F^\varepsilon)^{-1}(U)$ is an open
subset of~$\scrM$. Hence the Lebesgue measure of this
subset is non-zero, $\mu((F^\varepsilon)^{-1}(U))>0$.
By definition of the push-forward measure, it follows that~$\rho^\varepsilon(U)>0$. 
Hence every neighborhood of~$p$ has a non-zero measure, implying that~$p \in \supp \rho^\varepsilon$.
This concludes the proof.
\QED

In order to have a convenient notation, in what follows we always identify a point
in Minkowski space with the corresponding operator of the causal fermion system,
\beq \label{Midentify}
\text{identify} \quad x \in \scrM \qquad \text{with} \qquad F^\varepsilon(x) \in \F \:.
\eeq
In general, this identification is not one-to-one, because the mapping~$F^\varepsilon$
need not be injective. In the latter case, there are two points~$x,y \in \scrM$
such that the bilinear forms~$b_x$ and~$b_y$ coincide (see~\eqref{bxdef}).
In other words, all correlations between regularized wave functions coincide
at the points~$x$ and~$y$.
Using a more physical language, this means that the points~$x, y$ of Minkowski space
are not distinguishable by any measurements performed on the fermionic wave functions.
We take the point of view that in such situations, the points~$x$ and~$y$ should not
be distinguished physically, and that it is reasonable and desirable that the two points
are identified in the causal fermion system with the same space-time point~$F^\varepsilon(x)
= F^\varepsilon(y) \in M^\varepsilon := \supp \rho^\varepsilon$.
In philosophical terms, our construction realizes the principle of the 
identity of indiscernibles.
\sindex{identity of indiscernibles, principle of}%

We also remark that, due to the closure in~\eqref{suppprop}, it may happen that
the space-time~$M^\varepsilon$ contains a point~$z$ which does {\em{not}} lie in the image of~$F^\varepsilon$,
but is merely an accumulation point in~$F^\varepsilon(\scrM)$. In this case, the corresponding bilinear
form~$b(u,v) := \la u | z v \ra_\H$ can be approximated with
an arbitrarily small error by bilinear forms~$b_x$ with~$x \in \scrM$.
Since experiments always involve small imprecisions, we take the point of view that
it is again reasonable and desirable mathematically to include~$z$ to the space-time points.

Generally speaking, the just-discussed cases that~$F^\varepsilon$ is not injective or its
image is not closed seem mostly of academic interest. 
In all applications in this book,
the mapping~$F^\varepsilon$ will be injective and closed. In all these situations,
Proposition~\ref{prpscrMM} will give us a one-to-one correspondence between
points~$x \in \scrM$ and points~$F^\varepsilon(x) \in M^\varepsilon$.

We finally note that, working with the push-forward measure~\eqref{pushforward},
the volume measure on space-time~$M^\varepsilon$ as defined by the universal measure~$d\rho^\varepsilon$
always agrees under the identification~\eqref{Midentify} with the Lebesgue measure~$d\mu$ on~$\scrM$.

\subsectionn{Correspondence of Spinors and Physical Wave Functions} \label{seccorsw}
We proceed by explaining the connection between the spinor space~$S_x\scrM$
at a point~$x \in \scrM$ of Minkowski space
and the corresponding spin space~$S_xM \subset \H$ of the causal fermion system
(where we use the identification~\eqref{Midentify}).
\nindex{ai2@$(S_x\scrM, \Sl . \vert . \Sr_x)$ -- spinor space}%
This will also make it possible to get a connection between Dirac wave functions
in Minkowski space and wave functions as defined in~\S\ref{secwave}.
In preparation, we derive useful explicit formulas for the local correlation
operators. To this end, for any~$x \in \scrM$ we define the {\em{evaluation map}}~$e_x^\varepsilon$ by
\sindex{evaluation map}%
\nindex{am0@$e_x^\varepsilon : \H \rightarrow S_x\scrM$ -- evaluation map}%
\beq \label{evalmap}
e^\varepsilon_x \::\: \H \rightarrow S_x\scrM \:,\qquad
e^\varepsilon_x \,\psi = ({\mathfrak{R}}_\varepsilon \psi)(x)\:.
\eeq
Its adjoint is defined as usual, taking into account the corresponding inner products on the
domain and the target space, i.e.
\[ \la (e^\varepsilon_x)^* \chi \,|\, \psi \ra_\H = \overline{\chi} \,\big( e^\varepsilon_x \,\psi)
\qquad \text{for all~$\chi \in S_x\scrM$}\:. \]
We denote this adjoint by~$\iota^\varepsilon_x$,
\[ \iota^\varepsilon_x := (e^\varepsilon_x)^* \::\: S_x \scrM \rightarrow \H\:. \]
\nindex{am2@$\iota^\varepsilon_x := (e^\varepsilon_x)^*$ -- adjoint of evaluation map}%
Multiplying~$e^\varepsilon_x$ by~$\iota^\varepsilon_x$ gives us
back the local correlation operator~$F^\varepsilon(x)$. Namely,
\begin{align*}
\la \psi \,|\, F^\varepsilon(x)\, \phi \ra_\H = 
- \overline{({\mathfrak{R}}_\varepsilon \,\psi)(x)} ({\mathfrak{R}}_\varepsilon \,\phi)(x)
= -\overline{\big( e^\varepsilon_x \psi \big)} \big(e^\varepsilon_x \phi \big)
= - \la \psi \,|\, \iota^\varepsilon_x e^\varepsilon_x \,\phi \ra_\H
\end{align*}
and thus
\beq \label{Fepsdef}
F^\varepsilon(x) = -\iota^\varepsilon_x \,e^\varepsilon_x
= -\iota^\varepsilon_x \,\big(\iota^\varepsilon_x)^* \::\: \H \rightarrow \H \:.
\eeq

The next proposition gives the desired connection between the spinor space~$S_x\scrM$
and the corresponding spin space~$S_xM$. We first state and prove the proposition
and explain it afterwards.
\sindex{correspondence to Minkowski space!$S_xM \leftrightarrow S_x\scrM$}%
\nindex{ac2@$(S_xM, \Sl . \vert . \Sr_x)$ -- spin space}%
\nindex{ai2@$(S_x\scrM, \Sl . \vert . \Sr_x)$ -- spinor space}%
\begin{Prp} \label{prpisometry}
The mapping
\sindex{evaluation map}%
\[ e^\varepsilon_x|_{S_x} \::\: S_xM \rightarrow S_x \scrM \quad
\text{is an isometric embedding}\:. \]

If the point~$x$ is regular (see Definition~\ref{defregular}),
then this mapping is an isomorphism. Moreover,
the physical wave function of a vector~$u$
at~$x$ is mapped to the regularized Dirac wave function at~$x$,
\beq \label{waveagree}
e^\varepsilon_x|_{S_x}\, \psi^u(x) = \big({\mathfrak{R}}_\varepsilon u \big)(x) \:.
\eeq
Finally, the inverse of~$e^\varepsilon_x|_{S_x}$ is given by
\beq \label{invform}
\big(e^\varepsilon_x|_{S_x}\big)^{-1} =  -\big( x|_{S_x} \big)^{-1}
\iota^\varepsilon_x\::\: S_x\scrM \rightarrow S_x M \:.
\eeq
\end{Prp}
\Proof Let~$\psi, \phi \in S_xM$. Then
\begin{align*}
\overline{\big( e^\varepsilon_x \psi \big)} \big(e^\varepsilon_x \phi \big)
&= \la \psi \:|\: (e^\varepsilon_x)^* \,e^\varepsilon_x \,\phi \ra_\H
= \la \psi \:|\: \iota^\varepsilon_x \,e^\varepsilon_x \,\phi \ra_\H
\overset{\eqref{Fepsdef}}{=} -\la \psi \:|\: x \,\phi \ra_\H = \Sl \psi | \phi \Sr \:,
\end{align*}
proving that~$e^\varepsilon_x|_{S_x}$ is an isometry.
Since the inner product~$\Sl .|. \Sr_x$ on~$S_xM$ is non-degenerate, it follows that
the mapping~$e^\varepsilon_x|_{S_x}$ is injective. Hence it is an isometric embedding.

If~$x$ is regular, a dimensional argument shows that~$e^\varepsilon_x|_{S_x}$
is an isomorphism. Again a dimensional argument yields that the image of the operator
product~$F^\varepsilon(x) = -\iota^\varepsilon_x (\iota^\varepsilon_x)^*$ coincides with the image of the
operator~$\iota^\varepsilon_x$, i.e.
\[ \iota^\varepsilon_x(S_x\scrM) = S_xM \:. \]
As a consequence, for any vector~$u \in S_x^\perp$ and any~$\chi \in S_x \scrM$,
\[ 0 = \la \iota^\varepsilon_x \chi | u \ra = \Sl \chi | e^\varepsilon_x u \Sr_x \:, \]
showing that the operator~$e^\varepsilon_x$ vanishes on the orthogonal complement of~$S_x$.
Therefore,
\[ e^\varepsilon_x|_{S_x}\, \psi^u(x) = e^\varepsilon_x|_{S_x}\, \pi_x\, u = e^\varepsilon_x\, u = 
\big({\mathfrak{R}}_\varepsilon u \big)(x) \:, \]
proving~\eqref{waveagree}. Finally,
\[ -\big( x|_{S_x} \big)^{-1} \iota^\varepsilon_x \:e^\varepsilon_x|_{S_xM}
\overset{\eqref{Fepsdef}}{=} \big( x|_{S_x} \big)^{-1} \:x|_{S_x} = \1_{S_x} \:, \]
so that that the inverse of~$e^\varepsilon_x|_{S_x}$ is indeed given by the expression in~\eqref{invform}.
\QED

This proposition makes it possible to identify the spin space~$S_xM \subset \H$ 
endowed with the inner product~$\Sl .|. \Sr_x$
with a subspace of the spinor space~$S_x\scrM$ with the inner product~$\overline{\psi} \phi$.
If the space-time point~$x$ is singular (see Definition~\ref{defregular}),
this is all we can expect, because in this case the spaces~$S_xM$
and~$S_x\scrM$ have different dimensions and are clearly not isomorphic.
As already mentioned after Definition~\ref{defregular}, in most situations of physical interest
the point~$x$ will be regular. In this case, we even obtain an isomorphism of~$S_xM$ and~$S_x\scrM$
which preserves the inner products on these spaces.
The identity~\eqref{waveagree} shows that, under the above 
identifications, the physical wave function~$\psi^u$
(as defined by~\eqref{psiudef}) goes over to the regularized Dirac wave
function~$({\mathfrak{R}}_\varepsilon u)(x)$. This shows again that the causal fermion system
involves the {\em{regularized}} objects. Moreover, one sees that the
abstract formalism introduced in Section~\ref{secframe} indeed gives agreement with
the usual objects in Minkowski space. We remark that the above isomorphism
of~$S_xM$ and~$S_x\scrM$ also makes it possible to use unambiguously the same notation for
the corresponding inner products. Indeed, it is convenient to denote the inner product~$\overline{\psi} \phi$
on the Dirac spinors at a time point~$x \in \scrM$ by
\sindex{spin scalar product}%
\nindex{ad4@$\Sl . \vert . \Sr_x$ -- spin scalar product}%
\sindex{adjoint spinor}%
\nindex{ai4@$\overline{\psi} \phi = \Sl \psi \vert \phi \Sr$ -- inner product on spinors}%
\beq \label{sspMink}
\Sl .|. \Sr_x \::\: S_x\scrM \times S_x\scrM \rightarrow \C \:,\qquad
\Sl \psi | \phi \Sr_x = \overline{\psi} \phi \:.
\eeq
\nindex{ai2@$(S_x\scrM, \Sl . \vert . \Sr_x)$ -- spinor space}%
In order to avoid confusion, we avoided this notation so far. But from now on, we will sometimes use it.

In the next proposition we compute the kernel of the fermionic projector~$P^\varepsilon(x,y)$
(as defined by~\eqref{Pxydef}, where the subscript~$\varepsilon$ clarifies the dependence on the
UV regularization) in Minkowski space.
\sindex{fermionic projector!kernel of}%
\sindex{fermionic projector!regularized kernel of}%
\nindex{an4@$P^\varepsilon(x,y)$ -- regularized kernel of fermionic projector}%
Moreover, we prove that
the limit~$\varepsilon \searrow 0$ exists in the distributional sense.
\sindex{correspondence to Minkowski space!kernel of the fermionic projector}%
\sindex{correspondence to Minkowski space!physical wave function}%
\begin{Prp} \label{lemma54} Assume that the points~$x$ and~$y$ are regular. Then, under
the above identification of~$S_xM$ with~$S_x\scrM$, 
the kernel of the fermionic projector has the representation
\sindex{evaluation map}%
\sindex{fermionic projector!kernel of}%
\sindex{fermionic projector!regularized kernel of}%
\nindex{an4@$P^\varepsilon(x,y)$ -- regularized kernel of fermionic projector}%
\[ P^\varepsilon(x,y) = -e^\varepsilon_x \,\iota^\varepsilon_y \::\: S_y\scrM \rightarrow S_x\scrM \:. \]
Moreover, choosing an orthonormal basis~$(u_\ell)$ of~$\H$,
the kernel of the fermionic projector can be written as
\beq \label{Pepsbase}
P^\varepsilon(x,y) = -\sum_\ell \big({\mathfrak{R}}_\varepsilon u_\ell \big)(x)\:
\overline{\big({\mathfrak{R}}_\varepsilon u_\ell \big)(y)} \:.
\eeq

In the limit~$\varepsilon \searrow 0$, the kernel of the fermionic projector~$P^\varepsilon(x,y)$
converges as a bi-distribution to the unregularized kernel defined by
\beq \label{Pxykernel}
P(x,y) := -\sum_\ell u_\ell(x)\: \overline{u_\ell(y)} \:.
\eeq
More precisely, for every compact subset~$K \subset \scrM$ and every~$\delta>0$,
there is a constant~$\varepsilon_0>0$ such that for all~$\varepsilon \in (0, \varepsilon_0)$
and for all test wave functions~$\eta, \tilde{\eta} \in C^\infty_0(K, S\scrM)$,
\beq
\bigg| \iint_{\scrM \times \scrM} \overline{\eta(x)}\, \big( P^\varepsilon(x,y)
- P(x,y) \big)\, \tilde{\eta}(y)\: d^4x\: d^4y \,\bigg| \leq \delta\: |\eta|_{C^1(K)}\, |\tilde{\eta}|_{C^1(K)} \:.
\label{Pest}
\eeq
\end{Prp} \noindent
We remark that, since~$\H$ is separable, we can always choose an at most countable orthonormal
basis~$(u_\ell)$ of~$\H$. The inequality~\eqref{Pest} is discussed
in Exercise~\ref{ex1}.

\Proof[Proof of Proposition~\ref{lemma54}] We first note that
\[ P^\varepsilon(x,y) = e^\varepsilon_x \, \pi_x\, y\, \big(e^\varepsilon_y|_{S_y} \big)^{-1}
= -e^\varepsilon_x \, \pi_x\, y\, 
\big( y|_{S_y} \big)^{-1}\, \iota^\varepsilon_y = - e^\varepsilon_x \:\pi_x\: \iota^\varepsilon_y = 
- e^\varepsilon_x \:\iota^\varepsilon_y \:. \]
In an orthonormal basis~$(u)_\ell$, the completeness relation yields for any spinor~$\chi \in S_y\scrM$
\begin{align*}
P^\varepsilon(x,y)\,\chi &= -e^\varepsilon_x \,\iota^\varepsilon_y\,\chi
= - \sum_\ell \big(e^\varepsilon_x \,u_\ell \big) \la u_\ell \,|\, \iota^\varepsilon_y\, \chi \ra_\H
= - \sum_\ell \big(e^\varepsilon_x \,u_\ell \big) \:\big( \overline{e^\varepsilon_x \,u_\ell} \:\chi \big)\:,
\end{align*}
and using~\eqref{evalmap} gives~\eqref{Pepsbase}.

In order to prove~\eqref{Pest}, we introduce the functionals
\begin{align*}
\hspace{1cm} \Phi^\varepsilon_\eta \:&:\: \H \rightarrow \C \:, \hspace*{-2cm} &
\Phi^\varepsilon_\eta u &= \int_\scrM \overline{\eta(x)} \big( {\mathfrak{R}}_\varepsilon u)(x)\: d^4x \\
\intertext{and similarly without UV regularization,}
\Phi_\eta \:&:\: \H \rightarrow \C \:, \hspace*{-2cm} &
\Phi_\eta u &= \int_\scrM \overline{\eta(x)} \,u(x)\: d^4x \:.
\end{align*}
Then the left side of~\eqref{Pest} can be written in the compact form
\[ \big| \Phi^\varepsilon_\eta \:\big( \Phi^\varepsilon_{\tilde{\eta}} \big)^*
- \Phi_\eta \:\big( \Phi_{\tilde{\eta}} \big)^* \big| \:, \]
which can be estimated with the triangle inequality by
\beq \label{zwischen}
\big| \Phi^\varepsilon_\eta \:\big( \Phi^\varepsilon_{\tilde{\eta}} \big)^*
- \Phi_\eta \:\big( \Phi_{\tilde{\eta}} \big)^* \big| \leq 
\|\Phi^\varepsilon_\eta \| \:\big\| \Phi^\varepsilon_{\tilde{\eta}} - \Phi_{\tilde{\eta}} \big\|
+ \big\| \Phi^\varepsilon_\eta - \Phi_\eta \big\| \: \|\Phi_{\tilde{\eta}}\| \:.
\eeq

It remains to estimate the operator norms in~\eqref{zwischen}. To this end, we
use property~(iii) of Definition~\ref{defreg} in the following way:
First, the norm of~$\Phi_\eta$ can be estimated by
\[ \big| \Phi_\eta u \big| = \int_\scrM \overline{\eta(x)} \,u(x)\: d^4x
\leq |\eta|_{C^0(K)} \sqrt{\mu(K)} \:\Big( \int_K |u(x)|\: d^4x \Big)^\frac{1}{2} \:, \]
and again by applying~\eqref{intes}. This gives
\[ \|\Phi_\eta\| \leq c\: |\eta|_{C^0(K)} \:. \]
Next, we use the triangle inequality together with~\eqref{reges2} to obtain the inequality
\[ \big\| \Phi^\varepsilon_\eta \big\| \leq \big\| \Phi^\varepsilon_\eta - \Phi_\eta \big\|
+ \big\| \Phi_\eta \big\|
\leq \delta\,|\eta|_{C^1(K)} + c\, |\eta|_{C^0(K)} \leq  2c\, |\eta|_{C^1(K)} \:, \]
valid uniformly for all~$\varepsilon \in (0, \varepsilon_0)$
(note that property~(i) cannot be used to obtain such a uniform estimate
because we have no control on how the constant~$c$ in~\eqref{reges0} depends on~$\varepsilon$).
Finally, again applying~\eqref{reges2}, we also know that
\[ \big\| \Phi^\varepsilon_\eta - \Phi_\eta \big\| \leq \delta\,|\eta|_{C^1(K)} \:. \]
Using these inequalities in~\eqref{zwischen} gives the result.
\QED

\subsectionn{Correspondence of the Causal Structure} \label{seccorcs}
We now explain how the causal structure of Minkowski space is related to the
corresponding notions of a causal fermion system (see Definition~\ref{def2}
and the time direction~\eqref{tdir}). To this end, we need to specify~$\H$ as a closed
subspace of the solution space of the vacuum Dirac equation~\eqref{Dirfree}.
Clearly, this Dirac equation can be solved by the plane-wave ansatz
\[ \psi(x) = e^{-i k x}\: \chi_k \]
with a constant spinor~$\chi_k$. Evaluating the resulting algebraic equation for~$\chi$
shows that the momentum~$k$ must lie on the mass shell~$k^2=m^2$
(where~$k^2 \equiv k^j k_j$ is the Minkowski inner product).
\sindex{mass shell}%
The solutions on the upper and lower mass shell are the solutions of positive and
negative energy, respectively. In order to avoid potential confusion with other notions of energy
(like energy densities or energy expectation values), we here prefer the notion of
solutions of positive and negative {\em{frequency}}.
\sindex{Dirac sea}%
Taking Dirac's original concept literally, we here describe
the vacuum in Minkowski space by the completely filled Dirac sea.
Thus we choose~$\H$ as the subspace of the solution space spanned
by all plane-wave solutions of negative frequency.
We refer to this choice as a {\em{Dirac sea configuration}}.
\sindex{Dirac sea configuration}%

\nindex{ac6@$P(x,y)$ -- kernel of fermionic projector}%
\nindex{an4@$P^\varepsilon(x,y)$ -- regularized kernel of fermionic projector}%
\sindex{fermionic projector!regularized kernel of}%
\begin{Lemma} \label{lemmaDiracsea}
If~$\H$ is the subspace of the solution space of the Dirac equation~\eqref{Dirfree}
spanned by all negative-frequency solutions, then the unregularized kernel of the fermio\-nic projector
as defined by~\eqref{Pxykernel} is the tempered bi-distribution
\beq \label{Pxyvac}
P(x,y) = \int \frac{d^4k}{(2 \pi)^4}\:(\slashed{k}+m)\: \delta(k^2-m^2)\: \Theta(-k_0)\: e^{-ik(x-y)} \:,
\eeq
where~$\delta$ is Dirac's delta distribution,
$\Theta$ is the Heaviside function, $k (x-y)$ is a short notation for the Minkowski
inner product~$k_j\,(x-y)^j$, and the slash in~$\slashed{k} = k^j \gamma_j$ denotes contraction with
the Dirac matrices (the ``Feynman dagger'').
\nindex{ao2@$\delta$ -- Dirac's delta distribution}%
\nindex{ao4@$\Theta$ -- Heaviside function}%
\sindex{Feynman dagger}%
\nindex{ao6@$\slashed{k}$ -- Feynman dagger}%
\end{Lemma}
\Proof The integrand in~\eqref{Pxyvac} clearly is a tempered distribution.
Hence its Fourier transform~$P(x,y)$ is also a tempered distribution (in the vector~$y-x$
and also in both vectors~$x$ and~$y$). In addition, one verifies by direct computation
that~$P(x,y)$ is a distributional solution of the Dirac equation,
\begin{align*}
(i \Pdd_x - m)\, P(x,y)
&= \int \frac{d^4k}{(2 \pi)^4}\:(\slashed{k}-m) (\slashed{k}+m)\: \delta(k^2-m^2)\: \Theta(-k_0)\: e^{-ik(x-y)} \\
&= \int \frac{d^4k}{(2 \pi)^4}\:\big(k^2 - m^2 \big)\: \delta(k^2-m^2)\: \Theta(-k_0)\: e^{-ik(x-y)} = 0 \:.
\end{align*}
Due to the factor~$\Theta(-k_0)$, the distribution~$P(x,y)$
is composed of solutions of negative frequency. Moreover, since the matrix~$(\slashed{k}+m)$
has rank two, one sees that~$P(x,y)$ is indeed composed of {\em{all}} negative-frequency solutions.
It remains to show that the normalization of~$P(x,y)$ is compatible with~\eqref{Pxykernel}, meaning that
\beq \label{Pnorm}
-2 \pi \int_{\R^3} P\big( x,(t,\vec{y}) \big) \,\gamma^0\, P \big( (t,\vec{y}), z \big)\: d^3y = P(x,z) \:.
\eeq
This identity follows by a straightforward computation: First,
\begin{align*}
\int_{\R^3} & P \big( x, (t, \vec{y}) \big) \:\gamma^0\: P \big( (t, \vec{y}), z \big)\: d^3y  \\
&= \int_{\R^3} d^3y \int \frac{d^4k}{(2 \pi)^4}\: e^{-i k(x-y)} \int \frac{d^4q}{(2 \pi)^4}\: e^{-i q(y-z)}
\: P_m(k)\:\gamma^0\: P_m(q) \\
&= \int \frac{d^4k}{(2 \pi)^4} \int_\R \frac{d \lambda}{2 \pi}\; e^{-i k x + i q z}
\: P_m(k)\:\gamma^0\: P_m(q) \Big|_{q = (\lambda, \vec{k})} \:.
\end{align*}
Setting~$k=(\omega, \vec{k})$, we evaluate the $\delta$-distributions inside the factors~$P_m$,
\begin{align*}
\delta(k^2-m^2)& \, \delta(q^2-m^2) \big|_{q = (\lambda, \vec{k})}
= \delta \big( \omega^2 - |\vec{k}|^2 -m^2 \big) \:
\delta \big( \lambda^2 - |\vec{k}|^2 -m^2 \big) \\
&= \delta(\lambda^2 - \omega^2)\: \delta \big( \omega^2 - |\vec{k}|^2 -m^2 \big) \:.
\end{align*}
This shows that we only get a contribution if~$\lambda=\pm \omega$.
Using this fact together with the mass shell property~$\omega^2-|\vec{k}|^2=m^2$, 
we can simplify the Dirac matrices according to
\begin{align*}
(\slashed{k}+m) &\:\gamma^0\: (\slashed{q} + m)
= (\omega \gamma^0 + \vec{k} \vec{\gamma} + m) \,\gamma^0\,
(\pm \omega \gamma^0 + \vec{k} \vec{\gamma} + m) \\
&= (\omega \gamma^0 + \vec{k} \vec{\gamma} + m) \,
(\pm \omega \gamma^0 - \vec{k} \vec{\gamma} + m)\,\gamma^0 \\
&= \Big( (\pm \omega^2 + |\vec{k}|^2 + m^2) \,\gamma^0
+(1 \pm 1) \,\omega\,(\vec{k} \vec{\gamma}) + (1 \pm 1)\, m \omega \Big) \\
&= \left\{ \begin{array}{cl}
2 \omega\, (\slashed{k}+m) & \text{in case~$+$} \\
0 & \text{in case~$-\:.$} \end{array} \right.
\end{align*}
Hence we only get a contribution if~$\lambda=\omega$, giving rise to the identity
\[  \delta(\lambda^2 - \omega^2) = \frac{1}{2 |\omega|}\: \delta(\lambda-\omega)\:. \]
Combining these formulas, we obtain
\begin{align*}
\int_{\R^3} & P \big( x, (t, \vec{y}) \big) \:\gamma^0\: P \big( (t, \vec{y}), z \big)\: d^3y \\
&= \int \frac{d^4k}{(2 \pi)^4} \int_\R \frac{d \lambda}{2 \pi}\; e^{-i k (x-z)}
\:\delta(\lambda - \omega)\: \delta( k^2 -m^2)\:
\frac{2 \omega}{2 |\omega|}\, (\slashed{k}+m)\: \Theta(-k_0) \\
&= -\frac{1}{2 \pi} \int \frac{d^4k}{(2 \pi)^4} \; e^{-i k (x-z)}
\: \delta( k^2 -m^2)\: (\slashed{k}+m)\: \Theta(-k_0) \:.
\end{align*}
This gives the result.
\QED

The Fourier integral~\eqref{Pxyvac} can be computed in closed form, giving
an expression involving Bessel functions.
Since the general structure of the resulting formula will be important later on, we give the computation in detail.
In preparation, it is useful to pull the Dirac matrices out of the Fourier integral.
To this end, one rewrites the factor~$(\slashed{k}+m)$ in~\eqref{Pxyvac}
in terms of a differential operator in position space,
\beq \label{Pdiff}
P(x,y) = (i \Pdd_x + m) \,T_{m^2}(x,y) \:,
\eeq
where~$T_{m^2}$ is the scalar bi-distribution
\beq \label{Tm2def}
T_{m^2}(x,y) := \int \frac{d^4k}{(2 \pi)^4}\: \delta(k^2-m^2)\: \Theta(-k_0)\: e^{-ik(x-y)} \:.
\eeq
\nindex{ao8@$T_a(x,y)$ -- Fourier transform of lower mass shell}%
In the next lemma, we determine the singular structure of this distribution.
The method is to subtract an explicit singular distribution and to show that the difference
is a {\em{regular distribution}} (i.e.\ a locally integrable function, denoted by~$L^1_\text{\rm{loc}}$).
\sindex{distribution!regular}%
\sindex{principal value}%
\nindex{ap0@$\PP$ -- principal value}%
The distribution~$\PP/\xi^2$, denoted by {\em{principal value}}, is defined by
evaluating weakly with a test function $\eta \in C^\infty_0(\scrM)$
and by removing the positive and negative parts of the pole in a symmetric way.
There are different equivalent ways of writing the principal part, each of which could
serve as a possible definition (for mathematical details see Exercises~\ref{ex201} and~\ref{ex21}):
\beq \label{PPdef}
\begin{split}
\int &\frac{\PP}{\xi^2} \: \eta(\xi)\: d^4\xi
= \lim_{\nu \searrow 0} \int \Theta\big( |\xi^2| - \nu \big)\;
\frac{1}{\xi^2} \: \eta(\xi)\: d^4\xi \\
&= \lim_{\nu \searrow 0} \frac{1}{2} \sum_{\pm}
\int \frac{1}{\xi^2 \pm i \nu} \: \eta(\xi)\: d^4\xi
= \lim_{\nu \searrow 0} \frac{1}{2} \sum_{\pm}
\int \frac{1}{\xi^2 \pm i \nu \xi^0} \: \eta(\xi)\: d^4\xi
\end{split}
\eeq
(here~$\xi^2 \equiv \xi^j \xi_j$ is again the Minkowski inner product).

\begin{Lemma} \label{lemmaTintro}
On the light cone, the distribution~$T_{m^2}$ has the singularity structure
\beq \label{Tsingular}
T_{m^2}(x,y) = -\frac{1}{8 \pi^3}
\left( \frac{\PP}{\xi^2} +i \pi\, \delta(\xi^2)\,\epsilon(\xi^0) \right)
+ r(x,y) \:,
\eeq
where we set~$\xi := y-x$, and~$r \in L^1_\text{\rm{loc}}(\scrM \times \scrM)$ is a regular distribution.
\nindex{ao8@$T_a(x,y)$ -- Fourier transform of lower mass shell}%
\nindex{ap4@$\xi$ -- short notation for Minkowski vector~$y-x$}%
Here~$\epsilon$ is the sign function $\epsilon(x)=1$
for $x \geq 0$ and $\epsilon(x)=-1$ otherwise. 
\nindex{ap8@$\epsilon$ -- sign function}%
Away from the light cone (i.e.\ for~$\xi^2 \neq 0$), $T_{m^2}(x,y)$ is a smooth function given by
\begin{align} \label{Taway}
T_{m^2}(x,y) = \left\{ \begin{array}{cl} 
\displaystyle \frac{m}{16 \pi^2} \:\frac{Y_1\big(m\sqrt{\xi^2} \,\big)}{\sqrt{\xi^2}}
+\frac{i m}{16 \pi^2}\: \frac{J_1 \big(m\sqrt{\xi^2} \,\big)}{\sqrt{\xi^2}}\: \epsilon(\xi^0)
& \text{if~$\xi$ is timelike} \\[1em]
\displaystyle \frac{m}{8 \pi^3} \frac{K_1 \big(m\sqrt{-\xi^2} \,\big)}{\sqrt{-\xi^2}} & \text{if~$\xi$ is spacelike}\:,
\end{array} \right. \hspace*{-0.3em}
\end{align}
where~$J_1$, $Y_1$ and~$K_1$ are Bessel functions.
\nindex{ap6@$J_1$, $Y_1$ and~$K_1$ -- Bessel functions}%
\end{Lemma}
\Proof The Fourier integral is computed most conveniently by inserting
a con\-ver\-gence-generating factor. Thus for any~$\varepsilon>0$ we consider the Fourier integral
\beq \label{Tepsreg}
T^\varepsilon_{m^2}(x,y) := \int \frac{d^4k}{(2 \pi)^4}\: \delta(k^2-m^2)\: \Theta(-k_0)\: e^{-ik(x-y)}\:
e^{-\varepsilon \,|k_0|} \:.
\eeq
This Fourier integral can be computed pointwise, showing that~$T^\varepsilon(x,y)$ is a regular distribution.
Taking the limit~$\varepsilon \searrow 0$ in the distributional sense, we will then obtain~$T_{m^2}(x,y)$.

Setting~$\xi=y-x$ and~$t=\xi^0$, we first carry out
the integral over~$k_0$ to obtain
\begin{align*}
T^\varepsilon_{m^2}(x,y)&= \int \frac{d^4k}{(2\pi)^4}\:
	\delta(k^2-m^2)\:\Theta(-k_0) \:e^{ik \xi} \:e^{-\varepsilon \,|k_0|} \nonumber \\
	&= \int_{\R^3} \frac{d^3k}{(2\pi)^4}\:\frac{1}{2\sqrt{\vec{k}^2+m^2}}
	\: e^{-i\sqrt{\vec{k}^2+m^2}\, t-i \vec{k}\vec{\xi}}\:e^{-\varepsilon \sqrt{\vec{k}^2+m^2}} \:.
\end{align*}
Next, for the spatial momentum~$\vec{k}$ we introduce polar coordinates~$(p=|\vec{k}|,
\vartheta, \varphi)$, where~$\vartheta$ is the angle between~$\vec{k}$ and~$\vec{\xi}$,
and~$\varphi$ is the azimuthal angle. Also setting~$r=|\vec{\xi}|$, we get
\begin{align}
T^\varepsilon_{m^2}(x,y)&= \int_0^\infty \frac{dp}{2(2\pi)^3} \int_{-1}^1d\cos\theta
	\: \frac{p^2}{\sqrt{p^2+m^2}} \:e^{-(\varepsilon+it) \sqrt{p^2+m^2}}\: e^{-ipr\cos\theta} \nonumber \\
	&= \frac{1}{r}\int_0^\infty \frac{dp}{(2\pi)^3}\:
	\frac{p}{\sqrt{p^2+m^2}}\: e^{-(\varepsilon+it) \sqrt{p^2+m^2}}\: \sin(pr) \nonumber \\
	&= \frac{m^2}{(2\pi)^3} \frac{K_1 \big(m\sqrt{r^2+(\varepsilon+it)^2}
		\,\big)}{m\sqrt{r^2+(\varepsilon+it)^2}}, \label{Tbessel}
\end{align}
where the last integral was carried out using~\cite[formula (3.961.1)]{gradstein}.
Here the square root and the Bessel function~$K_1$ are defined as usual using a branch cut along
the negative real axis.

When taking the limit~$\varepsilon \searrow 0$, one must be careful for two reasons.
First, a pole forms on the light cone~$t=\pm r$. Second, the Bessel function~$K_1$
involves logarithms, which must be evaluated in the complex plane using the
branch cut along the negative real axis. For clarity, we treat these two issues after each other.
The asymptotic expansion of the Bessel function (see~\cite[(10.31.1)]{DLMF})
\[ K_1(z) = \frac{1}{z} + \O\big( z \log z \big) \]
yields that the pole on the light cone is of the form
\[ T^\varepsilon_{m^2}(x,y) = \frac{1}{(2\pi)^3} \:\frac{1}{r^2+(\varepsilon+it)^2} 
+ \O\big( \log|\xi^2|) \:, \]
uniformly in~$\varepsilon$. Therefore, after subtracting the pole, we can take the
limit~$\varepsilon \searrow 0$ as a locally integrable function, i.e.
\[ \lim_{\varepsilon \searrow 0}
\bigg( T^\varepsilon_{m^2}(x,y) - \frac{1}{(2\pi)^3} \:\frac{1}{r^2+(\varepsilon+it)^2} \bigg)
\in L^1_\text{\rm{loc}}(\scrM \times \scrM)\:. \]
For the subtracted pole, the limit~$\varepsilon \searrow 0$ can be computed in the
distributional sense by
\beq \label{delta-highdim}
\lim_{\varepsilon \searrow 0} \frac{1}{r^2+(\varepsilon+it)^2}
= \lim_{\varepsilon \searrow 0} \frac{1}{r^2 - t^2 + i \varepsilon t }
= -\frac{\PP}{\xi^2} -i \pi\, \delta(\xi^2)\,\epsilon(\xi^0)\:,
\eeq
where we used the distributional equations
\begin{align}
\lim_{\varepsilon \searrow 0} \left( \frac{1}{x - i \varepsilon} - \frac{1}{x + i \varepsilon} \right)
&=\, 2 \pi i \: \delta(x) \label{eq:delta-formula} \\
\frac{1}{2} \lim_{\varepsilon \searrow 0} \left(
\frac{1}{x - i \varepsilon} + \frac{1}{x + i \varepsilon} \right) &=: \frac{\PP}{x}\:.
\label{eq:PP-formula}
\end{align}
(for details see Exercises~\ref{ex2}--\ref{ex21}).
Here~$\epsilon$ is again the sign function $\epsilon(x)=1$
for $x \geq 0$ and $\epsilon(x)=-1$ otherwise. 
This gives~\eqref{Tsingular}.

\begin{figure}
\psscalebox{1.0 1.0} 
{
\begin{pspicture}(0,-2.0034792)(11.932222,2.0034792)
\psdots[linecolor=black, dotsize=0.1](0.30722222,-0.20236807)
\psline[linecolor=black, linewidth=0.02, arrowsize=0.05291666666666668cm 4.0,arrowlength=1.4,arrowinset=0.0]{->}(0.0,-0.0037566868)(5.1,-0.0032016642)
\psline[linecolor=black, linewidth=0.02, arrowsize=0.05291666666666668cm 4.0,arrowlength=1.4,arrowinset=0.0]{->}(3.3,-2.0034792)(3.3,2.096521)
\psline[linecolor=black, linewidth=0.04](4.9,0.19652082)(0.3,0.19652082)
\psline[linecolor=black, linewidth=0.04](4.9,-0.20347917)(0.3,-0.20347917)
\psline[linecolor=black, linewidth=0.02](0.3,-0.013479176)(0.3,-0.65347916)
\rput[bl](0.19222222,-1.0257014){\normalsize{$t^2$}}
\rput[bl](1.9222223,-0.6257014){\normalsize{$t<0$}}
\rput[bl](1.9422222,0.3942986){\normalsize{$t>0$}}
\psdots[linecolor=black, dotsize=0.1](0.30222222,0.19763194)
\psline[linecolor=black, linewidth=0.02, arrowsize=0.05291666666666668cm 4.0,arrowlength=1.4,arrowinset=0.0]{->}(6.5,-2.0034792)(6.5,2.096521)
\psline[linecolor=black, linewidth=0.02, arrowsize=0.05291666666666668cm 4.0,arrowlength=1.4,arrowinset=0.0]{->}(6.3,-0.0034791755)(9.705,-0.0034791755)
\psline[linecolor=black, linewidth=0.02](6.5,1.8015474)(6.3,1.8014942)
\psbezier[linecolor=black, linewidth=0.04](6.62,1.7965208)(6.630015,1.3908274)(6.650103,0.61585486)(6.684794,0.47255844)(6.7194853,0.32926205)(6.845412,0.20600404)(6.9941616,0.17245044)(7.142912,0.13889685)(8.840294,0.09498181)(9.46,0.081520826)
\psbezier[linecolor=black, linewidth=0.04](6.625,-1.8034792)(6.6400146,-1.5110548)(6.6551027,-0.69972855)(6.689794,-0.4717969)(6.7244854,-0.24386524)(6.8004117,-0.19152978)(6.9991617,-0.169939)(7.1979117,-0.14834821)(8.210294,-0.103061244)(9.465,-0.07847918)
\psline[linecolor=black, linewidth=0.02](6.495,-1.7984526)(6.295,-1.7985058)
\rput[bl](5.232222,-1.9157014){\normalsize{$-\sqrt{t^2}$}}
\rput[bl](5.6222224,1.6892986){\normalsize{$\sqrt{t^2}$}}
\rput[bl](4.3922224,1.5742986){\normalsize{$\C$}}
\rput[bl](8.752222,1.5242985){\normalsize{$\C$}}
\rput[bl](7.487222,0.2692986){\normalsize{$t>0$}}
\rput[bl](7.4922223,-0.5457014){\normalsize{$t<0$}}
\psdots[linecolor=black, dotsize=0.1](6.632222,1.797632)
\psdots[linecolor=black, dotsize=0.1](6.6172223,-1.792368)
\end{pspicture}
}
\caption{The set~$\{r^2+(\varepsilon+it)^2 \,|\, r \in \R\}$ (left) and its square root (right)}
\label{figcontour}
\end{figure}
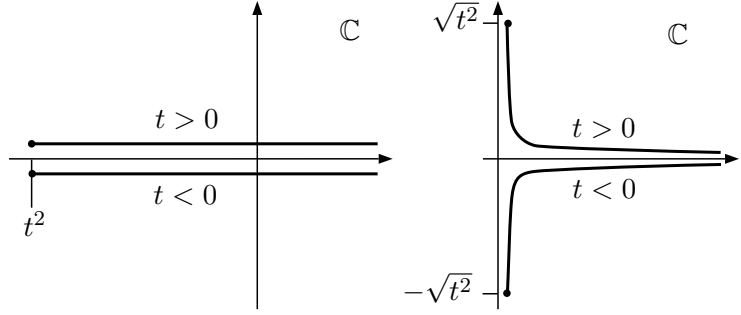

In order to compute the regular part of the distribution~$T_{m^2}$,
we may disregard the singularity on the light cone and may consider the case
that~$\xi$ is either spacelike or timelike. In the first case,
the argument~$m\sqrt{r^2+(\varepsilon+it)^2}$ of the Bessel function converges to
the positive real axis, where the Bessel function is analytic.
This gives the lower equation in~\eqref{Taway}.
In the remaining case that~$\xi$ is timelike, the
argument~$m\sqrt{r^2+(\varepsilon+it)^2}$ converges to the imaginary axis
(more precisely, to the upper imaginary axis if~$t>0$ and to the lower imaginary axis if~$t<0$; see
Figure~\ref{figcontour}).
Using the relations~\cite[(10.27.9) and~(10.27.10)]{DLMF}
\[ i \pi J_1(z) = -i K_1(-iz) -i K_1(iz) \qquad \text{and} \qquad 
-\pi Y_1(z) = -i K_1(-iz) + i K_1(iz) \]
(valid if~$|\arg z|<\frac{\pi}{2}$), one can express~$K_1$ near the upper and lower imaginary axis by
\[ K_1(\pm iz) = -\frac{\pi}{2} \big( J_1(z) \mp i Y_1(z) \big) \:. \]
Using these identities in~\eqref{Tbessel} and using that the Bessel functions~$J_1$ and~$K_1$
are analytic in a neighborhood of the positive real axis, one can take the limit~$\varepsilon \searrow 0$
to obtain the upper equation in~\eqref{Taway}.
\QED
We point out that the Bessel functions in~\eqref{Taway} are all real-valued.
In particular, one sees that~$T(x,y)$ is real-valued if the vector~$\xi$ is spacelike.
This fact can also be understood from a general symmetry argument (see Exercise~\ref{ex3}).

Using the result of Lemma~\ref{lemmaTintro} in~\eqref{Pdiff}, one
can derive corresponding formulas for~$P(x,y)$.
In particular, differentiating~\eqref{Tsingular}, one sees that~$P(x,y)$
has an even stronger singularity on the light cone which involves
terms of the form~$1/\xi^4$ and~$\delta'(\xi^2)$.
Differentiating~\eqref{Taway}, carrying out the derivatives
with the chain rule and using formulas for the
derivatives of Bessel functions (see~\cite[(10.6.6) and~(10.29.4)]{DLMF}),
one can also express the fermionic projector~$P(x,y)$ in terms of Bessel functions
(see Exercise~\ref{ex4}).
We do not give the resulting formulas, because we do not need the detailed form later on.
Instead, we here prefer to argue with structural properties of the distribution~$P(x,y)$.
This makes it possible to infer qualitative properties of the eigenvalues of~$A_{xy}$,
even without referring to the detailed form of the formulas in Lemma~\ref{lemmaTintro}.
From Lorentz symmetry, we know that for all~$x$ and~$y$ with spacelike or timelike separation,
$P(x,y)$ can be written as
\beq \label{Pxyrep}
P(x,y) = \alpha\, \xi_j \gamma^j + \beta\:\1
\eeq
with two complex-valued functions~$\alpha$ and~$\beta$ (where again~$\xi =y-x$).
Taking the adjoint with respect to the spin scalar product, we see that
\beq \label{Pyxrep}
P(y,x) = \overline{\alpha}\, \xi_j \gamma^j + \overline{\beta}\:\1 \:.
\eeq
As a consequence,
\beq \label{1}
A_{xy} = P(x,y)\, P(y,x) = a\, \xi_j \gamma^j + b\, \1
\eeq
with two real parameters~$a$ and $b$ given by
\beq \label{ab}
a = \alpha \overline{\beta} + \beta \overline{\alpha} \:,\qquad
b = |\alpha|^2 \,\xi^2 + |\beta|^2 \:.
\eeq
Applying the formula~$(A_{xy} - b \1)^2 = a^2\:\xi^2\,\1$,
the roots of the characteristic polynomial of~$A_{xy}$ are computed by
\beq \label{root}
b \pm \sqrt{a^2\: \xi^2} \:.
\eeq
Therefore, the eigenvalues of the closed chain are either real, or else they
form a complex conjugate pair\footnote{It is a general property of the
closed chain that if~$\lambda$ is an eigenvalue, then so is~$\overline{\lambda}$;
see Exercise~\ref{ex41}.}.
Which of the two cases occurs is determined by the sign of the factor~$\xi^2$.
This gives the agreement of the different notions of causality in the following sense:
\sindex{correspondence to Minkowski space!causal structure}%
\begin{Prp} \label{prpcorrespond}
Assume that~$P(x,y)$ is the unregularized kernel of the fermionic projector of
the vacuum~\eqref{Pxyvac}, and that the eigenvalues~$\lambda^{xy}_1,
\ldots, \lambda^{xy}_4$ are computed as the eigenvalues of the closed chain~\eqref{Axydef}.
Then the following statements hold:

If the points~$x, y \in \scrM$ have spacelike separation in Minkowski space,
then they are also spacelike separated in the sense of Definition~\ref{def2}.
If, on the other hand, the points~$x, y \in \scrM$ have timelike separation in Minkowski space,
then they are also timelike separated in the sense of Definition~\ref{def2}.
Even more, they are properly timelike separated (see Definition~\ref{defproptl})
in the sense that the closed chain~$A_{xy}$ has strictly positive eigenvalues and
definite eigenspaces.
Finally, if the points~$x, y \in \scrM$ have lightlike separation in Minkowski space,
then the causal structure of Definition~\ref{def2} is ill-defined.
\end{Prp} \noindent
The fact that the causal structure is ill-defined for lightlike separation
again explains why an UV regularization must be introduced.

\Proof[Proof of Proposition~\ref{prpcorrespond}]
If the vector~$\xi=y-x$ is spacelike, then the term~$\xi^2$ is negative. Thus
the eigenvalues in~\eqref{root} form a complex conjugate
pair, implying that they all have the same absolute value.
Thus the points are spacelike separated in the sense of Definition~\ref{def2}.

If the vector~$\xi$ is timelike, the term~$\xi^2$ in~\eqref{root} is positive, so that
the~$\lambda_j$ are all real.
In order to show that they do not have the same absolute value,
we need to verify that the parameters~$a$ and~$b$ are both non-zero.
This makes it necessary to refer to the explicit formula involving
Bessel functions~\eqref{Taway}: The Bessel functions~$Y_1$ and~$J_1$
do not have joint zeros on the positive real axis
(this can be understood abstractly from the fact that these Bessel functions
form a fundamental system of solutions of the Bessel equation; see
Exercise~\ref{ex5}).
As a consequence, the parameter~$\beta$ in~\eqref{Pxyrep} is non-zero.
Likewise, the derivatives~$Y_1'$ and~$J_1'$ do not have joint zeros
(as can again be verified from the fact that the Bessel functions form a fundamental system).
This implies that the parameter~$\alpha$ in~\eqref{Pxyrep} is non-zero.
We conclude that the parameter~$b$ in~\eqref{ab} is non-zero.
The combination of~$\alpha$ and~$\beta$ in the formula for~$a$ in~\eqref{ab}
can be rewritten in terms of a Wronskian of the Bessel function.
This Wronskian can be computed explicitly using~\cite[(10.5.2)]{DLMF},
implying that~$a$ is non-zero (see Exercise~\ref{ex4}).
We conclude that the points~$x$ and~$y$ are timelike separated in the sense of Definition~\ref{def2}.

In order to get the connection to proper timelike separation,
recall that if~$\xi$ is a timelike vector of Minkowski space, then
the closed chain has the form~\eqref{ab} with~$a,b \neq 0$.
A direct computation shows that this matrix is diagonalizable and that
the eigenspaces are definite with respect to the spin scalar product (see Exercise~\ref{ex6}).
Moreover, applying the Schwarz inequality to the explicit formulas~\eqref{ab},
one obtains
\beq \label{starin}
|a| \, \sqrt{\xi^2} = 2 \re \Big( \alpha \, \sqrt{\xi^2} \: \overline{\beta} \Big)
\overset{(\star)}{\leq} |\alpha|^2 \xi^2 + |\beta|^2 = b \:,
\eeq
proving that the eigenvalues in~\eqref{root} are non-negative. 
It remains to show that none of these eigenvalues vanishes.
To this end, it suffices to show that the inequality~($\star$) in~\eqref{starin}
is strict, which in turn is equivalent to proving that
\[ \im \big( \alpha \overline{\beta} \big) \neq 0 \:. \]
This inequality follows by a detailed analysis of the Bessel functions
(see~\cite[proof of Lemma~4.3]{lqg}).
We conclude that~$x$ and~$y$ are indeed properly timelike separated.

If the vector~$\xi$ is lightlike, then~$P(x,y)$ is not defined pointwise.
As a consequence, the closed chain is ill-defined.
\QED

This proposition cannot be applied directly to causal fermion systems because,
as explained in~\S\ref{secuvintro} and~\S\ref{secuvreg},
constructing a causal fermion system makes it necessary to introduce an
UV regularization. Nevertheless, the above proposition also gives the
correspondence of the different notions of causality for
causal fermion systems describing the Minkowski vacuum, as we now explain.
Thus let us consider the causal fermion system corresponding to the regularized
fermionic projector of the vacuum~$P^\varepsilon(x,y)$.
In the limit~$\varepsilon \searrow 0$, the kernel of the fermionic
projector~$P^\varepsilon(x,y)$ converges to the unregularized kernel~$P(x,y)$
(see~\eqref{Pest} in Proposition~\ref{prpisometry}).
If this convergence is pointwise, i.e.\ if for given space-time points~$x,y \in \scrM$,
\beq \label{pointwise}
\lim_{\varepsilon \searrow 0} P^\varepsilon(x,y) = P(x,y) \:,
\eeq
then the results of Proposition~\ref{prpcorrespond} also apply to the causal
fermion system, up to error terms which tend to zero as~$\varepsilon \searrow 0$
(for the stability of the notions of causality see Exercises~\ref{ex8} and~\ref{ex81}).
\sindex{Planck length}%
\nindex{aj9@$\ell_P$ -- Planck length}%
Thinking of~$\varepsilon$ as the Planck scale, this means physically that
the notion of causality of Definition~\ref{def2} agrees with the usual notion of
causality in Minkowski space, up to corrections which are so small that they
cannot be observed.
The subtle point of this argument is that it requires pointwise convergence~\eqref{pointwise}.
Clearly, such a pointwise convergence cannot hold if~$x$ and~$y$ are lightlike separated,
because the right side of~\eqref{pointwise} is ill-defined pointwise.
Expressed for a causal fermion system for fixed~$\varepsilon$ on the Planck scale,
this means that the notion of causality of Definition~\ref{def2} does {\em{not}} agree
with the usual notion of
causality if the vector~$\xi$ is almost lightlike in the sense that~$\big| |\xi^0| - |\vec{\xi}| \big|
\lesssim \varepsilon$. This is not surprising because we cannot expect
that the notion of causality in Minkowski space holds with a higher resolution than
the regularization scale~$\varepsilon$.
The remaining question is whether we have pointwise convergence~\eqref{pointwise}
if the points~$x$ and~$y$ have timelike or spacelike separation.
The answer is yes for a large class of regularizations (like for example the
regularization by mollification in Example~\ref{exmollify}).
However, the general notion of Definition~\ref{defreg}
only gives weak convergence of the kernels~\eqref{Pest}.
This shortcoming could be removed by adding a condition to Definition~\ref{defreg}
which ensures pointwise convergence away from the light cone.
On the other hand, such an additional condition
will not be needed for the constructions in this book,
and therefore it seems preferable not to impose it.
Nevertheless, the physical picture is that the regularized kernel should 
converge pointwise, at least for generic points~$x$ and~$y$ which lie
sufficiently far away from the light cone.
With this in mind, Proposition~\ref{prpcorrespond} indeed
shows that the notion of causality of Definition~\ref{def2} corresponds to the usual notion of
causality in Minkowski space, up to corrections which
are so small that they are irrelevant in most situations of interest.

We next make a remark on the notion of {\em{lightlike}} separation. At first sight,
Definition~\ref{def2} leaves many possibilities for lightlike separation:
two eigenvalues could be real and two could form a complex conjugate pair,
or there could be two complex conjugate pairs with different abolute values, etc.
However, in the above example of Minkowski space, we saw that lightlike
separation appears only as the boundary case between timelike and spacelike separation
when all eigenvalues are degenerate. More generally, all known examples and
numerical studies suggest that lightlike separation
is not favorable when minimizing the causal action. This can be understood
intuitively as follows: Since
the Lagrangian vanishes for spacelike separation, the causal action principle
aims at arranging spacelike separation for as many pairs or points as possible.
But in view of the constraints, it is impossible to arrange that all pairs of points
are spacelike separated. It seems preferable to satisfy the constraints by
arranging timelike separation for certain pairs or points, whereas the other pairs
of points are arranged to be spacelike separated. In this mechanism, lightlike separation
appears typically only for few pairs of points
as the boundary case between timelike and spacelike separation.

We conclude this section by explaining why the functional~${\mathscr{C}}$
\sindex{time direction}%
\nindex{ab8@${\mathscr{C}}(x, y)$ -- distinguishes time direction}%
introduced in~\eqref{Cform} gives information on the time direction.
Our first task is to rewrite this functional in terms of the regularized kernel of the fermionic
projector~$P^\varepsilon(x,y)$.

\begin{Lemma} Assume that the operator~$P^\varepsilon(x,x) : S_x\scrM \rightarrow S_x\scrM$
is invertible. Then, setting
\beq \label{nuxdef}
\nu(x)= P^\varepsilon(x,x)^{-1} \::\: S_x\scrM \rightarrow S_x\scrM \:,
\eeq
the functional~${\mathscr{C}}$, \eqref{Cform}, can be written as
\beq \label{Ccomm}
{\mathscr{C}}(x, y) = i \Tr_{S_x} \!\Big(  P^\varepsilon(x,y) \:\nu(y)\:P^\varepsilon(y,x)\:
\big[ \nu(x), A_{xy} \big] \Big)\:.
\eeq
\end{Lemma}
\Proof Since~$P(x,x) = \pi_x x|_{S_x} = x|_{S_x}$, we know that~$\nu(x) = ( x|_{S_x} )^{-1}$. Thus
\begin{align*}
\pi_x\, y\,x \,\pi_y\, \pi_x|_{S_x} &=
\pi_x y\: \pi_y x\: \pi_x y\: \nu(y)\: \pi_y x\: \nu(x) |_{S_x} \\
&= P^\varepsilon(x,y)\: P^\varepsilon(y,x)\: P^\varepsilon(x,y) \: \nu(y)\: P^\varepsilon(y,x)\: \nu(x) |_{S_x} \:.
\end{align*}
Using this formula in~\eqref{Cform}, we obtain
\begin{align*}
{\mathscr{C}}(x, y) &= i \Tr_{S_x} \!\big( y\,x \,\pi_y\, \pi_x|_{S_x} - y\,\pi_x \,\pi_y\,x|_{S_x}
\big) \\
&= i \Tr_{S_x} \!\Big(  P^\varepsilon(x,y)\, P^\varepsilon(y,x) \: P^\varepsilon(x,y) \:\nu(y)\:P^\varepsilon(y,x)\: \nu(x) \\
&\qquad \quad\;\, - P^\varepsilon(x,y)\, P^\varepsilon(y,x) \:\nu(x)\:P^\varepsilon(x,y) \:\nu(y)\:P^\varepsilon(y,x) \Big) \\
&= i \Tr_{S_x} \!\Big(  P^\varepsilon(x,y) \:\nu(y)\:P^\varepsilon(y,x)\: \nu(x)\: P^\varepsilon(x,y)\:P^\varepsilon(y,x) \\
&\qquad \quad\;\, - P^\varepsilon(x,y)\: \nu(y)\:P^\varepsilon(y,x) \:P^\varepsilon(x,y) \:P^\varepsilon(y,x)\: \nu(x)\:  \Big) \:.
\end{align*}
This gives the result.
\QED

We point out that the operator~$\nu(x)$ in~\eqref{nuxdef} is ill-defined without
UV regularization because evaluating the distribution~$P(x,y)$ on the diagonal~$x=y$
has no mathematical meaning. As a consequence, the functional~${\mathscr{C}}$
is ill-defined without UV regularization, even if~$x$ and~$y$ have timelike separation.
This makes the following computation somewhat delicate.
In order to keep the analysis reasonably simple, we assume that the
regularized kernel of the fermionic projector has {\em{vector-scalar structure}},
meaning that it is of the general form
\beq \label{vectorscalar}
P^\varepsilon(x,y) = v^\varepsilon_j(x,y) \:\gamma^j \:+\: \beta^\varepsilon(x,y) \:\1
\eeq
\sindex{fermionic projector!vector-scalar structure}%
with a vector and a scalar component. Here~$v^\varepsilon(x,y)$ is a complex vector field
(i.e.\ it can be written as~$v^\varepsilon = u^\varepsilon + i w^\varepsilon$
with Minkowski vectors~$u^\varepsilon$ and~$w^\varepsilon$ which need not be
collinear).
Moreover, we only consider the case that~$x$ and~$y$ have timelike separation
(for points with spacelike separation see Exercise~\ref{ex7}).
Then, evaluating~\eqref{vectorscalar} for~$x=y$, one sees that~$P^\varepsilon(x,x)$
can be written as
\[ P^\varepsilon(x,x) = v^\varepsilon_j(x) \:\gamma^j \:+\: \beta^\varepsilon(x) \:\1 \]
(where we set~$v^\varepsilon(x)=v^\varepsilon(x,x)$ and~$\beta^\varepsilon(x)=\beta^\varepsilon(x,x)$).
Since~$P^\varepsilon(x,x)$ is a symmetric operator on~$S_x\scrM$, it follows that~$v^\varepsilon$
is a real vector field, and~$\beta$ a real-valued function.
For a large class of regularizations, the matrix~$P^\varepsilon(x,x)$ is invertible because the
vectorial component dominates the scalar component (see Exercise~\ref{ex9}).
With this in mind, we
here assume that~$\nu(x)$ exists. Then it is given by
\beq \label{nuform}
\nu(x) = \frac{1}{\rho(x)} \:\Big( v^\varepsilon_j(x) \:\gamma^j - \beta^\varepsilon(x) \:\1 \Big) \:,
\eeq
where~$\rho :=  v^\varepsilon_j  \,(v^\varepsilon)^j - (\beta^\varepsilon)^2$.
Now we can compute the composite expression in~\eqref{Ccomm}, working
for all other terms with the unregularized formulas
(which is again justified if we have pointwise convergence~\eqref{pointwise}).
This gives the following result.
\sindex{time direction}%
\nindex{ab8@${\mathscr{C}}(x, y)$ -- distinguishes time direction}%

\begin{Prp} Using~\eqref{nuform} and replacing~$P^\varepsilon(x,y)$, $P^\varepsilon(y,x)$
and~$A_{xy}$ by the unregularized expressions~\eqref{Pxyrep}, \eqref{Pyxrep} and~\eqref{1},
the functional~${\mathscr{C}}$ is given by
\beq \label{Cfinal}
{\mathscr{C}}(x, y) = \frac{16 a}{\rho(x)\, \rho(y)}\: \im \big(\alpha \overline{\beta}\big)\;
\Big(v^\varepsilon(x)^j\, \xi_j\: v^\varepsilon(y)^k\, \xi_k - \xi^2\:
v^\varepsilon(x)^j \,v^\varepsilon(y)_j \Big) \:.
\eeq
\end{Prp}
\Proof Using~\eqref{nuform} and~\eqref{1} in~\eqref{Ccomm} gives
\begin{align*}
{\mathscr{C}}(x, y) &= i \Tr_{S_x} \!\Big(  P(x,y) \:\nu(y)\:P(y,x)\:
\big[ \nu(x), A_{xy} \big] \Big) \\
&= \frac{ia}{\rho(x)} \Tr_{S_x} \!\Big(  P(x,y) \:\nu(y)\:P(y,x)\:
\big[  \slashed{v}^\varepsilon(x), \slashed{\xi} \big] \Big) \:,
\end{align*}
where in the last step we used that the scalar components of~$A_{xy}$ and~$\nu(x)$ drop
out of the commutator.
Taking the scalar component of~$\nu(y)$, the two factors~$P(x,y)$ and~$P(y,x)$ combine
to the closed chain, which according to~\eqref{1} has no bilinear component, so that
the trace vanishes. Therefore, we only need to take into account the vectorial component of~$\nu(y)$.
Using~\eqref{Pxyrep} and~\eqref{Pyxrep}, we obtain
\begin{align*}
{\mathscr{C}}(x, y)
&= \frac{ia}{\rho(x)\, \rho(y)} \Tr_{S_x} \!\Big(  \big(\alpha \slashed{\xi}+\beta\,\1\big) \: \slashed{v}^\varepsilon(y) \:\big(\overline{\alpha} \slashed{\xi}+\overline{\beta}\,\1\big)\:
\big[  \slashed{v}^\varepsilon(x), \slashed{\xi} \big] \Big) \\
&= -\frac{a}{\rho(x)\, \rho(y)}\: \im \big(\alpha \overline{\beta}\big)\:
 \Tr_{S_x} \!\Big(  \big[ \slashed{\xi}, \slashed{v}^\varepsilon(y) \big]\:
\big[  \slashed{v}^\varepsilon(x), \slashed{\xi} \big] \Big) \:.
\end{align*}
Computing the trace of the product of Dirac matrices gives the result.
\QED
The critical reader may wonder why the functional which distinguishes the time direction
has the specific form~\eqref{Cform}. This question is addressed in Exercise~\ref{ex9a}, where
another similar functional is analyzed.

For the interpretation of the formula~\eqref{Cfinal}, we first consider
the case that~$y$ and~$x$ have space-like separation.
In this case, it turns out that the prefactor~$\im (\alpha \overline{\beta})$ vanishes,
so that~\eqref{Cfinal} gives no information on a time direction. This is consistent with the fact that
for points in Minkowski space with space-like separation, the notions of future- and past-directed
depend on the observer and cannot be defined in a covariant manner.
However, if~$y$ and~$x$ have timelike separation, then the factors~$a$ and~$\im (\alpha \overline{\beta})$
are indeed both non-zero (see the proof of Proposition~\ref{prpcorrespond}).
Therefore, the functional~${\mathscr{C}}$ is non-zero, provided that the
vector~$\xi$ is {\em{non-degenerate}} in the sense that it is linearly independent
of both~$v^\varepsilon(x)$ and~$v^\varepsilon(y)$.
Since the set of directions~$\xi$ for which these vectors are linearly
dependent has measure zero, we may always restrict attention to
non-degenerate directions.
Moreover, the formula~\eqref{Cfinal} shows that the functional~${\mathscr{C}}$ does not change sign
for~$\xi$ inside the upper or lower light cone.
On the other hand, ${\mathscr{C}}$
is antisymmetric under sign flips of~$\xi$ because interchanging~$x$ and~$y$
in~\eqref{Cform} obviously gives a minus sign.

\sindex{correspondence to Minkowski space!time direction}%
We conclude that for the regularized Dirac sea vacuum,
the sign of the functional~${\mathscr{C}}$ distinguishes a time direction.
Asymptotically as~$\varepsilon \searrow 0$, this time direction agrees with the distinction of the causal
past and causal future in Minkowski space.

To summarize, in this section we saw how the intrinsic structures of a causal fermion system
correspond to the usual structures in Minkowski space.
To this end, we constructed causal fermion systems from a regularized
Dirac sea configuration and analyzed the asymptotics as the UV regularization
is removed. For brevity, we only considered the topological and causal structure of space-time
as well as spinors and wave functions. The reader interested in geometric structures
like connection and curvature is referred to the detailed exposition in~\cite{lqg}.
Moreover, in Section~\ref{seclimit} below we shall explain how the methods and results
introduced in this section can be generalized to interacting systems.
\sindex{correspondence to Minkowski space|)}%

\section{Underlying Physical Principles} \label{secprinciples}
In order to clarify the physical concepts, we now briefly discuss the
underlying physical principles.
Causal fermion systems evolved from an attempt to combine several physical principles
in a coherent mathematical framework. As a result, these principles appear in the
framework in a specific way:
\begin{itemize}[leftmargin=1.3em, itemsep=0.2em]
\itemD The {\bf{principle of causality}} is built into a causal fermion system in a specific way,
as was explained in~\S\ref{seccausal} above.
\sindex{causality}%
\itemD The {\bf{Pauli exclusion principle}} is incorporated in a causal fermion system,
as can be seen in various ways. 
\sindex{Pauli exclusion principle}%
One formulation of the Pauli exclusion principle states that every fermionic one-particle state
can be occupied by at most one particle. In this formulation, the Pauli exclusion principle
is respected because every wave function can either be represented in the form~$\psi^u$
(the state is occupied) with~$u \in \H$ or it cannot be represented as a physical wave function
(the state is not occupied). 
Via these two conditions, the fermionic projector encodes for every state
the occupation numbers~$1$ and~$0$, respectively, but it is
impossible to describe higher occupation numbers.
More technically, one may obtain the connection to the fermionic Fock space formalism
by choosing an orthonormal basis~$u_1, \ldots, u_f$ of~$\H$ and forming the $f$-particle Hartree-Fock state
\[ \Psi := \psi^{u_1} \wedge \cdots \wedge \psi^{u_f} \:. \]
Clearly, the choice of the orthonormal basis is unique only up to the unitary transformations
\[ u_i \rightarrow \tilde{u}_i = \sum_{j=1}^f U_{ij} \,u_j \quad \text{with} \quad U \in \U(f)\:. \]
Due to the anti-symmetrization, this transformation changes the corresponding Hart\-ree-Fock state
only by an irrelevant phase factor,
\[ \psi^{\tilde{u}_1} \wedge \cdots \wedge \psi^{\tilde{u}_f} = \det U \;
\psi^{u_1} \wedge \cdots \wedge \psi^{u_f} \:. \]
Thus the configuration of the physical wave functions can be described by a
fermionic multi-particle wave function.
The Pauli exclusion principle becomes apparent
in the total anti-symmetrization of this wave function.
\itemD A {\bf{local gauge principle}} becomes apparent once we choose
\sindex{local gauge principle}%
\sindex{local gauge freedom}%
\sindex{local gauge transformation}%
basis representations of the spin spaces and write the wave functions in components.
Denoting the signature of~$(S_x, \Sl .|. \Sr_x)$ by~$(p(x),q(x))$, we choose
a pseudo-orthonormal basis~$(\mathfrak{e}_\alpha(x))_{\alpha=1,\ldots, p+q}$ of~$S_x$.
Then a wave function~$\psi$ can be represented as
\[ \psi(x) = \sum_{\alpha=1}^{p+q} \psi^\alpha(x)\: \mathfrak{e}_\alpha(x) \]
with component functions~$\psi^1, \ldots, \psi^{p+q}$.
The freedom in choosing the basis~$(\mathfrak{e}_\alpha)$ is described by the
group~$\U(p,q)$ of unitary transformations with respect to an inner product of signature~$(p,q)$.
This gives rise to the transformations
\[ \mathfrak{e}_\alpha(x) \rightarrow \sum_{\beta=1}^{p+q} U^{-1}(x)^\beta_\alpha\;
\mathfrak{e}_\beta(x) \qquad \text{and} \qquad
\psi^\alpha(x) \rightarrow  \sum_{\beta=1}^{p+q} U(x)^\alpha_\beta\: \psi^\beta(x) \]
with $U \in \U(p,q)$.
As the basis~$(\mathfrak{e}_\alpha)$ can be chosen independently at each space-time point,
one obtains {\em{local gauge transformations}} of the wave functions,
where the gauge group is determined to be the isometry group of the spin scalar product.
The causal action is
{\em{gauge invariant}} in the sense that it does not depend on the choice of spinor bases.
\itemD The {\bf{equivalence principle}} is incorporated in the following general way.
\sindex{equivalence principle}%
Space-time $M:= \supp \rho$ together with the universal measure~$\rho$ form a topological
measure space, being a more general structure than a Lorentzian manifold.
Therefore, when describing~$M$ by local coordinates, the freedom in choosing such
coordinates generalizes the freedom in choosing general reference frames in a space-time manifold.
Therefore, the equivalence principle of general relativity is respected. The causal action is {\em{generally
covariant}} in the sense that it does not depend on the choice of coordinates.
\end{itemize}

\section{The Dynamics of Causal Fermion Systems}
Similar to the Einstein-Hilbert action in general relativity, in the causal action principle
one varies space-time as well as all structures therein globally.
This global viewpoint implies that it is not obvious what the causal action principle tells us about the
dynamics of the system. The first step for clarifying the situation is to derive the
Euler-Lagrange (EL) equations corresponding to the causal action principle (\S\ref{secvary}).
Similar to the Einstein or Maxwell equations, these EL equations should
describe how the system evolves in time. Additional insight is gained by
studying Noether-like theorems which specify the quantities which are conserved in the dynamics
(\S\ref{secnoether}). Finally, we review results on the initial value problem (\S\ref{secinitial}).
We remark that more explicit information on the dynamics is obtained by
considering limiting cases in which the EL equations corresponding to the
causal action reduce to equations of a structure familiar from classical field theory
and quantum field theory (see Section~\ref{seclimit}).

\subsectionn{The Euler-Lagrange Equations} \label{secvary}
We return to the abstract setting of Section~\ref{secframe}.
Our goal is to derive the EL equations corresponding to the causal action principle
in the form most useful for our purposes. Let~$(\H, \F, \rho)$ be a causal fermion system.
We assume that~$\rho$ is a minimizer of the causal action principle.
However, we do not want to assume that the total volume~$\rho(\F)$ be finite.
Instead, we merely assume that~$\rho$ is {\em{locally finite}} in the sense that~$\rho(K)< \infty$
for every compact subset~$K \subset \F$.
\sindex{measure!locally finite}%
Moreover, we only consider variations of~$\rho$ of finite total variation (see the inequality in~\eqref{totvol}).
We treat the constraints with Lagrange multipliers (this procedure is justified in~\cite{lagrange}).
Thus for each constraint~\eqref{volconstraint}--\eqref{Tdef},
we add a corresponding Lagrange Lagrange multiplier term to the action.
We conclude that first variations of the functional
\beq \label{Skaplam}
\Sact_{\kappa,\lambda,\nu}
:= \Sact + \kappa \,\big( \T - C_1 \big) - \lambda \left( \int_\F \tr (x) \: d\rho - C_2 \right)
- \nu \big( \rho(\F) - C_3 \big)
\eeq
\nindex{aq4@$\Sact_{\kappa,\lambda,\nu}$ -- causal action including Lagrange parameters}%
vanish for suitable values of the Lagrange parameters~$\kappa, \lambda, \nu \in \R$,
where the constants~$C_1$, $C_2$ and~$C_3$ are the prescribed values of the constraints.
For clarity, we point out that the boundedness constraint merely is an inequality.
The method for handling this inequality constraint is to choose~$\kappa=0$ if~$\T(\rho)<C$, whereas
in the case~$\T(\rho)=C$ the Lagrange multiplier~$\kappa$ is in general non-zero
(for details see again~\cite{lagrange}). Introducing the short notation
\beq \label{Lkappadef}
\L_\kappa(x,y) := \L(x,y) + \kappa \, |xy|^2 \:,
\eeq
we can write the effective action as
\beq \label{Ssum}
\begin{split}
\Sact_{\kappa,\lambda,\nu}(\rho)
&= \iint_{\F \times \F} \L_\kappa(x,y) \: d\rho(x)\, d\rho(y)
- \lambda \int_\F \tr (x)\: d\rho(x) - \nu \:\rho(\F) \\
&\qquad - \kappa \,C_1 + \lambda \,C_2 + \nu \,C_3 \:.
\end{split}
\eeq

When considering first variations of the measure~$\rho$, it is useful to distinguish between two
types of variations. One possible variation is to multiply~$\rho$ by a positive
function~$f_\tau \::\: M \rightarrow \R^+$,
\beq \label{weight}
\rho_\tau = f_\tau\, \rho \:.
\eeq
\sindex{variation of universal measure!by multiplication}%
Clearly, this does not change the support of the measure. In order to change the support,
one can consider a function~$F_\tau \::\: M \rightarrow \F$ and take the push-forward measure,
\beq \label{pushtau}
\rho_\tau = (F_\tau)_* \rho \:.
\eeq
\sindex{variation of universal measure!by push-forward}%
Combining these two variations, we are led to considering the family of measures
\beq \label{rhoFf}
\rho_\tau = (F_\tau)_* \big( f_\tau \,\rho \big) \:.
\eeq
Clearly, $\rho_0$ should coincide with our minimizing measure~$\rho$, leading to the condition
\beq \label{fFinit}
f_0 \equiv 1 \qquad \text{and} \qquad F_0 \equiv \1 \:.
\eeq
Moreover, in order to ensure that the variation has finite total variation, we assume
that it is trivial outside a compact set~$K \subset M$, i.e. for all~$\tau \in (-\delta, \delta)$,
\beq \label{fFtrivial}
f_\tau \big|_{M \setminus K} \equiv 1 \qquad \text{and} \qquad
F_\tau \big|_{M \setminus K} \equiv \1 \:.
\eeq
Finally, we assume that the functions~$f_\tau$ and~$F_\tau$ are defined and smooth in~$\tau$
for all~$\tau \in (-\delta, \delta)$ for some~$\delta>0$.
Variations of the form~\eqref{rhoFf} are sufficiently general for all the purposes of this book
(more general variations will be discussed in Remark~\ref{remgenEL} below).

Choosing the function~$F_\tau$ in a specific way, one gets the following result.
\begin{Prp} \label{prptrxconst}
If~$\rho$ is a minimizing measure of the causal action principle, then
there is a real constant~$c$ such that
\beq \label{trxconst}
\tr(x) = c \qquad \text{for all~$x \in M$}\:.
\eeq
\end{Prp} \noindent
We often refer to~$\tr(x)$ as the {\em{local trace}} at the point~$x$.
\sindex{local trace}%
Then the above proposition can be stated that for a minimizing measure,
the local trace is constant in space-time.
\Proof[Proof of Proposition~\ref{prptrxconst}]
Using the definition of the push-forward measure and the fact that
the variation is trivial outside~$K$, the integral over a function~$\phi$ on~$\F$
can be written conveniently as
\begin{align*}
\int_\F \phi(x)\: d\rho_\tau(x) &= \int_\F \phi\big( F_\tau(x) \big)\: f_\tau(x)\: d\rho(x) \\
&= \int_K \phi\big( F_\tau(x) \big)\: f_\tau(x)\: d\rho(x)
+ \int_{M \setminus K} \phi(x)\: d\rho(x) \:.
\end{align*}

We choose the mapping~$F_\tau$ as
\sindex{local correlation operator!rescaling of}%
\beq \label{Ftauf}
F_\tau(x) = \frac{x}{\sqrt{f_\tau(x)}} \:.
\eeq
Using that~$\L_\kappa(x,y)$ is homogeneous in~$y$ of degree two, it follows that
\begin{align*}
\int_K &\L_\kappa\big( x, F_\tau(y) \big)\: f_\tau(y)\: d\rho(y)
= \int_K \L_\kappa\bigg( x, \frac{y}{\sqrt{f_\tau(y)}} \bigg)\: f_\tau(y)\: d\rho(y) \\
&= \int_K \L_\kappa(x,y) \frac{1}{f_\tau(y)} \: f_\tau(y)\:d\rho(y) 
= \int_K \L_\kappa(x,y)\: d\rho(y) \:.
\end{align*}
Arguing similarly in the variable~$x$, one sees that
the variation does not change the integrals over~$\L_\kappa$
in~\eqref{Ssum}. Hence it remains to consider the variation of the other terms in~\eqref{Ssum},
\begin{align*}
\Sact_{\kappa,\lambda,\nu}(\rho_\tau) - \Sact_{\kappa,\lambda,\nu}(\rho)
&= - \lambda \int_K \Big( \tr \big( F_\tau(x) \big) \, f_\tau(x) - \tr (x) \Big)\: d\rho(x) - \nu
\int_K (f_\tau(x) - 1)\: d\rho(x) \\
&= - \lambda \int_K \Big( \sqrt{f_\tau(x)} - 1 \Big)\: \tr(x)\, d\rho(x)
- \nu \int_K (f_\tau(x) - 1)\: d\rho(x) \:,
\end{align*}
where in the last step we used the linearity of the trace.
Choosing~$f_\tau = 1 + \tau g$ (where~$g$ is a bounded function supported in~$K$),
the first order variation, denoted by
\[ \delta \Sact_{\kappa,\lambda,\nu} = \frac{d}{d\tau}\, \Sact_{\kappa,\lambda,\nu}\Big|_{\tau=0} \:, \]
is computed by
\[ \delta \Sact_{\kappa,\lambda,\nu} = - \frac{\lambda}{2} \int_K g(x) \: \tr(x)\, d\rho(x)
- \nu \int_K g(x) \: d\rho(x)
= - \int_K g(x) \: \bigg[ \frac{\lambda}{2} \:\tr(x) + \nu \bigg]\, d\rho(x) \:. \]
Since~$g$ is arbitrary, it follows that the square brackets vanish identically.
This gives the result.
\QED

The result of this proposition is important because it tells us that seeking for
minimizers of the causal action, we should always arrange that the local trace
is constant on~$M$. If this constant is zero, then the measure supported at one point~$x$
given by~\eqref{xtrivial} is a trivial minimizer. Therefore, we shall always
restrict attention to the case~$c \neq 0$. Then we can arrange~\eqref{trxconst}
by the scaling
\sindex{local correlation operator!rescaling of}%
\sindex{local trace!arrange to be constant}%
\beq \label{rhorescale}
\rho \rightarrow F_* \rho \qquad \text{with} \qquad F(x) = \frac{c}{\tr(x)}\: x \:.
\eeq
Clearly, this transformation can be used only if the local trace has no zeros in~$M$.
In order to avoid the analysis of the zeros of the local trace,
we note that if the local trace has zeros, then the measure
cannot be a minimizer because the condition~\eqref{trxconst} is violated, and we cannot
arrange this condition by rescaling. Thus we may take the point of view
that this measure is not useful for us and should be discarded.
In other words, our strategy for constructing minimizers is to start from a measure~$\rho$
for which the local trace has no zeros in~$M$, and to perform the rescaling~\eqref{rhorescale}.
The resulting measure satisfies~\eqref{trxconst}.
With this in mind, in what follows we shall always assume that~\eqref{trxconst} holds.

In Proposition~\ref{prptrxconst} we considered variations
of the form~\eqref{rhoFf} with an arbitrary function~$f_\tau$.
Therefore, in what follows we may restrict attention to
variations obtained by taking the push-forward~\eqref{pushtau}
(more precisely, every linear perturbation can be decomposed uniquely
into the sum of a variation of the form~\eqref{rhoFf} with~$F_\tau$ given by~\eqref{Ftauf}
and a variation of the form~\eqref{pushtau}).
Variations of the form~\eqref{pushtau} can be described conveniently by working with
so-called {\em{variations of the physical wave functions}}, which we now introduce.
\sindex{variation of physical wave functions}%
\sindex{variation of universal measure!by variation of physical wave functions|see{variation of physical wave functions}}%
Our starting point is the wave evaluation operator~$\Psi$ introduced in~\eqref{weo},
\[ \Psi \::\: \H \rightarrow C^0(M, SM)\:, \qquad u \mapsto \psi^u \:. \]
We want to vary the wave evaluation operator.
Thus for given~$\delta>0$ and any~$\tau \in (-\delta, \delta)$ we consider a
mapping~$\Psi_\tau : \H \rightarrow C^0(M, SM)$.
For~$\tau=0$, this mapping should coincide with the wave evaluation operator~$\Psi$.
The family~$(\Psi_\tau)_{\tau \in (-\delta, \delta)}$ can be regarded as a simultaneous variation
of all physical wave functions of the system.
In fact, for any~$u \in \H$, the variation of the corresponding physical wave function is given by
\[ \psi^u_\tau := \Psi_\tau(u) \in C^0(M, SM) \:. \]
Next, we introduce the corresponding local correlation operators~$F_\tau$ by
\beq \label{Ftauvary}
F_\tau(x) := - \Psi_\tau(x)^* \Psi_\tau(x) \qquad \text{so that} \qquad
F_\tau \::\: M \rightarrow \F \:.
\eeq
In view of~\eqref{Fid}, we know that~$F_0(x)=x$. Therefore, the family~$(F_\tau)_{\tau \in (-\delta, \delta)}$
is a variation of the local correlation operators. Taking the push-forward measure~\eqref{pushtau}
gives rise to a variation~$(\rho_\tau)_{\tau \in (-\delta, \delta)}$ of the universal measure.
Indeed, if all points of~$K$ are regular (see Definition~\ref{defregular}),
every variation of the universal measure of the form~\eqref{pushtau} can be
realized by a variation of the wave functions (see Exercise~\ref{ex9b}).

We now work out the EL equations for the resulting class of variations of the universal
measure. In order for the constructions to be mathematically well-defined, we need
a few technical assumptions which are summarized in the following definition.
\begin{Def} \label{defvarc}
The {\bf{variation of the physical wave functions}}
is {\bf{smooth}} and {\bf{compact}} if the family of operators~$(\Psi_\tau)_{\tau \in (-\delta, \delta)}$
has the following properties:
\sindex{variation of universal measure!by variation of physical wave functions|see{variation of physical wave functions}}%
\sindex{variation of physical wave functions!smooth and compact}%
\begin{itemize}
\item[(a)] The variation is trivial on the orthogonal complement of a finite-dimensional
subspace~$I \subset \H$, i.e.
\[ \Psi_\tau |_{I^\perp} = \Psi \qquad \text{for all~$\tau \in (-\delta, \delta)$} \:. \]
\item[(b)] There is a compact subset~$K \subset M$ outside which the variation is trivial, i.e.
\[ \big( \Psi_\tau(u) \big) \big|_{M \setminus K} = \big( \Psi(u) \big) \big|_{M \setminus K}
\qquad \text{for all~$\tau \in (-\delta, \delta)$ and~$u \in \H$} \:. \]
\item[(c)] The Lagrangian is continuously differentiable in the sense that the derivative
\beq \label{ccond}
\frac{d}{d\tau} \L\big( x, F_\tau(y) \big) \big|_{\tau=0}
\eeq
exists and is continuous on~$M \times M$.
\end{itemize}
\end{Def} \noindent
With the conditions~(a) and~(b) we restrict attention to variations
which are sufficiently well-behaved (similar as
in the classical calculus of variations, where one restricts attention to
smooth and compactly supported variations).
It is a delicate point to satisfy the condition~(c), because (due to the absolute values of the eigenvalues in~\eqref{Lagrange}) the Lagrangian is only Lipschitz continuous on~$\F \times \F$.
Therefore, the derivative in~\eqref{ccond} does not need to exist, even if~$F_\tau(y)$ is smooth.
This means that in the applications, one must verify that the condition~(c) holds
(for details see Sections~\ref{s:sec5}, \ref{s:sec7} and many computations in subsequent sections).
Right now, we simply assume that the variation of the wave functions is smooth and compact.

By definition of the push-forward measure~\eqref{pushtau}, we know that for any integrable
function~$f$ on~$\F$,
\beq \label{pushid}
\int_\F f(x)\: d\rho_\tau = \int_\F f(F_\tau(x) \big)\: d\rho \:.
\eeq
In this way, the variation of the measure can be rewritten as a variation of the
arguments of the integrand. In particular, the variation of the action
can be written as
\[ \iint_{M \times M} \L\big(F_\tau(x), F_\tau(y) \big) \: d\rho(x)\, d\rho(y) \]
(and similarly for the other integrals).
Another benefit of working with the push-forward measure~\eqref{pushtau}
is that the total volume is preserved. Namely, combining the identity~\eqref{pushid}
with the assumption in Definition~\ref{defvarc}~(b), one readily verifies that
the volume constraint~\eqref{volconstraint} is satisfied in the sense that~$\rho_\tau$
satisfies the conditions~\eqref{totvol}.

Now we can compute the first variation by differentiating with respect to~$\tau$.
It is most convenient to express the causal action and the
constraints in terms of the kernel of the fermionic projector
(just as explained at the beginning of~\S\ref{secker}).
Moreover, it is preferable to consider the Lagrangian~$\L_\kappa(x,y)$ as a function only
of~$P_\tau(x,y)$ by writing the closed chain as
\beq \label{Atauex}
A^\tau_{xy} = P_\tau(x,y)\, P_\tau(x,y)^*
\eeq
(where the index~$\tau$ clarifies the dependence on the parameter~$\tau \in (-\delta, \delta)$,
and~$P_\tau(x,y)^*$ denotes similar to~\eqref{Pxysymm} the adjoint with respect to the spin scalar
product). When computing the variation of the Lagrangian, one must keep in mind that~$\L_\kappa(x,y)$
depends both on~$P_\tau(x,y)$ and on its adjoint~$P_\tau(x,y)^*$ (cf.~\eqref{Atauex}).
Therefore, when applying the chain rule, we obtain contributions which are complex linear
and complex anti-linear in~$\delta P(x,y)$. We write the first variation in terms of traces as
\[ \delta \L_\kappa(x,y) = \Tr_{S_y} \big( B\, \delta P(x,y) \big) + \Tr_{S_x} \!\big( C\, \delta P(x,y)^* \big) \]
with linear operators~$B : S_x \rightarrow S_y$ and~$C : S_y \rightarrow S_x$.
Since~$\delta P(x,y)$ can be chosen arbitrarily, this equation uniquely defines
both~$B$ and~$C$.
Since the variation of the Lagrangian is always real-valued, it follows that~$C=B^*$.
Using furthermore the
symmetry of the Lagrangian in the arguments~$x$ and~$y$, we conclude that
the first variation of the Lagrangian can be written as (see also~\cite[Section~5.2]{PFP})
\beq \label{delLdef}
\delta \L_\kappa(x,y) = 
\Tr_{S_y} \big( Q(y,x)\, \delta P(x,y) \big) + \Tr_{S_x} \!\big( Q(x,y)\, \delta P(x,y)^* \big)
\eeq
with a kernel~$Q(x,y) : S_y \rightarrow S_x$ which is symmetric in the sense that
\beq \label{Qsymm}
Q(x,y)^* = Q(y,x)\:.
\eeq
\nindex{aq6@$Q(x,y)$ -- first variation of the Lagrangian}%

The EL equations are expressed in terms of the kernel~$Q(x,y)$ as follows.
\sindex{Euler-Lagrange equations}%
\sindex{causal action!Euler-Lagrange equations of}%
\sindex{action!causal|see{causal action}}%
\begin{Prp} {\bf{(Euler-Lagrange equations)}} \label{prpEL}
Let~$\rho$ be a minimizer of the causal action principle. Then for a suitable choice of the
Lagrange parameters~$\lambda$ and~$\kappa$, the integral operator~$Q$ with kernel defined by~\eqref{delLdef}
satisfies the equations
\beq \label{Qrel}
\int_M Q(x,y)\, \psi^u(y)\: d\rho(y) = \frac{\lambda}{2}\: \psi^u(x) \qquad \text{for all~$u \in \H$ and~$x \in M$}\:.
\eeq
\end{Prp} \noindent
We note for clarity that by writing the equation~\eqref{Qrel} we imply that
the integral must exist and be finite.
\Proof[Proof of Proposition~\ref{prpEL}] Using~\eqref{delLdef}, the first variation of~$\Sact_{\kappa, \lambda, \nu}$
is computed by
\begin{align*}
\delta \Sact_{\kappa,\lambda,\nu}
=\;& \iint_{M\times M} \Big(
\Tr_{S_y} \big( Q(y,x)\, \delta P(x,y) \big) + \Tr_{S_x} \!\big( Q(x,y)\, \delta P(x,y)^* \big) \Big) \:d\rho(x) \: d\rho(y) \\
&- \lambda \int_M \Tr \big(\delta P(x,x) \big)\: d\rho(x) \:.
\end{align*}
Noting that~$\delta P(x,y) = \delta P(y,x)^*$, after renaming the integration variables in the first summand
of the double integral, we obtain
\beq \begin{split}
\delta \Sact_{\kappa,\lambda,\nu}
&= 2 \iint_{M\times M} \Tr_{S_x} \!\big( Q(x,y)\, \delta P(y,x) \big)\: d\rho(x)\, d\rho(y) \\
&\qquad - \lambda \int_M \Tr_{S_x} \!\big(\delta P(x,x) \big)\: d\rho(x) \:.
\end{split} \label{delS}
\eeq

Next, we express~$\delta P$ in terms of the variation of the physical wave
functions. By Lemma~\ref{lemmaPxyrep}, we know that
\[ P_\tau(y,x) = -\Psi_\tau(y) \Psi_\tau(x)^* \:. \]
Differentiating this relation gives
\[ \delta P(y,x) = -(\delta \Psi)(y) \:\Psi(x)^* -\Psi(y) \:(\delta \Psi)(x)^* \:. \]

We now specialize to the case that the variation is trivial on the
orthogonal complement of a one-dimensional subspace~$I=\text{span}(u) \subset \H$.
Then for any~$\phi \in S_y$,
\[ \delta P(y,x) \, \phi = -\delta \psi^u(y) \;\Sl \:\psi^u(x) \,|\, \phi \Sr_x
-\psi^u(y) \;\Sl \delta \psi^u(x) \,|\, \phi \Sr_x \:. \]
By inserting a phase factor according to
\[ \delta \psi^u \rightarrow e^{i \varphi}\: \delta \psi^u \:, \]
one sees that~$\delta \psi^u$ can be varied independently inside and outside the spin scalar product
(more precisely, denoting the variation of the
action~\eqref{delS} corresponding to~$\delta \psi^u$ by~$\delta \Sact_{\kappa,\lambda,\nu}[\delta \psi^u]$,
the linear combination~$\delta \Sact_{\kappa,\lambda,\nu}[\delta \psi^u]
+ i \,\delta \Sact_{\kappa,\lambda,\nu}[i \,\delta \psi^u]$
involves only the complex conjugate of~$\delta \psi^u$,
whereas~$\delta \psi^u$ without complex conjugation drops out).
We conclude that it suffices to consider variations inside the spin scalar product.
Thus the vanishing of the first variation~\eqref{delS} yields the condition
\begin{align*}
0 = 2 \iint_{M\times M} \Sl \delta \psi^u(x) \,|\, Q(x,y) \,\psi^u(y) \Sr_x
- \lambda \int_M \Sl \delta \psi^u(x) \,|\, \psi^u(x) \Sr_x \:.
\end{align*}
Since~$\delta \psi^u$ is
arbitrary (within the class of smooth and compactly supported variations), the result follows.
\QED

We remark that the kernel~$Q(x,y)$ also gives rise to an operator on the
one-particle Krein space~$(\K, \bra .|. \ket)$ as introduced in~\S\ref{secKrein}.
Thus, in analogy to~\eqref{Pdef}, one sets
\nindex{aq8@$Q$ -- integral operator on Krein space~$\K$}%
\[ Q \::\: \D(Q) \subset \K \rightarrow \K \:,\qquad (Q \psi)(x) =
\int_M Q(x,y)\, \psi(y)\, d\rho(y)\:, \]
where the domain~$\D(Q)$ can be chosen for example as the continuous
wave functions with compact support.
The symmetry property of the kernel~\eqref{Qsymm} implies that the operator~$Q$
is symmetric on the Krein space~$(\K, \bra .|. \ket)$.
The equation~\eqref{Qrel} can be written in a compact form as the operator equation
\sindex{Euler-Lagrange equations}%
\sindex{causal action!Euler-Lagrange equations of}%
\beq \label{Qrel2}
\boxed{\quad \big( 2 Q -\lambda \1 \big)\, \Psi = 0 \quad }
\eeq
(where~$\Psi$ is again the wave evaluation operator~\eqref{weo}).
In words, this equation means that the operator~$(2 Q - \lambda \1)$ vanishes
on the physical wave functions.
However, the operator equation~\eqref{Qrel2} is not satisfying mathematically
because the physical wave functions in the image of~$\Psi$ are in general not vectors
of the Krein space~$(\K, \bra .|. \ket)$ (see~\S\ref{secKrein}).
Nevertheless, \eqref{Qrel2} is useful as a short notation for the EL equations~\eqref{Qrel}.

\begin{Remark} {\bf{(more general variations)}} \label{remgenEL} {\em{
Clearly, the ansatz~\eqref{rhoFf} only covers a certain class of variations of the
universal measure. As a consequence, the resulting EL equations~\eqref{trxconst} and~\eqref{Qrel}
are only {\em{necessary}} conditions for~$\rho$ to be a critical point of the action~\eqref{Skaplam}.
We now explain how these necessary conditions are related to the
stronger EL equations as derived in~\cite{lagrange}.

Variations of the form~\eqref{rhoFf} have the property that the support of the universal measure
changes continuously (in the sense that for every compact set~$K \subset \F$ and every
open neighborhood~$U$ of~$K \cap \supp \rho$ there is~$\varepsilon>0$ such
that~$\supp \rho_\tau \cap K \subset U$ for all~$\tau$ with~$|\tau|<\varepsilon$).
In fact, up to regularity and smoothness issues which we shall not enter here,
{\em{every}} variation of~$\rho$ which changes its support continuously can be
written in the form~\eqref{rhoFf} (this could be proved abstractly using arguments as
in~\cite[Lemma~1.4]{continuum}).
Such variations can be regarded as the analogs of variations of the potentials, the metric or the wave functions
in classical field theory or quantum mechanics.
However, in the setting of causal fermion systems there are also more general smooth variations
for which the support of the measure~$\rho_\tau$
changes discontinuously. A typical example is to let~$\rho$ be a bounded measure and to set
\beq \label{rhosc}
\rho_\tau = (1-\tau^2) \:\rho + \tau^2 \,\rho(\F)\: \delta_x \:,
\eeq
where~$\delta_x$ is the Dirac measure supported at~$x \not \in \supp \rho$.
The EL equations corresponding to such variations have a different mathematical structure,
which we cannot explain in detail here.
Generally speaking, for interacting systems in Minkowski space,
the EL equations of Proposition~\ref{prpEL} give rise to an effective interaction
via {\em{classical}} gauge fields (this so-called {\em{continuum limit}} will be discussed in~\S\ref{seccl}).
\sindex{classical field!bosonic}%
\sindex{bosonic field!classical}%
\sindex{gauge field!classical}%
The EL equations corresponding to more general variations
like~\eqref{rhosc}, however, give rise to an effective interaction via bosonic {\em{quantum}} fields
(see Exercise~\ref{ex10}).
\sindex{quantum field!bosonic}%
\sindex{bosonic field!quantized}%
\sindex{gauge field!quantized}%
We will come back to this point in~\S\ref{secQFT}.
}} \QEDrem
\end{Remark} 

\begin{Remark} {\bf{(unitary variations in Krein space)}} \label{remvarKrein} {\em{
Rather than generalizing~\eqref{rhoFf}, one can also proceed in the opposite way
and restrict attention to a more special class of variations of the universal measure.
If this is done, one obtains weaker equations, meaning that the resulting EL equations
are only necessary conditions for the EL equations~\eqref{Qrel} of Proposition~\ref{prpEL} to hold.
Nevertheless, this procedure has its benefits in cases when the weaker EL equations
are easier to handle and/or if the weaker EL equations capture the essence
of~\eqref{Qrel} in certain limiting cases.
A specific class of variations which is of interest in this context
are so-called {\em{unitary variations in the Krein space}}.
\sindex{universal measure!variation of|see{variation of universal measure}}%
\sindex{variation of universal measure!by unitary variation in Krein space}%
\sindex{unitary variation in Krein space}%
Such variations were first considered in~\cite[Section~3.5]{PFP}.
It turns out that in the continuum limit, the resulting EL equations are equivalent to~\eqref{Qrel}.
The advantage of working with unitary variations in Krein spaces is that the volume and trace constraints
are respected by the variation, making it unnecessary to treat these constraints with
Lagrange multipliers. This method is also used in Chapter~\ref{sector}
(see Section~\ref{s:sec2} and~\S\ref{s:secELC}).
We now briefly outline the method and put it into the context of the variations~\eqref{rhoFf}.

We let~$U_\tau$ be a family of unitary operators on the Krein space~$(\K, \bra .|. \ket)$.
Setting
\beq \label{UtauPsi}
\Psi_\tau = U_\tau \circ \Psi \:,
\eeq
we obtain a corresponding variation of the physical wave functions.
Following~\eqref{Ftauvary} and~\eqref{pushtau} gives a corresponding
variation of the universal measure, i.e.\
\[ \rho_\tau = (F_\tau)_* \rho \qquad \text{with} \qquad
F_\tau(x) := - \Psi_\tau(x)^* \Psi_\tau(x) \:. \]
Since~$\rho_\tau$ is the push-forward of the measure~$\rho$, the volume constraint
is clearly satisfied. In order to verify the trace constraint, we note that formally,
\beq \begin{split}
\int_\F \tr(x)\: d\rho_\tau(x) &= \int_M \tr \big( F_\tau(x) \big)\: d\rho_\tau(x) \\
&= -\int_M \tr \big(\Psi_\tau(x)^* \Psi_\tau(x) \big)\: d\rho
= -\tr (\Psi_\tau^* \Psi_\tau \big) \:,
\end{split} \label{Minttr}
\eeq
where the adjoint~$\Psi_\tau^* : \K \rightarrow \H$ is defined using the respective inner products, i.e.
\[ \bra \Psi_\tau u \,|\, \phi \ket =  \la u \,|\, \Psi_\tau^* \,\phi \ra_\H \qquad \text{for~$u \in \H$,
$\phi \in \K$}\:. \]
Therefore, using~\eqref{UtauPsi} together with the fact that the operators~$U_\tau$ are unitary,
we conclude that
\[ \int_\F \tr(x)\: d\rho_\tau(x) = -\tr (\Psi_\tau^* \Psi_\tau \big)
= -\tr (\Psi^* U_\tau^*\, U_\tau \Psi_\tau \big) = 
-\tr (\Psi^* \Psi_\tau \big) = \int_M \tr(x)\: d\rho(x) \:, \]
showing that the trace constraint is indeed respected.
Clearly, this computation has the shortcoming that
the integral in~\eqref{Minttr} may diverge (see before~\eqref{totvol}),
and that~$\Psi$ does not necessarily map to~$\K$ (see~\S\ref{secKrein}).
But the above consideration can be given a mathematical meaning
when assuming that the operators~$\1-U_\tau$ can be represented
as integral operators with integral kernels which vanish outside
a compact subset of~$M \times M$. We refer to the details
to Definition~\ref{s:def21} and the constructions in Section~\ref{s:sec2}.  }} \QEDrem
\end{Remark} 

\subsectionn{Symmetries and Conserved Surface Layer Integrals} \label{secnoether}
In~\cite{noether} it is shown that symmetries of the Lagrangian give rise to conservation
laws. These results can be understood as adaptations of Noether's theorem
to the causal action principle. Since the mathematical structure of the causal action principle
is quite different from that of the Lagrangian formulation of classical field theory,
these adaptations are not straightforward.
We now explain a few concepts and results from~\cite{noether}
which are important for understanding the general physical picture.

\begin{figure}
\psscalebox{1.0 1.0} 
{
\begin{pspicture}(0,-1.511712)(10.629875,1.511712)
\definecolor{colour0}{rgb}{0.8,0.8,0.8}
\definecolor{colour1}{rgb}{0.6,0.6,0.6}
\pspolygon[linecolor=black, linewidth=0.002, fillstyle=solid,fillcolor=colour0](6.4146066,0.82162136)(6.739051,0.7238436)(6.98794,0.68384355)(7.312384,0.66162133)(7.54794,0.67939913)(7.912384,0.7593991)(8.299051,0.8705102)(8.676828,0.94162136)(9.010162,0.9549547)(9.312385,0.9371769)(9.690162,0.8571769)(10.036829,0.7371769)(10.365718,0.608288)(10.614607,0.42162135)(10.614607,-0.37837866)(6.4146066,-0.37837866)
\pspolygon[linecolor=black, linewidth=0.002, fillstyle=solid,fillcolor=colour1](6.4146066,1.2216214)(6.579051,1.1616213)(6.770162,1.1127324)(6.921273,1.0905102)(7.103495,1.0816213)(7.339051,1.0549546)(7.530162,1.0638436)(7.721273,1.0993991)(7.8857174,1.1393992)(8.10794,1.2060658)(8.299051,1.2549547)(8.512384,1.3038436)(8.694607,1.3260658)(8.890162,1.3305103)(9.081273,1.3393991)(9.379051,1.3216213)(9.659051,1.2593992)(9.9746065,1.1705103)(10.26794,1.0460658)(10.459051,0.94384354)(10.614607,0.82162136)(10.610162,0.028288014)(10.414606,0.1660658)(10.22794,0.26828802)(10.010162,0.37051025)(9.663495,0.47273245)(9.356829,0.53051025)(9.054606,0.548288)(8.814607,0.54384357)(8.58794,0.5171769)(8.387939,0.48162135)(8.22794,0.44162133)(7.90794,0.34828803)(7.6946063,0.29939914)(7.485718,0.26828802)(7.272384,0.26828802)(7.02794,0.28162134)(6.82794,0.3171769)(6.676829,0.35273245)(6.543495,0.38828802)(6.4146066,0.42162135)
\pspolygon[linecolor=black, linewidth=0.002, fillstyle=solid,fillcolor=colour0](0.014606438,0.82162136)(0.3390509,0.7238436)(0.5879398,0.68384355)(0.9123842,0.66162133)(1.1479398,0.67939913)(1.5123842,0.7593991)(1.8990508,0.8705102)(2.2768288,0.94162136)(2.610162,0.9549547)(2.9123843,0.9371769)(3.290162,0.8571769)(3.6368287,0.7371769)(3.9657176,0.608288)(4.2146063,0.42162135)(4.2146063,-0.37837866)(0.014606438,-0.37837866)
\psbezier[linecolor=black, linewidth=0.04](6.4057174,0.8260658)(7.6346064,0.45939913)(7.8634953,0.8349547)(8.636828,0.92828804)(9.410162,1.0216213)(10.165717,0.7927325)(10.614607,0.42162135)
\psbezier[linecolor=black, linewidth=0.04](0.005717549,0.8260658)(1.2346064,0.45939913)(1.4634954,0.8349547)(2.2368286,0.92828804)(3.0101619,1.0216213)(3.7657175,0.7927325)(4.2146063,0.42162135)
\rput[bl](2.0101619,0.050510235){$\Omega$}
\rput[bl](8.759051,0.0016213481){\normalsize{$\Omega$}}
\psline[linecolor=black, linewidth=0.04, arrowsize=0.09300000000000001cm 1.0,arrowlength=1.7,arrowinset=0.3]{->}(1.9434953,0.85495466)(1.8057176,1.6193991)
\rput[bl](2.0946064,1.1705103){$\nu$}
\psbezier[linecolor=black, linewidth=0.02](6.4146066,0.42384356)(7.6434956,0.057176903)(7.872384,0.43273246)(8.645718,0.52606577)(9.419051,0.61939913)(10.174606,0.39051023)(10.623495,0.019399125)
\psbezier[linecolor=black, linewidth=0.02](6.410162,1.2193991)(7.639051,0.8527325)(7.86794,1.228288)(8.6412735,1.3216213)(9.414606,1.4149547)(10.170162,1.1860658)(10.619051,0.8149547)
\rput[bl](8.499051,0.9993991){\normalsize{$y$}}
\rput[bl](7.8657174,0.49273247){\normalsize{$x$}}
\psdots[linecolor=black, dotsize=0.06](8.170162,0.65273243)
\psdots[linecolor=black, dotsize=0.06](8.796828,1.1327325)
\psline[linecolor=black, linewidth=0.02](6.1146064,1.2216214)(6.103495,0.82162136)
\rput[bl](5.736829,0.8993991){\normalsize{$\delta$}}
\rput[bl](3.6146064,0.888288){$\scrN$}
\rput[bl](1.1146064,-1.4117119){$\displaystyle \int_\scrN \cdots\, d\mu_\scrN$}
\rput[bl](5.7146063,-1.511712){$\displaystyle \int_\Omega d\rho(x) \int_{M \setminus \Omega} d\rho(y)\: \cdots\:\L(x,y)$}
\psline[linecolor=black, linewidth=0.02](6.0146065,1.2216214)(6.2146063,1.2216214)
\psline[linecolor=black, linewidth=0.02](6.0146065,0.82162136)(6.2146063,0.82162136)
\end{pspicture}
}
\caption{A surface integral and a corresponding surface layer integral}
\label{fignoether1}
\end{figure}
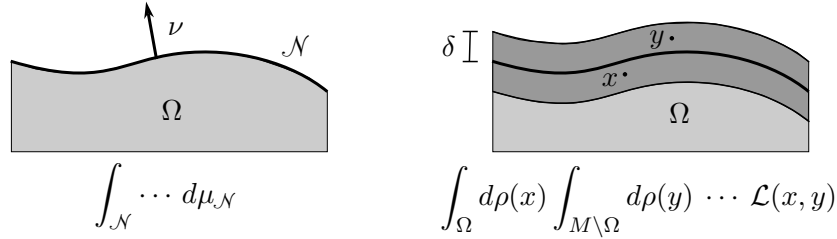

We first recall that the conservation laws obtained from the classical Noether theorem
state that the integral of a certain density over a Cauchy surface~$\scrN$ does not
depend on the choice of~$\scrN$. For example, charge conservation states that
the spatial integral of the charge density gives a constant. As another example, energy conservation
states that in a static space-time background, the integral of the energy density is a constant.
In general terms, the conserved quantities are surface integrals over a Cauchy surface~$\scrN$
(see the left of Figure~\ref{fignoether1}).
In the setting of causal fermion systems, it is unclear how such surface integrals should be
defined, in particular because we do not have a measure on hypersurfaces and because it
is not clear what the normal~$\nu$ on the hypersurface should be.
This is the reason why in the Noether-like theorems in~\cite{noether} one works instead 
of surface integrals with so-called {\em{surface layer integrals}}
\sindex{surface layer integral}%
where one integrates over a boundary layer
of a set~$\Omega \subset M$ (see the right of Figure~\ref{fignoether1}
and Exercise~\ref{ex11} for an illustration of how such double integrals arise).
The width~$\delta$ of this layer is the length scale on which~$\L(x,y)$
decays. For a system composed of Dirac particles (similar as explained in Section~\ref{secmink}
for the Minkowski vacuum and in~\S\ref{seccl} for interacting systems),
this length scale can be identified with the {\em{Compton scale}}~$\sim m^{-1}$
\sindex{Compton scale}%
of the Dirac particles. Thus the width of the surface layer is a non-zero macroscopic
length scale. In particular, the surface layer integrals cannot be identified with
or considered as a generalization of the surface integrals of the classical Noether theorem.
However, in many situations of interest the surface~$N$ is almost flat on the
Compton scale (the simplest example being a spatial hyperplane in Minkowski space). Then the surface layer
integral can be well-approximated by a corresponding surface integral.
It turns out that in this limiting case, the conservation laws obtained from the
Noether-like theorems in~\cite{noether} go over to corresponding classical
conservation laws.

From the conceptual point of view, the most interesting conservation law
is {\em{charge conservation}}.
\sindex{conservation law!for charge}%
In order to construct the underlying symmetry, we let~$\scrA$
be a bounded symmetric operator on~$\H$ and let
\[ \scrU_\tau := \exp(i \tau \scrA) \qquad \text{for~$\tau \in \R$} \]
be the corresponding one-parameter family of unitary transformations.
We introduce the family of transformations
\[ \Phi_\tau \,:\, \F \rightarrow \F\:,\qquad \Phi_\tau(x) = \scrU_\tau \,x\, \scrU_\tau^{-1} \:. \]
Since the Lagrangian is defined via the spectrum of operators on~$\H$, it clearly remains unchanged
if all operators are unitarily transformed, i.e.
\beq \label{Lsymm}
\L\big( \Phi_\tau(x), \Phi_\tau(y) \big) = \L(x,y) \:.
\eeq
In other words, the transformations~$\Phi_\tau$ describe a {\em{symmetry of the Lagrangian}}.
\sindex{symmetry!of the Lagrangian}%
Next, one constructs a corresponding one-parameter family of universal measures by taking the
push-forward,
\[ \rho_\tau := (\Phi_\tau)_* \rho \:. \]
As a consequence of the symmetry~\eqref{Lsymm}, this variation of the universal measure
leaves the action invariant. Under suitable differentiability assumptions, this
symmetry gives rise to the identity
\beq \label{conserve}
\frac{d}{d\tau} \int_\Omega d\rho(x) \int_{M \setminus \Omega} d\rho(y) \:\Big( \L\big( \Phi_\tau(x), y\big) -
\L\big(\Phi_{-\tau}(x), y \big) \Big) \Big|_{\tau=0} = 0 \:,
\eeq
valid for any compact subset~$\Omega \subset M$.

We now explain how the identity~\eqref{conserve} is related to a conservation law.
To this end, for simplicity we consider a system in Minkowski space
(similar as explained for the vacuum in Section~\ref{secmink}) and choose a sequence
of compact sets~$\Omega_n$ which exhaust the region between two Cauchy surfaces at times~$t=t_0$
and~$t=t_1$. Then the surface layer integral~\eqref{conserve} reduces to the difference
of integrals over surface layers at times~$t \approx t_0$ and~$t \approx t_1$.
Next, we choose~$\scrA =\pi_{\la u \ra}$ as the projection operator on the one-dimensional subspace
generated by a vector~$u \in \H$. Then in the limit~$\varepsilon \searrow 0$ in which the
UV regularization is removed, the resulting surface layer integral at time~$t \approx t_0$
reduces to the integral
\[ \int_{\R^3} \Sl u(t_0,\vec{x}) \,|\, \gamma^0 u(t_0, \vec{x}) \Sr_{(t_0, \vec{x})}\: d^3x\:, \]
thereby reproducing the probability integral in Dirac theory.
As a consequence, the representation of the scalar product~$\la .|. \ra_\H$ as an integral over a
Cauchy surface~\eqref{sprodMin} has a natural generalization to the setting of causal fermion systems,
if the surface integral is replaced by a corresponding surface layer integral.
This result also shows that the spatial normalization of the fermionic projector
(where one works with spatial integrals of the form~\eqref{Pnorm}; for details see~\cite{norm})
really is the correct normalization method which reflects the intrinsic conservation laws of the
causal fermion system.

The conservation laws in~\cite{noether} also give rise to the
{\em{conservation of energy and momentum}}, as we now outline.
\sindex{conservation law!for energy and momentum}%
In the classical Noether theorem, these conservation
laws are a consequence of space-time symmetries as described most conveniently using
the notion of Killing fields. Therefore, one must extend this notion to the setting of causal
fermion systems. Before explaining how this can be accomplished, we recall the
procedure in the classical Noether theorem:
In the notion of a Killing field, one distinguishes the background geometry from the
additional particles and fields. The background geometry must have a symmetry
as described by the Killing equation. The additional particles and fields, however, do not
need to have any symmetries. Nevertheless, one can construct a symmetry of the whole system
by actively transporting the particles and fields along the flow lines of the Killing field.
The conservation law corresponding to this symmetry transformation gives rise to
the conservation of energy and momentum.

In a causal fermion system, there is no clear-cut distinction between the background geometry
and the particles and fields of the system, because all of these structures are encoded in the
underlying causal fermion system and mutually depend on each other.
Therefore, instead of working with a symmetry of the background geometry,
we work with the notion of an approximate symmetry. By actively transforming
those physical wave functions which do not respect the symmetry,
such an approximate symmetry again gives rise to an exact
symmetry transformation, to which the Noether-like theorems in~\cite{noether} can be applied.
More precisely, one begins with a $C^1$-family of transformations~$(f_\tau)_{\tau \in (-\delta, \delta)}$ of space-time,
\beq \label{ftdef}
f_\tau \::\: M \rightarrow M \qquad \text{with} \qquad f_0 = \1 \:,
\eeq
which preserve the universal measure in the sense that~$(f_\tau)_* \rho = \rho$.
The family~$(f_\tau)$ can be regarded as the analog of a flow in space-time along a classical Killing field.
Moreover, one considers a family of unitary transformations~$(\scrU_\tau)_{\tau \in (-\delta, \delta)}$
on~$\H$ with the property that
\[ \scrU_{-\tau} \,\scrU_\tau = \1 \qquad \text{for all~$\tau \in (-\delta, \delta)$}\:. \]
Combining these transformations should give rise to an
{\em{approximate symmetry}} of the wave evaluation operator~\eqref{weo}
\sindex{symmetry!approximate}%
in the sense that if we compare the transformation of the space-time point with the unitary transformation
by setting
\beq \label{Edef}
E_\tau(u,x) := (\Psi u)\big(f_\tau(x) \big) - (\Psi \scrU^{-1}_\tau u)(x) \qquad (x \in M, u \in \H) \:,
\eeq
then the operator~$E_\tau : \H \rightarrow C^0(M, SM)$ should be sufficiently small. Here ``small'' means for example
that~$E$ vanishes on the orthogonal complement of a finite-dimensional subspace of~$\H$; for details see~\cite[Section~6]{noether}. Introducing the variation~$\Phi_\tau$ by
\[ \Phi_\tau \::\: M \rightarrow \F \:,\qquad \Phi_\tau(x) = \scrU_\tau \,x\, \scrU^{-1}_\tau \:, \]
we again obtain a symmetry of the Lagrangian~\eqref{Lsymm}.
This gives rise to conserved surface layer integrals of the form~\eqref{conserve}.
In order to bring these surface layer integrals into a computable form, one decomposes the first variation
of~$\Phi_\tau$ as
\beq \label{decompose}
\delta \Phi(x) := \partial_\tau \Phi_\tau(x) \big|_{\tau=0} = \delta f(x) + v(x) \:,
\eeq
where~$\delta f$ is the first variation of~$f_\tau$, \eqref{ftdef}, and~$v(x)$ is a vector field on~$\F$ 
along~$M$ which is transversal to~$M \subset \F$. Expressing~$v$ in terms of the operator~$E$
in~\eqref{Edef} shows that~$v$ is again small, making it possible to compute
the corresponding variation of the Lagrangian in~\eqref{conserve}.
We remark that in the decomposition~\eqref{decompose}, the vector field~$\delta f$ describes
a transformation of the space-time points. The vector field~$v$, however,
can be understood as an active transformation of all the objects in space-time which
do {\em{not}} have the space-time symmetry (similar as described above for
the parallel transport of the particles and fields along the flow
lines of the Killing field in the classical Noether theorem).

In order to get the connection to classical conservation laws, one again
studies a system in Minkowski space and considers the limiting case where a sequence~$\Omega_n$
exhausts the region between two Cauchy surfaces
at times~$t=t_0$ and~$t=t_1$. In this limiting case, the conserved surface layer integral
reduces to the surface integral
\[ \int_{\R^3} T_{i0} \,K^i\: d^3x\:, \]
where~$T_{ij}$ is the energy-momentum tensor of the Dirac particles and~$K= \delta f$ is a Killing field.
This shows that the conservation of energy and momentum is a 
special case of more general conservation laws which are intrinsic to causal fermion systems.


\subsectionn{The Initial Value Problem and Time Evolution} \label{secinitial}
\sindex{Cauchy problem}%
\sindex{initial value problem}%
\sindex{time evolution}%
In order to get a better understanding of the dynamics described by the causal
action principle, it is an important task to analyze the initial value problem.
The obvious questions are: What is the initial data? Is it clear that a solution exists? Is the solution unique?
How do solutions look like? Giving general answers to these questions is a difficult
mathematical problem. In order to evaluate the difficulties, one should recall that~$\rho$ describes
space-time as well as all structures therein. Therefore, similar as in the Cauchy problem for the Einstein equations, solving the initial value problem involves finding the geometry of space-time
together with the dynamics of all particles and fields.
In view of the complexity of this problem, at present there are only a few partial results.
First, in the paper~\cite{cauchy} an initial value problem is formulated and some existence and
uniqueness theorems are proven. We now review a few methods and results
of this paper. Moreover, at the end of this section we mention an approach
proposed in~\cite{jet} for obtaining more explicit information on the dynamics by
analyzing perturbations of a given minimizing measure.

Since the analysis of the causal action principle is technically demanding,
in~\cite{cauchy} one considers instead so-called {\em{causal variational principles in the
compact setting}}.
\sindex{causal variational principle!in the compact setting}%
\sindex{variational principle!causal}%
\sindex{causal variational principle|see{causal action principle}}%
In order to get into this simplified setting, one replaces~$\F$ by a compact metric space
(or a smooth manifold).
The Lagrangian is replaced by a non-negative Lipschitz-continuous function~$\L \in C^{0,1}(\F \times \F, \R^+_0)$
which is symmetric in its two arguments. Similar to~\eqref{Sdef} one minimizes the action
\[ \Sact(\rho) = \iint_{\F \times \F} \L(x,y)\: d\rho(x)\: d\rho(y) \]
in the class of all normalized regular Borel measures on~$\F$, but now
leaving out the constraints~\eqref{trconstraint} and~\eqref{Tdef}.
Space-time is again defined by~$M:= \supp \rho$.
The resulting {\em{causal structure}} is defined by saying that
two space-time points~$x, y \in M$ are called
{\em{timelike}} separated if $\L(x,y)>0$, and {\em{spacelike}} separated if~$\L(x,y)=0$.
\sindex{causality}%
\sindex{causal structure}%
\sindex{timelike}%
\sindex{spacelike}%
The principle of causality is again incorporated in the sense that pairs of points with spacelike
separation do no enter the action. But clearly, in this setting there are no wave functions. Nevertheless,
causal variational principles in the compact setting have many features of the causal action principle and
are therefore a good starting point for the analysis
(for a more detailed introduction and structural results on the minimizing measures see~\cite{support}).

When solving the classical Cauchy problem, instead of searching for a global solution for all
times, it is often easier to look for a local solution around a given initial value surface.
This concept of a local solution also reflects the
common physical situation where the physical system under consideration is only a small
subsystem of the whole universe. With this in mind, we would like to
``localize'' the variational principle to a subset~$\mathfrak{I} \subset \F$, referred to
as the {\em{inner region}}.
\nindex{ar0@$\mathfrak{I} \subset \F$ -- inner region}%
There is the complication that the Lagrangian~$\L(x,y)$ is nonlocal in the sense that
it may be non-zero for points~$x \in \mathfrak{I}$ and~$y \in \F \setminus \mathfrak{I}$.
In order to take this effect into account, one describes the influence
of the ``outer region'' $\F \setminus \mathfrak{I}$ by a so-called {\em{external potential}}
\sindex{external potential}%
$\phi : \F \rightarrow \R^+_0$. In the limiting case when the
outer region becomes large, this gives rise to the so-called {\em{inner variational principle}},
where the action defined by
\nindex{ar2@$\Sact_{\mathfrak{I}}[\rho, \phi]$ -- inner variational principle}%
\sindex{variational principle!inner}%
\beq \label{SIdef}
\Sact_{\mathfrak{I}}[\rho, \phi] = \iint_{\mathfrak{I} \times \mathfrak{I}} \L(x,y)\: d\rho(x)\: d\rho(y)
+ 2 \int_{\mathfrak{I}} \big( \phi(x) - \mathfrak{s} \big) \: d\rho(x)
\eeq
is minimized under variations of~$\rho$ in the class of regular Borel measures on~$\mathfrak{I}$
(not necessarily normalized because the volume constraint is now taken care of 
by the corresponding Lagrange parameter~$\mathfrak{s}>0$).

The {\em{initial values}} are described by a regular Borel measure~$\rho_0$
(which is to be thought of as the universal measure restricted to a time slice around
the initial value surface in space-time). The {\em{initial conditions}} are implemented
by demanding that
\beq \label{initial}
\rho \geq \rho_0 \:.
\eeq
The naive method of minimizing~\eqref{SIdef} under the constraint~\eqref{initial}
is not a sensible concept because the constraint~\eqref{initial} would give rise to undesirable
Lagrange multiplier terms in the EL equations.
Instead, one minimizes~\eqref{SIdef} without constraints, but chooses the external potential~$\phi$ in such
a way that the minimizing measure satisfies the initial values~\eqref{initial}.
It turns out that this procedure does not determine the external potential uniquely.
Therefore, the method proposed in~\cite{cauchy} is to {\em{optimize the external potential}}
by making it in a suitable sense ``small.''
As is made precise in~\cite{cauchy} in various situations, the resulting
interplay between minimizing the action and optimizing the external potential
gives rise to unique solutions of the initial-value problem with an optimal external potential.

We point out that, due to the mathematical simplifications made, the
results in~\cite{cauchy} do not apply to physically interesting situations
like the initial value problem for interacting Dirac sea configurations.
Moreover, the methods in~\cite{cauchy} do not seem to give
explicit information on the dynamics of causal fermion systems.
Therefore, it is a promising complementary approach to consider perturbations of a given minimizing measure
(which should describe the ``vacuum configuration'') and to analyze
the dynamics of the perturbations by studying the resulting EL equations.
This approach is pursued in~\cite{jet} in the following way.
In order to describe the perturbations of the minimizing measure~$\rho$,
one considers smooth variations for which the support of~$\rho$ changes
continuously. Combining~\eqref{pushtau} and~\eqref{weight}, 
these variations can be written as
\[ \tilde{\rho}_\tau = (F_\tau)_* \big( f_\tau \, \rho \big) \]
with a family of mappings~$F_\tau : M \rightarrow \F$ and
a family of non-negative functions~$f_\tau$.
Expanding in powers of~$\tau$, these variations can be described conveniently
in terms of sections of {\em{jet bundles}} over~$M$.
\sindex{jet bundle}%
The EL equations yield conditions
on the jets, which can be rewritten as dynamical equations in space-time.

\section{Limiting Cases} \label{seclimit}
We now discuss different limiting cases of causal fermion systems.
\subsectionn{The Quasi-Free Dirac Field and Hadamard States} \label{secquasifree}
We now turn attention to interacting systems. The simplest interaction is
obtained by inserting an {\em{external potential}} into the Dirac equation~\eqref{Dirfree},
\beq \label{Cont:DiracEq}
\big( i \gamma^j \partial_j + \B - m \big) \,\psi(x) = 0 \:.
\eeq
\sindex{external potential}%
\nindex{ar4@$\B$ -- external potential}%
\sindex{potential!external}%
Another situation of physical interest is to consider the Dirac equation in an external
classical gravitational field as described mathematically by a globally hyperbolic
Lorentzian manifold~$(\scrM, g)$.
\sindex{Lorentzian manifold!globally hyperbolic}%
\sindex{gravitational field}%
In this section, we explain how the methods and results of Section~\ref{secmink}
generalize to the situation when an external field is present.
This will also give a connection to quasi-free Dirac fields and Hadamard states.
In order to keep the explanations as simple as possible, we here restrict attention to
an external potential~$\B$ in Minkowski space, but remark that many methods and results could or have
been worked out also in the presence of a gravitational field.

The obvious conceptual difficulty when extending the constructions of Section~\ref{secmink}
is that one no longer has the notion of ``negative-frequency solutions''
which were essential for introducing Dirac sea configurations (see Lemma~\ref{lemmaDiracsea}).
In order to overcome this difficulty, 
one needs to decompose the solution space of the Dirac equation~\eqref{Cont:DiracEq}
into two subspaces, in such a way that without external potential, the two subspaces
reduce to the subspaces of positive and negative frequency.
This {\em{external field problem}} was solved perturbatively in~\cite{sea, grotz}
and non-perturbatively in~\cite{finite, infinite, hadamard} (for a more detailed
exposition see~\S\ref{secextfield} or~\cite[Section~2.1]{PFP}).
\sindex{external field problem}%

We now briefly outline the non-perturbative treatment, which relies on the
construction on the so-called {\em{fermionic signature operator}}.
\sindex{fermionic signature operator}%
Choosing again the scalar product~\eqref{sprodMin}, the solution space of the
Dirac equation~\eqref{Cont:DiracEq} forms a Hilbert space denoted by~$(\H_m, (.|.)_m)$.
\nindex{ar6@$(\H_m, (. \vert .)_m)$ -- Hilbert space of Dirac solutions of mass~$m$}%
Moreover, on the Dirac wave functions (not necessarily solutions of the Dirac equations)
one may introduce a dual pairing by integrating the spin scalar product over all of space-time,
\beq
\bra .|. \ket \::\: C^\infty(\scrM, S\scrM) \times C^\infty_0(\scrM, S\scrM) \rightarrow \C \:, \quad
\bra \psi|\phi \ket = \int_\scrM \Sl \psi | \phi \Sr_x \: d^4x\:. \label{stip} 
\eeq
\nindex{ar8@$\bra . \vert . \ket$ -- inner product on Dirac solutions}%
The basic idea is to extend this dual pairing to a bilinear form on the Hilbert space~$\H_m$
and to represent this bilinear form in terms of the Hilbert space scalar product
\[ \bra \phi_m | \psi_m \ket = ( \phi_m \,|\, \Sig\, \psi_m)_m \:. \]
\nindex{as0@$\Sig$ -- fermionic signature operator}%
If~$\scrM$ is a space-time of {\em{finite lifetime}}, this construction can indeed be carried out
and defines the {\em{fermionic signature operator}}~$\Sig$
being a bounded symmetric operator on~$\H_m$ (see~\cite{finite}).
The positive and negative spectral subspaces of~$\Sig$ give the desired
decomposition of~$\H_m$ into two subspaces.
We remark that the fermionic signature operator makes it possible to study
{\em{spectral geometry}} for Lorentzian signature
(see~\cite{drum} and~\cite{index} for the connection to index theory).
\sindex{spectral geometry}%
\sindex{index theory}%

In space-times of infinite lifetime like Minkowski space, the above method does not work
because~\eqref{stip} does not extend to a continuous bilinear form on~$\H_m \times \H_m$.
The underlying problem is that the time integral in~\eqref{stip} in general diverges for
solutions of the Dirac equation. In order to circumvent this problem, one considers
families of Dirac solutions~$(\psi_m)_{m \in I}$ (for an open interval~$I=(m_a, m_b)
\subset (0, \infty)$) and makes use of the fact that integrating over the mass parameter
generates decay of the wave functions for large times
(the {\em{mass oscillation property}}; for details see~\cite{infinite}).
\sindex{mass oscillation property}%
As a result, one can make sense of the equation
\[ \bra \int_I \psi_m \,dm \,|\, \int_I \psi_{m'} \,dm' \ket = \int_I (\psi_m \,|\, \Sig_m \,\phi_m)_m\: dm \:, \]
which uniquely defines a family of bounded symmetric operators~$(\Sig_m)_{m \in I}$.
Now the positive and negative spectral subspaces of the operator~$\Sig_m$ again give the desired
decomposition of~$\H_m$ into two subspaces.

Having decomposed the solution space, one may choose the Hilbert space~$\H$
of the causal fermion system as one of the two subspaces of the solution space.
Choosing an orthonormal basis~$(u_\ell)$ of~$\H$ and introducing
the unregularized kernel of the fermionic projector again by~\eqref{Pxykernel},
one obtains a two-point distribution~$P(x,y)$. Using that this two-point distribution
comes from a projection operator in the Hilbert space~$\H_m$, 
there is a canonical construction which gives a quasi-free
Dirac field together with a Fock representation such that the two-point distribution
coincides with~$P(x,y)$. In the canonical formalism, this result can be stated as follows
(for a formulation in the language of algebraic quantum field theory see~\cite[Theorem~1.4]{hadamard}):
\begin{Thm} \label{thmstate}
There are fermionic field operators~$\hat{\Psi}^\alpha(x)$ and~$\hat{\Psi}^\beta(y)^*$
together with a ground state~$|0\ket$ with the following properties: \\[0.3em]
(a) The canonical anti-commutation relations hold:\footnote{In order to avoid confusion, we note
that the operators~$\hat{\Psi}(x)^\dagger$ which appear in the usual
equal-time canonical commutation relations~$\{ \hat{\Psi}^\alpha(t,\vec{x}), \hat{\Psi}^\beta(t,\vec{y})^\dagger \} = 
\delta^\alpha_\beta\: \delta^3(\vec{x}-\vec{y})$ are related to the above operators
by~$\hat{\Psi}^\alpha(x)^\dagger = 2 \pi \sum_{\beta=1}^4 \hat{\Psi}^\beta(x)^* \:(\gamma^0)^\beta_\alpha$.}
\[ \big\{\hat{\Psi}^\alpha(x),\hat{\Psi}^\beta(y)^* \big\} = \tilde{k}_m(x,y)^\alpha_\beta\:,\qquad
\big\{ \hat{\Psi}^\alpha(x),\hat{\Psi}^\beta(y) \big\} = 0 = \big\{ \hat{\Psi}^\alpha(x)^*,\hat{\Psi}^\beta(y)^* \big\} \:. \]
(b) The two-point function is given by
\[ \bra 0 | \,\hat{\Psi}^\alpha(x) \,\hat{\Psi}^\beta(y)^* \,| 0 \ket = -P(x,y)^\alpha_\beta \:. \]
\end{Thm} \noindent
This theorem means that before introducing an UV regularization, the description
of the Dirac system using the fermionic projector is equivalent to the usual description
of a non-interacting Dirac field in quantum field theory.
\sindex{Hadamard state}%
\sindex{distribution!of Hadamard form}%
\nindex{as1@$\hat{\Psi}(x), \hat{\Psi}(x)^*$ -- fermionic field operators}%

Moreover, it is shown in~\cite{hadamard} that the two-point distribution~$P(x,y)$ is of {\em{Hadamard form}},
provided that~$\B$ is smooth, not too large and decays faster than quadratically for large times
(for details see~\cite[Theorem~1.3]{hadamard} and the references in this paper).
This result implies that the representation of the quasi-free Dirac field as obtained
from the fermionic projector is a suitable starting point for a perturbative treatment
of the resulting interacting theory (see for example~\cite{brunetti+dutsch+fredenhagen}).

In our context, the fact that~$P(x,y)$ is of Hadamard form implies that the results
in~\S\ref{seccausal} also apply in the presence of an external potential,
as we now explain. The Hadamard property means in words that the
bi-distribution~$P(x,y)$ in the presence of the external potential has the same singularity structure
as in the Minkowski vacuum. As a consequence, the arguments in~\S\ref{seccausal} remain true
if the points~$x$ and~$y$ are sufficiently close to each other. More precisely, the relevant length
scale is given by the inverse of the amplitude~$|\B(x)|^{-1}$ of the external
potential. On the other hand, the separation of the points~$x$ and~$y$ must be larger than
the scale~$\varepsilon$ on which regularization effects come into play.
Therefore, the causal structure of a causal fermion system agrees with that of Minkowski
space on the scale~$\varepsilon \ll \big|x^0-y^0\big| +\big|\vec{x}-\vec{y} \big| \ll |\B|^{-1}$
(where~$|\B|$ is any matrix norm).
Thinking of~$\varepsilon$ as being at least as small as the Planck length,
\sindex{Planck length}%
\nindex{aj9@$\ell_P$ -- Planck length}%
in most situations
of interest the lower bound is no restriction. The upper bound is also unproblematic because
the causal structure on the macroscopic scale can still be recovered by considering
paths in space-time and subdividing the path on a scale~$\delta \ll |\B|^{-1}$
(similar as explained in~\cite[Section~4.4]{lqg} for the spin connection).
With this in mind, we conclude that the causal structure of a causal fermion system
indeed agrees with that of Minkowski space, even in the presence of an external potential.

\subsectionn{Effective Interaction via Classical Gauge Fields} \label{seccl}
We now outline how to describe interacting systems in Minkowski space
by analyzing the EL equations corresponding to the causal action principle
as worked out in Proposition~\ref{prpEL}. In this so-called {\em{continuum limit}},
the interaction is described by classical gauge fields.
\sindex{continuum limit}%
Working out the details of this procedure is the
main objective of this book (see Sections~\ref{secreg} and~\ref{secspeccc} for the general formalism
and Chapters~\ref{sector}--\ref{quark} for the explicit analysis of different models).
Therefore, we here merely explain a few basic concepts.

Let us begin with the Minkowski vacuum. As shown in~\S\ref{secuvreg},
regularizing a vacuum Dirac sea configuration gives rise to a
causal fermion system~$(\H, \F, \rho^\varepsilon)$.
Moreover, we saw in the following sections~\S\ref{seccorst}--\S\ref{seccorsw} that
the inherent structures of the causal fermion system can be identified with those
of Minkowski space (in particular, see~\eqref{Midentify} as well as Propositions~\ref{prpisometry}
and~\ref{lemma54}). This makes it possible to write the EL equations~\eqref{Qrel} as
\beq \label{ELreg}
\int_\scrM Q^\varepsilon(x,y)\, \big({\mathfrak{R}}_\varepsilon u_\ell \big)(y)\: d^4y
= \frac{\lambda}{2}\: \big({\mathfrak{R}}_\varepsilon u_\ell \big)(x) \qquad \text{for all~$u \in \H$}\:,
\eeq
where the regularized kernel~$Q^\varepsilon(x,y)$ is again defined via~\eqref{delLdef}
as the derivative of the Lagrangian.
\nindex{as2@$Q^\varepsilon(x,y)$ -- first variation of the Lagrangian with regularization}%
Next, one chooses the Hilbert space~$\H$ as in~\S\ref{seccorcs} as the Dirac sea configuration
formed of all negative-energy solutions of the Dirac equation.
Then~$P^\varepsilon(x,y)$ can be computed explicitly by regularizing the
distribution~$P(x,y)$ as given in momentum space by~\eqref{Pxyvac} and
in position space by~\eqref{Pdiff} and Lemma~\ref{lemmaTintro}.
Computing~$Q^\varepsilon(x,y)$, it turns out that the
EL equations are mathematically well-defined if the convolution integral in~\eqref{ELreg} is
rewritten with the help of Plancherel's theorem as a multiplication in momentum space.
The analysis of the {\em{continuum limit}} gives a procedure for studying these equations in the
asymptotics~$\varepsilon \searrow 0$ when the regularization is removed.
The effective equations obtained in this asymptotic limit are evaluated most conveniently in
a formalism in which the unknown microscopic structure of space-time (as described by the regularization)
enters only in terms of a finite (typically small) number of so-called {\em{regularization parameters}}.
\sindex{regularization parameter}%
According to the method of variable regularization (see Remark~\ref{remmvr}),
one needs to analyze the dependence of the regularization parameters in detail.
It turns out that the causal fermion systems obtained from the vacuum Dirac sea configuration satisfy
the EL equations in the continuum limit, for any choice of the regularization parameters.

The first step towards interacting systems is to consider systems involving {\em{particles}}
and/or {\em{anti-particles}}. To this end, one simply modifies the constructions in~\S\ref{seccorcs}
by choosing the Hilbert space~$\H$ differently. Namely, instead of choosing
all negative-energy solutions, one chooses~$\H$ as a subspace of the solution space which
differs from the space of all negative-energy solutions by a finite-dimensional subspace.
In other words, $\H$ is obtained from the space of all negative-energy solutions by taking
out a finite number~$\na$ of states and by adding a finite number of states~$\np$ of positive energy.
Thus, denoting the regularized kernel of the fermionic projector of the Minkowski vacuum for
clarity by~$P^\varepsilon_\text{sea}(x,y)$, the kernel of the fermionic projector~\eqref{Pepsbase} can be written as
\beq \label{Pepsmod}
P^\varepsilon(x,y) = P^\varepsilon_\text{sea}(x,y)
-\sum_{k=1}^{\np} \big({\mathfrak{R}}_\varepsilon \psi_k \big)(x)
\overline{\big({\mathfrak{R}}_\varepsilon \psi_k \big)(y)}
+\sum_{l=1}^{\na} \big({\mathfrak{R}}_\varepsilon \phi_l \big)(x)
\overline{\big({\mathfrak{R}}_\varepsilon \phi_l \big)(y)} \:,
\eeq
where~$\psi_k$ and~$\phi_l$ are suitably normalized bases of the particle and anti-particle states,
respectively. 
\nindex{as3@$\np, \na$ -- number of particles and anti-particle states}%
\nindex{as4@$\psi_k, \phi_l$ -- particle and anti-particle states}%
\sindex{particles and anti-particles}%
\sindex{anti-particles}%
In this procedure, we again take Dirac's concept of a ``sea'' of particles literally
and describe particles and anti-particles by occupying positive-energy states and creating ``holes'' in the
Dirac sea, respectively. We also remark that the construction~\eqref{Pepsmod}
modifies the kernel of the fermionic projector only by smooth contributions and thus
preserves the singularity structure of~$P^\varepsilon(x,y)$ as~$\varepsilon \searrow 0$.
As a consequence, the correspondence of the inherent structures of the causal fermion systems
to the structures in Minkowski space remains unchanged (just as explained at the end of~\S\ref{secquasifree}
for an external potential).

According to~\eqref{Pepsmod}, the particle and anti-particle states modify the
kernel of the fermionic projector. It turns out that this has the effect that
the EL equations in the continuum limit no longer hold.
In order to again satisfy these equations, we need to introduce an interaction.
In mathematical terms, this means that the universal measure~$\rho$ must be modified.
The basic question is how to modify the universal measure in such a way that the
EL equations in the continuum limit again hold.
It turns out that it is a useful first step
to insert an external potential~$\B$ into the Dirac equation~\eqref{Dirfree}
by going over to the Dirac equation~\eqref{Cont:DiracEq}.
\nindex{ar4@$\B$ -- external potential}%
Choosing~$\H$ as a subspace of the solution space of this Dirac equation,
the constructions of Section~\ref{secmink} again apply and give rise to causal fermion
systems~$(\H, \F, \rho^\varepsilon)$. The potential~$\B$ modifies the dynamics of all physical wave functions
in a collective way. Now one can ask the question whether the resulting causal fermion systems satisfy
the EL equations in the continuum limit. It turns out that this is the case
if and only if the potential~$\B$ satisfies certain equations, which can be identified
with classical field equations for the potential~$\B$. In this way, the causal action principle
gives rise to classical field equations.
In order to make our concepts clear, we point out that the 
potential~$\B$ merely is a convenient device in order to describe the collective
behavior of all physical wave functions. It should not be considered as a fundamental
object of the theory. We also note that, in order to describe variations of the physical wave functions,
the potential in~\eqref{Cont:DiracEq} can be chosen arbitrarily (in particular, the potential
does not need to satisfy any field equations). Each choice of~$\B$
describes a different variation of the physical wave functions.
It is the EL equations in the continuum limit which single out the physically admissible potentials
as being those which satisfy the field equations.

Before going on, we briefly explain how the subspace~$\H$ is chosen.
Clearly, the Dirac equation~\eqref{Cont:DiracEq}
cannot in general be solved in closed form. Therefore, for an explicit analysis
one must use perturbative methods. When performing the perturbation expansion,
one must be careful about the proper normalization of the fermionic states
(in the sense that spatial integrals of the form~\eqref{Pnorm} should be preserved).
Moreover, one must make sure that the singular structure of~$P(x,y)$ in
position space is compatible with the causal action principle
(meaning that the light-cone expansion of~$P(x,y)$ only involves bounded
integrals of~$\B$ and its derivatives).
Satisfying these two requirements leads to the {\em{causal perturbation expansion}}
(see~\S\ref{secspatial} or~\cite{norm} and the references therein).
\sindex{perturbation expansion!causal}%
\sindex{causal perturbation expansion}%
We also mention that regularizing the perturbation expansion is a delicate issue.
This can already be understood for the simple regularization by mollification
in Example~\ref{exmollify}, in which case it is not clear whether one should first
mollify and then introduce the interaction or vice versa.
The correct method for regularizing the perturbation expansion is obtained
by demanding that the behavior under gauge transformations should be preserved
by the regularization. This leads to the {\em{regularized causal perturbation expansion}}
as developed in~\cite[Appendix~D]{PFP} and Appendix~\ref{l:appA}.
\sindex{perturbation expansion!regularized causal}%
\sindex{causal perturbation expansion!regularized}%

We proceed with a brief overview of the results of the analysis of the continuum limit.
In the following Chapters~\ref{sector}--\ref{quark}
the continuum limit is worked out in several steps beginning from simple systems
and ending with a system realizing the fermion configuration of the standard model.
For each of these systems, the continuum limit gives rise to effective equations for
second-quantized fermion fields
coupled to classical bosonic gauge fields (for the connection to second-quantized bosonic
fields see~\S\ref{secQFT} below).
\sindex{quantum field!fermionic}%
\sindex{fermionic field!quantized}%
\sindex{classical field!bosonic}%
\sindex{bosonic field!classical}%
\sindex{gauge field!classical}%
To explain the structure of the obtained results, it is preferable to first describe the
system modelling the leptons as analyzed in Chapter~\ref{lepton}.
The input to this model is the configuration of the leptons in the standard model
without interaction. Thus the fermionic projector of the vacuum is assumed to be
composed of three generations of Dirac particles of masses~$m_1, m_2, m_3>0$
(describing~$e$, $\mu$, $\tau$)
as well as three generations of Dirac particles of masses~$\tilde{m}_1, \tilde{m}_2,
\tilde{m}_3 \geq 0$ (describing the corresponding neutrinos).
Furthermore, we assume that the regularization of the neutrinos breaks the chiral
symmetry (implying that we only see their left-handed components).
We point out that the definition of the model does not involve any assumptions on the interaction.

The detailed analysis in Chapter~\ref{lepton} reveals that the effective interaction in the continuum limit
has the following structure.
The fermions satisfy the Dirac equation coupled to a left-handed $\SU(2)$-gauge
potential~$A_L=\big( A_L^{ij} \big)_{i,j=1,2}$,
\[ \left[ i \Pdd + \begin{pmatrix} \slashed{A}_L^{11} & \slashed{A}_L^{12}\, \UMNS^* \\[0.2em]
\slashed{A}_L^{21}\, \UMNS & -\slashed{A}_L^{11} \end{pmatrix} \chi_L
- m Y \right] \!\psi = 0 \:, \]
\sindex{potential!bosonic}%
\sindex{gauge potential}%
\sindex{potential!gauge}%
where we used a block matrix notation (in which the matrix entries are
$3 \times 3$-matrices). Here~$mY$ is a diagonal matrix composed of the fermion masses,
\beq \label{mY}
mY = \text{diag} (\tilde{m}_1, \tilde{m}_2, \tilde{m}_3,\: m_1, m_2, m_3)\:,
\eeq
and~$\UMNS$ is a unitary $3 \times 3$-matrix (taking the role of the
Maki-Nakagawa-Sakata matrix in the standard model).
The gauge potentials~$A_L$ satisfy a classical Yang-Mills-type equation, coupled
to the fermions. More precisely, writing the isospin dependence of the gauge potentials according
to~$A_L = \sum_{\alpha=1}^3 A_L^\alpha \sigma^\alpha$ in terms of Pauli matrices,
we obtain the field equations
\beq \label{l:YM}
\partial^k \partial_l (A^\alpha_L)^l - \Box (A^\alpha_L)^k - M_\alpha^2\, (A^\alpha_L)^k = c_\alpha\,
\overline{\psi} \big( \chi_L \gamma^k \, \sigma^\alpha \big) \psi\:,
\eeq
\sindex{field equations in the continuum limit}%
valid for~$\alpha=1,2,3$ (for notational simplicity, we wrote the Dirac current for one Dirac
particle; for a second-quantized Dirac field, this current is to be replaced by the expectation value
of the corresponding fermionic field operators). Here~$M_\alpha$ are the bosonic masses and~$c_\alpha$
the corresponding coupling constants.
The masses and coupling constants of the two off-diagonal components are
equal, i.e.\ $M_1=M_2$ and~$c_1 = c_2$,
but they may be different from the mass and coupling constant
of the diagonal component~$\alpha=3$. Generally speaking, the mass ratios~$M_1/m_1$, $M_3/m_1$ as
well as the coupling constants~$c_1$, $c_3$ depend on the regularization. For a given regularization,
they are computable.

Finally, our model involves a gravitational field described by the Einstein equations
\beq \label{l:Einstein}
R_{jk} - \frac{1}{2}\:R\: g_{jk} + \Lambda\, g_{jk} = \kappa\, T_{jk} \:,
\eeq
\sindex{Einstein equations}%
where~$R_{jk}$ denotes the Ricci tensor, $R$ is scalar curvature, and~$T_{jk}$
is the energy-momentum tensor of the Dirac field. Moreover, $\kappa$ and~$\Lambda$ denote the
gravitational and the cosmological constants, respectively.
We find that the gravitational constant scales like~$\kappa \sim \delta^2$,
where~$\delta \geq \varepsilon$ is the length scale on which the chiral symmetry is broken.
We remark that the regularization is not
necessarily constant in space-time but may have a dynamical behavior, in which case also the gravitational
constant would become dynamical. The resulting effect, referred to as dynamical gravitational coupling,
\sindex{dynamical gravitational coupling}%
will not be covered in this book, but we refer the interested reader to~\cite{dgc}.

In Chapter~\ref{quark}, a system is analyzed which realizes the
configuration of the leptons and quarks in the standard model.
The result is that the field equation~\eqref{l:YM} is replaced by
field equations for the electroweak and strong interactions after spontaneous
symmetry breaking (the dynamics of the corresponding Higgs field has not yet been analyzed).
Furthermore, the system again involves gravity~\eqref{l:Einstein}.

A few clarifying remarks are in order. First, the above field equations come with
corrections which for brevity we cannot discuss here (see Sections~\ref{s:secfield}, \ref{l:sec4}
and~\ref{l:secstructure}). Next, it is worth noting that, although
the states of the Dirac sea are explicitly taken into account in our analysis, they do not
enter the field equations.
\sindex{Dirac sea}%
More specifically, in a perturbative treatment,
the divergences of the Feynman diagram
describing the vacuum polarization drop out of the EL equations of the causal action.
Similarly, the naive ``infinite negative energy density'' of the
sea drops out of the Einstein equations, making it unnecessary to subtract any counter terms.
We finally remark that the only free parameters of the theory are the masses in~\eqref{mY}
as well as the parameter~$\delta$ which determines the gravitational constant.
The coupling constants, the bosonic masses and the mixing matrices
are functions of the regularization parameters
which are unknown due to our present lack of knowledge on the microscopic structure of space-time.
The regularization parameters cannot be chosen arbitrarily because they must satisfy certain
relations. But except for these constraints, the regularization parameters are currently treated
as free empirical parameters.

To summarize, the dynamics in the continuum limit is described by Dirac spinors
coupled to classical gauge fields and gravity. The effective continuum theory is manifestly covariant
under general coordinate transformations.
The only limitation of the continuum limit is that the bosonic fields are merely classical.
We shall come back to second-quantized bosonic fields in~\S\ref{secQFT} below.

\subsectionn{Effective Interaction via Bosonic Quantum Fields} \label{secQFT}
In~\S\ref{seccl} it was outlined that and in which sense
the regularized Dirac sea vacuum satisfies the EL equations~\eqref{Qrel}.
In simple terms, these results mean that the regularized Dirac sea vacuum is a critical
point of the causal action under variations of the physical wave functions
(see Definition~\ref{defvarc}).
We now explain why the regularized Dirac sea vacuum is  {\em{not a minimizer}} of the causal action
principle. This argument will lead us to a method for further decreasing the causal action.
It also gives some insight on the structure of the minimizing measure.
Working out this method systematically reveals that the resulting interaction
is to be described effectively by {\em{bosonic quantum fields}}.
\sindex{quantum field!bosonic}%
\sindex{bosonic field!quantized}%
\sindex{gauge field!quantized}%

Our argument is based on the general observation that a {\em{convex combination}} of
universal measures is again a universal measure.
\sindex{measure!convex combination of}%
More precisely, let~$\rho_1, \ldots, \rho_L$ be positive measures on~$\F$.
Choosing coefficients~$c_\ell$ with
\[ c_\ell \geq 0 \qquad \text{and} \qquad \sum_{\ell=1}^L c_\ell = 1 \:, \]
the convex combination
\beq \label{convexcombi}
\tilde{\rho} := \sum_{\ell=1}^L c_\ell\: \rho_\ell
\eeq
is again a positive measure on~$\F$. Moreover, if the~$\rho_\ell$ satisfy the
linear constraints (i.e.\ the volume constraint~\eqref{volconstraint}
and the trace constraint~\eqref{trconstraint}), then these constraints
are again respected by~$\tilde{\rho}$.
Taking convex combinations of universal measures resembles
superpositions of quantum states in quantum field theory.
However, as a major difference, the coefficients in
the convex combination~\eqref{convexcombi} must be real-valued
and non-negative.

Taking convex combinations of measures is a useful method
for decreasing the causal action.
Thus we want to choose the measures~$\rho_\ell$ and the coefficients~$c_\ell$
in~\eqref{convexcombi} in such a way that the boundedness constraint~\eqref{Tdef} is
satisfied and the causal action~\eqref{Sdef} is smaller than that of~$\rho$.
A simple but useful method for constructing the measures~$\rho_\ell$ is to
work with unitary transformations.
For a unitary operator~$V \in \U(\H)$, we define the measure~$V(\rho)$ by
\beq \label{Vrho}
(V \rho)(\Omega) = \rho \big(V \Omega V^{-1} \big) \:.
\eeq
Choosing unitary transformations~$V_1, \ldots, V_L$
we set~$\rho_\ell = V_\ell \rho$ and introduce~$\tilde{\rho}$ as a convex combination~\eqref{convexcombi}
where for simplicity we choose~$c_\ell=1/L$,
\[ \tilde{\rho} = \frac{1}{L} \sum_{\as=1}^L V_\as \rho \:. \]
The action becomes
\begin{align}
\Sact(\tilde{\rho}) &= \frac{1}{L^2} \sum_{\as, \bs=1}^L \iint_{\F \times \F} \L(x,y)\: d(V_\as \rho)(x)\:
d(V_\bs \rho)(y) \notag \\
&= \frac{\Sact(\rho)}{L} + \frac{1}{L^2} \sum_{\as \neq \bs} \iint_{\F \times \F} \L(x,y)\: d(V_\as \rho)(x)\:
d(V_\bs \rho)(y) \:. \label{Smix}
\end{align}
In order to analyze this equation more concretely, we consider the situation that~$(\H, \F, \rho)$ is a causal
fermion system describing a regularized Dirac sea configuration (see~\S\ref{seccorcs}).
Then, due to the factor~$1/L$, the first summand becomes small as~$L$ increases.
The second summand involves all the contributions for~$\as \neq \bs$.
If we can arrange that these contributions become small, then the action
of the new measure~$\tilde{\rho}$ will indeed be smaller than the action of~$\rho$.

Let us consider the contributions for~$\as \neq \bs$ in more detail. In order to simplify the explanations,
it is convenient to assume that the measures~$V_\as \rho$ have mutually disjoint supports
(this can typically be arranged by a suitable choice of the unitary transformations~$V_\as$).
Then the space-time~$\tilde{M} := \supp \tilde{\rho}$ can be decomposed into~$L$ ``sub-space-times''
$M_\as := \supp \rho_\as$,
\[ \tilde{M} = M_1 \cup \cdots \cup M_L \qquad \text{and} \qquad M_\as \cap M_\bs = \varnothing \quad
\text{if $\as \neq \bs$}\:. \]
Likewise, a physical wave function~$\psi^u$ can be decomposed into the contributions in
the individual sub-space-times,
\[ \psi^u = \sum_{\as=1}^L \psi^u_\as \qquad \text{with} \qquad
\psi^u_\as := \chi_{M_\as} \,\psi^u \]
(and~$\chi_{M_\as}$ is the characteristic function). This also gives rise to a corresponding
decomposition of the fermionic projector:
\begin{Lemma} Every sub-space-time~$M_\as$ of~$\tilde{M}$ is homeomorphic to~$M$,
with a homeomorphism given by
\[ \phi_\as \::\: M \rightarrow M_\as \:,\qquad
\phi_\as(x) := V_\as^* \,x\, V_\as \:. \]
Moreover, the mapping
\beq \label{Vasiso}
V_\as^* \big|_{S_x} \,:\, S_x \rightarrow S_{\phi_\as(x)}
\eeq
is an isomorphism of the corresponding spin spaces.
Identifying the spin spaces in different sub-space-times via this isomorphism,
the fermionic projector can be written as
\begin{align}
P(x,y) &= -\sum_{\as, \bs=1}^L \chi_{M_\as}(x) \: P_{\as, \bs}(x,y)\:\chi_{M_\bs}(y) \qquad \text{with} \\
P_{\as, \bs}(x,y) \:&\!:= \Psi(x) \:V_\as \,V_\bs^* \:\Psi(y)^* \:. \label{Pxymix}
\end{align}
\end{Lemma}
\Proof The definition of~$V \rho$, \eqref{Vrho}, immediately implies that the
transformation~\eqref{psiudef} maps~$M$ to~$M_\as$ and is a homeomorphism.
By definition of the physical wave function~\eqref{psiudef},
\[ \psi^u(\phi_\as(x)) = \pi_{\phi_\as(x)} = \pi_{V_\as^* x V_\as} u
= V_\as^* \,\pi_x\, V_\as u \:. \]
The identification~\eqref{Vasiso} makes it possible to leave out the factor~$V_\as^*$.
Then we can write the wave evaluation operator~\eqref{weo} as
\[ \tilde{\Psi}(x) = \sum_{\as=1}^L \chi_{M_\as}(x)\: \Psi(x)\: V_\as \:. \]
Applying~\eqref{Pid} gives the result.
\QED

This lemma makes it possible to rewrite the action~\eqref{Smix} as
\beq \label{SPmix}
\Sact(\tilde{\rho}) = \frac{\Sact(\rho)}{L} + \frac{1}{L^2} \sum_{\as \neq \bs}
\iint_{M \times M} \L\big[P_{\as, \bs}(x,y)\big] \: d\rho(x)\: d\rho(y) \:,
\eeq
where the square bracket means that the Lagrangian is computed
as a function of the kernel of the fermionic projector~$P_{\as, \bs}(x,y)$
(just as explained after~\eqref{Axydef} for the kernel~$P(x,y)$).
The identities~\eqref{Pxymix} and~\eqref{SPmix} give a good intuitive understanding of how
the action depends on the unitary operators~$V_\as$.
We first note that in the case~$\as = \bs$, the unitary operators in~\eqref{Pxymix} drop out,
so that~$P_{\as,\as}(x,y) = P(x,y)$. This also explains why the first summand in~\eqref{SPmix}
involves the original action~$\Sact(\rho)$.
In the case~$\as \neq \bs$, however, the unitary operators in~\eqref{Pxymix} do not drop out.
In particular, this makes it possible to introduce phase factors into the fermionic projector.
For example, one may change the phase of each physical wave function~$\psi^u_\as$
arbitrarily while keeping the physical wave functions~$\psi^u_\bs$ for~$\bs \neq \as$ unchanged.
Choosing the resulting phases randomly, one gets destructive interference, implying that
the kernel~$P_{\as,\bs}(x,y)$ becomes small.
Making use of this dephasing effect, one can make the summands
in~\eqref{SPmix} for~$\as \neq \bs$ small. A detailed analysis of the involved scalings
reveals that this indeed makes it possible to decrease the causal action while respecting
all constraints (see~\cite{qft}).

We refer to the measures~$\rho_\as = V_\as \rho$ as {\em{decoherent replicas}}
\sindex{decoherent replica of the universal measure}%
of~$\rho$. Thus in the above example, the universal measure~$\tilde{\rho}$
consists of a convex combination of many decoherent replicas of~$\rho$.
Likewise, space-time~$M$ is decomposed many sub-space-times
$M_1, \ldots, M_L$. Each of these sub-space-times describes the same physical
system because the geometric structures are identical
for all sub-space-times. However, the physical wave functions in the different sub-space-times
involve relative phases, with the effect that the correlations between the sub-space-times
(as described by the kernels~$P_{\as,\bs}(x,y)$) become small.
This means physically that the decoherent replicas do not interact with each other.
The resulting picture is that space-time looks effectively like a ``superposition'' of
the different sub-space-times. The dephasing can be understood similar to
decoherence effects in standard quantum field theory (see for example~\cite{joos}).

Instead of taking decoherent replicas of the same measure~$\rho$, one
can consider the situation that each of the measures~$\rho_\ell$
describes a sub-space-time which involves a different
classical bosonic field. In this way, one obtains effectively a superposition of classical field configurations.
This makes it possible to describe second-quantized bosonic fields
(see~\cite{entangle}). However, as the different sub-space-times do not interact with each other,
each sub-space-time has it own independent dynamics. This dynamics is described by the
classical bosonic field in the corresponding sub-space-time.

In order to obtain an interaction via second-quantized bosonic fields,
\sindex{quantum field!bosonic}%
\sindex{bosonic field!quantized}%
\sindex{gauge field!quantized}%
one needs to consider another limiting case in which
the dephasing involves only some of the physical
wave functions. In this case, the contributions~$P_{\as,\bs}$ with~$\as \neq \bs$ to the fermionic projector
are not necessarily small.
This also implies that relations which arise as a consequence of the collective behavior
of all physical wave functions (like the causal relations or classical bosonic fields)
still exist between the sub-space-times~$M_\as$ and~$M_\bs$.
In more physical terms, the sub-space-times still interact with each other.
This scenario is studied in~\cite{qft} and is referred to as the {\em{microscopic mixing
of physical wave functions}}.
\sindex{microscopic mixing of physical wave functions}%
In order to describe the effective interaction, one describes the
unitary operators~$V_\as$ by random matrices. Taking averages over the random
matrices, one finds that the effective interaction can be described perturbatively
in the Fock space formalism working with fermionic and bosonic field operators.
Working out the detailed combinatorics and the implications of the resulting quantum
field theory is work in progress (for the first step in this program see~\cite{qftlimit}).

\section*{Exercises}

\begin{Exercise} \label{exm4} (a causal fermion system on~$\ell_2$) {\em{
Let~$(\H = \ell^2, \la .|. \ra)$ be the Hilbert space of square-summable complex-valued
sequences. Thus, writing the vectors of~$\H$ as~$u=(u_i)_{i \in \N}$,
the scalar product is defined by
\[  \langle u | v \rangle = \sum_{i=1}^\infty \: \overline{u_i} \, v_i \: . \]
For any~$k \in \N$, we let $x_k$ be an operator on~$\H$ defined by
\[ \big( x_k \,u \big)_k = u_{k+1} \:,\qquad \big( x_k \,u \big)_{k+1} = u_{k} \]
and~$(x_k \,u)_i=0$ for all~$i \not \in \{ k, k+1\}$. In other words,
\beq \label{xkdef}
x_k \,u = \big( \underbrace{0,\ldots, 0}_{\text{$k-1$ entries}}, u_{k+1}, u_k, 0, \ldots \big) \:.
\eeq
Finally, we let~$\mu$ be the counting measure on~$\N$ (i.e.\ $\mu(X) =  \#X$ equals the
number of elements of~$X \subset \N$).
\begin{itemize}[leftmargin=2em]
\item[(a)] Show that every operator~$x_k$ has rank two, is symmetric, and has one positive and one
negative eigenvalue. Make yourself familiar with the concept that every operator is a point in~$\F$
(as introduced in Definition~\ref{defparticle}) for spin dimension~$n=1$.
\item[(b)] Let~$F : \N \rightarrow \F$ be the mapping which to every~$k$ associates the corresponding
operator~$x_k$. Show that the push-forward measure~$\rho = F_* \mu$ defined by~$\rho(\Omega)
= \mu(F^{-1}(\Omega))$ defines a measure on~$\F$. Show that this measure can also be characterized by
\[ \rho(\Omega) = \# \{ k \in \N \:|\: x_k \in \Omega \} \:. \]
(Clearly, we could also have taken this equation as the definition of~$\rho$. But
working with the push-forward measure is a good preparation for the constructions
in Section~\ref{secmink}.)
\item[(c)] Show that~$(\H, \F, \rho)$ is a causal fermion system of spin dimension one.
\end{itemize}
}} \end{Exercise}

\begin{Exercise} \label{exm3} {\em{
This exercise shows that the {\em{trace constraint}} ensures that the action is non-zero.
Let~$(\H, \F, \rho)$ be a causal fermion system of spin dimension~$n$.
\begin{itemize}[leftmargin=2em]
\item[(a)] Assume that~$\tr(x) \neq 0$. Show that~$\L(x,x)>0$.
(For a quantitative statement of this fact in the setting of discrete space-times
see~\cite[Proposition~4.3]{discrete}.)
\item[(b)] Assume that~$\int_\F \tr(x)\: d\rho \neq 0$.
Show that~$\Sact(\rho) > 0$.
\end{itemize}
}} \end{Exercise}

\begin{Exercise} \label{exm2} {\em{
This exercise explains why the causal action principle is ill-posed
in the case~$\dim \H=\infty$ and~$\rho(\F)<\infty$.
The underlying estimates were first given in the setting of discrete space-times
in~\cite[Lemma~5.1]{discrete}.
\begin{itemize}[leftmargin=2em]
\item[(a)] Let~$\H_0$ be a finite-dimensional Hilbert space of dimension~$n$
and~$(\H_0, \rho_0, \F_0)$ be a causal fermion system of finite total volume~$\rho_0(\F_0)$.
Let~$\iota : \H_0 \rightarrow \H$ be an isometric embedding.
Construct a causal fermion system~$(\H, \rho, \F)$
which has the same action, the same total volume
and the same values for the trace and boundedness constraints as the causal fermion
system~$(\H_0, \rho_0, \F_0)$.
\item[(b)] Let~$\H_1 = \H_0 \oplus \H_0$. Construct a causal fermion system~$(\H_1, \rho_1, \F_1)$
which has the same total volume and the same value of the trace constraint as~$(\H_0, \rho_0, \F_0)$ but a smaller action
and a smaller value of the boundedness constraint.
{\em{Hint:}} Let~$F_{1\!/\!2} : \Lin(\H_0) \rightarrow \Lin(\H_1)$ be the linear mappings
\[ \big(F_1(A)\big)(u \oplus v) = (Au) \oplus 0 \:,\qquad
\big(F_2(A)\big)(u \oplus v) = 0 \oplus (Av) \:. \]
Show that~$F_{1\!/\!2}$ map~$\F_0$ to~$\F_1$. Define~$\rho_1$ by
\[ \rho_1 = \frac{1}{2} \Big( (F_1)_* \rho + (F_2)_* \rho \Big) \:. \]
\item[(c)] Iterate the construction in~(b) and apply~(a)
to obtain a series of universal measures on~$\F$ of fixed total volume
and with fixed value of the trace constraint, for which the action and the values of the boundedness
constraint tend to zero. Do these universal measures converge? If yes, what is the limit?
\end{itemize}
}} \end{Exercise}

\begin{Exercise} \label{exm1} {\em{
The following example explains why the {\em{boundedness constraint}}~\eqref{Tdef} is needed
to ensure the existence of minimizers.
This example was first given in~\cite[Example~2.9]{continuum}.
Let~$\H=\C^4$. For a given parameter~$\tau>1$ consider the following mapping from
the sphere~$S^3 \subset \R^4$ to the linear operators on~$\H$,
\[ F \::\: S^3 \rightarrow \Lin(\H) \:,\quad F(x) = \sum_{i=1}^4 \tau\: x^i \gamma^i + \1\:. \]
Here~$\gamma^i$ are the four matrices
\[ \gamma^\alpha = \begin{pmatrix} \sigma^\alpha & 0 \\ 0 & -\sigma^\alpha \end{pmatrix},
\quad \alpha=1,2,3  \qquad \text{and} \qquad
\gamma^4 = \begin{pmatrix} 0 & \1 \\ \1 & 0 \end{pmatrix} \]
(and~$\sigma^\alpha$ are the Pauli matrices).
\begin{itemize}[leftmargin=2em]
\item[(a)] Verify by explicit computation that~$F(x)$ has two positive and two negative eigenvalues.
{\em{Hint:}} To simplify the computation one can make use of the fact that
the matrices~$\gamma^i$ satisfy the anti-commutation relations~$\{\gamma^i, \gamma^j\} = 2 \delta^{ij}\:\1$
(in other words, these matrices generate the Clifford algebra on Euclidean~$\R^4$).
\item[(b)] Let~$\mu$ be the normalized Lebesgue measure on~$S^3 \subset \R^4$.
Show that setting~$\rho = F_* \mu$ defines a causal fermion system of spin dimension two
and total volume one. Show that~$M:=\supp \rho$ is homeomorhic to~$S^3$.
\item[(c)] Compute the eigenvalues of~$F(x)\, F(y)$.
What is the causal structure of the causal fermion system?
\item[(d)] We now analyze the dependence on the parameter~$\tau$.
Show that the value of the trace constraint is independent of~$\tau$, whereas
\[ \lim_{\tau \rightarrow \infty} \Sact = 0 \qquad \text{and} \qquad \lim_{\tau \rightarrow \infty} \T = \infty\:. \]
Do the universal measures converge in the limit~$\tau \rightarrow \infty$? If yes, what is the limit?
\end{itemize}
}} \end{Exercise}

\begin{Exercise} \label{exm41} (support of a measure) {\em{
\begin{itemize}[leftmargin=2em]
\item[(a)] We return to the example of Exercise~\ref{exm4}.
Show that the support of~$\rho$ consists precisely of all the operators~$x_k$.
\item[(b)] In order to illustrate how to encode geometric information in the support of a measure,
let~$\scrM \subset \R^3$ be a smooth surface described in a parametrization~$\Phi$. Thus given
an open subset~$\Omega \subset \R^2$, we consider a smooth injective map
\[ \Phi \::\: \Omega \rightarrow \R^3 \]
with the property that~$D\Phi|_p : \R^2 \rightarrow \R^3$ has rank two for all~$p \in \Omega$.
Then the surface~$\scrM$ is defined as the image~$\Phi(\Omega) \subset \R^3$.
We now introduce the measure~$\rho$ as the {\em{push-forward measure}} of the
Lebesgue measure on~$\R^2$: Let~$\mu$ be the Lebesgue measure on~$\R^2$.
We define a set~$U \subset \R^3$ to be $\rho$-measurable if and only if its preimage~$\Phi^{-1}(U) \subset \R^2$
is $\mu$-measurable. On the $\rho$-measurable sets we define the measure~$\rho$ by
\[ \rho(U) = \mu\big( \Phi^{-1}(U) \big) \:. \]
Verify that the $\rho$-measurable sets form a $\sigma$-algebra, and that~$\rho$ is a measure.
What are the sets of $\rho$-measure zero? What is the support of the measure~$\rho$?

Suppose that~$\Phi$ is no longer assumed to be injective. Is~$\rho$ still a well-defined measure?
Is~$\rho$ well-defined if~$\Phi$ is only assumed to be continuous?
What are the minimal regularity assumptions on~$\Phi$ needed for the
push-forward measure to be well-defined? What is the support of~$\rho$ in this general setting?
\end{itemize}
}} \end{Exercise}

\begin{Exercise}  \label{exm5} (space-time and causal structure of the causal fermion system on~$\ell_2$) {\em{
We return to the example of Exercise~\ref{exm4}.
What is space-time~$M$? ({\em{Hint:}} See Exercise~\ref{exm41}~(a).)
What is the causal structure on~$M$? What is the resulting causal action?
Discuss the last result in the context of the trace constraint and Exercise~\ref{exm3}.
}} \end{Exercise}

\begin{Exercise}  \label{ex0} {\em{This exercise is devoted to the inequality
\beq \label{ineqsqrt}
\Big\| \sqrt{|y|} - \sqrt{|x|} \Big\| \leq \|y-x\|^\frac{1}{4} \:\|y+x\|^\frac{1}{4} \:,
\eeq
used in~\eqref{ineqstar} (the solution to this exercise can be found in~\cite{cfsrev}).
\begin{itemize}[leftmargin=2em]
\item[(a)] Let~$A$ and~$B$ be symmetric linear operators of finite rank.
Construct an explicit counter example for $2 \times 2$-matrices to the inequality
\[ \big\| |A| - |B| \big\| \leq \| A-B \| \]
(a similar exercise can be found in~\cite[Exercise~7 on page~217]{reed+simon}).
\item[(b)] Prove the inequality
\beq \label{ineqiter}
\big\| |A| - |B| \big\|^2 \leq \big\| A^2 - B^2 \big\| \:.
\eeq
{\em{Hint:}} First show that there is a vector~$u \in \H$ such that
\beq \label{ineqprep}
\big(|A| - |B| \big) u = \pm \big\||A| - |B| \big\|\,u
\eeq
and deduce the inequality
\[ \big\||A| - |B| \big\|
\leq \big\la u \,\big|\, \big(|A| + |B| \big) u \big\ra \:. \]
Then use~\eqref{ineqprep} once again.
\item[(c)] Iterate~\eqref{ineqiter} to obtain~\eqref{ineqsqrt}.
\end{itemize}
}}
\end{Exercise}

\begin{Exercise} \label{ex01} (Krein structure of the causal fermion system on~$\ell_2$) {\em{
We return to the example of Exercise~\ref{exm4} and Exercise~\ref{exm5}.
\begin{itemize}
\item[(a)] For any~$k \in \N$, construct the spin space~$S_{x_k}$ and its spin scalar product.
\item[(b)] Given a vector~$u \in \H$, what is the corresponding wave function~$\psi^u(x_k)$?
What is the Krein inner product~$\bra .|. \ket$?
\item[(c)] What is the topology on the Krein space~$\K$?
Does the wave evaluation operator~\eqref{weo} give rise to a well-defined and
continuous mapping~$\Psi: \mathcal \H \rightarrow \mathcal \K$?
If yes, is it an embedding? Is it surjective?
\item[(d)] Repeat part~(c) of this exercise for the causal fermion system obtained
if the operators~$x_k$ in~\eqref{xkdef} are multiplied by~$k$, i.e.
\[ x_k \,u = \big( \underbrace{0,\ldots, 0}_{\text{$k-1$ entries}}, k \,u_{k+1}, k \,u_k, 0, \ldots \big) \:. \]
\end{itemize}
}}
\end{Exercise}

\begin{Exercise} \label{ex1} {\em{
The goal of this exercise is to explore possible modifications
of the definition of {\em{regularization operators}}.
\begin{itemize}[leftmargin=2em]
\item[(a)]
Show that for the regularization operators in Example~\ref{exmollify}, the
estimate~\eqref{Pest} can be improved to
\beq \label{stronger}
\begin{split}
\bigg| \iint_{\scrM \times \scrM} & \overline{\eta(x)}\, \big( P^\varepsilon(x,y)
- P(x,y) \big)\, \tilde{\eta}(y)\: d^4x\: d^4y \,\bigg| \\
&\leq \delta\: 
\Big( |\eta|_{C^0(K)}\, |\tilde{\eta}|_{C^1(K)} + |\eta|_{C^1(K)}\, |\tilde{\eta}|_{C^0(K)}  \Big) \:.
\end{split}
\eeq
{\em{Hint:}} Using the notation in the proof of Proposition~\ref{lemma54},
one should first prove that
\[ \|\Phi^\varepsilon_\eta\| \leq c\, |\eta|_{C^0(K)} \:. \]
\item[(b)] Can Definition~\ref{defreg} be modified so that the stronger estimate~\eqref{stronger}
holds? Is there a natural way of doing so?
\end{itemize} }}
\end{Exercise}

\begin{Exercise} \label{ex2} {\em{ This exercise is devoted to a clean proof
of the distributional relation~\eqref{eq:delta-formula} in one dimension. More precisely, we want to prove the slightly
more general statement that for any function~$\eta \in C^1(\R) \cap L^1(\R)$,
\beq \label{deltaint}
\lim_{\varepsilon \searrow 0} \int_\R \eta(x) \left( \frac{1}{x - i \varepsilon} - \frac{1}{x + i \varepsilon} \right) dx
= 2 \pi i\: \eta(0) \:.
\eeq
\begin{itemize}[leftmargin=2em]
\item[(a)] Let~$\eta \in C^1(\R) \cap L^1(\R)$ with~$\eta(0)=0$. Show with the help of Lebesgue's dominated convergence theorem that~\eqref{deltaint} holds.
\item[(b)] Show with residues that~\eqref{deltaint} holds for the function~$\eta(x)=1/(x^2+1)$.
\item[(c)] Combine the results of~(a) and~(b) to prove~\eqref{deltaint} for general~$\eta \in C^1(\R) \cap L^1(\R)$.
\end{itemize} }}
\end{Exercise}

\begin{Exercise} \label{ex201} {\em{ This exercise recalls basics on the principal value
in one dimension~\eqref{eq:PP-formula}.
\begin{itemize}[leftmargin=2em]
\item[(a)] Repeat the method in Exercise~\ref{ex2} to show that the limit of the left side
of~\eqref{eq:PP-formula} exist for any~$\eta \in C^1(\R) \cap L^1(\R)$.
Derive a corresponding estimate which shows that~$\PP$ is a well-defined tempered distribution.
\item[(b)] Show that for any~$\eta \in C^1(\R) \cap L^1(\R)$,
\[ \PP(\eta) = \lim_{\varepsilon \searrow 0} \left(
\int_{-\infty}^{-\varepsilon} + \int_{\varepsilon}^\infty \right) \frac{\eta(x)}{x} \:dx \:. \]
\end{itemize} }}
\end{Exercise}

\begin{Exercise} \label{ex21} {\em{  The goal of this exercise is to justify that the one-dimensional
relations~\eqref{eq:delta-formula} and~\eqref{eq:PP-formula} can be used in the
four-dimensional setting~\eqref{delta-highdim}.
\begin{itemize}[leftmargin=2em]
\item[(a)] Let~$T$ be a distribution on~$\R$, $\Omega \subset \scrM$ be an open subset of Minkowski space
 and~$f : \Omega \rightarrow \R$ a smooth function with nowhere vanishing gradient. Show that the relation
\[ \big(f^* T)(\eta) := T \big( \phi_f(\eta) \big) \:, \qquad \eta \in C^\infty_0(\Omega) \]
with
\[ \phi_f(\eta)(t) := \frac{\partial}{\partial t} \int_{\Omega} \Theta\big( t-f(x) \big)\: \eta(x)\: d^4x \]
(where~$\Theta$ is the Heaviside function)
defines $f^* T$ as a distribution on~$\Omega$ (this is the so-called
{\em{pullback}} of~$T$ under~$f$; for details see~\cite[Section~7.2]{friedlander2}).
\item[(b)] Choosing~$\Omega$ as the half space in the future, $\Omega = \{ x\in \scrM, x^0>0\}$,
one can rewrite the expression on the left of~\eqref{delta-highdim} as
\[ \lim_{\varepsilon \searrow 0} \frac{1}{r^2 - t^2 + i \varepsilon} \:. \]
Use~(a) to conclude that this expression is a well-defined distribution for any~$\varepsilon>0$.
Show that the limit~$\varepsilon \searrow 0$ exist in the distributional sense.
\item[(c)] Repeating the procedure of~(b) for the half space in the past,
one obtains a distribution on~$\scrM \setminus \{t=0\}$.
Show that this distribution coincides with the limit in~\eqref{delta-highdim}.
{\em{Hint:}} Similar as in Exercise~\ref{ex2}, one can estimate the behavior at the origin
with Lebesgue's dominated convergence theorem
(to this end, first consider a test function which vanishes to a certain order at the origin).
\end{itemize}
}}
\end{Exercise}

\begin{Exercise} \label{ex3} {\em{
Show with a symmetry argument (without explicit computation of Fourier integrals!)
that the imaginary part of the distribution~$T(x,y)$ vanishes if~$x$ and~$y$ have space-like separation.
}}
\end{Exercise}

\begin{Exercise} \label{ex4} {\em{
This exercise is concerned with the {\em{Bessel functions}} in Lemma~\ref{lemmaTintro}.
\begin{itemize}[leftmargin=2em]
\item[(a)] Express the vectorial component of~$P(x,y)$
similar to~\eqref{Taway} in terms of Bessel functions.
{\em{Hint:}} Use~\eqref{Pdiff} together with the relations for the derivatives
of Bessel functions (see~\cite[(10.6.6) and~(10.29.4)]{DLMF}).
\item[(b)] Use~\eqref{Taway} together with the results of~(a)
to compute the parameters~$a$ and~$b$ in~\eqref{ab} 
in the case that~$x$ and~$y$ have timelike separation.
Simplify the formula for~$a$ using the relations for the
Wronskians of Bessel functions (see~\cite[(10.5.2)]{DLMF}).
\end{itemize}
}}
\end{Exercise}

\begin{Exercise} \label{ex41} {\em{
This exercise is devoted to analyzing general properties of the spectrum of
the closed chain.
\begin{itemize}[leftmargin=2em]
\item[(a)] As in Definition~\ref{def2}, we let~$x$ and~$y$ be symmetric operators
of finite rank on a Hilbert space~$(\H, \la .|. \ra_\H)$.
Show that there is a finite-dimensional subspace~$I \subset \H$ on which both~$x$
and~$y$ are invariant.
By choosing an orthonormal basis of~$I$ and restricting the operators to~$I$,
we may represent both~$x$ and~$y$ by Hermitian matrices.
Therefore, the remainder of this exercise is formulated for simplicity in terms of Hermitian matrices.
\item[(b)] Show that for any matrix~$Z$, the characteristic polynomials of~$Z$ and of its
adjoint~$Z^*$ (being the transposed complex conjugate matrix) are related by complex
conjugation, i.e. $\det(Z^*- \overline{\lambda} \:\1) = \overline{\det(Z-\lambda  \:\1)}$.
\item[(c)] Let~$X$ and~$Y$ be symmetric matrices. Show that the
characteristic polynomials of the matrices~$XY$ and~$YX$ coincide.
\item[(d)] Combine~(b) and~(c) to conclude that the characteristic polynomial
of~$XY$ has real coefficients, i.e.~$\det(XY- \overline{\lambda}  \:\1) = \overline{\det(XY-\lambda \:\1)}$.
Infer that the spectrum of the matrix product~$XY$ is symmetric about the real axis, i.e.\
\beq \label{XYsymm}
\det(XY- \lambda  \:\1)=0 \;\;\Longrightarrow\;\; \det(XY- \overline{\lambda} \:\1)=0 \:.
\eeq
\item[(e)] For the closed chain~\eqref{Axydef}, the mathematical setting is somewhat
different, because~$A_{xy}$ is a symmetric operator on the indefinite inner product
space $(S_x, \Sl .|. \Sr_x)$. On the other hand, we saw after~\eqref{Axydef}
that~$A_{xy}$ is isospectral to~$xy$. Indeed, the symmetry result~\eqref{XYsymm}
can be used to prove a corresponding statement for~$A_{xy}$,
\beq \label{Axysymm}
\det(A_{xy}- \lambda \:\1)=0 \;\;\Longrightarrow\;\; \det(A_{xy} - \overline{\lambda} \:\1)=0 \:.
\eeq
This result is well-known in the theory of indefinite inner product spaces
(see for example the textbooks~\cite{bognar, GLR}
or~\cite[Section~3]{discrete}). In order to derive it from~\eqref{XYsymm}, one
can proceed as follows: First, represent the indefinite inner product in the
form~$\Sl .|. \Sr =\la .|S\, x \ra$, where~$\la .|. \ra$ is a scalar product
and~$S$ is an invertible operator which is symmetric (with respect to this scalar product).
Next, show that the operator~$B:=A_{xy} S$ is symmetric (again with
respect to this scalar product). Finally, write the closed chain as~$A_{xy} = B S^{-1}$
and apply~\eqref{XYsymm}.
\end{itemize}
}}
\end{Exercise}

\begin{Exercise} \label{ex5} {\em{
This exercise recalls a few basic facts from the theory
of ordinary differential equations which are relevant
in the context of the Bessel functions in Lemma~\ref{lemmaTintro}
(for more material in this direction see for example the textbook~\cite{coddington}).
Let~$\phi_1$ and~$\phi_2$ be two linearly independent real-valued
solutions of the linear ordinary differential equation of second order
\[ \phi''(x) + a(x)\, \phi'(x) + b(x)\, \phi(x) = 0 \:, \]
where~$a$ and~$b$ are two smooth, real-valued functions on an open interval~$I$.
\begin{itemize}[leftmargin=2em]
\item[(a)] Show that at every~$x \in I$, either~$\phi_1(x)$ or~$\phi_2(x)$ is non-zero.
Moreover, either~$\phi_1'(x)$ or~$\phi_2'(x)$ is non-zero.
{\em{Hint:}} Combine the statement of the Picard-Lindel\"of theorem with
the fact that a general solution can be written as a linear combination of~$\phi_1$
and~$\phi_2$.
\item[(b)] Show that the Wronskian defined by~$ w(\phi_1, \phi_2) = \phi_1'(x)\, \phi_2(x) - \phi_1(x)\, \phi_2'(x)$ is independent of~$x$ and non-zero.
\end{itemize}
}}
\end{Exercise}

\begin{Exercise} \label{ex6} {\em{
Let~$\xi$ be a timelike vector, for simplicity normalized to~$\xi^2=1$.
Let~$A$ be the $4 \times 4$-matrix~$A = a \slashed{\xi} + b$.
Show that the operators
\[ F_\pm := \frac{1}{2} \left( \1 \pm \slashed{\xi} \right) \]
have rank two and map to eigenspaces of~$A$.
What are the corresponding eigenvalues? Show that the operators~$F_\pm$
are idempotent and symmetric with respect to the spin scalar product.
Show that the image of the operators~$F_\pm$ is positive or negative
definite. Moreover, the image of~$F_+$ is orthogonal to that of~$F_-$
(again with respect to the spin scalar product).
The results of this exercise can be summarized by saying that the~$F_\pm$ are the
spectral projection operators of~$A$.
}}
\end{Exercise}

\begin{Exercise} \label{ex8} {\em{
The goal of this exercise is to analyze in which sense the notion of causality
is stable under perturbations.
\begin{itemize}[leftmargin=2em]
\item[(a)] Show by a counter example with $3 \times 3$-matrices
that the notion of timelike separation (see Definition~\ref{def2}) is {\em{not}}
stable under perturbations.
\item[(b)] Show that the notion of properly timelike separation
(see Definition~\ref{defproptl}) is stable under perturbations. 
\item[(c)] We now analyze a setting in which the
notion of spacelike separation (see Definition~\ref{def2}) is stable under perturbations:
Assume that the regularized kernel~$P^\varepsilon(x,y)$ converges to the unregularized
kernel~\eqref{pointwise}. Moreover, assume that the eigenvalues of the regularized closed
chain are at least two-fold degenerate for every~$\varepsilon>0$. Finally,
assume that the eigenvalues of the
unregularized closed chain form a complex conjugate pair. 
Show that under these assumptions, the eigenvalues of the regularized closed chain
also form a complex conjugate pair for sufficiently small~$\varepsilon$.
\end{itemize}
In Exercise~\ref{ex81} a setting is given in which the assumptions in~(c) are
satisfied. 
}}
\end{Exercise}

\begin{Exercise} \label{ex81} {\em{
This exercise explains why the assumptions in Exercise~\ref{ex8}~(c) are reasonable.
It is a good preparation for the computation of the
eigenvalues of the closed chain in the vacuum to be carried out in~\S\ref{s:sec71}.
Assume that the regularized kernel~$P^\varepsilon(x,y)$ has vector-scalar structure~\eqref{vectorscalar}.
Compute the eigenvalues of the closed chain.
Why are they always two-fold degenerate? 
Explain why the bilinear contribution to the closed chain tends to gives rise to complex conjugate pairs
of eigenvalues.

In order to put these results into context, we remark that the picture in~\S\ref{s:sec71} is that
in space-like directions, the bilinear contribution gives rise to complex conjugate pairs of eigenvalues.
These are stable under perturbations according to Exercise~\ref{ex8}~(c).
}}
\end{Exercise}

\begin{Exercise} \label{ex7} {\em{
The goal of this exercise is to analyze the functional~${\mathscr{C}}$
for the regularized Dirac sea vacuum in spacelike directions.
\begin{itemize}[leftmargin=2em]
\item[(a)] Let~$\xi$ be a spacelike vector. Show that in the representation~\eqref{Pxyrep}
of the kernel of the fermionic projector, the parameter~$\alpha$ is imaginary,
whereas~$\beta$ is real.
{\em{Hint:}} Use~\eqref{Pdiff} with either the formula~\eqref{Taway} or the result of Exercise~\ref{ex3}.
\item[(b)] Deduce that the parameter~$a$ in~\eqref{1} vanishes if~$\xi$ is spacelike.
\item[(c)] What do these findings imply for the size of the functional~${\mathscr{C}}$?
{\em{Hint:}} Discuss the commutator in~\eqref{Ccomm}.
\end{itemize}
}}
\end{Exercise}

\begin{Exercise} \label{ex9} {\em{
Consider the kernel of the fermionic projector regularized by a con\-ver\-gence-generating
factor~$e^{\varepsilon \,|k_0|}$, i.e.\ similar to~\eqref{Tepsreg},
\[ P^\varepsilon(x,y) = \int \frac{d^4k}{(2 \pi)^4}\: (\slashed{k}+m)\:
\delta(k^2-m^2)\: \Theta(-k_0)\: e^{-ik(x-y)}\: e^{-\varepsilon \,|k_0|} \:. \]
Compute~$P^\varepsilon(x,x)$. How do the scalar and vectorial components
scale in~$\varepsilon$?
}}
\end{Exercise}

\begin{Exercise} \label{ex9a} {\em{
We now explore a functional which at first sight might seem a promising
alternative to~\eqref{Cform} for distinguishing a time direction.
Clearly, for the sign of a functional to distinguish a time direction, the functional should
be anti-symmetric in its arguments~$x$ and~$y$. The simplest functional with this property
is given by
\[ {\mathscr{B}} \::\: M \times M \rightarrow \R\:,\qquad {\mathscr{B}}(x,y) := \tr \big( y\,\pi_x - x\, \pi_y \big) \:. \]
\begin{itemize}[leftmargin=2em]
\item[(a)] Write the functional~${\mathscr{B}}$ similar to~\eqref{Ccomm} in the form
\beq \label{Btrace}
{\mathscr{B}}(x,y) = \Tr_{S_x} \!\big( \nu(x)\, A^\varepsilon_{xy} \big) - \Tr_{S_y} \!\big( \nu(y)\, A^\varepsilon_{yx} \big) \:.
\eeq
\item[(b)] Now assume that in a given spinor basis, the fermionic projector has vector-scalar structure~\eqref{vectorscalar}.
Show that only the scalar and vectorial components of~$A^\varepsilon_{xy}$ 
contribute to the trace in~\eqref{Btrace} (whereas the bilinear component drops out).
Deduce that, in the chosen spinor basis, the relations~$A^\varepsilon_{xy} = A^\varepsilon_{yx}$ holds
and
\[ {\mathscr{B}}(x,y) = \Tr \Big( \big( \nu(x) - \nu(y) \big)\, A^\varepsilon_{xy} \Big) \:. \]
Show that the last equation vanishes if the fermionic projector is {\em{homogeneous}} in the
sense that~$P^\varepsilon(x,x)=P^\varepsilon(y,y)$ for all~$x,y \in M$.
\end{itemize}
In non-technical terms, these results show that the functional~${\mathscr{B}}$ gives information on the
``deviation from homogeneity.'' But this functional cannot be used to distinguish a time direction.
In particular, it vanishes for regularized Dirac sea configurations in Minkowski space.
We remark that seeking for an anti-symmetric functional which does not vanish for homogeneous
fermionic projectors with vector-scalar structure leads directly to the functional~\eqref{Cform}. }}
\end{Exercise}

\begin{Exercise} \label{ex9b} {\em{ This exercise explains how a variation
of the universal measure described by a push-forward~\eqref{pushtau}
can be realized by a variation of the physical wave functions.
Thus we let~$F_\tau : M \rightarrow \F$ with~$\tau \in (-\delta, \delta)$
be a family of functions which satisfy the conditions
in~\eqref{fFinit} and~\eqref{fFtrivial}. Moreover, we assume that~$F_\tau$ is differentiable
in~$\tau$ and that all points in~$K$ are regular (see Definition~\ref{defregular}).
\begin{itemize}[leftmargin=2em]
\item[(a)] We first fix~$x \in K$.
Show that by decreasing~$\delta$, one can arrange that the operators~$F_\tau(x)$
have maximal rank~$2n$ for all~$\tau \in (-\delta, \delta)$. 
{\em{Hint:}} Make use of the fact that the spectrum of the operators~$F_\tau(x)$ depends
continuously on~$\tau$.
\item[(b)] We introduce the spin spaces~$S_x^\tau$ endowed with
corresponding inner products~$\Sl .|. \Sr_x^\tau$ in analogy to~\eqref{ssp} by
\[ S_x^\tau = \big(F_\tau(x)\big)(\H) \:,\qquad \Sl .|. \Sr_x^\tau = -\la . | F_\tau(x) . \ra_\H
\big|_{S_x^\tau \times S_x^\tau} \:. \]
Construct a family of isometries
\[ V_\tau(x) \::\: (S_x, \Sl .|. \Sr_x) \rightarrow (S^\tau_x, \Sl .|. \Sr^\tau_x) \]
which is differentiable in~$\tau$ (where ``isometric'' refers to the corresponding spin scalar product).
{\em{Hint:}} For example, one can work with the orthogonal projections
in~$\H$ and take the polar decomposition with respect to the spin scalar products.
\item[(c)] Consider the variation of the wave evaluation operator given by
\[ \Psi_\tau(x) = \big( V_\tau(x) \big)^{-1} \,\pi_{F_\tau(x)} \::\: \H \rightarrow S_x \]
(where~$\pi_{F_\tau(x)}$ is the orthogonal projection in~$\H$ on~$S^\tau_x$ as defined above).
Show that the relation~\eqref{Ftauvary} gives us back the family of functions~$F_\tau(x)$ we started with.
\item[(d)] So far, the point~$x \in K$ was fixed. We now extend the construction so that~$x$
can be varied: Use a compactness argument to show that there is~$\delta>0$ such that
the operators~$F_\tau(x)$ have maximal rank~$2n$ for all~$\tau \in (-\delta, \delta)$
and all~$x \in K$. Show that the mappings~$V_\tau(x)$ can be introduced such that they
depend continuously in~$x$ and are differentiable in~$\tau$.
\end{itemize}
}} \end{Exercise}

\begin{Exercise} \label{ex10} {\em{
The goal of this exercise is to illustrate the more general EL equations as derived
in~\cite{lagrange}. In order to simplify the setting, we leave out the constraints and replace~$\F$
by a compact manifold. Thus
let~$\F$ be a smooth {\em{compact}} manifold and~$\L \in C^{0,1}(\F \times \F, \R^+_0)$
be a non-negative Lipschitz-continuous function which is symmetric, i.e.
\beq \label{symmLcompact}
\L(x,y) = \L(y,x) \qquad \text{for all~$x,y \in \F$}\:.
\eeq
The {\em{causal variational principle}} is to minimize the action~$\Sact$ defined by
\beq \label{Sdefcompact}
\Sact(\rho) = \iint_{\F \times \F} \L(x,y)\: d\rho(x)\: d\rho(y)
\eeq
under variations of~$\rho$ in the class of (positive) normalized regular Borel measures.
Let~$\rho$ be a minimizer.
\begin{itemize}[leftmargin=2em]
\item[(a)] Show by analyzing variations of the form~\eqref{rhosc}
that the function~$\ell \in C^{0,1}(\F)$ defined by
\beq
\ell(x) = \int_\F \L(x,y)\: d\rho(y) \label{ldef}
\eeq
is minimal on the support of~$\rho$,
\beq \label{EL1}
\ell|_{\supp \rho} \,\equiv\, \inf_\F \ell \:.
\eeq
\item[(b)] We now consider second variations. Let~$(\H_\rho, \la .,. \ra_\rho)$ be the Hilbert space~$L^2(\F, d\rho)$.
Show that the operator~$\L_\rho$ defined by
\[ \L_\rho \::\: \H_\rho \rightarrow \H_\rho\:, \qquad (\L_\rho \psi)(x) =
\int_\F \L(x,y)\: \psi(y)\: d\rho(y) \]
is Hilbert-Schmidt. Show that it is non-negative.
{\em{Hint:}} Consider suitable variations of the form~$d\tilde{\rho}_\tau = d\rho + \tau \psi \,d\rho$
with~$\psi \in \H_\rho$.
\end{itemize}
We refer the reader interested in the analysis of the
causal variational principle in this compact setting to~\cite{continuum, support}.
}}
\end{Exercise}

\begin{Exercise} \label{ex11} {\em{
The goal of this exercise is to illustrate the Noether-like theorems
mentioned in~\S\ref{secnoether}.
In order to simplify the problem as far as possible, we again consider the compact
setting of Exercise~\ref{ex10} and assume furthermore that the Lagrangian is
smooth, i.e.\ $\L \in C^\infty(\F \times \F, \R^+_0)$.
Let~$\rho$ be a minimizer of the action~\eqref{Sdefcompact}
under variations of~$\rho$ in the class of (positive) normalized regular Borel measures.
Let~$u$ be a vector field on~$\F$. Assume that~$u$ is a {\em{symmetry of the
Lagrangian}} in the sense that
\beq \label{Lsymm2}
\left( u(x)^j \:\frac{\partial}{\partial x^j} + u(y)^j\: \frac{\partial}{\partial y^j} \right) \L(x,y) = 0 \qquad
\text{for all~$x,y \in \F$}\:.
\eeq
Prove that for any measurable set~$\Omega \subset \F$,
\[ \int_\Omega d\rho(x) \int_{\F \setminus \Omega} d\rho(y)\:
u(x)^j \:\frac{\partial}{\partial x^j} \L(x,y) = 0 \:. \]
{\em{Hint:}} Integrate~\eqref{Lsymm2} over~$\Omega \times \Omega$.
Transform  the integrals using the symmetry of the Lagrangian~\eqref{symmLcompact}.
Finally, make use of the EL equations~\eqref{EL1} and the smoothness of the function~$\ell$.
}}
\end{Exercise}

\chapter{Computational Tools} \label{tools}
In this chapter we introduce the computational methods needed for the analysis of the causal action principle
in the continuum limit. These methods are the backbone of the analysis given in Chapters~\ref{sector}--\ref{quark}.
Nevertheless, in order to facilitate the reading of the book, we made the
subsequent chapters accessible even without a detailed knowledge of the computational tools.
To this end, all the technical computations are given in the appendices, whereas in the
main Chapters~\ref{sector}--\ref{quark} these results are merely stated and explained.
Therefore, a reader who is willing to accept the results of the detailed computations
may skip the present chapter in a first reading.

Our main objective is to construct the fermionic projector in the presence of an external potential
and to analyze it in position space.
The first task is to define the unregularized fermionic projector~$P(x,y)$ in the presence of the external potential.
In this setting, the fermionic projector was constructed in a perturbation expansion in~$\B$
in~\cite{sea, grotz, norm}. More recently, a non-perturbative construction was given in~\cite{finite, infinite, hadamard}
(see also the brief review in~\S\ref{secquasifree}).
For the explicit analysis of the causal action principle to be carried out in this book, we need the
detailed formulas of the perturbation expansion. In order to focus on what is really needed in this book,
we here restrict attention to the perturbative treatment (Section~\ref{secfpext}).
The reader interested in non-perturbative methods is referred
to the introduction in~\cite{intro} or to the research papers~\cite{finite, infinite, hadamard}.

Our next task is to derive detailed formulas for the fermionic projector in position space.
Such formulas are most conveniently obtained using the so-called light-cone expansion
as first developed in~\cite{firstorder, light}. In Section~\ref{seclight} we give a self-contained
introduction to the light-cone expansion.

In Section~\ref{seclingrav} the causal perturbation expansion and the light-cone expansion
are adapted to the description of linearized gravity.

In Section~\ref{secreg} we turn attention to the ultraviolet regularization of the fermionic
projector. This leads us to the so-called formalism of the continuum limit, which makes
it possible to analyze how the different contributions to the causal action depend on the
regularization. In order to make the presentation easily accessible, we begin with
the example of an $i \varepsilon$-regularization (\S\ref{secieps}).
Then we consider linear combinations of such regularizations (\S\ref{seclineps})
and explain further regularization effects (\S\ref{secfurthereffects}).
Then the formalism of the continuum is introduced (\S\ref{sec73}),
and its derivation is outlined (\S\ref{secdercl}).
Our presentation is not as general as the original derivation as given
in~\cite[Chapter~4]{PFP}, but instead it aims at clarifying the main points of the construction.

In Section~\ref{secloctrace} we explain how to compute the local trace.
This is important in view of the rescaling procedure explained in \S\ref{secvary}
(see~\eqref{rhorescale}).

Finally, in Section~\ref{secspeccc} it is explained how the EL equations 
as derived in~\S\ref{secvary} can 
can be analyzed in the formalism of the continuum limit.

\section{The Fermionic Projector in an External Potential} \label{secfpext}

\subsectionn{The Fermionic Projector of the Vacuum}
Our starting point is the unregularized kernel of the fermionic projector of the vacuum which we already
encountered in~\S\ref{seccorcs} (see Lemma~\ref{lemmaDiracsea}, \eqref{Pdiff} and
Lemma~\ref{lemmaTintro}). For the later constructions, it is convenient
to clarify that we are in the Minkowski vacuum by adding an index ``vac.''
Moreover, we denote the mass by an additional index~$m$. Thus we define the
{\em{kernel of the fermionic projector of the vacuum}} as the bi-distribution
\beq \label{Fourier2}
P^\text{vac}_m(x,y) = \int \frac{d^4k}{(2 \pi)^4}\: P^\text{vac}_m(k)\: e^{-ik(x-y)}\:,
\eeq
where~$P^\text{vac}_m(k)$ is the distribution in momentum space
\beq \label{Fourier1}
P^\text{vac}_m(k) = (\slashed{k}+m)\: \delta(k^2-m^2)\: \Theta(-k^0)
\eeq
(and~$\Theta$ denotes the Heaviside function).
\sindex{fermionic projector!of the vacuum}%
\nindex{ba0@$P^\text{vac}_m(x,y)$ -- fermionic projector corresponding to a vacuum Dirac sea of mass~$m$}%
We also consider the distribution~$P^\text{vac}_m(x,y)$ as the integral kernel of an operator acting on
wave functions in space-time, i.e.
\beq \label{Pmop}
P^\text{vac}_m \::\: C^\infty_0(\scrM, S\scrM) \rightarrow C^\infty(\scrM, S\scrM) \:,\qquad
(P^\text{vac}_m \psi)(x) = \int_M P^\text{vac}_m(x,y)\: \psi(y)\: d^4y \:.
\eeq
This operator is the so-called fermionic projector of the vacuum.

Before going on, we briefly recall the physical picture. In~\eqref{Fourier2} we integrate over all the plane-wave
solutions of the Dirac equation of negative frequency
(the decomposition into plane-wave solutions was explained in detail
in Chapter~\ref{introduction}; see~\eqref{Pxykernel} and Lemma~\ref{lemmaDiracsea}).
Thus~$P^\text{vac}_m$ describes the ensemble of all
negative-frequency solutions of the Dirac equation.
As already mentioned in~\S\ref{seccorcs}, we use this Dirac sea configuration to describe the vacuum in Minkowski space.
\sindex{Dirac sea}%
\sindex{Dirac sea configuration}%
In order to describe a system with an additional particle, we simply add the corresponding
bra/ket-combination by setting
\[ P(x,y) = P_m^\text{vac}(x,y) -\frac{1}{2 \pi} \psi(x)\, \overline{\psi(y)} \:, \]
where~$\psi$ is a positive-frequency solution of the Dirac equation
(for the prefactor~$-1/(2 \pi)$ and the normalization of the wave function see~\S\ref{secnorm} below).
Similarly, we occupy several states by adding the bra/ket-combinations of several
particle states,
\[ P(x,y) = P_m^\text{vac}(x,y) - \frac{1}{2 \pi} \sum_{k=1}^{\np} \psi_k(x) \overline{\psi_k(y)} \]
(which need to be suitably ortho-normalized; see again~\S\ref{secnorm} below).
In order to introduce anti-particles, we similarly subtract bra/ket-combinations
\beq \label{Pvacparticle}
P(x,y) = P_m^\text{vac}(x,y)
- \frac{1}{2 \pi} \sum_{k=1}^{\np} \psi_k(x) \overline{\psi_k(y)}
+ \frac{1}{2 \pi} \sum_{l=1}^{\na} \phi_l(x) \overline{\phi_l(y)}\:,
\eeq
where~$\phi_1, \ldots, \phi_{\na}$ are the wave functions of negative-frequency solutions.
\nindex{as3@$\np, \na$ -- number of particles and anti-particle states}%
\nindex{as4@$\psi_k, \phi_l$ -- particle and anti-particle states}%
Thus in simple terms, we take Dirac's concept of the Dirac sea literally and describe particles by additional
occupied states and anti-particles by ``holes'' in the sea.

With the methods introduced so far, this description of particles and anti-particles by occupying states
and creating ``holes'' can only be performed in the non-interacting situation in which we can work
with plane-wave solutions of the Dirac equation. But it is not
obvious how the construction should be carried out if an external potential
is present. In order to tackle this problem, we first analyze how to describe the
completely filled Dirac sea in the presence of an external potential
(see~\S\ref{secextfield}--\S\ref{secspatial}). Afterwards, we will come back to the
description of systems involving particles and anti-particles (see~\S\ref{secnorm}).

\subsectionn{The External Field Problem} \label{secextfield}
We now return to the Dirac equation in the presence of an external potential~\eqref{Cont:DiracEq},
\beq \label{direx}
(i \Pdd + \B - m) \,\psi(x) = 0 \:,
\eeq
where~$\B$ is a smooth potential with suitable decay properties at spatial infinity and for
large times (to be specified in Lemma~\ref{l:lemma0} below).
\sindex{potential!bosonic}%
\sindex{potential!external}%
\nindex{ar4@$\B$ -- external potential}%
We now explain the basic problem in defining the fermionic projector in the presence of an external
potential.

The definition of the fermionic projector of the vacuum~\eqref{Fourier2}
and~\eqref{Fourier1} makes essential use of the fact that the solution space of the Dirac equation
splits into two subspaces of negative and positive frequency, respectively.
Indeed, this made it possible in~\eqref{Fourier1} to integrate only over the solutions of negative frequency.
In order to extend the definition of the fermionic projector to the case when an external potential
is present~\eqref{direx}, one needs to again decompose the solution space into two subspaces.
In the special case that~$\B$ is static, one can still separate the time dependence
by the plane wave ansatz~$\psi(t, \vec{x}) = e^{-i \omega t} \, \psi_\omega(\vec{x})$, so that
the sign of~$\omega$ gives a canonical splitting of the solution space.
This procedure is often referred to as the {\em{frequency splitting}}.
In the general time-dependent setting, however, no plane wave ansatz can be used,
so that the frequency splitting breaks down. Therefore, it is no longer obvious if there still is a
canonical decomposition of the solution space into two subspaces.

This problem is sometimes referred to as the {\em{external field problem}}
(for more details see Exercise~\ref{ex2.0} or the exposition in~\cite[Section~2.1]{PFP}).
\sindex{external field problem}%
It is a common belief that in the presence of a general time-dependent external potential,
there no longer exists a canonical decomposition of the solution space into two subspaces.
Nevertheless, it is still possible to decompose the solution space into two subspaces,
for example by using the sign of the spectrum of the Dirac Hamiltonian on a
distinguished Cauchy surface. But the decomposition is no longer canonical in the
sense that it involves an arbitrariness.
This arbitrariness is often associated to an observer, so that the choice of the subspaces
depends on the observer. As a consequence, the interpretation of the fermionic many-particle
state in terms of particles and anti-particles also depends on the observer.
This {\em{observer dependence of the particle interpretation}} becomes most apparent in the
Unruh effect in which the vacuum of the observer at rest is described by
a uniformly accelerated observer in terms of a thermal state involving particles and anti-particles.

Nevertheless, this reduction to particles and anti-particles as being objects associated to observers
only tells part of the truth. Namely, as shall be developed in what follows, even in the presence of a
time-dependent external potential there is a {\em{canonical decomposition of the solution space
into two subspaces}}. In the static situation, this decomposition reduces to the frequency splitting.
In the time-dependent situation, however, this decomposition depends on the global behavior of~$\B$
in space-time. In particular, this decomposition cannot be associated to a local observer.
Starting from the canonical decomposition of the solution space, one can again generate particles
and holes, giving rise to an interpretation of the many-particle state in terms of particles and anti-particles.
This particle interpretation is again independent of the choice of an observer. All constructions are
explicitly covariant.

\subsectionn{Main Ingredients to the Construction}
Before entering the constructions, we explain a few ingredients and ideas.
Generally speaking, we shall make use of additional properties of the fermionic
projector of the vacuum, which are not immediately apparent in the Fourier decomposition~\eqref{Fourier2}
and~\eqref{Fourier1}.
One ingredient is to use that {\em{causality}} is built into~$P^\text{vac}(x,y)$.
\sindex{causality!of the perturbation expansion}%
To see how this comes about, we decompose~$P^\text{vac}_m$ as
\beq
P^\text{vac}_m(x,y) = \frac{1}{2}\,\Big(p_m(x,y)-k_m(x,y)\Big) \:, \label{sea-pk}
\eeq
where~$p_m(x,y)$ and~$k_m(x,y)$ are the Fourier transforms of the distributions in momentum space
\begin{align}
p_m(q)&= (\slashed{q}+m)\:\delta(q^2-m^2) \label{defp} \\
k_m(q)&= (\slashed{q}+m)\:\delta(q^2-m^2)\:\epsilon(q^0) \label{defk}
\end{align}
(and~$\epsilon$ in~\eqref{defk} is again the sign function $\epsilon(x)=1$
for $x \geq 0$ and $\epsilon(x)=-1$ otherwise).
All these Fourier integrals are well-defined tempered distributions,
which are also distributional solutions of the vacuum Dirac equation.
\nindex{ba0@$P^\text{vac}_m(x,y)$ -- fermionic projector corresponding to a vacuum Dirac sea of mass~$m$}%
\nindex{ba8@$p_m, k_m$ -- fundamental solutions of the vacuum Dirac equation}%
The point is that the distribution~$k_m(x,y)$ is {\em{causal}} in the sense that it vanishes
if~$x$ and~$y$ have spacelike separation. In order to see this, it is
useful to introduce the {\em{advanced}} and the {\em{retarded Green's functions}} by
\beq
        s^{\lor}_{m}(q) = \lim_{\nu \searrow 0}
            \frac{\slashed{q} + m}{q^{2}-m^{2}-i \nu q^{0}}
            \qquad {\mbox{and}} \qquad
           s^{\wedge}_{m}(q) = \lim_{\nu \searrow 0}
            \frac{\slashed{q} + m}{q^{2}-m^{2}+i \nu q^{0}} \:,
        \label{8b}
\eeq
\sindex{Green's function!advanced}%
\sindex{Green's function!retarded}%
\nindex{bb0@$s_m^\vee, s_m^\wedge$ -- causal Green's functions of the vacuum Dirac equation}%
respectively (with the limit~$\nu \searrow 0$ taken
in the distributional sense). Taking their Fourier transform
\beq \label{Ft}
s_m(x,y) = \int \frac{d^4q}{(2 \pi)^4} \:s_m(q)\: e^{-iq (x-y)} \:,
\eeq
we obtain corresponding bi-distributions~$s^{\lor}_{m}(x,y)$ and~$s^\wedge_m(x,y)$. 
By direct computation one verifies that these Green's functions satisfy
the distributional equation
\beq
(i \Pdd_x - m) \: s_m(x,y) = \delta^4(x-y) \:. \label{Greendef}
\eeq
Moreover, computing the Fourier integral~\eqref{Ft}
with residues, one sees that the support of these Green's functions
lies in the upper respectively lower light cone, i.e.
\beq \label{GF}
\supp s^\lor_m(x,.) \subset J_x^\lor \;,\qquad
\supp s^\wedge_m(x,.) \subset J_x^\wedge \:,
\eeq
\nindex{bb2@$\supp$ -- support of distribution}%
where~$J_x^\vee$ and~$J_x^\wedge$ denote the points in the causal future respectively
past of $x$,
\begin{align*}
J_x^\vee &= \{ y \in M \,|\, (y-x)^2 \geq 0,\;
(y^0-x^0) \geq 0 \} \\
J_x^\wedge &= \{ y \in M \,|\, (y-x)^2 \geq 0, \;
(y^0-x^0) \leq 0 \}
\end{align*}
(for details see Exercise~\ref{ex2.1} or~\cite[Chapter~4]{intro}).
\nindex{bb4@$J_x^\vee, J_x^\wedge$ -- causal future and past}%
In view of~\eqref{Greendef}, the difference of the advanced and retarded Greens' functions
is a solution of the homogeneous Dirac equation. In order to compute it
in detail, we again make use of~\eqref{eq:delta-formula} to obtain
\beq \label{sm2calc}
\begin{split}
s_m^\vee(q) - s_m^\wedge(q) &= (\slashed{q} + m) \; \lim_{\nu
\searrow 0} \left[ \frac{1}{q^{2}-m^{2}-i\nu q^{0}} -
\frac{1}{q^{2}-m^{2}+i\nu q^{0}} \right] \\
&= (\slashed{q} + m) \; \lim_{\nu
\searrow 0} \left[ \frac{1}{q^{2}-m^{2}-i\nu} -
\frac{1}{q^{2}-m^{2}+i\nu} \right] \epsilon(q^{0}) \\
&= 2 \pi i\, (\slashed{q} + m) \: \delta(q^{2}-m^{2})\: \epsilon(q^{0})
\end{split}
\eeq
(for details see Exercise~\ref{ex2.2}).
Comparing with~\eqref{defk}, we conclude that the difference of the advanced
and retarded Green's functions is a multiple of~$k_m$
\beq \label{kmdef}
k_m(x,y) = \frac{1}{2\pi i} \left( s^{\vee}(x,y)-s^{\wedge}(x,y) \right) .
\eeq
In particular, this shows that~$k_m$ is indeed causal, i.e.
\beq \label{kcausal}
\supp k_m(x,.) \subset J_x \:,
\eeq
where~$J_x := J_x^\vee \cup J_x^\wedge$.
We refer to~$k_m$ as the {\em{causal fundamental solution}}.
\sindex{causal fundamental solution}%

Now~\eqref{sea-pk} can be understood as the decomposition of
the vacuum fermionic projector into a causal part (the distribution~$k_m$)
and a part which is not causal (the distribution~$p_m$; note that the
explicit formulas in~\eqref{Pdiff} and Lemma~\ref{lemmaTintro} show that~$p_m(x,y)$
is indeed non-zero for spacelike distances).
One idea behind our constructions is to perform the perturbation expansion
in such a way that the decomposition of~$P(x,y)$ in to a causal and
a non-causal part is preserved.

Another ingredient to our constructions is that the distributions~$p_m$ and~$k_m$
are related to each other by a {\em{functional calculus}}, as we now explain.
We first point out that for the space-time integral in~\eqref{Pmop} to exist,
we had to assume that the wave function~$\psi$ has suitable decay properties at infinity.
More specifically, the time integral in~\eqref{Pmop} in general diverges if~$\psi$ is a physical wave function,
being a solution of the Dirac equation.
In particular, the operator in~\eqref{Pmop} cannot be defined as an operator from a vector space
to itself, but it necessarily maps one function space to another function space.
As a consequence, it is impossible to multiply the operator~$P_m$ by itself.
This is obvious because the formal integral
\beq \label{PPint}
\int P^\text{vac}_m(x,z)\, P^\text{vac}_m(z,y) \,d^4z
\eeq
is ill-defined. This problem can be understood similarly in momentum space.
Namely, using that convolution in position space corresponds to multiplication in momentum space,
the integral in~\eqref{PPint} corresponds to the formal product
\[ 
P^\text{vac}_m(q)\, P^\text{vac}_m(q) \:, \]
which is again ill-defined because the square of the $\delta$-distribution in~\eqref{Fourier1}
makes no mathematical sense.
As we shall see, these obvious problems in the naive treatment of the fermionic projector are not
only a mathematical subtlety. On the contrary, the methods for overcoming these problems
will involve a careful analysis of the causal structure of the fermionic projector and of its
proper normalization.

It is important to observe that the above operator product 
does make sense if we consider two different mass parameters. Namely,
\begin{align*}
P^\text{vac}_m(q) \: P^\text{vac}_{m^\prime}(q) &= (\slashed{q} + m) \: \delta(q^2 - m^2) \:\Theta(-q^0)\;
   (\slashed{q} + m^\prime) \: \delta(q^2 - (m^\prime)^2) \:\Theta(-q^0) \\
&= \big( q^2 + (m+m^\prime)\, \slashed{q} + m m^\prime \big) \: \delta(m^2 -
   (m^\prime)^2) \; \delta(q^2 - m^2) \:\Theta(-q^0) \\
&= \big( q^2 + (m+m^\prime)\, \slashed{q} + m m^\prime \big) \: \frac{1}{2m} \:\delta(m -
   m^\prime) \; \delta(q^2 - m^2) \:\Theta(-q^0) \\
&= \delta(m-m^\prime) \: (\slashed{q} + m) \: \delta(q^2 - m^2) \:\Theta(-q^0)\: .
\end{align*}
giving rise to the distributional identity
\beq \label{massnorm}
P^\text{vac}_m \: P^\text{vac}_{m^\prime} = \delta(m-m^\prime)\: P^\text{vac}_m \:.
\eeq
This resembles idempotence, but it involves a $\delta$-distribution in the mass
parameter. We remark that this $\delta$-normalization in the mass parameter can be treated
in a mathematically convincing way using the notion of the mass oscillation property as introduced
in~\cite{infinite}. For brevity, we shall not enter these constructions here. Instead, we are content with the fact
that~\eqref{massnorm} is well-defined if we test in both~$m$ and~$q$.

This calculus can be used similarly for the operators~$p_m$ and~$k_m$ obtained
by considering the distributions~\eqref{defp} and~\eqref{defk} as multiplication operators
in momentum space. In particular, this gives rise to the relation
\beq \label{km2}
k_m\,k_{m'}=\delta(m-m')\:p_m
\eeq
(for details see Lemma~\ref{lemma21} below).
This identity is very useful because it allows us to deduce~$p_m$ from~$k_m$.
Therefore, our strategy is to first construct~$k_m$ in the presence of an external potential
using the underlying causal structure~\eqref{kmdef}.
Then we take~\eqref{km2} to define~$p_m$ in the presence of the external potential.
Finally, we use~\eqref{sea-pk} to define the fermionic projector.

There is one subtle point in the construction which we want to mention here:
the proper normalization of the states of the fermionic projector.
The most obvious method is to interpret and use the identity~\eqref{massnorm}
as a normalization condition. This so-called {\em{mass normalization}} was used
in~\cite{sea, grotz}; see also~\cite[Chapter~2]{PFP}.
More recently, the non-perturbative construction in~\cite{infinite} revealed that on
a general globally hyperbolic manifold, the mass normalization cannot be used
and should be replaced by the so-called {\em{spatial normalization}}.
\sindex{normalization!spatial}%
\sindex{normalization!mass}%
In~\cite{norm} the causal perturbation expansion is worked out for both the mass and
the spatial normalizations, and the methods and results are compared.
In~\cite[Section~2.2]{norm} the advantages of the spatial normalization are discussed,
but no decisive argument in favor of one of the normalization methods is given.
Finally, the Noether-like theorems in~\cite{noether} showed that the spatial normalization
is the proper normalization method, because it reflects the intrinsic conservation laws of the
causal fermion system (see~\cite[Remark~5.13]{noether} or the brief outline in~\S\ref{secnoether}).

With these results in mind, we here restrict attention to the spatial normalization, which we
now introduce. Recall that for a Dirac wave function~$\psi$, the quantity~$(\overline{\psi} \gamma^0 \psi)(t_0, \vec{x})$
has the interpretation as the probability density for the particle at time~$t_0$ to be at position~$\vec{x}$.
Integrating over space and polarizing, we obtain the scalar product~\eqref{sprodMin}, which we
also denote by
\beq \label{print}
(\psi | \phi)_{t_0} = 2 \pi \int_{\R^3} \overline{\psi(t_0, \vec{y})} \gamma^0 \phi(t_0,\vec{y})\: d^3y \:.
\eeq
\nindex{ai8@$(. \vert .)$ -- scalar product on Dirac wave functions}%
\nindex{bb8@$(. \vert .)_{t_0}$ -- scalar product on Dirac wave functions, computed at time~$t_0$}%
It follows from current conservation that for any solutions~$\psi, \phi$ of the Dirac equation,
this scalar product is independent of the choice of~$t_0$.
This is the case even in the presence of an external potential~\eqref{direx},
provided that the potential is symmetric with respect to the inner product on the spinors~\eqref{sspMink}, i.e.\
\beq \label{Bsymm}
\Sl \psi | \B \phi\Sr = \Sl \B \psi | \phi \Sr
\eeq
(see Exercise~\ref{ex2.3}).
\nindex{ar4@$\B$ -- external potential}%
\sindex{potential!external}%
Since the kernel of the fermionic projector is a solution of the Dirac equation,
one is led to evaluating the integral in~\eqref{print} for~$\phi(y)=P(y,z)$
and~$\overline{\psi(y)} = P(x,y)$. In the vacuum, the resulting integral can
be computed, giving a simple result.

\begin{Lemma} For any~$t \in \R$, there is the distributional relation
\beq \label{spatialnorm}
2 \pi \int_{\R^3} P^\text{\rm{vac}}_m \big( x, (t, \vec{y}) \big) \:\gamma^0\: P^\text{\rm{vac}}_m \big( (t, \vec{y}), z \big)\: d^3y
= -P^\text{\rm{vac}}_m(x,z)\:.
\eeq
\end{Lemma}
\nindex{ba0@$P^\text{vac}_m(x,y)$ -- fermionic projector corresponding to a vacuum Dirac sea of mass~$m$}%
\Proof The identity follows by a straightforward computation, which was already given in
the proof of Lemma~\ref{lemmaDiracsea} (see~\eqref{Pnorm} and the computation thereafter).
\QED
We refer to~\eqref{spatialnorm} as the {\bf{spatial normalization}} of the fermionic projector.
\sindex{normalization!spatial}%
It has the advantage that it is well-defined even for fixed~$m$.
Moreover, the normalization method is closely related to the probabilistic interpretation of the Dirac equation.

In the following sections~\S\ref{secpertgreen}--\S\ref{secspatial}, we shall carry out the
construction of the fermionic projector describing the completely filled Dirac sea
in the presence of the external potential~$\B$. Our method will make essential use
of generalizations of the underlying causal structure (as is apparent in~\eqref{sea-pk} and~\eqref{GF}), of
the relation between~$k_m$ and~$p_m$ as expressed by~\eqref{km2}, and of
the spatial normalization~\eqref{spatialnorm}.
Finally, in~\S\ref{secnorm} we shall extend the construction to allow for particles and anti-particles.

\subsectionn{The Perturbation Expansion of the Causal Green's Functions} \label{secpertgreen}
Using the causal support property, the advanced and retarded Green's functions~$\tilde{s}_m^{\vee}$
and~$\tilde{s}_m^{\wedge}$ are uniquely defined even in the presence of an external potential~\eqref{direx}.
They can be constructed non-perturbatively using the theory of symmetric hyperbolic
systems (see~\cite{john} or~\cite[Chapter~5]{intro}).
For our purposes, it is sufficient to work out their perturbation expansions:
The retarded Green's function is characterized by the conditions
\[ 
(i \Pdd + \B - m) \: \tilde{s}^\wedge_m(x,y) = \delta^4(x-y) \qquad \text{and} \qquad
{\mbox{supp}}\: \tilde{s}^\wedge_m(x,.) \subset J_x^\wedge \:. \]
\nindex{bc4@$\tilde{s}_m^\vee, \tilde{s}_m^\wedge$ -- causal Green's functions of the Dirac equation in 
an external potential}%
Employing the perturbation ansatz
\[ 
\tilde{s}^\wedge_m = \sum_{n=0}^\infty s^\wedge_{(n)} \qquad \text{with} \qquad
s^\wedge_{(0)} = s^\wedge_m \]
(where the subscript~$(n)$ denotes the order of perturbation theory),
we obtain for~$n=1,2,\ldots$ the inductive conditions
\beq \label{inductive}
(i \Pdd - m)\: s^\wedge_{(n)} = - \B\: s^\wedge_{(n-1)}
\qquad \text{and} \qquad
{\mbox{supp}}\, \tilde{s}^\wedge_{(n)}(x,.) \subset J_x^\wedge \:.
\eeq
Using the defining property of the Green's function~\eqref{Greendef}, one sees that
the left equation in~\eqref{inductive} can be solved in the case~$n=1$ by
\beq \label{s1ans}
s^\wedge_{(1)} = - s_m \,\B\, s^\wedge_m \:,
\eeq
where the operator product is defined as follows,
\beq \label{eq:defof-bprods}
(s_m \,\B\, s_m^\wedge)(x,y) := \int d^4z \:s_m(x,z) \,\B(z)\, s_m^\wedge(z,y)
\eeq
(the analytic justification of this and all other operator products
in this section will be given in Lemma~\ref{l:lemma0} below).
The operator~$s_m$ in~\eqref{s1ans} is any Green's function
(like the advanced, retarded or the symmetric Green's function).
In order to determine which Green's function to choose, we evaluate the
condition on the right side of~\eqref{inductive}.
Namely, if we choose~$s_m$ in~\eqref{s1ans} again as the retarded
Green's function, then the integral in~\eqref{eq:defof-bprods} vanishes
if~$x$ lies in the past of~$y$ because in this case the supports of the
distributions~$s^\wedge_m(x,.)$ and~$s_m^\wedge(.,y)$ do not intersect.
This leads us to setting
\[ s^\wedge_{(1)} = - s^\wedge_m \,\B\, s^\wedge_m \:. \]
Now we can evaluate~\eqref{inductive} inductively to obtain
\[ s^\wedge_{(n)} = - s^\wedge_m \,\B\, s^\wedge_{(n-1)} = \big( -s_m^{\wedge} \,\mathscr{B} \big)^n \,s_m^{\wedge} \:. \]
Proceeding similarly for the advanced Green's function, we obtain the unique perturbation
series
\beq \tilde{s}_m^{\vee}=\sum_{n=0}^{\infty} \big( -s_m^{\vee}\, \mathscr{B} \big)^n \,s_m^{\vee}\;, \qquad
\tilde{s}_m^{\wedge}=\sum_{n=0}^{\infty} \big( -s_m^{\wedge}\, \mathscr{B} \big)^n \,s_m^{\wedge}\:.
\label{series-scaustilde}
\eeq
Having derived a perturbation series for the causal Green's functions, we can also
define the causal fundamental solution in generalization of~\eqref{kmdef} by
\beq
\tilde{k}_m:=\frac{1}{2\pi i}(\tilde{s}_m^{\vee}-\tilde{s}_m^{\wedge})\;,   \label{def-ktil}
\eeq
\nindex{bc6@$\tilde{k}_m$ -- causal fundamental solution of the Dirac equation in an external potential}%
\sindex{causal fundamental solution}%
We now specify a class of potentials for which all the
operator products appearing here and later in this book are all well-defined in the distributional sense:
\begin{Lemma}
\label{l:lemma0}
Let $(C_j)$, $0 \leq j \leq n$, be a choice of operators $C_j \in \{ k_m, p_m,
s_m^\vee, s_m^\wedge \}$ (and~$p_m$, $k_m$ according to~\eqref{defp} and~\eqref{defk}).
If the external potential~$\B$ is smooth and decays so fast at infinity that the functions
$\B(x)$, $x^i \B(x)$, and $x^i x^j
\B(x)$ are integrable, then the operator product
\beq \label{l:2}
        (C_n \:\B \:C_{n-1}\: \B \cdots \B \:C_0)(x,y)
\eeq
is a well-defined tempered distribution on $\R^4 \times \R^4$.
\end{Lemma}
\Proof
Calculating the Fourier transform of~\eqref{l:2} gives the formal
expression
\begin{eqnarray}
\lefteqn{M(q_2,q_1) := \int \frac{d^4 p_1}{(2 \pi)^4}
\cdots \int \frac{d^4 p_{n-1}}{(2 \pi)^4} C_n(q_2) \;\hat\B(q_2-p_{n-1}) } \notag \\
&\times&\!\!\!\!\!
\:C_{n-1}(p_{n-1})\: \hat\B(p_{n-1}-p_{n-2})
\:\cdots\: C_1(p_1) \:\hat\B(p_1-q_1) \:C_0(q_1) \; , \label{l:3}
\end{eqnarray}
where we consider the $C_j$ as multiplication operators in momentum
space and where $\hat\B$ denotes the Fourier transform of
the function~$\B$ (it is more convenient to work in momentum
space because the operators $C_j$ are then diagonal).
We will show that $M(q_2,q_1)$ is a well-defined tempered distribution;
the Lemma then immediately follows by transforming back to position space.

The assumptions on $\B$ yield
that $\hat\B$ is $C^2$ and has rapid decay at infinity, i.e.
\[ \sup_{q \in \R^4, \;|\kappa| \leq 2} |q^{i_1} \cdots q^{i_n} \:
\partial_\kappa {\hat\B}(q)| \;<\; \infty \]
for all~$n$, all tensor indices $i_1,\ldots, i_n$ and all multi-indices
$\kappa$ (with $\kappa=(\kappa^1,\ldots,\kappa^q)$, $|\kappa|:=q$).
As is verified explicitly in momentum space, the
distributions $k_m$, $p_m$ and~$s_m$ are bounded in the Schwartz norms
of the test functions involving derivatives of only first order. More precisely,
\[ |C(f)| \leq {\mbox{const}}\: \|f\|_{4,1} \qquad \text{with} \qquad
{\text{$C=k_m$, $p_m$ or $s_m$ and $f \in {\mathcal{S}}(\R^4, \C^4)$}}\:, \]
where~${\mathcal{S}}(\R^4, \C^4)$ is the Schwartz space,
\nindex{bc8@${\mathcal{S}}(\R^4, \C^4)$ -- Schwartz space}%
\sindex{Schwartz space}%
and the Schwartz norms are defined as usual by
\sindex{distribution!tempered}%
\sindex{Schwartz norm}%
\nindex{bd0@$\NORM . \NORM_{p,q}$ -- Schwartz norm}%
\[ \|f\|_{p,q} = \max_{|I| \leq p,\; |J| \leq q} \;\;\sup_{x \in \R^4}
|x^I \:\partial_J f(x)| \]
(for basics on the Schwartz space and distributions see for example~\cite{friedlander2}).
As a consequence, we can apply the corresponding operators even to
functions with rapid decay which are only $C^1$. Furthermore, we can form the convolution of such
functions with~$C$; this gives continuous functions (which will no
longer have rapid decay, however). Since~$C$ involves first derivatives,
a convolution decreases the order of
differentiability of the function by one.

We consider the combination of multiplication and convolution
\beq \label{l:3a}
F(p_2) := \int \frac{d^4p_1}{(2 \pi)^4} \;
f(p_2-p_1) \:C(p_1) \:g(p_1) \:,
\eeq
where we assume that $f \in C^2$ has rapid decay and $g \in C^1$ is
bounded together with its first derivatives, $\|g\|_{0,1}<\infty$.
For any fixed $p_2$,
the integral in~\eqref{l:3a} is well-defined and finite because $f(p_2-.) \:g(.)$
is $C^1$ and has rapid decay. The resulting function $F$ is $C^1$ and
bounded together with its first derivatives, more precisely
\beq
\|F\|_{0,1} \leq {\mbox{const}} \;\|f\|_{4,2} \:\|g\|_{0,1} \:. \label{l:3c}
\eeq

After these preparations, we can estimate the integrals in \eqref{l:3} from
the right to the left: We choose two test functions $f, g \in
{\mathcal{S}}(\R^4, \C^{4})$ and introduce the functions
\begin{align}
F_1(p_1) &= \int \frac{d^4q_2}{(2 \pi)^4} \; \hat\B(p_1-q_1) \:C_0(q_1)\:g(q_1)\label{l:4z} \\
F_j(p_j) &= \int \frac{d^4p_{j-1}}{(2 \pi)^4} \:\hat\B(p_j-p_{j-1}) \:C_{j-1}(p_{j-1}) \:F_{j-1}(p_{j-1})
\:, \quad 1 < j \leq n \:. \label{l:4a}
\end{align}
The integral~\eqref{l:4z} is of the form~\eqref{l:3a} and satisfies the
above assumptions on the integrand. Using the bound
\eqref{l:3c}, we can proceed inductively in \eqref{l:4a}.
Finally, we perform the $q_2$-integration,
\beq \label{Mfg}
M(f,g) = \int \frac{d^4q_2}{(2 \pi)^4} \:f(q_2) \:C_n(q_2)
\:F_n(q_2) \:.
\eeq
We conclude that $M$ is a linear functional on
${\mathcal{S}}(\R^4,\C^{4}) \times {\mathcal{S}}(\R^4, \C^{4})$, which is
bounded in the Schwartz norm $\|.\|_{4,1}$ of the test functions.
\QED
We remark that the assumptions in this lemma are stronger than
what is needed for the operator products in~\eqref{series-scaustilde} and~\eqref{def-ktil}
to be well-defined:
First of all, the smoothness assumption for~$\B$ is unnecessarily strong; for example,
it would be sufficient to assume that~$\B$ is twice differentiable.
Moreover, using the causal structure, the contributions to the
above perturbation expansions are well-defined even without the decay assumptions
in Lemma~\ref{l:lemma0}. Namely, these perturbation expansions are all causal in the sense that
for any given~$x,y \in \scrM$, the distributions $\tilde{s}^\vee(x,y)$ and $\tilde{s}^\wedge(x,y)$ depend on the 
potential $\B$ only on in the so-called
\[ \text{causal diamond} \qquad \big( J^\vee_x \cap J^\wedge_y \big) \cup \big( J^\wedge_x \cap J^\vee_y \big) \:. \]
\sindex{causal diamond}%
Since the causal diamond is a bounded region of 
space-time, we may modify~$\B$ outside this bounded set to arrange the
decay assumptions without changing the contributions to the above perturbation expansions.

The reason why we prefer to impose with the stronger
assumptions in Lemma~\ref{l:lemma0} is that they will be needed later on.
Indeed, for the operator products appearing in the causal perturbation expansion of the
Dirac sea, the decay assumptions in Lemma~\ref{l:lemma0} will be required.
Moreover, the smoothness of~$\B$ will be needed for the light-cone expansion.

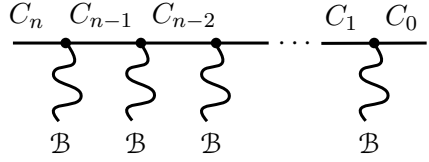
\begin{figure}
\psscalebox{1.0 1.0} 
{
\begin{pspicture}(0,-1.0062988)(8.4,1.0062988)
\psline[linecolor=black, linewidth=0.04](3.4527779,0.5368609)(0.05277778,0.5368609)
\rput[bl](0.545,-0.93036133){\normalsize{$\B$}}
\rput[bl](3.575,0.44463867){\normalsize{$\cdots$}}
\psdots[linecolor=black, dotsize=0.14](0.755,0.53797203)
\psbezier[linecolor=black, linewidth=0.04](0.6627778,-0.5031391)(0.4005642,-0.2025663)(0.9239167,-0.35638225)(0.9577778,-0.1831391)(0.99163884,-0.009895964)(0.55777776,-0.028139105)(0.5927778,0.13686089)(0.62777776,0.3018609)(1.0627778,0.101860896)(0.7777778,0.5168609)
\psbezier[linecolor=black, linewidth=0.04](1.6377778,-0.49313912)(1.3755642,-0.19256629)(1.8989167,-0.34638226)(1.9327778,-0.17313911)(1.9666388,0.0)(1.5327778,-0.018139105)(1.5677778,0.1468609)(1.6027777,0.3118609)(2.0377777,0.11186089)(1.7527778,0.5268609)
\psdots[linecolor=black, dotsize=0.14](1.755,0.53797203)
\psdots[linecolor=black, dotsize=0.14](2.755,0.53797203)
\psline[linecolor=black, linewidth=0.04](5.552778,0.5368609)(4.152786,0.5322594)
\psdots[linecolor=black, dotsize=0.14](4.855,0.53797203)
\rput[bl](5.0,0.71463865){\normalsize{$C_0$}}
\rput[bl](4.2,0.71463865){\normalsize{$C_1$}}
\rput[bl](0.8,0.71463865){\normalsize{$C_{n-1}$}}
\rput[bl](0.0,0.7396387){\normalsize{$C_n$}}
\rput[bl](1.9,0.71463865){\normalsize{$C_{n-2}$}}
\psbezier[linecolor=black, linewidth=0.04](2.6327777,-0.5031391)(2.3705642,-0.2025663)(2.8939166,-0.35638225)(2.9277778,-0.1831391)(2.961639,-0.009895964)(2.5277777,-0.028139105)(2.5627778,0.13686089)(2.5977778,0.3018609)(3.0327778,0.101860896)(2.7477777,0.5168609)
\psbezier[linecolor=black, linewidth=0.04](4.742778,-0.4981391)(4.480564,-0.1975663)(5.0039167,-0.35138226)(5.037778,-0.1781391)(5.071639,-0.0048959637)(4.637778,-0.023139106)(4.6727777,0.14186089)(4.707778,0.3068609)(5.142778,0.10686089)(4.8577776,0.5218609)
\rput[bl](1.545,-0.93036133){\normalsize{$\B$}}
\rput[bl](2.545,-0.93036133){\normalsize{$\B$}}
\rput[bl](4.645,-0.93036133){\normalsize{$\B$}}
\end{pspicture}
}
\caption{A Feynman tree diagram}
\label{figfeynman}
\end{figure}

The summands of the above perturbation expansions~\eqref{series-scaustilde}
and~\eqref{def-ktil} arise similarly in quantum field theory and are then
depicted by Feynman diagrams (see Figure~\ref{figfeynman}).
Using the language of quantum field theory, we also refer to the summands
of our perturbation expansions as {\em{Feynman diagrams}}.
Then the result of the last lemma can be understood from the fact that in the presence of an
external field one only encounters tree diagrams, which are all finite.
\sindex{Feynman diagram!tree diagram}%

\subsectionn{Computation of Operator Products}
We saw in~\eqref{massnorm} and~\eqref{km2} that operator products can be formed
if the mass is considered as a variable parameter. We now develop this method more
systematically. It is usually most convenient to work with the {\em{symmetric Green's function}} defined by
\beq \label{def-s}
s_m = \frac{1}{2}(s_m^{\vee}+s_m^{\wedge})\:.
\eeq
\sindex{Green's function!symmetric}%
\nindex{bd2@$s_m$ -- symmetric Green's functions of the vacuum Dirac equation}%

\begin{Lemma} \label{lemma21}
The following identities hold:
\begin{align}
p_m\,p_{m'}&=k_m\,k_{m'}=\delta(m-m')\:p_m\label{eq:pp-p} \\
p_m\,k_{m'}&=k_m\,p_{m'}=\delta(m-m')\:k_m\label{eq:pk-k} \\
p_m \,s_{m'}&=s_{m'}\,p_m=\frac{\PP}{m-m'}\:p_m\label{eq:ps-p} \\
k_m \,s_{m'}&=s_{m'}\,k_m=\frac{\PP}{m-m'}\:k_m\label{eq:ks-k} \\
s_m\,s_{m'}&=\frac{\PP}{m-m'}\:(s_m-s_{m'})+\pi^2\,\delta(m-m')\:p_m\;,\label{eq:ss-sp}
\end{align}
where $\PP$ denotes the principal value defined in analogy to~\eqref{PPdef}
alternatively by
\sindex{principal value}%
\nindex{ap0@$\PP$ -- principal value}%
\beq \label{PPdefmass}
\begin{split}
\int_{-\infty}^\infty \frac{\PP}{m} \: \eta(m)\: dm
&= \lim_{\nu \searrow 0} \left( \int_{-\infty}^{-\nu} + \int_{\nu}^\infty \right)
\frac{\eta(m)}{m} \: dm \\
&= \lim_{\nu \searrow 0} \frac{1}{2} \sum_{\pm}
\int_{-\infty}^\infty \frac{\eta(m)}{m \pm i \nu}\: dm \:.
\end{split}
\eeq
\end{Lemma}
\Proof
Calculating pointwise in momentum space, we obtain 
\begin{align*}
p_m(q)\,p_{m'}(q)&=(\slashed{q}+m)\: \delta(q^2-m^2)\;(\slashed{q}+m')\: \delta(q^2-m'^2)\\
&=\delta(m^2-m'^2)\:\delta(q^2-m^2) \;\Big(q^2+(m+m')\slashed{q}+mm' \Big)\\
&=\frac{1}{2m}\: \delta(m-m')\: \delta(q^2-m^2)\; \big(m^2+(m+m')\slashed{q}+mm' \big)\\
&=\frac{1}{2m}\: \delta(m-m')\: \delta(q^2-m^2)\;2m\, (m+\slashed{q})=\delta(m-m')\,p_m(q) \:.
\end{align*}
This gives the first part of \eqref{eq:pp-p}. The second part of this formula as well as formula \eqref{eq:pk-k} are obtained analogously. The formulas \eqref{eq:ps-p} and \eqref{eq:ks-k} are obtained as follows:
\begin{align*}
2 \,& p_m(q) \,s_{m'}(q)= \lim_{\nu \searrow 0}\:
\delta(q^2-m^2)(\slashed{q}+m) \left(\frac{\slashed{q}+m'}{q^2-m'^2-i\nu q^0}+\frac{\slashed{q}+m'}{q^2-m'^2+i\nu q^0}\right)\\
&=\lim_{\nu \searrow 0}\:\delta(q^2-m^2)\:\Big(q^2+(m+m')\slashed{q}+mm' \Big)
\left(\frac{1}{q^2-m'^2-i\nu q^0}+\frac{1}{q^2-m'^2+i\nu q^0}\right)\\
&=\lim_{\nu \searrow 0}\:\delta(q^2-m^2)\: \Big(m^2+(m+m')\slashed{q}+mm' \Big)
\left(\frac{1}{m^2-m'^2-i\nu q^0}+\frac{1}{m^2-m'^2+i\nu q^0}\right)\\
&=\lim_{\nu \searrow 0}\:\delta(q^2-m^2)(\slashed{q}+m) \left(\frac{(m+m')}{(m+m')(m-m')-i\nu q^0}+\frac{(m+m')}{(m+m')(m-m')+i\nu q^0}\right)\\
&=2 \:\frac{\PP}{m-m'}\: p_m(q)\:.
\end{align*}

The derivation of~\eqref{eq:ss-sp} is a bit more involved. Combining~\eqref{kmdef} and~\eqref{def-s}, we obtain
\beq \label{smkform}
s_m=s_m^{\vee}-i\pi k_m=s_m^{\wedge}+i\pi k_m \:.
\eeq
Thus we can express the product $s_m(q)\,s_{m'}(q)$ in two ways, namely as
\begin{align*}
s_m(q)\,s_{m'}(q) &=(s_m^{\vee}(q)-i\pi k_m(q))(s_{m'}^{\vee}(q)-i\pi k_{m'}(q))\\
&= s_m^{\vee}(q)\,s_{m'}^{\vee}(q) -\pi^2\delta(m-m')\,p_m(q) \\
&\qquad -i\pi \lim_{\nu \searrow 0} \left(k_{m'}(q)\frac{1}{m'-m-i\nu q^0}+k_{m}(q)
\frac{1}{m-m'-i\nu q^0}\right) ,
\end{align*}
or alternatively as
\begin{align*}
&s_m(q)\,s_{m'}(q) = (s_m^{\wedge}(q)+i\pi k_m(q))(s_{m'}^{\wedge}(q)+i\pi k_{m'}(q))\\
&= s_m^{\wedge}(q)\,s_{m'}^{\wedge}(q) -\pi^2\delta(m-m')\,p_m(q) \\
&\qquad +i\pi \lim_{\nu \searrow 0}
\left(k_{m'}(q)\frac{1}{m'-m+i\nu q^0}+k_{m}(q)\frac{1}{m-m'+i\nu q^0}\right) .
\end{align*}
Adding these two formulas yields
\begin{align*}
2\,&s_m(q)\,s_{m'}(q) - (s_m^{\vee}(q)\,s_{m'}^{\vee}(q)+s_m^{\wedge}(q)\,s_{m'}^{\wedge}(q))
+2\pi^2\delta(m-m')\,p_m(q) \\
&= i\pi \lim_{\nu \searrow 0} k_{m'}(q)\left(\frac{1}{m'-m+i\nu q^0}-\frac{1}{m'-m-i\nu q^0}\right)\\
&\qquad +i\pi \lim_{\nu \searrow 0}
k_{m}(q)\left(\frac{1}{m-m'+i\nu q^0}-\frac{1}{m-m'-i\nu q^0}\right)\\
&\!\overset{(*)}{=} i\pi k_{m'}(q)\epsilon(-q^0)\,2\pi i\,\delta(m'-m)
+i\pi k_{m}(q)\epsilon(-q^0)\,2\pi i\,\delta(m-m')\\
&= -2\pi^2\delta(m'-m)(-p_{m'}(q)) - 2\pi^2\delta(m-m')(-p_{m}(q))\:,
\end{align*}
where in~($*$) we applied~\eqref{eq:delta-formula}, and in the last line we used the
definitions of $p_m$ and $k_m$. We thus obtain
\beq 
s_m \,s_{m'} =  \frac{1}{2}
\left( s_m^{\vee}\, s_{m'}^{\vee} +s_m^{\wedge} \,s_{m'}^{\wedge} \right)
+ \pi^2\:\delta(m-m')\, p_m\:. \label{2ssprime}
\eeq

It remains to derive the relations
\beq \label{smmp}
s_m^{\vee}\,s_{m'}^{\vee}=\frac{\PP}{m-m'}(s_m^{\vee}-s_{m'}^{\vee})
\qquad \text{and} \qquad
s_m^{\wedge}\,s_{m'}^{\wedge}=\frac{\PP}{m-m'}(s_m^{\wedge}-s_{m'}^{\wedge})\:,
\eeq
which can be regarded as ``resolvent identities'' for the causal Green's functions.
It suffices to consider the case of the advanced Green's function.
Clearly, the operators on the right side of~\eqref{smmp}
satisfy the support condition~$\supp((s_m^{\vee}-s_{m'}^{\vee})(x,.))\subset J_x^{\vee}$, and from
\[ s_m^{\vee}\,s_{m'}^{\vee}(x,y)=\int d^4z\: s_m^{\vee}(x,z)\,s_{m'}^{\vee}(z,y) \]
we see that the operators on the left side of~\eqref{smmp} satisfy this support condition as well.
Moreover, the calculations
\[ (i\slashed{\partial}_x-m)\:s_m^{\vee}\,s_{m'}^{\vee}(x,y)=s_{m'}^{\vee}(x,y) \]
and
\begin{align*}
(i\slashed{\partial}_x&-m)\frac{\PP}{m-m'}(s_m^{\vee}-s_{m'}^{\vee})(x,y)\\
&=\frac{\PP}{m-m'} \:\Big( \delta(x-y)-(m'-m)s_{m'}^{\vee}(x,y)-\delta(x-y) \Big) =s_{m'}^{\vee}(x,y)
\end{align*}
show that both sides of~\eqref{smmp} satisfy the same inhomogeneous Dirac equation.
Hence their difference is a distributional solution of the homogeneous Dirac equation
which vanishes outside~$J_x^\vee$. The uniqueness of the solution of the Cauchy problem
for hyperbolic PDEs yields that this difference vanishes identically.
This proves~\eqref{smmp} and thus concludes the proof of~\eqref{eq:ss-sp}.
\QED

In the above operator products we get contributions of two different forms: those involving
a factor~$\delta(m-m')$ and those involving the principal value of~$1/(m-m')$.
In order to simplify the structure of the multiplication rules, it is useful to get rid of
the principal values by restricting attention to combinations in which all principal values
drop out in telescopic sums. To this end, we introduce the series of operator products
\begin{align}
b_m^<=\sum_{n=0}^{\infty} \big(-s_m\mathscr{B} \big)^n \:,\qquad
b_m=\sum_{n=0}^{\infty} \big(-\mathscr{B} s_m \big)^n\, \mathscr{B} \:,\qquad
b_m^>=\sum_{n=0}^{\infty} \big(-\mathscr{B} s_m \big)^n\:.
	\label{thm-defs}
\end{align}
\nindex{bd6@$b_m^<, b_m, b_m^>$ -- series of operator products}%

\begin{Corollary} \label{corol22}
Let $C\in \{p_m,k_m\}$ and $C^{\,\prime}\in\{p_{m'},k_{m'}\}$ as well as~$b_m^<$, $b_m^>$
as in~\eqref{thm-defs}. Then the following calculation rule holds:
\begin{align}
C\,b_m^>b_{m'}^<\,C^{\,\prime}=CC^{\,\prime}+\delta(m-m')\: \pi^2\: C\,b_m\, p_m\, b_m\,C^{\,\prime}.
\end{align}
\end{Corollary}
\Proof
Using the calculation rules of the previous lemma, we obtain
\begin{align*}
C \:\Big(\sum_{l=0}^{1} &(\mathscr{B}s_m)^l(s_{m'}\mathscr{B})^{n-l}\Big) \:C^{\,\prime}
= C s_{m'}\, \mathscr{B}\, C^{\,\prime} + C\, \mathscr{B}\, s_{m}\, C^{\,\prime} \\
&= \frac{\PP}{m-m'} \left( C \,\mathscr{B}\, C^{\,\prime} - C\, \mathscr{B}\, C^{\,\prime} \right) = 0\:.
\end{align*}
The same method also applies to higher order. We again get a telescopic sum,
but the last summand in~\eqref{eq:ss-sp} gives additional contributions. More precisely,
for any $n\geq2$,
\begin{align*}	
&C \:\Big(\sum_{l=0}^{n}(\mathscr{B}s_m)^l(s_{m'}\mathscr{B})^{n-l}\Big) \:C^{\,\prime} \\
&=C(\mathscr{B}s_m)^nC^{\,\prime}+C(s_{m'}\mathscr{B})^nC^{\,\prime}+C
\:\bigg[\sum_{l=1}^{n-1}(\mathscr{B}s_m)^l(s_{m'}\mathscr{B})^{n-l} \bigg] \:C^{\,\prime}\\
&=C\,\frac{\PP}{m-m'} \left[ -(\mathscr{B}s_m)^{n-1}\mathscr{B}+\mathscr{B}(s_{m'}\mathscr{B})^{n-1} \right]
C^{\,\prime} \\
&\quad+C\sum_{l=1}^{n-1}(\mathscr{B}s_m)^{l-1}\mathscr{B}\left(\frac{\PP}{m-m'}(s_m-s_{m'})+\delta(m-m')\pi^2p_m\right)\mathscr{B}(s_{m'}\mathscr{B})^{n-l-1}C^{\,\prime}\\
&=\frac{\PP}{m-m'}\,C \left[-(\mathscr{B}s_m)^{n-1}\mathscr{B}+\mathscr{B}(s_{m'}\mathscr{B})^{n-1} \right] C^{\,\prime}\\
&\quad+\frac{\PP}{m-m'}\,C\Big(\sum_{l=1}^{n-1}(\mathscr{B}s_m)^{l}(\mathscr{B}s_{m'})^{n-l-1}\mathscr{B}-
\sum_{l=0}^{n-2}(\mathscr{B}s_m)^{l}(\mathscr{B}s_{m'})^{n-l-1}\mathscr{B}\Big)C^{\,\prime}\\
&\quad+\delta(m-m')\:\pi^2\,C\sum_{l=1}^{n-1}(\mathscr{B}s_m)^{l-1}\mathscr{B}p_m\mathscr{B}(s_{m'}\mathscr{B})^{n-l-1}C^{\,\prime}\\
&=\delta(m-m')\:\pi^2\,C\sum_{l=1}^{n-1}(\mathscr{B}s_m)^{l-1}\mathscr{B}p_m\mathscr{B}(s_{m'}\mathscr{B})^{n-l-1}C^{\,\prime}\\
&=\delta(m-m')\:\pi^2\,C\sum_{l=0}^{n-2}(\mathscr{B}s_m)^{l}\mathscr{B}p_m\mathscr{B}(s_{m'}\mathscr{B})^{n-l-2}C^{\,\prime}.
\end{align*}
Thus, performing an index shift, we obtain
\begin{align*}
C\,b_m^> & b_{m'}^<\,C^{\,\prime} =C\sum_{n=0}^{\infty}(-\mathscr{B}s_m)^n\sum_{n'=0}^{\infty}(-s_{m'}\mathscr{B})^{n'}C^{\,\prime}\\
&=C\sum_{n=0}^{\infty}\sum_{l=0}^{n}(-\mathscr{B}s_m)^l(-s_{m'}\mathscr{B})^{n-l}C^{\,\prime}\\
&=CC^{\,\prime}+\delta(m-m')\pi^2\sum_{n=2}^{\infty}(-1)^nC\left(\sum_{l=0}^{n-2}(\mathscr{B}s_m)^{l}\mathscr{B}p_m\mathscr{B}(s_{m'}\mathscr{B})^{n-l-2}\right)C^{\,\prime}\\
&=CC^{\,\prime}+\delta(m-m')\pi^2\sum_{n=0}^{\infty}(-1)^nC\left(\sum_{l=0}^{n}(\mathscr{B}s_m)^{l}\mathscr{B}p_m\mathscr{B}(s_{m'}\mathscr{B})^{n-l}\right)C^{\,\prime}\\
&=CC^{\,\prime}+\delta(m-m')\pi^2C\,b_mp_mb_m\,C^{\,\prime}.
\end{align*}
This concludes the proof.
\QED

In what follows, we rewrite all operator products in terms of~$\B$ and~$p_m$, $k_m$
as well as the above combinations~$b_m^<$, $b_m$ and~$b_m^>$.
In order to explain how this can be done, we rewrite the perturbation expansion of~$\tilde{k}_m$
in this form.

\begin{Prp}\label{thm:kpert}
The perturbation expansion of~$\tilde{k}_m$ as given by~\eqref{def-ktil} 
can be written as
\beq \label{k-pert}
\tilde{k}_m=\sum_{\beta=0}^{\infty}(-i\pi)^{2\beta}b_m^<k_m(b_mk_m)^{2\beta}b_m^>\:,
\eeq
\nindex{bc6@$\tilde{k}_m$ -- causal fundamental solution of the Dirac equation in an external potential}%
\sindex{causal fundamental solution}%
where the factors~$b_m^<$, $b_m$ and $b_m^>$ are again the operator products in~\eqref{thm-defs},
\end{Prp}
\Proof
An explicit calculation shows that
\[ 
(i\slashed{\partial}+\mathscr{B}-m)\, b_m^< = 0\:. \]
As all operator products in \eqref{k-pert} have a factor $b_m^<$ at the left, the series in \eqref{k-pert} is a solution of the Dirac equation.

From \eqref{kmdef} and \eqref{def-s}, we have
\beq s_m^{\vee} = s_m+i\pi k_m\:, \qquad s_m^{\wedge} = s_m-i\pi k_m\:. \label{svw-sk}
\eeq
We substitute the series \eqref{series-scaustilde} into \eqref{def-ktil}, insert~\eqref{svw-sk} and expand. A reordering of the resulting sum gives the claim. The details of the reordering process can be found
in~\cite{sea}.
\QED \noindent

\subsectionn{The Causal Perturbation Expansion} \label{secspatial}
\sindex{causal perturbation expansion}%
$\quad$ We follow the constructions in~\cite{norm}.
Recall that, in the presence of an external potential~$\B$, the perturbation expansion of
the advanced and retarded Green's functions is unique by causality~\eqref{series-scaustilde}.
Moreover, Proposition~\ref{thm:kpert} gave us a unique perturbation expansion of the
causal fundamental solution~\eqref{k-pert}.

In the following constructions, we need to multiply the operator products in~\eqref{k-pert}.
These products have a mathematical meaning as distributions in the involved mass parameters.
Namely, according to Lemma~\ref{lemma21} and Corollary~\ref{corol22},
\begin{align}
p_m\,p_{m'}&=k_m\,k_{m'}=\delta(m-m')\:p_m \label{ppprod} \\
p_m\,k_{m'}&=k_m\,p_{m'}=\delta(m-m')\:k_m \\
k_m \,b_m^>\, b_{m'}^<\, k_{m'} &= \delta(m-m') \Big( p_m
+ \pi^2\,k_m\, b_m\, p_m\, b_m\, k_m \Big) \:. \label{bbprod}
\end{align}
Since these formulas all involve a common prefactor~$\delta(m-m')$,
we can introduce a convenient notation by leaving out this factor and omitting the
mass indices. For clarity, we denote this short notation with a dot, i.e.\ symbolically
\beq \label{sdotdef}
A \cdot B = C \qquad \text{stands for} \qquad
A_m \,B_{m'} = \delta(m-m')\: C_m \:.
\eeq
With this short notation, the above multiplication rules can be written in the compact form
\beq p \cdot p =k \cdot k =p\:, \qquad p \cdot k=k \cdot p=k\:,\qquad
k \,b^> \cdot b^< k =p+\pi^2 \,kbpbk \:. \label{rules}
\eeq
Writing~\eqref{k-pert} as
\beq \label{ktildef}
\tilde{k} =\sum_{\beta=0}^{\infty}(-i\pi)^{2\beta} \:b^<\, k\, (b k)^{2\beta}\, b^>\:,
\eeq
powers of the operator~$\tilde{k}$ with the product~\eqref{sdotdef}
are well-defined using the multiplication rules~\eqref{rules}.
This makes it possible to develop a spectral calculus for~$\tilde{k}$.
In particular, in~\cite{grotz} the operator~$P^\sea$
\nindex{be0@$P^\sea$ -- fermionic projector describing Dirac seas}%
is constructed as the projection operator
on the negative spectral subspace of~$\tilde{k}$.
We now give an equivalent construction using contour integrals, which gives a more systematic
procedure for computing all the contributions to the expansion
(for basics on the resolvent and contour integrals see Exercise~\ref{ex2.31}).

We introduce the resolvent by
\beq \label{Resdef}
\tilde{R}_\lambda = \big( \tilde{k} - \lambda \big)^{-1}\:.
\eeq
\sindex{resolvent!in causal perturbation expansion}%
\nindex{be2@$\tilde{R}_\lambda$ -- resolvent in presence of external potential}%
Writing~$\tilde{k}$ as
\beq \label{Delkdef}
\tilde{k} = k + \Delta k \:,
\eeq
(where~$k$ is the corresponding distribution in the vacuum), the resolvent~$\tilde{R}_\lambda$
can be written as a Neumann series,
\beq \label{tR}
\tilde{R}_\lambda = (k - \lambda + \Delta k)^{-1}
= (1 + R_\lambda \cdot \Delta k)^{-1} \cdot R_\lambda 
= \sum_{n=0}^\infty (-R_\lambda \cdot \Delta k)^n \cdot R_\lambda \:.
\eeq
\nindex{be4@$R_\lambda$ -- resolvent in Minkowski vacuum}%
The multiplication rules~\eqref{rules} imply that~$p$ is idempotent and
thus has the eigenvalues~$1$ and~$0$. Since the operator~$k$ commutes
with~$p$ and its square equals~$p$, it has the eigenvalues~$\pm 1$ and~$0$.
A short computation shows that the corresponding spectral projection
operators are~$(p \pm k)/2$ and~$\1-p$, respectively.
Hence we can write the unperturbed resolvent~$R_\lambda := (k - \lambda)^{-1}$ as
\beq \label{Rlamrel}
R_\lambda = \frac{p+k}{2} \left( \frac{1}{1-\lambda} \right) + \frac{p-k}{2} \left( \frac{1}{-1-\lambda} \right)
- \frac{\1-p}{\lambda} \:.
\eeq
Using this formula in~\eqref{tR}, to every order in perturbation theory we obtain
a meromorphic function in~$\lambda$ having poles only at~$\lambda=0$ and~$\lambda= \pm1$.

We now use contour integral methods to develop a spectral calculus. To this end,
we choose a contour~$\Gamma_-$ which encloses the point~$-1$ in counter-clockwise direction
and does not enclose the points~$1$ and~$0$. Similarly, $\Gamma_+$ is a contour which
encloses the point~$+1$ in counter-clockwise direction
and does not enclose the points~$-1$ and~$0$.
Moreover, we let~$f$ be a holomorphic function defined on
an open neighborhood of the points~$\pm 1$. We define~$f(\tilde{k})$ as the contour integral
\beq \label{fres}
f \big( \tilde{k} \big) := -\frac{1}{2 \pi i} \ointctrclockwise_{\Gamma_+ \cup \Gamma_-} f(\lambda)\:
\tilde{R}_\lambda\: d\lambda  \:.
\eeq
Using~\eqref{tR} together with the fact that to every order in perturbation theory, the integrand is
a meromorphic function in~$\lambda$ having poles only at~$\lambda=0$ and~$\lambda= \pm1$,
one sees that the operator~$f(\tilde{k})$ 
is well-defined to every order in perturbation theory and is independent of the choice of
the contours~$\Gamma_+$ and~$\Gamma_-$.

\begin{Thm} {\bf{(functional calculus)}} \label{thmfcalc}
\sindex{functional calculus!in causal perturbation expansion}%
For any functions~$f, g$ which are holomorphic in discs around~$\pm1$
which contain the contours~$\Gamma_\pm$,
\begin{align}
(i \Pdd+\B-m)\, f \big( \tilde{k} \big) &= 0 \label{fsol} \\
f \big( \tilde{k} \big) \cdot g \big( \tilde{k} \big) &= (fg)\big( \tilde{k} \big) \:. \label{fg}
\end{align}
\end{Thm}
\Proof 
Since the operator~$\tilde{k}$ maps to solutions of the Dirac equation, we know that
\[ (i \Pdd + \B - m)\: \tilde{R}_\lambda = (i \Pdd + \B - m)\, \big( -\lambda^{-1} \big) . \]
Taking the contour integral~\eqref{fres} gives~\eqref{fsol}.

The starting point for proving~\eqref{fg} is the resolvent identity
\beq \label{rr1}
\tilde{R}_\lambda \cdot \tilde{R}_{\lambda'} = \frac{1}{\lambda-\lambda'} \left(
\tilde{R}_\lambda - \tilde{R}_{\lambda'} \right) \:.
\eeq
We set~$\Gamma=\Gamma_+ \cup \Gamma_-$ and denote the corresponding
contour for~$\lambda'$ by~$\Gamma'$.
Since the integral~\eqref{fres} is independent of the precise choice of the contour,
we may choose
\[ \Gamma = \partial B_{\delta}(1) \cup \partial B_{\delta}(-1) \qquad \text{and} \qquad
\Gamma'=\partial B_{2 \delta}(1) \cup \partial B_{2 \delta}(-1) \]
for sufficiently small~$\delta<1/2$.
Then~$\Gamma$ does not enclose any point of~$\Gamma'$, implying that
\beq \label{rr2}
\ointctrclockwise_\Gamma \frac{f(\lambda)}{\lambda-\lambda'} \: d\lambda = 0 
\qquad \text{for all~$\lambda' \in \Gamma'$}\:.
\eeq
On the other hand, $\Gamma'$ encloses every point of~$\Gamma$, so that
\beq \label{rr3}
\ointctrclockwise_{\Gamma'} f(\lambda)\, g(\lambda')\: \frac{\tilde{R}_{\lambda}}{\lambda-\lambda'} \: d\lambda' =
-2 \pi i\,  f(\lambda)\, g(\lambda)\: \tilde{R}_\lambda \qquad \text{for all~$\lambda \in \Gamma$}\:.
\eeq
Combining~\eqref{rr1} with~\eqref{rr2} and~\eqref{rr3}, we obtain
\begin{align*}
f \big( \tilde{k} \big) \cdot g \big( \tilde{k} \big) 
&= -\frac{1}{4 \pi^2} \ointctrclockwise_{\Gamma} f(\lambda)\, d\lambda \ointctrclockwise_{\Gamma'} g(\lambda')\, d\lambda'\;
\frac{1}{\lambda-\lambda'} \left( \tilde{R}_\lambda - \tilde{R}_{\lambda'} \right) \\
&= -\frac{1}{2 \pi i} \ointctrclockwise_{\Gamma} f(\lambda)\, g(\lambda)\: \tilde{R}_\lambda \:d\lambda
= (fg)\big( \tilde{k} \big) \:.
\end{align*}
This concludes the proof.
\QED

The fermionic projector~$P^\sea$ is obtained by choosing a specific function~$f$, as we now
explain. First, the desired splitting of the solution space of the Dirac equation into two subspaces
(see~\S\ref{secextfield}) can now be obtained using the sign of the spectrum of~$\tilde{k}$.
More precisely, we choose~$P^\sea$ such that its image coincides with the negative spectral
subspace of~$\tilde{k}$. To this end, we choose a function~$f$ which vanishes identically in a
neighborhood of~$+1$. In a neighborhood of~$-1$, on the other hand, the form of~$f$ is
determined by the spatial normalization condition (see~\eqref{spatialnorm}).
Namely, the correct definition is
\nindex{be0@$P^\sea$ -- fermionic projector describing Dirac seas}%
\beq \label{Psea}
P^\sea = -\frac{1}{2 \pi i} \ointctrclockwise_{\Gamma_-} (-\lambda)\: \tilde{R}_\lambda\: d\lambda \:,
\eeq
as becomes clear in the next proposition.
\sindex{fermionic projector!in the presence of an external potential}%
\begin{Prp} \label{prpspatial}
The expansion~$P^\text{\rm{sea}}$ has the properties
\begin{align}
(i \Pdd+\B-m)\, P^\text{\rm{sea}} &= 0 \label{Psol2} \\
2 \pi \int_{\R^3} P^\text{\rm{sea}} \big( x, (t, \vec{y}) \big) \:\gamma^0\: P^\text{\rm{sea}}
\big( (t, \vec{y}), z \big)\: d^3y &= -P^\text{\rm{sea}}(x,z)\:. \label{Pspatial}
\end{align}
Moreover, $P^\sea$ is symmetric
\beq \label{Pseasymm}
(P^\sea)^* = P^\sea \:,
\eeq
where the star denotes the adjoint with respect to the space-time inner product~\eqref{stip}.
\end{Prp} \noindent
We note for clarity that for the kernel of the fermionic projector,
the symmetry property~\eqref{Pseasymm} means that
\beq \label{Pkersymm}
\big( P^\sea(x,y) \big)^* = P^\sea(y,x) \:,
\eeq
where the star denotes the adjoint with respect to the spin scalar product~\eqref{sspMink}.

In order to simplify the notation in the proof,
we abbreviate the spatial integral in~\eqref{Pspatial} by~$|_t$, i.e.
\nindex{be8@$\vert_t$ -- spatial normalization integral}%
\[ 
(A \,|_t\, B)(x,z) := 2 \pi \int_{\R^3} A \big( x, (t, \vec{y}) \big) \:\gamma^0\: B \big( (t, \vec{y}), z \big)\: d^3y \:. \]
We begin with a preparatory lemma.
\begin{Lemma} \label{lemmakmid}
For any~$t_0 \in \R$, the distribution~\eqref{def-ktil} has the property
\[ \tilde{k}_m \,|_{t_0}\, \tilde{k}_m = \tilde{k}_m \:. \]
\end{Lemma}
\Proof Clearly, it suffices to prove the relation when evaluated by a test function~$f$.
Then~$\tilde{\phi} := \tilde{k}_m(f)$ is a smooth solution of the Dirac equation with
spatially compact support. Therefore, it suffices to show that for any such solution,
\[ \tilde{\phi}(t, \vec{x}) = 2 \pi \int_{\R^3} \tilde{k}_m(t,\vec{x}; t_0, \vec{y})\,
\gamma^0\, \tilde{\phi}_0(\vec{y})\: d^3y\:. \]
Since~$\tilde{\phi}$ and~$\tilde{k}_m$ satisfy the Dirac equation, it suffices to prove
this equation in the case~$t>t_0$. In this case, the equation simplifies in view of~\eqref{def-ktil} to
\[ \tilde{\phi}(x) = i \int_{\R^3} \tilde{s}^\wedge_m(x,y)\, \gamma^0\, \tilde{\phi}_0(y) \big|_{y=(t_0, \vec{y})}
\: d^3y\:, \]
where we set~$x=(t, \vec{x})$.
This identity is derived as follows: We choose a non-negative function~$\eta \in C^\infty(\R)$
with~$\eta|_{[t_0, t]} \equiv 1$ and~$\eta_{(-\infty, t_0-1)} \equiv 0$.
We also consider~$\eta=\eta(x^0)$ as a function of the time variable in space-time. Then
\[ \tilde{\phi}(x) = (\eta \tilde{\phi})(x) = \tilde{s}^\wedge_m \big( (i \Pdd + \B - m) (\eta \tilde{\phi}) \big)
= \tilde{s}^\wedge_m \big( i \gamma^0 \,\dot{\eta}\, \tilde{\phi}) \big) \:, \]
where we used the defining equation of the Green's function~$\tilde{s}_m^\wedge (i \Pdd_x +\B- m)=\1$
together with the fact that~$\tilde{\phi}$ is a solution of the Dirac equation.
To conclude the proof, we choose a sequence~$\eta_l$ such that
the sequence of derivatives~$\dot{\eta}_l$ converges as~$l \rightarrow \infty$ in the
distributional sense to the $\delta$-distribution~$\delta_{t_0}$ supported at~$t_0$. Then
\begin{align*}
\tilde{s}^\wedge_m \big( i \gamma^0 \,\dot{\eta}\, \tilde{\phi}) \big)(x)
&= \int \left( \tilde{s}^\wedge_m(x,y) \big( i \gamma^0 \, \dot{\eta}(y^0)\, \tilde{\phi}(y) \big) \right) d^4y \\
&\rightarrow \int_{\R^3} \left( \tilde{s}^\wedge_m(x,y) \big( i \gamma^0 \tilde{\phi}) \right) 
\big|_{y=(t_0, \vec{y})} \:d^3y \:,
\end{align*}
giving the result.
\QED
An alternative, more computational proof of this lemma is sketched in Exercise~\ref{ex2.4}.

\Proof[Proof of Proposition~\ref{prpspatial}] $\;\;\;$
The Dirac equation~\eqref{Psol2} follows immediately from the identity~\eqref{fsol}.
In order to prove~\eqref{Pspatial}, we integrate the relations
\[ \tilde{R}_\lambda \cdot (\tilde{k}-\lambda) = \1 = (\tilde{k}-\lambda) \cdot \tilde{R}_\lambda \:, \]
to obtain
\[ \ointctrclockwise_{\Gamma_-} \tilde{R}_\lambda \cdot \tilde{k} \: d\lambda
= \ointctrclockwise_{\Gamma_-} \tilde{R}_\lambda\: \lambda\: d\lambda
= \ointctrclockwise_{\Gamma_-} \tilde{k}\, \tilde{R}_\lambda \: d\lambda \:. \]
As a consequence,
\[ P^\text{\rm{sea}} \,|_t\, P^\text{\rm{sea}}
= -\frac{1}{4 \pi^2} \ointctrclockwise_{\Gamma_-} d\lambda \ointctrclockwise_{\Gamma_-'} d\lambda' \:
\tilde{R}_\lambda \cdot \tilde{k} \,|_t\, \tilde{k} \cdot \tilde{R}_{\lambda'} \:, \]
and applying Lemma~\ref{lemmakmid} for~$t_0=t$ gives
\[ P^\text{\rm{sea}} \,|_t\, P^\text{\rm{sea}}
= -\frac{1}{4 \pi^2} \ointctrclockwise_{\Gamma_-} d\lambda \ointctrclockwise_{\Gamma_-'} d\lambda' \:
\tilde{R}_\lambda \cdot \tilde{k} \cdot \tilde{R}_{\lambda'}
= -\frac{1}{4 \pi^2} \ointctrclockwise_{\Gamma_-} \lambda\, d\lambda \ointctrclockwise_{\Gamma_-'} d\lambda' \:
\tilde{R}_\lambda \cdot \tilde{R}_{\lambda'} \:. \]
Now we can again apply~\eqref{rr1} and~\eqref{rr2} (which remains valid if the integrand involves
an additional factor~$\lambda$) as well as~\eqref{rr3}. We thus obtain
\[ P^\text{\rm{sea}} \,|_t\, P^\text{\rm{sea}} =  -\frac{1}{2 \pi i} \ointctrclockwise_{\Gamma_-} \lambda\,
\tilde{R}_\lambda \:d\lambda = -P^\text{\rm{sea}} \:. \]

It remains to prove the symmetry property~\eqref{Pseasymm}.
The operators~$p_m$, $k_m$ and~$s_m$ are obviously symmetric
(with respect to the inner product~\eqref{stip}).
According to~\eqref{k-pert}, the operator~$\tilde{k}_m$ is also symmetric. Hence
the resolvent~$\tilde{R}_\lambda$ defined by~\eqref{Resdef} has the property
\[ \tilde{R}_\lambda^* = \tilde{R}_{\overline{\lambda}}\:. \]
This property implies that if we consider the Laurent expansion of~$-\lambda \,\tilde{R}_\lambda$
around~$\lambda=-1$,
\[ -\lambda \:\tilde{R}_\lambda = \frac{A_{-1}}{\lambda+1} + A_0 + A_1\: (1+\lambda) + \cdots \:, \]
then the operators~$A_{-1}, A_0, \ldots$ are all symmetric with respect to~\eqref{stip}.
Since the contour integral~\eqref{Psea} simply gives the residue~$-A_{-1}$, we obtain~\eqref{Pseasymm}.
This concludes the proof.
\QED

In order to illustrate the above constructions, we now compute the first orders
of the perturbation expansion~\eqref{Psea}.
We first recall that in the computation rules~\eqref{ppprod}--\eqref{bbprod} no principal
values occur. Using these rules in~\eqref{tR} and~\eqref{Psea}, one
sees that also~$P^\sea$ involves no principal values.
With this in mind, we may omit all principal values in the computation,
even if we consider other operator products. In particular, we may write the
computation rules of Lemma~\ref{lemma21} as
\beq \label{rules2}
p \cdot s =s \cdot p = k \cdot s = s \cdot k = 0  \qquad \text{and} \qquad
s \cdot s = \pi^2 \,p \:.
\eeq
Combining~\eqref{rules} and~\eqref{rules2} with~\eqref{Rlamrel}, we obtain
\begin{align*}
R_\lambda \cdot s &= s \cdot R_\lambda = -\frac{1}{\lambda}\:s \\
R_\lambda \cdot k &= k \cdot R_\lambda
= \frac{p+k}{2} \left( \frac{1}{1-\lambda} \right) - \frac{p-k}{2} \left( \frac{1}{-1-\lambda} \right) 
\end{align*}

According to~\eqref{ktildef} and~\eqref{Delkdef},
\begin{align*}
\Delta k &=-s\mathscr{B}k-k\mathscr{B}s+k\mathscr{B}s\mathscr{B}s+s\mathscr{B}k\mathscr{B}s+s\mathscr{B}s\mathscr{B}k-\pi^2k\mathscr{B}k\mathscr{B}k +\O(\mathscr{B}^3)\:.
\intertext{Hence, using~\eqref{tR},}
\tilde{R}_\lambda &= \sum_{n=0}^\infty (-R_\lambda \cdot \Delta k)^n \cdot R_\lambda
= R_\lambda - R_\lambda \cdot \Delta k \cdot R_\lambda +
R_\lambda \cdot \Delta k \cdot R_\lambda \cdot \Delta k \cdot R_\lambda+ \O(\B^3) \\
&= R_\lambda - R_\lambda \cdot \left( -s\mathscr{B}k-k\mathscr{B}s+k\mathscr{B}s\mathscr{B}s+s\mathscr{B}k\mathscr{B}s+s\mathscr{B}s\mathscr{B}k-\pi^2k\mathscr{B}k\mathscr{B}k \right) \cdot R_\lambda \\
&\quad + R_\lambda \cdot \left( -s\mathscr{B}k-k\mathscr{B}s \right) \cdot R_\lambda \cdot
\left( -s\mathscr{B}k-k\mathscr{B}s \right)\cdot R_\lambda+ \O(\B^3) \:.
\end{align*}
Using~\eqref{Rlamrel} and computing the contour integrals, one obtains to first order
\begin{align}
P^\text{\rm{sea}} &= -\lambda\, \frac{p-k}{2} - s\,\mathscr{B} \:\frac{p-k}{2}
-\frac{p-k}{2}\: \mathscr{B} \,s \Big|_{\lambda=-1} + \O(\B^2) \notag \\
&= \frac{p-k}{2} - s \,\mathscr{B} \:\frac{p-k}{2} - \frac{p-k}{2}\: \mathscr{B} \,s + \O(\B^2) \:.
\label{delP1}
\end{align}
To second and higher orders, the resolvent~$\tilde{R}_\lambda$ involves higher poles at~$\lambda=-1$.
This gives rise to derivatives of the factor~$(-\lambda)$ in~\eqref{Psea},
having an influence of the combinatorics of the perturbation expansion (see Exercise~\ref{ex2.4-4}).
The reader interested in more details is referred to~\cite[Appendix~A]{norm}.
A few structural results of the causal perturbation expansion are treated in
Exercises~\ref{ex2.4-5}--\ref{ex2.4-52}.

\subsectionn{Introducing Particles and Anti-Particles} \label{secnorm}
We shall now make the me\-thod of occupying particle and anti-particle states~\eqref{Pvacparticle}
precise in the presence of an external potential. To this end, it is useful to construct
out of the kernel of the fermionic projector a projection operator on a Hilbert space, as
we now explain. On the smooth solutions of the Dirac equation~\eqref{direx} with spatially
compact support one can introduce the scalar product~\eqref{sprodMin}.
Due to current conservation, this scalar product is again independent of the choice of~$t$.
Taking the completion, the solution space of the Dirac equation becomes a Hilbert
space, which we denote by~$(\H_m, (.|.)_m)$.
We now introduce on the Dirac wave functions at time~$t$ the operator
\beq \label{Piseadef}
\begin{split}
&\Pi^\sea \::\: C^\infty_0(\scrN_t, S\scrM) \rightarrow C^\infty(\scrM, S\scrM) \:, \\
&(\Pi^\sea \psi)(x) = -2 \pi \int_{\R^3} P^\sea \big( x, (t, \vec{y}) \big)\: \gamma^0\, \psi(\vec{y})\: d^3 y \:,
\end{split}
\eeq
where~$\scrN_t := \{t\} \times \R^3 \subset \scrM$ denotes the spatial hyperplane at time~$t$.
\nindex{be9@$\scrN_t$ -- Cauchy surface at time~$t$}%
\nindex{bf0@$\Pi^\sea$ -- projection operator on Dirac sea states}%
According to~\eqref{Psol2}, this operator maps to the solutions of the Dirac equation.
Moreover, the spatial normalization property~\eqref{Pspatial} implies that~$\Pi^\sea$ can
be extended by continuity to a projection operator on~$\H_m$, i.e.
\[ \Pi^\sea : \H_m \rightarrow \H_m \qquad \text{with} \qquad (\Pi^\sea)^* = \Pi_\sea = \Pi_\sea^2 \]
(where the star now denotes the adjoint with respect to the scalar product~\eqref{sprodMin};
note that the last equation follows from the symmetry of the kernel~\eqref{Pkersymm}).

Now we can form another operator by adding and subtracting projection operators. More precisely,
the operator
\[ \Pi := \Pi^\sea + \Pi_{\text{span}(\psi_1,\ldots, \psi_{\np})} - \Pi_{\text{span}(\phi_1,\ldots, \phi_{\na})} \]
(where~$\Pi_U : \H_m \rightarrow \H_m$ denotes the orthogonal projection to a subspace~$U \subset \H_m$)
is again a projection operator, provided that the functions~$\phi_l$ are vectors in~$\H_m$
which lie in the image of~$\Pi^\sea$, whereas the vectors~$\psi_k \in \H_m$ are in the orthogonal
complement of the image of~$\Pi^\sea$. In order to comply with the usual normalization of
wave functions in quantum mechanics, we orthonormalize these vectors as follows,
\nindex{as4@$\psi_k, \phi_l$ -- particle and anti-particle states}%
\sindex{particles and anti-particles}%
\sindex{anti-particles}%
\beq \label{orthonorm}
(\psi_k | \psi_{k'})_m = 2 \pi\, \delta_{k,k'} \qquad \text{and} \qquad
(\phi_l | \phi_{l'})_m = 2 \pi\, \delta_{l,l'}
\eeq
(we included the factor~$2 \pi$ in order to account for the factor~$2 \pi$ in~\eqref{sprodMin}).
Then we can write~$\Pi$ more explicitly as
\[ \Pi \psi := \Pi^\sea \psi + \frac{1}{2 \pi} \sum_{k=1}^{\np} \psi_k \: (\psi_k | \psi)_m
- \frac{1}{2 \pi} \sum_{l=1}^{\na} \phi_l\: (\phi_l | \psi)_m \:. \]
This new projection operator can again be written in the form~\eqref{Piseadef} with
the distribution
\[ P(x,y) = P_m^\text{vac}(x,y)
- \frac{1}{2 \pi} \sum_{k=1}^{\np} \psi_k(x) \overline{\psi_k(y)}
+ \frac{1}{2 \pi} \sum_{l=1}^{\na} \phi_l(x) \overline{\phi_l(y)}\:. \]

This relation gives a mathematical justification for~\eqref{Pvacparticle}
in the presence of an external potential. Note that the wave functions~$\psi_k$ and~$\phi_l$
must be solutions of the Dirac equation~\eqref{direx}. Moreover, the~$\phi_l$ must be in
the image of~$\Pi^\sea$, whereas the~$\psi_k$ must be in the orthogonal complement of
the image of~$\Pi^\sea$. Finally, the normalization conditions~\eqref{orthonorm} can be
written as
\[ 
\int_{\R^3} (\overline{\psi_k} \gamma^0 \psi_{k'})(t, \vec{x}) \, d^3x = \delta_{k,k'} \:,\quad
\int_{\R^3} (\overline{\phi_l} \gamma^0 \phi_{l'})(t, \vec{x}) \, d^3x = \delta_{l,l'} \:. \]

\section{The Light-Cone Expansion} \label{seclight}
The light-cone expansion is a powerful tool for analyzing
the fermionic projector in position space. We now outline the constructions and results
as first given in~\cite{firstorder} and~\cite{light}.
Before beginning, we point out that the light-cone expansion is closely tied to the
causal perturbation expansion.
Namely, we shall see that the ``causality'' of the perturbation expansion
(as built in via~\eqref{def-ktil} into the resolvent~\eqref{Resdef}) will become apparent in the light-cone expansion
of~$P(x,y)$ in the fact that all appearing line integrals will be bounded integrals along
the line segment~$\overline{xy}$.
\sindex{causality!of the perturbation expansion}%
This specific feature of the light-cone expansion is of central importance
for the analysis of the continuum limit.

\subsectionn{Basic Definition}
We first give the basic definition of the light-cone expansion and explain it afterwards.
\begin{Def} \label{deflce}
A distribution~$A(x,y)$ on~$M \times M$ is
of the order~$\O((y-x)^{2p})$ for~$p \in \Z$ if the product
\[ (y-x)^{-2p} \: A(x,y) \]
\nindex{bf4@$\O((y-x)^{2p})$ -- order on the light cone}%
is a regular distribution (i.e.\ a locally integrable function).
An expansion of the form
\beq
A(x,y) = \sum_{j=g}^{\infty} A^{[j]}(x,y) \label{l:6a}
\eeq
with $g \in \Z$ is called {\bf{light-cone expansion}}
\sindex{light-cone expansion}%
if the~$A^{[j]}(x,y)$ are distributions of the order
$\O((y-x)^{2j})$ and if~$A$ is approximated by the partial sums
in the sense that for all~$p \geq g$,
\beq \label{l:6b}
A(x,y) - \sum_{j=g}^p A^{[j]}(x,y) \qquad {\text{is of the order~$\O\big( (y-x)^{2p+2} \big)$}}\:.
\eeq
\end{Def} \noindent
The parameter~$g$ gives the leading order of the singularity of~$A(x,y)$ on the light cone.
We point out that we do not demand that the infinite series in~\eqref{l:6a} converges. Thus, similar
to a formal Taylor series, the series in~\eqref{l:6a} is defined only via the approximation by the
partial sums~\eqref{l:6b}. The notion of the light-cone expansion is illustrated in Exercise~\ref{ex2.4-21}.

As a simple example for a light-cone expansion, we consider the distribution~$T_{m^2}(x,y)$
as introduced in~\eqref{Tm2def} and analyzed in Lemma~\ref{lemmaTintro}.
Expanding the Bessel functions in~\eqref{Taway} in a power series, one obtains
(see~\cite[(10.2.2), (10.8.1) and~(10.25.2), (10.31.1)]{DLMF})
\begin{align}
T_{m^2}(x,y) &= -\frac{1}{8 \pi^3} \:\bigg( \frac{\PP}{(y-x)^2} + i \pi \delta \big( (y-x)^2 \big) \: \epsilon \big( (y-x)^0 \big)
\bigg) \nonumber \\
&\quad +\frac{m^2}{32 \pi^3}\sum_{j=0}^\infty \frac{(-1)^j}{j! \: (j+1)!} \: \frac{\big( m^2 (y-x)^2 \big)^j}{4^j} \notag \\
&\qquad\qquad \times  \Big( \log \big|m^2 (y-x)^2 \big| + c_j
+ i \pi \:\Theta \big( (y-x)^2 \big) \:\epsilon\big( (y-x)^0 \big) \Big) \label{l:3.1}
\end{align}
\nindex{ao8@$T_a(x,y)$ -- Fourier transform of lower mass shell}%
with real coefficients~$c_j$ (here~$\Theta$ and~$\epsilon$ are again the Heaviside and the sign function,
respectively).
Due to the factors $(y-x)^{2j}$, this series representation is a light-cone expansion.
The term with the leading singularity becomes integrable after multiplying by~$(y-x)^2$,
showing that~$g=-1$.

The light-cone expansion of the kernel of the fermionic projector of the vacuum $P^\text{vac}(x,y)$
(see~\eqref{Fourier2} and~\eqref{Fourier1}) is readily obtained using the relation~\eqref{Pdiff}.
To this end, one simply applies the differential operator~$i \Pdd+m$ to the above series expansion of~$T_{m^2}$
and computes the derivatives term by term. Since differentiation
increases the order of the singularity on the light cone by one, we thus obtain a light-cone expansion
of the form~\eqref{l:6a} with~$g=-2$.

\subsectionn{Inductive Light-Cone Expansion of the Green's Functions} \label{seclcgreen}
We now return to the perturbation series for the causal Green's functions~\eqref{series-scaustilde}
derived in~\S\ref{secpertgreen}. Our goal is to develop a method
for performing the light-cone expansion of each summand of this perturbation series.
In order to get a first idea for how to proceed, we begin by considering the
free advanced Green's function $s^\vee_m$ of a the Dirac equation of mass~$m$ in position space:
Similar to~\eqref{Pdiff}, it is again convenient to pull the Dirac matrices out of~$s^\vee_m$ by setting
\sindex{Green's function!retarded}%
\nindex{bb0@$s_m^\vee, s_m^\wedge$ -- causal Green's functions of the vacuum Dirac equation}%
\beq
s^\vee_m(x,y) = (i \Pdd_x + m) \: S^\vee_{m^2}(x,y) \:, \label{l:10}
\eeq
where $S^\vee_{m^2}$ is the advanced Green's function of the 
Klein-Gordon operator,
\beq \label{l:11}
S^\vee_{m^2}(x,y) = \lim_{\nu \searrow 0} \int \frac{d^4p}{(2 \pi)^4}
\:\frac{1}{p^2-m^2-i \nu p^0} \:e^{-ip(x-y)} \:.
\eeq
\nindex{bg0@$S_{m^2}^\vee, S_{m^2}^\wedge$ -- causal Green's functions of the Klein-Gordon equation}%
Computing this Fourier integral and expanding the
resulting Bessel function in a power series gives (for details see Exercise~\ref{ex2.4-2})
\begin{align}
S^\vee_{m^2}(x,y) &= -\frac{1}{2 \pi} \:\delta \big( (y-x)^2 \big) \:
\Theta \big( y^0 - x^0 \big) \notag \\
&\quad\:+ \frac{m^2}{4 \pi} \:\frac{J_1 \Big( \sqrt{m^2 
\:(y-x)^2} \Big)}{\sqrt{m^2 \:(y-x)^2}} \:\Theta\big( (y-x)^2 \big) \:\Theta \big(y^0 - 
x^0 \big) \label{l:121} \\
&= -\frac{1}{2 \pi} \:\delta \big( (y-x)^2 \big) \: \Theta\big(y^0 - x^0 \big) \notag \\
&\quad\:+ \frac{m^2}{8 \pi}
\sum_{j=0}^\infty \frac{(-1)^j}{j! \:(j+1)!} \: \frac{\big( m^2 (y-x)^2 \big)^j}{4^j} \:
\Theta \big( (y-x)^2 \big) \:\Theta \big(y^0 - x^0 \big) \:.
\label{l:12}
\end{align}
This computation shows that $S^\vee_{m^2}(x,y)$ has a $\delta((y-x)^2)$-like
singularity on the light cone. Furthermore, one sees that $S^\vee_{m^2}$
is a power series in $m^2$. The important point for what follows is
that the higher order contributions in $m^2$ contain more 
factors $(y-x)^2$ and are thus of higher order on the light cone. 
More precisely,
\beq \label{l:24b}
\left( \frac{d}{dm^2} \right)^n S^\vee_{m^2 }(x,y) \Big|_{m=0} \qquad
\text{is of the order $\O\big((y-x)^{2n-2} \big)$}\:.
\eeq
According to \eqref{l:10}, the Dirac Green's function is obtained by 
computing the first partial derivatives of \eqref{l:12}. Therefore,
$s^\vee_m(x,y)$ has a singularity on the light cone which is even
$\sim \delta^\prime((y-x)^2)$.
The higher order contributions in $m$ are again of increasing order on 
the light cone. This means that we can view the Taylor expansion of 
\eqref{l:10} in $m$,
\[ s^\vee_m(x,y) = \sum_{n=0}^\infty (i \Pdd + m) \;\frac{1}{n!}
\left( \frac{d}{dm^2} \right)^n S^\vee_{m^2}(x,y)  \Big|_{m=0} \: , \]
as a light-cone expansion of the free Green's function. Our idea is to
generalize this formula to the case with interaction. More precisely, we want
to express the perturbed Green's function in the form
\beq
        \tilde{s}^\vee(x,y) = \sum_{n=0}^\infty F_n(x,y) \: \left( 
        \frac{d}{dm^2} \right)^n S^\vee_{m^2}(x,y)  \Big|_{m=0} 
        \label{l:14a}
\eeq
with factors $F_n$ which depend on the external potential.
We will see that this method is very convenient; especially, we can in 
this way avoid working with the rather complicated explicit formula \eqref{l:12}.
Apart from giving a motivation for the desired form \eqref{l:14a} of the
formulas of the light-cone expansion, the mass expansion \eqref{l:12}
\sindex{mass expansion}%
leads to the conjecture
that even the higher order contributions in the mass to the {\em{perturbed}}
Green's functions might be of higher order on the light cone.
If this conjecture was true, it would be a good idea to expand the
perturbation expansion of~$\tilde{s}$ with respect to the parameter $m$.
Therefore, our strategy is to first expand \eqref{series-scaustilde} with respect to
the mass and to try to express the contributions to the resulting expansion
in a form similar to \eqref{l:14a}.

The expansion of \eqref{series-scaustilde} with respect to $m$ gives a double 
sum over the orders in the mass parameter and in the external 
potential. It is convenient to combine these two expansions in a single 
perturbation series. To this end, we rewrite the Dirac operator as
\beq \label{l:18a}
i \Pdd + \B - m = i \Pdd + B \qquad {\mbox{with}} \qquad B:=\B-m \:.
\eeq
\nindex{bg2@$B=\B-m$ -- external potential combined with the mass}%
For the light-cone expansion of the Green's functions, we will always
view $B$ as the perturbation of
the Dirac operator. This has the advantage that the unperturbed objects
are massless. Expanding in powers of~$B$
gives the mass expansion and the perturbation expansion in one step.
\sindex{mass expansion}%
In order to simplify the notation, for the massless objects we usually
omit the index~$m$. Thus we write the Green's function of the
massless Dirac equation in the Minkowski vacuum as
\beq
        s^\vee(x,y) = i \Pdd_x \:S^\vee_{m^2}(x,y) \big|_{m=0}\:,\qquad
        s^\wedge(x,y) = i \Pdd_x \:S^\wedge_{m^2}(x,y) \big|_{m=0}\:.
        \label{l:11a}
\eeq
\sindex{Green's function!advanced}%
\sindex{Green's function!retarded}%
\nindex{bb0@$s_m^\vee, s_m^\wedge$ -- causal Green's functions of the vacuum Dirac equation}%
Then the interacting Green's functions are given by the perturbation series
\beq \tilde{s}^\vee = \sum_{k=0}^\infty (-s^\vee B)^k 
        s^\vee \:,\qquad \tilde{s}^\wedge = \sum_{k=0}^\infty
        (-s^\wedge B)^k s^\wedge \: .
        \label{l:11b}
\eeq
\nindex{bc4@$\tilde{s}_m^\vee, \tilde{s}_m^\wedge$ -- causal Green's functions of the Dirac equation in 
an external potential}%
The constructions of the following subsections are exactly the same for
the advanced and retarded Green's functions. In order to treat both 
cases at once, in the remainder of this section we will omit all 
superscripts `$^\vee$', `$^\wedge$'. The formulas for the advanced and 
retarded Green's functions are obtained by either adding `$^\vee$' or
`$^\wedge$' to all factors $s$, $S$.

We now explain how the individual contributions 
to the perturbation expansion \eqref{l:11b} can be written similar to 
the right side of \eqref{l:14a} as a sum of terms of increasing order 
on the light cone. For the mass expansion of $S_{m^2}$, we set $a=m^2$ and
use the notation
\beq \label{l:23b}
S^{(l)} = \left( \frac{d}{da} \right)^l S_a \big|_{a=0} \: .
\eeq
\nindex{bg8@$S^{(l)}$ -- mass expansion of $S_a$}%
In preparation, we derive some computation rules for the $S^{(l)}$:
$S_a$ satisfies the defining equation of a Klein-Gordon Green's function
\[ (-\Box_x - a) \:S_a(x,y) = \delta^4(x-y) \: . \]
Differentiating with respect to $a$ and setting~$a=0$ gives
\beq
-\Box_x S^{(l)}(x,y) = \delta_{l,0} \:\delta^4(x-y)
+ l \:S^{(l-1)}(x,y) \:,\qquad l \geq 0 . \label{l:5}
\eeq
(For $l=0$, this formula does not seem to make sense because $S^{(-1)}$ 
is undefined. The expression is meaningful, however, if one keeps 
in mind that in this case the factor $l$ is zero, and thus the whole 
second summand vanishes. We will also use this convention in the following 
calculations.) Next, we differentiate the formulas for $S_a$ in momentum space,
\beq
S_a^\vee(p) = \frac{1}{p^2-a-i \nu p^0} \:,\qquad
        S_a^\wedge(p)= \frac{1}{p^2-a+i \nu p^0}
        \label{l:21x}
\eeq
with respect to both $p$ and $a$. Comparing the results gives the 
relation
\[ \frac{\partial}{\partial p^k} S_{a}(p) = -2p_k \:\frac{d}{da} S_a(p) \: , \]
or, after expanding in the parameter $a$,
\beq
\frac{\partial}{\partial p^k} S^{(l)}(p) = -2 p_k \:S^{(l+1)}(p) \:,\qquad l \geq 0 .
        \label{l:21a}
\eeq
This formula also determines the derivatives of $S^{(l)}$ in position space; namely
\begin{align}
\frac{\partial}{\partial x^k} & S^{(l)}(x,y) =
\int \frac{d^4p}{(2 \pi)^4} \:S^{(l)}(p) \:(-i p_k) \:e^{-ip(x-y)} \notag \\
&\!\!\!\!\!\stackrel{\eqref{l:21a}}{=} \frac{i}{2} \int \frac{d^4p}{(2 \pi)^4} \: 
\frac{\partial}{\partial p^k} S^{(l-1)}(p) \; e^{-ip(x-y)} \notag \\
&= -\frac{i}{2} \int \frac{d^4p}{(2 \pi)^4} \: 
         S^{(l-1)}(p) \;\frac{\partial}{\partial p^k} e^{-ip(x-y)} \notag \\
&=\frac{1}{2} \: (y-x)_k \: S^{(l-1)}(x,y) \:,\qquad l \geq 1 .
        \label{l:7}
\end{align}
We iterate this relation to calculate the Laplacian,
\begin{align*}
        -\Box_x S^{(l)}(x,y) &= -\frac{1}{2} \:\frac{\partial}{\partial 
        x^k} \left( (y-x)^k \:S^{(l-1)}(x,y) \right) \\
        &= 2 \:S^{(l-1)}(x,y) + \frac{1}{4} \:(y-x)^2 \:S^{(l-2)}(x,y)
        \:,\qquad l \geq 2 .
\end{align*}
After comparing with \eqref{l:5}, we conclude that
\beq \label{l:22a}
(y-x)^2 \:S^{(l)}(x,y) = -4l\: S^{(l+1)}(x,y)\:,\qquad l \geq 0 \:.
\eeq
Finally, $S^{(l)}(x,y)$ is only a function of $(y-x)$, which 
implies that
\beq \label{l:20a}
\frac{\partial}{\partial x^k} S^{(l)}(x,y) =
-\frac{\partial}{\partial y^k} S^{(l)}(x,y) \:,\qquad l \geq 0 \:.
\eeq

The following lemma gives the light-cone expansion of an operator product which
is linear in the external potential. We will later use it for the 
iterative light-cone expansion of more complicated operator products;
in this case, the potential will be a composite 
expression in $B$ and its partial derivatives. In order to 
avoid confusion then, we denote the external potential by $V$.
\sindex{light-cone expansion!of the Green's function!to first order}%

\begin{Lemma}{\bf{(light-cone expansion to first order)}}
\label{l:lemma1} For any~$l,r \geq 0$, the operator product $S^{(l)} \:V\: S^{(r)}$ has the light-cone expansion
\begin{align}
(&S^{(l)} \:V\: S^{(r)})(x,y)\notag \\
&= \sum_{n=0}^\infty 
        \frac{1}{n!} \int_0^1 \alpha^{l} \:(1-\alpha)^{r} \:
        (\alpha - \alpha^2)^n \: (\Box^n V)_{|\alpha y + (1-\alpha) x} \:d\alpha \;
        S^{(n+l+r+1)}(x,y) \:.
        \label{l:4}
\end{align}
\end{Lemma}
\Proof The method of proof is to first compute the Laplacian of both 
sides of \eqref{l:4}. The resulting formulas will have a
similar structure, making it possible to proceed inductively.

On the left side of \eqref{l:4}, we calculate the Laplacian with the 
help of \eqref{l:5} to
\beq \label{l:6}
-\Box_x (S^{(l)} \:V\: S^{(r)})(x,y) = \delta_{l,0} 
        \:V(x) \: S^{(r)}(x,y) +
        l \: (S^{(l-1)} \:V\: S^{(r)})(x,y) \:.
\eeq

The Laplacian of the integral on the right side of \eqref{l:4} can be 
computed with \eqref{l:7} and \eqref{l:5},
\begin{align}
-&\Box_x \int_0^1 \alpha^{l} \:(1-\alpha)^{r} \:(\alpha-\alpha^2)^n
\: (\Box^n V)_{|\alpha y + (1-\alpha) x} \:d\alpha \;
S^{(n+l+r+1)}(x,y) \label{l:8} \\
         & = -\int_0^1 \alpha^{l} \:(1-\alpha)^{r+2} \:(\alpha-\alpha^2)^n \:
         (\Box^{n+1} V)_{|\alpha y + (1-\alpha) x} \:d\alpha \;
S^{(n+l+r+1)}(x,y) \notag \\
&\quad-\int_0^1 \alpha^{l} \:(1-\alpha)^{r+1} \:(\alpha-\alpha^2)^n \:
(\partial_k \Box^{n} V)_{|\alpha y + (1-\alpha) x}
\:d\alpha \; (y-x)^k \:
S^{(n+l+r)}(x,y) \notag \\
&\quad +(n+l+r+1) \int_0^1 \alpha^{l} \:(1-\alpha)^{r} \:(\alpha-\alpha^2)^n \:
(\Box^{n} V)_{|\alpha y + (1-\alpha) x}
\:d\alpha \;S^{(n+l+r)}(x,y) \: . \notag
\end{align}
In the second summand, we rewrite the partial derivative as a
derivative with respect to $\alpha$,
\[  (y-x)^k (\partial_k \Box^{n} V)_{|\alpha y + (1-\alpha) x}
= \frac{d}{d\alpha}  (\Box^{n} V)_{|\alpha y + (1-\alpha) x} \]
(as is verified immediately by computing the right side with the chain rule).
This makes it possible to integrate in~$\alpha$ by parts. We thus obtain
\begin{align*}
\int_0^1 &\alpha^{l} \:(1-\alpha)^{r+1} \:(\alpha-\alpha^2)^n \:
(\partial_k \Box^{n} V)_{|\alpha y + (1-\alpha) x}
\:d\alpha \; (y-x)^k \\
&= \int_0^1 \alpha^{l} \:(1-\alpha)^{r+1} \:(\alpha-\alpha^2)^n \:
\frac{d}{d\alpha} \Big( (\Box^{n} V) \big|_{\alpha y + (1-\alpha) x} \Big)
d\alpha \\
&= -\delta_{n,0} \:\delta_{l,0}\:V(x) - (n+l)
         \int_0^1 \alpha^{l} \:(1-\alpha)^{r+2} \:(\alpha-\alpha^2)^{n-1} \:
(\Box^{n} V)_{|\alpha y + (1-\alpha) x} \:d\alpha \\
&\qquad +(n+r+1) \int_0^1 \alpha^{l} \:(1-\alpha)^{r} \:(\alpha-\alpha^2)^n \:
(\Box^{n} V)_{|\alpha y + (1-\alpha) x} \:d\alpha \\
&= -\delta_{n,0} \:\delta_{l,0}\:V(x) \\
&\qquad -n \int_0^1 \alpha^{l} \:(1-\alpha)^{r+2} \:(\alpha-\alpha^2)^{n-1} \:
(\Box^{n} V)_{|\alpha y + (1-\alpha) x} \:d\alpha \\
&\qquad+(n+l+r+1) \int_0^1 \alpha^{l} \:(1-\alpha)^{r} \:(\alpha-\alpha^2)^n \:
(\Box^{n} V)_{|\alpha y + (1-\alpha) x} \:d\alpha \\
&\qquad-l \int_0^1 \alpha^{l-1} \:(1-\alpha)^{r} \:(\alpha-\alpha^2)^{n} \:
(\Box^{n} V)_{|\alpha y + (1-\alpha) x} \:d\alpha \: .
\end{align*}
We substitute back into the original equation to obtain
\begin{align*}
&\eqref{l:8} = \delta_{n,0} \:\delta_{l,0} \: V(x) \: S^{(r)}(x,y) \\
&\quad+l \int_0^1 \alpha^{l-1} \:(1-\alpha)^{r} \:(\alpha-\alpha^2)^{n} \:
(\Box^{n} V)_{|\alpha y + (1-\alpha) x} 
\:d\alpha \; S^{(n+l+r)}(x,y) \\
&\quad-\int_0^1 \alpha^{l} \:(1-\alpha)^{r+2} \:(\alpha-\alpha^2)^{n} \:
(\Box^{n+1} V)_{|\alpha y + (1-\alpha) x} 
\:d\alpha \; S^{(n+l+r+1)}(x,y) \\
&\quad+n\int_0^1 \alpha^{l} \:(1-\alpha)^{r+2} \:(\alpha-\alpha^2)^{n-1} \:
(\Box^{n} V)_{|\alpha y + (1-\alpha) x} 
\:d\alpha \; S^{(n+l+r)}(x,y) \: .
\end{align*}
After dividing by $n!$ and summation over $n$, the last two summands 
are telescopic and cancel each other. Thus one gets
\begin{align}
&-\Box \sum_{n=0}^\infty \frac{1}{n!} \int_0^1 
\alpha^{l} \:(1-\alpha)^{r} \:(\alpha-\alpha^2)^n \: (\Box^n 
V)_{|\alpha y + (1-\alpha) x} \:d\alpha \; S^{(n+l+r+1)}(x,y) \notag \\
&= \delta_{l, 0} \:V(x)\:S^{(r)}(x,y) \notag \\
&\qquad +l \sum_{n=0}^\infty \frac{1}{n!} \int_0^1 
\alpha^{l-1} \:(1-\alpha)^{r} \:(\alpha-\alpha^2)^n \: (\Box^n 
V)_{|\alpha y + (1-\alpha) x} \:d\alpha \; S^{(n+l+r)}(x,y) \:.\label{l:9a}
\end{align}

We now compare the formulas \eqref{l:6} and \eqref{l:9a} for the Laplacian of 
both sides of \eqref{l:4}. In the special case $l=0$, these formulas
coincide, and we can use a uniqueness argument for the solutions of the 
wave equation to prove \eqref{l:4}: We assume that
we consider the advanced Green's function (for the 
retarded Green's function, the argument is analogous). For given $y$,
we denote the difference of both sides of \eqref{l:4} by $F(x)$.
Since the support of $F(x)$ is in the past light cone $x \in
L^\wedge_y$, $F$ vanishes in a neighborhood of the 
hypersurface ${\mathcal{H}}=\{z \in \R^4 \:|\: z^0 = y^0 + 1\}$. 
Moreover, the Laplacian of $F$ is identically zero according to 
\eqref{l:6} and \eqref{l:9a}. We conclude that
\[      \Box F = 0 \qquad {\mbox{and}} \qquad F_{|{\mathcal{H}}}= \partial_k 
        F_{|{\mathcal{H}}} = 0 \: . \]
Since the wave equation has a unique solution for given initial data 
on the Cauchy surface ${\mathcal{H}}$, $F$ vanishes identically.

The general case follows by induction in $l$: Suppose that 
\eqref{l:4} holds for given $\hat{l}$ (and arbitrary $r$).
Then, according to \eqref{l:6}, \eqref{l:9a}, and the induction hypothesis,
the Laplacian of both sides of \eqref{l:4} coincides for $l = \hat{l} + 1$.
The above uniqueness argument for the solutions of the wave equation again 
gives \eqref{l:4}.
\QED
We recall for clarity that, according to \eqref{l:24b}, the higher 
$a$-derivatives of $S_a(x,y)$ are of higher order on the light cone. 
Thus the summands in \eqref{l:4} are of increasing order on the light 
cone, and the infinite sum is mathematically well-defined in the sense of Definition~\ref{deflce}
via the approximation by the partial sums \eqref{l:6b}.

Lemma \ref{l:lemma1} can be used for the light-cone expansion of 
more complicated operator products. To explain the method, we look at 
the simplest example of three factors~$S^{(0)}$ and two 
potentials~$V$ and~$W$,
\[ 
(S^{(0)} \:V\: S^{(0)} \:W\: S^{(0)})(x,y) = \int d^4z \: 
S^{(0)}(x,z) \: V(z)\; (S^{(0)}\:W\:S^{(0)})(z,y) \:. \]
Having split up the operator product in this form, we can apply 
Lemma \ref{l:lemma1} to the factor $S^{(0)} W S^{(0)}$,
\[ = \sum_{n=0}^\infty \frac{1}{n!} \int d^4z \;
S^{(0)}(x,z) \left\{ V(z) \int_0^1 (\alpha-\alpha^2)^n \:
(\Box^n W)_{|\alpha y + (1-\alpha) z} \:d\alpha \right\} S^{(n+1)}(z,y) \:. \]
Now we rewrite the $z$-integral as the operator product
$(S^{(0)} g_y S^{(0)})(x,y)$, where $g_y(z)$ is the function in the 
curly brackets. The $y$-dependence of $g_y$ causes no problems because we can 
view $y$ as a fixed parameter throughout the expansion. Thus we can simply
apply Lemma \ref{l:lemma1} once again to obtain
\begin{align*}
&= \sum_{m,n=0}^\infty \frac{1}{m!\:n!} \int_0^1 d\beta 
\;(1-\beta)^{n+1} \:(\beta-\beta^2)^m \:\int_0^1 d\alpha 
\;(\alpha-\alpha^2)^n \\
&\hspace*{1.5cm}
\times\; \Box^m_z \left( V(z) \:(\Box^n W)_{|\alpha y + (1-\alpha) 
z} \right)_{|z=\beta y + (1-\beta) x} \; S^{(m+n+2)}(x,y) \:.
\end{align*}
The Laplacian $\Box^m_z$ could be computed further with the Leibniz
rule. Notice that the manipulations of the infinite sums are
unproblematic because to every order on the light cone, the number of terms is
actually finite (the situation would be more difficult if we studied the 
convergence of the sum \eqref{l:6a}, but, as pointed out earlier, the 
light-cone expansion is defined merely via the partial sums).

We want to iteratively perform the light-cone expansion 
of the operator products in \eqref{l:11b}. This is not possible directly 
with the method just described, because \eqref{l:11b} contains the Dirac 
Green's function $s$ (instead of $S$). We must think about how to deal 
with this complication. Relation \eqref{l:11a} allows us to replace the
factors $s$ by $S$, but this gives additional partial derivatives in 
the operator product. These derivatives can be
carried out after each iteration step by applying the Leibniz rule and using
the differentiation rule~\eqref{l:7}. In the simplest example, we have
\begin{align*}
\lefteqn{ (s^{(0)} \:V\: S^{(0)})(x,y) = (i \Pdd_x) 
(S^{(0)}\:V\:S^{(0)})(x,y) } \\
&= i \Pdd_x \sum_{n=0}^\infty \frac{1}{n!} \int_0^1 
(\alpha-\alpha^2)^n \:(\Box^nV)_{|\alpha y + (1-\alpha) x} \:d\alpha\; 
S^{(n+1)}(x,y) \\
&= i \sum_{n=0}^\infty \frac{1}{n!} \int_0^1 
(1-\alpha)\:(\alpha-\alpha^2)^n \:(\Pdd\:\Box^nV)_{|\alpha y + (1-\alpha) x} \:d\alpha
\; S^{(n+1)}(x,y) \\
&\quad+\frac{i}{2}  \:\sum_{n=0}^\infty \frac{1}{n!} \int_0^1 
(\alpha-\alpha^2)^n \:(\Box^nV)_{|\alpha y + (1-\alpha) x}  \:d\alpha\;
(y-x)_j \gamma^j \; S^{(n)}(x,y) \: .
\end{align*}
The only problem with this method is that the partial derivatives
might hit a factor $S^{(0)}$, in which case the rule~\eqref{l:7} cannot be 
applied. In order to resolve this problem, we extend our 
constructions in a way which allows us to use all previous formulas 
also in this special case. To this end, we take \eqref{l:7} as the defining
equation for $(y-x)_k \:S^{(-1)}(x,y)$,
\beq
(y-x)_k \:S^{(-1)}(x,y) := 2 \:\frac{\partial}{\partial x^k}
        S^{(0)}(x,y) \label{l:29z}
\eeq
\nindex{bg8@$S^{(l)}$ -- mass expansion of $S_a$}%
(notice that $S^{(-1)}$ itself remains undefined, only the combination
$(y-x)_k \:S^{(-1)}(x,y)$ makes mathematical sense as the partial 
derivative of the distribution $S^{(0)}$). It turns out that with this definition, all our computation
rules as well as the light-cone expansion of Lemma~\ref{l:lemma1} remain valid
for~$S^{(-1)}$:
\begin{Lemma} {\bf{(light-cone expansion to first order for $r=-1$)}}
\label{l:lemma2}
The operator product $(S^{(l)} \:.\: S^{(-1)})$, $l \geq 0$, has the 
light-cone expansion
\begin{align*}
&\int d^4z \; S^{(l)}(x,z) \:V(z) \:(y-z)_k \:S^{(-1)}(z,y) \notag \\
&= \sum_{n=0}^\infty \frac{1}{n!} \int_0^1 \alpha^{l} \:(1-\alpha)^{-1} \:(\alpha-\alpha^2)^n
\: \Box^n_z \big( V(z) \: (y-z)_k \big)
\big|_{z=\alpha y + (1-\alpha) x} \: d\alpha \; S^{(n+l)}(x,y) \:.
\end{align*}
\end{Lemma} \noindent
Since the proof is straightforward, we omit it here
but refer to Exercise~\ref{ex2.5} or~\cite[proof of Lemma~2.2]{light}.
We note for clarity that the pole of the factor~$(1-\alpha)^{-1}$ at~$\alpha=1$
in the formula of the above lemma does not cause any problems.
Namely, in the case~$n=0$ it disappears
because~$(1-\alpha)^{-1} (y-z)=y-x$, whereas in the case~$n>0$
it is compensated by the zero of the factor $(\alpha-\alpha^2)^n$.

\subsectionn{Structural Results for Chiral Potentials} \label{secstructure}
In the previous section, we gave a constructive procedure for performing the
light-cone expansion of each summand of the perturbation expansion
for the causal Green's functions~\eqref{l:11b}. In this and the next section, we shall
explain how to use this method to uncover the structure of the Green's functions in position space.
To this end, we need to specify the form of the external potential~$\B$ in the
Dirac equation~\eqref{direx}. We are mostly interested in the situation that~$\B$
is composed of left- or right-handed potentials, i.e.\
\beq \label{Bchiral}
\B = \chi_L \:\slashed{A}_R + \chi_R \: \slashed{A}_L \:.
\eeq
\sindex{potential!chiral}%
\nindex{bh0@$\chi_{\LR}$ -- chiral projectors}%
\nindex{bh1@$\pseudo$ -- pseudoscalar matrix}%
\nindex{bh2@$A_L, A_R$ -- chiral potentials}%
(here~$\chi_{L\!/\!R} = \frac{1}{2}(\1 \mp \pseudo)$ are the chiral projectors,
and~$\pseudo = i \gamma^0 \gamma^1 \gamma^2 \gamma^3$ is the usual pseudoscalar matrix).
Such so-called {\em{chiral potential}} are of central interest because they allow for the description
of {\em{gauge fields}}.
\sindex{gauge potential}%
\sindex{potential!gauge|see{gauge potential}}%
\sindex{gauge field}%
For example, an electromagnetic field is described by choosing~$A_L=A_R=A$,
where~$A$ is the electromagnetic potential.
A left-handed potential is needed for example for describing the weak interaction in the standard model.
In this context, it is important to describe {\em{non-abelian}} gauge fields.
\sindex{gauge field!non-abelian}%
In this case, the potentials~$A_L$ and~$A_R$ take values in a Lie algebra.
For simplicity, we here always represent the potentials by matrices acting on~$\C^g$
with~$g \in \N$. In order to describe the coupling of the gauge gauge fields to the fermions, the Dirac
wave functions must also carry an index~$a=1,\ldots, g$.
Moreover, we want to allow for the situation that the system involves Dirac matrices
of different rest masses, which we label again by an index~$a$.
This leads to the following setup. We define the fermionic projector of the vacuum
and the Green's functions as direct sums of the corresponding operators with
rest masses~$m_1, \ldots, m_g$, i.e.
\beq \label{Pdirsum}
P^\text{vac} = \bigoplus_{a=1}^g P^\text{vac}_{m_a} \qquad \text{and} \qquad
s = \bigoplus_{a=1}^g s_{m_a}
\eeq
with~$P^\text{vac}_{m_a}$ and~$s_m$ according to~\eqref{Fourier1} and~\eqref{8b}.
We write the Dirac equation as
\beq \label{direxY}
(i \Pdd + \B - m Y) \,\psi(x) = 0
\eeq
with~$\B$ as in~\eqref{Bchiral}. Here~$Y$ is the {\em{mass matrix}} defined by
\[ Y = \frac{1}{m} \: \text{diag} \big(m_1, \ldots, m_a) \]
\sindex{mass matrix}%
\nindex{bh3@$m$ -- parameter used for mass expansion}%
\nindex{bh4@$Y$ -- mass matrix}%
(here~$m$ is introduced merely as an expansion parameter; the picture is that~$Y$
is dimensionless, whereas~$m$ carries the dimension of inverse length).
For later use, it is also convenient to allow for {\em{scalar}} and {\em{pseudoscalar potentials}}.
\sindex{potential!scalar}%
\sindex{potential!pseudoscalar}%
In order to built these potentials into the Dirac equation~\eqref{direxY}, it is most convenient
to replace the mass matrix by a space-time dependent matrix,\footnote{To avoid confusion,
we point out that our convention differs from that used in~\cite{light, PFP},
where the dynamical mass matrix is defined instead by~$Y=\chi_L Y_R + \chi_R Y_L$.
Our convention fits to our general rule that left- and right-handed potentials should couple to
the left- and right-handed component of the Dirac spinors, respectively
(see also~\eqref{psichiral} and the explanation thereafter).}
\beq \label{YLRdef}
Y=Y(x) := \chi_L Y_L(x) + \chi_R Y_R(x) \:,
\eeq
referred to as the {\em{dynamical mass matrix}}.
\sindex{mass matrix!dynamical}%

In analogy to~\eqref{l:18a}, we combine the mass term with the potential by setting
\beq \label{Bchiral2}
B = \chi_L \:\slashed{A}_R + \chi_R \: \slashed{A}_L - m Y\:.
\eeq
Then the perturbation expansion of the causal Green's functions can again be written
in the form~\eqref{l:11b}. The light-cone expansion can be carried out exactly as
explained in the previous section. The only point to keep in mind is that the chiral potentials
at different space-time points do not necessarily commute. Moreover, the chiral potentials
in general do not commute with the mass matrix. Therefore, in what follows we need
to be careful in keeping track of the order of multiplication.

Before going on, we explain our convention for the chiral indices of
potentials in~\eqref{YLRdef} and~\eqref{Bchiral2}.
We follow the usual rule that a left-handed potential couples to the
left-handed component of the Dirac wave function, whereas the right-handed potential couples to the
right-handed component of the wave function. Indeed, decomposing the Dirac wave function as
\beq \label{psichiral}
\psi = \chi_L\: \psi_L + \chi_R\: \psi_R \:,
\eeq
the Dirac equation~\eqref{direxY} becomes
\begin{align*}
0 &= \Big( i \Pdd + \chi_L \:\slashed{A}_R + \chi_R \: \slashed{A}_L - m \chi_L Y_L(x) - m \chi_R Y_R(x) \Big)
\Big( \chi_L\: \psi_L + \chi_R\: \psi_R \Big) \\
&= \chi_L \,\Big( \big(i \Pdd + \slashed{A}_R \big) \psi_R - m Y_L \psi_L \Big)
+ \chi_R \,\Big( \big(i \Pdd + \slashed{A}_L \big) \psi_L - m Y_R \psi_R \Big) \:.
\end{align*}
(here we use that the chirality is reversed at each Dirac matrix).
This shows that our conventions~\eqref{YLRdef} and~\eqref{Bchiral2} indeed
imply that left-handed potentials couple to~$\psi_L$ and right-handed potentials to~$\psi_R$.

The next theorem gives a structural result on the contributions to the light-cone expansion
of the Green's functions. For the line integrals, we introduce the short notation
\beq \label{l:29x}
\int_x^y [l,r\:|\: n] \:dz \; f(z) := \int_0^1 d\alpha \;
\alpha^{l}\:(1-\alpha)^{r} \:(\alpha-\alpha^2)^n \; f(\alpha y + (1-\alpha) x) \:.
\eeq
\nindex{bh6@$\int_x^y [l,r \:\vert\: n] \cdots$ -- short notation for line integrals}%
Furthermore, we abbreviate the following products with multi-indices,
\[ \partial_z^J := \frac{\partial}{\partial z^{j_1}} \cdots
\frac{\partial}{\partial z^{j_l}} \:,\qquad
(y-x)^J := (y-x)^{j_1} \cdots (y-x)^{j_l} \:,\quad
\gamma^J \;:=\; \gamma^{j_1} \cdots \gamma^{j_l} \:, \]
where $J=(j_1, \ldots, j_l)$.
\nindex{bh8@$\partial_z^J, (y-x)^J$ -- multi-index notation}%
\sindex{multi-index}%

\begin{Thm} \label{l:thm1}
In the presence of chiral potentials~\eqref{Bchiral2}, the light-cone expansion of the $k^{\mbox{\scriptsize{th}}}$ order 
contribution $((-s B)^k \:s)(x,y)$ to the perturbation series 
\eqref{l:11b} can be written as an infinite sum of expressions, each of which 
has the form
\sindex{light-cone expansion!of the Green's function}%
\begin{align}
\chi_{c_0} \:C \:(y-x)^I &\int_x^y [l_1, r_1 \:|\: n_1] \:dz_1 \; 
\partial_{z_1}^{I_1}\: \Box_{z_1}^{p_1} \:V^{(1)}_{J_1, c_1}(z_1)
\int_{z_1}^y [l_2, r_2 \:|\: n_2] \:dz_2 \; \partial_{z_2}^{I_2}\:
\Box_{z_2}^{p_2}\: V^{(2)}_{J_2, c_2}(z_2) \notag \\
&\cdots \int_{z_{k-1}}^y [l_k, r_k \:|\: n_k] \:dz_k \; 
\partial_{z_k}^{I_k}\: \Box_{z_k}^{p_k}\: V^{(k)}_{J_k, c_k}(z_k) \;
\gamma^J \;S^{(h)}(x,y) \:. \label{l:l1}
\end{align}
In this formula, $C$ is a complex number and the parameters
$l_a$, $r_a$, $n_a$, and $p_a$ are non-negative integers; the indices
$c$ and $c_a$ can take the two values $L$ or $R$.
The functions $V^{(a)}_{J_a, c_a}$ (where~$J_a$ is a multi-index and~$c_a \in \{L,R\}$ is a chiral index)
coincide with any of the individual potentials in~\eqref{Bchiral2} 
and~\eqref{YLRdef} with chirality $c_a$, i.e.
\beq
\begin{split}
V^{(a)}_{c_a} &= A_{c_a} \qquad\;\;\,
{\mbox{(in which case $|J_a|=1$)}} \qquad {\mbox{or}} \\
V^{(a)}_{c_a} &= m Y_{c_a} \qquad {\mbox{(in which case 
$|J_a|=0$)}} \:.
\end{split}  \label{l:42e}
\eeq
The chirality $c_a$ of the potentials is determined by the following rule:
\begin{itemize}[leftmargin=2em]
\item[(i)] The chirality is reversed precisely at every mass matrix, i.e.
\[ {\mbox{$c_{a-1}$ and $c_a$}}\;\;\left\{ \begin{array}{cl} 
\mbox{coincide} & \mbox{if $V^{(a)}_{c_a}=A_{c_a}$} \\
\mbox{are opposite} & \mbox{if $V^{(a)}_{c_a}=m Y_{c_a}$}
\end{array} \right. \]
for all~$a=1,\ldots, k$.
\end{itemize}
The tensor indices of the multi-indices in \eqref{l:l1} are all
contracted with each other, according to the following rules:
\begin{itemize}[leftmargin=2em]
\item[(a)] No two tensor indices of the same multi-index are
contracted with each other.
\item[(b)] The tensor indices of the factor $\gamma^{J}$ are 
all contracted with different multi-indices, in the order of their appearance in
the product \eqref{l:l1} (i.e., for $J=(j_1,\ldots,j_l)$ and $1 \leq a < b 
\leq l$, the multi-index with which $j_a$ is contracted must stand to the 
left of the multi-index corresponding to $j_b$).
\end{itemize}
The parameter $h$ is given by
\beq \label{l:l3}
2h = k - 1 - |I| + \sum_{a=1}^k \Big( |I_a| + 2 p_a \Big) \:.
\eeq       
The number of factors $(y-x)$ is bounded by
\beq \label{l:l3a}
|I| \leq k+1-\sum_{a=1}^k |I_a| \:.
\eeq        
\end{Thm}
Basically, this theorem states that the light-cone expansion of the
$k^{\mbox{\scriptsize{th}}}$ order Feynman diagrams can be written with $k$ 
nested line integrals. Notice that the potentials
$V^{(a)}(z_a)$ do in general not commute with each other, so that the order
of multiplication is important in \eqref{l:l1}.
In order to avoid misunderstandings, we point out that the derivatives
$\partial^{I_a}_{z_a}$ and $\Box_{z_a}^{p_a}$ do not only act on $V^{(a)}(z_a)$, but also on 
all the following factors $V^{(a+1)}(z_{a+1})$, $V^{(a+2)}(z_{a+2})$,\ldots\
(note that the variables $z_{a+1}$, $z_{a+2}$,\ldots\ implicitly depend 
on $z_a$ via the inductive definition of the line integrals).
Clearly, these derivatives could be carried out further with the
Leibniz rule, but it is easier not to do this at the moment.
The restrictions {\em{(a)}} and {\em{(b)}} on the possible contractions of
the tensor indices were imposed in order to avoid an abuse of our
multi-index notation. More precisely, {\em{(a)}} prevents factors
$(y-x)^2$ in $(y-x)^I$, an unnecessary
large number of $\gamma$-matrices in $\gamma^J$, and ``hidden'' Laplacians in
the partial derivatives $\partial_{z_a}^{I_a}$.
The rule {\em{(b)}}, on the other hand, prevents factors $(y-x)^2$ and
hidden Laplacians in combinations of the form
$(y-x)_i \:(y-x)_j \: \gamma^i \:\gamma^j$ and
$\partial_{ij} V^{(a)}_{J_a} \: \gamma^i \:\gamma^j$, respectively.
Our ordering condition for the $\gamma$-matrices is just a matter of
convenience. Relation \eqref{l:l3} is very useful because it
immediately tells for any configuration of the line integrals and 
potentials in \eqref{l:l1} what the corresponding order on the light cone is.
Notice that \eqref{l:l3} and \eqref{l:l3a} imply the inequality
\beq \label{l:45o}
h \geq -1 + \sum_{a=1}^k \left( |I_a| + p_a \right) \:.
\eeq
In particular, one sees that $h \geq -1$. In the case $h=-1$, \eqref{l:l3} yields
that $|I|>0$, so that \eqref{l:l1} must contain at least one factor $(y-x)$.
Therefore, the factor $S^{(h)}$ in \eqref{l:l1} is always well-defined
by either \eqref{l:23b} or \eqref{l:29z}.

We point out that, although the total number of summands \eqref{l:l1} 
is infinite, the number of summands for any given value of the 
parameter $h$ is finite. This is clear because, for fixed $h$, the 
relations \eqref{l:l3} and \eqref{l:l3a} only allow for a finite number of 
possibilities to choose the parameters $|I|$, $|I_a|$, and $p_a$,
giving rise to only a finite number of expressions of the 
form \eqref{l:l1}. Since, according to \eqref{l:24b}, the contributions
for higher values of $h$ are of higher order on the light cone, we 
conclude that the number of summands \eqref{l:l1} is finite to every 
order on the light cone. Therefore, the light-cone expansion of 
Theorem \ref{l:thm1} makes mathematical sense in terms of Definition~\ref{deflce}.
\\[.5em]
\Proof[Proof of Theorem \ref{l:thm1}.]
We proceed inductively in $k$.
For $k=0$, the assumption is true 
because in view of~\eqref{l:11a} and~\eqref{l:29z} we can write the free Dirac Green's function as
\beq
s(x,y) = (\chi_L + \chi_R) \:\frac{i}{2} \:(y-x)^j \gamma_j \:S^{(-1)}(x,y) \:, \label{l:233a}
\eeq
which is of the desired form~\eqref{l:l1}.
The conditions {\em{(i)}}, {\em{(a)}}, {\em{(b)}}, and the
relations \eqref{l:l3}, \eqref{l:l3a} are clearly satisfied.

Assume that the theorem holds for a given $k$. With the formula
\beq \label{l:i4}
\big( (-s B)^{k+1}\:s \big)(x,y) = -i \Pdd_x \int d^4 z \; S^{(0)}(x,z) \:B(z) \: \big((-s B)^k\:s \big)(z,y) \:,
\eeq
we can express the $(k+1)^{\mbox{\scriptsize{st}}}$ order 
contribution to the perturbation series
\eqref{l:11b} in terms of the $k^{\mbox{\scriptsize{th}}}$ order
contribution. We must show that \eqref{l:i4} can again 
be written as a sum of expressions of the form \eqref{l:l1} (with $k$ 
replaced by $k+1$), and that {\em{(i)}}, {\em{(a)}}, {\em{(b)}}, 
and \eqref{l:l3}, \eqref{l:l3a} are satisfied.
This is done in several construction steps:
\begin{itemize}[leftmargin=2em]
\item[\rm{\em{(1)}}] Chiral decomposition: \\
We substitute the induction hypothesis \eqref{l:l1} into \eqref{l:i4}.
This gives a sum of expressions of the form
\begin{eqnarray}
&& C \:i \Pdd_x \int d^4z \; S^{(0)}(x,z) \left\{ (y-z)^I \:
B(z) \:\chi_c \int_z^y [l_1, r_1 \:|\: n_1] \:dz_1 \;
\partial_{z_1}^{I_1} \:\Box^{p_1} \:V^{(1)}_{J_1, c_1}(z_1) \right.
\notag \\
&& \hspace*{1cm}
\cdots \left. \int_{z_{k-1}}^y [l_k, r_k \:|\: n_k] \:dz_k \;
\partial_{z_k}^{I_k}\: \Box^{p_k}\: V^{(k)}_{J_k, c_k}(z_k)
\;\gamma^J \right\} S^{(h)}(z,y) \:.
\label{l:45x}
\end{eqnarray}
We insert the specific form of the potential~$B$, \eqref{Bchiral2}, and expand.
Using the commutation rule $\gamma^i \:\chi_{L\!/\!R} =
\chi_{R\!/\!L} \:\gamma^i$, we bring all chiral projectors to the 
very left, where they can be combined with the formula $\chi_c 
\chi_d=\delta_{cd}\:\chi_c$ to a single chiral projector.
Next, we bring the $\gamma$-matrices of $B$ to the right and write 
them together with the factor $\gamma^J$ in
\eqref{l:45x} (notice that the Dirac matrices commute with the potentials
$V^{(a)}_{c_a}$, which act non-trivially only on the Dirac sea index).
Denoting the individual potentials of the factor~$B$ in~\eqref{l:45x}
by~$V^{(0)}_{J_0, c_0}$, we thus get for~\eqref{l:45x} a sum of expressions of the form
\begin{eqnarray}
&& \chi_c\:C \:i \Pdd_x \int d^4z \; S^{(0)}(x,z) \left\{ (y-z)^I
\: V^{(0)}_{J_0, c_0}(z) \int_z^y [l_1, r_1 \:|\: n_1] \:dz_1 \;
\partial_{z_1}^{I_1} \:\Box^{p_1} \:V^{(1)}_{J_1, c_1}(z_1) \right.
\notag \\
&& \hspace*{1cm}
\cdots \left. \int_{z_{k-1}}^y [l_k, r_k \:|\: n_k] \:dz_k \;
\partial_{z_k}^{I_k}\: \Box^{p_k}\: V^{(k)}_{J_k, c_k}(z_k)
\;\gamma^J \right\} S^{(h)}(z,y) \:.
\label{l:35u}
\end{eqnarray}
The chiral decomposition in~\eqref{Bchiral2} and~\eqref{YLRdef} 
imply that the chiralities in~\eqref{l:35u} satisfy the rule~{\em{(i)}}
(after relabeling the indices in an obvious way).
The chirality of the potentials will not be 
affected in all the following construction steps; to simplify the
notation, we will omit the indices $c_a$ from now on.
\item[\rm{\em{(2)}}] Light-cone expansion: \\
Since $y$ can be considered as a fixed parameter, we can in \eqref{l:35u}
apply Lemma~\ref{l:lemma1} with $V$ given by the expression in the curly brackets,
\begin{align}
\eqref{l:35u} =\;& \chi_c\:C \:i \Pdd_x \sum_{n=0}^\infty
\frac{1}{n!} \int_x^y [0, h \:|\: n] \:dz \notag \\
&\times\; \Box^n_z \left( (y-z)^I \:
V^{(0)}_{J_0}(z) \int_z^y [l_1, r_1 \:|\: n_1] \:dz_1 \;
\partial_{z_1}^{I_1} \:\Box^{p_1}\: V^{(1)}_{J_1}(z_1)
\right. \notag \\
& \hspace*{.7cm} \left. \cdots
\int_{z_{k-1}}^y [l_k, r_k \:|\: n_k] \:dz_k \;
\partial_{z_k}^{I_k} \:\Box^{p_k} \:V^{(k)}_{J_k}(z_k) \right)
\:\gamma^J \; S^{(n+h+1)}(x,y) \: . \label{l:33n}
\end{align}
\item[\rm{\em{(3)}}] Computation of the Laplacian $\Box^n_z$: \\
We carry out the $z$-derivatives in \eqref{l:33n} inductively with the
Leibniz rule. Each derivative can act either on the factors $(y-z)^I$
or on the functions $V^{(a)}$. In the first case, one of the 
factors $(y-z)$ disappears. Thus we get a sum of expressions of the form
\begin{align}
\chi_c\:C \:i \Pdd_x &\int_x^y [0, h \:|\: n] \:dz \; (y-z)^{\hat{I}}
\:\partial_{z}^{I_0} \:\Box^{p_0}_z \:V^{(0)}_{J_0}(z)
\int_{z}^y [l_1, r_1 \:|\: n_1] \:dz_1 \;
\partial_{z_1}^{I_1} \:\Box^{p_1} \:V^{(1)}_{J_1}(z_1) \notag \\
& \cdots
\int_{z_{k-1}}^y [l_k, r_k \:|\: n_k] \:dz_k \;
\partial_{z_k}^{I_k} \:\Box^{p_k} \:V^{(k)}_{J_k}(z_k)
\:\gamma^J \; S^{(n+h+1)}(x,y) \label{l:34j}
\end{align}
with $|\hat{I}| \leq |I|$ and
\beq \label{l:47x}
2n=|I|-|\hat{I}|+|I_0|+2p_0 \:.
\eeq
We can assume that no tensor indices of $\partial_z^{I_0}$ are contracted 
with each other (otherwise we rewrite the corresponding partial 
derivatives as additional Laplacians).
Then all the partial derivatives $\partial_z$ in \eqref{l:34j} were 
generated in the case when one derivative of a Laplacian
$\Box_z$ in \eqref{l:33n} hit a factor $(y-z)$ whereas the other derivative
acted on the $V^{(a)}$.
Thus the number of factors $(y-z)$ which disappeared by carrying out 
the Laplacians in \eqref{l:33n} is larger or equal than the number of partial 
derivatives $\partial_z$,
\beq \label{l:45a}
|I| - |\hat{I}| \geq |I_0| \:.
\eeq
\item[\rm{\em{(4)}}] Extraction of the factors $(y-x)$: \\
In \eqref{l:34j}, we iteratively apply the identity
\[ 
\int_x^y [0, r \:|\: n] \:dz \; (y-z) \cdots =
(y-x) \int_x^y [0, r+1 \:|\: n] \:dz \; \cdots \: . \]
This gives $(k+1)$ nested line integrals of the form
\begin{eqnarray}
\eqref{l:34j} &=& \chi_c\:C \:i \Pdd_x (y-x)^{\hat{I}} \; S^{(\hat{h})}(x,y)
\int_x^y [l_0, r_0 \:|\: n_0] \:dz_0 \;
\partial_{z_0}^{I_0} \:\Box^{p_0} \:V^{(0)}_{J_0}(z_0) \notag \\
&& \hspace*{1cm} \cdots
\int_{z_{k-1}}^y [l_k, r_k \:|\: n_k] \:dz_k \;
\partial_{z_k}^{I_k} \:\Box^{p_k} \:V^{(k)}_{J_k}(z_a)
\:\gamma^J \label{l:34g}
\end{eqnarray}
with
\begin{align}
l_0 &= 0 \:,\qquad r_0 = h+|\hat{I}| \:,\qquad n_0=n \label{l:42f} \\
0 &\leq 2 \hat{h} = 2(n + h + 1) \stackrel{\eqref{l:47x}}{=}
2 h + 2 + |I| - |\hat{I}| + |I_0| + 2 p_0 \:.\label{l:46a}
\end{align}
We can arrange that the parameters $l_0$, $r_0$, and $n_0$ are all 
positive: The only parameter which might be negative is $r_0$; in this 
case, $h=-1$, $|\hat{I}|=0$, and thus $r_0=-1$. The induction 
hypothesis \eqref{l:l3} yields that $|I|>0$. Thus $|I|>|\hat{I}|$, and 
relation \eqref{l:47x} gives that $(n_0=)n>0$. Therefore, we can 
apply the identity
\[ [l_0, r_0 \:|\: n_0] = [l_0+1, r_0+1 \:|\: n_0-1] \]
to make all the parameters in this bracket positive.
\item[\rm{\em{(5)}}] Computation of the partial derivative $\Pdd_x$: \\
The $x$-derivative in \eqref{l:34g} can act on the factors
$S^{(\hat{h})}$, $(y-x)^{\hat{I}}$, or $V^{(a)}(z_a)$. The first case can be
computed with the rules \eqref{l:7} or \eqref{l:29z}; it decreases $\hat{h}$ 
by one and gives one additional factor $(y-x)$. In the second case, 
one factor $(y-x)$ disappears, and thus $|\hat{I}|$ is decremented.
The last case can be handled with the  rule
\beq \label{l:34h}
\frac{\partial}{\partial x^k} \int_x^y [l,r \:|\: n] \:dz\; f(z,y) 
= \int_x^y [l,r+1 \:|\: n] \:\frac{\partial}{\partial z^k} f(z,y) \:,
\eeq
which increases $|I_0|$ by one. As is immediately verified in each of these
cases, equation \eqref{l:46a} transforms into
\beq
2 \hat{h} = 2h + 1 + |I| - |\hat{I}| + |I_0| + 2 p_0 \:, \label{l:47a}
\eeq
whereas inequality \eqref{l:45a} must be weakened to
\beq
|\hat{I}| \leq 1 + |I| - |I_0| \:. \label{l:47b}
\eeq
Finally, we combine the $\gamma$-matrix of the factor $\Pdd_x$ with $\gamma^J$.
\end{itemize}
After these transformations, the $(k+1)^{\mbox{\scriptsize{st}}}$ order 
Feynman diagram consists of a sum of terms of the form
\begin{align}
\chi_c\:C \:(y-x)^{\hat{I}} & \int_x^y [l_0, r_0 \:|\: n_0]
\:dz_0 \; \partial_{z_0}^{I_0}\: \Box_{z_0}^{p_0} \:V^{(0)}_{J_0}(z_0) \notag \\
& \cdots \int_{z_{k-1}}^y \!\!\!\![l_k, r_k \:|\: n_k] \:dz_k
\; \partial_{z_k}^{I_k}\: \Box_{z_k}^{p_k}\: V^{(k)}_{J_k}(z_k) \;
\gamma^J \;S^{(\hat{h})}(x,y) \:.
\label{l:54a}
\end{align}
Notice that the parameters $I_a, p_a$, $a=1,\ldots, k$, were not 
changed by the above construction steps; they are still the same as 
in the induction hypothesis \eqref{l:l1}.
After renaming the indices and the integration variables, \eqref{l:54a} 
is of the required form \eqref{l:l1}. The conditions {\em{(a)}} and {\em{(b)}}
for the contractions of the tensor indices, however, will in general be
violated. Therefore we need two further computation steps:
\begin{itemize}[leftmargin=2em]
\item[\rm{\em{(6)}}] Simplification of the Dirac matrices: \\
If any two of the tensor indices of the factor $\gamma^J$ 
are contracted with each other, we reorder the $\gamma$-matrices with 
the anti-commutation relations
\beq \label{l:ar1}
\{ \gamma^i,\:\gamma^j \} = 2 \:g^{ij}\:\1
\eeq
until the corresponding matrices are next to each other. Applying the
identity $\gamma^i \gamma_i = 4\:\1$, both Dirac matrices disappear.
We iterate this procedure until no tensor indices of $\gamma^J$ are 
contracted with each other (notice that the iteration comes to an end
because the number of $\gamma$-factors is decreased by two in each 
step). Again using the anti-commutation rule \eqref{l:ar1}, we reorder the
Dirac matrices until they are in the same order in which the factors 
to which their tensor indices are contracted appear in the product
\eqref{l:54a}. If any two of the $\gamma$-matrices are 
contracted with the same multi-index, these $\gamma$-matrices are
next to each other, and we can use the symmetry in the tensor
indices to eliminate them both, more precisely
\begin{align}
(y-x)_i \:(y-x)_j \cdots \gamma^i \gamma^j &= (y-x)^2 \cdots \1 \\
\partial_{ij} V^{(a)} \cdots \gamma^i \gamma^j &= \Box V^{(a)} \cdots \1 \:.
        \label{l:58e}
\end{align}
After all these transformations, condition {\em{(b)}} is satisfied.

Notice that the parameters $|I_a|$ and $p_a$ are in general changed in 
this construction step. More precisely, each transformation 
\eqref{l:58e} modifies the parameters according to
\beq \label{l:25t}
|I_a| \rightarrow |I_a| - 2 \qquad \text{and} \qquad
p_a \rightarrow p_a + 1 \:.
\eeq
\item[\rm{\em{(7)}}] Handling of the new contractions: \\
If any two tensor indices of a factor $\partial_{z_a}^{I_a}$ are contracted
with each other, we rewrite the corresponding partial derivatives as a 
Laplacian; this changes the parameters $|I_a|$ and $p_a$ according to
\eqref{l:25t}.
If two tensor indices of the factor $(y-x)^{\hat{I}}$ are contracted with
each other, this gives a factor $(y-x)^2$. Using the identity
\eqref{l:22a}, we inductively absorb the factors $(y-x)^2$
into $S^{(\hat{h})}(x,y)$, which transforms $\hat{h}$ and $|\hat{I}|$ as
\beq \label{l:25u}
\hat{h} \:\rightarrow\: \hat{h}+1 \qquad \text{and} \qquad
|\hat{I}| \:\rightarrow\: |\hat{I}| - 2 \:.
\eeq
After these transformations, condition {\em{(a)}} is also satisfied.
\end{itemize}
After all these construction steps, the 
$(k+1)^{\mbox{\scriptsize{st}}}$ order Feynman diagram is a sum of 
terms of the form \eqref{l:54a} satisfying the conditions {\em{(a)}} and 
{\em{(b)}}. It remains to show that the relations \eqref{l:l3} and \eqref{l:l3a}
remain valid in our inductive construction: As mentioned earlier, the 
parameters $I_a$, $p_a$, $a=1,\ldots,k$ are not changed in the 
construction steps {\em{(1)}} to {\em{(5)}}. In the steps {\em{(6)}} and 
{\em{(7)}}, the transformations \eqref{l:25t} and \eqref{l:25u} preserve 
both the induction hypothesis \eqref{l:l3},\eqref{l:l3a} and the relations 
\eqref{l:47a},\eqref{l:47b}, as is immediately verified. By substituting \eqref{l:47a}
and \eqref{l:47b} into \eqref{l:25t},\eqref{l:25u}, we obtain
\[ 2 \hat{h} = (k+1) - 1 - |\hat{I}| + \sum_{a=0}^k |I_a| + 2 p_a 
\:,\qquad |\hat{I}| \leq (k+1) + 1 - \sum_{a=0}^k |I_a| \:. \]
This concludes the proof. \QED

\subsectionn{Reduction to the Phase-Free Contribution} \label{secresum}
The shortcoming of the constructions of the previous section is that the resulting formulas become
more and more involved to higher order in perturbation theory. Moreover, to any order on the light cone,
one gets an infinite number of contributions. In order to clarify the structure of
the singularities on the light-cone, it is therefore essential to collect and rearrange the different
contribution to the light-cone expansion. This procedure is called {\em{resummation}}
of the light-cone expansion.
\sindex{light-cone expansion!resummation of}%
\sindex{resummation!of light-cone expansion}%
After the resummation, the light-cone expansion of~$\tilde{s}(x,y)$ will,
to every order on the light cone, consist of only a finite number of terms.
Before beginning, we remark that the resummation technique can also be understood
from underlying gauge symmetries. In order no to mix mathematical constructions with
physical considerations, we postpone the explanation of gauge phases and gauge transformations
to~\S\ref{s:sec72} (however, the idea of working with local transformations will be used in
our constructions; see~\eqref{spinunit} and the computations thereafter).

In order to give a first idea of how the resummation works,
we consider the leading singularity on the light cone
by neglecting all terms of the order~$\O((y-x)^{-2})$.
According to~\eqref{l:24b}, we need to take into account only
the contributions~\eqref{l:l1} with~$h=-1$. The inequality~\eqref{l:45o} implies that no
derivatives of the potentials appear. Moreover, we obtain from~\eqref{l:l3}
that $|I|=k+1$. Using the rules~{\em{(a)}} and~{\em{(b)}},
we conclude that one tensor index of the multi-index~$I$ is
contracted with a Dirac matrix, whereas all the remaining $k$ indices of~$I$
are contracted with chiral potentials.
Therefore, all~$k$ potentials are chiral, and
no dynamical mass matrices appear.
A detailed calculation yields for the~$k^{\mbox{\scriptsize{th}}}$ order
Feynman diagram a term of precisely this structure,
\begin{align*}
\chi_c &\:\big((-s B)^k s\big)(x,y) = \chi_c \:(-i)^k \int_x^y dz_1 \;
(y-x)_{j_1} \:A^{j_1}_c(z_1) \\
& \times \int_{z_1}^y dz_2 \:(y-z_1)_{j_2}\: A^{j_2}_c(z_2)
\cdots \int_{z_{k-1}}^y dz_k \:(y-z_k)_{j_k}\: A^{j_k}_c(z_k) \:
s(x,y) + \O\big( (y-x)^{-2} \big)\:.
\end{align*}
The obtained nested line integrals can be identified with the
summands of the familiar Dyson series.
This allows us to carry out the sum over all Feynman diagrams,
\beq \label{sho}
\chi_c \:\tilde{s}(x,y) = \chi_c \:\Pexp \left( -i \int_x^y
(y-x)_j \:A^j_c(z)\:dz \right) s(x,y) \:+\:
\O((y-x)^{-2}) \:,
\eeq
where~$\Pexp$ is defined as follows.
\begin{Def} \label{2.5.4}
For a smooth one-parameter family of matrices $F(\alpha)$, $\alpha
\in \R$, the {\bf{ordered exponential}}~$\Pexp (\int F(\alpha) \:d\alpha)$
is given by the Dyson series
\begin{align*}
\Pexp \bigg( \int_a^b F(\alpha) \:d\alpha \bigg)
&= \1 + \int_a^b F(t_0) \:dt_0 \:+\: \int_a^b dt_0\:F(t_0) \int_{t_0}^b dt_1 \: F(t_1) \notag \\
&\qquad \!+ \int_a^b dt_0\:F(t_0) \int_{t_0}^b dt_1 \: F(t_1) \int_{t_1}^b dt_2 \:F(t_2) + \cdots \:. 
\end{align*}
For ordered exponentials over the chiral potentials, we use the short notations
\begin{align*}
&\Pexp \bigg( -i \int_x^y (y-x)_j \:A^j_c(z)\:dz \bigg) =
\Pexp \bigg(-i \int_x^y A^j_c \:(y-x)_j \bigg) = \Pe^{-i \int_x^y A^j_c \:(y-x)_j} \\
&\quad:= \Pexp \bigg( -i \int_0^1 A^j_c \big|_{\alpha y + (1-\alpha) x} \: (y-x)_j \: d\alpha \bigg) .
\end{align*}        
\end{Def} \noindent
Sometimes, we shall find it more convenient to write~$\Pexp(\cdots)$ as~$\Pe^{(\cdots)}$.
\nindex{bi0@$\Pexp, \Pe$ -- ordered exponential}%
\sindex{ordered exponential}%
For elementary properties of the ordered exponentials we refer
to Exercise~\ref{ex2.4-3}. For the general background on the ordered exponential
we refer to~\cite[X.12]{reed+simon2} or to the closely related time-ordered or path-ordered
exponential in the physics literature (see for example~\cite[Section~4.2]{peskin+schroeder}).
The connection to local gauge transformations is explained in Exercise~\ref{ex2.7}.

To lower order on the light cone, the situation clearly is more complicated.
The idea is to rearrange the contributions of the light-cone expansion in a such a way
that certain subseries can be summed up to again obtain ordered exponentials of the chiral
potentials. This idea is made precise in the following proposition and theorem,
which we state and explain before giving their proofs.

Note that the partial derivatives in~\eqref{l:l1} may be contracted with the
factors~$y-x$. If this is the case, the corresponding combination
\beq \label{tangential}
(y-x)^j\: \frac{\partial}{\partial z_k^j}
\eeq
is a derivative in the direction of the vector~$y-x$.
Since the direction~$y-x$ is tangential to the corresponding line integral,
such so-called {\em{tangential derivatives}} can be rewritten as
\sindex{tangential derivative}%
derivatives with respect to the corresponding integration variable
(for details see Exercise~\ref{ex2.6} or the proof of Proposition~\ref{l:thm2} below). Integrating by parts,
the tangential derivatives disappear. Proceeding in this way, one can
in fact eliminate all tangential derivatives, as is made precise in the
following Proposition.

\begin{Prp} {\bf{(elimination of tangential derivatives)}} \label{l:thm2} \\
Every contribution~\eqref{l:l1} to the light cone expansion of Theorem~\ref{l:thm1} can be written
as a finite sum of expressions of the form
\begin{align}
\chi_c\:C \:(y-x)^K \:W^{(0)}(x) &\int_x^y [l_1, r_1 \:|\: n_1] \:dz_1 \;
W^{(1)}(z_1) \int_{z_1}^y [l_2, r_2 \:|\: n_2] \:dz_2 \; W^{(2)}(z_2) \notag \\
& \cdots \int_{z_{\alpha-1}}^y [l_\alpha, r_\alpha \:|\: 
n_\alpha] \:dz_\alpha \; W^{(\alpha)}(z_\alpha) \;
\gamma^J \; S^{(h)}(x,y) \label{l:70}
\end{align}
with~$\alpha \leq k$, where the factors $W^{(\beta)}$ are composed of the potentials and 
their partial derivatives,
\beq \label{l:71}
W^{(\beta)} = (\partial^{K_{a_\beta}} \Box^{p_{a_\beta}} 
V^{(a_\beta)}_{J_{a_\beta}, c_{a_\beta}}) \cdots (\partial^{K_{b_\beta}} \Box^{p_{b_\beta}} 
V^{(b_\beta)}_{J_{b_\beta}, c_{b_\beta}})
\eeq
with $a_1=1$, $a_{\beta+1}=b_\beta+1$, $b_\beta \geq a_\beta-1$ (in 
the case $b_\beta=a_\beta-1$, $W^{(\beta)}$ is identically one),
and $b_\alpha=k$.
The parameters $l_a$, $r_a$, and $n_a$ are non-negative integers,
$C$ is a complex number, and $c=L\!/\!R$, $c_a=L\!/\!R$ are chiral 
indices. The potentials~$V^{(a)}$ are again given by~\eqref{l:42e};
their chirality is determined by the rule~(i) in Theorem \ref{l:thm1}.
The tensor indices of the multi-indices $J$, $K$, $J_a$, and 
$K_a$ are all contracted with each other, according to the rules (a),(b) 
of Theorem \ref{l:thm1} and
\begin{itemize}[leftmargin=2em]
\item[(c)] The tensor indices of $(y-x)^K$ are all 
contracted with the tensor indices of the factors $V^{(a)}_{J_a}$ or $\gamma^J$ 
(but not with the factors $\partial^{K_a}$).
\end{itemize}
 We have the relation
\beq \label{l:71b}
2h = k - 1 - |K| + \sum_{a=1}^k \big( |K_a| + 2 p_a \big) \:.
\eeq
\end{Prp}
Before coming to the proof, we make precise how this proposition can be used to
simplify the light-cone expansion.
\begin{Def} \label{l:def_pf}
A contribution of the form~\eqref{l:l1} to the light-cone expansion of Theorem~\ref{l:thm2}
is called {\bf{phase-free}} if all the tangential potentials
\sindex{light-cone expansion!phase-free contribution}%
$V^{(a)}_{J_a}$ are differentiated, i.e.
\[ |K_a|+2 p_a > 0 \qquad {\mbox{whenever}}\qquad {\mbox{$J_a$ is contracted with
$(y-x)^K$.}} \]
From every phase-free contribution the corresponding
{\bf{phase-inserted}} contribution is obtained as follows:
\sindex{light-cone expansion!phase-inserted contribution}%
We insert ordered exponentials according to the replacement rule
\beq \label{l:p30}
W^{(\beta)}(z_\beta) \;\longrightarrow\;
W^{(\beta)}(z_\beta) \:\Pexp \left(-i \int_{z_\beta}^{z_{\beta+1}} A^{j_\beta}_{c_\beta}
\: (z_{\beta+1} - z_\beta)_{j_\beta} \right),\quad \beta=0,\ldots,\alpha\:,
\eeq
where we set~$z_0=x$ and~$z_{\alpha+1}=y$.
The chiralities~$c_\beta$
are determined by the relations~$c_0 = c$ and
\beq \label{crule}
\begin{split}
{\text{$c_{\beta-1}$ and $c_\beta$}} & \left\{ \!\!\!\begin{array}{c}
{\text{coincide}} \\ {\text{are opposite}} \end{array} \!\!\right\} \\
&{\text{if $W^{(\beta-1)}$ contains an}} \left\{ \!\!\!\begin{array}{c}
{\text{even}} \\ {\text{odd}} \end{array} \!\!\right\}
{\text{number of factors $Y_.$.}}
\end{split}
\eeq
\end{Def}

\begin{Thm}
\label{l:thm3}
To every order on the light cone, the number of phase-free contributions
is finite.
The light-cone expansion of the Green's function $\tilde{s}(x,y)$
is given by the sum of the corresponding phase-inserted contributions.
\end{Thm} 

This theorem gives a convenient procedure for performing the light-cone expansion of the Green's function.
The only task is to compute to any order on the light cone
the finite number of phase-free contributions. Then one inserts
ordered exponentials according to Definition~\ref{l:def_pf}.
Note that this method is constructive in the sense that it gives a procedure with which the light-cone
expansion of every Feynman diagram can be carried out explicitly.
\sindex{computer algebra}%
Indeed, this procedure is implemented in the C++-program {\textsf{class\_commute}}\footnote{The
C++ program {\textsf{class\_commute}} and its computational output as well as the resulting
Mathematica worksheets were included as ancillary files to the arXiv submission arXiv:1211.3351 [math-ph].}.
These computations are illustrated in Exercise~\ref{ex2.71}.

The remainder of this section is devoted to the proof of Proposition~\ref{l:thm2} and Theorem~\ref{l:thm3}.
We begin with a preparatory lemma which controls the number of tangential derivatives
in the contributions~\eqref{l:l1} in Theorem~\ref{l:thm1}.
\begin{Lemma} \label{l:lemma3}
For any~$a \in \{1,\ldots, k\}$, we let~$t_a$ be the number of tensor indices of the
multi-index~$I_a$ in~\eqref{l:l1} which are contracted with the factor~$(y-x)^I$.
Then the following inequalities hold for all~$a=1,\ldots, k$:
\beq \label{l:73f}
l_a + n_a \geq t_a-1 \qquad {\mbox{and}} \qquad r_a + n_a \geq \sum_{b=a}^k t_b \:.
\eeq
\end{Lemma}
\Proof As in the proof of Theorem~\ref{l:thm1}, we proceed inductively in the
order $k$ of the perturbation theory.
For $k=0$, the inequalities~\eqref{l:73f} are trivially satisfied according
to~\eqref{l:233a}.
Assume that~\eqref{l:73f} is true for a given $k$. We go through the construction 
steps {\em{(1)}} to {\em{(7)}} of Theorem \ref{l:thm1} and check that the
inequalities~\eqref{l:73f} then also hold in~\eqref{l:54a} for $a=0,\ldots, k$.

We first consider the case $a>0$. The parameters $l_a$, $r_a$, and $n_a$ remain
unchanged in all the construction steps of Theorem \ref{l:thm1}.
Furthermore, it is obvious that the 
parameters $t_a$ are not affected in the steps~{\em{(1)}}, {\em{(2)}}, 
{\em{(4)}} and {\em{(7)}}. In the steps~{\em{(3)}} and~{\em{(5)}}, the computation 
of the derivatives $\Box^n_z$ and $\Pdd_x$ might annihilate some of 
the factors $(y-x)$ which were contracted with the factors 
$\partial_{z_a}^{I_a}$; this may decrease the parameters $t_a$. For 
the analysis of step~{\em{(6)}}, note that all $\gamma$-matrices which
are contracted with factors $(y-x)$ stand to the left of those 
$\gamma$-matrices which are contracted with the 
$\partial_{z_a}^{I_a}$, $a=1,\ldots, k$ (this follows from the ordering 
condition~{\em{(b)}} in the induction hypothesis and the fact that
additional factors $(y-x)^j \cdots\gamma_j$ are only generated during the 
construction if the partial derivative $\Pdd_x$ 
hits $S^{(\hat{h})}$ in step~{\em{(5)}}; in this case, the corresponding 
$\gamma$-matrix stands at the very left in $\gamma^J$). Therefore the 
commutations of the Dirac matrices do not lead to additional contractions
between factors $(y-x)$ and $\partial_{z_a}^{I_a}$, which implies that
the parameters $t_a$ remain unchanged in step~{\em{(6)}}.
We conclude that the $l_a$, $r_a$, and $n_a$ remain unchanged whereas the $t_a$
may only decrease, and thus~\eqref{l:73f} holds for $a=1,\ldots, k$
throughout all the construction steps.

It remains to show that the inequalities~\eqref{l:73f} hold in~\eqref{l:54a} for
$a=0$. We first look at the situation after step~{\em{(4)}}
in~\eqref{l:34g}: The values~\eqref{l:42f} for $l_0$, $r_0$, and $n_0$ give in 
combination with~\eqref{l:47x} the equations
\begin{align}
l_0 + n_0 &= \frac{1}{2} \left( |I| - |\hat{I}| + |I_0| + 2 p_0 \right) \label{l:ln1} \\
r_0 + n_0 &= h + \frac{1}{2} \left( |I| + |\hat{I}| + |I_0| + 2 p_0 \right) . \label{l:rn1}
\end{align}
Moreover, the number of tangential derivatives $t_0$ at the first 
potential is clearly bounded by the total number of derivatives there,
\beq \label{l:bn1}
|I_0| \geq t_0 \:.
\eeq
Furthermore, the total number of tangential derivatives is smaller 
than the number of factors $(y-x)$,
\beq \label{l:bn2}
|\hat{I}| \geq \sum_{a=0}^k t_a \:.
\eeq
Substituting~\eqref{l:45a} and~\eqref{l:bn1} into~\eqref{l:ln1} yields the 
inequalities
\beq \label{l:5t}
l_0 + n_0 \geq |I_0| + p_0 \geq t_0 \:.
\eeq
In order to get a bound for $r_0+n_0$, we must distinguish two cases. 
If $h \geq 0$, we substitute~\eqref{l:45a} into~\eqref{l:rn1} and get 
with~\eqref{l:bn2} the inequality
\beq \label{l:5u}
r_0 + n_0 \geq |\hat{I}| + |I_0| + p_0 \geq |\hat{I}| \geq \sum_{a=0}^k t_a \:.
\eeq
In the case $h=-1$, \eqref{l:45o} shows that $|I_a|$, and consequently 
also $t_a$, vanish for $1 \leq a \leq k$. Furthermore, \eqref{l:l3}
yields that $|I| \neq 0$. Thus~\eqref{l:rn1} and~\eqref{l:bn1}, 
\eqref{l:bn2} give the bound
\[ r_0 + n_0 \geq h + \frac{|I|}{2} + \frac{1}{2} 
\:\sum_{a=0}^k t_a + \frac{1}{2} \:t_0
\geq \frac{1}{2} \:\sum_{a=0}^k t_a 
+ \frac{1}{2} \:t_0 \:, \]
where we used in the last inequality that $h+|I|/2 \geq -1/2$ and that all the 
other terms are integers. Since $t_0=\sum_{a=0}^k t_a$, we conclude 
that inequality~\eqref{l:5u} also holds in the case $h=-1$.

We finally consider how the bounds~\eqref{l:5t} and~\eqref{l:5u} for $l_0 + n_0$ and
$r_0 + n_0$ must be modified in the subsequent construction steps.
In step~{\em{(5)}}, the partial derivative $\Pdd_x$ may annihilate a factor 
$(y-x)$, in which case the parameters $t_a$ might decrease. On the 
other hand, the partial derivatives $\Pdd_x$ may produce an additional 
factor $\partial_{z_0}$; in this case, $r_0$ is incremented according 
to~\eqref{l:34h}. In step~{\em{(6)}}, only this additional factor 
$\partial_{z_0}$ may be contracted with $(y-x)^{\hat{I}}$. Step~{\em{(7)}}
does not change $l_0$, $r_0$, $n_0$, and $t_0$. Putting 
these transformations together, we conclude that the inequality~\eqref{l:5t} for $l_0+n_0$ must be weakened by one, whereas the bound~\eqref{l:5u} for $r_0 + n_0$ remains valid as it is.
This gives precisely the inequalities~\eqref{l:73f} for $a=0$.
\QED

\Proof[Proof of Proposition~\ref{l:thm2}]
The basic method for the proof is to iteratively eliminate those partial
derivatives~$\partial_{z_a}^{I_a}$ in~\eqref{l:l1} which are contracted with
a factor $(y-x)$. This is accomplished with the integration-by-parts formula
\begin{align*}
&(y-x)^j \int_x^y [l, r \:|\: n] \:dz \;\partial_j f(z)
\stackrel{\eqref{l:29x}}{=} \int_0^1 d\alpha \; \alpha^l 
\:(1-\alpha)^r \:(\alpha-\alpha^2)^n \:\frac{d}{d\alpha} f(\alpha 
y + (1-\alpha) x) \notag \\
&= \delta_{r+n, 0} \:f(y) \:-\: \delta_{l+n, 0} \:f(x) \notag \\
&\qquad -(l+n) \int_x^y [l-1, r \:|\: n] \:dz \; f(z) \;+\;
(r+n) \int_x^y [l, r-1 \:|\: n] \:dz \; f(z) \:. 
\end{align*}
In order to see the main difficulty, we consider the example of two
nested line integrals with two tangential derivatives
\begin{align}
(y&-x)^j \:(y-x)^k \int_x^y [0,1 \:|\: 0] \:dz_1 \; 
V^{(1)}(z_1) \int_{z_1}^y [0,1 \:|\: 0] \:dz_2 \; \partial_{jk} 
V^{(2)}(z_2) \label{l:80} \\
&= (y-x)^j \int_x^y [0,0 \:|\: 0] \:dz_1 \; 
V^{(1)}(z_1) \: (y-z_1)^k \int_{z_1}^y [0,1 \:|\: 0] \:dz_2 \; \partial_{jk} 
V^{(2)}(z_2) \notag \\
&= -(y-x)^j \int_x^y dz_1 \; V^{(1)}(z_1) \: \partial_j V^{(2)}(z_1)
\label{l:81} \\
&\qquad+(y-x)^j \int_x^y dz_1 \; V^{(1)}(z_1) \int_{z_1}^y dz_2 \; 
\partial_j V^{(2)}(z_2) \:. \label{l:82}
\end{align}
Although the line integrals in~\eqref{l:80} satisfy the conditions 
of Theorem \ref{l:thm1}, the expression cannot be transformed into the 
required form~\eqref{l:70}. Namely, in~\eqref{l:81} we cannot eliminate 
the remaining tangential derivative (because partial integration would
yield a term $(y-x)^j \:\partial_j V^{(1)}(z_1)$).
In~\eqref{l:82}, on the other hand, we can successfully perform a second partial 
integration
\[ \eqref{l:82} = \int_x^y [0,-1 \:|\: 0] \:dz_1 \;V^{(1)}(z_1) 
\;(V^{(2)}(y) - V^{(2)}(z_1)) \:, \]
but then the second parameter in the bracket $[.,.\:|\:.]$ becomes 
negative. More generally, we must ensure that the boundary terms contain no 
tangential derivatives, and that the parameters $l_a, r_a$, and $n_a$ 
stay positive in the construction.

Since the chirality of the potentials is not affected by the partial 
integrations, it is obvious that the rule~{\em{(i)}} in Theorem \ref{l:thm1} will remain valid. For ease
in notation, in the remainder of the proof we usually omit the indices~$c_a$.

First of all, we split up the factor $(y-x)^I$ in~\eqref{l:l1} in the form
$(y-x)^I = (y-x)^K \:(y-x)^L$, where $L$ are those tensor indices which are 
contracted with the partial derivatives $\partial^{I_a}_{z_a}$, $a=1,\ldots, k$.
Setting $b=1$ and $z_0=x$, the first line integral in~\eqref{l:l1} can 
be written as
\beq \label{l:81a}
\cdots \:(y-z_{b-1})^L \int_{z_{b-1}}^y [l_b, r_b \:|\: n_b] \:dz_b 
\; \partial^{I_b}_{z_b} \:\Box^{p_b}_{z_b} \:V^{(b)}_{J_b}(z_b) \:\cdots\:.
\eeq
We rewrite the tangential derivatives in this line integral as derivatives in
the integration variable,
\beq \label{l:82a}
= \cdots \:(y-z_{b-1})^N \int_0^1 d\alpha \: \alpha^l 
\:(1-\alpha)^r \; \left( \frac{d}{d\alpha} \right)^q
\partial^{K_b}_{z_b} \:\Box^{p_b}_{z_b} \:V^{(b)}_{J_b}(z_b) \:\cdots
\eeq
with $|L|=|N|+q$ and $l=l_b+n_b$, $r=r_b+n_b$.
Lemma \ref{l:lemma3} gives the bounds
\beq \label{l:83}
l \geq q-1 \qquad \text{and} \qquad r \geq q + |N| \:.
\eeq
More generally, we use~\eqref{l:82a} and~\eqref{l:83} as our induction 
hypothesis, where the left factor `$\cdots$' stands for all 
previous line integrals (which contain {\em{no}} tangential 
derivatives), and the right factor `$\cdots$' stands for subsequent 
line integrals. The tensor indices of the factor $(y-z_{a-1})^N$ must all be
contracted with the partial derivatives $\partial^{I_a}_{z_a}$ for $a>b$ and
thus give tangential derivatives in the subsequent line integrals.
The induction step is to show that all the $\alpha$-derivatives 
in~\eqref{l:82a} can be eliminated, and that we can write the 
resulting expressions again in the form~\eqref{l:82a} and~\eqref{l:83} with $b$ 
replaced by $b+1$. Under the assumption that this induction step
holds, we can eliminate all tangential derivatives in $k$ steps.
The resulting expressions are very similar to~\eqref{l:70} and~\eqref{l:71}. 
The only difference is that the derivatives $\partial^{K_a}$ and 
$\Box^{p_a}$ in the resulting expressions are differential operators 
acting on all the following factors $V^{(a)}$, $V^{(a+1)}$,\ldots; 
in~\eqref{l:71}, on the other hand, the partial derivatives act only on the 
adjacent potential $V^{(a)}$. In order to bring the resulting 
expressions into the required form, we finally carry out all the 
derivatives with the Leibniz rule and the chain rule~\eqref{l:34h}.

For the proof of the induction step, we integrate in~\eqref{l:82a} $q$ times 
by parts (if $q$ is zero, we can skip the partial integrations; our 
expression is then of the form~\eqref{l:84}). Since the powers of 
the factors $\alpha$ and $(1-\alpha)$ are decreased at most by one in 
each partial integration step,~\eqref{l:83} implies that the boundary values
vanish unless in the last step for $\alpha=0$.
We thus obtain a sum of terms of the form
\beq \label{l:83b}
\cdots (y-z_{b-1})^N \;\partial^{K_b}_{z_b} \:\Box^{p_b}_{z_b} \:
V^{(b)}_{J_b}(z_b) \:\cdots_{|z_b \equiv z_{b-1}}
\eeq
and
\beq
\begin{split} \label{l:84}
\cdots (y-z_{b-1})^N \int_{z_{b-1}}^y [l, r \:|\: n=0] \:dz_b \;
\partial^{K_b}_{z_b} \:\Box^{p_b}_{z_b} & \:V^{(b)}_{J_b}(z_b) \\
& {\text{with $l \geq 0$, $r \geq |N|$}}\:.
\end{split}
\eeq
In~\eqref{l:84}, we iteratively use the relation
\[ 
(y-x)^j \int_x^y [l, r \:|\: n] \:dz \;\cdots = \int_x^y [l, r-1 
        \:|\: n] \:dz \; (y-z)^j \;\cdots \]
to bring all factors $(y-z_{b-1})$ to the right. We thus obtain expressions of the form
\beq
\eqref{l:84} = \cdots \int_{z_{b-1}}^y [l, r \:|\: n=0] \:dz_b \;
(y-z_{b})^N \:\partial^{K_b}_{z_b} \:\Box^{p_b}_{z_b} \:V^{(b)}_{J_b}(z_b) 
\cdots \quad {\mbox{with $l, r \geq 0$}} \:.  \label{l:85}
\eeq

In both cases~\eqref{l:83b} and~\eqref{l:85}, we have an expression of 
the form
\beq \label{l:86}
\cdots (y-z_b)^N \;\partial^{K_b}_{z_b} \:\Box^{p_b}_{z_b}\:
        V^{(b)}_{J_b}(z_b) \cdots \: ,
\eeq
where the first factor `$\cdots$' stands for line integrals 
without tangential derivatives, and where none of the factors $(y-z_b)$ 
are contracted with $\partial^{K_b}_{z_b}$. Applying the ``inverse
Leibniz rules''
\begin{align*}
(y-x)^j \:\frac{\partial}{\partial x^k} &= \frac{\partial}{\partial x^k}\: (y-x)^j + \delta^j_k 
\\
(y-x)_j \:\Box_x &= \Box_x \: (y-x)_j + 2 \frac{\partial}{\partial x^j} \:, 
\end{align*}
we iteratively commute all factors $(y-z_b)$ in~\eqref{l:86} to the right.
This gives a sum of expressions of the form
\beq \label{l:87}
\cdots \partial^{K_b}_{z_b} \:\Box^{p_b}_{z_b}\:
        V^{(b)}_{J_b}(z_b) \; (y-z_b)^L \cdots \:,
\eeq
where the factors $(y-z_b)$ are all contracted with the partial derivatives
$\partial^{I_a}_{z_a}$, $a=b+1,\ldots, k$. The Leibniz rules may have 
annihilated some factors $(y-z_b)$ (i.e., $|L|$ might be smaller 
than $|N|$); in this case, the parameters $t_a$, $a=b+1,\ldots,k$ 
have decreased. As a consequence, the inequalities of Lemma 
\ref{l:lemma3} are still valid for all expressions~\eqref{l:87}.
If we write~\eqref{l:87} in the form~\eqref{l:81a} with $b$ replaced 
by $b+1$, we can thus split up the tangential derivatives in the form~\eqref{l:82a}
and~\eqref{l:83}. This concludes the proof of the induction step.

It remains to derive equation~\eqref{l:71b}: Note that each integration by parts
decreases both the number of factors $(y-z_{a-1})$ and the 
total number of partial derivatives by one. If we carry out the 
remaining derivatives with the Leibniz rule (in the last step of the 
proof), this does not change the total order $\sum_{a=1}^k |K_a|+2 
p_a$ of the derivatives. Therefore, relation~\eqref{l:l3} in Theorem~\ref{l:thm1} transforms into~\eqref{l:71b}.
\QED

We come to the proof of Theorem~\ref{l:thm3}.
A possible method would be to rearrange all the
contributions to the light-cone expansion of Theorem~\ref{l:thm1}
until recovering the Dyson 
series of the ordered exponentials in~\eqref{l:p30}. However, this method
has the disadvantage of being rather involved. It is more elegant
to use a particular form of {\em{local gauge invariance}} of the Green's function
for the proof (for basics see Exercise~\ref{ex2.7}). To this end, for given $x$ and $y$ we will transform the
spinors locally. The transformation will be such that the light-cone expansion of the 
transformed Green's function $\hat{s}(x,y)$ consists precisely of all 
phase-free contributions. Using the transformation law of 
the Green's function, we then show that the light-cone expansion of 
$\tilde{s}(x,y)$ is obtained from that of $\hat{s}(x,y)$ by inserting 
unitary matrices into the line integrals. Finally, we prove that 
these unitary matrices coincide with the ordered exponentials in Definition~\ref{l:def_pf}.

In preparation, we consider the transformation law of the Dirac 
operator and the Green's function under generalized local phase 
transformations of the spinors. We let $U_L(x)$ and $U_R(x)$ be two unitary matrices
acting on the Lie algebra index of the gauge potential.
We transform the wave functions according to
\beq \label{spinunit}
\psi(x) \rightarrow \hat{\psi}(x) = U(x)\:\psi(x) \qquad \text{with} \qquad
        U(x) = \chi_L \:U_L(x) + \chi_R \:U_R(x) \:.
\eeq
Thus~$U_L$ and~$U_R$ transform the left and right handed 
component of the wave functions, respectively.
We point out that transformation~$U$ is {\em{not}} unitary
with respect to the spin scalar product because~$\chi_L^* = \chi_R$
and therefore
\begin{align*}
V &:= U^{-1} = \chi_L \:U_L^{-1} + \chi_R \:U_R^{-1} \qquad
{\mbox{but}} \\
U^* &\:= \gamma^0 \:U^\dagger \:\gamma^0 =
\chi_R \:U_L^{-1} + \chi_L \:U_R^{-1} \: .
\end{align*}
Therefore, in what follows we carefully distinguish between~$U$, 
$U^*$ and their inverses~$V$ and~$V^*$. As an immediate consequence of 
the Dirac equation $(i \Pdd + \B - m) \psi =0$, the transformed
wave functions $\hat{\psi}$ satisfies the equation
\[ V^* (i \Pdd + B) V \:\hat{\psi} = 0 \:. \]
A short computation yields for the transformed Dirac operator
\[ V^* (i \Pdd + B) V = i \Pdd + \hat{B} \]
with
\[ \hat{B} = \chi_L \:(\hat{A}_R \!\!\!\!\!\!\!\slash \;\:- m \:\hat{Y}_L) \:+\:
\chi_R \:(\hat{A}_L \!\!\!\!\!\!\!\slash \;\;- m \:\hat{Y}_R) \:, \]
where~$\hat{A}_{L\!/\!R}$ and $\hat{Y}_{L\!/\!R}$ are the potentials
\begin{align}
\hat{A}_{L\!/\!R}^j &= U_{L\!/\!R} \:A_{L\!/\!R}^j \:U_{L\!/\!R}^{-1} 
\:+\: i U_{L\!/\!R} (\partial^j U_{L\!/\!R}^{-1}) \label{l:p8} \\
\hat{Y}_{L\!/\!R} &= U_{R\!/\!L} \:Y \:U_{L\!/\!R}^{-1} \:. \label{l:p9}
\end{align}
We denote the advanced and retarded Green's functions of the transformed
Dirac operator $i \Pdd + \hat{B}$ by $\hat{s}$. They satisfy the
equation
\beq
\big( i \Pdd_x \:+\: \hat{B}(x) \big)\, \hat{s}(x,y) = \delta^4(x-y) \:. \label{l:p0}
\eeq
Since we can view~$\hat{B}$ as the perturbation of the Dirac 
operator, the Green's function~$\hat{s}$ has, in analogy to~\eqref{l:11b},
the perturbation expansion
\beq \label{l:p10}
\hat{s} = \sum_{n=0}^\infty (-s \hat{B})^n \:s \:.
\eeq
The important point for what follows is that the Green's functions 
$\tilde{s}$ and $\hat{s}$ are related to each other by the local 
transformation
\beq \label{l:p1}
\hat{s}(x,y) = U(x) \:\tilde{s}(x,y) \:U(y)^* \:.
\eeq
This is verified as follows: The right side of~\eqref{l:p1} also satisfies the
defining equation~\eqref{l:p0} of the Green's functions; namely
\begin{align*}
(i &\Pdd_x + \hat{B}(x)) \: U(x) 
\:\tilde{s}(x,y)\:U(y)^* = V(x)^* \:(i \Pdd_x + B(x)) \:V(x) \;U(x)\:\tilde{s}(x,y)\:U(y)^* \\
&= V(x)^* \:(i \Pdd_x + B(x)) \:\tilde{s}(x,y)\:U(y)^* = V(x)^* \:\delta^4(x-y) \:U(y)^* \\
&=  V(x)^* \:U(x)^* \:\delta^4(x-y) = \delta^4(x-y) \:.
\end{align*}
Furthermore, the supports of both sides of~\eqref{l:p1} lie (depending 
on whether we consider the advanced or retarded Green's functions) 
either in the upper or in the lower light cone. A uniqueness argument 
for the solutions of hyperbolic differential equations yields that 
both sides of~\eqref{l:p1} coincide.

We next specify the unitary transformations $U_L$ and $U_R$: We fix the
points $x$ and $y$. For any point $z$ on the line segment $\overline{xy}$,
we chose $U_{L\!/\!R}(z)$ as
\beq
U_{L\!/\!R}(z) = \Pexp \left(-i \int_x^z A^j_{L\!/\!R} \;
(z-x)_j \right) \:. \label{l:p3}
\eeq
Using the differential equation for the ordered exponential
(see Exercise~\ref{ex2.4-3})
\beq \label{l:71x}
(y-x)^k \frac{\partial}{\partial x^k} \Pe^{-i \int_x^y A^j_c \:(y-x)_j}
= i (y-x)_k \:A^k_c(x) \; \Pe^{-i \int_x^y A^j_c \:(y-x)_j} \:,
\eeq
we obtain
\begin{align*}
(y-x)^j \;U_c(z) \:(\partial_j U_c(z)^{-1}) &= \Pe^{-i \int_x^z 
A^k_c \: (z-x)_k} \:(y-x)^j \frac{\partial}{\partial z^j}
\Pe^{-i \int_z^x A^k_c \: (x-z)_k} \\
&= \Pe^{-i \int_x^z A^k_c \: (z-x)_k} \:i (y-x)_j\:A^j_c(z) \:
\Pe^{-i \int_z^x A^k_c \: (x-z)_k} \\
&= i (y-x)_j \:U_c(z) \:A^j_c(z) \: U_c(z)^{-1} \:.
\end{align*}
Using this formula in~\eqref{l:p8} gives
\beq \label{l:p11}
\hat{A}^j_{L\!/\!R}(z) \:(y-x)_j = 0 \qquad
{\mbox{for $z \in \overline{xy}$}}\:.
\eeq
Thus our choice of $U_L$ and $U_R$ makes the potentials $\hat{A}_L(z)$ 
and $\hat{A}_R(z)$ for~$z \in \overline{xy}$ orthogonal to the vector~$(y-x)$. 
Before going on, we point out that we did not specify $U_{L\!/\!R}(z)$ away from the
line segment $z \in \overline{xy}$; the unitary transformation $U_{L\!/\!R}$ 
may be arbitrary there. This also implies that also $\hat{A}_{L\!/\!R}$ is 
undetermined outside the line segment $\overline{xy}$. In particular, all the 
non-tangential derivatives of $\hat{A}_{L\!/\!R}(z)$ for~$z \in \overline{xy}$
are undetermined. However, \eqref{l:p3} does give
constraints for the tangential derivatives. For example, 
differentiating~\eqref{l:p11} in the direction $(y-x)$ yields
\[ 
(y-x)^j \:(y-x)_k \:\partial_j \hat{A}^k_{L\!/\!R}(z) =0
\qquad \text{for} \qquad z \in \overline{xy} \:. \]

We now consider the perturbation expansion~\eqref{l:p10}. The light-cone 
expansion of all Feynman diagrams according to Theorem~\ref{l:thm1} 
gives a sum of terms of the form
\begin{align}
\chi_c\:&C \:(y-x)^K \:\hat{W}^{(0)}(x) \int_x^y [l_1, r_1 \:|\: n_1]
\:dz_1 \; \hat{W}^{(1)}(z_1) \int_{z_1}^y [l_2, r_2 \:|\: n_2] \:dz_2 \;
\hat{W}^{(2)}(z_2) \notag \\
&\cdots \int_{z_{\alpha-1}}^y [l_\alpha, r_\alpha \:|\: 
n_\alpha] \:dz_\alpha \; \hat{W}^{(\alpha)}(z_\alpha) \;
\gamma^J \; S^{(h)}(x,y) \:, \label{l:t1}
\end{align}
where the factors $\hat{W}^{(\beta)}$ are of the form
\beq \label{l:t2}
\hat{W}^{(\beta)} = (\partial^{K_{a_\beta}} \Box^{p_{a_\beta}} 
\hat{V}^{(a_\beta)}_{J_{a_\beta}, c_{a_\beta}})
\cdots (\partial^{K_{b_\beta}} \Box^{p_{b_\beta}} 
\hat{V}^{(b_\beta)}_{J_{b_\beta}, c_{b_\beta}}) \:.
\eeq
Because of~\eqref{l:p11}, all the contributions which are not 
phase-free vanish. Furthermore, according to Theorem~\ref{l:thm1}, the 
contributions~\eqref{l:t1} and~\eqref{l:t2} contain no tangential derivatives. 
Clearly, the derivatives in these formulas may have a component in
direction of $(y-x)$. But the contribution of the derivatives transversal 
to $(y-x)$ uniquely determines the form of each derivative term. 
Therefore, all the phase-free contributions of the form~\eqref{l:t1} and~\eqref{l:t2}
are independent in the sense that we have no 
algebraic relations between them. We conclude that, as long as the 
potentials $\hat{A}_{L\!/\!R}$ and $\hat{Y}_{L\!/\!R}$ are only specified 
by~\eqref{l:p8}, \eqref{l:p9}) and~\eqref{l:p3}, the light-cone expansion~\eqref{l:t1} and~\eqref{l:t2}
consists precisely of all phase-free contributions.

Next, we exploit the local transformation law~\eqref{l:p1} of the Green's 
functions: We solve this equation for $\tilde{s}$,
\beq \label{l:t0}
\tilde{s}(x,y) = V(x) \:\hat{s}(x,y) \:V(y)^* \:.
\eeq
The transformation~$U_{L\!/\!R}$ does not enter on the 
left side of this equation. Thus the right side of~\eqref{l:t0}
is also independent of $U_{L\!/\!R}$. In particular, we conclude that the 
light-cone expansion of $\hat{s}(x,y)$ must be independent of the 
derivatives of $U_{L\!/\!R}$ along the line segment $\overline{xy}$.
At first sight, this might seem inconsistent because the individual 
contributions~\eqref{l:t1} and~\eqref{l:t2} do depend on the derivatives of 
$U_{L\!/\!R}$ (this is obvious if one substitutes~\eqref{l:p8} and~\eqref{l:p9}
into~\eqref{l:t2} and carries out the derivatives with the 
Leibniz rule). The right way to understand the independence of 
$\hat{s}(x,y)$ on the derivatives of $U_{L\!/\!R}$ is that
all derivative terms of $U_{L\!/\!R}$ cancel each other to every 
order on the light cone if the (finite) sum over all contributions~\eqref{l:t1}
to the light-cone expansion of $\hat{s}(x,y)$ is carried out.
Since we will form the sum over all contributions to the light-cone 
expansion in the end, it suffices to consider only those contributions to the
light-cone expansion which contain no derivatives of $U_{L\!/\!R}$.
This means that we can substitute~\eqref{l:p8} and~\eqref{l:p9} into~\eqref{l:t2},
forget about the derivative term $i 
U_{L\!/\!R} (\partial^j U_{L\!/\!R}^{-1})$ in~\eqref{l:p8}, and pull the unitary
transformations $U_{L\!/\!R}, U_{L\!/\!R}^{-1}$
out of the derivatives. In other words, we can replace $\hat{W}^{(\beta)}$, \eqref{l:t2}, by
\beq
\hat{W}^{(\beta)} = U_{d_{a_\beta}}
(\partial^{K_{a_\beta}} \Box^{p_{a_\beta}} 
V^{(a_\beta)}_{J_{a_\beta}, c_{a_\beta}}) U_{c_{a_\beta}}^{-1} \:
\cdots U_{d_{b_\beta}} (\partial^{K_{b_\beta}} \Box^{p_{b_\beta}} 
V^{(b_\beta)}_{J_{b_\beta}, c_{b_\beta}}) U_{c_{b_\beta}}^{-1} \label{l:t3}
\eeq
with chiral indices $c_a, d_a = L\!/\!R$.
The light-cone expansion of $\hat{s}(x,y)$ consists precisely of the sum of 
all phase-free contributions of the form~\eqref{l:t1} and~\eqref{l:t3}.

The chiralities $c_a$, $d_a$ of the unitary transformations
$U_{L\!/\!R}$, $U_{L\!/\!R}^{-1}$ in~\eqref{l:t3} are determined by the rule~{\em{(i)}}
in Theorem~\ref{l:thm1} and by~\eqref{l:p8} and~\eqref{l:p9}.
According to this rule, the indices $c_{a-1}$ and $c_a$ 
coincide iff $V^{(a)}$ is a chiral potential. According to~\eqref{l:p8}
and~\eqref{l:p9}, on the other hand, the indices $d_a$ and $c_a$ 
coincide iff $V^{(a)}=A_{L\!/\!R}$. We conclude that the indices $c_{a-1}$ and 
$d_a$ always coincide. Thus all the intermediate factors
$U_{c_{a-1}}^{-1} U_{d_a}$ give the identity, and~\eqref{l:t3} simplifies to
\beq \label{l:t4}
\hat{W}^{(\beta)} = U_{d_\beta} \:W^{(\beta)} 
\:U_{c_\beta}^{-1} \:.
\eeq
Furthermore, the chiralities~$d_\beta$ and~$c_\beta$ coincide if and only if 
$W^{(\beta)}$ contains an even number of dynamic mass matrices.

Finally, we substitute the light-cone expansion~\eqref{l:t1} for $\hat{s}(x,y)$ as well as~\eqref{l:t4}
into~\eqref{l:t0}. This gives for the light-cone 
expansion of $\tilde{s}(x,y)$ a sum of expressions of the form
\begin{align}
\chi_c\:C & \:(y-x)^K \:U_c^{-1}(x)
\:(U_{d_0} W^{(0)} U_{c_0}^{-1})(x) \int_x^y [l_1, r_1 \:|\: n_1] \:dz_1 \;
(U_{d_1} W^{(1)} U_{c_1}^{-1})(z_1) \notag \\
&\cdots \int_{z_{\alpha-1}}^y [l_\alpha, r_\alpha \:|\: 
n_\alpha] \:dz_\alpha \; (U_{d_\alpha} W^{(0)} U_{c_\alpha}^{-1})(z_\alpha)  \;U_{c_{\alpha+1}}(y) \;
\gamma^J \; S^{(h)}(x,y) \:,
\label{l:t5}
\end{align}
where the sum runs over all phase-free contributions of this type. 
Similar to the considerations before~\eqref{l:t4}, one sees that
adjacent unitary transformations always have the same 
chirality. Therefore, renaming the chiral indices, the expressions~\eqref{l:t5} can be written
in the simpler form
\begin{align*}
\chi_c\:C &\:(y-x)^K \:W^{(0)}(x) \int_x^y [l_1, r_1 \:|\: n_1]
\:dz_1 \; U_{c_1}(x)^{-1} \: U_{c_1}(z_1) \: W^{(1)} \\
&\cdots \int_{z_{\alpha-1}}^y [l_\alpha, r_\alpha \:|\: 
n_\alpha] \:dz_\alpha \; U_{c_\alpha}(z_{\alpha-1})^{-1} \: 
U_{c_\alpha}(z_\alpha) \:W^{(0)}(z_\alpha) \;U_{c_{\alpha+1}}(z_\alpha)^{-1} \\
& \times \: U_{c_{\alpha+1}}(y) \; \gamma^J \; S^{(h)}(x,y) \:,
\end{align*}
where the chiral indices $c_a$ satisfy the rule~\eqref{crule}. 
According to~\eqref{l:p3}, the factors $U_c^{-1}(.) \:U_c(.)$ coincide 
with the ordered exponentials in~\eqref{l:p30}.
This concludes the proof of Theorem~\ref{l:thm3}.

\subsectionn{The Residual Argument} \label{l:sec_31}
In the previous sections, the light-cone expansion was performed for the
causal Green's functions. We now want to extend our methods and results to
the fermionic projector.
We begin by describing how the light-cone expansion of the Green's 
functions can be understood in momentum space. Apart from giving a 
different point of view, this will make it possible to get a connection
to the light-cone expansion of the fermionic projector. For notational
simplicity, we restrict attention to the case~$g=1$ where in~\eqref{Pdirsum}
there is only one direct summand (the generalization to several direct summands
is obtained in a straightforward way by replacing all vacuum operators
as in~\eqref{Pdirsum} by corresponding direct sums).
As in~\eqref{l:18a}, we again combine the rest mass and the external potential
in a potential~$B$. Furthermore, we only consider the advanced Green's function; for the
retarded Green's function, the calculation is analogous.

Suppose that we want to perform the light-cone expansion of the $k^{\mbox{\scriptsize{th}}}$ order
contribution to the perturbation series~\eqref{l:11b}. Using that the Green's function
is diagonal in momentum space and that multiplying by~$B$ in position space
corresponds to a convolution in momentum space, we can write the contribution as a multiple
Fourier integral,
\begin{align}
\big( (&-s^\vee B)^k s^\vee \big)(x,y) \notag \\
&= \int \frac{d^4p}{(2 \pi)^4}
        \int \frac{d^4q_1}{(2 \pi)^4} \cdots \int \frac{d^4q_k}{(2 \pi)^4}
        \: \Delta s^\vee(p;q_1,\ldots,q_k) \:e^{-i(p+q_1+\cdots+q_k)x + i p y} \;,
        \label{l:31}
\end{align}
where the distribution $\Delta s^\vee(p;q_1,\ldots,q_k)$ is the Feynman 
diagram in momentum space,
\begin{align}
\Delta s^\vee(p;q_1,\ldots,q_k) &= (-1)^k \:s^\vee(p+q_1+\cdots+q_k) 
\:\hat{B}(q_k)\: s^\vee(p+q_1+\cdots+q_{k-1}) \:\hat{B}(q_{k-1}) 
\notag \\
& \qquad\qquad \cdots\; \hat{B}(q_2) \:s^\vee(p+q_1) \:\hat{B}(q_1)\:
s^\vee(p) \label{l:A2}
\end{align}
(here~$\hat{B}$ denotes the Fourier transform of the potential $B$, and 
$s^\vee(p)$ is the multiplication operator in momentum space).
For the arguments of the Green's functions, we introduce the 
abbreviation
\beq
p_0 := p \qquad {\mbox{and}} \qquad p_l\ := p+q_1+\cdots+q_l 
\;,\quad 1 \leq l \leq k \:. \label{l:32x}
\eeq
Substituting the explicit formulas \eqref{l:10} and 
\eqref{l:21x} into \eqref{l:A2}, we obtain
\begin{align*}
\Delta s^\vee(p; &\;q_1,\ldots,q_k) = (-1)^k \;
\slashed{p}_k \:\hat{B}(q_k) \:\slashed{p}_{k-1}
\;\cdots\; \slashed{p}_1 \:\hat{B}(q_1)\: \slashed{p}_0 \notag \\
&\times \lim_{\nu_0, \ldots, \nu_k \searrow 0} 
\frac{1}{(p_k)^2 - i \nu_k p^0_k} \:
\frac{1}{(p_{k-1})^2- i \nu_{k-1} p^0_{k-1}} \:\cdots\:
\frac{1}{(p_0)^2- i \nu_0 p^0_0} \:.
\end{align*}
We already know that the limits~$\nu_0, \ldots, \nu_k \searrow 0$ exist in the
distributional sense. This can be understood directly from the fact that, fixing
the momenta~$q_1,\ldots q_k$ as well as~$\vec{p}$, the above expression for~$\Delta s^\vee$
is a meromorphic function in~$p^0$ having poles only in the lower half plane.
Computing the Fourier transform with residues, we obtain a well-defined
expression which remains finite as~$\nu_0, \ldots, \nu_k \searrow 0$.
This consideration also shows that we may choose the~$\nu_0, \ldots, \nu_k$
to be equal, i.e.
\begin{align}
\Delta s^\vee(p; &\;q_1,\ldots,q_k) = (-1)^k \;
\slashed{p}_k \:\hat{B}(q_k) \:\slashed{p}_{k-1}
\;\cdots\; \slashed{p}_1 \:\hat{B}(q_1)\: \slashed{p}_0 \notag \\
&\times \lim_{\nu \searrow 0} 
\frac{1}{(p_k)^2 - i \nu p^0_k} \:
\frac{1}{(p_{k-1})^2- i \nu p^0_{k-1}} \:\cdots\:
\frac{1}{(p_0)^2- i \nu p^0_0} \:. \label{l:A22}
\end{align}
We now expand the Klein-Gordon Green's functions in \eqref{l:A22}
with respect to the 
momenta $p_l - p$. If we expand the terms $i \nu p^0_l$
with a geometric series,
\[ \frac{1}{(p_l)^2 - i \nu p^0_l} = \sum_{n=0}^\infty 
\frac{(i \nu \:(p^0_l - p^0))^n}{((p_l)^2 - i \nu 
p^0)^{1+n}} \:, \]
all contributions with $n \geq 1$ contain factors $\nu$ and 
vanish in the limit $\nu \searrow 0$. Therefore, we must only
expand with respect to the parameters $((p_l)^2 - p^2)$. This gives, 
again with geometric series,
\begin{align*}
\Delta s^\vee(p; &\;q_1,\ldots,q_k) = (-1)^k \;
\slashed{p}_k \:\hat{B}(q_k) \:\slashed{p}_{k-1}
\;\cdots\; \slashed{p}_1 \:\hat{B}(q_1)\: \slashed{p}_0 \\
&\times \sum_{n_1,\ldots,n_k=0}^\infty (p^2 - p_k^2)^{n_k}
\:\cdots\: (p^2 - p_1^2)^{n_1} \; \lim_{\nu \searrow 0} 
\frac{1}{(p^2 - i \nu p^0)^{1+k+n_1+\cdots+n_k}} \: .
\end{align*}
Rewriting the negative power of $(p^2-i \nu p^0)$ as a mass-derivative,
\begin{align}
&\frac{1}{(p^2 - i \nu p^0)^{1+k+n_1+\cdots+n_k}} \notag \\
&\qquad = \frac{1}{(k+n_1+\cdots+n_k)!} \left( \frac{d}{da} 
\right)^{k+n_1+\cdots+n_k} \frac{1}{p^2 - a - i \nu 
p^0} \bigg|_{a=0} \:, \label{l:34z}
\end{align}
we obtain a formula containing only one Green's function. Namely,
using the notation~\eqref{l:23b}, we get
\beq \begin{split}
&\Delta s^\vee(p; q_1,\ldots,q_k) = (-1)^k \;
\slashed{p}_k \:\hat{B}(q_k) \:\slashed{p}_{k-1}
\;\cdots\; \slashed{p}_1 \:\hat{B}(q_1)\: \slashed{p}_0 \\
&\times \!\!\!\!\! \sum_{n_1,\ldots,n_k=0}^\infty \frac{1}{(k+n_1+\cdots+n_k)!}
\;(p^2 - p_k^2)^{n_k} \:\cdots\: (p^2 - p_1^2)^{n_1} \;
S^{\vee (k+n_1+\cdots+n_k)}(p) \:.
\end{split} \label{l:A3}
\eeq
This is the basic equation for the light-cone expansion of the Green's 
functions in momentum space.
\sindex{light-cone expansion!in momentum space}%
Similar to the light-cone expansion of the previous section, \eqref{l:A3} involves the mass derivatives
of the Green's functions $S^{\vee (.)}$. In order to get a connection to the nested line integrals of, say, Theorem~\ref{l:thm1},
it remains to transform the polynomials in the momenta $p_0,\ldots,p_k$ as follows:
Using~\eqref{l:32x}, we rewrite \eqref{l:A3} in terms 
of the momenta $p$, $q_1,\ldots,q_k$ and multiply out. 
Furthermore, we simplify the Dirac matrices with the anti-commutation 
rules \eqref{l:ar1}. This gives for \eqref{l:A3} a sum of terms of the form
\beq
\chi_c \:C\: \gamma^I \:q_k^{I_k} \cdots q_1^{I_1} 
\;\tilde{V}^{(k)}_{J_k, c_k}(q_k) \cdots \tilde{V}^{(1)}_{J_1, c_1}(q_1)
\:p^L \;S^{\vee (h)}(p) \qquad \big(h \geq \big[ |L|/2 \big] \big) \:,
        \label{l:A4}
\eeq
where the tensor indices of the multi-indices
$I$, $I_l$, $J_l$, and $L$ are contracted 
with each other (similar to the notation of Theorem~\ref{l:thm1}, the 
factors $\tilde{V}^{(l)}_{J_l, c_l}$ stand for the individual 
potentials of $\hat{B}$). If tensor indices of the power $p^L$ are 
contracted with each other, we can eliminate the 
corresponding factors $p^2$ iteratively with the rule \eqref{l:5}, more precisely
\beq
p^2 \:S^{\vee (h)}(p) = h \:S^{\vee (h-1)}(p) \qquad (h \geq 1) \:.
        \label{l:37x}
\eeq
In this way, we can arrange that the tensor indices of $p^L$ in 
\eqref{l:A4} are all contracted with tensor indices of the 
factors $\gamma^I$, $q_l^{I_l}$, or $\tilde{V}^{(l)}_{J_l, c_l}$.
By iteratively applying the differentiation rule \eqref{l:21a}, we can
now rewrite the power $p^L$ in \eqref{l:A4} with $p$-derivatives, e.g.
\begin{align*}
p_j\: p_k\: S^{\vee (2)}(p) &= -\frac{1}{2} 
\:p_j\:\frac{\partial}{\partial p^k} S^{\vee (1)}(p) = -\frac{1}{2} 
\:\frac{\partial}{\partial p^k} (p_j \:S^{\vee (1)}(p)) + \frac{1}{2} 
\:g_{jk}\: S^{\vee (1)}(p) \\
&= \frac{1}{4} \:\frac{\partial^2}{\partial p^j \:\partial p^k} 
S^{(0)}(p) + \frac{1}{2} \:g_{jk} \:S^{(1)}(p) \:.
\end{align*}
In this way, we obtain for $\Delta s^\vee(p; q_1,\ldots,q_k)$ a sum of terms
of the form
\beq
\chi_c \:C\: \gamma^I \:q_k^{I_k} \cdots q_1^{I_1} 
\;\tilde{V}^{(k)}_{J_k, c_k}(q_k) \cdots \tilde{V}^{(1)}_{J_1, c_1}(q_1)
\:\partial_p^K \;S^{\vee (h)}(p) \:,
        \label{l:A5}
\eeq
where no tensor indices of the derivatives $\partial_p^K$ are 
contracted with each other.
We substitute these terms into \eqref{l:31} and transform them to
position space. Integrating the derivatives 
$\partial_p^K$ by parts gives factors $(y-x)^K$. The factors 
$q_l^{I_l}$, on the other hand, can be written as partial 
derivatives $\partial^{I_l}$ acting on the potentials $V^{(l)}$.
More precisely, substituting into \eqref{l:31}, the term \eqref{l:A5} gives
the contribution
\beq
\chi_c \:C\:i^{|I_1|+\cdots+|I_k|} \:(-i)^{|K|} \;\gamma^I
\;(\partial^{I_k} V^{(k)}_{J_k, c_k}(x)) \cdots
(\partial^{I_1} V^{(1)}_{J_1, c_1}(x))
\:(y-x)^K \;S^{\vee (h)}(x,y) \: ,
        \label{l:A6}
\eeq
where the tensor indices of the factor $(y-x)^K$ are all contracted 
with tensor indices of the multi-indices $I$, $I_l$, or $J_l$.
The Feynman diagram $((-s B)^k s)(x,y)$ coincides with the sum of
all these contributions.

This expansion has much similarity with the light-cone expansion of 
Theorem~\ref{l:thm1}. Namely, if one expands the nested line 
integrals in \eqref{l:l1} in a Taylor series around $x$, one gets 
precisely the expansion into terms of the form \eqref{l:A6}. Clearly, 
the light-cone expansion of Theorem~\ref{l:thm1} goes far beyond 
the expansion \eqref{l:A6}, because the dependence on the external potential is 
described by non-local line integrals. Nevertheless, the 
expansion in momentum space \eqref{l:A3} and subsequent Fourier 
transformation give an easy way of understanding in principle how the 
formulas of the light-cone expansion come about. We remark that, after going through the
details of the combinatorics and rearranging the 
contributions \eqref{l:A6}, one can indeed recover the Taylor series of the 
line integrals in \eqref{l:l1}. This gives an alternative method for 
proving Theorem~\ref{l:thm1}. However, it is obvious that this becomes 
complicated and does not yield the most elegant approach (the reader 
interested in the details of this method is referred to \cite{firstorder}, 
where a very similar technique is used for the light-cone expansion to first 
order in the external potential).

Next, we want to generalize the previous construction to other types of Green's functions.
Since, similar to \eqref{l:34z}, we must rewrite a product of Green's functions as 
the mass derivative of a single Green's function, we can only expect 
the construction to work if all Green's functions in the product 
\eqref{l:A2} are of the same type (e.g.\ the construction breaks down 
for a ``mixed'' operator product containing both advanced and retarded 
Green's functions). But we need not necessarily
work with the advanced or retarded Green's functions. Instead, we can use
Green's functions with a different location of the poles in the complex
$p^0$-plane: We consider the Green's functions
\beq \label{l:G}
s^\pm(p) = \slashed{p} \: S^\pm_{a \:|\: a=0}(p) \qquad {\mbox{with}} 
\qquad S^\pm_a(p) = \lim_{\nu \searrow 0} \frac{1}{p^2-a \mp i \nu}
\eeq
\nindex{bi2@$s^\pm$ -- Feynman propagator of the vacuum Dirac equation}%
\nindex{bi4@$S^\pm_a$ -- Feynman propagator of the vacuum Klein Gordon equation}%
and again use the notation \eqref{l:23b},
\[ S^{\pm \:(l)} = \left( \frac{d}{da} \right)^l S^\pm_{a \:|\: a=0} \: . \]
The distribution~$s^-$ is referred to as the {\em{Feynman propagator}} (see Exercise~\ref{ex2.2-1}).
\sindex{Feynman propagator}%
The perturbation expansion of these Dirac Green's functions is,
similar to \eqref{series-scaustilde} or \eqref{l:11b}, given by the formal series
\beq \label{l:F}
\tilde{s}^+ := \sum_{n=0}^\infty (-s^+ \:B)^n 
s^+ \qquad \text{and} \qquad \tilde{s}^- := \sum_{n=0}^\infty (-s^- \:B)^n s^- \:.
\eeq
\nindex{bi6@$\tilde{s}^\pm$ -- Feynman propagator of the Dirac equation in an external potential}%
The light-cone expansion in momentum space is performed exactly as for 
\sindex{light-cone expansion!in momentum space}%
the advanced and retarded Green's functions. In analogy to 
\eqref{l:31} and \eqref{l:A3}, we thus obtain the formula
\begin{align*}
((-&s^\pm \:B)^k \:s^\pm)(x,y) \notag \\
&= \int \frac{d^4p}{(2 \pi)^4}
        \int \frac{d^4q_1}{(2 \pi)^4} \cdots \int \frac{d^4q_k}{(2 \pi)^4}
        \: \Delta s^\pm(p;q_1,\ldots,q_k) \:e^{-i(p+q_1+\cdots+q_k)x + i p y}
\end{align*}
with
\begin{align*}
\Delta s^\pm&(p; q_1,\ldots,q_k) = (-1)^k \;
\slashed{p}_k \:\hat{B}(q_k) \:\slashed{p}_{k-1}
\;\cdots\; \slashed{p}_1 \:\hat{B}(q_1)\: \slashed{p}_0 \notag \\
&\times \sum_{n_1,\ldots,n_k=0}^\infty \frac{1}{(k+n_1+\cdots+n_k)!}
\;(p^2 - p_k^2)^{n_k} \:\cdots\: (p^2 - p_1^2)^{n_1} \;
S^{\pm\: (k+n_1+\cdots+n_k)} \:.
\end{align*}
Since $S^\pm$ are Green's functions of the Klein-Gordon equation, 
they clearly also satisfy the identity \eqref{l:37x}.
Furthermore, the differentiation rule \eqref{l:21a} is also valid 
for $S^\pm$; namely
\begin{align*}
\frac{\partial}{\partial p^j} S^{\pm \:(l)}(p) &= \left( 
\frac{d}{da} \right)^l \:\lim_{\nu \searrow 0} 
\frac{\partial}{\partial p^j} \left( \frac{1}{p^2 - a \mp i 
\nu} \right) \bigg|_{a=0} \\
&= \left( \frac{d}{da} \right)^l \:\lim_{\nu \searrow 0} 
\frac{-2 p_j}{(p^2 - a \mp i \nu)^2} \bigg|_{a=0} = -2 p_j 
\:S^{\pm\: (l+1)}(p) \: .
\end{align*}
Therefore we can, exactly as in \eqref{l:A5}, rewrite the power $p^L$ 
with $p$-derivatives. Thus the expansion \eqref{l:A6} is valid in 
the same way for the Green's functions $s^\pm$ if one only replaces 
the index ``$^\vee$'' in \eqref{l:A6} by ``$^\pm$''. As explained 
before, the expansion \eqref{l:A6} can be obtained from the light-cone 
expansion of Theorem~\ref{l:thm1} by expanding the potentials around 
the space-time point $x$. Since the formulas of the light-cone 
expansion are uniquely determined by this Taylor expansion, we 
immediately conclude that the statement of Theorem~\ref{l:thm1} is also 
valid for the $k^{\mbox{\scriptsize{th}}}$ order contribution to the 
perturbation expansion \eqref{l:F} if the factor $S^{(h)}$ in \eqref{l:l1}
stands more generally for $S^{+\:(h)}$ or $S^{-\:(h)}$, respectively. 
This simple analogy between the formulas of the light-cone expansions 
of the Feynman diagrams $((-s^{\vee / \wedge} \:B)^k \:s^{\vee / 
\wedge})$ and $((-s^\pm \:B)^k \:s^\pm)$, which are obtained by changing the 
location of the poles of the vacuum Green's functions in momentum space, 
is called the {\em{residual argument}}
(the name is motivated by the fact that the effect of changing the location of the poles
becomes apparent when taking the Fourier integral with residues).
\sindex{residual argument}%

Having other Green's functions to our disposal, one can also form more general
solutions of the homogeneous equation. Namely, taking the difference of~$s^+$
and~$s^-$, we obtain similar to~\eqref{sm2calc},
\beq \label{pm2calc}
s^+(q) - s^-(q) = \slashed{q} \; \lim_{\nu
\searrow 0} \left[ \frac{1}{q^{2}-i\nu} -
\frac{1}{q^{2}+i\nu} \right] = 2 \pi i\, \slashed{q} \: \delta(q^{2}) = 2 \pi i\, p(q)
\eeq
with~$p$ according to~\eqref{defp}.
Replacing the Green's functions by those in the external potential,
one gets a canonical perturbation series for~$p$.
As we shall see below (see~\S\ref{sechec}), this perturbation series does {\em{not}} 
agree with the causal perturbation expansion~\eqref{Psea}.
Therefore, we denote the obtained operator with an additional index~$\res$.
Similar to~\eqref{def-ktil}, we thus introduce the
{\em{residual fundamental solution}}~$\tilde{p}^\res$ by
\nindex{bj0@$\tilde{p}^\res(x,y)$ -- residual fundamental solution}%
\beq \label{l:E1}
\tilde{p}^\res := \frac{1}{2 \pi i} \:\big( \tilde{s}^+ - \tilde{s}^- \big) \:.
\eeq
We now introduce the residual fermionic projector by replacing
the operators~$p_m$ and~$k_m$ in~\eqref{sea-pk} by the corresponding
perturbation series.
\begin{Def}
\label{l:def_res}
The {\bf{residual fermionic projector}}
\sindex{fermionic projector!residual}%
$\tilde{P}^\res(x,y)$ is defined by
\beq \label{l:E0}
\tilde{P}^\res(x,y) = \frac{1}{2}\,
\big(\tilde{p}^\res - \tilde{k} \big)(x,y) \:,
\eeq
\nindex{bi8@$\tilde{P}^\res(x,y)$ -- residual fermionic projector}%
where the operator~$\tilde{p}^\res$ is defined in~\eqref{l:E1},
and~$\tilde{k}$ is again given by~\eqref{ktildef}.
\end{Def} \noindent
Similar to~\eqref{Psea}, the residual fermionic projector also has a contour integral representation
(see Exercise~\ref{ex2.8}).

Applying the residual argument, the light-cone expansion of the Green's functions immediately 
carries over to $\tilde{P}^\res$:
As in~\eqref{Tm2def} we denote the lower mass shell by $T_a$, i.e.\ in momentum space
\beq \label{l:31s}
T_a(q) = \Theta(-q^0)\: \delta(q^2-a) \:.
\eeq
In analogy to the mass expansion of the Green's functions~\eqref{l:23b}, we set
\beq \label{l:F2}
T^{(l)}_\text{\tiny{formal}} = \left( \frac{d}{da} \right)^l T_a \big|_{a=0} \:.
\eeq
In order not to distract from the main idea, we postpone the analysis of whether
these derivatives exist to~\S\ref{l:sec_33}. This is why we added the index ``formal.''

\begin{Prp} \label{l:prp4}
The light-cone expansion of the residual fermionic projector $\tilde{P}^\res(x,y)$
is obtained from that of the causal Green's functions 
by the replacement
\[ S^{(l)} \rightarrow T^{(l)}_\text{\tiny{\rm{formal}}} \:. \]
\end{Prp}
\sindex{light-cone expansion!for the residual fermionic projector}%
\Proof The starting point is the light-cone expansion of the causal Green's functions
(see Theorem~\ref{l:thm1}, Theorem~\ref{l:thm2} and Theorem~\ref{l:thm3}).
By linearity, this light-cone expansion also hold for~$\tilde{k}$
defined by~\eqref{def-ktil}, after the replacements
\[ S^{(l)} \rightarrow \frac{1}{2 \pi i} \:\Big( S^{\vee\, (l)} - S^{\wedge\, (l)} \Big) \:. \]
Using the residual argument, the light-cone expansion of the Green's functions~$\tilde{s}^\pm$
is obtained by the replacements~$S^{(l)} \rightarrow S^{\pm\, (l)}$.
It follows by linearity that~$\tilde{p}^\res$ as defined by~\eqref{l:E1} also has a light-cone
expansion obtained by the replacements
\[ S^{(l)} \rightarrow \frac{1}{2 \pi i} \:\Big( S^{+\, (l)} - S^{-\, (l)} \Big) \:. \]
Finally, again by linearity, we obtain the light-cone expansion of
residual fermionic projector~\eqref{l:E0} by the replacements
\[ S^{(l)} \rightarrow \frac{1}{4 \pi i} \:\Big( S^{+\, (l)} - S^{-\, (l)} - S^{\vee\, (l)} + S^{\wedge\, (l)} \Big) \:. \]

A direct computation in analogy to~\eqref{sm2calc} and~\eqref{pm2calc} shows that
\[ \frac{1}{4 \pi i} \:\Big( S^+ - S^- - S^\vee + S^\wedge \Big) = T_a \:. \]
This concludes the proof.
\QED
We point out that the result of Proposition~\ref{l:prp4} is only formal because
we have not yet analyzed whether the factors~$T^{(l)}_\text{\tiny{formal}}$ are mathematically
well-defined. This will be done in the next section.

\subsectionn{The Non-Causal Low Energy Contribution} \label{l:sec_33}
We now want to put the residual argument and the 
formal light-cone expansion of Proposition~\ref{l:prp4} on a satisfying 
mathematical basis.
In order to explain what precisely we need to do, we first
recall how the light-cone expansion of the Green's functions makes 
mathematical sense: Theorem~\ref{l:thm1} gives a representation of every 
Feynman diagram of the perturbation series \eqref{l:11b} as an infinite 
sum of contributions of the form \eqref{l:l1}. According to the bound 
\eqref{l:45o}, there are, for any given $h$, only a finite number of
possibilities to choose $I_a$ and $p_a$; as a consequence, we get, for
fixed $h$, only a finite number of contributions \eqref{l:l1}.
Thus we can write the light-cone expansion in the symbolic form
\beq \label{l:m1}
\big( (-s B)^k \:s \big)(x,y) = \sum_{h=-1}^\infty 
\:\sum_{\mbox{\scriptsize{finite}}} \:\cdots\: S^{(h)}(x,y) \:,
\eeq
where `$\cdots$' stands for a configuration of the $\gamma$-matrices 
and nested line integrals in \eqref{l:70}. According to the
explicit formula \eqref{l:12}, the higher $a$-derivatives of $S_a(x,y)$ 
contain more factors $(y-x)^2$ and are thus of higher order on the 
light cone. This makes it possible to make mathematical sense of the infinite 
series in~\eqref{l:m1} as a light-cone expansion.

According to Proposition~\ref{l:prp4}, all the results for the Green's function
are, on a formal level, also valid for the residual fermionic projector.
We begin by considering the light-cone expansion of the individual
Feynman diagrams in more detail. Similar to 
\eqref{l:m1}, the $k^{\mbox{\scriptsize{th}}}$ order contribution 
$\Delta P^\res$ to the residual fermionic projector has an 
expansion of the form
\beq \label{l:m2}
\Delta P^\res(x,y) = \sum_{h=-1}^\infty 
\:\sum_{\mbox{\scriptsize{finite}}} \:\cdots\: T^{(h)}_{\text{\tiny{\rm{formal}}}}(x,y) \:,
\eeq
where $T^{(h)}_{\text{\tiny{formal}}}$ is the $a$-derivative \eqref{l:F2} of the lower mass 
shell $T_a$, \eqref{l:31s}. In position space, $T_a$ is given explicitly
in~\eqref{l:3.1}.
The basic difference between the light-cone expansions 
\eqref{l:m1} and \eqref{l:m2} is related to the logarithmic pole $\log |a|$
in \eqref{l:3.1}. Namely, as a consequence of this logarithm, the higher 
$a$-derivatives of $T_a$ are {\em{not}} of higher order on the light 
cone. To the order $\O((y-x)^2)$, for example, one has
\beq \label{l:m3}
\left( \frac{d}{da} \right)^n T_a(x,y) = \frac{1}{32 \pi^3} \left( 
\frac{d}{da} \right)^n (a \:\log |a|) + \O((y-x)^2)
\qquad (n \geq 2) \:.
\eeq
In our context of an expansion around $a=0$, the situation is even worse, because the 
$a$-derivatives of $T_a$ are singular for $a \rightarrow 0$ (as one 
sees e.g.\ in \eqref{l:m3}). Thus not even the individual contributions 
to the light-cone expansion make mathematical sense. These 
difficulties arising from the logarithm in \eqref{l:3.1} are called the 
{\em{logarithmic mass problem}}
\sindex{logarithmic mass problem}%
(see \cite{firstorder} for a more detailed 
discussion in a slightly different setting). Since we know from Lemma~\ref{l:lemma0}
that the Feynman diagrams are all well-defined, the 
logarithmic mass problem is not a problem of the perturbation 
expansion, but shows that something is wrong with the light-cone 
expansion of Proposition~\ref{l:prp4}.

In order to resolve the logarithmic mass problem, we first 
``regularize'' the formal light-cone expansion by taking out the 
problematic $\log |a|$ term. By resumming the formal light-cone 
expansion, we then show that the difference between the residual 
Dirac sea and the ``regularized'' Dirac sea is a smooth function in 
position space. We introduce the notation
\begin{align}
T_a^\reg(x,y) &= T_a(x,y) - \frac{a}{32 \pi^3} \:\log|a| \: \sum_{j=0}^\infty
\frac{(-1)^j}{j! \: (j+1)!} \: \frac{(a \xi^2)^j}{4^j} \label{Tregdef} \\
T^{(l)} &= \left( \frac{d}{da} \right)^l
T_{a \:|\: a=0}^\reg \label{l:3zz}
\end{align}
(where~$\xi^2 \equiv \xi^j \xi_j$ denotes again the Minkowski inner product).
\nindex{bj1@$T^\reg_a$ -- $T_a$ with $\log$-terms in~$a$ removed}%
\nindex{bj2@$T^{(l)}$ -- mass expansion of $T^\reg_a$}%

\begin{Def}
\label{l:def_le}
The {\bf{causal contribution}} 
$\tilde{P}^{\mbox{\scriptsize{\rm{causal}}}}$ to the fermionic projector is obtained 
from the residual Dirac sea $\tilde{P}^\res$ by 
replacing all factors~$T^{(h)}_{\text{\tiny{\rm{formal}}}}$ in the formal light-cone expansion by~$T^{(h)}$.
\sindex{fermionic projector!causal contribution}%
\nindex{bj4@$\tilde{P}^{\mbox{\scriptsize{\rm{causal}}}}$ -- causal contribution to fermionic projector}%
\sindex{fermionic projector!non-causal low energy contribution}%
\nindex{bj6@$\tilde{P}^\lec$ -- non-causal low energy contribution to fermionic projector}%
The {\bf{non-causal low energy contribution}}
$\tilde{P}^\lec$ to the fermionic projector is given by
\[ \tilde{P}^\lec(x,y) =
\tilde{P}^\res(x,y) -
\tilde{P}^{\mbox{\scriptsize{\rm{causal}}}}(x,y) \: . \]
\end{Def}
By the replacement $T^{(h)}_{\text{\tiny{formal}}} \rightarrow 
T^{(h)}$, the formal light-cone expansion of 
Proposition~\ref{l:prp4} becomes mathematically meaningful in the sense of 
Definition~\ref{deflce}. Thus we can restate this result as a theorem,
leaving out the word ``formal.''
\begin{Thm}
The light-cone expansion of the causal Green's functions also holds for the
causal contribution~$\tilde{P}^{\mbox{\scriptsize{\rm{causal}}}}$ to the 
fermionic projector if one simply replaces~$S^{(l)} \rightarrow T^{(l)}$
with~$T^{(l)}$ according to~\eqref{l:3zz}.
\end{Thm} 

Since~$T_a - T_a^\reg$ is a smooth function in~$x$ and~$y$,
it is natural to expect that the non-causal low energy contribution should also be smooth.
This is indeed the case, in the following sense.
\begin{Thm} \label{l:thm6}
To every order in the external potential~$\B$, the non-causal low energy contribution 
$\tilde{P}^\lec(x,y)$ is a smooth function in $x$ and $y$.
\end{Thm}
\sindex{fermionic projector!smooth contributions to}%
The subtle point in the proof is that, to every order in perturbation theory,
the non-causal low energy contribution involves an infinite number of summands.
Although each summand is smooth, it is not clear whether the infinite sum converges
and gives rise to a smooth function.
This makes it necessary to use a {\em{resummation technique for the smooth
contributions to the light-cone expansion}}.
\sindex{resummation!of smooth contributions}%
For brevity, we do not enter these constructions here but instead refer the interested reader
to~\cite[Proof of Theorem~3.8]{light}. The resummation technique will also be introduced
and applied in Appendix~\ref{s:appresum}.

\subsectionn{The Non-Causal High Energy Contribution} \label{sechec}
In the previous sections (\S\ref{l:sec_31} and~\S\ref{l:sec_33}) we performed the light-cone expansion of
the residual fermionic projector~$\tilde{P}^\res$
(see Definition~\ref{l:def_res}). The remaining task is to deduce
the light-cone expansion of the fermionic projector~$P^\sea$ with spatial
normalization (as defined by~\eqref{Psea}).
We now prove that~$P^\sea$ and~$\tilde{P}^\res$ have the
same light-cone expansion.

We begin by giving the difference between the fermionic projector and the 
residual fermionic projector a name.
\begin{Def}
\label{l:def_he}
The {\bf{non-causal high energy contribution}}
$\tilde{P}^\hec(x,y)$ to the fermionic projector is given by
\[ \tilde{P}^\hec(x,y) = P^\sea(x,y) - \tilde{P}^\res(x,y) \:. \]
\sindex{fermionic projector!non-causal high energy contribution}%
\nindex{bj8@$\tilde{P}^\hec$ -- non-causal high energy contribution to fermionic projector}%
\end{Def}

\begin{Thm} \label{l:thm4}
To every order in the external potential~$\B$, the non-causal high energy contribution 
$\tilde{P}^\hec(x,y)$ is a smooth function in $x$ and $y$.
\end{Thm}
\sindex{fermionic projector!smooth contributions to}%
\Proof Our first task is to rewrite the perturbation expansion of~$\tilde{P}^\res$
in terms of the potential~$\B$. To this end, one combines the rest masses
of the Dirac particles with the unperturbed Green's functions.
Thus for the advanced and retarded Green's functions, we return to the
perturbation expansions~\eqref{series-scaustilde}.
Similarly, for the Green's functions~$\tilde{s}^\pm$, we rewrite~\eqref{l:F} as
\[ \tilde{s}_m^+ = \sum_{n=0}^\infty (-s^+_m \:\B)^n 
s^+_m \qquad \text{and} \qquad \tilde{s}^-_m = \sum_{n=0}^\infty (-s^-_m \:\B)^n s^-_m \:. \]
Then~$\tilde{k}$ and~$\tilde{p}^\res$ are defined again by~\eqref{def-ktil}
and~\eqref{l:E1}, respectively. As a result, the operators~$\tilde{k}$ and~$\tilde{p}^\res$
are defined as sums of operator products of the form
\beq \label{l:n7}
C_n \:\B\:C_{n-1} \:\B\: \cdots \: \B\:C_0 \:,
\eeq
where the factors $C_l$ coincide with either $k$, $p$ or $s$.

Next, we need a few structural properties of the causal perturbation expansion.
These results are derived in Exercises~\ref{ex2.4-5}--\ref{ex2.4-52}.
Alternatively, these results are obvious from the detailed formulas in the research papers~\cite{grotz, norm}.
First, the operator~$\tilde{k}$ has the contour integral representation
(see Exercise~\ref{ex2.4-5}~(a))
\[ \tilde{k} = -\frac{1}{2 \pi i} \ointctrclockwise_{\Gamma_+ \cup \Gamma_-} \lambda\:
\tilde{R}_\lambda\: d\lambda \:. \]
As a consequence, the fermionic projector~$P^\sea$, \eqref{Psea}, can be represented as
\[ P^\sea = \frac{1}{2} \,\big( \tilde{p} - \tilde{k} \big) \:, \]
where~$\tilde{p}$ is defined by
\[ \tilde{p} := -\frac{1}{2 \pi i} \left( \ointctrclockwise_{\Gamma_+} - \ointctrclockwise_{\Gamma_-} \right)
\lambda\: \tilde{R}_\lambda\: d\lambda \]
(see Exercise~\ref{ex2.4-5}~(b)). Comparing with~\eqref{l:E0} and Definition~\ref{l:def_he}, we conclude that
\[ \tilde{P}^\hec = \frac{1}{2}\:\big( \tilde{p} - \tilde{p}^\res \big) \:. \]
Next, the operator~$\tilde{p}$ has the following properties:
\begin{itemize}[leftmargin=2em]
\item[(i)] Every contribution to the perturbation expansion of~$\tilde{p}$ contains an even number of factors~$k$.
\item[(ii)] If in the perturbation series for~$\tilde{p}$ one replaces all factors~$k$ by factors~$p$,
one gets precisely the perturbation series for~$\tilde{p}^\res$.
\end{itemize}
These properties can be read off from the explicit
formulas for~$\tilde{p}$ and~$\tilde{p}^\res$ given in~\cite{grotz, norm}.
For abstract proofs, one can proceed as follows.
Property~(i) is shown in Exercise~\ref{ex2.4-51}.
In order to prove~(ii), we first bring the perturbation expansion of the residual fundamental solution into a more
explicit form. Comparing~\eqref{l:E1} with~\eqref{def-ktil} and noting that in view of~\eqref{pm2calc},
the Green's functions~$s^\pm_m$ satisfy in analogy to~\eqref{smkform} the relations
\[ s =s^+-i\pi p = s^- +i \pi p \:, \]
we find that the perturbation expansion of~$\tilde{p}^\res$
is obtained from that for~$\tilde{k}$, \eqref{k-pert}, simply by replacing all factors~$k$ by factors~$p$,
\[ \tilde{p}^\res = \sum_{\beta=0}^\infty (-i \pi)^{2 \beta} \;b^< \:p\: (b \:p)^{2 \beta} \:b^> \:. \]
In Exercise~\ref{ex2.4-52} it is shown that exactly the same perturbation series is obtained
if in the perturbation series for~$\tilde{p}$ one replaces all factors~$k$ by factors~$p$. This proves~(ii).

Using the above properties~(i) and~(ii), we can convert the perturbation series for $\tilde{p}$ into that for
$\tilde{p}^\res$ by iteratively replacing pairs 
of factors $k$ in the operator products by pairs of factors $p$. Thus the 
difference $\tilde{p} - \tilde{p}^\res$ can, 
to every order in perturbation theory, be written as a finite sum of 
expressions of the form
\beq \label{l:n6}
\begin{split}
C_n \:\B \cdots C_{b+1} \:&\B \:\Big( p \:\B\: 
C_{b-1} \cdots C_{a+1} \:\B\: p \\
&\:- k \:\B\: C_{b-1} \cdots C_{a+1} \:\B\: k\Big) 
\:\B\:C_{a-1} \cdots \B\:C_0 \:,
\end{split}
\eeq
where the factors $C_l$ again stand for $k$, $p$ or $s$.
Therefore, it remains to show that~\eqref{l:n6} is a smooth function in position 
space.

We first simplify our problem: Once we have shown that 
the bracket in~\eqref{l:n6} is smooth and bounded in position space, 
the additional multiplications to the very left and right can be carried out 
by iteratively multiplying with $\B$ and forming the 
convolution with $C_l$, which again gives a smooth and bounded 
function in each step (notice that, according to the assumptions of 
Lemma \ref{l:lemma0}, $\B$ decays sufficiently fast at infinity). 
Thus we must only consider the bracket in~\eqref{l:n6}. We rewrite this 
bracket with the projectors $\frac{1}{2}(p-k)$ and $\frac{1}{2}(p+k)$ 
on the lower and upper mass shells,
\begin{align*}
p \:\B\: & C_{n-1} \cdots C_1 \:\B\: p
\:-\: k \:\B\: C_{n-1} \cdots C_1 \:\B\: k \\
&= \frac{1}{2}\:(p+k) \:\B\: C_{n-1} \cdots C_1 
\:\B\: (p-k) \:+\:\frac{1}{2}\:(p-k) \:\B\: C_{n-1} \cdots C_1 
\:\B\: (p+k) \:.
\end{align*}
For symmetry reasons, it suffices to consider the first summand of this 
decomposition,
\beq \label{l:p-k}
((p+k) \:\B\: C_{n-1} \cdots C_1 \:\B\: (p-k))(x,y) \:,
\eeq
where the factors $C_l$ again stand for $k$, $p$, or $s$. Our task is
to show that \eqref{l:p-k} is a smooth function in position space.

We proceed in momentum space. We say that a function $f(q)$ has 
{\em{rapid decay for positive frequency}} if it is $C^1$, bounded 
together with its first derivatives (i.e.\ $\sup |f|, \sup|\partial_l 
f| < \infty$), and satisfies for every $\alpha>0$ the bounds
\beq \label{l:n12}
\sup_{\omega >0,\; \vec{k} \in \R^3} |\omega^\alpha \:f(\omega, \vec{k})|,\;
\sup_{\omega >0,\; \vec{k} \in \R^3} |\omega^\alpha \:\partial_l
f(\omega, \vec{k})| \;<\; \infty \:.
\eeq
After setting $C_0=p-k$ and $C_n = p+k$,
the operator product \eqref{l:p-k} is of the form \eqref{l:2}. We choose 
a function $g$ with rapid decay for positive frequency and decompose 
the operator product in the form \eqref{l:4z},\eqref{l:4a}. It follows by 
induction that the functions $F_j$ all have rapid decay for positive 
frequency: The induction hypothesis is obvious by setting $F_0=g$. The 
induction step is to show that for a function $F_{j-1}$ with rapid 
decay for positive frequency, the convolution
\beq \label{l:n13}
F_j(\omega, \vec{k}) = \int \frac{d\omega^\prime}{2 \pi}
\int\frac{ d\vec{k}^\prime}{(2 \pi)^3} \; 
\hat{B}(\omega - \omega^\prime, \vec{k}-\vec{k}^\prime) 
\:C_{j-1}(\omega^\prime, \vec{k}^\prime) \:F_{j-1}(\omega^\prime, 
\vec{k}^\prime)
\eeq
also has rapid decay for positive frequency. In Lemma~\ref{l:lemma0}, 
it was shown that $F_j$ is $C^1$ and bounded together with its first 
derivatives. As a consequence, we must only establish the bounds \eqref{l:n12}
for $\omega>1$. Moreover, because of the monotonicity $\omega^\alpha < 
\omega^\beta$ for $\alpha<\beta$ (and $\omega >1$), it suffices
to show that there are arbitrarily large numbers $\alpha$ satisfying 
the bounds \eqref{l:n12}; we only consider $\alpha=2n$ with $n \in \N$.
For $\omega>1$ and $\omega^\prime \in \R$, we have the inequality
\[ \omega^{2n} \leq (2 \omega^\prime)^{2n} 
\:\Theta(\omega^\prime) + (2 (\omega-\omega^\prime))^{2n} \: , \]
as is immediately verified by checking the three regions 
$\omega^\prime \leq 0$, $0 < \omega^\prime \leq \omega/2$, and 
$\omega^\prime>\omega/2$. We combine this inequality with \eqref{l:n13} and
obtain for $\omega>1$ the estimate
\beq \label{E12int}
|\omega^{2n} \:F_j(\omega, \vec{k})| \leq \bigg| \int \frac{d\omega^\prime}{(2 \pi)} 
\int \frac{d\vec{k}^\prime}{(2 \pi)^3} \: \big( E_1 + E_2 \big) \bigg| \:,
\eeq
where~$E_1$ and~$E_2$ are given by
\begin{align}
E_1 &= \hat{B}(\omega-\omega^\prime, 
\vec{k}-\vec{k}^\prime) \:C_{j-1}(\omega^\prime, \vec{k}^\prime) 
\left[ (2 \omega^\prime)^{2n} \:\Theta(\omega^\prime) 
\:F_{j-1}(\omega^\prime, \vec{k}) \right] \label{l:n14} \\
E_2 &= \left[ (2(\omega-\omega^\prime))^{2n} 
\:\hat{B}(\omega-\omega^\prime, \vec{k}-\vec{k}^\prime) \right]
\:C_{j-1}(\omega^\prime, \vec{k}^\prime) 
\:F_{j-1}(\omega^\prime, \vec{k}) \:. \label{l:n15}
\end{align}
According to the induction hypothesis, the square bracket in 
\eqref{l:n14} is bounded together with its first derivatives. Since 
$\hat{B}$ has rapid decay at infinity, the square bracket in \eqref{l:n15} also 
has rapid decay at infinity. As a consequence, the integral in~\eqref{E12int}
satisfies the hypothesis considered in Lemma~\ref{l:lemma0} for \eqref{l:3a}
and is therefore bounded. In order to estimate the expression~$|\omega^{2n} \partial_l F_j|$, we
differentiate \eqref{l:n13} and obtain similar to~\eqref{l:n14} and~\eqref{l:n15} the inequality
\begin{align*}
&|\omega^{2n} \:\partial_l F_j(\omega, \vec{k})| \\
&\leq \bigg| \int \frac{d\omega^\prime}{2 \pi} 
\int \frac{d\vec{k}^\prime}{(2 \pi)^3} \;
\partial_l \hat{B}(\omega-\omega^\prime, 
\vec{k}-\vec{k}^\prime) \:C_{j-1}(\omega^\prime, \vec{k}^\prime) 
\left[ (2 \omega^\prime)^{2n} \:\Theta(\omega^\prime) 
\:F_{j-1}(\omega^\prime, \vec{k}) \right] \bigg| \\
&\quad+\bigg| \int \frac{d\omega^\prime}{d \omega} \int 
\frac{d\vec{k}^\prime}{(2 \pi)^3} \:
\left[ (2(\omega-\omega^\prime))^{2n} 
\:\partial_l \hat{B}(\omega-\omega^\prime, \vec{k}-\vec{k}^\prime) \right]
\:C_{j-1}(\omega^\prime, \vec{k}^\prime) 
\:F_{j-1}(\omega^\prime, \vec{k}) \bigg| \:.
\end{align*}
This concludes the proof of the induction step.

We just showed that for a function $g$ with rapid decay for 
positive frequency, the function
\beq \label{l:n15a}
F_n(q) = \int \frac{d^4q_1}{(2 \pi)^4} \left( \B\:C_{n-1}\:\B
\cdots \B\:C_1\:\B\:C_0 \right)(q,q_1) \:g(q_1)
\eeq
has rapid decay for positive frequency. We now consider what this 
means for our operator product \eqref{l:p-k} in position space. For a 
given four-vector $y=(y^0, \vec{y})$, we choose
\[ g(\omega, \vec{k}) = \eta(\omega)\; e^{-i 
(\omega y^0 - \vec{k} \vec{y})} \: , \]
where $\eta$ is a smooth function with $\eta(\omega)=1$ for 
$\omega \leq 0$ and $\eta(\omega)=0$ for $\omega>1$ (this choice of $g$ clearly 
has rapid decay for positive frequency). Since the support of the 
factor $C_0=(p-k)$ is the lower mass cone $\{q^2 \geq 0,\: q^0 \leq 
0\}$, $g(\omega, \vec{k})$ enters into the integral \eqref{l:n15a} only 
for negative $\omega$. But for $\omega \leq 0$, the cutoff function $\eta$ is 
identically one. Thus the integral \eqref{l:n15a} is simply a Fourier 
integral; i.e., with a mixed notation in momentum and position space,
\[ F_n(q) = \left( \B\:C_{n-1}\:\B \cdots 
\B\:C_1\:\B\:(p-k) \right)(q,y) \: . \]
Next, we multiply from the left with the operator $(p+k)$,
\beq \label{l:n16}
\left( (p+k)\:\B\:C_{n-1}\:\B \cdots 
\B\:C_1\:\B\:(p-k) \right)(q,y) =
(p+k)(q) \:F_n(q) \: .
\eeq
Since~$F_n$ has rapid decay for positive frequency and $(p+k)$ has 
its support in the upper mass cone $\{q^2 \geq 0,\:q^0>0\}$, their product 
decays fast at infinity. More precisely,
\[ \left| q^I \:(p+k)(q) \:F_n(q) \right| \leq {\mbox{const}}(I) 
\: (p+k)(q) \]
for any multi-index $I$. As a consequence, the Fourier transform of 
\eqref{l:n16} is even finite after multiplying with an arbitrary number 
of factors $q$, i.e.
\[ \left| \int \frac{d^4q}{(2 \pi)^4} \:q^I
\;(p+k)(q) \:F_n(q) \; e^{-i q x} \right| \leq
{\mbox{const}}(I) \;<\; \infty \]
for all $x$ and $I$.
This shows that our operator product in position space \eqref{l:p-k} is 
bounded and, for fixed $y$, a smooth function in $x$ (with derivative 
bounds which are uniform in $y$). Similarly, one 
obtains that \eqref{l:p-k} is, for fixed $x$, a smooth function in $y$. We 
conclude that the distribution \eqref{l:p-k} is a smooth and bounded function.
\QED

\subsectionn{The Unregularized Fermionic Projector in Position Space} \label{seckernunreg}
The previous constructions give a representation of the fermionic projector 
in the presence of chiral and scalar/pseudoscalar potentials (see~\eqref{direxY},
\eqref{Bchiral} and~\eqref{YLRdef}) of the form
\label{Ple}
\beq \label{fprep}
\begin{split}
P^{\mbox{\scriptsize{sea}}}(x,y) &= \sum_{n=-1}^\infty
{\mbox{(phase-inserted line integrals)}} \times  T^{(n)}(x,y) \\
&\qquad + \tilde{P}^\lec(x,y) + \tilde{P}^\hec(x,y) \:.
\end{split}
\eeq
\nindex{be0@$P^\sea$ -- fermionic projector describing Dirac seas}%
\sindex{light-cone expansion!of fermionic projector}%
\sindex{fermionic projector!light-cone expansion of}%
\sindex{fermionic projector!unregularized in position space}%
\sindex{fermionic projector!in the presence of an external potential}%
Here the series is a light-cone expansion which describes
the singular behavior of the fermionic projector on the light cone
non-perturbatively. It is obtained from the light-cone expansion of the
Green's functions by the simple replacement rule
\[ S^{(n)} \longrightarrow T^{(n)} \]
(with~$T^{(n)}$ as defined in~\eqref{l:3zz}).
In particular, the phase-inserted line integrals are exactly
the same as those for the Green's functions (see Definition~\ref{l:def_pf}).
The contributions $\tilde{P}^\lec$ and~$\tilde{P}^\hec$, on the other hand,
are both given perturbatively by a series of terms which are all smooth on the light cone.
The ``causality'' of the causal perturbation expansion
can be understood from the fact that the phase-inserted line integrals
in~\eqref{fprep} are all bounded integrals along the line segment joining
the points~$x$ and~$y$ (whereas the light-cone expansion of general operator
products involves unbounded line integrals).
\sindex{causality!of the perturbation expansion}%
\sindex{causality!of the light-cone expansion}%
In particular, when~$y$ lies in the causal future or past of~$x$,
the light-cone expansion in~\eqref{fprep} depends
on the external potential only inside the causal diamond
$(J^\lor_x \cap J^\wedge_y) \cup (J^\wedge_x \cap J^\lor_y)$.
Nevertheless, the light-cone expansion is not causal in this strict
sense because there are contributions for~$y \not \in J_x$.
Furthermore, the low and high energy contributions cannot be described with
line integrals and violate locality as well as causality.
This non-locality can be understood from the fact that
the fermionic projector is a global object in space-time.
We conclude that the singular behavior of the
fermionic projector on the light-cone can be described explicitly by
causal line integrals, whereas the smooth contributions to the fermionic projector are governed by
non-local effects.

We finally remark that the decomposition~\eqref{fprep} is also a suitable
starting point for analyzing the smooth contributions to the fermionic projector.
Indeed, the low energy contribution~$\tilde{P}^\lec$ can be computed effectively by resumming the
perturbation expansion, as is explained in Appendix~\ref{s:appresum}.
The high energy contribution~$\tilde{P}^\hec$, on the other hand,
is given in terms of operator products, which can be analyzed with Fourier methods.

\section{Description of Linearized Gravity} \label{seclingrav}
We now outline how our computational tools apply in the presence of a gravitational field.
Note that so far, the external potential~$\B$ in the Dirac equation~\eqref{direx}
was assumed to be a multiplication operator. When describing a gravitational field, however,
the derivative terms in the Dirac equation are modified. The gravitational field can still
be described by the Dirac equation~\eqref{direx} if we allow for~$\B$
to be a first order differential operator. This means that the causal perturbation expansion of~\S\ref{secspatial} still applies. An analysis similar to that in Lemma~\ref{l:lemma0} shows that
the contributions to the perturbation series are again all well-defined and finite, provided that the
gravitational field is smooth and decays sufficiently fast at infinity.
In order to perform the light-cone expansion of the Green's functions,
it is convenient to commute the differential operators contained in~$\B$ to the very left
to obtain operator products of the form
\[ \frac{\partial}{\partial x^I} \big[ s \,Z_1\, \cdots \,Z_n\, s \big](x,y) \:, \]
where the~$Z_1,\ldots, Z_n$ are again multiplication operators (which contain tensor
indices contracted with the multi-index~$I$).
This makes it possible to perform the light-cone expansion of the square brackets
with the inductive procedure described in~\S\ref{seclcgreen}.
Carrying out the derivatives~$\partial_x^I$ gives the desired light-cone expansion of
the Green's function.

The basic difficulty with this construction is that, due to the additional derivatives,
the contributions to higher order in perturbation theory become more and more
singular on the light cone. In particular, the structural results of~\S\ref{secstructure}
no longer hold, and the resummation method of~\S\ref{secresum} no longer applies.
These difficulties are closely related to the fact that in the presence of a gravitational field,
the light cone is no longer the light cone of Minkowski space, but it is generated by
the null geodesics of the Lorentzian metric. This ``deformation of the light cone''
by the gravitational field is an effect which cannot be properly described by
a light-cone expansion in Minkowski space.
A possible way out is to use to use the non-perturbative construction in~\cite{infinite, hadamard}.
The structure of the singularities on the light-cone can then be analyzed with the
so-called Hadamard expansion (for explicit computations for the fermionic
projector we refer to~\cite[Appendix~A]{lqg}).
Since we do not want to enter these techniques here, we simply describe how
{\em{linearized gravity}} can be described with our methods.
For more details we refer to~\cite[Appendix~B]{firstorder}.

For the metric, we consider a first order perturbation~$h_{jk}$ of the Minkowski
metric~$\eta_{jk}={\mbox{diag}}(1,-1,-1,-1)$,
\[ g_{jk}(x) = \eta_{jk} + h_{jk}(x) \:. \]
\nindex{bk2@$g_{jk}$ -- Lorentzian metric}%
\nindex{bk4@$\eta_{jk}$ -- Minkowski metric}%
\nindex{bk6@$h_{jk}$ -- linear perturbation of metric}%
\sindex{gravitational field!linearized}%
As in the usual formalism (see for example~\cite[\S105 and~\S107]{landau2}), we raise and lower 
tensor indices with the Minkowski metric.
Using the transformation of~$h_{jk}$ under infinitesimal coordinate 
transformations, we can assume \cite[\S105]{landau2} that
\[ \partial^k h_{jk} = \frac{1}{2} \:\partial_j h \qquad \text{with} \qquad h := h^k_k \:. \]
A straightforward computation (using for example the formalism introduced in~\cite{U22})
shows that in the so-called symmetric gauge, the Dirac operator takes the form
\[ i \Pdd_x - \frac{i}{2} \:\gamma^j \:h_{jk}\:\eta^{kl} 
\:\frac{\partial}{\partial x^l} +\frac{i}{8} \:(\Pdd h) \:. \]
In contrast to~\eqref{direx}, now the perturbation itself is a differential operator.

One complication arises from the fact that the integration measure in 
curved space is $\sqrt{|g|} \:d^4x = (1+\frac{h}{2}) \:d^4x$, whereas 
the formula~\eqref{delP1} for the perturbation of the fermionic projector is
valid only if one has the integration measure~$d^4 x$ of Minkowski space. 
Therefore we first transform the system such that the integration 
measure becomes~$d^4x$, then apply~\eqref{delP1}, and finally transform back to 
the original integration measure $\sqrt{|g|} \:d^4x$. Rewriting the space-time
inner product~\eqref{stip} as
\[ \int_{\scrM} \Sl \psi | \phi \Sr\: d\mu(x) 
= \int_{\R^4} \Sl \psi | \phi \Sr \:\sqrt{|g|} \:d^4x = \int_{\R^4} 
\Sl (|g|^{\frac{1}{4}} \psi) \:| \: (|g|^{\frac{1}{4}} \:\phi \Sr \:d^4x \:, \]
the transformation to the measure $d^4x$ is accomplished by
\begin{eqnarray*}
\psi(x) & \rightarrow & \hat{\psi}(x) =
|g|^{\frac{1}{4}}(x) \:\psi(x)  \\
i \Pdd_x - \frac{i}{2} \:\gamma^j \:h_j^k\: \partial_k
+ \frac{i}{8} \:(\Pdd h) & \rightarrow & |g|^{\frac{1}{4}} \left(
i \Pdd_x - \frac{i}{2} \:\gamma^j \:h_j^k\: \partial_k
+ \frac{i}{8} \:(\Pdd h) \right) |g|^{-\frac{1}{4}} \\
&& = i \Pdd_x - \frac{i}{2} \:\gamma^j \:h_j^k\: \partial_k
- \frac{i}{8} \:(\Pdd h) \:.
\end{eqnarray*}
The perturbation $\Delta P^{(d^4x)}$ of the transformed system is given by~\eqref{delP1},
\begin{align}
\Delta P^{(d^4x)}(x,y) &= -\int d^4z \:
\bigg\{ s(x,z) \Big( -\frac{i}{2}\:\gamma^j\:h_j^k \frac{\partial}{\partial z^k}
        - \frac{i}{8} (\Pdd h)(z) \Big) P(z,y) \notag \\
&\qquad \qquad\;\; + P(x,z) \Big( -\frac{i}{2}\:\gamma^j\:h_j^k
\frac{\partial}{\partial z^k} - \frac{i}{8} (\Pdd h)(z) \Big) s(z,y) \bigg\} \:.
        \label{pd4x}
\end{align}
The formula for the transformation of the Dirac sea to the original 
integration measure $\sqrt{|g|} \:d^4x$ is
\[ P(x,y) \:+\: \Delta P(x,y) = |g|^{-\frac{1}{4}}(x) 
\:|g|^{-\frac{1}{4}}(y) \left( P(x,y) + \Delta P^{(d^4x)}(x,y) \right) . \]
Thus
\[ \Delta P(x,y) = \Delta P^{(d^4x)}(x,y) -\frac{1}{4} \:( h(x) + h(y)) \:P(x,y) \:. \]
Since the factors $P(z,y)$ and $s(z,y)$ in~\eqref{pd4x} only depend on the difference
vector~$z-y$, we can rewrite the $z$-derivatives as $y$-derivatives,
\[ \frac{\partial}{\partial z^k} P(z,y) =
-\frac{\partial}{\partial y^k} P(z,y) \:,\qquad
\frac{\partial}{\partial z^k} s(z,y) =
-\frac{\partial}{\partial y^k} s(z,y) \:, \]
which can be  pulled out of the integral. Furthermore, the relations
\begin{align*}
\int d^4z \; P(x,z) \:\big(i \Pdd_z h(z) \big)\: s(z,y)
&= \int d^4z \; P(x,z) \:\big[(i \Pdd_z-m),\: h(z) \big]\: s(z,y) \\
&=-P(x,y) \:h(y) \\
\int d^4z \; s(x,z) \:\big(i \Pdd_z h(z) \big)\: P(z,y) &= h(x) \:P(x,y)
\end{align*}
make it possible to simplify the factors $(\Pdd h)$ in the integral. In the 
resulting formula for $\Delta P(x,y)$, one recovers the perturbation
by an electromagnetic potential. More precisely,
\beq \label{dA}
\Delta P(x,y) = \Big(-\frac{1}{8}\:h(x) \:-\: \frac{3}{8}\:h(y) \Big)\,
P(x,y) \:-\: \frac{i}{2}\, \frac{\partial}{\partial y^k} \Delta P[\gamma^j 
h^k_j](x,y) \:,
\eeq
\sindex{fermionic projector!in the presence of linearized gravity}%
where $\Delta P[\gamma^j h^k_j](x,y)$ is the perturbation~\eqref{delP1} of 
the Dirac sea corresponding to the electromagnetic potential $\B=\gamma^j h^k_j$.
The light-cone expansion of $\Delta P(x,y)$ is obtained by substituting the
light-cone expansion of $\Delta P[\gamma^j h^k_j](x,y)$ into~\eqref{dA}
and computing the $y$-derivative.

\section{The Formalism of the Continuum Limit} \label{secreg}
In Section~\ref{seclight} we developed a method for analyzing the
unregularized kernel of the fermionic projector in position space
(see the summary in~\S\ref{seckernunreg}).
Our next goal is to extend these methods in order to include an {\em{ultraviolet
regularization}}.
\sindex{regularization!ultraviolet (UV)}%
Following the method of variable regularization
(see Remark~\ref{remmvr}), the allowed class of regularizations should be as large
as possible. Moreover, we need to analyze in detail how the causal action
and the corresponding EL equations depend on the regularization.
As we shall see, these issues can be treated conveniently in the so-called
{\em{formalism of the continuum limit}}, which is also most suitable for explicit computations.
\sindex{continuum limit!formalism of the}%

The formalism of the continuum limit was first introduced in~\cite[Chapter~4]{PFP}, based on earlier
considerations in the unpublished preprint~\cite{endlich}. In particular, the analysis
in~\cite[Sections~4.3--4.5]{PFP} puts the formalism on a rigorous basis.
For better readability, we here follow the original ideas in~\cite{endlich} and develop
the formalism from a more computational perspective.
This makes it possible to explain the main points of the formalism in a non-technical way.
Generalizing the concepts, we then obtain the formalism of the continuum limit.
In order avoid repetitions, we only outline the general derivation and refer the reader
interested in the details to~\cite[Sections~4.3--4.5]{PFP} and Appendix~\ref{l:appA}.

\subsectionn{Example: The $i \varepsilon$-Regularization} \label{secieps}
In Section~\ref{secmink} we introduced the UV regularization in Minkowski space
using general regularization operators (see Definition~\ref{defreg} and the resulting
regularized kernel in Proposition~\ref{lemma54}). In order to get a better idea of
what the effect of the regularization is, we now consider an explicit example.
To this end, we assume that the regularized kernel of the fermionic projector,
denoted again by~$P^\varepsilon(x,y)$, is {\em{homogeneous}} in the sense
that it depends only on the difference vector~$\xi:=y-x$.
\sindex{fermionic projector!homogeneous}%
Then the kernel can be written as a Fourier integral
\beq P^\varepsilon(x,y) = \int \frac{d^4k}{(2 \pi)^4}\: \hat{P}^\varepsilon(k)\:
e^{-ik(x-y)} \label{B}
\eeq
\nindex{an4@$P^\varepsilon(x,y)$ -- regularized kernel of fermionic projector}%
\sindex{fermionic projector!regularized kernel of}%
\sindex{regularization!$i \varepsilon-$}%
with a distribution~$\hat{P}^\varepsilon(k)$.
From the computational point of view, the simplest possible regularization method is to
modify the unregularized kernel~\eqref{Pxyvac} by inserting a convergence-generating
exponential factor. This leads us to choosing
\beq \label{convergence-generate}
\hat{P}^\varepsilon(k) = (\slashed{k} + m) \,\delta(k^2-m^2)\, \Theta(-k^0)
\: \exp \big( \varepsilon k^0 \big)\:,
\eeq
where~$\varepsilon>0$ is the regularization length.
The convergence-generating factor ensures that the Fourier integral~\eqref{B}
converges pointwise for any vector~$\xi \in \scrM$.
Moreover, differentiating~\eqref{B} with respect to~$x$ or~$y$ gives
rise to powers of~$k$. Since these polynomial factors are dominated
by the convergence-generating exponential factor, the Fourier integral again
converges pointwise. We thus conclude that~$P^\varepsilon(x,y)$ is a smooth function,
\[ P^\varepsilon(.,.) \in C^\infty(\scrM \times \scrM) \:. \]
Therefore, all composite expressions in the kernel of the fermionic projector are well-defined
(like the closed chain~\eqref{Axydef}, its eigenvalues~$\lambda^{xy}_1,\ldots \lambda^{xy}_{2n}$,
the Lagrangian~\eqref{Ldiff}, the integrands in~\eqref{trconstraint} and~\eqref{Tdef}
as well as the kernel~$Q(x,y)$ in~\eqref{delLdef}).
But clearly, the singularities on the light-cone reappear in the limit~$\varepsilon \searrow 0$,
and the composite expressions will diverge.
In other words, the limit~$\varepsilon \searrow 0$ is a {\em{singular limit}}.
Our goal is to analyze this singular limit in detail.

The effect of the convergence-generating factor in~\eqref{convergence-generate}
can be described conveniently in position space. Namely, introducing the short notations
\[ \omega = k^0 \qquad \text{and} \qquad \xi = (t, \vec{x}) \:, \]
one can combine the exponential with the phase factor of the Fourier transform,
\[ \exp(\varepsilon k^0)\: e^{i k \xi} = e^{i \omega (t - i \varepsilon) - i \vec{k} \vec{x}} \:. \]
This shows that the regularization amounts to the replacement
\beq \label{regreplace}
t \rightarrow t - i \varepsilon \:.
\eeq
This simple replacement rule motivates the name {\bf{$i \varepsilon$-regularization}}.

In order to illustrate how to work with this regularization, we next derive explicit
formulas for the fermionic projector of the vacuum with this regularization.
Our starting point is the light-cone expansion of the unregularized fermionic
projector~\eqref{fprep}. More specifically, pulling the Dirac matrices out of
the Fourier integral~\eqref{Pdiff} and expanding in a Taylor series in the mass
parameter using~\eqref{l:3zz} and~\eqref{Tregdef}, we obtain
\[ P^\text{vac}(x,y) = (i \Pdd + m)\: T_{m^2}
= (i \Pdd + m)\: \bigg( \sum_{n=0}^\infty \frac{m^{2n}}{n!}\:  T^{(n)} + \text{(smooth contributions)} \bigg) . \]
Since we are again interested mainly in the behavior of the singularities, for simplicity we shall disregard
the smooth contributions. Clearly, such smooth contributions are important, and they also affect the singularities
of composite expressions on the light cone (for example if multiplied by a singular contribution
when forming the closed chain). But of course, smooth contributions can be treated in composite
expressions in a straightforward way. Therefore, we now focus on the singularities and do
all computations modulo smooth contributions.
\sindex{error term!smooth contribution}%
Then the residual argument shows that the~$T^{(n)}$ satisfy the same computation
rules as the Green's functions in~\eqref{l:7} and~\eqref{l:20a},
\beq \label{Trules}
\frac{\partial}{\partial x^k} T^{(l)}(x,y) = -\frac{\partial}{\partial y^k} T^{(l)}(x,y)
= \frac{1}{2} \: \xi_k \: T^{(l-1)}(x,y)
\eeq
(again valid up to smooth contributions; for an explicit derivation see Exercise~\ref{ex2.10}).
We thus obtain the light-cone expansion
\beq \label{Pvaclight}
P^\text{vac}(x,y) = \frac{i \slashed{\xi}}{2} \sum_{n=0}^\infty \frac{m^{2n}}{n!}\: T^{(-1+n)}
+ \sum_{n=0}^\infty \frac{m^{2n+1}}{n!}\: T^{(n)}
\eeq
(where in analogy to~\eqref{l:29z} we use~\eqref{Trules} to {\em{define}}~$T^{(-1)}$).

The next step is to apply the replacement rule~\eqref{regreplace}.
The factor~$\slashed{\xi}$ becomes
\beq \label{xiprule}
\slashed{\xi} \rightarrow \slashed{\xi}^\varepsilon := (t-i \varepsilon) \gamma^0 - \vec{\xi} \vec{\gamma} \:.
\eeq
\nindex{bl0@$\xi^\varepsilon$ -- Minkowski vector~$y-x$ with $i \varepsilon$-regularization}%
In order to regularize the factors~$T^{(l)}$, we first note that,
applying the replacement rule~\eqref{regreplace} to the distribution~$T_a$ computed in Lemma~\ref{lemmaTintro},
one really obtains a smooth function. Moreover, using the series expansion~\eqref{l:3.1},
one can compute the factors~$T^{(n)}$ as defined by~\eqref{l:3zz} and~\eqref{Tregdef}.
When doing so, it is most convenient to combine the principal part with the $\delta$-contribution
as well as the logarithm with the Heaviside function by using the identities
\begin{align*}
\frac{\PP}{\xi^2} + i \pi \delta \big( \xi^2 \big) \: \epsilon \big( \xi^0 \big) 
&= \lim_{\nu \searrow 0} \: \frac{1}{\xi^2-i \nu \xi^0} \\
\log \big|(y-x)^2 \big| + i \pi \:\Theta \big( \xi^2 \big)\: \epsilon \big( \xi^0 \big)
&= \lim_{\nu \searrow 0} \Big( \log \big( \xi^2-i \nu \xi^0 \big) - i \pi \Big) ,
\end{align*}
where the logarithm is understood in the complex plane which is cut along the positive real axis
such that~$\lim_{\nu \searrow 0} \log(x+i \nu)$ is real for $x>0$.
This gives
\begin{align}
T^{(0)} &\rightarrow -\frac{1}{8 \pi^3}\: \frac{1}{(t-i \varepsilon)^2 - \big|\vec{\xi} \big|^2} \label{rule0} \\
T^{(1)} &\rightarrow \frac{1}{32 \pi^3}\: \log \Big( (t-i \varepsilon)^2 - \big|\vec{\xi} \big|^2 \Big) \label{rule1} \:,
\end{align}
and similar for the other distributions~$T^{(n)}$.
These replacement rules are compatible with our earlier computation rules like~\eqref{Trules}.
These rules can also be used to compute~$T^{(-1)}$ via~\eqref{l:29z} to obtain
\beq \label{rulem1}
T^{(-1)} \rightarrow -\frac{1}{2 \pi^3}\: \frac{1}{\big( (t-i \varepsilon)^2 - r^2 \big)^2} \:,
\eeq
where we set~$r=|\vec{\xi}|$.

Next, in~\eqref{Pvaclight} we apply the replacement rule~\eqref{xiprule}
and replace the factors~$T^{(l)}$ according to rules like~\eqref{rule0}--\eqref{rulem1}.
We thus obtain the regularized fermionic projector of the vacuum~$P^\varepsilon(x,y)$.
The kernel~$P^\varepsilon(y,x)$ is obtained by taking the conjugate with respect to
the spin scalar product (see~\eqref{Pxysymm} or~\eqref{Pkersymm}).
Then one can form the closed chain~$A_{xy}$ by~\eqref{Axydef} and compute all other quantities
of interest. In order to give a concrete example, let us consider the massless case.
Then
\begin{align*}
P(x,y) &= \frac{i}{2}\: \slashed{\xi} \: T^{(-1)} \qquad \text{and thus} \\
P^\varepsilon(x,y) &=-\frac{i}{4 \pi^3}\: \frac{(t-i \varepsilon) \gamma^0 - \vec{\xi} \vec{\gamma}}
{\big( (t-i \varepsilon)^2 - r^2 \big)^2} \\
P^\varepsilon(y,x) &= P^\varepsilon(x,y)^*
= \frac{i}{4 \pi^3}\: \frac{(t+i \varepsilon) \gamma^0 - \vec{\xi} \vec{\gamma}}
{\big( (t+i \varepsilon)^2 - r^2 \big)^2} \\
A^\varepsilon_{xy} &= P^\varepsilon(x,y)\, P^\varepsilon(y,x) \\
&= \frac{1}{16 \pi^6}\: \frac{1}{\big| (t-i \varepsilon)^2 - r^2 \big|^4}\;
\Big( (t-i \varepsilon) \gamma^0 - \vec{\xi} \vec{\gamma} \Big) 
\Big( (t+i \varepsilon) \gamma^0 - \vec{\xi} \vec{\gamma} \Big) \:.
\end{align*}
Simplifying the Dirac matrices according to
\beq \label{iecont}
\big(\slashed{\xi} - i \varepsilon \gamma^0 \big) \big(\slashed{\xi} + i \varepsilon \gamma^0 \big)
= \xi^2 - i \varepsilon [\gamma^0, \slashed{\xi}] + \varepsilon^2 \:,
\eeq
we obtain
\beq \label{Axyvac}
A^\varepsilon_{xy}
= \frac{1}{16 \pi^6}\: \frac{(t^2-r^2) - i \varepsilon [\gamma^0, \slashed{\xi}] + \varepsilon^2}
{\big| (t-i \varepsilon)^2 - r^2 \big|^4}\:.
\eeq
In order to compute the eigenvalues of this matrix, the task is to diagonalize the
bilinear contribution~$i \varepsilon [\gamma^0, \slashed{\xi}]$. The calculation
\[ \big( i \varepsilon [\gamma^0, \slashed{\xi}] \big)^2
= -4 \varepsilon^2 \:\gamma^0 (\vec{\xi} \vec{\gamma}) \gamma^0 (\vec{\xi} \vec{\gamma})
= 4 \varepsilon^2 \:\gamma^0 \gamma^0 (\vec{\xi} \vec{\gamma}) (\vec{\xi} \vec{\gamma}) 
= -4 \varepsilon^2 \, |\vec{\xi}|^2 < 0 \]
shows that this bilinear contribution has complex eigenvalues.
Thus the {\em{regularization makes the spacelike region larger}}.
As we shall see below, this happens in a much more general setting.
It is a desirable effect because it decreases the causal action.

Clearly, the singular behavior of the resulting expressions in the limit~$\varepsilon \searrow 0$ is rather complicated.
However, one limiting case, which will be important later on, is relatively easy to handle.
This limiting case is to consider the region {\em{close to the light cone}} and {\em{away from the origin}}.
For simplicity, we restrict attention to the {\em{upper light cone}}~$t \approx r$ (but clearly, the lower
light cone can be treated similarly). Then ``close to the light cone'' means that~$t-r$
is much smaller than~$r$, whereas ``away from the origin'' means that~$\varepsilon$ is much smaller
that~$r$. Under these assumptions, we have approximately
\[ (t-i \varepsilon)^2 - r^2 = (t+r-i \varepsilon) (t-r-i \varepsilon) \approx 2r\, (t-r-i \varepsilon)\:. \]
In order to make the approximation precise, we write the error term as
\beq \label{errorterms}
(t-i \varepsilon)^2 - r^2 = 2r\, (t-r-i \varepsilon) \left(1+ \O \Big( \frac{t-r}{r} \Big)
+ \O \Big( \frac{\varepsilon}{r} \Big) \right) .
\eeq
Computing up to error terms of this type, the above formulas~\eqref{rule0}--\eqref{rulem1}
can be simplified to
\begin{align}
T^{(0)} &\rightarrow -\frac{1}{8 \pi^3}\: \frac{1}{2r\: (t-r-i \varepsilon)} \label{T0app} \\
T^{(1)} &\rightarrow \frac{1}{32 \pi^3}\: \log \big( 2r \:(t-r-i \varepsilon) \big) \label{T1app} \\
T^{(-1)} &\rightarrow -\frac{1}{8 \pi^3\,r^2}\: \frac{1}{(t-r-i \varepsilon)^2} \:. \label{Tm1app}
\end{align}
Using this approximation, the closed chain~\eqref{Axyvac} simplifies to
\[ A^\varepsilon_{xy}
= \frac{1}{256 \pi^6\,r^4}\: \frac{(t^2-r^2) - i \varepsilon [\gamma^0, \slashed{\xi}] + \varepsilon^2}{|t-r-i \varepsilon|^4}\:. \]
Moreover, the numerator can be further simplified.
We first note that, since~$\xi$ is close to the light cone, the factor~$\xi^2$ can be arbitrarily small.
Therefore, despite the factor~$\varepsilon$, the summand~$\varepsilon [\gamma^0, \slashed{\xi}]$ cannot be left out.
But the summand~$\varepsilon^2$ is of higher order in~$\varepsilon/r$ and can be omitted.
We conclude that
\beq \label{Axyfinal}
A^\varepsilon_{xy}
= \frac{1}{256 \pi^6\,r^4}\: \frac{(t^2-r^2) - i \varepsilon [\gamma^0, \slashed{\xi}]}{|t-r-i \varepsilon|^4}
\left(1+ \O \Big( \frac{t-r}{r} \Big) + \O \Big( \frac{\varepsilon}{r} \Big) \right) .
\eeq

Clearly, composite expressions diverge in the limit~$\varepsilon \searrow 0$.
In order to analyze this singular behavior, the proper method is
to evaluate weakly in~$t$ for fixed~$r$. Thus one considers integrals of the form
\beq \label{weval}
\int_{-\infty}^\infty \eta(t) \: \big(\cdots\big) \: dt
\eeq
for a smooth test function~$\eta$, where~``$\cdots$'' stands for a composite expression
in the~$T^{(n)}$ and~$\overline{T^{(n)}}$. Then ``$\cdots$'' is a meromorphic function in~$t$ with poles
at~$t=\pm r \pm i \varepsilon$. This makes it possible to compute the integral with the help
of residues. The reader interested in an explicit example is referred to Exercise~\ref{ex2.9}.
Here we proceed by compiling and explaining a few general conclusions which will be important later on.

\begin{itemize}[leftmargin=2em]
\item[(a)] The integrand in~\eqref{weval} has poles at~$t=\pm r \pm i \varepsilon$.
Again restricting attention to the upper light cone, we only need to consider the
poles at~$t=r \pm i \varepsilon$. When computing the residues at these points,
the variable~$t-r$ is of the order~$\varepsilon$. Therefore, the two error terms
in~\eqref{errorterms} become the same. For convenience, we usually write the error terms as
\beq \label{hot}
\cdots + \text{(higher orders in~$\varepsilon/|\vec{\xi}|$)}\:.
\eeq
\sindex{error term!higher order in $\varepsilon / \vert \vec{\xi} \vert$}%
Moreover, the theorem of residues gives rise to contributions where the test function~$\eta$
is differentiated. Every such derivative gives rise to an additional factor of~$\varepsilon$.
In order to keep the dimensions of length, we write the resulting error terms in the form
\beq \label{hotmacro}
+ \text{(higher orders in~$\varepsilon/\ell_\text{macro}$)}\:,
\eeq
\nindex{bl2@$\ell_\text{macro}$ -- macroscopic length scale}%
\sindex{error term!higher order in $\varepsilon/\ell_\text{macro}$}%
where~$\ell_\text{macro}$ denotes the ``macroscopic'' length scale on which~$\eta$ varies.
\item[(b)] The scaling of the integral~\eqref{weval} in~$\varepsilon$ and~$r$
can be described by
\beq \label{weakscale}
T^{(n)} \sim \big( \varepsilon \,|\vec{\xi}| \big)^{n-1}
\qquad \text{and} \qquad dt \sim \varepsilon \:.
\eeq
The resulting scaling of a composite expression in powers of~$1/(\varepsilon \,|\vec{\xi}|)$ is
referred to as the {\em{degree}} of the expression.
\sindex{degree on the light cone}%
One should carefully distinguish
the powers of~$1/(\varepsilon \,|\vec{\xi}|)$ defining the degree from the 
factors~$\varepsilon/|\vec{\xi}|$ appearing in the error terms in~\eqref{hot}.
To make this distinction, it is important that we have two independent variables~$\varepsilon$ and~$|\vec{\xi}|$,
and that we consider the scaling behavior in both variables.
In this way, when evaluating a sum of expressions of different degrees,
our methods make it possible to evaluate each degree separately, each with error
terms of the form~\eqref{hot} and~\eqref{hotmacro}.
\item[(c)] The scaling behavior of the factors~$\xi^\varepsilon$ is more subtle, as we now explain.
If a factor~$\xi^\varepsilon$ is contracted to Dirac matrices or to a macroscopic function
(like a gauge potential or the Dirac current), we may simply disregard the regularization~\eqref{xiprule}, i.e.
\begin{align*}
\slashed{\xi}^\varepsilon &= \slashed{\xi} + \text{(higher orders in~$\varepsilon/|\vec{\xi}|$)} \\
\xi^\varepsilon_j \, f^j &= \xi_j  \, f^j + \text{(higher orders in~$\varepsilon/|\vec{\xi}|$)}
\end{align*}
(where~$f_j$ is a macroscopic vector field). We refer to such factors~$\xi^\varepsilon$ as {\em{outer factors}}.
\sindex{outer factor~$\xi$}%
\item[(d)] Two factors~$\xi^\varepsilon$ which are contracted to each other are called {\em{inner factors}}.
\sindex{inner factor~$\xi$}%
Since the resulting function~$\xi^2$ is very small on the light cone, the
factor~$\varepsilon$ in~\eqref{xiprule} must be taken into account, i.e.
\beq \label{xieps}
\big(\xi^\varepsilon \big)^2 = (t- i \varepsilon)^2 - |\vec{\xi}|^2 = t^2 - |\vec{\xi}|^2
- 2 i \varepsilon t - \varepsilon^2 \:.
\eeq
But similar as in~\eqref{errorterms}, the quadratic term in~$\varepsilon$ may be dropped, i.e.
\beq \label{contract1}
\big(\xi^\varepsilon \big)^2 = t^2 - |\vec{\xi}|^2 - 2 i \varepsilon t + \text{(higher orders in~$\varepsilon/|\vec{\xi}|$)} \:.
\eeq
The general rule is that in every contraction, the factors~$i \varepsilon$ must be taken into account
linearly. This means in particular that the regularized factors~$\xi^\varepsilon$ are no longer real,
but must be treated as complex-valued vectors. Taking their complex conjugate corresponds
to flipping the sign of~$\varepsilon$, i.e.
\[ \overline{\xi}^\varepsilon =  (t+i \varepsilon, \vec{\xi}) \:. \]
Taking the adjoint of~$\slashed{\xi}^\varepsilon$ (with respect to the spin scalar product),
we need to take the complex conjugate of~$\xi^\varepsilon$, i.e.
\[ \big(\overline{\xi}^\varepsilon \big)^* = \overline{\slashed{\xi}^\varepsilon} \:. \]
One must carefully distinguish~$\xi^\varepsilon$ and~$\overline{\xi^\varepsilon}$ in all computations.
\item[(e)] Clearly, a factor~$\xi^\varepsilon$ may also be contracted to a factor~$\overline{\xi}^\varepsilon$,
or two factors~$\overline{\xi}^\varepsilon$ may be contracted to each other.
In these cases, we again refer to the factors~$\xi^\varepsilon$ and~$\overline{\xi}^\varepsilon$
as {\em{inner factors}}.
\sindex{inner factor~$\xi$}%
Since we only take into account~$\varepsilon$ linearly, we get
\begin{align*}
\big(\overline{\xi}^\varepsilon \big)^2 &= t^2 - |\vec{\xi}|^2
+ 2 i \varepsilon t + \text{(higher orders in~$\varepsilon/|\vec{\xi}|$)} \\
(\xi^\varepsilon)_j \:(\overline{\xi}^\varepsilon)^j
&= t^2 - |\vec{\xi}|^2 + \text{(higher orders in~$\varepsilon/|\vec{\xi}|$)} \:.
\end{align*}
Comparing these formulas with~\eqref{contract1}, one sees that
\beq \label{contract2}
(\xi^\varepsilon)_j \,(\overline{\xi}^\varepsilon)^j = \frac{1}{2} \Big( \big( \xi^\varepsilon \big)^2 +
\big( \overline{\xi}^\varepsilon \big)^2 \Big) + \text{(higher orders in~$\varepsilon/|\vec{\xi}|$)} \:.
\eeq
This identity will appear later in a much more general context as the
so-called {\em{contraction rule}}.
\sindex{contraction rule}%

After applying this contraction rule, one gets products of the form~$(\xi^\varepsilon)^2\, T^{(l)}$.
We remark that such  products can be further simplified.
Namely, according to the residual argument, the rule~\eqref{l:22a} also holds for~$S^{(l)}$
replaced by~$T^{(l)}$, up to smooth contributions. In fact, this rule even holds with
regularization, i.e.\ for all~$l \geq 0$
\beq \label{l:20reg}
\big(\xi^{(p)})^2 \:T^{(l)} = -4 p \, T^{(l-1)} + \text{(smooth contributions)}
\eeq
(the smooth contributions are of course important, but they can be treated together with the
other smooth contributions to the fermionic projector as outlined in~\S\ref{seckernunreg}).
The reader interested in the details of the derivation of the identity~\eqref{l:20reg} is referred to
Exercise~\ref{ex2.10}.
\item[(f)] We mention one more structure which in the present example is easy to understand,
and which will come up in a more general context later on. Namely, suppose that
the composite expression ``$\cdots$'' in~\eqref{weval} can be written as a time derivative.
Then we can integrate by parts,
\[ \int_{-\infty}^\infty \eta(t) \: \frac{\partial F}{\partial t}\: dt
= -\int_{-\infty}^\infty \big( \partial_t \eta(t) \big) \: F(t)\: dt \:. \]
Since derivatives of the test function scale like factors~$1/\ell_\text{macro}$, this
contribution is much smaller than expected from the scalings~\eqref{weakscale}. We write
\beq \label{zerorel}
\int_{-\infty}^\infty \eta(t) \: \frac{\partial F}{\partial t}\: dt
= 0 \;+\; \text{(higher orders in~$\varepsilon/\ell_\text{macro}$)} \:.
\eeq
This relation shows that certain composite expressions in the factors~$T^{(n)}$ and~$\overline{T^{(n)}}$
vanish when evaluated weakly on the light cone. In other words, there are relations between
composite expressions.

These relations are expressed most conveniently in terms of so-called {\em{integration-by-parts rules}}.
\sindex{integration-by-parts rule}%
The starting point for deriving these rules is the identity~\eqref{Trules}
which holds up to smooth contributions, i.e.\ for all~$l \geq 0$
\beq \label{l:7T}
\frac{\partial}{\partial x^k} T^{(l)}(x,y) =
\frac{1}{2} \: (y-x)_k \: T^{(l-1)}(x,y) + \text{(smooth contributions)}
\eeq
(recall that in the case~$l=0$, this relation serves as the {\em{definition}} of~$T^{(-1)}$).
For an explicit derivation of the identity~\eqref{l:7T} we again refer to Exercise~\ref{ex2.10}.
Considering a derivative in time direction (and noting that~$\partial_t = -\partial_{x^0}$), we obtain
\[ \frac{\partial}{\partial t} T^{(l)}(x,y) =
-\frac{1}{2} \: t \: T^{(l-1)}(x,y) + \text{(smooth contributions)} \:. \]
Near the upper light cone, we can write this identity as
\begin{align*}
\frac{1}{r}\: &\frac{\partial}{\partial t} T^{(l)}(x,y) = -\frac{1}{2} \: T^{(l-1)}(x,y) \\
&+ \text{(smooth contributions)}
+ \text{(higher orders in~$\varepsilon/|\vec{\xi}|$)} \:.
\end{align*}
Introducing the abbreviation
\beq \label{nabladef}
\nabla := \frac{1}{t}\: \frac{\partial}{\partial t} \:,
\eeq
\nindex{bl4@$\nabla$ -- derivation on the light cone}%
we thus obtain the relations
\beq \label{nablarel}
\nabla T^{(l)} = -\frac{1}{2} \: T^{(l-1)} \:.
\eeq
Moreover, the identity~\eqref{zerorel} can be written in the short symbolic form
\beq \label{nablazero}
\nabla \big( \cdots \big) = 0
+ \text{(smooth contributions)}
+ \text{(higher orders in~$\varepsilon/|\vec{\xi}|$)} \:,
\eeq
where~``$\cdots$'' again stands for a composite expression in the~$T^{(n)}$ and~$\overline{T^{(n)}}$.
\end{itemize}
We finally remark that, at this stage, neglecting all terms of the order~\eqref{hot}
merely is a matter of convenience. In fact, one can also take into account
the higher orders in~$\varepsilon/|\vec{\xi|}|$ by performing an expansion
in powers of~$\varepsilon/|\vec{\xi|}$. Such an expansion is called
{\em{regularization expansion}}.
\sindex{regularization expansion}%
We will come back to the regularization
expansion in~\S\ref{secdercl}. But before, we analyze the situation for
more general regularizations.

\subsectionn{Example: Linear Combinations of $i \varepsilon$-Regularizations} \label{seclineps}
Clearly, the $i \varepsilon$-regularization is very special and ad-hoc.
In order to get a first idea on what happens for more general regularizations, it
is instructive to consider linear combinations of $i \varepsilon$-regularizations. To this end,
we choose an integer~$N$ and generalize~\eqref{convergence-generate} to
\beq \label{lincomb}
\hat{P}^\varepsilon(k) = (\slashed{k} + m) \,\delta(k^2-m^2)\, \Theta(-k^0)\:
\bigg( \sum_{a=1}^N c_a \,\exp \big( \varepsilon \,d_a \,k^0 \big) \bigg)
\eeq
with positive parameters~$d_1,\ldots, d_N$ and real numbers~$c_1,\ldots, c_N$ which add up to one,
\[ c_1 + \cdots + c_N = 1 \:. \]
\sindex{regularization!linear combination of $i \varepsilon$-}%
In fact, by choosing~$N$ sufficiently large, with this ansatz one can approximate any regularization
of the form
\beq \label{regmult}
\hat{P}^\varepsilon(k) = (\slashed{k} + m) \,\delta(k^2-m^2)\, \Theta(-k^0)\:
\hat{h} \big( k^0 \big) \:,
\eeq
corresponding to a regularization by convolution with a function~$h(t)$
(being a special case of the regularizations in Example~\ref{exmollify}).

For regularizations of the form~\eqref{lincomb}, we can again evaluate weakly
on the light cone~\eqref{weval}. It turns out that the scalings in~$\varepsilon$
and~$|\vec{\xi}|$ are exactly the same as for the $i \varepsilon$-regularization.
In order to see this in a simple setting, one can consider a polynomial in~$T^{(n)}$
and~$\overline{T^{(n)}}$,
\[ T^{(l_1)} \cdots T^{(l_\alpha)} \:
\overline{T^{(n_1)} \cdots T^{(n_\beta)}} \:. \]
When evaluating weakly on the light cone, one can pull the sums 
of the linear combinations in~\eqref{lincomb} out of the integral, i.e.
\beq \begin{split} \label{Tdint}
&\int_{-\infty}^\infty \eta(t)\: T^{(l_1)} \cdots T^{(l_\alpha)} \:
\overline{T^{(n_1)} \cdots T^{(n_\beta)}}\: dt \\
&= \sum_{a_1, \ldots, a_\alpha, b_1, \ldots, b_\beta=1}^N \!\!\!\!\!\!\!\!\!\!
c_{a_1} \cdots c_{a_\alpha}\; c_{b_1} \cdots c_{b_\beta}
\int_{-\infty}^\infty \eta(t)\: T^{(l_1)}_{d_{a_1}} \cdots T^{(l_\alpha)}_{d_{a_\alpha}} \:
\overline{T^{(n_1)}_{d_{b_1}} \cdots T^{(n_\beta)}_{d_{b_\beta}}}\: dt \:,
\end{split}
\eeq
where~$T^{(n)}_d$ denotes the $i\varepsilon$-regularization with~$\varepsilon$
replaced by~$\varepsilon d$.
Again computing up to the error terms~\eqref{hot} and~\eqref{hotmacro},
one can again use the explicit formulas for~$T^{(n)}$ like~\eqref{T0app}--\eqref{Tm1app}
and analyze the integral with residues.
The only difference compared to the analysis of the $i \varepsilon$-regularization is that
one has many poles at positions~$t=r \pm i \varepsilon d_a$, and the residue theorem gives
sums over these poles. But obviously, this has no effect on all scalings.

The contraction of the inner factors must be handled with care, as we now explain.
Using~\eqref{xiprule} and forming linear combinations, one sees that
the factor~$\slashed{\xi} T^{(n)}$ is to be regularized according to
\beq \label{xiTreg}
\slashed{\xi}\, T^{(n)} \rightarrow \sum_{a=1}^N c_a \Big( \big(t - i \varepsilon d_a \big) \gamma^0
-\vec{\xi} \vec{\gamma} \Big) \:T^{(n)}_{d_a}
\eeq
(with~$T^{(n)}_d$ again as in~\eqref{Tdint}). When forming composite expressions, one must
take into account that the regularized factors~$\xi$ and~$T^{(n)}$ both carry the same summation index.
Therefore, one should regard the factors~$T^{(n)}$ and~$\slashed{\xi}$
as belonging together. It is useful to make this connection explicit in the notation. Therefore, we
discard~\eqref{xiprule} and introduce instead the more general rule
\[ \slashed{\xi} \,T^{(n)} \rightarrow \slashed{\xi}^{(n)} \, T^{(n)} \:, \]
where the right side is a short notation for the sum in~\eqref{xiTreg}.

Contracting two inner factors~$\xi$ in this formalism gives
\begin{align}
\big( \xi^{(l)} \big)_j &\,T^{(l)} \; \big( \xi^{(n)} \big)^j \,T^{(n)} =
\sum_{a,b=1}^N c_a c_b \; \Big( \big(t - i \varepsilon d_a \big)
,\vec{\xi} \,\Big)_j \:T^{(l)}_{d_a} \;\Big( \big(t - i \varepsilon d_b \big)
,\vec{\xi} \,\Big)^j \:T^{(n)}_{d_b} \notag \\
&= \sum_{a,b=1}^N c_a c_b \:T^{(l)}_{d_a} \,T^{(n)}_{d_b}
\left( t^2 - i \varepsilon t d_a - i \varepsilon t d_b -\varepsilon^2 d_a d_b - |\vec{\xi}|^2 \right) .
\label{epsquadrat}
\end{align}
This is considerably more complicated than~\eqref{xieps}. However, if as in~\eqref{contract1}
we drop the term quadratic in~$\varepsilon$, the formula can be simplified to
\begin{align}
\big( &\xi^{(p)} \big)_j \,T^{(l)} \; \big( \xi^{(q)} \big)^j \,T^{(n)} \notag \\
&= \sum_{a,b=1}^N c_a c_b \:T^{(l)}_{d_a} \,T^{(n)}_{d_b}
\left( t^2 - i \varepsilon t d_a - i \varepsilon t d_b- |\vec{\xi}|^2 \right)
+ \text{(higher orders in~$\varepsilon/|\vec{\xi}|$)} \\
&= \frac{1}{2} \sum_{a,b=1}^N c_a c_b \:T^{(l)}_{d_a} \,T^{(n)}_{d_b}
\left( \Big( \big(t - i \varepsilon d_a \big),\vec{\xi} \,\Big)^2
+ \Big( \big(t - i \varepsilon d_b \big) ,\vec{\xi} \,\Big)^2 \right) \label{squares} \\
&\qquad + \text{(higher orders\ in~$\varepsilon/|\vec{\xi}|$)} \notag \\
&= \frac{1}{2} \Big(\big(\xi^{(l)})^2 + \big(\xi^{(n)})^2 \Big) \, T^{(l)} \, T^{(n)}
+ \text{(higher orders in~$\varepsilon/|\vec{\xi}|$)}\:,
\end{align}
where the squares in~\eqref{squares} denote the Minkowski inner product,
and where in the last step we introduced the notation
\beq \label{prodsimp}
\big(\xi^{(l)})^2 \:T^{(l)} = \sum_{a=1}^N c_a \,\Big( \big(t - i \varepsilon d_a \big)^2 -
\big|\vec{\xi} \big|^2 \Big) \:T^{(l)}_{d_a} \:.
\eeq
In this way, the contraction rules~\eqref{contract1} can be generalized to
\begin{align}
(\xi^{(l)})^j \, (\xi^{(n)})_j &= \frac{1}{2} \left( (\xi^{(l)})^2 + (\xi^{(n)})^2 \right) \label{contract3}
\intertext{Similarly the contraction rule~\eqref{contract2} becomes}
(\xi^{(l)})^j \, \overline{(\xi^{(n)})_j} &=
\frac{1}{2} \left( (\xi^{(l)})^2 + \overline{(\xi^{(n)})^2} \right) . \label{contract4}
\end{align}
We remark that this product can again be simplified using~\eqref{l:20reg}, giving
rise to the computation rule
\[ \big(\xi^{(p)})^2 \:T^{(l)} = -4 p \, T^{(l-1)} + \text{(smooth contributions)} \:. \]
We also remark that the integration-by-parts rules~\eqref{nablarel}
and~\eqref{nablazero} with~$\nabla$ according to~\eqref{nabladef} remain valid,
as one sees immediately by applying~\eqref{nablarel} to each summand in~\eqref{lincomb}
and by noting that~\eqref{zerorel} holds for any regularization.

Working with linear combinations of $i \varepsilon$-regularization gives a first
hint why one should disregard error terms of the form~\eqref{hot} and~\eqref{hotmacro},
as we now explain. Using the method of variable regularization (see Remark~\ref{remmvr}), we must
show that the structure of the effective equations in the continuum limit does not depend
on the details of the regularization. Evaluating weakly on the light cone and neglecting
error terms of the form~\eqref{hot} and~\eqref{hotmacro}, one gets relatively simple
computation rules (like~\eqref{contract3}, \eqref{contract4} or~\eqref{l:20reg}), giving
rise to a formalism which captures the structure of the EL equation independent of
regularization details. However, for example the quadratic term in~$\varepsilon$ in~\eqref{epsquadrat}
\beq \label{quterm}
-\varepsilon^2 \sum_{a,b=1}^N c_a c_b \:d_a d_b \:T^{(l)}_{d_a} \,T^{(n)}_{d_b}
\eeq
has a different structure. Namely,
even after prescribing linear moments as they appear in~\eqref{xiTreg}, there is
a lot of freedom to give the quadratic term in~\eqref{quterm} an arbitrary value.
More generally, if we computed the terms~\eqref{hot} or~\eqref{hotmacro},
these contributions would depend on the regularization in a complicated way,
so much so that without knowing the regularization in detail, it would be impossible to evaluate these
contributions. This is the reason why we shall disregard these contributions.
Clearly, at this stage, the above argument is not quite satisfying because notions like
``complicated'' and ``knowing the regularization in detail'' are somewhat vague.
The argument will be made more precise in~\S\ref{secdercl} using Fourier methods.

\subsectionn{Further Regularization Effects} \label{secfurthereffects}
Working with linear combinations of $i \varepsilon$-regularizations,
one is still in the restrictive class of regularizations of the form~\eqref{regmult}
where the unregularized distribution is multiplied in momentum space by a convergence-generating
function~$\hat{h}(k^0)$. Considering more general regularizations gives rise to
additional effects. We now list those regularization effects will be important later on:
\begin{itemize}[leftmargin=2em]
\itemD The support of the distribution in~\eqref{regmult} can be slightly deformed
from the hyperboloid to another hypersurface.
It turns out that in this case, one can still perform a mass expansion of the form~\eqref{Pvaclight}.
But the regularization of the factors~$T^{(n)}$ also depends on the power of the mass
in the corresponding contribution to the fermionic projector. In order to implement this
effect into our formalism, one adds a subscript~$[.]$ to the factors~$T^{(n)}$ which counts the power in~$m$.
Thus we regularize the contributions to the light-cone expansion according to the rule
\[ m^p \,T^{(n)} \rightarrow m^p \,T^{(n)}_{[p]} \:. \]
\nindex{bl6@$T_{[p]}^{(n)}$ -- ultraviolet regularized $T^{(n)}$ }%
For example, the regularization of the light-cone expansion of the vacuum~\eqref{Pvaclight} now takes the form
\[ P^\varepsilon(x,y) = \sum_{n=0}^\infty \frac{m^{2n}}{n!}\: \frac{i \slashed{\xi}^{(-1+n)}}{2} \:T^{(-1+n)}_{[2n]}
+ \sum_{n=0}^\infty \frac{m^{2n+1}}{n!}\: T^{(n)}_{[2n+1]} \:. \]
Regularizing the fermionic projector in the presence of an external potential,
one gets contributions involving factors~$T^{(n)}_{[p]}$ with the same~$n$ but different values of~$p$.
These factors must be treated as being different (although they clearly coincide without regularization).
\itemD The direction of the vector~$k$ which appears in the factor~$\slashed{k}$ in~\eqref{regmult}
can be slightly changed by the regularization. This leads to the notion of
the {\em{shear of surface states}}.
\sindex{shear of surface states}%
This effect is of importance when inner factors are contracted.
More precisely, one needs to modify the calculation rule~\eqref{l:20reg} to
\[ \big(\xi^{(p)})^2 \:T^{(n)}_{[p]} = -4 \left( n \:T^{(n+1)}_{[p]}
+ T^{(n+2)}_{\{p \}} \right) + \text{(smooth contributions)} \:, \]
where the factors~$T^{(l)}_{\{p \}}$ with curly brackets have the same scaling behavior
as the corresponding factors with square brackets but are regularized differently.
\nindex{bl8@$T^{(n)}_{\{p\}}$ -- factor in continuum limit describing the shear of surface states}%
\itemD There may be additional contributions to~$\hat{P}(k)$ which lie outside the
hyperboloid in~\eqref{regmult} or the deformation thereof.
It turns out that the resulting contributions can be absorbed into the error terms~\eqref{hot}
and~\eqref{hotmacro} (for details see~\S\ref{secdercl}).
\end{itemize}
We also remark that the regularization of neutrinos is more involved because
the regularization must break the chiral symmetry and because the corresponding Dirac sea
can ``mimic'' a Dirac sea of a different mass. In order not to distract from the
main points of our construction, these extensions of the formalism will be introduced
later when we need them (see Section~\ref{l:sec2}).

\subsectionn{The Formalism of the Continuum Limit} \label{sec73}
\sindex{continuum limit}%
After the above motivation and preparations, we now present the formalism of the
continuum limit. In~\S\ref{secdercl} we shall outline the derivation of this formalism
as first given in~\cite[Chapter~4]{PFP}.

Before beginning, we point out that we work {\em{modulo smooth contributions}}
throughout. The reason for this procedure is that the smooth contributions can be
computed in a straightforward manner by first evaluating composite expressions
away from the light cone (where they are smooth) and taking the limit when~$y-x$
approaches the light cone. Clearly, computing the smooth contributions is 
important and not always easy (for details see Appendix~\ref{s:appresum}).
But these computations are not related to the problem of the singularities on the
light cone to be considered here.
\sindex{error term!smooth contribution}%

Our starting point is the light-cone expansion of the unregularized fermionic
projector~$P(x,y)$ (as given in~\S\ref{seckernunreg}).
In order to regularize the light-cone expansion on the length scale~$\varepsilon$, we proceed as follows.
\sindex{light-cone expansion!regularization}%
\sindex{regularization!of the light-cone expansion}%
The smooth contributions are all left unchanged. For the regularization
of the factors~$T^{(n)}$, we employ the replacement rule
\beq \label{contri}
m^p \,T^{(n)} \rightarrow m^p \,T^{(n)}_{[p]}\:,
\eeq
where the factors~$T^{(n)}_{[p]}$ are smooth functions of~$\xi$.
\nindex{bl6@$T_{[p]}^{(n)}$ -- ultraviolet regularized $T^{(n)}$ }%
Fortunately, the rather complicated detailed form of the factors~$T^{(n)}_{[p]}$
will not be needed here, because these functions can be treated symbolically using the
following simple calculation rules. In computations one may treat the~$T^{(n)}_{[p]}$
like complex functions. However, one must be careful when tensor indices of factors~$\slashed{\xi}$
are contracted with each other. Naively, this gives a factor~$\xi^2$ which vanishes on the
light cone and thus changes the singular behavior on the light cone. In order to describe this
effect correctly, we first write every summand of the light cone expansion~\eqref{fprep}
such that it involves at most one factor~$\slashed{\xi}$ (this can always be arranged using
the anti-commutation relations of the Dirac matrices).
We now associate every factor~$\slashed{\xi}$ to the corresponding factor~$T^{(n)}_{[p]}$.
In short calculations, this can be indicated by putting brackets around the two factors,
whereas in the general situation we add corresponding indices
to the factor~$\slashed{\xi}$, giving rise to the replacement rule
\beq \label{xicontri}
m^p \,\slashed{\xi} T^{(n)} \rightarrow m^p \,\slashed{\xi}^{(n)}_{[p]}\, T^{(n)}_{[p]}\:.
\eeq
\nindex{bm2@$\xi^{(n)}_{[p]}$ -- ultraviolet regularized factor $\xi$}%
For example, we write the regularized fermionic projector of the vacuum as
\[ P^\varepsilon = \frac{i}{2} \sum_{n=0}^\infty \frac{m^{2n}}{n!}\: \slashed{\xi}^{(-1+n)}_{[2n]}\, T^{(-1+n)}_{[2n]}
+ \sum_{n=0}^\infty \frac{m^{2n+1}}{n!}\: T^{(n)}_{[2n+1]} \:. \]
\nindex{an4@$P^\varepsilon(x,y)$ -- regularized kernel of fermionic projector}%
\sindex{fermionic projector!regularized kernel of}%

The kernel~$P(y,x)$ is obtained by taking the conjugate (see~\eqref{Pkersymm}).
The conjugates of the factors~$T^{(n)}_{[p]}$ and~$\xi^{(n)}_{[p]}$ are the complex conjugates,
\[ \overline{T^{(n)}_{[p]}} := \big(T^{(n)}_{[p]} \big)^* \qquad \text{and} \qquad
\overline{\xi^{(n)}_{[p]}} := \big(\xi^{(n)}_{[p]} \big)^* \:. \]
One must carefully distinguish between these factors with and without complex conjugation.
In particular, the factors~$\slashed{\xi}^{(n)}_{[p]}$ need not be symmetric,
\[ \big( \slashed{\xi}^{(n)}_{[p]} \big)^* \neq \slashed{\xi}^{(n)}_{[p]} \qquad \text{in general}\:. \]

When forming composite expressions, the tensor indices of the factors~$\xi$ are 
contracted to other tensor indices.
The factors~$\xi$ which are contracted to other factors~$\xi$ are called {\em{inner factors}}.
\sindex{inner factor~$\xi$}%
The contractions of the inner factors are handled with the so-called {\em{contraction rules}}
\sindex{contraction rule}%
\begin{align}
(\xi^{(n)}_{[p]})^j \, (\xi^{(n')}_{[p']})_j &=
\frac{1}{2} \left( z^{(n)}_{[p]} + z^{(n')}_{[p']} \right) \label{eq52} \\
(\xi^{(n)}_{[p]})^j \, \overline{(\xi^{(n')}_{[p']})_j} &=
\frac{1}{2} \left( z^{(n)}_{[p]} + \overline{z^{(n')}_{[p']}} \right) \label{eq53} \\
z^{(n)}_{[p]} \,T^{(n)}_{[p]} &= -4 \left( n \:T^{(n+1)}_{[p]}
+ T^{(n+2)}_{\{p \}} \right) , \label{eq54}
\end{align}
\nindex{bl8@$T^{(n)}_{\{p\}}$ -- factor in continuum limit describing the shear of surface states}%
which are to be complemented by the complex conjugates of these equations.
Here the factors~$z^{(n)}_{[p]}$
\nindex{bm8@$z^{(n)}_{[p]}$ -- abbreviation for $(\xi^{(n)}_{[p]})^2$}%
 can be regarded simply as a book-keeping device
to ensure the correct application of the rule~\eqref{eq54}.
The factors~$T^{(n)}_{\{p\}}$
\nindex{bl6@$T_{[p]}^{(n)}$ -- ultraviolet regularized $T^{(n)}$ }%
have the same scaling behavior as the~$T^{(n)}_{[p]}$,
but their detailed form is somewhat different; we simply treat them as a new class of symbols.
In cases where the lower index does not need to be specified we write~$T^{(n)}_\circ$.
\nindex{bn2@$T^{(n)}_\circ$ -- stands for~$T^{(n)}_{\{p\}}$ or~$T^{(n)}_{[p]}$}%
After applying the contraction rules, all inner factors~$\xi$ have disappeared.
The remaining so-called {\em{outer factors}}~$\xi$
\sindex{outer factor~$\xi$}%
need no special attention and are treated
like smooth functions.

Next, to any factor~$T^{(n)}_\circ$ we associate the {\em{degree}} $\deg T^{(n)}_\circ$ by
\nindex{bn4@$\deg$ -- degree on light cone}%
\sindex{degree on the light cone}%
\[ \deg T^{(n)}_\circ = 1-n \:. \]
The degree is additive in products, whereas the degree of a quotient is defined as the
difference of the degrees of numerator and denominator. The degree of an expression
can be thought of as describing the order of its singularity on the light cone, in the sense that a
larger degree corresponds to a stronger singularity (for example, the
contraction rule~\eqref{eq54} increments~$n$ and thus decrements the degree, in
agreement with the naive observation that the function~$z=\xi^2$ vanishes on the light cone).
Using formal Taylor series, we can expand in the degree. In all our applications, this will
give rise to terms of the form
\beq \label{sfr}
\eta(x,y) \:
\frac{ T^{(a_1)}_\circ \cdots T^{(a_\alpha)}_\circ \:
\overline{T^{(b_1)}_\circ \cdots T^{(b_\beta)}_\circ} }
{ T^{(c_1)}_\circ \cdots T^{(c_\gamma)}_\circ \:
\overline{T^{(d_1)}_\circ \cdots T^{(d_\delta)}_\circ} } \qquad \text{with~$\eta(x,y)$ smooth}\:.
\eeq
The quotient of the two monomials in this equation is referred to as a {\em{simple fraction}}.
\sindex{simple fraction}%

A simple fraction can be given a quantitative meaning by considering one-dimensional integrals
along curves which cross the light cone transversely away from the origin~$\xi=0$.
This procedure is called {\em{weak evaluation on the light cone}}.
\sindex{evaluation on the light cone!weak}%
For our purpose, it suffices to integrate over the time coordinate~$t=\xi^0$ for fixed~$\vec{\xi} \neq 0$.
Moreover, using the symmetry under reflections~$\xi \rightarrow -\xi$, it suffices to consider the upper
light cone~$t \approx |\vec{\xi}|$. The resulting integrals diverge if the regularization
is removed. The leading contribution for small~$\varepsilon$ can be written as
\beq
\int_{|\vec{\xi}|-\varepsilon}^{|\vec{\xi}|+\varepsilon} dt \; \eta(t,\vec{\xi}) \:
\frac{ T^{(a_1)}_\circ \cdots T^{(a_\alpha)}_\circ \:
\overline{T^{(b_1)}_\circ \cdots T^{(b_\beta)}_\circ} }
{ T^{(c_1)}_\circ \cdots T^{(c_\gamma)}_\circ \:
\overline{T^{(d_1)}_\circ \cdots T^{(d_\delta)}_\circ} }
\;\approx\; \eta(|\vec{\xi}|,\vec{\xi}) \:\frac{c_\reg}{(i |\vec{\xi}|)^L}
\;\frac{\log^r (\varepsilon |\vec{\xi}|)}{\varepsilon^{L-1}}\:, \label{asy}
\eeq
where~$L$ is the degree of the simple fraction and~$c_\reg$, the so-called {\em{regularization parameter}},
\nindex{bn6@$L$ -- degree of simple fraction}%
\sindex{regularization parameter}%
\nindex{bn8@$c_{\text{reg}}$ -- regularization parameter}%
is a real-valued function of the spatial direction~$\vec{\xi}/|\vec{\xi}|$ which also depends on
the simple fraction and on the regularization details
(the error of the approximation will be specified below). The integer~$r$ describes
a possible logarithmic divergence. Apart from this logarithmic divergence, the
scalings in both~$\xi$ and~$\varepsilon$ are described by the degree.

When analyzing a sum of expressions of the form~\eqref{sfr}, one must
know if the corresponding regularization parameters are related to each other.
In this respect, the {\em{integration-by-parts rules}}
\sindex{integration-by-parts rule}%
are important, which are described
symbolically as follows. On the factors~$T^{(n)}_\circ$ we introduce a derivation~$\nabla$ by
\[ \nabla T^{(n)}_\circ = T^{(n-1)}_\circ \:. \]
\nindex{bl4@$\nabla$ -- derivation on the light cone}%
Extending this derivation with the Leibniz and quotient rules to simple fractions, the
integration-by-parts rules state that
\beq \label{ipart}
\nabla \left( \frac{ T^{(a_1)}_\circ \cdots T^{(a_\alpha)}_\circ \:
\overline{T^{(b_1)}_\circ \cdots T^{(b_\beta)}_\circ} }
{ T^{(c_1)}_\circ \cdots T^{(c_\gamma)}_\circ \:
\overline{T^{(d_1)}_\circ \cdots T^{(d_\delta)}_\circ} }
\right) = 0 \:.
\eeq
These rules give relations between simple fractions. The name is motivated by the
integration-by-parts method as explained for the $i \varepsilon$-regularization in~\eqref{zerorel}.
Simple fractions which are not related
to each other by the integration-by-parts rules are called {\em{basic fractions}}. As shown
in~\cite[Appendix~E]{PFP}, there are no further relations between the basic fractions.
Thus the corresponding {\em{basic regularization parameters}} are linearly independent.
\sindex{regularization parameter!basic}%

The above symbolic computation rules give a convenient procedure to evaluate composite expressions
in the fermionic projector, referred to as the {\em{analysis in the continuum limit}}:
\sindex{continuum limit!analysis in the}%
After applying the contraction rules and expanding in the degree, 
the EL equations can be rewritten as equations involving
a finite number of terms of the form~\eqref{sfr}. By applying the integration-by-parts rules,
we can arrange that all simple fractions are basic fractions.
We evaluate weakly on the light cone~\eqref{asy} and collect the terms according to their
scaling in~$\xi$. Taking for every given scaling in~$\xi$ only the leading pole in~$\varepsilon$,
we obtain equations which involve linear combinations of smooth functions and basic regularization parameters.
We consider the basic regularization parameters as empirical parameters describing the
unknown microscopic structure of space-time.
\sindex{regularization parameter!basic}%
We thus end up with equations involving
smooth functions and a finite number of free parameters. We point out that these free parameters
cannot be chosen arbitrarily because they might be constrained by inequalities
(see the discussion after~\cite[Theorem~E.1]{PFP}). Also, the values of the basic regularization
parameters should ultimately be justified by an analysis of vacuum minimizers of the causal
action principle.

We finally specify the error of the above expansions. By not regularizing the bosonic potentials
and fermionic wave functions, we clearly disregard the
\beq \label{ap1}
\text{higher orders in~$\varepsilon/\ell_\text{macro}$}\:.
\eeq
\sindex{error term!higher order in $\varepsilon/\ell_\text{macro}$}%
Furthermore, in~\eqref{asy} we must stay away from the origin, meaning that we neglect the
\beq \label{ap2}
\text{higher orders in~$\varepsilon/|\vec{\xi}|$}\:.
\eeq
\sindex{error term!higher order in $\varepsilon / \vert \vec{\xi} \vert$}%
The higher oder corrections in~$\varepsilon/|\vec{\xi}|$ depend on the fine structure of
the regularization and thus seem unknown for principal reasons. Neglecting the terms in~\eqref{ap1}
and~\eqref{ap2} also justifies the formal Taylor expansion in the degree.
Clearly, leaving out the terms~\eqref{ap2} is justified only if~$|\vec{\xi}| \gg \varepsilon$.
Therefore, whenever using the above formalism, we must always ensure that~$|\vec{\xi}|$
is much larger than~$\varepsilon$
(we will come back to this point in~\S\ref{secspecQ}, \S\ref{s:secELC}
and Appendix~\ref{s:appnull}).

\subsectionn{Outline of the Derivation} \label{secdercl}
We now outline the derivation of the formalism of the continuum limit 
(for more details see~\cite[Chapter~4]{PFP}).
The method relies on an asymptotic analysis of the Fourier integral~\eqref{B},
\beq \label{p:2k}
P^\varepsilon(x,y) = \int \frac{d^4k}{(2 \pi)^4}\: \hat{P}^\varepsilon(k)\:e^{ik \xi}\:.
\eeq
\nindex{an4@$P^\varepsilon(x,y)$ -- regularized kernel of fermionic projector}%
\sindex{fermionic projector!regularized kernel of}%
For simplicity, we begin the analysis for the scalar component, i.e.\ we consider the case
\beq \label{p:25s}
\hat{P}^\varepsilon(p) = \phi(p) \:f(p)
\eeq
(the vector component will be treated after~\eqref{p:25v} below).
We may assume that the spatial component of the vector
$\xi$ points in the direction of the $x$-axis of our
Cartesian coordinate system, i.e.\ $y-x=(t,r,0,0)$ with $r>0$.
Choosing cylindrical coordinates $\omega$, $k$, $\rho$ and $\varphi$
in momentum space, defined by $p=(\omega, \vec{p})$ and $\vec{p}=(k,\:
\rho \:\cos \varphi,\: \rho \:\sin \varphi)$, the Fourier integral becomes
\beq \label{p:2pq}
P(x,y) = \frac{1}{(2 \pi)^4} \int_{-\infty}^\infty d\omega
\int_{-\infty}^\infty dk \int_0^\infty \rho\: d\rho \int_0^{2 \pi}
d\varphi \; \hat{P}^\varepsilon(\omega, k, \rho, \varphi) \;e^{i \omega t - i k r} \: .
\eeq
Since the exponential factor in this formula is independent of $\rho$
and $\varphi$, we can write the fermionic projector as the
two-dimensional Fourier transform
\beq \label{p:2pp}
P(x,y) = 2 \int_{-\infty}^\infty d\omega \int_{-\infty}^\infty dk \;
h(\omega, k) \:e^{i \omega t - i k r}
\eeq
of a function $h$ defined by
\beq \label{p:2q}
h(\omega, k) = \frac{1}{2 \:(2 \pi)^4} \int_0^\infty \rho \:d\rho
\int_0^{2 \pi} d\varphi \;(\phi \:f)(\omega, k, \rho, \varphi) \:.
\eeq

We want to analyze $P(x,y)$ close to the light cone $(y-x)^2=0$ away from
the origin $y=x$. Without loss of generality, we may restrict
attention to the upper light cone $t=r$. Thus we are interested in
the region $t \approx r >0$. The ``light-cone coordinates''
\sindex{light-cone coordinates}%
\nindex{bo4@$s, l$ -- light-cone coordinates}%
\beq \label{p:2q1}
s = \frac{1}{2} \:(t-r) \:,\qquad l = \frac{1}{2} \:(t+r)
\eeq
are well-suited to this region, because the ``small'' variable $s$
vanishes for $t=r$, whereas the ``large'' variable $l$ is positive and
non-zero. Introducing also the associated momenta
\beq \label{p:2q2}
u = -k-\omega \:,\qquad v = k-\omega \:,
\eeq
we can write the fermionic projector as
\beq \label{p:2r}
P(s,l) = \int_{-\infty}^\infty du \int_{-\infty}^\infty dv
\;h(u,v) \:e^{-i(us + vl)} \:.
\eeq
\nindex{bo6@$u, v$ -- momenta associated to light-cone coordinates}%

Let us briefly discuss the qualitative form of the function $h$, \eqref{p:2q}.
Without regularization, the scalar component is given by the $\delta$-distribution on the lower
mass shell~$\hat{P} = m \:\delta(p^2-m^2) \:\Theta(-p^0)$. In this case, the integral~\eqref{p:2q}
can be evaluated to be
\begin{align}
h &= \frac{m}{2 \:(2 \pi)^4} \int_0^\infty \rho\:d\rho
\int_0^{2 \pi} d\varphi \;\delta(\omega^2 - k^2 - \rho^2 - m^2)
\:\Theta(-\omega) \notag \\
&= \frac{m}{4 \:(2 \pi)^3} \:\Theta(\omega^2 - k^2 - m^2)
     \:\Theta(-\omega) = \frac{m}{32 \pi^3} \:\Theta(uv - m^2)
     \:\Theta(u) \:.
    \label{p:2rr}
\end{align}
Thus integrating over $\rho$ and $\varphi$ yields a constant function
in the interior of the two-di\-men\-sio\-nal ``lower mass shell''
$\omega^2 - k^2 = m^2$, $\omega<0$.
From this we conclude that for small momenta, where the regularization
should play no role, the function~$h$ should have a
discontinuity along the hyperbola $\{ uv=m^2,\: u>0\}$, be zero
below (i.e.\ for $uv<m^2$) and be nearly constant above. 
The precise form of $h$ for large energy or momentum can be arbitrary.
We only know that~$h$ decays at infinity.

It is instructive to discuss the energy scales. Clearly,
one scale is given by the regularization length~$\varepsilon$.
In momentum space, this corresponds to the high energy
scale~$\varepsilon^{-1}$.
We sometimes refer to the region~$|\omega| + |k| \gtrsim \varepsilon^{-1}$
as the high energy region.
The relevant low energy scale, on the other hand, is $\varepsilon m^2$ (it is zero
for massless fermions). This is because the hyperbola $uv=m^2$ comes as close
to the $v$-axis as as $v \sim \varepsilon m^2$ before entering the high energy region.
Finally, the Compton scale~$m$ lies between the low- and high energy scales,
\[ \varepsilon m^2 \lesssim m \lesssim \varepsilon^{-1} \:. \]
Since we want to analyze the situation close to the light cone, we choose the
``small'' light-cone parameter $s$ on the regularization scale, i.e.
\beq \label{p:2z}
s \lesssim \varepsilon \:.
\eeq
The ``large'' light-cone parameter $l$, on the other hand, is non-zero. We shall always choose this
scale between the regularization scale and the Compton scale,
\beq \label{p:2s}
\varepsilon \ll l \ll \frac{1}{m}\:.
\eeq
Since~$\varepsilon m \ll 1$, the inequalities in~\eqref{p:2s}
still leave us the freedom to vary~$l$ on many orders of magnitude.

Our task is to evaluate the Fourier
integral \eqref{p:2r} using the scales \eqref{p:2z} and \eqref{p:2s}.
In preparation, we discuss and specify the function $h(u,v)$ for fixed $u$,
also denoted by $h_u(v)$. Without regularization~\eqref{p:2rr}, the function~$h_u$
has  a discontinuous ``jump'' from zero to a finite value on the hyperbola.
Therefore, we cannot expect that~$h_u$ is continuous when a regularization is present.
On the contrary, the decay for large~$v$ suggests that~$h_u$ might have another
discontinuity for large~$v$, where it might ``jump'' to zero.
In order to keep the presentation reasonably simple,
we assume that $h_u$ is always of this general form, i.e.
\beq \label{p:2t}
h_u(v) = \left\{ \begin{array}{cl} 0 & {\mbox{for $v<\alpha_u$ or
$v>\beta_u$}} \\
{\mbox{smooth}} & {\mbox{for $\alpha_u \leq v \leq \beta_u$}}
\end{array} \right.
\eeq
with parameters $\alpha_u<\beta_u$. The case of less than two
discontinuities can be obtained from \eqref{p:2t} by setting
$h_u(\alpha_u)$ or $h_u(\beta_u)$ equal to zero, or alternatively by
moving the position of the discontinuities $\alpha_u$ or $\beta_u$ to
infinity. We remark that the discontinuity at $v=\beta_u$ will become
irrelevant later; it is taken into account only in order to explain why the
behavior of the fermionic projector
on the light cone is independent of many regularization details.

Without regularization~\eqref{p:2rr}, the function~$h_u(v)$ is
constant for~$v \geq \alpha_u$. Thus the $v$-dependence
of $h_u(v)$ for $\alpha_u \leq v \leq \beta_u$ merely is a
consequence of the regularization, and it is therefore reasonable to
assume that the $v$-derivatives of $h_u(v)$ scale in powers of the
regularization length~$\varepsilon$. More precisely, we assume that there is a
constant $c_1 \ll l / \varepsilon$ such that
\beq \label{p:2v}
|h_u^{(n)}(v)| \leq \big(c_1 \,\varepsilon \big)^n \:\max
|h_u|  \qquad {\mbox{for $\alpha_u \leq v \leq \beta_u$}} \:,
\eeq
where the derivatives at $v=\alpha_u$ and $\beta_u$ are understood as
the right- and left-sided limits, respectively. This regularity
condition is typically satisfied for polynomial, exponential and
trigonometric functions, but it excludes the case that the function~$h_u$
has small-scale fluctuations. Clearly, we could also consider a more general ansatz for
$h_u$ with more than two discontinuities or weaker regularity
assumptions. But this does not seem to be the point because all
interesting effects, namely the influence of discontinuities for
small and large $v$ as well as of smooth regions, can already be
studied in the setting \eqref{p:2t}, \eqref{p:2v}.

Let us analyze the $v$-integral of the Fourier transform \eqref{p:2r},
\beq \label{p:2w}
P_u(l) \;:=\; \int_{-\infty}^\infty h_u(v) \: e^{-ivl} \;dv \:.
\eeq
According to the left inequality in \eqref{p:2s}, the exponential factor
in \eqref{p:2w} is highly oscillatory on the scale $v \sim 1/\varepsilon$. Thus we can
expect that the smooth component of $h_u$ only gives a small
contribution to the integral \eqref{p:2w}, so that the discontinuities
at $\alpha_u$ and $\beta_u$ should play the dominant role.
In order to make this picture mathematically
precise, in~\eqref{p:2w} we iteratively integrate~$K$ times by parts,
\begin{align}
P_u(l) &= \int_{\alpha_u}^{\beta_u} h_u(v) \:e^{-ivl} \:dv =
-\frac{1}{il} \int_{\alpha_u}^{\beta_u} dv\; h_u(v) \:\frac{d}{dv} e^{-ivl} \notag \\
&= -\frac{1}{il} \left. h_u(v) \:e^{-ivl}
\right|_{\alpha_u}^{\beta_u} \:+\: \frac{1}{il}
\int_{\alpha_u}^{\beta_u} h_u^\prime(v) \:e^{-ivl} \:dl
= \cdots = \notag \\
&= -\frac{1}{il} \sum_{n=0}^{K-1}
\left(\frac{1}{il}\right)^n \: \left. h_u^{(n)}(v) \:e^{-ivl}
\right|_{\alpha_u}^{\beta_u} \:+\: \left( \frac{1}{il} \right)^K
\int_{\alpha_u}^{\beta_u} h_u^{(K)}(v) \:e^{-ivl} \:dl \:. \label{p:2x}
\end{align}
If we bound all summands in~\eqref{p:2x} using the first inequality in
\eqref{p:2s} and the regularity condition~\eqref{p:2v}, each
$v$-derivative appears in combination with a power of~$l^{-1}$, and
giving a factor $c_1 \varepsilon/l \ll 1$. Thus in the limit
$K \to \infty$, we may drop the integral in \eqref{p:2x} to obtain
\beq \label{p:2y}
P_u(l) = -\frac{1}{il} \sum_{n=0}^\infty
\left(\frac{1}{il}\right)^n \: \left. h_u^{(n)}(v) \:e^{-ivl}
\right|_{\alpha_u}^{\beta_u} \: .
\eeq
This expansion converges, and its summands decay like $(c_1 \varepsilon/ l)^n$.

Using \eqref{p:2w}, we can write the Fourier transform \eqref{p:2r} as
\beq \label{p:21}
P(s,l) = \int_{-\infty}^\infty P_u(l) \:e^{-ius} \:du \:.
\eeq
Notice that, apart from the constraints \eqref{p:2s}, the ``large''
variable $l$ can be freely chosen. We want to study the functional
dependence of \eqref{p:21} on the parameter $l$. In preparation, we
consider an integral of the general form
\beq \label{p:22}
\int_a^b f(u) \:e^{-i \gamma(u) \:l} \:du \:,
\eeq
where we assume that $(u, \gamma(u))$ is a curve in the high energy
region in the sense that~$\gamma \sim 1/\varepsilon$. Furthermore, we assume that $\gamma$ is
monotone with $|\gamma^\prime| \sim 1$ and that $(b-a) \sim 1/\varepsilon$. By
transforming the integration variable, we can then write \eqref{p:22} as
the Fourier integral
\beq \label{p:23}
\int_{\gamma(a)}^{\gamma(b)} f \:|\gamma^\prime|^{-1} \:e^{-i \gamma
l} \:d\gamma \:.
\eeq
If the function $f \:|\gamma^\prime|^{-1}$ is smooth,
its Fourier transform \eqref{p:23} has rapid decay in
the variable $l$. Under the stronger assumption that $f \:|\gamma^\prime|^{-1}$
varies on the scale~$1/\varepsilon$, we conclude that the length scale for
this rapid decay is of the order $l \sim \varepsilon$. As a consequence, the
rapid decay can be detected even under the constraint
$l<l_{\mbox{\scriptsize{max}}}$ imposed by \eqref{p:2s}, and we say
that \eqref{p:23} has {\em{rapid decay in $l$}}. The reader who feels
uncomfortable with this informal definition can immediately make this
notion mathematically precise by an integration by parts argument
similar to \eqref{p:2x} imposing for $f \:|\gamma^\prime|^{-1}$ a
condition of type \eqref{p:2v}. The precise mathematical meaning of
rapid decay in $l$ for the integral \eqref{p:22} is that for every
integer $k$ there should be constants $c \sim 1$ and parameters~$l_{\min}, l_{\max}$
in the range~$\varepsilon \ll l_{\min} \ll l_{\max} \ll 1/m$ such
that for all $l \in (l_{\min}, l_{\max})$,
\[ \int_a^b f(u) \:e^{-i \gamma(u) \:l} \:du \;\leq\; c \: \Big( \frac{\varepsilon}{l} \Big)^k
\:\int_a^b |f(u)| \:du \:. \]

We return to the analysis of the integral~\eqref{p:21}.
The boundary terms in~\eqref{p:2y} at $\beta_u$ yield
contributions to $P(s,l)$ of the form
\beq \label{p:24}
-\left(\frac{1}{il}\right)^{n+1} \int_{-\infty}^\infty
h_u^{(n)} \big( \beta_u \big) \:e^{-i \beta_u l - i u s} \:du \:.
\eeq
According to \eqref{p:2z}, the length scale for the oscillations of
the factor $\exp (-ius)$ is $u \sim 1/\varepsilon$. Under the reasonable assumption
that $\beta_u$ is monotone and that the functions $|\beta^\prime(u)|^{-1}$
and $h_u^{(n)}(\beta_u)$ vary on the scale $1/\varepsilon$, the
integral~\eqref{p:24} is of the form~\eqref{p:23}, and the above
consideration yields that~\eqref{p:24} has rapid decay in $l$.
We conclude that it suffices to consider the boundary terms
in~\eqref{p:2y} at~$\alpha_u$. Using this result in~\eqref{p:21}, we obtain
\beq 
P(s,l) = \sum_{n=0}^\infty \left(\frac{1}{il}\right)^{n+1}
\int_{-\infty}^\infty h_u^{(n)}(\alpha_u) \:e^{-i \alpha_u l - ius}
\:du \;+\; {\mbox{(rapid decay in $l$)}} \:. \label{p:25a}
\eeq

The integral~\eqref{p:25a} cannot be estimated again
using the ``oscillation argument'' after~\eqref{p:22},
because, according to~\eqref{p:2rr}, the function~$\alpha_u$ tends asymptotically to
zero for large~$u$, so that the factor $\exp(-i \alpha_u l)$ is non-oscillating in this region.
Instead, we expand this factor in a Taylor series,
\beq \label{p:25}
P(s,l) = \sum_{n,k=0}^\infty \frac{1}{k!} \:(il)^{k-n-1}
\int_{-\infty}^\infty h_u^{(n)}(\alpha_u) \:(-\alpha_u)^k \: e^{-ius}
\:du \: .
\eeq
Let us discuss this expansion. Without regularization~\eqref{p:2rr},
the function~$\alpha_u = m^2/u$ involves the mass.
Therefore, expanding in powers of~$\alpha_u$ corresponds precisely to
the expansion in the mass expansion as considered earlier
(see~\eqref{l:24b} and~\eqref{l:F2} and the explanations thereafter).
With this in mind, we can regard~\eqref{p:25} as a generalization
of the {\em{mass expansion}} to the setting with regularization.
This expansion is clearly justified if~$\alpha_u l \ll 1$.
However, as the function~$m^2/u$ has a pole at~$u=0$,
the function~$\alpha_u$ becomes large for small~$u$,
so that it is not clear whether the mass expansion is sensible.
Indeed, this issue is closely related to the logarithmic mass
problem which was mentioned in~\S\ref{l:sec_33} and was resolved by
working with the ``regularized'' distribution~$T^\reg_a$, \eqref{Tregdef}.
In the present setting, this ``regularization procedure'' can be understood as follows:
For small momenta~$u \ll 1/\varepsilon$,
our oscillation argument after \eqref{p:22} again applies and shows
that the resulting contribution to~$P(s,l)$ decays rapidly in~$l$.
Therefore, disregarding contributions with rapid decay in~$l$, we
may restrict attention to the region~$u \gtrsim \varepsilon$ where
\beq \label{p:25n}
\alpha_u < \alpha_{\mbox{\scriptsize{max}}}
\ll l_{\mbox{\scriptsize{max}}}^{-1} \:.
\eeq \label{alphamax}
Then~$\alpha_u l \ll 1$, justifying the mass expansion~\eqref{p:25}.

For a fixed value of $k-n$, all summands in \eqref{p:25} have the
same $l$-dependence. Let us compare the relative size of these terms.
According to our regularity assumption \eqref{p:2v}, the derivatives
of $h$ scale like $h_u^{(n)} \sim \varepsilon^n$. Using the bound
\eqref{p:25n}, we conclude that, for a fixed
power of $l$, the summands in \eqref{p:25} decrease like
$(\varepsilon \alpha_{\mbox{\scriptsize{max}}})^n$.
Thus it is a very good approximation to drop the summands for large $n$.
At first sight, it might seem admissible to take into account only the first
summand $n=0$. But the situation is not quite so simple.
For example, it may happen that, when restricted to the curve $(u, \alpha_u)$,
the function $h(u,v)$ is so small that the summands for $n=0$ in
\eqref{p:25} are indeed not dominant. More generally, we need to know that
for some $n_0 \geq 0$, the function $h^{(n_0)}_u(\alpha_u)$ is really of the
order given in \eqref{p:2v}, i.e.
\beq \label{p:27h}
|h^{(n_0)}_u(\alpha_u)| \geq c \:\big( c_1\, \varepsilon \big)^{n_0} \:
\max |h_u| \eeq
with a positive constant $c$ which is of the order one.
If this condition is satisfied, we may neglect
all summands for $n>n_0$, and collecting the terms in powers of $l$, we
conclude that
\begin{eqnarray}
\lefteqn{ P(s,l) } \notag \\
&=& \frac{1}{(il)^{n_0+1}} \sum_{k=0}^\infty (-il)^k \!\!\!\!\!\!\!
\sum_{n=\max (n_0-k, 0)}^{n_0} \frac{(-1)^{n_0 - n}}{(k-n_0+n)!}
\int_{-\infty}^\infty h_u^{(n)}(\alpha_u) \:\alpha_u^{k-n_0+n}
\:e^{-ius} \:du \notag \\
&&+\sum_{n=n_0+1}^\infty \frac{1}{(il)^{n+1}} \int_{-\infty}^\infty
h_u^{(n)}(\alpha_u) \:e^{-ius} \:du
\:+\:{\mbox{(rapid decay in $l$)}} \notag \\
&&+\: {\mbox{(higher orders in
$\varepsilon \alpha_{\mbox{\scriptsize{max}}}$)}}\:.
    \label{p:26}
\end{eqnarray}
We point out that, according to \eqref{p:25n},
\[ \varepsilon \alpha_{\mbox{\scriptsize{max}}} \ll
\varepsilon / l_{\mbox{\scriptsize{max}}}\:, \]
and this explains why we disregard the higher orders in~$\varepsilon \alpha_{\mbox{\scriptsize{max}}}$.
In our case, the function $h_u$ has in the low energy region
according to \eqref{p:2rr} the form $h_u(\alpha_u) = m/(32 \pi^3)\:
\Theta(u)$. Hence it is natural to assume that
\eqref{p:27h} is satisfied for $n_0=0$. Introducing the shorter notation
\beq \label{p:27x}
h(u) := h_u(\alpha(u)) \:,\quad h^{[n]}(u) :=
h_u^{(n)}(\alpha_u) \:,\quad \alpha(u) :=
\alpha_u \:,
\eeq
we have thus derived the following result.\\[.5em]
{\bf{Expansion of the scalar component:}} {\em{Close to the light cone
\eqref{p:2z}, \eqref{p:2s}, the scalar component \eqref{p:25s} of
the fermionic projector of the vacuum has the expansion}}
\begin{align}
P(s,l) &= \frac{1}{il} \sum_{k=0}^\infty \frac{(-il)^k}{k!}
\int_{-\infty}^\infty h \:\alpha^k \:e^{-ius} \:du \label{p:27a} \\
&\qquad+\sum_{n=1}^\infty \frac{1}{(il)^{n+1}} \int_{-\infty}^\infty
h^{[n]} \:e^{-ius} \:du \label{p:27b} \\
&\qquad +\:{\mbox{(rapid decay in $l$)}} \:+\: {\mbox{(higher orders in
$\varepsilon \alpha_{\mbox{\scriptsize{max}}}$)}} \label{p:27c}
\end{align}
{\em{with suitable regularization functions $h$, $h^{[n]}$ and $\alpha$.
In the low energy region $u \ll 1/\varepsilon$, the regularization functions are}}
\beq \label{p:2rs}
h(u) = \frac{m}{32 \pi^3} \:\Theta(u) \:,\qquad
h^{[n]}(u) = 0 \:,\qquad \alpha(u) = \alpha_u = \frac{m^2}{u} \:.
\eeq

In this expansion, the $l$-dependence is written out
similar to a Laurent expansion. The main simplification compared to
our earlier Fourier representation is that the
dependence on the regularization is now described by functions of only
one variable, denoted by $h$, $h^{[n]}$ and $\alpha$. In composite
expressions in $P(s,l)$, we will typically get convolutions of these
functions; such one-dimensional convolutions can be easily analyzed. The simplification to one-dimensional
regularization functions became possible because many details of the
regularization affect only the contribution with rapid decay in $l$,
which we do not consider here. Notice that the summands in
\eqref{p:27a} and \eqref{p:27b} decay like $(l
\:\alpha_{\mbox{\scriptsize{max}}})^k/k! \ll
(l/l_{\mbox{\scriptsize{max}}})^k/k!$ and $(\varepsilon/l)^n$,
respectively. In the low energy limit \eqref{p:2rs}, the expansion
\eqref{p:27a} goes over to a power series in $m^2$, and we thus refer to
\eqref{p:27a} as the {\em{mass expansion}}.
\sindex{mass expansion}%
In the mass expansion, the regularization is
described by only two functions $h$ and $\alpha$. The series
\eqref{p:27b}, on the other hand, is a pure regularization effect and
is thus called the {\em{regularization expansion}}.
\sindex{regularization expansion}%
It involves an
infinite number of regularization functions $h^{[n]}$. Accordingly, we
will use the notions of mass and regularization expansions also for
other expansions of type \eqref{p:26}.

We now outline how to extend the previous analysis to the
vector component. More precisely, we will analyze
the Fourier integral \eqref{p:2k} for
\beq \label{p:25v}
    \hat{P}^\varepsilon(p) = v_j(p) \:\gamma^j \:f(p)
\eeq
close to the light cone. We again choose light-cone coordinates $(s,l,x_2,
x_3)$ with $y-x=(s,l,0,0)$ ($s$ and $l$ are given by \eqref{p:2q1},
while $x_2$ and $x_3$ are Cartesian coordinates in the orthogonal
complement of the $sl$-plane). The associated momenta are denoted by
$p=(u,v,p_2, p_3)$ with $u$ and $v$ according to \eqref{p:2q2}.
As in \eqref{p:2pp}, we integrate out the coordinates
perpendicular to $u$ and $v$,
\beq \label{p:28a}
h_j(u,v) := \frac{1}{2 \:(2 \pi)^4} \int_{-\infty}^\infty dp_2
\int_{-\infty}^\infty dp_3 \; (v_j \:f)(u,v,p_2,p_3) \:.
\eeq
We thus obtain a representation of the fermionic projector involving
two-dimensional Fourier integrals
\[ P(s,l) = \gamma^j \:P_j(s,l) \]
with
\beq
P_j(s,l) := \int_{-\infty}^\infty du \int_{-\infty}^\infty dv \;
h_j(u,v) \:e^{-i(us + vl)} \:. \label{p:28b}
\eeq
The tensor indices in \eqref{p:28a} and \eqref{p:28b} refer to the
coordinate system $(s,l,x_2,x_3)$. For clarity, we denote the range of
the indices by $j=s,l,2,3$; thus
\beq \label{p:28c}
\gamma^s = \frac{1}{2}\:(\gamma^0 - \gamma^1) \:,\qquad
\gamma^l = \frac{1}{2}\:(\gamma^0 + \gamma^1) \:,
\eeq
where $\gamma^0,\ldots,\gamma^3$ are the usual Dirac matrices of
Minkowski space. Since without regularization, $\hat{P}=\slashed{p}
 \:\delta(p^2 - m^2) \:\Theta(-p^0)$, the functions~$h_j$ can be computed
similar to \eqref{p:2rr} to be
\beq \label{p:29}
\gamma^j \:h_j(u,v) = \frac{1}{32 \pi^3} (-u \gamma^s - v \gamma^l)
\:\Theta(uv - m^2) \:\Theta(u) \:.
\eeq
This limiting case specifies the regularized $h_j(u,v)$ for small
energy-momentum $u,v \ll 1/\varepsilon$. In order to keep the form of the
functions $h_j$ in the high energy region sufficiently general, we
merely assume in what follows that the functions~$h_j$ have all the properties
which se assumed for the function $h$ above. This gives the following result.\\[.5em]
{\bf{Expansion of the vector component:}} {\em{Close to the light cone
\eqref{p:2z}, \eqref{p:2s}, the vector component \eqref{p:25v} of
the fermionic projector of the vacuum has the expansion $P = \gamma^j
P_j$ with}}

\begin{eqnarray}
\lefteqn{ P_s(s,l) = \frac{1}{il} \sum_{k=0}^\infty \frac{(-il)^k}{k!}
\int_{-\infty}^\infty -u \:g_s \: \:\alpha^k
\:e^{-ius} \:du } \notag \\
&&+\sum_{n=1}^\infty \frac{1}{(il)^{n+1}} \int_{-\infty}^\infty
- u \:g_s^{[n]} \:e^{-ius} \:du \notag \\
&&+\:{\mbox{(rapid decay in $l$)}} \:+\: {\mbox{(higher orders in
$\varepsilon \alpha_{\mbox{\scriptsize{max}}}$)}} \label{p:210} \\
\lefteqn{ P_l(s,l) = \frac{1}{(il)^2} \sum_{k=0}^\infty \frac{(-il)^k}{k!}
\int_{-\infty}^\infty \left[ (k-1) \: \alpha^k \:+\: k
\:\frac{b}{u} \: \alpha^{k-1} \right] \:g_l
\:e^{-ius} \:du } \notag \\
&&+\sum_{n=1}^\infty \frac{1}{(il)^{n+2}} \int_{-\infty}^\infty
- (n+1)\: g_l^{[n]} \:e^{-ius} \:du \notag \\
&&+\:{\mbox{(rapid decay in $l$)}} \:+\: {\mbox{(higher orders in
$\varepsilon \alpha_{\mbox{\scriptsize{max}}})$}}\label{p:211} \\
\lefteqn{ P_{2\!/\!3}(s,l) = \frac{1}{(il)^2} \sum_{k=0}^\infty \frac{(-il)^k}{k!}
\int_{-\infty}^\infty \left[ \alpha^k \:+\: k
\:\frac{b_{2\!/\!3}}{u} \: \alpha^{k-1} \right] \:g_{2\!/\!3}
\:e^{-ius} \:du } \notag \\
&&+\sum_{n=1}^\infty \frac{1}{(il)^{n+2}} \int_{-\infty}^\infty
g_{2\!/\!3}^{[n]} \:e^{-ius} \:du \notag \\
&&+\:{\mbox{(rapid decay in $l$)}} \:+\: {\mbox{(higher orders in
$\varepsilon \alpha_{\mbox{\scriptsize{max}}}$)}} \qquad \quad \label{p:212}
\end{eqnarray}
{\em{and suitable regularization functions $g_j$, $g_j^{[n]}$, $b$,
$b_{2\!/\!3}$ and the mass regularization function $\alpha$ as in
\eqref{p:27a} and~\eqref{p:2rs}. In the low energy region $u \ll 1/\varepsilon$, the
regularization functions have the form}}
\begin{align}
g_s(u) &= \frac{1}{32 \pi^3} \:\Theta(u) \:,\qquad
g_s^{[n]}(u) = 0 \label{p:regs} \\
g_l(u) &= \frac{1}{32 \pi^3} \:\Theta(u) \:,\qquad
g_l^{[n]}(u) = b(u) = 0 \label{p:2rv2} \\
g_{2\!/\!3}(u) &= g_{2\!/\!3}(u) = b_{2\!/\!3}(u) = 0 \:.
    \label{p:59c}
\end{align}

In order to explain these formulas,
we consider the situation where, like in the case without
regularization, the vector $v(p)$ in \eqref{p:25v} points into the
direction $p$. In this case, we can write the vector component as
\beq \label{p:2pr}
\hat{P}^\varepsilon(p) = p_j \gamma^j \:(\phi f)(p) \:,
\eeq
where $(\phi f)$ has the form of the scalar component as considered
above. Since multiplication in momentum space
corresponds to differentiation in position space, we obtain for \eqref{p:28b}
\[ P(s,l) = -i \left( \gamma^s \frac{\partial}{\partial s} +
\gamma^l \frac{\partial}{\partial l} +
\gamma^2 \frac{\partial}{\partial x^2} +
\gamma^3 \frac{\partial}{\partial x^3} \right)
P_{\mbox{\scriptsize{scalar}}}(s,l) \:, \]
where $P_{\mbox{\scriptsize{scalar}}}$ is the scalar component
\eqref{p:2r} with $h$ as in \eqref{p:2q}. We now substitute for
$P_{\mbox{\scriptsize{scalar}}}$ the expansion on the light cone
\eqref{p:27a}--\eqref{p:27c} and carry out the partial
derivatives. For the $s$- and $l$-components, this gives
exactly the expansions \eqref{p:210}, \eqref{p:211} with
\beq \label{p:2M1}
g_s = g_l = h \:,\quad g_s^{[n]} = g_l^{[n]} = h^{[n]}
\:,\quad b = 0 \:. 
\eeq
For the components $j=2,3$, the calculation of the partial
derivatives is not quite so straightforward because the expansion of
the scalar component \eqref{p:27a}--\eqref{p:27c} was carried out
for fixed $x_2$ and $x_3$. Nevertheless, one can deduce also the
expansion \eqref{p:212} from \eqref{p:27a}--\eqref{p:27c} if one
considers $x_2$ and $x_3$ as parameters of the regularization
functions $h$, $h^{[n]}$ and $\alpha$, and differentiates through,
keeping in mind that differentiation yields a factor $1/l$ (to get the
scaling dimensions right). In this way, the simple example \eqref{p:2pr}
explains the general structure of the expansions
\eqref{p:210}--\eqref{p:212}. We point out that the regularization
function $b$ vanishes identically in \eqref{p:2M1}. This means that
$b$ is non-zero only when the direction of the vector field $v$ is modified
by the regularization. Thinking in terms of the decomposition into the
one-particle states, we refer to this regularization effect as the
{\em{shear of the surface states}}.
\sindex{shear of surface states!}%

The derivation of these formulas uses the same methods as for
the scalar components. The analysis is a bit more subtle because
one must carefully analyze the scaling of the different components.
We refer the interested reader to~\cite[Section~4.4]{PFP}.

Computing composite expressions using the above Fourier representations,
one readily verifies the calculations rules stated in~\S\ref{sec73}.
The details can be found in~\cite[Section~4.5]{PFP}.

\section{Computation of the Local Trace} \label{secloctrace}
When deriving the EL equations in~\S\ref{secvary}, we showed in Proposition~\ref{prptrxconst}
that for every minimizer of the causal action principle, the local trace is constant in space-time.
We also argued that this condition should be satisfied by the rescaling~\eqref{rhorescale}.
In the Minkowski vacuum, the local trace is obviously constant because the kernel of the fermionic projector
is translation invariant (see our ansatz~\eqref{p:2k}). But in the presence of an external potential,
the local trace will in general no longer be constant, making it necessary to perform the
rescaling~\eqref{rhorescale}. We now explain how to compute the local trace and discuss the effect
of the rescaling~\eqref{rhorescale}.

We begin by noting that, using the abstract definition of the kernel of the fermionic projector~\eqref{Pxydef},
we know that the local trace can be computed by
\[ \tr(x) = \Tr_{S_x} \!\big( P^\varepsilon(x,x) \big) \:. \]
\sindex{local trace}%
In what follows, we usually omit the subscript~$S_x$ and regard $\Tr$ as the trace
of a $4 \times 4$-matrix. In the vacuum, one can compute this trace from~\eqref{p:2k} to conclude the
scaling
\beq \label{epsloctrace}
\Tr_{S_x} \!\big( P^\varepsilon(x,x) \big) = c\;\frac{m}{\varepsilon^2} \;\Big(1 + \O \big( m \varepsilon \big) \Big) \:,
\eeq
where the constant~$c$ depends on the regularization method
(for an explicit computation in the $i \varepsilon$-regularization see Exercise~\ref{ex2.101}).

In the next proposition we specify how the local trace is affected by the external potential.
\begin{Prp} \label{prploctr}
In the presence of a smooth external chiral potential~\eqref{Bchiral}
with the properties as in Lemma~\ref{l:lemma0}, the contribution~$\Delta P$ to the fermionic projector
to order~$n$ in perturbation theory influences the local trace only by an error term of the form
\beq \label{loctrbound}
\bigg| \Tr_{S_x} \!\big( \Delta P^\varepsilon(x,x) \big) \bigg| \leq \frac{C}{\varepsilon} \:,
\eeq
\sindex{local trace!effect of chiral potentials on}%
where the constant~$C$ depends on~$m$, $n$ as well as on the potential~$\B$ and its partial derivatives.
Moreover, the function~$\Tr_{S_x} ( \Delta P^\varepsilon(x,x) )$ is smooth in~$x$.
\end{Prp} \noindent
This result implies that, when rescaling the causal fermion system according to~\eqref{rhorescale},
we only pick up smooth error terms of the order~$\varepsilon/\ell_\text{macro}$. Since such error terms
are neglected in the continuum limit (see~\eqref{ap1}), we may disregard the rescaling~\eqref{rhorescale}.
This is the reason why the rescaling~\eqref{rhorescale} will not be considered further in this book.

Before coming to the proof of the above proposition, we note that for a {\em{gravitational field}}, the situation
is more involved. Namely, for linear gravity as considered in Section~\ref{seclingrav}, the
change of the local trace is typically of the order
\beq \label{trtyp}
\Tr_{S_x} \!\big( \Delta P^\varepsilon(x,x) \big) \sim \frac{m}{\varepsilon^2}\: \O(h)\:.
\eeq
\sindex{gravitational field!effect on local trace}%
Clearly, this is sufficient in order to treat a weak gravitational field.
However, when constructing causal fermion systems non-perturbatively
in curved space-time (as is done in~\cite[Section~4]{finite}), the macroscopic space-time dependence
of the local trace must be taken into account, meaning that 
the rescaling procedure~\eqref{rhorescale} will change the causal fermion system
substantially. The same is true if a {\em{scalar potential}} is considered,
\sindex{potential!scalar!effect on local trace}%
because in this case the local trace takes the form
\beq \label{epsloctrace2}
\Tr_{S_x}\! \big( P^\varepsilon(x,x) \big)  = \frac{c}{\varepsilon^2} \:\frac{\Tr_{S_x} \!\big(B(x) \big)}{\dim(S_x)}
+ \O \Big(\frac{1}{\varepsilon} \Big) \:,
\eeq
where the potential~$B$ again includes the mass~\eqref{l:18a} (for the derivation
see Exercise~\ref{ex2.102}).

\Proof[Proof of Proposition~\ref{prploctr}]
As shown in Theorem~\ref{l:thm4}, to every order in perturbation theory,
the non-causal high energy contribution~$\tilde{p}-\tilde{p}^\res$
is a smooth function in~$x$ and~$y$. Therefore, it is even bounded for~$x=y$, and we do
not need to consider it here. Hence it suffices to consider the perturbation expansions of~$\tilde{k}$
and~$\tilde{p}^\res$. These perturbation expansions must be regularized on the scale~$\varepsilon$.
The procedure for this is explained in the appendix (see Appendix~\ref{l:appA}).
In order to keep the presentation as simple as possible, here we shall not enter the
regularized causal perturbation theory. Instead, we consider the unregularized perturbation
expansion and make use of the fact that the regularization gives rise to a decay in momentum
space on the scale~$\varepsilon^{-1}$. This simplified procedure will be justified by a short remark
at the end of the proof.

In view of~\eqref{def-ktil} and~\eqref{l:E1}, instead of~$\tilde{k}$
and~$\tilde{p}^\res$ we can just as well consider the causal Green's functions~$s^\wedge$ and~$s^\vee$
(see~\eqref{series-scaustilde}) as well as the Green's functions~$s^+$ and~$s^-$ (see~\eqref{l:F}).
For the causal Green's function, we can apply the structural results on the light-cone expansion
stated in Theorem~\ref{l:thm1}. Using the residual argument, this theorem holds just as well for
the Green's functions~$s^\pm$. With this in mind, we may restrict attention to the causal Green's
functions, which we again simply denote by~$s$.

The formula~\eqref{trtyp} can also be expressed by saying that~$S^{(0)} \sim \varepsilon^{-2}$.
Since increasing the upper index gives a scaling factor~$\xi^2$, which for~$x=y$ is translated
to a scaling factor~$\varepsilon^2$, we have
\beq \label{Shscale}
S^{(h)} \sim \varepsilon^{-2+2h} \:.
\eeq
Moreover, every factor~$\xi$ in the light-cone expansion gives rise to a scaling factor
\beq \label{xiscale}
\xi \sim \varepsilon \:.
\eeq
Applying these scalings to a contribution of the light-cone expansion in Theorem~\ref{l:thm1}, we find that
\[ \eqref{l:11} \sim \varepsilon^{-2+2h+|I|} \:. \]
Therefore, our task is to show that all expressions of the form~\eqref{l:11} which contribute to the local trace
satisfy the inequality
\beq \label{toshow}
2h+|I| > 0 \:.
\eeq

Using the identity~\eqref{l:l3}, the inequality~\eqref{toshow} is equivalent to
\[ k - 1 + \sum_{a=1}^k \Big(|I_a| + 2 p_a \Big) > 0 \:. \]
Obviously, it suffices to consider the cases~$k=0$ and~$k=1$.
If~$k=0$, the fermionic projector is odd (i.e.\ it contains an odd number of Dirac matrices),
so that the local trace vanishes.
In the case~$k=1$, on the other hand, the contribution involving the chiral potential is
again odd and vanishes. The contribution involving the mass matrix~$mY$, on the other
hand, is precisely the term~$m Y S^{(0)}$ whose local trace was computed in~\eqref{epsloctrace}.
This concludes the proof, provided that the scalings~\eqref{Shscale} and~\eqref{xiscale} hold.

The scalings~\eqref{Shscale} and~\eqref{xiscale} are justified by the regularized causal
perturbation theory developed in Appendix~\ref{l:appA}.
We here explain the reason for the scalings: In the regularized causal perturbation calculation,
the ``causality'' is built in by {\em{demanding}} that the resulting regularized light-cone expansion
again only involves integrals along the line segment~$\overline{xy}$ (and not integrals along
the whole straight line through~$x$ and~$y$). In more technical terms, this is achieved by
demanding that the contributions to the perturbation expansion remain bounded in the
limit when the momentum of the external potential tends to zero
(this method was first used in~\cite[Appendix~D]{PFP}). This procedure ensures that
a factor~$\xi$ in the unregularized light-cone expansion really gives a scaling factor~$\varepsilon$,
\eqref{xiscale}. The scaling~\eqref{Shscale}, on the other hand, follows immediately from the
fact that the local trace is obtained by integrating over the momentum variables
(similar as in Exercise~\ref{ex2.101}), and that the regularization gives decay in momentum space
on the scale~$\varepsilon^{-1}$.
\QED

\section{Spectral Analysis of the Closed Chain} \label{secspeccc}
In this section we explain how to analyze the EL equations corresponding to
the causal action in the continuum limit.
Since the Lagrangian involves the eigenvalues of the closed chain, the main task is
to compute the spectral decomposition of~$A_{xy}^\varepsilon = P^\varepsilon(x,y)\, P^\varepsilon(y,x)$.
We first compute this spectral decomposition in the vacuum (\S\ref{secregvac}).
This spectral decomposition has the special properties that the eigenvalues are non-real and
form complex conjugate pairs, and that the corresponding eigenvectors are null
(with respect to the spin scalar product). In order to simplify the subsequent computations,
it is very convenient to choose a spinor basis which reflects these special properties of
the closed chain of the vacuum. This so-called {\em{double null spinor frame}}
is introduced in~\S\ref{secdnsf}.
In~\S\ref{secpertspec} we proceed by describing the interaction perturbatively
using contour integral methods.
In~\S\ref{secspecA} we derive a few general properties of the spectral representation
of the closed chain.
Finally, in~\S\ref{secspecQ} we use the obtained spectral representation
of the closed chain to rewrite the EL equations in a form suitable for an explicit analysis.

\subsectionn{Spectral Decomposition of the Regularized Vacuum} \label{secregvac}
In order to analyze the causal action principle,
we clearly need to know the eigenvalues~$\lambda^{xy}_i$ of the closed chain. Moreover, in order to
bring the EL equations into a tractable form, we also need to know the corresponding eigenspaces.
We now compute the spectral decomposition of the closed chain for the regularized
fermionic projector of the vacuum. We first do the computation in general, and then
rewrite it using the formalism of the continuum limit.

As in~\S\ref{seccorcs} we assume that the regularized fermionic projector of the vacuum
is {\em{homogeneous}} and has a {\em{vector-scalar structure}}~\eqref{vectorscalar}.
These assumptions are reasonable and sufficiently general for our purposes.
Thus we assume that~$P^\varepsilon(x,y)$ can again be written
again as the Fourier integral~\eqref{B}, where~$\hat{P}^\varepsilon$ now
is a distribution of the form
\beq \label{C}
\hat{P}^\varepsilon(k) = \hat{g}_j(k)\: \gamma^j + \hat{h}(k)
\eeq
with real-valued distributions~$\hat{g}_j$ and~$\hat{h}$.
Here the parameter~$\varepsilon>0$ denotes the length scale of the regularization.
Thus, expressed in momentum space, the distributions~$\hat{g}_j$
and~$\hat{h}$ should decay at infinity on the scale~$k \sim \varepsilon^{-1}$.
This means in position space that the kernel of the fermionic projector has the form
\beq \label{e:2g}
P^\varepsilon(x,y) = g_j(x,y)\: \gamma^j + h(x,y)
\eeq
with smooth functions~$g_j$ and~$h$ whose derivatives scale at most in powers of~$\varepsilon^{-1}$.
As~$\varepsilon$ tends to zero, the regularized fermionic
projectors should go over to the unregularized fermionic projector,
\beq
\lim_{\varepsilon \searrow 0} P^\varepsilon(x,y) = P(x,y)
\qquad {\mbox{as a distribution.}} \label{Z}
\eeq

According to~\eqref{Axydef}, we introduce the corresponding closed chain by
\beq \label{chain}
A^\varepsilon_{xy} = P^\varepsilon(x,y)\: P^\varepsilon(y,x)\:.
\eeq
In the next lemma we compute the roots of the characteristic polynomial of this matrix.
For ease in notation we shall often omit the subscripts ``$xy$.''
\begin{Lemma} \label{lemma11} The characteristic polynomial of the closed chain~$A^\varepsilon_{xy}$
has two roots~$\lambda_\pm$.
Either the~$\lambda_\pm$ form a complex conjugate pair, $\overline{\lambda_+}=\lambda_-$,
or else they are both real and have the same sign.
The roots are given explicitly by
\beq \label{e:2e}
\lambda_\pm = g \overline{g} \:+\: h \overline{h} \:\pm\:
\sqrt{(g \overline{g})^2 \:-\: g^2\:\overline{g}^2 \:+\:
(g \overline{h} + h \overline{g})^2 } \:.
\eeq
\end{Lemma}
\Proof We write the fermionic projector in position space as
\[ P^\varepsilon(x,y) = g_j(x,y)\: \gamma^j + h(x,y) \:,\qquad
P^\varepsilon(y,x) = \overline{g_j(x,y)}\: \gamma^j + \overline{h(x,y)} \:. \]
Thus, omitting the arguments~$x$ and~$y$,
\[ A^\varepsilon_{xy} = (\slashed{g} + h)(\overline{\slashed{g}} + \overline{h})\:. \]
Omitting the superscript~$\varepsilon$ and the subscript~$xy$, we obtain
\beq
A = \slashed{g} \:\overline{\slashed{g}} \:+\: h \:\overline{\slashed{g}} \:+\:
\slashed{g} \:\overline{h} \:+\: h \overline{h} \:. \label{e:2nn}
\eeq
It is useful to decompose $A$ in the form
\[ A = A_1 \:+\: A_2 \:+\: \mu \]
with
\[ A_1 = \frac{1}{2} \:[\slashed{g}, \overline{\slashed{g}}] \:, \quad
A_2 = h \:\overline{\slashed{g}} \:+\: \slashed{g} \:\overline{h} \:,\quad
\mu = g \overline{g} \:+\: h \overline{h} \]
and $g \overline{g} \equiv g_j \:\overline{g^j}$. Then the matrices~$A_1$
and $A_2$ anti-commute, and thus
\beq \label{e:2d2}
(A-\mu)^2 = A_1^2 + A_2^2 = (g \overline{g})^2 \:-\: g^2
\:\overline{g}^2 \:+\: (g \overline{h} + h \overline{g})^2 \:.
\eeq
The right side of~\eqref{e:2d2} is a multiple of the identity matrix, and
so~\eqref{e:2d2} is a quadratic equation for $A$.
The roots~$\lambda_\pm$ of this equation as given by~\eqref{e:2e}
are the zeros of the characteristic polynomial of $A$.
If the discriminant is negative,
the~$\lambda_\pm$ form a complex conjugate pair. If conversely the discriminant is positive,
the~$\lambda_\pm$ are both real. In order to show that they have the same sign, we compute
their product,
\begin{eqnarray*}
\lambda_+ \lambda_- &=& (g \overline{g} + h \overline{h})^2 - \left[ (g \overline{g})^2
- g^2\: \overline{g}^2 + (g \overline{h} + h \overline{g})^2 \right] \\
&=& 2\:(g \overline{g})\: |h|^2 + |h|^4 + g^2\: \overline{g}^2 -
(g \overline{h} + h \overline{g})^2 \\
&=& |h|^4 + g^2\: \overline{g}^2 - g^2\: \overline{h}^2 - h^2\: \overline{g}^2 \\
&=& (g^2-h^2)(\overline{g}^2 - \overline{h}^2) \geq 0\:.
\end{eqnarray*}
This concludes the proof.
\QED

In the degenerate case that the two eigenvalues~$\lambda_+$ and~$\lambda_-$ coincide,
the relation~\eqref{e:2d2} shows that the matrix~$A-\mu$ is nilpotent.
However, in this case the matrix~$A-\mu$ need not vanish
(as one sees from~\eqref{e:2nn}), giving examples where the
matrix~$A$ is {\em{not diagonalizable}}.
Except for this degenerate case, the matrix~$A$ is indeed diagonalizable and has two-dimensional
eigenspaces:
\begin{Lemma} \label{lemmanondeg}
In the case~$\lambda_+ \neq \lambda_-$, the matrix~$A_{xy}$
is diagonalizable and has two-dimensional eigenspaces. It has the spectral representation
\beq \label{e:2m}
    A_{xy} = \sum_{s=\pm} \lambda_s^{xy}\: F_s^{xy} \:,
\eeq
where the spectral projections are given by
\beq \label{e:2f}
F_\pm^{xy} = \frac{\1}{2} \:\pm\:\frac{\frac{1}{2} \:[\slashed{g}, \overline{\slashed{g}}] \:+\:
h \overline{\slashed{g}} \:+\: \slashed{g} \overline{h}}{2 \:
\sqrt{(g \overline{g})^2 \:-\: g^2\:\overline{g}^2 \:+\:
(g \overline{h} + h \overline{g})^2 } } \:.
\eeq
\sindex{closed chain!spectral analysis of}%
\end{Lemma}
\Proof If we {\em{assume}} that $A$ is
diagonalizable, then $\lambda_\pm$ are the two eigenvalues of $A$, and the
corresponding spectral projectors $F_\pm$ are given by
\beq \label{e:2f1}
F_\pm = \frac{\1}{2} \:\pm\: \frac{1}{\lambda_+ - \lambda_-} \left(
A \:-\: \frac{1}{2}\:(\lambda_+ + \lambda_-)\:\1 \right) \:.
\eeq
Applying~\eqref{e:2e} gives~\eqref{e:2f}. Taking their trace, one sees that
the matrices~$F_+$ and~$F_-$ both have rank two.

In order to prove that~$A$ is diagonalizable, one takes formulas~\eqref{e:2f}
and shows by direct computation that (see Exercise~\ref{ex2.11})
\beq \label{AFformula}
A \:F_\pm = \lambda_\pm \:F_\pm \qquad{\mbox{and}}\qquad
F_+ + F_- = \1 \:.
\eeq
This shows that the images of $F_+$ and $F_-$ are indeed eigenspaces
of~$A$ which span $\C^4$.
\QED

Our next step is to rewrite the spectral representation using the formalism
of the continuum limit. Let us compute the leading singularity on the light cone. Then
\beq \label{Peq}
P(x,y) = \frac{i}{2}\: \slashed{\xi}\: T^{(-1)}_{[0]} + (\deg < 2 )\:,
\eeq
where for notational convenience we omitted the indices~$^{-1}_{[0]}$ of the factor~$\xi$,
and where the bracket~$(\deg < 2)$ stands for terms of degree at most one.
Using this formula for the fermionic projector, the closed chain becomes
\beq \label{clcvac}
A_{xy} = \frac{1}{4}\: (\slashed{\xi} T^{(-1)}_{[0]}) (\overline{\slashed{\xi} T^{(-1)}_{[0]}}) 
+ \slashed{\xi} (\deg \leq 3) + (\deg < 3)\:,
\eeq
where~$\overline{\slashed{\xi}} := \overline{\xi_j} \gamma^j$.
Its trace can be computed with the help of the contraction rules~\eqref{eq53},
\[ \Tr (A_{xy}) = (\xi_j \overline{\xi^j})\: T^{(-1)}_{[0]}\, \overline{T^{(-1)}_{[0]}}
= \frac{1}{2} \left(z + \overline{z} \right) T^{(-1)}_{[0]}\, \overline{T^{(-1)}_{[0]}}
+ (\deg < 3)\:. \]
We next compute the square of the trace-free part of the closed chain,
\begin{align*}
\Big( &A_{xy} - \frac{1}{4}\: \Tr (A_{xy})\,\1 \Big)^2 =
\frac{1}{16} \left( \slashed{\xi} \overline{\slashed{\xi}} -\frac{z+\overline{z}}{2} \right)^2 \left(T^{(-1)}_{[0]} \overline{T^{(-1)}_{[0]}} \right)^2 \\
&= \frac{1}{16} \left( \slashed{\xi} \overline{\slashed{\xi}} \slashed{\xi} \overline{\slashed{\xi}}- (z+\overline{z})\:
\slashed{\xi} \overline{\slashed{\xi}} + \frac{1}{4}\: (z+\overline{z})^2 \right) \left(T^{(-1)}_{[0]} \overline{T^{(-1)}_{[0]}} \right)^2 \\
&= \frac{1}{64}\:(z-\overline{z})^2 \:
\Big(T^{(-1)}_{[0]} \overline{T^{(-1)}_{[0]}} \Big)^2.
\end{align*}
Combining these formulas, we see that to leading degree,
the closed chain is a solution of the polynomial equation
\beq \label{poly}
\left( A_{xy} - \frac{1}{8}\, (z+\overline{z})\:T^{(-1)}_{[0]} \overline{T^{(-1)}_{[0]}} \right)^2
= \left( \frac{1}{8}\, (z-\overline{z})\:T^{(-1)}_{[0]} \overline{T^{(-1)}_{[0]}} \right)^2 .
\eeq
We point out that the calculations so far are only formal, but they have a well-defined meaning
in the formalism of the continuum, because to all our end formulas we will be able to apply
the weak evaluation formula~\eqref{asy}. Having this in mind, we can interpret the roots
of the polynomial in~\eqref{poly}
\[ \lambda_+ = \frac{1}{4}\: \big(z\, T^{(-1)}_{[0]} \big) \,\overline{T^{(-1)}_{[0]}} \qquad \text{and} \qquad
\lambda_- = \frac{1}{4}\: T^{(-1)}_{[0]} \,\overline{\big(z \,T^{(-1)}_{[0]} \big)} \]
as the eigenvalues of the closed chain. Using the contraction rule~\eqref{eq54}, these eigenvalues
simplify to (see also~\cite[eq.~(5.3.20)]{PFP})
\beq \label{lpm}
\lambda_+ = T^{(0)}_{[0]} \,\overline{T^{(-1)}_{[0]}} + (\deg < 3) \:,\qquad
\lambda_- = T^{(-1)}_{[0]} \,\overline{T^{(0)}_{[0]}} + (\deg < 3)\:.
\eeq
The corresponding spectral projectors become (see also~\cite[eq.~(5.3.21)]{PFP})
\beq \label{Fpm}
F_\pm = \frac{1}{2} \Big( \1 \pm \frac{[\slashed{\xi}, \overline{\slashed{\xi}}]}{z-\overline{z}} \Big)
+ \slashed{\xi} (\deg \leq 0) + (\deg < 0) \:.
\eeq
Since in the formalism of the continuum limit, the factors~$z$ and~$\overline{z}$ are treated as
two different functions, we do not need to worry about the possibility that the eigenvalues~$\lambda_+$
and~$\lambda_-$ might coincide or that the denominator in~\eqref{Fpm} might vanish.
Similarly, we can treat~$\xi$ and~$\overline{\xi}$
simply as two different vectors. Then the methods and results of Lemma~\ref{lemmanondeg}
apply and show that the matrices~$F_+$ and~$F_-$ have rank two,
so that the eigenvalues~$\lambda_+$ and~$\lambda_-$ are both two-fold degenerate.
By direct computation, one finds that (see Exercise~\ref{ex2.12})
\beq
F_\pm \:P(x,y) = \left\{ \begin{array}{cc}
0 & {\mbox{for ``$+$''}} \\
\frac{i}{2}\:\slashed{\xi}\:T^{(-1)}_{[0]} & {\mbox{for ``$-$''}}
\end{array} \right. \:+\: (\deg <2)\:. \label{FPasy}
\eeq
From~\eqref{lpm} and~\eqref{Fpm} one sees that
the eigenvalues of the closed chain form a {\em{complex conjugate pair}}
and are both {\em{two-fold degenerate}}.
Using this result in~\eqref{Ldiff}, one comes to the important
conclusion that the Lagrangian vanishes identically,
implying that, using the formalism of the continuum limit,
the fermionic projector of the vacuum is a minimizer of the
causal action. We will return to this point in a more general context in~\S\ref{secspecQ}.

The lower degrees on the light cone can be computed in a straightforward way by
expanding the formulas~\eqref{e:2f}. To give an impression, we here list a few formulas:
\begin{align*}
\lambda_\pm &= \frac{1}{4} \: \times\: \left\{ \begin{array}{cc}
(z\: T^{(-1)}_{[0]})\: \overline{T^{(-1)}_{[0]}} \:+\:
(z\: T^{(0)}_{[2]})\: \overline{T^{(-1)}_{[0]}} \:+\:
(z\: T^{(-1)}_{[0]})\: \overline{T^{(0)}_{[2]}} & {\mbox{for ``$+$''}} \\[.5em]
T^{(-1)}_{[0]}\: (\overline{z\: T^{(-1)}_{[0]}}) \:+\:
T^{(-1)}_{[0]}\: (\overline{z\: T^{(0)}_{[2]}}) \:+\:
T^{(0)}_{[2]}\: (\overline{z\: T^{(-1)}_{[0]}}) & {\mbox{for ``$-$''}} \end{array} \right. \notag \\[.1em]
&\quad+\: T^{(0)}_{[1]}\: \overline{T^{(0)}_{[1]}}
\:\mp\:\frac{T^{(0)}_{[1]}\: \overline{T^{(-1)}_{[0]}}
-T^{(-1)}_{[0]} \:\overline{T^{(0)}_{[1]}}}
{ T^{(0)}_{[0]}\: \overline{T^{(-1)}_{[0]}}
- T^{(-1)}_{[0]} \:\overline{T^{(0)}_{[0]}}} \:
(T^{(0)}_{[1]}\: \overline{T^{(0)}_{[0]}}
-T^{(0)}_{[0]} \:\overline{T^{(0)}_{[1]}}) \notag \\
&\quad+\: (\deg < 2) \:. \\
F_\pm\: P(x,y) &= \frac{i}{4}\:(\slashed{\xi}\:T^{(-1)}_{[0]}) \:+\:
(\deg < 2) \notag \\
&\quad\pm \frac{i}{4}\: \frac{(\slashed{\xi} \:T^{(-1)}_{[0]}) (T^{(0)}_{[0]}\:
\overline{T^{(-1)}_{[0]}} + T^{(-1)}_{[0]}\:
\overline{T^{(0)}_{[0]}} ) \:-\:
2\:(\overline{\slashed{\xi} \:T^{(-1)}_{[0]}})\: T^{(-1)}_{[0]}\:
T^{(0)}_{[0]}} {T^{(0)}_{[0]}\: \overline{T^{(-1)}_{[0]}}
- T^{(-1)}_{[0]} \: \overline{T^{(0)}_{[0]}}} \:.
\end{align*}
These formulas can be obtained more systematically with the perturbation expansion of
the spectral decomposition which we now describe.

\subsectionn{The Double Null Spinor Frame} \label{secdnsf}
Before entering the perturbation calculation, it is convenient to choose a specific eigenvector
basis of the closed chain of the vacuum. This basis is referred to as the
{\em{double null spinor frame}} and is denoted by~$(\mathfrak{f}^{L\!/\!R}_\pm)$.
\sindex{double null spinor frame}%
\nindex{bo8@$(\mathfrak{f}^{L \back / \back R}_\pm)$ -- double null spinor frame}%
Performing computations in the double null spinor frame is an improvement of
the method of ``factorizing matrix traces'' as introduced in~\cite[Appendix~G.2]{PFP}.
Following~\eqref{clcvac}, we introduce the matrix
\[ A^0_{xy} = \frac{1}{4}\: (\slashed{\xi} T^{(-1)}_{[0]}) (\overline{\slashed{\xi} T^{(-1)}_{[0]}}) \:. \]
\nindex{bp0@$A^0_{xy}$ -- unperturbed closed chain}%
According to~\eqref{lpm} and~\eqref{Fpm}, in the formalism of the
continuum limit the corresponding eigenvalues and spectral projectors are given by
\begin{gather}
\lambda_+ = T^{(0)}_{[0]} \,\overline{T^{(-1)}_{[0]}} \:,\qquad
\lambda_- = T^{(-1)}_{[0]} \,\overline{T^{(0)}_{[0]}} \label{l0pm} \\
F_\pm = \frac{1}{2} \Big( \1 \pm \frac{[\slashed{\xi}, \overline{\slashed{\xi}}]}{z-\overline{z}} \Big) ,
\label{F0def}
\end{gather}
\nindex{bp2@$\lambda_\pm, F_\pm$ -- eigenvalues and spectral projections of unperturbed closed chain}%
and they satisfy the relations
\[ F_+ \, \slashed{\xi} \overline{\slashed{\xi}} = z \,F_+ \:,\qquad \text{and} \qquad
F_- \, \slashed{\xi} \overline{\slashed{\xi}} = \overline{z} \,F_- \:. \]
Furthermore, the matrix~$A^0_{xy}$ is invariant on the left- and right-handed components, and thus we
may choose joint eigenvectors of the matrices~$A_0$ and~$\pseudo$.
This leads us to introduce the four eigenvectors~$\mathfrak{f}^{L\!/\!R}_\pm$ by the relations
\beq \label{zetaeigen}
\boxed{ \quad \chi_c \,F_s\, \mathfrak{f}^c_s = \mathfrak{f}^c_s \quad } 
\eeq
with $c \in \{L, R\}$ and~$s \in \{ +,- \}$,
which define each of these vectors up to a complex factor. For clarity in notation, we again write the
inner product on Dirac spinors~$\overline{\psi} \phi \equiv \psi^\dagger \gamma^0 \phi$
as~$\Sl \psi | \phi \Sr$, and refer to it as the {\em{spin scalar product}}.
\nindex{ad4@$\Sl . \vert . \Sr_x$ -- spin scalar product}%
\sindex{adjoint spinor}%
\nindex{ai4@$\overline{\psi} \phi = \Sl \psi \vert \phi \Sr$ -- inner product on spinors}%
\sindex{spin scalar product}%
Then the calculation
\[ \Sl \mathfrak{f}^L_+ \,|\, \mathfrak{f}^L_+ \Sr =  \Sl \chi_L \mathfrak{f}^L_+ \,|\, \chi_L \mathfrak{f}^L_+ \Sr
= \Sl \mathfrak{f}^L_+ \,|\, \chi_R \,\chi_L \,\mathfrak{f}^L_+ \Sr  = 0 \]
(and similarly for the other eigenvectors) shows that these vectors are indeed all null with respect
to the spin scalar product.
Moreover, taking the adjoint of~\eqref{F0def} with respect to the spin scalar product, one
sees that
\beq \label{F0adj}
(F_+)^* = F_-\:.
\eeq
As a consequence, the inner products vanish unless the lower indices are different, for example
\[ \Sl \mathfrak{f}^L_+ \,|\, \mathfrak{f}^R_+ \Sr = \Sl F_+ \mathfrak{f}^L_+ \,|\, F_+ \mathfrak{f}^R_+ \Sr
= \Sl \mathfrak{f}^L_+ \,|\, F_-\, F_+\, \mathfrak{f}^R_+ \Sr = 0\:. \]
We conclude that all inner products between the basis vectors vanish except
for the inner products~$\Sl \mathfrak{f}^L_+ | \mathfrak{f}^R_- \Sr$, $\Sl \mathfrak{f}^R_+ | \mathfrak{f}^L_- \Sr$ as well as their complex
conjugates~$\Sl \mathfrak{f}^R_- | \mathfrak{f}^L_+ \Sr$ and~$\Sl \mathfrak{f}^L_- | \mathfrak{f}^R_+ \Sr$. We assume that all the non-vanishing inner products are equal to one,
\beq \label{absone}
\boxed{ \quad
|\Sl \mathfrak{f}^L_+ \,|\, \mathfrak{f}^R_- \Sr| = 1 = |\Sl \mathfrak{f}^R_+ \,|\, \mathfrak{f}^L_- \Sr|\:.
\quad }
\eeq
In order to specify the phases and relative scalings of the basis vectors, we
introduce a space-like unit vector $u$ which is orthogonal to both~$\xi$ and~$\overline{\xi}$.
Then the imaginary vector $v=i u$ satisfies the relations
\beq \label{vrel}
\langle v, \xi \rangle = 0 = \langle v, \overline{\xi} \rangle \:,\quad
 \langle v, v \rangle = 1 \quad \text{and} \quad
\overline{v} = -v \:.
\eeq
As a consequence, the operator~$\slashed{v}$ commutes with~$F_+$ and~$F_-$, and since
it flips parity, we may set~$\mathfrak{f}^R_+ = \slashed{v} \,\mathfrak{f}^L_+$.
\nindex{bp8@$\slashed{v}$ -- operator commuting with~$F_\pm$}%
Next, a straightforward computation using~\eqref{F0def} gives the identities
\beq \label{Fident}
F_- \,\slashed{\xi} = \slashed{\xi}\, F_+ \qquad \text{and} \qquad
F_- \,\overline{\slashed{\xi}} = \overline{\slashed{\xi}} \, F_+\:.
\eeq
These identities can be used as follows. The first identity implies that
\[ \big( \chi_R F_- \,\slashed{\xi} \big) \,\mathfrak{f}^L_+
= \slashed{\xi}\,\chi_L F_+ \,\mathfrak{f}^L_+ \sim \slashed{\xi}\,\mathfrak{f}^L_+ \:, \]
showing that the vectors~$\slashed{\xi} \,\mathfrak{f}^L_+$ and~$\mathfrak{f}^R_-$ are linearly dependent. 
The calculation
\[ \Sl \mathfrak{f}^L_+ \,|\, \slashed{\xi} \mathfrak{f}^L_+ \Sr
= \Sl \mathfrak{f}^L_+ \,|\, \slashed{\xi} \:\frac{\slashed{\xi} \overline{\slashed{\xi}}}{z}\:\mathfrak{f}^L_+ \Sr 
= \Sl \mathfrak{f}^L_+ \,|\, \frac{\slashed{\xi} \slashed{\xi} }{z}\:\overline{\slashed{\xi}} \:\mathfrak{f}^L_+ \Sr 
= \Sl \mathfrak{f}^L_+ \,|\, \overline{\slashed{\xi}} \mathfrak{f}^L_+ \Sr \]
(where we used~\eqref{zetaeigen} and~\eqref{F0def}) shows that the vector~$\slashed{\xi} \,\mathfrak{f}^L_+$
is in fact a real multiple of~$\mathfrak{f}^R_-$. Hence by normalizing~$\mathfrak{f}^L_+$
appropriately, we can arrange\footnote{Let us explain why we do not consider
the opposite sign~$\mathfrak{f}^R_- = -\slashed{\xi} \,\mathfrak{f}^L_+$. To this end,
we must show that~$\Sl \mathfrak{f}^L_+ | \slashed{\xi} \mathfrak{f}^L_+ \Sr > 0$.
Since for any given positive or definite spinor~$\zeta$, the vector~$\chi_L F_+ \zeta$ is a multiple of~$\mathfrak{f}^L_+$, it suffices to compute instead the sign of the
combination~$\Sl \chi_L F_+ \zeta | \slashed{\xi} \chi_L F_+ \zeta \Sr$.
Applying~\eqref{F0adj} and~\eqref{Fident}, this inner product simplifies
to~$\Sl \zeta | \chi_R F_-\slashed{\xi} \zeta \Sr$. With the help of~\eqref{FPasy} and~\eqref{Peq}, we
can treat the factor~$\slashed{\xi}$ as an outer factor. Then our inner product simplifies to the expectation
value~$\Sl \zeta | \chi_R \slashed{\xi} \zeta \Sr$. This expectation value is positive if we follow
the convention introduced before~\eqref{asy} that~$\xi^0>0$.}
that~$\mathfrak{f}^R_- = \slashed{\xi} \,\mathfrak{f}^L_+$.
Using the second identity in~\eqref{Fident}, we also find that~$\mathfrak{f}^R_- =
\overline{\slashed{\xi}} \,\mathfrak{f}^L_+$. Similarly, we may also set~$\mathfrak{f}^L_-
= \slashed{\xi} \,\mathfrak{f}^R_+ = \overline{\slashed{\xi}} \,\mathfrak{f}^R_+$.
The resulting relations between our basis vectors are summarized in the following diagram:
\beq \label{commute}
\begin{CD}
\mathfrak{f}^L_+ @> \slashed{v} > \phantom{\quad\;\; \quad} > \mathfrak{f}^R_+ \\
@ V \displaystyle \slashed{\xi} V \displaystyle \overline{\slashed{\xi}} V @
V \displaystyle \slashed{\xi} V \displaystyle \overline{\slashed{\xi}} V \\
\mathfrak{f}^R_- @> -\slashed{v}>\phantom{\quad\;\; \quad} > \mathfrak{f}^L_-
\end{CD} 
\eeq
With~\eqref{zetaeigen}, \eqref{absone} and~\eqref{commute} we have introduced the
double null spinor frame~$(\mathfrak{f}^{L\!/\!R}_\pm)$. The construction involves the
freedom in choosing the operator~$\slashed{v}$ according to~\eqref{vrel};
for given~$\slashed{v}$, the basis vectors are unique up to an irrelevant common phase.
The construction of the double null spinor frame is illustrated in Exercise~\ref{ex2.13}.

We next explain how we can represent a given linear operator~$B$ on the spinors in the double null
frame~$(\mathfrak{f}^{L\!/\!R}_\pm)$. Following the notation in~\cite[Appendix~G]{PFP},
we denote the matrix element in the column~$(c,s)$ and row~$(c',s')$ by~$\mathfrak{F}^{c c'}_{s s'}(B)$.
These matrix entries are obtained by acting with~$B$ on the vector~$\mathfrak{f}^{c'}_{s'}$
and taking the inner product with the basis vector which is conjugate to~$\mathfrak{f}^{c}_{s}$, i.e.\
\beq \label{matrixelement}
\mathfrak{F}^{c c'}_{s s'}(B) = \Sl \mathfrak{f}^{\overline{c}}_{\overline{s}} \,|\, B\, \mathfrak{f}^{c'}_{s'} \Sr\:,
\eeq
where the conjugation flips the indices according to~$L \leftrightarrow R$
and~$+ \leftrightarrow -$.
\nindex{bq0@$\mathfrak{F}^{c c'}_{s s'}$ -- matrix elements in double null spinor frame}%
Similarly, we can also express the projectors~$\chi_c F_s$ in terms of the basis vectors,
for example
\beq \label{LF0rel}
\chi_L F_+ = |\mathfrak{f}^L_+ \Sr \Sl \mathfrak{f}^R_- | \:.
\eeq
For computing~\eqref{matrixelement},
we use the relations in~\eqref{commute} to express
the vector~$\mathfrak{f}^{c'}_{s'}$ in terms of~$\mathfrak{f}^L_+$, choosing the
relations which do not involve factors of~$\overline{\slashed{\xi}}$.
Similarly, we express the vector~$\mathfrak{f}^{\overline{c}}_{\overline{s}}$ in
terms of~$\mathfrak{f}^R_-$, avoiding factors of~$\slashed{\xi}$.
Applying~\eqref{LF0rel}, we can then rewrite the inner product as a
trace involving the operator~$F_+$. More precisely, a straightforward calculation yields
\beq
\left. \begin{array}{lcl}
\!\!\!\!\!\mathfrak{F}^{LL}_{++}(B) = \Tr (F_+ \:\chi_L\: B)
&\!\!\!\!,\quad& \displaystyle \mathfrak{F}^{LR}_{++}(B) = \Tr(F_+ \:\slashed{v}\: \chi_L\: B)
\\[.4em]
\!\!\!\!\!\mathfrak{F}^{LL}_{+-}(B) = \Tr (\slashed{\xi}\: F_+ \:\slashed{v}\:\chi_L\: B)
&\!\!\!\!,\quad& \mathfrak{F}^{LR}_{+-}(B) = \Tr(\slashed{\xi}\: F_+ \:\chi_L\: B) \\[.4em]
\!\!\!\!\!\mathfrak{F}^{LL}_{-+}(B) = \displaystyle \frac{1}{z} \,\Tr (F_+ \:\slashed{v}\:\slashed{\xi}\chi_L\: B)
&\!\!\!\!,\quad& \mathfrak{F}^{LR}_{-+}(B) =
\displaystyle \frac{1}{z} \,\Tr(F_+ \:\slashed{\xi}\: \chi_L\: B) \\[.7em]
\!\!\!\!\!\mathfrak{F}^{LL}_{--}(B) = \displaystyle \frac{1}{z}\,\Tr (\slashed{\xi}\: F_+ \:\slashed{\xi} \:\chi_L\: B)
&\!\!\!\!,\quad& \mathfrak{F}^{LR}_{--}(B) = \displaystyle \frac{1}{z}\,\Tr(\slashed{\xi}\:
F_+ \:\slashed{v}\:\slashed{\xi}\: \chi_L\: B)
\end{array} \right\} \label{e:Ce2}
\eeq
(see also~\cite[eq.~(G.19)]{PFP}, where these relations are derived with a different method).
Indeed, it suffices to compute the given eight matrix elements, because the other eight
matrix elements are obtained by the replacements~$L \leftrightarrow R$.
Moreover, the matrix elements of the adjoint (with respect to the spin scalar product)
are obtained by
\[ \mathfrak{F}^{c c'}_{s s'}(B^*) = \Sl \mathfrak{f}^{\overline{c}}_{\overline{s}} \,|\, B^*
\,\mathfrak{f}^{c'}_{s'} \Sr
= \overline{\Sl \mathfrak{f}^{c'}_{s'} \,|\, B \,\mathfrak{f}^{\overline{c}}_{\overline{s}} \Sr}
=  \overline{\mathfrak{F}^{\overline{c' c}}_{\overline{s' s}}(B)}\:. \]
A simple example for how to compute the matrix elements in the double null
spinor frame is given in Exercise~\ref{ex2.14}.

\subsectionn{Perturbing the Spectral Decomposition} \label{secpertspec}
Omitting the arguments~$(x,y)$, we decompose the fermionic projector as
\[ P = P_0 + \Delta P \:, \]
where~$P_0$ is the vacuum fermionic projector (possibly modified by gauge phases).
This gives rise to the decomposition of $A$
\beq
A = A_0 + \Delta A \label{e:Cb1}
\eeq
with
\begin{eqnarray}
\hspace*{-0.5cm}A_0 &=& P_0(x,y)\: P_0(y,x) \label{e:Cc1} \\
\hspace*{-0.5cm}\Delta A &=& \Delta P(x,y) \: P_0(y,x) \:+\: P_0(x,y) \:\Delta P(y,x) \:+\:
\Delta P(x,y) \:\Delta P(y,x) \:. \label{e:Cd1}
\end{eqnarray}
The eigenvalues and spectral projectors of $A_0$ were computed
explicitly in~\S\ref{secregvac}. In view of later generalizations, we write
the obtained spectral decomposition as
\[ A_0 = \sum_{k=1}^K \lambda_k\: F_k \]
with~$K=2$, where~$\lambda_k$ are distinct eigenvalues with corresponding spectral
projections~$F_k$. Since the perturbation~$\Delta A$ will in general remove the degeneracies,
we cannot expect that by perturbing $F_k$ we again obtain spectral projection operators.
But we can form projectors $G_k$ on the space spanned
by all eigenvectors of $A$ whose eigenvalues are sufficiently close to
$\lambda_k$. The $G_k$ are most conveniently introduced using contour
integrals. We choose $\varepsilon>0$ such that
\[ |\lambda_i-\lambda_j| \;<\; 2 \varepsilon \qquad {\mbox{for all
$i,j=1,\ldots,K$ and $i \neq j$}}. \]
Then we set
\beq \label{e:Ba}
G_k = \frac{1}{2 \pi i} \oint_{|z-\lambda_k|=\varepsilon}
(z-A)^{-1}\: dz\:,
\eeq
Combining the resolvent identity with the Cauchy integral formula,
one sees that~$G_k$ is indeed an idempotent operator
whose image is the invariant subspace corresponding to the eigenvalues near~$\lambda_k$
(for details see Exercise~\ref{ex2.31}).

The integral formula~\eqref{e:Ba} is very useful for a perturbation expansion.
To this end, we substitute~\eqref{e:Cb1} into~\eqref{e:Ba} and compute the
inverse with the Neumann series,
\begin{eqnarray*}
G_k &=& \frac{1}{2 \pi i} \oint_{|z-\lambda_k|=\varepsilon} \big( z-A_0-\Delta A \big)^{-1}\: dz \\
&=& \frac{1}{2 \pi i} \oint_{|z-\lambda_k|=\varepsilon}
\left(\1 - (z-A_0)^{-1}\: \Delta A \right)^{-1}\:(z-A_0)^{-1}\: dz \\
&=& \frac{1}{2 \pi i} \oint_{|z-\lambda_k|=\varepsilon}
\sum_{n=0}^\infty \left((z-A_0)^{-1}\: \Delta A \right)^n \:(z-A_0)^{-1}\:
dz\: .
\end{eqnarray*}
Interchanging the integral with the infinite sum gives the perturbation
expansion,
\beq \label{e:Bb}
G_k = \sum_{n=0}^\infty\: \frac{1}{2 \pi i}
\oint_{|z-\lambda_k|=\varepsilon}
\left((z-A_0)^{-1}\: \Delta A \right)^n \:(z-A_0)^{-1}\: dz\:,
\eeq
where $n$ is the order in perturbation theory. After substituting in the
spectral representation for $(z-A_0)^{-1}$,
\beq
(z-A_0)^{-1} = \sum_{l=1}^K \frac{F_l}{z-\lambda_l}\:, \label{e:BzA0}
\eeq
the contour integral in~\eqref{e:Bb} can be carried out with residues.
For example, we obtain to second order,
\begin{align}
G_k &= F_k \:+\: \sum_{l \neq k} \frac{1}{\lambda_k-\lambda_l}
\left( F_k\: \Delta A\: F_l \:+\: F_l \:\Delta A\: F_k \right)
\:+\: \O((\Delta A)^3) \notag \\
&\qquad +\sum_{l, m \neq k} \frac{1}{(\lambda_k-\lambda_{l})
(\lambda_k-\lambda_{m})}  \notag \\
&\hspace*{1.5cm} \times
\left(
F_k\: \Delta A\: F_{l} \:\Delta A\: F_{m} +
F_{l}\: \Delta A\: F_k \:\Delta A\: F_{m} +
F_{l}\: \Delta A\: F_{m} \:\Delta A\: F_k \right) \notag \\
&\qquad -\sum_{l \neq k} \frac{1}{(\lambda_k-\lambda_l)^2}
\notag \\
&\hspace*{1.5cm} \times
\left( F_k\: \Delta A\: F_k \:\Delta A\: F_l +
F_k\: \Delta A\: F_l \:\Delta A\: F_k +
F_l\: \Delta A\: F_k \:\Delta A\: F_k \right). \label{e:Bcc}
\end{align}
To order $n>2$, the corresponding formulas are clearly more complicated,
but even then they involve matrix products which are all of the form
\beq
F_{k_1} \:\Delta A\: F_{k_2} \:\Delta A\: \cdots \:F_{k_n} \:\Delta A\:
F_{k_{n+1}}\:. \label{e:Bd}
\eeq
An example of a first order perturbation computation is given in Exercise~\ref{ex2.15}.

\subsectionn{General Properties of the Spectral Decomposition} \label{secspecA}
We now derive a few general properties of the spectral decomposition of the closed chain.

\begin{Lemma} \label{lemma71}
Assume that for a one-parameter family of fermionic projectors~$P(\tau)$ and fixed~$x,y \in M$,
the matrices~$A_{xy}$ and~$A_{yx}$ are diagonalizable for all~$\tau$ in a neighborhood
of~$\tau=0$, and that the eigenvalues of the matrix~$A_{xy}|_{\tau=0}$ are all non-real.
Then the unperturbed closed chain $A_{xy}$ has a spectral representation
\beq \label{Arep}
A_{xy} \big|_{\tau=0} = \sum_{k=1}^4 \lambda^{xy}_k F^{xy}_k
\eeq
with the following properties.
\sindex{closed chain!spectral analysis of}%
The last two eigenvalues and spectral projectors are related
to the first two by
\beq \label{lorder}
\lambda^{xy}_3 = \overline{\lambda^{xy}_1}\:,\quad F^{xy}_3 = (F^{xy}_1)^*
\qquad \text{and} \qquad
\lambda^{xy}_4 = \overline{\lambda^{xy}_2}\:, \quad F^{xy}_4 = (F^{xy}_2)^*\:.
\eeq
The first order perturbation~$\delta A_{xy} = \partial_\tau A_{xy}|_{\tau=0}$
of the closed chain is diagonal in the bases of the non-trivial degenerate subspaces, i.e.
\beq \label{Fdiag}
F^{xy}_k (\delta A_{xy}) F^{xy}_l = 0 \qquad \text{if~$k \neq l$
and~$\lambda^{xy}_k = \lambda^{xy}_l$.}
\eeq
The closed chain~$A_{yx}$ has a corresponding spectral representation
satisfying~\eqref{Arep}--\eqref{Fdiag} with all indices `$xy$' are replaced by `$yx$'.
The spectral representations of~$A_{xy}$ and~$A_{yx}$ are related to each other by
\beq \label{eq77}
\lambda^{xy}_k =  \lambda^{yx}_k \qquad \text{and} \qquad
F_k^{xy}\, P(x,y) = P(x,y)\, F_k^{yx}\:.
\eeq
\end{Lemma}
\Proof By continuity, the eigenvalues of the matrix~$A_{xy}$ are non-real
in a neighborhood of~$\tau=0$. Moreover, by direct computation
one sees that the matrix~$A_{xy}$ is symmetric
in the sense that~$A_{xy} = A_{xy}^* = \gamma^0 A_{xy}^\dagger \gamma^0$.
Hence, using the idempotence of the matrix~$\gamma^0$ together with the multiplicity of the
determinant, we find that
\[ \det(A_{xy} - \lambda)  = 
\det( \gamma^0(A_{xy}^\dagger - \lambda) \gamma^0) = 
\det(A_{xy}^\dagger - \lambda) = \overline{\det(A_{xy} - \overline{\lambda})} \:. \]
Hence if~$\lambda$ is an eigenvalue of the matrix~$A_{xy}$, so is~$\overline{\lambda}$.
Thus the eigenvalues must form complex conjugate pairs.

We first complete the proof in the case that there are no degeneracies.
For any eigenvalue~$\lambda$ of~$A_{xy}$ we choose a
polynomial~$p_\lambda(z)$ with~$p_\lambda(\lambda)=1$
and~$p_\lambda(\mu)=0$ for all other spectral points~$\mu$.
Then the spectral projector on the eigenspace corresponding to~$\lambda$,
denoted by~$F^{xy}_\lambda$, is given by
\beq \label{Frel}
F_\lambda^{xy} = p_\lambda(A_{xy}) \:.
\eeq
Taking the adjoint and possibly after reordering the indices~$k$, we
obtain the relations~\eqref{Arep} and~\eqref{lorder}.
The general matrix relation $\det(BC-\lambda) = \det(CB-\lambda)$ (see for
example~\cite[Section~3]{discrete}) shows that the closed chains~$A_{xy}$
and~$A_{yx}$ have the same spectrum.
Multiplying~\eqref{Frel} by~$P(x,y)$ and iteratively applying the relation
\[ A_{xy}\, P(x,y) = P(x,y)\, P(y,x)\, P(x,y) = P(x,y)\, A_{yx} \:, \]
we find that~$F_\lambda^{xy}\, P(x,y) = P(x,y)\, F_\lambda^{yx}$.
Thus we can label the eigenvalues of the matrix~$A_{yx}$ such that~\eqref{eq77} holds.

In the case with degeneracies, the assumption that~$A_{xy}$ is diagonalizable in
a neighborhood of~$\tau=0$ allows us to diagonalize~$\delta A_{xy}$ on the
degenerate subspaces (see for example~\cite{baumgaertel}
or the similar method for self-adjoint operators in~\cite[Section~11.1.2]{schwabl1}).
This yields~\eqref{Fdiag}, whereas~\eqref{lorder} can be
arranged by a suitable ordering of the spectral projectors~$F^{xy}_k$.
In the degenerate subspaces of~$A_{yx}$ we can choose the bases such
that~\eqref{Arep} and~\eqref{lorder} hold (with `$xy$' replaced by `$yx$')
and that~\eqref{eq77} is satisfied. It remains to prove that~\eqref{Fdiag}
also holds for~$A_{yx}$:
From~\eqref{Fdiag} we know that for any pair~$l,k$ with~$\lambda^{xy}_l =\lambda^{xy}_k$,
\begin{align*}
0 &=  F^{xy}_k (\delta A_{xy}) F^{xy}_l = F^{xy}_k \Big(\delta P(x,y)\, P(y,x)
+ P(x,y)\, \delta P(y,x) \Big) F^{xy}_l \\
&=  F^{xy}_k (\delta P(x,y)) F^{yx}_l \, P(y,x) + P(x,y) F^{yx}_k (\delta P(y,x)) F^{xy}_l \:,
\end{align*}
where in the last line we applied the second equation in~\eqref{eq77}.
Multiplying by $P(y,x)$ from the left and by~$P(x,y)$ on the right, we find
\[ 0 = P(y,x) F^{xy}_k (\delta P(x,y)) F^{yx}_l \,\lambda^{yx}_l + \lambda^{yx}_k  F^{yx}_k (\delta P(y,x)) F^{xy}_l P(x,y) \:. \]
After dividing by~$\lambda^{yx}_l=\lambda^{yx}_k$ (note that the eigenvalues are non-zero
because they are assumed to form complex conjugate pairs), we can again use the
second equation in~\eqref{eq77} to obtain
\begin{align*}
0 &= P(y,x) F^{xy}_k (\delta P(x,y)) F^{yx}_l  + F^{yx}_k (\delta P(y,x)) F^{xy}_l P(x,y) \\
&= F^{yx}_k \Big(P(y,x) \:\delta P(x,y) + \delta P(y,x)\: P(x,y) \Big) F^{yx}_l
= F^{yx}_k (\delta A_{yx}) F^{yx}_l\:,
\end{align*}
concluding the proof.
\QED

\subsectionn{Spectral Analysis of the Euler-Lagrange Equations} \label{secspecQ}
We now explain how the spectral decomposition of the closed chain can be used
to analyze the causal action principle introduced in~\S\ref{secbasicdef}
as well as the corresponding EL equations as worked out in~\S\ref{secvary}.
For the regularized Dirac sea vacuum as considered in~\S\ref{secregvac},
the situation is quite simple. Namely, according to Lemma~\ref{lemma11}
(or more explicitly in~\eqref{l0pm}), the closed chain has two eigenvalues which form
a complex conjugate pair. As a consequence, the eigenvalues all have the same
absolute value. Writing the Lagrangian in the form~\eqref{Ldiff}, one sees
that the Lagrangian vanishes identically. We come to the following conclusion:
\beq
\begin{tabular}{l}
\text{In the formalism of the continuum limit, the regularized Dirac sea vacuum} \\
\text{is a minimizer of the causal action.}
\end{tabular}
\eeq
\sindex{causal action!minimizer}%
If the fermionic projector of the vacuum is perturbed (for example by an external potential
or by additional particle or antiparticle states), the degeneracy of the eigenvalues will in general
disappear, so that the spectrum will consist of two complex conjugate pairs.
As a consequence, the causal action will no longer vanish.
In order to analyze whether we still have a critical point of the causal action,
one needs to analyze the corresponding EL equations in Proposition~\ref{prpEL}.
To this end, it is very convenient to rewrite these EL equations using the spectral
decomposition of the closed chain, as we now explain.

For simplicity, we again restrict attention to Dirac spinors and spin dimension two.
Moreover, we only consider the case that the
Lagrange multipliers~$\kappa$ and~$\lambda$ in Proposition~\ref{prpEL} are both equal to zero.
The generalization to higher spin dimension and to non-trivial $\kappa$ and~$\lambda$ are
straightforward and will be carried out later on (see Lemma~\ref{lemma73},
Lemma~\ref{s:lemma81} and the similar results in~\S\ref{l:sec40}).
Writing the Lagrangian in the form~\eqref{Ldiff}, we have
\beq
\L(x,y) = \frac{1}{8} \sum_{i,j=1}^{4} \Big( |\lambda^{xy}_i| - |\lambda^{xy}_j| \Big)^2 \:. \label{t:Lex}
\eeq
The relation~\eqref{Fdiag} allows us to compute the
variation of the eigenvalues by a standard first order perturbation calculation
without degeneracies,
\beq \label{t:firstper}
\delta \lambda^{xy}_k = \Tr(F^{xy}_k \, \delta A_{xy})\:.
\eeq
Using that that~$\delta |\lambda| = \re(\overline{\lambda}\, \delta \lambda/|\lambda|)$, we
can compute the first variation of~\eqref{t:Lex} by
\beq \label{t:dL}
\delta \L(x,y) = \frac{1}{2}\, \re \sum_{j,k=1}^{4}
\left( |\lambda^{xy}_k| - |\lambda^{xy}_j| \right)\:
\frac{\overline{\lambda^{xy}_k}}{|\lambda^{xy}_k|}\: \Tr(F^{xy}_k \, \delta A_{xy})\:.
\eeq
We now insert the identity
\[ \delta A_{xy} = \delta P(x,y)\, P(y,x) + P(x,y)\, \delta P(y,x) \:. \]
Cyclically commuting the arguments of the trace, we obtain
\begin{align*}
\delta \L(x,y) &= \frac{1}{2}\, \sum_{j,k=1}^{4}
\Big( |\lambda^{xy}_k| - |\lambda^{xy}_j| \Big)\\
&\quad\; \times \re 
\Tr \bigg[ \frac{\overline{\lambda^{xy}_k}}{|\lambda^{xy}_k|}\: P(y,x) \,F^{xy}_k \,
\delta P(x,y) + 
\frac{\overline{\lambda^{xy}_k}}{|\lambda^{xy}_k|}\: F^{xy}_k
P(x,y)\, \delta P(y,x) \bigg]\:.
\end{align*}
Using~\eqref{lorder} and~\eqref{eq77}, one sees that the first summand in the
square bracket is the adjoint of the second summand. Therefore, the trace of the
square bracket is real-valued, so that it is unnecessary take the real part.
Comparing with~\eqref{delLdef}, we conclude that
\beq \label{t:Qspec}
Q(x,y) = \frac{1}{2} \sum_{j,k=1}^{4}
\Big( |\lambda^{xy}_k| - |\lambda^{xy}_j| \Big)\:
\frac{\overline{\lambda^{xy}_k}}{|\lambda^{xy}_k|}\:F^{xy}_k \,P(x,y)
\eeq
(where we again used~\eqref{eq77}). In the vacuum, when the eigenvalues of the closed chain
form a complex conjugate pair~\eqref{l0pm}, the kernel~$Q(x,y)$ vanishes identically
in the formalism of the continuum limit. If the fermionic projector of the vacuum is perturbed,
the first order perturbation of~$Q(x,y)$ can be computed easily with the help of~\eqref{t:firstper}.
The higher orders in perturbation theory can be treated systematically by using the contour method
in~\S\ref{secpertspec} and by evaluating the resulting expressions in the formalism of the continuum limit.

The above methods give a mathematical meaning to~$Q(x,y)$ in the formalism of the
continuum limit. The remaining difficulty is that in the EL equations worked out
in Proposition~\ref{prpEL}, the kernel~$Q(x,y)$ appears inside an integral~\eqref{Qrel},
and one must control the error terms~\eqref{ap1} and~\eqref{ap2} inside this integral.
The method is to choose a vector~$u \in \H$ such that its
physical wave function~$\psi^u$ is supported away from~$x$, up to a small
error. This method is referred to as {\em{testing on null lines}}.
\sindex{testing on null lines}%
In a more physical picture, one chooses~$\psi^u$ as an {\em{ultrarelativistic
wave packet}} localized near a null curve which does not meet the space-time point~$x$.
\sindex{ultrarelativistic wave packet}%
Applying this method to~\eqref{Qrel}, the left side is evaluated weakly on the
light cone, whereas the right side vanishes. In this way, the
EL equations in the continuum limit reduce to
\[ Q(x,y) = 0 \qquad \text{evaluated weakly on the light cone}\:. \]
We refer for details to~\S\ref{s:secELC}. The estimates of all the error
terms are worked out in Appendix~\ref{s:appnull}.

\section*{Exercises}
\begin{Exercise} \label{ex2.0}  (external field problem) {\em{
In physics textbooks, the notions of a ``particle'' and ``anti-particle'' are often associated to the frequency
(or equivalently the energy) of the solutions: solutions of positive frequency are called particles, whereas
the negative-frequency solutions are reinterpreted as describing anti-particle states.
The aim of this exercise is to explain why these notions are ill-defined in the presence of a
time-dependent potential. To this end, we consider the Dirac equation
\begin{align}\label{DiracExt}
(i \slashed{\partial}+\B-m) \psi = 0 \:,
\end{align}
where~$\B$ is a ``step potential in time'' i.e.\
\[ \B(t,\vec{x}) = V \gamma^0\:\Theta(t) \:\Theta(1-t) \]
with a real parameter~$V$.
\begin{itemize}[leftmargin=2em]
\item[(a)] Separate out the spatial dependence for any given~$\vec{k} \in \R^3$
with the plane-wave ansatz
\[ \psi(t,\vec{x}) = e^{i \vec{k} \vec{x}} \,\phi(t) \]
(where~$\phi$ is a spinor-valued function). Derive the resulting ordinary differential equation for~$\phi(t)$.
\item[(b)] Clearly, the potential has discontinuities at~$t=0$ and~$t=1$. Show that there are
two fundamental solutions~$\phi_1, \phi_2 \in C^0(\R, \C^4)$ which are smooth solutions of the
ODE except at the points~$t=0$ and~$t=1$.
{\em{Remark:}} This procedure is familiar to physics students from quantum mechanics textbooks
where wave functions are ``glued together'' at discontinuities of step potentials.
From the mathematical point of view, the ``glueing'' of the solutions can be justified by saying
that~$\phi_1$ and~$\phi_2$ are a fundamental system of {\em{weak solutions}} of the ODE.
To the reader who is not familiar with these concepts, it might be
instructive to verify that the notion of ``weak solution''
really gives rise to a continuity condition for~$\phi$. (Likewise, for a second order equation
like the Schr\"odinger equation, the notion of ``weak solution'' gives rise to $C^1$-solutions
whose second derivatives are discontinuous.)
\item[(c)] Consider a ``scattering process'' where for negative times the solution is of the form
\[ \phi(t) = e^{-i \omega t} \:\chi \:, \]
where~$\chi$ is a constant spinor and~$\omega := \sqrt{\vec{k}^2 - m^2}$. Show that for
time~$t>1$, this solution can be written as
\[ \phi(t) = e^{-i \omega t} \,\chi_+ +  e^{+i \omega t} \,\chi_- \]
with constant spinors~$\chi_+$ and~$\chi_-$. Compute~$\chi_+$ and~$\chi_-$ explicitly
as functions of~$\chi$ and~$V$. Verify in particular that~$\chi_-$ in general does not vanish.
\item[(d)] What does this mean for the interpretation of the solution in terms of ``particles''
and ``anti-particles''? Why can the frequency of the solutions not be used for a
global concept of particles and anti-particles?
How can a pair creation/annihilation process be understood in our example?
{\em{Remark:}} In order to avoid misunderstandings, we point out that the above arguments
only show that the {\em{frequency}} cannot be used to obtain a global particle interpretation.
They do not rule out the possibility that there may be a well-defined global particle interpretation
using other properties of the solutions. In fact, such a global particle interpretation
is provided by the causal perturbation expansion (or the
corresponding functional analytic constructions in~\cite{finite, infinite, hadamard}).
However, this global particle interpretation in general does not coincide with the ``particles''
and ``anti-particles'' as experienced by a local observer.
\end{itemize}
}} \end{Exercise}

\begin{Exercise} \label{ex2.1} {\em{
This exercise is devoted to the {\em{advanced Green's function}}~$s^\vee_m$
(for a more computational exercise on the advanced Green's function
see Exercise~\ref{ex2.4-21} below).
\begin{itemize}[leftmargin=2em]
\item[(a)] Assume that~$m>0$. Show that the limit~$\nu \searrow 0$
in~\eqref{8b} exist in the distributional sense.
\item[(b)] Show that the limit~$\nu \searrow 0$ in~\eqref{8b} also exists in the
massless case~$m=0$ and that
\[ \lim_{m \searrow 0} s_m^\vee(k) = s_0^\vee(k) \qquad \text{as a distribution}\:. \]
{\em{Hint:}} Proceed similar as in Exercise~\ref{ex21}.
\item[(c)] Consider the Fourier integral in the $q^0$-variable
\[ \int_{-\infty}^\infty \frac{1}{q^{2}-m^{2}-i \nu q^{0}} \:e^{i q^0 t}\: dq^0 \:. \]
Show with residues that this integral vanishes for sufficiently small~$\nu$ if~$t<0$.
\item[(d)] Argue with Lorentz invariance to prove the left side of~\eqref{GF}.
\end{itemize}
}} \end{Exercise}

\begin{Exercise} \label{ex2.2-1} {\em{
Modifying the location of the poles in~\eqref{8b} gives rise to the distribution
\[ s^F_m(k) := \lim_{\nu \searrow 0}
\frac{\slashed{k} + m}{k^{2}-m^{2}+i \nu} \:. \]
This is the well-known {\em{Feynman propagator}}, which is often
described intuitively by saying that ``positive frequencies move to the future and
negative frequencies move to the past.'' Make this sentence precise by
a computation similar to that in Exercise~\ref{ex2.1}~(c).
}} \end{Exercise}

\begin{Exercise} \label{ex2.2} {\em{
\begin{itemize}[leftmargin=2em]
\item[(a)] Assume that~$m>0$. Give a detailed proof of the distributional
relation~\eqref{kmdef}. {\em{Hint:}} Argue similar as in Exercise~\ref{ex21}.
\item[(b)] Prove that~\eqref{kmdef} also holds in the case~$m=0$.
{\em{Hint:}} The subtle point is to analyze the behavior at~$q=0$.
To this end, apply Lebesgue's dominated convergence theorem.
\end{itemize}
}} \end{Exercise}

\begin{Exercise} \label{ex2.3} (probability integral and current conservation) {\em{
Let~$\psi, \phi$ be two solutions of the Dirac equation~\eqref{direx} with a
smooth potential~$\B$ which is symmetric~\eqref{Bsymm}.
Moreover, assume that~$\psi$ and~$\phi$ are smooth and have spatially
compact support.
\begin{itemize}[leftmargin=2em]
\item[(a)] Show that the integral~\eqref{print} is independent of~$t_0$.
\item[(b)] More generally, let~$\scrN$ be a Cauchy surface in Minkowski space
with future-directed normal~$\nu$. Show that the integral
\[ \int_\scrN \overline{\psi} (\slashed{\nu} \phi)\: d\mu_\scrN \]
is independent of the choice of the Cauchy surface
(where~$d\mu_\scrN$ is the volume measure corresponding to the induced
Riemannian metric on~$\scrN$).
{\em{Hint:}} Show that the vector field~$\overline{\psi} \gamma^j \phi$ is divergence-free
and apply the Gau{\ss} divergence theorem.
\end{itemize}
}} \end{Exercise}

\begin{Exercise} \label{ex2.31}  (resolvent and contour integrals) {\em{
The aim of this exercise is to make the reader familiar with the notion of the resolvent and the
contour integral representation of spectral projectors in the finite-dimensional setting.
More details and generalizations to infinite dimensions can be found
in the book by Kato~\cite{kato}.
\begin{itemize}[leftmargin=2em]
\item[(a)] Let~$A \in \Lin(\C^k)$ be a $k \times k$-matrix.
The {\em{resolvent set}} is the set of all~$\lambda \in \C$ for which the matrix~$(A-\lambda)$
is invertible. The {\em{spectrum}} is the complement of the resolvent set.
For any~$\lambda$ in the resolvent set, we define the {\em{resolvent}} $R_\lambda$ by
\[ R_\lambda = (A-\lambda\1)^{-1} \]
(we use this sign convention consistently, although some authors use the opposite sign convention).
Prove the {\em{resolvent identity}}
\[ R_\lambda \: R_{\lambda'} = \frac{1}{\lambda-\lambda'} \:\big(
R_\lambda - R_{\lambda'} \big) \:, \]
valid for any~$\lambda, \lambda'$ in the resolvent set.
{\em{Hint:}} Multiply the identity~$\lambda'-\lambda = (A-\lambda) - (A-\lambda')$
from the left and right by a resolvent.
\item[(b)] Assume that~$A$ is a Hermitian matrix. Let~$\Gamma$ be a contour which encloses
only one eigenvalue~$\lambda_0$ with winding number one. Show that the contour integral
\beq \label{cintmatrix}
-\frac{1}{2 \pi i} \ointctrclockwise_{\Gamma} R_\lambda\: d\lambda
\eeq
is an orthogonal projection onto the corresponding eigenspace.
{\em{Hint:}} Choose an eigenvector basis and apply the Cauchy integral formula.
\item[(c)] Now let~$A$ be any matrix. Let~$\Gamma$ be a contour which encloses
a point~$\lambda_0$ in the spectrum with winding number one. Show that the contour integral~\eqref{cintmatrix}
is an idempotent operator whose image is the corresponding invariant subspace.
{\em{Hint:}} Choose a Jordan representation of the matrix.
Restrict attention to one Jordan block. Then the resolvent can be written as a Neumann
series, which reduces to a finite sum. The resulting integral can be computed
with residues.
\item[(d)] Derive the idempotence relation in~(c) directly from the resolvent identity.
{\em{Hint:}} A very similar computation is given in the proof of Theorem~\ref{thmfcalc}.
\end{itemize}
}} \end{Exercise}

\begin{Exercise} \label{ex2.4}  {\em{ In this exercise we explore an alternative
and more computational proof of Lemma~\ref{lemmakmid}.
\begin{itemize}[leftmargin=2em]
\item[(a)] Show by direct computation in momentum space that~$k_m \,|_{t_0}\, k_m = k_m$.
{\em{Hint:}} Proceed similarly as in the derivation of~\eqref{Pnorm}
in the proof of Lemma~\ref{lemmaDiracsea}.
\item[(b)] Show that due to current conservation (see Exercise~\ref{ex2.3} above),
the operator~$\tilde{k}_m \,|_{t_0}\, \tilde{k}_m$ is independent of~$t_0$.
Therefore, it suffices to compute the limit~$t_0 \rightarrow -\infty$.
In order to study this limit, assume for technical simplicity that~$\B$ has compact support.
Show with the help of~\eqref{kmdef}, \eqref{series-scaustilde} and~\eqref{def-ktil}
that for sufficiently small~$t_0<0$,
\begin{align*}
\tilde{k}_m \,|_{t_0}\, \tilde{k}_m &= \frac{1}{4 \pi^2}
\sum_{n,n'=0}^\infty (-s^\wedge_m \B)^n \,s^\wedge_m \,|_{t_0}\, s^\vee_m \,(-\B s^\vee_m)^{n'} \\
&=\sum_{n,n'=0}^\infty (-s^\wedge_m \B)^n \,k_m \,|_{t_0}\, k_m \,(-\B s^\vee_m)^{n'} \:.
\end{align*}
\item[(c)] Apply the result of~(a) together with~\eqref{kmdef} to conclude
that~$\tilde{k}_m \,|_{t_0}\, \tilde{k}_m = \tilde{k}_m$.
\end{itemize}
}} \end{Exercise}

\begin{Exercise} \label{ex2.4-4}  (causal perturbation expansion to second order) {\em{
\begin{itemize}[leftmargin=2em]
\item[(a)] Compute~$P^\sea$ to second order in~$\B$. {\em{Hint:}} Use~\eqref{Psea}
as well as the perturbation series for~$\tilde{k}$.
The resulting formulas are also listed in~\cite[Appendix~A]{norm}.
\item[(b)] The so-called {\em{residual fermionic projector}} is defined by modifying the
integrand in~\eqref{Psea} to
\[ P^\sea_\res = -\frac{1}{2 \pi i} \ointctrclockwise_{\Gamma_-} \tilde{R}_\lambda\: d\lambda \:. \]
Show that to first order in~$\B$, the operators~$P^\sea$ and~$P^\sea_\res$ coincide.
However, there is a difference to second order in~$\B$. Compute it.
{\em{Hint:}} In order to simplify the computation, it is helpful to write the difference as
\[ P^\sea - P^\sea_\res = \frac{1}{2 \pi i} \ointctrclockwise_{\Gamma_-} (\lambda+1)\:
\tilde{R}_\lambda\: d\lambda \:, \]
and to use that the factor~$\lambda+1$ decreases the order of the pole at~$\lambda=-1$.
\end{itemize}
}} \end{Exercise}

\begin{Exercise} \label{ex2.4-5} (the fundamental solution~$\tilde{p}$) {\em{
\begin{itemize}[leftmargin=2em]
\item[(a)] Show that the operator~$\tilde{k}$ has the contour integral representation
\[ \tilde{k} = -\frac{1}{2 \pi i} \ointctrclockwise_{\Gamma_+ \cup \Gamma_-} \lambda\:
\tilde{R}_\lambda\: d\lambda  \:. \]
{\em{Hint:}} Use~\eqref{Resdef} or the functional calculus of Theorem~\ref{thmfcalc}.
\item[(b)] Conclude that the fermionic projector~$P^\sea$, \eqref{Psea}, can be represented as
\[ P^\sea = \frac{1}{2} \,\big( \tilde{p} - \tilde{k} \big) \:, \]
where~$\tilde{p}$ is defined by
\beq \label{tilpint}
\tilde{p} := -\frac{1}{2 \pi i} \left( \ointctrclockwise_{\Gamma_+} - \ointctrclockwise_{\Gamma_-} \right)
\lambda\: \tilde{R}_\lambda\: d\lambda \:.
\eeq
\end{itemize}
}} \end{Exercise}

\begin{Exercise} \label{ex2.4-51} (structural properties of~$\tilde{p}$: even number of factors~$k$) {\em{
The goal of this exercise is to show that every contribution to the perturbation expansion of~$\tilde{p}$
contains an even number of factors~$k$.
\begin{itemize}[leftmargin=2em]
\item[(a)] Read off from~\eqref{ktildef} and~\eqref{Delkdef} that every contribution to~$\Delta k$
involves an odd number of factors~$k$.
\item[(b)] Carry out the products $\cdot$ in the perturbation series for~$\tilde{R}_\lambda$ in~\eqref{tR} 
with the help of~\eqref{Rlamrel} and the multiplication rules~\eqref{rules} and~\eqref{rules2}.
Show that this gives rise to a sum of terms of the form
\beq \label{pcombi}
\frac{c\, \lambda^r}{(1-\lambda^2)^q}\;
C^{[p]} \:\B \:C^{[p-1]}\: \B \cdots \B \:C^{[0]}
\eeq
with operators $C^{[j]} \in \{ k, p, s \}$, parameters~$q \geq 0$ and~$r \in \Z$ as well as a combinatorial factor~$c$.
Show that the total number of factors~$p$ and~$k$ in the above operator product is always odd.
Show that the number of factors~$p$ in the above operator product is even if and only if the
parameter~$r$ is even. {\em{Hint:}} Use the result of~(a) and analyze how the number
of factors~$p$ and~$k$ changes when the different multiplication rules are applied.
\item[(c)] Substitute~\eqref{pcombi} into~\eqref{tilpint} and analyze the
symmetry of the resulting contour integral under the transformation~$\lambda \rightarrow -\lambda$. 
Use the results of~(b) to deduce that every contribution to the perturbation expansion of~$\tilde{p}$
contains an even number of factors~$k$.
\end{itemize}
}} \end{Exercise}

\begin{Exercise} \label{ex2.4-52} (structural properties of~$\tilde{p}$: replacing~$k$ by~$p$) {\em{
In this exercise we compute what one gets if in the perturbation series for~$\tilde{p}$ one replaces
all factors~$k$ by~$p$.
\begin{itemize}[leftmargin=2em]
\item[(a)] Show that replacing all factors~$k$ by~$p$, the formula~\eqref{ktildef} becomes
\beq \label{extilkrel}
\tilde{k} = \sum_{\beta=0}^{\infty}(-i\pi)^{2\beta} \:b^<\, p\, (b p)^{2\beta}\, b^> \:.
\eeq
Show that this gives rise to the simple multiplication rule
\[ \tilde{k} \cdot \tilde{k} = \tilde{k} \:. \]
{\em{Hint:}} Use the multiplication rules~\eqref{rules} and~\eqref{rules2}.
It might be helpful to insert the identity~$\1 = p + (\1-p)$.
\item[(b)] Use the result of~(a) to deduce that~$\tilde{p} = \tilde{p}^\res = \tilde{k}$.
{\em{Hint:}} Compute the contour integrals with residues.
\item[(c)] Show that the perturbation expansion of~$\tilde{p}^\res$ involves no operators~$k$.
{\em{Hint:}} Use the explicit formulas~\eqref{l:E1}, \eqref{l:F} and~\eqref{pm2calc}.
Indeed, the perturbation expansion of~$\tilde{p}^\res$ coincides precisely with
the series in~\eqref{extilkrel}.
\item[(d)] Combine the results of~(b) and~(c) to conclude that replacing all factors~$k$ by~$p$,
the perturbation series of~$\tilde{p}$ goes over to that of~$\tilde{p}^\res$.
\end{itemize}
}} \end{Exercise}

\begin{Exercise} \label{ex2.4-21} {\em{ This exercise 
explains the notion of the {\em{light-cone expansion}} in simple examples.
\begin{itemize}[leftmargin=2em]
\item[(a)] What is the light-cone expansion of a smooth function on~$M \times M$?
In which sense is it trivial? In which sense is it non-unique?
\item[(b)] Show that~$A(x,y) = \log \big(|y-x|^2 \big)$ is a well-defined distribution on~$M \times M$.
What is the order on the light cone? Write down a light-cone expansion.
\item[(c)] Now consider the distributional derivatives
\[ \Big( \frac{\partial}{\partial x^0} \Big)^p A(x,y) \qquad \text{with} \qquad p \in \N \]
and~$A(x,y)$ as in part~(b). What is the order on the light cone? Write down a light-cone expansion.
\item[(d)] Consider the function
\[ E(x,y) = \sin\big( (y-x)^2 \big)\: \log \big(|y-x|^2 \big)\:. \]
Determine the order on the light cone and give a light cone expansion.
\item[(e)] Consider the function
\[ E(x,y) = \left\{ \begin{array}{cl} \displaystyle e^{-\frac{1}{(y-x)^2}} & \text{if~$(y-x)^2 \geq 0$} \\[0.15em]
0 & \text{otherwise .} \end{array} \right. \]
Determine the order on the light cone and give a light cone expansion.
\item[(f)] Show that the expression
\[ \lim_{\varepsilon \searrow 0} \frac{\log \big(|y-x|^2 \big)}{(y-x)^4 + i \varepsilon} \]
is a well-defined distribution on~$M \times M$. Derive its light-cone expansion.
\end{itemize}
}} \end{Exercise}

\begin{Exercise} \label{ex2.4-2}  {\em{ This exercise is devoted to
computing the Fourier transform of the {\em{advanced Green's function}}~\eqref{l:11}
and deriving the series expansion~\eqref{l:12}.
\begin{itemize}[leftmargin=2em]
\item[(a)] As in Lemma~\ref{lemmaTintro}, we set~$\xi=y-x$ and~$\xi=(t,\vec{\xi})$ with~$t>0$.
Moreover, we choose polar coordinates~$r=(|\vec{\xi}|, \vartheta, \varphi)$. Carry out the $\omega$-integration
with residues and compute the angular integrals to obtain
\[ S^\vee_{m^2}(x,y) = \frac{i}{8 \pi r} \int_0^\infty \frac{p}{\omega(p)} \, \big( e^{-ipr} - e^{ipr}\big) \big(
e^{i \omega(p)\, t} - e^{-i \omega(p)\, t}\big)\, dp\:, \]
where~$p=|\vec{p}|$ and~$\omega(p):= \sqrt{|\vec{p}^2| + m^2}$.
Justify this integral as the Fourier transform of a distribution and show that
\[ \qquad S^\vee_{m^2}(x,y) =  \frac{i}{8 \pi r} \lim_{\varepsilon \searrow 0}\int_0^\infty e^{-\varepsilon p}\:
\frac{p}{\omega(p)}\, \big( e^{-ipr} - e^{ipr}\big) \big( e^{i \omega(p) \,t} - e^{-i \omega(p) \,t}\big)\, dp \]
with convergence as a distribution.
\item[(b)] Verify~\eqref{l:121} in the case~$m=0$ by setting~$\omega(p)=p$ and using~\eqref{eq:delta-formula}.
\item[(c)] In order to analyze the behavior away from the light cone, it is most convenient to
take the limit~$r \searrow 0$ and use Lorentz invariance. Show that in this limit,
\begin{align}
S^\vee_{m^2}(x,y) &=  \frac{1}{4 \pi} \lim_{\varepsilon \searrow 0}\int_0^\infty e^{-\varepsilon p}\:
\frac{p^2}{\omega(p)} \:\big( e^{i \omega(p) \,t} - e^{-i \omega(p) \,t}\big)\, dp \label{Spform} \\
&=  \frac{1}{4 \pi} \lim_{\varepsilon \searrow \omega}\int_m^\infty e^{-\varepsilon p}\:
\sqrt{\omega^2-m^2}\:\big( e^{i \omega t} - e^{-i \omega t}\big)\, d\omega\:.
\end{align}
Compute this integral using~\cite[formula (3.961.1)]{gradstein} (similar as in the proof
of Lem\-ma~\ref{lemmaTintro}. Use the relations between Bessel functions~\cite[(10.27.6), (10.27.11)]{DLMF}
to obtain~\eqref{l:121} away from the light cone.

As an alternative method for computing the Fourier integral, one can begin
from the integral representation for~$J_0$ in~\cite[(10.9.12)]{DLMF}, differentiate with respect to~$x$
and use~\cite[(10.6.3)]{DLMF}.
\item[(d)] Combine the results of~(b) and~(c) to prove~\eqref{l:121}. Why is there no additional
contribution at~$\xi=0$?
\item[(e)] Use the series expansion~\cite[(10.2.2)]{DLMF} to derive~\eqref{l:12}.
\item[(f)] The series expansion~\eqref{l:12} can also be derived without using Bessel functions.
To this end, one expands~\eqref{Spform} in powers of~$m^2$ and computes the
Fourier transform term by term. Verify explicitly that this procedure really gives~\eqref{l:12}.
\end{itemize}
}} \end{Exercise}

\begin{Exercise} \label{ex2.5}  {\em{This exercise is devoted to the proof of Lemma~\ref{l:lemma2}
as given in~\cite[Lemma~2.2]{light}.
\begin{itemize}[leftmargin=2em]
\item[(a)] Use~\eqref{l:29z} to derive the identity
\beq \label{yder}
\qquad \int d^4z \:S^{(l)}(x,z) \:V(z) \:(y-z)_k\:S^{(-1)}(z,y) = -2\:\frac{\partial}{\partial y^k} (S^{(l)} \:V\: 
S^{(0)})(x,y) \:.
\eeq
\item[(b)] Apply Lemma~\ref{l:lemma1} and carry out the $y$-derivative in~\eqref{yder} to obtain
the formula in Lemma~\ref{l:lemma2}. {\em{Hint:}} Use the identity
\[ \qquad \partial_k \Box^n V(z) = -\frac{1}{2 (n+1)} \:\Box^{n+1}_z \Big(V(z) \:(y-z)_k \Big) +
\frac{1}{2 (n+1)} \:\Big(\Box^{n+1}_z V(z) \Big) \:(y-z)_k \]
and shift the summation index.
\end{itemize}
}} \end{Exercise}

\begin{Exercise} \label{ex2.4-3}  {\em{ In this exercise we collect
elementary properties of the {\em{ordered exponential}}.
\begin{itemize}[leftmargin=2em]
\item[(a)] Assume that the matrix-valued function~$F$ in Definition~\ref{2.5.4} is commutative in the sense that
\[ \big[F(\alpha), F(\beta) \big] = 0 \qquad \text{for all~$\alpha, \beta \in [a,b]$}\:. \]
Show that the ordered exponential reduces to the ordinary exponential,
\[ \Pexp \bigg( \int_a^b F(\alpha) \:d\alpha \bigg) = \exp \bigg( \int_a^b F(\alpha) \:d\alpha \bigg) \:. \]
{\em{Hint:}} Show inductively that
\[ \int_a^b dt_0\:F(t_0) \int_{t_0}^b dt_1 \: F(t_1)
\cdots \int_{t_{n-1}}^b dt_n \:F(t_n) = \frac{1}{(n+1)!} \bigg(  \int_a^b F(t)\:dt \bigg)^{n+1} \:. \]
\item[(b)] Assume that~$F$ is continuous on~$[a,b]$. Show that the Dyson series converges
absolutely and that
\[ \bigg\| \Pexp \bigg( \int_a^b F(\alpha) \:d\alpha \bigg) \bigg\|
\leq \exp \bigg( \int_a^b \big\|F(\alpha)\big\| \:d\alpha \bigg) \:. \]
{\em{Hint:}} Estimate the integrals and apply~(a).
\item[(c)] Show by direct computation that the ordered exponential satisfies the equations
\begin{gather}
\frac{d}{da} \Pexp \bigg( \int_a^b F(\alpha) \:d\alpha \bigg)
= -F(a)\: \Pexp \bigg( \int_a^b F(\alpha) \:d\alpha \bigg) \label{Fadiff} \\
\Pexp \bigg( \int_a^a F(\alpha) \:d\alpha \bigg) = \1\:. \label{Fgroup2}
\end{gather}
Use the uniqueness theorem for solutions of ordinary differential equations to
give an alternative definition in terms of the solution of an initial-value problem.
Use this reformulation to show the group property
\beq \label{Fgroup}
\Pexp \bigg( \int_a^b F(\alpha) \:d\alpha \bigg) \Pexp \bigg( \int_b^c F(\alpha) \:d\alpha \bigg)
= \Pexp \bigg( \int_a^c F(\alpha) \:d\alpha \bigg) \:.
\eeq
\item[(d)] Show that
\beq \label{Fbdiff}
\frac{d}{db} \Pexp \bigg( \int_a^b F(\alpha) \:d\alpha \bigg)
= \Pexp \bigg( \int_a^b F(\alpha) \:d\alpha \bigg) \: F(b) \:.
\eeq
{\em{Hint:}} Differentiate the identity~\eqref{Fgroup} in the case~$c=a$
and use the group properties~\eqref{Fgroup2} and~\eqref{Fgroup}.
\item[(e)] Show that
\[ \Pexp \bigg( \int_a^b F(\alpha) \:d\alpha \bigg)^*
= \Pexp \bigg( \int_b^a \big(-F(\alpha)^*\big) \:d\alpha \bigg) \:. \]
Deduce that if~$F(\alpha)$ is an anti-Hermitian matrix, then the
ordered exponential is a unitary matrix.
{\em{Hint:}} There are two alternative methods.
One method is to argue using the differential equations~\eqref{Fadiff} and~\eqref{Fbdiff}
or with the group property. A more computational approach is to take the adjoint of the Dyson series
and reparametrize the integrals.
\end{itemize}
}} \end{Exercise}

\begin{Exercise} \label{ex2.7} {\em{
This exercise recalls the concept of {\em{local gauge transformations}} and gets the connection
to the {\em{ordered exponential}}.
\begin{itemize}[leftmargin=2em]
\item[(a)] An electromagnetic potential~$A$ of the form~$A_j=\partial_j \Lambda$ with
a real-valued function~$\Lambda$ is called a pure gauge potential. Show that~$(i \Pdd + \slashed{A} - m)
= U (i \Pdd - m) U^{-1}$, where~$U$ is the phase factor~$U=e^{i \Lambda}$.
Conclude that every solution of the Dirac equation~$(i \Pdd + \slashed{A} - m) \tilde{\psi}=0$
can be written in the form~$\tilde{\psi} = U \psi$, where~$\psi$ is a solution of the vacuum Dirac equation.
In other words, pure gauge potentials merely describe local phase transformations of the wave functions.
\item[(b)] Generalize the argument of~(a) to the case of non-abelian gauge fields and an additional
gauge potential using the relation
\[ U (i \Pdd +\slashed{A} - m \,\1) U^{-1} = i \Pdd + U \slashed{A} U^{-1} + iU
\big(\Pdd U^{-1} \big)- m\,\1 \:, \]
where now~$U(x)$ is a unitary matrix (the mass matrix was left out for simplicity).
How does the gauge potential transform under local unitary transformations of the spinors?
\item[(c)] Prove that for a pure gauge potential~$A= iU (\Pdd U^{-1})$ the
ordered exponential of Definition~\ref{2.5.4} simplifies to
\[ \Pexp \bigg(-i \int_x^y A^j \:(y-x)_j \bigg) = U(x)\: U(y)^{-1} \:. \]
{\em{Hint:}} Apply the integration-by-parts method of Exercise~\ref{ex2.6}
to the Dyson series. Alternatively, one can make use of the differential equation~\eqref{Fadiff}
with initial conditions~\eqref{Fgroup2}.
\end{itemize}
}} \end{Exercise}

\begin{Exercise} \label{ex2.6} {\em{This exercise illustrates the handling of the {\em{tangential derivatives}}
mentioned before Proposition~\ref{l:thm2}.
Let~$z=\beta y + (1-\beta) x$ be a point on the line segment~$\overline{xy}$. Show that
\[ \int_z^y [p,q|0]\: f(z')\: dz' =
\int_0^1 \alpha^p\, (1-\alpha)^q\: f\Big( \alpha \, (1-\beta)(y-x) + z \Big) \]
Deduce the identity
\begin{align*}
(y-&x)^j \int_z^y [p,q|0]\: (\partial_j f)(z')\: dz' \\
&= \frac{1}{1-\beta} \int_0^1 \alpha^p\, (1-\alpha)^q\: \frac{d}{d\alpha}
f\Big( \alpha \, (1-\beta)(y-x) + z \Big) \:d\alpha\:.
\end{align*}
In the case~$p,q>0$, integrate by parts to derive the computation rule
\[ (y-x)^j \int_z^y [p,q|0]\: (\partial_j f)(z')\: dz'
= -\frac{1}{1-\beta} \int_z^y \Big(p \: [p-1,q \,|\, 0] - q\:[p,q-1 \,|\, 0] \Big)\: f \:. \]
What is the analogous computation rule in the cases~$p=0$ and/or $q=0$?
}} \end{Exercise}

\begin{Exercise} \label{ex2.71} {\em{This exercise explains how the Maxwell field tensor 
and the Maxwell current arise in the light cone expansion. To this end, we consider the first order
perturbation of the massless Green's function by an electromagnetic potential~$A$,
\[ \Delta s_0 := - s_0 \,\slashed{A}\, s_0 \:. \]
\begin{itemize}[leftmargin=2em]
\item[(a)] Show that the leading contributions to the light-cone expansion of~$\Delta s_m$ have the form
\begin{align}
(\Delta s_0)(x,y)  &= \frac{1}{2} \int_x^y A_i(z)\, \xi^i \; \slashed{\xi}\, S^{(-1)}(x,y) \\
&\qquad +\int_x^y dz\:[0,1\:|\: 0]\: (\Pdd A_{i})(z)\, \xi^i \; S^{(0)}(x,y) \label{fieldterm} \\
&\qquad -\int_x^y dz\:[0,0\:|\: 0]\: \slashed{A}(z)\: S^{(0)}(x,y) \\
&\qquad+ \slashed{A}(x) \: S^{(0)}(x,y) + \slashed{\xi}\, \O\big(\xi^{-2} \big) + \O\big(\xi^0 \big) \:,
\end{align}
where~$\xi := y-x$. {\em{Hint:}} First compute~$s_0$ using~\eqref{l:4} and~\eqref{l:29z}. Then
perform the light-cone expansion of the first order perturbation by using Lemma~\ref{l:lemma2}
and then by differentiating similar as done in the displayed computation before~\eqref{l:29z}.
Finally, the resulting formulas can be simplified by using~\eqref{l:22a} and by
integrating the tangential derivatives by part (see Exercise~\ref{ex2.6} or the proof of Proposition~\ref{l:thm2}).
\item[(b)] Which of the above contributions are phase-free? Show that the contribution which
is {\em{not}} phase-free can be understood as the first-order contribution to the gauge phase
in~\eqref{sho}.
\item[(c)] Rewrite the phase-free contributions in an explicitly gauge-invariant way. \\
{\em{Hint:}} In~\eqref{fieldterm} use the identity~$\Pdd A_{i})(z)\, \xi^i
= \gamma^j F_{ji} \xi^i - \xi^j \partial_j \slashed{A}$.
Note that this generates a tangential derivative (see~\eqref{tangential}).
Integrate it by parts as explained in Exercise~\ref{ex2.6} or in the proof of Proposition~\ref{l:thm2}.
\item[(d)] Compute the contributions to the above light-cone expansion of the form~$\sim \slashed{\xi}\,\cdots\, S^{(0)}$.
There is a term involving~$\Box A$. Rewrite it in an explicitly gauge-invariant way using the Maxwell
current~$j_i := \partial_{ik} A^k - \Box A_i$.
\item[(e)] The reader who wants to get more computational practice may find it instructive to carry
out the light-cone expansion up to the order~$\O(\xi^2)$. In particular, there is
a term~$\sim (\Box \slashed{A})\, S^{(1)}$. Rewriting the contributions again
an explicitly gauge-invariant form, one thus obtains a contribution~$\sim \gamma^k j_k\, S^{(1)}$.
In fact, this contribution gives rise to the Maxwell current in the field equations
in the continuum limit.
\end{itemize}
We note that all these computations are explained in more detail in~\cite[Appendix~A]{light}.
}} \end{Exercise}

\begin{Exercise} \label{ex2.8} (contour integral representation of the residual
fermionic projector) {\em{
In Exercise~\ref{ex2.4-5}~(a) we derived a contour integral representation for
the operator~$\tilde{k}$ in~\eqref{l:E0}. Thus it remains to derive a contour
integral representation for the operator~$\tilde{p}^\res$ as defined by
defined by~\eqref{l:E1}.
Verify to second order in perturbation theory (see Exercise~\ref{ex2.4-4})
that~$\tilde{p}^\res$ has the contour integral representation
\[ \tilde{p}^\res = -\frac{1}{2 \pi i} \ointctrclockwise_{\Gamma_+ \cup \Gamma_-}
\tilde{R}_\lambda\: d\lambda  \:. \]
{\em{Remark:}} This equation indeed holds to every order in perturbation theory.
This is a consequence of an underlying symmetry of the perturbation expansions
with mass and spatial normalizations as explained in~\cite[Section~3.4]{norm}.
}} \end{Exercise}

\begin{Exercise} \label{ex2.9} {\em{The goal of this exercise is to
explore {\em{weak evaluation on the light cone}}
in the example of the massless closed chain of the vacuum~\eqref{Axyvac}.
Thus in view of~\eqref{weval}, we want to analyze the integral
\beq \label{exweakint}
\int_{-\infty}^\infty \eta(t)\: 
\frac{(t^2-r^2) - i \varepsilon [\gamma^0, \slashed{\xi}] + \varepsilon^2}
{\big| (t-i \varepsilon)^2 - r^2 \big|^4}\: dt
\eeq
for a test function~$\eta \in C^\infty_0(\R)$ asymptotically as~$\varepsilon \searrow 0$.
\begin{itemize}[leftmargin=2em]
\item[(a)] Choose~$r>0$. Show that, changing the integral only by contributions which are bounded uniformly
in~$\varepsilon$, we may replace~$\eta(t)$ by a test function supported in the interval~$(r/2, 2r)$
around the upper light cone.
\item[(b)] Use the identity
\[ \frac{1}{(t-i \varepsilon)^2 - r^2} =
\frac{1}{(t-i \varepsilon-r)(t-i \varepsilon+r)} = \frac{1}{2r} \left(
\frac{1}{t-i \varepsilon-r} - \frac{1}{t-i \varepsilon+r} \right) \]
to rewrite the integrand in~\eqref{exweakint} in the form
\[ \sum_{p,q=0}^2 \frac{\eta_{p,q}(t,r, \varepsilon)}{(t-i \varepsilon-r)^p\, (t+i \varepsilon-r)^q} \:, \]
with functions~$\eta_{p,q}(t,r, \varepsilon)$ which in the limit~$\varepsilon \searrow 0$
converge in~$C^\infty$ to smooth functions~$\eta_{p,q}(t,r)$, i.e.
\[ \lim_{\varepsilon \searrow 0} \partial_t^\alpha \partial_r^\beta
\eta_{p,q}(t,r, \varepsilon) = \partial_t^\alpha \partial_r^\beta \eta_{p,q}(t,r) \qquad \text{for all~$\alpha, \beta \geq 0$}\:. \]
Compute the functions~$\eta_{p,q}$.
Verify that the contribution for~$p=q=2$ agrees with the approximation~\eqref{Axyfinal}.
\item[(c)] We now compute the leading contributions and specify what we mean by ``leading.''
First compute the following integrals with residues:
\begin{align*}
I_0(\varepsilon) &:= \int_{-\infty}^\infty \frac{1}{(t-i \varepsilon-r)^2\, (t+i \varepsilon-r)^2}\: dt \\
I_1(\varepsilon) &:= \int_{-\infty}^\infty \frac{t-r}{(t-i \varepsilon-r)^2\, (t+i \varepsilon-r)^2}\: dt \:.
\end{align*}
Show that
\begin{align*}
\int_{-\infty}^\infty & \frac{\eta_{2,2}(t,r)}{(t-i \varepsilon-r)^2\, (t+i \varepsilon-r)^2}\: dt \\
&= I_0(\varepsilon)\: \eta_{2,2}(r,r) + I_1(\varepsilon) \big( \partial_t \eta_{2,2}\big)(r,r) +\O(\varepsilon) \:.
\end{align*}
{\em{Hint:}} To estimate the error term, proceed similar as in Exercise~\ref{ex2}~(a).
\item[(d)] We now analyze the dependence of the resulting terms on~$r$.
To this end, first compute~$\eta_{2,2}(r,r)$ and~$(\partial_t \eta_{2,2})(r,r)$.
Verify the rules~\eqref{weakscale}.
Verify the scaling of the error terms~\eqref{hot} and~\eqref{hotmacro}, where
we use the convention that every derivative of~$\eta$ gives rise to a factor~$1/\ell_\text{macro}$.
\item[(e)] Show that the integrals for~$p<2$ or~$q<2$ can be absorbed into the error terms.
Also show that the term~$\sim \varepsilon^2$ in~\eqref{Axyvac} can be absorbed into the error terms.
\item[(f)] So far we analyzed the integrals with the simplified test functions~$\eta_{p,q}(t,r)$.
Show that replacing them by~$\eta_{p,q}(t,r, \varepsilon)$ changes the integrals only by
error terms of the form~\eqref{hot} and~\eqref{hotmacro}.
\end{itemize}
}} \end{Exercise}

\begin{Exercise} \label{ex2.10} {\em{This exercise explains how the identities~\eqref{l:20reg}
and~\eqref{l:7T} can be derived by explicit computation.
\begin{itemize}[leftmargin=2em]
\item[(a)] Use~\eqref{l:3zz} together with~\eqref{Tregdef} and the series expansion~\eqref{l:31}
to derive explicit formulas for~$T^{(l)}$ for all~$l \geq 0$.
Use the relation~\eqref{Trules} in the case~$l=0$ to also compute~$T^{(-1)}$.
\item[(b)] Show that for all~$n \geq 0$,
\beq \label{xi2T}
\xi^2 \, T^{(l)}(x,y) = -4\, T^{(l-1)} + \text{(smooth contributions)}\:.
\eeq
Why do the ``smooth contributions'' arise?
\item[(c)] Verify that the relation~\eqref{xi2T} remains valid for the $i \varepsilon$-regularization.
{\em{Hint:}} One can argue without computations directly with a meromorphic extension
using~\eqref{regreplace}.
\item[(d)] Verify the identities~\eqref{l:7T} by explicit computation. What are the ``smooth contributions''?
Show that these identities remain valid for the $i \varepsilon$-regularization.
\end{itemize}
}} \end{Exercise}

\begin{Exercise} \label{ex2.101} (computation of the local trace) {\em{
Compute~$P^\varepsilon(x,x)$ in the Minkowski vacuum with $i \varepsilon$-regulari\-za\-tion (see~\eqref{B}
and~\eqref{convergence-generate}). How do the vector and scalar components scale in~$m$
and~$\varepsilon$? Verify the scaling of the local trace~\eqref{epsloctrace}.
}} \end{Exercise}

\begin{Exercise} \label{ex2.102} (scalar potentials and the local trace) {\em{
Consider a potential~$\B$ composed of chiral potentials and a scalar
potential, i.e.\ in generalization of~\eqref{Bchiral},
\[ \B = \chi_L \:\slashed{A}_R + \chi_R \: \slashed{A}_L
+ \Phi(x) \:. \]
\begin{itemize}[leftmargin=2em]
\item[(a)] Show that the scalar potential can be
combined with the mass terms to obtain a Dirac equation of the form~\eqref{l:18a}
with~$B$ as in~\eqref{Bchiral2}, but now with~$Y(x)$ depending on~$x$.
We remark that this so-called {\em{dynamical mass matrix}} was first introduced
in~\cite[Section~2]{light} (also including a pseudoscalar potential); see
also~\cite[Section~2.5]{PFP}.
\item[(b)] Go through the proof of Theorem~\ref{l:thm1} and convince yourself
that the statement of the theorem remains valid in the presence of a scalar potential
if in~\eqref{l:42e} the matrix~$Y$ is replaced by~$Y(x)$.
{\em{Remark:}} This generalization of Theorem~\ref{l:thm1} is given in~\cite[Theorem~2.3]{light}.
\item[(c)] Use this generalization of Theorem~\ref{l:thm1} together with the scaling argument in
the proof of Proposition~\ref{prploctr} to derive the formula for the local trace~\eqref{epsloctrace2}.
\end{itemize}
}} \end{Exercise}

\begin{Exercise} \label{ex2.11} (spectral representation of~$A_{xy}$) {\em{
Derive the formulas~\eqref{AFformula} by a straightforward computation using~\eqref{e:2nn} and~\eqref{e:2f}.
}} \end{Exercise}

\begin{Exercise} \label{ex2.12} (spectral representation in the continuum limit) {\em{
Derive~\eqref{FPasy} by using~\eqref{Fpm}, \eqref{Peq} as well as the
contraction rules~\eqref{eq52}--\eqref{eq54}.
}} \end{Exercise}

\begin{Exercise} \label{ex2.13} {\em{ In this exercise we
consider the {\em{double null spinor frame}} in the example of the $i \varepsilon$-regularization.
\begin{itemize}[leftmargin=2em]
\item[(a)] Consider a point~$(t,\vec{\xi})$ on the upper light cone, i.e.\ $t=|\vec{\xi}|$
(more specifically one may choose~$\vec{x}=(t,0,0)$). Use~\eqref{xiprule} to
and compute~$z$ (up to errors of the form~\eqref{hot}). Compute the spectral projectors~\eqref{F0def}.
Verify the relations~\eqref{Fident}.
\item[(b)] Compute the solutions~$\f^c_s$ of the eigenvector equations~\eqref{zetaeigen}.
Normalize them according to~\eqref{absone}. What is the remaining freedom to modify
the eigenvectors.
\item[(c)] Choose a space-like unit vector~$u$ which is orthogonal to~$\xi$ and~$\bar{\xi}$.
What is the freedom in choosing this vector?
Show that by suitably choosing the phases of the eigenvectors~$\f^c_s$ one can arrange
that the relations in~\eqref{commute} hold. What is the remaining freedom in choosing
the frame~$(\f^c_s)$?
\item[(d)] The diagram~\eqref{commute} implies in particular
that~$\slashed{\xi} \,\f^L_+ = \overline{\slashed{\xi}} \,\f^L_+$.
Explain how this identity can be understood in view of the error terms~\eqref{hot}.
\end{itemize}
}} \end{Exercise}

\begin{Exercise} \label{ex2.14} (matrix elements in the double null spinor frame) {\em{
Compute the matrix elements~$\mathfrak{F}^{LL}_{++}(B)$, $\mathfrak{F}^{LL}_{+-}(B)$,
$\mathfrak{F}^{LR}_{+-}(B)$ and $\mathfrak{F}^{RR}_{+-}(B)$ for~$B$ given by
\[ B = \frac{i}{2} \:\chi_L \, \slashed{\xi} \,T^{(-1)}_{[0]} \:. \]
Simplify the expression as far as possible.
{\em{Hint:}} Use the cyclic property of the trace, the anti-commutation relations of the Dirac
matrices and the contraction rules.
}} \end{Exercise}

\begin{Exercise} \label{ex2.15} (Perturbation of the eigenvalues of the closed chain) {\em{
The light-cone expansion can be understood as giving corrections to the fermionic projector
of lower order on the light cone. We now explore how these corrections affect the eigenvalues
of the closed chain, and which of them are compatible with the EL equations.
In order to work in a specific example, we assume that the unperturbed
fermionic projector is
\[ P(x,y) = \frac{i}{2} \:\slashed{\xi} \,T^{(-1)}_{[0]} \]
(similar as considered in Exercise~\ref{ex2.14}), whereas the perturbation
has a left- and right-handed component,
\[ \Delta P(x,y) = \chi_L \: \slashed{\nu}_L +  \chi_R \: \slashed{\nu}_R \:, \]
where~$\nu_L$ and~$\nu_R$ are given vectors in Minkowski space.
\begin{itemize}[leftmargin=2em]
\item[(a)] Compute the corresponding perturbation~$\Delta \lambda^{xy}_k$ to leading order
in the degree on the light cone. What is the leading degree? Which eigenvalues change, which remain the same?
{\em{Hint:}} Use the usual formula for first order perturbations (see~\eqref{t:firstper})
and rewrite it in the double null spinor frame.
\item[(b)] For which vectors $\nu_L$ and $\nu_R$ does the relation~$| \lambda^{xy}_k| = | \lambda^{xy}_l|$
hold for all~$k,l \in \{1,\ldots, 4\}$?
Show that these relations are a sufficient condition for the EL equations to be satisfied.
What would one need to verify in order to conclude that these relations are necessary?
{\em{Hint:}} Consider~\eqref{t:firstper} and~\eqref{t:Qspec}. Keep in mind that the EL equations are
evaluated weakly on the light cone.
\end{itemize}
}} \end{Exercise}


\chapter[A System of one Sector]
{An Action Principle for an Interacting Fermion System and its Analysis in the Continuum Limit} \label{sector}

\begin{abstract}
We introduce and analyze a system of relativistic fermions in a space-time continuum,
which interact via an action principle as previously considered in a discrete space-time.
The model is defined by specifying the vacuum as a sum of Dirac seas
corresponding to several generations of elementary particles.
The only free parameters entering the model are the fermion masses.
We find dynamical field equations if and only if the number of generations
is at least three. If the number of generations equals three,
the dynamics is described by a massive axial potential coupled to the
Dirac spinors. The coupling constant and the rest mass of the axial field depend on the
regularization; for a given regularization method they can be computed
as functions of the fermion masses.
The bosonic mass term arises as a consequence of a symmetry breaking effect, giving
an alternative to the Higgs mechanism.
In addition to the standard loop corrections of quantum field theory, we find new types of
correction terms to the field equations which violate causality. These non-causal corrections
are too small for giving obvious contradictions to physical observations,
but they might open the possibility to test the approach in future experiments.
\end{abstract}

\section{Introduction}
In~\cite{PFP} it was proposed to formulate physics based on a new action principle in space-time.
On the fundamental level, this action principle is defined in so-called discrete space-time for
a finite collection of projectors in an indefinite inner product space (see also~\cite{discrete}).
An effect of spontaneous symmetry breaking~\cite{osymm} leads to the emergence of a
discrete causal structure (see~\cite{small} for an explanation in simple examples), which for many
space-time points and many particles should go over to the usual causal structure
of Minkowski space (for the connection between discrete and continuum space-times we also refer
to~\cite{ssymm, osymm, discrete} and the survey article~\cite{lrev}).
Furthermore, on a more phenomenological level, it is shown in~\cite[Chapters~4--8]{PFP} that
the action can also be analyzed in Minkowski space in the so-called {\em{continuum limit}}, where
the interaction is described effectively by classical gauge fields coupled to second-quantized Dirac fields.
Finally, generalizing our approach has led to the mathematical framework of
so-called causal fermion systems (cf.~\cite{rrev} and the references therein).

Apart from deriving the general formalism of the continuum limit, in~\cite[Chapters~4--8]{PFP} it
is shown that for a suitable system involving 24 Dirac seas, the resulting effective gauge group as well
as the coupling of the effective gauge fields to the Dirac fields have striking similarities to
the standard model. However, the detailed form of the effective interaction so far has not
been worked out.

This work is the first of a series of papers devoted to the detailed analysis of our action
principle in the continuum limit and to the derivation of the resulting field equations.
In order to make the presentation as clear and easily accessible
as possible, our procedure is to begin with small systems, which are composed of only a few Dirac
seas, and then to gradually build up larger and more complicated systems.
In the present paper, we consider a system of several Dirac seas (corresponding to
several ``generations'' of elementary particles) in the simplest possible configuration
referred to as a {\em{single sector}}. 
The only free parameters entering the model are the masses of the Dirac particles of each generation.
However, we do not specify the form of the interaction, which is completely determined by
our action principle. Also, we do not put in coupling constants nor the masses of gauge bosons.
The analysis of the model in the continuum limit reveals that we get dynamical field equations
if and only if the number of generations is at least three. If the number of generations equals three,
the dynamics can be described by a {\em{massive axial potential}}~$A_\text{\rm{a}}$ coupled to the
Dirac equation. The corresponding Dirac and field equations (stated for notational
simplicity for one Dirac particle) become
\[ \boxed{ \quad (i \Pdd + \pseudo \slashed{A}_\text{\rm{a}} - m) \psi = 0 \:,\qquad
C_0 \,j^k_\text{\rm{a}} - C_2\, A^k_\text{\rm{a}} = 12 \pi^2\, 
\overline{\psi} \pseudo \gamma^k \psi \:,\quad } \]
where~$j^k_\text{\rm{a}} = \partial^k_{\;\:l} A^l_\text{\rm{a}} - \Box A^k_\text{\rm{a}}$ is the corresponding
axial current (here~$\pseudo$ is the pseudoscalar matrix, which is often denoted by~$\gamma^5$).
\nindex{bh1@$\pseudo$ -- pseudoscalar matrix}%
\sindex{field equations in the continuum limit}%
\sindex{coupling constant!of axial field}%
The coupling constant and the rest mass of the axial gauge field are described
by the constants~$C_0$ and~$C_2$, which for a given regularization method can be computed
as functions of the fermion masses.
The mass term of the gauge field arises as a consequence of a symmetry breaking effect, giving
an alternative to the Higgs mechanism.
The field equations involve surprising corrections which challenge
the standard model of elementary particle physics: First, the field equations involve
additional convolution terms of the form
\beq \label{s:fconvolve}
- f_{[0]}* j^k_\text{\rm{a}} + 6 f_{[2]}* A^k_\text{\rm{a}} \:,
\eeq
where~$f_{[p]}$ are explicit Lorentz invariant distributions. These convolution
terms give rise to small corrections which violate causality. Moreover, we get new types of higher order
corrections to the field equations. We also find additional potentials which are non-dynamical in the
sense that they vanish away from the sources.

In order to make the paper self-consistent, we introduce our fermion
systems and the continuum limit from the basics. However, to avoid an excessive
overlap with previous work, we present a somewhat different point of view,
where instead of considering a discrete space-time or a space-time continuum of finite volume,
we work exclusively in Minkowski space.
Furthermore, we always restrict attention to a single sector.
For clarity, we omit the more technical aspects of the regularization, relying instead on results from
the corresponding chapters of the book~\cite{PFP}.

The paper is organized as follows. In Section~\ref{s:sec2} we introduce our action principle
in a space-time continuum. In Sections~\ref{s:sec3}--\ref{s:sec5} we review and adapt
the methods for analyzing this action principle in the continuum as developed in~\cite{PFP}.
More precisely, in Section~\ref{s:sec3} we describe the vacuum by a system of regularized Dirac seas.
We list all the assumptions on the vacuum state, either motivating them or explaining
how they can be justified.
In Section~\ref{s:sec4} we construct more general fermion configurations in Minkowski space
by modifying and perturbing the vacuum state, also introducing particles and gauge fields.
We also outline the mathematical methods for analyzing the unregularized fermionic projector
with interaction. In Section~\ref{s:sec5}, we explain how interacting systems are to be regularized,
and how to treat the regularization in an effective way. This leads us to the formalism of the
continuum limit, which allows us to analyze our action principle in the continuum,
taking into account the unknown regularization details by a finite number of free parameters.
In the following Sections~\ref{s:sec7}--\ref{s:secnonlocal} the continuum limit of our action
principle is worked out in detail; this is the main part of the paper where we present our new results.
Section~\ref{s:sec7} is devoted to the leading singularities of the Euler-Lagrange equations
on the light cone, where the vacuum contributions (\S\ref{s:sec71})
are modified by phases coming from the chiral gauge potentials (\S\ref{s:sec72}).
The next lower orders of singularities are analyzed in Section~\ref{s:sec8}.
Then the currents of the gauge fields come into play, and we also get a mass term corresponding to
the axial gauge field (\S\ref{s:sec81}). Furthermore, we find a corresponding contribution of
the Dirac current (\S\ref{s:sec82}). A priori, the different current terms are not comparable,
because the gauge currents have logarithmic poles on the light cone (\S\ref{s:sec83}).
But provided that the number of generations is at least three, these logarithmic poles can be
compensated by a local axial transformation, as is developed in~\S\ref{s:sec84}--\S\ref{s:sec86}.
After considering more general local transformations (\S\ref{s:secgenlocal}),
in~\S\ref{s:secprobaxial} we explain a basic shortcoming of local transformations.
This motivates us to work instead with so-called microlocal transformations,
which are developed in~\S\ref{s:secnonlocaxial}--\S\ref{s:secshearmicro}.

Section~\ref{s:secfield} is devoted to the derivation and analysis of the field equations.
In~\S\ref{s:secfield1} we show that the Euler-Lagrange equations corresponding to our action
principle give rise to relations between the Dirac and gauge currents.
If the number of generations equals three, we thus obtain field equations for the
axial gauge potential (see Theorem~\ref{s:thmfield}).
These field equations involve non-causal correction terms, which are analyzed and discussed
in~\S\ref{s:secnocausal} and~\S\ref{s:sechighorder}.
In~\S\ref{s:secquantcorr} we explain schematically how the standard loop corrections of quantum field theory (QFT)
appear in our framework, and how loop corrections of the non-causal terms could be obtained.
In~\S\ref{s:secnohiggs} we get a connection to the Higgs mechanism and explain why
our model involves no Higgs particle. We finally compute the coupling
constant and the rest mass of the axial field for a few simple regularizations (\S\ref{s:secexample}).

In Section~\ref{s:secthree} we analyze and discuss further potentials and fields, including
scalar and pseudoscalar potentials, bilinear potentials, as well as
the gravitational field and a conformal axial field.
In Section~\ref{s:secnonlocal} we consider nonlocal potentials, which can be used to
satisfy the Euler-Lagrange equations to higher order in an expansion near the origin.

In order not to interrupt the explanations in the main sections by longer calculations,
the more technical parts are worked out in the appendices.
Appendix~\ref{s:appnull} supplements the estimates needed for the derivation of the
Euler-Lagrange equations in the continuum limit in~\S\ref{s:secELC}.
All the calculations in the formalism of the continuum limit
as needed in Sections~\ref{s:sec7}--\ref{s:secthree} are combined in Appendix~\ref{s:appspec},
which also reviews the general method as developed in~\cite[Appendix~G]{PFP}.
All the formulas given in this appendix have been obtained with the help of computer algebra.
In Appendix~\ref{s:applocaxial} we give a general argument which explains why local transformation cannot
be used to compensate the logarithmic poles of the current terms.
In Appendix~\ref{s:appresum} we compute and analyze the smooth contributions to the fermionic projector
as needed in~\S\ref{s:secfield1}; this is done by modifying a resummation technique first
introduced in~\cite{firstorder}.
Finally, in Appendix~\ref{s:apprho} we outline how our constructions and results can be extended
to the setting where the Dirac seas involve weight factors, as was proposed in~\cite{reg}
and~\cite{vacstab}.

\section{An Action Principle for Fermion Systems in Minkowski Space} \label{s:sec2}
In relativistic quantum mechanics, a fermionic particle is described by a Dirac wave
function~$\psi$ in Minkowski space~$(\scrM, \langle .,. \rangle)$.
In order to describe a many-particle system, we consider an operator~$P$
\nindex{ag1@$P$ -- fermionic projector}%
on the Dirac wave functions and interpret
the vectors in the image of~$P$ as the occupied fermionic states
of the system (for a discussion of the Pauli exclusion principle and the connection to the
fermionic Fock space formalism see~\cite[Chapter~3 and Appendix~A]{PFP}). We assume that~$P$ has
an integral representation
\sindex{fermionic projector!kernel of}%
\beq \label{s:Pdef}
(P \psi)(x) = \int_\scrM P(x,y)\: \psi(y)\: d^4y
\eeq
with an integral kernel~$P(x,y)$. Moreover, we assume for technical simplicity that $P(x,y)$  is
continuous in both arguments~$x$ and~$y$; then the integral in~\eqref{s:Pdef} is clearly well-defined if
for the domain of definition of~$P$ we choose for example the space $C^\infty_0(\scrM, S\scrM)$ of
smooth wave functions with compact support. Moreover, we assume that~$P$ is symmetric with
respect to the Lorentz invariant inner product
\beq \label{s:iprod}
\bra \psi | \phi \ket = \int_\scrM \overline{\psi(x)} \phi(x)\: d^4x\:,
\eeq
\nindex{af4@$\bra . \vert . \ket$ -- inner product on Krein space}%
where~$\overline{\psi} \equiv \psi^\dagger \gamma^0$ is the usual adjoint spinor
($\psi^\dagger$ is the complex conjugate spinor). 
\sindex{adjoint spinor}%
\nindex{ai4@$\overline{\psi} \phi = \Sl \psi \vert \phi \Sr$ -- inner product on spinors}%
In other words, we demand that
\beq \label{s:Psymm1}
\bra P \psi | \phi \ket = \bra \psi | P \phi \ket \quad
{\text{for all~$\psi, \phi \in C^\infty_0(\scrM, S\scrM)$}}\:.
\eeq
This condition can also be expressed in terms of the kernel by
\beq \label{s:Padjoint}
P(x,y)^* \equiv \gamma^0 P(x,y)^\dagger \gamma^0 = P(y,x)
\quad {\text{for all~$x,y \in \scrM$}}\:,
\eeq
where the dagger denotes the transposed, complex conjugate matrix.
We refer to~$P$ as the {\em{fermionic projector}}.
\sindex{fermionic projector}%
The vectors in the image of~$P$
are referred to as the {\em{physical wave functions}}.
\sindex{wave function!physical}%
 We point out that for the moment,
these wave functions do not need to be solutions of a Dirac equation.

For any space-time points~$x$ and~$y$, we next introduce the {\em{closed chain}} $A_{xy}$ by
\sindex{closed chain}%
\nindex{ad0@$A_{xy}$ -- closed chain}%
\beq \label{s:Adef}
A_{xy} = P(x,y)\, P(y,x)\:.
\eeq
It is a $4 \times 4$-matrix which can be considered as a linear operator on the wave functions
at~$x$. For any such linear operator~$A$ we define the {\em{spectral weight}} $|A|$ by
\nindex{aa1@$\vert \,.\, \vert$ -- spectral weight}%
\sindex{spectral weight}%
\beq \label{s:swdef}
|A| = \sum_{i=1}^4 |\lambda_i|\:,
\eeq
where~$\lambda_1, \ldots, \lambda_4$ are the eigenvalues of~$A$ counted with algebraic
multiplicities. For any~$x, y \in \scrM$ we define the {\em{Lagrangian}} $\L$ by
\nindex{aa2@$\L(x,y)$ -- Lagrangian}%
\sindex{Lagrangian}%
\beq \label{s:Ldef}
\L_{xy}[P] = |A_{xy}^2| - \frac{1}{4}\: |A_{xy}|^2 \:.
\eeq
Integrating over space-time, we can furthermore introduce the functionals
\beq \boxed{ \quad \begin{split}
\Sact[P] \;&\stackrel{\text{formally}}{=}\; \iint_{\scrM \times \scrM} \L_{xy}[P] \:d^4 x\: d^4y \label{s:STdef} \\
\T[P] \;&\stackrel{\text{formally}}{=}\; \iint_{\scrM \times \scrM} |A_{xy}|^2 \:d^4 x\: d^4y\:.
\end{split} \quad }
\eeq
\nindex{aa4@$\Sact(\rho)$ -- causal action}%
\nindex{aa8@$\T(\rho)$ -- functional in boundedness constraint}%
These expressions are only formal because the integrands need not decay for large~$x$ or~$y$,
and thus the integrals may be infinite
(similar as in classical field theory, where the space-time integral over the Lagrangian
diverges without imposing suitable decay properties at infinity).
The functional~$\Sact$ is referred to as the {\em{causal action}}.
\sindex{causal action}%
Our variational principle is to minimize~$\Sact$ under the constraint that~$\T$ is kept
fixed\footnote{Clearly, the constraint of keeping~$\T$ fixed is stronger than
the boundedness constraint~\eqref{Tdef} which merely imposes that~$\T$ must be bounded from above.
Working with~\eqref{Tdef} is preferable when working out the existence theory~\cite{continuum}.
However, for what follows here, it makes no difference if~$\T$ is kept fixed or only stays bounded,
because both variational principles give rise to the same Euler-Lagrange equations.}.
For this minimization, we vary the fermionic projector in the following sense.
In order to prevent trivial minimizers, the variation should preserve the
normalization of the wave functions. This normalization should be performed with
respect to the Lorentz invariant inner product~\eqref{s:iprod}
(in a more abstract language, we thus consider {\em{unitary variations in the Krein space}}
as introduced in Remark~\ref{remvarKrein}).
\sindex{variation of universal measure!by unitary variation in Krein space}%
\sindex{unitary variation in Krein space}%
However, we do not want to assume that this inner product is finite for the wave
functions~$\psi$ in the image of~$P$ (indeed, for physical wave functions, the inner product~$\bra \psi | \psi \ket$ is in general infinite because the time integral diverges). Our method for avoiding
the divergences in~\eqref{s:iprod} and~\eqref{s:STdef} is to consider
variations which outside a compact set are the identity.
\begin{Def} \label{s:def21}
An operator~$U$ on the Dirac wave functions is called
{\bf{unitary in a compact region}} if
\sindex{unitary in a compact region}%
\begin{itemize}
\item[(i)] $\bra U \psi \,|\, U \psi \ket = \bra \psi \,|\, \psi \ket$ for all compactly supported~$\psi$.
\item[(ii)] The operator~$V := U-\1$ has the representation
\[ (V \psi)(x) = \int_\scrM v(x,y)\, \psi(y)\, d^4y \]
with a smooth integral kernel~$v(x,y)$ which has compact support, i.e.\ there is a
compact set~$K \subset \scrM$ such that
\[ v(x,y) = 0 \qquad {\text{unless $x \in K$ and $y \in K$}}\:. \]
\end{itemize}
\end{Def} \noindent
Thus introducing a variation of the wave functions by the transformation~$\psi \rightarrow
U \psi$, all the wave functions are changed only in the compact region~$K \subset \scrM$,
in such a way that all inner products in this region, i.e.\ all the integrals
\[ \int_K \overline{\psi(x)} \,\phi(x) \: d^4x \:, \]
remain unchanged. Having introduced a well-defined notion of  ``varying the fermionic projector
while respecting the inner product~\eqref{s:iprod},'' we can now specify what we mean by
a minimizer.
\begin{Def} \label{s:def12}
A fermionic projector~$P$ of the
form~\eqref{s:Pdef} is a {\bf{minimizer}} of the variational principle
\beq \label{s:actprinciple}
\text{minimize $\Sact$ for fixed~$\T$}
\eeq
if for any operator~$U$ which is unitary in a compact region and satisfies the constraint
\beq \label{s:intdelT}
\int_\scrM d^4x \int_\scrM d^4y \:\Big( \big|A_{xy}[P] \big|^2 - \big| A_{xy}[U P U^{-1}] \big|^2 \Big)
= 0\:,
\eeq
the functional~$\Sact$ satisfies the inequality
\beq
\int_\scrM d^4x \int_\scrM d^4y \:\Big(\L_{xy}[U P U^{-1}] -  \L_{xy}[P] \Big)
\;\geq\; 0\:. \label{s:intdelL}
\eeq
\end{Def} \noindent
We point out that, since~$U$ changes the wave functions only inside a compact set~$K$,
the integrands in~\eqref{s:intdelT} and~\eqref{s:intdelL} clearly vanish if~$x$ {\em{and}} $y$ are
outside~$K$.
However, it is not obvious that the integrals over the region~$x \in K$ and~$y \in \scrM \setminus K$
(and similarly~$x \in \scrM \setminus K$ and~$y \in \scrM$) exist. 
By writing~\eqref{s:intdelT} and~\eqref{s:intdelL} we implicitly demand that
the integrand in~\eqref{s:intdelT} and the negative part of the integrand in~\eqref{s:intdelL}
should be in~$L^1(\scrM \times \scrM, \R)$.

Before going on, we briefly discuss this action principle and bring it into the context of
previous work. We first remark that, in contrast to~\cite{PFP, discrete},
we here ignore the condition that~$P$ should be idempotent. This is done merely
to simplify the presentation, anticipating that the idempotence condition will not be of relevance
in this paper. The action principle~\eqref{s:actprinciple} was first introduced in a discrete
space-time in~\cite[Section~3.5]{PFP}. Apart from the obvious replacement of sums by integrals,
the action here differs from
that in~\cite[Section~3.5]{PFP} only by an irrelevant multiple of the constraint~$\T$.
This has the advantage that the Lagrangian~\eqref{s:Ldef} coincides with the so-called critical case
of the auxiliary Lagrangian as introduced in~\cite{discrete}; this is the case relevant in our
setting of one sector. Note that this Lagrangian is symmetric (see~\cite[eq.~(13)]{discrete}) and
non-negative,
\[ \L_{xy}[P] = \L_{yx}[P] \qquad {\text{and}} \qquad
\L_{xy}[P] \geq 0 \:. \]
Moreover, the action principle~\eqref{s:actprinciple} can be regarded as an infinite volume
limit of the variational principle in~\cite[Section~3]{continuum} (possibly also in the
limit where the number of particles tends to infinity).
In the special case of homogeneous systems, our variational principle is closely related
to the variational principle in infinite volume as considered in~\cite[Section~4]{continuum}.
Working with unitary transformations in a compact region, we can make sense of the action principle
even in infinite space-time volume without assuming homogeneity;
this procedure can be seen in analogy to considering variations of compact support
in the Lagrangian formulation of classical field theory (like a variation~$\delta A \in
C^\infty_0(\scrM, \R^4)$ of the electromagnetic potential in classical electrodynamics).

\section{Assuming a Vacuum Minimizer} \label{s:sec3}
Apart from the general existence results in~\cite{discrete, continuum}
and the simple examples in~\cite{small, continuum}, almost nothing is known about the minimizers of
our action principle. Therefore, before we can do physics, we need to assume the existence of a special minimizer which describes a physically meaningful vacuum. In this section, we compile our assumptions
on this vacuum minimizer, and we outline in which sense and to what extent these assumptions
have been justified in~\cite{reg, vacstab}. At the end of this section, we will explain how to work
with these assumptions in practice.

Taking Dirac's original concept seriously, we want to describe the vacuum by ``completely
filled Dirac seas'' corresponding to the masses $m_1, \ldots, m_g$ of $g$ generations of
elementary particles
\nindex{ca2@$g$ -- number of generations}%
(later we will set~$g=3$, but for the moment it is preferable not to
specify the number of generations). Thus our first ansatz for the integral kernel of the fermionic
projector of the vacuum is the Fourier transform of the projectors~$\frac{1}{2 m_\beta}
(\slashed{k} +m_\beta)$ on the Dirac states on the lower mass shells,
\beq \label{s:A}
P(x,y) = \sum_{\beta=1}^g \int \frac{d^4k}{(2 \pi)^4}\: (\slashed{k} +m_\beta)\:
\delta(k^2-m_\beta^2)\: \Theta(-k^0)\: e^{-ik(x-y)}\:.
\eeq
\sindex{fermionic projector!of the vacuum}%
\nindex{ca3@$m_\beta$ -- masses of charged fermions}%
(Here~$\Theta$ is the Heaviside function, and $k (x-y)$ is a short notation for the Minkowski
inner product~$\langle k, x-y \rangle$. The slash denotes contraction with the Dirac matrices,
thus~$\slashed{k} = k_j \gamma^j$.
We always work in natural units
$\hbar=c=1$, and for the signature of the Minkowski inner product we use the
convention~$(+ - -\, -)$.)
We always index the masses in increasing order,
\beq \label{s:morder}
m_1 < m_2 < \ldots < m_g \:.
\eeq
The Fourier integral~\eqref{s:A} is well-defined as a distribution. If the vector $y-x$
is spacelike or timelike, the integral~\eqref{s:A} exists even pointwise.
However, if the vector $y-x$ is null, the distribution~$P(x,y)$ is singular
(for details see~\cite[Section~2.5]{PFP} or~\S\ref{seccorcs}). In physical terms, these singularities occur if
$y$ lies on the light cone centered at~$x$.
Thus we refer to the singularities on the set where~$(x-y)^2=0$
as the {\em{singularities on the light cone}}.
As a consequence of these singularities, the pointwise product in~\eqref{s:Adef}
is ill-defined, and the Lagrangian~\eqref{s:Ldef} has no mathematical meaning.
In order to resolve this problem, one needs to introduce an {\em{ultraviolet regularization}}.
\sindex{regularization!ultraviolet (UV)}%
In position space, this regularization can be viewed as a ``smoothing'' on a microscopic
length scale. It seems natural to identify this microscopic length scale with the
{\em{Planck length}}~$\ell_P$, although it may be even smaller.
\sindex{Planck length}%
\nindex{aj9@$\ell_P$ -- Planck length}%
Thus we always assume that~$\varepsilon \lesssim \ell_P$.
Likewise, in momentum space the regularization corresponds to a cutoff or decay
on the scale~$\varepsilon^{-1}$, which is at least as large as the {\em{Planck energy}}~$E_P=\ell_P^{-1}$.
\sindex{Planck energy}%
\nindex{ca4@$E_P$ -- Planck energy}%
Clearly, the regularization scale is extremely small compared to the length scale $\ell_{\text{macro}}$
\nindex{bl2@$\ell_\text{macro}$ -- macroscopic length scale}%
of macroscopic physics, and thus it seems reasonable to expand in powers
of~$\varepsilon/\ell_{\text{macro}}$. However, such an expansion would not be mathematically
meaningful, because Taylor series can be performed only in continuous variables
(but not in a constant, no matter how small). Therefore, it is preferable to denote the regularization length by the variable~$\varepsilon$,
which may vary in the range~$0<\varepsilon \ll \ell_{\text{macro}}$.
\nindex{aj8@$\varepsilon$ -- regularization length}%
\sindex{regularization length}%
We are thus led to a one-parameter family of regularizations. We assume that
these regularized Dirac sea configurations are all minimizers. We also compile all assumptions
on the regularization as introduced in~\cite[Chapter~4]{PFP}.
\begin{Assumption} {\bf{(regularized Dirac sea vacuum)}} \label{s:assumption}
{\em{ There is a family~$(P^\varepsilon)_{\varepsilon>0}$ of fermionic projectors
\nindex{an4@$P^\varepsilon(x,y)$ -- regularized kernel of fermionic projector}%
\sindex{fermionic projector!regularized kernel of}%
whose kernels~$P^\varepsilon(x,y)$ (as defined by~\eqref{s:Pdef}) have the following properties:
\begin{itemize}[leftmargin=2.5em]
\item[(i)] Every~$P^\varepsilon(x,y)$ is a {\em{minimizer}} in the sense of Definition~\ref{s:def12}.
\item[(ii)] Every~$P^\varepsilon(x,y)$ is {\em{homogeneous}}, i.e.\ it depends only on the
\sindex{fermionic projector!homogeneous}%
variable~$\xi := y-x$.%
\item[(iii)] Taking its Fourier transform,
\beq \label{s:PFT}
P^\varepsilon(x,y) = \int \frac{d^4k}{(2 \pi)^4} \:\hat{P}^\varepsilon(k) \: e^{-i k (x-y)}\:,
\eeq
$\hat{P}^\varepsilon$ is a distribution with a {\em{vector-scalar structure}}, i.e.
\beq \label{s:Peps}
\hat{P}^\varepsilon(k) = (v^\varepsilon_j(k) \:\gamma^j \:+\: \phi^\varepsilon(k) \:\1) \:f^\varepsilon(p)
\eeq
\sindex{fermionic projector!vector-scalar structure}%
with a vector field $v^\varepsilon$, a scalar field $\phi^\varepsilon$ and a
distribution~$f^\varepsilon$, which are all real-valued.
\item[(iv)] If the regularization is removed, $P^\varepsilon$ goes over to~$P$
(as given by~\eqref{s:A}), i.e.
\[ \lim_{\varepsilon \searrow 0} \hat{P}^\varepsilon(k) = \hat{P}(k)
:= \sum_{\beta=1}^g  (\slashed{k}+m_\beta)\: \delta(k^2-m_\beta^2)\: \Theta(-k^0) \]
with convergence in the distributional sense.
\end{itemize}
The assumptions so far seem natural and are easy to state. In order to understand the following
assumptions, one should notice that the singularities of~$P(x,y)$ on the light cone arise
because its Fourier transform~$\hat{P}(k)$ is supported on the mass shells~$k^2=m_\beta^2$,
\sindex{mass shell}%
which are hypersurfaces being asymptotic to the mass cone~$k^2=0$
\sindex{mass cone}%
(for details see~\cite[Section~4.2]{PFP}). Thus in order to control the behavior of~$P^\varepsilon$
near the light cone, we need to make suitable assumptions on~$P^\varepsilon(\omega, \vec{k})$
for~$\omega \approx -|\vec{k}| \sim \varepsilon^{-1}$.
\begin{itemize}[leftmargin=2.5em]
\item[(v)] We assume that the distribution~$\hat{P}^\varepsilon$ is supported on hypersurfaces
described by graphs, i.e.\ the distribution~$f^\varepsilon$ in~\eqref{s:Peps} should be of the form
\beq \label{s:hypersurf}
f^\varepsilon(\omega, \vec{k}) = \sum_{\beta=1}^g \delta \!\left( \omega + |\vec{k}| + \alpha_\beta(\vec{k}) \right) \:.
\eeq
These hypersurfaces should be asymptotic to the mass cone in the sense that
\[ \alpha_\beta(\vec{k}) \sim \varepsilon \qquad \text{if~$|\vec{k}| \sim \varepsilon^{-1}$}\:. \]
Except for these singularities, $\hat{P}^\varepsilon(k)$ is so regular that the singularities
as~$\varepsilon \searrow 0$ of~$P^\varepsilon(x,y)$ on the light cone are completely
described by the behavior of~$\hat{P}^\varepsilon(k)$ on the hypersurfaces~\eqref{s:hypersurf},
up to corrections of higher order in~$\varepsilon$.
We refer to this assumption as the {\em{restriction to surfaces states}}.
\sindex{surface state!restriction to}%
\item[(vi)] On the hypersurfaces~\eqref{s:hypersurf} and for~$|\vec{k}| \sim \varepsilon^{-1}$,
the vector field~$v^\varepsilon$ in~\eqref{s:Peps} should be parallel to~$k$, up to a small
error term. More precisely, decomposing~$v^\varepsilon$ as
\beq \label{s:vepsrel}
v^\varepsilon = s^\varepsilon(k) \, k + \vec{w}^\varepsilon(k)
\eeq
with a scalar function~$s^\varepsilon$, the vector field~$\vec{w}^\varepsilon$ should be bounded by
\beq \label{s:shear}
|\vec{w}^\varepsilon(k)| < \varepsilon_\text{shear} \qquad \text{where} \qquad
\varepsilon_\text{shear} \ll 1\:.
\eeq
We refer to~$\varepsilon_\text{shear}$ as the {\em{shear parameter}}.
\nindex{ca6@$\varepsilon_\text{shear}$ -- shear parameter}%
Considering the effect of this assumption in position space, we say that the {\em{vector component is null
on the light cone}}.
\sindex{vector component!is null on the light cone}%
\item[(vii)] The functions in~\eqref{s:Peps} either vanish, $\phi^\varepsilon(k)=0=v^\varepsilon(k)$,
or else~$\phi^\varepsilon(k)>0$ and the vector field~$v^\varepsilon$ is time-like and past-directed.
Furthermore,
\[ v^\varepsilon(p)^2 = \phi^\varepsilon(p)^2\:. \]
\end{itemize} }}
\end{Assumption} \noindent
For a discussion of the assumptions~(v) and~(vi) we refer to~\cite[Chapter~4]{PFP}.
The condition~(vii) requires a brief explanation. This assumption is clearly satisfied without
regularization~\eqref{s:A} (in which case we choose~$v(p)=p/(2\omega)$ and~$\phi$ a positive function
which on the mass shells takes the values~$m_\alpha/(2 \omega)$).
A closely related condition was first proposed in~\cite[Chapter~4]{PFP} as the
assumption of {\em{half-occupied surface states}}.
\sindex{surface state!half-occupied}%
This condition was motivated
by the wish to realize the Dirac sea configurations with as few occupied states as possible,
noting that the condition~(vii) implies that the matrix~$\hat{P}^\varepsilon(k)$ has rank
at most two. Furthermore, the condition~(vii) implies that the image of the matrix~$\hat{P}^\varepsilon(k)$
is negative definite with respect to the inner product~$\overline{\psi} \phi$.
From the mathematical point of view, this definiteness is crucial for
our action principle to be mathematically well-defined
(see the reformulation as a causal variational principle in~\cite{continuum, rrev}
as well as the general compactness result~\cite[Theorem~4.2]{continuum}).
Thus the physical intuition and
the mathematical requirements
fit together. Moreover, in the case when~$\hat{P}^\varepsilon(k)$ does not vanish, we can choose
a suitably normalized orthogonal basis~$(\psi_{p,1}, \psi_{p,2})$ of the image of~$\hat{P}^\varepsilon(k)$
such that~$(2 \pi)^4 \hat{P}^\varepsilon(k)=-\psi_{k,1} \overline{\psi_{k,1}}
-\psi_{k,2} \overline{\psi_{k,2}}$.
Substituting this representation into the Fourier integral~\eqref{s:PFT} and using~\eqref{s:hypersurf},
we obtain
\beq \label{s:Perep}
P^\varepsilon(x,y) = -\sum_{\beta=1}^g \int_{\R^3} d\vec{k} \sum_{a=1,2}
\psi_{\vec{k} \beta a}(x) \:\overline{\psi_{\vec{k} \beta a}(y)}\:,
\eeq
where~$\psi_{\vec{k} \beta a(x)} = \psi_{p,a} \,e^{-i p x}$ for~$p=(-|\vec{k}|-\alpha_\beta(\vec{k}),
\vec{k})$. This representation is helpful because it shows that the regularized fermionic projector of the
vacuum is composed of negative-energy wave functions; the index~$a$ can be thought of as describing
the two spin orientations.

We next outline the approach taken to justify the above assumptions. In~\cite{reg} a class
of regularizations is constructed for which the action remains finite when the regularization
is removed (more precisely, this is done by proving that the constructed regularizations
satisfy the so-called assumption of a distributional~${\mathcal{M}}P$-product).
These regularizations are spherically symmetric, but they break the Lorentz symmetry.
However, after suitably removing the regularization, we obtain a well-defined
Lorentz invariant action principle.
This Lorentz invariant action principle is analyzed in~\cite{vacstab}, and it is shown that
for certain values of the masses and the so-called weight factors
(which for simplicity we do not consider in the main text of this paper; see however
Appendix~\ref{s:apprho}), the Dirac sea configuration~\eqref{s:A}
is indeed a minimizer, in a sense made precise using the notion of state stability.
Following these results, ``good candidates'' for satisfying the above assumptions are
obtained by regularizing the state stable Dirac sea configurations from~\cite{vacstab} according
to the regularization scheme in~\cite{reg}. The remaining task for giving a rigorous justification of
Assumption~\ref{s:assumption} is to use the freedom in choosing the regularization such as
to obtain a minimizer in the sense of Definition~\ref{s:def12}.
This task seems difficult and has not yet been accomplished.
In~\cite[Theorem~4.2]{continuum} the existence of minimizers is proved within the class
of homogeneous fermionic projectors; but this is considerably weaker than being
a minimizer in the sense of Definition~\ref{s:def12}. In technical terms,
the main difficulty is to quantify the influence of the spherically symmetric
regularization on the action, even taking into account
contributions which remain finite when the regularization is removed.
Despite this difficult and technically challenging open problem, it is fair to say that the results
of~\cite{reg, vacstab} show that Dirac sea configurations tend to make our action small, thus
explaining why Assumption~\ref{s:assumption} is a reasonable starting point for the
continuum analysis.

We finally explain how to work with the above assumptions in practice. Ideally, 
the fields~$v^\varepsilon$, $\phi^\varepsilon$
and the distribution~$f^\varepsilon$ in~\eqref{s:Peps} could be determined
by minimizing our action~\eqref{s:actprinciple}, thus giving detailed information on~$P^\varepsilon$.
Such a minimization process is indeed possible (see~\cite[Theorem~4.2]{continuum} for a
general existence result and~\cite{ssymm} for a lattice formulation), but so far has not been
analyzed in sufficient depth. Thus for the time being, there is a lot of freedom to choose
the functions in~\eqref{s:Peps}.
Our program is not to make a specific choice but to consider instead general
functions~$v^\varepsilon$, $\phi^\varepsilon$ and~$f^\varepsilon$.
Our subsequent analysis will clearly depend on the choice of these functions, and our task is to look
for conclusions which are robust to regularization details.
This so-called {\em{method of variable regularization}}
\sindex{regularization!method of variable}%
(which is worked out in detail in~\cite[Section~4.1]{PFP}) leads to the formalism of the continuum
limit which will be explained in Section~\ref{s:sec5} below.

\section{Introducing an Interaction} \label{s:sec4}
Our next goal is to generalize the regularized fermionic projector~$P^\varepsilon$
of the previous section such as to include an interaction. Postponing the treatment of the
regularization to Section~\ref{s:sec5}, we shall now extend the definition of the
fermionic projector of the vacuum~\eqref{s:A} to the case with interaction.
We outline the methods developed in~\cite{sea, firstorder, light}; see also~\cite[Chapter~2]{PFP}
or Chapter~\ref{tools} in this book.

\subsectionn{A Dirac Equation for the Fermionic Projector} \label{s:sec41}
First, it is useful to recover~\eqref{s:A} as a solution of a Dirac equation: 
Replacing the ordinary sum in~\eqref{s:A} by a direct sum, we introduce
the so-called {\em{auxiliary fermionic projector}} $P^\text{aux}$ by
\beq \label{s:Pauxvac}
P^\text{aux}(x,y) = \bigoplus_{\beta=1}^g \int \frac{d^4k}{(2 \pi)^4}\: (\slashed{k}+m_\beta)\:
\delta(k^2-m_\beta^2)\: \Theta(-k^0)\: e^{-ik(x-y)}
\eeq
(thus~$P^\text{aux}(x,y)$ is represented by a $4g \times 4g$-matrix).
\sindex{fermionic projector!auxiliary}%
\nindex{ca8@$P^\text{aux}$ -- auxiliary fermionic projector}%
It is a solution of the free Dirac equation
\beq \label{s:D}
(i \Pdd_x - m Y)\, P^\text{aux}(x,y) = 0\:,
\eeq
where the {\em{mass matrix}} $Y$ is composed of the rest masses corresponding to the~$g$
generations,
\[ m Y = \bigoplus_{\beta=1}^g m_\beta \]
\sindex{mass matrix}%
\nindex{bh3@$m$ -- parameter used for mass expansion}%
\nindex{bh4@$Y$ -- mass matrix}%
(here~$m>0$ is an arbitrary mass parameter which makes~$Y$ dimensionless and will be useful for
expansions in the mass parameter; see also~\cite[Section~2.3]{PFP} or Section~\ref{seclight}).
The fermionic projector of the vacuum is obtained
from~$P^\text{aux}$ by summing over the generation indices,
\begin{equation} \label{s:pt}
P = \sum_{\alpha, \beta=1}^g (P^\text{aux})^\alpha_\beta\:.
\end{equation}
This summation removes the generation indices, leaving us with the configuration of one sector.
In~\cite{PFP} this operation is referred to as the {\em{partial trace}}.
However, this notion might be confusing because it suggests that in~\eqref{s:pt} one
should set~$\alpha=\beta$ and sum over one index (for a more detailed discussion
see~\cite[paragraph after Lemma~2.6.1]{PFP}). In order to avoid this potential source of confusion,
in this book we always refer to the operator in~\eqref{s:pt} as the 
the {\em{sectorial projection}}.
\sindex{sectorial projection}%

The obvious idea for introducing an interaction is to replace the free Dirac equation~\eqref{s:D}
by a Dirac equation with interaction,
\beq
(i \Pdd + \B - m Y)\, P^\text{aux}(x,y) = 0\:, \label{s:diracPaux}
\eeq
where~$\B$ is a general perturbation operator,
\nindex{ar4@$\B$ -- external potential}%
\sindex{potential!bosonic}%
\sindex{potential!external}%
and to introduce the fermionic projector again forming the sectorial projection~\eqref{s:pt}.
In order to ensure that the resulting fermionic projector is again symmetric~\eqref{s:Psymm1},
we generalize the inner product~\eqref{s:iprod} to the wave functions of the auxiliary Dirac
equation by setting
\beq \label{s:iprodaux}
\bra \psi_\text{aux} | \phi_\text{aux} \ket =
\sum_{\beta=1}^g \int_\scrM \overline{\psi_\text{aux}^\beta(x)} \phi_\text{aux}^\beta(x)\: d^4x\:,
\eeq
and demand that the auxiliary fermionic projector should be symmetric with respect to this
new inner product,
\[ \bra P^\text{aux} \psi_\text{aux} | \phi_\text{aux} \ket = \bra \psi_\text{aux} | P^\text{aux}
\phi_\text{aux} \ket \quad
{\text{for all~$\psi_\text{aux}, \phi_\text{aux} \in C^\infty_0(\scrM, S\scrM)^{g}$}}\:. \]
In order to obtain a coherent framework, we shall always assume that the Dirac operator
is symmetric with respect to this inner product,
\[ \bra (i \Pdd + \B - m Y) \psi_\text{aux} | \phi_\text{aux} \ket = \bra \psi_\text{aux} | (i \Pdd + \B - m Y)
\phi_\text{aux} \ket . \]
This equation gives a condition for the operator~$\B$ describing the interaction.
Apart from this condition and suitable regularity and decay assumptions, the operator~$\B$
can be chosen arbitrarily; in particular, it can be time dependent.
In typical applications, $\B$ is a multiplication or differential operator composed of
{\em{bosonic potentials}} and fields.
\sindex{potential!bosonic}%
\sindex{potential!external}%
\sindex{classical field!bosonic}%
\sindex{bosonic field!classical}%
\sindex{gauge field!classical}%
The choices of~$\B$ relevant for this work will be discussed in~\S\ref{s:discussion} below.

\subsectionn{The Interacting Dirac Sea} \label{s:sec42}
Clearly, the Dirac equation~\eqref{s:diracPaux} has many different solutions, and thus in
order to determine~$P^\text{aux}$, we need to specify of which one-particle
states~$P^\text{aux}$ should be composed. In the vacuum~\eqref{s:Pauxvac}, this can be done
by taking all the negative-energy solutions, i.e.\ all states on the lower mass shells
$\{k^2=m^2_\beta, k^0<0\}$. Unfortunately, the concept of negative energy
does not carry over to the situation of a time-dependent interaction~\eqref{s:diracPaux}, because
in this case the energy of the Dirac wave functions is not conserved; this is
the so-called {\em{external field problem}} (see~\cite[Section~2.1]{PFP} or Section~\ref{secfpext}).
\sindex{external field problem}%
The clue for resolving this problem is the observation that the negative-energy states in~\eqref{s:Pauxvac}
can be characterized alternatively using the causality of the Dirac Green's functions in a specific way.
This causal approach generalizes to the situation~\eqref{s:diracPaux}
and makes it possible to extend the concept of the Dirac sea to the time-dependent setting.
It gives rise to a unique definition of the fermionic projector~$P^\sea$
in terms of a power series in~$\B$. More precisely, the so-called {\em{causal perturbation expansion}}
\sindex{causal perturbation expansion}%
expresses~$P^\sea$ as sums of operator products
\beq \label{s:cpower}
P^\sea = \sum_{k=0}^\infty \sum_{\alpha=0}^{\alpha_{\max}(k)} c_\alpha\;
C_{1,\alpha} \, \B\,C_{2,\alpha} \,\B\, \cdots \,\B\, C_{k+1, \alpha} \:,
\eeq
\nindex{be0@$P^\sea$ -- fermionic projector describing Dirac seas}%
where the factors~$C_{l,\alpha}$ are the Green's functions or fundamental solutions of the
free Dirac equation~\eqref{s:D}, and the~$c_\alpha$ are combinatorial factors
(for details see~\cite{sea} and~\cite[Sections~2.2--2.3]{PFP}; for a more recent account on
idempotence and unitarity questions see~\cite{grotz, norm}).
In the language of Feynman diagrams,
\sindex{Feynman diagram!tree diagram}%
each summand in~\eqref{s:cpower} is a tree diagram.
These tree diagrams are all finite, provided
that~$\B$ satisfies suitable regularity and decay assumptions at infinity
(see~\cite[Lemma~2.2.2.]{PFP} or Lemma~\ref{l:lemma0}).

\subsectionn{Introducing Particles and Anti-Particles} \label{s:sec43}
The fermionic projector~$P^\sea$ is interpreted as a generalization of
completely filled Dirac seas to the interacting situation~\eqref{s:diracPaux}.
In order to bring particles and anti-particles into the system,
we add the projectors
on states $\psi_1, \ldots, \psi_{\np}$ which are {\em{not}} contained in the image of the operator~$P^\sea$ (the particle states)
and subtract the projectors on states $\phi_1, \ldots, \phi_{\na}$
which are in the image of~$P^\sea$ (the anti-particle states),
\beq \label{s:particles}
P^\text{aux}(x,y) = P^\sea(x,y)
-\frac{1}{2 \pi} \sum_{k=1}^{\np} \psi_k(x) \overline{\psi_k(y)}
+\frac{1}{2 \pi} \sum_{l=1}^{\na} \phi_l(x) \overline{\phi_l(y)}\:.
\eeq
\nindex{as3@$\np, \na$ -- number of particles and anti-particle states}%
\nindex{as4@$\psi_k, \phi_l$ -- particle and anti-particle states}%
\sindex{particles and anti-particles}%
\sindex{anti-particles}%
Then the fermionic projector is again obtained by forming the sectorial projection~\eqref{s:pt}. Here the wave functions in~\eqref{s:particles} are to be normalized such that
they are orthonormal with respect to the usual integral over the probability density, i.e.\
\beq \label{s:normalize}
\int_{\R^3} (\overline{\psi_k} \gamma^0 \psi_{k'})(t, \vec{x}) \, d^3x = \delta_{k,k'} \:,\quad
\int_{\R^3} (\overline{\phi_l} \gamma^0 \phi_{l'})(t, \vec{x}) \, d^3x = \delta_{l,l'} \:.
\eeq
The factors~$\pm \frac{1}{2 \pi}$ in~\eqref{s:particles} are needed for the
proper normalization of the fermionic states (for details see~\cite{norm}
or~\S\ref{secnorm}).

\subsectionn{The Light-Cone Expansion and Resummation} \label{s:sec44}
We now outline the methods for analyzing the fermionic projector in position space (for
details see~\cite{firstorder, light} or Section~\ref{seclight}). The following notion is very useful for describing
the structure of the singularities on the light cone.
\begin{Def} \label{s:l:def1}
A distribution~$A(x,y)$ on~$\scrM \times \scrM$ is of the order~$\O((y-x)^{2p})$,
$p \in \Z$, if the product
\[ (y-x)^{-2p} \: A(x,y) \]
is a regular distribution (i.e.\ a locally integrable function).
It has the {\bf{light-cone expansion}}
\nindex{bf4@$\O((y-x)^{2p})$ -- order on the light cone}%
\sindex{light-cone expansion}%
\[ A(x,y) = \sum_{j=g_0}^{\infty} A^{[j]}(x,y) \]
with $g_0 \in \Z$ if the distributions $A^{[j]}(x,y)$ are of the order
$\O((y-x)^{2j})$ and if $A$ is approximated by the partial sums
in the sense that for all $p \geq g$,
\[ A(x,y) - \sum_{j=g_0}^p A^{[j]}(x,y) \qquad
        \text{is of the order~$\O((y-x)^{2p+2})$}\:. \]
\end{Def} \noindent
Thus the light-cone expansion is an expansion in
the orders of the singularity on the light cone. As the main difference to a
Taylor expansion, for any fixed~$x$ the expansion parameter~$(y-x)^2$ vanishes for all~$y$
in an unbounded set, namely the whole light cone centered at~$x$.
In this sense, the light-cone expansion is a {\em{nonlocal}} expansion.

For a convenient formulation of the light-cone expansion of the fermionic projector,
it is helpful to work with a {\em{generating function}}, i.e.\ a power series in a real parameter
$a > 0$ whose coefficients are functions in~$(y-x)^2$ which are of increasing 
order on the light cone. The first ansatz for such a generating function is
the Fourier transform~$T_a(x,y)$ of the lower mass shell with $k^2=a$,
\beq \label{s:Tadef}
T_a(x,y) = \int \frac{d^4k}{(2 \pi)^4} \: \delta(k^2-a)\: \Theta(-k^0) \:e^{-ik(x-y)} \:.
\eeq
\nindex{ao8@$T_a(x,y)$ -- Fourier transform of lower mass shell}%
Carrying out the Fourier integral and expanding the resulting Bessel functions, one obtains
\beq \label{s:Taser}
\begin{split}
T_a(x,y)
=& -\frac{1}{8 \pi^3} \:
\left( \frac{\text{PP}}{\xi^2} \:+\: i \pi \delta (\xi^2) \:
\epsilon(\xi^0) \right) \\
&+\: \frac{a}{32 \pi^3}\sum_{j=0}^\infty  \left( \log |a \xi^2| + c_j
+ i \pi \:\Theta(\xi^2) \:\epsilon(\xi^0) \right)
\frac{(-1)^j}{j! \: (j+1)!} \: \frac{(a
\xi^2)^j}{4^j} \:,
\end{split}
\eeq
where we again used the abbreviation~$\xi = y-x$, and~$\epsilon$ denotes the sign function
(i.e.\ $\epsilon(x)=1$ if~$x \geq 0$ and~$\epsilon(x)=-1$ otherwise).
The real coefficients~$c_j$ are given explicitly in~\cite[Section~2.5]{PFP}.
Unfortunately, due to the factor~$\log|a \xi^2|$, the expression~\eqref{s:Taser} is not a power series
in~$a$. In order to bypass this problem, we simply remove the logarithms in~$a$
by subtracting suitable counter terms,
\beq \label{s:Tacounter}
T_a^\reg(x,y) := T_a(x,y) - \frac{a}{32 \pi^3} \:\log |a| \: \sum_{j=0}^\infty
\frac{(-1)^j}{j! \: (j+1)!} \: \frac{(a \xi^2)^j}{4^j} \:.
\eeq
\nindex{bj1@$T^\reg_a$ -- $T_a$ with $\log$-terms in~$a$ removed}%
The resulting distribution~$T_a^\reg$ is a power series in~$a$,
and it is indeed the right choice for our generating function. We
denote its coefficients by
\beq \label{s:Tndef}
T^{(n)} =  \left( \frac{d}{da} \right)^n 
T^{\text{reg}}_a \Big|_{a=0} \qquad (n=0,1,2,\ldots)
\eeq
\nindex{bj2@$T^{(l)}$ -- mass expansion of $T^\reg_a$}%
and also introduce~$T^{(-1)}$ via the distributional equation
\beq \label{s:Tm1def}
\frac{\partial}{\partial x^k} T^{(0)}(x,y) = \frac{1}{2} \: (y-x)_k \: T^{(-1)}(x,y) \: .
\eeq
We remark for clarity that removing the logarithmic poles in~$a$ has similarity to
an infrared regularization, because infrared problems also appear when the mass
parameters tend to zero. This is the motivation for using the superscript ``reg.''
But clearly, this ``regularization'' is not related to the ultraviolet regularization
in Assumption~\ref{s:assumption}.

Combining Fourier techniques with methods of hyperbolic partial differential equations, one can perform
the light-cone expansion of each summand of the perturbation series~\eqref{s:cpower}.
After suitably rearranging all the resulting contributions, one can partially
carry out the infinite sums.
\sindex{light-cone expansion!resummation of}%
\sindex{resummation!of light-cone expansion}%
 This so-called {\em{resummation}} gives rise to
an expansion of the interacting fermionic projector of the form
\begin{align}
P^{\text{sea}}(x,y) = & \sum_{n=-1}^\infty
\sum_{k} m^{p_k} 
{\text{(phase-inserted nested line integrals)}} \times  T^{(n)}(x,y) \nonumber \\
&+ \tilde{P}^\lec(x,y) + \tilde{P}^\hec(x,y) \:.
\label{s:fprep}
\end{align}
\nindex{be0@$P^\sea$ -- fermionic projector describing Dirac seas}%
\sindex{light-cone expansion!of fermionic projector}%
\sindex{fermionic projector!light-cone expansion of}%
\sindex{fermionic projector!unregularized in position space}%
\sindex{fermionic projector!in the presence of an external potential}%
\nindex{bj6@$\tilde{P}^\lec$ -- non-causal low energy contribution to fermionic projector}%
\nindex{bj8@$\tilde{P}^\hec$ -- non-causal high energy contribution to fermionic projector}%
Here the $n$-summands describe the different orders of the singularities on the light cone,
whereas the $k$-sum describes all contributions to a given order on the light cone.
The phase-inserted nested line integrals involve~$\B$ and its partial derivatives,
possibly sandwiched between time-ordered exponentials of chiral potentials.
Since these nested line integrals are smooth functions in~$x$ and~$y$,
the series in~\eqref{s:fprep} is a light-cone expansion in the sense of Definition~\ref{s:l:def1},
provided that the $k$-sum is finite for every~$n$. This is indeed the case
if~$\B$ is composed of scalar, pseudoscalar and chiral potentials~\cite{light},
whereas for a more general perturbation operator~$\B$ this condition still needs to be verified.
This expansion is {\em{causal}} in the sense that it depends on~$\B$ and its
partial derivatives only along the line segment~$\overline{xy}$.
The contributions~$\tilde{P}^\lec$ and~$\tilde{P}^\hec$, on the other hand, are
not causal but depend instead on the global behavior of~$\B$ in space-time.
They can be written as a series of functions which are all smooth in~$x$ and~$y$.
Their different internal structure gives rise to the names {\em{non-causal
low energy contribution}} and {\em{non-causal high energy contribution}},
respectively.

For an introduction to the light-cone expansion and the required mathematical methods
we refer to~\cite{firstorder} and~\cite{light}, the exposition in~\cite[Section~2.5]{PFP}
or Chapter~\ref{tools} in this book.
The formulas of the light-cone expansion needed in this work are compiled in Appendix~\ref{s:appspec}.

\subsectionn{Clarifying Remarks} \label{s:discussion}
The above constructions require a few explanations. We first point out that, although we are
working with one-particle wave functions, the ansatz for the fermionic projector~\eqref{s:particles}
describes a many-particle quantum state. In order to get a connection to the Fock space
formalism, one can take the wedge product of the wave
functions~$\psi_k$ and~$\phi_l$ to obtain a vector in the fermionic Fock space
(for details see~\cite[Appendix~A]{PFP}).
\sindex{Fock space}%
We conclude that~\eqref{s:particles} describes {\em{second-quantized fermions}}.
For the description of entangled states see~\cite{entangle}.
\sindex{fermionic field!quantized}%
\sindex{quantum field!fermionic}%

One should keep in mind that at this stage, the form of the potential~$\B$ has not been specified;
it can be an arbitrary operator. Indeed, we regard the operator~$\B$
merely as a device for modifying or perturbing the fermionic projector. We do not want to preassume which of
these perturbations are physically relevant; instead, we want to select the relevant perturbations
purely on the basis of whether they are admissible for minimizers of our action
principle~\eqref{s:actprinciple}.
In other words, our action principle should decide how the physical interaction looks like, even
quantitatively in the sense that our action principle should determine the corresponding field equations.
Following this concept, we should
choose~$\B$ as general as possible, even allowing for potentials which are usually not considered
in physics. We now give a brief overview over the potentials which will be of relevance in the present
work. The most obvious choice is an electromagnetic potential\footnote{\label{s:units}
\sindex{units}%
For convenience we shall always omit the coupling constant~$e$
\nindex{cb0@$e$ -- coupling constant}%
 in the Dirac equation. Our convention
is obtained from the usual choice~$\B = e \slashed{A}$ by the transformation $A \rightarrow e^{-1} A$.
The coupling constant clearly reappears in the Maxwell equations, which we write in natural units 
and with the Heaviside-Lorentz convention as
$\partial_{jk} A^k - \Box A_k = e^2 \overline{\psi} \gamma_k \psi$. As usual, the fine structure
constant is given by~$\alpha = e^2/(4 \pi)$.},
\beq \label{s:EM}
\B = \slashed{A}\:.
\eeq
More generally, one can choose {\em{chiral potentials}},
\sindex{potential!chiral}%
which may be non-diagonal in the
generations,
\beq \label{s:chiral}
\B = \chi_L\: \slashed{A}_R + \chi_R\: \slashed{A}_L\:,
\eeq
where~$A_{L\!/\!R} = (A^i_{L\!/\!R})^\alpha_\beta$
\nindex{bh2@$A_L, A_R$ -- chiral potentials}%
with generation indices~$\alpha, \beta=1,\ldots, g$ and a vector index~$i=0,\ldots,3$
(here~$\chi_{L\!/\!R} = \frac{1}{2}(\1 \mp \pseudo)$ are the chiral projectors,
and~$\pseudo = i \gamma^0 \gamma^1 \gamma^2 \gamma^3$ is the usual pseudoscalar matrix).
\sindex{potential!chiral}%
\nindex{bh0@$\chi_{\LR}$ -- chiral projectors}%
\nindex{bh1@$\pseudo$ -- pseudoscalar matrix}%
\nindex{bh2@$A_L, A_R$ -- chiral potentials}%
To describe a {\em{gravitational field}},
\sindex{gravitational field}%
one needs to choose~$\B$ as a differential operator of
first order; more precisely,
\beq \label{s:grav}
\B = \Dir - i \Pdd\:,
\eeq
where~$\Dir$ is the Dirac operator in the presence of a gravitational field.

The above choices of~$\B$ are of course motivated by physical fields
observed in nature. However, we point out that we do not assume any field equations.
Thus the electromagnetic potential in~\eqref{s:EM} does not need to satisfy Maxwell's equations,
in~\eqref{s:chiral} we do not assume Yang-Mills-type equations for the chiral gauge fields, and in~\eqref{s:grav} the Einstein equations
are not imposed. This is because, as already pointed out above, our goal is to derive the classical field equations from our action principle~\eqref{s:actprinciple}.

Apart from the above choices of~$\B$ motivated from physics, one can also
choose other physically less obvious operators, like for example
{\em{scalar}} or {\em{pseudoscalar potentials}},
\sindex{potential!scalar}%
\sindex{potential!pseudoscalar}%
\beq \label{s:pseudoscalar}
\B = \Phi + i \pseudo \Xi
\eeq
\nindex{cb2@$\Phi$ -- scalar potential}%
\nindex{cb4@$\Xi$ -- pseudoscalar potential}%
with~$\Phi = \Phi^\alpha_\beta$, $\Xi = \Xi^\alpha_\beta$ and~$\alpha, \beta=1,\ldots,g$.
Furthermore, one can consider a {\em{scalar differential operator}},
\[ \B = i \Phi^j \partial_j \:, \]
or a higher order differential operator. More specifically, we will find a {\em{pseudoscalar
differential potential}} useful,
\sindex{potential!pseudoscalar differential}%
\[ \B = \pseudo \left( v^j \partial_j + \partial_j v^j \right) \:. \]
It is worth noting that one does not need to restrict attention to differential operators.
Indeed, $\B$ can also be an integral operator, in which case we talk of {\em{nonlocal potentials}}.
Clearly, one can also take linear combinations of all the above operators~$\B$.

Next, it is worth noting that for the moment, we consider~$\B$ as a-priori given,
and thus at this stage, our system consists of Dirac particles moving in an {\em{external field}}.
\sindex{potential!external}%
\sindex{external potential}%
However, our action principle~\eqref{s:actprinciple} will give relations between the potentials
contained in~$\B$ and the Dirac wave functions in~\eqref{s:particles}, and thus these
potentials will be influenced by the Dirac wave functions.
This leads to a mutual coupling of the potentials to the Dirac wave functions,
giving rise to a fully interacting system. We also point out that the potentials and fields contained
in~$\B$ should be regarded as {\em{classical}}.
\sindex{bosonic field!classical}%
\sindex{classical field!bosonic}%
Indeed, in this book we will always work
with classical bosonic fields. However, as is worked out in~\cite{entangle, qft, qftlimit}, 
the framework of the fermionic projector also allows for the description of second-quantized bosonic
fields.
\sindex{bosonic field!quantized}%
\sindex{quantum field!bosonic}%

\subsectionn{Relation to Other Approaches}
Having outlined our approach, we can now give a short review of related works.
In order to get a connection to our description of the Dirac sea in~\S\ref{s:sec42},
we begin with the construction of quantum fields in an external field.
\sindex{external field problem}%
\sindex{potential!external}%
\sindex{external potential}%
Historically, this problem was first analyzed in the {\em{Fock space formalism}}.
\sindex{Fock space}%
Klaus and Scharf~\cite{klaus+scharf1, klaus+scharf2} considered
the Fock representation of the electron-positron field in the presence of a static external field.
They noticed that the Hamiltonian needs to be regularized by subtracting suitable counter terms which
depend on the external field. They also noticed that the electron-positron field operators
in the external field form a Fock representation on the standard Fock space of free fields
only if the external field satisfies a certain regularity condition. This regularity condition
is quite restrictive and excludes many cases of physical interest (like a magnetic
field~\cite{nenciu+scharf} and a Coulomb potential~\cite{klaus}). In particular,
these results show that different external fields in general give rise to nonequivalent Fock representations
of the electron-positron field operators.
More recently, in~\cite{hainzl+sere1, hainzl+sere2} the vacuum state was constructed
for a system of Dirac particles with electrostatic interaction
in the Bogoliubov-Dirac-Fock approximation. The conclusion of this analysis is
that for mathematical consistency, one must take into account all the states forming the
Dirac sea. Furthermore, the interaction mixes the states in such a way that
it becomes impossible to distinguish between the particle states and the states of the Dirac sea.

In the time-dependent setting, Fierz and Scharf~\cite{fierz+scharf} proposed
that the Fock representation should be adapted to the external field as measured by a local observer.
Then the Fock representation becomes time and observer dependent.
This implies that the distinction between particles and anti-particles no longer has an invariant meaning,
but it depends on the choice of an observer. In this formulation, the usual particle interpretation of
quantum states only makes sense for the in- and outgoing scattering states, but it has no invariant meaning
for intermediate times.
For a related approach which allows for the construction of quantum fields in the
presence of an external magnetic field see~\cite{merkl}.
In all the above approaches, the Dirac sea leads to divergences, which must be treated by
an ultraviolet regularization and suitable counter terms.
\sindex{renormalization!by counter terms}%
\sindex{regularization}%

As an alternative to working with Fock spaces, one can use the 
so-called {\em{point splitting renormalization method}},
\sindex{renormalization!point splitting method}%
which is particularly useful for renormalizing the expectation value of the energy-momentum
tensor~\cite{christensen}. The idea is to replace a function of one variable~$T(x)$ by
a two-point distribution~$T(x,y)$, and to take the limit~$y \rightarrow x$ after subtracting
suitable singular distributions which take the role of counter terms.
Analyzing the singular structure of the counter terms leads to the so-called {\em{Hadamard condition}}
(see for example~\cite{fulling+sweeny+wald}).
\sindex{Hadamard condition}%
Reformulating the Hadamard condition for the two-point function as a
local spectral condition for the wave front set~\cite{radzikowski}
turns out to be very useful for the axiomatic formulation of free quantum field theory in curved space-time.
As in the Fock space formalism, in the point splitting approach the particle interpretation depends
on the observer. This is reflected mathematically by the fact that the Hadamard condition
specifies the two-point distribution only up to smooth contributions, thus leaving the
smooth particle wave functions undetermined.
For a good introduction to free quantum fields in curved space-time we refer to the
recent book~\cite{baer+fredenhagen}.

As mentioned at the beginning of~\S\ref{s:discussion},
in our approach the connection to the Fock space formalism is obtained by choosing a basis
of the image of the fermionic projector and taking the wedge product of the basis vectors
(for details see~\cite[Appendix~A]{PFP} or~\cite{entangle}). If in this construction the states
of the Dirac sea are taken into account, we get precisely the framework in~\cite{fierz+scharf}.
The connection to the Hadamard
condition is even closer. Indeed, considering the light-cone expansion locally for~$y$ near~$x$,
the summands in~\eqref{s:fprep} coincide precisely with the
singular distributions in the Hadamard construction. Since the
non-causal contributions~$\tilde{P}^\hec$ and~$\tilde{P}^\lec$ are smooth functions, 
we conclude that the integral kernel of the fermionic projector satisfies the Hadamard condition,
provided that the perturbation expansions of~$\tilde{P}^\hec$ and~$\tilde{P}^\lec$ converge
(a subtle technical problem which we do not want to enter here).
Thus in a given external field, $P^\sea(x,y)$ can be interpreted as
the two-point function, and using the methods of~\cite{radzikowski, baer+fredenhagen}
one could construct the corresponding free QFT.
This construction has been carried out in~\cite{hadamard} in the presence of
an external potential.

A major difference of our approach is that our framework allows for the description of
an {\em{interacting theory}}, where the coupling of the fermions to bosonic fields
and the back-reaction of the bosonic fields to the fermions is taken into account.
In this setting, the interaction is described by our action principle~\eqref{s:STdef}.
The mathematical framework is no longer equivalent to standard QFT.
In particular, {\em{$P(x,y)$ cannot be interpreted as the two-point function}} of a corresponding
QFT, simply because the notions of QFT can no longer be used.
But we still get a connection to the Feynman diagrams of QFT
(as will be explained in~\S\ref{s:secquantcorr} below).

Another major difference of our approach is that the distribution~$P^\sea$ as defined by the
causal perturbation expansion~\eqref{s:cpower} distinguishes a unique state which can
be interpreted as the fermionic vacuum state where all Dirac seas are
completely filled. Thus working relative to this distinguished state, there is a unique
{\em{observer independent particle interpretation}}, even at intermediate times
(see~\cite[Section~5]{sea} for a discussion of this point). At first sight, this
distinguished particle interpretation might seem of purely academic interest, because~$P^\sea$
is defined globally in space-time and is thus not accessible to a local observer.
However, our action principle~\eqref{s:STdef} does have access to quantities
defined globally in space-time, and in this way the distinguished particle interpretation
enters the physical equations. More precisely, $P^\sea$ drops out of the
Euler-Lagrange equations corresponding to our action principle, up to terms which
are well-defined and explicitly computable, even including a uniquely determined
smooth contribution.
In this way, the arbitrariness of working modulo smooth contributions (in the Hadamard condition)
or modulo regular counter terms (in the Fock space formalism) is removed.
The corresponding smooth contributions to the physical equations
will be analyzed in~\S\ref{s:secfield1} and Appendix~\ref{s:appresum}. They are
nonlocal and violate causality, as will be explained in~\S\ref{s:secnocausal}.

A frequently asked question is how our approach relates to Connes' {\em{noncommutative
geometry}}~\cite{connes}.
\sindex{noncommutative geometry}%
In particular, can our approach be thought of as a Lorentzian version of noncommutative geometry?
Clearly, both approaches have in common that the Dirac operator plays a central role.
Moreover, the light-cone expansion is the Lorentzian analog of local expansions of the resolvent or
the heat kernel near the diagonal. A major difference is that instead of considering the whole spectrum of
the Dirac operator, we only consider the eigenspaces corresponding to the masses~$m_\alpha$ of
the Dirac particles of our system. Furthermore, we only take ``half the eigenspaces'' by
constructing Dirac seas, and we also build in additional particle and anti-particle states~\eqref{s:particles}.
Another major difference concerns the mathematical structure of our action principle~\eqref{s:STdef}.
Namely, this action cannot be thought of as a spectral action, because it is impossible to express
it in terms of spectral properties of the Dirac operator. This is obvious from the fact that
in~\eqref{s:Ldef} and~\eqref{s:STdef} we perform a nonlinear (and even non-analytic) transformation
of the kernel~$P(x,y)$ before integrating over~$x$ and~$y$. As a consequence, there is no connection
to a regularized trace or Hilbert-Schmidt norm of~$P$. The specific form of our action principle
makes it possible to regard the structures of Minkowski space as emerging from a self-organization
of the wave functions in discrete space time (see~\cite{lrev}), an idea which has no correspondence in
noncommutative geometry. On the other hand, noncommutative geometry has deep connections
to Riemannian geometry, index theory and number theory. We conclude that despite superficial
similarities, the aims, ideas and methods of our approach are quite different from those in
noncommutative geometry.

\section{The Continuum Limit} \label{s:sec5}
\sindex{continuum limit}%
In Section~\ref{s:sec3} we described the vacuum by a family of regularized fermionic projectors~$P^\varepsilon$.
Our next goal is to use the information on the regularized vacuum to also regularize the fermionic
projector with interaction. We cannot expect that this information will suffice to determine the
interacting fermionic projector in all details, because it is unknown how the interaction affects the
fermionic projector on the microscopic scale. But as shown in~\cite[Chapter~4 and Appendix~D]{PFP},
there is a canonical method to regularize the formulas of the light-cone expansion~\eqref{s:fprep}.
This method also gives a meaning to composite expressions as needed for the analysis of
the action principle introduced in Section~\ref{s:sec2}. In particular, it allows us to analyze
the corresponding Euler-Lagrange equations in the continuum,
taking into account the unknown regularization details by a finite number of free parameters.
We now outline this method, relying for all technical issues on the detailed analysis in~\cite{PFP}.
The method in~\S\ref{s:secELC} is a major improvement and simplification of the techniques
in~\cite[Appendix~F]{PFP}. An introduction to the methods is given in Section~\ref{secreg}.

\subsectionn{Weak Evaluation on the Light Cone} \label{s:sec51}
Our method relies on the physically reasonable {\em{assumption of macroscopic potentials and
wave functions}} which states that both the bosonic potentials in~\eqref{s:diracPaux} and the
fermionic wave functions in~\eqref{s:particles} vary only on the macroscopic scale and
are thus almost constant on the regularization scale~$\varepsilon$. Then the
idea is to regularize the perturbation expansion~\eqref{s:cpower} in such a way that the
interaction modifies the fermionic projector also only on the macroscopic scale.
As exemplified in~\cite[Appendix~D]{PFP} in the perturbation expansion to first order,
this idea can be realized by demanding that the perturbation expansion should be gauge invariant
and should satisfy a causality condition. Performing the light-cone expansion of the thus regularized
perturbation expansion and using the form of the regularized vacuum minimizers as
specified in Assumption~\ref{s:assumption}, one obtains a simple regularization scheme for
the continuum fermionic projector~\eqref{s:fprep}, which we now describe.

The non-causal contributions~$\tilde{P}^\lec$ and~$\tilde{P}^\hec$, which are already smooth
in~$x$ and~$y$, are not regularized. Likewise, the smooth nested line-integrals are not regularized.
Thus we only regularize the singularities of the factors~$T^{(n)}$ on the light cone, and this
is done by the replacement rule
\sindex{light-cone expansion!regularization}%
\sindex{regularization!of the light-cone expansion}%
\beq \label{s:contri}
m^p \,T^{(n)} \rightarrow m^p \,T^{(n)}_{[p]}\:,
\eeq
\nindex{bl6@$T_{[p]}^{(n)}$ -- ultraviolet regularized $T^{(n)}$ }%
where the factors~$T^{(n)}_{[p]}(\xi)$ are smooth functions defined as Fourier integrals
involving the functions~$v^\varepsilon$, $\phi^\varepsilon$ and~$f^\varepsilon$
in the ansatz~\eqref{s:Peps}. If the sectorial projection is formed, we clarify the handling of the
generation index by accents, where~$\acute{\;\:}$ and~$\grave{\;\:}$
denote the summation over an upper and lower generation index, respectively
\sindex{sectorial projection}%
\nindex{cb6@$\hat{\;},\: \acute{\;} \ldots \grave{\;}$ -- short notation for sectorial projection}%
More precisely, we extend the replacement rule~\eqref{s:contri} to (see also~\cite[Section~7.1]{PFP})
\beq \label{s:tildedef}
\sum_{\alpha,\beta, \gamma_1, \ldots, \gamma_{p-1}=1}^g \!\!\!\!\! m^p \:
\underbrace{Y^\alpha_{\gamma_1} \cdots Y^{\gamma_1}_{\gamma_2}
\cdots Y^{\gamma_{p-1}}_\beta}_{\text{$p$ factors~$Y$}}
\,T^{(n)} \rightarrow m^p \:\acute{Y} Y \cdots \grave{Y} \,T^{(n)}_{[p]}\:,
\eeq
and use the notation\footnote{ \label{s:footg}
In contrast to the convention in~\cite{PFP}, here we always write out the 
factors~$g$ which count the number of generations (in~\cite{PFP}, the factor~$g$
\nindex{ca2@$g$ -- number of generations}%
was absorbed into the
factors~$T^{(n)}_{[0]}$ and~$\overline{T^{(n)}_{[0]}}$). The shorter notation in~\cite{PFP} has
the disadvantage that reinserting the factors of~$g$ in the end is a potential source of confusion and 
may lead to computational errors. In the convention here, the factors~$T^{(n)}_\circ$ without regularization
always coincide with the distributions~\eqref{s:Tndef} and~\eqref{s:Tm1def}.}
\[ \sum_{\alpha,\beta=1}^g m^0 \:\delta^\alpha_\beta
\,T^{(n)} \rightarrow m^0\,g\,T^{(n)}_{[0]} \qquad \text{and} \qquad
\sum_{\alpha,\beta=1}^g m \:Y^\alpha_\beta
\,T^{(n)} \rightarrow m\, \hat{Y} \,T^{(n)}_{[1]}\:. \]
Fortunately, the rather complicated detailed form of the factors~$T^{(n)}_{[p]}$
will not be needed here, because these functions can always be treated symbolically using the
following simple calculation rules. In computations one may treat the~$T^{(n)}_{[p]}$
like complex functions. However, one must be careful when tensor indices of factors~$\slashed{\xi}$
are contracted with each other. Naively, this gives a factor~$\xi^2$ which vanishes on the
light cone and thus changes the singular behavior on the light cone. In order to describe this
effect correctly, we first write every summand of the light cone expansion~\eqref{s:fprep}
such that it involves at most one factor~$\slashed{\xi}$ (this can always be arranged using
the anti-commutation relations of the Dirac matrices).
We now associate every factor~$\slashed{\xi}$ to the corresponding factor~$T^{(n)}_{[p]}$.
In simple calculations, this can be indicated by putting brackets around the two factors,
whereas in the general situation we add an index to the factor~$\slashed{\xi}$, giving rise to the replacement rule
\[ m^p \,\slashed{\xi} \,T^{(n)} \rightarrow m^p \,\slashed{\xi}^{(n)}_{[p]} \, T^{(n)}_{[p]} \:. \]
\nindex{bm2@$\xi^{(n)}_{[p]}$ -- ultraviolet regularized factor $\xi$}%
The factors~$\slashed{\xi}$ which are contracted to other factors~$\slashed{\xi}$ are called {\em{inner factors}}.
\sindex{inner factor~$\xi$}%
The contractions of inner factors can be handled with the so-called {\em{contraction rules}}
\sindex{contraction rule}%
\begin{align}
(\xi^{(n)}_{[p]})^j \, (\xi^{(n')}_{[p']})_j &=
\frac{1}{2} \left( z^{(n)}_{[p]} + z^{(n')}_{[p']} \right) \label{s:eq52} \\
(\xi^{(n)}_{[p]})^j \, \overline{(\xi^{(n')}_{[p']})_j} &=
\frac{1}{2} \left( z^{(n)}_{[p]} + \overline{z^{(n')}_{[p']}} \right) \label{s:eq53} \\
z^{(n)}_{[p]} \,T^{(n)}_{[p]} &= -4 \left( n \:T^{(n+1)}_{[p]}
+ T^{(n+2)}_{\{p \}} \right) , \label{s:eq54}
\end{align}
which are to be complemented by the complex conjugates of these equations.
Here the factors~$z^{(n)}_{[p]}$
\nindex{bm8@$z^{(n)}_{[p]}$ -- abbreviation for $(\xi^{(n)}_{[p]})^2$}%
 can be regarded simply as a book-keeping device
to ensure the correct application of the rule~\eqref{s:eq54}.
The factors~$T^{(n)}_{\{p\}}$
\nindex{bl8@$T^{(n)}_{\{p\}}$ -- factor in continuum limit describing the shear of surface states}%
have the same scaling behavior as the~$T^{(n)}_{[p]}$,
but their detailed form is somewhat different; we simply treat them as a new class of symbols\footnote{We
remark that the functions~$T^{(n)}_{\{p\}}$ will be of no relevance in this chapter, because
they contribute to the EL equations only to degree three and lower; see~\S\ref{s:sec810}.}.
In cases where the lower index does not need to be specified we write~$T^{(n)}_\circ$.
\nindex{bn2@$T^{(n)}_\circ$ -- stands for~$T^{(n)}_{\{p\}}$ or~$T^{(n)}_{[p]}$}%
After applying the contraction rules, all inner factors~$\xi$ have disappeared.
The remaining so-called {\em{outer factors}}~$\xi$
\sindex{outer factor~$\xi$}%
need no special attention and are treated like smooth functions.

Next, to any factor~$T^{(n)}_\circ$ we associate the {\em{degree}} $\deg T^{(n)}_\circ$ by
\sindex{degree on the light cone}%
\nindex{bn4@$\deg$ -- degree on light cone}%
\[ \deg T^{(n)}_\circ = 1-n \:. \]
The degree is additive in products, whereas the degree of a quotient is defined as the
difference of the degrees of numerator and denominator. The degree of an expression
can be thought of as describing the order of its singularity on the light cone, in the sense that a
larger degree corresponds to a stronger singularity (for example, the
contraction rule~\eqref{s:eq54} increments~$n$ and thus decrements the degree, in
agreement with the naive observation that the function~$z=\xi^2$ vanishes on the light cone).
Using formal Taylor expansions, we can expand in the degree. In all our applications, this will
give rise to terms of the form
\beq \label{s:sfr}
\eta(x,y) \:
\frac{ T^{(a_1)}_\circ \cdots T^{(a_\alpha)}_\circ \:
\overline{T^{(b_1)}_\circ \cdots T^{(b_\beta)}_\circ} }
{ T^{(c_1)}_\circ \cdots T^{(c_\gamma)}_\circ \:
\overline{T^{(d_1)}_\circ \cdots T^{(d_\delta)}_\circ} } \qquad \text{with~$\eta(x,y)$ smooth}\:.
\eeq
Here the quotient of the two monomials is referred to as a {\em{simple fraction}}.
\sindex{simple fraction}%

A simple fraction can be given a quantitative meaning by considering one-dimensional integrals
along curves which cross the light cone transversely away from the origin~$\xi=0$.
This procedure is called {\em{weak evaluation on the light cone}}.
\sindex{evaluation on the light cone!weak}%
For our purpose, it suffices to integrate over the time coordinate~$t=\xi^0$ for fixed~$\vec{\xi} \neq 0$.
Moreover, using the symmetry under reflections~$\xi \rightarrow -\xi$, it suffices to consider the upper
light cone~$t \approx |\vec{\xi}|$. The resulting integrals will diverge if the regularization
is removed. The leading contribution for small~$\varepsilon$ can be written as
\beq
\int_{|\vec{\xi}|-\varepsilon}^{|\vec{\xi}|+\varepsilon} dt \; \eta(t,\vec{\xi}) \:
\frac{ T^{(a_1)}_\circ \cdots T^{(a_\alpha)}_\circ \:
\overline{T^{(b_1)}_\circ \cdots T^{(b_\beta)}_\circ} }
{ T^{(c_1)}_\circ \cdots T^{(c_\gamma)}_\circ \:
\overline{T^{(d_1)}_\circ \cdots T^{(d_\delta)}_\circ} }
\;\approx\; \eta(|\vec{\xi}|,\vec{\xi}) \:\frac{c_{\text{reg}}}{(i |\vec{\xi}|)^L}
\;\frac{\log^r (\varepsilon |\vec{\xi}|)}{\varepsilon^{L-1}}\:, \label{s:asy}
\eeq
where~$L$ is the degree and~$c_{\text{reg}}$, the so-called {\em{regularization parameter}},
\nindex{bn6@$L$ -- degree of simple fraction}%
\sindex{regularization parameter}%
\nindex{bn8@$c_{\text{reg}}$ -- regularization parameter}%
is a real-valued function of the spatial direction~$\vec{\xi}/|\vec{\xi}|$ which also depends on
the simple fraction and on the regularization details
(the error of the approximation will be specified below). The integer~$r$ describes
a possible logarithmic divergence; we postpone its
discussion until when we need it (see~\S\ref{s:sec83}). Apart from this logarithmic divergence, the
scalings in both~$\xi$ and~$\varepsilon$ are described by the degree.

When analyzing a sum of expressions of the form~\eqref{s:sfr}, one must
know if the corresponding regularization parameters are related to each other.
In this respect, the {\em{integration-by-parts rules}}
\sindex{integration-by-parts rule}%
are important, which are described
symbolically as follows. On the factors~$T^{(n)}_\circ$ we introduce a derivation~$\nabla$ by
\[ \nabla T^{(n)}_\circ = T^{(n-1)}_\circ \:. \]
\nindex{bl4@$\nabla$ -- derivation on the light cone}%
Extending this derivation with the Leibniz and quotient rules to simple fractions, the
integration-by-parts rules states that
\beq \label{s:ipart}
\nabla \left( \frac{ T^{(a_1)}_\circ \cdots T^{(a_\alpha)}_\circ \:
\overline{T^{(b_1)}_\circ \cdots T^{(b_\beta)}_\circ} }
{ T^{(c_1)}_\circ \cdots T^{(c_\gamma)}_\circ \:
\overline{T^{(d_1)}_\circ \cdots T^{(d_\delta)}_\circ} }
\right) = 0 \:.
\eeq
These rules give relations between simple fractions (the name is motivated by the fact that
when evaluating~\eqref{s:ipart} weakly on the light cone~\eqref{s:asy}, the rules state that the integral
over a derivative vanishes). Simple fractions which are not related
to each by the integration-by-parts rules are called {\em{basic fractions}}. As shown
in~\cite[Appendix~E]{PFP}, there are no further relations between the basic fractions.
Thus the corresponding {\em{basic regularization parameters}} are linearly independent.
\sindex{regularization parameter!basic}%

We next specify the error of the above expansions. By not regularizing the bosonic potentials
and fermionic wave functions, we clearly miss the
\beq \label{s:ap1}
\text{higher orders in~$\varepsilon/\ell_\text{macro}$}\:.
\eeq
\sindex{error term!higher order in $\varepsilon/\ell_\text{macro}$}%
Furthermore, in~\eqref{s:asy} we must stay away from the origin, meaning that we neglect the
\beq \label{s:ap2}
\text{higher orders in~$\varepsilon/|\vec{\xi}|$}\:.
\eeq
\sindex{error term!higher order in $\varepsilon / \vert \vec{\xi} \vert$}%
The higher oder corrections in~$\varepsilon/|\vec{\xi}|$ depend on the fine structure of
the regularization and thus seem unknown for principal reasons.
Neglecting the terms in~\eqref{s:ap1} and~\eqref{s:ap2} also justifies the
formal Taylor expansion in the degree. Neglecting the terms~\eqref{s:ap2} clearly makes it
necessary to choose~$|\vec{\xi}| \gg \varepsilon$.
Finally, we disregard the higher order
corrections in the parameter~$\varepsilon_\text{shear}$ in~\eqref{s:shear}.
\nindex{ca6@$\varepsilon_\text{shear}$ -- shear parameter}%

The above symbolic computation rules give a convenient procedure to evaluate composite expressions
in the fermionic projector, referred to as the {\em{analysis in the continuum limit}}:
\sindex{continuum limit!analysis in the}%
After applying the contraction rules and expanding in the degree, we obtain equations involving
a finite number of terms of the form~\eqref{s:sfr}. By applying the integration-by-parts rules,
we can arrange that all simple fractions are basic fractions.
We evaluate weakly on the light cone~\eqref{s:asy} and collect the terms according to their
scaling in~$\xi$. Taking for every given scaling in~$\xi$ only the leading pole in~$\varepsilon$,
we obtain equations which involve linear combinations of smooth functions and basic regularization parameters. We consider the basic regularization parameters as empirical parameters describing the
unknown microscopic structure of space-time. We thus end up with equations involving
smooth functions and a finite number of free parameters. We point out that these free parameters
cannot be chosen arbitrarily because they might be constrained by inequalities
(see the discussion after~\cite[Theorem~E.1]{PFP}). Also, the values of the basic regularization
parameters should ultimately be justified by an analysis of vacuum minimizers of our
variational principle (as discussed at the end of Section~\ref{s:sec3}).

In view of the later considerations in~\S\ref{s:secfield1}, we point out that the above calculation rules
are valid only {\em{modulo smooth contributions}} to the fermionic projector.
\sindex{error term!smooth contribution}%
This can be understood
from the fact that these rules only deal with the terms of the series in~\eqref{s:fprep}, but they do not take into
account the smooth non-causal high and low energy contributions. But the above calculation rules
affect these smooth contributions as well.
To give a simple example, we consider the distribution~$T^{(0)}$, which according
to~\eqref{s:Taser}--\eqref{s:Tndef} is given by
\[ T^{(0)} = -\frac{1}{8 \pi^3} \: \left( \frac{\text{PP}}{\xi^2} \:+\: i \pi \delta (\xi^2) \:
\varepsilon (\xi^0) \right) . \]
Multiplying by~$z=\xi^2$ in the distributional sense gives a constant
\beq \label{s:multdist}
z T^{(0)} = -\frac{1}{8 \pi^3}\: .
\eeq
On the other hand, the contraction rule~\eqref{s:eq54} yields
\beq \label{s:multweak}
z^{(0)}_{[0]} T^{(0)}_{[p]} = - 4 T^{(2)}_{\{p\}} \:.
\eeq
The last relation gives much finer information than the distributional equation~\eqref{s:multdist},
which is essential when we want to evaluate composite expressions weakly on the light cone~\eqref{s:asy}.
However, the constant term in~\eqref{s:multdist} does not appear in~\eqref{s:multweak}. The way
to think about this shortcoming is that this
constant term is smooth and can thus be taken into account by modifying the corresponding
low energy contribution~$\tilde{P}^\lec(x,y)$ in~\eqref{s:fprep}.
Indeed, this situation is not as complicated as it might seem at first sight.
Namely, the smooth contributions to the fermionic projector need special attention
anyway and must be computed using the resummation technique explained in
Appendix~\ref{s:appresum}. When performing this resummation, we can in one step also
compute all the smooth contributions which were not taken into account by the formalism of
the continuum limit. Thus altogether we have a convenient method where we first concentrate
on the singularities on the light cone, whereas the neglected smooth contributions
will be supplemented later when performing the resummation.

We note that the above procedure needs to be modified for the description of {\em{gravity}},
\sindex{gravitational field}%
because in this case the gravitational constant makes it necessary to have
relations between terms involving different powers of a fundamental length scale.
These generalizations are worked out in Chapter~\ref{lepton}.

\subsectionn{The Euler-Lagrange Equations in the Continuum Limit} \label{s:secELC}
We now return to the action principle of Section~\ref{s:sec2}. Our goal is to bring the
conditions for a minimizer~\eqref{s:intdelT} and~\eqref{s:intdelL} into a form suitable for the
analysis in the continuum limit. We begin by considering a smooth family~$P(\tau)$ of fermionic
projectors and compute the corresponding first variation of the action. We differentiate~\eqref{s:intdelL}
with respect to~$\tau$, treating the constraint~\eqref{s:intdelT} with a Lagrange multiplier
(for the mathematical justification of this procedure see the related paper~\cite{lagrange}).
For convenience, we introduce the functional
\beq \label{s:Smudef}
\Sact_\mu[P] \;\stackrel{\text{formally}}{=}\; \iint_{\scrM \times \scrM}\L_\mu[A_{xy}]\:d^4x\: d^4y
\qquad \text{with} \qquad
\L_\mu[A] = |A^2| - \mu |A|^2 \:.
\eeq
\nindex{cc2@$\Sact_\mu$ -- causal action involving one Lagrange multiplier}%
\nindex{cc4@$\L_\mu$ -- Lagrangian involving one Lagrange multiplier}%
Choosing~$\mu=\frac{1}{4}$ gives precisely the action~\eqref{s:STdef}, whereas
by allowing a general~$\mu \in \R$ we take into account the Lagrange multiplier.
We thus obtain the condition
\beq \label{s:EL1}
0 = \delta \Sact_\mu[P]
= \iint_{\scrM\times \scrM} \re \Tr \Big\{ \nabla \L_\mu[A_{xy}]
\: \delta P(x,y) \Big\}  \:d^4x \: d^4y\:,
\eeq
where~$\delta P :=  P'(0)$. Here we consider~$P(y,x)$ via
\[ P(y,x) = P(x,y)^* \equiv \gamma^0 P(x,y)^\dagger \gamma^0 \]
as a function of~$P(x,y)$, and~$\nabla$ denotes the gradient where the real and imaginary
parts of~$P(x,y)$ are treated as independent variables, i.e.
\beq \label{s:grad}
(\nabla f)^\alpha_\beta := \frac{\partial f}{\partial \re P(x,y)^\beta_\alpha} - i
\frac{\partial f}{\partial \im P(x,y)^\beta_\alpha} \:,
\eeq
and~$\alpha,\beta=1,\ldots, 4$ are spinor indices.
Introducing the integral operator~$R$ with kernel
\beq \label{s:Rdef}
R(y,x) := \nabla \L_\mu[A_{xy}] \:,
\eeq
we can write~\eqref{s:EL1} as a trace of an operator product, 
\[ \delta \Sact_\mu[P] = \re \tr \big( R\; \delta P \big) \:. \]
In order to get rid of the real part, it is convenient to replace~$R$ by its
symmetric part. More precisely, introducing the symmetric operator~$Q$ with kernel
\beq \label{s:Qdef}
Q(x,y) = \frac{1}{4} \Big( R(x,y) + R(y,x)^* \Big) \:,
\eeq
\nindex{aq6@$Q(x,y)$ -- first variation of the Lagrangian}%
we can write the variation as
\beq \label{s:QdP}
\delta \Sact_\mu[P] = 2 \tr \big( Q\; \delta P \big) \:.
\eeq

As explained before Definition~\ref{s:def21}, we want to vary the fermionic projector by
unitary transformations in a compact region. Thus the family of fermionic projectors~$P(\tau)$
should be of the form
\beq \label{s:UPU}
P(\tau) = U^{-1}(\tau) \,P\, U(\tau)
\eeq
with a smooth family~$U(\tau)$ of
unitary transformations in a fixed compact region~$K$ (see Definition~\ref{s:def21}) with~$U(0)=\1$.
Then the operator~$B=-i U'(0)$ has the integral representation
\[ (B \psi)(x) = \int_\scrM B(x,y)\, \psi(y)\, d^4y \]
with a smooth compactly supported integral kernel $B \in C^\infty_0(K \times K, \C^{4 \times 4})$.
Differentiating~\eqref{s:UPU} yields that~$\delta P = i[P,B]$, and substituting this identity
into~\eqref{s:QdP} and cyclically commuting the operators inside the trace, we
can rewrite the condition~\eqref{s:EL1} as
\[ 0 = \tr \big( [P,Q] B \big) \:. \]
Since~$B$ is arbitrary, we obtain the Euler-Lagrange (EL) equations
\beq \label{s:ELeqns}
\boxed{ \quad [P,Q] = 0 \:, \quad }
\eeq
\sindex{Euler-Lagrange equations}%
\sindex{causal action!Euler-Lagrange equations of}%
stating that two operators in space-time should commute. For more details on the derivation
of the EL equations we refer to~\cite[Section~3.5]{PFP} and \S\ref{secvary}.

When analyzing the commutator~\eqref{s:ELeqns} in the continuum limit, the
kernel~$Q(x,y)$ can be evaluated weakly using the formula~\eqref{s:asy}. The subtle point is that,
according to~\eqref{s:ap2}, this weak evaluation formula only applies if~$x$ and~$y$
stay apart. But writing the commutator in~\eqref{s:ELeqns} with integral kernels,
\beq \label{s:comm}
[P,Q](x,y) = \int_\scrM \Big( P(x,z)\, Q(z,y) - Q(x,z)\, P(z,y) \Big) \,d^4z\:,
\eeq
we also integrate over the regions~$z \approx y$ and~$z \approx x$ where the kernels~$Q(z,y)$
and~$Q(x,z)$ are ill-defined. There are several methods to resolve this difficulty,
which all give the same end result. The cleanest method is the method of {\em{testing
on null lines}}.
\sindex{testing on null lines}%
We now explain the ideas and results of this last method, referring for the rigorous
derivation to Appendix~\ref{s:appnull} (for other methods of testing see~\cite[Appendix~F]{PFP}).
The idea is to take the expectation value of the commutator in~\eqref{s:comm}
for two wave functions~$\psi_1$ and~$\psi_2$, one being in the kernel and one in the image of the operator~$P$. Thus
\beq \label{s:P12}
P \psi_1 = 0 \qquad \text{and} \qquad \psi_2 = P \phi
\eeq
for a suitable wave function~$\phi$. Then, using the symmetry of~$P$
with respect to the indefinite inner product~\eqref{s:iprod}, we find
\beq \label{s:PQeval}
\bra \psi_1 \,|\, [P,Q] \,\phi \ket = \bra P \psi_1 \,|\, Q \phi \ket - \bra \psi_1 \,|\, Q P \phi \ket 
= - \bra \psi_1 \,|\, Q \psi_2 \ket \:.
\eeq
Now the commutator has disappeared, and the EL equations~\eqref{s:ELeqns} give rise to the condition
\beq \label{s:Q12}
0 = \bra \psi_1 | Q \,\psi_2 \ket = \iint_{\scrM \times \scrM} Q(x,y)\: \psi_1(x)\: \psi_2(y)\: d^4x\, d^4y  \:.
\eeq
The hope is that by choosing suitable wave functions~$\psi_1$ and~$\psi_2$
of the form~\eqref{s:P12} having disjoint supports, we can evaluate the expectation value~\eqref{s:Q12}
weakly on the light cone~\eqref{s:asy}, thus making sense of the EL equations in the continuum
limit.

The key question is to what extent the constraints~\eqref{s:P12} restrict the freedom in
choosing the wave functions~$\psi_1$ and~$\psi_2$. For clarity, we here explain the situation
in the simplified situation where~$P$ is composed of one free Dirac sea of mass~$m$,
\beq \label{s:onesea}
P(x,y) = \int \frac{d^4k}{(2 \pi)^4}\: (\slashed{k} +m)\:
\delta(k^2-m^2)\: \Theta(-k^0)\: e^{-ik(x-y)}
\eeq
(but~$Q$ can be a general operator for which the methods of Section~\ref{s:sec5} apply).
The generalization to several generations and a~$P$ with general interaction
is worked out in Appendix~\ref{s:appnull}.
In order to extract information  from~\eqref{s:Q12}
and~\eqref{s:asy}, it is desirable that the wave functions~$\psi_1$ and~$\psi_2$ are as much as
possible localized in space-time. For the wave function~$\psi_1$, this requirement is easy to fulfill by
removing a strip of width~$\Delta \omega$ around the lower mass shell in momentum space.
For example, we can construct a wave function supported near the origin
by choosing for a given parameter~$\delta>0$ a smooth function~$\eta$ supported in the ball of
radius~$\delta$ in Euclidean~$\R^4$ and setting
\beq \label{s:psi1}
\psi_1(x) = \int \frac{d^4k}{(2 \pi)^4}\: \hat{\eta}(k) 
\:\chi_{\R \setminus [-\Delta \omega, \Delta \omega]} \Big(k^0+\sqrt{|\vec{k}|^2+m^2} \Big)\:
e^{-ikx} \:,
\eeq
where~$\hat{\eta}$ is the Fourier transform of~$\eta$, and~$\chi_I$ is the characteristic function
defined by~$\chi_I(x)=1$ if~$x \in I$ and~$\chi_I(x)=0$ otherwise.
In the limit~$\Delta \omega \searrow 0$, the characteristic function in~\eqref{s:psi1}
becomes the identity, so that~$\psi_1$ goes over to~$\eta$. Moreover, for any~$\Delta \omega >0$, the function~$\psi_1$
is indeed in the kernel of the operator~$P$, because it vanishes on the lower mass shell.
Thus by choosing~$\Delta \omega$ sufficiently small, we can arrange that~$\psi_1$ is arbitrarily close
to~$\eta$ and satisfies the condition in~\eqref{s:P12} (indeed, in finite space-time volume one cannot
choose~$\Delta \omega$ arbitrarily small, leading to small corrections which will be specified in
Appendix~\ref{s:appnull}; see Remark~\ref{s:remlife}).

The construction of~$\psi_2$ is a bit more difficult because~$\psi_2$ must lie in the image of~$P$,
and thus it must be a negative-energy solution of the Dirac equation $(i \Pdd -m) \psi_2=0$.
Due to current conservation, it is obviously not possible to choose~$\psi_2$ to be localized in
space-time; the best we can do is to localize in space by considering a wave packet.
According to the Heisenberg Uncertainty Principle, localization in a small spatial region requires
large momenta, and thus we are led to considering an {\em{ultrarelativistic wave packet}}
\sindex{ultrarelativistic wave packet}%
of negative energy moving along a null line~$\mathfrak{L}$, which does not intersect the
ball~$B_\delta(0) \subset \R^4$ where~$\psi_1$ is localized.
By a suitable rotation and/or a Lorentz boost of our reference frame $(t, \vec{x})$, we can arrange that
\[ \mathfrak{L} = \{ (\tau, -\tau+\ell, 0, 0) \text{ with } \tau \in \R \} \]
with~$\ell>0$. For~$\psi_2$ we take the ansatz
\beq \label{s:psi2}
\psi_2 = (i \Pdd + m) \left( e^{-i \Omega (t+x)} \:\phi(t+x-\ell,y,z) \right) + \text{(small corrections)}\:,
\eeq
where the smooth function~$\phi$ is supported in~$B_\delta(\vec{0}) \subset \R^3$,
and the frequency~$\Omega<0$ as well as the length scales~$\delta$ and~$\ell$ are chosen in the range
\beq \label{s:scales}
\varepsilon \ll |\Omega|^{-1} \ll \delta \ll \ell, \ell_\text{macro}, m^{-1} \:.
\eeq
The small corrections in~\eqref{s:psi2} are due to the non-zero rest mass, the dispersion and the condition
that~$\psi_2$ must have no contribution of positive energy (for details see Appendix~\ref{s:appnull}).

\begin{figure}
\begin{picture}(0,0)%
\includegraphics{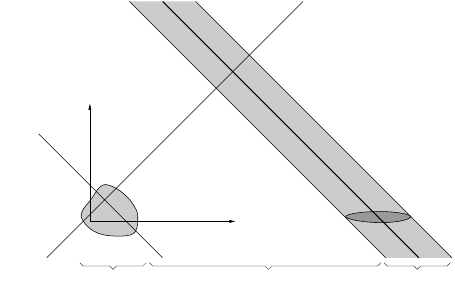}%
\end{picture}%
\setlength{\unitlength}{1160sp}%
\begingroup\makeatletter\ifx\SetFigFontNFSS\undefined%
\gdef\SetFigFontNFSS#1#2#3#4#5{%
  \reset@font\fontsize{#1}{#2pt}%
  \fontfamily{#3}\fontseries{#4}\fontshape{#5}%
  \selectfont}%
\fi\endgroup%
\begin{picture}(12322,8100)(-5324,-1378)
\put(2926,3959){\makebox(0,0)[lb]{\smash{{\SetFigFontNFSS{11}{13.2}{\rmdefault}{\mddefault}{\updefault}$\text{supp}\, \psi_2$}}}}
\put(2656,5939){\makebox(0,0)[lb]{\smash{{\SetFigFontNFSS{11}{13.2}{\rmdefault}{\mddefault}{\updefault}$Q(x,y)$}}}}
\put(6046,839){\makebox(0,0)[lb]{\smash{{\SetFigFontNFSS{11}{13.2}{\rmdefault}{\mddefault}{\updefault}$\text{supp}\, \phi(\vec{x})$}}}}
\put(-2399,-1171){\makebox(0,0)[lb]{\smash{{\SetFigFontNFSS{11}{13.2}{\rmdefault}{\mddefault}{\updefault}$\delta$}}}}
\put(1831,-1186){\makebox(0,0)[lb]{\smash{{\SetFigFontNFSS{11}{13.2}{\rmdefault}{\mddefault}{\updefault}$\ell$}}}}
\put(5881,-1201){\makebox(0,0)[lb]{\smash{{\SetFigFontNFSS{11}{13.2}{\rmdefault}{\mddefault}{\updefault}$\delta$}}}}
\put(1486,4664){\makebox(0,0)[b]{\smash{{\SetFigFontNFSS{11}{13.2}{\rmdefault}{\mddefault}{\updefault}$y$}}}}
\put(3466,2399){\makebox(0,0)[lb]{\smash{{\SetFigFontNFSS{11}{13.2}{\rmdefault}{\mddefault}{\updefault}$\mathfrak{L}$}}}}
\put(-2564,3629){\makebox(0,0)[b]{\smash{{\SetFigFontNFSS{11}{13.2}{\rmdefault}{\mddefault}{\updefault}$t$}}}}
\put(811, 74){\makebox(0,0)[b]{\smash{{\SetFigFontNFSS{11}{13.2}{\rmdefault}{\mddefault}{\updefault}$\vec{x}$}}}}
\put(-5309,974){\makebox(0,0)[lb]{\smash{{\SetFigFontNFSS{11}{13.2}{\rmdefault}{\mddefault}{\updefault}$\text{supp}\, \psi_1$}}}}
\put(-2069,1109){\makebox(0,0)[b]{\smash{{\SetFigFontNFSS{11}{13.2}{\rmdefault}{\mddefault}{\updefault}$x$}}}}
\end{picture}%
\caption{Intersection of the null line~$\mathfrak{L}$ with the singular set of~$Q(x,y)$}
\label{s:fig1}
\end{figure}
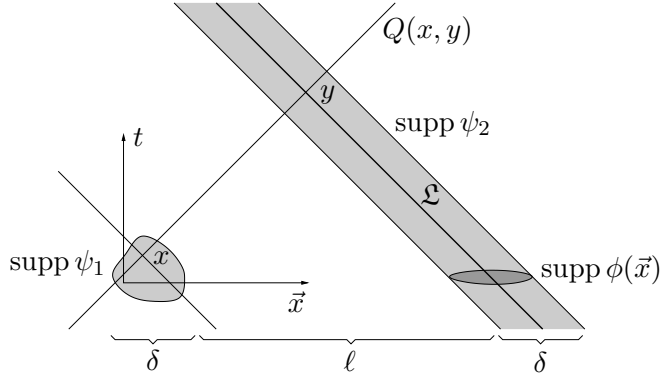

Except for the small corrections to be specified in Appendix~\ref{s:appnull},
the support of the wave function~$\psi_1$  in~\eqref{s:psi1} lies in~$B_\delta(0)$,
and thus it is disjoint from the support~$B_\delta(\mathfrak{L})$ of the wave
function~$\psi_2$ in~\eqref{s:psi2}.
Hence the integrals in~\eqref{s:Q12} only involve the region $x \neq y$
where~$Q(x,y)$ is well-defined in the continuum limit.
Furthermore, the null line~$\mathfrak{L}$ intersects the null cone around~$x$ in precisely one
point~$y$ for which~$|\xi^0|=|\vec{\xi}| \sim \ell$ (see Figure~\ref{s:fig1}).
Since this intersection is transverse, we can evaluate the expectation value~\eqref{s:Q12}
with the help of~\eqref{s:asy}. In view of the freedom in choosing
the parameter~$\ell$ and the direction of~$\mathfrak{L}$, we conclude that~\eqref{s:asy} itself must vanish,
\beq \label{s:ELcl}
\boxed{ \quad Q(x,y) = 0 \quad \text{if evaluated weakly on the light cone}\:. \quad }
\eeq
\sindex{Euler-Lagrange equations}%
\sindex{causal action!Euler-Lagrange equations of}%

The above consideration is made rigorous in Appendix~\ref{s:appnull}. More precisely,
in Proposition~\ref{s:prpnull}, the above arguments are extended to the setting involving
several generations and a general interaction, and the scaling of the correction terms in~\eqref{s:psi2}
is specified to every order in perturbation theory. This proposition applies
to our action principle~\eqref{s:actprinciple} and all interactions to
be considered here, thus justifying~\eqref{s:ELcl} in all cases of interest in this book.
Moreover, in Remark~\ref{s:remlife}
we consider the corrections to~\eqref{s:psi1} which arise if the lifetime of the
universe is finite. Using that this lifetime can be estimated by the
time from the big bang as known from experiments,
we show that the correction to~\eqref{s:psi1} can indeed be neglected for our universe.

To summarize, we saw that within the formalism of the continuum limit, the commutator
in~\eqref{s:ELeqns} vanishes only if~$Q(x,y)$ itself is zero. This result is the strongest condition
we could hope for, because in view of~\eqref{s:QdP} it implies that arbitrary first variations of the
action vanish, even  if we disregard the constraint that~$P$ must be a projector.
We refer to~\eqref{s:ELcl} as the {\em{Euler-Lagrange equations in the continuum limit}}.
\sindex{Euler-Lagrange equations!in the continuum limit}%

We finally remark that by replacing the null lines by null geodesics, the above method
could immediately be generalized to situations involving a gravitational field. However, the
estimates of Appendix~\ref{s:appnull} would become more demanding.

\section{The Euler-Lagrange Equations to Degree Five} \label{s:sec7}
We proceed with the analysis of the EL equations in the continuum limit~\eqref{s:ELcl}
using the methods outlined in Sections~\ref{s:sec4} and~\ref{s:sec5}.
For clarity, we begin in the vacuum and then introduce more and more interaction terms.
Furthermore, we consider the contributions to the EL equations to decreasing degree on the
light cone. In this section, we consider the most singular contributions of degree five.
The contributions of degree four will be analyzed in Section~\ref{s:sec8}, whereas
the contributions to even lower degree are discussed in Section~\ref{s:secthree}.

We point out that many results of this section were already obtained in~\cite[Chapters~5 and~6]{PFP}
for more general systems, which however involve only one generation.
In order to lay consistent foundations for the new calculations of Sections~\ref{s:sec8}--\ref{s:secthree},
we here present all calculations in a self-contained way.

\subsectionn{The Vacuum} \label{s:sec71}
In order to perform the light-cone expansion of the fermionic projector of the vacuum, we first
pull the Dirac matrices out of the Fourier integral~\eqref{s:Pauxvac} and use~\eqref{s:Tadef}
to obtain
\beq \label{s:6.0}
P^\text{aux}(x,y) = \bigoplus_{\beta=1}^g \big(i \Pdd_x + m_\beta \big) \,T_a(x,y) \Big|_{a=m_\beta^2}\:.
\eeq
\nindex{ca8@$P^\text{aux}$ -- auxiliary fermionic projector}%
After removing the logarithmic mass terms by the replacement $T_a \rightarrow T_a^\reg$,
the light-cone expansion reduces to a Taylor expansion in the mass parameter~$a$.
Restricting attention to the leading degree on the light cone, it suffices to consider the
first term of this expansion. Using~\eqref{s:Tm1def} and forming the sectorial projection~\eqref{s:pt},
we obtain for the regularized fermionic projector (for the factors of~$g$ see Footnote~\ref{s:footg}
on page~\pageref{s:footg})
\beq \label{s:Peq}
P(x,y) = \frac{ig}{2}\: \slashed{\xi}\: T^{(-1)}_{[0]} + (\deg < 2 )\:,
\eeq
where for notational convenience we omitted the indices~$^{-1}_{[0]}$ of the factor~$\xi$,
and where the bracket~$(\deg < 2)$ stands for terms of degree at most one.

Using this formula for the fermionic projector, the closed chain~\eqref{s:Adef} becomes
\beq \label{s:clcvac}
A_{xy} = \frac{g^2}{4}\: (\slashed{\xi} T^{(-1)}_{[0]}) (\overline{\slashed{\xi} T^{(-1)}_{[0]}}) 
+ \slashed{\xi} (\deg \leq 3) + (\deg < 3)\:,
\eeq
where~$\overline{\slashed{\xi}} := \overline{\xi_j} \gamma^j$.
Its trace can be computed with the help of the contraction rules~\eqref{s:eq53},
\[ \Tr (A_{xy}) = g^2 \,(\xi_j \overline{\xi^j})\: T^{(-1)}_{[0]}\, \overline{T^{(-1)}_{[0]}}
= \frac{g^2}{2} \left(z + \overline{z} \right) T^{(-1)}_{[0]}\, \overline{T^{(-1)}_{[0]}}
+ (\deg < 3)\:. \]
Next we compute the square of the trace-free part of the closed chain,
\begin{align*}
\Big( &A_{xy} - \frac{1}{4}\: \Tr (A_{xy})\,\1 \Big)^2 =
\frac{g^4}{16} \left( \slashed{\xi} \overline{\slashed{\xi}} -\frac{z+\overline{z}}{2} \right)^2 \left(T^{(-1)}_{[0]} \overline{T^{(-1)}_{[0]}} \right)^2 \\
&= \frac{g^4}{16} \left( \slashed{\xi} \overline{\slashed{\xi}} \slashed{\xi} \overline{\slashed{\xi}}- (z+\overline{z})\:
\slashed{\xi} \overline{\slashed{\xi}} + \frac{1}{4}\: (z+\overline{z})^2 \right) \left(T^{(-1)}_{[0]} \overline{T^{(-1)}_{[0]}} \right)^2 \\
&= \frac{g^4}{64}\:(z-\overline{z})^2 \:
\left(T^{(-1)}_{[0]} \overline{T^{(-1)}_{[0]}} \right)^2.
\end{align*}
Combining these formulas, we see that to leading degree,
the closed chain is a solution of the polynomial equation
\beq \label{s:poly}
\left( A_{xy} - \frac{g^2}{8}\, (z+\overline{z})\:T^{(-1)}_{[0]} \overline{T^{(-1)}_{[0]}} \right)^2
= \left( \frac{g^2}{8}\, (z-\overline{z})\:T^{(-1)}_{[0]} \overline{T^{(-1)}_{[0]}} \right)^2 .
\eeq
We point out that the calculations so far are only formal, but they have a well-defined meaning
in the formalism of the continuum, because to our end formulas we will be able to apply
the weak evaluation formula~\eqref{s:asy}. Having this in mind, we can interpret the roots
of the polynomial in~\eqref{s:poly}
\[ \lambda_+ = \frac{g^2}{4}\: (z\, T^{(-1)}_{[0]}) \,\overline{T^{(-1)}_{[0]}} \qquad \text{and} \qquad
\lambda_- = \frac{g^2}{4}\: T^{(-1)}_{[0]} \,\overline{(z \,T^{(-1)}_{[0]})} \]
as the eigenvalues of the closed chain. Using the contraction rule~\eqref{s:eq54}, these eigenvalues
simplify to (see also~\cite[eq.~(5.3.20)]{PFP} or~\eqref{lpm})
\beq \label{s:lpm} \boxed{ \quad
\lambda_+ = g^2\, T^{(0)}_{[0]} \,\overline{T^{(-1)}_{[0]}} + (\deg < 3) \:,\qquad
\lambda_- = g^2\,T^{(-1)}_{[0]} \,\overline{T^{(0)}_{[0]}} + (\deg < 3)\:. \quad }
\eeq
The corresponding spectral projectors, denoted by~$F_\pm$, are given by
\[ F_+ = \frac{A_{xy} - \lambda_-}{\lambda_+-\lambda_-} \:,\qquad
F_+ = \frac{A_{xy} - \lambda_+}{\lambda_--\lambda_+}\:; \]
\nindex{bp2@$\lambda_\pm, F_\pm$ -- eigenvalues and spectral projections of unperturbed closed chain}%
a short calculation yields (see also~\cite[eq.~(5.3.21)]{PFP} or~\eqref{Fpm})
\beq \label{s:Fpm} \boxed{ \quad
F_\pm = \frac{1}{2} \Big( \1 \pm \frac{[\slashed{\xi}, \overline{\slashed{\xi}}]}{z-\overline{z}} \Big)
+ \slashed{\xi} (\deg \leq 0) + (\deg < 0) \:. \quad }
\eeq
Since in the formalism of the continuum limit, the factors~$z$ and~$\overline{z}$ are treated as
two different functions, we do not need to worry about the possibility that the
denominator in~\eqref{s:Fpm} might vanish. Similarly, we can treat~$\xi$ and~$\overline{\xi}$
simply as two different vectors. Then the matrices~$F_+$ and~$F_-$ have rank two,
so that the eigenvalues~$\lambda_+$ and~$\lambda_-$ are both two-fold degenerate.
A straightforward calculation yields
\beq \label{s:Aspec}
A_{xy} = \lambda_+ F_+ + \lambda_- F_- + \slashed{\xi} (\deg \leq 3) + (\deg < 3) \:,
\eeq
showing that our spectral decomposition is indeed complete.
An important general conclusion from~\eqref{s:lpm} and~\eqref{s:Fpm} is
that in the vacuum, the eigenvalues of the closed chain form a {\em{complex conjugate pair}},
and are both {\em{two-fold degenerate}}.

We now give the corresponding operator~$Q$ which appears in the EL equations of the continuum
limit~\eqref{s:ELcl}.
\begin{Prp} \label{s:prpQ}
For the fermionic projector of the vacuum~\eqref{s:Peq}, the operator~$Q$
as defined by~\eqref{s:Qdef} and~\eqref{s:Rdef} takes the form
\beq \label{s:Qvac}
Q(x,y) = i \slashed{\xi}  \, g^3\,(1-4 \mu)\:
T^{(0)}_{[0]} T^{(-1)}_{[0]}\,\overline{T^{(-1)}_{[0]}} + (\deg < 5)\:.
\eeq
\end{Prp} \noindent
In order not to distract from the main points, we first discuss the consequences of this result and
derive it afterwards. According to the EL equations in the continuum limit~\eqref{s:ELcl},
the expression~\eqref{s:Qvac} must vanish. This determines the value of the Lagrange
multiplier~$\mu=\frac{1}{4}$. Thus the action~\eqref{s:Smudef} reduces to the action in~\eqref{s:STdef},
and we conclude that
\beq \label{s:Scrit}
\text{$P$ is a critical point of $\Sact$} \:,
\eeq
disregarding the constraint~$\T=\text{const}$. This result can be understood immediately
from the form of the Lagrangian~\eqref{s:Ldef} and the fact that the eigenvalues of~$A_{xy}$
form a complex conjugate pair. Namely, writing the spectral weights in~\eqref{s:Ldef}
via~\eqref{s:swdef} as sums over the eigenvalues~$\lambda^{xy}_\pm$ (both of multiplicity two),
we obtain
\[ \L_{xy}[P] = \left( |\lambda_+| - |\lambda_-| \right)^2\:. \]
The expression~$|\lambda_+|-|\lambda_-|$ clearly vanishes for a complex conjugate pair,
and the fact that it appears quadratically is the reason why even first variations of~$\L_{xy}[P]$
vanish, explaining~\eqref{s:Scrit}.

In the last argument we only used that the eigenvalues of~$A_{xy}$ form a complex
conjugate pair. Therefore, we can use this argument to show that~$Q$ vanishes in a more general sense:
\label{s:ccpgen}
First, a straightforward calculation yields that the eigenvalues of the closed chain~$A_{xy}$
form a complex conjugate pair to every degree on the light cone (for details see~\cite[Section~5.3]{PFP}
or Section~\ref{secspeccc}),
and thus~$Q$ vanishes identically in the formalism of the continuum limit.
Moreover, going beyond the formalism of the continuum limit, in~\cite{reg} it is shown that there
are regularizations of the vacuum for which the operator~$Q$ vanishes up to
contributions which stay finite in the limit~$\varepsilon \searrow 0$. Furthermore, in~\cite{reg} it is
shown that restricting attention to such regularizations does not give any constraints for the
regularization parameters~$c_\text{reg}$ in~\eqref{s:asy}.
Since we are here interested in the singularities of~$Q(x,y)$ in the limit~$\varepsilon \searrow 0$
as described by the weak evaluation formula~\eqref{s:asy}, we can in what follows assume that in the
vacuum, the operator~$Q$ vanishes identically.

The remainder of this section is devoted to deriving the result of Proposition~\ref{s:prpQ}.
For the derivation it is preferable to bypass the computation of the gradient~\eqref{s:Rdef}
by determining~$Q$ directly from~\eqref{s:QdP}.
For later use, we assume a more general spectral decomposition of~$A_{xy}$
with eigenvectors~$\lambda^{xy}_1, \ldots, \lambda^{xy}_4$ and corresponding
one-dimensional spectral projectors~$F^{xy}_1, \ldots, F^{xy}_4$.
This setting can be obtained from~\eqref{s:Aspec} by choosing pseudo-orthonormal bases in the
degenerate eigenspaces and letting~$F^{xy}_k$ be the projectors onto the span of these
basis vectors. It is convenient to choose these bases according to Lemma~\ref{lemma71}.

For later use, we next compute the operator~$Q$ in the general setting of the previous lemma.
Noting that the function~$\L_\mu$ in~\eqref{s:Smudef} depends only on the absolute values of the
eigenvalues, we can write
\[ \L_\mu[A_{xy}] = \L_\mu(|\lambda^{xy}_1|, \ldots, |\lambda^{xy}_4|)\:. \]
The partial derivatives of the function~$\L_\mu(|\lambda^{xy}_1|, \ldots, |\lambda^{xy}_4|)$
will be denoted by~$D_k$.
\nindex{cc8@$D$ -- partial derivative of~$\L_\mu$}%
\begin{Lemma} \label{lemma73}
Under the assumptions of Lemma~\ref{lemma71}, the operator~$Q$ in~\eqref{s:QdP} is given by
\beq \label{Qgen}
Q(x,y) = \sum_{k=1}^4 D_k \L_\mu \big(|\lambda^{xy}_1|, \ldots, |\lambda^{xy}_4| \big)\:
\frac{\overline{\lambda^{xy}_k}}{|\lambda^{xy}_k|}\: F^{xy}_k \, P(x,y) \:.
\eeq
\nindex{aq6@$Q(x,y)$ -- first variation of the Lagrangian}%
\end{Lemma}
\Proof
The relation~\eqref{Fdiag} allows us to compute the
variation of the eigenvalues by a standard first order perturbation calculation
without degeneracies,
\beq \label{firstper}
\delta \lambda^{xy}_k = \Tr(F^{xy}_k \, \delta A_{xy})\:.
\eeq
Using that that~$\delta |\lambda| = \re(\overline{\lambda}\, \delta \lambda/|\lambda|)$, we
can compute the first variation of this function with the help of~\eqref{firstper},
\beq \label{dL}
\delta \L_\mu[A_{xy}]
= \re \sum_{k=1}^4 D_k \L_\mu \big(|\lambda^{xy}_1|, \ldots, |\lambda^{xy}_4|
\big)\: \frac{\overline{\lambda^{xy}_k}}{|\lambda^{xy}_k|}\: \Tr(F^{xy}_k \, \delta A_{xy})\:.
\eeq
In the last trace we substitute the identity
\[ \delta A_{xy} = \delta P(x,y)\, P(y,x) + P(x,y)\, \delta P(y,x) \]
and cyclically commute the arguments to obtain
\begin{align*}
\Tr(F^{xy}_k \, \delta A_{xy}) &=
\Tr \big( F^{xy}_k \,P(x,y) \:\delta P(y,x) + P(y,x) F^{xy}_k\: \delta P(x,y) \big) \\
&= \Tr \big( F^{xy}_k \,P(x,y) \:\delta P(y,x) + F^{yx}_k\, P(y,x) \:\delta P(x,y) \big) \,,
\end{align*}
where in the last step we applied~\eqref{eq77}.
Substituting this formula into~\eqref{dL} and integrating over~$x$ and~$y$, we can exchange
the names of~$x$ and~$y$ such that only~$\delta P(y,x)$ appears. We thus obtain
\beq \label{delSmu}
\delta \Sact_\mu[P] = 2 \,\re \iint_M d^4x \:d^4y \Tr \left( Q(x,y)\: \delta P(y,x) \right)
\eeq
with the integral kernel~$Q(x,y)$ given by~\eqref{Qgen}.
Using Lemma~\ref{lemma71}, one sees that the operator corresponding to this
integral kernel is symmetric (i.e.\ $Q(x,y)^*=Q(y,x)$). As a consequence,
the integral in~\eqref{delSmu} is real, so that it is unnecessary to take the real part.
Comparing with~\eqref{s:QdP}, we conclude that the operator with kernel~\eqref{Qgen} indeed
coincides with the operator~$Q$ in~\eqref{s:QdP}. We note that due to the sum in~\eqref{Qgen},
it is irrelevant how the bases were chosen on the degenerate subspaces of~$A_{xy}$.
\QED

\Proof[Proof of Proposition~\ref{s:prpQ}]
Let us specialize the general formula~\eqref{Qgen} to our spectral representation with
eigenvalues~\eqref{s:lpm} and spectral projectors~\eqref{s:Fpm}. First, from~\eqref{s:Smudef} we
readily obtain that
\[ D_k \L_\mu \big(|\lambda^{xy}_1|, \ldots, |\lambda^{xy}_4| \big) = 2 |\lambda_k|
- 2 \mu \sum\nolimits_{l=1}^4 |\lambda_l| = 2 (1-4 \mu)\: |\lambda_-| \:. \]
The product~$F^{xy}_k \, P(x,y)$ can be computed with the help of~\eqref{s:Peq} and~\eqref{s:Fpm}
as well as the relations
\[ [\slashed{\xi}, \overline{\slashed{\xi}}]\, \slashed{\xi} = 2 \la \overline{\xi}, \xi \ra\: \slashed{\xi} - 2 \xi^2\,
\overline{\slashed{\xi}} = -(z-\overline{z}) \:\slashed{\xi} \:, \]
where in the last step we treated the factors~$\slashed{\xi}$ and~$\overline{\slashed{\xi}}$
as outer factors and applied the contraction rules~\eqref{s:eq52} and~\eqref{s:eq53}. We thus obtain
(see also~\cite[eq.~(5.3.23)]{PFP} or~\eqref{FPasy})
\beq \label{s:FPasy}
F^{xy}_+ P(x,y) = (\deg < 2) \:,\qquad
F^{xy}_- P(x,y) = \frac{i g}{2}\:\slashed{\xi}\, T^{(-1)}_{[0]} + (\deg < 2)\:.
\eeq
Substituting these formulas into~\eqref{Qgen} and using~\eqref{s:lpm}, the result follows.
\QED

\subsectionn{Chiral Gauge Potentials} \label{s:sec72}
We now begin the study of interacting systems by introducing {\em{chiral potentials}}.
\sindex{potential!chiral}%
Thus we choose the operator~$\B$ in the auxiliary Dirac equation with
interaction~\eqref{s:diracPaux} according to~\eqref{s:chiral} with two real vector fields~$A_L$
and~$A_R$. Sometimes it is convenient to write~$\B$ in the form
\nindex{bh2@$A_L, A_R$ -- chiral potentials}%
\beq \label{s:va}
\B = \slashed{A}_\text{\rm{v}} + \pseudo \slashed{A}_\text{\rm{a}}
\eeq
with a {\em{vector potential}}~$A_\text{\rm{v}}$ and an {\em{axial
potential}}~$A_\text{\rm{a}}$ defined by
\sindex{potential!vector}%
\sindex{potential!axial}%
\beq \label{s:axialdef}
A_\text{\rm{v}}=(A_L+A_R)/2 \qquad \text{and} \qquad A_\text{\rm{a}}=(A_L-A_R)/2\:.
\eeq
\nindex{cd0@$A_\text{\rm{v}}$ -- vector potential}%
\nindex{cd2@$A_\text{\rm{a}}$ -- axial potential}%
To the considered highest degree on the light cone, the chiral gauge potentials merely
describe phase transformations of the left- and right-handed components of the fermio\-nic
projector (for details see~\cite{light}, \cite[Section~2.5]{PFP} or Section~\ref{seclight}). More precisely, the fermionic projector
is obtained from~\eqref{s:Peq} by inserting the phase factors
\beq \label{s:Pchiral}
P(x,y) = \frac{ig}{2}\: \left( \chi_L \,e^{-i \Lambda^{xy}_L} + \chi_R \,e^{-i \Lambda^{xy}_R} 
\right) \slashed{\xi}\: T^{(-1)}_{[0]} + (\deg < 2 )\:,
\eeq
where the functions~$\Lambda^{xy}_{L\!/\!R}$ are integrals of the chiral potentials along the
line segment~$\overline{xy}$,
\beq \label{s:Lambda}
\Lambda^{xy}_{L\!/\!R} = \int_x^y A^j_{L\!/\!R} \,\xi_j :=
\int_0^1 A^j_{L\!/\!R}|_{\tau y + (1-\tau) x}\: \xi_j \: d\tau \:.
\eeq
\nindex{cd4@$\Lambda^{xy}_{\LR}$ -- integrated chiral potentials}%
Consequently, the closed chain is obtained from~\eqref{s:clcvac} by inserting phase factors,
\beq \label{s:clc}
A_{xy} = \frac{g^2}{4} \left( \chi_L \,\nu_L + \chi_R \,\nu_R
\right) (\slashed{\xi} T^{(-1)}_{[0]}) (\overline{\slashed{\xi} T^{(-1)}_{[0]}}) 
+ \slashed{\xi} (\deg \leq 3) + (\deg < 3)\:,
\eeq
where
\beq \label{s:nudef}
\nu_L = \overline{\nu_R} = e^{-i (\Lambda_L^{xy} - \Lambda_R^{xy})} 
= \exp \Big(-2i \int_x^y A_\text{\rm{a}}^j \,\xi_j \Big) \:.
\eeq
\nindex{cd6@$\nu_{\LR}$ -- chiral phases}%
From~\eqref{s:clc} one sees that the matrix~$A_{xy}$ is invariant on the left- and right-handed
subspaces (i.e.\ on the image of the operators~$\chi_L$ and~$\chi_R$).
On each of these invariant subspaces, it coincides up to a phase with the closed chain
of the vacuum~\eqref{s:clcvac}. Using these facts, the eigenvalues~$(\lambda^c_s)_{c \in \{L,R\},
s \in \{+,-\}}$ and corresponding spectral projectors~$F^c_s$ are immediately computed by
\beq \label{s:unperturb}
\boxed{ \quad \lambda^{L\!/\!R}_\pm = \nu_{L\!/\!R}\: \lambda_\pm \qquad \text{and} \qquad
F^{L\!/\!R}_\pm = \chi_{L\!/\!R} \:F_\pm \quad }
\eeq
\nindex{cd8@$\lambda^\LR_\pm, F^\LR_\pm$ -- eigenvalues and corresponding spectral projectors
of closed chain}%
with~$\lambda_s$ and~$F_s$ as in~\eqref{s:lpm} and~\eqref{s:Fpm}.
We conclude that the eigenvalues of the closed chain are again complex, but in general
they now form two complex conjugate pairs. Since the eigenvalues~$\lambda^L_c$ and~$\lambda^R_c$
differ only by a phase, we see that all eigenvalues have the same absolute value,
\beq \label{s:labs}
|\lambda^L_+| = |\lambda^R_+| = |\lambda^L_-| = |\lambda^R_-| \:.
\eeq
Writing the Lagrangian~\eqref{s:Ldef} as
\beq
\L_{xy}[P] = \sum_{c ,s } |\lambda^{c}_s |^2 - \frac{1}{4}
\Big( \sum_{c ,s } |\lambda^{c}_s| \Big)^2
= \frac{1}{8} \sum_{c,c' \in \{L,R\}}  \;\sum_{s,s' \in \{\pm\} }
\left( |\lambda^c_s| - |\lambda^{c'}_{s'}| \right)^2 \label{s:Lcrit}
\eeq
(where we sum over~$c \in \{L,R\}$ and~$s \in  \{\pm\}$),
we find that~$\L$ vanishes identically. Since the Lagrangian is quadratic
in~$|\lambda^c_s| - |\lambda^{c'}_{s'}|$, also first variations of~$\L$ vanish,
suggesting that the operator~$Q(x,y)$ should again vanish identically. This is indeed the
case, as is verified immediately by applying Lemmas~\ref{lemma71} and~\ref{lemma73}.
We conclude that for chiral potentials, the EL equations in the continuum limit~\eqref{s:ELcl}
are satisfied to degree five on the light cone.

We end this section by explaining
how the line integrals in~\eqref{s:Lambda} and the phase factors in~\eqref{s:Pchiral}
and~\eqref{s:clc} can be understood from an underlying local gauge symmetry
\sindex{gauge symmetry}%
(for more details in the general context of non-abelian gauge fields see~\cite[Section~6.1]{PFP}).
The local phase transformation~$\psi(x) \rightarrow e^{i \Lambda(x)} \psi(x)$
with a real function~$\Lambda$ describes a unitary transformation of the wave functions (with respect
to the inner product~\eqref{s:iprod}). Transforming all objects unitarily, we obtain the
transformation laws
\begin{align}
i \Pdd + \B - m Y &\rightarrow 
e^{i \Lambda(x)} \,( i \Pdd + \B - m Y)\, e^{-i \Lambda(x)} =  i \Pdd + \B - m Y + (\Pdd \Lambda) \\
P(x,y) &\rightarrow e^{i \Lambda(x)} \, P(x,y) e^{- i \Lambda(y)} \label{s:Pgauge} \\
A_{xy} &\rightarrow e^{i \Lambda(x)} \, P(x,y) e^{- i \Lambda(y)}\;
e^{i \Lambda(y)} \, P(y,x) e^{- i \Lambda(x)} = A_{xy}\:. \label{s:gauge}
\end{align}
The transformation of the Dirac operator corresponds to a transformation of the vector and
axial potentials by
\beq \label{s:Agauge}
A_\text{\rm{v}} \rightarrow A_\text{\rm{v}} + \partial \Lambda \qquad \text{and} \qquad A_\text{\rm{a}} \rightarrow A_\text{\rm{a}}\:.
\eeq
These are the familiar gauge transformations of electrodynamics.
\sindex{gauge transformation}%
Using the formula
\[ \Lambda(y)-\Lambda(x) = \int_0^1 \frac{d}{d\tau} \Lambda|_{\tau y + (1-\tau) x}\:  d\tau
= \int_x^y (\partial_j \Lambda)\: \xi^j\: d\tau \:, \]
the phases in~\eqref{s:Pgauge} can be described similar to~\eqref{s:Lambda} in terms of
line integrals. This explains why the phase factors in~\eqref{s:Pchiral} describe the correct
behavior under gauge transformations.
According to~\eqref{s:gauge}, the closed chain~$A_{xy}$ is gauge invariant.
This is consistent with the fact that in~\eqref{s:clc} and~\eqref{s:nudef} only the axial potential
enters, which according to~\eqref{s:Agauge} is also gauge invariant.

In order to transform the axial potential, one can consider the local
transformation~$\psi(x) \rightarrow e^{-i \pseudo \Lambda(x)} \,\psi(x)$.
In contrast to the above gauge transformation, this transformation is {\em{not unitary}}
(with respect to the inner product~\eqref{s:iprod}), and the requirement that the Dirac operator and the fermionic projector must be symmetric operators leads us to the transformations
\begin{align}
i \Pdd + \B - m Y &\rightarrow 
e^{i \pseudo \Lambda(x)} \,( i \Pdd + \B - m Y)\, e^{i \pseudo \Lambda(x)} \label{s:axialgauge} \\
&=  i \Pdd + e^{i \pseudo \Lambda(x)} (\B - m Y) e^{i \pseudo \Lambda(x)} +
\pseudo (\Pdd \Lambda) \nonumber \\
P(x,y) &\rightarrow e^{-i \pseudo \Lambda(x)} \, P(x,y) e^{-i \pseudo \Lambda(x)}\:. \nonumber
\end{align}
Thus the vector and axial potentials transform as desired by
\[ A_\text{\rm{v}} \rightarrow A_\text{\rm{v}}  \qquad \text{and} \qquad A_\text{\rm{a}} \rightarrow A_\text{\rm{a}} +
\partial \Lambda \]
(and also the term~$mY$ is modified, but this is of no relevance for the argument here).
The point is that when we now consider the transformation of the closed chain,
\beq \label{s:axialchain}
A_{xy} \rightarrow e^{i \pseudo \Lambda(x)} P(x,y) \,e^{i \pseudo \Lambda(y)}\;
e^{i \pseudo \Lambda(y)} P(y,x) \,e^{i \pseudo \Lambda(x)} \:,
\eeq
the local transformations do {\em{not}} drop out. This explains why in~\eqref{s:clc} phases involving
the axial potentials appear.

For clarity, we point out that the field tensors and the currents of the chiral gauge potentials
also affect the fermionic projector, in a way which cannot be understood from the simple gauge
transformation laws considered above. The corresponding contributions to the operator~$Q$
will be of degree four, and we shall consider them in the next section.

\section{The Euler-Lagrange Equations to Degree Four}  \label{s:sec8}
We come to the analysis of the EL equations to the next lower degree four on the light cone.
In preparation, we bring the EL equations into a convenient form.
\begin{Lemma} \label{s:lemma81}
To degree four, the EL equations in the continuum limit~\eqref{s:ELcl}
are equivalent to the equation
\beq \label{s:RRdef}
{\mathcal{R}} := \frac{\Delta ( |\lambda^L_-| - |\lambda^R_-| )}{|\lambda_-|}\;g^3\:
T^{(0)}_{[0]} T^{(-1)}_{[0]} \:\overline{T^{(-1)}_{[0]}}  = 0 + (\deg<4)\:,
\eeq
\nindex{ce0@${\mathcal{R}}$ -- appears in EL equations to degree four}%
where $\Delta$ denotes the perturbation of the eigenvalues~\eqref{s:unperturb} to degree two.
\end{Lemma}
\Proof According to~\eqref{s:unperturb}, the eigenvalues to degree three are all non-real.
Since this property is stable under perturbations of lower degree,
we can again apply Lemmas~\ref{lemma71} and~\ref{lemma73}. Noting that
before~\eqref{s:Scrit}, we fixed the Lagrange multiplier to~$\mu=\frac{1}{4}$, we
consider the Lagrangian~\eqref{s:Ldef}, which we now write in analogy to~\eqref{s:Lcrit} as
\[ \L_{xy}[P] = \frac{1}{8} \sum_{k,l=1}^4 \big( |\lambda^{xy}_k| - |\lambda^{xy}_l| \big)^2 \:. \]
Then the relation~\eqref{Qgen} can be written as
\[ Q(x,y) = \frac{1}{2}
\sum_{k,l=1}^4 \Big\{ |\lambda^{xy}_k| - |\lambda^{xy}_l| \Big\} \:
\frac{\overline{\lambda^{xy}_k}}{|\lambda^{xy}_k|}\: F^{xy}_k \, P(x,y) \:. \]
According to~\eqref{s:labs}, the curly brackets vanish for the unperturbed eigenvalues.
This has the convenient consequence that to degree four, it suffices
to take into account the perturbation of the curly brackets, whereas everywhere else we may
work with the unperturbed spectral decomposition~\eqref{s:unperturb},
\[ Q(x,y) = \frac{1}{2}
\sum_{k,l=1}^4 \Delta\Big( |\lambda^{xy}_k| - |\lambda^{xy}_l| \Big) \:
\frac{\overline{\lambda^{xy}_k}}{|\lambda^{xy}_k|}\: F^{xy}_k \, P(x,y) + (\deg<4) \:. \]
Using~\eqref{s:FPasy}, we see that we only get a contribution if~$\lambda_k$
equals~$\lambda^L_-$ or~$\lambda^R_-$. Furthermore, we can apply~\eqref{lorder},
numbering the eigenvalues such that~$\overline{\lambda^\pm_L} = \lambda^\mp_R$.
We thus obtain
\beq \label{s:Qxidef}
Q(x,y) =  \sum_{c \in \{L,R\}}
\Delta\Big( |\lambda^c_-| - |\lambda^c_+| \Big) 
\frac{\overline{\lambda^c_-}}{|\lambda^c_-|}\; \chi_c\: \frac{i \slashed{\xi}}{2}\:g\, T^{(-1)}_{[0]}
+ (\deg<4)\:.
\eeq
The EL equations~\eqref{s:ELcl} imply that the left- and
right-handed components of this expression must vanish separately. Thus, again applying~\eqref{lorder},
we obtain the sufficient and necessary condition
\[ \Delta\Big( |\lambda^L_-| - |\lambda^R_-| \Big) \frac{\overline{\lambda_-}}{|\lambda_-|}
\:g\, T^{(-1)}_{[0]} + (\deg<4) = 0\:. \]
The explicit formulas~\eqref{s:unperturb} and~\eqref{s:lpm} yield the result.
\QED

It is important to observe that the EL equations only involve the difference of the
absolute values of the left- and right-handed eigenvalues. This can immediately be understood
as follows. To the leading degree three, the eigenvalues of~$A_{xy}$ form two
complex conjugate pairs (see~\eqref{s:unperturb}). Since this property is preserved
under perturbations, we can again write the Lagrangian in the form~\eqref{s:Lcrit}.
Hence the Lagrangian vanishes identically unless the absolute values of the
eigenvalues are different for the two pairs. This explains the term~$\Delta
( |\lambda^L_-| - |\lambda^R_-| )$ in~\eqref{s:RRdef}.

As explained on page~\pageref{s:ccpgen}, the expression~$\Delta
( |\lambda^L_-| - |\lambda^R_-| )$ vanishes in the vacuum.
Furthermore, the phase factors in~\eqref{s:unperturb} drop out of this expression.
But new types of contributions to the interacting fermionic projector come into play,
as we now explain.

\subsectionn{The Axial Current Terms and the Mass Terms} \label{s:sec81}
An interaction by chiral potentials~\eqref{s:chiral} as introduced in~\S\ref{s:sec72} affects the
fermionic projector in a rather complicated way. For clarity, we treat the different
terms in succession, beginning with the contributions near the origin~$\xi =0$
(the contributions away from the origin will be considered in Section~\ref{s:secnonlocal}).
For the Taylor expansion around~$\xi=0$ we note that when evaluated weakly on the light
cone~\eqref{s:asy}, a simple fraction of degree~$L$ has a pole~$|\vec{\xi}|^{-L}$.
This leads us to say that a term of the form~\eqref{s:sfr} is of the order~$k$
at the origin if the smooth function~$\eta$ vanishes at the origin to the order~$k+L$.
\begin{Def} \label{s:def82} An expression of the form~\eqref{s:sfr} is said to be
of {\bf{order~$o (|\vec{\xi}|^k)$ at the origin}} if the function~$\eta$ is
in the class~$o((|\xi^0| + |\vec{\xi}|)^{k+L})$.
\end{Def} \noindent
\nindex{ce2@$o( \vert \vec{\xi} \vert^k)$ -- order at the origin}%
\sindex{order!at the origin}%
In the next lemma we specify the contributions to the EL equations
to degree four on the light cone, to leading order at the origin.
\begin{Lemma} \label{s:lemmalc1}
For an interaction described by vector and axial potentials~\eqref{s:chiral}, the
expression~${\mathcal{R}}$ as defined by~\eqref{s:RRdef} takes the form
\begin{align} \label{s:eq82}
{\mathcal{R}} = -i \xi_k \left( j^k_\text{\rm{a}}\: N_1
- m^2 A^k_\text{\rm{a}} \:N_2 \right) + (\deg < 4) + o \big( |\vec{\xi}|^{-3} \big)\:,
\end{align}
where~$j_\text{\rm{a}}$ is the axial current
\beq \label{s:jadef}
j_\text{\rm{a}}^k = \partial^k_{\;j}A^j_\text{\rm{a}} - \Box A_\text{\rm{a}}^k \:,
\eeq
\nindex{ce4@$j_\text{\rm{a}}$ -- axial current}%
\sindex{current!axial}%
and~$N_1, N_2$ are the simple fractions
\begin{align}
N_1 =& \frac{g^3}{6\,\overline{T^{(0)}_{[0]}}}
\Big[  \Big( T^{(0)}_{[0]} T^{(0)}_{[0]} - 2 \, T^{(1)}_{[0]} T^{(-1)}_{[0]} \Big)
\overline{T^{(0)}_{[0]} T^{(-1)}_{[0]}} - c.c. \Big] \label{s:N1def} \\
N_2 =& -\frac{2}{\overline{T^{(0)}_{[0]}}}
\left[   \Big( g\, \hat{Y}^2\: T^{(-1)}_{[0]} T^{(0)}_{[0]} \overline{ T^{(0)}_{[1]} T^{(0)}_{[1]}}
+ g^2\, \acute{Y} \grave{Y}\: T^{(-1)}_{[0]} T^{(1)}_{[2]}  \overline{T^{(-1)}_{[0]} T^{(0)}_{[0]}} \Big)
- c.c. \right] . \label{s:N2def}
\end{align}
Here ``$c.c.\!$'' denotes the complex conjugate of the preceding simple fraction; the
accents were defined in~\eqref{s:tildedef}.
\nindex{ce6@$c.c.$ -- complex conjugate simple fraction}%
\end{Lemma} \noindent
In order not to distract from the main ideas, we postpone the proof of this lemma to
Appendix~\ref{s:appspec} and proceed right away with the physical discussion.
From the mathematical point of view, the appearance of the axial current~$j_\text{\rm{a}}$ is not
surprising, because the light-cone expansion of the fermionic projector involves derivatives of
the potentials. In physical terms, this shows that the axial potential affects the
fermionic projector not only via the phases in~\eqref{s:Pchiral}, but also via
the axial current. The term~$-i \xi_k \:j^k_a N_1$ is referred to as the {\em{current term}}.
\sindex{fermionic projector!axial current term}%
The other term~$-i \xi_k \:m^2 A^k_a N_2$ could not appear in ordinary Yang-Mills theories
because it would not be gauge invariant.
However, as pointed out after~\eqref{s:axialgauge}, the axial $U(1)$-transformations
do {\em{not}} correspond to a local gauge symmetry, because they are not unitary.
Instead, they describe relative phase transformations of the left- and right-handed components of
the fermionic projector, thereby changing the physics of the system.
Only the phase transformations~\eqref{s:gauge} correspond to a local gauge symmetry,
and in view of~\eqref{s:Agauge}, the term~$-i \xi_k \:m^2 A^k_a N_2$ is indeed consistent with
this local $U(1)$-symmetry.

Since the direction~$\xi$ can be chosen arbitrarily on the light cone, the condition~\eqref{s:RRdef}
implies that the bracket in~\eqref{s:eq82} must vanish,
\beq \label{s:YM1}
j^k_\text{\rm{a}}\: N_1 - m^2 A^k_\text{\rm{a}} \:N_2 = 0 \:.
\eeq
If~$N_1$ and~$N_2$ could be treated as constants, this equation would go over to
field equations for the axial potential~$A_\text{\rm{a}}$ with rest mass $m^2 N_2/N_1$.
For this reason, we refer to the term~$-i \xi_k \:m^2 A^k_a N_2$ in~\eqref{s:eq82} as the
{\em{mass term}}.
\sindex{mass term}%
It is remarkable that in our framework, the bosonic mass term appears naturally,
without the need for the Higgs mechanism of spontaneous symmetry breaking
(for a detailed discussion of this point see~\S\ref{s:secnohiggs}).
\sindex{Higgs mechanism}%
\sindex{spontaneous symmetry breaking}%
We also point out that the simple fraction~$N_2$ involves the mass matrix~$Y$, and thus
the mass term in~\eqref{s:YM1} depends on the masses of the fermions of the system.

In order to make the argument after~\eqref{s:YM1} precise, we need to analyze the simple fractions~$N_1$
and~$N_2$ weakly on the light cone. Before this will be carried out in~\S\ref{s:sec83},
we specify how the Dirac current enters the EL equations.

\subsectionn{The Dirac Current Terms} \label{s:sec82}
As explained in~\S\ref{s:sec43}, the particles and anti-particles of the system
enter the auxiliary fermionic projector via~\eqref{s:particles}, where we orthonormalize
the wave functions according to~\eqref{s:normalize}. Introducing the left- and right-handed
component of the Dirac current by
\[ J^i_{L\!/\!R} = \sum_{k=1}^{\np} \overline{\psi_k} \chi_{R\!/\!L} \gamma^i \psi_k
- \sum_{l=1}^{\na} \overline{\phi_l} \chi_{R\!/\!L} \gamma^i \phi_l \:, \]
\nindex{ce8@$J_{\LR}$ -- chiral Dirac current}%
\sindex{current!chiral Dirac}%
a decomposition similar to~\eqref{s:axialdef} leads us to define the
{\em{axial Dirac current}} by
\beq \label{s:Jadef}
J^i_\text{\rm{a}} = \sum_{k=1}^{\np} \overline{\psi_k} \pseudo \gamma^i \psi_k
- \sum_{l=1}^{\na} \overline{\phi_l} \pseudo \gamma^i \phi_l \:.
\eeq
\nindex{cf0@$J_\text{\rm{a}}$ -- axial Dirac current}%
\sindex{current!axial Dirac}%
The next lemma gives the corresponding contribution to the EL equations,
to leading order at the origin.

\begin{Lemma} \label{s:lemmalc2}
Introducing the axial Dirac current by
the particle and anti-particle wave functions in~\eqref{s:particles} leads to
a contribution to~${\mathcal{R}}$ of the form
\[ {\mathcal{R}} \asymp i \xi_k \:
J^k_\text{\rm{a}}\: N_3 +(\deg < 4)  + o \big( |\vec{\xi}|^{-3} \big)\:, \]
where
\beq \label{s:N3def}
N_3 = \frac{g^2}{8 \pi}\:\frac{1}{\overline{T^{(0)}_{[0]}}}
\Big[T^{(-1)}_{[0]}\: \overline{T^{(0)}_{[0]}T^{(-1)}_{[0]}}  - c.c. \Big] .
\eeq
\end{Lemma} \noindent
Here the symbol ``$\asymp$''
\nindex{cf2@$\asymp$ -- denotes a contribution}%
\sindex{fermionic projector!Dirac current term}%
means that we merely give the contribution to~${\mathcal{R}}$
by the Dirac current, but do not repeat the earlier contributions given in Lemma~\ref{s:lemmalc1}.
The proof of this lemma is again postponed to Appendix~\ref{s:appspec}.

\subsectionn{The Logarithmic Poles on the Light Cone} \label{s:sec83}
Combining the results of Lemmas~\ref{s:lemma81}, \ref{s:lemmalc1} and~\ref{s:lemmalc2},
the Euler-Lagrange equations give rise to the equation
\[ \xi_k \left( j^k_\text{\rm{a}} \: N_1- m^2 A_\text{\rm{a}}^k\:N_2 - J^k_a\: N_3 \right) = 0 \:, \]
which involves the axial potential~$A_\text{\rm{a}}$ (see~\eqref{s:va}), the corresponding axial bosonic
current~\eqref{s:jadef} and the axial Dirac current~\eqref{s:Jadef}.
At first sight, this equation resembles a bosonic field equation, which describes the coupling
of the Dirac spinors to the bosonic field and involves a bosonic mass term. However, the situation
is not quite so simple, because the factors~$N_1$, $N_2$ and~$N_3$ (see~\eqref{s:N1def},
\eqref{s:N2def} and~\eqref{s:N3def}) have a mathematical meaning only when evaluated
weakly on the light cone~\eqref{s:asy}. Let us analyze the weak evaluation in more detail.
The simple fraction~$N_3$ is composed of the functions~$T^{(0)}_{[0]}$, $T^{(-1)}_{[0]}$
and their complex conjugates, which according to~\eqref{s:Taser}--\eqref{s:Tm1def}
all have poles of the order~$\xi^{-2}$ or~$\xi^{-4}$. In particular, no logarithmic poles
appear, and thus we may apply~\eqref{s:asy} with~$r=0$ to obtain
\[ \int_{|\vec{\xi}|-\varepsilon}^{|\vec{\xi}|+\varepsilon} dt \; \eta \:
\xi_k\:J^k_a\: N_3 = \frac{c^{\text{reg}}_3}{\varepsilon^3 |\vec{\xi}|^4}\:
\eta(x)\:\xi_k\:J^k_a(x) + (\deg < 4) + o \big( |\vec{\xi}|^{-3} \big) \]
with a regularization parameter~$c^{\text{reg}}_3$,
where we omitted error terms of the form~\eqref{s:ap1} and~\eqref{s:ap2}.
The simple fractions~$N_1$ and~$N_2$, on the other hand, involve in addition
the functions~$T^{(1)}_\circ$ and~$\overline{T^{(1)}_\circ}$, which according
to~\eqref{s:Taser}--\eqref{s:Tndef} involve a factor $\log |\xi^2|$ and thus
have a {\em{logarithmic pole on the light cone}}.
\sindex{logarithmic pole on the light cone}%
As a consequence, in~\eqref{s:asy} we also obtain
contributions with~$r=1$,
\begin{align*}
&\int_{|\vec{\xi}|-\varepsilon}^{|\vec{\xi}|+\varepsilon} dt \; \eta \:
\xi_k \left( j^k_\text{\rm{a}} \: N_1- m^2 A^k\:N_2 \right)
= (\deg < 4) + o \big( |\vec{\xi}|^{-3} \big) \\
&+ \frac{1}{\varepsilon^3 |\vec{\xi}|^4}\:
\eta(x)\:\xi_k \left[ j^k_\text{\rm{a}} \: \left(c^{\text{reg}}_1 + d^{\text{reg}}_1\:
\log(\varepsilon |\vec{\xi}|) \right)
- m^2 A_\text{\rm{a}}^k\left(c^{\text{reg}}_2 + d^{\text{reg}}_2\:
\log(\varepsilon |\vec{\xi}|) \right) \right] ,
\end{align*}
involving four regularization parameters~$c^{\text{reg}}_{1\!/\!2}$
and~$d^{\text{reg}}_{1\!/\!2}$.
Combining the above weak evaluation formulas, the freedom in choosing the radius~$|\vec{\xi}|$
and the spatial direction~$\vec{\xi}/|\vec{\xi}|$ implies that the logarithmic and
non-logarithmic terms must vanish separately,
\begin{align}
j^k_\text{\rm{a}} \: d^{\text{reg}}_1
- m^2 A_\text{\rm{a}}^k \:d^{\text{reg}}_2 &=0 \label{s:fe1} \\
j^k_\text{\rm{a}} \: c^{\text{reg}}_1
- m^2 A_\text{\rm{a}}^k \:c^{\text{reg}}_2 &= J^k_a\: c^{\text{reg}}_3\:, \label{s:fe2}
\end{align}
where~$c^{\text{reg}}_{1\!/\!2}$ and~$d^{\text{reg}}_{1\!/\!2}$ are
constants depending on the particular regularization.

Unfortunately, the system of equations~\eqref{s:fe1} into~\eqref{s:fe2} is overdetermined.
Thus turns out to be too restrictive for physical applications, as we now explain.
We begin with the case of a generic regularization for which the
constants~$c^{\text{reg}}_1$,  $c^{\text{reg}}_3$ and~$d^{\text{reg}}_1$ are non-zero.
Thus solving~\eqref{s:fe1} for~$j_\text{\rm{a}}$ and substituting into~\eqref{s:fe2}, one
obtains an algebraic equation involving~$J_\text{\rm{a}}$ and~$A_\text{\rm{a}}$.
This means that either~$J_\text{\rm{a}}$ must vanish identically, or else the gauge potential~$A_\text{\rm{a}}$
is fixed to a constant times~$J_\text{\rm{a}}$ and thus cannot be dynamical.
Both cases are not interesting from a physical point of view.
The basic reason for this shortcoming is that the bosonic current and mass terms
have logarithmic poles on the light cone, whereas the Dirac current terms
involve no such logarithms. Our method for overcoming this problem is
to insert additional potentials into the Dirac equation, with the aim of compensating the
logarithmic poles of the bosonic current and mass terms.
Before entering these constructions in~\S\ref{s:sec84},
we now briefly discuss alternative methods for treating the logarithmic poles.

An obvious idea for reducing the system~\eqref{s:fe1} and~\eqref{s:fe2} to a single equation is to
restrict attention to non-generic regularizations
\sindex{regularization!non-generic}%
where the constants~$c^\reg_i$ and/or~$d^\reg_i$
take special values. In particular, it seems tempting to demand that
$d^\reg_1= d^\reg_2= 0$, so that~\eqref{s:fe1} is trivially satisfied, leaving
us with the field equations~\eqref{s:fe2}. This method does not work, as the following consideration
shows. Differentiating~\eqref{s:Tacounter} and using~\eqref{s:Taser}, one sees that
\beq \label{s:logpole}
32 \pi^3 \:T^{(1)}_\circ = \log |\xi^2| + c_0 + i \pi \Theta(\xi^2)\: \epsilon(\xi^0)\: .
\eeq
Evaluating near the upper light cone~$\xi^0 \approx |\vec{\xi}|$, we can apply the
relation~$ \log |\xi^2| = \log \left| \xi^0+|\vec{\xi}| \right| + \log \left| \xi^0-|\vec{\xi}| \right|$ to obtain
\beq \label{s:T1ex}
32 \pi^3 \:T^{(1)}_\circ = \log |2 \vec{\xi}| + \log \left|\xi^0 - |\vec{\xi}| \right| 
+ i \pi \Theta(\xi^0 - |\vec{\xi}|) + c_0 + {\mathscr{O}}(\xi^0 - |\vec{\xi}|) \: .
\eeq
When evaluating the corresponding simple fraction weakly~\eqref{s:asy}, the first term in~\eqref{s:T1ex}
gives rise to the $\log|\vec{\xi}|$-dependence,
the second term gives the $\log \varepsilon$-dependence, whereas all the other terms
do not involve logarithms or are of higher order in~$\varepsilon$.
Obviously, the same is true for the complex conjugate~$\overline{T^{(1)}_\circ}$.
Since in~\eqref{s:asy} the vector~$\vec{\xi}$ is fixed, the vanishing of the
$\log|\vec{\xi}|$-dependent contribution to the integral~\eqref{s:asy}
implies that the simple fraction still vanishes if the factors~$T^{(1)}_\circ$
and~$\overline{T^{(1)}_\circ}$ are replaced by constants.
Inspecting the~$T^{(1)}$-dependence of~\eqref{s:N1def} and~\eqref{s:N2def} and
comparing with~\eqref{s:N3def}, we find that
\[ d^\reg_1 = 0 \; \Longleftrightarrow \; d^\reg_2 =0  \qquad \text{and} \qquad
d^\reg_1 = 0 \; \Longrightarrow \; c^\reg_3 = 0 \:. \]
Thus if the constants~$d^\reg_1$ and~$d^\reg_2$ in~\eqref{s:fe1} vanish,
then~\eqref{s:fe1} becomes trivial as desired. But then the constant~$c^\reg_3$ in~\eqref{s:fe2}
is also zero, so that the Dirac current drops out of the field equation.
Again, we do not end up with physically reasonable equations.

Sticking to the idea of considering regularizations where the regularization constants
have special values, the remaining method is to assume that {\em{all}} regularization constants
in~\eqref{s:fe1} and~\eqref{s:fe2} vanish. Then the EL equations would be trivially satisfied to degree
four on the light cone, and one would have to proceed to the analysis to degree three on the light cone.
This method does not seem to be promising for the following reasons.
First, it is not clear whether there exist regularizations for which all the
regularization constants in~\eqref{s:fe1} and~\eqref{s:fe2} vanish. In any case, it
seems difficult to satisfy all these conditions, and the resulting regularizations would have to
be of a very special form. This would not be fully convincing, because one might prefer 
not to restrict the class of admissible regularizations at this point.
Secondly, there is no reason to believe that the situation to degree three would be better,
at least not without imposing additional relations between regularization constants,
giving rise to even more constraints for the admissible regularizations.

We conclude that assuming special values for the regularization constants in~\eqref{s:fe1}
and~\eqref{s:fe2} does not seem to be a promising strategy.
Thus in what follows we shall {\em{not}} impose any constraints on the regularization constants,
which also has the advantage that our constructions will apply to {\em{any}} regularization.
Then the only possible strategy is to try to compensate the logarithmic poles by 
a suitable transformation of the fermionic projector.

\subsectionn{A Pseudoscalar Differential Potential} \label{s:sec84}
Compensating the logarithmic poles of the bosonic current and mass terms
by a suitable transformation of the fermionic projector is not an easy task,
because it is not at all obvious how such a transformation should look like.
We approach the problem in several steps, following the original path
which eventually led us to the microlocal transformation to be introduced in~\S\ref{s:secgennonloc}.
The most obvious method is to inserting additional potentials into the auxiliary Dirac
equation~\eqref{s:diracPaux} and to analyze the effect on the fermionic projector.
In order to get contributions of comparable structure, these potentials should involve a vector field~$v$,
which should be equal either to the axial potential~$A_\text{\rm{a}}$ or to the corresponding axial
current~$j_\text{\rm{a}}$ (see Lemma~\ref{s:lemmalc1}).
Since contracting the vector index of~$v$ with the Dirac matrices would again give rise to
chiral potentials~\eqref{s:chiral}, we now prefer to contract~$v$
with partial derivatives. Moreover, since we want to compensate contributions
which are odd under parity transformations (i.e.\ which change signs if we flip the left-
and right-handed components), the resulting operator must involve
the pseudoscalar matrix~$\pseudo$. The requirement that the Dirac operator should be
symmetric with respect to the inner product~\eqref{s:iprod} leads us to the
ansatz involving an anti-commutator
\beq \label{s:pdp}
{\mathscr{B}} = \pseudo \left\{ v^j, \partial_j \right\} = 
2 \pseudo v^j \partial_j + \pseudo \big(\partial_j v^j \big)\:.
\eeq
We refer to this ansatz as a {\em{pseudoscalar differential potential}}.
\sindex{potential!pseudoscalar differential}%
Our ansatz seems unusual because such differential potentials do not occur in
the standard model nor in general relativity. However, as explained in~\S\ref{s:discussion},
we are free to modify the Dirac equation arbitrarily.

The corresponding leading contribution to the fermionic projector is of the form (for details
see equation~\eqref{s:sdlight} in Appendix~\ref{s:appspec})
\beq \label{s:sdlead} \begin{split}
P(x,y) \asymp&\: \frac{g}{2} \: \pseudo \xi_i \left(v^i(y) + v^i(x) \right) T^{(-1)} \\
&+ \frac{g}{2}\: \pseudo \xi_i  \int_x^y \left[ \slashed{\xi}, (\Pdd v^i) \right] T^{(-1)} 
\:+\: (\deg<2)\:.
\end{split}
\eeq
This contribution has a pole of order~$\xi^{-4}$ on the light cone and is therefore much more
singular than the desired logarithmic pole. 
A straightforward calculation shows that~\eqref{s:sdlead} does contribute to the expression
${\mathcal{R}}$ in Lemma~\ref{s:lemma81}, and thus we conclude that~\eqref{s:sdlead}
is not suitable for compensating the logarithmic pole.

The key for making use of the pseudoscalar differential potential~\eqref{s:pdp}
is to observe that the required logarithmic poles do appear to higher order in a mass expansion.
More precisely, to leading order at the origin, the cubic contribution to the fermionic projector is
\beq \label{s:v3}
P(x,y) \asymp  \frac{m^3}{4} \pseudo \left[ v_j^{(3)}(x) + \O \big( |\xi^0|+|\vec{\xi}| \big) \right]
\left( \slashed{\xi} \xi^j \, T^{(0)} - 2 \gamma^j\, T^{(1)} \right) + (\deg < -1)\:,
\eeq
where~$v^{(3)}$ is a Hermitian matrix composed of~$v$ and~$Y$, 
\beq \label{s:v3def}
v^{(3)} = i \left( v Y Y Y - Y v Y Y + Y Y v Y - Y Y Y v \right)
\eeq
(for details see equation~\eqref{s:sdlight3} in Appendix~\ref{s:appspec}).
Thus there is hope that the logarithmic poles can be compensated,
provided that we can arrange that the contributions by~\eqref{s:pdp} to~${\mathcal{R}}$
of order~$m^0$, $m$ and~$m^2$ in a mass expansion vanish.
The last requirement cannot be met if we consider one Dirac sea, because
the term~\eqref{s:sdlead} does contribute to~${\mathcal{R}}$.
But if we consider several Dirac seas, we have more freedom, as the pseudoscalar
differential potential~\eqref{s:pdp} can be chosen differently for each Dirac sea. For example, we
can multiply the potentials acting on the different Dirac seas by real constants~$\g_\alpha$,
\beq \label{s:van1}
({\mathscr{B}})^\alpha_\beta = \g_\alpha \,\delta_{\alpha \beta}\: \pseudo \left\{ v^j, \partial_j \right\} 
\qquad \text{with} \qquad \alpha, \beta=1,\ldots, g\:.
\eeq
Using this additional freedom, it is indeed possible to arrange that the contribution~\eqref{s:sdlead}
drops out of~${\mathcal{R}}$. This consideration explains why we must consider
{\em{several generations}} of elementary particles.
\sindex{generations!why several}%

The critical reader might object that there might be other choices of the operator~${\mathscr{B}}$
which could make it possible to compensate the logarithmic poles without the need
for several generations. However, the following consideration shows that~\eqref{s:pdp}
is indeed the only useful ansatz, provided that we work with local operators
(for nonlocal operators see~\S\ref{s:secgennonloc} 
and Section~\ref{s:secnonlocal}). The only zero order operator
are the chiral potentials~\eqref{s:chiral}, which were already considered in~\S\ref{s:sec72}.
Apart from~\eqref{s:pdp}, the only first order differential operator involving the vector field~$v$
and the pseudoscalar matrix~$\pseudo$ is the operator
\[ \pseudo \left\{ v_j \sigma^{jk},  \partial_k \right\} \:, \]
where~$\sigma^{jk} = \frac{i}{2} [\gamma^j, \gamma^k]$ are the bilinear covariants.
This ansatz can be shown to be useless, basically because the calculations in the continuum limit
give rise to contractions with the vector~$\xi$, which vanish (see also~\S\ref{s:sec88}).
Differential operators of higher order must involve the wave operator~$\Box$, which
applied to the Dirac wave functions gives rise to lower order operators. This shows that it is
not useful to consider differential operators of order higher than one.
We conclude that~\eqref{s:pdp} and its generalizations to several generations (like~\eqref{s:van1})
are indeed the only possible ans\"atze for compensating the logarithmic poles.

We end the discussion by having a closer look at the matrix~$v^{(3)}$, \eqref{s:v3def}.
Note that the ansatz~\eqref{s:van1} is diagonal in the generation index and thus
commutes with the mass matrix~$Y$. As a consequence, the matrix~$v^{(3)}$ vanishes.
This means that for compensating the logarithmic poles, the ansatz~\eqref{s:van1} is not
sufficient, but we must allow for non-zero off-diagonal elements in the generation index.
Thus we replace the factors~$b_\alpha$ in~\eqref{s:van1} by a Hermitian
matrix~$\g=(\g^\alpha_\beta)_{\alpha, \beta=1,\ldots, g}$, the so-called
{\em{generation mixing matrix}}.
\sindex{generation mixing matrix}%
\nindex{cf4@$\g$ -- generation mixing matrix}%
Later on, the generation mixing matrix will depend on the space-time point~$x$. This leads us to
generalize~\eqref{s:van1} by the ansatz
\beq \label{s:van2}
({\mathscr{B}})^\alpha_\beta =\pseudo \left\{ \g^\alpha_\beta(x) \:v^j(x), \partial_j
\right\} \:,
\eeq
thus allowing that the pseudoscalar differential potential mixes the generations.

\subsectionn{A Vector Differential Potential} $\quad$
Modifying the auxiliary Dirac equation~\eqref{s:diracPaux} by a first oder operator~\eqref{s:pdp}
or~\eqref{s:van2} changes the behavior of its solutions drastically. In particular, it is not clear whether
the operator~$\B$ can be treated perturbatively~\eqref{s:cpower}. In order to analyze and
resolve this problem, we begin by discussing the case when the potential~$v$ in~\eqref{s:pdp}
is a {\em{constant}} vector field, for simplicity for one Dirac sea of mass~$m$. Then taking the
Fourier transform, the Dirac equation reduces to the algebraic equation
\beq \label{s:Div}
(\slashed{k} -2 i \pseudo v^j k_j - m) \,\hat{\psi}(k) = 0 \:.
\eeq
Multiplying from the left by the matrix~$(\slashed{k} -2 i \pseudo v^j k_j + m)$, we find that
the momentum of a plane-wave solution must satisfy the dispersion relation
\[ k^2 - 4 (v^j k_j)^2 - m^2 = 0 \:. \]
Rewriting this equation as
\[ g^{ij} k_i k_j - m^2 = 0 \qquad \text{with} \qquad
g^{ij} := \eta^{ij} - 4 v^i v^j \:, \]
where~$\eta^{ij}=\text{diag}(1,-1,-1,-1)$ is again the Minkowski metric,
we see that the new dispersion relation is the same as that for the Klein-Gordon equation in
a space-time with Lorentzian metric~$g^{ij}$. In particular, the characteristics
of the Dirac equation become the null directions of the metric~$g^{ij}$.
In other words, the light cone is ``deformed'' to that of the new metric~$g^{ij}$.

This deformation of the light cone leads to a serious problem when we want to
compensate the logarithmic poles, as we now discuss. Suppose that we
introduce a pseudoscalar differential potential which according to~\eqref{s:van1}
or~\eqref{s:van2} depends on the generation index. In the case~\eqref{s:van1},
the Dirac seas feel different dispersion relations. In particular, the singularities of the
fermionic projector~$P(x,y)$ will no longer be supported on one light cone, but will be distributed
on the union of the light cones corresponding to the Lorentzian metrics~$g^{ij}_\alpha =
\eta^{ij} - 4 v^i_\alpha v^j_\alpha$. 
The ansatz~\eqref{s:van2} leads to a similar effect of a ``dissociation of the light cone.''
In the EL equations, this would lead to large additional contributions, which are
highly singular on the light cone and can certainly not compensate the logarithmic
poles.

Our method for bypassing this problem is to introduce another differential potential
which transforms the dispersion relation back to that of the Klein-Gordon equation
in Minkowski space. In the case of a constant vector field~$v$ and one generation,
this can be achieved by choosing matrices~$G^j$ which satisfy the anti-commutation relations
\[ \left\{ G^i, G^j \right\} = 2 \eta^{ij} + 8 v^i v^j \qquad \text{and} \qquad
\left\{ \pseudo, G^i(x) \right\} = 0 \:, \]
and by modifying~\eqref{s:Div} to
\[ (G^j  k_j -2 i \pseudo v^j k_j - m) \,\hat{\psi}(k) = 0 \:. \]
This modification of the Dirac matrices can be interpreted as introducing a constant
gravitational potential corresponding to the metric~$\eta^{ij} + 4 v^i_\alpha v^j_\alpha$.
This construction is extended to the general case~\eqref{s:van2} as follows.
We choose $(4g \times 4g)$-matrices $G^j(x)$ which 
are symmetric with respect to the inner product $\overline{\psi} \phi$ on the Dirac spinors
and satisfy the anti-commutation relations
\beq \label{s:acr}
\left\{ G^i(x), G^j(x) \right\} = 2 \eta^{ij} + 8 \,\g(x)^2 \,v^i(x)\, v^j(x) \qquad \text{and} \qquad
\left\{ \pseudo, G^i(x) \right\} = 0 \:.
\eeq
In the auxiliary Dirac equation~\eqref{s:diracPaux} we insert the additional operator
\beq \label{s:vdp}
\B = i \big(G^j(x)-\gamma^j \big) \partial_j + G^j(x) \,E_j(x) \:,
\eeq
where the matrices~$E_j$ involve the spin connection coefficients and are not of
importance here (for details see for example~\cite[Section~1.5]{PFP}).
We refer to~\eqref{s:vdp} as a {\em{vector differential potential}}.
\sindex{potential!vector differential}%
In the case~\eqref{s:van1}, this construction can be understood as introducing
for each Dirac sea a gravitational potential corresponding to the
metric~$\eta^{ij} + 4 \g_\alpha^2 v^i_\alpha v^j_\alpha$, whereas in case~\eqref{s:van2}, the 
interpretation is bit more complicated due to the off-diagonal terms.

\subsectionn{Recovering the Differential Potentials by a Local Axial Transformation} \label{s:sec86}
By introducing the differential potentials~\eqref{s:van2} and~\eqref{s:vdp} with~$G^j$
according to~\eqref{s:acr}, we inserted differential operators into the auxiliary Dirac
equation~\eqref{s:diracPaux}. We will now show that the effect of these operators on the 
solutions of the auxiliary Dirac equation can be described by a local
transformation
\beq \label{s:lut}
\psi_\text{aux}(x) \rightarrow U(x)\, \psi_\text{aux}(x)\:,
\eeq
\nindex{cf6@$U(x)$ -- local axial transformation}%
which is unitary with respect to the inner product~\eqref{s:iprodaux}.

Recall that we introduced the vector differential potential~\eqref{s:vdp} with the goal of
transforming the dispersion relation back to the form in the vacuum.
Thus if~$v$ is a constant vector field, the combination~\eqref{s:van2}+\eqref{s:vdp} leaves the momenta
of plane-wave solutions unchanged. This suggests that the sum~\eqref{s:van2}+\eqref{s:vdp} might
merely describe a unitary transformation of the Dirac wave functions. Thus 
we hope that there might be a unitary matrix~$U(x)$ such that
\[ U (i \Pdd - m Y) U^{-1} = i \Pdd + \eqref{s:pdp} + \eqref{s:vdp} \:. \]
Let us verify that there really is such a unitary transformation. The natural ansatz for~$U$
is an exponential of an axial matrix involving the vector field~$v$ and the generation mixing
matrix,
\beq \label{s:Udef}
U(x) = \exp \left(-i {\g}(x) \,\pseudo \gamma^j v_j(x) \right) .
\eeq
Writing out the exponential series and using that~$(\pseudo \gamma^j v_j)^2 = -v^2$,
we obtain
\beq \label{s:Udef2}
U(x) = \cos( {\g} \varphi)\: \1 -  i\,\frac{\sin({\g} \varphi)}{\varphi} \:\pseudo \slashed{v}\:,\qquad
U(x)^{-1} = \cos( {\g} \varphi)\: \1 +  i\,\frac{\sin({\g} \varphi)}{\varphi} \:\pseudo \slashed{v} \:,
\eeq
\nindex{cf8@$U(x)$ -- local axial transformation}%
where the angle~$\varphi := \sqrt{-v^2}$ is real or imaginary
(note that~\eqref{s:Udef2} is well-defined even in the limit~$\varphi \rightarrow 0$).
A short calculation yields
\[ U \gamma^j - \gamma^j U =
-i\,\frac{\sin ({\g} \varphi)}{\varphi} \left[ \pseudo \slashed{v}, \gamma^j \right]
= -2i \pseudo\, v^j\: \frac{\sin ({\g} \varphi)}{\varphi} \]
and thus
\begin{align}
U &(i \Pdd - m Y) \,U^{-1} = i U \gamma^j U^{-1} \partial_j + U \gamma^j (i \partial_j U^{-1})
- m U Y U^{-1} \nonumber \\
&=  i \Pdd + 2 \pseudo v^j\: \frac{\sin ({\g} \varphi)}{\varphi}\:U^{-1} \partial_j +
U \gamma^j (i \partial_j U^{-1}) - m U Y U^{-1} \nonumber \\
&=i \Pdd + \pseudo\: \frac{\sin (2 {\g} \varphi)}{\varphi}\:v^j \partial_j   +
2i \: \frac{\sin^2 ({\g} \varphi)}{\varphi^2}\: \slashed{v}\: v^j \partial_j
+ U \gamma^j (i \partial_j U^{-1}) - m U Y U^{-1}\:. \label{s:Dtrans}
\end{align}
In order to verify that the resulting Dirac operator allows us to recover both~\eqref{s:van2} and~\eqref{s:vdp},
we assume that~$v^2$ is so small that~$\sin (2 {\g} \varphi) \approx 2
{\g} \varphi$ and~$\sin^2 ({\g} \varphi) \approx \g^2
\varphi^2$. Then the second summand in~\eqref{s:Dtrans}
reduces precisely to the differential operator in the relation~\eqref{s:van2}.
The third summand in~\eqref{s:Dtrans} gives precisely the differential operator in~\eqref{s:vdp},
noting that~\eqref{s:acr} has the solution~$G^j = \gamma^j + 2 \g^2 \slashed{v} v^j + \O(v^4)$.
Likewise, a direct calculation shows that the multiplication operators in~\eqref{s:van2} and~\eqref{s:vdp}
are contained in the fourth summand in~\eqref{s:Dtrans}.
Writing out the fourth and fifth summands in~\eqref{s:Dtrans}, one finds a
rather complicated combination of additional chiral, scalar, pseudoscalar and even bilinear potentials.
These additional potentials do not cause any problems; on the contrary, they guarantee that
the total transformation of the Dirac wave functions simply is the local transformation~\eqref{s:lut}.
We conclude that with~\eqref{s:Dtrans} we have found a Dirac operator which
includes the differential potentials in~\eqref{s:van2} and~\eqref{s:vdp}. It has the nice property that
it can easily be treated non-perturbatively by the simple local transformation~\eqref{s:lut}.
We refer to the transformation~\eqref{s:lut} with~$U$ according to~\eqref{s:Udef}
as the {\em{local axial transformation}}.
\sindex{transformation of the fermionic projector!local axial}%

The local axial transformation was analyzed in detail in a previous version of the present
work (see~arXiv:0908.1542v3 [math-ph]). There are two reasons why the local axial
transformation will no longer be used here. First, the unitarity of~\eqref{s:lut} turns out not to be essential,
as it can be dropped in the more general construction given in~\S\ref{s:secgenlocal} below.
Second and more importantly, compensating the logarithmic poles of the current and and mass terms
by a local axial transformation leads to additional contributions to the fermionic projector
of higher order in the local transformation (for details see Appendix~C in~arXiv:0908.1542v3 [math-ph]).
It turns out that the gauge phases which appear in these additional contributions
(which were not considered in~arXiv:0908.1542v3 [math-ph]) enter the EL equations
in a way which makes it impossible to satisfy these equations.
This problem was first observed when working out the follow-up paper on systems involving
neutrinos (see Chapter~\ref{lepton}). The method for overcoming this problem also led to major revisions
of the present paper. The basic problem will be explained in~\S\ref{s:secprobaxial} and resolved
in~\S\ref{s:secnonlocaxial}.

Before entering the generalizations and discussing the shortcomings of the local axial transformation,
we now briefly review how the local axial transformation can be used to
compensate the logarithmic poles of the current and mass terms.
For simplicity, we only consider a perturbation expansion to first order in~$v$.
Then the transformation~\eqref{s:Udef} simplifies to
\beq \label{s:Uexp}
U(x) = \1 - i \g \pseudo \, \slashed{v}(x) + \O(v^2)\:.
\eeq
Transforming the auxiliary fermionic projector~\eqref{s:Pauxvac} by~$U$ and forming the
sectorial projection~\eqref{s:pt}, we obtain for the perturbation of the fermionic projector the expression
\beq \label{s:Pform}
P \asymp -i  \pseudo \slashed{v} \,\acute{\g} \grave{P} + i \acute{P} \grave{\g} \,\pseudo \slashed{v} 
+ \O(v^2)\:,
\eeq
where we denoted the sectorial projection similar to~\eqref{s:tildedef} by accents.
Here we always sum over one index of the generation mixing matrix.
Thus it is convenient to introduce real functions~$c_\alpha$ and~$d_\alpha$ by
\beq \label{s:cddef}
\sum_{\alpha=1}^g \g^\alpha_\beta = c_\beta + i d_\beta \qquad \text{and} \qquad
\sum_{\beta=1}^g \g^\alpha_\beta = c_\alpha - i d_\alpha \:,
\eeq
\nindex{cg0@$c_\alpha, d_\alpha$ -- sectorial projection of $\g$}%
where the last equation is verified by taking the adjoint of the first and using
that~$\g$ is Hermitian. Combining these equations with the fact that the auxiliary fermionic
projector of the vacuum is diagonal on the generations, we can write~\eqref{s:Pform} as
\beq \label{s:Pabasy}
P \asymp \sum_{\beta=1}^g \Big( -i \left[ c_\beta \pseudo \slashed{v}, P_\beta \right] 
+ \left\{ d_\beta \pseudo \slashed{v}, P_\beta \right\} \Big)  + \O(v^2)\:,
\eeq
where the~$P_\beta$ stand for the direct summands in~\eqref{s:Pauxvac}.
The next lemma shows that the functions~$c_\beta$ drop out of the EL equations;
the proof is again given in Appendix~\ref{s:appspec}.
\begin{Lemma} \label{s:lemmalogterm1} $\quad$
The perturbation of the fermionic projector
by the functions~$c_\beta$ in~\eqref{s:Pabasy}
does not contribute linearly to the EL equations.
\end{Lemma} \noindent
This leaves
us with the~$g$ real coefficients~$d_\beta$. In order to study their effect,
we expand~$P_\beta$ in~\eqref{s:Pabasy} in powers of the mass. The zeroth order
in the mass expansion vanishes in view of the identity
\beq \label{s:cdm0}
\sum_{\beta=1}^g d_\beta  = 0
\eeq
(which follows immediately from~\eqref{s:cddef} by summing over the free generation index).
The first order in the mass expansion yields instead of a logarithmic pole a stronger singularity~$\sim \xi^{-2}$
and contributes to the EL equations even to degree five on the light cone.
In order for these contributions to vanish, we need to impose that
\beq \label{s:cdm1}
\sum_{\beta=1}^g m_\beta\, d_\beta = 0\:.
\eeq
The remaining contributions to EL equations are indeed of degree four. They
involve logarithmic poles, making it possible to compensate the corresponding poles
of the current and mass terms.

It is worth noting that this procedure only works if the number of generations equals three.
Namely, in the case~$g <3$, the equations~\eqref{s:cdm0} and~\eqref{s:cdm1} only have the
trivial solution~$d_\beta \equiv0$. 
In the case~$g=3$, the conditions~\eqref{s:cdm0} and~\eqref{s:cdm1}
leave one free constant, which is uniquely fixed by compensating the logarithmic pole.
In the case~$g>3$, however, there are free constants even after compensating the logarithmic pole,
so that the resulting field equations are underdetermined.

In order to avoid confusion, we finally point out that it is crucial that the local axial transformation
is performed {\em{before}} taking the sectorial projection.
Namely, if we performed a similar transformation after taking the sectorial projection,
\beq \label{s:Pgauge2}
P(x,y) \rightarrow V(x)\, P(x,y)\, V(y)^* \qquad \text{with} \qquad
V(x) = e^{- i \pseudo \slashed{B}(x)}
\eeq
with a real vector field~$V$, then~$V(x)$ would be a unitary transformation on the spinors.
Such a transformation can be regarded as a {\em{local gauge transformation}}
(see~\cite[Sections~1.5 and~3.1]{PFP}). Since our action is gauge invariant (see~\cite[Section~3.5]{PFP}
or~\cite{discrete}), it would have no effect on our physical system.
By performing a gauge transformation~\eqref{s:Pgauge2} with~$B = -\hat{g}\, v$, we
can always arrange that the matrix~$\g$ in the local axial transformation~\eqref{s:Udef}
has the property
\beq \label{s:E0rel}
\hat{g}(x) = 0 \qquad \text{for all~$x$}\:.
\eeq

\subsectionn{General Local Transformations} \label{s:secgenlocal}
We now generalize the local axial transformation~\eqref{s:lut} and bring it in connection
to the causal perturbation expansion of~\S\ref{s:sec42}.
\sindex{transformation of the fermionic projector!general local}%
Thus suppose that~$U(x)$ is a linear transformation of the spinors at every space-time point.
For simplicity, we assume that~$U(x)$ is invertible and depends smoothly on~$x$,
but it need not be a symmetric operator.
\nindex{cg2@$U(x)$ -- general local transformation}%
We then transform the Dirac operator with interaction
locally by~$U(x)^{-1}$,
\beq \label{s:dirloc}
i \Pdd + \B - m Y \longrightarrow \big( U(x)^{-1} \big)^* \,\big( i \Pdd_x + \B - m Y \big) \,
U^{-1}(x) \:.
\eeq
Writing the transformed Dirac operator in the form~$i \Pdd + {\mathcal{B}}- mY$
with a new perturbation operator~${\mathcal{B}}$, we can again introduce the
corresponding fermionic projector as outlined in Section~\ref{s:sec4}
with the causal perturbation expansion followed by the light-cone expansion and a resummation
of the perturbation expansion.

Introducing the perturbation operator by the local transformation~\eqref{s:dirloc} has the
advantage that the effect on the fermionic projector can be described explicitly. In preparation,
we consider the advanced and retarded Green's functions~$s^\vee$ and~$s^\wedge$, which
are characterized by the equations
\[ (i \Pdd + \B - m Y) \, s^\wedge(x,y) = \delta^4(x-y) = (i \Pdd + \B - m Y) \, s^\vee(x,y) \]
and the fact that they are supported in the upper and lower light cone, respectively
(for details see~\cite[Section~2.2]{PFP}, \cite{light} or Section~\ref{secfpext}). A direct calculation shows that
under the transformation~\eqref{s:dirloc}, the Green's functions simply transform locally by
\[ s^{\vee\!/\!\wedge}(x,y) \longrightarrow U(x)\, s^{\vee \!/\! \wedge}(x,y)\, U(y)^{-1}\:. \]
The transformation of the auxiliary fermionic projector is more involved, because
one has nonlocal contributions and must satisfy normalization conditions
(for details see~\cite{grotz, norm} and~\cite{light}). However, the residual argument
(see~\cite[Section~3.1]{light}, \cite[Section~6]{grotz} or Section~\ref{seclight}) shows that the
poles of the auxiliary fermionic projector have the same structure as those of the
causal Green's functions, i.e.
\beq \label{s:lct}
\tilde{P}^\text{aux}(x,y) = U(x)\, \tilde{P}^\text{aux}_\B(x,y)\, U(y)^* +
\text{(smooth contributions)}\:,
\eeq
where~$\tilde{P}^\text{aux}_\B$ denotes the auxiliary fermionic projector
in the presence of the external field~$\B$.
Thus as long as we are concerned with the poles of the fermionic projector,
we can work with the simple local transformation~\eqref{s:lct} of the auxiliary fermionic projector.
The fermionic projector is again obtained by forming the sectorial projection~\eqref{s:pt}.
Again using the notation~\eqref{s:tildedef}, we thus obtain
\beq \label{s:UPxy}
\tilde{P}(x,y) = \acute{U}(x)\, (x,y)\, \tilde{P}^\text{aux}_\B(x,y)\, \grave{U}(y)^* +
\text{(smooth contributions)} \:.
\eeq

This construction shows that the unitarity condition for~$U(x)$ used in~\eqref{s:lut}
is not needed. Moreover, since the construction is based on the causal perturbation
expansion for the new perturbation~${\mathcal{B}}$ (which also incorporates the local
transformation), we do not need to worry about the normalization of the fermionic states.
In other words, the auxiliary fermionic projector defined by~\eqref{s:lct} is
idempotent (as is made precise in~\cite{norm}).
The prize we pay is that the additional smooth contributions in~\eqref{s:UPxy} are not
known explicitly. But they could be computed with the resummation technique
developed in Appendix~\ref{s:appresum}.

\subsectionn{The Shear Contributions by the Local Axial Transformation} \label{s:secprobaxial}
In order to explain the problem of the local axial transformation~\eqref{s:Uexp}, we need to analyze
its effect on the fermionic projector to higher order in the vector field~$v$.
The nonlinear dependence has two reasons: First, $U$ depends nonlinearly on~$v$
(see~\eqref{s:Udef}) and second, the transformation~\eqref{s:UPxy} involves two factors of~$U$.
For clarity, we want to begin with the second effect. To this end, we simplify~$U$ to a
transformation which is linear in~$v$,
\beq \label{s:Ulin}
U(x) = \1 + i E(x) \qquad \text{with} \qquad E(x) = -\g \pseudo \, \slashed{v}(x) \:.
\eeq
Note that, in contrast to~\eqref{s:Uexp}, we no longer have an error term.
As a consequence, the transformation~$U(x)$ is no longer unitary. Nevertheless,
the general construction in~\S\ref{s:secgenlocal} applies (see~\eqref{s:dirloc}--\eqref{s:UPxy}).
Clearly, the linear ansatz~\eqref{s:Ulin} leaves us with the freedom to perform additional local transformations
which are of higher order in~$v$; these will be considered in Appendix~\ref{s:applocaxial}.

Using~\eqref{s:E0rel}, the effect of the local axial transformation~\eqref{s:UPxy}, \eqref{s:Ulin} on
the vacuum fermionic projector~\eqref{s:Pauxvac} is described by the transformation
\[ P(x,y) \rightarrow P(x,y) + \Delta P(x,y) \:, \]
where~$\Delta P$ is computed by
\begin{align}
\Delta P(x,y) \:=\:&i m \big( \acute{E}(x) \grave{Y} - \acute{Y} \grave{E}(y)^* \big) \,T^{(0)}_{[1]}
\label{s:Y1} \\
&+ \frac{i}{2}\: \acute{E}(x)\, \slashed{\xi}\, \grave{E}(y)^*\: T^{(-1)}_{[0]} \label{s:E20} \\
&+ m \:\acute{E}(x) \,Y \grave{E}(y)^* \:T^{(0)}_{[1]} \:+\: \slashed{\xi} (\deg < 2) + (\deg < 1)\:. \label{s:E21}
\end{align}
The contribution~\eqref{s:Y1} was already considered in~\S\ref{s:sec86};
it vanishes in view of~\eqref{s:cdm1}. The logarithmic poles on the light cone are
contained in the error terms in~\eqref{s:E21}.
To higher order in~$v$, we get the additional contributions~\eqref{s:E20} and~\eqref{s:E21},
which need to be taken into account.

Let us analyze the effect of the most singular contribution~\eqref{s:E20} in more detail.
Using~\eqref{s:Ulin}, we obtain
\begin{align}
\Delta P(x,y) &\asymp \frac{i}{2}\: \acute{\g} \grave{\g} \, \slashed{v} \slashed{\xi} \slashed{v}\: T^{(-1)}_{[0]} 
+ (\deg < 2) \label{s:Pcont} \\
\Delta A_{xy} &\asymp \Delta P(x,y) \, P(y,x) + P(x,y)\, \Delta P(y,x) + \O(v^4) \notag \\
&\!\!\overset{\eqref{s:Peq}}{=} \frac{g}{4}\: \acute{\g} \grave{\g}\:
\left( \slashed{v} \slashed{\xi} \slashed{v} \overline{\slashed{\xi}} + \slashed{\xi} \slashed{v} \overline{\slashed{\xi}} \slashed{v} \right)\:
T^{(-1)}_{[0]} \,\overline{T^{(-1)}_{[0]}} + \O(v^4) \notag \\
&= g\: \acute{\g} \grave{\g}\: \la v, \xi \ra^2 \:
T^{(-1)}_{[0]} \,\overline{T^{(-1)}_{[0]}} + \O(v^4) + (\deg < 4) \label{s:Acont} \\
\Delta \lambda_\pm &\asymp g\: \acute{\g} \grave{\g}\: \la v, \xi \ra^2 \:
T^{(-1)}_{[0]} \,\overline{T^{(-1)}_{[0]}} + \O(v^4) + (\deg < 4) \:. \label{s:lambdacont}
\end{align}
Thus we get a contribution to the closed chain which is of degree four on the light cone
and is thus more singular than the closed chain of the vacuum (see~\eqref{s:clcvac}).
We call this contribution the {\em{shear contribution of the local axial transformation}}
\sindex{shear contribution!by the local axial transformation}%
(the connection to the parameter~$\varepsilon_\text{shear}$ in~\eqref{s:shear} will
become clear in~\S\ref{s:secnonlocaxial} below).
The resulting contribution to the eigenvalues is also of degree four on the light cone.
This by itself is not a problem because the
contribution to~$A_{xy}$, \eqref{s:Acont}, is a real multiple of the identity matrix.
Thus it changes both eigenvalues by the same real amount (see~\eqref{s:lambdacont}).
As a consequence, the eigenvalues~$\lambda_\pm$ remain complex conjugate pairs~\eqref{s:labs},
so that the Lagrangian is still zero.

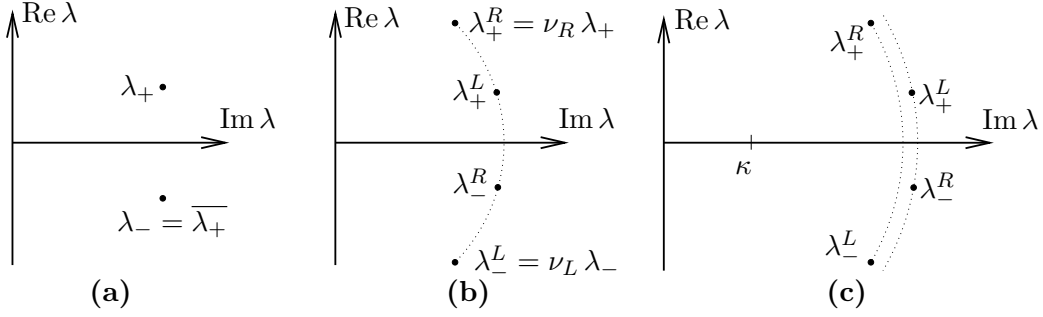
\begin{figure}
\begin{picture}(0,0)%
\includegraphics{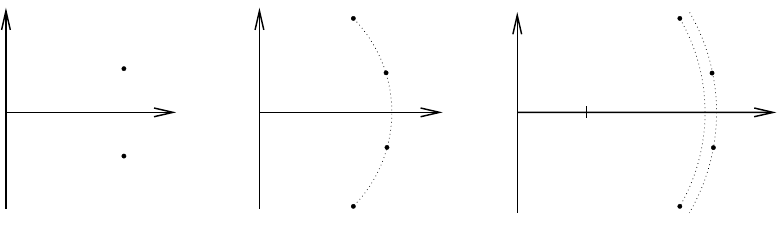}%
\end{picture}%
\setlength{\unitlength}{1533sp}%
\begingroup\makeatletter\ifx\SetFigFont\undefined%
\gdef\SetFigFont#1#2#3#4#5{%
  \reset@font\fontsize{#1}{#2pt}%
  \fontfamily{#3}\fontseries{#4}\fontshape{#5}%
  \selectfont}%
\fi\endgroup%
\begin{picture}(16025,4992)(-5701,-4567)
\put(-5429, 86){\makebox(0,0)[lb]{\smash{{\SetFigFont{11}{13.2}{\rmdefault}{\mddefault}{\updefault}$\re \lambda$}}}}
\put(1530,-1166){\makebox(0,0)[lb]{\smash{{\SetFigFont{11}{13.2}{\rmdefault}{\mddefault}{\updefault}$\lambda^L_+$}}}}
\put(1537,-2682){\makebox(0,0)[lb]{\smash{{\SetFigFont{11}{13.2}{\rmdefault}{\mddefault}{\updefault}$\lambda^R_-$}}}}
\put(-3839,-1092){\makebox(0,0)[lb]{\smash{{\SetFigFont{11}{13.2}{\rmdefault}{\mddefault}{\updefault}$\lambda_+$}}}}
\put(-3884,-3286){\makebox(0,0)[lb]{\smash{{\SetFigFont{11}{13.2}{\rmdefault}{\mddefault}{\updefault}$\lambda_-=\overline{\lambda_+}$}}}}
\put(6106,-2360){\makebox(0,0)[lb]{\smash{{\SetFigFont{11}{13.2}{\rmdefault}{\mddefault}{\updefault}$\kappa$}}}}
\put(-216, 11){\makebox(0,0)[lb]{\smash{{\SetFigFont{11}{13.2}{\rmdefault}{\mddefault}{\updefault}$\re \lambda$}}}}
\put(5124,-19){\makebox(0,0)[lb]{\smash{{\SetFigFont{11}{13.2}{\rmdefault}{\mddefault}{\updefault}$\re \lambda$}}}}
\put(9066,-1240){\makebox(0,0)[lb]{\smash{{\SetFigFont{11}{13.2}{\rmdefault}{\mddefault}{\updefault}$\lambda^L_+$}}}}
\put(9089,-2750){\makebox(0,0)[lb]{\smash{{\SetFigFont{11}{13.2}{\rmdefault}{\mddefault}{\updefault}$\lambda^R_-$}}}}
\put(-2229,-1617){\makebox(0,0)[lb]{\smash{{\SetFigFont{11}{13.2}{\rmdefault}{\mddefault}{\updefault}$\im \lambda$}}}}
\put(3253,-1624){\makebox(0,0)[lb]{\smash{{\SetFigFont{11}{13.2}{\rmdefault}{\mddefault}{\updefault}$\im \lambda$}}}}
\put(10108,-1646){\makebox(0,0)[lb]{\smash{{\SetFigFont{11}{13.2}{\rmdefault}{\mddefault}{\updefault}$\im \lambda$}}}}
\put(-4333,-4426){\makebox(0,0)[lb]{\smash{{\SetFigFont{11}{13.2}{\rmdefault}{\mddefault}{\updefault}{\bf{(a)}}}}}}
\put(1435,-4426){\makebox(0,0)[lb]{\smash{{\SetFigFont{11}{13.2}{\rmdefault}{\mddefault}{\updefault}{\bf{(b)}}}}}}
\put(7585,-4426){\makebox(0,0)[lb]{\smash{{\SetFigFont{11}{13.2}{\rmdefault}{\mddefault}{\updefault}{\bf{(c)}}}}}}
\put(1797,-72){\makebox(0,0)[lb]{\smash{{\SetFigFont{11}{13.2}{\rmdefault}{\mddefault}{\updefault}$\lambda^R_+=\nu_R \,\lambda_+$}}}}
\put(1880,-3927){\makebox(0,0)[lb]{\smash{{\SetFigFont{11}{13.2}{\rmdefault}{\mddefault}{\updefault}$\lambda^L_-=\nu_L \,\lambda_-$}}}}
\put(7632,-371){\makebox(0,0)[lb]{\smash{{\SetFigFont{11}{13.2}{\rmdefault}{\mddefault}{\updefault}$\lambda^R_+$}}}}
\put(7551,-3684){\makebox(0,0)[lb]{\smash{{\SetFigFont{11}{13.2}{\rmdefault}{\mddefault}{\updefault}$\lambda^L_-$}}}}
\end{picture}%
\caption{The eigenvalues of~$A_{xy}$ to degree five on the light cone
in the vacuum {\bf{(a)}}, in the presence of chiral gauge fields {\bf{(b)}},
and after a local axial transformation {\bf{(c)}}}
\label{s:figabslambda}
\end{figure}

However, the situation becomes more involved when the gauge phases are taken into
account. Namely, according to~\eqref{s:Pchiral} and~\eqref{s:Lambda}, we need to
multiply the fermionic projector of the vacuum by the phase
factor~$(\chi_L \,e^{-i \Lambda^{xy}_L} + \chi_R \,e^{-i \Lambda^{xy}_R})$.
As the local axial transformation is odd and thus flips the chirality, \eqref{s:Pcont} becomes
\beq \label{s:badphase}
\Delta P(x,y) \asymp
\frac{i}{2}\: \acute{\g} \grave{\g} \,
\left( \chi_L \,e^{-i \Lambda^{xy}_R} + \chi_R \,e^{-i \Lambda^{xy}_L}  \right)
\left( \slashed{v} \slashed{\xi} \slashed{v} \right) \: T^{(-1)}_{[0]} \:.
\eeq
In view of~\eqref{s:Pchiral}, the phases cancel in the closed chain~\eqref{s:Acont}.
We conclude that the contribution to the eigenvalues~\eqref{s:lambdacont} does not
involve any gauge phases, i.e.
\[ \Delta \lambda^{L\!/\!R}_\pm \asymp \kappa \qquad \text{with} \qquad
\kappa := g\: \acute{\g} \grave{\g}\: \la v, \xi \ra^2 \:
T^{(-1)}_{[0]} \,\overline{T^{(-1)}_{[0]}} + \O(v^4) + (\deg < 4) \:. \]
Combining this result with~\eqref{s:unperturb}, we obtain
\beq \label{s:lambdak}
\lambda^{L\!/\!R}_\pm = \nu_{L\!/\!R}\: \lambda_\pm + \kappa \:.
\eeq
As a consequence, the eigenvalues~$\lambda^{L\!/\!R}_\pm$ will in general no longer
all have the same absolute value (see Figure~\ref{s:figabslambda}). 
Thus the Lagrangian~\eqref{s:Lcrit} does not vanish, and the EL equations are no longer
satisfied to degree five on the light cone.

As a possible method to bypass this problem, one could hope to compensate
the contribution~\eqref{s:E20} by other local transformations of the form~\eqref{s:lct}.
For example, one could work with a local transformation involving a bilinear
potential~$B_{ij}$ with
\[ v_i \,v_j = B_{ik} \,B_j^{\;\,k}\:. \]
Another potential method is to work with the contributions to~\eqref{s:Udef2} of higher order
in~$v$ (which were disregarded in~\eqref{s:Ulin}). It turns out that all these methods
necessarily fail. We now merely explain the underlying reason and refer for the detailed analysis
to Appendix~\ref{s:applocaxial}.
The basic problem of the local axial transformation is that it generates
a shear contribution to the closed chain (see~\eqref{s:Acont}).
The appearance of such contributions can be understood as follows.
The vector~$\xi = y-x$ is null on the light cone. As a consequence, contracting
the vectors~$\xi$ and~$\overline{\xi}$ gives factors~$z$ and~$\overline{z}$,
which vanish on the light cone (see the calculation~\eqref{s:clcvac}--\eqref{s:lpm}).
If, as a consequence of the local axial transformation, the vector~$\xi$ is no longer
null on the light cone, then the inner product~$\la \xi, \overline{\xi} \ra$ no longer
vanishes on the light cone. This gives rise to contributions to the closed chain
which are more singular on the light cone. These are precisely the shear contributions.
Thus in order to prevent shear contributions, one must make sure that the
factors~$\xi$ remain lightlike on the light-cone.

A lightlike vector~$\xi$ has the special property that the bilinear
form~$\overline{\psi} \slashed{\xi} \phi$ is positive semi-definite, but degenerate (i.e.\ 
it is {\em{not}} positive definite).
Indeed, these two properties encode the fact that~$\slashed{\xi}^2=0$.
A local transformation~$\slashed{\xi} \rightarrow V^* \slashed{\xi} V$
preserves the definiteness of the corresponding bilinear form,
because~$\overline{\psi} (V^* \slashed{\xi} V) \phi = \overline{V \psi} \,\slashed{\xi}\, (V \phi) \geq 0$.
If~$V$ is invertible, the inner product also remains degenerate.
However, the transformation~$\grave{U}^* \::\: \C^4 \rightarrow \C^{4g}$
in~\eqref{s:UPxy} is certainly not invertible. As a consequence, the bilinear form~$\overline{\psi} \acute{U} \slashed{\xi}
\grave{U}^* \phi$ will in general be positive definite, giving rise to the undesirable shear contributions.
The condition that the bilinear form~$\overline{\psi} \acute{U} \slashed{\xi}
\grave{U}^* \phi$ be degenerate gives rise to strong constraints for the local transformation.
As worked out in Appendix~\ref{s:applocaxial}, these constraints are not compatible with
local axial transformations.

\subsectionn{Homogeneous Transformations in the High-Frequency Limit} \label{s:secnonlocaxial}
In order to resolve the problems caused by the shear contributions,
\sindex{shear contribution!by the local axial transformation}%
we now reconsider the local axial transformation in the special case of a constant transformation.
We again write~$U$ in the form~\eqref{s:Ulin}. Transforming the vacuum fermionic projector~\eqref{s:Pauxvac}
and disregarding the smooth contributions in~\eqref{s:UPxy}, we obtain
\beq \label{s:Pconst}
\tilde{P} = \acute{U} P^\text{aux} \,\grave{U}^*
= P - i \pseudo \slashed{v} \:(\acute{\g} \,\grave{P}^\text{aux})
+ i (\acute{P}^\text{aux} \,\grave{\g}) \:\pseudo \slashed{v}
+ \pseudo \slashed{v} \:(\acute{\g} \,P^\text{aux} \,\grave{\g}) \:\pseudo \slashed{v} \:.
\eeq
As outlined in~\S\ref{s:sec86}, the contributions linear in~$v$ give rise to the desired
axial terms with logarithmic poles on the light cone. The problematic shear contribution, however,
is quadratic in~$v$ (see~\eqref{s:Pcont} and the discussion thereafter).
For the unregularized fermionic projector~\eqref{s:A}, this shear
contribution can be written in momentum space as
\begin{align}
\tilde{P} &\asymp \sum_{\beta=1}^g \acute{\g}_\alpha \slashed{v} \slashed{k} \slashed{v}\,\grave{\g}^\alpha
\: \delta(k^2-m_\alpha^2)\: \Theta(-k^0) \\
&= \sum_{\beta=1}^g \acute{\g}_\alpha \Big( 2 \la v, k \ra\, \slashed{v} - v^2 \, \slashed{k} \Big) \grave{\g}^\alpha\;
\delta(k^2-m_\alpha^2)\: \Theta(-k^0)\:. \label{s:Pt2}
\end{align}
(With regularization, one transforms similarly the distribution~\eqref{s:Peps}.
This has no effect on the following consideration if one keeps in mind that
the vector field~$v^\varepsilon$ is time-like and past-directed
according to our assumption~(vii).)
Since the momentum~$k$ is timelike and past-directed, a short calculation shows that
the vector~$2 \la v, k \ra\, v - v^2 \, k$ is non-spacelike and past-directed (no matter
how~$v$ is chosen). As a consequence, adding up the contributions by different Dirac seas,
we necessarily obtain a timelike and past-directed vector, even if by the replacement~$v \rightarrow v_\beta$
we allowed the vector field in~\eqref{s:Pt2} to depend on the generation index~$\beta$.
This means that the contributions by different Dirac seas cannot compensate each other.
Transforming to position space, we necessarily get a contribution to~$\tilde{P}$ which is
highly singular on the light cone, leading to the problems discussed in~\S\ref{s:secprobaxial}. 
Using the decomposition~\eqref{s:vepsrel}, one sees that~$\tilde{P}$ violates the condition~\eqref{s:shear}.
\nindex{ca6@$\varepsilon_\text{shear}$ -- shear parameter}%
We can thus say that the local axial transformation introduces a shear of the surface states
(see also~\cite[Section~4.4]{PFP}). This also explains why we referred to~\eqref{s:Acont} as the
shear contribution.

In order to bypass the problems caused by the shear contributions, we need to introduce
an axial contribution to~$\tilde{P}$ without generating a vectorial contribution
of the form~\eqref{s:Pt2}. To explain the method, we consider
in generalization of~\eqref{s:Pconst} the transformation
\beq \label{s:Pnonloc}
\tilde{P}(k) = \acute{U}(k)\: P^\text{aux}(k)\: \grave{U}(k)^* 
= \sum_{\beta=1}^g  U_\beta(k)\: P_\beta(k)\: U_\beta(k)^* \:,
\eeq
where~$U(k)$ is a multiplication operator in momentum space.
\nindex{cg6@$U(k)$ -- nonlocal homogeneous transformation}%
Before going on, we
briefly discuss this ansatz. We first point out that the operator~$U(k)$ can be an arbitrary
multiplication operator in momentum space. Representing it in position space gives rise
to a convolution operator with an integral kernel~$U(x-y)$ which will in general be non-zero
if~$x \neq y$. Thus~\eqref{s:Pnonloc} describes a {\em{nonlocal homogeneous transformation}}.
\sindex{homogeneous transformation!in the high-frequency limit}%
We shall see that the freedom to choose~$U$ as a function of~$k$ will be essential for
our construction to work. Clearly, \eqref{s:Pnonloc} gives a lot of freedom to modify
the fermionic projector. On the other hand, we will see in the subsequent calculations that~\eqref{s:Pnonloc}
is quite restrictive is the sense that it imposes inequalities on the vector and axial components
of~$\tilde{P}(k)$. The basic reason for these restrictions can be understood from the fact
that~$-\tilde{P}(k)$ is necessarily a {\em{positive operator}}, meaning that
\[ - \overline{\psi} \tilde{P}(k) \psi \geq 0 \qquad \text{for every spinor~$\psi$} \:. \]
Namely, taking the expectation value of~\eqref{s:Pnonloc}, we obtain
\[ \overline{\psi} \tilde{P}(k) \psi = \sum_{\beta=1}^g \overline{(U_\beta^* \psi)} P_\beta(k) (U_\beta^* \psi)
\leq 0 \:, \] 
where in the last step we used the fact that the operators~$-(\slashed{k}+m_\beta)$ are obviously positive
for~$k$ on the lower mass shell. We remark that this positivity of~$(-\tilde{P})$ in momentum space
gives rise to corresponding positivity properties of the fermionic projector in position space;
these are worked out in Appendix~\ref{s:applocaxial}. For general properties of positive operators
in indefinite inner product space we refer to~\cite[Section~4]{discrete}.

We first analyze the effect of the above transformations in the
{\em{high-frequency limit}}. Thus we consider a momentum~$k=(\omega, \vec{k})$
on the mass cone and consider the asymptotics as~$\Omega:=|\omega| \rightarrow \infty$.
Then the projector onto the states of momentum~$k$ of rest mass~$m$ converges after
suitable rescaling,
\beq \label{s:Oscale}
\frac{1}{\Omega} \: (\slashed{k}+m) \xrightarrow{\Omega \rightarrow \infty}
\hat{\slashed{k}} \qquad \text{with} \qquad \hat{k} := (-|\vec{k}|, \vec{k})\:.
\eeq
Note that the operator~$\hat{\slashed{k}}$ is nilpotent and negative semi-definite
(in the sense that $\Sl \psi | \hat{\slashed{k}} \psi \Sr \leq 0$ for every spinor~$\psi$).
We want to generate an axial contribution of the
form~$\pseudo \slashed{v}$ (independent of~$k$). Thus in view of the scaling~\eqref{s:Oscale},
our goal is to perturb the operator~$\hat{k}$ such as to generate an axial contribution of the form
\beq \label{s:axialwanted}
\frac{1}{\Omega}\: \pseudo \slashed{v} \qquad \text{with a timelike vector~$v$}\:.
\eeq
If this perturbation is described by a unitary transformation, the resulting operator will
again be nilpotent and negative semi-definite. According to the methods in~\S\ref{s:secgenlocal},
however, the perturbation can also be described by a non-unitary transformation.
But even then the semi-definiteness is preserved, and the operator will also remain
close to a nilpotent operator provided that we want to vary its eigenvalues only slightly.
Therefore, in order to to get an idea for how to perturb the states,
it is helpful to characterize the general form of nilpotent and negative semi-definite operators.
\begin{Lemma} Suppose that the linear operator~$A$ is nilpotent and negative semi-definite. Then~$A$
has the representation
\beq \label{s:Arep2}
A = \left(\slashed{q} + a\, i \pseudo \right) \Big(\1 + i \, \big( \slashed{u} + b\, i \pseudo \big) \Big)
\eeq
with a non-spacelike past-directed vector~$q$ and a non-spacelike vector~$u$ as well as real
parameters~$a$ and~$b$ such that
\[ \la q, u \ra = ab \]
and moreover one of the following two alternative conditions holds:
\beq \label{s:cases}
\left\{ \begin{split} q^2 &=a^2 &&\quad\text{and}\quad&  -1 &\leq u^2-b^2 \leq 0 \\
q^2 &> a^2 &&\quad \text{and} \quad& -1 &= u^2 - b^2 \:.
\end{split} \right.
\eeq
\end{Lemma}
\Proof It is convenient to interpret the component proportional to~$i \pseudo$ as
an additional spatial coordinate. To this end, we introduce the matrices~$\Gamma^0, \ldots, \Gamma^4$
by~$\Gamma^i = \gamma^i$ if~$i=0,\ldots, 3$ and~$\Gamma^4 = i \pseudo$. These
matrices generate a $5$-dimensional Clifford algebra of signature~$(1,4)$.
Introducing the vectors~$\underline{q} = (q, a), \underline{u} = (u, b) \in \R^{1,4}$,
the lemma can be restated that there is a non-spacelike past-directed vector~$\underline{q} \in \R^{1,4}$ and
a vector~$\underline{u} \in \R^{1,4}$ with~$-1 \leq \underline{u}^2 \leq 0$ such that
\beq \label{s:Aform}
A = \Gamma^a q_a \left(\1 + i \Gamma^b u_b \right) \qquad \text{with} \qquad
\la \underline{q}, \underline{u} \ra=0 \quad \text{and} \quad
\underline{q}^2 = 0 \;\text{ or }\; \underline{u}^2 = -1 \:.
\eeq

The calculation
\beq \label{s:Acalc} \begin{split}
A^2 &= \Gamma^a q_a \left(\1 + i \Gamma^b u_b \right)
\Gamma^c q_c \left(\1 + i \Gamma^d u_d \right) \\
&= \Gamma^a q_a \Gamma^c q_c  \left(\1 - i \Gamma^b u_b \right) \left(\1 + i \Gamma^d u_d \right)
= \underline{q}^2\, (1 + \underline{u}^2) = 0 
\end{split}
\eeq
shows that every matrix of the form~\eqref{s:Aform} is indeed nilpotent. In order to show 
conversely that any nilpotent matrix~$A$ can be written in the form~\eqref{s:Aform}, we introduce the bilinear
covariants by~$\Sigma^{ab} = \frac{i}{2}\,[\Gamma^a, \Gamma^b]$. Then
the matrices~$\1$, $\Gamma^a$ and~$\Sigma^{ab}$ form a basis of the symmetric linear operators.
Since a nilpotent matrix is trace-free, we know that~$A$ has the basis representation
\[ A = \Gamma^a q_a + \Sigma^{ab}\, B_{ab} \]
with a vector~$\underline{q} \in \R^{1,4}$ and an anti-symmetric tensor~$B$.
Using the anti-commutation relations, it follows that
\begin{align*}
0 = A^2 &= \underline{q}^2 + \left( \Sigma^{ab}\, B_{ab} \right)^2 + \left\{\Gamma^a q_a, \Sigma^{ab}\, B_{ab} \right\} \\
&= \underline{q}^2 + B_{ab} \,B^{ab} + \frac{1}{4}\: \epsilon^{abcde} \,B_{ab}\, B_{cd}\, \Gamma_e
+ \epsilon^{abcde} \,B_{ab}\, q_c\, \Sigma_{de} \:,
\end{align*}
where~$\epsilon^{abcde}$ is the totally antisymmetric symbol in~$\R^{1,4}$.
\nindex{cg7@$\epsilon_{ijkl}$ -- totally antisymmetric tensor in Minkowski space}%
It follows that the expression~$\epsilon^{abcde} \,B_{ab}\, q_c=0$ vanishes, which
implies that~$B$ can be written as the anti-symmetrized tensor product of~$\underline{q}$ with another
vector~$\underline{u}$,
$B_{ab} = q_{[a} u_{b]}$. Then the vectorial component of~$A^2$ also vanishes, i.e.
\[ 0 = A^2 = \underline{q}^2 + B_{ab} \,B^{ab} = \underline{q}^2 + \la \underline{q},\underline{u} \ra^2
- \underline{u}^2 \underline{q}^2\:. \]
In the case~$\underline{q}^2=0$, it follows that~$\la \underline{q}, \underline{u} \ra=0$.
On the other hand if~$\underline{q}^2 \neq 0$, by adding a multiple of~$\underline{q}$ to~$\underline{u}$
we can arrange that again~$\la \underline{q}, \underline{u} \ra=0$. We conclude that~$\Sigma^{ab} B_{ab} = i \Gamma^a q_a \Gamma^b u_b$,
giving the desired representation in~\eqref{s:Aform}. Finally, the relations~$\underline{q}^2=0$
or~$\underline{u}^2=-1$ again follow from~\eqref{s:Acalc}.

It remains to show that~$\underline{q}$ is non-spacelike and past-directed and
that~$-1 \leq \underline{u}^2 \leq 0$. To verify these inequalities, it is most convenient to
arrange by a Lorentz transformation in~$R^{1,4}$ that the subspace spanned by~$q$ and~$u$
lies in the~$(0,1)$ or~$(1,2)$ plane or is a null surface. A straightforward calculation in each of these
cases using that~$A$ is negative semi-definite gives the result.
\QED
Let us discuss the result of this lemma. Multiplying out,
the representation~\eqref{s:Arep2} becomes
\beq \label{s:vpba}
A = \underbrace{\slashed{q}}_{\text{{vectorial}}} + \underbrace{a\, i \pseudo}_{\text{{pseudoscalar}}}
+ \underbrace{\frac{i}{2} \:[\slashed{q}, \slashed{u}]}_{\text{{bilinear}}}
- \underbrace{a\, \pseudo \slashed{u} + b\, \pseudo \slashed{q}}_{\text{{axial}}}\:.
\eeq
The operator~$\hat{k}$ can be realized by choosing~$q=\hat{k}$, $u=0$ and~$a=b=0$.
Then we are clearly in the first case in~\eqref{s:cases}.
The simplest operator in the second case is obtained by choosing~$u=0$, $a=0$ and~$b=\mp 1$,
giving $A=\chi_{L\!/\!R} \,\slashed{q}$, where~$q$ is any non-spacelike past-directed vector.
In view of this example, we refer to the second case in~\eqref{s:cases} as the {\em{chiral limit}}.
Because of the equation~$u^2-b^2=-1$, the chiral limit cannot be obtained by continuously
deforming the operator~$\hat{k}$. Since we have a perturbation of~$\hat{k}$ in mind, in what follows we
always restrict attention to the first case in~\eqref{s:cases}.
If we choose~$a=0$, the vector~$q$ must be light-like. Moreover, the axial
contribution in~\eqref{s:vpba} is proportional to~$\pseudo \slashed{q}$, making it impossible to
generate a contribution of the form~\eqref{s:axialwanted}. We conclude that we must choose~$a \neq 0$.
This implies that we necessarily get a pseudoscalar contribution and a timelike vectorial contribution.
Furthermore, in order for the axial contribution to point into a general direction~$v$, we
also need to choose~$u \neq 0$. This implies that we also get a bilinear contribution.
We conclude that when generating the desired axial contribution~\eqref{s:axialwanted},
we necessarily generate error terms having a vectorial, a pseudoscalar and a bilinear contribution.
The appearance of such error terms can be understood similar to the shear contribution in~\eqref{s:Pt2}.
The good news is that now the error terms can be arranged to be much smaller than the shear contributions,
as we now make precise by specifying the scalings.

Choosing a unitary matrix~$U$ with an axial and a vector component,
\begin{align}
U &= \exp \left( \frac{i}{\sqrt{\Omega}}\:Z \right)
\qquad \text{with} \qquad
Z = \nu\, \pseudo \slashed{v} + \lambda\, \slashed{v}\:,\quad \lambda, \nu \in \R\:, \label{s:Zansatz} \\
\intertext{we obtain}
Z^2 &= (-\nu^2+\lambda^2)\, v^2 \label{s:Z2rel} \\
Z \slashed{k} Z &= (\nu\, \pseudo + \lambda)\,\slashed{v} \slashed{k} \slashed{v} \,
(-\nu\, \pseudo + \lambda ) 
= (\nu\, \pseudo + \lambda)^2 \,\slashed{v} \slashed{k} \slashed{v} \notag \\
&= (\nu^2 +\lambda^2 + 2 \lambda \nu\, \pseudo)\: (2 \la v,k \ra\, \slashed{v} - v^2\, \slashed{k}) \notag \\
U \slashed{k} U^{-1} &= \slashed{k} + \frac{i}{\sqrt{\Omega}}\: [Z, \slashed{k}] -\frac{1}{2 \Omega}\:
\{Z^2, \slashed{k} \} + \frac{1}{\Omega}\: Z \slashed{k} Z + \O \big( \Omega^{-\frac{3}{2}} \big) \notag \\
&=  (\slashed{k}+m) + \frac{1}{\sqrt{\Omega}} \Big( \nu\: 2 \la k, v \ra\: i \pseudo  + \lambda\: i [\slashed{v}, \slashed{k}] \Big)
+ \frac{1}{\Omega}\: (\nu^2-\lambda^2)\, v^2\: \slashed{k} \notag \\
&\quad + \frac{1}{\Omega}\: (\nu^2 +\lambda^2 + 2 \lambda \nu\, \pseudo)\: (2 \la v,k \ra\, \slashed{v} - v^2\, \slashed{k})
+ \O \big( \Omega^{-\frac{3}{2}} \big) \notag
\end{align}
and thus
\begin{align}
U \,(\slashed{k}+m)\, U^{-1} 
&=  (\slashed{k}+m) + \frac{1}{\sqrt{\Omega}} \Big( \nu\: 2 \la v,k \ra\: i \pseudo  + \lambda\: i [\slashed{v}, \slashed{k}] \Big)
\label{s:Uk1} \\
&\quad +  \frac{2}{\Omega}\: (\nu^2 +\lambda^2 + 2 \lambda \nu\, \pseudo)\: \la v,k \ra\: \slashed{v} \label{s:Uk2} \\
&\quad - \frac{2}{\Omega}\: (\lambda^2 + \lambda \nu\, \pseudo)\: v^2\, \slashed{k}+ \O \big( \Omega^{-\frac{3}{2}} \big) \label{s:Uk3} \:.
\end{align}
The last contribution~\eqref{s:Uk3} points into the direction~$\slashed{k}$. It can be compensated by
the subsequent transformation
\[ U (\slashed{k}+m) U^{-1} \longrightarrow V U (\slashed{k}+m) U^{-1} V^* \]
with
\[ V = \1 + \frac{1}{\Omega} \left( \lambda^2\, v^2 + \lambda \nu \, v^2\: \pseudo \right) \:. \]
(this transformation describes a unitary pseudoscalar transformation combined with
a scaling by the factor~$(1+\lambda^2 v^2/\Omega)^2$).
In view of this transformation, we can drop~\eqref{s:Uk3}.
The remaining contributions~\eqref{s:Uk1} and~\eqref{s:Uk2} can be understood in analogy to
our discussion of~\eqref{s:vpba}.
Namely, the desired axial contribution is contained in~\eqref{s:Uk2}, which also involves a vectorial error term.
The second summand in~\eqref{s:Uk1} is the error term composed of the pseudoscalar and the
bilinear components.
By inspecting~\eqref{s:vpba}, one verifies that the above scaling is optimal in the sense that
the error terms must be at least of the order~$\Omega^{-\frac{1}{2}}$.
By repeating the above calculation for a more general matrix~$Z$, one can verify that
the form of the contributions in~\eqref{s:Uk1} and~\eqref{s:Uk2} is uniquely determined,
so that the only arbitrariness is to choose the two free parameters~$\lambda$
and~$\nu$\footnote{We remark that the ansatz~\eqref{s:Zansatz} for~$Z$ can be
can be generalized to~$Z = \pseudo \slashed{g} + \slashed{h}$ involving two
vectors~$g$ and~$h$. This gives additional freedom to modify the vector component,
but without any influence on the axial component to be considered here.
For details we refer to~\S\ref{l:seccurrent}.}.
For what follows, it is important that the error terms in~\eqref{s:Uk1} do not have a fixed sign.
Thus when considering several generations, it will be possible to arrange that
the contributions~\eqref{s:Uk1} cancel each other. The vectorial error term
in~\eqref{s:Uk2}, on the other hand, has a definite sign, leading to a timelike and past-directed
contribution. But it is by a factor~$1/\Omega$ smaller than the corresponding shear terms
in~\eqref{s:Pt2}.

Let us extend the last computation to a system of~$g$ Dirac seas
and generate the desired logarithmic pole on the light cone.
Our goal is to generate an axial contribution of the form 
\beq \label{s:poswanted}
\tilde{P}(x,y) \asymp \pseudo \slashed{v}\, \log(\xi^2)
\eeq
with a timelike vector~$v$.
Taking the Fourier transform, one sees that the contribution should be
proportional to the derivative of the $\delta$~distribution,
\beq \label{s:wanted}
\pseudo \slashed{v}\,\delta'(k^2)\: \Theta(-k^0)\:.
\eeq
Similar to~\eqref{s:Pt2}, we can transform each sea independently. Thus
\beq \label{s:Pseatrans}
\tilde{P} = \sum_{\beta=1}^g U_\beta(k)\: (\slashed{k} + m_\beta)\: U_\beta(k)^{-1}\:
\delta(k^2 - m_\beta^2)\:,
\eeq
where the operators~$U_\beta$ are again of the form~\eqref{s:Zansatz}.
Using~\eqref{s:Uk1} and~\eqref{s:Uk2}, we obtain
\beq \begin{split}
\tilde{P}(k) = \sum_{\beta=1}^g &\bigg[ (\slashed{k} + m_\beta)
+ \frac{1}{\sqrt{\Omega}}\:\Big( \nu_\beta\,2 \la v,k \ra\, i \pseudo + \lambda_\beta \,i [\slashed{v}, \slashed{k}]  \Big) \\
&\;\;+ \frac{4 \la v, k \ra}{\Omega}\: \Big( \lambda_\beta \nu_\beta\, \pseudo \slashed{v}\
+ \text{(vectorial)} \Big) \bigg]\: \delta(k^2 - m_\beta^2)\: \Theta(-k^0)\:.
\end{split} \label{s:Pktrans}
\eeq
Expanding the $\delta$-distributions according to~$\delta(k^2 - m_\beta^2)
= \delta(k^2) -m_\beta^2\: \delta'(k^2) + (\deg<0)$, we obtain the conditions
\begin{align}
\sum_{\beta=1}^g \lambda_\beta &= 0\:, &\hspace*{-1cm} \sum_{\beta=1}^g \nu_\beta &= 0 \label{s:nonc1} \\
\sum_{\beta=1}^g \lambda_\beta \,\nu_\beta &= 0\:, &\hspace*{-1cm}
\sum_{\beta=1}^g m_\beta^2 \:\lambda_\beta \,\nu_\beta &= -\frac{\Omega}{4 \la v, k \ra} \neq 0\:. \label{s:nonc2}
\end{align}
Since~$k$ scales like~$\Omega$, we see that these equations have the correct scaling
if we assume that the parameters~$\nu_\beta$ and~$\lambda_\beta$
all scale~$\sim \Omega^0$.

Let us verify that the equations~\eqref{s:nonc1} and~\eqref{s:nonc2} do not admit solutions
if the number of generations~$g < 3$. In the case~$g=1$, the equations~\eqref{s:nonc1} only
have the trivial solution, in contradiction to the right equation in~\eqref{s:nonc2}.
In the case~$g=2$, the equations~\eqref{s:nonc1} imply that~$\nu_2=-\nu_1$
and~$\lambda_2=-\lambda_1$. Using these relations on the left of~\eqref{s:nonc2},
it follows that~$\lambda_1=\nu_1=0$, again in contradiction to the right equation in~\eqref{s:nonc2}.

In order to analyze the case of three generations,
it is convenient to arrange with the transformation~$\nu_\beta \rightarrow -\nu_\beta \,\Omega/ (4 \la w,k \ra)$
that the inhomogeneity in~\eqref{s:nonc2} equals one. Moreover, introducing the parameters
\[ a_\beta = \frac{1}{2} \left( \lambda_\beta + \nu_\beta \right) \:,\qquad
b_\beta = \frac{1}{2} \left( \lambda_\beta - \nu_\beta \right) \:, \]
the equations~\eqref{s:nonc1} and~\eqref{s:nonc2} become
\begin{align}
\sum_{\beta=1}^3 a_\beta &= 0\:, &\hspace*{-1cm} \sum_{\beta=1}^g b_\beta &= 0 \label{s:n1} \\
\sum_{\beta=1}^3 (a_\beta^2-b_\beta^2) &= 0\:, &\hspace*{-1cm}
\sum_{\beta=1}^3 m_\beta^2 \:(a_\beta^2-b_\beta^2) &= 1\:. \label{s:n2}
\end{align}
We solve the linear equations~\eqref{s:n1} for~$a_3=a_3(a_1, a_2)$
and~$b_3 = b_3(b_1, b_2)$. In order to analyze the remaining quadratic equations~\eqref{s:n2} 
for~$a_1, a_2$ and~$b_1, b_2$, it is most convenient to introduce the
scalar products~$\la .,. \ra_0$ and~$\la .,. \ra_2$ by
\[ \sum_{\beta=1}^3 a_\beta^2 = \Big\la \bigg( \begin{matrix} a_1 \\ a_2 \end{matrix} \bigg),
\bigg( \begin{matrix} a_1 \\ a_2 \end{matrix} \bigg) \Big\ra_0 \:,\qquad
\sum_{\beta=1}^3 m_\beta^2 \:a_\beta^2 = \Big\la \bigg( \begin{matrix} a_1 \\ a_2 \end{matrix} \bigg),
\bigg( \begin{matrix} a_1 \\ a_2 \end{matrix} \bigg) \Big\ra_2 \:, \]
making it possible to write~\eqref{s:n2} as
\beq \label{s:rew1}
\la a, a \ra_0 = \la b, b \ra_0 \qquad \text{and} \qquad
\la a, a \ra_2 - \la b, b \ra_2 = 1
\eeq
(where~$a=(a_1, a_2)$ and~$a=(b_1, b_2)$). Representing the
scalar products with signature matrices~$S_0$ and~$S_2$,
\[ \la a, a \ra_0 = \la a, S_0 \,a \ra_{\R^2} \:, \qquad \la a, a \ra_2 = \la a, S_2\, a \ra_{\R^2} \:, \]
we can rewrite~\eqref{s:rew1} with the scalar product~$\la .,. \ra_0$,
\[ \la a, a \ra_0 = \la b, b \ra_0 \qquad \text{and} \qquad
\big\la a, (S_0^{-1} S_2)\, a \big\ra_0 - \big\la b, (S_0^{-1} S_2)\, b \big\ra_0 = 1\:. \]
Thus we seek for vectors~$a$ and~$b$ having the same norm, such that the
difference of their expectation values of the operator~$(S_0^{-1} S_2)$ equals one.
Having two equations for four unknowns, we can clearly not expect a unique solution.
But we get a unique solution by imposing that the error terms should be as small as possible,
which means that we want to the norms on the left to be minimal.
For the computation, it is useful that the operator~$(S_0^{-1} S_2)$ is symmetric with respect to~$\la ., \ra_0$.
Thus we can choose an eigenvector basis~$e_1, e_2$ of~$(S_0^{-1} S_2)$ which is orthonormal
with respect to~$\la .,. \ra_0$. Representing~$a=\alpha_k e_k$ and~$b=\beta_k e_k$ in this basis,
we obtain the equations
\[ \alpha_1^2 +\alpha_2^2 = \beta_1^2 +\beta_2^2 \qquad \text{and} \qquad
\mu_1 \, (\alpha_1^2-\beta_1^2)  + \mu_2\, (\alpha_2^2 - \beta_2^2) = 1\:, \] 
where the eigenvalues~$\mu_{1\!/\!2}$ are computed by
\beq \label{s:772}
\mu_{1\!/\!2} = \frac{1}{3} \left( m_1^2 + m_2^2 + m_3^2 \pm \sqrt{
m_1^4 + m_2^4 + m_3^4 - m_1^2 \,m_2^2 - m_2^2 \,m_3^2 - m_1^2 \,m_3^2} \right) .
\eeq
Now our minimization problem leads us to choose~$\alpha_1^2$ as large
as possible and~$\beta_1^2$ as small as possible, giving the unique solution
\[ \alpha_1 = \beta_2 = \frac{1}{\sqrt{\mu_1 - \mu_2}} \:, \qquad
\alpha_2 = \beta_1 = 0\:. \]
We thus obtain the following result.
\begin{Prp} \label{s:prpaxial1}
Considering homogeneous transformations of the vacuum fer\-mio\-nic projector,
it is impossible to generate a logarithmic pole~\eqref{s:poswanted} if the number of generations
is smaller than three. If the number of generations equals three,
by minimizing the error terms we get a unique solution
of the resulting equations~\eqref{s:nonc1} and~\eqref{s:nonc2}.
Thus for sufficiently large~$\Omega$, there are transformations~$U_\beta(k)$ such that
the transformed fermionic projector~\eqref{s:Pseatrans} is of the form
\beq
\begin{split}
\tilde{P}(k) &= P(k) + \pseudo \slashed{v} \sum_{\beta=1}^3 C_\beta\, \delta(k^2-m_\beta^2)
+ \text{(vectorial)} \:\delta(k^2) \Big(1+ \O(\Omega^{-1}) \Big) \\
&\quad + \text{(pseudoscalar or bilinear)} \:\sqrt{\Omega} \:\delta'(k^2) \Big(1+ \O(\Omega^{-1}) \Big) \:.
\end{split} \label{s:PasyO}
\eeq
The coefficients~$C_\beta$ satisfy the relations
\begin{align}
\sum_{\beta=1}^3 C_\beta &= 0 \:, \qquad
\sum_{\beta=1}^3 m_\beta^2 \,C_\beta = -1 \:, \label{s:nologeq} \\
\sum_{\beta=1}^3 m_\beta^2 \log(m_\beta^2)\: C_\beta
&= -\frac{\sum_{\beta=1}^3 m_\beta^2 \log(m_\beta^2)
\Big( 2 m_\beta^2 - \sum_{\alpha \neq \beta} m_\alpha^2 \Big)}
{\sum_{\beta=1}^3 m_\beta^2 \Big( 2 m_\beta^2 - \sum_{\alpha \neq \beta} m_\alpha^2 \Big)} \:. \label{s:logeq}
\end{align}
\end{Prp}
\Proof Introducing the abbreviation~$C_\beta = \lambda_\beta \nu_\beta \: 4 \la v, k \ra/\Omega$,
the relations~\eqref{s:nologeq} follow immediately from~\eqref{s:nonc2}.
The identity~\eqref{s:logeq} is obtained by a straightforward calculation using the
explicit form of the parameters~$\lambda_\beta$ and~$\nu_\beta$.
\QED

We remark that our analysis could be carried out similarly in the case of more than three
generations. We expect that, similar as explained after~\eqref{s:cdm1} for the local axial
transformation, the equation should be underdetermined for than three generations.
But we will not enter the details here. Instead, we shall always choose~$g=3$, noting
that this is the smallest number for which one can compensate the logarithmic poles on the light-cone.

\subsectionn{The Microlocal Chiral Transformation} \label{s:secgennonloc}
The setting in the previous section was rather special because we only considered
homogeneous transformations in the high-frequency limit.
But the methods and results can be generalized in a straightforward way, leading to
the so-called microlocal chiral transformation. We now explain these generalizations.

We first recall that with the ansatz~\eqref{s:Zansatz} we built in a specific scaling
which was suitable to describe the asymptotic behavior in the high-frequency limit~$\Omega \rightarrow \infty$.
More precisely, we showed that there are
transformations~$U_\beta(k)$ such that the transformed fermionic projector~\eqref{s:PasyO}
involves the desired axial contribution~\eqref{s:wanted} as well as vectorial, pseudoscalar and bilinear
error terms. By adding lower order terms in~$\Omega$
to~\eqref{s:Zansatz}, one could extend the analysis to also include correction terms
which decay faster for large~$\Omega$.
More generally, one can consider transformations of the form~$U(k)=e^{iZ(k)}$ with~$Z(k)$ as in~\eqref{s:Zansatz} without taking the high-energy limit
(for ease in notation, we now combined the factor~$1/\sqrt{\Omega}$ in~\eqref{s:Zansatz}
with the operator~$Z$). Then using~\eqref{s:Pnonloc}, a state of the fermionic projector
transforms to
\beq \label{s:Ukform}
\begin{split}
U&(k)\, (\slashed{k}+m)\, U(k)^* = e^{iZ} \,(\slashed{k}+m)\, e^{-i Z} \\
&= \Big( \cos(\alpha) + \frac{i Z}{\alpha} \:\sin(\alpha) \Big) \,\slashed{k}\,
\Big( \cos(\alpha) - \frac{i Z}{\alpha} \:\sin(\alpha) \Big) + m \:,
\end{split}
\eeq
where we set~$\alpha=\sqrt{Z^2}$ and used that, according to~\eqref{s:Z2rel}, $Z^2$
is a multiple of the identity matrix. A straightforward computation shows that by choosing~$Z$
appropriately, one can generate an arbitrary axial contribution, at the cost of generating vectorial, bilinear and
pseudoscalar error terms.
\sindex{homogeneous transformation!in the low-frequency region}%
Such homogeneous transformations in the low-frequency region
will be analyzed systematically in~\S\ref{s:sec110} below.
Here we simply choose~$Z(k)$ in the low-frequency region 
such that the coefficients~$C_\beta$ in~\eqref{s:PasyO} are constants.
Taking the Fourier transform, we thus obtain the following result.
\begin{Lemma} \label{s:lemmahom}
There is a homogeneous transformation~$U(k)$ such that 
the transformed fermionic projector~\eqref{s:Pnonloc} is of the form
\begin{align*}
\tilde{P}(x,y) &= P(x,y) + \pseudo \slashed{v}\, T^{(1)}_{[3]} + \pseudo \slashed{v}\, (\deg < 0) + \text{(smooth contributions)} \\
&\quad+ \slashed{v}\, (\deg<2) + \text{(pseudoscalar or bilinear)} \: (\deg<1)  \:.
\end{align*}
\end{Lemma} \noindent
We note that the smooth axial contributions can be computed explicitly from~\eqref{s:PasyO}
and~\eqref{s:logeq}. The smooth vectorial, pseudoscalar and bilinear contributions,
however, are undetermined.

Our next step is to extend our methods to the non-homogeneous setting where
the vector~$v$ in~\eqref{s:poswanted} is a smooth vector field.
In order to determine the relevant length scales, we first read off from~\eqref{s:Zansatz} (for constant~$\lambda$ and~$\nu$) that~$U(k)$ varies on the scale~$m$ of the rest masses of the Dirac seas. Thus in position
space, the distribution~$U(x,y)$ decays for~$x-y$ on the Compton scale.
Moreover, the vector field~$v$ varies on the scale~$\ell_\text{macro}$ of macroscopic physics.
We want to treat~$v$ as a slowly varying function in space-time. To this end, we need
to assume that the difference vector~$x-y$ of the fermionic projector
is much smaller than the macroscopic length scale. This leads us to evaluate
weakly~\eqref{s:asy} for~$\vec{\xi}$ on the scale
\beq \label{s:xiscale}
\varepsilon \ll |\vec{\xi}| \ll \ell_\text{macro}\:.
\eeq
Under this assumption, we can write~\eqref{s:poswanted} as
\[ \tilde{P}(x,y) \asymp \pseudo \slashed{v} \Big( \frac{x+y}{2} \Big)\, \log(\xi^2) 
+ \text{(higher orders in~$| \vec{\xi} |/\ell_\text{macro}$})\:. \]
In order to clarify the dependence on the vector~$v$, we now denote the homogeneous
nonlocal transformation of Lemma~\ref{s:lemmahom} by~$U(k, v)$.
We introduce the distribution~$U(x,y)$ by
\beq \label{s:Umicro}
U(x,y) = \int \frac{d^4k}{(2 \pi)^4}\: U\Big( k, v \Big(\frac{x+y}{2} \Big) \Big)\: e^{-i k(x-y)}
\eeq
\nindex{cg8@$U(x,y)$ -- microlocal chiral transformation}%
\sindex{quasi-homogeneous ansatz}%
and consider it as the integral kernel of a corresponding operator~$U$.
We generalize the transformation~\eqref{s:Pnonloc} by
\beq \label{s:Pmicroloc}
\tilde{P} := \acute{U} P^\text{aux} \,\grave{U}^* \:.
\eeq
The Fourier integral in~\eqref{s:Umicro} resembles the so-called Weyl map which is used to transform the
Wigner distribution to an integral kernel in position space.
More generally, the so-called {\em{quasi-homogeneous ansatz}}~\eqref{s:Umicro}
\sindex{quasi-homogeneous ansatz}%
is frequently used in microlocal analysis in order to approximately localize functions simultaneously
in position and in momentum space. For this reason, we refer to~$U$ as the
{\em{microlocal chiral transformation}}.
\sindex{transformation of the fermionic projector!microlocal chiral}%

Let us specify the error of the quasi-homogeneous approximation.
As the kernel~$U(x,y)$ decays on the Compton scale, at first sight one might expect
a relative error of the order~$(m \,\ell_\text{macro})^{-1}$. In fact, the situation is improved
by using~\eqref{s:xiscale}, as the following argument shows.
The kernel~$U(x,y)$ will in general be singular at~$x=y$, but it can be chosen
to be smooth otherwise. Hence we can decompose~$U(x,y)$ as
\beq \label{s:Usmooth}
U(x,y) = U_{|\vec{\xi}|}(x,y) + U_\text{smooth}(x,y)\:,
\eeq
where~$U_\text{smooth}$ is smooth and~$U_{|\vec{\xi}|}$ is supported for~$|\vec{x}-\vec{y}| \leq |\vec{\xi}|/2$.
Using this decomposition in~\eqref{s:Pmicroloc}, one sees that the error is of the
order~$|\vec{\xi}|/\ell_\text{macro}$, up to smooth contributions. We thus obtain the following result.
\begin{Prp} \label{s:prpaxial2}
If the number of generations equals three, then for any macroscopic
vector field~$v$ there is a microlocal chiral transformation~\eqref{s:Umicro} such that
the transformed fermionic projector~\eqref{s:Pmicroloc} is of the form
\beq \label{s:Ptres}
\begin{split}
\tilde{P}(x,y) &= P(x,y) + \pseudo \slashed{v}\, T^{(1)}_{[3]} \,\big( 1 + \O(|\vec{\xi}|/\ell_\text{macro}) \big) 
+ \pseudo \slashed{v}\, (\deg < 0) + \slashed{v}\, (\deg<2) \\
&\quad + \text{(pseudoscalar or bilinear)} \: (\deg<1) + \text{(smooth contributions)} \:.
\end{split}
\eeq
\end{Prp} \noindent
We now give the corresponding contribution to~${\mathcal{R}}$; the derivation is
postponed to Appendix~\ref{s:appspec}.

\begin{Lemma} \label{s:lemmalogterm2}
The perturbation of the fermio\-nic projector by the microlocal chiral transformation
of Proposition~\ref{s:prpaxial2} leads to a contribution to~${\mathcal{R}}$ of the form
\[ {\mathcal{R}} \asymp i \xi_k \:
v^k\: N_4 \:\big( 1 + \O(|\vec{\xi}|/\ell_\text{macro}) \big) +(\deg < 4)  + o \big(|\vec{\xi}|^{-3} \big)\:, \]
where
\beq
N_4 = -\frac{g}{\overline{T^{(0)}_{[0]}}} 
\Big[ T^{(1)}_{[3]} T^{(-1)}_{[0]} \overline{T^{(-1)}_{[0]} T^{(0)}_{[0]}} - c.c. \Big] .
\label{s:N4def2}
\eeq
\end{Lemma} \noindent
We note that the smooth contributions in~\eqref{s:Ptres} also affect~${\mathcal{R}}$;
this will be analyzed in detail in~\S\ref{s:secfield1}.

It is a shortcoming of the transformation~\eqref{s:Pmicroloc} that it does not necessarily
preserve the normalization of the fermionic projector
(note that the operator defined by the quasi-homogeneous ansatz~\eqref{s:Umicro}
is in general not unitary). We now explain how this problem
can be fixed by adapting the methods in~\S\ref{s:secgenlocal}: Similar to~\eqref{s:dirloc}, we
transform the Dirac operator by~$U^{-1}$, and rewrite the transformed Dirac operator as
\beq \label{s:dirnonloc}
\big( U^{-1} \big)^* \,\big( i \Pdd_x + \B - m Y \big) \,
U^{-1} = i \Pdd_x + \nf - m Y \:.
\eeq
As~$U$ is non-local, the operator~$\nf$ will in general be a {\em{nonlocal potential}}
(i.e.\ it can be written as an integral operator with a nonlocal integral kernel~$\nf(x,y)$).
\sindex{potential!nonlocal}%
Now we can again perform the causal perturbation expansion.
The resulting fermionic projector is properly normalized (for details see~\cite{norm}).
As explained in~\eqref{s:lct} and~\eqref{s:UPxy}, the residual argument yields that
the fermionic projector defined by the causal perturbation expansion coincides
with~\eqref{s:Pmicroloc} up to smooth contributions.

We conclude by pointing out that the previous constructions involve several correction terms
which will not be analyzed further in this book:
\sindex{field equations in the continuum limit!corrections!due to microlocal chiral transformation}%
First, there are the corrections of the order~$|\vec{\xi}|/\ell_\text{macro}$ in~\eqref{s:Ptres}.
Moreover, \eqref{s:Ptres} involves corrections of the order~$(m \,\ell_\text{macro})^{-1}$ to
the smooth contributions, generated by the operator~$U_\text{smooth}$ in~\eqref{s:Usmooth}.
Finally, the causal perturbation expansion corresponding to the nonlocal Dirac operator~\eqref{s:dirnonloc}
involve smooth corrections to~\eqref{s:Pmicroloc}. Since these corrections vanish for
homogeneous perturbations (because in this case, the transformation~$U(k)$ is unitary),
they should again be of the order~$(m \,\ell_\text{macro})^{-1}$.
We thus expect that all these correction terms are smaller than the
quantum corrections~\eqref{s:fconvolve} to be considered in~\S\ref{s:secfield1},
which are of the order~$k/m$ (where~$k$ is the momentum of the bosonic field).
This is the reason why in this paper, we shall not enter the analysis of these correction terms.

\subsectionn{The Shear Contributions by the Microlocal Chiral Transformation} \label{s:secshearmicro}
With the constructions in~\S\ref{s:secnonlocaxial} and~\S\ref{s:secgennonloc}, we could
avoid the shear contribution~\eqref{s:E20} caused by the local axial transformation.
Instead, the microlocal chiral transformation gives rise to the vectorial contribution in~\eqref{s:Uk2}.
We now analyze the effect of this contribution in detail and compare the situation to that of the shear terms
in~\S\ref{s:secprobaxial}. The vectorial contribution in~\eqref{s:Uk2} can be generated by
the transformation of the Dirac states
\[ (\slashed{k} + m) \rightarrow \slashed{k} + \frac{2}{\Omega}\: (\nu^2 +\lambda^2)\: \la v,k \ra\: \slashed{v} + m\:. \]
Writing the fermionic projector in the form~\eqref{s:Peps}, this transformation describes
a change of the direction of the vector field~$v^\varepsilon$. 
Using the notion in~\eqref{s:vepsrel} and~\eqref{s:shear}, this transformation again introduces
a shear of the surface states. We thus refer to the resulting contributions to the fermionic projector
as the {\em{shear contributions by the microlocal chiral transformation}}.
\sindex{shear contribution!by the microlocal chiral transformation}%
In Proposition~\ref{s:prpaxial1}, Lemma~\ref{s:lemmahom} and Proposition~\ref{s:prpaxial2},
the shear contributions were included in the error term~$\slashed{v} (\deg<2)$.
A short computation shows that they can be written as
\beq \label{s:DelPmicro}
\Delta P(x,y) = \frac{c}{m^2}\: \slashed{v}\: T^{(0)}_{[1]} \,\big( 1 + \O(|\vec{\xi}|/\ell_\text{macro}) \big) ,
\eeq
where~$c$ is a real-valued, dimensionless constant depending on the ratios of the fermionic masses
(the reason for the subscript~$[1]$ is that, according to~\eqref{s:Uk2}, this contribution
can be obtained form the contribution~$\sim m$ to~$\slashed{k}+m$
by multiplication with a function which is constant as~$\Omega \rightarrow \infty$).
Perturbing the fermionic projector of the vacuum, we obtain 
similar to~\eqref{s:Acont}
\begin{align*}
\Delta A_{xy} &= \frac{igc}{2 m^2} \Big( \slashed{\xi} \slashed{v} \:T^{(-1)}_{[0]} \overline{T^{(0)}_{[1]}}
- \slashed{v} \slashed{\xi} \: T^{(0)}_{[1]} \overline{T^{(-1)}_{[0]}} \Big) + (\deg < 3)\:.
\end{align*}
As this contribution is invariant under the replacements~$L \leftrightarrow R$,
it clearly drops out of the EL equations.

The situation becomes more subtle when the gauge phases are taken into account.
Similar to~\eqref{s:badphase}, the microlocal chiral transformation flips the chirality.
Thus~\eqref{s:DelPmicro} becomes
\beq \label{s:DelPphase}
\Delta P(x,y) = \frac{c}{m^2}\:
\Big( \chi_L \,e^{-i \Lambda^{xy}_R} + \chi_R \,e^{-i \Lambda^{xy}_L}  \Big)\:
\slashed{v}\: T^{(0)}_{[1]} \,\big( 1 + \O(|\vec{\xi}|/\ell_\text{macro}) \big) .
\eeq
Also using~\eqref{s:Pchiral} and~\eqref{s:clc}, we obtain
\beq \begin{split}
A_{xy} &= \frac{g^2}{4} \:(\chi_L\: \nu_L + \chi_R\: \nu_R)\:
(\slashed{\xi} T^{(-1)}_{[0]}) (\overline{\slashed{\xi} T^{(-1)}_{[0]}}) \\
&\quad
+ \frac{igc}{2 m^2} \Big( \slashed{\xi} \slashed{v} \:T^{(-1)}_{[0]} \overline{T^{(0)}_{[1]}}
- \slashed{v} \slashed{\xi} \: T^{(0)}_{[1]} \overline{T^{(-1)}_{[0]}} \Big) 
+ \slashed{\xi} (\deg \leq 3) + (\deg < 3)\:.
\end{split} \label{s:Amic}
\eeq
We now give the corresponding eigenvalues of the closed chain; the proof is
again given in  Appendix~\ref{s:appspec}.
\begin{Lemma} \label{s:lemmalammicro} The closed chain~\eqref{s:Amic} has the eigenvalues
\[ \lambda^{L\!/\!R}_+ =
\nu_{L\!/\!R}\: g^2\, T^{(0)}_{[0]} \overline{T^{(-1)}_{[0]}}
- \frac{igc}{m^2} \: v_k \xi^k\: T^{(0)}_{[1]} \overline{T^{(-1)}_{[0]}} + (\deg < 3)\:, \]
and the eigenvalues~$\lambda^{L\!/\!R}_-$ are obtained by complex conjugation
(see~\eqref{lorder}).
\end{Lemma} \noindent
This lemma shows that, similar to~\eqref{s:lambdak}, the eigenvalues of the closed chain are
perturbed by a contribution which does not involve the chiral phases.
This means that, just as shown in Figure~\ref{s:figabslambda}, in general the eigenvalues
will no longer have the same absolute value.
In other words, the fermionic projector again involves shear contributions which violate
the EL equations to degree five on the light cone.

However, now the situation is much better, because the perturbation~\eqref{s:DelPphase}
is of degree one on the light cone (and not of degree two as~\eqref{s:badphase}).
This makes it possible to ``modify the phase'' of~\eqref{s:DelPphase} by the following construction.
The auxiliary fermionic projector~$\tilde{P}$  in the presence of chiral gauge fields and including the microlocal
chiral transformation is defined by (cf.~\eqref{s:Pmicroloc})
The auxiliary fermionic projector~$\tilde{P}$ in the presence of chiral gauge fields and including the microlocal
chiral transformation is a solution of the Dirac equation
\beq \label{s:Dunflip}
\Dir \, \tilde{P}^\text{aux} = 0 \:,
\eeq
where the operator~$\Dir$ is obtained from
the Dirac operator with chiral gauge fields (as given in~\eqref{s:diracPaux} and~\eqref{s:chiral})
by performing the nonlocal transformation~\eqref{s:dirnonloc},
\beq \label{s:scrDdef}
\Dir :=  (U^{-1})^* \,\big(i \Pdd_x + \chi_L \slashed{A}_R + \chi_R \slashed{A}_L - m Y \big)\, U^{-1} \:.
\eeq
As explained after~\eqref{s:dirnonloc}, one can take this Dirac equation as the
starting point and introduce the fermionic projector by the causal perturbation expansion as outlined
in Section~\ref{s:sec4}. The resulting fermionic projector coincides agrees with~\eqref{s:dirnonloc},
up to smooth contributions which guarantee the desired normalization.
We now modify the operator~$U$, which implicitly changes the potentials in the Dirac operator~\eqref{s:scrDdef},
and, via the causal perturbation expansion, also the fermionic projector.
More precisely, we introduce the new Dirac operator
\beq \label{s:Dflip}
\Dir_\text{flip} :=  (U_\text{flip}^{-1})^* \,\big(i \Pdd_x + \chi_L \slashed{A}_R + \chi_R \slashed{A}_L - m Y \big)\, U_\text{flip}^{-1} \:,
\eeq
where for~$U_\text{flip}$ we make the ansatz
\beq \label{s:Uflip}
U_\text{flip} = \1 + (U - \1) \, V \:,
\eeq
and~$V$ is defined similar to~\eqref{s:cpower} by a perturbation series
\beq \label{s:Vansatz}
V = \1 + \sum_{k=1}^\infty \sum_{\alpha=0}^{\alpha_{\max}(k)} c_\alpha\;
\B_{1,\alpha} \,C_{1,\alpha} \,\B_{2,\alpha}\, \cdots \,\B_{k,\alpha}\, C_{k, \alpha}
\eeq
with combinatorial factors~$c_\alpha$ and Green's functions or fundamental solutions~$C_{\ell, \alpha}$.
Here the factors~$\B_{\ell, \alpha}$ are chiral potentials which can be freely chosen, giving
a lot of freedom to modify the gauge phases in the fermionic projector.
However, our ansatz~\eqref{s:Uflip} ensures that these modified phases only enter the
contributions generated by the microlocal chiral transformation
(using the notation~\eqref{s:Zansatz}, $V$ only affects the orders~$1/\sqrt{\Omega}$
or higher of~$P$, whereas the leading order~$\sim \Omega^0$ is not altered by~$V$).
Taking the Dirac equation
\beq \label{s:Dirflip}
\Dir_\text{flip} \, \tilde{P}^\text{aux} = 0
\eeq
as the starting point, one can again introduce the fermionic projector
by the causal perturbation expansion as outlined in Section~\ref{s:sec4}.
The next proposition shows that by a suitable choice of the operator~$V$, one can arrange that
the gauge phases in~\eqref{s:DelPphase} are flipped.
The proof will again be given in Appendix~\ref{s:appspec}.
\begin{Prp} \label{s:prpflip}
By a suitable choice of the perturbation series~\eqref{s:Vansatz},
one can arrange that the auxiliary fermionic projector~$\tilde{P}^\text{aux}$ defined by~\eqref{s:Dirflip} differs from
that defined by~\eqref{s:Dunflip} in that the contribution~\eqref{s:DelPphase} is modified to
\[ \Delta P(x,y) = \frac{c}{m^2}\:
\Big( \chi_L \,e^{-i \Lambda^{xy}_L} + \chi_R \,e^{-i \Lambda^{xy}_R}  \Big)\:
\slashed{v}\: T^{(0)}_{[1]} \,\big( 1 + \O(|\vec{\xi}|/\ell_\text{macro}) \big) . \]
All the other contributions to the fermionic projector remain unchanged,
up to error terms of the order~$o(|\vec{\xi}|) \,(\deg <2)$.
\end{Prp} \noindent
Following this result, the closed chain~\eqref{s:Amic} and the eigenvalue~$\lambda_+$ become
\begin{align*}
A_{xy} &= (\chi_L\: \nu_L + \chi_R\: \nu_R) \bigg[ \frac{g^2}{4}\:
(\slashed{\xi} T^{(-1)}_{[0]}) (\overline{\slashed{\xi} T^{(-1)}_{[0]}}) 
+ \frac{igc}{2 m^2} \Big( \slashed{\xi} \slashed{v} \:T^{(-1)}_{[0]} \overline{T^{(0)}_{[1]}}
- \slashed{v} \slashed{\xi} \: T^{(0)}_{[1]} \overline{T^{(-1)}_{[0]}} \Big) \bigg] \\
&\qquad + \slashed{\xi} (\deg \leq 3) + (\deg < 3) \\
\lambda^{L\!/\!R}_+ &=
\nu_{L\!/\!R} \Big[ g^2\, T^{(0)}_{[0]} \overline{T^{(-1)}_{[0]}}
- \frac{igc}{m^2} \: v_k \xi^k\: T^{(0)}_{[1]} \overline{T^{(-1)}_{[0]}} \Big] + (\deg < 3)\:.
\end{align*}
In particular, one sees that the eigenvalues all have the same absolute value,
so that the EL equations are satisfied to degree five.
To degree four, the error terms~$\tilde{P}(x,y) \sim \O(|\vec{\xi}|^2) \, (\deg \leq 1)$
appear. But as they involve at least two factors~$\xi$, they are of one order higher in~$|\vec{\xi}|$
than the current and mass terms. For this reason, we shall not consider them here.
We conclude that the shear contributions drop out of the EL equations.

\section{The Field Equations} \label{s:secfield}
Having developed a method for compensating the logarithmic poles on the light cone,
we are now in the position to derive and analyze the field equations.
We again point out that our method only works if the number of generations
is at least three (see Proposition~\ref{s:prpaxial1}). In what follows, we assume that the
number of generations equals three.

\subsectionn{The Smooth Contributions to the Fermionic Projector at the Origin} \label{s:secfield1}
We add the contributions from Lemmas~\ref{s:lemmalc1}, \ref{s:lemmalc2} and~\ref{s:lemmalogterm2}
and collect all the terms which involve factors of~$T^{(1)}_\circ$ or~$\overline{T^{(1)}_\circ}$.
Using~\eqref{s:logpole}, we find that the contribution to~${\mathcal{R}}$ involving
factors of~$\log |\xi^2|$ has the form
\begin{align}
{\mathcal{R}} \asymp & -\frac{i \xi_k}{16 \pi^3}\:
\bigg\{ \frac{j^k_\text{\rm{a}}}{6} - m^2\, \acute{Y} \grave{Y}\, A^k_\text{\rm{a}} -  \frac{v^k}{2}
\bigg\}\:
\frac{\log |\xi^2|}{\overline{T^{(0)}_{[0]}}}\:g^2\,
T^{(-1)}_{[0]} \overline{T^{(-1)}_{[0]}} \Big(T^{(0)}_{[0]} - \overline{T^{(0)}_{[0]}} \Big)
\nonumber \\
&+ (\deg < 4) + o \big( |\vec{\xi}|^{-3} \big)\:. \label{s:fieldprep}
\end{align}
As explained after~\eqref{s:T1ex}, this term must vanish.
This leads us to choose~$v$ as
\beq \label{s:vspec}
v = \frac{j_\text{\rm{a}} }{3} - 2 m^2\, \acute{Y} \grave{Y} A_\text{\rm{a}} \:.
\eeq
Then the logarithmic poles of~${\mathcal{R}}$ have disappeared.

Before analyzing the remaining contributions to~${\mathcal{R}}$,
we must have a closer look at the non-causal low- and high energy
contributions~$\tilde{P}^\lec$ and~$\tilde{P}^\hec$ in the light cone
expansion~\eqref{s:fprep}.
\nindex{bj6@$\tilde{P}^\lec$ -- non-causal low energy contribution to fermionic projector}%
\nindex{bj8@$\tilde{P}^\hec$ -- non-causal high energy contribution to fermionic projector}%
\sindex{fermionic projector!non-causal low energy contribution}%
\sindex{fermionic projector!non-causal high energy contribution}%
These smooth contributions to the fermionic
projector were disregarded in the formalism of the continuum limit as outlined
in Section~\ref{s:sec5}. 
\sindex{fermionic projector!smooth contributions to}%
This is justified as long as singular contributions to the
fermionic projector are considered.
In particular, contributions to~$P(x,y)$ involving
the functions~$T^{(-1)}_\circ$, $T^{(0)}_\circ$ or
their complex conjugates have poles on the light cone, and therefore smooth
corrections would be of lower degree on the light cone,
meaning that the corresponding contributions to the EL equations
would be negligible corrections of the form~\eqref{s:ap1}.
However, the factors~$T^{(1)}_\circ$ and~$\overline{T^{(1)}_\circ}$ only
have a logarithmic pole, and after the above cancellations of the logarithmic poles,
the remaining leading contributions are indeed bounded functions.
Thus smooth corrections become relevant. We conclude that it is necessary to
determine the {\em{smooth contributions}} to the fermionic projector~$P(x,y)$ at the
origin~$x=y$.
\sindex{field equations in the continuum limit!corrections!due to smooth, noncausal contributions to fermionic projector}%
This analysis is carried out in Appendix~\ref{s:appresum} to first
order in the bosonic potentials using a resummation technique. 
In what follows, we use these results and explain them.

In order to introduce a convenient notation, we write the factors~$T^{(1)}_{[p]}$
in generalization of~\eqref{s:logpole} as
\beq \label{s:logpole2}
T^{(1)}_{[p]} = \frac{1}{32 \pi^3} \Big( \log |\xi^2| + i \pi \,\Theta(\xi^2)\,
\epsilon(\xi^0) \Big) +  s_{[p]} \:,
\eeq
\nindex{ch0@$s_{[p]}$ -- smooth contribution to~$T^{(1)}_{[p]}$}%
where the real-valued functions~$s_{[p]}$, which may depend on the
masses and the bosonic potentials, will be specified below.
Taking the complex conjugate of~\eqref{s:logpole}, we get a similar
representation for~$\overline{T^{(1)}_{[p]}}$. Substituting these formulas
into~${\mathcal{R}}$, the factors~$\log |\xi^2|$ cancel each other
as a consequence of~\eqref{s:fieldprep} and~\eqref{s:vspec}.
Moreover, a short calculation shows that
the factors~$i \pi \,\Theta(\xi^2)\, \epsilon(\xi^0)$ in~\eqref{s:logpole2} also
drop out. Applying Lemma~\ref{s:lemma81}, the EL equations to degree four
yield the vector equation
\beq \label{s:field0}
j_a\, N_5 -m^2 A_a \,N_6 = J_a \,N_3
\eeq
with~$N_3$ as in~\eqref{s:N3def} and
\begin{align}
N_5 =& \frac{g^3}{6\, \overline{T^{(0)}_{[0]}}} \Big[
T^{(0)}_{[0]} T^{(0)}_{[0]} \overline{T^{(0)}_{[0]} T^{(-1)}_{[0]}} - c.c.  \Big] \label{s:N51} \\
&+ g \,(s_{[0]}-s_{[3]})\,\frac{g^2}{3\, \overline{T^{(0)}_{[0]}}}\:
T^{(-1)}_{[0]} \overline{T^{(-1)}_{[0]}} \Big( T^{(0)}_{[0]} - \overline{T^{(0)}_{[0]}} \Big) \\
N_6 =& -\frac{2g \, \hat{Y}^2}{\overline{T^{(0)}_{[0]}}}
\Big[  T^{(-1)}_{[0]} T^{(0)}_{[0]} \overline{ T^{(0)}_{[1]} T^{(0)}_{[1]}} - c.c. \Big] \\
&+ (s_{[2]}-s_{[3]})\, \frac{2g^2 \,\acute{Y} \grave{Y}}{\overline{T^{(0)}_{[0]}}}\:
 T^{(-1)}_{[0]} \overline{T^{(-1)}_{[0]}} \Big( T^{(0)}_{[0]} - \overline{T^{(0)}_{[0]}} \Big) \,.
\label{s:N64} 
\end{align}
By direct inspection one verifies that the integration-by-parts rules~\eqref{s:ipart} do not yield
relations between the simple fractions. In other words, the appearing
simple fractions are all basic fractions. When evaluating weakly on the light cone~\eqref{s:asy},
all basic fractions are of degree four, thus producing the same
factor~$\varepsilon^{-3} (i |\vec{\xi}|)^{-4}$. Using furthermore that no
logarithmic divergences appear, we conclude that~\eqref{s:field0} must hold if the basic
fractions are replaced by the corresponding regularization parameters.
We can thus rewrite~\eqref{s:field0} as
\beq \label{s:field1}
\Big(c_0 - c_1 (s_{[0]}-s_{[3]})\Big) j_a
 -m^2 \left( c_2 \hat{Y}^2 + c_3 \acute{Y} \grave{Y}- 2 c_1 (s_{[2]}-s_{[3]})
 \acute{Y} \grave{Y} \right) A_a = \frac{c_1}{8 \pi}\: J_a \:,
\eeq
where the constants~$c_0, \ldots, c_3$ are the four regularization parameters
corresponding to the basic fractions appearing in~\eqref{s:N3def} and~\eqref{s:N51}--\eqref{s:N64}.
Since we are free to multiply~\eqref{s:field1} by a non-zero constant,
our field equations~\eqref{s:field1} involve {\em{three regularization parameters}}.
For a given regularization scheme, these parameters can be computed to obtain
numerical constants, as will be explored further in~\S\ref{s:secexample}.
Alternatively, these parameters can be regarded as empirical constants which
take into account the unknown microscopic structure of space-time.
Apart from these three constants, all the quantities in~\eqref{s:field1} are objects of
macroscopic physics, defined independent of the regularization.

It remains to determine~$s_{[0]}$, $s_{[2]}$ and~$s_{[3]}$.
As in~\eqref{s:fieldprep}, we again consider the leading order at the origin,
and thus it suffices to compute the functions~$s_{[p]}(x,y)$ at~$x=y$.
Let us begin with the calculation of~$s_{[3]}$.
Since the factors~$T^{(1)}_{[3]}$ and~$\overline{T^{(1)}_{[3]}}$
only appear in~\eqref{s:N4def2}, the function~$s_{[3]}$ is obtained by computing
the smooth contributions to the fermionic projector generated by the microlocal chiral transformation.
The vector contribution clearly drops out of~\eqref{s:RRdef}.
The pseudoscalar and bilinear contributions are even. As a consequence, the leading
contribution to~${\mathcal{R}}$ involves a factor of~$m$ and is thus of lower degree on the
light cone (for details see~\S\ref{s:sec87} and~\S\ref{s:sec88} and Lemma~\ref{s:lemmascal}).
Thus it remains to consider the axial contribution to the fermionic projector.
According to~\eqref{s:Pseatrans}, we obtain contributions from the different Dirac seas.
Using the formulas~\eqref{s:Taser}--\eqref{s:Tm1def} for the Fourier transform of the lower mass shell,
we find that the relevant contribution of the~$\beta^\text{th}$ Dirac sea
involving the logarithmic pole and the constant term is given by
\[ P_\beta(x,y) \asymp \frac{m_\beta^3}{32 \pi^3}\: \log(m_\beta^2 |\xi^2|) + c \:, \]
where~$c$ equals the constant~$c_0$ in~\eqref{s:Taser}.
Since in~\eqref{s:field1} only the differences of the functions~$s_{[p]}$ appear,
we may always disregard this constant. Then the smooth contribution at the origin
is given by~\eqref{s:logeq}. More precisely, collecting all prefactors
and comparing with~\eqref{s:logpole2}, we obtain
\beq \label{s:s3def}
s_{[3]} = \frac{1}{32 \pi^3}\: \frac{\sum_{\beta=1}^3 m_\beta^2 \log(m_\beta^2)
\Big( 2 m_\beta^2 - \sum_{\alpha \neq \beta} m_\alpha^2 \Big)}
{\sum_{\beta=1}^3 m_\beta^2 \Big( 2 m_\beta^2 - \sum_{\alpha \neq \beta} m_\alpha^2 \Big)}\:.
\eeq

The functions~$s_{[0]}$ and~$s_{[2]}$ are more difficult to compute. Therefore, we first state
the result and discuss it afterwards.
\begin{Lemma} \label{s:lemmasmooth}
The operators~$s_{[0]}$ and~$s_{[2]}$ appearing in~\eqref{s:logpole2} and~\eqref{s:field1} have the form
\begin{align}
s_{[0]} j_\text{\rm{a}} &= \frac{1}{3 \!\cdot\! 32 \pi^3}  \sum_{\beta=1}^3
\left( \log(m_\beta^2) \,j_a + f^\beta_{[0]}*j_\text{\rm{a}}  \right) + \O(A_a^2) \label{s:s0def} \\
s_{[2]} A_\text{\rm{a}} &= \frac{1}{32 \pi^3\, m^2 \acute{Y} \grave{Y}}
 \sum_{\beta=1}^3 m_\beta^2 \Big( \log(m_\beta^2) \, A_\text{\rm{a}} + f^\beta_{[2]}*A_\text{\rm{a}}  \Big)
 + \O(A_a^2) , \label{s:s2def}
\end{align}
\nindex{ch2@$f^\beta_{[p]}$ -- contribution to~$s_{[p]}$}%
where the star denotes convolution, i.e.
\[ (f^\beta_{[p]}*h)(x) = \int f^\beta_{[p]}(x-y)\: h(y)\: d^4y\:. \]
\nindex{ch4@$*$ -- convolution}%
The convolution kernels are the Fourier transforms of the distributions
\begin{align}
\hat{f}^\beta_{[0]}(q) &= 6\, \int_0^1 (\alpha-\alpha^2)\: \log \bigg| 1 - (\alpha-\alpha^2)
\,\frac{q^2}{m_\beta^2} \bigg| \, d\alpha \label{s:fb0def} \\
\hat{f}^\beta_{[2]}(q) &= \int_0^1 \log \bigg| 1 - (\alpha-\alpha^2) \,\frac{q^2}{m_\beta^2} \bigg|
\, d\alpha \:. \label{s:fb2def}
\end{align}
\end{Lemma} \noindent
\nindex{ch6@$\hat{f}^\beta_{[p]}$ -- Fourier transform of~$f^\beta_{[p]}$}%
Postponing the proof of this lemma to Appendix~\ref{s:appresum}, we here merely discuss
the result. For clarity, we first remark that the convolution operators in~\eqref{s:s0def}
and~\eqref{s:s2def} can also be regarded as multiplication operators in momentum space,
defined by
\[ f^\beta_{[p]} * e^{-i q x} = \hat{f}^\beta_{[p]}(q)\, e^{-i q x} \]
with the functions~$\hat{f}^\beta_{[p]}$ as in~\eqref{s:fb0def} and~\eqref{s:fb2def}.
Next, we note that the integrands in~\eqref{s:fb0def} and~\eqref{s:fb2def} only
have logarithmic poles, so that the integrals are finite. In Appendix~\ref{s:appresum},
these integrals are even computed in closed form (see Lemma~\ref{s:lemmaclosed} and
Figure~\ref{s:fig3}).
Next, we point out that these integrals vanish
if~$q^2=0$, because then the logarithm in the integrand is zero. Therefore,
the convolutions by~$f^\beta_{[p]}$ can be regarded as higher order corrections in~$q^2$ to the
field equations. Thus we can say that~$s_{[0]}$ and~$s_{[2]}$ are composed of
constant terms involving logarithms of the Dirac masses, correction terms~$f^\beta_{[p]}$
taking into account the dependence on the momentum~$q^2$ of the bosonic potential,
and finally correction terms of higher order in the bosonic potential.

The constant term in~$s_{[0]}$ can be understood from the following simple consideration
(the argument for~$s_{[2]}$ is similar). The naive approach to determine the constant
contribution to the $\beta^\text{th}$ Dirac sea is to differentiate~\eqref{s:Taser}
at~$a=m_\beta^2$ to obtain
\[ T^{(1)}_{[0]} \asymp \frac{1}{32 \pi^3} \left( \log(m_\beta^2) + 1 \right) , \]
where we again omitted the irrelevant constant~$c_0$.
Forming the sectorial projection and comparing with~\eqref{s:logpole2}, we obtain the contribution
\beq \label{s:sguess}
s_{[0]} \asymp \frac{1}{3 \!\cdot\! 32 \pi^3}  \sum_{\beta=1}^3 \left( \log(m_\beta^2) + 1 \right)\:.
\eeq
This naive guess is wrong because there is also a contribution to the fermionic projector of the form
$\sim \slashed{\xi} \xi_k j^k_\text{\rm{a}} T^{(0)}_{[0]}$,
which when contracted with~$\slashed{\xi}$ yields another constant term which is not taken
into account by the formalism of Section~\ref{s:sec5}
(see the term~\eqref{s:xij} in Appendix~\ref{s:appspec}). This additional contribution
cancels the summand~$+1$ in~\eqref{s:sguess}, giving the desired constant term in~\eqref{s:s0def}.

Next, it is instructive to consider the scaling behavior of the functions~$s_{[p]}$ in the fermion
masses. To this end, we consider a
joint scaling~$m_\beta \rightarrow L m_\beta$ of all masses.
Sine the expressions~\eqref{s:s3def}, \eqref{s:s0def} and~\eqref{s:s2def} have the same
powers of the masses in the numerator and denominator, our scaling amounts to
the replacement~$\log(m_\beta^2) \rightarrow \log(m_\beta^2) + 2 \log L$.
Using the specific form of the operators~$s_{[p]}$, one easily verifies that
the transformation of the constant terms can be described by the
replacement~$s_{[p]} \rightarrow s_{[p]}+ 2/(32 \pi^3) \log L$. We conclude
that for differences of these operators as appearing in~\eqref{s:field1}, the constant terms
are indeed scaling invariant. In other words, the constant terms in the expressions~$s_{[0]}-s_{[3]}$
and~$s_{[2]}-s_{[3]}$ depend only on quotients of the masses~$m_1$, $m_2$ and~$m_3$.

Before discussing the different correction terms in~\eqref{s:s0def} and~\eqref{s:s2def}, it is convenient
to combine all the constant terms in~\eqref{s:s0def}, \eqref{s:s2def} and~\eqref{s:field1}.
More precisely, multiplying~\eqref{s:field1} by~$96 \pi^3/c_0$ gives the following result.
\begin{Thm} \label{s:thmfield}
The EL equations to degree four on the light cone give rise to the condition
\beq \label{s:field2}
(C_0 - f_{[0]}*) j_\text{\rm{a}} - (C_2 - 6 f_{[2]}*) A_\text{\rm{a}} = 12 \pi^2\, J_\text{\rm{a}} 
+ \O(A_\text{\rm{a}}^2)
\eeq
\nindex{ch2@$f^\beta_{[p]}$ -- contribution to~$s_{[p]}$}%
\nindex{ci0@$C_0, C_2$ -- regularization parameters}%
involving the axial bosonic potential~$A_\text{\rm{a}}$, the corresponding axial current~$j_\text{\rm{a}}$
and the axial Dirac current~$J_\text{\rm{a}}$ (see~\eqref{s:axialdef}, \eqref{s:jadef} and~\eqref{s:Jadef}).
Here the convolution kernels are the Fourier transforms of the distributions
\begin{align*}
\hat{f}_{[0]}(q) &= \sum_{\beta=1}^3 \hat{f}^\beta_{[0]}(q)\,
= \sum_{\beta=1}^3 \:6\! \int_0^1 (\alpha-\alpha^2)\: \log \bigg| 1 - (\alpha-\alpha^2)
\,\frac{q^2}{m_\beta^2} \bigg| \, d\alpha \\
\hat{f}_{[2]}(q) &= \sum_{\beta=1}^3 m_\beta^2 \,\hat{f}^\beta_{[0]}(q) =
\sum_{\beta=1}^3 m_\beta^2 \int_0^1 \log \bigg| 1 - (\alpha-\alpha^2) \,\frac{q^2}{m_\beta^2} \bigg|
\, d\alpha
\end{align*}
(with the functions~$\hat{f}^\beta_{[p]}$ as defined by~\eqref{s:fb0def} and~\eqref{s:fb2def}).
The constants~$C_0$ and~$m_1^2 C_2$ depend only on the regularization and on the
ratios of the masses of the fermions.
\end{Thm} \noindent
With this theorem, we have derived the desired field equations for the axial potential~$A_\text{\rm{a}}$.
They form a linear hyperbolic system of equations involving a mass term, with corrections in the
momentum squared and of higher order in the potential. It is remarkable that the corrections in
the momentum squared are described by explicit convolutions, which do not involve any free constants.
In order to make the effect of the convolution terms smaller, one must choose the
constants~$C_0$ and~$C_2$ larger, also leading to a smaller coupling of the Dirac current.
Thus the effect of the convolution terms decreases for a smaller coupling constant,
but it cannot be arranged to vanish completely.

We proceed by explaining and analyzing the above theorem, beginning with the convolution
operators~$f_{[p]}$ (\S\ref{s:secnocausal}) and the higher orders in the potential
(\S\ref{s:sechighorder}). In~\S\ref{s:secquantcorr} we explain how the standard loop corrections
of QFT appear in our model.
 In~\S\ref{s:secnohiggs} we explain why the Higgs boson does not appear
in our framework. Finally, in~\S\ref{s:secexample} we compute the coupling constants and
the bosonic rest mass for a few simple regularizations.

\subsectionn{Violation of Causality and the Vacuum Polarization} \label{s:secnocausal}
\sindex{causality violation}%
\sindex{non-causal correction!by convolution terms}%
In this section we want to clarify the significance of the convolution operators in the field
equations~\eqref{s:field2}.
Our first step is to bring the convolution kernels into a more suitable form. For any~$a>0$,
we denote by~$S_a$ the following Green's function of the Klein-Gordon equation,
\begin{align}
S_a(x,y) &= \int \frac{d^4 k}{(2 \pi)^4}\: \frac{\text{PP}}{k^2-a}\: e^{-ik(x-y)} \label{s:Samom} \\
&= -\frac{1}{2 \pi} \:\delta(\xi^2) + \frac{a}{4 \pi} \:\frac{J_1\big(\sqrt{a 
\:\xi^2} \big)}{\sqrt{a \xi^2}} \:\Theta(\xi^2)\:, \label{s:Sapos}
\end{align}
\sindex{Green's function!symmetric}%
\nindex{bg0@$S_{m^2}$ -- symmetric Green's functions of the Klein-Gordon equation}%
where in the last step we again set~$\xi=y-x$ and
computed the Fourier integral using the Bessel function~$J_1$.
This Green's function is obviously causal in the sense that it vanishes for spacelike~$\xi$.
Due to the principal part, it is the mean of the advanced and retarded Green's function;
this choice has the advantage that~$S_a$ is symmetric, meaning
that~$\overline{S_a(x,y)} = S_a(y,x)$.
Expanding the Bessel function in a power series, the square roots drop out, showing
that~$S_a$ is a power series in~$a$.
In view of the explicit and quite convenient formula~\eqref{s:Sapos},
it seems useful to express the convolution kernels in terms of~$S_a$. This is done in the
next lemma.
\begin{Lemma} \label{s:lemma93}
The distributions~$f^\beta_{[p]}$ as defined by~\eqref{s:fb0def} and~\eqref{s:fb2def} can
be written as
\begin{align}
f^\beta_{[0]}(x-y) &= \int_{4 m_\beta^2}^\infty 
\left(S_a(x,y) + \frac{\delta^4(x-y)}{a} \right) \sqrt{a-4m_\beta^2} \:(a+2m_\beta^2)\:
\frac{da}{a^\frac{3}{2}} \label{s:f0rep} \\
f^\beta_{[2]}(x-y) &= \int_{4 m_\beta^2}^\infty \left(S_a(x,y) + \frac{\delta^4(x-y)}{a} \right)
\sqrt{a-4m_\beta^2}\: \frac{da}{\sqrt{a}}\:.
\end{align}
\end{Lemma}
\nindex{ch2@$f^\beta_{[p]}$ -- contribution to~$s_{[p]}$}%
\sindex{field equations in the continuum limit!corrections!due to smooth, noncausal contributions to fermionic projector}%
\Proof We first compute the Fourier transform of the distribution $\log|1-q^2/b|$ for given~$b>0$.
Using that $\lim_{a \rightarrow \infty} \log|1-q^2/a|=0$ with convergence as a distribution,
we have
\[ \log \left| 1 - \frac{q^2}{b} \right| = -\int_b^\infty \frac{d}{da} \log \left| 1 - \frac{q^2}{a} \right| da
= \int_b^\infty \left( \frac{\text{PP}}{q^2-a} + \frac{1}{a} \right) da \:. \]
Now we can compute the Fourier transform with the help of~\eqref{s:Samom}.
Setting~$b=m_\beta^2/(\alpha-\alpha^2)$, we obtain
\[ \int \frac{d^4q}{(2 \pi)^4}\: \log \left| 1 - (\alpha-\alpha^2)\, \frac{q^2}{m_\beta^2} \right|
e^{-iq(x-y)}
= \int_{\frac{m_\beta^2}{\alpha-\alpha^2}}^\infty
\left( S_a(x,y) + \frac{\delta^4(x-y)}{a} \right) da \:. \]
We finally integrate over~$\alpha$, interchange the orders of integration,
\[ \int_0^1 d\alpha\: (\alpha-\alpha^2)^r \int_{\frac{m_\beta^2}{\alpha-\alpha^2}}^\infty
da\:(\cdots) = \int_{4 m_\beta^2}^\infty da\: (\cdots) \int_0^1 d\alpha\: (\alpha-\alpha^2)^r\:
\Theta \Big(a - \frac{m_\beta^2}{\alpha-\alpha^2} \Big) , \]
and compute the last integral.
\QED

Qualitatively speaking, this lemma shows that the distributions~$f_{[p]}(x,y)$
can be obtained by integrating the Green's function~$S_a$ over the mass parameter~$a$
and by subtracting a suitable counter term localized at~$\xi=0$.
The interesting conclusion is that the convolution kernels~$f_{[p]}(x,y)$
in the field equations~\eqref{s:field2} are weakly causal in the sense that they vanish for
spacelike~$\xi$. But they are not strictly causal in the sense that
the past influences the future and also {\em{the future influences the past}}.

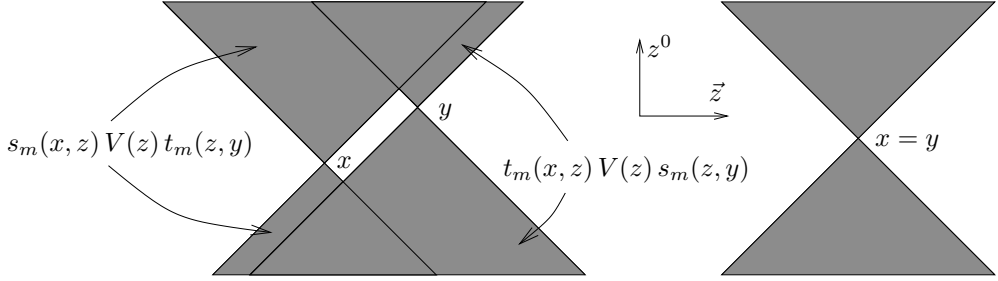
\begin{figure}
\begin{picture}(0,0)%
\includegraphics{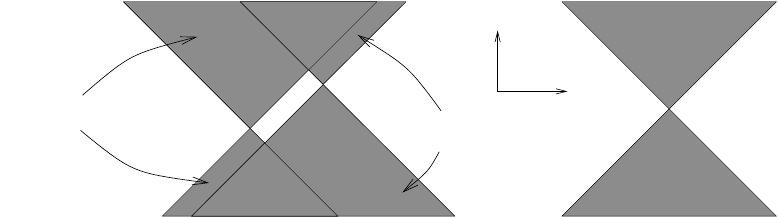}%
\end{picture}%
\setlength{\unitlength}{1367sp}%
\begingroup\makeatletter\ifx\SetFigFontNFSS\undefined%
\gdef\SetFigFontNFSS#1#2#3#4#5{%
  \reset@font\fontsize{#1}{#2pt}%
  \fontfamily{#3}\fontseries{#4}\fontshape{#5}%
  \selectfont}%
\fi\endgroup%
\begin{picture}(17937,4974)(-9374,-3898)
\put(6391,-1546){\makebox(0,0)[lb]{\smash{{\SetFigFontNFSS{10}{12.0}{\rmdefault}{\mddefault}{\updefault}$x=y$}}}}
\put(-1529,-991){\makebox(0,0)[lb]{\smash{{\SetFigFontNFSS{10}{12.0}{\rmdefault}{\mddefault}{\updefault}$y$}}}}
\put(-3374,-1951){\makebox(0,0)[lb]{\smash{{\SetFigFontNFSS{10}{12.0}{\rmdefault}{\mddefault}{\updefault}$x$}}}}
\put(2251, 74){\makebox(0,0)[lb]{\smash{{\SetFigFontNFSS{10}{12.0}{\rmdefault}{\mddefault}{\updefault}$z^0$}}}}
\put(3376,-736){\makebox(0,0)[lb]{\smash{{\SetFigFontNFSS{10}{12.0}{\rmdefault}{\mddefault}{\updefault}$\vec{z}$}}}}
\put(-359,-2086){\makebox(0,0)[lb]{\smash{{\SetFigFontNFSS{10}{12.0}{\rmdefault}{\mddefault}{\updefault}$t_m(x,z)\, V(z)\, s_m(z,y)$}}}}
\put(-9359,-1636){\makebox(0,0)[lb]{\smash{{\SetFigFontNFSS{10}{12.0}{\rmdefault}{\mddefault}{\updefault}$s_m(x,z)\, V(z)\, t_m(z,y)$}}}}
\end{picture}%
\caption{The support of the integrand in~\eqref{s:DelT}}
\label{s:fig2}
\end{figure}

Before discussing whether and how such a violation of causality could be observed in experiments,
we give a simple consideration which conveys an intuitive understanding for
how the non-causal contributions to the field
equations come about. For simplicity, we consider the linear perturbation~$\Delta P$ of a
Dirac sea of mass~$m$ by a potential~$\B$,
\beq \label{s:DelT}
\Delta P(x,y)
= -\int d^4z \:\Big(s_m(x,z) \:V(z)\: t_m(z,y) \:+\: t_m(x,z) \:V(z)\: s_m(z,y) \Big) \: ,
\eeq
where~$s_m$ is the Dirac Green's function and~$t_m$ denotes the Dirac sea of the vacuum, i.e.
\beq \label{s:smdef1}
t_m = (i \Pdd + m) T_{m^2} \qquad \text{and} \qquad s_m = (i \Pdd + m) S_{m^2} \:,
\eeq
\nindex{bd2@$s_m$ -- symmetric Green's functions of the vacuum Dirac equation}%
\nindex{ci4@$t_m$ -- vacuum Dirac sea}%
and~$T_{m^2}$ and~$S_{m^2}$ as defined by~\eqref{s:Tadef} and~\eqref{s:Samom}
(for details see~\cite[eqs~(2.4) and~(2.5)]{firstorder} or Section~\ref{secfpext}). 
Let us consider the support of the integrand in~\eqref{s:DelT}.
The Green's function~$s_m$ vanishes outside the light cone (see~\eqref{s:Sapos}),
whereas the distribution~$t_m$ is non-causal (see~\eqref{s:Taser}). Thus
in~\eqref{s:DelT} we integrate over the union of the double light cone
(meaning the interior of the light cones and their boundaries) centered at the points~$x$
and~$y$; see the left of Figure~\ref{s:fig2}.
In the limit~$x \rightarrow y$, the integral in~\eqref{s:DelT} will diverge, as becomes
apparent in the poles of light cone expansion. But after subtracting these divergent contributions,
we can take the limit~$x \rightarrow y$ to obtain a well-defined integral over the
double light cone centered at the point~$x=y$; see the right of Figure~\ref{s:fig2}.
Indeed, the finite contribution at the origin described by this integral corresponds precisely to
the smooth contribution to the fermionic projector as considered in~\S\ref{s:secfield1}.
This consideration explains why the distributions~$f^\beta_{[p]}(x-y)$ vanish for
spacelike~$\xi$. Moreover, one sees that the distributions~$f^\beta_{[p]}(x,y)$ are closely related to
the pointwise product in position space of the Bessel functions appearing in the 
distributions~$T_{m^2}(x,y)$ and~$S_{m^2}(x,y)$ for timelike~$\xi$. 
Going into more details, this argument could even be elaborated to an alternative method
for computing the convolution kernels. However, for actual computations this alternative
method would be less convenient than the resummation technique of Appendix~\ref{s:appresum}.

One might object that the above violation of causality occurs simply because
in~\eqref{s:DelT} we are working with the wrong Green's functions. Indeed, if
in~\eqref{s:DelT} the first and second factors~$s_m$ were replaced by the regarded and
advanced Green's function, respectively, the support of the integral would become
strictly causal in the sense that~$z$ must lie in the causal past of~$x$ or~$y$.
However, modifying the Green's functions in this way is not admissible, as it would
destroy the property that the Dirac sea is composed only of half of the solutions of the
Dirac equation.
More generally, the uniqueness of the perturbation expansion of the fermionic projector
follows from a causality argument (see~\cite[Section~2.2]{PFP} or Section~\ref{secfpext}).
Thus there is no freedom in modifying the perturbation expansion, and thus the above violation
of causality cannot be avoided.

The violation of causality in the field equations breaks with one of the most fundamental
physical principles.
\sindex{causality violation}%
The immediate question is whether and how this effect could be verified
in experiments.
\sindex{causality violation!in experiments}%
We conclude this section by discussing this question. Before beginning,
we point out that the present paper is concerned with a simple fermion system, and one should
be careful to draw physical conclusions from this oversimplified physical model.
Also, the author has no expertise to address experimental issues.
Nevertheless, it seems worth exploring the potential consequences of the causality
violation in a few ``Gedanken experiments,'' just to make sure that we do not get immediate
contradictions to physical observations. In order to be closer to everyday physics, let us consider
what happened if we inserted the nonlocal convolution term into Maxwell's equations.
For simplicity, we consider one Dirac wave function~$\psi$ of mass~$m$.
Thus dropping the mass term in~\eqref{s:field2} and choosing for convenience the Lorentz gauge,
the modified Dirac-Maxwell equations become
\sindex{Dirac-Maxwell equations!modified by convolution terms}%
\beq \label{s:diracmaxwell}
\boxed{ \quad (i \Pdd + \slashed{A}-m) \psi = 0 \:, \qquad
-\Big[ 1 - \frac{e^2}{12 \pi^2}\:f_{[0]}* \Big] \Box A_k = e^2\, \overline{\psi} \gamma_k \psi  \:,
\quad }
\eeq
where we chose the constant~$C_0$ such that without the convolution terms, the Maxwell equations
take the familiar form~$-\Box A_k = e^2 \overline{\psi} \gamma_k \psi$
(note that we again use the convention where the Dirac equation involves no
coupling constants; see also Footnote~\ref{s:units} on page~\pageref{s:units}).
In view of Lemma~\ref{s:lemma93}, the square bracket is an integral operator which vanishes
for spacelike distances. Furthermore, we see from~\eqref{s:fb0def} and~\eqref{s:fb2def}
(for more details see Lemma~\ref{s:lemmaclosed}) that the functions~$\hat{f}^\beta_{[0]}(q)$
diverge for large~$q^2$ only logarithmically. Thus in view of the smallness of the
fine structure constant~$e^2/4 \pi \approx 1/137$, for the energy range accessible by experiments
the square bracket in~\eqref{s:diracmaxwell} is an invertible operator. Thus we may write our
modified Dirac-Maxwell equations as
\beq \label{s:maxsource}
(i \Pdd + \slashed{A}-m) \psi = 0 \:, \qquad -\Box A_k = \left[ 1 - \frac{e^2}{12 \pi^2}\:f_{[0]}* \right]^{-1} e^2\, \overline{\psi} \gamma_k \psi  \:,
\eeq
showing that the convolution term can be regarded as a modification of the source term.
Alternatively, one may write the Maxwell equation in the standard form
\beq \label{s:reform1}
-\Box \tilde{A}_k = e^2\, \overline{\psi} \gamma_k \psi
\eeq
with a so-called {\em{auxiliary potential}} $\tilde{A}$
\sindex{potential!auxiliary}%
\nindex{ci6@$\tilde{A}$ -- auxiliary potential}%
and take the point of view that
the convolution term only affects the coupling of the electromagnetic potential to the Dirac equation,
\beq \label{s:reform2}
(i \Pdd + \slashed{A}-m) \psi = 0 \qquad \text{with} \qquad
A := \left[ 1 - \frac{e^2}{12 \pi^2}\:f_{[0]}* \right]^{-1} \tilde{A}
\eeq
(note that the wave and convolution operators commute, as they are both
multiplication operators in momentum space).
Both the ``source form''~\eqref{s:maxsource} and the ``coupling form''~\eqref{s:reform1} and~\eqref{s:reform2} are useful; they give different points of view
on the same system of equations.
We point out that, as the inverse of a causal operator, the operator on the right of~\eqref{s:maxsource}
and~\eqref{s:reform2} is again causal in the sense that its integral kernel vanishes for spacelike distances.
Moreover, for large timelike distances, the kernel~$f_{[0]}$ is oscillatory and decays.
More specifically, writing the Green's function~$S_a$ in~\eqref{s:f0rep} with Bessel functions
and using their asymptotic expansion for large~$\xi^2$, one finds that
\beq \label{s:f0asy}
f^\beta_{[0]}(x-y) \sim m_\beta\: (\xi^2)^{-\frac{3}{2}}\: \cos \left( \sqrt{4 m_\beta^2 \xi^2}
+ \varphi \right) \qquad \text{if $\xi^2 \gg m_\beta^2$}
\eeq
(where~$\varphi$ is an irrelevant phase).

The formulation~\eqref{s:reform1} and~\eqref{s:reform2} reveals that
our modified Dirac-Maxwell equations are of {\em{variational form}}.
\sindex{modified Dirac-Maxwell equations!in variational form}%
More precisely, they can be recovered as the EL equations corresponding to the modified
Dirac-Maxwell action
\[ \Sact_\text{DM} = \int_\scrM \bigg\{ \overline{\psi} \Big(i \Pdd + \left[ 1 - \frac{e^2}{12 \pi^2}\:f_{[0]}* \right]^{-1} 
\;\tilde{\!\!\slashed{A}} - m \Big) \psi
- \frac{1}{4 e^2} \tilde{F}_{ij}\: \tilde{F}^{ij} \bigg\} \,d^4x \:, \]
where~$\tilde{F}$ is the field tensor corresponding to the auxiliary potential.
Hence by applying Noether's theorem, we obtain corresponding conserved quantities,
in particular the total {\em{electric charge}} and the total {\em{energy}} of the system.
Thus all conservation laws of the classical Dirac-Maxwell system still hold, but
clearly the form of the conserved quantities must be modified by suitable convolution terms.

The simplest idea for detecting the convolution term is to expose an electron to a {\em{laser pulse}}.
Then the convolution term in the Dirac equation~\eqref{s:reform2} might seem to imply that
the electron should ``feel'' the electromagnetic wave at a distance, or could even be influenced
by a laser beam flying by in the future, at a time when the electron may already have moved away.
However, such obvious violations of causality are impossible for the following reason:
An electromagnetic wave satisfies the vacuum Maxwell equations~$\Box \tilde{A}=0$
(see~\eqref{s:reform1}). Thus the momentum squared of the electromagnetic wave vanishes,
implying that~$f_{[0]}*\tilde{A}=0$, so that the convolution term in~\eqref{s:reform2} drops out.
In more general terms, the convolution terms are constant if the bosonic field is on-shell.
We conclude that the convolution terms can be detected only by {\em{off-shell}} bosonic fields,
which according to~\eqref{s:reform1} occur only at the electromagnetic sources.

Another idea for observing the convolution term is that, according to~\eqref{s:maxsource},
it modifies the way the Dirac current generates an electromagnetic field.
Due to the prefactor~$e^2/(12 \pi^2)$ and in view of the fact that the kernel~$f_{[0]}$
decays and has an oscillatory behavior~\eqref{s:f0asy},
this effect will not be large, but it could nevertheless be observable.
In particular, one may ask whether the positive and negative charges of protons and electrons
still compensate each other in such a way that macroscopic objects appear neutral.
If this were not the case, this would have drastic consequences, because then the electromagnetic
forces would dominate gravity on the large scale.
To analyze this question we consider for example a crystal containing exactly as many positive and
negative charges. Then the corresponding auxiliary potential~$\tilde{A}$ vanishes outside the crystal
(except for dipole effects, which fall off rapidly with increasing distance).
As a consequence, the potential~$A$ defined by~\eqref{s:reform2} also vanishes outside the
crystal, and thus there are no observable electrostatic forces outside the crystal,
in agreement with physical observations.

More generally, the above considerations show that the convolution term can lead to
observable effects only if the sources of the electromagnetic field and the Dirac particles
on which it acts are very close to each other, meaning that the whole interaction must take
place on the scale of the Compton length of the electron.
One conceivable way of measuring this effect is by considering {\em{electron-electron scattering}}.
In order to concentrate on the violation of causality, it seems preferable to
avoid the noise of the usual electromagnetic interactions by considering two wave packets
which stay causally separated, but nevertheless come as close as the Compton length.
In this case, an electron in the future could even affect the motion of an electron in the past.
However, due to the Heisenberg uncertainty principle, localizing a wave packet on the Compton scale
implies that the energy uncertainty is of the order of the rest mass, so that pair creation becomes
a relevant effect. Therefore, arranging such wave packets seems a very difficult task.

Another potential method for observing the convolution term is to get a connection
to the high-precision measurements of atomic spectra. Thus we conclude the discussion
by considering the {\em{static}} situation.
Integrating the Green's function~\eqref{s:Samom} over time, we can compute the
remaining spatial Fourier integral with residues to obtain the familiar Yukawa potential,
\begin{align*}
V_a(\vec{\xi}) &:=\int_{-\infty}^\infty S_a(x,y) \,d\xi^0
= -\int_{\R^3} \frac{d\vec{k}}{(2 \pi)^3}\: \frac{e^{-i\vec{k}\vec{\xi}}}{|\vec{k}|^2+a} \\
&\:= -\frac{1}{(2 \pi)^2} \int_0^\infty \frac{k^2 dk}{k^2+a}
\int_{-1}^1 e^{-i k r \cos \vartheta} d\cos \vartheta \\
&\:= \frac{1}{(2 \pi)^2\, ir} \int_{-\infty}^\infty \frac{k}{k^2+a}\:e^{-i k r}\:dk
= -\frac{1}{4 \pi}\: \frac{e^{- \sqrt{a} r}}{r}\:,
\end{align*}
where we set~$r=|\vec{\xi}|$.
\nindex{ci8@$r$ -- radius $\vert \vec{\xi} \vert$}%
Hence in the static case, the convolution operator reduces to the three-dimensional integral
\[ (f_{[0]} * h)(\vec{x}) = \int_{\R^3} f_{[0]}(\vec{x}-\vec{y})\, h(\vec{y})\: d\vec{y} \]
involving the kernel
\beq
f_{[0]}(\vec{\xi}) = \frac{1}{3} \sum_{\beta=1}^3 \int_{4 m_\beta^2}^\infty 
\bigg[V_a(\vec{\xi}) + \frac{\delta^3(\vec{\xi})}{a} \bigg]
\sqrt{a-4m_\beta^2} \:(a+2m_\beta^2)\:
\frac{da}{a^\frac{3}{2}} \:. \label{s:fker}
\eeq
We now consider a classical point charge~$Ze$ located at the origin. In order to compute the
corresponding electric field~$A_0$, we consider the corresponding Maxwell equation~\eqref{s:maxsource},
\[ \Delta A_0(\vec{x}) = \left[ 1 - \frac{e^2}{12 \pi^2}\:f_{[0]}* \right]^{-1} Ze^2\, \delta^3(\vec{x})
= Ze^2\, \delta^3(\vec{x}) + \frac{Z e^4}{12 \pi^2}\:f_{[0]}(\vec{x}) + \O(e^6)\:. \]
In order to solve for $A_0$, we convolute both sides with the Newtonian
potential~$V_0(\xi)=-1/(4 \pi r)$. To compute the resulting convolution of the Newtonian potential
with~$f_{[0]}$, we first observe that~\eqref{s:fker} involves the Yukawa potential~$V_a$.
Since convolution corresponds to multiplication in momentum space, we can use the
simple transformation
\[ \frac{1}{|\vec{k}|^2}\: \frac{1}{|\vec{k}|^2+a} = \frac{1}{a} \left(- \frac{1}{|\vec{k}|^2+a} 
+\frac{1}{|\vec{k}|^2} \right) \]
to conclude that
\[ (V_0 * V_a)(\vec{x}) = \frac{1}{a} \left( V_a(\vec{x}) - V_0(\vec{x}) \right)\:. \]
We thus obtain
\[ A_0(\vec{x}) = -\frac{Ze^2}{4 \pi r}+ \frac{Ze^4}{12 \pi^2} \sum_{\beta=1}^3 \int_{4 m_\beta^2}^\infty 
\bigg[-\frac{e^{- \sqrt{a} |\vec{\xi}|}}{4 \pi r} \: \frac{1}{a} \bigg]
\sqrt{a-4m_\beta^2} \:(a+2m_\beta^2)\:
\frac{da}{a^\frac{3}{2}} \:.  \]
Here the first summand is the Coulomb potential, whereas the second summand is an
additional short-range potential. This is very similar to the situation for the relativistic correction described
by the Darwin term (a relativistic correction to the Schr\"odinger equation;
see for example~\cite[Section~3.3]{sakuraiadv}).
Concentrating the short range potential at the origin by the replacement
\[ \frac{e^{- \sqrt{a}\, r}}{4 \pi r} \rightarrow \frac{1}{a}\: \delta^3(\vec{x}) \:, \]
we can carry out the $a$-integral to obtain
\[ A_0(\vec{x}) = -\frac{Ze^2}{4 \pi r} - \frac{Ze^4}{60 \pi^2} \sum_{\beta=1}^3 
\frac{1}{m_\beta^2}\:  \delta^3(\vec{x})\:. \]
We thus end up with a correction to the Dirac Hamiltonian of the form
\beq \label{s:Hnonlocal}
H_\text{noncausal} = \frac{Ze^4}{60 \pi^2} \sum_{\beta=1}^3 
\frac{1}{m_\beta^2}\:  \delta^3(\vec{x})\:.
\eeq

This coincides precisely with the Uehling potential which describes the  one-loop vacuum
polarization in the static situation (see for example~\cite[eq.~(7.94) and~(7.95)]{peskin+schroeder}).
At first sight, it might be surprising that we get the same
result as in perturbative QFT, although we did not consider a fermionic loop
diagram and did not encounter the usual ultraviolet divergences. In order to see the
connection, it is preferable to reconsider the original derivation by Uehling and Serber~\cite{uehling,
serber}, which was based on earlier papers by Dirac~\cite{dirac3} and Heisenberg~\cite{heisenberg2}.
Similar to~\eqref{s:Perep}, Dirac considers the sum over all sea states,
\[ R(t, \vec{x}; t', \vec{x}') = \sum_{l \text{ occupied}} \psi_l(t, \vec{x}) \:\overline{\psi_l(t', \vec{x}')} \:, \]
where the wave functions~$\psi_l$ are solutions of the Dirac equation
\[ \big( i \Pdd + e \slashed{A}(t, \vec{x}) - m \big) \psi_l(t, \vec{x}) = 0 \:. \]
Thus up to an irrelevant overall constant, $R$ coincides precisely with the kernel of the fermionic
projector~$P(x,y)$.
Dirac realizes that~$R$ has singularities on the light cone and discusses their form.
Heisenberg pushes the calculation a bit further and, using physical
conservation laws and the requirement of gauge invariance, he brings the singular contribution to~$R$
into a canonical form. The he argues that this singular contribution to~$R$ should simply be
dropped. Uehling and Serber took the remaining regular contribution to~$R$ as the starting point
to derive the corresponding correction to the Maxwell equations.
Since the regular contribution to~$R$ coincides precisely with the non-causal contributions
in~\eqref{s:fprep} (albeit in a less explicit form where the underlying causal structure is not apparent),
it is clear that~\eqref{s:Hnonlocal} coincides with the usual Uehling potential.

We conclude that the non-causal correction reproduces the usual vacuum polarization
as described by the Uehling potential. The main difference in our approach is that
the singular contributions to the fermionic projector are not disregarded or removed, but they are
carried along in our analysis.
These singular contributions then drop out of the Euler-Lagrange equations corresponding to our
action principle~\eqref{s:actprinciple}. In this way, all divergences disappear.
The remaining finite contributions to the fermionic projector give rise to
the bosonic current and mass terms in the resulting field equations~\eqref{s:field2},
and also yield the convolution terms which describe the vacuum polarization.
The main advantage of the fermionic projector approach is that no counter terms are
needed. The back-reaction of the Dirac sea on the electromagnetic field is finite,
no divergences occur. Moreover, as we do not need counter terms computed from the
Minkowski vacuum, the
setting immediately becomes background independent. It is to be expected (although it has not yet
been worked out in detail) that the singularities of the fermionic projector will also drop out of the Euler-Lagrange equations if one sets up the theory in curved space-time.

We finally remark that the connection to Feynman diagrams will be explained in more detail
in~\S\ref{s:secquantcorr}.

\subsectionn{Higher Order Non-Causal Corrections to the Field Equations} \label{s:sechighorder}
\sindex{non-causal correction!of higher order}%
The non-causal convolution terms in the previous section were obtained by computing the
non-causal contributions in~\eqref{s:fprep} at the origin, considering the first order of the perturbation
expansion~\eqref{s:cpower}.
Likewise, the higher orders of this expansion also contribute to~$\tilde{P}^\lec$ and~$\tilde{P}^\hec$,
giving rise to higher order non-causal corrections to the field equations. In this section we briefly
discuss the structure of these correction terms (computing them in detail goes beyond the scope
of this book).
\sindex{field equations in the continuum limit!corrections!due to smooth, noncausal contributions to fermionic projector}%

It is natural to distinguish between the low and high energy contributions.
The non-causal {\em{low energy contribution}} $\tilde{P}^\lec$ in~\eqref{s:fprep} can be computed at
the origin to every order in~$\B$ by extending the resummation technique of Appendix~\ref{s:appresum}
to higher order (more precisely, according to the residual argument, we again
get sums of the form~\eqref{s:Tlight}, but with nested line integrals and multiple series,
which are to be carried out iteratively).
\sindex{fermionic projector!non-causal low energy contribution}%
\nindex{bj6@$\tilde{P}^\lec$ -- non-causal low energy contribution to fermionic projector}%
Similar as explained to first order after Lemma~\ref{s:lemma93}, the resulting corrections to the field
equation are weakly causal in the sense that they can be described by convolutions with integral kernels
which vanish for spacelike distances.
Thus they have the same mathematical structure, but are clearly much smaller than the
convolution terms in~\S\ref{s:secnocausal}.

The non-causal {\em{high energy contribution}} $\tilde{P}^\hec$ in~\eqref{s:fprep} is more interesting,
because it gives rise to corrections of different type.
\sindex{fermionic projector!non-causal high energy contribution}%
\nindex{bj8@$\tilde{P}^\hec$ -- non-causal high energy contribution to fermionic projector}%
For simplicity, we explain their mathematical structure
only in the case of one generation and only for the
leading contribution to~$\tilde{P}^\hec$ (see~\cite{grotz} for details)
\[ \tilde{P}^\hec = -\frac{\pi^2}{4} \Big( t_m \,\B\, \overline{t_m} \,\B\, t_m -
\overline{t_m} \,\B\, t_m \,\B\, \overline{t_m} \Big) + \O(\B^3) \:, \]
where we set
\[ t_m = (i \Pdd + m)\, T_{m^2} \qquad \text{and} \qquad \overline{t_m} = 
(i \Pdd + m)\, \overline{T_{m^2}}\:, \]
\nindex{cj0@$\overline{t_m}$ -- upper Dirac mass shell}%
and~$\overline{T_a}$ is the complex conjugate of the distribution~$T_a$, \eqref{s:Tadef}.
Thus the distributions~$t_m$ and~$\overline{t_m}$ are supported on the lower and upper
mass shell, respectively. Evaluating this expression at the origin gives
\beq \label{s:Pheor}
\begin{split}
\tilde{P}^\hec(x,x) = -\frac{\pi^2}{4} \int_\scrM \! d^4 z_1 &\int_\scrM \! d^4 z_2 \,
\Big(   t_m(x,z_1) \,\B(z_1)\, \overline{t_m}(z_1, z_2) \,\B(z_2)\, t_m(z_2, x) \\
- &\overline{t_m}(x,z_1) \,\B(z_1)\, t_m(z_1, z_2) \,\B(z_2)\, \overline{t_m}(z_2, x) \Big)  + \O(\B^3) \:.
\end{split}
\eeq
This is similar to a second order tree diagram, but instead of Green's functions it involves
the projectors onto the lower and upper mass shells, which appear in alternating order.
The expression is well-defined and finite (see~\cite[Lemma~2.2.2]{PFP} or Lemma~\ref{l:lemma0}).
Similar to the correction terms in Theorem~\ref{s:thmfield}, our expression is a convolution,
but now it involves two integrals, each of which contains one factor of~$\B$.
Consequently, the integral kernel depends on two arguments~$z_1$ and~$z_2$.
The interesting point is that this integral kernel does not vanish even if the vectors~$z_1-x$
or~$z_2-x$ are space-like. Thus the corresponding corrections to the field equations
violate causality even in the strong sense that in addition to an influence of the future on the past,
there are even {\em{interactions for spacelike distances}}.
\sindex{causality violation!for spacelike distances}%
This surprising result is in
sharp contrast to conventional physical theories. However, since for space-like separation the
kernels~$t_m$ decay exponentially fast on the Compton scale, the effect is extremely small.
In particular, describing this exponential decay by the Yukawa potential,
this effect could be described similar to the correction~\eqref{s:fker} and~\eqref{s:Hnonlocal}.
But compared to the latter first order correction, the second order correction by~$\tilde{P}^\hec$ would
be smaller by a factor~$e^2$. In view of the discussion in~\S\ref{s:secnocausal},
measuring this correction is at present out of reach. Thus it seems that
the only promising approach for detecting an effect of the high energy contribution
is to look for an experiment which is sensitive to interactions between regions of space-time
with spacelike separation, without being affected by any causal interactions.

\subsectionn{The Standard Quantum Corrections to the Field Equations} \label{s:secquantcorr}
\sindex{quantum corrections}%
We now explain how the quantum corrections due to the Feynman loop diagrams
arise in our model. We will recover all the standard quantum corrections. Moreover, we will
obtain quantum corrections of the previously described non-causal terms
(see~\S\ref{s:secnocausal} and~\S\ref{s:sechighorder}). 
For clarity, we proceed in several steps and
begin by leaving out the non-causal convolution terms in the field equations~\eqref{s:field2}.
Furthermore, we consider only one Dirac particle of mass~$m$. Under this simplifying assumption,
the interaction is described by the coupled Dirac-Yang/Mills equations
\beq \label{s:DYM}
(i \Pdd + \pseudo \slashed{A}-m) \psi = 0 \:, \qquad
\partial_{kl} A^l - \Box A_k - M^2 A_k = e^2 \,\overline{\psi} \pseudo \gamma_k \psi \:,
\eeq
\sindex{Dirac-Yang/Mills equations}%
where~$A$ is the axial potential, and the bosonic rest mass~$M$ and the coupling constant~$e$
are determined from~\eqref{s:field2} by setting~$M^2=C_2/C_0$ and~$e^2=12 \pi^2/C_0$.
We point out that the wave function~$\psi$ and the bosonic field~$A$ in~\eqref{s:DYM} are classical
in the sense that no second quantization has been performed.

The equations~\eqref{s:DYM} form a coupled system of nonlinear hyperbolic partial differential
equations. For such a system, standard methods give local existence and uniqueness
results (see for example~\cite[Section~5.3]{john} or~\cite[Chapter~16]{taylor3}),
but constructing global solutions is a very difficult task. 
Therefore, we must rely on a perturbative treatment, giving us a connection to Feynman diagrams.
\sindex{Feynman diagram}%
Although this connection is quite elementary, it does not seem to be well-known to mathematicians
working on partial differential equations. In physics, on the other hand, Feynman diagrams
are usually derived from second quantized fields, where the connection to nonlinear partial
differential equations is no longer apparent. Therefore, we now explain the procedure schematically
from the basics, hopefully bridging a gap between the mathematics and physics communities.
In order to be in a simpler and more familiar setting, we consider instead of~\eqref{s:DYM}
the Dirac-Maxwell equations
\sindex{Dirac-Maxwell equations}%
in the Lorentz gauge, as considered in quantum electrodynamics
(see for example~\cite{bjorken})\footnote{
In order to bring the system~\eqref{s:DYM} into a comparable form, one first takes the
divergence of the Yang/Mills equation to obtain
\[ - M^2 \partial_k A^k = e^2 \, \overline{\partial_k \psi} \pseudo \gamma^k \psi
+ e^2 \,\overline{\psi} \pseudo \gamma^k \partial_k \psi = 
-2 i e^2 m \: \overline{\psi} \pseudo \psi\: , \]
where in the last step we used the Dirac equation.
In particular, the divergence of~$A$ in general does not vanish.
It seems convenient to subtract from~$A$ the gradient of a scalar field~$\Phi$,
\[ B_k := A_k - \partial_k \Phi \:, \]
in such a way that the new potential~$B$ becomes divergence-free. This leads to the
system of equations
\[ \big( i \Pdd -m+\pseudo \slashed{B} + \pseudo (\Pdd \Phi) \big) \psi = 0\:,\;\;\;
-\Box \Phi = -\frac{2 i e^2 m}{M^2} \: \overline{\psi} \pseudo \psi \:,\;\;\;
(-\Box-M^2) B_k = e^2 \,\overline{\psi} \pseudo \gamma_k \psi + M^2 \partial_k \Phi \:. \]
This system has the same structure as~\eqref{s:DM}, and it can be analyzed with
exactly the same methods. For the handling of the factors~$e$ see Footnote~\ref{s:units}
on page~\pageref{s:units}.}
\beq \label{s:DM}
(i \Pdd + e \slashed{A}-m) \psi = 0 \:, \qquad
- \Box A_k = e \,\overline{\psi} \gamma_k \psi \:. 
\eeq
The natural question in the theory of hyperbolic partial differential equations is the Cauchy problem,
where we seek for solutions of~\eqref{s:DM} for given initial values
\sindex{Cauchy problem}%
\beq \label{s:DMinit}
\psi(t,\vec{x})|_{t=0} = \psi_0(\vec{x}) \:,\qquad A(t,\vec{x})|_{t=0}=A_0(\vec{x}) \:,\quad \partial_t A(t,\vec{x})|_{t=0}=A_1(\vec{x})\:.
\eeq
In preparation, we formulate the equations as a system which is of first order in time. To this end,
we introduce the field~$\Phi$ with components
\beq \label{s:Phidef}
\Phi = \begin{pmatrix} \psi \\ A \\ i \partial_t A \end{pmatrix} ,
\eeq
and write the system in the Hamiltonian form
\beq \label{s:hamilt}
i \partial_t \Phi(t,\vec{x}) = H \big( \Phi(t,\vec{x}) \big) := H_0 \Phi + e B(\Phi) \:,
\eeq
\nindex{cj2@$H$ -- full Hamiltonian}%
where in the last step we decomposed the Hamiltonian into its linear and non-linear parts given by
\beq \label{s:H0Bdef}
H_0 = \begin{pmatrix} -i \gamma^0 \,\vec{\gamma} \vec{\nabla} + \gamma^0 m & 0 & 0 \\
0 & 0 & 1 \\ 0 & -\Delta & 0 \end{pmatrix} , \qquad
B(\Phi) = \begin{pmatrix} -\gamma^0 \slashed{A} \psi \\ 0 \\ \overline{\psi} \gamma \psi \end{pmatrix} .
\eeq
\nindex{cj4@$H_0$ -- free Hamiltonian}%
\nindex{cj6@$B$ -- perturbation of Hamiltonian}%
In the case~$e=0$, we have a linear equation,
 which is immediately solved formally by exponentiation,
\[ \Phi(t) = e^{-i t H_0} \Phi_0 \:, \]
where we set~$\Phi_0 = \Phi|_{t=0}$. This equation is given a rigorous meaning by writing
the so-called time evolution operator~$e^{-i t H_0}$ as an integral operator in the spatial variables.
\begin{Lemma} \label{s:lemmafree}
For any~$t \geq 0$, the operator~$e^{-i t H_0}$ can be written as
\beq \label{s:Rprop}
(e^{-i t H_0} \Phi)(\vec{x}) = \int_{\R^3} R_t(\vec{x} - \vec{y})\: \Phi(\vec{y})\: d\vec{y}\:,
\eeq
where the integral kernel is the distribution
\beq \label{s:Rtdef}
R_t(\vec{x}) =   \begin{pmatrix} s_m^\wedge(t, \vec{x}) \:(i \gamma^0) & 0 & 0 \\
0 & -\partial_t S_0^\wedge(t, \vec{x}) & i S_0^\wedge(t, \vec{x}) \\
0 & -i \Delta S_0^\wedge(t, \vec{x}) & -\partial_t S_0^\wedge(t, \vec{x})
\end{pmatrix} ,
\eeq
which involves the retarded Green's functions defined by
\sindex{Green's function!retarded}
\nindex{bg0@$S_{m^2}^\vee, S_{m^2}^\wedge$ -- causal Green's functions of the Klein-Gordon equation}%
\begin{align}
S^\wedge_a(x) &= \lim_{\varepsilon \searrow 0}
\int \frac{d^4 k}{(2 \pi)^4}\;
\frac{e^{-i k x}}{k^2-m^2 + i \varepsilon k^0}  \\
s^\wedge_m(t,\vec{x}) &= (i \Pdd_x + m)\: S^\wedge_{m^2}(x)\:. \label{s:smret}
\end{align}
\end{Lemma}
\Proof Using that for any~$t>0$, the Green's function~$S^\wedge_a$ is a solution of the
Klein-Gordon equation~$(-\Box - a) S_a(x)=0$, a short calculation using~\eqref{s:smret}
shows that~\eqref{s:Rprop} is a solution of the equation~$(i \partial_t - H_0) (e^{-i t H_0} \Phi)=0$.
In order to verify the correct initial conditions, we differentiate~$S^\wedge_a$ with
respect to time and carry out the $t$-integration with residues to obtain
\begin{align*}
\lim_{t \searrow 0} &\: \partial_t^n S^\wedge_a(t,\vec{x}) = \frac{1}{(2 \pi)^4} \int_{\R^3}
d \vec{k} \:e^{i \vec{k} \vec{x}} \: \lim_{\varepsilon, t \searrow 0} \int_{-\infty}^\infty
\frac{(-i \omega)^n}{\omega^2- |\vec{k}|^2-m^2 + i \varepsilon \omega}\: e^{-i \omega t} \:d\omega \\
&= \frac{1}{(2 \pi)^4} \int_{\R^3} d \vec{k} \:e^{i \vec{k} \vec{x}}
\; (-2 \pi i)\: \frac{(-i \omega)^n}{2 \omega} \Big|_{\omega= \pm \sqrt{|\vec{k}|^2+m^2}}
= \left\{ \begin{array}{cl} 0 & \text{if~$n=0$} \\
-\delta^3(\vec{x}) & \text{if~$n=1\:.$} \end{array} \right. 
\end{align*}
Using this result in~\eqref{s:smret} and~\eqref{s:Rtdef} shows that indeed~$\lim_{t \searrow 0}
R_t(\vec{x}) = \delta^3(\vec{x})$.
\QED
In the nonlinear situation~$e \neq 0$, it is useful to work in the so-called ``interaction
picture'' (see for example~\cite[Section~8.5]{schwabl1}).
\sindex{interaction picture}%
We thus employ the ansatz
\beq \label{s:intpic}
\Phi(t) = e^{-i t H_0} \Phi_\text{int}(t) \:,
\eeq
giving rise to the nonlinear equation
\beq \label{s:Hinteq}
i \partial_t \Phi_\text{int} = e B_\text{int}(\Phi_\text{int}(t))\:,
\eeq
\nindex{cj8@$B_\text{int}$ -- perturbation in interaction picture}%
where
\[ B_\text{int}(\Phi_\text{int}(t)) = e^{i t H_0} B \big( e^{-i t H_0} \Phi_\text{int}(t) \big) . \]
We regard~\eqref{s:Hinteq} as an ordinary differential equation in time, which 
in view of~\eqref{s:Rprop} is nonlocal in space.
From~\eqref{s:intpic} one sees that~$\Phi_\text{int}$ comes with
the initial data~$\Phi_\text{int}|_{t=0} = \Phi_0$. Taking a power ansatz in~$e$,
\[ \Phi_\text{int}(t) = \Phi_\text{int}^{(0)}(t) + e \,\Phi_\text{int}^{(1)}(t) + e^2 \,\Phi_\text{int}^{(2)}(t)
+ \cdots \:, \]
a formal solution of the Cauchy problem for~$\Phi_\text{int}$ is obtained 
by integrating~\eqref{s:Hinteq} inductively order by order,
\begin{align*}
\Phi_\text{int}^{(0)}(t) &= \Phi_0 \:,\qquad
\Phi_\text{int}^{(1)}(t) = -i \int_0^t B_\text{int} \Big( \Phi_\text{int}^{(0)}(\tau) \Big)\, d\tau \\
\Phi_\text{int}^{(2)}(t) &= -i \int_0^t \nabla B_\text{int} \Big(\Phi_\text{int}^{(0)}(\tau) \Big)  \cdot
\Phi_\text{int}^{(1)}(\tau)\: d\tau \\
&= (-i)^2 \int_0^t d\tau \:\nabla B_\text{int} \Big( \Phi_\text{int}^{(0)}(\tau) \Big)
\int_0^\tau d\sigma \:B_\text{int} \Big( \Phi_\text{int}^{(0)}(\sigma) \Big) , \quad \ldots
\end{align*}
(here~$\nabla B$ denotes the Jacobi matrix of~$B$, where as in~\eqref{s:grad}
we consider the real and imaginary parts of the arguments as independent variables).
Substituting these formulas into~\eqref{s:intpic}, we obtain the desired solution~$\Phi$
of the original Cauchy problem expressed as a sum of iterated time integrals,
involving intermediate factors of the free time evolution operator~$e^{-i \tau H_0}$.
In particular, we obtain to second order
\begin{align*}
\Phi(t) =&\: e^{-i t H_0} \Phi_0 - ie \int_0^t e^{-i (t-\tau) H_0} B \big( e^{-i \tau H_0}
\Phi_0 \big) \,d\tau \\
& - e^2 \int_0^t d\tau \:e^{-i (t-\tau) H_0} \:\nabla B \big( e^{-i \tau H_0} \Phi_0 \big)
\int_0^\tau d\sigma \: e^{-i (\tau-\sigma) H_0} \:B \big( e^{-i \sigma H_0} \big)
+ \O(e^3)\:.
\end{align*}
We remark that in the case when~$B(\Phi)$ is linear in~$\Phi$, this expansion simplifies
to the well-known Dyson series (also referred to as the time-ordered exponential).
\sindex{Dyson series}%
In view of~\eqref{s:Phidef}, we have derived a unique formal solution of the Cauchy
problem~\eqref{s:DM} and~\eqref{s:DMinit}.

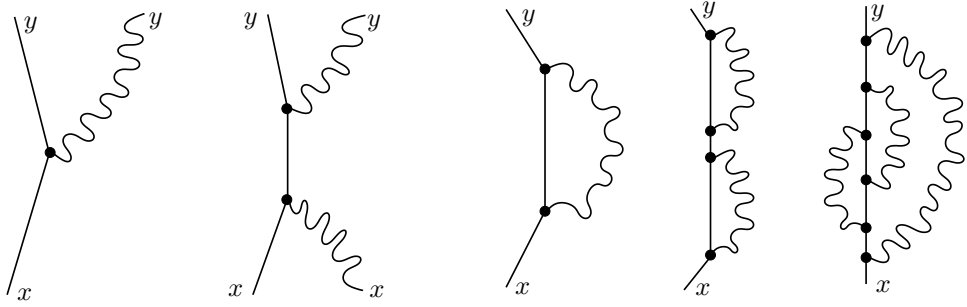
\begin{figure}
\begin{picture}(0,0)%
\includegraphics{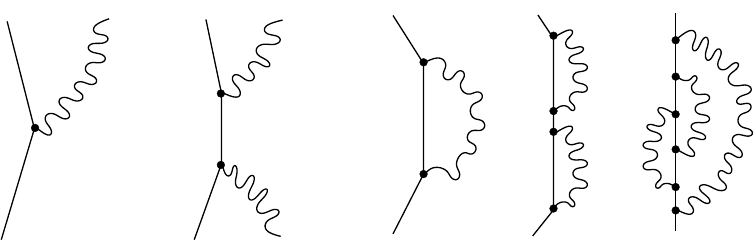}%
\end{picture}%
\setlength{\unitlength}{1367sp}%
\begingroup\makeatletter\ifx\SetFigFontNFSS\undefined%
\gdef\SetFigFontNFSS#1#2#3#4#5{%
  \reset@font\fontsize{#1}{#2pt}%
  \fontfamily{#3}\fontseries{#4}\fontshape{#5}%
  \selectfont}%
\fi\endgroup%
\begin{picture}(17408,5743)(-8492,-5027)
\put(-8279,-4891){\makebox(0,0)[lb]{\smash{{\SetFigFontNFSS{10}{12.0}{\rmdefault}{\mddefault}{\updefault}$x$}}}}
\put(-8159, 59){\makebox(0,0)[lb]{\smash{{\SetFigFontNFSS{10}{12.0}{\rmdefault}{\mddefault}{\updefault}$y$}}}}
\put(-5894,134){\makebox(0,0)[lb]{\smash{{\SetFigFontNFSS{10}{12.0}{\rmdefault}{\mddefault}{\updefault}$y$}}}}
\put(-4454,-4816){\makebox(0,0)[lb]{\smash{{\SetFigFontNFSS{10}{12.0}{\rmdefault}{\mddefault}{\updefault}$x$}}}}
\put(-4169,119){\makebox(0,0)[lb]{\smash{{\SetFigFontNFSS{10}{12.0}{\rmdefault}{\mddefault}{\updefault}$y$}}}}
\put(-1934, 89){\makebox(0,0)[lb]{\smash{{\SetFigFontNFSS{10}{12.0}{\rmdefault}{\mddefault}{\updefault}$y$}}}}
\put(751,-4771){\makebox(0,0)[lb]{\smash{{\SetFigFontNFSS{10}{12.0}{\rmdefault}{\mddefault}{\updefault}$x$}}}}
\put(4021,-4801){\makebox(0,0)[lb]{\smash{{\SetFigFontNFSS{10}{12.0}{\rmdefault}{\mddefault}{\updefault}$x$}}}}
\put(7201,269){\makebox(0,0)[lb]{\smash{{\SetFigFontNFSS{10}{12.0}{\rmdefault}{\mddefault}{\updefault}$y$}}}}
\put(7291,-4741){\makebox(0,0)[lb]{\smash{{\SetFigFontNFSS{10}{12.0}{\rmdefault}{\mddefault}{\updefault}$x$}}}}
\put(4141,329){\makebox(0,0)[lb]{\smash{{\SetFigFontNFSS{10}{12.0}{\rmdefault}{\mddefault}{\updefault}$y$}}}}
\put(871,179){\makebox(0,0)[lb]{\smash{{\SetFigFontNFSS{10}{12.0}{\rmdefault}{\mddefault}{\updefault}$y$}}}}
\put(-1889,-4831){\makebox(0,0)[lb]{\smash{{\SetFigFontNFSS{10}{12.0}{\rmdefault}{\mddefault}{\updefault}$x$}}}}
\end{picture}%
\caption{Feynman tree diagrams (left) and bosonic loop diagrams (right)}
\label{s:figbosonic}
\end{figure}

Combining the above expansion of~$\Phi(t)$ with the formula for the time evolution
operator in Lemma~\ref{s:lemmafree}, one can write the above perturbation expansion in a
manifestly covariant form. Namely, when multiplying the operators~$R_t$ with~$B$
(or similarly~$\nabla B$ or higher derivatives), the factors~$\gamma^0$ in the first component
of~\eqref{s:Rtdef} and in the formula for~$B$ in~\eqref{s:H0Bdef} cancel each other, giving the
Lorentz invariant expression~$s^\wedge \slashed{A}$. Likewise, the Dirac current in~\eqref{s:H0Bdef}
multiplies the retarded Green's function~$S^\wedge_0$.
Moreover, we can combine the spatial and time time integrals to integrals over Minkowski space.
In this way,  we can identify the contributions to the perturbation expansion with the familiar
Feynman diagrams. More precisely, every integration variable corresponds to a vertex of
the diagram, whereas the bosonic and fermionic Green's functions~$S^\wedge_0$ and~$s^\wedge_m$
are written as wiggled and straight lines, respectively. Denoting the argument of the
solution~$\Phi(t, \vec{y})$ by~$y$, whereas~$x=(0, \vec{x})$ stands for the argument of the
initial values, we obtain all {\em{tree diagrams}}
\sindex{Feynman diagram!tree diagram}%
as exemplified in Figure~\ref{s:figbosonic} (left).
We come to the following conclusion:
\begin{itemize}[leftmargin=2em]
\itemD All tree diagrams are obtained from the nonlinear system of partial
differential equations~\eqref{s:DM}, working purely with classical fields.
\end{itemize}
We point out that the {\em{bosonic loop diagrams}}
\sindex{Feynman diagram!bosonic loop}
are missing as a consequence of the strict time ordering in the solution
of the Cauchy problem. However, if one introduces a microscopic background
field or takes into account the mechanism of {\em{microscopic mixing}},
then one also gets bosonic loop diagrams (as for example in Figure~\ref{s:figbosonic} (right)).
We shall not enter these constructions here but refer the interested reader
to~\cite{loop, qft}.

In order to make the connection to QFT clearer, we point out that in
quantum physics one usually does not consider the initial value problem~\eqref{s:DMinit}.
Instead, one is interested in the $n$-point functions, which give information about the
correlation of the fields at different space-time points. The {\em{two-point function}}
is obtained by choosing initial values involving $\delta^3$-distributions. Similarly, all the $n$-point
functions can be recovered once the solution of the Cauchy problem is known.
Thus from a conceptual point of view, the only difference between our expansion
and the Feynman diagrams in QFT is that, since in quantum physics the Feynman
diagrams do not come from an initial value problem, there is a {\em{freedom in choosing the
Green's function}}. Note that in the setting of the Cauchy problem, one necessarily gets the
retarded Green's function (see~\eqref{s:Rtdef}). In contrast, in QFT one is
free to work instead with any other Green's function. Indeed, different choices lead to
different approaches for handling the perturbation series. The most common choice is
the so-called {\em{Feynman propagator}} (see for example~\cite{bjorken}), which is
motivated from the physical picture that the positive frequencies (describing particles)
move to the future, whereas the negative frequencies (corresponding to anti-particles) move
to the past. In this standard approach,
the loop diagrams diverge. This problem is bypassed in the renormalization program
by first regularizing the diagrams, and then removing the regularization while
simultaneously adjusting the masses and coupling constants (see for example~\cite{collins}).
\sindex{renormalization}%
A QFT is called {\em{renormalizable}} if this renormalization procedure works to all
orders in perturbation theory, involving only a finite number of effective constants.
There are different equivalent renormalization procedures, the most common
being dimensional renormalization (see for example~\cite{peskin+schroeder}).
But the Feynman propagator is not a canonical choice, and indeed this choice suffers from the problem of
not being invariant under general coordinate transformations (for more details see~\cite[Section~2.1]{PFP}).
An alternative method, which seems natural but has not yet been worked out, would be to extend the
choice of Green's functions in the causal perturbation expansion~\eqref{s:cpower}
(see also~\cite{grotz, norm}) to the loop diagrams.
Yet another method is the so-called {\em{causal approach}} based on ideas of
Epstein and Glaser~\cite{epstein+glaser}, which uses the freedom in choosing the Green's function
to avoid the divergences of QFT (see also~\cite{scharf}).
We also mention that our above derivation of Feynman diagrams is certainly not the
most sophisticated or most elegant method. Maybe the cleanest method for the formal
perturbation expansion is obtained in the framework of path integrals (see for
example~\cite{kleinert, pokorski}).

\begin{figure}
\begin{picture}(0,0)%
\includegraphics{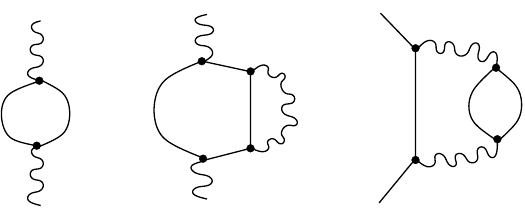}%
\end{picture}%
\setlength{\unitlength}{1367sp}%
\begingroup\makeatletter\ifx\SetFigFontNFSS\undefined%
\gdef\SetFigFontNFSS#1#2#3#4#5{%
  \reset@font\fontsize{#1}{#2pt}%
  \fontfamily{#3}\fontseries{#4}\fontshape{#5}%
  \selectfont}%
\fi\endgroup%
\begin{picture}(12076,4933)(-8888,-4727)
\put(-7859,-4591){\makebox(0,0)[lb]{\smash{{\SetFigFontNFSS{10}{12.0}{\rmdefault}{\mddefault}{\updefault}$x$}}}}
\put(-7829,-481){\makebox(0,0)[lb]{\smash{{\SetFigFontNFSS{10}{12.0}{\rmdefault}{\mddefault}{\updefault}$y$}}}}
\put(-4019,-4591){\makebox(0,0)[lb]{\smash{{\SetFigFontNFSS{10}{12.0}{\rmdefault}{\mddefault}{\updefault}$x$}}}}
\put(-3899,-331){\makebox(0,0)[lb]{\smash{{\SetFigFontNFSS{10}{12.0}{\rmdefault}{\mddefault}{\updefault}$y$}}}}
\put(121,-4591){\makebox(0,0)[lb]{\smash{{\SetFigFontNFSS{10}{12.0}{\rmdefault}{\mddefault}{\updefault}$x$}}}}
\put(211,-181){\makebox(0,0)[lb]{\smash{{\SetFigFontNFSS{10}{12.0}{\rmdefault}{\mddefault}{\updefault}$y$}}}}
\end{picture}%
\caption{Typical Feynman diagrams involving fermion loops}
\label{s:figfermionic}
\end{figure}
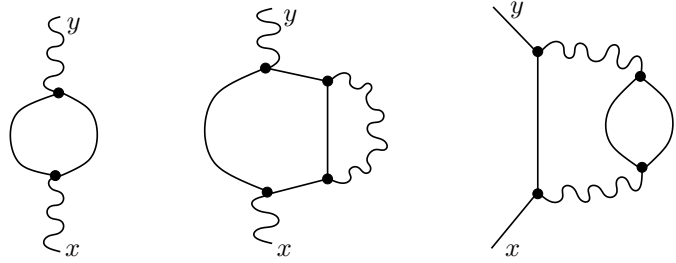

Recall that one simplification of the system~\eqref{s:DYM} was that we considered
only one Dirac particle and disregarded the interaction of this particle with the states
of the Dirac sea. In particular, we did not allow for the creation of a particle/anti-particle
pair. This shortcoming is reflected in our perturbation expansion in that
the {\em{fermionic loop diagrams}} are missing (see Figure~\ref{s:figfermionic} for a few examples).%
\sindex{Feynman diagram!fermionic loop}%
However, we already encountered the loop diagram in Figure~\ref{s:figfermionic} (left) when
discussing the vacuum polarization in~\S\ref{s:secnocausal}. This shows that
the fermion loops appear once we take into account the convolution terms in~\eqref{s:field2}
and the other non-causal corrections mentioned in~\S\ref{s:sechighorder}. In order to
include these corrections in the above expansion, we simply add them to the perturbation
operator~$e B$ in~\eqref{s:hamilt}. We conclude that
\begin{itemize}[leftmargin=2em]
\itemD The framework of the fermionic projector in the continuum limit yields
all tree diagrams and all fermionic loop diagrams.
\end{itemize}
We point out that this statement does not imply that the framework of the fermionic projector
is equivalent to perturbative QFT. As a major difference, the perturbation expansion of
the fermionic projector involves a non-trivial combinatorics of operator products involving different types
of Green's functions and fundamental solutions (see~\cite{grotz} and Section~\ref{secfpext}
for details or~\eqref{s:Pheor} for an example of a contribution which is absent in standard QFT).
This difference has no influence on the singularities of the resulting Feynman diagrams, 
and thus we expect that the renormalizability of the theory is not affected.
But the higher-loop radiative corrections should depend on the detailed combinatorics, giving the
hope to obtain small deviations from standard QFT which might be tested experimentally.

We close this section with two remarks. First, we again point out that
in the above derivation of the Feynman diagrams, we worked with classical
bosonic fields. This raises the question why a quantization of the bosonic fields
is at all needed, and what this ``quantization'' actually means.
Here we shall not enter a discussion of this point, but refer the
reader to~\cite[Section~4]{rev} and to the constructions in~\cite{entangle, qft}.
The second remark is that the described method of first taking the continuum limit, then
expanding the resulting equations in terms of Feynman diagrams and renormalizing
these diagrams should be considered as preliminary. This method has the great advantage that
it gives a simple connection to Feynman diagrams and to the renormalization program,
making it easier to compare our approach to standard QFT.
But ultimately, a fully convincing theory should work exclusively with the regularized fermionic projector,
thereby completely avoiding the ultraviolet divergences of QFT.
Before one can attack this program, one needs to have a better understanding
of our variational principle in discrete space-time.

\subsectionn{The Absence of the Higgs Boson} \label{s:secnohiggs}
In this section we compare the mechanism leading to the mass term in the field equations
(see~\eqref{s:YM1} and~\eqref{s:field2}) with the Higgs mechanism of the standard model.
\sindex{Higgs mechanism}%
\sindex{Higgs field}%
Clearly, our framework is considerably different from that of the standard model,
so that no simple comparison is possible. But in order to make the connection between the
formalisms as close as possible, we can consider the effective action of the continuum limit and compare
it to the action of a corresponding model involving a Higgs field.
\sindex{effective action}%
\sindex{action!effective}%
For simplicity leaving out the non-causal convolution terms and considering only one particle of
mass~$m$, the field equations~\eqref{s:field2} coupled to the Dirac equation are recovered
as the EL equations corresponding to the action
\beq \label{s:Seff}
{\mathcal{S}}_\text{DYM} =  \int_\scrM \left\{ \overline{\psi} (i \Pdd + \pseudo \slashed{A} - m) \psi
-\frac{1}{4e^2}\: F_{ij} F^{ij} + \frac{M^2}{2 e^2}\: A_j A^j 
\right\} d^4x\:,
\eeq
where~$A$ denotes the axial potential and~$F_{ij} = \partial_i A_j - \partial_j A_i$ is the
corresponding field tensor. 
\nindex{ck0@$F_{ij}$ -- field tensor}%
\nindex{ck2@$M$ -- bosonic mass}%
The coupling constant~$e$ and the bosonic mass~$M$ are
related to the constants in~\eqref{s:field2} by~$e^2 = 12 \pi^2/C_0$ and~$M^2 = C_2/C_0$.
We point out that this action is {\em{not}} invariant under the axial gauge transformation
\beq \label{s:axial}
\psi(x) \rightarrow e^{-i \pseudo \Lambda(x)} \psi(x) \:, \qquad
A \rightarrow A + \partial \Lambda\:,
\eeq
because both the fermionic mass term~$m \overline{\psi} \psi$ and the bosonic mass
term~$M^2 A_j A^j/(2 e^2)$ have no axial symmetry. 
As explained in~\S\ref{s:sec72}, the absence of an axial symmetry can be understood
from the fact that the transformation of the wave function in~\eqref{s:axial} is not unitary,
and thus it does not correspond to a local symmetry of our functionals in~\eqref{s:STdef} (see
also~\eqref{s:axialgauge} and~\eqref{s:axialchain}).

The axial gauge transformation~\eqref{s:axial}
\sindex{gauge transformation!axial}%
can be realized as a local symmetry
by adding a Higgs field~$\phi$,
\sindex{Higgs field}%
in complete analogy to the procedure in the standard
model. More precisely, we introduce~$\phi$ as a complex scalar field which behaves under
axial gauge transformations as
\beq \label{s:axialHiggs}
\phi(x) \rightarrow e^{-2i \Lambda(x)} \phi(x)\:.
\eeq
The fermionic mass term can be made gauge invariant by inserting suitable factors of~$\phi$.
Moreover, in view of~\eqref{s:axial}, we can introduce a corresponding gauge-covariant derivative~$D$ by
\[ D_j = \partial_j + 2 i A_j \:. \]
Thus the {\em{Dirac-Yang/Mills-Higgs action}} defined by
\sindex{Dirac-Yang/Mills-Higgs action}%
\beq \label{s:SDYMH} \begin{split}
{\mathcal{S}}_\text{DYMH} &=  \int_\scrM \left\{ \overline{\psi} (i \Pdd + \pseudo \slashed{A}) \psi
- m \overline{\psi} (\phi \chi_L + \overline{\phi} \chi_R) \psi \right. \\
&\qquad\quad\; \left. -\frac{1}{4e^2}\: F_{ij} F^{ij} + \frac{M^2}{8 e^2}\, (\overline{D_j \phi})
(D^j \phi) - V\! \left(|\phi|^2 \right) \right\} d^4x \end{split}
\eeq
is invariant under the axial gauge transformation~\eqref{s:axial} and~\eqref{s:axialHiggs}.
We now follow the construction of spontaneous symmetry breaking in the standard model.
\sindex{spontaneous symmetry breaking}%
For~$V$ we choose a double well potential having its minimum at~$|\phi|^2=1$.
Then the Higgs field~$\phi$ has a non-trivial vacuum with~$|\phi|=1$. Thus choosing an axial gauge
where~$\phi$ is real and positive, we can write~$\phi$ as
\[ \phi(x) = 1 + h(x) \]
with a real-valued field~$h$. Since~$h$ vanishes in the vacuum, we may expand
the action in powers of~$h$. Taking the leading orders in~$h$, we obtain the
{\em{action after spontaneous symmetry breaking}}
\[ {\mathcal{S}}_\text{DYMH} =  {\mathcal{S}}_\text{DYM} + {\mathcal{S}}_\text{Higgs} \]
with
\beq \label{s:SHiggs}
{\mathcal{S}}_\text{Higgs} = 
\int_\scrM \left\{- m h\: \overline{\psi} \psi 
+ \frac{M^2 h}{e^2}\: A_j A^j  + \frac{M^2}{8 e^2} \,(\partial_j h) (\partial^j h) - 2 V''(1)\:
h^2 \right\} d^4x \:.
\eeq
We conclude that for the action~\eqref{s:SDYMH}, the Higgs mechanism yields an action
which reproduces the effective action of the continuum limit~\eqref{s:Seff}, but gives rise
to additional terms involving a real Higgs boson~$h$. The Higgs boson has a rest mass
as determined by the free parameter~$V''(1)$. It couples to both the wave function~$\psi$
and the axial potential~$A$.

The Higgs field~$h$ can also be described in the setting of the fermionic projector,
as we now explain. Note that the coupling terms of the Higgs field to~$\psi$ and~$A$ can be
obtained from~\eqref{s:Seff} by varying the masses according to
\[ m \rightarrow (1+h(x)) \,m \:,\qquad M \rightarrow (1+h(x))\, M \:. \]
Taking into account that in our framework, the bosonic masses are given in terms of
the fermion masses (see~\eqref{s:field1}), this variation is described simply by
inserting a {\em{scalar potential}} into the Dirac equation.
\sindex{potential!scalar}%
Likewise, for a system involving
several generations, we must scale all fermion masses by a factor~$1+h$.
This is implemented in the auxiliary Dirac equation~\eqref{s:diracPaux} by choosing
\beq \label{s:scalpert}
\B = -m h(x) \,Y\:.
\eeq

The remaining question is whether scalar perturbations of the form~\eqref{s:scalpert}
occur in the setting of the fermionic projector, and whether our action principle~\eqref{s:actprinciple}
reproduces the dynamics of the Higgs field as described by~\eqref{s:SHiggs}.
A-priori, {\em{any}} symmetric perturbation of the Dirac equation is admissible, and thus we
can certainly consider the scalar perturbation~\eqref{s:scalpert}.
Since~\eqref{s:scalpert} is even under parity transformations, the corresponding leading perturbations of
the eigenvalues~$\lambda^L_s$ and~$\lambda^R_s$ will be the same.
In view of~\eqref{s:RRdef} and the formulas for the unperturbed eigenvalues~\eqref{s:unperturb},
we find that the leading contributions by~\eqref{s:scalpert} drop out of the EL
equations. We thus conclude that, although a Higgs field can be described in our framework,
the action principle~\eqref{s:actprinciple} does not describe a dynamics of this field,
but instead predicts that the Higgs field must vanish identically
(for more details on scalar perturbations see Lemma~\ref{s:lemmascal}).

We point out that for systems involving the direct sum of several sectors
as studied in Chapters~\ref{lepton} and~\ref{quark}, it is conceivable
that scalar/pseudoscalar potentials indeed give rise to dynamical degrees of freedom which
can be identified with the Higgs field. The corresponding field equations could be obtained
by analyzing the EL equations to degree three on the light cone
(see also Section~\ref{q:sechiggs}).

\subsectionn{The Coupling Constant and the Bosonic Mass in Examples} \label{s:secexample}
The regularization parameters~$c_0, \ldots, c_3$ in the field equations~\eqref{s:field1} are given in terms
of the simple fractions in~\eqref{s:N3def} and~\eqref{s:N51}--\eqref{s:N64}.
For a given regularization method, we can evaluate these simple
fractions and compute the coupling constant and the bosonic rest mass.
\sindex{coupling constant!of axial field}%
We now exemplify the procedure by considering the two simplest methods of regularization: \\[-0.8em]

\noindent {\bf{(A)}}  The $i \varepsilon$-regularization:
\sindex{regularization!$i \varepsilon-$}%
Exactly as in~\S\ref{secieps}, we define
the distribution~$\hat{P}^\varepsilon$ in~\eqref{s:PFT} by
inserting an exponential convergence generating factor
into the integrand of~\eqref{s:A},
\[ \hat{P}^\varepsilon(k) = \sum_{\beta=1}^g (\slashed{k} + m_\beta) \,\delta(k^2-m^2)\, \Theta(-k^0)
\: \exp(\varepsilon k^0)\: \:. \]
To the considered leading degree on the light cone, this regularization corresponds to the
simple replacements (cf.~\eqref{rule0} and~\eqref{rulem1})
\beq \label{s:Tregsim}
T^{(0)}_{[p]} \rightarrow -\frac{1}{8 \pi^3}\: \frac{1}{2r\: (t-r-i \varepsilon)} \:,\qquad
T^{(-1)}_{[p]} \rightarrow -\frac{2}{r}\: \frac{\partial}{\partial t} T^{(0)}_{[p]}
= -\frac{1}{8 \pi^3\,r^2}\: \frac{1}{(t-r-i \varepsilon)^2} \:,
\eeq
and similarly for the complex conjugates. Using these formulas in~\eqref{s:field1},
the basic fractions all coincide up to constants, giving the equation
\[ \Big( -\frac{3}{2} - 96 \pi^3 \:(s_{[0]}-s_{[3]}) \Big) j_a
+ 2 m^2 \left(\hat{Y}^2 + 96 \pi^3 (s_{[2]}-s_{[3]}) \acute{Y} \grave{Y}
\right) A_a = 12 \pi^2\: J_a \:. \]
According to Lemma~\ref{s:lemmasmooth} and~\eqref{s:s3def}, the functions~$s_{[p]}$
involve the masses of the fermions and also the convolution terms~$f^\beta_{[p]}$.
For clarity, we here leave out the convolution terms (which are analyzed in detail
in~\S\ref{s:secnocausal} and Appendix~\ref{s:appresum}). Then we can write the field equation
in the usual form
\[ j_a - M^2 A_a = e^2 J_a \:, \]
\nindex{ck2@$M$ -- bosonic mass}%
\nindex{cb0@$e$ -- coupling constant}%
where the coupling constant~$e$ and the mass~$M$ are given functions
of the ratios~$m_2/m_1$ and~$m_3/m_1$ of the fermion masses.
\begin{figure}
\begin{center}
\includegraphics[width=7cm]{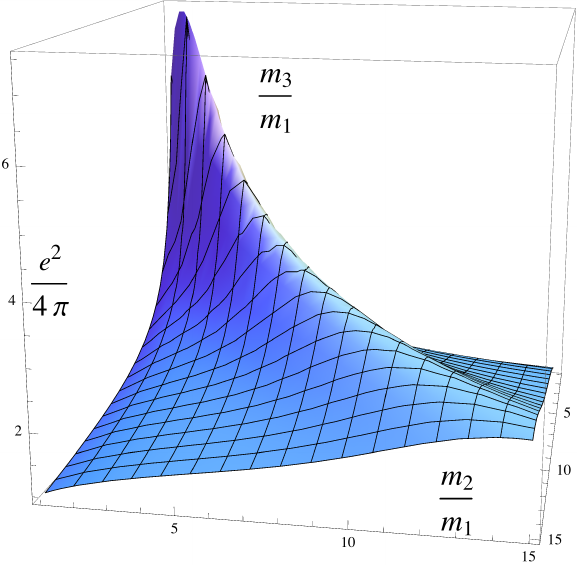} 
\includegraphics[width=7cm]{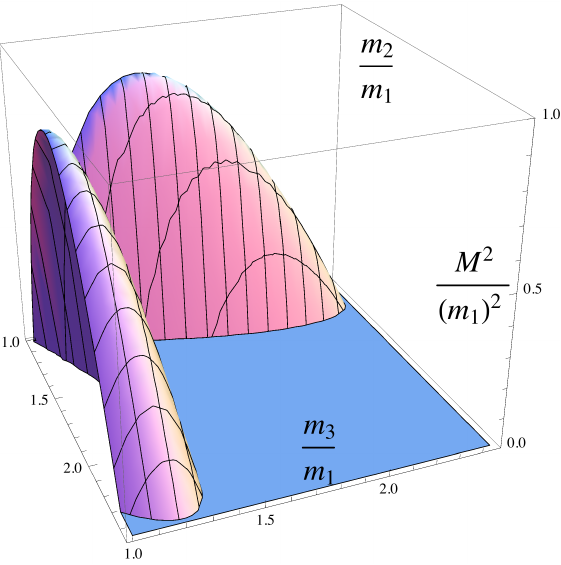}
\end{center} 
\caption{The coupling constant and the bosonic mass for the regularization~\eqref{s:Tregsim}}
\label{s:figreg1}
\end{figure}
In Figure~\ref{s:figreg1}, these constants are plotted as functions of the mass ratios.
The coupling constant is of the order one; it is largest if the fermion masses are close to each other.
The term~$M^2$ is always positive, so that the mass term has the correct sign.
The bosonic mass has the same order of magnitude as the fermion masses.

\noindent {\bf{(B)}} A {\em{cutoff}} in momentum space:
\sindex{regularization!by cutoff}%
In~{\bf{(A)}} we considered a regularization which was ``soft'' in the sense that it was smooth in momentum space. To give a complementary
example, we now consider the ``hard'' regularization obtained by
inserting a Heaviside function into the integrand of~\eqref{s:A}. Thus we
define the distribution~$\hat{P}^\varepsilon$ in~\eqref{s:PFT} by
\[ \hat{P}^\varepsilon(k) = \sum_{\beta=1}^g (\slashed{k} + m_\beta) \,\delta(k^2-m^2)\, \Theta(-k^0)
\; \Theta(1 + \epsilon k^0) \:. \]
This regularization is described in analogy to~\eqref{s:Tregsim} by
\[ T^{(0)}_{[p]} \rightarrow -\frac{1}{16 \pi^3}\: \frac{1-e^{-\frac{i (t-r)}{\varepsilon}}}{2r\: (t-r)} \:,\qquad T^{(-1)}_{[p]} \rightarrow -\frac{2}{r}\: \frac{\partial}{\partial t} T^{(0)}_{[p]} \:. \]
Thus as expected, the cutoff in momentum space gives rise to rapid oscillations in position space.
Using these formulas in~\eqref{s:field1}, the resulting basic fractions are no longer multiples of
each other. But the weak evaluation integrals~\eqref{s:asy} can still be computed in closed form.
We thus obtain the field equation
\[ \Big( -\frac{9}{4} - 96 \pi^3 \:(s_{[0]}-s_{[3]}) \Big) j_a
+ 3 m^2 \left(\hat{Y}^2 +64 \pi^3 (s_{[2]}-s_{[3]}) \acute{Y} \grave{Y}
\right) A_a = 12 \pi^2\: J_a \:. \]
This equation can be analyzed exactly as in example~{\bf{(A)}}, giving the same qualitative
results. It is remarkable that the constants in the field equations in example~{\bf{(A)}}
and~{\bf{(B)}} differ at most by a factor~$3/2$, indicating that our results do not depend
sensitively on the method of regularization.

\section{The Euler-Lagrange Equations to Degree Three and Lower} \label{s:secthree}
The wave functions in~\eqref{s:particles} do not only have a vector and axial component
as considered in~\S\ref{s:sec82}, but they also have scalar, pseudoscalar and bilinear
components. We will now analyze the effect of these contributions on the EL equations\
(\S\ref{s:sec87} and~\S\ref{s:sec88}). 
Moreover, we will insert further potentials into the Dirac equation and analyze the
consequences. More precisely, in~\S\ref{s:sec88} we consider bilinear potentials,
whereas in~\S\ref{s:sec89} we consider scalar and pseudoscalar potentials
and discuss the remaining possibilities in choosing other potentials and fields.
We conclude this section with a discussion of the structure of the EL equations to degree
three and lower (\S\ref{s:sec810}).

\subsectionn{Scalar and Pseudoscalar Currents} \label{s:sec87}
In analogy to the Dirac currents in~\S\ref{s:sec82}, we introduce the
{\em{scalar Dirac current}}~$J_\text{s}$ and the
{\em{pseudoscalar Dirac current}}~$J_\text{p}$ by
\sindex{current!scalar Dirac}%
\sindex{current!pseudoscalar Dirac}%
\nindex{ck8@$J_\text{s}$ -- scalar Dirac current}
\nindex{cl0@$J_\text{p}$ -- pseudoscalar Dirac current}
\beq \label{s:Jspdef}
J_\text{s} = \sum_{k=1}^{\np} \overline{\psi_k} \psi_k - \sum_{l=1}^{\na} \overline{\phi_l} \phi_l
\qquad \text{and} \qquad
J_\text{p} = \sum_{k=1}^{\np} \overline{\psi_k} \,i \pseudo \psi_k - \sum_{l=1}^{\na} \overline{\phi_l}
\,i \pseudo  \phi_l \:.
\eeq
According to~\eqref{s:particles}, these currents lead to a perturbation of the fermionic projector.
In view of the fact that the scalar and pseudoscalar currents involve no Dirac matrix which could be
contracted with a factor of~$\xi$, one expects that the resulting contribution to the
EL equations should be one degree lower than that of the axial current (see Lemma~\ref{s:lemmalc2}).
Thus to leading order at the origin, one might expect contributions of the form
\beq \label{s:Qnaiv}
Q(x,y) \asymp J_\text{p}\: \text{(monomial of degree three)} + (\deg < 3) \:.
\eeq
However, perturbing the fermionic projector of the
vacuum by~\eqref{s:Jspdef}, one sees that the corresponding contribution to~$Q(x,y)$ of
degree three on the light cone vanishes (see Lemma~\ref{s:lemmascal}, where it also explained
how the cancellations come about).
Taking into account the axial potentials, we do get contributions to~$Q(x,y)$ of degree three
on the light cone, which are of the form
\beq \label{s:QAaJs}
Q(x,y) \asymp A_a^j \,\xi_j \: J_{\text{s}}\: \text{(monomial of degree three)} + (\deg < 3) \:.
\eeq
However, these contribution have the same tensor structure as the contributions of the
axial potentials and currents of degree three, which will be discussed in~\S\ref{s:sec810} below.

\subsectionn{Bilinear Currents and Potentials} \label{s:sec88}
The particles and anti-particle currents in~\eqref{s:particles} also have a bilinear component, leading
us to introduce the {\em{bilinear Dirac current}}~$J_\text{b}$ by
\sindex{current!bilinear Dirac}%
\nindex{cl2@$J_\text{b}$ -- bilinear Dirac current}

\[ J_\text{b}^{ij} = \sum_{k=1}^{\np} \overline{\psi_k} \sigma^{ij} \psi_k
- \sum_{l=1}^{\na} \overline{\phi_l} \sigma^{ij} \phi_l \]
(here~$\sigma^{jk} = \frac{i}{2} [\gamma^j, \gamma^k]$ are again the bilinear covariants).
Likewise, one may want to insert a {\em{bilinear potential}}
\[ \B = H_{ij}(x) \,\sigma^{ij} \]
into the auxiliary Dirac equation~\eqref{s:diracPaux} (where~$H$ is an anti-symmetric tensor field).

Let us briefly discuss the effect of a bilinear current and a bilinear potential on the EL equations.
The bilinear current corresponds to a perturbation of the fermionic projector of the form
\[ \Delta P(x,y) = -\frac{1}{8 \pi}\, J_\text{b}^{ij} \sigma_{ij} + o \big( |\vec{\xi}|^0 \big)\:. \]
The bilinear potential, on the other hand, gives rise to different types of contributions
to the light-cone expansion which involve~$H$ and its partial derivatives
(for details see~\cite[Appendix~A.5]{firstorder}). 
When computing the perturbations of the eigenvalues~$\lambda^{L\!/\!R}_\pm$ of the closed chain
(cf.~\cite[Appendix~G]{PFP} or Appendix~\ref{s:appspec}), all Dirac matrices in~$\Delta P$
are contracted with outer factors~$\xi$. As a consequence, the contribution of the bilinear
current drops out. Moreover, due to the anti-symmetry of~$H$, all the bilinear contributions
of the bilinear potential to the EL equations vanish. What remains are terms involving the
divergence~$\partial_j H^{ij}$ of the bilinear
potential or derivatives of the divergence. All these terms can be interpreted as effective
vector or axial potentials or derivatives thereof. Furthermore, the resulting contributions
to the EL equations are of degree three on the light cone. Such contributions will
again be discussed in~\S\ref{s:sec810} below.

\subsectionn{Further Potentials and Fields} \label{s:sec89}
Having considered many different perturbations of the Dirac operator,
we are now in the position to draw a few general conclusions, and to discuss a
few potentials and fields which are not covered by the previous analysis.
First of all, we point out that in Lemmas~\ref{s:lemmalc1} and~\ref{s:lemmalc2} the vector
components dropped out, so that we got no contribution by an electromagnetic field.
\sindex{electromagnetic field}%
This cancellation can be understood from the general structure of our action principle, namely from the fact
that the Lagrangian~\eqref{s:Lcrit} involves
the differences of the absolute values of the left- and right-handed eigenvalues.
As a consequence, any perturbation which affects the eigenvalues~$\lambda^L_s$ and~$\lambda^R_s$
in the same way necessarily drops out of the EL equations.
In other words, the EL equations are only affected by perturbations of the fermionic projector
which have {\em{odd parity}}.

The fact that the electromagnetic field does not enter the EL equations does not necessarily imply
that the electromagnetic field must vanish. But it means that no electromagnetic fields are generated
in the system, so that the only possible electromagnetic field must be radiation coming from infinity.
Having isolated systems without incoming radiation in mind, we conclude that our system
involves {\em{no electromagnetic field}}.

The above consideration for the electromagnetic field also applies to the  {\em{gravitational field}},
\sindex{gravitational field}%
as we now explain.
Introducing gravitational fields (see for example~\cite[Section~1.5]{PFP}) gives rise to
contributions to the light-cone expansion which involve the metric, the curvature and
the derivatives of curvature (see~\cite[AppendixB]{firstorder}). The main effect of the
gravitational field can be understood as a ``deformation'' of the light cone corresponding to the
fact that the light cone is now generated by the null geodesics. The corresponding contributions to the
light-cone expansion drop out of the EL equations if we make the action principle diffeomorphism
invariant simply by replacing the measure~$d^4x$ in~\eqref{s:STdef} by~$\sqrt{|\det g_{ij}|}\, d^4x$.
In addition, there are terms involving the curvature of space-time,
whose singularities on the light cone are of so small degree that the corresponding closed chain
can again be treated perturbatively. Since these curvature terms are even under parity
transformations, they drop out of the EL equations. We conclude that our model involves
no gravitational field.

From the physical point of view, it might seem disappointing that our model involves no electromagnetic
and gravitational fields. However, the simple explanation is that our system of one sector
is too small to involve these fields. If one considers systems of several sectors, the equation
analogous to~\eqref{s:Lcrit} will involve the differences of the eigenvalues~$\lambda^c_s$
and~$\lambda^{c'}_{s'}$ in different sectors. Then potentials no longer drop out even if they
have even parity, provided that they are not the same in all sectors. Only a detailed analysis of
systems involving several sectors will show whether
the electromagnetic and gravitational fields will appear in the physically correct way
(see the Chapters~\ref{lepton} and~\ref{quark}).

Having understood why gravitational fields drop out of the EL equations, one might want to
consider instead an {\em{axial gravitational field}} as described by a perturbation of the form
\sindex{potential!axial gravitational}%
\beq \label{s:axialgrav}
\B = i \pseudo \gamma_i \,h^{ij}\, \partial_j + \text{(lower order terms)} \:.
\eeq
Such a field cannot occur for the following reason. As just mentioned, a gravitational field
describes a deformation of the light cone, which due to the diffeomorphism invariance of
our action principle does not enter the EL equations. Similarly, an axial gravitational
field~\eqref{s:axialgrav} describes a deformation of the light cone, but now differently
for the left- and right-handed components of the Dirac sea. Thus considering a
perturbation of the form~\eqref{s:axialgrav}, the light cone would ``split up'' into
two separate light cones being the singular sets of the left- and right-handed 
components of the fermionic projector, respectively.
As a consequence, the leading singularities of the closed chain could no longer
compensate each other in the EL equations, so that the EL equations would be violated
to degree five on the light cone.
Thus for the EL to be satisfied, the axial gravitational field must vanish.

In order to avoid the problem of the axial deformations of the light cone, one may want to consider
a so-called {\em{axial conformal field}}
\sindex{potential!axial conformal}%
\beq \label{s:axialconf}
\B = i  \Xi(x) \pseudo \gamma^j \partial_j + \text{(lower order terms)} \:.
\eeq
But this field is of no use, as the following consideration shows. Let us consider 
for a given real function~$\Lambda$ the so-called {\em{axial scaling transformation}}~$U = 
e^{\pseudo \Lambda(x)}$. This transformation is unitary, and thus it has no
effect on the EL equations. Transforming the Dirac operator according to
\begin{align*}
(i \Pdd - m) \rightarrow & U (i \Pdd - m) U^{-1} =  i U^2 \Pdd - m + i \pseudo (\Pdd \Lambda) \\
&=   \cosh(2 \Lambda(x))\, i \Pdd + \sinh(2 \Lambda(x))\, i \pseudo \Pdd - m
 + i \pseudo (\Pdd \Lambda) \:,
\end{align*}
the summand~$\sinh(2 \Lambda(x))\, i \pseudo \Pdd$ can be identified with the
first-order term in~\eqref{s:axialconf}. The summand~$\cosh(2 \Lambda(x))\, i \Pdd$,
on the other hand, is a conformal gravitational field, and we already saw that gravitational fields
do not enter the EL equations. Thus in total, we are left with a perturbation of the Dirac
operator by chiral potentials, as considered earlier in this chapter.

Next, we briefly consider a {\em{scalar}} or {\em{pseudoscalar potential}}~\eqref{s:pseudoscalar}.
\sindex{potential!scalar}%
\sindex{potential!pseudoscalar}%
The leading contributions to the fermionic projector involve the potentials~$\Xi$ and~$\Phi$,
whereas to lower degree on the light cone also derivatives of these potentials appear.
Since~$\Phi$ has even parity, its leading contribution to the EL equations vanishes.
For the potential~$\Xi$, the leading contribution cancels in analogy to~\eqref{s:Qnaiv}
(see also Lemma~\ref{s:lemmascal}). But to degree four on the light cone, one gets
cross terms similar to~\eqref{s:QAaJs} which also involve the axial potential.
In order for these additional terms to vanish, we are led to setting the scalar and pseudoscalar
potentials equal to zero. Thus there seems no point in considering scalar or pseudoscalar potentials.
Nevertheless, scalar or pseudoscalar perturbations might enter the EL equations to degree three
and lower, as will be discussed in~\S\ref{s:sec810} below.

In the analysis of the axial potential we made one assumption which requires a brief
explanation. Namely, when introducing the axial potential~$A_\text{\rm{a}}$ in~\S\ref{s:sec72}, we assumed
that it couples to all generations in the same way (see~\eqref{s:va}).
In view of the constructions in~\S\ref{s:sec84}--\S\ref{s:sec86}, where it was essential that
the potentials were different for each generation, the ansatz~\eqref{s:va} seems rather special,
and one might wonder what would happen if we replaced the potential~$A_\text{\rm{a}}$ in~\eqref{s:va} by
a matrix potential acting non-trivially on the generations. Indeed, this scenario was already discussed
in~\cite[Remark~6.2.3]{PFP}, and thus here we briefly repeat the main argument. Suppose that
the potential~$A_\text{\rm{a}}$ in~\eqref{s:va} were a matrix. Then the exponentials in~\eqref{s:Pchiral}
would have to be replaced by ordered exponentials of the form
\[ \text{Texp} \left( -i \int_x^y A^j_\text{\rm{a}} \xi_j \right) . \]
This is a unitary matrix whose eigenvalues can be regarded as different phase factors.
Thus when forming the sectorial projection, we do not get a single phase factor, but instead a linear combination
of different phases. As a consequence, the relations~\eqref{s:labs} will in general be violated,
so that we get a contribution to the EL equations to degree five on the light cone.
Reverting this argument, we can say that the EL equations to degree five imply
that the eigenvalues~\eqref{s:unperturb} should involve only one phase, meaning that the
axial potential must in fact be of the form~\eqref{s:va} with a vector field~$A_\text{\rm{a}}$.

\subsectionn{The Non-Dynamical Character of the EL Equations to Lower Degree} \label{s:sec810}
In our analysis of the EL equations we began with the leading degree five on the light cone.
The analysis to degree four revealed the field equations and thus described the dynamics.
Generally speaking, to degree three and lower on the light cone, we get many more conditions,
but on the other hand, we also get much more freedom to modify the fermionic projector.
Namely, to degree three the Dirac currents as well as vector and axial potentials give rise to
many terms, which in general do not cancel each other, even if the field equations
of Section~\ref{s:secfield} are satisfied. For example, the factors~$T^{(n)}_{ \{p\} }$ involving
curly brackets come into play (see~\eqref{s:eq54}), and we also get contributions involving
higher derivatives of the potentials. To degree two and one on the light cone, we
get even more terms, involving the cross terms of different potentials and the terms
generated by the mass expansion of the fermionic projector. In order to satisfy
the EL equations to lower degree, all these terms must cancel each other.
The good news is that we also get more and more free parameters. For example, we can consider
scalar and pseudoscalar potentials~\eqref{s:pseudoscalar}, or bilinear potentials, or the other
potentials discussed in~\S\ref{s:sec89}. All these potentials can be chosen independently for each
generation. Taking into account that to lower degree on the light cone,
more and more cross terms between different perturbations come into play, we obtain a
very complicated structure involving a large number of free parameters to modify the EL equations
to degree three and lower.

In view of this complexity, it is not clear whether the EL equations can
be satisfied to every degree on the light cone or not. The analysis becomes so complicated that it
seems impossible to answer this question even with more computational effort.
A possible philosophy to deal with this situation would be to take the pragmatic point of view
that one should simply satisfy the EL equations as far as possible, but stop once the equations
can no longer be handled. Since we do not find this point of view convincing, we now go one step
further and explain why the analysis of the EL to lower degree (no matter what the results of this
analysis would or will be) will have no influence on the dynamics of the system.

The field equations of Section~\ref{s:secfield} were {\em{dynamical}} in the sense that
they involved partial differential equations of the potentials, and by solving these equations
one finds that the potential is non-trivial even away from the sources.
This property of the field equations is a consequence of the fact that the leading
contributions to the fermionic projector (namely the phase factors in~\eqref{s:Pchiral})
dropped out of the EL equations. This made is possible that the derivative terms became
relevant (although they were of lower degree on the light cone), leading to dynamical field equations.
The set of perturbations which can lead to
dynamical field equations is very limited, because this requires that the potential itself
must drop out of the EL equations, meaning that the potential must correspond to a local symmetry
of the system. All in this sense dynamical perturbations have been considered in this chapter.
This implies that all further potentials and fields will be {\em{non-dynamical}}
in the sense that the potentials themselves (and not their derivatives) would enter the EL equations.
\sindex{potential!non-dynamical}%
This would give rise to algebraic relations between these potentials.
In particular, these potentials would vanish away from the sources. Thus they do not
describe a dynamical interaction, also making it difficult to observe them experimentally. 

A possible idea for avoiding non-dynamical potentials is to choose the potentials
differently for each generation, in such a way that the most singular contribution vanishes
when forming the sectorial projection.
This would open the possibility that the leading contribution to the EL equations might involve
derivatives, thereby giving rise to dynamical field equations.
We now give an argument which explains why this idea does not seem to work:
Suppose that the potential is chosen as a matrix on the generations. Then for the
potential to drop out of the EL equations, we must impose that a certain sectorial projection
involving the potential must vanish at every space-time point.
But this implies that this sectorial projection also vanishes if the potential is replaced by its
derivatives. In particular, these derivative terms again drop out of the EL equations, making it
impossible to get dynamical field equations.
We remark that the last argument fails if the derivative terms also involve factors of the mass matrix,
in which case the sectorial projection might be non-zero. But this situation seems rather
artificial, and we shall not enter its analysis here.

This concludes our analysis of {\em{local}} potentials. Our treatment was exhaustive in the sense
that we considered all multiplication operators and considered all relevant first order operators.
Our results gave a good qualitative picture of how the fermionic projector is affected
by different kinds of perturbations of the Dirac operator, and what the resulting contributions
to the EL equations are. But clearly, the present paper cannot cover all possible perturbations,
and some details still remain unsettled. Nevertheless, as explained above,
our analysis covers all perturbations which should be of relevance for the dynamics of
the fermions in our model.

\section{Nonlocal Potentials} \label{s:secnonlocal}
So far, we only analyzed the EL equations at the origin, i.e.\ for the leading contribution in
an expansion in the parameter~$\xi = y-x$. But clearly, the EL equations
should also be satisfied away from the origin. In this section, we will explore
whether this can be accomplished by introducing {\em{nonlocal}}
potentials into the Dirac equation.
\sindex{potential!nonlocal}%
We shall see that this method indeed makes it possible
to satisfy the EL equations to every order in an expansion around~$\xi=0$
(see Theorem~\ref{s:thmloc}). However, it will not become possible to satisfy the EL equations
globally for all~$x$ and~$y$. This will lead us to the conclusion that all nonlocal potentials
should vanish in the continuum limit.

In order to introduce the problem, we consider
the perturbation of the fermionic projector by a particle wave function~$\psi$, i.e.\ in
view of~\eqref{s:particles}
\beq \label{s:particle}
P(x,y) \asymp -\frac{1}{2 \pi}\: \psi(x) \overline{\psi(y)}\:.
\eeq
In the variable~$y-x$, this contribution oscillates on the scale of the Compton wave length,
whereas in the variable~$y+x$, it varies typically on the larger atomic or macroscopic scale.
For this reason, it is appropriate to begin by analyzing {\em{homogeneous}} perturbations
which depend only on the variable~$y-x$ (see~\S\ref{s:sec110}--\S\ref{s:sec112}).
In~\S\ref{s:sec113} we shall extend our constructions to build in an additional dependence on
the variable~$y+x$. The results and physical significance of our analysis will be discussed
in~\S\ref{s:secnonlocconclude}.

\subsectionn{Homogeneous Transformations in the Low-Frequency Regime} \label{s:sec110}
We now return to the homogeneous transformation of the fermionic projector~\eqref{s:Ukform}.
In \S\ref{s:secgennonloc} we analyzed this transformation, choosing~$Z(k)$ such that
the coefficients~$C_\beta$ in~\eqref{s:PasyO} were constants. This had the advantage that
the corresponding smooth contribution to the fermionic projector was easily computable
(see~\eqref{s:s3def}). However, $Z(k)$ can be chosen more generally as an arbitrary
function of~$k$. In order to complement the analysis of~\S\ref{s:secnonlocaxial}, where
the emphasis was on the singularities on the light cone, we now concentrate on smooth
contributions to the fermionic projector.
\sindex{homogeneous transformation!in the low-frequency region}%
We thus assume that~$Z(k)$ has rapid decay.
For technical simplicity, we assume that~$Z$ is a Schwartz function in momentum space,
$Z \in {\mathcal{S}}(\hscrM)$.
\nindex{cl4@${\mathcal{S}}(\hscrM)$ -- Schwartz functions in momentum space}%
As a consequence, the function~$\delta P(x,y)$ is smooth.
Clearly, more general functions could be realized by approximation.

In order to explain the basic idea in the homogeneous setting, we again consider
the system~\eqref{s:onesea} of one Dirac sea of mass~$m$. Transforming each state
according to~\eqref{s:Ukform} gives rise to a perturbation of the fermionic projector of the form
\beq \label{s:delPuni}
\begin{split}
\delta P(x,y) &= \int \frac{d^4k}{(2 \pi)^4} \:\pseudo \slashed{v}(k)\, \delta(k^2-m^2)\:
\Theta(-k^0) \: e^{-ik(x-y)} \\
&\qquad + \text{(smooth vectorial, pseudoscalar or bilinear contributions)}\:,
\end{split}
\eeq
where~$v$ is any vector field of rapid decay.
Exactly as explained in the derivation of~\eqref{s:s3def}, the vectorial contribution drops out of~\eqref{s:RRdef},
whereas the pseudoscalar and bilinear contributions drop out of the EL equations to degree four.
Hence it suffices to consider the axial contribution. Moreover, in the EL equations, $\delta P(x,y)$
is always contracted with a factor~$\xi$. Thus it suffices to consider the expression
\beq
\frac{1}{4} \,\Tr \Big( \pseudo \slashed{\xi}\: \delta P(x,y) \Big) 
= -\int \frac{d^4k}{(2 \pi)^4} \:
\big\la \xi, v(k) \big\ra \:T_{m^2}(k) \: e^{ik \xi}\:, \label{s:axFourier0}
\eeq
where the distribution~$T_{m^2}(k) := \delta(k^2-m^2)\, \Theta(-k^0)$ is supported on the
lower mass shell. For what follows, it will be useful to write the inner product~$\big\la \xi, v(k) \big\ra
= \xi^i v_i$ in components and to analyze the components separately. Dividing by~$\xi^i$,
we can reduce~\eqref{s:axFourier0} to four scalar Fourier integrals of the form
\beq \label{s:sFourier}
f(\xi) = \int \frac{d^4k}{(2 \pi)^4} \:\hat{f}(k)\:T_{m^2}(k) \: e^{ik \xi}\:.
\eeq

\subsectionn{Homogeneous Perturbations by Varying the Momenta} \label{s:sec111}
Another homogeneous perturbation which was not considered so far is to 
{\em{vary the momenta}} of the states, dropping the mass shell condition.
For simplicity, we begin with the system~\eqref{s:onesea} of one
Dirac sea for a given rest mass~$m > 0$.
This system is composed of states~$\psi_k(x) = \chi(k) \,e^{-i k x}$ which are plane waves
of momentum~$k$ on the lower mass shell.
Considering a variation~$\delta k$ of the momentum leads us to replace~$\psi_k(x)$
by the plane wave
\sindex{homogeneous perturbation!by varying the momenta}%
\beq \label{s:pwsol}
\chi(k)  \,e^{-i (k+\delta k) x} \:.
\eeq
Then to first order in~$\delta k$, the individual states are varied by
\[ \delta \psi_k(x) = -i \:\la x, \delta k \ra \: \psi_k(x) + \O((\delta k)^2)\:, \]
and this yields the following perturbation of the fermionic projector,
\beq \label{s:homperturb}
\delta P(x,y) = i \int \frac{d^4k}{(2 \pi)^4} \big\la \xi, \delta k \big\ra  (\slashed{k} +m)\, \delta(k^2-m^2)\:
\Theta(-k^0) \: e^{-ik(x-y)} \:.
\eeq
The vector field~$\delta k(k)$ can be chosen arbitrarily on the lower mass shell, giving us
a lot of freedom to vary~$P(x,y)$. However, due to the factor~$\xi$ inside the
inner product, the variation necessarily vanishes at the origin~$x=y$.
For technical simplicity we again assume that~$\delta k$ has rapid decay.

We want to explore whether such a variation of the momenta allows us to modify
the EL equations. First of all, we remind the reader that only the axial component of the currents
enters the EL equations to degree four (whereas the scalar, pseudoscalar and bilinear components
drop out; see~\S\ref{s:sec87} and~\S\ref{s:sec88}). But the perturbation~\eqref{s:homperturb}
does not have an axial component, and thus we must generalize~\eqref{s:homperturb}
such as to include a perturbation which is odd under parity transformations.
To this end, we choose a vector field~$q$ in momentum space with the properties
\beq \label{s:qprop}
\la k, q(k) \ra = 0 \qquad \text{and} \qquad q(k)^2 = -1 \:.
\eeq
Then the operators
\beq \label{s:Pipmdef}
\Pi_{\pm}(k) := \frac{1}{2} \left( 1 \mp \pseudo \slashed{q}(k) \right)
\eeq
are projectors which commute with the fermionic projector of the vacuum;
they project onto the two spin orientations in the direction of~$q$ and~$-q$, respectively.
Thus multiplying the fermionic projector by~$\Pi_\pm$,
\beq \label{s:Ppmdef}
P_\pm(k) = \Pi_{\pm}(k)\, (\slashed{k}+m) \:\delta(k^2-m^2)\, \Theta(-k^0) \:,
\eeq
we decompose the Dirac sea into two ``subseas'' $P_\pm$, which are still
composed of solutions of the free Dirac equation. We remark that this decomposition
was already used in~\cite[Appendix~C.1]{PFP}, where it was shown that
in a suitable limit~$m \searrow 0$, the projectors~$P_\pm(k)$ go over to chiral Dirac seas
composed of left- or right-handed states. Here the above decomposition gives us the freedom to vary
the momenta of each subsea independently. Thus we generalize~\eqref{s:homperturb} by
\beq \label{s:homperturb2}
\delta P(x,y) = i \sum_{s =\pm}
\int \frac{d^4k}{(2 \pi)^4} \big\la \xi, \delta k_s \big\ra P_s(k) \: e^{-ik(x-y)} \:,
\eeq
\nindex{cl6@$P_\pm$ -- half filled Dirac sea}%
where~$\delta k_+$ and~$\delta k_-$ are two vector fields on the lower mass shell.
For our purposes, it will be sufficient to always assume that these vector fields
are smooth and have rapid decay. Thus we can extend them to 
Schwartz functions in momentum space.
As a consequence, the function~$\delta P(x,y)$ is smooth (but due to the restriction
to the mass shell in~\eqref{s:homperturb2}, it will in general not have rapid decay).

In the EL equations, $\delta P(x,y)$ is contracted with a factor~$\slashed{\xi}$. The corresponding
vector and axial components are computed by
\begin{align}
\frac{1}{4} \, \Tr \Big( \slashed{\xi}\: \delta P(x,y) \Big) 
&= \frac{i}{2} \int \frac{d^4k}{(2 \pi)^4} \:\big\la \xi, \delta k_+ + \delta k_- \big\ra\,
\big\la \xi, k \big\ra \:T_{m^2}(k) \: e^{ik \xi} \label{s:vecFourier} \\
\frac{1}{4} \, \Tr \Big( \pseudo \slashed{\xi}\: \delta P(x,y) \Big) 
&= -\frac{i}{2} \int \frac{d^4k}{(2 \pi)^4} \:\big\la \xi, \delta k_+ - \delta k_- \big\ra\,
\big\la \xi, mq \big\ra \:T_{m^2}(k) \: e^{ik \xi} \label{s:axFourier}
\end{align}
(where again~$T_{m^2}(k) = \delta(k^2-m^2)\, \Theta(-k^0)$).
Collecting the factors of~$\xi$, these expressions can be written in the
form~$\xi_j \xi_l A^{jl}$ with a symmetric tensor
field~$A^{jl}(\xi)$. We first want to eliminate the tensor indices,
leaving us with scalar Fourier transforms. Thus suppose that for a given smooth tensor field~$A^{jl}(\xi)$
we want to find the corresponding vector fields~$\delta k_\pm$ and~$q$ in~\eqref{s:vecFourier}
or~\eqref{s:axFourier}. In the vector component~\eqref{s:vecFourier}, we can rewrite
the factor~$\big\la \xi, k \big\ra$ as a $\xi$-derivative, leading to the equation
\beq \label{s:xiODE}
\xi^j \frac{\partial}{\partial \xi^j} f_l(\xi) = \xi^j A_{jl}(\xi)\:,
\eeq
where~$f_l$ is the Fourier integral
\[ f_l = \frac{1}{2} \int \frac{d^4k}{(2 \pi)^4} \:(\delta k_+ + \delta k_-)_l \:T_{m^2}(k) \: e^{ik \xi} \:. \]
Integrating the ordinary differential equation~\eqref{s:xiODE} gives the solution
\[ f_l(\xi) = \xi^j \int_0^1 A_{jl}(\tau \xi)\: d\tau \:, \]
being a smooth vector field. Thus for every choice of the vector index~$l$, we must
again solve the equation~\eqref{s:sFourier}, if 
we set~$f=f_l$ and~$\hat{f} = (\delta k_+ + \delta k_-)_l/2$.
In this way, we have reduced~\eqref{s:vecFourier} to scalar Fourier integrals of the form~\eqref{s:sFourier}.
The same can be accomplished for the axial component~\eqref{s:axFourier} with the
following construction. We write the factor~$\xi$ in~\eqref{s:axFourier}, which is contracted with~$q$,
as a $k$-derivative of the factor~$e^{i k \xi}$ and integrate by parts. Using the relation
\[ q^j \frac{\partial}{\partial k^j} T_{m^2}(k) 
= q^j \frac{\partial}{\partial k^j} \left( \delta(k^2-m^2)\, \Theta(-k^0) \right)
= 2 q^j k_j \: \delta'(k^2-m^2)\, \Theta(-k^0) = 0\:, \]
where in the last step we applied the orthogonality relation in~\eqref{s:qprop}, we obtain
the equation
\[ \xi^j A_{jl}(\xi) = \frac{1}{2} \int \frac{d^4k}{(2 \pi)^4} \left[
\frac{\partial}{\partial k^j} \Big( (\delta k_+ - \delta k_- )_l \:mq^j \Big)  \right]
T_{m^2}(k) \: e^{ik \xi} \:. \]
We again fix the index~$l$, but now set~$f(\xi) = \xi^j A_{jl}(\xi)$. Furthermore,
we introduce the vector field~$v^j = (\delta k_+ - \delta k_- )_l \:mq^j$. Suppose that
the smooth scalar function~$f$ can be represented in the form~\eqref{s:sFourier} with a
suitable Schwartz function~$\hat{f}$. Then the remaining task is to satisfy on the lower
mass shell the equation
\beq \label{s:divM}
\frac{\partial}{\partial k^j} v^j(k) = \hat{f} \:.
\eeq
In order to verify that this equation always has a solution, it is useful to rewrite
it as a geometric PDE defined intrinsically on the hyperbola
\[ \mathcal{H} := \{k \in \hscrM \,|\, k^2=m^2, \:k^0<0\} \:. \]
Namely, the orthogonality condition
in~\eqref{s:qprop} implies that the vector~$v$ is tangential to~$\mathcal{H}$,
and the derivatives in~\eqref{s:divM} can be rewritten as the covariant divergence on~$\mathcal{H}$,
\beq \label{s:covv}
\nabla_j v^j = \hat{f} \in {\mathcal{S}}(\mathcal{H})
\eeq
(where~$\nabla$ is the Levi-Civita connection on~${\mathcal{H}}$).
Conversely, for a given vector field~$v$ on~${\mathcal{H}}$ which satisfies~\eqref{s:covv}, 
extending~$v$ to a vector field on~$\hscrM$ which is everywhere orthogonal to~$k$
gives the desired solution of~\eqref{s:divM}.
A simple solution of~\eqref{s:covv} is obtained by first solving the Poisson equation
$\Delta_{\mathcal{H}} \phi = \hat{f}$ (for example using the explicit form of the Green's function
on the hyperbola) and setting~$v=\nabla \phi$.
However, this solution has the disadvantage that~$v$ has no rapid decay at infinity.
In order to do better, we must exploit that our function~$f(\xi)= \xi^j A_{jl}(\xi)$ vanishes at the origin,
and thus the integral of~$\hat{f}$ vanishes,
\beq \label{s:zmean}
\int_{\mathcal{H}} \hat{f} \,d\mu_{\mathcal{H}} = 0 \:.
\eeq
Combining this fact with the freedom to add to~$v$ an arbitrary divergence-free vector field,
one can indeed construct a solution~$v$ of~\eqref{s:covv} within the Schwartz class,
as is shown in the following lemma\footnote{I thank Bernd Ammann for 
the idea of solving the equation on the leaves of a foliation, after subtracting the mean
value of~$f$.}.

\begin{Lemma} \label{s:lemmaintegrate} For every function~$\hat{f} \in {\mathcal{S}}({\mathcal{H}})$
whose integral vanishes~\eqref{s:zmean}, there is a vector field~$v \in {\mathcal{S}}({\mathcal{H}})$
which satisfies~\eqref{s:covv}.
\end{Lemma}
\Proof We parametrize the hyperbola~${\mathcal{H}}$ by
\[ \left(-\sqrt{m^2+\rho^2}, \rho \cos \vartheta, \rho \sin \vartheta \cos \varphi,
\rho \sin \vartheta \sin \varphi \right) \in \hscrM , \]
where~$\rho:=|\vec{k}|$, and~$(\vartheta, \varphi)$ are the standard polar coordinates on the $2$-sphere.
Then the metric on~${\mathcal{H}}$ is diagonal,
\[ g_{ij} = \text{diag} \left( \frac{m^2}{\rho^2+m^2}, \rho^2, \rho^2 \sin^2 \vartheta \right) . \]
For the vector field~$v$ we take the ansatz as the sum of a radial part~$u^j=(u^\rho,0,0)$ and
an angular part~$w^j=(0,w^\vartheta, w^\varphi)$. We also regard~$w$ as a vector field on the sphere.
Then the equation~\eqref{s:covv} can be written as
\beq \label{s:combeq}
\text{div}_{\mathcal{H}}(u) + \text{div}_{S^2}(w) = \hat{f}\:.
\eeq

Taking the average of~$\hat{f}$ over spheres defines a spherically
symmetric Schwartz function~$\overline{f}$,
\[ \overline{f}(\rho) := \frac{1}{4 \pi} \int_{S^2} \hat{f}(\rho, \vartheta, \varphi) \:d\varphi\: d\cos \vartheta
\; \in\; {\mathcal{S}}({\mathcal{H}})\:. \]
Writing the integration measure as~$d\mu_{\mathcal{H}} = \sqrt{\det g} \,d\rho \,d\vartheta \,d\varphi$
and using Fubini, the condition~\eqref{s:zmean} implies that
\beq \label{s:wholeint}
\int_0^\infty \overline{f}(\rho)\: \frac{\rho^2\, d\rho}{\sqrt{\rho^2+m^2}} = 0\:.
\eeq
Solving the Poisson equation~$\Delta_{\mathcal{H}} \phi = \overline{f}$, the resulting
function~$\phi$ is smooth and again spherically symmetric. Setting~$u = \nabla \phi$,
we obtain a smooth radial vector field, being a solution of the equation
\beq \label{s:ueq}
\text{div}_{\mathcal{H}}(u) = \overline{f} \:.
\eeq
Writing the covariant divergence as
\[ \text{div}_{\mathcal{H}}(u) = \frac{1}{\sqrt{\det g}} \frac{\partial}{\partial x^j} \left( \sqrt{\det g}
\: u^j \right) = \frac{\sqrt{\rho^2+m^2}}{\rho^2} \frac{\partial}{\partial \rho} \left(
\frac{\rho^2\, u^\rho}{\sqrt{\rho^2+m^2}} \right) , \]
the divergence condition~\eqref{s:ueq} becomes an ordinary differential equation, having
the explicit solution
\[ u^\rho(\rho) = \frac{\sqrt{\rho^2+m^2}}{\rho^2}
\int_0^\rho \overline{f}(\tau)\: \frac{\tau^2\, d\tau}{\sqrt{\tau^2+m^2}} \:. \]
Using~\eqref{s:wholeint}, one immediately verifies that the vector field~$u$
and all its derivatives have rapid decay at infinity. Hence~$u$ is in the
desired Schwartz class.

Using~\eqref{s:ueq} in~\eqref{s:combeq}, it remains to consider the differential equation
\beq \label{s:angular}
\text{div}_{S^2} (w) = \hat{f} - \overline{f}\:.
\eeq
Since the function~$\hat{f} - \overline{f}$ has mean zero on every sphere, it
can be expanded in terms of spherical harmonics starting at~$l=1$,
\[ (\hat{f} - \overline{f})(\rho, \vartheta, \varphi) = 
\sum_{l=1}^\infty \sum_{k=-l}^l c_{lk}(\rho) \:Y_{lk}(\vartheta, \varphi)\:. \]
Since~$\hat{f} - \overline{f}$ is a Schwartz function, 
the coefficients~$c_{lk}$ are all smooth in~$\rho$,
and these coefficients together with all their $\rho$-derivatives
have rapid decay in both~$\rho$ and~$l$, uniformly in~$k$.
Hence the Poisson equation~$\Delta_{S^2} \phi = \hat{f} - \overline{f}$ can
be solved explicitly by
\[ \phi(\rho, \vartheta, \varphi) = - \sum_{l=1}^\infty \sum_{k=-l}^l \;\frac{c_{lk}(\rho)}{l(l+1)}
\:Y_{lk}(\vartheta, \varphi) \:, \]
defining again a Schwartz function on~${\mathcal{H}}$. Introducing the vector field~$w$
by~$w^j = \nabla_{S^2}^j \phi = (0, \partial_\vartheta \phi, \cos^{-2} \vartheta\, \partial_\varphi \phi)$
gives the desired solution of~\eqref{s:angular} in the Schwartz class.
\QED

\subsectionn{The Analysis of Homogeneous Perturbations on the Light Cone} \label{s:sec112}
\sindex{homogeneous perturbation!analysis on the light cone}%
With the above constructions, we have reduced the analysis of the homogeneous
transformation~\eqref{s:axFourier} as well as the perturbation by varying of the
momenta~\eqref{s:homperturb2} to the scalar Fourier transform~\eqref{s:sFourier}.
Having a weak evaluation on the light cone~\eqref{s:asy} in mind, we may restrict attention
to the light cone~$L = \{ \xi \:|\: \xi^2=0 \}$. Thus our task is to analyze the Fourier integral
\beq \label{s:sF2}
f(\xi) = \int \frac{d^4k}{(2 \pi)^4} \:\hat{f}(k)\: \delta(k^2-m^2)\, \Theta(-k^0)\: e^{ik \xi}
\qquad \text{for~$\xi \in L$}\:.
\eeq
More precisely, for a given smooth function~$f \in C^\infty(\scrM)$ we want to find a Schwartz
function~$\hat{f} \in {\mathcal{S}}(\hscrM)$ such that the Fourier integral~\eqref{s:sF2} coincides
on the light cone with~$f$. The question is for which~$f$ such a function~$\hat{f}$ exists.
We begin the analysis in the simple case that the function~$\hat{f}$ when
restricted to the mass shell depends only on the variable~$\omega = k^0$
(by linearity, we can later realize more general functions~$\hat{f}$ by superposition).
Then the resulting Fourier integral is spherically symmetric, so that the
function~$f(\xi)$ will only depend on the time and radial variables~$t=\xi^0$
and~$r=|\vec{\xi}|$. Restricting attention to the light cone~$t=\pm r$, we end up with
a one-dimensional problem. More precisely, setting~$p=|\vec{k}|$ and denoting the
angle between~$\vec{\xi}$ and~$\vec{k}$ by~$\vartheta$, the Fourier integral~\eqref{s:sF2} becomes
\begin{align*}
f(t,r) &= \frac{1}{8 \pi^3} \int_{-\infty}^0 d\omega\: \hat{f}(\omega)\,e^{i \omega t}
\int_0^\infty p^2\, dp \;\delta(\omega^2-p^2-m^2) \int_{-1}^1 d\cos \vartheta\:  e^{-i p r \cos \vartheta} \\
&= \frac{i}{8 \pi^3\, r} \int_{-\infty}^0 d\omega\: \hat{f}(\omega)\,e^{i \omega t}
\int_0^\infty p\, dp \: \delta(\omega^2-p^2-m^2) \left( e^{-i p r} - e^{i p r} \right) \\
&= \frac{i}{16 \pi^3\, r} \int_{-\infty}^{-m} d\omega\: \hat{f}(\omega)\,e^{i \omega t}\:
\Big( e^{-i \sqrt{\omega^2-m^2} r} - e^{i \sqrt{\omega^2-m^2} r} \Big) \:.
\end{align*}
Hence on the light cone~$t=\pm r$ we obtain the representation
\beq \label{s:ft}
i t \,f(t) = \frac{1}{16 \pi^3} \int_{-\infty}^{-m} d\omega\: \hat{f}(\omega)
\left( e^{i \omega_+ t} - e^{i \omega_- t} \right) ,
\eeq
where we set
\[ \omega_\pm := \omega \pm \sqrt{\omega^2-m^2} \:. \]
The right side of~\eqref{s:ft} differs from an ordinary Fourier integral in two ways:
First, the integrand does not involve one plane wave, but the difference of the
two plane waves~$e^{i \omega_\pm t}$, whose frequencies are related to each other
by~$\omega_+ \omega_- = m^2$. Second, in~\eqref{s:ft} only negative frequencies
appear. Let us discuss these two differences after each other.
The appearance of the combination $( e^{i \omega_+ t} - e^{i \omega_- t})$
in~\eqref{s:ft} means that the coefficients of the plane waves cannot be chosen arbitrarily, but a
contribution for a frequency~$\omega_-<-m$ always comes with a corresponding contribution
of frequency~$\omega_+>-m$. This frequency constraint makes it impossible to
represent a general negative-frequency function~$f$; for example, 
it is impossible to represent a function~$t f(t)$ whose frequencies are supported in the
interval~$[-\infty, -m)$. However, this {\em{frequency constraint}}
\sindex{frequency!constraint of negative}%
can be regarded as a shortcoming
of working with a single Dirac sea. If we considered instead a realistic system of
several Dirac seas~\eqref{s:A}, the Fourier integral~\eqref{s:ft} would involve a sum over the
generations,
\beq \label{s:ft2}
i t \,f(t) = \frac{1}{16 \pi^3} \sum_{\beta=1}^g \int_{-\infty}^{-m_\beta} d\omega\: \hat{f}^\beta(\omega)
\left( e^{i \omega^\beta_+ t} - e^{i \omega^\beta_- t} \right)
\eeq
with~$\omega_\pm^\beta(\omega) := \omega \pm (\omega^2-m_\beta^2)^{-\frac{1}{2}}$ and~$g>1$.
Then the freedom in choosing~$g$ independent functions~$\hat{f}^\beta$ would indeed make it
possible to approximate any negative-frequency function, as the following lemma shows.

\begin{Lemma} \label{s:lemmaapprox}
Assume that the number of generations~$g \geq 2$. Assume furthermore that
the one-dimensional Fourier transform~$\hat{f}$ of a given Schwartz
function~$f \in {\mathcal{S}}(\R)$ is supported in the interval~$(-\infty, 0)$.
Then there is a sequence of Schwartz functions~$\hat{f}_n^\beta \in {\mathcal{S}}(\R)$ such that
the corresponding functions~$f_n(t)$ defined by the Fourier integrals~\eqref{s:ft2} 
as well as all their derivatives converge uniformly to~$f(t)$,
\beq \label{s:converge}
\sup_{t \in \R} \left| \partial_t^K (f_n(t) - f(t)) \right| \xrightarrow{n \rightarrow \infty} 0
\qquad \text{for all~$K \geq 0$}\:.
\eeq
\end{Lemma}
\Proof
We want to find functions~$\hat{f}^\beta$ in~\eqref{s:ft2} such that
the right side of~\eqref{s:ft2} gives the plane wave~$e^{i \Omega t}$ with~$\Omega<0$.
Again ordering the masses according to~\eqref{s:morder},
we choose~$\omega_1$ such that~$\omega_+^g(\omega_1)=\Omega$ or~$\omega_-^g(\omega_1)
=\Omega$, i.e.
\[ \omega_1 = \frac{\Omega^2+m_g^2}{2 \Omega}\:. \]
Then choosing~$\hat{f}^g(\omega)=\pm \delta(\omega-\omega_1)$, we obtain the desired
plane wave~$e^{i \Omega t}$, but as an error term we get the plane wave~$-e^{i \Omega_1 t}$
with~$\Omega_1 = m_g^2/\Omega$. In order to compensate the error, we next
choose~$\omega_2$ such that~$\omega_+^1(\omega_2)=\Omega_1$ or~$\omega_-^1(\omega_2)
=\Omega_1$.
Choosing~$\hat{f}^1(\omega) = \delta(\omega-\omega_2)$, the plane wave~$-e^{i \Omega_1 t}$
drops out, but we obtain instead the plane wave~$e^{i \Omega_2 t}$
with~$\Omega_2=m_1^2/\Omega_1 = m_1^2 \Omega/m_g^2$.
We proceed by compensating the plane waves in turns by the last Dirac sea and the
first Dirac sea. After $n$ iteration steps, the functions~$\hat{f}^1$ and~$\hat{f}^g$
take the form
\beq \label{s:hfex}
\hat{f}^1_n(\omega) = -\sum_{l=1}^{n} \delta \bigg(
\omega - \frac{\Omega_{2n+1}^2 +m_1^2}{2 \Omega_{2n+1}} \bigg) \:,\qquad
\hat{f}^g_n(\omega) = \sum_{l=0}^{n-1} \delta \bigg(
\omega - \frac{\Omega_{2n}^2 +m_g^2}{2 \Omega_{2n}} \bigg) \:,
\eeq
where
\[ \Omega_{2n} = \frac{m_1^{2n}}{m_g^{2n}}\: \Omega \qquad \text{and} \qquad
 \Omega_{2n+1} = \frac{m_g^{2n+2}}{m_1^{2n}\, \Omega}\:. \]
The Fourier integral~\eqref{s:ft2} gives rise to the plane waves
\[ e^{i \Omega t} - e^{i \lambda^n \Omega t} \qquad \text{where} \qquad
\lambda := \frac{m_1^2}{m_g^2} < 1\:. \]

In order to form superpositions of these plane waves, we next multiply by a Schwartz
function~$\hat{h}(\Omega)$ and integrate over~$\Omega$. Then the Fourier integral~\eqref{s:ft2}
becomes
\[ i t f_n(t) = \int_{-\infty}^0 \hat{h}(\Omega) \left( e^{i \Omega t} -  e^{i \lambda^n \Omega t}
\right) d\Omega \:. \]
Choosing~$\hat{h}(\omega) = -\partial_\omega \hat{f}(\omega)/(2 \pi)$, we can extend the
integration to the whole real line. After integrating by parts, we can carry out the Fourier integral to obtain
\[ f_n(t) = \frac{1}{2 \pi} \int_{-\infty}^\infty \hat{f}(\Omega) \left( e^{i \Omega t} -  \lambda^n
e^{i \Omega \,(\lambda^n  t)} \right) d\Omega
= f(t) - \lambda^n f(\lambda^n t)\:. \]
From this explicit formula it is obvious that the functions~$f_n$ converge in the limit~$n \rightarrow \infty$
in the sense~\eqref{s:converge}.
\QED
As is immediately verified, the functions~$f_n$ as well as all their derivatives also converge in~$L^2(\R)$.
However, we point out that for the functions~$\hat{f}^1_n$ and~$\hat{f}^g_n$,
the convergence is a bit more subtle. Namely, from~\eqref{s:hfex} one sees that
for large~$n$, these functions involve more and more contributions for large~$\omega$. A
direct calculation shows that in the limit~$n \rightarrow \infty$, these functions converge to
smooth functions which decay at infinity only~$\sim 1/\omega$.

Let us now discuss the consequences of the fact that~\eqref{s:ft} 
only involves negative frequencies.
\sindex{frequency!constraint of negative}%
This restriction is already obvious in the Fourier
integral~\eqref{s:homperturb2}, before the reduction to scalar Fourier integrals~\eqref{s:sF2} or~\eqref{s:ft}.
Since contractions with factors~$\xi$ merely correspond to differentiations
in momentum space which preserve the sign of the frequencies,
the following considerations apply in the same way before or after the
contractions with~$\xi$ have been performed.
We point out that the contribution by a Dirac wave function~\eqref{s:particle} can be composed of positive
frequencies (=particles) or negative frequencies (=anti-particles), and thus in~\eqref{s:particle}
we cannot restrict attention to negative frequencies.
This raises the question whether a contribution to~\eqref{s:particle} of positive frequency
can be compensated by a contribution to~\eqref{s:sF2} of negative frequency.
The answer to this question is not quite obvious, because the EL equations involve
both~$P(x,y)$ and its adjoint~$P(y,x)=P(x,y)^*$. Since taking the adjoint reverses the sign of the
frequencies, a negative-frequency contribution to~$P(x,y)$ affects the EL equations
by contributions of both positive and negative frequency. Thus one might hope
that perturbations of~$P(x,y)$ of positive and negative frequency could compensate each other
in the EL equations. However, such a compensation is impossible, as the following lemma shows.

\begin{Lemma} \label{s:lemmanocomp} Assume that~$\hat{f}$ and~$\hat{g}$ are the Fourier
transforms of chiral perturbations of the fermionic projector, such that~$\hat{f}$ has a
non-vanishing contribution inside the upper mass cone, whereas~$\hat{g}$ is supported
inside the lower mass cone. Then the linear contributions of~$f$ and~$g$ to the EL equations
to degree four cannot compensate each other for all~$\xi$.
\end{Lemma}
\Proof By linearity, we may restrict attention to the spherically symmetric situation,
so that~$f$ and~$g$ restricted to the light cone are functions of one variable~$t$.
Since the negative-frequency component of~$f$ can clearly be compensated by~$g$,
we can assume that~$f$ and~$g$ are composed purely of positive and negative frequencies, 
respectively. The perturbation~$g$ affects the EL equations to degree four
by (see Lemma~\eqref{s:lemmalc2} and its proof in Appendix~\ref{s:appspec})
\[ {\mathcal{R}} \asymp c \left( M_1 \,g(\xi) + M_2 \, \overline{g(\xi)} \right)
+(\deg < 4) , \]
where
\beq \label{s:M12}
M_1 = T^{(-1)}_{[0]} \overline{T^{(-1)}_{[0]}} \qquad \text{and} \qquad
M_2 = -T^{(-1)}_{[0]} \overline{T^{(-1)}_{[0]}}\: \frac{T^{(0)}_{[0]}}{\overline{T^{(0)}_{[0]}}}\:,
\eeq
and~$c>0$ is an irrelevant constant.
Evaluating weakly on the light cone~\eqref{s:asy}, we obtain the contribution
\beq \label{s:gasy}
c_1 \, g(t) + c_2\, \overline{g(t)}\:,
\eeq
where~$c_1$ and~$c_2$ are real regularization parameters (real because the degree is even).
Similarly, the perturbation~$f$ yields the contribution
\beq \label{s:fasy}
c_1 \, f(t) + c_2\, \overline{f(t)}\:.
\eeq
In order for~\eqref{s:fasy} to compensate~\eqref{s:gasy}, both the negative and positive
frequencies must cancel each other, leading to the conditions
\beq \label{s:comprel}
c_1 \, g(t) = c_2\, \overline{f(t)}  \qquad \text{and} \qquad
c_2\, \overline{g(t)} = c_1 \, f(t) \:.
\eeq
Taking the complex conjugate of the first equation, multiplying it by~$c_2$
and subtracting $c_1$ times the second equation, we get
\[ \left( c_1^2 - c_2^2 \right)  f(t) = 0\:. \]
We thus obtain the condition
\beq \label{s:c12}
c_1 = \pm c_2 \:.
\eeq
If this condition holds, we can indeed satisfy~\eqref{s:comprel} by
setting~$g(t) = \pm \overline{f(t)}$.

We conclude that~$f$ and~$g$ can compensate each other if and only
if we impose the relation~\eqref{s:c12} between the regularization parameters corresponding
to the basic fractions~$M_1$ and~$M_2$. 
Imposing relations between the regularization parameters was not used previously in
this paper, and one could simply reject~\eqref{s:c12} by saying that we do not want to
restrict the class of admissible regularization by introducing such relations.
\sindex{regularization!non-generic}%
However, this argumentation would not be fully convincing, as it would not allow for
the possibility that the microscopic structure of space-time on the regularization scale~$\varepsilon$ corresponds
to a regularization which does have the special property~\eqref{s:c12}. This possibility is ruled out by the
following argument which shows that there are in fact no regularizations which satisfy~\eqref{s:c12}:
As only the real parts of basic fractions enter~\eqref{s:asy},
it suffices to consider the real parts of~$M_1$ and~$M_2$. Using the specific form of these monomials
in~\eqref{s:M12}, we obtain
\[ \frac{\re(M_1 + M_2)}{2} =
\left| \frac{T^{(-1)}_{[0]}}{T^{(0)}_{[0]}} \right|^2 \left( \im T^{(0)}_{[0]} \right)^2
\:,\quad
\frac{\re(M_1 - M_2)}{2} = 
\left| \frac{T^{(-1)}_{[0]}}{T^{(0)}_{[0]}} \right|^2 \left( \re T^{(0)}_{[0]} \right)^2 . \]
Both these expressions are non-negative, and thus there cannot be cancellations between positive
and negative contributions, no matter how we regularize.
Without a regularization, we know from~\eqref{s:Taser}--\eqref{s:Tndef} that
\[ \re T^{(0)} = -\frac{1}{8 \pi^3}\: \frac{\text{PP}}{\xi^2} \qquad \text{and} \qquad
\im T^{(0)}_{[0]}=-\frac{i}{8 \pi^2}\: \delta(\xi^2)\, \epsilon(\xi^0)\:. \]
Regularizing these terms, we find that for any regularization, both~$M_1+M_2$
and~$M_1-M_2$ are non-zero to degree four on the light cone. Hence~\eqref{s:c12} is violated.
\QED
We come to the definitive conclusion that using perturbations of the form~\eqref{s:homperturb2},
it is in general impossible to satisfy the EL equations to degree four globally for all~$\xi$.
But, as we will now show, it is possible to satisfy the EL equations {\em{locally}} near~$\xi=0$,
in the sense that we can compensate all contributions in a Taylor expansion in~$\xi$ to
an arbitrarily high order. We consider the obvious generalization of~\eqref{s:homperturb2} to several
generations
\[ \delta P(x,y) = i \sum_{\beta=1}^g \sum_{s =\pm}
\int \frac{d^4k}{(2 \pi)^4} \big\la \xi, \delta k^\beta_s \big\ra \:P^\beta_\pm(k) \: e^{-ik(x-y)} \:, \]
where~$P^\pm_\beta$ is obtained from~\eqref{s:Ppmdef} if one replaces the
vector field~$q$ in~\eqref{s:Pipmdef} by a vector field~$q^\beta$
for the corresponding generation.

\begin{Prp} \label{s:prphom} Suppose that the number of generations~$g \geq 2$.
Then for any given smooth functions~$h_\text{\rm{v}}, h_\text{\rm{a}} \in C^\infty(\scrM)$ and
every parameter~$L>2$, there are vector fields~$\delta k_{\pm}^\beta$ and~$q^\beta$ in the
Schwartz class such that for all multi-indices~$\kappa$ with~$1 \leq |\kappa| \leq L$,
\[ \partial_\xi^\kappa \left[
\Tr \Big( \slashed{\xi}\: \delta P(x,y) \Big) - h_\text{\rm{v}}(\xi) \right]\! \Big|_{\xi=0} = 0 =
\partial_\xi^\kappa \left[
\Tr \Big( \pseudo \slashed{\xi}\: \delta P(x,y) \Big) - h_\text{\rm{a}}(\xi) \right] \! \Big|_{\xi=0}\: . \]
\end{Prp} \noindent
\Proof
Following the arguments after~\eqref{s:axFourier0} as well as after~\eqref{s:homperturb2} and
applying Lem\-ma~\ref{s:lemmaintegrate}, it again suffices to analyze scalar Fourier integrals.
Furthermore, using the polarization formula for the multi-index~$\kappa$, 
we may restrict attention to the spherically symmetric situation~\eqref{s:ft2}
with Schwartz functions~$\hat{f}^\beta \in {\mathcal{S}}(\R)$. Keeping track of the factors~$\xi$
in the arguments after~\eqref{s:axFourier0} and~\eqref{s:homperturb2}, it remains to show that for every smooth
function~$h(t)$ there are functions~$\hat{f}_\beta \in {\mathcal{S}}(\hscrM)$ such that
the corresponding function~$f(t)$ defined by~\eqref{s:ft2} satisfies the conditions
\beq \label{s:scalcond}
\frac{d^l}{dt^l} \big( f(t)-h(t) \big) \big|_{t=0} = 0 \qquad \text{for all~$l=0,\ldots, L$}\:.
\eeq

We choose a test function~$\hat{\eta} \in C^\infty_0((-1,1))$ and denote its
Fourier transform by~$\eta$. Furthermore, we choose a parameter~$\Omega_0 < -4 m_g$ and introduce
the function~$\hat{\eta}_{\Omega_0} \in C^\infty_0((-5 m_g, -3 m_g))$ by
\[ \hat{\eta}_{\Omega_0}(\Omega) = \frac{1}{| 4 \Omega_0|}\:
\hat{\eta} \Big( \frac{\Omega - \Omega_0}{4 \Omega_0} \Big) \:. \]
Thus~$\hat{\eta}_{\Omega_0}$ is supported for large negative frequencies.
For any such frequency~$\Omega \in \text{supp} \,\hat{\eta}_{\Omega_0}$, we want to
construct the plane wave~$e^{i \Omega t}$, with an error term which is again of large
negative frequency. To this end, we proceed similar as in the proof of Lemma~\ref{s:lemmaapprox}
by iteratively perturbing the first and last Dirac seas, but now beginning with the first sea.
Thus we first construct the plane wave~$e^{i \Omega t}$ by perturbing the first sea, and compensate the
error term by perturbing the last Dirac sea. This gives in analogy to~\eqref{s:hfex}
\[ \hat{f}^1(\omega) = \delta \bigg(
\omega - \frac{\Omega^2 +m_1^2}{2 \Omega} \bigg) \:,\qquad
\hat{f}^g(\omega) = -\delta \bigg(
\omega - \frac{\Omega \,m_g^2}{2 m_1^2} -  \frac{m_1^2}{2 \Omega} \bigg) \:, \]
giving rise to the plane wave
\beq \label{s:pwave}
e^{i \Omega t} - e^{i \Omega t \:m_g^2/m_1^2} \:.
\eeq
Multiplying by~$\hat{\eta}_{\Omega_0}$ and integrating over~$\Omega$, we find that the
function
\[ f(t) := \frac{1}{it} \Big(
e^{i \Omega_0 t}\, \eta(4 \Omega_0 t) - e^{i \Omega_1 t}\, \eta(4 \Omega_1 t)  \Big)
\qquad \text{with} \qquad \Omega_1 = \frac{m_g^2}{m_1^2}\: \Omega_0 \]
has the desired Fourier representation~\eqref{s:ft2}.
Since differentiating~\eqref{s:ft2} with respect to~$t$ merely generates factors of~$\omega^\beta_\pm$,
the functions~$f^{(l)}(t) := t^{-1} \partial_t^l (t f(t))$, $l=1,\ldots, L$, can again be represented
in the form~\eqref{s:ft2}.
By a suitable choice of the function~$\hat{\eta}$, we can clearly arrange that the
parameters~$f(0), f^{(1)}(0), \ldots, f^{(L)}(0)$ are linearly independent.
Thus by adding to~$f$ a suitable linear combination of the functions~$f^{(l)}$,
we can arrange~\eqref{s:scalcond}.
\QED
For clarity, we point out that the function~$f$ in~\eqref{s:scalcond} will in general not
be a good global approximation to~$h$. In particular, it is impossible to pass to the
limit~$L \rightarrow \infty$.
We also remark that this proposition also holds in the case~$g=1$ of only Dirac sea.
However, in this case it would not be possible to compensate the error term, so that
instead of~\eqref{s:pwave} we would have to work with the combination~$
e^{i \Omega t} - e^{i t \,m^2/\Omega}$.
This has the disadvantage that in the limit~$\Omega \rightarrow -\infty$, we would get
contributions of low frequency, making it impossible to generalize the result to the
non-homogeneous situation (see the proof of Theorem~\ref{s:thmloc} below).
This is why Proposition~\ref{s:prphom} was formulated only in the case~$g \geq 2$.

\subsectionn{Nonlocal Potentials, the Quasi-Homogeneous Ansatz} \label{s:sec113}
We now want to extend the previous results to the non-homogeneous situation.
For notational simplicity, we will write all formulas only for one generation.
But as we only consider perturbations which are diagonal on the generations,
all constructions immediately carry over to several generations by
taking sums. Our method is to first describe our previous perturbations of the
fermionic projector~\eqref{s:homperturb2} by homogeneous
perturbations of the Dirac operator. Replacing this perturbation operator by a nonlocal
operator will then make it possible to describe the desired non-homogeneous perturbations of the
fermionic projector.

We first note that the perturbation of the fermionic projector~\eqref{s:delPuni}
came about by the unitary transformation~\eqref{s:Ukform}. We can describe it alternatively
by perturbing the Dirac operator in momentum space to
\[ \slashed{k} + \B(k) - m \qquad \text{with} \qquad \B(k) = U(k) \,\slashed{k} \,U(k)^* - \slashed{k} \:. \]
Similarly, the plane wave~\eqref{s:pwsol} is a solution
of the Dirac equation~$(i \Pdd - \delta \slashed{k} - m) \psi = {\mathcal{O}}((\delta k)^2)$.
Thus the linear perturbation~\eqref{s:homperturb} can be described equivalently by
working with the perturbed Dirac operator in momentum space
\[ \slashed{k} - \delta \slashed{k}(k) - m \:. \]
Likewise, for the perturbation~\eqref{s:homperturb2}, we must find a perturbation~$\nf$ of the
Dirac operator which is symmetric and, when restricted to the image of the operators~$P_\pm(k)$,
reduces to the operators~$-\delta \slashed{k}_\pm$. In order to determine~$\nf$, it is convenient to decompose
the vector fields~$\delta k_\pm$ as
\beq \label{s:VApdef}
\delta k_\pm = V \pm (\phi\, q + A) \qquad \text{where} \qquad
\la A, q \ra = 0
\eeq
(thus~$V$ is the vector part, whereas~$\phi$ and~$A$ describe the axial components
parallel and orthogonal to~$q$, respectively).
Then using~\eqref{s:qprop} together with the relations
\[ \pseudo \slashed{q}\, \Pi_\pm = \mp \Pi_\pm \:, \]
we obtain
\begin{align*}
-\delta \slashed{k}_\pm P_\pm &= \left( -\slashed{V} \mp (\phi\, \slashed{q} + \slashed{A}) \right) P_\pm
=  -\slashed{V} P_\pm  + (\phi\, \slashed{q} + \slashed{A}) (\pseudo \slashed{q}) P_\pm \\
&= \left(-\slashed{V} + \phi \pseudo + \pseudo \slashed{q}\, \slashed{A} \right) P_\pm
= \Big[-\slashed{V} + \frac{\phi}{m} \pseudo \slashed{k} + \pseudo \slashed{q}\, \slashed{A} \Big] P_\pm\:,
\end{align*}
where in the last step we used that~$(\slashed{k}-m) P_\pm = 0$. The square bracket has
the desired properties of~$\nf$. Thus the perturbation~\eqref{s:homperturb2} is equivalently
described by the Dirac operator in momentum space
\beq \label{s:dirhom}
\slashed{k} + \nf + m \qquad \text{with} \qquad
\nf(k) = -\slashed{V}(k) + \frac{\phi(k)}{m} \pseudo \slashed{k} + \pseudo \slashed{q}(k)\, \slashed{A}(k) \:,
\eeq
and the fields~$V$, $A$ and~$\phi$ as defined by~\eqref{s:VApdef}.
Note that~$\nf$ is composed of three terms, which can be regarded as a vector, an axial and
a bilinear perturbation of the Dirac operator. The perturbations in~\eqref{s:dirhom} are
all homogeneous. The reader interested in the perturbation expansion for the corresponding
Dirac solutions is referred to~\cite[Appendix~C.1]{PFP}.

In order to generalize to non-homogeneous perturbations, we write the operator~$\nf$ as the
convolution operator in position space
\beq \label{s:nonloc}
(\nf \,\psi)(x) = \int_\scrM \nf(x,y) \: \psi(y)\, d^4y
\eeq
with the integral kernel
\[ \nf(x,y) = \int \frac{d^4 k}{(2 \pi)^4}\:\nf(k) \, e^{i k \xi} \:. \]
Now we can replace~$\nf(x,y)$ by a general nonlocal kernel.
\nindex{cl8@$\nf(x,y)$ -- nonlocal kernel}%
\sindex{potential!nonlocal}%
We refer to the operator~$\nf$ as a {\em{nonlocal potential}} in the Dirac equation.
For technical convenience, it seems appropriate to make suitable decay assumptions at infinity,
for example by demanding that
\beq \label{s:ndecay}
\nf(x,y) \in {\mathcal{S}}(\scrM \times \scrM)\:.
\eeq
\nindex{cm0@${\mathcal{S}}(\scrM \times \scrM)$ -- Schwartz kernel in Minkowski space}%
Then the corresponding fermionic projector can be introduced perturbatively exactly as 
outlined in~\S\ref{s:sec44}. Before we can make use of the general ansatz~\eqref{s:nonloc}
and~\eqref{s:ndecay}, we must specify~$\nf(x,y)$. To this end, we fix the variable~$\zeta := y+x$ and consider
the fermionic projector as a function of the variable~$\xi$ only. Then we are again in the homogeneous
setting, and we can choose the operator~$\nf$ as in~\eqref{s:dirhom}. In order to clarify the dependence
on the parameter~$\zeta$, we denote this operator by~$\nf(k, \zeta)$. As explained at the beginning
of this section, we consider~$y+x$ as a macroscopic variable, whereas~$k$ is the momentum of
the quantum mechanical oscillations. This motivates us to introduce the kernel~$\nf(x,y)$ similar
to~\eqref{s:Umicro} by the {\em{quasi-homogeneous ansatz}}
\beq \label{s:adiabatic}
\nf(x,y) = \int \frac{d^4k}{(2 \pi)^4}\: \nf \big(k, y+x \big) \:e^{i k (y-x)}\: .
\eeq
\sindex{quasi-homogeneous ansatz}%
We remark for clarity that this procedure generalizes our method~\eqref{s:dirnonloc} of rewriting the microlocal
chiral transformation in terms of a perturbation of the Dirac operator. Now~$\nf$ may involve additional
nonlocal potentials which describe variations of the momenta of the plain-wave solutions.

The quasi-homogeneous ansatz
makes it possible to satisfy the EL equations to degree four locally around every space-time point,
as is made precise in the following theorem.
\begin{Thm} \label{s:thmloc}
Suppose that the number of generations~$g \geq 2$, and that we are given
an integer~$L \geq 2$ and a parameter~$\delta>0$.
Then for any given smooth functions~$h_\text{\rm{v}}, h_\text{\rm{a}} \in {\mathcal{S}}(\scrM \times \scrM)$,
there is a nonlocal potential~$\nf$ of the form~\eqref{s:nonloc} with a kernel~$\nf(x,y) \in
{\mathcal{S}}(\scrM \times \scrM)$ of the form~\eqref{s:adiabatic}
such that for all multi-indices~$\kappa$ with~$1 \leq |\kappa| \leq L$,
\[ \bigg| \partial_\xi^\kappa \big[ \Tr \big( \slashed{\xi}\: \Delta P(x,y) \big) - h_\text{\rm{v}}(x,y) \big] \Big|_{x=y} \bigg|
+ \bigg| \partial_\xi^\kappa \left[
\Tr \big( \pseudo \slashed{\xi}\: \Delta P(x,y) \big) - h_\text{\rm{a}}(x,y) \right]  \Big|_{x=y} \bigg| < \delta\: . \]
Here~$\Delta P$ denotes the perturbation of the fermionic projector 
to first order in the nonlocal potential~$\nf$.
\end{Thm}
\Proof For a fixed choice of the parameter~$\Omega_0$ and for any given~$\zeta$,
we choose the homogeneous perturbation~$\delta P$
as in the proof of Proposition~\ref{s:prphom} and rewrite it according to~\eqref{s:dirhom}
as a homogeneous perturbation~$\nf(k,\zeta)$ of the Dirac equation.
Introducing the nonlocal potential by~\eqref{s:nonloc} and~\eqref{s:adiabatic},
the rapid decay of the functions~$h_\text{\rm{v}}$ and~$h_\text{\rm{a}}$ implies that the
kernel~$\nf(x,y)$ has rapid decay also in~$\zeta$ (the rapid decay in~$\xi$ is obvious
because~$\nf(.,\zeta) \in {\mathcal{S}}(\hscrM)$). Since the same is true for all derivatives,
we conclude that~$\nf(x,y) \in {\mathcal{S}}(\scrM \times \scrM)$.

The corresponding perturbation of the fermionic projector~$\Delta P$ can be analyzed
with the methods introduced in~\cite{firstorder} (see also Section~\ref{seclight}).
We first pull out the Dirac matrices to obtain
\beq \label{s:DelP2}
\Delta P(x,y) = (i \Pdd_x + m) \left(-i \frac{\partial}{\partial y^k}+m \right) \Delta T_{m^2}[\nf](x,y)\: \gamma^k \:,
\eeq
where~$\Delta T_{m^2}$ is the perturbation of the corresponding solution of the
inhomogeneous Klein-Gordon equation (see~\cite[eqs~(2.4) and~(2.5)]{firstorder})
\begin{align*}
\Delta T_{m^2}&[\nf](x,y) =
 - \int_\scrM d^4z_1 \int_\scrM d^4 z_2 \\
& \times \Big( S_{m^2}(x,z_1) \:\nf(z_1, z_2)\:
	T_{m^2}(z_2,y) \:+\: T_{m^2}(x,z_1) \:\nf(z_1, z_2)\: S_{m^2}(z_2,y)  \Big)
\end{align*}
(where~$T_a$ and~$S_a$ are again given by~\eqref{s:Tadef} and~\eqref{s:Samom}).
We next transform to momentum space. Setting
\[ \nf(p,q) = \int_\scrM \nf(p, \zeta) \, e^{\frac{i q \zeta}{2}}\: d^4 \zeta \]
with the ``macroscopic'' momentum vector~$q$, 
the above formula for~$\Delta T_{m^2}$ becomes (see~\cite[eqs~(3.8) and~(3.9)]{firstorder})
\begin{align*}
\Delta &T_{m^2}[\nf] \Big( p+\frac{q}{2}, p-\frac{q}{2} \Big) \\
&= -S_{m^2} \Big( p+\frac{q}{2} \Big)\, \nf(p,q)\, T_{m^2} \Big( p-\frac{q}{2} \Big)
-T_{m^2} \Big( p+\frac{q}{2} \Big)\, \nf(p,q)\, S_{m^2} \Big( p-\frac{q}{2} \Big) \:.
\end{align*}
Now we can perform the light-cone expansion exactly as in~\cite[Section~3]{firstorder}.
This gives (cf.~\cite[eq.~(3.21)]{firstorder})
\begin{align*}
\Delta &T_{m^2}[\nf] \Big( p+\frac{q}{2}, p-\frac{q}{2} \Big) = - \nf(p,q)\\
& \times \sum_{n=0}^\infty
        \frac{(-1)^n}{n!} \left( \frac{q^2}{4} \right)^n \sum_{k=0}^\infty
        \frac{1}{(2k+1)!} \sum_{l=0}^k
        \left[\!\! \begin{array}{c} 2k \\ l
        \end{array} \!\!\right] \:\left( \frac{q^2}{2} \right)^l \:
        \left(\frac{q^j}{2} \frac{\partial}{\partial p^j}\right)^{2k-2l}
        T_{m^2}^{(n+1+l)}(p) \;,
\end{align*}
where the curly brackets are combinatorial factors whose detailed form is not needed here
(see~\cite[eq.~(3.13)]{firstorder}).

Let us discuss how the Fourier transform of this expansion behaves
in the limit~$\Omega_0 \rightarrow -\infty$. According to the construction
of the functions~$\hat{f}_\beta$ in the proof of Proposition~\ref{s:prphom},
the function~$\nf(p,q)$ is supported in the region~$p^2 \sim \Omega_0^2$.
Moreover, the $p$-derivatives of~$\nf$ scale in powers of~$1/\Omega_0$.
This implies that every derivative of the factor~$T_{m^2}$
gives a scaling factor of~$\Omega_0^{-2}$. Since every such derivative
comes with factor $q^2$, we obtain a scaling factor~$(q^2/\Omega_0^2)^{n+l}$.
Thus in the limit~$\Omega_0 \rightarrow -\infty$, it suffices to consider
the lowest summand in~$n+l$,
\[ \Delta T_{m^2}[\nf] = - \nf(p,q)
\sum_{k=0}^\infty
        \frac{1}{(2k+1)!} \:
        \left(\frac{q^j}{2} \frac{\partial}{\partial p^j}\right)^{2k}
        T_{m^2}^{(1)}(p) \,\left[ 1 + \O \Big( \frac{q^2}{\Omega_0^2} \Big) \right] . \]
When transforming to position space, the $p$-derivatives can be integrated by parts.
If they act on the function~$\nf(p,q)$, this generates scale factors of the
order~$\O(|q|/|\Omega_0|)$ which again tend to zero as~$\Omega_0 \rightarrow -\infty$.
Thus it remains to consider the case when these derivatives act on the
plane wave~$e^{i p \xi}$. We thus obtain
\[ \Delta T_{m^2}[\nf](x,y) = - \int \frac{d^4p}{(2 \pi)^4} \int \frac{d^4q}{(2 \pi)^4} \:
\nf(p,q) \sum_{k=0}^\infty \frac{(-q \xi)^{2k}}{2^k (2k+1)!} \:
T_{m^2}^{(1)}(p) \:e^{i p \xi - \frac{i q \zeta}{2}} + \O \Big( \frac{1}{\Omega_0} \Big) \,. \]
We now consider a Taylor expansion in~$\xi$ around~$\xi=0$ up to the given order~$L$.
This amounts to replacing the factor~$e^{e^{i p \xi}}$ by its power series and collecting
the powers of~$\xi$. The remaining task is to compare the factors~$p \xi$ with~$q \xi$.
This is a subtle point, because the fact that~$p^2 \sim \Omega_0^2$ does not imply
that the inner product~$p \xi$ is large. Indeed, this effect was responsible for the
appearance of low frequencies in the Fourier integral~\eqref{s:ft}. However, in the proof of
Proposition~\ref{s:prphom} we arranged by a suitable choice of~$\hat{f}_1$ and~$\hat{f}_g$
that these low-frequency contributions cancel, so that we were working only with
the high-frequency terms~\eqref{s:pwave}. Restating this fact in the present context,
we can say that for the leading contribution to~$\Delta T_{m^2}$, the
factor~$p \xi$ is larger than $q \xi$ by a factor of the order~$\O(|q|/|\Omega_0|)$.
Thus it suffices the consider the summand~$k=0$,
\[ \Delta T_{m^2}[\nf](x,y) = - \int \frac{d^4p}{(2 \pi)^4} \int \frac{d^4q}{(2 \pi)^4} \:
\nf(p,q) \:T_{m^2}^{(1)}(p) \:e^{i p \xi - \frac{i q \zeta}{2}} + \O \Big( \frac{1}{\Omega_0} \Big) \,. \]
Now we can carry out the $q$-integration to obtain
\[ \Delta T_{m^2}[\nf](x,y) = - 16 \int \frac{d^4q}{(2 \pi)^4} \:
\nf(p,y+x) \:T_{m^2}^{(1)}(p) \:e^{i p \xi} + \O \Big( \frac{1}{\Omega_0} \Big) \,. \]
Using this result in~\eqref{s:DelP2}, we can carry out the derivatives to recover
precisely the homogeneous perturbation~\eqref{s:homperturb2} for fixed~$\zeta$.
\QED
The scaling argument used in the last proof can be understood non-technically as follows.
It clearly suffices to consider the region where~$y$ lies
in a small neighborhood of~$x$. Thus we may perform the rescaling~$\xi \rightarrow \xi/\lambda$
with a scale factor~$\lambda \gg 1$, leaving~$\zeta$ unchanged.
This corresponds to changing the momentum scale by~$\Omega_0 \rightarrow \lambda \Omega_0$.
In the limiting case $\lambda \rightarrow \infty$, the fermionic projector depends on~$\xi$
on a smaller and smaller scale. On this scale, the dependence on the variable~$x+y$
drops out, so that the quasi-homogeneous ansatz~\eqref{s:adiabatic} becomes exact.
For this argument to work, one must ensure that in the homogeneous setting all frequencies
scale like~$\Omega_0$, as was arranged in~\eqref{s:pwave}.

\subsectionn{Discussion and Concluding Remarks} \label{s:secnonlocconclude}
The previous analysis puts us into the position to discuss the scope and significance
of nonlocal potentials. It is remarkable that by choosing suitable nonlocal potentials,
one can satisfy the EL equations to any order in an expansion in powers of~$\xi$
(see Theorem~\ref{s:thmloc}).
However, this is not sufficient because, as shown in~\S\ref{s:secELC}, the EL equations~\eqref{s:ELcl}
must be satisfied globally (i.e.\ for any~$x, y$ with~$|\vec{\xi}| \gg \varepsilon$; see~\eqref{s:ap2}).
Such global solutions of the EL equations, however, cannot be constructed with the help
of nonlocal potentials (see Lemma~\ref{s:lemmanocomp}
and the paragraph before this lemma). More specifically, as one sees from~\eqref{s:pwave},
every plane wave~$e^{i \Omega t}$ comes with an error term of the same magnitude.
This raises the question whether nonlocal potentials are at all useful for fulfilling the EL equations.

In order to answer this question, it is helpful to first consider the EL equations
on the more restrictive scale~\eqref{s:xiscale}. On this scale, we could satisfy the EL equations
by working with local potentials and the microlocal chiral transformation.
Introducing additional nonlocal potentials does not seem helpful because in view of~\eqref{s:pwave},
it becomes impossible to satisfy the EL equation for {\em{all}}~$x,y$ in the range~\eqref{s:xiscale}.
More specifically, by arranging that the constants~$C_\beta$ in~\eqref{s:PasyO} were constant,
we achieved that the contribution by the microlocal chiral transformation was the
desired logarithmic pole plus smooth contributions which varied on the scale~$\ell_\text{macro}$.
According to~\eqref{s:pwave}, additional nonlocal potentials would introduce contributions
which oscillate on the macroscopic scale, thus violating the EL equations on the scale~\eqref{s:xiscale}.
We conclude that in order to satisfy the EL equations
in a weak evaluation for any~$|\vec{\xi}|$ in the range~\eqref{s:xiscale},
the nonlocal potentials considered in this section must necessarily vanish.

We point out that with our methods, it seems impossible to satisfy the EL equations
globally (i.e.\ to fulfill~\eqref{s:ELcl} for~$x, y$ with~$|\vec{\xi}| \gtrsim \ell_\text{macro}$).
At first sight, this might seem to imply that the EL equations are overdetermined and cannot be
solved. On the other hand, the general compactness results in~\cite{continuum} indicate
that our action principle does have non-trivial minimizers, so that the EL equations are expected
to admit solutions. Thus there should be a way to compensate the above nonlocal error terms.
A possible method is to modify the wave functions globally in space-time. Whether and how
in detail this is supposed to work is a difficult question which we cannot answer here.
Instead, we explain what this situation means physically: Suppose that a physical system is described
by a minimizer of our action principle. Then the corresponding EL equations to degree four
do not only yield the field equations, but they give rise to additional conditions which are nonlocal
and can therefore not be specified by a local observer.
In~\cite[Section~7]{rev} such so-called {\em{nonlocal quantum conditions}} were proposed to
explain phenomena which in ordinary quantum mechanics are probabilistic.

More specifically, the fact that the EL equations cannot in general be satisfied globally might explain
the tendency for quantum mechanical wave functions to be localized, as we now outline.
Suppose that the fermionic projector is perturbed by a fermionic
wave function~\eqref{s:particle}. At the origin~$\xi=0$, this
perturbation leads to the field equations as worked out in Section~\ref{s:secfield}.
The higher orders in~$\xi$ can be compensated by nonlocal potentials.
But the contribution for large~$\xi$ cannot be compensated, thereby increasing
our action. Thus seeking for minimizers, our action principle
should try to arrange that the contribution~\eqref{s:particle} vanishes
for large~$\xi$. This might explain why quantum mechanical wave functions
are usually not spread out over large distances, but are as much as possible
localized, even behaving as point particles. This idea is explained further in~\cite{dice2010}.


\chapter[A System Involving Neutrinos]{The Continuum Limit of a Fermion System Involving Neutrinos:
Weak and Gravitational Interactions} \label{lepton}

\begin{abstract}
We analyze the causal action principle for a system of relativistic fer\-mions
composed of massive Dirac particles and neutrinos.
In the continuum limit, we obtain an effective interaction described by
a left-handed, massive $\SU(2)$ gauge field and a gravitational field.
The off-diagonal gauge potentials involve a unitary mixing matrix, 
which is similar to the Maki-Nakagawa-Sakata matrix in the standard model.
\end{abstract}

\section{Introduction} \label{l:sec1}
In~\cite{PFP} it was proposed to formulate physics based on a new action principle in space-time.
In Chapter~\ref{sector}, this action principle was worked out in detail in the so-called continuum limit
for a simple model involving several generations of massive Dirac particles.
We now extend this analysis to a model which includes neutrinos.
In the continuum limit, we shall obtain an effective interaction described by a left-handed massive
$\SU(2)$ gauge field and a gravitational field.

More specifically, we again consider the causal action principle introduced in~\cite{PFP}. Thus we define the
causal Lagrangian by
\beq \label{l:Ldef}
\L[A_{xy}] = |A_{xy}^2| - \frac{1}{8}\: |A_{xy}|^2 \:,
\eeq
\nindex{aa2@$\L(x,y)$ -- Lagrangian}%
\sindex{Lagrangian!causal}%
where~$A_{xy} = P(x,y)\, P(y,x)$ denotes the closed chain corresponding to the
fermionic projector~$P(x,y)$,
\sindex{closed chain}%
\nindex{ad0@$A_{xy}$ -- closed chain}%
and~$|A| = \sum_{i=1}^8 |\lambda_i|$ is the spectral weight
(where~$\lambda_i$ are the eigenvalues of~$A$ counted with algebraic multiplicities).
We introduce the action~$\Sact$ and the constraint~$\T$ by
\[ \Sact[P] = \iint_{\scrM \times \scrM} \L[A_{xy}] \:d^4 x\: d^4y \:,\qquad
\T[P] = \iint_{\scrM \times \scrM} |A_{xy}|^2 \:d^4 x\: d^4y\:, \]
\nindex{aa4@$\Sact(\rho)$ -- causal action}%
\sindex{causal action}%
\nindex{aa8@$\T(\rho)$ -- functional in boundedness constraint}%
where~$(\scrM, \langle .,. \rangle)$ denotes Minkowski space.
The causal action principle is to
\beq \label{l:actprinciple}
\text{minimize $\Sact$ for fixed~$\T$}\:.
\eeq
This action principle is given a rigorous meaning in Section~\ref{s:sec2}.
Every minimizer is a critical point of the so-called auxiliary action
\beq \label{l:auxact}
\Sact_\mu[P] = \iint_{\scrM \times \scrM} \L_\mu[A_{xy}] \:d^4 x\: d^4y \:,\qquad
\L_\mu[A_{xy}] = |A_{xy}^2| - \mu \: |A_{xy}|^2 \:,
\eeq
\nindex{cc2@$\Sact_\mu$ -- causal action involving one Lagrange multiplier}%
\nindex{cc4@$\L_\mu$ -- Lagrangian involving one Lagrange multiplier}%
which involves a Lagrange multiplier~$\mu \in \R$.

We model the configuration of the fermions  by a system consisting of a doublet of two sectors,
\sindex{sector}%
each composed of three generations. Thus we describe the vacuum by the fermionic projector
\beq \label{l:Pvac}
P(x,y) = P^N(x,y) \oplus P^C(x,y) \:,
\eeq
\nindex{da0@$P^N(x,y)$ -- neutrino sector of the vacuum fermionic projector}%
\nindex{da2@$P^C(x,y)$ -- charged component of the vacuum fermionic projector}%
\sindex{fermionic projector!of the vacuum}%
where the {\em{charged sector}} $P^C$ is formed exactly as the fermionic projector in
Chapter~\ref{sector} as a sum of Dirac seas, i.e.
\beq \label{l:PC}
P^C(x,y) = \sum_{\beta=1}^3 P^\text{vac}_{m_\beta}(x,y) \:,
\eeq
\sindex{sector!charged}%
where~$m_\beta$ are the masses of the fermions and~$P^\text{vac}_m$ is the distribution
\beq \label{l:Pmdef}
P^\text{vac}_m(x,y) = \int \frac{d^4k}{(2 \pi)^4}\: (\slashed{k}+m)\: \delta(k^2-m^2)\: \Theta(-k^0)\: e^{-ik(x-y)}\:.
\eeq
\nindex{ca3@$m_\beta$ -- masses of charged fermions}%
\nindex{ba0@$P^\text{vac}_m(x,y)$ -- fermionic projector corresponding to a vacuum Dirac sea of mass~$m$}%
For the {\em{neutrino sector}}~$P^N$ we consider two different ans\"atze.
\sindex{sector!neutrino}%
The first ansatz of {\em{chiral neutrinos}} is to take a sum of left-handed, massless
Dirac seas,
\beq \label{l:PN}
P^N(x,y) = \sum_{\beta=1}^3 \chi_L \,P^\text{vac}_0(x,y) \:.
\eeq
\sindex{neutrino!chiral}%
The configuration of Dirac seas~\eqref{l:Pvac}, \eqref{l:PC} and~\eqref{l:PN} models precisely the
leptons in the standard model. It was considered earlier in~\cite[Section~5.1]{PFP}.
The chiral ansatz~\eqref{l:PN} has the shortcoming that the neutrinos are necessarily massless,
in contradiction to experimental observations. In order to describe massive neutrinos, we
proceed as follows. As the mass mixes the left- and right-handed spinor components in the
Dirac equation, for massive Dirac particles it is impossible to to restrict attention to one chirality.
This leads us to the ansatz of {\em{massive neutrinos}}
\beq \label{l:massneutrino}
P^N(x,y) = \sum_{\beta=1}^3 P^\text{vac}_{\tilde{m}_\beta}(x,y)\:.
\eeq
\sindex{neutrino!massive}%
\nindex{da4@$\tilde{m}_\beta$ -- neutrino masses}%
Here the neutrino masses~$\tilde{m}_\beta \geq 0$ will in general be different from
the masses~$m_\beta$ in the charged sector.
Except for the different masses, the ans\"atze~\eqref{l:PC} and~\eqref{l:massneutrino} are exactly the same.
In particular, it might seem surprising that~\eqref{l:massneutrino} does not
distinguish the left- or right-handed component, in contrast to the observation that neutrinos are
always left-handed. In order to obtain consistency with experiments, if working with~\eqref{l:massneutrino}
we need to make sure that the {\em{interaction}} distinguishes one chirality.
For example, if we described massive neutrinos by~\eqref{l:massneutrino} and found that
the neutrinos only couple to left-handed gauge fields, then the right-handed neutrinos,
although being present in~\eqref{l:massneutrino}, would not be observable.
With this in mind, working with~\eqref{l:massneutrino} seems a possible approach, provided that
we find a way to break the chiral symmetry in the interaction. It is a major goal of this paper
to work out how this can be accomplished.

Working out the continuum limit for the above systems gives the following results.
First, we rule out the chiral ansatz~\eqref{l:PN} by showing that it does not admit a
global minimizer of the causal action principle. Thus in the fermionic projector approach,
we must necessarily work with the massive ansatz~\eqref{l:massneutrino}.
We find that at least one of the neutrino masses~$\tilde{m}_\beta$ must be strictly
positive. In order to break the chiral symmetry, we introduce additional
right-handed states into the neutrino sector. It is a delicate question how
this should be done. We discuss different approaches, in particular the so-called shear
states and general surface states. The conclusion is that if the right-handed states
and the regularization are introduced suitably, then the continuum limit is well-defined.
Moreover, the structure of the effective interaction in the continuum limit is described as follows.
The fermions satisfy the Dirac equation coupled to a left-handed $\SU(2)$-gauge potential~$A_L$,
\beq \label{l:Dirac}
\left[ i \Pdd + \chi_R \begin{pmatrix} \slashed{A}_L^{11} & \slashed{A}_L^{12}\, \UMNS^* \\[0.2em]
\slashed{A}_L^{21}\, \UMNS & -\slashed{A}_L^{11} \end{pmatrix}
- m Y \right] \!\psi = 0 \:,
\eeq
where we used a block matrix notation (where the matrix entries are
$3 \times 3$-matrices). Here~$mY$ is a diagonal matrix composed of the fermion masses,
\[ mY = \text{diag} (\tilde{m}_1, \tilde{m}_2, \tilde{m}_3,\: m_1, m_2, m_3)\:, \]
\sindex{mass matrix}%
\nindex{bh3@$m$ -- parameter used for mass expansion}%
\nindex{bh4@$Y$ -- mass matrix}%
\nindex{da6@$\UMNS$ -- Maki-Nakagawa-Sakata (MNS) matrix}%
\sindex{mixing matrix!Maki-Nakagawa-Sakata (MNS) matrix}%
and~$\UMNS$ is a unitary $3 \times 3$-matrix. In analogy to the standard model,
we refer to~$\UMNS$ as the Maki-Nakagawa-Sakata (MNS) matrix.
The gauge potentials~$A_L$ satisfy a classical Yang-Mills-type equation, coupled
to the fermions. More precisely, writing the isospin dependence of the gauge potentials according
to~$A_L = \sum_{\alpha=1}^3 A_L^\alpha \sigma^\alpha$ in terms of Pauli matrices,
we obtain the field equations
\beq \label{i:YM}
\partial^k_{\;\:l} (A^\alpha_L)^l - \Box (A^\alpha_L)^k - M_\alpha^2\, (A^\alpha_L)^k = c_\alpha\,
\overline{\psi} \big( \chi_L \gamma^k \, \sigma^\alpha \big) \psi\:,
\eeq
valid for~$\alpha=1,2,3$. Here~$M_\alpha$ are the bosonic masses and~$c_\alpha$
the corresponding coupling constants.
\nindex{da8@$M_\alpha$ -- bosonic masses}%
\nindex{db0@$c_\alpha$ -- coupling constant}%
The masses and coupling constants of the two off-diagonal components are
equal, i.e.\ $M_1=M_2$ and~$c_1 = c_2$,
but they may be different from the mass and coupling constant
of the diagonal component~$\alpha=3$.

Moreover, our model involves a gravitational field described by the Einstein equations
\beq \label{i:Einstein}
R_{jk} - \frac{1}{2}\:R\: g_{jk} + \Lambda\, g_{jk} = \kappa\, T_{jk} \:,
\eeq
\sindex{Einstein equations}%
\nindex{db2@$R_{jk}$ -- Ricci tensor}%
\nindex{db4@$R$ -- scalar curvature}%
\nindex{db6@$T_{jk}$ -- energy-momentum tensor}%
\sindex{cosmological constant}%
\nindex{db8@$\Lambda$ -- cosmological constant}%
\nindex{dc0@$\kappa$ -- gravitational constant}%
\sindex{coupling constant!gravitational}%
where~$R_{jk}$ denotes the Ricci tensor, $R$ is scalar curvature, and~$T_{jk}$
is the energy-momentum tensor of the Dirac field. Moreover, $\kappa$ and~$\Lambda$ denote the
gravitational and the cosmological constants, respectively.
We find that the gravitational constant scales like~$\kappa \sim \delta^2$, where~$\delta$ is
the length scale on which the shear and general surface states become relevant.
The dynamics in the continuum limit is described by the coupled 
Dirac-Yang/Mills-Einstein equations~\eqref{l:Dirac}, \eqref{i:YM} and~\eqref{i:Einstein}.
These equations are of variational form, meaning that they
can be recovered as Euler-Lagrange equations corresponding to an
``effective action.'' The effective continuum theory is manifestly covariant
under general coordinate transformations.

For ease in notation, the field equations~\eqref{i:YM} (and similarly
the Einstein equations~\eqref{i:Einstein}) were written only for one
fermionic wave function~$\psi$. But clearly, the equations hold similarly for many-fermion
systems (see Theorem~\ref{l:thmfield}). In this context, it is worth noting that, although
the states of the Dirac sea are explicitly taken into account in our analysis, they do not
enter the Einstein equations. Thus the naive ``infinite negative energy density'' of the
sea drops out of the field equations, making it unnecessary to subtract any counter terms.

Similar as explained in Chapter~\ref{sector} for an axial field, we again obtain corrections to the field equations
which are nonlocal and violate causality in the sense that the future may influence the past.
Moreover, for a given regularization one can compute
the coupling constant, the bosonic mass, and the gravitational constant.

We remark that in this book, we always assume that the regularization
and the corresponding scales~$\varepsilon$ and~$\delta$ are
constant in space-time. Although this seems a good approximation locally,
it is conceivable that the regularization does change on the astrophysical
or cosmological scale. In this case, the gravitational constant would 
no longer be constant in space-time, but would become dynamical.
The resulting effect, referred to as dynamical gravitational coupling,
\sindex{dynamical gravitational coupling}%
will not be covered in this book, but we refer the interested reader to~\cite{dgc}.

We note that in this paper, we restrict attention to explaining our computations and results;
for all conceptual issues and more references
we refer to Chapter~\ref{sector} and the survey article~\cite{srev}.

\section{Regularizing the Neutrino Sector} \label{l:sec2}
In this section, we explain how the neutrino sector is to be regularized. We begin in~\S\ref{l:sec20}
by reviewing the regularization method used in~\cite{PFP}
(see also Section~\ref{secreg}). Then we give an argument why this method
is not sufficient for our purposes (see~\S\ref{l:sec21}). This leads us to extending our methods (see~\S\ref{l:sec22}),
and we will explain why these methods only work for the ansatz of massive neutrinos
(see~\S\ref{l:secmassive}). In~\S\ref{l:sec23} we introduce the resulting general
regularization scheme for the vacuum neutrino sector.
In~\S\ref{l:sec24} we explain how to introduce an interaction, relying for the more technical aspects
on Appendix~\ref{l:appA}. Finally, in~\S\ref{l:seciota} we introduce a modification of the
formalism of the continuum limit which makes some computations more transparent.

\subsectionn{A Naive Regularization of the Neutrino Sector} \label{l:sec20}
As in Section~\ref{s:sec3} we denote the regularized fermionic projector of the vacuum
by~$P^\varepsilon$,
\nindex{aj8@$\varepsilon$ -- regularization length}%
\nindex{an4@$P^\varepsilon(x,y)$ -- regularized kernel of fermionic projector}%
where the parameter~$\varepsilon$ is the length scale of the regularization.
This regularization length can be thought of as the Planck length, but it could be even smaller.
Here we shall always assume that~$P^\varepsilon$ is {\em{homogeneous}}, meaning
that it depends only on the difference vector~$\xi:=y-x$. 
This is a natural physical assumption
as the vacuum state should not distinguish a specific point in space-time.
The simplest regularization method for the vacuum neutrino sector is to replace the above
distribution~$P^N(x,y)$ (see~\eqref{l:PN}) by a function~$P^N_\varepsilon$ which is again left-handed,
\beq \label{l:Pnaive}
P^N_\varepsilon(x,y) = \chi_L\: g_j(\xi)\: \gamma^j \:.
\eeq
Such a regularization, in what follows referred to as a {\em{naive regularization}},
\sindex{regularization!naive}%
was used in~\cite{PFP} (see~\cite[eq.~(5.3.1)]{PFP}). It has the effect that the corresponding
closed chain vanishes due to so-called {\em{chiral cancellations}} (see~\cite[eq.~5.3.2]{PFP}),
\sindex{chiral cancellation}%
\[ A^N_{xy} := P^N_\varepsilon(x,y)\, P^N_\varepsilon(y,x) = \chi_L \:\slashed{g}(x,y) \;\chi_L \:\slashed{g}(y,x) =
    \chi_L\: \chi_R \:\slashed{g}(x,y)\:\slashed{g}(y,x) = 0 \:. \]

Regularizing the charged sector as explained in~\cite[Chapter~4]{PFP} or Chapter~\ref{sector},
the closed chain of the regularized fermionic projector~$P^\varepsilon$ of the whole system
is of the form
\[ A_{xy} = P^\varepsilon(x,y)\, P^\varepsilon(y,x) = 0 \oplus A^C_{xy} \:. \]
Hence the closed chain has the eigenvalue zero with multiplicity four
as well as the non-trivial eigenvalues~$\lambda_+$ and~$\lambda_-$, both with multiplicity two
(see \cite[Section~5.3]{PFP}). Let us recall from~\cite[Chapter~5]{PFP} how by a specific choice
of the Lagrange multiplier~$\mu$ we can arrange that
the EL equations are satisfied: The operator~$Q$ corresponding to the
action~\eqref{l:auxact} is computed by (see~\cite[Section~3.5]{PFP} or~Section~\ref{s:sec7})
\[ Q(x,y) = (-4 \mu) \oplus \Big[ (1-4 \mu) \Big] \:\sum_{s = \pm} \overline{\lambda_s} F_s \:P(x,y) . \]
In order for the operator~$Q$ to vanish on the charged sector, we must choose
\beq \label{l:muchoice}
\mu = \frac{1}{4}\:.
\eeq
Then
\[ Q(x,y) = -\sum_{s = \pm} \overline{\lambda_s} F_s \:P^N_\varepsilon(x,y) \:\oplus\: 0 , \]
and multiplying by~$P(y,z)$, we again get chiral cancellations to obtain
\[ Q(x,y) P(y,z) = - \sum_{s = \pm} \overline{\lambda_s} F_s \;\chi_L \:\slashed{g}(x,z)\,
\chi_L \:\slashed{g}(z,y) \:\oplus\: 0 = 0\:. \]
Similarly, the pointwise product $P(x,y) \,Q(y,z)$ also vanishes, showing that the EL
equations~$[P,Q]$ are indeed satisfied in the vacuum.

Before going on, we note for clarity that in~\cite{PFP}, the chiral regularization
ansatz~\eqref{l:Pnaive} was overridden on the large scale
in order to arrange a suitable normalization of the chiral
fermionic states (see~\cite[Appendix~C]{PFP}). More precisely, $P^N_\varepsilon$ was constructed
by projecting out half of the states of a Dirac sea of mass~$m$. The formula~\eqref{l:Pnaive} was
recovered in the limit~$m \searrow 0$. In this so-called {\em{singular mass limit}}, the normalization
integrals did not converge, making it possible to arrange a proper normalization, although for the
limit~\eqref{l:Pnaive} the normalization integral would vanish due to chiral
cancellations.
\sindex{singular mass limit}%
However, in~\cite[Appendix~C.1]{PFP} it was explained that the formalism of the continuum
limit is well-behaved in the singular mass limit, thus justifying why we were allowed to describe
the regularized chiral Dirac seas by~\eqref{l:Pnaive}.

\subsectionn{Instability of the Naively Regularized Neutrino Sector} \label{l:sec21}
\sindex{regularization!naive}%
We now give an argument which shows that
if the neutrino sector is regularized in the neutrino sector according to~\eqref{l:Pnaive},
the system~\eqref{l:Pvac} cannot be an absolute minimum of the causal action principle~\eqref{l:actprinciple}.
Suppose conversely that a fermionic projector~$P^\varepsilon$, which in the neutrino sector
is regularized according to~\eqref{l:Pnaive}, is an absolute minimum of the action
principle~\eqref{l:actprinciple}. Then any variation of the fermionic projector can only increase
the action. Evaluating this condition for specific variations leads to the notion of
{\em{state stability}}, which we now recall (for details see~\cite[Section~5.6]{PFP} or~\cite{vacstab}).
\sindex{state stability}%
This notion makes it necessary to assume that our regularization is
{\em{macroscopic away from the light cone}},
\sindex{regularization!macroscopic away from light cone}%
meaning that the difference~$P^\varepsilon(x,y)-P(x,y)$ should be small
pointwise except if the vector~$y-x$ is close to the light cone
(see~\cite[Section~5.6]{PFP}). This condition seems to be fulfilled for any reasonable regularization, and
thus we shall always assume it from now on.
Suppose that the state~$\psi$ is occupied by a particle (i.e.\ that~$\psi$ lies in the image of the
operator~$P^\varepsilon$), whereas the state~$\phi$ is not occupied. We assume that~$\psi$ and~$\phi$
are suitably normalized and negative definite with respect to the indefinite inner product
\beq \label{l:iprod}
\bra \psi | \phi \ket = \int_\scrM \overline{\psi(x)} \phi(x)\: d^4x\:.
\eeq
Then the ansatz
\beq \label{l:delP}
\delta P^\varepsilon(x,y) = \psi(x) \overline{\psi(y)} - \phi(x) \overline{\phi(y)}
\eeq
describes an admissible perturbation of~$P^\varepsilon$.
Since the number of occupied states is very large, $\delta P^\varepsilon$
is a very small perturbation (which even becomes infinitesimally small in the infinite volume limit).
Thus we may consider~$\delta P$ as a first order variation and treat the constraint in~\eqref{l:actprinciple} with a Lagrange multiplier. We point out that the set of possible variations~$\delta P^\varepsilon$ does
not form a vector space, because it is restricted by additional conditions. This is seen most easily from the
fact that~$-\delta P^\varepsilon$ is not an admissible variation, as it does not preserve the rank
of~$P^\varepsilon$. The fact that possible variations~$\delta P^\varepsilon$ are restricted
has the consequence that we merely get the variational {\em{in}}equality
\beq \label{l:ss1}
{\mathcal{S}}_\mu[P^\varepsilon +\delta P^\varepsilon] \geq {\mathcal{S}}_\mu[P^\varepsilon]\:,
\eeq
valid for all admissible variations of the form~\eqref{l:delP}.

Next, we consider variations which are {\em{homogeneous}}, meaning that~$\psi$ and~$\phi$ are
plane waves of momenta~$k$ respectively~$q$,
\beq \label{l:pw}
\psi(x) = \hat{\psi} \: e^{-i k x}\:, \qquad \phi(x) = \hat{\phi} \: e^{-i q x} \:.
\eeq
Then both~$P^\varepsilon$ and the variation~$\delta P$ depend only on the difference
vector~$\xi = y-x$. Thus after carrying out one integral in~\eqref{l:auxact}, we obtain a constant,
so that the second integral diverges. Thinking of the infinite volume limit of a system in finite $4$-volume,
we can remove this divergence simply by omitting the second integral. Then~\eqref{l:ss1} simplifies
to the {\em{state stability condition}}
\beq \label{l:ss2}
\int_\scrM \delta \L_\mu[A(\xi)] \:d^4 \xi \geq 0 \:.
\eeq

In order to analyze state stability for our system~\eqref{l:Pvac}, we first choose the
Lagrange multiplier according to~\eqref{l:muchoice}. Moreover, we assume that~$\psi$ is
a state of the charged sector, whereas~$\phi$ is in the neutrino sector,
\beq \label{l:hPP}
\hat{\psi} = 0 \oplus \hat{\psi}^C \:,\qquad \hat{\phi} = \hat{\phi}^N \oplus 0 \:.
\eeq
Since~$\psi$ should be an occupied state, it must clearly be a solution of one of the Dirac
equations~$(i \Pdd - m_\alpha) \psi=0$ with~$\alpha \in \{1,2,3\}$. The state~$\phi$, on the other hand,
should be unoccupied; we assume for simplicity that its momentum~$q$ is outside the support
of~$P^N_\varepsilon$,
\beq \label{l:qsupp}
q \not \in \text{supp} \,\hat{g}
\eeq
(where~$\hat{g}$ is the Fourier transform of the vector field~$g$ in~\eqref{l:Pnaive}).
Thus our variation removes a state from a Dirac sea in the charged sector and
occupies instead an unoccupied state in the neutrino sector with arbitrary momentum~$q$
(in particular, $\phi$ does not need to satisfy any Dirac equation).
Let us compute the corresponding variation of the Lagrangian. First, using that the spectral weight
is additive on direct sums, we find that
\begin{align*}
\delta \L_\frac{1}{4} &= \delta \Big( |A^2| - \frac{1}{4} |A|^2 \Big)
= \delta |A^2| - \frac{1}{2} \: |A| \: \delta |A| \\
&= \delta \big| (A^C)^2 \big| + \delta \big| (A^N)^2 \big| - \frac{1}{2} \left( |A^C|+|A^N| \right)
\left( \delta |A^C|+ \delta |A^N| \right)\:.
\end{align*}
This formula simplifies if we use that~$A^N$ vanishes due to chiral cancellations. Moreover,
the first order variation of~$(A^N)^2$ vanishes because
\[ \delta \left( (A^N)^2 \right) = (\delta A^N) A^N + A^N (\delta A^N) = 0\:. \]
Finally, $\delta |A^N| = |(A^N + \delta A^N)| - |A^N| = |\delta A^N|$.
This gives
\beq \label{l:delL}
\delta \L_\frac{1}{4} = 
\delta \Big( \big| (A^C)^2 \big| - \frac{1}{4}\: |A^C|^2 \Big)
- \frac{1}{2}\: |A^C|\: |\delta A^N| \:.
\eeq
Note that~$\psi$ only affects the first term, whereas~$\phi$ influences only the second term.
In the first term the neutrino sector does not appear, and thus the state stability analysis for
one sector as carried out in~\cite[Section~5.6]{PFP} and~\cite{vacstab} applies.
From this analysis, we know that the charged sector should be regularized
in compliance with the condition of a distributional ${\mathcal{M}} P$-product
(see also~\cite{reg}). Then the first term in~\eqref{l:delL} leads to a finite variation of our action.
The point is that the second term in~\eqref{l:delL} is {\em{negative}}. In the next lemma we show
that it is even unbounded below, proving that our system indeed violates the state stability
condition~\eqref{l:ss2}.

\begin{Lemma} Suppose that~$P^\varepsilon$ is a regularization of the distribution~\eqref{l:Pvac}
which is macroscopic away from the light cone and which in the neutrino sector is of the
form~\eqref{l:Pnaive}. Then for any constant~$C>0$ there is a properly normalized, negative definite
wave function~$\phi$ satisfying~\eqref{l:pw}, \eqref{l:hPP} and~\eqref{l:qsupp} such that the corresponding
variation of the fermionic projector
\beq \label{l:varPhi}
\delta P^\varepsilon(x,y) = - \phi(x) \overline{\phi(y)}
\eeq
satisfies the inequality
\[ \int_\scrM |A^C|\: |\delta A^N| \, d^4 \xi > C\:. \]
\end{Lemma}
\Proof For convenience, we occupy two fermionic states of the same momentum~$q$ such that
\beq \label{l:2state}
\delta P^N_\varepsilon(x,y) =  (\slashed{p} + m)\: e^{-i q (y-x)} \:,
\eeq
where~$p$ is a vector on the lower
hyperboloid~${\mathcal{H}}_m := \{p \:|\: p^2=m^2  \text{ and } p^0 < 0\}$, and~$m$ is a positive parameter which involves the normalization constant. For this simple ansatz one easily verifies that
the image of~$\delta P^N$ is indeed two-dimensional and negative definite. By occupying the two
states in two separate steps, one can decompose~\eqref{l:2state} into two variations of the
required form~\eqref{l:varPhi}. Therefore, it suffices to prove the lemma for the variation~\eqref{l:2state}.

Using~\eqref{l:Pnaive} and~\eqref{l:2state}, the variation of~$A^N$ is computed to be
\[ \delta A^N = \chi_L \slashed{g}(x,y) (\slashed{p} + m)\: e^{i q \xi} + \chi_R (\slashed{p} + m)\: \slashed{g}(y,x)\:
e^{-i q \xi} \:. \]
To simplify the notation, we omit the arguments~$x$ and~$y$ and write~$g(\xi)=g(x,y)$.
Then~$g$ is a complex vector field with~$\overline{g(\xi)} = g(y,x)$.
Using that our regularization is macroscopic away from the light cone, there clearly is
a set~$\Omega \subset \scrM$ of positive Lebesgue measure such that both the vector field~$g$
and the function~$|A^C|$ are non-zero for all~$\xi \in \Omega$. Then we can
choose a past directed null vector~$\nf$ such that~$\langle \nf, g \rangle$ is non-zero on a set
$\Omega' \subset \Omega$ again of positive measure. We now consider a sequence of
vectors~$p_l \in {\mathcal{H}}_m$ which converge to the ray~$\R^+ \nf$
in the sense that there are coefficients~$c_l$ with
\[ p_l - c_l \,\nf \rightarrow 0 \quad \text{and} \quad c_l \rightarrow \infty\:. \]
Then on~$\Omega'$, the inner product~$\langle p_l , g \rangle$ diverges as~$l \rightarrow \infty$.
A short computation shows that in this limit, the eigenvalues of the matrix~$\delta A_l^N$ also diverge.
Computing these eigenvalues asymptotically, one finds that
\[ |\delta A_l^N| \geq 4 \,|\langle p_l, g \rangle| + \O(l^0) \:. \]
Hence for large~$l$,
\[ \int_\scrM |A^C|\: |\delta A_l^N| \geq
\int_{\Omega'} |A^C|\: |\langle p_l, g \rangle| \;\xrightarrow{l \rightarrow \infty}\; \infty\:, \]
completing the proof.
\QED
It is remarkable that the above argument applies independent of any regularization details.
We learn that regularizing the neutrino sector by a left-handed function~\eqref{l:Pnaive} necessarily
leads to an instability of the vacuum. The only way to avoid this instability is to consider more
general regularizations where~$P^N_\varepsilon$ also involves a right-handed component.

\subsectionn{Regularizing the Vacuum Neutrino Sector -- Introductory Discussion} \label{l:sec22}
We begin by explaining our regularization method for one massless left-handed Dirac sea,
\[ P(x,y) = \chi_L\: P^\text{vac}_0(x,y) \]
(several seas and massive neutrinos will be considered later in this section).
Working with a left-handed Dirac sea is motivated by the fact that right-handed neutrinos
have never been observed in nature. To be precise, this physical observation only tells us that
there should be no right-handed neutrinos in the low-energy regime. However, on the regularization
scale~$\varepsilon^{-1}$, which is at least as large as the Planck energy~$E_P$ and
therefore clearly inaccessible to experiments, there might well be right-handed
neutrinos. Thus it seems physically admissible to regularize~$P$ by
\beq \label{l:Pepsneutrino}
P^\varepsilon(x,y) = \chi_L \,\slashed{g}_L(x,y) + \chi_R \,\slashed{g}_R(x,y)\:,
\eeq
provided that the Fourier transform~$\hat{g}_R(k)$ vanishes if~$|k^0|+|\vec{k}| \ll \varepsilon^{-1}$.

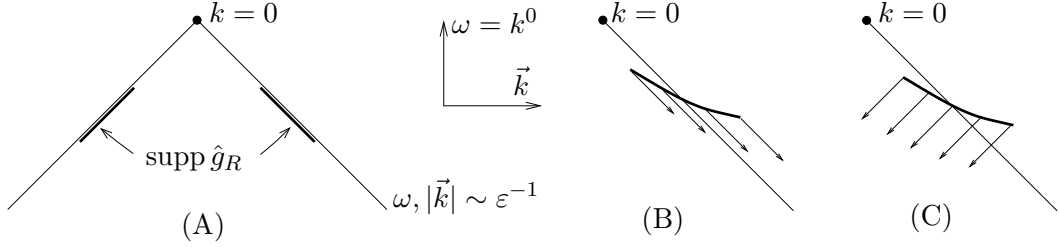
\begin{figure} %
\begin{picture}(0,0)%
\includegraphics{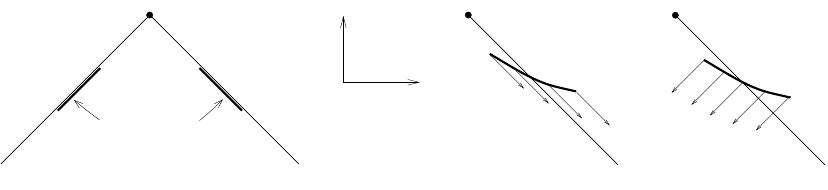}%
\end{picture}%
\setlength{\unitlength}{1492sp}%
\begingroup\makeatletter\ifx\SetFigFontNFSS\undefined%
\gdef\SetFigFontNFSS#1#2#3#4#5{%
  \reset@font\fontsize{#1}{#2pt}%
  \fontfamily{#3}\fontseries{#4}\fontshape{#5}%
  \selectfont}%
\fi\endgroup%
\begin{picture}(17480,4005)(-921,-2857)
\put(1396,-1576){\makebox(0,0)[lb]{\smash{{\SetFigFontNFSS{11}{13.2}{\rmdefault}{\mddefault}{\updefault}$\text{supp}\, \hat{g}_R$}}}}
\put(5506,-2266){\makebox(0,0)[lb]{\smash{{\SetFigFontNFSS{11}{13.2}{\rmdefault}{\mddefault}{\updefault}$\omega, |\vec{k}| \sim \varepsilon^{-1}$}}}}
\put(7501,-406){\makebox(0,0)[lb]{\smash{{\SetFigFontNFSS{11}{13.2}{\rmdefault}{\mddefault}{\updefault}$\vec{k}$}}}}
\put(6481,629){\makebox(0,0)[lb]{\smash{{\SetFigFontNFSS{11}{13.2}{\rmdefault}{\mddefault}{\updefault}$\omega=k^0$}}}}
\put(2446,809){\makebox(0,0)[lb]{\smash{{\SetFigFontNFSS{11}{13.2}{\rmdefault}{\mddefault}{\updefault}$k=0$}}}}
\put(9121,809){\makebox(0,0)[lb]{\smash{{\SetFigFontNFSS{11}{13.2}{\rmdefault}{\mddefault}{\updefault}$k=0$}}}}
\put(13561,809){\makebox(0,0)[lb]{\smash{{\SetFigFontNFSS{11}{13.2}{\rmdefault}{\mddefault}{\updefault}$k=0$}}}}
\put(1981,-2716){\makebox(0,0)[lb]{\smash{{\SetFigFontNFSS{11}{13.2}{\rmdefault}{\mddefault}{\updefault}(A)}}}}
\put(9631,-2626){\makebox(0,0)[lb]{\smash{{\SetFigFontNFSS{11}{13.2}{\rmdefault}{\mddefault}{\updefault}(B)}}}}
\put(14086,-2566){\makebox(0,0)[lb]{\smash{{\SetFigFontNFSS{11}{13.2}{\rmdefault}{\mddefault}{\updefault}(C)}}}}
\end{picture}%
\caption{Plots of~$\hat{g}_R$ exemplifying different regularization mechanisms in the neutrino sector}
\label{l:fig1}
\end{figure}

In order to explain the effect of such a {\em{right-handed high-energy component}}, we
begin with the simplest example where~$\hat{g}_R$ is supported on the lower mass cone,
\beq \label{l:gR0}
\hat{\slashed{g}}_R(k) = 8 \pi^2\, \slashed{k} \:  \hat{h}(\omega)\, \delta(k^2)\:,
\eeq
where~$\omega \equiv k^0$, and the non-negative
function~$\hat{h}$ is supported in the high-energy region~$\omega \sim \varepsilon^{-1}$
(see Figure~\ref{l:fig1}~(A)).
We compute the Fourier integrals by
\sindex{regularization!spherically symmetric}%
\begin{align*}
\slashed{g}_R(\xi) &= 8 \pi^2\int \frac{d^4k}{(2 \pi)^4}\: \slashed{k}\: \hat{h}(\omega) \,\delta(k^2)\: e^{i k \xi} 
= -8 i \pi^2 \,\Pdd_\xi \int \frac{d^4k}{(2 \pi)^4}\: \hat{h}(\omega)\: \delta(k^2)\: e^{i k \xi} \\
&= -2 i \:\Pdd_\xi \int_{-\infty}^0 \frac{d\omega}{2 \pi} \: \hat{h}(\omega)\: e^{i \omega t}
\int_0^\infty p^2\: dp \: \delta(\omega^2 - p^2) \int_{-1}^1 d\cos \vartheta \: e^{-i p r \cos \vartheta} \\
&= 2 \:\Pdd_\xi \left[ \frac{1}{r} \int_{-\infty}^0 \frac{d\omega}{2 \pi} \: \hat{h}(\omega)\: e^{i \omega t}
\int_0^\infty p\: dp \: \delta(\omega^2 - p^2) \big( e^{-i p r} - e^{i p r} \big) \right] \\
&= -\Pdd_\xi \left[ \frac{1}{r} \int_{-\infty}^0 \frac{d\omega}{2 \pi} \: \hat{h}(\omega)\: e^{i \omega t}\,
\big( e^{-i \omega r} - e^{i \omega r} \big) \right] ,
\end{align*}
where we set $t=\xi^0$, $r=|\vec{\xi}|$ and chose polar coordinates~$(p=|\vec{k}|, \vartheta, \varphi)$.
This gives the simple formula
\[ \slashed{g}_R(\xi) = -\Pdd_\xi \: \frac{h(t-r) - h(t+r)}{r}\:, \]
where~$h$ is the one-dimensional Fourier transform of~$\hat{h}$.
Under the natural assumption that the derivatives of~$\hat{h}$ scale in powers of~$\varepsilon$,
the function~$h$ decays rapidly on the regularization scale. Then~$\slashed{g}_R$ vanishes
except if~$\xi$ is close to the light cone, so that the regularization is {\em{again macroscopic
away from the light cone}}.
\sindex{regularization!macroscopic away from light cone}%
But the contribution~\eqref{l:gR0} does affect the singularities on the
light cone, and it is thus of importance in the continuum limit. More specifically, on the upper light
cone away from the origin~$t \approx r \gg \varepsilon$, we obtain the contribution
\beq \label{l:gRpos0} \begin{split}
\slashed{g}_R(\xi) =&\: -\Pdd_\xi \: \frac{h(t-r)}{r} =
-(\gamma^0 - \gamma^r)\: \frac{h'(t-r)}{r} + \gamma^r\: \frac{h(t-r)}{r^2} \\
&+ \text{(rapid decay in~$r$)} \:,
\end{split}
\eeq
where we set~$\gamma^r=(\vec{\xi} \vec{\gamma})/r$. This contribution is compatible
with the formalism of the continuum limit, because it has a similar structure and
the same scaling as corresponding contributions by a regularized Dirac sea  (see~\cite{reg}, where
the same notation and sign conventions are used).

Regularizing the neutrino sector of our fermionic projector~\eqref{l:Pvac} using
a right-handed high-energy component has the consequence that {\em{no chiral cancellations}} occur.
\sindex{chiral cancellation}%
Hence the EL equations become
\beq \label{l:ELreg}
\sum_i \Big( |\lambda_i| - \mu \sum_l |\lambda_l| \Big) \frac{\overline{\lambda_i}}{|\lambda_i|}
\: F_i \,P(x,y) = 0\:,
\eeq
where~$i$ labels the eigenvalues of~$A_{xy}$. For these equations to be satisfied, we must
choose
\beq \label{l:muchoice2}
\mu = \frac{1}{8} \:,
\eeq
and furthermore we must impose that the eigenvalues of~$A_{xy}$ all have the same
absolute values in the sense that
\[ \big( |\lambda_i| - |\lambda_j| \big) \frac{\overline{\lambda_i}}{|\lambda_i|}
\: F_i \,P(x,y) = 0 \qquad \text{for all~$i,j$.} \]
In simple terms, the matrix~$A^N$ must have the {\em{same spectral properties}}
as~$A^C$.

This consideration points to a shortcoming of the regularization~\eqref{l:gR0}. Namely, 
the expression~\eqref{l:gRpos0} does not involve a mass parameter, and thus the corresponding
contribution to the closed chain~$A^N$ cannot have the same spectral properties as~$A^C$,
which has a non-trivial mass expansion. A possible solution to this problem is to
consider states on a more {\em{general hypersurface}}, as we now explain again in the example of
a spherically symmetric regularization. We choose
\sindex{regularization!spherically symmetric}%
\beq \label{l:gR1}
\hat{\slashed{g}}_R(k) = -4 \pi^2\, (\gamma^0 + \gamma^k) \:  \hat{h}(\omega)\, \delta \big(
|\vec{k}|-K(\omega) \big) \:,
\eeq
where~$\gamma^k = \vec{k} \vec{\gamma}/k$, and~$h$ is chosen as in~\eqref{l:gR0}.
We again assume that~$\hat{\slashed{g}}$ is supported in the high-energy region, meaning that
\beq \label{l:highenergy}
\hat{h}(\omega)=0 \qquad \text{if~$|\omega| \ll \varepsilon^{-1}$}\:.
\eeq
Setting~$K=-\omega$, we get back to~\eqref{l:gR0}; but now the function~$K$ gives a more
general dispersion relation (see Figure~\ref{l:fig1}~(B)). Carrying out the Fourier integrals, we obtain
\begin{align*}
g^0_R(\xi) &= -4 \pi^2 \int \frac{d^4k}{(2 \pi)^4}\: 
\hat{h}(\omega)\, \delta \big( |\vec{k}|-K(\omega) \big)\: e^{i k \xi} \\
&= -\int_{-\infty}^0 \frac{d\omega}{2 \pi} \:\hat{h}(\omega)\, e^{i \omega t} \int_0^\infty p^2 dp\:
\delta(p-K(\omega)) \int_{-1}^1 d\cos \vartheta \: e^{-i p r \cos \vartheta} \\
&= -\frac{i}{r} \int_{-\infty}^0 \frac{d\omega}{2 \pi} \:\hat{h}(\omega)\, e^{i \omega t} \int_0^\infty p\, dp\:
\delta(p-K(\omega)) \left( e^{-i p r} - e^{i p r} \right) \\
&= \frac{i}{r} \int_{-\infty}^0 \frac{d\omega}{2 \pi} \:\hat{h}(\omega)\, K(\omega) 
\:e^{i \omega t} \,\Big( e^{i K r} - e^{-i K r} \Big) \\
(\vec{\gamma} \vec{g}_R)(\xi) &= -4 \pi^2\: (i \vec{\gamma} \vec{\nabla})  \int \frac{d^4k}{(2 \pi)^4}\: 
\hat{h}(\omega)\, \delta(k-K(\omega))
\:\frac{1}{|\vec{k}|} \: e^{i k \xi} \\
&= -\vec{\gamma} \vec{\nabla} \left[ \frac{1}{r} \int_{-\infty}^0 \frac{d\omega}{2 \pi}
\:\hat{h}(\omega)\: e^{i \omega t}\, \Big( e^{i K r} - e^{-i K r} \Big) \right] \\
&= -\frac{i \gamma^r}{r} \int_{-\infty}^0 \frac{d\omega}{2 \pi}
\:\hat{h}(\omega)\:K(\omega)\: e^{i \omega t}\, \Big( e^{i K r} + e^{-i K r} \Big) \\
&\quad+ \frac{\gamma^r}{r^2} \int_{-\infty}^0 \frac{d\omega}{2 \pi}
\:\hat{h}(\omega)\: e^{i \omega t}\, \Big( e^{i K r} - e^{-i K r} \Big) \:.
\end{align*}
Evaluating as in~\eqref{l:gRpos0} on the upper light cone away from the origin, we conclude that
\begin{align*}
\slashed{g}_R(\xi) &=\: \int_{-\infty}^0 \frac{d\omega}{2 \pi} \:\hat{h}(\omega)
\left( i \:\frac{\gamma^0-\gamma^r}{r}\: K(\omega)
+ \frac{\gamma^r}{r^2} \right) e^{i (\omega t+ K r)} \\
&\qquad + \text{(rapid decay in~$r$)} \:.
\end{align*}
For ease in notation, from now on we will omit the rapidly decaying error term.
Rearranging the exponentials, we obtain
\[ \slashed{g}_R(\xi) = \int_{-\infty}^0 \frac{d\omega}{2 \pi} \:e^{i (\omega + K) r}\:\hat{h}(\omega)
\left( i \:\frac{\gamma^0-\gamma^r}{r}\: K(\omega)
+ \frac{\gamma^r}{r^2} \right) e^{i \omega (t-r)} \:. \]
Now the {\em{mass expansion}} can be performed by expanding the factor~$\exp(i (\omega + K) r)$,
\sindex{mass expansion}%
\begin{align}
\slashed{g}_R(\xi) &= \sum_{n=0}^\infty \frac{(ir)^n}{n!} \int_{-\infty}^0 \frac{d\omega}{2 \pi} \:\hat{h}(\omega)\,
(\omega+K)^n \left( i \:\frac{\gamma^0-\gamma^r}{r}\: K(\omega)
+ \frac{\gamma^r}{r^2} \right) e^{i \omega (t-r)} \label{l:massexp}  \\
&= \int_{-\infty}^0 \frac{d\omega}{2 \pi} \:\hat{h}(\omega)
\left( \frac{\gamma^r}{r^2} + i \:\frac{K \gamma^0+ \omega \gamma^r}{r} + \cdots \right)
e^{i \omega (t-r)} \:. \nonumber
\end{align}
We conclude that the general ansatz~\eqref{l:gR1} gives rise to a mass expansion
which is similar to that for a massive Dirac sea (see~\cite[Chaper~4]{PFP}
or Section~\ref{seclight}). By modifying
the geometry of the hypersurface~$\{|\vec{k}|=K(\omega)\}$, we have a lot of freedom to modify
the contributions to the mass expansion.
We point out that, in contrast to
the mass expansion for a massive Dirac sea, the mass expansion in~\eqref{l:massexp}
involves {\em{no logarithmic poles}}. This is because here we only consider
high-energy states~\eqref{l:highenergy}, whereas the logarithmic poles are a
consequence of the low-frequency behavior of the massive Dirac seas
(for details see the discussion of the logarithmic mass problem in~\cite[Sections~2.5 and~4.3]{PFP}).

We now come to another regularization effect. The regularizations~\eqref{l:gR0} and~\eqref{l:gR1}
considered so far have the property that~$\slashed{g}_R$ is a multiple of the
matrix~$\chi_L(\gamma^0 + \gamma^k)$, as
is indicated in Figure~\eqref{l:fig1}~(B) by the arrows (to avoid confusion with the signs, we
note that on the lower mass shell, $\slashed{k} = \omega \gamma^0 - \vec{k} \vec{\gamma}
= \omega \,( \gamma^0 + \gamma^k)$). Clearly, we could also have flipped the
sign of~$\gamma^k$, i.e.\ instead of~\eqref{l:gR1},
\beq \label{l:gR2}
\hat{\slashed{g}}_R(k) = -4 \pi^2\, (\gamma^0 - \gamma^k) \:  \hat{h}(\omega)\, \delta \big(
|\vec{k}|-K(\omega) \big)
\eeq
(see Figure~\ref{l:fig1}~(C)). In order to explain the consequence of this sign change in the simplest
possible case, we consider the two functions
\[ \hat{\slashed{g}}_\pm(k) = 8 \pi^2\: \omega\, (\gamma^0 \pm \gamma^k)\:
\hat{h}(\omega)\, \delta(k^2)\:, \]
whose Fourier transforms are given in analogy to~\eqref{l:gRpos0} on the upper light cone by
\beq \label{l:gR2pos}
\slashed{g}_\pm(\xi) = -(\gamma^0 \mp \gamma^r)\: \frac{h'(t-r)}{r} \mp \gamma^r\: \frac{h(t-r)}{r^2} \:.
\eeq
When multiplying~$\slashed{g}_+$ by itself, the identity~$(\gamma^0+\gamma^r)^2=0$ gives
rise to a cancellation. For example, in the expression
\beq \label{l:cancel}
\frac{1}{4}\,\Tr \left( \slashed{g}_+(\xi) \,\slashed{g}_+(\xi)^* \right)
= \frac{2 {\mbox{Re}}(h'(t-r)\, \overline{h(t-r)})}{r^3}\:-\: \frac{|h(t-r)|^2}{r^4}
\eeq
the term~$\sim r^{-2}$ has dropped out. The situation is different if we multiply~$\slashed{g}_+$ by~$\slashed{g}_-$.
For example, in
\beq \label{l:nocancel}
\frac{1}{4}\,\Tr \left( \slashed{g}_+(\xi) \,\slashed{g}_-(\xi)^* \right)
= \frac{2 |h'(t-r)|^2}{r^2} -
 \frac{2 i \,{\mbox{Im}}(h'(t-r)\, \overline{h(t-r)})}{r^3}\:+\: \frac{|h(t-r)|^2}{r^4}
\eeq
no cancellation occurs, so that the term~$\sim r^{-2}$ is present.
From this consideration we learn that by flipping the sign of~$\gamma^r$
as in~\eqref{l:gR2}, we can generate terms in the closed chain which have a different scaling
behavior in the radius.

In order to clarify the last construction, it is helpful to describe the situation in
terms of the general notions introduced in~\cite[Section~4.4]{PFP}. The fact that the leading term in~\eqref{l:gRpos0} is proportional to~$(\gamma^0-\gamma^r)$ can be expressed by saying that the
{\em{vector component is null on the light cone}}. When forming the closed chain, the
term quadratic in the leading terms drops out, implying that~$A_{xy} \sim r^{-3}$.
In momentum space, this situation
corresponds to the fact that the vector~$\hat{g}(k)$ points almost in the
same direction as~$k$. In other words, the {\em{shear of the surface states}} is small.
Thus in~\eqref{l:gR0} and~\eqref{l:gR1} as well as in~$g_+$, the shear is small, implying that the vector
component is null on the light cone, explaining the cancellation of the term~$\sim r^{-2}$
in~\eqref{l:cancel}. The states in~\eqref{l:gR2} and~$g_-$, however, have a large shear. Thus the corresponding
vector component is not null on the light cone, explaining the term~$\sim r^{-2}$ in~\eqref{l:nocancel}.
We point out that states of large shear have never been considered before, as in~\cite{PFP} we
always assumed the shear to be small. For simplicity, we refer to the states in~\eqref{l:gR2}
and~$g_-$ as {\em{shear states}}.
\sindex{shear state}%

We next outline how the above considerations can be adapted to the general an\-s\"atze~\eqref{l:PN}
and~\eqref{l:massneutrino}. In order to describe {\em{several chiral Dirac seas}}, one simply
adds regularized Dirac seas, each of which might involve a right-handed high-energy component
and/or shear states. In other words, in the chiral ansatz~\eqref{l:PN} one replaces each summand
by a Dirac sea regularized as described above.
In the massive ansatz~\eqref{l:massneutrino}, we regularize every massive Dirac sea
exactly as described in~\cite[Chapter~4]{PFP}. Moreover, in order to distinguish the neutrino
sector from a massive sector, we add one or several right-handed high-energy contributions.
In this way, the regularization breaks the chiral symmetry.

We finally make a few remarks which clarify our considerations and bring them
into the context of previous work.
\begin{Remark} {\em{
\begin{itemize}
\item[(1)] We point out that the above assumption of spherical symmetry was merely a technical simplification. But this assumption is not crucial for the arguments, and indeed it will be
relaxed in~\S\ref{l:sec23}. We also point out that in all previous regularizations,
the occupied states formed a hypersurface in momentum space. In this paper, we will
always restrict attention to such {\em{surface states}} (see~\cite[Section~4.3]{PFP}). The underlying
guiding principle is that one should try to build up the regularized fermionic projector with as
few occupied states as possible. This can be understood from the general framework
of causal variational principles as introduced in~\cite{discrete, continuum}. Namely, in this
framework the minimum of the action decreases if the number of particles gets larger\footnote{To be
precise, this results holds for operators in the class~${\mathcal{P}}^f$ 
(see~\cite[Definition~2.7]{discrete}) if the fermionic operator is rescaled such that
its trace is independent of~$f$. In the formulation with local correlation matrices
(see~\cite[Section~3.2]{continuum}) and under the trace constraint, the canonical
embedding~$\C^f \hookrightarrow \C^{f+1}$ allows one to regard a system of~$f$ particles
as a special system of~$f+1$ particles. Since varying within the set of~$f+1$-particle systems gives more
freedom, it is obvious that the action decreases if~$f$ gets larger.}. Thus to construct minimizers, one
should always keep the number of particles fixed.
Conversely, one could also construct minimizers by keeping the action fixed and decreasing the
number of particles. With this in mind, a regularization involving fewer particles
corresponds to a smaller action and is thus preferable. 

\item[(2)]
It is worth mentioning that in all the above regularizations we worked with {\em{null states}},
\sindex{state!null}%
meaning that for every~$k$, the image of the operator~$\hat{P}(k)$ is null with respect to the
spin scalar product. Such null states can be obtained from properly normalized negative definite states
by taking a singular mass limit, similar as worked out in~\cite[Appendix~C]{PFP}.

\item[(3)]
At first sight, our procedure for regularizing might seem very special and ad-hoc. However,
it catches all essential effects of more general regularizations, as we now outline.
First, states of large shear could be used just as well for the regularization of massive Dirac seas,
also in the charged sector. However, our analysis in Section~\ref{l:sec5} will reveal that
the EL equations will only involve the difference in the regularization used in the
charged sector compared to that in the neutrino sector. Thus it is no loss
of generality to regularize the charged sector simply according to~\cite[Chapter~4]{PFP},
and to account for shear states only in the neutrino sector.
Next, in the high-energy region one could also work with {\em{massive states}}.
\sindex{state!massive}%
In order to
break the chiral symmetry, one could project out one spin state with the ansatz
\beq \label{l:rhosea}
\slashed{g}(p) = 
\frac{1}{2}\:(\1 -\rho \slashed{q})\: (\slashed{k}+m) \: \hat{h}(k)
\eeq
with~$p^2=m^2, q^2=-1$ and~$\langle q, k \rangle =0$
(see~\cite[eq.~(C.1.5)]{PFP}, where a corresponding Dirac sea is considered before taking the
singular mass limit).
However, this procedure would have two disadvantages. First, massive states
would yield additional contributions to the fermionic projector, whereas~\eqref{l:rhosea}
even gives rise to bilinear and pseudoscalar contributions, which would all cause technical
complications. Secondly, massive states involve both left- and right-handed components,
which are coupled together in such a way that it would be more difficult to
introduce a general interaction. Apart from these disadvantages,
working with massive states does not seem to lead to any interesting effects.
This is why we decided not to consider them in this paper.

\item[(4)]
We mention that for a fully convincing justification of the vacuum fermionic projector~\eqref{l:Pvac}
and of our regularization method, one should extend the {\em{state stability analysis}} from~\cite{vacstab}
\sindex{state stability}%
to a system of a charged sector and a neutrino sector. Since this analysis only takes into account the
behavior of the fermionic projector away from the light cone, the high-energy behavior
of~$P^\varepsilon$
plays no role, so that one could simply work with the explicit formula for the unregularized
fermionic projector~\eqref{l:Pvac}. Then the methods of~\cite{vacstab} apply to each of the
sectors. However, the two sectors are coupled by the term~$|A|^2$ in the Lagrangian.
The results of this analysis will depend on the value of the Lagrange multiplier~\eqref{l:muchoice2} as well as
on the choice of all lepton masses (including the neutrino masses).
Clearly, the details of this analysis are too involved for predicting results.
For the moment, all one can say is that there is no general counter argument
(in the spirit of~\S\ref{l:sec21}) which might prevent state stability.
\end{itemize}
}} \end{Remark}

\subsectionn{Ruling out the Chiral Neutrino Ansatz} \label{l:secmassive}
In this section, we give an argument which shows that for
chiral neutrinos there is no regularization which gives rise to a stable minimum of the causal
action principle.
\sindex{neutrino!chiral}%
More precisely, we will show that even taking into account the
regularization effects discussed in the previous section, it is impossible to arrange that
the vacuum satisfies the EL equations in the continuum limit~\eqref{l:ELreg} and~\eqref{l:muchoice2}.
Our argument applies in such generality (i.e.\ without any specific assumptions on
the regularization) that it will lead us to drop the ansatz of chiral neutrinos~\eqref{l:PN},
leaving us with the ansatz of massive neutrinos~\eqref{l:massneutrino}.

Considering massive neutrinos is clearly consistent with the experimental observation of neutrino
oscillations. Based on these experimental findings, we could also have
restricted attention to the ansatz~\eqref{l:massneutrino} right away. On the other hand,
considering also chiral neutrinos~\eqref{l:PN} has the advantage that we can conclude that
massive neutrinos are needed even for mathematical consistency.
This conclusion is of particular interest because in the neutrino experiments, the mass of
the neutrinos is observed indirectly from the fact that different generations of neutrinos
are converted into each other. This leaves the possibility that neutrinos might be massless, and
that the neutrino oscillations can be explained instead by modifying the weak interaction.
The following argument rules out this possibility by giving an independent reason why
there must be massive neutrinos.

Recall that the Dirac seas in the charged sector~$P^C$, \eqref{l:PC}, can be written as
\beq \label{l:Ta}
P^\text{vac}_m(x,y) = (i \Pdd_x + m)\: T_{m^2}(x,y) \:,
\eeq
where~$T_{m^2}$ is the Fourier transform of the lower mass shell,
\[ T_{m^2}(x,y) = \int \frac{d^4k}{(2 \pi)^4} \:
\delta(k^2-m^2)\: \Theta(-k^0) \:e^{-ik(x-y)} \: . \]
Computing this Fourier integral and expanding the resulting
Bessel functions gives the expansion in position space
\beq \label{l:logdiv} \begin{split}
T_{m^2}(x,y) &= -\frac{1}{8 \pi^3} \:
\left( \frac{\mbox{PP}}{\xi^2} \:+\: i \pi \delta (\xi^2) \:
\varepsilon (\xi^0) \right) \\
&\qquad +\: \frac{m^2}{32 \pi^3} \left( \log |m^2 \xi^2| + c
+ i \pi \:\Theta(\xi^2) \:\epsilon(\xi^0) \right) + \O(\xi^2 \, \log(\xi^2))\:.
\end{split}
\eeq
(see~\cite[Section~2.5]{PFP} or~\S\ref{s:sec44}). The point for what follows is
that the light-cone expansion of~$P^\text{vac}_m(x,y)$ involves a logarithmic pole~$\sim \log(\xi^2)$.
As a consequence, in the EL equations~\eqref{l:ELreg} we get contributions
to~\eqref{l:ELreg} which involve the logarithm of the radius~$|\vec{x} - \vec{y}|$
(for details see~\S\ref{s:sec51} or the weak evaluation formula~\eqref{l:asy} below).
In order to satisfy the EL equations, these logarithmic contributions in the
charged sector must be compensated by corresponding logarithmic contributions in the neutrino
sector.

Now assume that we consider the chiral neutrino ansatz~\eqref{l:PN}.
Then the light-cone expansion of~$T_N$ does not involve logarithmic poles
(indeed, the distribution~$P^\text{vac}_0$ can be given explicitly in position space by
taking the limit~$m \searrow 0$ in~\eqref{l:Ta} and~\eqref{l:logdiv}).
Thus the logarithmic contributions in the radius must come from the
high-energy component to the fermionic projector. However, as one sees
explicitly from the formulas~\eqref{l:massexp} and~\eqref{l:gR2pos},
the high-energy component is a Laurent series in the radius and does not involve
any logarithms. This explains why with chiral neutrinos alone it is impossible to satisfy
the EL equations.

This problem can also be understood in more general terms as follows.
The logarithmic poles of~$P^\text{vac}_m(x,y)$ are an infrared effect related to the
fact that the square root is not an analytic function
(see the discussion of the so-called logarithmic mass problem in~\cite[Sections~2.5
and~4.5]{PFP}). Thus in order to arrange logarithmic contributions in the
high-energy region, one would have to work with states on a surface
with a singularity. Then the logarithm in the radius would show up in the
next-to leading order on the light cone. Thus in order to compensate
the logarithms in~\eqref{l:logdiv}, the contribution by the high-energy states
would be just as singular on the light cone as the contribution
by the highest pole in~\eqref{l:logdiv}. Apart from the fact that it seems
difficult to construct such high-energy contributions,
such constructions could no longer be regarded as regularizations of
Dirac sea structures. Instead, one would have to put in specific additional
structures ad hoc, in contrast to the concept behind the method of variable
regularization (see~\cite[Section~4.1]{PFP} or Remark~\ref{remmvr}).

The above arguments show that at least one generation
of neutrinos must be massive. In particular, we must give up the ansatz~\eqref{l:PN}
of chiral neutrinos. Instead, we shall always work with massive
neutrinos~\eqref{l:massneutrino}, and we need to assume that at least one
of the masses~$\tilde{m}_\beta$ is non-zero.

For clarity, we finally remark that our arguments also leave the possibility
to choose another ansatz which involves a combination of both chiral and
massive neutrinos, i.e.
\beq \label{l:PNmix}
P^N(x,y) = \sum_{\beta=1}^{\beta_0} \chi_L P^\text{vac}_0(x,y) +
\sum_{\beta=\beta_0+1}^3 P^\text{vac}_{m_\beta}(x,y) \qquad \text{with} \qquad
\beta_0 \in \{1,2\}\:.
\eeq
The only reason why we do not consider this ansatz here is that it seems more natural
to describe all neutrino generations in the same way.
All our methods could be extended in a straightforward way to the
ansatz~\eqref{l:PNmix}.

\subsectionn{A Formalism for the Regularized Vacuum Fermionic Projector} \label{l:sec23}
In the following sections~\S\ref{l:sec23} and~\S\ref{l:sec24}, we
incorporate the regularization effects discussed in~\S\ref{l:sec22}
to the formalism of the continuum limit. Beginning with the vacuum, we recall that
in~\cite[Section~4.5]{PFP} we described the regularization by complex
factors~$T^{(n)}_{[p]}$ and~$T^{(n)}_{\{ p \}}$ (see also~\S\ref{s:sec51}).
The upper index~$n$ tells about the order of the singularity
on the light cone, whereas the lower index keeps track of the orders in a mass expansion.
In~\S\ref{l:sec22}, we considered a chiral decomposition~\eqref{l:Pepsneutrino} and
chose the left- and right-handed components independently. This can be indicated in our formalism
by a chiral index~$c \in \{L, R\}$, which we insert into the subscript. Thus we write the
regularization~\eqref{l:Pepsneutrino} and~\eqref{l:gR0} symbolically as
\[ P^\varepsilon(x,y) = \frac{i}{2} \left( \chi_L \,\slashed{\xi} T^{(-1)}_{[L, 0]} +
\chi_R \,\slashed{\xi} T^{(-1)}_{[R, 0]} \right) . \]
If the regularization effects of the previous section are {\em{not}} used in the left-
or right-handed component, we simply omit the chiral index. Thus if
we work with general surface states or shear states only in the right-handed component,
we leave out the left-handed chiral index,
\[ P^\varepsilon(x,y) = \frac{i}{2} \left( \chi_L \,\slashed{\xi} T^{(-1)}_{[0]} +
\chi_R \,\slashed{\xi} T^{(-1)}_{[R, 0]} \right) . \]
When using the same notation as in the charged sector, we always indicate that we assume the corresponding
regularizations to be compatible. Thus for factors~$T^{(n)}_\circ$ without a chiral index,
we shall use the same calculation rules in the neutrino and in the charged sector. This will also make it
possible to introduce an interaction between these sectors
(for details see~\S\ref{l:sec24} and Appendix~\ref{l:appA}). If we consider a sector of
massive neutrinos~\eqref{l:massneutrino}, we first perform the mass expansion of every
Dirac sea
\sindex{mass expansion}%
\beq \label{l:smass}
P_m^\varepsilon = \frac{i \slashed{\xi}}{2} \sum_{n=0}^\infty \frac{m^{2n}}{n!}\: T^{(-1+n)}_{[2n]}
+ \sum_{n=0}^\infty \frac{m^{2n+1}}{n!}\: T^{(n)}_{[2n+1]}
\eeq
and then add the chiral index to the massless component,
\beq \label{l:neureg1} \begin{split}
P^\varepsilon_m(x,y) &= \frac{i}{2} \left( \chi_L \,\slashed{\xi} T^{(-1)}_{[0]} +
\chi_R \,\slashed{\xi} T^{(-1)}_{[R, 0]} \right) \\
&\quad\; + \frac{i \slashed{\xi}}{2} \sum_{n=1}^\infty \frac{m^{2n}}{n!}\: T^{(-1+n)}_{[2n]}
+ \sum_{n=0}^\infty \frac{m^{2n+1}}{n!}\: T^{(n)}_{[2n+1]} \:.
\end{split} \eeq

Now the regularization effects of the previous section can be incorporated by
introducing more general factors~$T^{(n)}_{[c,p]}$ and~$T^{(n)}_{\{ c,p \} }$
and by imposing suitable computation rules. Before beginning, we point out that the more
general factors should all comply with our weak evaluation rule
\sindex{evaluation on the light cone!weak}%
\beq \label{l:asy}
\int_{|\vec{\xi}|-\varepsilon}^{|\vec{\xi}|+\varepsilon} dt \; \eta(t,\vec{\xi}) \:
\frac{ T^{(a_1)}_\circ \cdots T^{(a_\alpha)}_\circ \:
\overline{T^{(b_1)}_\circ \cdots T^{(b_\beta)}_\circ} }
{ T^{(c_1)}_\circ \cdots T^{(c_\gamma)}_\circ \:
\overline{T^{(d_1)}_\circ \cdots T^{(d_\delta)}_\circ} }
= \eta(|\vec{\xi}|,\vec{\xi}) \:\frac{c_{\reg}}{(i |\vec{\xi}|)^L}
\;\frac{\log^k (\varepsilon |\vec{\xi}|)}{\varepsilon^{L-1}} \:,
\eeq
which holds up to
\beq \label{l:neglect}
\text{(higher orders in~$\varepsilon/\ell_\text{macro}$ and~$\varepsilon/|\vec{\xi}|$)}\:.
\eeq
Here~$L$ is the degree defined by~$\deg T^{(n)}_\circ = 1-n$,
and~$c_{\reg}$ is a so-called {\em{regularization parameter}}
(for details see again~\cite[Section~4.5]{PFP} or~\S\ref{s:sec51}).
\nindex{bn6@$L$ -- degree of simple fraction}%
\sindex{regularization parameter}%
\nindex{bn8@$c_{\text{reg}}$ -- regularization parameter}%
\sindex{degree on the light cone}%
The quotient of products of factors~$T^{(n)}_\circ$ and~$\overline{T^{(n)}_\circ}$
in~\eqref{l:asy} is referred to as a {\em{simple fraction}}.
\sindex{simple fraction}%
In order to take into account the mass expansion~\eqref{l:massexp}, we
replace every factor~$T^{(-1)}_{[c,0]}$ by the formal series
\sindex{mass expansion}%
\beq \label{l:Texp}
\sum_{n=0}^\infty \frac{1}{n!}\: \frac{1}{\delta^{2n}} T^{(-1+n)}_{[c,2n]} \:.
\eeq
This notation has the advantage that it resembles the even part of the standard mass expansion~\eqref{l:smass}.
In order to get the scaling dimensions right, we inserted a factor~$\delta^{-2n}$,
where the parameter~$\delta$ has the dimension of a length.
\nindex{dc2@$\delta$ -- length scale of shear and general surface states}%
The scaling of~$\delta$ will be specified later (see~\eqref{l:deltascale}, 
Section~\ref{l:seccurv} and Section~\ref{l:secgrav}).
For the moment, in order to make sense of the mass expansion, we only need to assume that
the
\beq \label{l:delscale}
\text{length scale } \delta \gg \varepsilon \:.
\eeq
But~$\delta$ could be much smaller than the Compton wave length
of the fermions of the system. It could even be on the same scale as the regularization
length~$\varepsilon$.
We thus replace~\eqref{l:neureg1} by
\nindex{bl6@$T_{[p]}^{(n)}$ -- ultraviolet regularized $T^{(n)}$ }%
\nindex{bl8@$T^{(n)}_{\{p\}}$ -- factor in continuum limit describing the shear of surface states}%
\beq \label{l:neureg2} \begin{split}
P^\varepsilon_m(x,y) &= \chi_L\: \frac{i \slashed{\xi}}{2} \: T^{(-1)}_{[0]}
+ \chi_R\: \frac{i \slashed{\xi}}{2} 
\sum_{n=0}^\infty \frac{1}{n!}\: \frac{1}{\delta^{2n}} T^{(-1+n)}_{[R,2n]} \\
&\quad\; + \frac{i \slashed{\xi}}{2} \sum_{n=1}^\infty \frac{m^{2n}}{n!}\: T^{(-1+n)}_{[2n]}
+ \sum_{n=0}^\infty \frac{m^{2n+1}}{n!}\: T^{(n)}_{[2n+1]} \:.
\end{split} \eeq

The effect of large shear can be incorporated in our {\em{contraction rules}}, as we now explain.
Recall that our usual contraction rules read
\begin{align}
(\slashed{\xi}^{(n)}_{[p]})^j \, (\slashed{\xi}^{(n')}_{[p']})_j &=
\frac{1}{2} \left( z^{(n)}_{[p]} + z^{(n')}_{[p']} \right) + 
\big( \text{higher orders in~$\varepsilon/|\vec{\xi}|$} \big) \label{l:contract} \\
z^{(n)}_{[p]} \,T^{(n)}_{[p]} &= -4 \left( n \:T^{(n+1)}_{[p]}
+ T^{(n+2)}_{\{p \}} \right) \label{l:ocontract}
\end{align}
\nindex{bm2@$\xi^{(n)}_{[p]}$ -- ultraviolet regularized factor $\xi$}%
\nindex{bm8@$z^{(n)}_{[p]}$ -- abbreviation for $(\xi^{(n)}_{[p]})^2$}%
(and similarly for the complex conjugates, cf.~\cite[Section~4.5]{PFP} or~\S\ref{s:sec51}).
We extend the first rule in the obvious way by inserting lower chiral indices.
In the second rule we insert a factor~$\delta^{-2}$,
\sindex{contraction rule}%
\beq \label{l:ncontract}
z^{(n)}_{[c,p]} \,T^{(n)}_{[c,p]} = -4 \left( n \:T^{(n+1)}_{[c,p]}
+ \frac{1}{\delta^2}\: T^{(n+2)}_{\{c, p \}} \right) .
\eeq
The factor~$\delta^{-2}$ has the advantage that it ensures that the factors with square
and curly brackets have the same scaling dimension (as one sees by comparing~\eqref{l:ncontract}
with~\eqref{l:Texp} or~\eqref{l:smass}; we remark that this point was not taken care of in~\cite{PFP}
and Chapter~\ref{sector}, simply because the factors with curly brackets played no role).
The term~$\delta^{-2} T^{(n+2)}_{\{c, p \}}$ can be associated precisely to the
shear states. For example, in the expression
\[ \frac{1}{8}\, \Tr \left( (\slashed{\xi} T^{(-1)}_{[0]})\: (\slashed{\xi} T^{(-1)}_{[R, 0]}) \right)
= T^{(0)}_{[0]} T^{(-1)}_{[R, 0]} + T^{(-1)}_{[0]} T^{(0)}_{[R, 0]}
- T^{(1)}_{\{ 0 \}} T^{(-1)}_{[R, 0]} - \frac{1}{\delta^2}\: T^{(-1)}_{[0]} T^{(1)}_{\{ R, 0 \}} \:, \]
the last summand involves an additional scaling factor of~$r$ and can thus be used to describe
the effect observed in~\eqref{l:nocancel}. Using again~\eqref{l:delscale}, we can reproduce
the scaling of the first summand in~\eqref{l:nocancel}.

In the weak evaluation formula~\eqref{l:asy}, one can integrate by parts.
This gives rise to the following {\em{integration-by-parts rules}}.
\sindex{integration-by-parts rule}%
On the factors~$T^{(n)}_\circ$ we introduce a derivation~$\nabla$ by
\[ \nabla T^{(n)}_\circ = T^{(n-1)}_\circ \:. \]
\nindex{bl4@$\nabla$ -- derivation on the light cone}%
Extending this derivation with the Leibniz and quotient rules, the
integration-by-parts rules states that
\beq \label{l:ipart}
\nabla \left( \frac{ T^{(a_1)}_\circ \cdots T^{(a_\alpha)}_\circ \:
\overline{T^{(b_1)}_\circ \cdots T^{(b_\beta)}_\circ} }
{ T^{(c_1)}_\circ \cdots T^{(c_\gamma)}_\circ \:
\overline{T^{(d_1)}_\circ \cdots T^{(d_\delta)}_\circ} }
\right) = 0 \:.
\eeq
As shown in~\cite[Appendix~E]{PFP}, there are no further relations between the 
factors~$T^{(a)}_\circ$.

We finally point out that the chiral factors~$T^{(n)}_{[c,p]}$ and~$T^{(n)}_{\{ c, p \}}$
were introduced in such a way that the weak evaluation formula~\eqref{l:asy}
remains valid. However, one should keep in mind that these chiral factors do not
have logarithmic singularities on the light cone, which implies that they have no influence
on the power~$k$ in~\eqref{l:asy}. This follows from the fact that the chiral factors
only describe high-energy effects, whereas the logarithmic poles are a consequence
of the low-frequency behavior of the massive Dirac seas (see also the explicit example~\eqref{l:massexp}
and the explanation thereafter).

\sindex{degree on the light cone}%
\subsectionn{Interacting Systems, Regularization of the Light-Cone Expansion} \label{l:sec24}
We now extend the previous formalism such as to include a general interaction; for the derivation see
Appendix~\ref{l:appA}. For simplicity, we restrict attention to the system~\eqref{l:Pvac}
with massive neutrinos~\eqref{l:massneutrino} and a non-trivial regularization of the neutrino
sector by right-handed high-energy states.
But our methods apply to more general systems as well (see Remark~\ref{l:remgen} below).
In preparation, as in~\cite[Section~2.3]{PFP} and~\S\ref{s:sec41} it is helpful to introduce the
{\em{auxiliary fermionic projector}} as the direct sum of all Dirac seas.
\sindex{fermionic projector!auxiliary}%
In order to allow
the interaction to be as general as possible, it is preferable to describe the right-handed
high-energy states by a separate component of the auxiliary fermionic projector.
Thus we set
\nindex{ca8@$P^\text{aux}$ -- auxiliary fermionic projector}%
\beq \label{l:Paux0}
P^\text{aux} = P^N_\text{aux} \oplus P^C_\text{aux}\:,
\eeq
where
\beq \label{l:Paux}
P^N_\text{aux} = \Big( \bigoplus_{\beta=1}^3 P^\text{vac}_{\tilde{m}_\beta} \Big) \oplus 0
\qquad \text{and} \qquad
P^C_\text{aux} = \bigoplus_{\beta=1}^3 P^\text{vac}_{m_\beta} \:.
\eeq
\nindex{dc31@$P^N_\text{aux}$ -- neutrino sector of auxiliary fermionic projector}%
\nindex{dc32@$P^C_\text{aux}$ -- charged component of auxiliary fermionic projector}%
Note that~$P^\text{aux}$ is composed of seven direct summands, four in the neutrino
and three in the charged sector.
As the fourth component of the neutrino sector is reserved for right-handed
high-energy neutrinos (possibly occupying shear or general surface states), the corresponding
component vanishes without regularization~\eqref{l:Paux}.

In order to recover~$P^\text{aux}$ from a solution of the Dirac equation, we introduce the
{\em{chiral asymmetry matrix}}~$X$ by
\sindex{chiral asymmetry matrix}%
\nindex{dc4@$X$ -- chiral asymmetry matrix}%
\beq \label{l:treg1}
X = \left( \1_{\C^3} \oplus \tau_\reg \,\chi_R \right) \oplus \1_{\C^3} \:.
\eeq
Here~$\tau_\reg$ is a dimensionless parameter, which we always assume to take values in the range
\nindex{dc6@$\tau_\reg$ -- dimensionless parameter for high-energy states}%
\[ 0 < \tau_\reg \leq 1 \:. \]
It has two purposes. First, it indicates that the corresponding direct summand involves a non-trivial
regularization. This will be useful below when we derive constraints for the interaction.
Second, it can be used to modify the amplitude of the regularization effects.
In the limit~$\tau_\reg \searrow 0$, the general surface states and shear states are absent,
whereas in the case~$\tau_\reg=1$, they have the same order of magnitude as the regular states.

Next, we introduce the {\em{mass matrix}}~$Y$ by
\sindex{mass matrix}%
\nindex{bh3@$m$ -- parameter used for mass expansion}%
\nindex{bh4@$Y$ -- mass matrix}%
\beq \label{l:Ydef1}
Y = \frac{1}{m}\, \text{diag} \big( \tilde{m}_1, \tilde{m}_2, \tilde{m}_3, 0, m_1, m_2, m_3 \big)
\eeq
(here~$m$ is an arbitrary mass parameter which makes~$Y$ dimensionless and is useful
for the mass expansion; see also~\cite[Section~2.3]{PFP} or~\S\ref{s:sec41}).
In the limiting case~$\tau_\reg \searrow 0$, we can then write~$P^\text{aux}$ as
\beq \label{l:Pauxdef}
P^\text{aux} = X t = t X^* \qquad \text{with} \qquad t := \bigoplus_{\beta=1}^7 P^\text{vac}_{m Y^\beta_\beta} \:.
\eeq
\nindex{dc8@$t$ -- distribution composed of vacuum Dirac seas}%
In the case~$\tau_\reg > 0$, the fourth direct summand will contain additional
states. We here model these states by a massless Dirac sea
(the shear, and general surface states will be obtained later from these
massless Dirac states by building in a non-trivial regularization).
Thus we also use the ansatz~\eqref{l:Pauxdef} in the case~$\tau_\reg>0$.
Since~$t$ is composed of Dirac seas, it is a solution of the Dirac equation
\beq \label{l:Dfree}
(i \Pdd - m Y) \,t = 0 \:.
\eeq

In order to introduce the interaction, we insert an operator~$\B$ into the Dirac equation,
\beq \label{l:Dinteract}
(i \Pdd + \B - m Y) \,\tilde{t} = 0 \:.
\eeq
\nindex{dd0@$\tilde{t}$ -- distribution composed of Dirac seas in the presence of an external potential}%
\nindex{ar4@$\B$ -- external potential}%
\sindex{potential!bosonic}%
\sindex{potential!external}%
Just as explained in~\cite[Section~2.2]{PFP} and~\cite{grotz}, the {\em{causal perturbation theory}}
defines~$\tilde{t}$ in terms of a unique perturbation series (see also Section~\ref{secfpext}).
The {\em{light-cone expansion}}
\sindex{causal perturbation expansion}%
\sindex{light-cone expansion}%
(see~\cite[Section~2.5]{PFP} and the references therein or Section~\ref{seclight}) is a method for analyzing
the singularities of~$\tilde{t}$ near the light cone. This gives a representation of~$\tilde{t}$
of the form
\begin{align}
\tilde{t}(x,y) = & \sum_{n=-1}^\infty
\sum_{k} m^{p_k} 
{\text{(nested bounded line integrals)}} \times  T^{(n)}(x,y) \nonumber \\
&+ \tilde{P}^\lec(x,y) + \tilde{P}^\hec(x,y) \:, \label{l:tlight}
\end{align}
where~$\tilde{P}^\lec(x,y)$ and~$\tilde{P}^\hec(x,y)$ are smooth to every order in perturbation theory.
The remaining problem is to insert the chiral asymmetry matrix~$X$ into the perturbation series
to obtain the auxiliary fermionic projector with interaction~$\tilde{P}^\text{aux}$.
As is shown in Appendix~\ref{l:appA}, the operator~$\tilde{P}^\text{aux}$ can be uniquely defined
in full generality, without any assumptions on~$\B$. However, for the resulting
light-cone expansion to involve only {\em{bounded}} line integrals, we need to assume the
{\em{causality compatibility condition}}
\beq \label{l:ccc}
(i \Pdd + \B - m Y)\, X = X^* \,(i \Pdd + \B - m Y) \qquad \text{for all~$\tau_\reg \in (0,1]$.}
\eeq
\sindex{causality compatibility condition}%
A similar condition is considered in~\cite[Definition~2.3.2]{PFP}. Here the additional
parameter~$\tau_\reg$ entails the further constraint that the right-handed neutrino
states must not interact with the regular sea states. This constraint can be understood from
the fact that gauge fields or gravitational fields should change space-time only on the macroscopic
scale, but they should leave the microscopic space-time structure unchanged. This gives rise
to conditions for the admissible interactions of the high-energy states. As is worked out in
Appendix~\ref{l:appA}, the gauge fields and the gravitational field must not lead to a  ``mixing'' of
the right-handed high-energy states with other states.

Assuming that the causality compatibility condition holds, the auxiliary fermionic
projector of the sea states~$P^\sea$ is obtained similar to~\eqref{l:Pauxdef} by
multiplication with the chiral asymmetry matrix. Incorporating the mass expansion similar
to~\eqref{l:Texp} leads to the following formalism. We multiply the formulas of the light-cone
expansion by~$X$ from the left or by~$X^*$ from the right (which as a consequence of~\eqref{l:ccc}
gives the same result). The regularization is built in by the formal replacements
\begin{align}
m^p \,T^{(n)} &\rightarrow m^p \,T^{(n)}_{[p]}\:, \label{l:frep1} \\
\tau_\reg \,T^{(n)} &\rightarrow \tau_\reg \sum_{k=0}^\infty \frac{1}{k!}\: \frac{1}{\delta^{2k}}
T^{(k+n)}_{[R,2n]} \:. \label{l:frep2}
\end{align}
Next, we introduce particles and anti-particles by occupying additional states or removing
states from the sea, i.e.\
\beq \label{l:particles}
P^\text{aux}(x,y) = P^\sea(x,y)
-\frac{1}{2 \pi} \sum_{k=1}^{\np} \psi_k(x) \overline{\psi_k(y)}
+\frac{1}{2 \pi} \sum_{l=1}^{\na} \phi_l(x) \overline{\phi_l(y)}\:.
\eeq
For the normalization of the particle and anti-particle states we refer to~\cite[Section~2.8]{PFP}
and~\S\ref{s:sec43}.
Finally, we introduce the regularized fermionic projector~$P$ by forming the {\em{sectorial projection}}
(see also~\cite[Section~2.3]{PFP} or~\eqref{s:pt}),
\beq \label{l:partrace0}
(P)^i_j = \sum_{\alpha, \beta} (\tilde{P}^\text{aux})^{(i,\alpha)}_{(j, \beta)} \:,
\eeq
where~$i,j \in \{1,2\}$ is the sector index, whereas the indices~$\alpha$ and~$\beta$ run over
the corresponding generations (i.e., $\alpha \in \{1, \ldots 4\}$ if~$i=1$ and~$\alpha \in \{1, 2, 3 \}$
if~$i=2$). We again indicate the sectorial projection of the mass matrices by accents
\sindex{sectorial projection}%
\nindex{cb6@$\hat{\;},\: \acute{\;} \ldots \grave{\;}$ -- short notation for sectorial projection}%
(see~\cite[Section~7.1]{PFP} or~\eqref{s:tildedef}),
\beq \label{l:accents}
\hat{Y} = \sum_{\alpha} Y^\alpha_\alpha\:, \qquad
\acute{Y} Y \cdots \grave{Y} = \sum_{\alpha,\beta, \gamma_1, \ldots, \gamma_{p-1}}
Y^\alpha_{\gamma_1} \cdots Y^{\gamma_1}_{\gamma_2}
\cdots Y^{\gamma_{p-1}}_\beta .
\eeq

\begin{Remark} (Regularizing general systems with interaction) \label{l:remgen} {\em{
We now outline how the above construction fits into a general framework for describing
interacting fermion system with chiral asymmetry. Suppose we consider a system which in
the vacuum is composed of a direct sum of sums of Dirac seas, some of which involve
non-trivial regularizations composed of right- or left-handed high-energy shear or general surface states.
Then the interaction can be introduced as follows: To obtain the auxiliary fermionic projector,
we replace the sums by direct sums. For each Dirac sea which should involve a non-trivial
regularization, we add a direct summand involving a left- or right-handed massless Dirac sea.
After reordering the direct summands, we thus obtain
\beq \label{l:dirsum2}
P^\text{aux} = \Big( \bigoplus_{\ell=1}^{\ell_1} P^\text{vac}_{m_j} \Big) \oplus \Big( \bigoplus_{\ell
=\ell_1+1}^{\ell_2} \chi_L P^\text{vac}_0 \Big) \oplus \Big( \bigoplus_{\ell=\ell_2+1}^{\ell_\text{max}}
\chi_R P^\text{vac}_0 \Big)
\eeq
with parameters~$1 \leq \ell_1 \leq \ell_2 \leq \ell_\text{max}$. In order to keep track of which
direct summand belongs to which sector, we form a partition~$L_1, \ldots, L_N$
of~$\{1, \ldots, \ell_\text{max} \}$ such that~$L_i$ contains all the seas in the~$i^\text{th}$
sector. Then the fermionic projector of the vacuum is obtained by forming the sectorial projection as follows,
\beq \label{l:partrace}
P^i_j = \sum_{\alpha \in L_i} \,\sum_{\beta \in L_j}\, (P^\text{aux})^\alpha_\beta \:,\qquad
i,j=1, \ldots, N\:.
\eeq

The next step is to specify the intended form of the regularization by
parameters~$\tau^\reg_1, \ldots, \tau^\reg_p$ with~$p \in \N_0$.
The rule is that to every left- or right-handed massless Dirac sea which
corresponds to a non-trivial regularization we associate a parameter~$\tau^\reg_k$.
Regularizations which we consider to be identical are associated the same parameter;
for different regularizations we take different parameters. Introducing the chiral asymmetry matrix~$X$,
the mass matrix~$Y$, and the distribution~$t$ by
\begin{align}
m Y &= \left( m_1, \ldots, m_{\ell_1} \right) \oplus \left(0, \ldots, 0 \right) \oplus \left(0, \ldots, 0 \right)
\label{l:Ydef2} \\
X &= \left( 1, \ldots, 1 \right) \oplus \chi_L \!\left(1, \ldots, 1, \tau^\reg_{k_1},
\ldots \tau^\reg_{k_a} \right) \oplus \chi_R \!\left(1, \ldots, 1,
\tau^\reg_{k_{a+1}}, \ldots \tau^\reg_{k_b} \right) \label{l:treg2} \\
t &=  \Big( \bigoplus_{\ell=1}^{\ell_1} P^\text{vac}_{m_\ell} \Big) \oplus \Big( \bigoplus_{\ell
=\ell_1+1}^{\ell_2} P^\text{vac}_0 \Big) \oplus \Big( \bigoplus_{\ell=\ell_2+1}^{\ell_\text{max}} P^\text{vac}_0 \Big) ,
\end{align}
the interaction can again be described by inserting an operator~$\B$ into the Dirac
equation~\eqref{l:Dinteract}. Now the causality compatibility condition~\eqref{l:ccc} must hold
for all values of the regularization parameters~$\tau^\reg_1, \ldots, \tau^\reg_k$,
thus allowing for an interaction only between seas with identical regularization.
Using the causal perturbation expansion and the light-cone expansion, we can again
represent~$\tilde{t}$ in the form~\eqref{l:tlight}. The regularization is again introduced
by setting~$P^\sea = t X^*$ and applying the replacement rules~\eqref{l:frep1} as well as
\[ \chi_{L\!/\!R} \:\tau_j^\reg \,T^{(n)} \rightarrow  \chi_{L\!/\!R}\: \tau_j^\reg
\sum_{k=0}^\infty \frac{1}{k!}\: \frac{1}{\delta^{2k}} T^{(k+n)}_{[R\!/\!L,2n, j]} \:, \]
\sindex{regularization!of the light-cone expansion}%
where the additional index~$j$ in the subscript~$[R\!/\!L,2n, j]$
indicates that the factors~$T^{(n)}_\circ$ corresponding to different parameters~$\tau_j$ must
be treated as different functions. This means that the basic fractions formed of these functions
are all linearly independent in the sense made precise in~\cite[Appendix~E]{PFP}. Finally, we
introduce particles and anti-particles again by~\eqref{l:particles} and obtain the fermionic projector
by forming the sectorial projection~\eqref{l:partrace}.
\QEDrem }}
\end{Remark}

\subsectionn{The $\iota$-Formalism} \label{l:seciota}
\sindex{$\iota$-formalism}%
In the formalism of the continuum limit reviewed in~\S\ref{l:sec23}, the regularization is
described in terms of contraction rules.
While this formulation is most convenient for most computations, it has the disadvantage
that the effect of the regularization on the inner factors~$\slashed{\xi}^{(n)}_\circ$ is not explicit.
The $\iota$-formalism remedies this shortcoming by providing more detailed formulas for the
regularized fermionic projector in position space.
The formalism will be used in~\S\ref{l:sec32}, \S\ref{l:seccurv} and~\S\ref{l:secfield}.
It will also be important for the derivation of the Einstein equations in Section~\ref{l:secgrav}.
Here we introduce the formalism and illustrate its usefulness in simple examples.

We begin for clarity with one Dirac sea in the charged sector. Then the mass expansion gives
(cf.~\eqref{l:smass}; see also~\cite[Section~4.5]{PFP})
\sindex{mass expansion}%
\[ P_m^\varepsilon = \frac{i}{2} \sum_{n=0}^\infty \frac{m^{2n}}{n!}\: \slashed{\xi}\, T^{(-1+n)}_{[2n]}
+ \sum_{n=0}^\infty \frac{m^{2n+1}}{n!}\: T^{(n)}_{[2n+1]} \:. \]
We choose a vector~$\check{\xi}$ which is real-valued, lightlike and approximates~$\xi$, i.e.
\nindex{dd2@$\check{\xi}$ -- real lightlike vector in $\iota$-formalism}%
\beq \label{l:xirel}
\check{\xi}^2 = 0 \:,\qquad \overline{\check{\xi}} = \check{\xi} \qquad \text{and} \qquad
\check{\xi} = \xi + \text{(higher orders in~$\varepsilon/|\vec{\xi}|$)}\:.
\eeq
Replacing all factors~$\xi$ in~$P_m^\varepsilon$ by~$\check{\xi}$, we obtain
the function~$\check{P}^\varepsilon_m$,
\[ \check{P}_m^\varepsilon := \frac{i}{2} \sum_{n=0}^\infty \frac{m^{2n}}{n!}\:  \check{\slashed{\xi}}\, T^{(-1+n)}_{[2n]}
+ \sum_{n=0}^\infty \frac{m^{2n+1}}{n!}\: T^{(n)}_{[2n+1]} \:. \]
\nindex{dd4@$\check{P}_m^\varepsilon$ -- lightlike component of vacuum fer\-mio\-nic projector in $\iota$-formalism}%
Clearly, this function differs from~$P^\varepsilon_m$ by vectorial contributions.
We now want to determine these additional contributions by using that the
contraction rules~\eqref{l:contract} and~\eqref{l:ocontract} hold.
It is most convenient to denote the involved vectors by~$\iota^{(n)}_{[p]}$, which we
always normalize such that
\nindex{dd6@$\iota^{(n)}_\circ$ -- vector describing regularization in $\iota$-formalism}%
\beq \label{l:iotac}
\langle \check{\xi}, \iota^{(n)}_\circ \rangle = 1 \:.
\eeq
Then the contraction rules~\eqref{l:contract} and~\eqref{l:ocontract}
are satisfied by the ansatz
\beq \label{l:Piota}
P_m^\varepsilon = \check{P}_m^\varepsilon
-i \sum_{n=0}^\infty \frac{m^{2n}}{n!}\:  \iotaslsh^{(-1+n)}_{[2n]}
\left( (n-1) \,T^{(n)}_{[2n]} + T^{(n+1)}_{\{2n\}} \right) \:,
\eeq
as is verified by a straightforward calculation.
To explain the essence of this computation, let us consider only the leading contribution in the mass expansion,
\beq \label{l:Pexa}
P^\varepsilon = \frac{i}{2}\: \check{\slashed{\xi}} T^{(-1)}_{[0]} + i \iotaslsh^{(-1)}_{[0]}\: T^{(0)}_{[0]} 
+ (\deg<-1) + \O(m) \:.
\eeq
Taking the square, we obtain
\begin{align*}
(P^\varepsilon)^2 &= -\langle \check{\xi}, \iota^{(-1)}_{[0]} \rangle\: T^{(-1)}_{[0]} T^{(0)}_{[0]}
- \langle \iota^{(-1)}_{[0]}, \iota^{(-1)}_{[0]} \rangle\: T^{(0)}_{[0]} T^{(0)}_{[0]} + (\deg<-2) + \O(m) \\
&= - T^{(-1)}_{[0]} T^{(0)}_{[0]} - \langle \iota^{(-1)}_{[0]}, \iota^{(-1)}_{[0]} \rangle\: T^{(0)}_{[0]} T^{(0)}_{[0]} 
+ (\deg<-2) + \O(m) \:.
\end{align*}
The first summand reproduces the contraction rules~\eqref{l:contract} and~\eqref{l:ocontract}.
Compared to this first summand, the second summand is of higher order in~$\varepsilon/ |\vec{\xi}|$.
It is thus omitted in the formalism of the continuum limit, where only the leading
contribution in~$\varepsilon/|\vec{\xi}|$ is taken into account (for details see~\cite[Chapter~4]{PFP}).
More generally, when forming composite expressions of~\eqref{l:Piota} in the formalism
of the continuum limit, only the mixed products~$\la \xi^{(n)}_\circ, \iota^{(n')}_\circ \ra$
need to be taken into account, whereas the products~$\la \iota^{(n)}_\circ, \iota^{(n')}_\circ \ra$
involving two factors~$\iota^{(\cdot)}_\circ$ may be disregarded.
With this in mind, one easily sees that the ansatz~\eqref{l:Piota} indeed incorporates the contraction
rules~\eqref{l:contract} and~\eqref{l:ocontract}.
Concerning the uniqueness of the representation~\eqref{l:Piota},
there is clearly the freedom to change the vectors~$\iota^{(n)}_\circ$,
as long as the relations~\eqref{l:iotac} are respected. 
Apart from this obvious arbitrariness, the representation~\eqref{l:Piota}
is unique up to contributions of higher order in~$\varepsilon/|\vec{\xi}|$,
which can be neglected in a weak evaluation on the light cone.

In order to extend the above formalism to include the regularization effects in the neutrino sector,
we define~$\check{P}^\varepsilon_m$
by replacing all factors~$\xi$ in~\eqref{l:neureg2} by~$\check{\xi}$. Writing
\begin{align}
P^\varepsilon_m(x,y) = \check{P}_m^\varepsilon &- i \chi_L \iotaslsh^{(-1)}_{[0]}
\left( - T^{(0)}_{[0]} + T^{(1)}_{\{0 \}} \right) \label{l:iota1} \\
&-i \chi_R \sum_{n=0}^\infty \frac{1}{n!}\: \frac{1}{\delta^{2n}}\: \iotaslsh^{(-1+n)}_{[2n]}
\left( (n-1) \,T^{(n)}_{[R, 2n]} + \frac{1}{\delta^2}\: T^{(n+1)}_{\{R, 2n\}} \right) \label{l:iota2} \\
&-i \sum_{n=1}^\infty \frac{m^{2n}}{n!}\:  \iotaslsh^{(-1+n)}_{[2n]}
\left( (n-1) \,T^{(n)}_{[2n]} + T^{(n+1)}_{\{2n\}} \right) , \label{l:iota3}
\end{align}
a direct calculation shows that the contraction rules~\eqref{l:contract},
\eqref{l:ocontract} and~\eqref{l:ncontract} are indeed respected.

Clearly, the $\iota$-formalism is equivalent to the standard formalism of~\S\ref{l:sec23}.
However, it makes some computations more transparent, as we now explain.
For simplicity, we again consider the leading order in the mass expansion~\eqref{l:Pexa}
and omit all correction terms, i.e.
\beq \label{l:Pepsiota} \begin{split}
P^\varepsilon(x,y) &= \frac{i}{2}\: \check{\slashed{\xi}} T^{(-1)}_{[0]} + i \iotaslsh^{(-1)}_{[0]}\: T^{(0)}_{[0]} \\
P^\varepsilon(y,x) &= P^\varepsilon(x,y)^* = -\frac{i}{2}\: \check{\slashed{\xi}} \overline{T^{(-1)}_{[0]}}
- i \overline{\iotaslsh^{(-1)}_{[0]}\: T^{(0)}_{[0]}} \:.
\end{split}
\eeq
Suppose we want to compute the eigenvalues of the closed chain.
As we already saw in the example~\eqref{l:Pexa}, contractions between
two factors~$\iota^{(n)}_\circ$ are of higher order in~$\varepsilon/|\vec{\xi}|$. Thus,
in view of the relations~\eqref{l:xirel}, it suffices to take into account the mixed terms, i.e.
\beq \label{l:Axyiota}
A_{xy} = \frac{1}{2}\: \iotaslsh^{(-1)}_{[0]} \check{\slashed{\xi}}\: T^{(0)}_{[0]} \overline{T^{(-1)}_{[0]}}
+ \frac{1}{2}\: \check{\slashed{\xi}}\, \overline{\iotaslsh^{(-1)}_{[0]}} \;T^{(-1)}_{[0]}
\overline{T^{(0)}_{[0]}} + \text{(higher orders in~$\varepsilon/|\vec{\xi}|$)}\:.
\eeq
When taking powers of~$A_{xy}$, any product of the first summand in~\eqref{l:Axyiota} with
the second summand in~\eqref{l:Axyiota} vanishes, because we get two adjacent factors~$\check{\slashed{\xi}}$.
Similarly, we also get zero when the second summand is multiplied by the first summand,
because in this case we get two adjacent factors~$\iotaslsh$. We thus obtain
\beq \label{l:Axyp}
(A_{xy})^p =  \Big( \frac{1}{2}\: \iotaslsh^{(-1)}_{[0]}
\check{\slashed{\xi}}\: T^{(0)}_{[0]} \overline{T^{(-1)}_{[0]}} \Big)^p
+ \Big( \frac{1}{2}\: \check{\slashed{\xi}}\, \overline{\iotaslsh^{(-1)}_{[0]}} \;T^{(-1)}_{[0]}
\overline{T^{(0)}_{[0]}} \Big)^p ,
\eeq
where we again omitted the higher orders in~$\varepsilon/|\vec{\xi}|$.
Moreover, powers of products of~$\slashed{\xi}$ and~$\iotaslsh$ can be
simplified using the anti-commutation relations; for example,
\[ \Big( \check{\slashed{\xi}}\, \overline{\iotaslsh^{(-1)}_{[0]}} \Big)^2 = 
\check{\slashed{\xi}} \,\overline{\iotaslsh^{(-1)}_{[0]}} \check{\slashed{\xi}} \overline{\iotaslsh^{(-1)}_{[0]}}
= 2 \,\check{\slashed{\xi}} \:\big\la \overline{\iota^{(-1)}_{[0]}}, \check{\xi}
\big\ra \:\overline{\iotaslsh^{(-1)}_{[0]}}\:, \]
and applying~\eqref{l:iotac} together with the fact that~$\check{\xi}$ is real, we obtain
\[ \Big( \check{\slashed{\xi}}\, \overline{\iotaslsh^{(-1)}_{[0]}} \Big)^2 = 2 \,\check{\slashed{\xi}}\,
\overline{\iotaslsh^{(-1)}_{[0]}} \:. \]
This shows that the Dirac matrices in~\eqref{l:Axyp} in the first and second summand in~\eqref{l:Axyp}
both have the eigenvalues two and zero.
From this fact we can immediately read off the eigenvalues of~\eqref{l:Axyiota} to be
\[ \lambda_+ = T^{(0)}_{[0]} \overline{T^{(-1)}_{[0]}} \qquad \text{and} \qquad
\lambda_- = T^{(-1)}_{[0]} \overline{T^{(0)}_{[0]}}\:. \]
Clearly, these formulas were obtained earlier in the usual formalism (for details see~\cite[Sections~5.3 and~6.1]{PFP}
or~\S\ref{s:sec71}). But the above consideration gives a more direct understanding
for how these formulas come about.

Another advantage is that it becomes clearer how different contributions to the fermionic projector
influence the eigenvalues. We explain this in the example of a left-handed contribution of the form
\beq \label{l:chiralex}
P(x,y) \asymp \chi_L \slashed{u} \:.
\eeq
The corresponding contribution to the left-handed component of the closed chain is given by
\[ \chi_L A_{xy} \asymp \chi_L \slashed{u} \: P^\varepsilon(y,x) \:. \]
If we substitute~$P^\varepsilon(y,x)$ according to~\eqref{l:Pepsiota},
the factor~$\iota$ will be contracted in any composite expression
either with~$u$ or with another factor~$\iota$. In both cases, we get contributions
of higher order in~$\varepsilon/|\vec{\xi}|$. Hence we can disregard the factor~$\iota$,
\[ \chi_L A_{xy} \asymp
-\frac{i}{2}\: \chi_L \: \slashed{u} \:\check{\slashed{\xi}} \:\overline{T^{(-1)}_{[0]}} \:. \]
When multiplying with~\eqref{l:Axyiota}, the product with the second summand vanishes.
Even more, using the anti-commutation relations, one finds that
\[ (A_{xy})^p \:\slashed{u} \:\check{\slashed{\xi}}\: (A_{xy})^q = 
\la u, \check{\xi} \ra \:\Big( \frac{1}{2}\: \iotaslsh^{(-1)}_{[0]}
\check{\slashed{\xi}}\: T^{(0)}_{[0]} \overline{T^{(-1)}_{[0]}} \Big)^{p+q} \:. \]
This implies that only the eigenvalue~$\lambda_{L+}$ is influenced; more precisely,
\[ \lambda_{L+} \asymp -\frac{i}{2}\: u_j \xi^j \:\overline{T^{(-1)}_{[0]}}
\qquad \text{and} \qquad \lambda_{L-} \asymp 0 \:. \]
Of course, this result is consistent with earlier computations
(see for example the proof of Lemma~\ref{s:lemmalc2} in Appendix~\ref{s:appspec}).

\section{The Euler-Lagrange Equations to Degree Five} \label{l:sec3}
Before entering the analysis of the EL equations, we briefly recall the basics.
Counting with algebraic multiplicities,
the closed chain~$A_{xy}$ has eight eigenvalues, which we denote by~$\lambda^{xy}_{ncs}$,
where~$n \in \{1,2\}$, $c \in \{L,R\}$ and~$s \in \{+,-\}$.
\nindex{dd8@$\lambda^{xy}_{ncs}, F^{xy}_{ncs}$ -- eigenvalues and corresponding spectral projectors
of closed chain}%
The corresponding spectral projectors are denoted by~$F^{xy}_{ncs}$.
In case of degeneracies, we usually omit the lower indices on which the eigenvalues do not depend.
For example, in the case of the four-fold degeneracy $\lambda_{1L+}=\lambda_{2L+}=\lambda_{1R+}
=\lambda_{2R+}$, we simply denote the corresponding eigenvalue by~$\lambda_+$ and the
spectral projector onto the four-dimensional eigenspace by~$F_+$.

The considerations in the previous section led us to choosing the Lagrange multiplier~$\mu=\frac{1}{8}$
(see~\eqref{l:muchoice2}), and thus a minimizer~$P$ is a critical point of the auxiliary action
\[ \Sact[P] = \iint_{\scrM \times \scrM} \L[A_{xy}]\: d^4x\, d^4y \]
with~$\L$ according to~\eqref{l:Ldef},
\[ \L[A_{xy}] = \sum_{n,c,s} |\lambda^{xy}_{ncs}|^2 - \frac{1}{8}
\left( \sum_{n,c,s} |\lambda^{xy}_{ncs}| \right)^2 
= \frac{1}{16}  \sum_{n,c,s}  \;\sum_{n',c',s'} \Big( |\lambda^{xy}_{ncs}|
- |\lambda^{xy}_{n'c's'}|\Big)^2 . \]
Considering first order variations of~$P$, one gets the EL equations
(see~\cite[Section~3.5]{PFP} or for more details~\eqref{s:ELeqns})
\beq \label{l:ELcommutator}
[P,Q] = 0 \:,
\eeq
where the operator~$Q$ has the integral kernel (see~\cite[Sections~3.5 and~5.4]{PFP})
\begin{align}
Q(x,y) &= \frac{1}{2} \sum_{ncs} \frac{\partial \L}{\partial \lambda^{xy}_{ncs}}
\:F_{ncs}^{xy}\: P(x,y) \nonumber \\
&= \sum_{n,c,s} \bigg[  |\lambda^{xy}_{ncs}| - \frac{1}{8} \sum_{n',c',s'}
|\lambda^{xy}_{n'c's'}| \bigg] \: \frac{\overline{\lambda^{xy}_{ncs}}}{|\lambda^{xy}_{ncs}|}
\: F_{ncs}^{xy}\: P(x,y) \:. \label{l:Qform}
\end{align}
\nindex{aq6@$Q(x,y)$ -- first variation of the Lagrangian}%
By testing on null lines (see~\S\ref{s:secELC} and Appendix~\ref{s:appnull}), one sees that the
commutator~\eqref{l:ELcommutator} vanishes if and only if~$Q$ itself is zero.
We thus obtain the EL equations in the continuum limit
\beq \label{l:EL}
\boxed{ \quad Q(x,y) = 0 \quad \text{if evaluated weakly on the light cone}\:. \quad }
\eeq
By relating the spectral decomposition of~$A_{xy}$ to that
of~$A_{yx}$ (see~\cite[Lemma~3.5.1]{PFP}), one sees that the operator~$Q$ is symmetric, meaning that
\beq \label{l:Qsymm}
Q(x,y)^* = Q(y,x)\:.
\eeq

As in Chapter~\ref{sector} we shall analyze the EL equations~\eqref{l:EL}
degree by degree on the light cone.
In this section, we consider the leading degree five, both in the vacuum
and in the presence of gauge potentials.
In Section~\ref{l:sec4} we then consider the next degree four.

\subsectionn{The Vacuum} \label{l:sec31} $\;\;\;$
Applying the formalism of~\S\ref{l:sec23} and~\S\ref{l:sec24} to
the an\-satz~\eqref{l:Pvac}, \eqref{l:PC} and~\eqref{l:massneutrino} and forming the sectorial projection,
we obtain according to~\eqref{l:Paux0} and~\eqref{l:Paux} for the vacuum fermionic
projector the expression
\beq \label{l:Pfreeex}
P(x,y) = \frac{i}{2} \begin{pmatrix}
3 \,\slashed{\xi} T^{(-1)}_{[0]} +
\chi_R \,\tau_\reg \,\slashed{\xi} T^{(-1)}_{[R, 0]} & 0 \\
0 & 3 \,\slashed{\xi} T^{(-1)}_{[0]} \end{pmatrix} + (\deg < 2) \:,
\eeq
where we used a matrix notation in the isospin index. Thus
\begin{align*}
\chi_L A_{xy} &= 
\frac{3}{4} \: \chi_L \begin{pmatrix}
3 \,\slashed{\xi} T^{(-1)}_{[0]} \overline{\slashed{\xi} T^{(-1)}_{[0]}}
+ \tau_\reg \:\slashed{\xi} T^{(-1)}_{[0]} \overline{\slashed{\xi} T^{(-1)}_{[R, 0]}} & 0 \\
0 & 3 \,\slashed{\xi} T^{(-1)}_{[0]} \overline{\slashed{\xi} T^{(-1)}_{[0]}} \end{pmatrix} \\
&\quad + \slashed{\xi}\, (\deg < 3) + (\deg < 2) \:,
\end{align*}
and the right-handed component is obtained by taking the adjoint.
The eigenvalues can be computed in the charged and neutrino sectors
exactly as in~\S\ref{s:sec71} to obtain
\beq
\lambda_{2+L} = \lambda_{2+R} = 
\overline{\lambda_{2-R}} = \overline{\lambda_{2-L}} = 9\, T^{(0)}_{[0]} \overline{T^{(-1)}_{[0]}} + (\deg<3)
\label{l:lamM}
\eeq
and
\begin{align*}
\lambda_{1+L} &= \overline{\lambda_{1-R}}
= 3\, T^{(0)}_{[0]} \Big( 3 \,\overline{T^{(-1)}_{[0]}} + \tau_\reg\, \overline{T^{(-1)}_{[R,0]}} \Big) + (\deg<3) \\
\lambda_{1+R} &= \overline{\lambda_{1-L}}
= \Big( 3 \,T^{(0)}_{[0]} +  \tau_\reg\, T^{(0)}_{[R,0]} \Big)\, 3\, \overline{T^{(-1)}_{[0]}} + (\deg<3) \:.
\end{align*}
The corresponding spectral projectors can be computed
exactly as in~\cite[Sections~5.3 and~6.1]{PFP} or~Section~\ref{s:sec7} to
\beq \label{l:Fpm1}
F_{1cs} = \begin{pmatrix} \chi_c F_s & 0 \\ 0 & 0 \end{pmatrix} \:,\qquad
F_{2cs} = \begin{pmatrix} 0 & 0 \\ 0 & \chi_c F_s \end{pmatrix} ,
\eeq
where~$F_\pm$ are given by
\beq \label{l:Fpm2}
F_\pm := \frac{1}{2} \Big( \1 \pm \frac{[\slashed{\xi}, \overline{\slashed{\xi}}]}{z-\overline{z}} \Big)
+ \slashed{\xi} (\deg \leq 0) + (\deg < 0)\:.
\eeq
Here the omitted indices of the factors~$\xi$, $z$ and their complex conjugates
are to be chosen in accordance with the corresponding factors~$T^{(-1)}_\circ$
and~$\overline{T^{(-1)}_\circ}$, respectively. In the charged sector, this simply 
amounts to adding indices~$^{(-1)}_{[0]}$ to all such factors.
In the neutrino sector, however, one must keep in mind the contributions involving~$\tau_\reg$,
making it necessary to keep track of the factors~$T^{(n)}_{[R, \circ]}$. More precisely, setting
\beq \label{l:Lndef}
L^{(n)}_{[p]} = T^{(n)}_{[p]} + \frac{1}{3}\:\tau_\reg\, T^{(n)}_{[R,p]}\:,
\eeq
\nindex{de0@$L^{(n)}_{[p]} = T^{(n)}_{[p]} + \tau_\reg T^{(n)}_{[R,p]}/3$}%
we obtain
\begin{align*}
2 \,\chi_R F_\pm
&= \1 \pm \frac{1}{4 L^{(0)}_{[0]} - L^{(-1)}_{[0]} \,\overline{z^{(-1)}_{[0]}}}
\left[ \slashed{\xi}^{(-1)}_{[0]} T^{(-1)}_{[0]} + \frac{1}{3}\,
\tau_\reg\, \slashed{\xi}^{(-1)}_{[R,0]} T^{(-1)}_{[R,0]},\: \overline{\slashed{\xi}^{(-1)}_{[0]}} \right] \\
2 \,\chi_L F_\pm
&= \1 \pm \frac{1}{z^{(-1)}_{[0]}\: \overline{L^{(-1)}_{[0]}} - 4 \overline{L^{(0)}_{[0]}}}
\left[ \slashed{\xi}^{(-1)}_{[0]},\: \overline{3\, \slashed{\xi}^{(-1)}_{[0]} T^{(-1)}_{[0]} +
\tau_\reg\, \slashed{\xi}^{(-1)}_{[R,0]} T^{(-1)}_{[R,0]}} \right]
\end{align*}
with the error terms as in~\eqref{l:Fpm1}.
Moreover, a direct computation shows that (cf.~\cite[eq.~(5.3.23)]{PFP})
\begin{align}
F_{nc+}\,P(x,y) &= (\deg < 0)  \label{l:Fxirel1} \\
F_{1c-}\, P(x,y) &= \chi_c \begin{pmatrix} 1  & 0 \\ 0 & 0 \end{pmatrix} P(x,y) + (\deg < 0) \label{l:Fxirel2} \\
F_{2c-}\, P(x,y) &= \chi_c \begin{pmatrix} 0 & 0 \\ 0 & 1 \end{pmatrix} P(x,y) + (\deg < 0) \:. \label{l:Fxirel3}
\end{align}

Evaluating the EL equations~\eqref{l:EL} by substituting the above formulas into~\eqref{l:Qform},
we obtain the three conditions
\begin{align}
\left( 2 \left| \overline{ T^{(-1)}_{[0]} } T^{(0)}_{[0]} \right|
- \left| \overline{ L^{(-1)}_{[0]} } T^{(0)}_{[0]} \right|
- \left| \overline{T^{(-1)}_{[0]}}  L^{(0)}_{[0]} \right|
\right) \frac{\overline{T^{(-1)}_{[0]}} T^{(0)}_{[0]}}{
\big| \overline{T^{(-1)}_{[0]}} T^{(0)}_{[0]} \big|}\: T^{(-1)}_{[0]} &= 0 \label{l:cone} \\
\left( 3 \left| \overline{ L^{(-1)}_{[0]} } T^{(0)}_{[0]} \right|
- 2 \left| \overline{ T^{(-1)}_{[0]} } T^{(0)}_{[0]} \right|
- \left| \overline{T^{(-1)}_{[0]}}  L^{(0)}_{[0]} \right|
\right) \frac{\overline{L^{(-1)}_{[0]}} T^{(0)}_{[0]}}{
\big| \overline{L^{(-1)}_{[0]}} T^{(0)}_{[0]} \big|}\: L^{(-1)}_{[0]} &= 0 \\
\left( 3 \left| \overline{ T^{(-1)}_{[0]} } L^{(0)}_{[0]} \right|
- 2 \left| \overline{T^{(-1)}_{[0]}}  T^{(0)}_{[0]} \right|
- \left| \overline{L^{(-1)}_{[0]} } T^{(0)}_{[0]} \right|
\right) \frac{\overline{T^{(-1)}_{[0]}} L^{(0)}_{[0]}}{
\big| \overline{T^{(-1)}_{[0]}} L^{(0)}_{[0]} \big|}\: T^{(-1)}_{[0]} &= 0\:. \label{l:cthree}
\end{align}
These three equations must be satisfied in a weak evaluation on the light cone.

To summarize, evaluating the EL equations for the fermionic projector of the vacuum~\eqref{l:Pfreeex},
we obtain a finite hierarchy of equations to be satisfied in a weak evaluation on the light cone.
As the detailed form of these equations is quite lengthy and will not be needed later on,
we omit the explicit formulas.

\subsectionn{The Gauge Phases} \label{l:sec32}
Let us introduce chiral gauge potentials. As the auxiliary fermionic projector~\eqref{l:Paux0}
has seven components, the most general ansatz for chiral potentials would correspond
to the gauge group~$\U(7)_L \times \U(7)_R$.
\sindex{potential!chiral}%
\nindex{bh2@$A_L, A_R$ -- chiral potentials}%
However, the causality
compatibility conditions~\eqref{l:ccc} reduce the gauge group to
\beq \label{l:GG1}
\U(6)_L \times \U(6)_R \times \U(1)_R \:,
\eeq
where the groups~$\U(6)_L$ and~$\U(6)_R$ act on the first and last three components,
whereas the group~$\U(1)_R$ acts on the fourth component.
Similar as in~\S\ref{s:sec72}, to degree five the gauge potentials describe phase
transformations of the left- and right-handed components of the fermionic projector,
\beq
P^\text{aux}(x,y) \rightarrow \big( \chi_L \,U_L(x,y) + \chi_R \,U_R(x,y)
\big) P^\text{aux}(x,y) + (\deg < 2 )\:. \label{l:Pchiral}
\eeq
However, as the gauge group~\eqref{l:GG1} is non-abelian, the unitary
operators~$U_{L\!/\!R}$ now involve the ordered exponential
(for details see~\cite[Section~2.5]{PFP} or~\cite[Section~2.2]{light})
\nindex{de2@$U_c$ -- unitary matrix involving gauge pha\-ses}%
\beq
U_{L\!/\!R} = \Pexp \Big( -i \int_x^y A^j_{L\!/\!R} \,\xi_j \Big) \:.  \label{l:Lambda}
\eeq
Substituting the gauge potentials corresponding to the gauge group~\eqref{l:GG1}
and forming the sectorial projection, we obtain
\beq \begin{split} \label{l:Ptransform}
\chi_L P(x,y) &=  \chi_L \:\frac{i \slashed{\xi}}{2}\: T^{(-1)}_{[0]} \begin{pmatrix}
\hat{U}_L^{11} & \hat{U}_L^{12} \\[0.2em]
\hat{U}_L^{21} & \hat{U}_L^{22} \end{pmatrix}  + (\deg < 2) \\
\chi_R P(x,y) &=  \chi_R \:\frac{i \slashed{\xi}}{2}
\left[  T^{(-1)}_{[0]} \begin{pmatrix}
\hat{U}_R^{11} & \hat{U}_R^{12} \\[0.2em]
\hat{U}_R^{21} & \hat{U}_R^{22} \end{pmatrix}
+ \begin{pmatrix}
V \,T^{(-1)}_{[R, 0]} & 0 \\ 0 & 0 \end{pmatrix} \right] + (\deg < 2) \:,
\end{split}
\eeq
where
\[ U_{L\!/\!R} = \begin{pmatrix}
U_{L\!/\!R}^{11} & U_{L\!/\!R}^{12} \\[0.2em]
U_{L\!/\!R}^{21} & U_{L\!/\!R}^{22} \end{pmatrix} \in \U(6) \:,\qquad
V \in \U(1)\:, \]
and the hat denotes the sectorial projection,
\beq \label{l:ptrace}
\hat{U}^{ij}_L = \sum_{\alpha, \beta=1}^3 (U^{ij}_L)^\alpha_\beta \:, \qquad
\hat{U}_R = \sum_{\alpha, \beta=1}^3 (U_R)^\alpha_\beta \:.
\eeq

At this point it is important to observe that our notation in~\eqref{l:Ptransform} is oversimplified
because it does not make manifest that the four matrices~$U^{11}_{L\!/\!R}$ and~$U^{22}_{L\!/\!R}$
on the block diagonal describe a mixing of three regularized Dirac seas. Thus when the
sectorial projection is formed,
one gets new linear combinations of the regularized Dirac seas, which are then
described effectively by the factor~$T^{(-1)}_{[0]}$.
The analysis in~\cite{reg} gives a strong indication that an admissible regularization can be obtained
only by taking a sum of several Dirac seas and by delicately adjusting their regularizations
(more precisely, the property of a distributional ${\mathcal{M}} P$-product can be arranged only
for a sum of at least three Dirac seas).
This means that if we take a different linear combination of our three regularized Dirac seas,
we cannot expect that the resulting regularization is still admissible.
In order to avoid this subtle but important problem, we must impose that each of the
four matrices~$U_L^{11}$, $U_L^{22}$, $U_R^{11}$ and~$U_R^{22}$ is a multiple
of the identity matrix, because only in this case we get up to a constant the same linear combination
of regularized Dirac seas as in the vacuum
(for more details and similar considerations see~\cite[Remark~6.2.3]{PFP} and~\S\ref{s:sec89}).
This argument shows that the matrices~$U_L^{11}$, $U_L^{22}$, $U_R^{11}$ and~$U_R^{22}$
must be multiples of the identity matrix.
The following lemma tells us what these conditions mean for~$U_L$ and~$U_R$.
\begin{Lemma} \label{l:lemmablock}
Suppose that~$\G \subset \U(6)$ is a Lie subgroup such that in the standard 
representation on~$\C^6$, every~$g \in \G$ is of the form
\beq \label{l:gcond}
g = \begin{pmatrix} a \1_{\C^3} & * \\ * & c \1_{\C^3} \end{pmatrix} \qquad
\text{with~$a, c \in \R\:,$}
\eeq
where we used a $(3 \times 3)$ block matrix notation, and the stars stand for arbitrary
$3 \times 3$-matrices. Then there is a matrix~$U \in \U(3)$ such that
every~$g \in \G$ has the representation
\beq \label{l:gform}
g = \begin{pmatrix} a \1_{\C^3} & \overline{b} \,U^* \\ b \,U & c \1_{\C^3} \end{pmatrix}
\qquad \text{with} \qquad
\begin{pmatrix} a  & \overline{b} \\ b & c \end{pmatrix} \in \U(2)\:.
\eeq
In particular, $\G$ is isomorphic to a Lie subgroup of~$\U(2)$.
\end{Lemma}
\Proof For any~$A \in T_e \G$ we consider the one-parameter subgroup~$V(\tau) = e^{i \tau A}$
($\tau \in \R$). Evaluating~\eqref{l:gcond} to first order in~$\tau$, we find that
\[ A = \begin{pmatrix} a \1_{\C^3} &  Z^* \\ Z & c \1_{\C^3} \end{pmatrix} \]
with a $3 \times 3$-matrix~$Z$. Considering~\eqref{l:gcond} for the quadratic terms in~$\tau$,
we find that the matrices~$Z Z^*$ and~$Z^* Z$ are multiples of the identity matrix.
Taking the polar decomposition of~$Z$, we find that there is a unitary matrix~$U$ such that
\beq \label{l:Arep}
A = \begin{pmatrix} a \1_{\C^3} &  \overline{b} U^* \\ b U & c \1_{\C^3} \end{pmatrix} 
\qquad \text{with~$a,c \in \R$ and~$b \in \C$}\:.
\eeq
Exponentiating, one finds that~$V(\tau)$ is of the required form~\eqref{l:gform},
but with~$U$ depending on~$A$.

We next choose two matrices~$A, \tilde{A} \in T_e \G$ and represent them in the
form~\eqref{l:Arep} (where tildes always refer to~$\tilde{A}$). It remains to show
that~$U$ and~$\tilde{U}$ coincide up to a phase,
\beq \label{l:VVtrel}
\tilde{U} = e^{i \varphi} \,U \qquad \text{with~$\varphi \in \R$}\:.
\eeq
To this end, we consider the one-parameter
subgroup~$V(\tau) = e^{i \tau (A+\tilde{A})}$. Evaluating~\eqref{l:gcond} to second order in~$\tau$,
we obtain the condition
\[ \{ A, \tilde{A} \} = \begin{pmatrix} d \1_{\C^3} & * \\ * & e \1_{\C^3} \end{pmatrix} 
\qquad \text{with~$d,e \in \R$}. \]
Writing out this condition using~\eqref{l:Arep}, we find that
\beq \label{l:VVtcond}
a \tilde{a} + \overline{b} U^*\: \tilde{b} \tilde{U} = d \1_{\C^3}  \:.
\eeq
Let us show that there is a parameter~$\varphi \in \R$ such that~\eqref{l:VVtrel} holds.
If~$b$ or~$\tilde{b}$ vanish, there is nothing to prove. Otherwise, we know from~\eqref{l:VVtcond}
that the matrix~$U^* \tilde{U}$ is a multiple of the identity matrix. Since this matrix is unitary, it
follows that~$U^* \tilde{U} = e^{i \varphi} \1_{\C^3}$, proving~\eqref{l:VVtrel}.
\QED
We point out that the matrix~$U \in \U(3)$ is the same for all~$g \in \G$; this means that~$U$
will be a constant matrix in space-time.

Using the representation~\eqref{l:gform} in~\eqref{l:Ptransform}, the left-handed component of
the fermio\-nic projector becomes
\beq \label{l:Pleft}
\chi_L P(x,y) =  \chi_L \:\frac{i \slashed{\xi}}{2}\: T^{(-1)}_{[0]} \begin{pmatrix}
U_L^{11} & U_L^{12}\,\UMNS^* \\[0.2em]
U_L^{21}\,\UMNS & U_L^{22} \end{pmatrix}  + (\deg < 2) \:,
\eeq
where~$U_L \in U(2)$, and~$\UMNS \in \U(3)$ is a constant matrix.
The matrix~$\UMNS$ can be identified with the {\em{MNS matrix}} in
the electroweak theory.
\sindex{Maki-Nakagawa-Sakata (MNS) matrix}%
\nindex{da6@$\UMNS$ -- Maki-Nakagawa-Sakata (MNS) matrix}%
\sindex{mixing matrix!Maki-Nakagawa-Sakata (MNS) matrix}%
In~\eqref{l:Pleft}, we still need to make sense of the expressions
\beq \label{l:remain}
\hat{U}_\text{\tiny{MNS}} \,T^{(-1)}_{[0]} \qquad \text{and} \qquad
\hat{U}^*_\text{\tiny{MNS}} \,T^{(-1)}_{[0]} \:.
\eeq
Again, the matrix~$\UMNS$ describes a mixing of regularized Dirac seas, now even combining the
seas with different isospin. Since~$\UMNS$ is constant, one can take the point
of view that we should adjust the regularizations of all six Dirac seas in such a way that
the expressions in~\eqref{l:remain} are admissible (in the sense that the fermionic projector
has the property of a distributional ${\mathcal{M}} P$-product; see~\cite{reg}).

For the right-handed component, the high-energy component~$T^{(-1)}_{[R,0]}$ makes the
argument a bit more involved. Applying Lemma~\ref{l:lemmablock} to the
right-handed component, we obtain a representation of the form
\[ \chi_R P(x,y) =  \chi_R \:\frac{i \slashed{\xi}}{2}
\left[  T^{(-1)}_{[0]} \begin{pmatrix}
U_R^{11} & \!\!\! U_R^{12}\,U^* \\[0.3em]
U_R^{21}\,U & \!\!\! U_R^{22} \end{pmatrix}
+ \tau_\reg\, T^{(-1)}_{[R, 0]} \begin{pmatrix}
V & 0 \\ 0 & 0 \end{pmatrix} \right] \!+ (\deg < 2) \]
with~$(U_R, V) \in U(2) \times U(1)$ and a fixed matrix~$U \in U(3)$.
As explained after~\eqref{l:ptrace}, our notation is again a bit too simple in that
it does not make manifest that the three Dirac seas and the right-handed high-energy states
will in general all be regularized differently, and that only their linear combination
is described effectively by the factors~$T^{(-1)}_\circ$.
With this in mind, we can repeat the argument after~\eqref{l:ptrace} to conclude that the
relative prefactor of the regularization functions in the upper left
matrix entry should not be affected by the gauge potentials, i.e.
\[  U_R^{11} T^{(-1)}_{[0]} + \tau_\reg \,V\,  T^{(-1)}_{[R, 0]} = \kappa \,\big(  T^{(-1)}_{[0]} + 
\tau_\reg\, T^{(-1)}_{[R, 0]}
\big) \qquad \text{with~$\kappa \in \C$}\:. \]
In particular, one sees that~$U_R^{11}$ must be a phase factor, and this implies that~$U_R$
must be a diagonal matrix. Moreover, we find that~$V = U_R^{11}$.

Putting our results together, we conclude that the admissible gauge group is
\beq \label{l:ggroup}
\G = \U(2)_L \times \U(1)_R \times \U(1)_R \:.
\eeq
Choosing a corresponding potential~$(A_L, A_R^C, A_R^N)
\in \u(2) \times \u(1) \times \u(1)$, the interaction is described by the operator
\beq \label{l:Bform}
\B = \chi_R \begin{pmatrix} \slashed{A}_L^{11} & \slashed{A}_L^{12}\, \UMNS^* \\[0.2em]
\slashed{A}_L^{21}\, \UMNS & \slashed{A}_L^{22} \end{pmatrix}
+ \chi_L \begin{pmatrix} \slashed{A}_R^N & 0 \\
0 & \slashed{A}_R^C \end{pmatrix}  .
\eeq
\nindex{de6@$A_L$ -- left-handed gauge potential}%
\nindex{de8@$A_R^N$ -- right-handed potential in neutrino sector}%
\nindex{df0@$A_R^C$ -- right-handed potential in charged sector}%
Thus the~$\U(1)$-potentials~$A_R^N$ and~$A_R^C$ couple to the right-handed
component of the two isospin components. The $\U(2)$-potential $A_L$, on the other
hand, acts on the left-handed components, mixing the two isospin components.
The $\UMNS$-matrix describes a mixing of the generations in the off-diagonal isospin
components of~$A_L$.

In order to analyze the EL equations to degree five
in the presence of the above gauge potentials, we need to compute the eigenvalues
of the closed chain (see~\eqref{l:EL} and~\eqref{l:Qform}).
Combining~\eqref{l:Ptransform} with the form of the gauge potentials as specified in~\eqref{l:ggroup}
and~\eqref{l:Bform}, we obtain
\begin{align}
\chi_L P(x,y) &=  \frac{3}{2}\:\chi_L \:i \slashed{\xi}\: T^{(-1)}_{[0]} 
\begin{pmatrix} U_L^{11} & \overline{c} \:U_L^{12} \\[0.2em]
c\: U_L^{21} & U_L^{22} \end{pmatrix} + (\deg < 2) \label{l:PLtransform} \\
\chi_R P(x,y) &=  \frac{3}{2}\:\chi_R \:i \slashed{\xi}
\begin{pmatrix} V_R^N\,L^{(-1)}_{[0]}  & 0 \\ 0 & V_R^C\, T^{(-1)}_{[0]}
\end{pmatrix} + (\deg < 2) \label{l:PRtransform}
\end{align}
with~$U_L \in U(2)$ and~$V_R^N, V_R^C \in U(1)$, where we again used
the notation~\eqref{l:Lndef} and introduced the complex number
\beq \label{l:cdef}
c = \frac{1}{3} \: \hat{U}_\text{\tiny{MNS}} \:.
\eeq
It follows for the closed chain that
\begin{align}
\chi_L A_{xy} &= \frac{9}{4} \: \chi_L
\begin{pmatrix} U_L^{11} & \overline{c} \:U_L^{12} \\[0.2em]
c\: U_L^{21} & U_L^{22} \end{pmatrix}
\begin{pmatrix} V_R^N & 0 \\ 0 & V_R^C \end{pmatrix}
\begin{pmatrix} \slashed{\xi} T^{(-1)}_{[0]} \overline{\slashed{\xi} L^{(-1)}_{[0]}} & 0 \\
0 & \slashed{\xi} T^{(-1)}_{[0]} \overline{\slashed{\xi} T^{(-1)}_{[0]}} \end{pmatrix}  \label{l:chain5} \\
& \qquad + \slashed{\xi}\, (\deg < 3) + (\deg < 2) \:. \nonumber
\end{align}

When diagonalizing the matrix~\eqref{l:chain5}, the
factor~$\overline{L^{(-1)}_{[0]}}$ causes major difficulties because it leads to
microscopic oscillations of the eigenvectors.
Let us explain this problem in detail. First, it is convenient to use the $\iota$-formalism, because
then, similar as explained after~\eqref{l:Axyiota}, the contributions~$\sim \iotaslsh \check{\slashed{\xi}}$
and~$\sim \check{\slashed{\xi}} \iotaslsh$ act on complementary subspaces. Thus it remains to
diagonalize the $2 \times 2$-matrices
\begin{align*}
&\begin{pmatrix} U_L^{11} & \overline{c} \:U_L^{12} \\[0.2em]
c\: U_L^{21} & U_L^{22} \end{pmatrix}
\begin{pmatrix} V_R^N \:T^{(0)}_{[0]} \overline{L^{(-1)}_{[0]}} & 0 \\[-0.3em]
0 & V_R^C\: T^{(0)}_{[0]} \overline{T^{(-1)}_{[0]}}\end{pmatrix} \\[-1em]
\intertext{and} \\[-3em]
&\begin{pmatrix} U_L^{11} & \overline{c} \:U_L^{12} \\[0.2em]
c\: U_L^{21} & U_L^{22} \end{pmatrix}
\begin{pmatrix} V_R^N \:T^{(-1)}_{[0]} \overline{L^{(0)}_{[0]}} & 0 \\[-0.3em]
0 & V_R^C\: T^{(-1)}_{[0]} \overline{T^{(0)}_{[0]}}\end{pmatrix} .
\end{align*}
The characteristic polynomial involves square roots of linear combinations of the inner matrix elements,
describing non-trivial fluctuations of the eigenvalues on the regularization scale~$\varepsilon$.
Such expressions are ill-defined in the formalism of the continuum limit.
A first idea for overcoming this problem would be to
extend the formalism such as to include square roots of linear combinations of simple fractions. 
However, even if one succeeded in extending the continuum limit in this way,
it would be unclear how the resulting square root expressions
after weak evaluation would depend on the smooth parameters~$U_L^{ij}$ and~$V_R^{N\!/\!C}$.
The basic difficulty is that integrating over the microscopic oscillations will in general not
preserve the square root structure (as a simple example, an integral of the
form~$\int_0^\infty \sqrt{a+x}\: f(x)\, dx$ cannot in general be written again as a square root of say
the form~$\sqrt{a b+c}$). This is the reason why the complications related to the
factor~$L^{(-1)}_{[0]}$ in~\eqref{l:chain5} seem to arise as a matter of principle.

In order to bypass this difficulty, we must restrict attention to a parameter range where
the eigenvalues of the above matrices can be computed perturbatively. In order to
make the scaling precise, we write~$\tau_\reg$ as
\begin{equation}
\tau_\reg = (m \varepsilon)^{p_\reg} \qquad \text{with} \qquad
0 < p_\reg < 2\:. \label{l:preg}
\end{equation}
\nindex{dc6@$\tau_\reg$ -- dimensionless parameter for high-energy states}%
\nindex{df4@$p_\reg$ -- determines scaling~$\tau_\reg = (m \varepsilon)^{p_\reg}$}%
Under this assumption, we know that that the relation
\beq \label{l:rcond}
T^{(n)}_{[p]} = L^{(n)}_{[p]} \:\big(1+\O \big((m \varepsilon)^{p_\reg} \big)\big) \qquad \text{holds pointwise}
\eeq
(by ``holds pointwise'' we mean that if we multiply~$T^{(n)}_{[p]} - L^{(n)}_{[p]}$ by
any simple fraction and evaluate weakly according to~\eqref{l:asy}, we get zero
up to an error of the specified order).
\sindex{evaluation on the light cone!pointwise}%
Making~$\tau_\reg$ small in this sense does not necessarily imply that the above matrices
can be diagonalized perturbatively, because we need to compare~$\tau_\reg$ to the
size of the off-diagonal matrix elements~$U_R^{12}$ and~$U_R^{21}$.
As they are given as line integrals over the chiral potentials (cf.~\eqref{l:Lambda}),
their size is described by
\[ \|A_L^{12}\| \cdot |\vec{\xi}| \qquad \text{and} \qquad  \|A_L^{21} \| \cdot |\vec{\xi}| \]
(where~$\|.\|$ is a Euclidean norm defined in the same reference frame as~$\vec{\xi}$).
This leads us to the following two cases:
\beq \label{l:casesiii}
\text{\bf{(i)}} \quad |\vec{\xi}| \gg \frac{(m \varepsilon)^{p_\reg}}{\|A_L^{12}\|+\|A_L^{21}\|}
\:,\qquad \qquad
\text{\bf{(ii)}} \quad |\vec{\xi}| \ll \frac{(m \varepsilon)^{p_\reg}}{\|A_L^{12}\|+\|A_L^{21}\|} \:.
\eeq
In fact, the computations are tractable in both cases, as we now explain. \\[1em]
{\bf{Case~(i).}} We expand in powers of~$\tau_\reg$. We begin with the case~$\tau_\reg=0$.
Then in the vacuum, \eqref{l:rcond} implies that the relations~\eqref{l:cone}--\eqref{l:cthree}
are trivially satisfied. If gauge potentials are present, in the above matrices we can
factor out the scalar functions~$T^{(0)}_{[0]} \overline{T^{(-1)}_{[0]}}$ and~$T^{(-1)}_{[0]} \overline{T^{(0)}_{[0]}}$,
respectively. Thus it remains to compute the eigenvalues and spectral projectors of the $2 \times 2$-matrix
\beq \label{l:maprod}
\begin{pmatrix} U_L^{11} & \overline{c} \:U_L^{12} \\[0.2em]
c\: U_L^{21} & U_L^{22} \end{pmatrix}
\begin{pmatrix} V_R^N & 0 \\ 0 & V_R^C \end{pmatrix} .
\eeq

\begin{Lemma} \label{l:lemma32}
The matrix in~\eqref{l:maprod} is normal (i.e.\ it commutes with its
adjoint). Moreover, its eigenvalues have the same absolute value.
\end{Lemma}
\Proof We denote the matrix in~\eqref{l:maprod} by~$B$ and
write the two factors in~\eqref{l:maprod} in terms of Pauli matrices as
\[ B = (a \1 + i \vec{v} \vec{\sigma})\; e^{i \varphi} (b \1 + i \vec{w} \vec{\sigma}) \]
with~$a,b, \varphi \in \R$ and~$\vec{v}, \vec{w} \in \R^3$. Using the multiplication rules of Pauli matrices,
one finds that
\beq \label{l:Pauliid}
e^{-i \varphi} \,B = (a b - \vec{v} \vec{w}) \1 + i (a \vec{w} + b \vec{v} + \vec{v} \wedge \vec{w}) \,\vec{\sigma} \:.
\eeq
A short calculation shows that this matrix is normal.
Moreover, the eigenvalues of~$B$ are computed by
\[ e^{i \varphi} \left( (a b - \vec{v} \vec{w}) \pm i \,\big| a \vec{w} + b \vec{v} + \vec{v} \wedge \vec{w} \big|
\right) \:. \]
Obviously, these eigenvalues have the same absolute value.
\QED

We denote the eigenvalues and corresponding spectral projection operators
of the matrix in~\eqref{l:maprod} by~$\nu_{nL}$ and~$I_n$. Then, according to the above lemma,
\nindex{df6@$\nu_{nc}, I_{nc}$ -- eigenvalues and corresponding spectral projectors
of matrix involving pha\-ses}%
\beq \label{l:nuabs}
|\nu_{1L}| = |\nu_{2L}| \qquad \text{and} \qquad I_n^* = I_n \:.
\eeq
For the left-handed component of the closed chain~\eqref{l:chain5} we thus obtain the
eigenvalues~$\lambda_{nLs}$ and spectral projectors~$F_{nLs}$ given by
\beq \label{l:lambdaL}
\lambda_{nLs} = \nu_{nL}\, \lambda_s \:,\qquad
F_{n L s} = \chi_L\: I_{n}\: F_s\:,
\eeq
where $\lambda_\pm$ and~$F_s$ are given by (cf.~\eqref{l:lamM} and~\eqref{l:Fpm2},
\begin{gather}
\lambda_+ = 9\, T^{(0)}_{[0]} \,\overline{T^{(-1)}_{[0]}} + (\deg < 3) \:,\qquad
\lambda_- = 9\,T^{(-1)}_{[0]} \,\overline{T^{(0)}_{[0]}} + (\deg < 3) \label{l:lpm} \\
F_\pm = \frac{1}{2} \Big( \1 \pm \frac{[\slashed{\xi}, \overline{\slashed{\xi}}]}{z-\overline{z}} \Big)
+ \slashed{\xi} (\deg \leq 0) + (\deg < 0)\:. \label{l:Fpm}
\end{gather}
The spectral decomposition of~$\chi_R A_{xy}$ is obtained by complex conjugation,
\beq \label{l:lambdaR}
\lambda_{n R \pm} = \nu_{nR}\, \lambda_\pm =
\overline{\lambda_{nL \mp}} = \overline{\nu_{nL}} \: \lambda_\pm\:,\qquad
F_{nL \pm} = F_{nR \mp}^* \:.
\eeq
Combining these relations with~\eqref{l:nuabs} and~\eqref{l:lpm}, 
we conclude that all the eigenvalues of the closed chain have the same absolute value.
Thus in view of~\eqref{l:Qform}, the EL equations are indeed satisfied for~$\tau_\reg=0$.
In order to treat the higher orders in~$\tau_\reg$, one performs a power expansion
up to the required order in the Planck length. The EL equations can be satisfied
to every order in~$\tau_\reg$ by imposing suitable conditions on the regularization functions.
Thus one gets a finite hierarchy of equations to be satisfied in a weak evaluation
on the light cone. \\[1em]
{\bf{Case~(ii).}} We perform a perturbation expansion in the off-diagonal elements~$U_L^{21}$
and~$U_L^{12}$. If we set these matrix elements to zero, we again get a spectral
representation of the form~\eqref{l:lambdaL}--\eqref{l:lambdaR}, but now with
\beq \label{l:I12diag}
I_1 = \begin{pmatrix} 1 & 0 \\ 0 & 0 \end{pmatrix} \:,\qquad
I_2 = \begin{pmatrix} 0 & 0 \\ 0 & 1 \end{pmatrix}
\eeq
and
\[ \nu_1 = U_L^{11}\, V_R^N \:,\qquad \nu_2 = U_R^{22}\, V_R^C \:. \]
Since the diagonal elements of any $U(2)$-matrix have the same absolute value,
it follows that~\eqref{l:nuabs} again holds. Hence the EL equations are again satisfied
in the case~$U_L^{21}=0=U_L^{12}$. Expanding in powers of~$U_L^{21}$ and~$U_L^{12}$
again gives a finite hierarchy of equations to be evaluated weakly on the light cone,
which can again be satisfied by imposing suitable conditions on the regularization functions.

We conclude that to degree five on the light cone, the EL equations can be satisfied
by a suitable choice of the regularization functions, whenever the EL equations have
a well-defined continuum limit. Clearly, the detailed computation of admissible regularizations
is rather involved. Fortunately, we do not need to work out the details, because
they will not be needed later on.

\section{The Euler-Lagrange Equations to Degree Four} \label{l:sec4}
We now come to the analysis of the EL equations to degree four on the light cone.
Before beginning, we clarify our scalings. Recall that the mass expansion
increases the upper index of the factors~$T^{(n)}_\circ$ and thus decreases the degree on
the light cone. In view of the weak evaluation formula~\eqref{l:asy}, the mass expansion gives scaling
factors~$m^2 \,\varepsilon |\vec{\xi}|$. Moreover, the parameter~$\tau_\reg$
gives scaling factors~$(m \varepsilon)^{p_\reg}$ (see~\eqref{l:preg}).
Unless stated otherwise, we shall only consider the leading order in~$(m \varepsilon)^{p_\reg}$,
meaning that we allow for an error term of the form
\begin{equation}
\big(1+\O \big((m \varepsilon)^{p_\reg} \big)\big) \:. \label{l:leadpreg}
\end{equation}
Finally, the weak evaluation formulas
involve error terms of the form~\eqref{l:neglect}.
Since the contributions to the EL equations to degree four
on the light cone involve at least one scaling factor~$m^2\, \varepsilon |\vec{\xi}|$ 
(from the mass expansion) or a factor
with the similar scaling~$\varepsilon |\vec{\xi}| /\ell_\text{macro}^2$ 
(from the light-cone expansion),
the factors~$\varepsilon/|\vec{\xi}|$ (which arise from the regularization expansion)
give rise to at least one factor~$m^2 \varepsilon^2$, which can be absorbed into the error
term~\eqref{l:leadpreg}. Hence, unless stated otherwise, in all the subsequent calculations we
neglect the
\beq \label{l:higherneglect}
\text{(higher orders in~$\varepsilon/\ell_\text{macro}$ and~$(m \varepsilon)^{p_\reg}$)}\:.
\eeq
For ease in notation, in most computations we omit to write out the corresponding error
term~$(1+\O(\varepsilon/\ell_\text{macro}) +\O((m \varepsilon)^{p_\reg}))$.

\subsectionn{General Structural Results} \label{l:sec40}
We again denote the eigenvalues of the closed chain~$A_{xy}$ by~$\lambda^{xy}_{ncs}$.
These eigenvalues will be obtained by perturbing the eigenvalues with gauge
phases as given in~\eqref{l:lambdaL} and~\eqref{l:lambdaR}. As a consequence, they will
again form complex conjugate pairs, i.e.
\beq \label{l:ccp}
\lambda^{xy}_{nR\pm} = \overline{\lambda^{xy}_{nL\mp}} \:.
\eeq
As the unperturbed eigenvalues all have the same absolute value (see~\eqref{l:lambdaL},
\eqref{l:nuabs} and~\eqref{l:lpm}), to degree four we only need to take into account the perturbation of the
square bracket in~\eqref{l:Qform}. Thus the EL equations reduce to the condition
\beq \label{l:EL4}
0 = \Delta Q(x,y) := \sum_{n,c,s} \bigg[  \Delta |\lambda^{xy}_{ncs}| - \frac{1}{8} \sum_{n',c',s'}
\Delta |\lambda^{xy}_{n'c's'}| \bigg] \: \frac{\overline{\lambda^{xy}_{ncs}}}{|\lambda^{xy}_{ncs}|}
\: F_{ncs}^{xy}\: P(x,y) \:,
\eeq
\nindex{df8@$\Delta Q(x,y)$ -- first order perturbation of~$Q(x,y)$}%
where we again evaluate weakly on the light cone and consider the perturbation of the eigenvalues
to degree two (also, the superscript~$xy$ clarifies the dependence of the eigenvalues
on the space-time points).

Here the unperturbed spectral projectors~$F_{ncs}$ were computed explicitly in~\eqref{l:lambdaL}
and~\eqref{l:Fpm}. Moreover, the relations~\eqref{l:Fxirel1}--\eqref{l:Fxirel3} can be written in the shorter form
\[ F^{xy}_+ \slashed{\xi} = (\deg < 0) \:,\qquad
F^{xy}_- \slashed{\xi} = \slashed{\xi} + (\deg < 0)\:. \]
Combining these relations with the explicit formulas for the
corresponding unperturbed eigenvalues (see~\eqref{l:lambdaL} and~\eqref{l:lpm})
as well as using~\eqref{l:ccp}, we can write~$\Delta Q(x,y)$ as
\beq \label{l:DelQrep}
\Delta Q(x,y) = \frac{i}{2}\: \sum_{n,s} \bigg[  {\mathscr{K}}_{nc}(x,y) - \frac{1}{4} \sum_{n',c'}
{\mathscr{K}}_{n'c'}(x,y) \bigg] \,I_n\: \chi_c\, \slashed{\xi} + (\deg < 4)\:,
\eeq
where
\beq \label{l:Kncdef}
{\mathscr{K}}_{nc}(x,y) := \frac{\Delta |\lambda^{xy}_{nc-}|}{|\lambda_-|}\;3^3\:
T^{(0)}_{[0]} T^{(-1)}_{[0]} \:\overline{T^{(-1)}_{[0]}}
\eeq
(for more details see the proof of Lemma~\ref{s:lemma81}).
Since the smooth factors in~\eqref{l:DelQrep}
are irrelevant, the EL equations~\eqref{l:EL4} reduce to the conditions
\beq \label{l:Knccond}
{\mathscr{K}}_{1L} = {\mathscr{K}}_{2L} = {\mathscr{K}}_{1R} = {\mathscr{K}}_{2R}
\quad \mod (\deg < 4) \:.
\eeq
\nindex{dg0@${\mathscr{K}}_{nc}$ -- matrices entering the EL equations to degree four}%

For all the contributions to the fermionic projector of interest in this paper, it will suffice
to compute~$\Delta |\lambda^{xy}_{nc+}|$ in a perturbation calculation of first or second order.
Then the complex numbers~${\mathscr{K}}_{nc}$ can be recovered as traces of~$I_n$ with
suitable $2 \times 2$-matrices, as the following lemma shows.
\begin{Lemma} \label{l:lemma41}
In a perturbation calculation to first order, there are $2 \times 2$-matrices
${\mathscr{K}}_L$ and~${\mathscr{K}}_R$ such that
\nindex{dg2@${\mathscr{K}}_{c}$ -- matrices entering the EL equations to degree four}%
\beq \label{l:Kdef}
{\mathscr{K}}_{nc} = \Tr_{\C^2} \left( I_n \,{\mathscr{K}}_c \right) + (\deg < 4)\:.
\eeq
In a second order perturbation calculation, one can again arrange~\eqref{l:Kdef},
provided that the gauge phases~$\nu_{nc}$ in the unperturbed eigenvalues~\eqref{l:lambdaL}
and~\eqref{l:lambdaR} must not to be taken into account and that the perturbation vanishes on the
degenerate subspaces in the sense that
\beq \label{l:secc}
F_+ \,(\Delta A)\, F_+ = 0 \:.
\eeq
\end{Lemma}
\Proof In view of~\eqref{l:Kncdef}, it clearly suffices to show that~$\Delta |\lambda_{nc+}|$
can be written as such a trace. Writing
\[ \Delta |\lambda_{nc+}| = \frac{1}{2 |\lambda_+|} \Big( (\Delta \lambda_{nc+})\, \overline{\lambda_+}
+ \lambda_+ \,\overline{(\Delta \lambda_{nc+})} \Big) \]
and using~\eqref{l:ccp}, one concludes that it suffices to show that
\beq \label{l:lamtr}
\Delta \lambda_{ncs} = \Tr_{\C^2} (I_n B)
\eeq
for a suitable $2 \times 2$-matrix~$B=B(c,s)$.

The linear perturbation is given by
\[ \Delta \lambda_{ncs} = \Tr (F_{ncs}\, \Delta A) \:. \]
As the unperturbed spectral projectors involve a factor~$I_n$ (see~\eqref{l:lambdaL} and~\eqref{l:lambdaR}),
this is obviously of the form~\eqref{l:lamtr}.

Using~\eqref{l:secc}, we have to second order
\beq \label{l:secondorder}
\Delta \lambda_{ncs} =
\sum_{n',c'} \frac{1}{\lambda_{ncs} - \lambda_{n'c' (-s)}}\: \Tr(F_{ncs} \,\Delta A\, F_{n'c' (-s)}
\,\Delta A)  \:.
\eeq
Disregarding the gauge phases~$\nu_{cs}$ in~\eqref{l:lambdaL} and~\eqref{l:lambdaR}, we get
\begin{align*}
\Delta \lambda_{ncs} &=
\sum_{n',c'} \frac{1}{\lambda_s - \lambda_{-s}}\: \Tr(F_{ncs} \,\Delta A\, F_{n'c' (-s)} \,\Delta A)  \\
&= \frac{1}{\lambda_s - \lambda_{-s}}\:\: \Tr( \chi_c \,I_n \,F_s \,\Delta A\, F_{-s} \,\Delta A) \:,
\end{align*}
where in the last line we used the form of the spectral projectors in~\eqref{l:lambdaL} and~\eqref{l:lambdaR}
and carried out the sums over~$n'$ and~$c'$. This is again of the form~\eqref{l:lamtr}.
\QED

Instead of analyzing the conditions~\eqref{l:Knccond}, we shall always analyze the stronger conditions
\beq \label{l:Kcond}
\boxed{ \quad {\mathscr{K}}_L(x,y) = {\mathscr{K}}_R(x,y) = c(\xi)\, \1_{\C^2}\:. \quad }
\eeq
This requires a detailed explanation, depending on the two cases in~\eqref{l:casesiii}.
In Case~{\bf{(i)}}, when the projectors~$I_n$ are determined by the chiral gauge potentials,
the condition~\eqref{l:Kcond} can be understood in two different ways. The first, more physical argument is
to note that the spectral projectors~$I_n$ of the
matrix product~\eqref{l:maprod} depend on the local gauge potentials~$A_L$ and~$A_R$.
In order for these potentials to be dynamical, the EL equations should not give
algebraic constraints for these potentials (i.e.\ constraints which involve the potentials
but not their derivatives). This can be achieved by demanding that the conditions~\eqref{l:Knccond}
should be satisfied for any choice of the potentials. In view of~\eqref{l:Kdef},
this implies that~\eqref{l:Kcond} must hold.

To give the alternative, more mathematical argument, let us assume conversely that
one of the matrices~${\mathscr{K}}_L$ or~${\mathscr{K}}_R$ is {\em{not}} a multiple of the identity
matrix. Then the perturbation calculation would involve terms mixing the free eigenspaces
corresponding to~$\lambda_{1cs}$ and~$\lambda_{2cs}$.
More precisely, to first order one would have to diagonalize the perturbation on the
corresponding degenerate subspace.
To second order, the resulting contribution to the perturbation calculation would
look similar to~\eqref{l:secondorder}, but it would also involve
factors of~$(\lambda_{1cs} - \lambda_{2cs})^{-1}$.
In both cases, the perturbed eigenvalues would no longer be a power series in the bosonic
potentials. Analyzing these non-analytic contributions in the EL equations~\eqref{l:Knccond},
one finds that they must all vanish identically. Working out this argument in more detail,
one could even derive~\eqref{l:Kcond} from the EL equations.

In Case~{\bf{(ii)}} in~\eqref{l:casesiii}, the projectors~$I_n$ are isospin-diagonal~\eqref{l:I12diag},
so that~\eqref{l:Knccond} only tests the diagonal elements of~${\mathscr{K}}_c$.
Thus at first sight, \eqref{l:Kcond} seems a too strong condition.
However, even in this case the condition~\eqref{l:Kcond} can be justified as follows.
The left-handed gauge potentials modify the left-handed component of the fermionic projector
by generalized phase transformations. If the involved gauge potential is off-diagonal,
it makes an off-diagonal components of~$P(x,y)$ diagonal and vice versa.
As a consequence, satisfying~\eqref{l:Knccond} in the presence of off-diagonal gauge potentials
is equivalent to satisfying~\eqref{l:Knccond}. 
We will come back to this argument in more detail in Section~\ref{l:sec6}.

We finally use~\eqref{l:Kdef} in~\eqref{l:DelQrep} to obtain a useful representation of~$\Delta Q$:
\begin{Corollary} \label{l:cor42} $\;\;\;$
Under the assumptions of Lemma~\ref{l:lemma41}, the kernel~$\Delta Q(x,y)$
in~\eqref{l:EL4} has the representation
\beq \label{l:DelQ}
\Delta Q(x,y) = \frac{i}{2} \sum_{n,c} \Tr_{\C^2} \!\big( I_n \,{\mathcal{Q}}_c \big)
\,I_n\: \chi_c\, \slashed{\xi} \:,
\eeq
where
\beq \label{l:Rdef}
{\mathcal{Q}}_L := {\mathscr{K}}_L - \frac{1}{4}\: \Tr_{\C^2} \!\big(
{\mathscr{K}}_L+{\mathscr{K}}_R \big)\: \1_{\C^2}
\eeq
(and~${\mathcal{Q}}_R$ is obtained by the obvious replacements~$L \leftrightarrow R$).
\nindex{dg4@${\mathcal{Q}}_{c}$ -- matrices entering the EL equations to degree four}%
\end{Corollary} \noindent
The stronger condition~\eqref{l:Kcond} is then equivalent to demanding that the relations
\beq \label{l:ELR}
{\mathcal{Q}}_L(x,y) = 0 = {\mathcal{Q}}_R(x,y)
\eeq
hold in a weak evaluation on the light cone.

\subsectionn{The Vacuum} \label{l:sec42}
We begin by analyzing the eigenvalues of the closed chain in the vacuum.
As the fermionic projector is diagonal in the isospin index, we can consider the charged sector and
the neutrino sector after each other. In the {\em{charged sector}}, the eigenvalues
can be computed exactly as in~\cite[Section~5.3]{PFP} or~\S\ref{s:sec71}. Using the notation and conventions
in Chapter~\ref{sector}, we obtain
\begin{align*}
P(x,y) \:=\:& \frac{3i}{2} \:\slashed{\xi}\, T^{(-1)}_{[0]} + \frac{i}{2}\:m^2 \,\acute{Y} \grave{Y}\,
T^{(0)}_{[2]} + m \hat{Y}\: T^{(0)}_{[1]} + (\deg < 1) \\
A_{xy} \:=\:& \frac{3}{4} \:\slashed{\xi} \overline{\slashed{\xi}}
\left( 3\,T^{(-1)}_{[0]} \overline{T^{(-1)}_{[0]}} + m^2 \,\acute{Y} \grave{Y} \Big(
T^{(0)}_{[2]} \overline{T^{(-1)}_{[0]}} + T^{(-1)}_{[0]} \overline{T^{(0)}_{[2]}} \Big) \right) \\
& + \frac{3i}{2} \:m \hat{Y} \left( \slashed{\xi}\, T^{(-1)}_{[0]} \overline{T^{(0)}_{[1]}}
- T^{(0)}_{[1]} \overline{\slashed{\xi}\, T^{(-1)}_{[0]}} \right) \\
& + m^2 \hat{Y}^2\: T^{(0)}_{[1]} \overline{T^{(0)}_{[1]}} + (\deg < 2)\:.
\end{align*}
A straightforward calculation shows that the closed chain has two eigenvalues~$\lambda_\pm$,
both with multiplicity two. They have the form
\beq \label{l:lampm} \begin{split}
\lambda_+ &= 9\: T^{(0)}_{[0]} \overline{T^{(-1)}_{[0]}} + m^2 \,(\cdots) +  (\deg < 2) \\
\lambda_- &= 9\: T^{(-1)}_{[0]} \overline{T^{(0)}_{[0]}} + m^2 \,(\cdots) +  (\deg < 2) \:,
\end{split} \eeq
where~$(\cdots)$ stands for additional terms, whose explicit form will not be needed here
(for details see~\cite[eq.~(5.3.24)]{PFP}).

In the {\em{neutrino sector}}, by using~\eqref{l:neureg2} in the
ansatz~\eqref{l:massneutrino} and~\eqref{l:Paux}, after forming the sectorial projection we obtain
\begin{align*}
P(x,y) \:=\:& \frac{3 i \slashed{\xi}}{2} \: T^{(-1)}_{[0]} + \chi_R\: \tau_\reg\: \frac{i \slashed{\xi}}{2}\,
\big( T^{(-1)}_{[R,0]} +\delta^{-2}\, T^{(0)}_{[R,2]} \big) \\
&+\frac{i}{2} \:\slashed{\xi}\: m^2 \,\acute{Y} \grave{Y}\, T^{(0)}_{[2]} + m \hat{Y}\: T^{(0)}_{[1]} + (\deg < 1) \\
\chi_L A_{xy} \:=\:& \frac{3}{4} \:\chi_L \: \slashed{\xi} \overline{\slashed{\xi}}
\: T^{(-1)}_{[0]}\overline{ \big(  3 \,T^{(-1)}_{[0]} + \tau_\reg\, T^{(-1)}_{[R, 0]}
+  \tau_\reg\, \delta^{-2}\, T^{(0)}_{[R,2]} \big) } \\
&+\frac{3}{4} \:\slashed{\xi} \overline{\slashed{\xi}}\:  m^2 \,\acute{Y} \grave{Y} \Big(
T^{(0)}_{[2]} \overline{T^{(-1)}_{[0]}} + T^{(-1)}_{[0]} \overline{T^{(0)}_{[2]}} \Big)
+ m^2 \hat{Y}^2\: T^{(0)}_{[1]} \overline{T^{(0)}_{[1]}}  \\
& + \frac{3i}{2} \:m \hat{Y} \Big( \slashed{\xi}\, T^{(-1)}_{[0]} \overline{T^{(0)}_{[1]}}
- T^{(0)}_{[1]} \overline{\slashed{\xi}\, T^{(-1)}_{[0]}} \Big) + (\deg < 2)\:.
\end{align*}
The contraction rules~\eqref{l:contract} and~\eqref{l:ncontract}
yield~$(\slashed{\xi} \overline{\slashed{\xi}})^2 = (z + \overline{z})\:
\slashed{\xi} \overline{\slashed{\xi}} + z \overline{z}$ and thus
\[ (\slashed{\xi} \overline{\slashed{\xi}} - z) (\slashed{\xi} \overline{\slashed{\xi}} - \overline{z}) = 0 \:. \]
This shows that the matrix~$\slashed{\xi} \overline{\slashed{\xi}}$ has the eigenvalues~$z$ and~$\overline{z}$.
Also applying~\eqref{l:rcond}, the eigenvalues of the closed chain are computed by
\begin{align}
\lambda_{L+} &= \frac{3}{4}\: z
\: T^{(-1)}_{[0]} \overline{\big( 3\, L^{(-1)}_{[0]} + \tau_\reg\, \delta^{-2}\,
T^{(0)}_{[R,2]} \big) } + m^2 \,(\cdots) \nonumber \\
&= 9 \,T^{(0)}_{[0]} \overline{L^{(-1)}_{[0]}}
+ 3 \,\tau_\reg\,\delta^{-2} \,T^{(0)}_{[0]} \overline{T^{(0)}_{[R, 2]}} + m^2\, (\cdots) + (\deg < 2) \label{l:lamL} \\
\lambda_{L-} &= \frac{3}{4}\:
T^{(-1)}_{[0]}\: \overline{z\, \big( 3 \,T^{(-1)}_{[0]} + \tau_\reg\, T^{(-1)}_{[R, 0]}
+  \tau_\reg\, \delta^{-2}\, T^{(0)}_{[R,2]} \big) } + m^2\, (\cdots) \nonumber \\
&= 9\, T^{(-1)}_{[0]} \overline{L^{(0)}_{[0]}}
- 3 \,\tau_\reg\,\delta^{-2}\, T^{(-1)}_{[0]} \overline{T^{(1)}_{\{R, 0 \}} } + m^2\,(\cdots) + (\deg < 2)\:, \label{l:lamR}
\end{align}
where~$L^{(n)}_\circ$ is again given by~\eqref{l:Lndef}, and~$m^2\,(\cdots)$ denotes the same contributions
as in~\eqref{l:lampm} with the masses~$m_\beta$
replaced by the corresponding neutrino masses~$\tilde{m}_\beta$. The two other eigenvalues are
again obtained by complex conjugation~\eqref{l:ccp}.

The first summands in~\eqref{l:lamL} and~\eqref{l:lamR} are of degree three on the light cone
and were already analyzed in Section~\ref{l:sec3}.
Thus the point of interest here are the summands involving~$\delta$.
Before analyzing them in detail, we point out that they arise for two different reasons:
The term in~\eqref{l:lamL} is a consequence of the mass expansion of
general surface states. The term in~\eqref{l:lamR}, on the other hand,
corresponds to the last term in the contraction rule~\eqref{l:ncontract}, which
takes into account the shear of the surface states.

Let us specify the scaling of the terms involving~$\delta$. Recall that
the parameter~$\tau_\reg$ scales according to~\eqref{l:preg}, whereas~$\delta$
is only specified by~\eqref{l:delscale}.
We want that the general surface and shear states make up for the fact that the
masses~$m_\beta$ of the charged fermions are different from the neutrino masses~$\tilde{m}_\beta$.
Therefore, it would be natural to impose that the summands involving~$\delta$
should have the same scaling as the contributions~$m^2\, (\cdots)$ arising in the standard
mass expansion. This gives rise to the scaling
\[ \frac{\tau_\reg}{\delta^2} \eqsim m^2 \:, \]
and thus~$\delta \eqsim m\: (m \varepsilon)^{\frac{p_\reg}{2}}$.
But~$\delta$ can also be chosen smaller. In this case, the terms involving~$\delta$
in~\eqref{l:lamL} and~\eqref{l:lamR} could dominate the contributions by the standard mass expansion.
But they do not need to, because their leading contributions may cancel when evaluated weakly on
the light cone. With this in mind, we allow for the scaling
\begin{equation}
\varepsilon \ll \delta \lesssim \frac{1}{m}\: (m \varepsilon)^{\frac{p_\reg}{2}}\:. \label{l:deltascale}
\end{equation}
Assuming this scaling, by choosing the regularization parameters corresponding to 
the factors~$T^{(0)}_{[R,2]}$ and~$T^{(1)}_{\{R, 0 \} }$ appropriately, we can arrange that~\eqref{l:EL4} holds.
This procedure works independent of the masses~$m_\beta$ and~$\tilde{m}_\beta$.

\subsectionn{The Current and Mass Terms} \label{l:seccurrent}
We now come to the analysis of the interaction. More precisely, we want to study the effect 
of the fermionic wave functions in~\eqref{l:particles} and of
the chiral potentials~\eqref{l:Bform} in the Dirac operator~\eqref{l:Dinteract} on the EL equations
to degree four. As in Section~\ref{s:sec8} we consider the contribution near the origin in a
Taylor expansion around~$\xi=0$.
\begin{Def} \label{l:def82} The integrand in~\eqref{l:asy} is said to be
of {\bf{order~$o(|\vec{\xi}|^k)$ at the origin}} if the function~$\eta$ is
in the class~$o((|\xi^0| + |\vec{\xi}|)^{k+L})$.
Likewise, a contribution to the fermionic projector of the form~$P(x,y) \asymp \eta(x,y)\: T^{(n)}$
is of the order~$o(|\vec{\xi}|^k)$ if~$\eta \in o((|\xi^0| + |\vec{\xi}|)^{k+1-n})$.
\end{Def}
\nindex{ce2@$o( \vert \vec{\xi} \vert^k)$ -- order at the origin}%
\sindex{order!at the origin}%

Before stating the main result, we define the bosonic current~$j_{L\!/\!R}$ and the
Dirac current~$J_{L\!/\!R}$ by
\begin{align}
j_{L\!/\!R}^k &= \partial^k_{\;j}A^j_{L\!/\!R} - \Box A_{L\!/\!R} \label{l:jdef} \\
(J^k_{L\!/\!R})^{(i,\alpha)}_{(j, \beta)}
&= \sum_{l=1}^{\np} \overline{\psi_l^{(j, \beta)}} \chi_{R\!/\!L} \gamma^k \psi_l^{(i,\alpha)}
- \sum_{l=1}^{\na} \overline{\phi_l^{(j, \beta)}} \chi_{R\!/\!L} \gamma^k \phi_l^{(i,\alpha)} \:. \label{l:Jdef}
\end{align}
\nindex{dg6@$j_{\LR}$ -- chiral bosonic current}%
\nindex{ce8@$J_{\LR}$ -- chiral Dirac current}%
\sindex{fermionic projector!Dirac current term}%
\sindex{fermionic projector!chiral bosonic current term}%
Note that, due to the dependence on the isospin and generation indices, these currents
are $6 \times 6$-matrices. We also point out that for the sake of brevity, in~\eqref{l:jdef} we omitted the terms
quadratic in the potentials which arise for a non-abelian gauge group.
But as the form of these quadratic terms is uniquely determined from the well-known behavior under
gauge transformations, they could be inserted into all our equations in an obvious way.
Similar to the notation~\eqref{l:accents}, we denote the
sectorial projection by~$\hat{\jmath}$ and~$\hat{J}$.
Moreover, we introduce the $2 \times 2$-matrix-valued vector field~${\mathfrak{J}}_L$ by
\nindex{dg8@${\mathfrak{J}}_c$ -- matrix composed of current and mass terms}%
\begin{align}
\!\!\!{\mathfrak{J}}^k_L =&\; \hat{J}_R^k\: K_1 + \hat{\jmath}_L^k \:K_2  + \hat{\jmath}_R^k \:K_3
\label{l:Jterm} \\ 
&-3 m^2  \left( \acute{A}^k_L Y \grave{Y} + \acute{Y} Y \grave{A}^k_L \right) 
K_4 \label{l:mterm1} \\ 
&+ m^2 \left( \hat{A}_L^k \: \acute{Y} \grave{Y} + \acute{Y} \grave{Y}\: \hat{A}_L^k \right)
K_4 \label{l:mterm2} \\ 
&-3 m^2 \left( \acute{A}_R^k Y \grave{Y} - 2 \acute{Y} A_L^k \,\grave{Y}
+ \acute{Y} Y \grave{A}_R^k \right) K_5 \label{l:mterm3} \\ 
&-6 m^2 \left( \acute{A}_L^k \grave{Y}\: \hat{Y} + \hat{Y}\: \acute{Y} \grave{A}_L^k \right)
K_6 \label{l:mterm4} \\ 
&+6 m^2 \left( \hat{Y} 
\acute{A}_L^k \grave{Y} + \acute{Y} \grave{A}_L^k \: \hat{Y} \right) K_7 \label{l:mterm5} \\ 
&+m^2 \left( \hat{A}_L^k \hat{Y} \hat{Y} + 2 \hat{Y} \hat{A}_R^k \hat{Y}
+ \hat{Y} \hat{Y} \hat{A}_L^k \right) K_6 \label{l:mterm6} \\ 
&-m^2 \left( \hat{A}_R^k \hat{Y} \hat{Y} + 2 \hat{Y} \hat{A}_L^k \hat{Y}
+ \hat{Y} \hat{Y} \hat{A}_R^k \right) K_7 \:, \label{l:mterm7} 
\end{align}
where~$K_1, \ldots, K_7$ are the expressions
\begin{align*}
K_1 &= -\frac{3}{16 \pi}\: \frac{1}{\overline{T^{(0)}_{[0]}}}
\left[T^{(-1)}_{[0]} T^{(0)}_{[0]}\: \overline{T^{(-1)}_{[0]}} - c.c. \right] \\
K_2 &= \frac{3}{4} \: \frac{1}{\overline{T^{(0)}_{[0]}}}
 \Big[ T^{(0)}_{[0]} T^{(0)}_{[0]} \:\overline{T^{(-1)}_{[0]} T^{(0)}_{[0]} }
- c.c. \Big] \\
K_3 &= \frac{3}{2}\: \frac{1}{\overline{T^{(0)}_{[0]}}}\:
\Big[ T^{(-1)}_{[0]} T^{(1)}_{[0]} \:\overline{T^{(-1)}_{[0]} T^{(0)}_{[0]}} - c.c. \Big] \\
K_4 &= \frac{1}{4}\: \frac{1}{\overline{T^{(0)}_{[0]}}}\:
\Big[\:T^{(0)}_{[0]} T^{(0)}_{[2]} \:\overline{T^{(-1)}_{[0]} T^{(0)}_{[0]}} - c.c. \Big] \\
K_5 &= \frac{1}{4}\: \frac{1}{\overline{T^{(0)}_{[0]}}}\:
\Big[ T^{(-1)}_{[0]} T^{(1)}_{[2]} \overline{T^{(-1)}_{[0]} T^{(0)}_{[0]}} - c.c. \Big] \\
K_6 &= \frac{1}{12} \: \frac{T^{(0)}_{[0]} \overline{T^{(0)}_{[0]}}}{\overline{T^{(0)}_{[0]}}}\:
\frac{\Big( T^{(0)}_{[1]} \overline{T^{(-1)}_{[0]}} - T^{(-1)}_{[0]}
\overline{T^{(0)}_{[1]}} \Big)^2}
{T^{(0)}_{[0]} \overline{T^{(-1)}_{[0]}} - T^{(-1)}_{[0]} \overline{T^{(0)}_{[0]}}} \\
K_7 &= \frac{1}{12} \: \frac{T^{(-1)}_{[0]} \overline{T^{(-1)}_{[0]}}}{\overline{T^{(0)}_{[0]}}}\:
\frac{ \Big( T^{(0)}_{[1]} \overline{T^{(0)}_{[0]}} - T^{(0)}_{[0]}
\overline{T^{(0)}_{[1]}} \Big)^2}
{T^{(0)}_{[0]} \overline{T^{(-1)}_{[0]}} - T^{(-1)}_{[0]} \overline{T^{(0)}_{[0]}}} \:,
\end{align*}
evaluated weakly on the light cone~\eqref{l:asy} (and $c.c.$ denotes the complex conjugate).
Similarly, the matrix~${\mathfrak{J}}_R$ is defined by the replacements~$L \leftrightarrow R$.

\begin{Lemma} \label{l:lemma44}
The contribution of the bosonic current~\eqref{l:jdef} and of the Dirac
current~\eqref{l:Jdef} to the order~$(\deg < 4) + o(|\vec{\xi}|^{-3})$ 
are of the form~\eqref{l:DelQ} and~\eqref{l:Rdef} with
\[ {\mathscr{K}}_{L\!/\!R} =  i \xi_k \:\mathfrak{J}_{L\!/\!R}^k \:+\: (\deg < 4) + o \big( |\vec{\xi}|^{-3} \big) \:. \]
\end{Lemma}
\Proof
The perturbation of the eigenvalues is obtained by a perturbation calculation to first and second
order (see~\cite[Appendix~G]{PFP} and Appendix~\ref{s:appspec}). The resulting matrix traces are
computed most conveniently in the double null spinor frame~$(\mathfrak{f}^{L\!/\!R}_\pm)$
with the methods described in Appendix~\ref{s:appspec}. One finds that~$\Delta A$ is diagonal on
the degenerate subspaces, so that the second order contribution is given by~\eqref{l:secondorder}.
Moreover, the gauge phases~$\nu_{nc}$ in the unperturbed eigenvalues~\eqref{l:lambdaL}
and~\eqref{l:lambdaR} only affect the error term~$o(|\vec{\xi}|^{-3})$.
We conclude that Lemma~\ref{l:lemma41} applies, and thus~${\mathscr{K}}_L$
and~${\mathscr{K}}_R$ are well-defined.

In order to compute~${\mathscr{K}}_{L\!/\!R}$, we need to take into account the following
contributions to the light-cone expansion of the fermionic projector:
\begin{align*}
\chi_L \,P(x,y) \asymp\:&
-\frac{1}{2} \:\chi_L\:\slashed{\xi}\, \xi_i  \int_x^y [0,0 \,|\, 1]\: j_L^i\: T^{(0)} \\
&-\chi_L\: \int_x^y [0,2 \,|\, 0]\, j_L^i\,\gamma_i\: T^{(1)} \\
&-i m\, \chi_L\:\xi_i \int_x^y Y A_R^i \: T^{(0)} \\
&+\frac{im}{2}\:\chi_L\:\slashed{\xi} \int_x^y (Y \slashed{A}_R - \slashed{A}_L Y)\: T^{(0)} \\
&+im\,\chi_L \int_x^y [0,1 \,|\, 0] \left( Y (\partial_j A_R^j) - (\partial_j A_L^j) \,Y \right) T^{(1)} \\
&+\frac{m^2}{2}\:\chi_L\:\slashed{\xi} \,\xi_i \int_x^y [1,0 \,|\, 0]\, Y Y A_L^i \: T^{(0)} \\
&+\frac{m^2}{2}\:\chi_L\:\slashed{\xi} \,\xi_i \int_x^y [0,1 \,|\, 0]\, A_L^i Y Y \: T^{(0)} \\
&+m^2\,\chi_L \int_x^y [1,0 \,|\, 0]\, Y Y \slashed{A}_L\: T^{(1)} \\
&-m^2\,\chi_L \int_x^y [0,0 \,|\, 0]\, Y \slashed{A}_R \,Y\: T^{(1)} \\
&+m^2\,\chi_L \int_x^y [0,1 \,|\, 0]\, \slashed{A}_L \,Y Y \: T^{(1)}
\end{align*}
(for the derivation see~\cite[Appendix~B]{PFP} and~\cite[Appendix~A]{light};
cf.\ also Appendix~\ref{s:appspec}).
A long but straightforward calculation (which we carried out with the help of
the C++ program {\textsf{class\_commute}}
\sindex{computer algebra}%
and an algorithm implemented
in {\textsf{Mathematica}}) gives the result\footnote{The
C++ program {\textsf{class\_commute}} and its computational output as well as the
Mathematica worksheets were included as ancillary files to the arXiv submission arXiv:1211.3351 [math-ph].}.

We finally mention a rather subtle point in the calculation:
According to~\eqref{l:lambdaL} and~\eqref{l:lambdaR}, our unperturbed eigenvalues
involve gauge phases and can thus be expanded in powers of~$A_c^k \xi_k$.
As a consequence, we must take into account
contributions of the form~\eqref{l:secondorder} where the factors~$\Delta A$
involve no gauge potentials, but the unperturbed eigenvalues~$\lambda_{ncs}$
are expanded linearly in~$A_c^k \xi_k$.
In this case, the corresponding contributions involving no factors of~$A_c^k \xi_k$
can be identified with contributions to the eigenvalues in
the vacuum in~\eqref{l:lampm} and~\eqref{l:lamL}, \eqref{l:lamR}. Using that
the vacuum eigenvalues all have the same absolute value, the contributions
linear in~$A_c^k \xi_k$ can be simplified to obtain the formulas for~${\mathfrak{J}}^k_c$ listed above.
Another, somewhat simpler method to get the same result is to use
that the operator~$Q$ is symmetric~\eqref{l:Qsymm} (see~\cite[Lemma~3.5.1]{PFP}).
Thus it suffices to compute the symmetric part $(\Delta Q(x,y) + \Delta Q(y,x)^*)/2$ of the
operator~$\Delta Q$ as defined by~\eqref{l:EL4}. This again gives the above formulas
for~${\mathfrak{J}}^k_c$, without using any relations between the vacuum eigenvalues.
\QED

Let us briefly discuss the obtained formula for~${\mathfrak{J}}_R$. The summands in~\eqref{l:Jterm}
involve the chiral gauge currents and Dirac currents; they can be understood in analogy to the current terms
in~\S\ref{s:sec81} and~\S\ref{s:sec82}. The contributions~\eqref{l:mterm1}-\eqref{l:mterm7} are the
mass terms. They are considerably more complicated than in~\S\ref{s:sec81}. These complications
are caused by the fact that we here consider left- and right-handed gauge potentials acting on two sectors,
involving a mixing of the generations. In order to clarify the structure of the mass terms, it is instructive
to look at the special case of a $U(1)$ vector potential, i.e.\ $A_L=A_R = A \!\cdot\! \1_{\C^2}$
(with a real vector field~$A$). In this case, the terms~\eqref{l:mterm1} and~\eqref{l:mterm2} cancel each
other (note that~$\hat{A} \acute{Y} \cdots = 3 \acute{A} Y \cdots$), and~\eqref{l:mterm3} vanishes.
Similarly, the summand~\eqref{l:mterm4} cancels~\eqref{l:mterm6}, and~\eqref{l:mterm5} cancels~\eqref{l:mterm7}. Thus the mass terms are zero, in agreement with local gauge invariance.

\subsectionn{The Microlocal Chiral Transformation} \label{l:secmicroloc}
\sindex{transformation of the fermionic projector!microlocal chiral}%
The simple fractions~$K_3$ and~$K_5$ involve factors~$T^{(1)}_\circ$ which have a logarithmic
pole on the light cone. Before working out the field equations, we must compensate these logarithmic
poles by a suitable transformation of the fermionic projector. 
We again work with a microlocal chiral transformation as developed
in~\S\ref{s:secprobaxial}--\S\ref{s:secshearmicro}.
As the generalizations to a system of two sectors is not straightforward, we
give the necessary constructions step by step.
Before beginning, we mention for clarity
that in the following sections~\S\ref{l:secmicroloc} and~\S\ref{l:secshear}
we will construct contributions to~$P(x,y)$ which enter
the EL equations to degree four only linearly. Therefore, it is obvious that Lemma~\ref{l:lemma41}
again applies.

As in~\S\ref{s:secnonlocaxial} we begin in the homogeneous setting and work in momentum space.
Then the logarithmic poles on the light cone correspond to a contribution to the fermionic projector
of the form
\beq \label{l:wanted}
\tilde{P}(k) \asymp \left( \chi_L \,\slashed{v}_L + \chi_R \,\slashed{v}_R \right) \delta'(k^2)\, \Theta(-k^0) \:,
\eeq
where the vector components~$v_L^j$ are Hermitian $2 \times 2$-matrices acting on the sector index.
In order to generate the desired contribution~\eqref{l:wanted}, we consider a
homogeneous transformation of the fermionic projector of the vacuum of the form
\beq \label{l:Pkans}
\tilde{P}(k) = \acute{U}(k)\, P^\text{aux}(k)\, \grave{U}(k)^*
\eeq
\nindex{dh0@$U(k)$ -- homogeneous transformation}%
with a multiplication operator in momentum space~$U(k)$.
With the operator~$U(k)$ we want to modify the states of vacuum Dirac sea with the
aim of generating a contribution which can compensate the logarithmic poles.
We denote the absolute value of the energy of the states by~$\Omega=|k^0|$.
\nindex{dh2@$\Omega$ -- absolute value of energy}%
We are mainly interested in the regime~$m \ll \Omega \ll \varepsilon^{-1}$ where regularization effects
play no role. Therefore, we may disregard the right-handed high-energy states
and write the vacuum fermionic projector according to~\eqref{l:Pvac}, \eqref{l:Pmdef} and
\eqref{l:massneutrino}. Expanding in the mass, we obtain
\beq \label{l:Pauxex}
P^\text{aux} = (\slashed{k} + mY)\, \delta(k^2)\, \Theta(-k^0) - (\slashed{k}+mY) \:m^2 Y^2\, \delta'(k^2)\, \Theta(-k^0)
+ (\deg < 0)\:.
\eeq
For the transformation~$U(k)$ in~\eqref{l:Pkans} we take the ansatz
\beq \label{l:Uans}
U(k) = \1 + \frac{i}{\sqrt{\Omega}} \:Z(k) \qquad \text{with} \qquad
Z = \chi_L\, L^j \gamma_j + \chi_R \,R^j \gamma_j \:,
\eeq
\nindex{dh4@$Z(k)$ -- generator of homogeneous transformation}%
\nindex{dh6@$L(k), R(k)$ -- chiral components of~$Z(k)$}%
where~$L^j$ and~$R^j$ are $6 \times 6$-matrices
(not necessarily Hermitian) which act on the generation and sector indices.
For simplicity, we assume that the dependence on the vector index can be written as
\beq \label{l:LRsimp}
L^j = L\, v_L^j \qquad \text{and} \qquad R^j = R\, v_R^j
\eeq
with real vector fields~$v_L$ and~$v_R$ (and $6 \times 6$-matrices~$L$ and~$R$).
The ansatz~\eqref{l:Uans} can be regarded as the linear Taylor expansion of the
exponential~$U = \exp(i Z/\sqrt{\Omega})$,
giving agreement to~\S\ref{s:secnonlocaxial} (in view of the fact that the quadratic and higher orders of
this Taylor expansion dropped out in~\S\ref{s:secnonlocaxial}, for simplicity we leave them out here).
Note that the operator~$U(k)$ is in general not unitary (for details see Remark~\ref{l:remunitary} below).

Applying the transformation~\eqref{l:Pkans} and~\eqref{l:Uans} to~\eqref{l:Pauxex}, only 
the isospin matrices are influenced. A short calculation gives
\begin{align}
\chi_L \,\acute{U} \,&(\slashed{k}+mY) \,\grave{U}^* = \chi_L (3 \slashed{k} +m \hat{Y}) \label{l:unper1} \\
&+ \frac{i}{\sqrt{\Omega}} \:\chi_L \left( \hat{\slashed{L}} \slashed{k} - \slashed{k} {\hat{\slashed{R}}}^* \right) 
+ \frac{i m}{\sqrt{\Omega}} \:\chi_L \left( \acute{\slashed{L}} \grave{Y} - \acute{Y} {\grave{\slashed{L}}}^* \right)
\label{l:forder} \\
&+ \frac{1}{\Omega} \:\chi_L \acute{\slashed{L}} \slashed{k} \grave{\slashed{L}}^*
+ \frac{m}{\Omega} \:\chi_L \acute{\slashed{L}} Y \grave{\slashed{R}}^* \label{l:sorder} \\
\chi_L \,\acute{U} \,&(\slashed{k}+mY)\,m^2 Y^2\, \grave{U}^* = \chi_L (\slashed{k}\: m^2 \acute{Y} \grave{Y} 
+m^3\, \acute{Y} Y \grave{Y}) \label{l:unper2} \\
&+ \frac{i m^2}{\sqrt{\Omega}} \:\chi_L \left( \acute{\slashed{L}} Y \grave{Y} \,\slashed{k}
- \slashed{k} \,\acute{Y} Y {\grave{\slashed{R}}}^* \right) 
+ \frac{i m^3}{\sqrt{\Omega}} \:\chi_L \left( \acute{\slashed{L}} YY \grave{Y} - \acute{Y} YY
{\grave{\slashed{L}}}^* \right) \label{l:ford} \\
& + \frac{m^2}{\Omega} \:\chi_L \acute{\slashed{L}} \slashed{k}\, Y^2 \grave{\slashed{L}}^*
+ \frac{m^3}{\Omega} \:\chi_L \acute{\slashed{L}} Y^3 \grave{\slashed{R}}^* \label{l:sord}
\end{align}
(and similarly for the right-handed component).
Let us discuss the obtained contributions. Clearly, the terms~\eqref{l:unper1} and~\eqref{l:unper2}
are the unperturbed contributions.
Generally speaking, due to the factor~$\delta(k^2)$ in~\eqref{l:Pauxex},
the contributions~\eqref{l:forder} and~\eqref{l:ford} are singular on the light cone
and should vanish, whereas the desired logarithmic contribution~\eqref{l:wanted}
must be contained in~\eqref{l:sorder} or~\eqref{l:sord}.
The terms~\eqref{l:forder} of order~$\Omega^{-\frac{1}{2}}$ contribute to the EL equations
to degree five on the light cone. Thus in order for them to vanish, we need to impose that
\begin{align}
\hat{L} &= 0 = \hat{R} \qquad \label{l:1cond} \\
\acute{L} \grave{Y} - \acute{Y} {\grave{L}}^* & = 0 =
\acute{R} \grave{Y} - \acute{Y} {\grave{R}}^* \:. \label{l:2cond}
\end{align}
The last summand in~\eqref{l:sorder} does not involve a factor~$\slashed{k}$ and is
even. As a consequence, it only enters the EL equations in combination with another factor of~$m$,
giving rise to a contribution of degree three on the light cone (for details see Lemma~\ref{s:lemmascal}).
With this in mind, we may disregard the last summand in~\eqref{l:sorder}.
Similarly, the last summand in~\eqref{l:sord} and the first summand in~\eqref{l:ford} are even
and can again be omitted. In order for the second summand in~\eqref{l:ford} to vanish,
we demand that
\beq
\acute{L} YY \grave{Y} - \acute{Y} YY {\grave{L}}^* = 0 =
\acute{R} YY \grave{Y} - \acute{Y} YY {\grave{R}}^* \:. \label{l:3cond}
\eeq
Then it remains to consider the first summand in~\eqref{l:sorder} and the first summand
in~\eqref{l:sord}. We thus end up with a left-handed (and similarly right-handed)
contribution to the fermionic projector of the form
\beq \label{l:effcont}
\chi_L \tilde{P}(k) \asymp
\frac{1}{\Omega} \:\chi_L \acute{\slashed{L}} \slashed{k} \grave{\slashed{L}}^*\:\delta(k^2)\, \Theta(-k^0)
- \frac{m^2}{\Omega} \:\chi_L \acute{\slashed{L}} \slashed{k}\, Y^2 \grave{\slashed{L}}^*\:
\delta'(k^2)\, \Theta(-k^0) \:.
\eeq
Note that the conditions~\eqref{l:1cond}--\eqref{l:3cond} are linear in~$L$ and~$R$,
whereas the contribution~\eqref{l:effcont} is quadratic.

Before going on, we remark that at first sight, one might want to replace
the conditions~\eqref{l:2cond} and~\eqref{l:3cond} by the weaker conditions
\beq \label{l:Yfalse}
\begin{split}
\acute{L} \grave{Y} - \acute{Y} {\grave{L}}^* & =
\acute{R} \grave{Y} - \acute{Y} {\grave{R}}^* = i v_1(k)\, \1_{\C^2} \\
\acute{L} YY \grave{Y} - \acute{Y} YY {\grave{L}}^* &=
\acute{R} YY \grave{Y} - \acute{Y} YY {\grave{R}}^* = i v_3(k)\, \1_{\C^2}
\end{split}
\eeq
involving two real-valued vector fields~$v_1$ and~$v_3$. Namely, as the resulting
contribution to the fermionic projector acts trivially on the isospin index and is symmetric under
the replacement~$L \leftrightarrow R$,
it perturbs the eigenvalues of the closed chain in a way that the absolute values
of all eigenvalues remain equal, so that the EL equations are still satisfied.
However, this argument is too simple because the gauge phases must be taken
into account. For the contributions in~\eqref{l:effcont}, the methods in~\S\ref{s:secshearmicro}
make it possible to arrange that the gauge phases enter in a way which is compatible with
the EL equations. For the contributions corresponding to~\eqref{l:Yfalse}, however, it is impossible
to arrange that the gauge phases drop out of the EL equations.
Hence~$v_1$ and~$v_3$ would necessarily enter the EL equations.
As the scaling factors~$1/\sqrt{\Omega}$ in~\eqref{l:forder} and~\eqref{l:ford} give rise to a different
$|\vec{\xi}|$-dependence, these contributions to the EL equations would have a different
scaling behavior in the radius. As a consequence, the EL equations would only be satisfied
if~$v_1 \equiv v_3 \equiv 0$.

For clarity, we want to focus our attention to the component of~\eqref{l:effcont}
which will give the dominant contribution to the EL equations.
For the moment, we only motivate in words how this component is chosen;
the detailed justification that the other components can really be neglected will be given
in the proof of Proposition~\ref{l:prphom} below.
In the EL equations, the chiral component~\eqref{l:effcont} is contracted with a factor~$\xi$.
This means in momentum space that the main contribution of~\eqref{l:effcont} to the
EL equations is obtained by contracting with a factor~$k$ (this will be justified in
detailed in the proof of Proposition~\ref{l:prphom} below). Therefore, we
use the anti-commutation relations to rewrite~\eqref{l:effcont} as
\[ \tilde{P}(k) = \chi_L\, P_L^j(k) \gamma_j + \chi_R\, P_R^j(k) \gamma_j \]
(here we use the specific form~\eqref{l:LRsimp} of our ansatz). We now contract with~$k$ to obtain
\beq \label{l:Pcontract}
\begin{split}
P_L[k] := P_L^j(k)\, k_j &= \frac{1}{\Omega} \left( 2 \acute{L}_i \grave{L}_j^*\: k^i k^j
- k^2\: \acute{L}^j \grave{L}^*_j \right) \delta(k^2)\, \Theta(-k^0) \\
&\quad - \frac{m^2}{\Omega} \left( 2 \acute{L}_i\,Y^2\, \grave{L}_j^* k^i k^j
- k^2\: \acute{L}^j \,Y^2\,\grave{L}^*_j \right) \delta'(k^2)\, \Theta(-k^0) \:.
\end{split}
\eeq
As the factor~$k^2$ vanishes on the mass shell, we may omit the resulting terms
(for details see again the proof of Proposition~\ref{l:prphom} below). We thus obtain
\beq \label{l:PLk}
P_L[k] =  \frac{2}{\Omega}\: L[k]\, L[k]^* \:\delta(k^2)\, \Theta(-k^0)
- \frac{2}{\Omega}\: L[k]\,m^2 Y^2\, L[k]^* \:\delta'(k^2)\, \Theta(-k^0) \:,
\eeq
where we set~$L[k] = \acute{L}_j(k)\, k^j$ (note that~$L[k]$ is a $2 \times 6$-matrix,
and the star simply denotes the adjoint of this matrix).
The right-handed component is obtained by the obvious replacements~$L \rightarrow R$.

Let us work out the conditions needed for generating a contribution of the desired form~\eqref{l:wanted}.
Similar as explained in~\S\ref{s:secnonlocaxial}, the first summand in~\eqref{l:PLk} necessarily gives a
contribution to the fermionic projector.
For this contribution to drop out of the EL equations,
we need to impose that it is vectorial and proportional to the identity matrix, i.e.\
\beq \label{l:4con}
L[k]\, L[k]^* = R[k]\, R[k]^* = {\mathfrak{c}}_0(k)\, \1_{\C^2}
\eeq
with some constant~${\mathfrak{c}}_0(k)$.
\nindex{dh7@${\mathfrak{c}}_0(k), {\mathfrak{c}}_2(k)$ -- parameters in microlocal chiral transformation}%
In order to better justify that~\eqref{l:4con} is a necessary condition, we remark that 
the contribution to the fermionic projector corresponding to the first summand in~\eqref{l:PLk}
is of the form~$P \simeq \chi_L\, \slashed{v}\, T^{[0]}_{[1]}$. The resulting contribution to the
eigenvalues of the closed chain is~$\Delta \lambda_{L+} \simeq i v_j \xi^j T^{(0)}_{[1]} \overline{T^{(-1)}_{[0]}}$,
and a direct computation shows that this gives rise to a non-trivial contribution to the
EL equations unless~\eqref{l:4con} holds.

The second summand in~\eqref{l:PLk} is of the desired form~\eqref{l:wanted}. 
Keeping in mind that we may again allow for
a vector contribution proportional to the identity, we get the conditions
\beq \label{l:5con} \begin{split}
L[k]\, m^2 Y^2\, L[k]^* &= \frac{\Omega}{2}\: v_L[k] + {\mathfrak{c}}_2(k)\, \1_{\C^2} \\
R[k]\, m^2 Y^2\, R[k]^* &=  \frac{\Omega}{2}\: v_R[k] + {\mathfrak{c}}_2(k)\, \1_{\C^2} \:,
\end{split}
\eeq
where we set~$v_{L\!/\!R}[k] = v_{L\!/\!R}^j(k)\, k_j$
(and~${\mathfrak{c}}_2$ is another free constant).
\nindex{dh7@${\mathfrak{c}}_0(k), {\mathfrak{c}}_2(k)$ -- parameters in microlocal chiral transformation}%
Our task is to solve the quadratic equations~\eqref{l:4con} and~\eqref{l:5con}
under the linear constraints~\eqref{l:1cond}--\eqref{l:3cond}.
Moreover, in order to compute the smooth contribution to the fermionic projector,
we need to determine the expectation values involving the logarithms of the masses
\beq \label{l:logcon}
L[k] \left( m^2 Y^2\, \log(m Y) \right) L[k]^* \qquad \text{and} \qquad
R[k] \left( m^2 Y^2\, \log(m Y) \right) R[k]^* \:.
\eeq

We next describe a method for treating the quadratic equations~\eqref{l:4con} and~\eqref{l:5con}
under the linear constraints~\eqref{l:1cond}
(the linear constraints~\eqref{l:2cond} and~\eqref{l:3cond} will be treated afterwards).
We first restrict attention to the left-handed component and consider the corresponding
equations in~\eqref{l:1cond}, \eqref{l:4con} and~\eqref{l:5con}
(the right-handed component can be treated similarly). We write the matrix $L[k]$ in components,
\beq \label{l:Lcomp}
L[k] = \begin{pmatrix} l_{11} & l_{12} & l_{13} & l_{14} & l_{15} & l_{16} \\
l_{21} & l_{22} & l_{23} & l_{24} & l_{25} & l_{26}
\end{pmatrix} ,
\eeq
where the matrix entries~$l_{ab}$ are complex numbers.
We use the linear relations~\eqref{l:1cond} to express the third and sixth columns
of the matrices by
\beq \label{l:linrel}
l_{a3} = -l_{a1} - l_{a2} \:,\quad l_{a6} = -l_{a4} - l_{a5} \qquad (a=1,2)\:.
\eeq
This reduces the number of free parameters to~$8$ complex
parameters, which we combine to the matrix
\beq \label{l:Psidef}
\psi = \begin{pmatrix} \psi_1 \\ \psi_2 \end{pmatrix} \qquad \text{with} \qquad
\psi_a = (l_{a1}, l_{a2}, l_{a4}, l_{a5}) \:.
\eeq
We introduce on~$\C^4$ the scalar product~$\la.,. \ra_0$ as well as the positive
semi-definite inner product~$\la .,. \ra_2$ by
\beq \label{l:scalrep}
\la \psi_a, \psi_b \ra_0 = (L[k] \,L[k]^*)^a_b \qquad \text{and} \qquad
\la \psi_a, \psi_b \ra_2 = (L[k] \,m^2 Y^2\, L[k]^*)^a_b
\eeq
(where we implicitly use~\eqref{l:linrel} to determine the third and sixth columns of~$L[k]$).
We represent these scalar products with signature matrices,
\[ \la \psi, \phi \ra_0 = \la \psi, S_0\, \phi \ra_{\C^4}\:,\qquad
\la \psi, \phi \ra_2 = \la \psi, S_2\, \phi \ra_{\C^4}\:. \]
\nindex{dh8@$S_0, S_2$ -- signature matrices}%
Expressing~$\la .,. \ra_2$ in terms of~$\la .,. \ra_0$,
\[ \la \psi, \phi \ra_2 = \la \psi , S_0^{-1} S_2 \,\phi \ra_0\:, \]
the resulting linear operator~$S_0^{-1} S_2$ is symmetric with respect to~$\la .,. \ra_0$.
Thus by diagonalizing the matrix~$S_0^{-1} S_2$, one can construct an eigenvector
basis~$e_1, \ldots, e_4$ which is orthonormal with respect to~$\la .,. \ra_0$, i.e.\
\beq \label{l:mudef}
\la e_a, e_b \ra_0 = \delta_{ab} \:,\qquad
\la e_a, e_b \ra_2 = \mu_a\: \delta_{ab} \:.
\eeq
As the matrices have real-valued entries, we can choose the eigenvectors~$e_a$
such that all their components are real.
Moreover, as the matrices~$S_0$ and~$S_2$ are block-diagonal in the isospin index,
we may choose the eigenvectors such that~$e_1$ and~$e_2$ have isospin up,
whereas~$e_3$ and~$e_4$ have isospin down, i.e.
\[ e_1, e_2 = (*,*,0,0) \:,\qquad e_3, e_4 = (0,0,*,*) \]
(where the star stands for an arbitrary real-valued entry).
Finally, we always order the eigenvalues and eigenvectors such that
\beq \label{l:signconv}
0 \leq \mu_1 \leq \mu_2 \qquad \text{and} \qquad \mu_3 \leq \mu_4\:.
\eeq
Writing the vectors~$\psi_a$ in~\eqref{l:Psidef} in this eigenvector basis,
\beq \label{l:psiadef}
\psi_a = \sum_{d=1}^4 \psi_a^d e_d\:,
\eeq
we can express~\eqref{l:scalrep} in the simpler form
\beq \label{l:scalrep2}
(L[k] \,L[k]^*)^a_b = \sum_{d=1}^4 \overline{\psi_a^d} \psi_b^d \:,\qquad
(L[k] \,m^2 Y^2 \,L[k]^*)^a_b = \sum_{d=1}^4 \mu_d \:\overline{\psi_a^d} \psi_b^d \:.
\eeq
Moreover, the linear condition~\eqref{l:1cond} is satisfied.

In order to treat the remaining linear constraints~\eqref{l:2cond} and~\eqref{l:3cond}, we
decompose the coefficients in~\eqref{l:psiadef} into their real and imaginary parts,
\[ \psi_{1\!/\!2}^d = a_{1\!/\!2}^d + i\, b_{1\!/\!2}^d \:,\qquad (d=1,\ldots, 4) \:. \]
Considering the diagonal entries of~\eqref{l:2cond} and~\eqref{l:3cond} shows that
\[ b_1^1=b_1^2 = b_2^3=b_2^4 = 0 \:. \]
The off-diagonal entries make it possible to express~$a_1^3$, $a_1^4$ in terms of~$a_2^1$, $a_2^2$
and~$b_1^3$, $b_1^4$ in terms of ~$b_2^1$, $b_2^2$, leaving us
with the eight real parameters~$a_1^1, a_1^2, a_2^3, a_2^4$
and~$a_2^1, a_2^2, b_2^1, b_2^2$.

In order to simplify the setting, it is useful to observe that all our constraints are invariant
if we multiply the rows of the matrix~$\psi$ in~\eqref{l:Psidef} by phase factors according to
\beq \label{l:phase}
\psi_1 \rightarrow e^{i \varphi}\, \psi_1\:, \qquad
\psi_2 \rightarrow e^{-i \varphi}\, \psi_2
\qquad \text{with} \qquad \varphi \in \R\:.
\eeq
These transformations only affect the off-diagonal isospin components of the left-handed
matrix in~\eqref{l:5con}. With this in mind, we can assume that
this matrix has real components and can thus be decomposed in terms of Pauli matrices as
\beq \label{l:m2Lrep}
L[k]\, m^2 Y^2 L[k]^* = t \,\1 + x\, \sigma^1 + z\, \sigma^3 \:.
\eeq
Using this equation in~\eqref{l:scalrep2} and evaluating the real part of the off-diagonal elements of~\eqref{l:4con},
one finds that~$b_2^1= 0 = b_2^2$, leaving us with the six real
parameters~$a_1^1, a_1^2, a_2^1, a_2^2, a_2^3, a_2^4$.
With these six parameters, we need to satisfy three quadratic relations in~\eqref{l:m2Lrep}
and three quadratic relations in~\eqref{l:4con}. This suggests that for given
parameters~${\mathfrak{c}}_0$ and~${\mathfrak{c}}_2$
as well as~$t, x, z$, there should be a discrete (possibly empty) set of solutions.

In preparation, we analyze the case when all potentials are diagonal in the isospin index.
\begin{Example} {\bf{(isospin-diagonal potentials)}} \label{l:example45} {\em{
Assume that the parameter~$x$ in~\eqref{l:m2Lrep} vanishes.
Evaluating the real part of the off-diagonal components of~\eqref{l:4con} and~\eqref{l:m2Lrep},
one finds that~$a_2^1=0=a_2^2$. The diagonal components of~\eqref{l:4con} and~\eqref{l:m2Lrep}
give the quadratic equations
\begin{align}
(a^1_1)^2 &= \frac{-t-z+{\mathfrak{c}}_0 \,\mu_2}{\mu_2 - \mu_1} \:,&
(a^1_2)^2 &= \frac{t+z-{\mathfrak{c}}_0 \,\mu_1}{\mu_2 - \mu_1} \label{l:qusys1} \\ 
(a^2_3)^2 &= \frac{-t+z+{\mathfrak{c}}_0 \,\mu_4}{\mu_4 - \mu_3} \:, &
(a^2_4)^2 &= \frac{t-z-{\mathfrak{c}}_0 \,\mu_3}{\mu_4 - \mu_3} \:. \label{l:qusys2}
\end{align}
For these equations to admit solutions, we need to assume the non-degeneracies
\[  \mu_2 \neq \mu_1 \qquad \text{and} \qquad \mu_3 \neq \mu_4 \:. \]
Then there are solutions if and only if all the squares are non-negative. In view of
our sign conventions~\eqref{l:signconv}, we obtain the conditions
\beq \label{l:lamineq}
{\mathfrak{c}}_0 \,\mu_1 \leq t+z \leq {\mathfrak{c}}_0 \,\mu_2 \qquad \text{and} \qquad
{\mathfrak{c}}_0 \,\mu_3 \leq t-z \leq {\mathfrak{c}}_0 \,\mu_4\:.
\eeq
Provided that these inequalities hold, the matrix entries~$a^1_1$, $a^1_2$, $a^2_3$
and~$a^2_4$ are uniquely determined up to signs. For any solution obtained in this way,
one can compute the logarithmic expectation value~\eqref{l:logcon}.

In order to analyze the conditions~\eqref{l:lamineq}, we first note that changing the constant~${\mathfrak{c}}_2$
corresponds to adding a constant to the parameter~$t$ (see~\eqref{l:m2Lrep} and~\eqref{l:5con}).
Hence we can always satisfy~\eqref{l:lamineq} by choosing~${\mathfrak{c}}_0$ and~${\mathfrak{c}}_2$ sufficiently large,
provided that
\beq \label{l:muin}
\mu_1 \leq \mu_4 \qquad \text{and} \qquad \mu_3 \leq \mu_2 \:.
\eeq
If conversely these conditions are violated, it is impossible to satisfy~\eqref{l:lamineq}
in the case~$z=0$. The physical meaning of the inequalities~\eqref{l:muin}
will be discussed in Remark~\ref{l:remlargemass} below.
\QEDrem
}} \end{Example}

In the next proposition, we use a perturbation argument to show that
the inequalities~\eqref{l:muin} guarantee the existence of the desired
homogeneous transformations even if off-diagonal isospin components are present.

\begin{Prp} \label{l:prphom}
Assume that the parameters~$\mu_1, \ldots, \mu_4$ as defined by~\eqref{l:mudef} and~\eqref{l:signconv}
satisfy the inequalities~\eqref{l:muin}.
Then for any choice of the chiral potentials~$v_L$ and~$v_R$ in~\eqref{l:wanted},
there is a homogeneous chiral transformation of the form~\eqref{l:Uans}
such that the transformed fermionic projector~\eqref{l:Pkans} is of the form
\begin{align}
\tilde{P}(k) &= P(k) + (\chi_L \slashed{v}_L + \chi_R \slashed{v}_R)\, T^{(1)}_{[3, \mathfrak{c}]} \label{l:des1} \\
&\quad+ \text{(vectorial)}\:\1_{\C^2} \:\delta(k^2) \Big(1+ \O(\Omega^{-1}) \Big) \label{l:des2} \\
&\quad+ \text{(vectorial)}\:\1_{\C^2} \:\delta'(k^2) \Big(1+ \O \big( \Omega^{-\frac{1}{2}} \big) \Big) \label{l:des3} \\
&\quad + \text{(pseudoscalar or bilinear)} \:\sqrt{\Omega} \:\delta'(k^2) \Big(1+ \O(\Omega^{-1}) \Big) \label{l:des4} \\
&\quad + \text{(higher orders in~$\varepsilon/|\vec{\xi}|$)}\:. \label{l:des5}
\end{align}
\end{Prp} \noindent
Before coming to the proof, we point out that the values of the parameters~${\mathfrak{c}}_0$
and~${\mathfrak{c}}_2$ are not determined by this proposition.
They can be specified similar as in~\S\ref{s:secnonlocaxial}
by choosing the homogeneous transformation
such that~${\mathfrak{c}}_0$ is minimal (see also Section~\ref{l:sec6}).
In order to clarify the dependence on~${\mathfrak{c}}_0$ and~${\mathfrak{c}}_2$,
we simply added a subscript~${\mathfrak{c}}$ to the factor~$T^{(1)}_{[3]}$.
Similar to~\eqref{s:logpole2}, this factor can be written in position space as
\[ T^{(1)}_{[3, {\mathfrak{c}}]} = \frac{1}{32 \pi^3} \Big( \log |\xi^2| + i \pi \,\Theta(\xi^2)\,
\epsilon(\xi^0) \Big) +  s_{[3, {\mathfrak{c}}]} \:, \]
where~$s_{[3, {\mathfrak{c}}]}$ is a real-valued
smooth function which depends on the choice of~${\mathfrak{c}}_0$ and~${\mathfrak{c}}_2$.
In fact, $s_{[p, {\mathfrak{c}}]}$ may even depend on the isospin components of~$v_L$ and~$v_R$;
but for ease in notation we shall not make this possible dependence explicit.

\Proof[Proof of Proposition~\ref{l:prphom}]
We first show that for sufficiently large~${\mathfrak{c}}_0$ and~${\mathfrak{c}}_2$,
there are solutions of~\eqref{l:m2Lrep} and of the left equation in~\eqref{l:4con}.
Evaluating the real part of the off-diagonal components of~\eqref{l:4con} and~\eqref{l:m2Lrep},
we get linear equations in~$a_2^1$ and~$a_2^2$, making it possible to express~$a_2^1$ and~$a_2^2$
in terms of~$a_1^1, a_1^2, a_2^3, a_2^4$. These relations do not involve~${\mathfrak{c}}_0$
nor~${\mathfrak{c}}_2$. As a consequence, the diagonal components of~\eqref{l:4con} and~\eqref{l:m2Lrep}
give a system of equations, which for large parameters~${\mathfrak{c}}_0$ and~${\mathfrak{c}}_2$ are a perturbation of the system~\eqref{l:qusys1} and~\eqref{l:qusys2}. Hence for sufficiently large~${\mathfrak{c}}_0$ and~${\mathfrak{c}}_2$, there are solutions by the implicit function theorem.

Repeating the above arguments for the right-handed potentials, we obtain matrices~$L[k]$
and~$R[k]$ such that~\eqref{l:4con} and~\eqref{l:5con} hold. Moreover, it is clear from our
constructions that~\eqref{l:1cond}, \eqref{l:2cond} and~\eqref{l:3cond} are satisfied.
It remains to go through all the contributions~\eqref{l:unper1}--\eqref{l:sord} and
to verify that they are of the form~\eqref{l:des1}--\eqref{l:des5}.
Clearly, \eqref{l:unper1} and~\eqref{l:unper2} combine to the summand~$P(k)$ in~\eqref{l:des1}.
The contributions in~\eqref{l:forder} vanish due to~\eqref{l:1cond} and~\eqref{l:2cond}.
The second summand in~\eqref{l:sorder} as well as the first summand in~\eqref{l:ford}
are of the form~\eqref{l:des4}. The second summand in~\eqref{l:ford} vanishes
in view of~\eqref{l:3cond}. Hence it really suffices to consider the first summand
in~\eqref{l:sorder} and the first summand in~\eqref{l:sord}, which were combined earlier
in~\eqref{l:effcont}.

It remains to justify the contraction with the momentum~$k$, which led us to analyze~\eqref{l:PLk}.
To this end, we need to consider the derivation of the weak evaluation formulas on the
light cone in~\cite[Chapter~4]{PFP}. More precisely, the expansion of
the vector component in~\cite[eq.~(4.4.6)--(4.4.8)]{PFP} shows that~$k$ and~$\xi$
are collinear, up to errors of the order~$\varepsilon/|\vec{\xi}|$.
Moreover, the terms in~\eqref{l:Pcontract} which involve a factor~$k^2$
are again of the order~$\varepsilon/|\vec{\xi}|$ smaller than the terms where the
factors~$k$ are both contracted to~$\acute{L}$ or~$\grave{L}^*$.
This explains the error term~\eqref{l:des5}.
\QED
We remark that the error term~\eqref{l:des5} could probably be improved
by analyzing those components of~$\acute{L}^j(k)$ which vanish
in the contraction~$\acute{L}^j(k)\, k^j$. Here we shall not enter this analysis
because errors of the order~$\varepsilon/|\vec{\xi}|$ appear anyway when
evaluating weakly on the light cone~\eqref{l:asy}.

Exactly as in~\S\ref{s:secgennonloc}, one can use a quasi-homogeneous ansatz
to extend the above methods to a microlocal chiral transformation of the form
\beq \label{l:Umicro}
U(x,y) = \int \frac{d^4k}{(2 \pi)^4}\: U\Big( k, v_{L\!/\!R} \Big(\frac{x+y}{2} \Big) \Big)\: e^{-i k(x-y)}\:,
\eeq
\nindex{cg8@$U(x,y)$ -- microlocal chiral transformation}%
and one introduced the auxiliary fermionic projector is defined via the Dirac equation
\beq \label{l:Dirnon}
(U^{-1})^* (i \Pdd - m Y) \,U^{-1} \tilde{P}^\text{aux} = 0\:.
\eeq
This gives the following result.
\begin{Prp} \label{l:prpmicroloc}
Assume that the parameters~$\mu_1, \ldots, \mu_4$ as defined by~\eqref{l:mudef} and~\eqref{l:signconv}
satisfy the inequalities~\eqref{l:muin}.
Then for any choice of the chiral potentials~$v_L$ and~$v_R$ in~\eqref{l:wanted},
there is a microlocal chiral transformation of the form~\eqref{l:Umicro} such that
the transformed fermionic projector~$\tilde{P} := \acute{U} P^\text{aux} \,\grave{U}^*$
is of the form
\begin{align}
\tilde{P}(x,y) &= P(x,y) + (\chi_L \slashed{v}_L + \chi_R \slashed{v}_R)\, T^{(1)}_{[3, \mathfrak{c}]} 
 \,\big( 1 + \O(|\vec{\xi}|/\ell_\text{macro}) \big) \label{l:microlog} \\
&\quad + \text{(vectorial)}\:\1_{\C^2} \:(\deg < 2)
+ \text{(pseudoscalar or bilinear)} \:(\deg < 1) \label{l:vpb} \\
&\quad + \text{(smooth contributions)} + \text{(higher orders in~$\varepsilon/|\vec{\xi}|$)} \:.
\end{align}
\end{Prp}

We conclude this section with two remarks.
\begin{Remark} {\bf{(Unitarity of~$U$)}} \label{l:remunitary} {\em{
We now explain why it would be preferable that the operator~$U$
in the microlocal transformation were unitary, and how and to which extent
this can be arranged.
We begin with the homogeneous setting~\eqref{l:Pkans} and~\eqref{l:Uans}.
As pointed out after~\eqref{l:Uans}, the operator~$U$ as given by~\eqref{l:Uans}
is in general not unitary. However, the following construction makes it possible to
replace~$U$ by a unitary operator without effecting out results:
We first consider the left-handed matrices~$L^j(k)$.
Note that our analysis only involved the sectorial projection~$\acute{L}[k]$
of these matrices contracted with~$k$. Moreover, by multiplying the 
columns by a phase~\eqref{l:phase} we could arrange that all the components in~\eqref{l:logcon}
were real. In this situation, a straightforward analysis shows that there is indeed
a Hermitian $6 \times 6$-matrix whose sectorial projection coincides with~\eqref{l:logcon}.
By choosing the other components of~$L^j(k)$ appropriately, one can arrange
that the matrices~$L^j(k)$ are all Hermitian, and~\eqref{l:logcon} still holds.
Similarly, one can also arrange that the matrices~$R^j(k)$ are Hermitian.
Replacing the ansatz~\eqref{l:Uans} by~$U = \exp(i Z/\sqrt{\Omega})$,
we get a unitary operator. A straightforward calculation shows that
expanding the exponential in a Taylor series, the second and higher orders
of this expansion only effect the error terms in Proposition~\ref{l:prphom}
(for a similar calculation see~\S\ref{s:secnonlocaxial}).

Having arranged that~$U$ is unitary has the advantage that the auxiliary fermionic
projector defined via the Dirac equation~\eqref{l:Dirnon}
is simply given by~$\tilde{P}^\text{aux} = U P U^*$
(whereas if~$U$ were not unitary, the auxiliary fermionic projector would involve
unknown smooth correction terms; see the similar discussion for local transformations
in~\S\ref{s:secgenlocal}).

In the microlocal setting~\eqref{l:Umicro}, the transformation~$U$ will no longer be
unitary, even if the used homogeneous transformations~$U(., v_{L\!/\!R})$
are unitary for every~$v_{L\!/\!R}$. Thus it seems unavoidable that
the fermionic projector defined via the Dirac equation~\eqref{l:Dirnon} will differ
from the operator~$U P U^*$ by smooth contributions on the light cone
(see also the discussion after~\eqref{s:N4def2}).
But even then it is of advantage to choose the homogeneous transformations~$U(., v_{L\!/\!R})$
to be unitary, because then the correction terms obviously vanish
in the limit~$\ell_\text{macro} \rightarrow \infty$. More precisely, a straightforward
analysis shows that these correction terms are of the order~$|\vec{\xi}|/\ell_\text{macro}$. \QEDrem
}} \end{Remark}

\begin{Remark} {\bf{(Lower bound on the largest neutrino mass)}} \label{l:remlargemass} {\em{
The inequalities~\eqref{l:lamineq} give constraints for the masses of the fermions, as
we now explain. Thinking of the interactions of the standard model, we want to
be able to treat the case when a left-handed gauge field but no right-handed gauge fields are present.
In this case, ${\mathfrak{c}}_0$ is non-zero, but the parameter~$z$ vanishes for the right-handed component.
In view of our sign conventions~\eqref{l:signconv}, the first inequality in~\eqref{l:lamineq}
implies that~${\mathfrak{c}}_0 > 0$. Then the inequalities~\eqref{l:lamineq} yield the necessary
conditions~\eqref{l:muin}. More precisely, the eigenvalues~$\mu_1, \ldots, \mu_4$
are given in terms of the lepton masses by (see also~\eqref{s:772})
\begin{align*}
\mu_{1\!/\!2} &= \frac{1}{3} \left( \tilde{m}_1^2 + \tilde{m}_2^2 + \tilde{m}_3^2 \mp \sqrt{
\tilde{m}_1^4 + \tilde{m}_2^4 + \tilde{m}_3^4 - \tilde{m}_1^2 \,\tilde{m}_2^2 - \tilde{m}_2^2 \,\tilde{m}_3^2
- \tilde{m}_1^2 \,\tilde{m}_3^2} \right) \\
\mu_{3\!/\!4} &= \frac{1}{3} \left( m_1^2 + m_2^2 + m_3^2 \mp \sqrt{
m_1^4 + m_2^4 + m_3^4 - m_1^2 \,m_2^2 - m_2^2 \,m_3^2
- m_1^2 \,m_3^2} \right) .
\end{align*}
The first inequality in~\eqref{l:muin} is satisfied once the mass~$m_3$ of the $\tau$-lepton
is much larger than the neutrino masses, as is the case for present experimental data.
However, the second inequality in~\eqref{l:muin} demands that the largest neutrino mass~$\tilde{m}_3$
must be at least of the same order of magnitude as~$m_2$.
In particular, our model does not allow for a description of the interactions in the standard model
if all neutrino masses are too small.

Before comparing this prediction with experiments, one should clearly take into account
that we are working here with the naked masses, which differ from the physical masses by the
contributions due to the self-interaction (with a natural ultraviolet cutoff given by the
regularization length~$\varepsilon$). Moreover, one should consider the possibility
of heavy and yet unobserved so-called sterile neutrinos.
\QEDrem
}} \end{Remark}

We finally point out that here our method was to compensate {\em{all}} the logarithmic
poles by a microlocal chiral transformation. Following the method of treating the
algebraic constraints which will be introduced in \S\ref{l:secalgebra} below,
one can take the alternative point of view that it suffices to compensate
the logarithmic poles in the direction of the {\em{dynamical}} gauge potentials.
This alternative method is preferable because it gives a bit more freedom in choosing the microlocal
chiral transformation. For conceptual clarity, we postpone this improved method to
Chapter~\ref{quark}, where a system involving quarks is analyzed
(see~\S\ref{q:secefflag}).

\subsectionn{The Shear Contributions} \label{l:secshear}
We proceed by analyzing the higher orders in an expansion in the chiral gauge potentials.
Qualitatively speaking, these higher order contributions describe generalized phase transformations of
the fermionic projector. Our task is to analyze how precisely the gauge phases come up and
how they enter the EL equations.
The most singular contributions to discuss are the error terms
\beq \label{l:errorshear}
\text{(vectorial)}\:\1_{\C^2} \:(\deg=1)
\eeq
in Proposition~\ref{l:prpmicroloc}. If modified by gauge phases, these error terms
give rise to the so-called {\em{shear contributions}} by the microlocal chiral transformation.
\sindex{shear contribution!by the microlocal chiral transformation}%
In the setting of one sector, these shear contributions were analyzed in detail in~\S\ref{s:secshearmicro}.
As the adaptation to the present setting of two sectors is not straightforward,
we give the construction in detail.

Recall that the gauge phases enter the fermionic projector to degree two according
to~\eqref{l:PLtransform} and~\eqref{l:PRtransform}. In order to ensure that the error term~\eqref{l:errorshear}
drops out of the EL equations, it must depend on the gauge phases exactly as~\eqref{l:PLtransform}, i.e.\
it must be modified by the gauge phases to
\beq \label{l:shearphase}
\left[ \chi_L 
\begin{pmatrix} U_L^{11} & \overline{c} \:U_L^{12} \\[0.2em] c\: U_L^{21} & U_L^{22} \end{pmatrix}
+ \chi_R \begin{pmatrix} V_R^N & 0 \\ 0 & V_R^C \end{pmatrix} \right]
\times \text{(vectorial)}\:\1_{\C^2} \:(\deg=1)\:.
\eeq
Namely, if~\eqref{l:shearphase} holds, then the corresponding contributions to the closed
chain involve the gauge phases exactly as in~\eqref{l:chain5}, and a straightforward
calculation using~\eqref{l:Pauliid} (as well as~\eqref{l:Lndef} and~\eqref{l:rcond})
shows that the eigenvalues of the closed chain all
have the same absolute value. If conversely~\eqref{l:shearphase} is violated, then
the eigenvalues of the closed chain are not the same, and the EL equations will
be violated (at least without imposing conditions on the regularization functions).
We conclude that the transformation law~\eqref{l:shearphase} is necessary and sufficient
for the EL equations to be satisfied to degree five on the light cone.

In order to arrange~\eqref{l:shearphase}, we follow the procedure in~\S\ref{s:secshearmicro}
and write down the Dirac equation for the auxiliary fermionic projector
\beq \label{l:Dunflip}
\Dir_\text{flip} \, \tilde{P}^\text{aux} = 0 \:,
\eeq
\nindex{di0@$\Dir_\text{flip}$ -- Dirac operator including microlocal chiral transformation}%
where~$\Dir_\text{flip}$ is obtained from the Dirac operator with chiral gauge fields by
\beq \label{l:Diroddeven}
\Dir_\text{flip} :=  (U_\text{flip}^{-1})^* \,\big(i \Pdd_x + \chi_L \slashed{A}_R + \chi_R \slashed{A}_L - m Y \big)\, U_\text{flip}^{-1} \:,
\eeq
where~$U_\text{flip}$ is obtained from the operator~$U$ in~\eqref{l:Dirnon} by
\[ U_\text{flip} = \1 + (\U-\1)\, V \:, \]
and~$V$ is the unitary perturbation flow which changes the gauge potentials from~$A_{L\!/\!R}$
to~$A^\even_{L\!/\!R}$,
\beq \label{l:Vflow}
V = U_\text{flow}[\chi_L \slashed{A}^\even_R + \chi_R \slashed{A}^\even_L]\;
U_\text{flow}[\chi_L \slashed{A}_R + \chi_R \slashed{A}_L]^{-1}
\eeq
(see~\eqref{s:scrDdef}, \eqref{s:Uflip} and~\eqref{a:Vflowdef}).
Exactly as in the proof of Proposition~\ref{s:prpflip}, one sees that
the component~$\sim \Omega^{-1}$ of~$P$ satisfies the Dirac equation
involving the chiral gauge potentials~$A_{L\!/\!R}^\even$. In view of~\eqref{l:effcont}
and~\eqref{l:4con}, we find that the left-handed contribution of~\eqref{l:errorshear} is modified by the
chiral gauge potentials to
\beq \label{l:Lgauge}
L[k]\, \Pexp \Big(-i \int_x^y (A^\even_R)_j \,\xi^j \Big) L[k]^*\:.
\eeq
Thus similar as in~\eqref{l:Pchiral}, gauge phases appear. The difference is that
the chirality is flipped, and moreover here the new potentials~$A^\even_{L\!/\!R}$ enter.
A-priori, these potentials can be chosen arbitrarily according to the gauge group~\eqref{l:GG1}.

For the right-handed component of the fermionic projector, we can use that~$A_R$ is
sector-diagonal (see~\eqref{l:Bform}). Thus we obtain agreement between the phase transformations
corresponding to~$A_R$ and the transformation law~\eqref{l:Lgauge} (with~$L$ exchanged by~$R$)
simply by choosing
\beq \label{l:AevenL}
A_L^\even = \begin{pmatrix} A_R^N & 0 \\ 0 & A_R^C \end{pmatrix} .
\eeq
For the left-handed component of the fermionic projector,
the basic difficulty is that the matrix~$L[k]$ is non-trivial in the generation index
(see~\eqref{l:Lcomp}--\eqref{l:Psidef}). Moreover, the gauge potential~$A_L$ involves
the MNS matrix~$\UMNS$ (see~\eqref{l:Pleft}).
Therefore, it is not obvious how~\eqref{l:Lgauge} can be related to~\eqref{l:Pleft}.
But the following construction shows that for a specific choice of~$A^\even_R$
the connection can be made:
We denote the two column vectors of~$L[k]^*$ by~$\ell_1, \ell_2 \in \C^6$.
In view of~\eqref{l:4con}, these vectors are orthogonal.
We set~${\mathfrak{e}}_1=\ell_1/\|\ell_1\|$ and~${\mathfrak{e}}_4 = \ell_2/\|\ell_2\|$
and extend these two vectors to an orthonormal basis~${\mathfrak{e}}_1, \ldots, {\mathfrak{e}}_6$
of~$\C^6$.
\nindex{di2@$\mathfrak{e}_1, \ldots, \mathfrak{e}_6$ -- orthonormal basis}%
We choose~$A^\even_R$ such that in this basis it has the form
\beq \label{l:AevenR}
A^\even_R(k,x) = \begin{pmatrix} A_L^{11}(x) & A_L^{12}(x)\, V(x)^* \\[0.2em]
A_L^{21}(x)\, V(x) & A_L^{22}(x) \end{pmatrix} ,
\eeq
where we used a block matrix representation in the
two subspaces $\text{span}({\mathfrak{e}}_1, {\mathfrak{e}}_2,
{\mathfrak{e}}_3)$ and $\text{span}( {\mathfrak{e}}_4, {\mathfrak{e}}_5, {\mathfrak{e}}_6)$.
Here the potentials~$\slashed{A}_L^{ij}$ are chosen as in~\eqref{l:Bform},
and~$V(x) \in \U(3)$ is an arbitrary unitary matrix.
We point out that the whole construction depends on the momentum~$k$ of the homogeneous
transformation in~\eqref{l:Lgauge}, as is made clear by the notation~$A^\even_R(k,x)$.
Substituting the ansatz~\eqref{l:AevenR} in~\eqref{l:Lgauge} and using that the columns of~$L[k]^*$
are multiples of~${\mathfrak{e}}_1$ and~${\mathfrak{e}}_4$, we obtain
\beq
\begin{split}
L[k] &\Pexp \Big(-i \int_x^y (A^\even_R)_j \,\xi^j \Big) L[k]^* \\
&=  \begin{pmatrix} U_L^{11} & \overline{d} \:U_L^{12} \\[0.2em]
d\: U_L^{21} & U_L^{22} \end{pmatrix} \:L[k]\, L[k]^*
\overset{\eqref{l:4con}}{=} \begin{pmatrix} U_L^{11} & \overline{d} \:U_L^{12} \\[0.2em]
d\: U_L^{21} & U_L^{22} \end{pmatrix} \:{\mathfrak{c}}_0(k)\, \1_{\C^2} \:,
\end{split}  \label{l:doccur}
\eeq
where~$U_L^{ij}$ as in~\eqref{l:PLtransform} and~$d= V^1_1$.
Choosing~$V$ such that~$d$ coincides with the parameter~$c$ in~\eqref{l:PLtransform},
we recover the transformation law of the left-handed component in~\eqref{l:shearphase}.
Repeating the above construction for the right-handed component (by flipping the chirality
and replacing~$L[k]$ by~$R[k]$), we obtain precisely the transformation law~\eqref{l:shearphase}.

In order to get into the microlocal setting, it is useful to observe that the $k$-dependence
of~$A^\even_R$ can be described by a unitary transformation,
\beq \label{l:prepmicro}
A^\even_R(k,x) = W(k)\: A_L(x)\: W(k)^* \qquad \text{with} \qquad
W(k) \in \U(6) \:.
\eeq
Interpreting~$W$ as a multiplication operator in momentum space and~$A_L$ as a multiplication
operator in position space, we can introduce~$A^\even_R$ as the operator product
\beq \label{l:Aevc}
A^\even_R = W A_L W^* \:.
\eeq
We point out that the so-defined potential~$A^\even_R$ is non-local.
As the microlocal chiral transformation is non-local on the Compton scale,
one might expect naively that the same should be true for~$A^\even_R$.
However, $A^\even_R$ can be arranged to be localized on the much smaller
regularization scale~$\varepsilon$, as the following argument shows:
The $k$-dependence of~$W$ is determined by the matrix entries of~$L[k]$.
The analysis in~\S\ref{l:secmicroloc} shows that the matrix entries of~$L[k]$
vary in~$k$ on the scale of the energy~$\varepsilon^{-1}$ (in contrast to the matrix~$Z$, which in view of
the factor~$1/\sqrt{\Omega}$ in~\eqref{l:Uans} varies on the scale~$m$).
Taking the Fourier transform, the operator~$W$ decays in position space on the regularization scale.

This improved scaling has the positive effect that the error term caused by
the quasilocal ansatz~\eqref{l:Aevc} is of the order~$\varepsilon/\ell_\text{macro}$.
Hence the gauge phases enter the left-handed component of the error term~\eqref{l:errorshear} as
\[ \chi_L \begin{pmatrix} U_L^{11} & \overline{c} \:U_L^{12} \\[0.2em]
c\: U_L^{21} & U_L^{22} \end{pmatrix}
\big( 1 + \O(\varepsilon/\ell_\text{macro}) \big) \text{(vectorial)}\:\1_{\C^2} \:(\deg=1) \:. \]
Carrying out a similar construction for the right-handed component, we obtain the following result.
\begin{Prp} Introducing the potentials~$A^\even_{L\!/\!R}$ in~\eqref{l:Vflow}
according to~\eqref{l:AevenL} and~\eqref{l:Aevc},
the error term~\eqref{l:errorshear} in Proposition~\ref{l:prpmicroloc} transforms to
\[ \left[ \chi_L 
\begin{pmatrix} U_L^{11} & \overline{c} \:U_L^{12} \\[0.2em] c\: U_L^{21} & U_L^{22} \end{pmatrix}
+ \chi_R \begin{pmatrix} V_R^N & 0 \\ 0 & V_R^C \end{pmatrix} \right]
\big( 1+\O(\varepsilon/\ell_\text{macro}) \big)\; \text{\rm{(vectorial)}}\:\1_{\C^2} \:(\deg=1)
\:. \]
\end{Prp}

In this way, we have arranged that the
EL equations are satisfied to degree five on the light cone.
Note that the above construction involves the freedom in choosing the
basis vectors~${\mathfrak{e}}_2, {\mathfrak{e}}_3, {\mathfrak{e}}_5, {\mathfrak{e}}_6$
as well as the unitary matrix~$V$ in~\eqref{l:AevenR}. This will be analyzed in more
detail in~\S\ref{l:seclogAA}.

We finally point out that in general, the Dirac operator~\eqref{l:Diroddeven} violates
the causality compatibility condition~\eqref{l:ccc}. This implies that the light-cone
expansion of the auxiliary fermionic projector may involve
unbounded line integrals. However, this causes no problems because these
unbounded line integrals drop out when taking the sectorial projection.
Thus the fermionic projector is again causal in the sense that its light-cone expansion only involves
bounded line integrals.

\section{The Energy-Momentum Tensor and the Curvature Terms} \label{l:seccurv}
In this section we compute other relevant contributions to the EL equations to degree four:
the contributions by the energy-momentum tensor and the curvature of space-time.
\subsectionn{The Energy-Momentum Tensor of the Dirac Field}
Considering the contribution of the particle and anti-particle wave functions in~\eqref{l:particles}
at the origin~$x=y$ gives rise to the Dirac current terms as considered in~\S\ref{l:seccurrent}
(for details see also~\S\ref{s:sec82}). We now go one order higher in
an expansion around the origin~$\xi=0$.
Setting~$z=(x+y)/2$ and expanding in powers of~$\xi$ according to
\begin{align*}
\psi(x) &= \psi(z- \xi/2) = \psi(z) - \frac{1}{2}\: \xi^j \partial_j \psi(z)
+ o(|\vec{\xi}|) \\
\psi(y) &= \psi(z+\xi/2) = \psi(z) + \frac{1}{2}\: \xi^j \partial_j \psi(z)
+ o(|\vec{\xi}|) \\
\psi(x) \overline{\psi(y)} &= \psi(z) \overline{\psi(z)}
- \frac{1}{2} \:\xi^j \left( \big( \partial_j \psi(z) \big) \overline{\psi(z)} - \psi(z) 
\big( \partial_j \overline{\psi}(z) \big) \right)
+ o(|\vec{\xi}|) \:,
\end{align*}
we can write the contribution by the particles and anti-particles as
\sindex{fermionic projector!energy-momentum term}%
\[ P(x,y) \asymp -\frac{1}{8 \pi} \sum_{c=L\!/\!R} \chi_c \gamma_k \left(  \hat{J}^k_c - i \xi_l\, \hat{T}^{kl}_c \right) 
+ o(|\vec{\xi}|) + \text{(even contributions)} \:, \]
where
\beq \label{l:Tdef}
(T^{kl}_{L\!/\!R})^{(i,\alpha)}_{(j, \beta)} = -\im \sum_{a=1}^{\np} 
\overline{\psi_a^{(j, \beta)}} \chi_{R\!/\!L} \gamma^k \partial^l \psi_a^{(i,\alpha)}
+ \im \sum_{b=1}^{\na} \overline{\phi_b^{(j, \beta)}} \chi_{R\!/\!L} \gamma^k \partial^l \phi_b^{(i,\alpha)} \:,
\eeq
and similar to~\eqref{l:accents}, the hat denotes the sectorial projection.
We denote the vectorial component by
\[ T^{kl} := T^{kl}_L + T^{kl}_R \:. \]
Taking the trace over the generation and isospin indices, we obtain the 
{\em{energy-mo\-men\-tum tensor}} of the particles and anti-particles.
We now compute the resulting contribution to the matrices~${\mathscr{K}}_{L\!/\!R}$
as introduced in Lemma~\ref{l:lemma41}. The proof of this lemma
is obtained similar to that of Lemma~\ref{l:lemma44} by a straightforward computation.
\sindex{energy-momentum tensor!of Dirac field}%
\nindex{db6@$T_{jk}$ -- energy-momentum tensor}%
\begin{Lemma} \label{l:lemmaT}
The tensors~$T^{kl}_{L\!/\!R}$, \eqref{l:Tdef}, give the following
contribution to the matrices~${\mathscr{K}}_{L\!/\!R}$ in~\eqref{l:DelQ} and~\eqref{l:Rdef},
\beq \label{l:JKL1}
{\mathscr{K}}_{L\!/\!R} \asymp  \hat{T}^{kl}_{R\!/\!L} \:\xi_k\, \xi_l\: K_8 \:+\: (\deg < 4) + o \big( |\vec{\xi}|^{-2}
\big) \:,
\eeq
where~$K_8$ is the simple fraction
\[ K_8 = \frac{3}{16 \pi}\: \frac{1}{\overline{T^{(0)}_{[0]}}}
\left[T^{(-1)}_{[0]} T^{(0)}_{[0]}\: \overline{T^{(-1)}_{[0]}} + c.c. \right] \]
(note that~$K_8$ differs from~$K_1$ on page~\pageref{l:mterm7} in that the
term~$-c.c.$ has been replaced by~$+c.c.$).
\end{Lemma} \noindent
The energy-momentum tensor of the gauge field will be computed in~\S\ref{secFFT} below.

\subsectionn{The Curvature Terms}
The obvious idea for compensating the above contributions to the EL equations is to
modify the Lorentzian metric. At first sight, one might want to introduce a metric
which depends on the isospin index.
However, such a dependence cannot occur, as the
following argument shows: The singular set of the fermionic projector~$P(x,y)$ is given by
the pair of points~$(x,y)$ with light-like separation. If the metric depended on the isospin
components, the singular set would be different in different isospin components.
Thus the light cone would ``split up'' into two separate light cones.
As a consequence, the leading singularities of the closed chain could no longer
compensate each other in the EL equations, so that the EL equations would be violated
to degree five on the light cone.

Strictly speaking, this argument leaves the possibility to introduce a conformal factor
which depends on the isospin (because a conformal transformation does not affect
the causal structure). However, as the conformal weight enters the closed chain
to degree five on the light cone, the EL equations will be satisfied only if the conformal factor is independent
of isospin.

The above arguments readily extend to a chiral dependence of the metric: If
the left- and right-handed component of the fermionic projector would feel a different
metric, then the singular sets of the left- and right-handed components of the closed chain
would again be different, thereby violating the EL equations to degree five
(for a similar argument for an axial gravitational field see the discussion in~\S\ref{s:sec89}).

Following these considerations, we are led to introducing a Lorentzian metric~$g_{ij}$.
Linear perturbations of the metric were studied in~\cite[Appendix~B]{firstorder}
(see also Section~\ref{seclingrav}).
The contributions to the fermionic projector involving the curvature tensor
were computed by
\begin{align}
P(x,y) \asymp\:& 
\frac{i}{48} \:R_{jk}\: \xi^j \xi^k \: \slashed{\xi}\: T^{(-1)} \label{l:curv1} \\
&+ \frac{i}{24} R_{jk}\: \xi^j \gamma^k\: T^{(0)}
+ \slashed{\xi} \,(\deg \leq 1) + (\deg < 1) \:, \label{l:curv2}
\end{align}
where~$R_{jk}$ denotes the Ricci tensor
\nindex{db2@$R_{jk}$ -- Ricci tensor}%
(we only consider the leading contribution in an expansion in powers of~$|\vec{\xi}|/\ell_\text{macro}$).
We refer to~\eqref{l:curv1} and~\eqref{l:curv2} as the {\em{curvature terms}}.
More generally, in~\cite[Appendix~A]{lqg} the singularity structure of the fermionic projector 
was analyzed on a globally hyperbolic Lorentzian manifold (for details see also~\cite{drgrotz}).
Transforming the formulas in~\cite{lqg, drgrotz} to the the coordinate system
and gauge used in~\cite{firstorder}, one sees that~\eqref{l:curv1} and~\eqref{l:curv2} also hold non-perturbatively.
In particular, the results in~\cite{lqg} show that, to the considered degree on the light cone,
no quadratic or even higher order curvature expressions occur.
In what follows, we consider~\eqref{l:curv1} and~\eqref{l:curv2} as a perturbation of the
fermionic projector in Minkowski space. This is necessary because at present, the formalism
of the continuum limit has only been worked out in Minkowski space. Therefore, strictly speaking,
the following results are perturbative. But after extending the formalism of the continuum limit
to curved space-time (which seems quite straightforward because the framework of the fermionic
projector approach is diffeomorphism invariant), our results would immediately carry over to
a globally hyperbolic Lorentzian manifold.

Let us analyze how the curvature terms enter the eigenvalues of the closed chain.
We first consider the case when we strengthen~\eqref{l:deltascale} by assuming that
\begin{equation}
\varepsilon \ll \delta \ll \frac{1}{m}\: (m \varepsilon)^{\frac{p_\reg}{2}} \label{l:deltascale2}
\end{equation}
(the case~$\delta \simeq m\: (m \varepsilon)^{\frac{p_\reg}{2}}$ will be discussed
in Section~\ref{l:secgrav}). The assumption~\eqref{l:deltascale2} makes it possible to
omit the terms~$\sim m^2 R_{ij}$.

\begin{Lemma} \label{l:lemmaR}
The curvature of the Lorentzian metric gives the following
contribution to the matrices~${\mathscr{K}}_{L\!/\!R}$ in~\eqref{l:Kdef},
\sindex{fermionic projector!curvature term}%
\begin{align}
{\mathscr{K}}_c \asymp\:&
\frac{5}{24}\: \frac{1}{48} \: R_{kl}\: \xi^k \xi^l\; A^{(0)}_{xy}\, P^{(0)}(x,y) \label{l:curv0} \\
&+ \delta_{cL}\: \frac{\tau_\reg}{\delta^2}\: R_{kl}\: \xi^k \xi^l
\begin{pmatrix} 1 & 0 \\ 0 & 0 \end{pmatrix} K_{16} \label{l:curv3} \\
&+ m^2\, R_{kl}\: \xi^k \xi^l\, (\deg=4) + (\deg<4) + o \big( |\vec{\xi}|^{-2} \big) \:, \nonumber
\end{align}
where~$K_{16}$ is the following simple fraction of degree four,
\beq \label{l:K16}
K_{16} = \frac{27}{32}\: \big| T^{(-1)}_{[0]}\big|^2 \:\overline{T^{(0)}_{[0]}}^{-1}
\left( T^{(1)}_{[R,2]} \overline{L^{(0)}_{[0]}} + L^{(0)}_{[0]} \overline{T^{(1)}_{[R,2]}} \right)
\eeq
(and~$P^{(0)}(x,y)$ and~$A^{(0)}_{xy}$ denote the vacuum fermionic projector and the closed chain
of the vacuum, respectively).
\nindex{di8@$P^{(0)}(x,y), A^{(0)}_{xy}, \lambda^{(0)}_{ncs}$ -- objects of the vacuum}%
\end{Lemma}
\Proof
The contribution~\eqref{l:curv1} multiplies the fermionic projector of the vacuum by a
scalar factor. Thus it can be combined with the vacuum fermionic projector~$P^{(0)}$ to the expression
\beq \label{l:cP0}
c_{xy} \:P^{(0)}(x,y) \qquad \text{with} \qquad
c_{xy} := 1 + \frac{1}{24} \:R_{jk}\: \xi^j \xi^k\:.
\eeq
Hence the closed chain and the eigenvalues are simply multiplied by a common prefactor,
\[ A_{xy} = c_{xy}^2 \:A^{(0)}_{xy} \:,\qquad
\lambda_{ncs}  = c_{xy}^2 \:\lambda^{(0)}_{ncs} \:. \]
As a consequence, the contribution~\eqref{l:curv1} can be written in the form~\eqref{l:curv0}.

The summand~\eqref{l:curv2} is a bit more involved, and we treat it in the $\iota$-formalism.
The closed chain is computed by
\begin{align}
A_{xy} \asymp\:& \frac{3}{16} \: R_{jk}\: \check{\xi}^j \gamma^k\: \check{\slashed{\xi}}\;
T^{(0)}_{[0]} \,\overline{T^{(-1)}_{[0]}} \label{l:curv21} \\
&+\frac{3}{16} \: R_{jk}\: \check{\slashed{\xi}} \:\check{\xi}^j \gamma^k\;
T^{(-1)}_{[0]} \,\overline{T^{(0)}_{[0]}} \:. \label{l:curv22}
\end{align}
Similar as explained for the chiral contribution after~\eqref{l:chiralex}, the eigenvalues~$\lambda_{nc+}$
are only perturbed by~\eqref{l:curv21}. More precisely,
\[ \lambda_{nc+} \asymp 
\frac{3}{16} \: R_{jk}\: \xi^j \xi^k\:
T^{(0)}_{[0]} \,\overline{T^{(-1)}_{[0]}} \:, \]
and the other eigenvalues are obtained by complex conjugation~\eqref{l:ccp}.
In particular, one sees that the eigenvalues are perturbed only by a common prefactor.
Combining the perturbation with the eigenvalues of the vacuum, we obtain
\[ \lambda_{nc+}  = \left(1 + \frac{1}{48} \: R_{jk}\: \xi^j \xi^k \right) \lambda^{(0)}_{nc+} \:. \]
In view of~\eqref{l:ccp}, this relation also holds for the eigenvalues~$\lambda_{nc-}$.
We conclude that~\eqref{l:curv2} can again be absorbed into~\eqref{l:curv0}.
A short calculation using~\eqref{l:Kncdef} shows that the contributions so far
combine precisely to~\eqref{l:curv0}.

It remains to consider the effects of shear and of the general surface terms.
The shear contribution is described by a homogeneous transformation
of the spinors which is localized on the scale~$\varepsilon$ (for details see Appendix~\ref{l:appA}).
Since this transformation does not effect the macroscopic prefactor~$c_{xy}$
in~\eqref{l:cP0}, the eigenvalues are again changed only by a common prefactor.
Hence~\eqref{l:curv1} drops out of the EL equations for the shear states.
The contributions~\eqref{l:curv21} and~\eqref{l:curv22}, on the other hand, do not involve~$\iota$,
and are thus absent for the shear states.
We conclude that also~\eqref{l:curv2} drops out of the EL equations for the shear states.

We finally consider the general surface states. As~\eqref{l:curv21} is a smooth factor times the
vacuum fermionic projector, the Ricci tensor again drops out of the EL equations.
For the remaining term~\eqref{l:curv22}, the replacement rule~\eqref{l:frep2} yields
the contribution of the general mass expansion
\[ \chi_R \,P^\varepsilon(x,y) \asymp \frac{i}{24} \: R_{jk}\: \xi^j \gamma^k\: 
\frac{\tau_\reg}{\delta^2} \begin{pmatrix} T^{(1)}_{[R,2]} & 0 \\ 0 & 0 \end{pmatrix} . \]
As a consequence,
\begin{align*}
\chi_R \,A_{xy} &\asymp \frac{3}{48}\: R_{jk}\: \xi^j \gamma^k\:
\overline{\slashed{\xi}} \; \frac{\tau_\reg}{\delta^2} \begin{pmatrix} T^{(1)}_{[R,2]} \overline{T^{(-1)}_{[0]}}
& 0 \\ 0 & 0 \end{pmatrix} \\
\lambda_{nR+} &\asymp \frac{3}{48}\: R_{jk}\: \xi^j \xi^k\:
\frac{\tau_\reg}{\delta^2}\: T^{(1)}_{[R,2]} \overline{T^{(-1)}_{[0]}}
\:\Tr_{\C^2} \left( I_n \begin{pmatrix} 1 & 0 \\ 0 & 0 \end{pmatrix} \right) \\
{\mathscr{K}}_L &\asymp 
\frac{27}{32}\: R_{jk}\: \xi^j \xi^k\:
\frac{\tau_\reg}{\delta^2}\: \left( T^{(1)}_{[R,2]} \overline{L^{(0)}_{[0]}}
+ L^{(0)}_{[0]}\overline{T^{(1)}_{[R,2]}} \right)
\:\frac{1}{\overline{T^{(0)}_{[0]}}}\: \big| T^{(-1)}_{[0]}\big|^2
\begin{pmatrix} 1 & 0 \\ 0 & 0 \end{pmatrix} .
\end{align*}
Similarly,
\begin{align*}
\chi_L \,A_{xy} &\asymp \frac{3}{48}\: \slashed{\xi} \:R_{jk}\: \xi^j \gamma^k
\; \frac{\tau_\reg}{\delta^2} \begin{pmatrix} T^{(-1)}_{[0]} \overline{T^{(1)}_{[R,2]}}
& 0 \\ 0 & 0 \end{pmatrix} ,
\end{align*}
and a straightforward computation yields~$\lambda_{nL+} \asymp 0$
and~${\mathscr{K}}_R \asymp 0$. This concludes the proof.
\QED

\subsectionn{The Energy-Momentum Tensor of the Gauge Field} \label{secFFT}
We now compute the effect of the energy-momentum tensor of the chiral gauge field.
We denote the chiral field tensor by~$F_c^{jk} = \partial^j A_c^k - \partial^k A_c^j$.
\nindex{di9@$F^{jk}_{\LR}$ -- chiral field tensor}%
\sindex{energy-momentum tensor!of gauge fields}%
\begin{Lemma} \label{lemmaFFT}
The field tensor of the gauge fields gives the following
contribution to the matrix~${\mathscr{K}}_L$ in~\eqref{l:DelQ} and~\eqref{l:Rdef},
\begin{align}
{\mathscr{K}}_L \asymp\;&
-\frac{g^2}{3} \: \acute{F}^L_{ki}\, \grave{F}_L^{kj}\: \xi^i\, \xi_j
\frac{\big|T^{(-1)}_{[0]} \big|^2}{\overline{T^{(0)}_{[0]}}} \left( T^{(1)}_{[0]} \overline{T^{(0)}_{[0]}}
+ T^{(0)}_{[0]} \overline{T^{(1)}_{[0]}} \right) \label{billog} \\
&-\frac{g^2}{24} \: \acute{F}^R_{ki}\, \grave{F}_R^{kj}\: \xi^i\, \xi_j\;
T^{(0)}_{[0]} \left( T^{(0)}_{[0]} \overline{T^{(-1)}_{[0]}} + T^{(-1)}_{[0]} \overline{T^{(0)}_{[0]}} \right) \\ 
&+\frac{g}{8}\: \hat{F}^L_{ki}\: \hat{F}_R^{kj} \:\xi^i\, \xi_j\: T^{(0)}_{[0]}\:
\frac{ \big( T^{(0)}_{[0]} T^{(0)}_{[0]} \overline{T^{(-1)}_{[0]} T^{(-1)}_{[0]}} - c.c. \big) }
{T^{(0)}_{[0]} \overline{T^{(-1)}_{[0]}} - T^{(-1)}_{[0]} \overline{T^{(0)}_{[0]}}} \\
&-\frac{g}{8}\: \epsilon_{ijkl}\: \xi^i\, \xi_a \: \hat{F}_R^{aj}\: \hat{F}_L^{kl}\,
\frac{T^{(0)}_{[0]}\: \big| T^{(-1)}_{[0]} T^{(0)}_{[0]} \big|^2}{T^{(0)}_{[0]} \overline{T^{(-1)}_{[0]}} - T^{(-1)}_{[0]} \overline{T^{(0)}_{[0]}}} \\
&-\frac{g}{8}\: \epsilon_{ijkl}\: \xi^i\, \xi_a \: \hat{F}_L^{aj}\: \hat{F}_R^{kl}\,
\frac{T^{(-1)}_{[0]} T^{(-1)}_{[0]} T^{(0)}_{[0]} \overline{T^{(0)}_{[0]} T^{(0)}_{[0]}}}
{T^{(0)}_{[0]} \overline{T^{(-1)}_{[0]}} - T^{(-1)}_{[0]} \overline{T^{(0)}_{[0]}}} \\
&-\frac{g}{8}\: \epsilon_{ijkl} \: \hat{F}_R^{ij}\: \xi^k\, \xi_a\: \hat{F}_L^{al}\,
\frac{T^{(0)}_{[0]} T^{(0)}_{[0]} T^{(0)}_{[0]} \overline{T^{(-1)}_{[0]} T^{(-1)}_{[0]}}}
{T^{(0)}_{[0]} \overline{T^{(-1)}_{[0]}} - T^{(-1)}_{[0]} \overline{T^{(0)}_{[0]}}} \\
&-\frac{g}{8}\: \epsilon_{ijkl} \: \hat{F}_L^{ij}\: \xi^k\, \xi_a\: \hat{F}_R^{al}\,
\frac{T^{(-1)}_{[0]} T^{(0)}_{[0]} T^{(0)}_{[0]} \overline{T^{(-1)}_{[0]} T^{(0)}_{[0]}}}
{T^{(0)}_{[0]} \overline{T^{(-1)}_{[0]}} - T^{(-1)}_{[0]} \overline{T^{(0)}_{[0]}}} \\
&+(\deg < 4) + o \big( |\vec{\xi}|^{-2} \big)\:. \notag
\end{align}
The contribution to~$\K_R$ is obtained by the replacements~$L \leftrightarrow R$.
\end{Lemma} \noindent
The proof of this lemma is given in Appendix~\ref{appEMT}.
The main reason why the formulas are rather complicated is
that the eigenvalues of the closed chain must be computed
to second order in perturbation theory.

The last lemma reveals the general problem that~\eqref{billog} involves
logarithmic poles on the light cone. This comes about because~$P(x,y)$
involves a term quadratic in the field tensor with a logarithmic pole on the light cone,
which when contracted with a factor~$\slashed{\xi}$ has the form
\beq \label{FFlog}
\frac{1}{4}\, \Tr \big( \slashed{\xi}\, \chi_{L\!/\!R}\: P(x,y) \big) \asymp
-\frac{i}{3} F^{L\!/\!R}_{ki}\: F_{L\!/\!R}^{kj}\, \xi^i\, \xi_j \: T^{(1)}_{[0]} + o \big(|\vec{\xi}|^2 \big)
\eeq
(for the detailed computations see again Appendix~\ref{appEMT}).
Similar as explained in \S\ref{s:sec82} for the logarithmic poles of the current terms,
the logarithmic pole must again be compensated by a suitable microlocal transformation.
This can indeed be accomplished, as we now explain. For clarity, we outline and discuss the
method before entering the details.
First, it suffices to consider the homogeneous situation in the high-frequency limit,
because the macroscopic space-time dependence of the energy-momentum tensor can
be taken into account just as in~\S\ref{s:secgennonloc} by a corresponding
quasi-homogeneous ansatz of the form~\eqref{s:Umicro}.
Generally speaking, our method is to generate a contribution of
the form as in Proposition~\ref{l:prphom}, but with the logarithmic
contributions in~\eqref{l:des1} and~\eqref{l:des3} of the more specific form that they are vectorial and
act trivially on the sector index, i.e.
\begin{align}
\tilde{P}(k) &= P(k) + \text{(vectorial)}\:\1_{\C^2}\, T^{(1)}_{[3, \mathfrak{c}]} \Big(1+ \O \big( \Omega^{-\frac{1}{2}} \big) \Big) \label{Pn1} \\
&\quad+ \text{(vectorial)}\:\1_{\C^2} \:\delta(k^2) \Big(1+ \O(\Omega^{-1}) \Big) \label{Pn2} \\
&\quad + \text{(pseudoscalar or bilinear)} \:\sqrt{\Omega} \:\delta'(k^2) \Big(1+ \O(\Omega^{-1}) \Big)
+ \text{(higher orders in~$\varepsilon/|\vec{\xi}|$)}\:. \notag
\end{align}
These contributions to the fermionic projector change the matrices~$\mathscr{K}_L$
and~$\mathscr{K}_R$ in~\eqref{l:Rdef} only by a multiple of the identity matrix
and thus drop out of the EL equations~\eqref{l:EL4}.
However, the above contributions~\eqref{Pn1} and~\eqref{Pn2} are modified by gauge phases,
and our strategy is to arrange these gauge phases in such a way that the modification linear in the
gauge potential will give the desired bilinear logarithmic contribution.
In order to understand the scaling in~$\xi$, one should keep in mind that the
term~\eqref{Pn1} gives a contribution to~${\mathscr{K}}_{L\!/\!R}$
of the form~$v_j \xi^j\: (K_2+K_3)$ (cf.~\eqref{l:Jterm}).
Therefore, the first order modification by a gauge potential~$A$ will be of the form
\beq \label{genform}
\sim v_j \xi^j\, A_k \xi^k \, (\deg=4) \:,
\eeq
involving as desired two factors of~$\xi$. Next, we need to explain how the potential~$A$
in~\eqref{genform} is to be chosen.
It is essential for our construction that the gauge phases of the contributions by the
microlocal chiral transformation are determined by the potentials~$A^\even_{L\!/\!R}$
in~\eqref{l:Vflow} (see~\eqref{l:Lgauge}).
It is very helpful that the potentials~$A^\even_{L\!/\!R}$ can be
chosen independent of the chiral potentials~$A_{L\!/\!R}$ in~\eqref{l:Bform}.
This makes it possible to arrange that the contribution~\eqref{genform} involves the desired
bilinear logarithmic term needed to compensate the contribution~\eqref{FFlog},
without affecting the gauge phases as analyzed in~\S\ref{l:sec32}.

The detailed construction is carried out in the following lemma.
\begin{Lemma} \label{lemmaTcomp}
Assume that the chiral potentials are of the form~\eqref{l:Bform}
with a left-handed $\SU(2)$-potential (i.e.~$A^N_R=A^C_R=0$ and~$A_L^{11}=A_L^{22}$).
Then the logarithmic pole of the contribution to the fermionic projector~\eqref{FFlog}
can be compensated by the shear contributions corresponding to a microlocal chiral transformation
for a suitable choice of the potentials~$A^\even_{L\!/\!R}$ in~\eqref{l:Vflow}.
\end{Lemma}
\Proof The first step is to arrange the contributions~\eqref{Pn1} and~\eqref{Pn2}
by specializing the transformation used in~\S\ref{l:secmicroloc}.
Thus we again choose the ansatz~\eqref{l:Uans}, but now with a pure vector component, i.e.\
\[ U(k) = \1 + \frac{i}{\sqrt{\Omega}}\: L^j \gamma_j \]
with $6 \times 6$-matrices~$L^j$. Next, we choose the matrices~$L^j$ as diagonal matrices
involving one vector field~$v$,
\beq \label{i:Lvec}
L^j \gamma_j = \slashed{v}\: \text{diag} \big( \lambda_1, \ldots,  \lambda_6 \big) \:.
\eeq
This ansatz has the advantage that the conditions~\eqref{l:1cond}--\eqref{l:3cond}
give rise to independent conditions in the two sectors. In the charged sector, we
satisfy these conditions by arranging that
\begin{gather}
\Big\la \begin{pmatrix} \lambda_4 \\ \lambda_5 \\ \lambda_6 \end{pmatrix},
\begin{pmatrix} 1 \\ 1 \\ 1 \end{pmatrix} \Big\ra_{\C^3} = 0 \label{ortho1} \\
\im \Big\la \begin{pmatrix} \lambda_4 \\ \lambda_5 \\ \lambda_6 \end{pmatrix},
\begin{pmatrix} m_1 \\ m_2 \\ m_3 \end{pmatrix} \Big\ra_{\C^3}
=0 = \im \Big\la \begin{pmatrix} \lambda_4 \\ \lambda_5 \\ \lambda_6 \end{pmatrix},
\begin{pmatrix} m_1^3 \\[0.2em] m_2^3 \\[0.2em] m_3^3 \end{pmatrix} \Big\ra_{\C^3} \:. \label{ortho2}
\end{gather}
In the neutrino sector, we obtain impose similar relations for the vector~$(\lambda_1, \lambda_2, \lambda_3)$.
Moreover, the formulas~\eqref{l:4con} and~\eqref{l:5con} give rise to the conditions
\begin{align*}
|\lambda_1|^2 + |\lambda_2|^2 + |\lambda_3|^2 &= |\lambda_4|^2 + |\lambda_5|^2 + |\lambda_6|^2 \\
\tilde{m}_1^2\, |\lambda_1|^2 + \tilde{m}_2^2\, |\lambda_2|^2 + \tilde{m}_3^2\, |\lambda_3|^2 &=
m_1^2\, |\lambda_4|^2 + m_2^2\, |\lambda_5|^2 + m_3^2\, |\lambda_6|^2 \:.
\end{align*}
By choosing the parameters~$\lambda_1, \ldots \lambda_6$ according to the above
conditions, we can arrange a contribution to the fermionic projector of the form~\eqref{Pn1}
and~\eqref{Pn2}.

Next, we need to specify the gauge potentials~$A^\even_{L\!/\!R}$.
Since the bilinear logarithmic contribution to be compensated~\eqref{FFlog} is sector diagonal,
we also choose the potentials sector diagonal, i.e.
\beq \label{Aevc}
A^\even_{L\!/\!R} = \begin{pmatrix}  A^N_{L\!/\!R} & 0 \\ 0 &  A^C_{L\!/\!R} \end{pmatrix} \:,
\eeq
where we used the same block matrix notation as in~\eqref{l:Bform}.
This makes it possible to again analyze the two sectors separately. We only consider
the charged sector, because the neutrino sector can be treated in exactly the same way.
It is convenient to choose the vectors in~$\C^3$
\[ f_1 = \begin{pmatrix} 1 \\ 1 \\ 1 \end{pmatrix}\:,\qquad
f_2 = \begin{pmatrix} \lambda_4 \\ \lambda_5 \\ \lambda_6 \end{pmatrix}\:,\qquad
f_3 =  \begin{pmatrix} m_1 \\ m_2 \\ m_3 \end{pmatrix} \]
(which to avoid trivialities we may assume to be all non-zero).
We need to ensure that the gauge phases do not give rise to contributions which
are more singular than the logarithmic poles. This is a rather subtle point, which we explain
in detail. First, the construction of~\S\ref{l:secshear} yields that the
gauge phases of the gauge potentials~$A^\even_{L\!/\!R}$ enter exactly as those
of the chiral potentials~$A_{L\!/\!R}$, except that the chirality is flipped (see~\eqref{l:Lgauge}).
Next, we must keep in mind that the chirality of the gauge phase flips at every factor of the mass
matrix~$Y$ (see the relations~\eqref{mT1} and~\eqref{mT2}).
We need to make sure that the gauge potentials drop out of the expressions in~\eqref{l:1cond}
and~\eqref{l:2cond}. More specifically, it is no problem if a vectorial gauge phase comes up,
because the corresponding phase factor is the same for the left- and right-handed components.
Arranging that it is also the same in both sectors, such a phase factor drops out of the EL equations.
Therefore, our task is to ensure that the expressions in~\eqref{l:1cond}
and~\eqref{l:2cond} only involve a vectorial phase factor. This leads to the conditions
\begin{align}
A^C_L \,f_1 = \!&\,\;0 = A^C_R \,f_1 \label{zero1} \\
\la f_2, \big( A^C_L \big)^p \,f_3 \ra_{\C^3} &= \la f_3, \big( A^C_R \big)^p \,f_2 \ra_{\C^3} 
\qquad \text{for all~$p \in \N$} \:. \label{zero2}
\end{align}
Next, we need to make sure that also the contribution~\eqref{Pn2} involves only
a vectorial phase factor. Here it suffices to consider the linear contribution in the potential,
because higher orders in the potential may be disregarded in the EL equations.
Thus we need to arrange that
\beq \label{zero3}
\la f_2, A^C_L \,f_2 \ra_{\C^3} = \la f_2, A^C_R \,f_2 \ra_{\C^3} \:.
\eeq
Finally, we need to arrange that the axial potential does modify the contribution~\eqref{Pn1},
because this will give rise to the desired term used to compensate the bilinear logarithmic term~\eqref{FFlog}.
By scaling we may impose that
\beq
\Big\la f_2, \Big\{ \big(A^C_L - A^C_R \big), Y^2  \Big\} f_2 \Big\ra_{\C^3} = 1 \:, \label{nonzero}
\eeq
were~$Y^2$ is the mass matrix in the charged sector,
i.e.\ $m Y = \text{diag}(m_1, m_2, m_3)$
(here the anti-commutator comes about in view of~\eqref{mT3} and~\eqref{mT4}).

In order to construct potentials satisfying the conditions~\eqref{zero1}--\eqref{nonzero},
we make the ansatz
\beq \label{Apsi}
A^C_{L\!/\!R} = |\psi_{L\!/\!R} \ra \la \psi_{L\!/\!R}| \qquad \text{with~$\psi_{L\!/\!R} \in \C^3$}\:.
\eeq
Then the conditions~\eqref{zero1} and~\eqref{zero3} become
\begin{align}
\la \psi_L , f_1 \ra_{\C^3} =\!&\,\;0 = \la \psi_R , f_1 \ra_{\C^3} \label{psicond1} \\
\big| \la \psi_L , f_2 \ra_{\C^3} \big|^2 &=  \big| \la \psi_R , f_2 \ra_{\C^3} \big|^2 \:. \label{psicond2}
\intertext{Next, the condition~\eqref{zero2} for~$p=1$ yields}
\la f_2 , \psi_L \ra_{\C^3}\: \la \psi_L , f_3 \ra_{\C^3} &= 
\la f_3 , \psi_R \ra_{\C^3}\: \la \psi_R , f_2 \ra_{\C^3} \:. \label{psicond3}
\intertext{Since the matrices~$A^C_{L\!/\!R}$ have rank one, their powers in~\eqref{zero2}
are scalar multiples times the matrices themselves. Therefore, the conditions~\eqref{zero2} for
general~$p$ are satisfied if and only if the vectors~$\psi_L$ and~$\psi_R$ have the same norm,}
\|\psi_L\|_{\C^3} &= \|\psi_R\|_{\C^3} \:. \label{psicond4}
\end{align}

In order to verify whether the above system of equations admits non-trivial solutions,
we first count the number of degrees of freedom.
Since the phases of~$\psi_L$ and~$\psi_R$ do not influence the
corresponding potentials~\eqref{Apsi}, we begin with ten real degrees of freedom
(five for~$\psi_L$ and five for~$\psi_R$).
The equations~\eqref{psicond1}--\eqref{psicond4} give $4+1+2+1=8$ conditions,
leaving us with two free parameters. This suggest that there should indeed be non-trivial solutions.
In order to construct these solutions, it is useful to choose an orthonormal basis~$(e_1, e_2)$ of the
orthogonal complement of the vector~$f_1 \in \C^3$.
Since~$f_2$ is orthogonal to~$f_1$ (see~\eqref{ortho1}), we can choose the first basis
vector~$e_1$ in the direction of~$f_2$, so that~$f_2 = \|f_2\|\, e_1$.
The conditions~\eqref{psicond1} mean that the vectors~$\psi_L$ and~$\psi_R$ are
in the span of~$e_1$ and~$e_2$.
Combining the conditions~\eqref{psicond2} and~\eqref{psicond4}
with the freedom in changing the phase of both~$\psi_L$ and~$\psi_R$, we can arrange
that in the basis~$(e_1, e_2)$, the vectors~$\psi_L$ and~$\psi_R$ have the form
\[ \psi_L = \begin{pmatrix} a \\ e^{i \varphi_L} b \end{pmatrix} \qquad \text{and} \qquad
\psi_R = \begin{pmatrix} a \\ e^{i \varphi_R} b \end{pmatrix} \]
with~$a,b \geq 0$ and suitable phases~$\varphi_{L\!/\!R}$.
Then~$\la f_2 , \psi_L \ra_{\C^3} = \la f_2 , \psi_R \ra_{\C^3} = a\,\|f_2\|$,
so that the condition~\eqref{psicond3} simplifies to
\beq \label{psifin1}
\la \psi_L, f_3 \ra_{\C^3} = \overline{ \la \psi_R, f_3 \ra_{\C^3} } \:.
\eeq
The relation~\eqref{nonzero} becomes
\[ \re \Big( \la f_2, \psi_L\ra_{\C^3}\: \la \psi_L, Y^2 f_2 \ra_{\C^3}
- \la f_2, \psi_R \ra_{\C^3}\: \la \psi_R, Y^2 f_2 \ra_{\C^3} \Big) = 1 \:, \]
which in our basis representation simplifies to
\beq \label{psifin2}
a\, \|f_2\|\: \re \big\la \psi_L-\psi_R, Y^2 f_2 \big\ra_{\C^3} = 1 \:.
\eeq

The previous construction make it possible to arrange a contribution to~$\K_{L\!/\!R}$
of the form~\eqref{genform} for an arbitrary choice of the vector fields~$v$ and~$A$.
This is not quite sufficient for compensating a contribution by a symmetric tensor field
of the form~$T_{jk} \xi^j \xi^k$ because the tensor field does not need to be
decomposable into a product of two vector fields. The tensor field can only be
written as a linear combination of such products, i.e.\
\beq \label{Tcomp}
T_{jk} = \sum_{p=1}^{p_{\max}} \left( v_j^{(p)}\, A_k^{(p)} + v_k^{(p)}\, A_j^{(p)} \right)
\eeq
for a suitable parameter~$p_{\max} \in \N$ and vector fields~$v^{(p)}$ and~$A^{(p)}$.
This leads us to generalize the ans\"atze~\eqref{i:Lvec} and~\eqref{Aevc} to
\[ L^j \gamma_j = \sum_{b=1}^B \slashed{v}^{(b)}\:
\text{diag} \big( \lambda_1^{(b)}, \ldots,  \lambda_6^{(b)} \big) \qquad \text{and} \qquad 
A^\even_{L\!/\!R} =
\sum_{b=1}^B \begin{pmatrix}  A^{N,(b)}_{L\!/\!R} & 0 \\ 0 &  A^{C,(b)}_{L\!/\!R} \end{pmatrix} \]
for a suitable parameter~$B$.
Then the linear conditions~\eqref{ortho1} and~\eqref{ortho2} can be satisfied for each~$b$ separately.
The quadratic conditions and relations, however, connect the vector fields for~$b$ and~$b'$
with~$b \neq b'$. A detailed analysis of the resulting equations shows that that it is indeed
possible to compensate the logarithmic terms for a general symmetric tensor~$T_{jk}$.
\QED

\section{Structural Contributions to the Euler-Lagrange Equations} \label{l:secstructure}
In this section, we analyze additional contributions to the EL equations to degree four
on the light cone. These contributions will not enter the field equations, but they
are nevertheless important because they give constraints for the form of the admissible
gauge fields and thus determine the structure of the interaction.
For this reason, we call them {\em{structural contributions}}.
\sindex{Euler-Lagrange equations!structural contributions}%

\subsectionn{The Bilinear Logarithmic Terms} \label{l:seclogAA}
We now return to the logarithmic singularities on the light cone.
In~\S\ref{l:seccurrent} we computed the corresponding contributions to the
EL equations to the order~$o(|\vec{\xi}|^{-3})$ at the origin.
In~\S\ref{l:secmicroloc}, we succeeded in compensating the logarithmic singularities
by a microlocal chiral transformation.
The remaining question is how the logarithmic singularities behave
in the next order in a Taylor expansion around~$\xi=0$.
It turns out that the analysis of this question yields constraints for the
form of the admissible gauge fields, as is made precise by the following proposition.
\begin{Prp} \label{l:prpbillog}
Assume that the parameter~${\mathfrak{c}}_2$ in~\eqref{l:5con} is sufficiently large and that
the chiral potentials in~\eqref{l:Bform} satisfy the conditions
\beq \label{l:LRmix}
\begin{split}
A^{11}_L - A^N_R &= \pm (A^{22}_L - A^C_R) \qquad \text{at all space-time points, and} \\
A^{11}_L - A^N_R &= -(A^{22}_L - A^C_R) \qquad \text{at all space-time points with~$A^{12}_L \neq 0$\:.}
\end{split}
\eeq
\nindex{de8@$A_R^N$ -- right-handed potential in neutrino sector}%
\nindex{df0@$A_R^C$ -- right-handed potential in charged sector}%
Moreover, in Case~{\bf{(i)}} in~\eqref{l:casesiii} we assume that the MNS matrix
and the mass matrix satisfy the relation
\beq \label{l:UYrel}
\begin{pmatrix} 0 &  \acute{U}_\text{\tiny{MNS}}^* \\ \acute{U}_\text{\tiny{MNS}} & 0 \end{pmatrix}
Y \grave{Y} = \acute{Y} Y \begin{pmatrix} 0 & \grave{U}_\text{\tiny{MNS}}^* \\
\grave{U}_\text{\tiny{MNS}} & 0 \end{pmatrix} \:.
\eeq
Then one can arrange by a suitable choice of the basis~${\mathfrak{e}}_1, \ldots, {\mathfrak{e}}_6$ and
the unitary matrix~$V$ in~\eqref{l:AevenR} that the contributions to the EL equations~$\sim |\vec{\xi}|^{-3}
\log |\vec{\xi}|$ vanish.

If conversely~\eqref{l:LRmix} does not hold and if we do not assume any relations between
the regularization parameters, then the EL equations of order~$|\vec{\xi}|^{-3} \log |\vec{\xi}|$
are necessarily violated at some space-time point.
\end{Prp} \noindent
The importance of this proposition is that it poses a further constraint on the form of the
chiral gauge potentials.

The remainder of this section is devoted to the proof of Proposition~\ref{l:prpbillog}.
Generally speaking, our task is to analyze how the gauge phases enter the
logarithmic singularities of the fermionic projector. We begin with the logarithmic current term
\[ \chi_L P^\text{aux}(x,y) \asymp -2 \,\chi_L \int_x^y [0,0 \,|\, 1]\, j_L^i\,\gamma_i\: T^{(1)} \:, \]
which gives rise to the last summand in~\eqref{l:Jterm}
(and similarly for the right-handed component; for details see~\eqref{s:dj }and~\eqref{s:jLi}
or~\cite[Appendix~B]{PFP}). According to the general rules for
inserting ordered exponentials (see~\cite[Definition~2.9]{light}, \cite[Definition~2.5.5]{PFP}
or Definition~\ref{l:def_pf}),
the gauge potentials enter the logarithmic current term according to
\beq \label{l:jphase}
-2 \,\chi_L \int_x^y 
[0,0 \,|\, 1]\, \Pe^{-i \int_x^z A_L^k (z-x)_k} j_L^i(z)\,\gamma_i \Pe^{-i \int_z^y A_L^l (y-z)_l}
\: T^{(1)}
\eeq
(where~$\Pe \equiv \Pexp$ again denotes the ordered exponential~\eqref{l:Lambda}).
Performing a Taylor expansion around~$\xi=0$ gives
\begin{align}
\chi_L P^\text{aux}(x,y) \asymp\;& -\frac{1}{3}\: \chi_L \,j_L^i \Big( \frac{x+y}{2} \Big) \gamma_i \: T^{(1)}
\label{l:cur1} \\
& +\frac{i}{6} \: \chi_L \left( A_L^j \xi_j \; j_L^i \gamma_i + j_L^i \gamma_i \;A_L^j \xi_j \right) \: T^{(1)} 
+ o(|\vec{\xi}|)  \label{l:cur2}
\end{align}
(note that in~\eqref{l:cur2} it plays no role if the functions are evaluated at~$x$ or~$y$
because the difference can be combined with the error term).

We arranged by the microlocal chiral transformation that the logarithmic singularity
of~\eqref{l:cur1} is compensated by the second summand in~\eqref{l:microlog}.
Since both~\eqref{l:cur1} and~\eqref{l:Umicro} involve the argument~$(x+y)/2$,
the logarithmic singularities compensate each other even if~$x$ and~$y$ are far apart
(up to the error terms as specified in~\eqref{l:cur2} and Proposition~\ref{l:prpmicroloc}).
Thus it remains to analyze how the gauge phases enter~\eqref{l:microlog}.
To this end, we adapt the method introduced after~\eqref{l:Lgauge} to the
matrix products in~\eqref{l:5con}.
Beginning with the left-handed component, the square of the mass matrix is modified by the
gauge phases similar to~\eqref{l:jphase} and~\eqref{l:cur1}, \eqref{l:cur2} by
\begin{align*}
\chi_L\: m^2 Y^2 &\rightarrow \chi_L
\int_x^y \Pe^{-i \int_x^z A_L^k (z-x)_k} m^2 Y^2 \Pe^{-i \int_z^y A_L^l (y-z)_l} \:dz \\
&\quad =\chi_L\: m^2 Y^2 - \frac{i}{2} \: \chi_L \: m^2\left( A_L^j \xi_j\, Y^2 + Y^2 \,A_L^j \xi_j \right)
+ o(|\vec{\xi}|)
\end{align*}
(for details see~\cite[Section~2 and Appendix~A]{light}). When using this transformation
law in~\eqref{l:5con}, we need to take into account that, similar to~\eqref{l:Lgauge}, the
chiral gauge potentials~$A_{L\!/\!R}$ must be replaced by~$A_{R\!/\!L}^\even$.
Thus we need to compute the expectation values
\beq \label{l:Pm2}
\chi_L P(x,y) \asymp -\frac{i}{2}\:m^2\: L[k] \Big( A^\even_R[\xi]\, Y^2 + Y^2 \,A^\even_R[\xi] \Big) L[k]^* \:,
\eeq
where the square bracket again denotes a contraction,
$A^\even_R[\xi] \equiv (A^\even_R)_k \,\xi^k$.

Again choosing the basis~${\mathfrak{e}}_1, \ldots, {\mathfrak{e}}_6$,
the potential~$A^\even_R$ is of the form~\eqref{l:AevenR}.
Now we must treat the diagonal and the off-diagonal elements of~$A_L$
separately. Obviously, the diagonal entries in~\eqref{l:AevenR} map
the eigenvectors~${\mathfrak{e}}_1$ and~${\mathfrak{e}}_4$ to each other.
Hence~\eqref{l:Pm2} gives rise to the anti-commutator
\begin{align}
\chi_L P(x,y) &\asymp -\frac{i}{2}\: \left\{ \begin{pmatrix}
A^{11}_L[\xi] & 0 \\ 0 & A^{22}_L[\xi]  \end{pmatrix},
L[k] \:m^2 Y^2 L[k]^* \right\} \nonumber \\
&\!\!\!\!\!\overset{\eqref{l:5con}}{=} -\frac{i}{2}\: \left\{ \begin{pmatrix}
A^{11}_L[\xi] & 0 \\ 0 & A^{22}_L[\xi]  \end{pmatrix},
\frac{\Omega}{2}\: v_L[k] + {\mathfrak{c}}_2(k)\, \1_{\C^2} \right\} . \label{l:PgD}
\end{align}
\nindex{de6@$A_L$ -- left-handed gauge potential}%
For the off-diagonal elements of~$A_L$, the matrices~$V$ and~$V^*$
in~\eqref{l:AevenR} make the situation more complicated. 
For example, the lower left matrix entry in~\eqref{l:AevenR}
maps~${\mathfrak{e}}_1$ to a non-trivial linear combination
of~${\mathfrak{e}}_4, {\mathfrak{e}}_5, {\mathfrak{e}}_6$, i.e.\ for
any $6 \times 6$-matrix~$B$,
\beq \label{l:Vfreedom}
L[k] \,B\, \begin{pmatrix} 0 & 0 \\ V & 0 \end{pmatrix} L[k]^* 
= \begin{pmatrix} \|\ell_1\|^2 \:(B^1_4 V^1_1 + B^1_5 V^2_1 + B^1_6 V^3_1) & 0 \\[0.2em]
\|\ell_2\|\,\|\ell_1\| \:(B^4_4 V^1_1 + B^4_5 V^2_1 + B^4_6 V^3_1) & 0
\end{pmatrix} .
\eeq
Similarly, the upper right matrix entry in~\eqref{l:AevenR} maps~${\mathfrak{e}}_4$ to a non-trivial
linear combination of~${\mathfrak{e}}_1, {\mathfrak{e}}_2, {\mathfrak{e}}_3$.
As a consequence, the off-diagonal elements of~$A_L$ yield a contribution to~\eqref{l:Pm2}
of the general form
\beq \label{l:PgO}
\chi_L P(x,y) \asymp A^{12}_L[\xi]\: G(k) + A^{21}_L[\xi]\: G(k)^*
\qquad \text{with} \qquad
G = \begin{pmatrix} G^{11} & 0 \\ G^{12} & G^{22} \end{pmatrix} .
\eeq
Here the $2 \times 2$-matrix $G(k)$ depends on~${\mathfrak{c}}_2$ and~$v_L[k]$
as well as on the choice of the basis vectors~${\mathfrak{e}}_2, {\mathfrak{e}}_3,
{\mathfrak{e}}_5, {\mathfrak{e}}_6$ and the matrix~$V$ in~\eqref{l:AevenR}.
Counting the number of free parameters, one sees that~$G(k)$ can be chosen
arbitrarily, up to inequality constraints which come from the fact that~$V$ must be unitary and that the
entries of the matrix~$m^2 Y^2$ in the basis~$({\mathfrak{e}}_1,
\ldots, {\mathfrak{e}}_6)$ cannot be too large due to the Schwarz inequality.
These inequality constraints can always be satisfied by suitably increasing
the parameter~${\mathfrak{c}}_2$. For this reason, we can treat~$G(k)$ as an arbitrary
matrix involving three free real parameters.

The right-handed component of the microlocal chiral transformation can be treated similarly.
The only difference is that the right-handed gauge potentials are already diagonal
in view of~\eqref{l:Bform}. Thus we obtain in analogy to~\eqref{l:PgD}
\beq
\chi_R P(x,y) \asymp -\frac{i}{2}\: \left\{ \begin{pmatrix}
A^N_R[\xi] & 0 \\ 0 & A^C_R[\xi]  \end{pmatrix},
\frac{\Omega}{2}\: v_R[k] + {\mathfrak{c}}_2(k)\, \1_{\C^2} \right\} , \label{l:PR}
\eeq
\nindex{de8@$A_R^N$ -- right-handed potential in neutrino sector}%
\nindex{df0@$A_R^C$ -- right-handed potential in charged sector}%
whereas~\eqref{l:PgO} has no correspondence in the right-handed component.

Comparing~\eqref{l:jphase} with~\eqref{l:PgD}, \eqref{l:PgO} and~\eqref{l:PR}, one sees
that the transformation laws are the same for the diagonal elements of~$A_L$ and~$A_R$.
For the off-diagonal elements of~$A_L$, we can always choose~$G(k)$ such as to get
agreement with~\eqref{l:jphase}. We conclude that by a suitable choice of~$G(k)$
we can arrange that the transformation law~\eqref{l:jphase} agrees
with~\eqref{l:PgD}, \eqref{l:PgO} and~\eqref{l:PR}. As a consequence, the logarithmic poles
of the current terms are compensated by the microlocal chiral transformation,
even taking into account the gauge phases to the order~$o(|\xi|)$.

We next consider the logarithmic mass terms
\begin{align}
\chi_L P^\text{aux}(x,y) &\asymp m^2\,\chi_L \int_x^y [1,0 \,|\, 0]\, Y Y \slashed{A}_L\: T^{(1)} \label{l:logm1} \\
&\quad -m^2\,\chi_L \int_x^y [0,0 \,|\, 0]\, Y \slashed{A}_R \,Y\: T^{(1)} \\
&\quad +m^2\,\chi_L \int_x^y [0,1 \,|\, 0]\, \slashed{A}_L \,Y Y \: T^{(1)} \:, \label{l:logm3}
\end{align}
which give rise to the summand~\eqref{l:mterm3} (for details see~\eqref{s:Pmass1}--\eqref{s:Pendgag}
or~\cite[Appendix~B]{PFP}). Here the gauge phases enter somewhat differently,
as we now describe.
\begin{Lemma} Contracting the logarithmic mass terms~\eqref{l:logm1}--\eqref{l:logm3} with
a factor~$\xi$ and including the gauge phases, we obtain
\begin{align}
\frac{1}{2} &\, \Tr \big( i \slashed{\xi}\, \chi_L \, P^\text{\rm{aux}}(x,y) \big) \nonumber \\
&\asymp \frac{i m^2}{2} \left( A_L^j \big(z_1)\, YY - 2 Y A_R^j(z_2)\, Y + YY A_L^j(z_3)
\right) \xi_j \:T^{(1)} \label{l:mp1} \\
&\quad +\frac{m^2}{8} \Big( (A_L^j \xi_j) (A_L^k \xi_k) YY + 2 (A_L^j \xi_j) YY (A_L^k \xi_k)
+ YY (A_L^j \xi_j) (A_L^k \xi_k) \Big)\, T^{(1)} \label{l:mp2} \\
&\quad -\frac{m^2}{2}\: Y (A_R^j  \xi_j) (A_R^k \xi_k)\: Y \:T^{(1)} +o \big( |\vec{\xi}|^2 \big)\:, \label{l:mp3}
\end{align}
where
\[ z_1 = \frac{3 x + y}{4} \:,\qquad z_2 = \frac{x+y}{2} \:,\qquad z_3 = \frac{x+3y}{4}\:. \]
The right-handed component is obtained by the obvious replacements~$L \leftrightarrow R$.
\end{Lemma}
\Proof Following the method of~\cite[proof of Theorem~2.10]{light} (see the proof of
Theorem~\ref{l:thm3}), we first choose a
special gauge and then use the behavior of the fermionic projector under chiral gauge transformations.
More precisely, with a chiral gauge transformation we can arrange that~$A_L$
and~$A_R$ vanish identically along the line segment~$\overline{xy}$.
In the new gauge, the mass matrix~$Y$ is no longer constant, but it is to be
replaced by dynamical mass matrices~$Y_{L\!/\!R}(x)$ (see~\cite[eq.~(2.8)]{light},
\cite[eq.~(2.5.9)]{PFP} or~\eqref{YLRdef}). Performing the light-cone expansion in this gauge,
a straightforward calculation yields
\[ \frac{1}{2} \, \Tr \big( i \slashed{\xi}\, \chi_L \, P^\text{aux}(x,y) \big)
\asymp m^2 \Big( Y_L(x)  \:Y_R(y) - \int_x^y (Y_L\, Y_R)(z)\:dz \Big) \:T^{(1)}\:. \]
Transforming back to the original gauge amounts to inserting ordered exponentials
according to the rules in~\cite[Definition~2.9]{light} (see also~\cite[Definition~2.5.5]{PFP}
or Definition~\ref{l:def_pf}).
We thus obtain
\beq \begin{split}
\frac{1}{2} \, \Tr \big( i \slashed{\xi}\, &\chi_L \, P^\text{\rm{aux}}(x,y) \big)
\asymp m^2 \:Y
\Pe^{-i \int_x^y A_R^j \xi_j} Y \:T^{(1)} \\
& -m^2 \int_x^y \Pe^{-i \int_x^z A_L^j \,(z-x)_j} \,Y Y\, \Pe^{-i \int_z^y A_L^j \,(y-z)_j} \:dz \:T^{(1)}\:.
\end{split} \label{l:phaselog}
\eeq
Expanding in powers of~$\xi$ gives the result.
\QED

The contribution~\eqref{l:mp1} is the mass term which we already encountered in~\eqref{l:mterm3}.
In contrast to~\eqref{l:cur1} and~\eqref{l:Umicro}, the term~\eqref{l:mp1} does not
only depend on the variable~$(x+y)/2$. However, forming the sectorial projection, for a
{\em{diagonal}} potential~$A_L$ we obtain in view of~\eqref{l:Bform} that
\beq \label{l:mittel}
\begin{split}
\left( \acute{A}^j_L(z_1)  \:Y \grave{Y} + \acute{Y} Y\: \grave{A}^j_L(z_3) \right) \xi_j
&= \left( A_L^j(z_1)+A_L^j(z_3) \right) \xi_j \, \acute{Y} \grave{Y} \\
&= 2 A_L^j \Big( \frac{x+y}{2} \Big)\, \xi_j\, \acute{Y} \grave{Y} + o(|\vec{\xi}|) \:.
\end{split}
\eeq
This makes it possible to compensate the logarithmic singularity of~\eqref{l:mp1}
by the second term in~\eqref{l:microlog}, up to the specified error terms.
For the {\em{off-diagonal potentials}}, the situation is more complicated
and depends on the two cases in~\eqref{l:casesiii}.
If we are in Case~{\bf{(i)}} and~\eqref{l:UYrel} is satisfied,
then~\eqref{l:mittel} also holds for the off-diagonal potentials.
As a consequence, the logarithmic singularity of~\eqref{l:mp1} can again be compensated
by the second term in~\eqref{l:microlog}.
However, if we do not impose~\eqref{l:UYrel}, then it seems impossible to
compensate the off-diagonal logarithmic terms by a microlocal chiral
transformation~\eqref{l:microlog}. If we assume instead that
we are in Case~{\bf{(ii)}} in~\eqref{l:casesiii}, then the spectral projectors~$I_n$
are diagonal~\eqref{l:I12diag}, so that off-diagonal potentials are irrelevant
as they do not enter the EL equations~\eqref{l:Knccond}.
We conclude that we can compensate the logarithmic singularities of~\eqref{l:mp1}
in Case~{\bf{(i)}} under the additional assumptions~\eqref{l:UYrel},
and in Case~{\bf{(ii)}} without any additional assumptions.

The terms~\eqref{l:mp2} and~\eqref{l:mp3}, on the other hand, are quadratic
in the chiral gauge potentials.
Analyzing whether these terms are compatible with the
transformation law~\eqref{l:PgD}, \eqref{l:PgO} and~\eqref{l:PR} of the microlocal
chiral transformation gives the following result.

\begin{Lemma} \label{l:lemmamass1}
Consider the component of the fermionic projector
which involves a bilinear tensor field and has a logarithmic pole on the light cone,
\beq \label{l:Pbillog}
P(x,y) \asymp \left( \chi_L \,h_L^{ij}(x,y) \,\gamma_i\, \xi_j + \chi_R \,h_R^{ij}(x,y) \,\gamma_i\, \xi_j \right)
 \:T^{(1)}
\eeq
\sindex{fermionic projector!bilinear logarithmic term}%
(where~$h^{ij}_{L\!/\!R}$ is a smooth tensor field acting as a $2 \times 2$-matrix on the isospin index).
If~\eqref{l:LRmix} holds and~${\mathfrak{c}}_2$ is sufficiently large,
then one can arrange by a suitable choice of the basis~${\mathfrak{e}}_1, \ldots, {\mathfrak{e}}_6$ and 
of the unitary matrix~$V$ in~\eqref{l:AevenR} that
\beq \label{l:bilcond}
\chi_L \,h_L^{ij} + \chi_R \,h_R^{ij} = h^{ij}\: \acute{Y} \grave{Y} \eeq
(where~$h^{ij}$ is a suitable tensor field which acts trivially on the isospin index).
If conversely~\eqref{l:LRmix} does not hold, then~\eqref{l:bilcond} is necessarily violated
at some space-time point.
\end{Lemma}
\Proof We first analyze the right-handed component.
If~\eqref{l:mp1} is transformed according to~\eqref{l:PR}, we could argue just as for the
logarithmic current terms to conclude that the contribution of the form~\eqref{l:Pbillog} vanishes.
Therefore, it suffices to consider the terms obtained by subtracting from~\eqref{l:mp1}--\eqref{l:mp3}
the term~\eqref{l:mp1} transformed according to~\eqref{l:Pm2}, giving rise to the expression
\beq \label{l:BLform}
\begin{split}
B_L :=\:& -\frac{m^2}{4} \Big\{ A_R^\even[\xi], \left( A_L[\xi] \big(z_1)\, YY - 2 Y A_R[\xi]\, Y + YY A_L[\xi]
\right) \Big\} \:T^{(1)} \\
&+\frac{m^2}{8} \Big( A_L[\xi]^2  YY + 2 A_L[\xi] YY A_L[\xi]
+ YY A_L[\xi]^2 \xi_k) \Big)\, T^{(1)} \\
&-\frac{m^2}{2}\: Y A_R[\xi]^2\: Y \:T^{(1)} \:,
\end{split}
\eeq
and similarly for the right-handed component. We must arrange that~$\hat{B}_L$
and~$\hat{B}_R$ are multiples of the identity matrix and coincide.
We first analyze~$B_R$.
Then according to~\eqref{l:AevenR}, the potential~$A_L^\even$ coincides with~$A_R$
and is sector diagonal. We thus obtain
\begin{align*}
B_R =\:&-\frac{m^2}{8} \Big( A_R[\xi]^2\, YY +2 A_R[\xi] \,YY A_R[\xi] + YY A_R[\xi]^2 \Big)\: T^{(1)} \\
&+\frac{m^2}{2} \Big( A_R[\xi]\, Y A_L[\xi]\, Y - Y A_L[\xi]^2\, Y
+ Y A_L[\xi]\, Y A_R[\xi] \Big)\: T^{(1)} \:.
\end{align*}
We decompose~$A_L$ into its diagonal and off-diagonal elements, denoted by
\[ A_L[\xi] = A_L^d[\xi] + A_L^o[\xi] \:. \]
A straightforward calculation using the identity
\[ A_L[\xi]^2 = A^d_L[\xi]^2 + A^o_L[\xi]^2 + \{ A^d_L[\xi], A^o_L[\xi] \} \]
gives
\begin{align}
B_R =\;&-\frac{m^2}{2} \: Y \,\Big( \big( A_R[\xi] - A^d_L[\xi] \big)^2 + A^o_L[\xi]^2 \Big)\, Y \:T^{(1)} \label{l:B1} \\
&+\frac{m^2}{2}\: Y \,\Big\{ A_R[\xi] - A_L^d[\xi], A^o_L[\xi] \Big\}\, Y \:T^{(1)} \:. \label{l:B2}
\end{align}
Clearly, the matrix~$A^o_L[\xi]^2$ is a multiple of the identity matrix.
The matrix~$(A_R[\xi] - A^d_L)^2$, on the other hand, is a multiple of the identity
matrix if and only if~\eqref{l:LRmix} holds.
The anti-commutator in~\eqref{l:B2} is zero on the diagonal.
It vanishes provided that~\eqref{l:LRmix} holds.
We conclude that the
contributions~\eqref{l:B1} and~\eqref{l:B2} act trivially on the isospin index
if and only if~\eqref{l:LRmix} holds. In this case,
\beq \label{l:Bres}
B_R = -\frac{m^2}{2} \: Y \,\Big( \big( A_R[\xi] - A^d_L[\xi] \big)^2 + A^o_L[\xi]^2 \Big)\, Y \:T^{(1)}\:.
\eeq

It remains to show that under the assumption~\eqref{l:LRmix}, we can arrange that
the corresponding left-handed contribution~$\hat{B}_L$
is also a multiple of the identity matrix, and that it coincides with~\eqref{l:Bres}.
Now~$A_R^\even$ is given by~\eqref{l:AevenR}. The diagonal entries of~$A_R^\even$
coincide with those of~$A_L$, giving rise to the contribution
\begin{align*}
B_L \asymp\:
&-\frac{m^2}{4} \Big\{ A^d_L[\xi], \big( A_L[\xi]\, YY - 2 Y A_R[\xi]\, Y + YY A_L[\xi] \big) \Big\}\: T^{(1)} \\
&+\frac{m^2}{8} \left( A_L[\xi]^2 \,YY + 2 A_L[\xi]\, YY A_L[\xi] + YY A_L[\xi]^2 \right) T^{(1)}
-\frac{m^2}{2}\: Y A_R[\xi]^2\, Y \:T^{(1)}\:.
\end{align*}
Similar as in~\eqref{l:PgO}, we can add contributions which involve~$A_L^{12}$
or~$A_L^{21}$. A short calculation shows that in this way,
we can indeed arrange that~$\hat{B}_L$ coincides with~$\hat{B}_R$ as given by~\eqref{l:Bres}.
\QED

\begin{Remark} {\em{  {\bf{(necessity of a mixing matrix)}} \label{l:rem54}
\sindex{mixing matrix!necessity of}%
The proof of Lemma~\ref{l:lemmamass1} even gives
an explanation why the mixing matrix~$\UMNS$ must occur. Namely, the 
following consideration shows that the method of proof fails if~$\UMNS$ is trivial:
Suppose that~$\UMNS=\1$. Then the parameter~$c$ in~\eqref{l:cdef} is equal to one.
As a consequence, the parameter~$d$ in~\eqref{l:doccur} also equals one,
implying that~$V^1_1=1$. Since~$V$ is unitary, it follows that~$V^2_1=V^3_1=0$.
As a consequence, the freedom in choosing~$V(x)$ does not make it possible
to modify~\eqref{l:Vfreedom}. Thus the matrix~$G(k)$ in~\eqref{l:PgO} can no longer
be chosen arbitrarily, in general making it impossible to arrange~\eqref{l:bilcond}.

We note that in~\S\ref{q:secproof}, a different argument is given which also yields that the
mixing matrix must be non-trivial (see Lemma~\ref{q:lemma5319}). }} \QEDrem
\end{Remark}

The next lemma gives the connection to the EL equations.
\begin{Lemma} \label{l:lemmamass2} The contributions to the EL equations~$\sim |\vec{\xi}|^{-3}
\log |\vec{\xi}|$ vanish if and only if the condition~\eqref{l:bilcond} holds.
\end{Lemma}
\Proof A direct computation shows that the terms of the form~\eqref{l:Pbillog}
contribute to the EL equations of the order~$|\vec{\xi}|^{-3}
\log |\vec{\xi}|$ unless~\eqref{l:bilcond} holds. Therefore, our task
is to show that it is impossible to compensate a term of the form~\eqref{l:Pbillog}
by a generalized microlocal chiral transformation.
It clearly suffices to consider the homogeneous setting in the high-frequency limit
as introduced in~\S\ref{s:secnonlocaxial}. Transforming to momentum space,
the contribution~\eqref{l:Pbillog} corresponds to the distribution
\beq \label{l:wanted2}
\gamma^i \,h_{ij}\,k^j \:\delta''(k^2)\: \Theta(-k^0)\:.
\eeq
Having only three generations to our disposal, such a contribution would necessarily give rise to
error terms of the form
\[ \frac{1}{m^2} \gamma^i \,h_{ij}\,k^j \:\delta'(k^2)\: \Theta(-k^0) \qquad \text{or} \qquad
\frac{1}{m^4} \gamma^i \,h_{ij}\,k^j \:\delta(k^2)\: \Theta(-k^0) \:. \]
These error terms are as large as the shear contributions by local axial transformation
as analyzed in \S\ref{s:secprobaxial}, causing problems in the EL equations
(for details see~ \S\ref{s:secprobaxial} and Appendix~\ref{s:applocaxial}).
Instead of going through these arguments again, we here rule out~\eqref{l:wanted2}
with the following alternative consideration: In order to generate the contribution~\eqref{l:wanted2},
at least one of the Dirac seas would have to be perturbed by a contribution with the scaling
\[ \frac{1}{m^4}\: \gamma^i \,h_{ij}\,k^j \: \delta(k^2-m_\alpha^2)\: \Theta(-k^0) \:. \]
Due to the factor~$k^j$, this perturbation is by a scaling factor~$\Omega$ larger than the
perturbations considered in~\S\ref{s:secnonlocaxial}. Thus one would have to consider a
transformation of the form (cf.~\eqref{s:Zansatz})
\[ U = \exp \left( i Z \right) \qquad \text{with} \qquad Z = \O(\Omega^0) \:. \]
This transformation does not decay in~$\Omega$ and thus cannot be treated perturbatively.
Treating it non-perturbatively, the resulting shear contributions violate the EL equations.
\QED
Combining Lemmas~\ref{l:lemmamass1} and~\ref{l:lemmamass2} gives Proposition~\ref{l:prpbillog}.

\subsectionn{The Field Tensor Terms} \label{l:secfield}
We now come to the analysis of the contributions to the fermionic projector
\sindex{fermionic projector!field tensor term}%
\begin{align}
\chi_L\, P(x,y) &\asymp \frac{1}{4} \:\chi_L\:\slashed{\xi} \int_x^y F_L^{ij} \, \gamma_i \gamma_j \: T^{(0)} 
- \chi_L \:\xi_i \int_x^y [0,1 \,|\, 0]\, F_L^{ij}\, \gamma_j \: T^{(0)} \nonumber \\
&= \frac{1}{2} \:\chi_L\: \int_x^y (2 \alpha-1)\: \xi_i \,F_L^{ij}\, \gamma_j\: T^{(0)}
+ \frac{i}{4} \:\chi_L \int_x^y \epsilon_{ijkl} \,F_L^{ij}  \,\xi^k\, \pseudo \gamma^l\: T^{(0)}\:,
\label{l:FT}
\end{align}
which we refer to as the {\em{field tensor terms}}
(see~\cite[Appendix~A]{firstorder}, \cite[Appendix~A]{light} and Appendix~\ref{s:appspec};
note that here we only consider the {\em{phase-free}} contributions,
to which gauge phases can be inserted according to the rules in~\cite{light}
or Definition~\ref{l:def_pf}).
In Chapter~\ref{sector}, the field tensor terms were disregarded because they vanish
when the Dirac matrices are contracted with outer factors~$\xi$. Now we will analyze the field tensor
terms in the $\iota$-formalism introduced in~\S\ref{l:seciota}. 
This will give additional constraints for the form of the admissible gauge fields
(see relation~\eqref{l:FTcond} below).

In this section, the corrections in~$\tau_\reg$ are essential. It is most convenient to
keep the terms involving~$\tau_\reg$ in all computations.
We assume that we evaluate weakly for such a small vector~$\xi$
that we are in Case~{\bf{(ii)}} in~\eqref{l:casesiii}
(this will be discussed in Section~\ref{l:sec6}).
It then suffices to consider the sector-diagonal elements of the closed chain.
Moreover, by restricting attention to the first or second isospin component, we
can compute the spectral decomposition of the closed chain in the neutrino 
sector ($n=1$) and the chiral sector ($n=2$) separately.
For a uniform notation, we introduce the notation
\nindex{dk0@$M^{(l)}_n$ -- short notation for factors~$T^{(l)}_{[0]}$ or~$L^{(l)}_{[0]}$}%
\[ M^{(l)}_n = \left\{ \begin{array}{cl} L^{(l)}_{[0]} & \text{if~$n=1$} \\[0.4em]
T^{(l)}_{[0]} & \text{if~$n=2$}
\end{array} \right. \]
with~$L^{(l)}_\circ$ as given by~\eqref{l:Lndef}.
Then the unperturbed eigenvalues are given by
\[ \lambda_{nL-} = 9\, T^{(-1)}_{[0]} \overline{M^{(0)}_n} \:,\qquad
\lambda_{nR-} = 9\, M^{(-1)}_n \overline{T^{(0)}_{[0]}} \:. \]
Moreover, using the calculations
\begin{gather*}
\frac{\overline{\lambda_{nL-}}}{|\lambda_{nL-}|}\:\chi_L P(x,y) =
3i\, \chi_L\, \slashed{\xi}\: \frac{ M^{(0)}_n \overline{T^{(-1)}_{[0]}} }
{\big| T^{(-1)}_{[0]} M^{(0)}_n \big|}\: T^{(-1)}_{[0]}
= 3i\, \chi_L\, \slashed{\xi}\: \frac{ \big| T^{(-1)}_{[0]} \big| }{\big| M^{(0)}_n \big|}\: M^{(0)}_n \\
\frac{\overline{\lambda_{nR-}}}{|\lambda_{nR-}|}\:\chi_R P(x,y) =
3i\, \chi_R\, \slashed{\xi}\: \frac{ \big| M^{(-1)}_n \big| }{\big| T^{(0)}_{[0]} \big|}\: T^{(0)}_{[0]}
\end{gather*}
in~\eqref{l:EL4}, we can write the EL equations as
\begin{align}
\bigg( \Delta |\lambda_{nL-}| - \frac{1}{4} \sum_{n',c'} \Delta |\lambda_{n' c' -}| \bigg)
\frac{ \big| T^{(-1)}_{[0]} \big| }{\big| M^{(0)}_n \big|}\: M^{(0)}_n &= 0 \label{l:ELl} \\
\bigg( \Delta |\lambda_{nR-}| - \frac{1}{4} \sum_{n',c'} \Delta |\lambda_{n' c' -}| \bigg)
\frac{ \big| M^{(-1)}_n \big| }{\big| T^{(0)}_{[0]} \big|}\: T^{(0)}_{[0]} &= 0 \:. \label{l:ELr}
\end{align}
Note that in the case~$\tau_\reg=0$, these equations reduce to
our earlier conditions~\eqref{l:Knccond} and~\eqref{l:Kncdef}.

Our task is to analyze how~\eqref{l:FT} influences the eigenvalues~$\lambda_{nc-}$ of the closed chain.
\begin{Lemma} \label{l:lemmaFT}
The field tensor terms~\eqref{l:FT} contribute to the eigenvalues~$\lambda_{nc-}$ by
\begin{align}
\lambda_{nL-} \asymp\:& \frac{3i}{2} \: \int_x^y (2 \alpha-1)\: \Tr_{\C^2} \Big(I_n\, \hat{F}_L^{ij}\,
\check{\xi}_i \, \big( \iota^{(-1)}_{[0]} \big)_j \Big) \: T^{(0)}_{[0]} \overline{M^{(0)}_n} \label{l:FTL1} \\
&+\frac{3}{4} \int_x^y \Tr_{\C^2} \Big( I_n \:\epsilon_{ijkl} \,\hat{F}_L^{ij}  \,\check{\xi}^k\,
\big( \iota^{(-1)}_{[0]} \big)^l \Big)\: T^{(0)}_{[0]} \overline{M^{(0)}_n} + (\deg < 2) \label{l:FTL2} \\
\lambda_{nR-} \asymp\:& \frac{3i}{2} \: \int_x^y (2 \alpha-1)\: \Tr_{\C^2} \Big(I_n\, \hat{F}_R^{ij}\,
\check{\xi}_i \, \big( \iota^{(-1)}_{[0]} \big)_j \Big) \:M^{(0)}_n \overline{T^{(0)}_{[0]}}  \label{l:FTR1} \\
&-\frac{3}{4} \int_x^y \Tr_{\C^2} \Big( I_n \:\epsilon_{ijkl} \,\hat{F}_R^{ij}  \,\check{\xi}^k\,
\big( \iota^{(-1)}_{[0]} \big)^l \Big)\:M^{(0)}_n \overline{T^{(0)}_{[0]}}  + (\deg < 2) \:. \label{l:FTR2}
\end{align}
\end{Lemma}
\Proof
We first consider the effect of a left-handed field on the left-handed eigenvalues.
Every summand in~\eqref{l:FT} involves a factor~$\xi T^{(0)}$.
As the factor~$\iota^{(0)}$ gives no contribution (see~\eqref{l:iota3}), we regularize~\eqref{l:FT} in the
$\iota$-formalism by
\beq \label{l:FTreg}
\chi_L\, P(x,y) \asymp \frac{1}{2} \:\chi_L \int_x^y (2 \alpha-1)\: \check{\xi}_i \,\hat{F}_L^{ij}\, \gamma_j\: T^{(0)}_{[0]}
+ \frac{i}{4} \:\chi_L \int_x^y \epsilon_{ijkl} \,\hat{F}_L^{ij}  \,\check{\xi}^k\, \pseudo \gamma^l\: T^{(0)}_{[0]}
\eeq
(where the hat again denotes the sectorial projection). For computing the effect on the eigenvalues, we
first multiply by the vacuum fermionic projector~$P^{(0)}(y,x)$ to form the closed chain.
Then we multiply by powers of the vacuum chain~\eqref{l:Axyp}
and take the trace. Since the number of factors~$\iota$ in~\eqref{l:Axyp} always equals the number
of factors~$\hat{\xi}$, and taking into account that~\eqref{l:FTreg} vanishes when contracted with
a factor~$\hat{\xi}$, we conclude that the factor~$P^{(0)}(y,x)$ must contain a factor~$\iota$.
In view of~\eqref{l:iota3}, this means that we only need to take into account the
contribution~$P^{(0)}(y,x) \asymp -3i \:\overline{\iotaslsh^{(-1)}_{[0]}\:L^{(0)}_{[0]}}$. We thus obtain
\begin{align*}
\chi_L\, A_{xy} \asymp\:&
\frac{3i}{2} \:\chi_L\: \int_x^y (2 \alpha-1)\: \check{\xi}_i \,\hat{F}_L^{ij}\, \gamma_j\: T^{(0)}_{[0]}\:
\overline{ \iotaslsh^{(-1)}_{[0]} \:M^{(0)}_n } \\
&- \frac{3}{4} \:\chi_L \int_x^y \epsilon_{ijkl} \,\hat{F}_L^{ij}  \,\check{\xi}^k\,
\pseudo \gamma^l\: T^{(0)}_{[0]} \:\overline{ \iotaslsh^{(-1)}_{[0]} \:M^{(0)}_n }\:.
\end{align*}
Since the last Dirac factor involves~$\iota$, this contribution vanishes when multiplied by
the first summand in~\eqref{l:Axyp}. Thus our field tensor term only influences
the eigenvalue~$\lambda_{nL-}$. A short calculation gives~\eqref{l:FTL1} and~\eqref{l:FTL2}.
Similarly, a right-handed field only influences the corresponding right-handed eigenvalues
by~\eqref{l:FTR1} and~\eqref{l:FTR2}. The result follows by linearity.
\QED
Before going on, we remark that the above contributions do not appear in the standard
formalism of the continuum limit, where all factors~$\xi$ which are contracted to macroscopic
functions are treated as outer factors. In order to get back to the standard formalism, one can simply
impose that~$F_{ij} \check{\xi}^i \iota^j=0$. However, this procedure, which was implicitly used in
Chapter~\ref{sector}, is not quite convincing because it only works if the regularization is adapted locally to the
field tensor. If we want to construct a regularization which is admissible for any field tensor
(which should of course satisfy the field equations), then the contributions
by Lemma~\ref{l:lemmaFT} must be taken into account.
\begin{Corollary} \label{l:corFT} Introducing the macroscopic functions
\begin{align}
a_{n\, L\!/\!R} =\:& \frac{3i}{4} \: \int_x^y (2 \alpha-1)\: \Tr_{\C^2} \Big(I_n\, \hat{F}_{L\!/\!R}^{ij}\,
\check{\xi}_i \, \big( \iota^{(-1)}_{[0]} \big)_j \Big) \label{l:anc1} \\
&\pm \frac{3}{8} \int_x^y \Tr_{\C^2} \Big( I_n \:\epsilon_{ijkl} \,\hat{F}_{L\!/\!R}^{ij}  \,\check{\xi}^k\,
\big( \iota^{(-1)}_{[0]} \big)^l \Big) \:, \label{l:anc2}
\end{align}
the absolute values of the eigenvalues are perturbed by the field tensor terms~\eqref{l:FT} according to
\begin{align*}
\Delta |\lambda_{nL-}| &= 
\frac{\big| M^{(0)}_n \big|}{\big| T^{(-1)}_{[0]} \big|}\:
\Big( a_{nL} \: T^{(0)}_{[0]} \overline{T^{(-1)}_{[0]}} + 
\overline{a_{nL}} \: T^{(-1)}_{[0]} \overline{T^{(0)}_{[0]}} \Big) \\
\Delta |\lambda_{nR-}| &= \frac{\big| T^{(0)}_{[0]} \big|}{\big| M^{(-1)}_n \big|}\:
\Big( a_{nR} \: M^{(0)}_n \overline{M^{(-1)}_n} 
+ \overline{a_{nR}} \: M^{(-1)}_n \overline{M^{(0)}_n} \Big) \:.
\end{align*}
\end{Corollary}
\Proof Writing the result of Lemma~\ref{l:lemmaFT} as
\[ \Delta \lambda_{nL-} = 2 a_{nL} \: T^{(0)}_{[0]} \overline{M^{(0)}_n} \:,\qquad
\Delta \lambda_{nR-} = 2 a_{nR} \: M^{(0)}_n \overline{T^{(0)}_{[0]}} \:, \]
we obtain
\[ \Delta |\lambda_{nL-}| = \frac{1}{\big| T^{(-1)}_{[0]} \overline{M^{(0)}_n} \big|}\:
\re \left( a_{nL} \: T^{(0)}_{[0]} \overline{M^{(0)}_n} M^{(0)}_n \overline{T^{(-1)}_{[0]}} \right)
= \frac{\big| M^{(0)}_n \big|}{\big| T^{(-1)}_{[0]} \big|}\:
\re \left( a_{nL} \: T^{(0)}_{[0]} \overline{T^{(-1)}_{[0]}} \right) \:. \]
The calculation for~$\Delta |\lambda_{nR-}|$ is analogous.
\QED

After these preparations, we are ready to analyze the EL equations~\eqref{l:ELl}, \eqref{l:ELr}.
We begin with the case~$\tau_\reg=0$. Then we can set~$M^{(l)}_n = T^{(l)}_{[0]}$,
giving the conditions~\eqref{l:Knccond}, where now
\[ {\mathscr{K}}_{nc} = \Delta |\lambda_{nL-}| \;
\frac{ \big| T^{(-1)}_{[0]} \big| }{\big| T^{(0)}_n \big|}\: M^{(0)}_n
= a_{nc} \: T^{(0)}_{[0]} T^{(0)}_{[0]} \overline{T^{(-1)}_{[0]}} + 
\overline{a_{nc}} \: T^{(0)}_{[0]} T^{(-1)}_{[0]}  \overline{T^{(0)}_{[0]}} \:. \]
This formula can be simplified further with the integration-by-parts rules.
Namely, applying~\eqref{l:ipart}, we obtain
\[ 0 = \nabla \left( T^{(0)}_{[0]}T^{(0)}_{[0]} \overline{T^{(0)}_{[0]}} \right)
= 2\, T^{(0)}_{[0]}T^{(-1)}_{[0]} \overline{T^{(0)}_{[0]}}
+ T^{(0)}_{[0]}T^{(0)}_{[0]} \overline{T^{(-1)}_{[0]}} \:. \]
Using this relation, we conclude that
\[ {\mathscr{K}}_{nc} = -\Big( 2 a_{nc} - \overline{a_{nc}} \Big)\, T^{(0)}_{[0]} T^{(-1)}_{[0]} \overline{T^{(0)}_{[0]}}
= -\Big( \re(a_{nc}) + 3 \im(a_{nc}) \Big)\, T^{(0)}_{[0]} T^{(-1)}_{[0]} \overline{T^{(0)}_{[0]}} \:. \]
If any non-trivial gauge field is present, the four macroscopic functions~$\re(a_{nc}) + 3 \im(a_{nc})$ will not
all be the same (note that even for a vectorial field which acts trivially on the isospin index,
the contribution~\eqref{l:anc2} has opposite signs for~$a_{nL}$ and~$a_{nR}$).
This implies that~\eqref{l:Knccond} can be satisfied only if we impose the regularization condition
\beq
T^{(0)}_{[0]} T^{(-1)}_{[0]} \overline{T^{(0)}_{[0]}} = 0 \qquad
\text{in a weak evaluation on the light cone} \:. \label{l:RC1}
\eeq

In order to compute the effect of~$\tau_\reg$, we first note that
the perturbations~$\Delta |\lambda_{2c-}|$ do not involve~$\tau_\reg$
(as is obvious from Corollary~\ref{l:corFT}).
Moreover, the contribution of these eigenvalues to~\eqref{l:ELl} and~\eqref{l:ELr}
for~$n=2$ is independent of~$\tau_\reg$. In view of~\eqref{l:RC1}, these contributions
drop out of the EL equations.
Next, the eigenvalue~$\lambda_{1L-}$ contributes to~\eqref{l:ELl} and~\eqref{l:ELr} for~$n=2$ by
\beq \label{l:L0ex}
-\frac{1}{4} \:\Delta |\lambda_{1L-}| \frac{ \big| T^{(-1)}_{[0]} \big| }{\big| T^{(0)}_{[0]} \big|}\: T^{(0)}_{[0]}
= \frac{\big| L^{(0)}_{[0]} \big|}{\big| T^{(0)}_{[0]} \big|}\: T^{(0)}_{[0]}
\Big( a_{nL} \: T^{(0)}_{[0]} \overline{T^{(-1)}_{[0]}} + 
\overline{a_{nL}} \: T^{(-1)}_{[0]} \overline{T^{(0)}_{[0]}} \Big) .
\eeq
This is in general non-zero. Thus in order to allow for left-handed gauge fields in the
neutrino sector, we need to impose additional conditions on the regularization functions.
The simplest method is to impose that
\beq \label{l:T0point}
\big| L^{(0)}_{[0]} \big| = \big| T^{(0)}_{[0]} \big| \left( 1 + \O \big( (m \varepsilon)^{2 p_\reg} \big) \right)
\qquad \text{pointwise}\:.
\eeq
Then~\eqref{l:L0ex} again vanishes as a consequence of~\eqref{l:RC1},
up to terms quadratic in~$\tau_\reg$. We note that this is compatible with~\eqref{l:rcond} and poses
an additional condition on the regularization in the case~$n=0$ and~$p=0$.
We also remark that~\eqref{l:T0point} could be replaced by a finite number of equations
to be satisfied in a weak evaluation on the light cone. But as these equations are rather
involved, we here prefer the stronger pointwise condition~\eqref{l:T0point}.
We also note that, in contrast to the condition~\eqref{l:RC1} for the regularization
of ordinary Dirac seas, the relation~\eqref{l:T0point} imposes a constraint only on the
right-handed high-energy states.

It remains to consider the terms involving~$T^{(-1)}_{[R,0]}$. These are~$\Delta |\lambda_{1R-}|$
as well as the factor~$|M^{(-1)}_n|$ in~\eqref{l:ELr} in case~$n=1$.
Collecting all the corresponding contributions to the EL equations, we get a finite
number of equations to be satisfied in a weak evaluation on the light cone.
Again, we could satisfy all these equations by imposing suitable conditions on the regularization.
However, these additional conditions would basically imply that~$T^{(-1)}_{[R,0]} = 0$
vanishes, meaning that there are no non-trivial regularization effects.
For this reason, our strategy is not to impose any more regularization conditions.
Then the EL equations~\eqref{l:ELl} and~\eqref{l:ELr} are satisfied if and only if
there is no right-handed gauge field in the neutrino sector and if the vectorial component
is trace-free, because only under these conditions all the equations involving~$T^{(-1)}_{[R,0]}$
or~$\overline{T^{(-1)}_{[R,0]}}$ are satisfied.
If the field tensor vanishes everywhere, we can arrange by a global gauge transformation
that also the corresponding potential vanishes globally.
We have thus derived the following result.

\begin{Prp} \label{l:prp58} Taking into account the contributions by the field tensor terms
in Lemma~\ref{l:lemmaFT}, the EL equations to degree four can be satisfied
only if the regularization satisfies the conditions~\eqref{l:RC1} and~\eqref{l:T0point}
(or a weaker version of~\eqref{l:T0point} involving weak evaluations on the light cone).
If no further regularization conditions are imposed, then the chiral potentials must
satisfy at all space-time points the conditions 
\beq \label{l:FTcond}
\Tr( {\mathfrak{I}}_1 A_R ) = 0 \qquad \text{and} \qquad \Tr( A_L + A_R ) = 0 \:,
\eeq
where~${\mathfrak{I}}_1$ is the projection on the neutrino sector.
If conversely the conditions~\eqref{l:RC1}, \eqref{l:T0point} and~\eqref{l:FTcond} are satisfied, then
the field tensor terms do not contribute to the EL equations of degree four.
\end{Prp} \noindent
We note for clarity that in the vacuum, the operator~${\mathfrak{I}}_1$ coincides
with the projection operator~$I_1$ in~\eqref{l:nuabs}. However, the operator~$I_1$ in general
depends on the gauge potentials. The operator~${\mathfrak{I}}_1$, however, is
a fixed matrix projecting on the neutrino sector. The reason why~${\mathfrak{I}}_1$
comes up is that the conditions~\eqref{l:FTcond} are derived in Case~{\bf{(ii)}} in~\eqref{l:casesiii} where the
matrix~$I_1$ is given by~\eqref{l:I12diag}.

\section{The Effective Action in the Continuum Limit} \label{l:sec5}
\subsectionn{Treating the Algebraic Constraints} \label{l:secalgebra}
Let us briefly review our general strategy for deriving the field equations in the continuum limit.
The starting point is the fermionic projector of the vacuum, being composed of solutions
of the free Dirac equation (cf.~\eqref{l:Dfree})
\[ (i \Pdd - m Y) \,\psi = 0 \:. \]
As explained in~\S\ref{l:seccurrent}, a wave function~$\psi$ gives rise to a contribution
to the fermionic projector of the form (cf.~\eqref{l:particles})
\beq \label{l:psiex0}
-\frac{1}{2 \pi} \overline{\psi}(x) \psi(x) \:,
\eeq
which enters the EL equations. Our method for satisfying the EL equations is to
introduce a suitable potential into the Dirac equation (see~\eqref{l:Dinteract})
\beq \label{l:direx0}
(i \Pdd + \B - m Y) \psi = 0 \:.
\eeq
After compensating the resulting logarithmic singularities by a microlocal chiral transformation
(see~\S\ref{l:secmicroloc}), we can hope that the remaining contributions to
the fermionic projector by the bosonic current can
compensate the contribution by the Dirac current~\eqref{l:psiex0}.
This general procedure was worked out for an axial potential in Chapter~\ref{sector}.

In the present setting of two sectors, the situation is more complicated
because the potential~$\B$ in~\eqref{l:direx0}
must satisfy additional equations which involve the chiral potentials without
derivatives (see~\eqref{l:Bform} and the
conditions arising from the structural contributions in Section~\ref{l:secstructure}).
We refer to these equations as the {\em{algebraic constraints}}.
\sindex{potential!algebraic constraint for}%
These constraints imply that the corresponding bosonic currents must be of
a specific form. As a consequence, we cannot expect to compensate an arbitrary contribution
of the form~\eqref{l:psiex0} in the EL-equations~\eqref{l:EL}.
Graphically speaking, we can only hope to compensate those contributions to~\eqref{l:psiex0} which
are ``parallel'' to the bosonic degrees of freedom.
The problem is that it is not obvious how to decompose the Dirac current~\eqref{l:psiex0}
into contributions ``parallel'' and ``orthogonal'' to the degrees of freedom of the bosonic current,
simply because there is no obvious scalar product on the contributions to~$Q$.
The goal of this section is to give a systematic procedure for deriving the
field equations in the continuum limit, taking into account the algebraic constraints.
These field equations will be recovered as the critical points of a corresponding
{\em{effective action}}.
\sindex{effective action}%
\sindex{action!effective|see{effective action}}%

In order to understand the concept behind the derivation of the effective action,
one should keep in mind that we regard only the fermionic states (including the states
of the Dirac sea) as the basic physical objects. The bosonic fields, however, are merely
auxiliary objects used for describing the collective behavior of the fermionic states.
With this in mind, in contrast to the usual Lagrangian formalism,
we do not need to derive the Dirac equation from an action principle. On the contrary,
the Dirac equation~\eqref{l:direx0} serves as the {\em{definition}} of~$\B$ introduced for describing
the behavior of the fermionic states.
Then the appearance of algebraic constraints like~\eqref{l:Bform} can be understood as
constraints for the admissible variations of the fermionic projector.
Such constraints are typically handled by demanding
that the action should be critical only under the admissible variations.
Thus in our setting, the natural idea is to demand that the first variation of
the action (see~\eqref{s:QdP})
\beq \label{l:varS}
\delta \Sact_\mu[P] = 2 \tr \left( Q\; \delta P \right)
\eeq
should vanish for all variations~$\delta P$ which are admissible in the sense that they are
described by variations of~$\B$ which satisfy the algebraic constraints.
Unfortunately, this method cannot be implemented directly because,
in order to obtain information independent of regularization details, the operator~$Q$
must be evaluated weakly on the light-cone (see~\S\ref{s:sec51} and~\S\ref{s:secELC}).
This means that the kernel~$\delta P(x,y)$ of the operator~$\delta P$ must satisfy the two
conditions that it be smooth and that it vanishes in a neighborhood of the diagonal~$x=y$,
\beq \label{l:twocond}
\delta P(x,y) \in C^\infty(\scrM \times \scrM) \qquad \text{and} \qquad
\delta P(x,y) = 0 \text{ unless~$|\vec{\xi}| \gg \varepsilon$}\:.
\eeq
However, the perturbation~$\delta P$ corresponding to a perturbation of the
bosonic potentials does not satisfy these two conditions, because in this case~$\delta P(x,y)$
is singular on the light cone and non-zero at~$x=y$.

Our method to overcome this shortcoming is to take~$\delta P$ as obtained from
a variation of~$\B$, and to arrange the additional requirements~\eqref{l:twocond}
by smoothing~$\delta P(x,y)$ and by setting it to zero in a neighborhood of~$x=y$.
This procedure can be understood as follows: The bosonic potentials satisfying
the algebraic constraints tell us about the admissible directions for varying~$P$.
But these variations need not necessarily be performed for all the Dirac states
simultaneously. Instead, it seems reasonable that only the low-energy states
(i.e.\ the states with frequencies $\ll \varepsilon^{-1}$) are varied. Then~$\delta P$ is smooth.
Moreover, by combining different such variations, one can arrange that~$\delta P$ vanishes
at the origin. The resulting variations satisfy~\eqref{l:twocond}, and we use them for
testing in~\eqref{l:varS}.

The goal of this section is to use the just-described method to derive EL equations in the continuum limit.
For clarity, we first treat the chiral gauge field, whereas the gravitational field will be considered
afterwards in a similar manner.
In preparation, we note that quantities like currents and fields take values in the Hermitian $6 \times 6$-matrices
and have a left- and right-handed component. Thus, taking the direct sum of the two chiral components,
it is useful to introduce the real vector space
\beq \label{l:S6def}
{\mathfrak{S}}_6 := \Symm(\C^6) \oplus \Symm(\C^6) \:,
\eeq
where~$\Symm(\C^6)$ denotes the Hermitian $6 \times 6$-matrices.
\nindex{dk2@$\Symm(\C^n)$ -- Hermitian $n \times n$-matrices}%
\nindex{dk4@${\mathfrak{S}}_n = \Symm(\C^6) \oplus \Symm(\C^6)$ --  left- and right-handed matrices}%
For example, the Dirac current~\eqref{l:Jdef} can be regarded as an element of~${\mathfrak{S}}_6$,
\[ {\mathscr{J}} := (J_L, J_R) \in {\mathfrak{S}}_6 \]
(here we disregard the tensor indices, which will be included later in a straightforward way).
In the EL equations, the Dirac current enters only after forming the sectorial projection.
In what follows, it is convenient to consider the sectorial projection as an operation
\beq \label{l:hatdef}
\hat \::\: {\mathfrak{S}}_6 \rightarrow {\mathfrak{S}}_2 \subset {\mathfrak{S}}_6 \:,
\eeq
where in the last inclusion we regard a symmetric $2 \times 2$-matrix
as a $6 \times 6$-matrix which acts trivially on the generation index
(and~${\mathfrak{S}}_2$ denotes similar to~\eqref{l:S6def} the chiral Hermitian $2 \times 2$-matrices).
The chiral gauge potential and current~\eqref{l:jdef} can also be regarded as elements of~${\mathfrak{S}}_6$.
However, they can take values only in a subspace of~${\mathfrak{S}}_6$, as we now make precise.
We denote the gauge group corresponding to the admissible gauge potentials
by~$\G \subset \U(6)_L \times \U(6)_R$ and refer to it as the {\em{dynamical gauge group}}
\sindex{gauge group!dynamical}%
\nindex{dk8@$\G$ -- dynamical gauge group}%
(recall that in~\S\ref{l:sec32} we found the group~\eqref{l:ggroup}, and taking into account
the additional constraints encountered in Section~\ref{l:secstructure}, the dynamical gauge group
is a proper subgroup of~\eqref{l:ggroup}).
\sindex{gauge group!dynamical}%
 The {\em{dynamical gauge potentials}}
\sindex{gauge potential!dynamical}%
are elements of the corresponding Lie algebra~$\g = T_e \G$, 
the so-called {\em{dynamical gauge algebra}}. It can be identified
with a subspace of~${\mathfrak{S}}_6$.
\sindex{gauge algebra!dynamical}%
\nindex{dl0@$\g$ -- dynamical gauge algebra}%
The dynamical potentials and corresponding bosonic currents take values in
the dynamical subspace,
\beq \label{l:scrA}
\A := (A_L, A_R) \in \g \qquad \text{and} \qquad (j_L, j_R) \in \g\:.
\eeq

We now evaluate~\eqref{l:varS} for~$\delta P$ being a variation in direction of the dynamical subspace.
In order to evaluate this equation, we need to analyze how the potentials
 and currents enter the EL equations~\eqref{l:Knccond}.
We consider the contributions to degree four on the light cone after compensating the logarithmic
poles and evaluate weakly on the light cone.
The corresponding contribution~$\Delta Q$ is given in Corollary~\ref{l:cor42}.
In order to determine the variation~$\delta P$, one should keep in mind that it is vectorial, and that the
the left- and right-handed gauge potentials affect the left- and right-handed components of~$\delta P$, respectively. Moreover, $\delta P$ involves a sectorial projection. We thus obtain
\beq \label{l:Qeff}
\Tr_{\C^8} \big( \Delta Q\; (\chi_L \hat{A}_L + \chi_R \hat{A}_R) \,\slashed{u} \big) = 0 \qquad
\text{for all~$\A=(A_L, A_R) \in \g$}\:.
\eeq
Here~$u$ is an arbitrary vector field, whose only purpose is to get a contraction
with the factor~$\slashed{\xi}$ in~\eqref{l:DelQ}.
As explained before~\eqref{l:Kcond}, we want to consider the stronger conditions
which are independent of the projectors~$I_n$.
Using the form of~$\Delta Q$ in Corollary~\ref{l:cor42}, we thus obtain the conditions
\beq \label{l:fieldeff1}
\Tr_{\C^2} \big( {\mathcal{Q}}_L\, \hat{A}_R + {\mathcal{Q}}_R\, \hat{A}_L \big) = 0 \qquad
\text{for all~$\A=(A_L, A_R) \in \g$}
\eeq
(see also~\eqref{l:ELR} and note that the chirality flips at the factor~$\slashed{\xi}$ in~\eqref{l:DelQ}).

In order to recover~\eqref{l:fieldeff1} from an effective variational principle, our
goal is to choose a Dirac Lagrangian~$\LDirac$ and
a Yang-Mills Lagrangian~$\LYM$ such that varying the gauge
potentials in the effective gauge algebra gives the left side of~\eqref{l:fieldeff1}
with~$\A$ replaced by the variation~$\delta \A$ of the potentials.
In order to keep track of the contractions of the tensor indices, it is useful to
again use the matrix-valued vector field~$\mathfrak{J}_{L\!/\!R}^k$  in Lemma~\ref{l:lemma44}.
Similar to~\eqref{l:Rdef} we set
\beq \label{l:Rckdef}
{\mathcal{Q}}_c^k := {\mathfrak{J}}_c^k - \frac{1}{4}\: \Tr_{\C^2} \!\big(
{\mathfrak{J}}^k_L+{\mathfrak{J}}^k_R \big)\: \1_{\C^2} \:.
\eeq
Then we would like to choose~$\LDirac$ and
a Yang-Mills Lagrangian~$\LYM$ such that
\beq \label{l:varEL}
K(\varepsilon, \xi)\, \frac{\delta}{\delta \A} \Big( \LDirac + \LYM \Big)
= \Tr_{\C^2} \big( {\mathcal{Q}}^k_L[\hat{\mathscr{J}}, \A]\: (\delta \hat{A}_R)_k + {\mathcal{Q}}^k_R[
\hat{\mathscr{J}}, \A]\: (\delta \hat{A}_L)_k \big)
\eeq
for any~$\delta \A = (\delta A_L, \delta A_R) \in \g$.
The square brackets~$[\hat{\mathscr{J}}, \A]$ clarify the dependence on the
chiral potentials and on the sectorial projection of the Dirac current.
Our notation also points out that for example the left-handed component~$\mathcal{Q}^k_L$
may depend on both the left- and right-handed components of the currents
(as becomes explicit in~\eqref{l:Jterm}--\eqref{l:mterm7}).
The way the equation~\eqref{l:varEL} is to be understood is that the right side is to be evaluated weakly
according to~\eqref{l:asy}. We demand that the dependence on the regularization length~$\varepsilon$
and on the direction~$\xi$ can be absorbed in the prefactor~$K$.
If this has been accomplished, 
the continuum limit of the EL equations corresponding to the causal action principle
can be recovered by seeking for critical points of the effective action
\beq \label{l:Seff0}
\Sact_\text{eff} = \int_{\R^4} \left( \LDirac + \LYM \right) d^4x \:.
\eeq
\sindex{effective action}%
\nindex{dl2@$\Sact_\text{eff}$ -- effective action}%

The above construction can be adapted in a straightforward way to the 
gravitational field. To this end, we consider the contribution to~$\Delta Q$
by the energy-momentum and the Ricci tensor as computed in Section~\ref{l:seccurv}.
Since the gravitational field couples to the right- and left-handed components of all fermions
in the same way, it corresponds to a variation in the direction~$(\1, \1) \in {\mathfrak{S}}_2$.
We thus obtain in analogy to~\eqref{l:Qeff} the condition
\beq \label{l:Qeffgrav}
\Tr_{\C^8} \big( \Delta Q\; \slashed{u} \big) = 0 \:.
\eeq
Similar to~\eqref{l:varEL} we want to recover this condition as the critical point of
an effective Lagrangian. In order to recover the Einstein equations, we want to add the Einstein-Hilbert action.
Moreover, in curved space-time one clearly replaces the integration
measure in~\eqref{l:Seff0} by~$\sqrt{-\det g} \:d^4x$, where~$g$ again denotes the Lorentzian metric.
Moreover, the Dirac action should clearly involve the Dirac operator in curved space-time.
In order to treat the tensor indices properly, we introduce a matrix-valued
symmetric $2$-tensor~${\mathcal{Q}}^{kl}$ by
\beq \label{l:Ikldef}
\Tr_{\C^8} \big( \Delta Q\; \slashed{u} \big) = i \xi_j u^j \, {\mathcal{Q}}^{kl}[\hat{T}, g] \,\xi_k \xi_l \:.
\eeq
(where the factors~$\xi_i \xi_j$ are precisely those in~\eqref{l:JKL1} and
similarly in Lemma~\ref{l:lemmaR}). 
The square bracket~$[\hat{T},g]$ clarifies the dependence on the energy-momentum tensor
(which involves a sectorial projection) and the
metric. Our goal is to find an effective action such that, in analogy to~\eqref{l:varEL},
\beq \label{l:varELgrav}
i K(\varepsilon, \xi)\; \frac{\delta}{\delta g} \Big(
\left( \LDirac + \LYM + \LEH \right)
\sqrt{-\deg g} \Big) = {\mathcal{Q}}^{kl}[\hat{T}, g] \;\delta g_{kl}
\eeq
with the Einstein-Hilbert action
\sindex{effective Lagrangian!Einstein-Hilbert}%
\sindex{Lagrangian!effective}%
\beq \label{l:Lcurv}
\LEH = \frac{1}{\kappa(\varepsilon, \delta)}\: (R+2 \Lambda)
\eeq
(where~$R$ denotes scalar curvature and~$\Lambda \in \R$ is the cosmological constant).
We point out that the gravitational coupling constant~$\kappa$
may depend on the length scales~$\varepsilon$ and~$\delta$
(recall that the parameter~$\delta$ gives the length scale for the shear contributions;
see~\eqref{l:Texp} and~\eqref{l:delscale}). The dependence on~$\varepsilon$ or~$\delta$ is needed 
in order to take into account that the gravitational constant is not dimensionless.
This procedure will also make it possible to link the Planck length to the regularization
lengths~$\varepsilon$ or~$\delta$.

We thus obtain the effective action
\beq \label{l:Seff}
\Sact_\text{eff} = \int_\scrM \left( \LDirac + \LYM + \LEH \right) \sqrt{-\deg g}\, d^4x \:.
\eeq
\nindex{dl2@$\Sact_\text{eff}$ -- effective action}%
Varying the chiral potentials in~$\g$ gives the bosonic field equations,
whereas varying the metric gives the equations for gravity.
We again point out that the variation of the effective action must always be performed under
the constraint that the Dirac equation~\eqref{l:direx0} holds. Thus we do not need to derive
the Dirac equation from the effective action. Instead, the Dirac equation holds a-priori and must
be respected by the variation. The resulting procedure for computing variations will be explained
in~\S\ref{l:secvary}.

\subsectionn{The Effective Dirac Action} \label{l:sec52}
Our goal is to find Lagrangians such that the equations~\eqref{l:varEL} and~\eqref{l:varELgrav} hold.
The main task is to choose the Dirac Lagrangian such that the coupling of the Dirac wave functions
to the chiral potentials and the gravitational fields as described by~$\Delta Q$
is compatible with the variations of the Dirac Lagrangian in~\eqref{l:varEL} and~\eqref{l:varELgrav}.
Usually, the coupling of the Dirac wave functions to the bosonic fields is described by the
{\em{Dirac Lagrangian}}, which in our context takes the form
\beq \label{l:LDir}
\LDirac = \re \overline{\psi} (i \Pdd + \B - mY) \psi
\eeq
\sindex{effective Lagrangian!Dirac}%
\sindex{Lagrangian!effective|see{effective Lagrangian}}%
(note that using the symmetry of the Dirac operator,
the real part can be omitted if one integrates over space-time).
The corresponding Dirac action has the nice feature that varying the Dirac wave functions
gives the Dirac equation~\eqref{l:Dinteract}.
The standard method would be to add to~\eqref{l:LDir} a Yang-Mills Lagrangian, in such a way that
varying the bosonic potentials gives the effective EL equations~\eqref{l:fieldeff1}.
However, in our situation this standard method does not work, because
according to~\eqref{l:varEL}, the effective EL equations involve the {\em{sectorial projection}}
of the Dirac current, whereas varying~$\B$ in~\eqref{l:LDir} yields the Dirac current without
a sectorial projection. A similar problem occurs when we try to recover the equations
for the gravitational field~\eqref{l:varELgrav} from a variational principle. The standard procedure
is to add the Einstein-Hilbert action. But then varying the metric would give the
energy-momentum tensor of the Dirac wave functions without a sectorial projection, in contrast
to the sectorial projection~$\hat{T}_{jk}$ in~\eqref{l:varELgrav}.

In order to resolve this problem, we need to modify the Dirac Lagrangian in such a way that the
sectorial projection is built in correctly. It is now convenient to describe the sectorial projection
by a projection operator~$\sproj$,
\nindex{dl6@$\sproj$ -- sectorial projection operator}%
\[ \sproj = \frac{1}{3} \begin{pmatrix} 1 & 1 & 1 \\ 1 & 1 & 1 \\ 1 & 1 & 1 \end{pmatrix} \::\:
\C^3 \rightarrow \C^3 \:, \]
acting on the generations. In agreement with our earlier notation, $\sproj$ acts on~$\C^6$
as the block-diagonal matrix
\[ \begin{pmatrix} \sproj & 0 \\ 0 & \sproj \end{pmatrix} \::\: \C^6 \rightarrow \C^6 \:. \]
Likewise, $\sproj$ may act on the left- and right-handed components.
Then the operation in~\eqref{l:hatdef} can be realized by acting with~$\sproj$ from the left and
from the right; for example
\[ \hat{\mathscr{J}} = 9\,\sproj {\mathscr{J}} \sproj \:. \]
The most obvious idea is to insert a sectorial projection into~\eqref{l:LDir},
\[ \re \Big( \overline{\psi} \:3 \sproj\, (i \Pdd + \B - mY) \psi \Big) \:. \]
Then varying the metric gives the desired energy-momentum term~$\hat{T}_{ij}$.
However, when varying the chiral potential, the mixing matrix~$\UMNS$
comes up in the wrong way. This leads us to also take the sectorial projection of~$\B$.
We thus choose
\beq \label{l:LDirpi}
\LDirac = \re \Big( \overline{\psi} \:3 \sproj\, (i \Pdd + \sproj \B \sproj - mY) \psi \Big) \:.
\eeq
Then varying the bosonic potentials also gives agreement with the sectorial projections
of the factors~$\hat{g}_L$ and~$\hat{g}_R$ in~\eqref{l:fieldeff1}.
We point out that varying the Dirac wave functions in the Dirac action corresponding to~\eqref{l:LDir}
does {\em{not}} give the Dirac equation~\eqref{l:LDir}. This is not a general problem
because, as explained in~\S\ref{l:secalgebra}, in our approach the Dirac equation holds trivially as
the defining equation for the bosonic potentials. Nevertheless, at first sight it might seem that the
Dirac Lagrangian~\eqref{l:LDir} should be inconsistent with the Dirac equation.
In~\S\ref{l:secvary} we will see that there are indeed no inconsistencies if the variations are
handled properly.

There is one more modification which we want to implement in the Dirac Lagran\-gian~\eqref{l:LDirpi}.
Namely, in order to have more freedom to modify the coupling of the right-handed neutrinos to the
gravitational field, we insert a parameter~$\tau$ into~$\sproj$ which modifies
the left-handed component of the upper isospin component,
\nindex{dl8@$\sproj_\tau$ -- sectorial projection with modified right-handed neutrino coupling}%
\[ \sproj_\tau := \begin{pmatrix} 1 + \tau \chi_L & 0 \\ 0 & 1 \end{pmatrix}
\: \sproj \qquad \text{with~$\tau \in \R$} \:. \]
We define our final Dirac Lagrangian by
\sindex{effective Lagrangian!Dirac}%
\beq \label{l:LDirpar}
\LDirac = \re \Big( \overline{\psi} \:3 \sproj_\tau (i \Pdd +
\sproj \B \sproj - mY) \psi \Big) \:.
\eeq
If a left-handed gauge field~$\B$ is varied, then
the parameter~$\tau$ drops out because the right-handed neutrinos
do not couple to the chiral gauge fields. However, the parameter~$\tau$ will make a difference
when considering variations of the metric. We will come back to this point in Section~\ref{l:secgrav} below.

The effective action is obtained as usual by adding to~\eqref{l:LDirpar} suitable Lagrangians involving
the chiral gauge field and scalar curvature.
They will be worked out in detail in Sections~\ref{l:sec6} and~\ref{l:secgrav}.

\subsectionn{Varying the Effective Dirac Action} \label{l:secvary}
We now explain how the effective action~\eqref{l:Seff0}
with the Dirac Lagrangian~\eqref{l:LDirpar} is to be combined with the Dirac equation~\eqref{l:direx0}
(or similarly the action~\eqref{l:Seff} with the corresponding Dirac equation in the gravitational field).

We again point out that in our approach, the Dirac equation~\eqref{l:direx0} is
trivially satisfied, because it serves as the definition of the bosonic potentials in~$\B$.
The bosonic potentials in~$\B$ are merely a device for describing the
behavior of the wave functions~$\psi$ in the fermionic projector.
With this concept in mind, the method of varying the bosonic potentials
for fixed wave functions (as used after~\eqref{l:Seff0}) is not the proper procedure.
The procedure is not completely wrong, because in many situations the wave functions do not
change much when varying the bosonic potentials, and in these cases it is admissible to
consider them as being fixed. But in general, it is not a consistent procedure to
vary~$\B$ for fixed~$\psi$, because then the Dirac equation~\eqref{l:direx0} will
be violated. Taking the Dirac equation as the definition of~$\B$, the only way to vary the bosonic
potentials is to also vary the wave functions according to~\eqref{l:direx0}, and to
consider the the effective Lagrangian under the resulting joint variations  of~$\B$ and~$\psi$.

Let us compute such variations, for simplicity for a variation of the bosonic potential
in Minkowski space (the method works similarly in the presence of a gravitational field
and for variations of the metric).
\begin{Prp} \label{l:prpvary}
Varying the potential~$\B$ in the Dirac action corresponding to the Dirac Lagrangian~\eqref{l:LDirpar}
under the constraint that the Dirac equation~\eqref{l:direx0} holds, we obtain the first variation
\begin{align}
\delta & \Sact_\text{\text{\tiny{\rm{Dirac}}}} = \re \int_\scrM \Sl \psi \,|\, X_\tau\, (\delta \B)\, \psi \Sr \:d^4x \label{l:naive} \\
&-\re \int_\scrM \Sl \psi \,|\, 3 \sproj_\tau \, (\delta \B) \,(\1-\sproj)\, \psi \Sr \:d^4x \label{l:correct0} \\
&-\re \int_\scrM \Big( \Sl \delta \psi \,|\,3 \sproj_\tau\, \B\, (\1-\sproj)\, \psi \Sr
+ \Sl \psi \,|\,3 \sproj_\tau\, (\1-\sproj)\, \B \,\delta \psi \Sr \Big) \,d^4x \label{l:correct1} \\
&-\re \int_\scrM \Sl \psi \,|\,\Big( (\B-m Y) \, X_\tau^*
-X_\tau\, (\B-m Y) \Big)\, \delta \psi \Sr \Big) \,d^4x \label{l:correct2}
\end{align}
(where~$\Sl \psi | \phi \Sr \equiv \overline{\psi} \phi$ denotes the spin scalar product).
Here~$X_\tau$ is the matrix
\beq \label{l:Xtdef}
X_\tau = \begin{pmatrix} 1+ \tau \chi_L & 0 \\ 0 & 1 \end{pmatrix} \otimes \1_{\C^3}\:,
\eeq
and the variation of the wave function~$\delta \psi$ is given by
\beq \label{l:delpsi}
\delta \psi = - \tilde{s} \,(\delta \B)\, \psi \:,
\eeq
where~$\tilde{s}$ is a Green's function of the Dirac equation~\eqref{l:direx0},
\[ (i \Pdd + \B - m Y) \,\tilde{s} = \1 \:. \]
\end{Prp}
\Proof Let~$\delta B$ be the variation of~$\B$. In order to
satisfy the Dirac equation, we must vary the wave function according to~\eqref{l:delpsi}.
The variation of the Dirac wave function does not have compact support,
making it necessary to take into account boundary terms when integrating by parts.
In order to treat these boundary terms properly, we multiply the variation
of the wave function by a test function~$\eta \in C^\infty_0(\scrM)$. Thus instead of~\eqref{l:delpsi}
we consider the variation
\[ \widetilde{\delta \psi} = - \eta \,\tilde{s} \,(\delta \B)\, \psi \:. \]
At the end, we will remove the test function by taking the limit~$\eta \rightarrow 1$ in which~$\eta$
goes over to the function constant one.

The resulting variation of the Dirac action is computed by
\begin{align*}
\delta \Sact_\text{\text{\tiny{\rm{Dirac}}}} &= \int_\scrM \delta \LDirac \,d^4x 
= \re \int_\scrM \Big( \Sl \widetilde{\delta \psi} \,|\,3 \sproj_\tau\, (i \Pdd + \sproj \B \sproj - mY) \psi \Sr \\
& \qquad
+ \Sl \psi \,|\,3 \sproj_\tau\, (\delta (\sproj \B \sproj))\, \psi \Sr
+ \Sl \psi \,|\,3 \sproj_\tau\, (i \Pdd + \sproj \B \sproj - mY) \,\widetilde{\delta \psi} \Sr \Big) \,d^4x \:.
\end{align*}
Using that~$\psi$ satisfies the Dirac equation, and that~$\widetilde{\delta \psi}$ satisfies the
inhomogeneous Dirac equation
\[ (i \Pdd + \B - mY) \,\widetilde{\delta \psi} 
= -i \big(\Pdd \eta \big) \: \tilde{s} \,(\delta \B)\, \psi - \eta (\delta \B)\, \psi \:, \]
we obtain
\begin{align*}
\delta \Sact_\text{\text{\tiny{\rm{Dirac}}}}
&= \re \int_\scrM \eta \:\Big( \Sl \delta \psi \,|\,3 \sproj_\tau\, (\sproj \B \sproj - \B)\, \psi \Sr
+ \Sl \psi \,|\,3 \sproj_\tau\, (\sproj \B \sproj - \B) \,\delta \psi \Sr \Big)\, d^4x \\
&\quad+ \re \int_\scrM \Big(
\Sl \psi \,|\,3 \sproj_\tau\, \big( \delta (\sproj \B \sproj)  - \eta\,(\delta \B) \big) \psi \Sr
-i  \:\Sl \psi \,|\,3 \sproj_\tau \big(\Pdd \eta \big) \: \big( \tilde{s} \,(\delta \B)\, \psi \big) \Sr \Big) \,d^4x \:.
\end{align*}
In the last term we decompose the matrix~$3 \sproj_\tau$ into its diagonal and off-diagonal parts,
\[ 3 \sproj_\tau = X_\tau + Z \qquad \text{with} \qquad Z = 3 \sproj - \1_{\C^3} \]
and~$X_\tau$ according to~\eqref{l:Xtdef}. Thus, using~\eqref{l:delpsi},
\begin{align*}
-i &\int_\scrM \Sl \psi \,|\,3 \sproj_\tau \big(\Pdd \eta \big) \: \big( \tilde{s} \,(\delta \B)\, \psi \big) \Sr \,d^4x \notag \\
&= i \int_\scrM (\partial_j \eta) \:\Sl \psi \,|\,X_\tau \gamma^j \,\delta \psi \Sr \,d^4x 
+i \int_\scrM (\partial_j \eta) \:\Sl \psi \,|\,Z \gamma^j \,\delta \psi \Sr \,d^4x \:.
\end{align*}
In the first integral we integrate by parts,
\begin{align*}
i &\int_\scrM (\partial_j \eta) \:\Sl \psi \,|\,X_\tau \gamma^j \,\delta \psi \Sr \,d^4x \\
&=  \int_\scrM \eta \:\Big( \Sl i \partial_j \psi \,|\,X_\tau \gamma^j \,\delta \psi \Sr 
-  \Sl \psi \,|\,X_\tau \gamma^j \,i \partial_j \,(\delta \psi) \Sr \Big)\,d^4x\:.
\end{align*}
The pseudoscalar matrix in~$X_\tau$ anti-commutes with the Dirac matrix~$\gamma^j$.
Since the pseudoscalar matrix is anti-symmetric with respect to the spin scalar product, we
can express this anti-commutation by
\[ X_\tau \gamma^j = \gamma^j X_\tau^* \:. \]
Then we can rewrite the partial derivatives~$i \partial_j$ with the Dirac equation~\eqref{l:direx0}
and the inhomogeneous Dirac equation equation for~$\delta \psi$,
\[ (i \Pdd + \B-m Y) \delta \psi = -(\delta \B)\, \psi \:. \]
This gives
\begin{align*}
i \int_\scrM & (\partial_j \eta) \:\Sl \psi \,|\,X_\tau \gamma^j \,\delta \psi \Sr \,d^4x \\
=& \int_\scrM \eta \:\Big( \Sl i \Pdd \psi \,|\,X^*_\tau \gamma^j \,\delta \psi \Sr 
- \Sl \psi \,|\,X_\tau \, i \Pdd \,(\delta \psi) \Sr \Big)\,d^4x \\
=& -\int_\scrM \eta \:\Sl \psi \,|\,\Big( (\B - m Y)\,X^*_\tau - X_\tau\,(\B-mY) \Big) \,(\delta \psi) \Sr \,d^4x \\
&+ \int_\scrM \eta \:\Sl \psi \,|\,X_\tau \, (\delta \B) \,\psi \Sr \,d^4x \:.
\end{align*}

Combining all the terms, we obtain
\begin{align}
\delta \Sact_\text{\text{\tiny{\rm{Dirac}}}}
&= \re \int_\scrM \eta \:\Big( \Sl \delta \psi \,|\,3 \sproj_\tau\, (\sproj \B \sproj - \B)\, \psi \Sr
+ \Sl \psi \,|\,3 \sproj_\tau\, (\sproj \B \sproj - \B) \,\delta \psi \Sr \Big)\, d^4x \\
&\quad+ \re \int_\scrM 
\Sl \psi \,|\,3 \sproj_\tau\, \big( \delta (\sproj \B \sproj)  - \eta\,(\delta \B) \big) \psi \Sr \,d^4x \\
&\quad+\re \int_\scrM i (\partial_j \eta) \:\Sl \psi \,|\,Z \gamma^j \,\delta \psi \Sr \,d^4x \label{l:boundary} \\
&\quad -\re \int_\scrM \eta \:\Sl \psi \,|\,\Big( (\B - m Y)\,X^*_\tau - X_\tau\,(\B-mY) \Big) \,(\delta \psi) \Sr \,d^4x \\
&\quad + \re \int_\scrM \eta \:\Sl \psi \,|\,X_\tau \, (\delta \B) \,\psi \Sr \,d^4x \:.
\end{align}
Now we may take the limit~$\eta \rightarrow 1$.
In this limit, the integral~\eqref{l:boundary} goes to zero, as will be justified
in Lemma~\ref{l:lemmaboundary} below. Rearranging the terms using the relation~$\sproj_\tau \sproj
= \sproj_\tau$ gives the result.
\QED
We now explain why the integral~\eqref{l:boundary} tends to zero if~$\eta \rightarrow 1$.
Since this is a rather subtle point, we give the details. However,
for technical simplicity we assume that the bosonic potential has compact support.
The result could be extended in a straightforward manner to
the case that the potential has suitable decay properties at infinity by estimating
the Lippmann-Schwinger equation (we refer the interested reader
to the exposition in~\cite{intro} and to similar methods in~\cite{hadamard}).

\begin{Lemma} \label{l:lemmaboundary}
Assume that the fermion masses are different in the generations, i.e.
\[ m_\alpha \neq m_\beta \quad \text{and} \quad \tilde{m}_\alpha \neq \tilde{m}_\beta
\qquad \text{for all~$\alpha, \beta \in \{1,2,3\}$ and~$\alpha \neq \beta\:.$} \]
Moreover, assume that the potential~$\B$ and its variation~$\delta \B$
are smooth and have compact support, and that~$\psi$ is smooth.
Then for any test function~$\eta \in C^\infty_0(\R^4)$ which
is constant in a neighborhood of the origin,
\beq \label{l:stint}
\lim_{L \rightarrow \infty}
\int_\scrM \Sl \psi \,|\,Z \gamma^j \,\big( \tilde{s} \,(\delta \B)\, \psi \big) \Sr \;
\frac{\partial}{\partial x^j} \eta \Big( \frac{x}{L} \Big) \: d^4x = 0 \:.
\eeq
\end{Lemma}
\Proof By choosing~$L$ sufficiently large, we can arrange that~$\eta$
is constant on the support of~$\B$ and~$\delta B$. Then
we may replace~$\psi$ and~$\phi := \tilde{s} \,(\delta \B)\, \psi$ by smooth solutions
of the vacuum Dirac equation~$(i \Pdd - mY) \psi = 0 = (i \Pdd - mY) \phi$.
Since the matrix~$\1-3 \sproj_\tau$ vanishes on the diagonal and only mixes the wave functions
within each sector, the integral~\eqref{l:stint} can be rewritten as a finite sum of integrals of the form
\beq \label{l:stint2}
\int_\scrM \Sl \psi_\alpha \,|\, \gamma^j \,\phi_\beta \Sr \; \partial_j \eta_L(x) \: d^4x \:,
\eeq
where~$\eta_L(x) := \eta(x/L)$, and where~$\psi_\alpha$ and~$\phi_\beta$ are solutions of the
Dirac equation for different masses,
\[ (i \Pdd - m_\alpha) \psi_\alpha = 0 = (i \Pdd - m_\beta) \phi_\beta \qquad
\text{and~$m_\alpha \neq m_\beta$}\:. \]
Writing the solutions as distributions in momentum space,
\[ \hat{\psi}_\alpha(k) = f(k)\: \delta(k^2-m_\alpha^2) \:,\qquad
\hat{\phi}_\beta(k) = g(k)\: \delta(k^2-m_\beta^2) \:, \]
the smoothness of~$\psi_\alpha$ and~$\phi_\beta$ implies that the functions~$f$
and~$g$ can be chosen to have rapid decay. Then the integral in~\eqref{l:stint2} can be rewritten
in momentum space as
\[ -i \int_{\hscrM} \Sl \hat{\psi}_\alpha \,|\, \slashed{k} \,(\hat{\eta}_L * \hat{\phi}_\beta) \Sr
\: \frac{d^4k}{(2 \pi)^4} \:, \]
where the star denotes the convolution of the distribution~$\hat{\phi}_\beta$ with
the test function~$\hat{\eta}_L$ giving a Schwartz function (note that the smoothness of~$\eta_L$
implies that~$\hat{\eta}_L$ has rapid decay), and the integral is to
be understood that the distribution~$\hat{\psi}_\alpha$ is applied to this Schwartz function.
Since the functions~$f$ and~$g$ have rapid decay, for any~$\varepsilon>0$ there is a
compact set~$K \subset \hscrM$ such that
\[ \Big| \int_{\hscrM \setminus K} \Sl \hat{\psi}_\alpha \,|\, \slashed{k} \,(\hat{\eta}_L * \hat{\phi}_\beta) \Sr
\: d^4k \Big| < \varepsilon \:, \]
uniformly in~$L$. For any fixed~$K$, the supports of the distributions~$\delta(k^2-m_\alpha^2)$
and~$\delta(k^2-m_\beta^2)$ have a finite separation (measured in the Euclidean norm
on~$\R^4$ in a chosen reference frame). Since~$\hat{\eta}_L(k) = L^4 \,\hat{\eta}(L k)$,
by increasing~$L$ we can arrange that the function~$\hat{\eta}_L$ decays on a smaller and smaller scale.
Since~$\hat{\eta}$ has rapid decay, this implies that the integral over~$K$ tends to zero,
\[ \lim_{L \rightarrow \infty} \int_{K} \Sl \hat{\psi}_\alpha \,|\, \slashed{k} \,(\hat{\eta}_L * \hat{\phi}_\beta) \Sr
\: d^4k =0 \:. \]
Since~$\varepsilon$ is arbitrary, the result follows.
\QED

Combining the result of Proposition~\ref{l:prpvary} with the 
variation of the Yang-Mills Lagrangian in~\eqref{l:Seff0} (which can be computed in the standard way,
see Section~\ref{l:sec6}),
one obtains field equations describing the dynamics of the chiral gauge field
and its coupling to the Dirac particles and anti-particles.
Together with the Dirac equation~\eqref{l:direx0}, one obtains a consistent set of
equations which we regard as the {\em{effective EL-equations in the continuum limit}}.

Let us discuss the structure of the resulting field equations: The first term~\eqref{l:naive}
differs from the standard contribution obtained by varying the bosonic potential in the Dirac
Lagrangian~\eqref{l:LDirpar} by the fact that the sectorial projection has disappeared.
This is desirable because the resulting contribution looks very much like the variation of the
standard Lagrangian~\eqref{l:LDir}.
The only difference is the additional factor~$X_\tau$. However, this factor comes into play only
if one considers gauge fields which couple to the right-handed neutrinos.
Such gauge fields will be ruled out in the present paper. They also do not appear in the standard
model. Therefore, the factor~$X_\tau$ in~\eqref{l:naive} seems consistent with observations.

The terms~\eqref{l:correct0}--\eqref{l:correct2} are additional contributions which are absent in the standard Lagrangian formulation. They can be understood as corrections which are needed in order to get consistency
with the Dirac equation~\eqref{l:direx0}. We refer to the terms~\eqref{l:correct0}--\eqref{l:correct2} as the
{\em{sectorial corrections}} to the field equations.
\sindex{field equations in the continuum limit!corrections!sectorial corrections}%
The term~\eqref{l:correct0} modifies the coupling of those chiral gauge potentials which involve a
non-trivial mixing matrix. The correction~\eqref{l:correct1} can be understood similarly.
As a difference, it involves the Green's function~$\tilde{s}$ and is therefore
nonlocal (we note that the choice of the Green's function~$\tilde{s}$
in~\eqref{l:delpsi} is uniquely determined by the causal perturbation expansion~\cite{norm};
see also Section~\ref{secfpext}).
The correction term~\eqref{l:correct2} is also nonlocal and comes into play when the neutrinos are
massive.
The appearance of these nonlocal correction terms are a prediction of the fermionic projector approach.
It is conceivable that these corrections are testable in experiments.
More specifically, the corrections vanish if the mixing matrices do not come into play
and if the Dirac wave functions are eigenstates
of the mass matrix. Thinking of the analogous situation for the standard model,
the corrections vanish for example for electrons with an electromagnetic interaction.
However, they come play in an interaction via $W$-bosons
if the wave function~$\psi$ is a non-trivial superposition of for example an electron and a muon.
The detailed mechanism triggered by the nonlocal effects is unclear and still needs to investigated.
All we can say for the moment is that the corrections~\eqref{l:correct1} and~\eqref{l:correct2}
cease to play any role as soon as the Dirac Lagrangian no longer involves cross terms of electrons and muons.

We finally point out that the effective action cannot be regarded as
some kind of ``continuum limit'' of the causal action principle.
It is merely a method for recovering the EL equations corresponding to the
causal action principle in the continuum limit from an effective variational principle.
The basic difference of the causal action principle and the effective action
can be understood already from the fact that the causal action is minimized,
whereas for the effective action one only seeks for critical points.
Thus the effective action should be regarded merely
as a convenient method for getting the connection to the standard Lagrangian
formalism. In particular, by applying Noether's theorem to the effective action, one can
immediately deduce conservation laws for the effective EL equations.

\section{The Field Equations for Chiral Gauge Fields} \label{l:sec6}
We now use the methods of Section~\ref{l:sec5} to compute the
effective action for the coupling of the Dirac field to the gauge fields.
In order to determine the dynamical gauge algebra~$\g \subset {\mathfrak{S}}_6$
(defined before~\eqref{l:scrA}), we first recall that in~\S\ref{l:sec32}
we derived the admissible gauge group~\eqref{l:ggroup}
together with the representation of the gauge potentials~\eqref{l:Bform}.
In Section~\ref{l:secstructure}, we obtained further restrictions for the
gauge potentials. Namely, the analysis of the bilinear logarithmic terms
in~\S\ref{l:seclogAA} revealed that the diagonal elements must satisfy the
constraint~\eqref{l:LRmix}. The field tensor terms in~\S\ref{l:secfield}, on the other hand,
gave us the two linear constraints~\eqref{l:FTcond} for the field tensor, which
due to gauge symmetry we can also regard as constraints for the potentials.
Putting these conditions together,
we conclude that the dynamical gauge potentials must be of one of the two
alternative forms
\begin{align}
\B &= \chi_R \begin{pmatrix} \slashed{A}_L^{11} & 0 \\[0.2em]
0 & 0 \end{pmatrix}
+ \chi_L \begin{pmatrix} 0 & 0 \\
0 & -\slashed{A}_L^{11} \end{pmatrix}  \quad \text{or} \label{l:nogroup} \\
\B &= \chi_R \begin{pmatrix} \slashed{A}_L^{11} & \slashed{A}_L^{12}\, \UMNS^* \\[0.2em]
\slashed{A}_L^{21}\, \UMNS & -\slashed{A}_L^{11}  \end{pmatrix}
+ \chi_L \begin{pmatrix} 0 & 0 \\
0 & 0 \end{pmatrix} . \label{l:SU2}
\end{align}
\nindex{ar4@$\B$ -- external potential}%
The potentials of the form~\eqref{l:nogroup} do not form a Lie algebra
(because taking a commutator, the resulting potential has the same sign
on the two isospin components, in contradiction to~\eqref{l:nogroup}).
This means that the structure of~\eqref{l:nogroup} is not preserved under local gauge
transformations corresponding to the potentials of the form~\eqref{l:nogroup}.
As this seems to be inconsistent, we disregard this case.
We thus restrict attention to the remaining case~\eqref{l:SU2}, where~$\g$
is the Lie algebra~$\text{su}(2)$, which acts on the left-handed component of the spinors
and involves the MNS mixing matrix.

Using the results of Section~\ref{l:sec4}, it is straightforward to compute
the right side of~\eqref{l:varEL}. Namely, applying Lemma~\ref{l:lemma44} together
with~\eqref{l:Jterm} and~\eqref{l:Rckdef} (and using that~$\delta A_c$ is traceless), for
the right side of~\eqref{l:varEL} we obtain the contribution
\[ \Tr_{\C^2} \big( {\mathcal{Q}}^k_R[\hat{\mathscr{J}}, \A]\: (\delta \hat{A}_L)_k \big)
\asymp K_1 \Tr_{\C^2} \big(J^k_L\: (\delta \hat{A}_L)_k \big) \:. \]
This is compatible with the variation of the Dirac Lagrangian~\eqref{l:LDirpar}
(for fixed wave functions) if we choose
\[ K(\varepsilon, \xi) = 3 \, K_1\:. \]
It is worth pointing out that this compatibility involves both the fact that
the left-handed gauge potentials couple only to the left-handed
component of the Dirac current and also that only the sectorial projection
of the potentials and currents appears. In particular, our method would fail if~\eqref{l:Jterm}
involved the left-handed component of the Dirac current.

The contributions by the bosonic current and mass terms as listed in~\eqref{l:Jterm}--\eqref{l:mterm7}
are a bit more difficult to handle because the logarithmic poles must be removed
with the help of the microlocal chiral transformation (see Proposition~\ref{l:prpmicroloc}).
If this is done, the resulting contributions have a rather complicated form.
However, the general structure is easy to understand:
First, writing the $\SU(2)_L$-gauge potentials in~\eqref{l:Bform} in components
\beq \label{l:ALrep}
A_L^\alpha = \frac{1}{2} \Tr(\sigma^\alpha A_L) \:,
\eeq
it is obvious from the symmetries that all contributions
involving~$A_L^\alpha \cdot \delta A_L^\beta$ 
and~$j_L^\alpha \cdot \delta A_L^\beta$  with~$\alpha \neq \beta$ vanish.
Second, in view of the symmetry under relative phase transformations of the isospin components
\[ \psi \rightarrow \begin{pmatrix} e^{i \varphi} & 0 \\ 0 & e^{-i \varphi} \end{pmatrix} \psi \:, \]
the contributions involving~$A_L^\alpha \cdot \delta A_L^\alpha$ 
and~$j_L^\alpha \cdot \delta A_L^\alpha$ coincide for~$\alpha=1$ and~$\alpha=2$.
Hence the equation~\eqref{l:varEL} can be satisfied for a bosonic Lagrangian of the form
\sindex{effective Lagrangian!Yang-Mills}%
\begin{align*}
\LYM &= a_1 \, \Big( (\partial_j A_L^1) (\partial^j A_L^1)
+ (\partial_j A_L^2) (\partial^j A_L^2) \Big) + a_3\, (\partial_j A_L^3) (\partial^j A_L^3) \\
&\qquad + b_1 \Big( (A_L^1)^2 + (A_L^2) \Big) + b_3\, (A_L^3)^2 )
\end{align*}
and suitable constants~$a_1, a_2$ and~$b_1, b_3$.
We thus obtain the following result.
\begin{Thm} \label{l:thmfield}
Expressing the $\SU(2)_L$-gauge potentials in Pauli matrices acting on
the isospin~\eqref{l:ALrep} (and similarly for the currents), the field equations read
\[ j_L^\alpha -  M_\alpha^2\, A^\alpha_L = c_\alpha \:J_L^\alpha
\:+\:( f_{[0]}* j_L)^\alpha + (f_{[2]}* A_L)^\alpha \:, \]
\sindex{field equations in the continuum limit!for gauge fields}%
where~$j_L$ and~$J_R$ are the currents~\eqref{l:jdef}
and~\eqref{l:Jdef}, respectively.
The mass parameters~$M_\alpha$ and the coupling constants~$c_\alpha$ satisfy the relations
\[ M_1=M_2 \qquad \text{and} \qquad c_1 = c_2 \:. \]
Finally, the distributions~$f_{[0]}$ and~$f_{[2]}$ are convolution kernels.
\end{Thm} \noindent
The convolution kernels take into account the following corrections:
\sindex{field equations in the continuum limit!corrections}%
\begin{itemize}[leftmargin=2em]
\itemD The corrections due to the smooth, noncausal contributions to the
fermionic projector. These corrections include the vacuum polarization due to the
fermion loops. These corrections are discussed further in~\S\ref{s:secfield1}--\S\ref{s:sechighorder}.
\sindex{field equations in the continuum limit!corrections!due to smooth, noncausal contributions to fermionic projector}%
\itemD The corrections due to the microlocal chiral transformation (see the last paragraph
in~\S\ref{s:secgennonloc}).
\sindex{field equations in the continuum limit!corrections!due to microlocal chiral transformation}%
\itemD The sectorial corrections (see Proposition~\ref{l:prpvary} and the explanation
after the proof of Lemma~\ref{l:lemmaboundary}).
\sindex{field equations in the continuum limit!corrections!sectorial corrections}%
\end{itemize}

Qualitatively speaking, this theorem can be understood similar to the results in Chapter~\ref{sector}.
Also, the calculations use exactly the same methods.
The appearance of bosonic masses and the connection to a spontaneous
breaking of the gauge symmetry is explained in~\S\ref{s:secnohiggs}.
In particular, the convolution kernels~$f_{[0]}$ and~$f_{[2]}$
are computed and interpreted just as in~\S\ref{s:secfield1} and~\S\ref{s:secnocausal}.
In view of these similarities, we here omit the detailed computations and only point to two
steps in the computations which are not quite straightforward.
First, as mentioned after Proposition~\ref{l:prphom},
the constants~${\mathfrak{c}}_0$ and~${\mathfrak{c}}_2$ are not determined
by this proposition. Following the strategy used in~\S\ref{s:secnonlocaxial}, we can
fix these constants by minimizing~${\mathfrak{c}}_0$.
Thus we choose the microlocal chiral transformation in such a way that the
vectorial contribution~\eqref{l:4con} to the fermionic projector is as small as possible.
Using this method, for a given regularization one can also compute
the coupling constants and the masses similar as in~\S\ref{s:secexample}.

The second step which requires an explanation concerns the computation of the
coupling constants and bosonic masses for a given regularization
in the spirit of~\S\ref{s:secexample}. Here one must distinguish the two
Cases~{\bf{(i)}} and~{\bf{(ii)}} in~\eqref{l:casesiii}.
Which of these cases applies depends crucially on the choice of the parameter~$p_\reg$
in~\eqref{l:preg}. In particular, by choosing~$p_\reg$ sufficiently small (and thus the
parameter~$\tau_\reg$ in~\eqref{l:treg1} sufficiently large), we can arrange that
we are in Case~{\bf{(ii)}}. In order to keep the setting as general as possible,
we deliberately left open which of the cases should be physically relevant.
We found that all our computations up to and including Section~\ref{l:sec4}
apply in the same way in both cases.
In the analysis of the bilinear logarithmic terms in~\S\ref{l:seclogAA}, however,
our constructions apply in Case~{\bf{(i)}} only under the additional assumption~\eqref{l:UYrel}.
The analysis of the field tensor terms in~\S\ref{l:secfield}
was carried out only in Case~{\bf{(ii)}} (and at present it is unclear how the results could
be extended to Case~{\bf{(i)}}). This gives a strong indication that the physically relevant
scaling should indeed be described by Case~{\bf{(ii)}}. This scaling can be realized by
choosing the parameter~$p_\reg$ in~\eqref{l:preg} sufficiently small.
Thus in a physical model, the the parameter~$\tau_\reg$ in~\eqref{l:treg1} should be chosen
sufficiently large.

Arranging in this way that we are in Case~{\bf{(ii)}}, it remains to justify
the transition from the EL equations~\eqref{l:Knccond} to the stronger conditions~\eqref{l:Kcond}.
We already indicated an argument before Corollary~\ref{l:cor42}. We are now
in the position to make this argument precise: Recall that in Case~{\bf{(ii)}},
the spectral projectors~$I_n$ are isospin-diagonal~\eqref{l:I12diag}.
The perturbation of these spectral projectors by the gauge phases
leads to a finite hierarchy of equations to be satisfied in a weak evaluation on the light cone.
With this in mind, it suffices to satisfy~\eqref{l:Knccond} with~$I_n$ according to~\eqref{l:I12diag}.
But clearly, we must take into account that the gauge phases enter the matrices~${\mathscr{K}}_{nc}$,
as we now explain.
We begin with the Dirac current terms. As the left-handed component of a wave function is
modified by the gauge phases in the obvious way by
\[ \chi_L \,\psi(y) \rightarrow \chi_L \,\exp \left( -i \int_x^y A_L^j \xi_j \right) \psi(y) \:, \]
the gauge potential enters the Dirac current term as described by the replacement
\[ J_L \rightarrow \left( 1 - i A_L^j \Big( \frac{x+y}{2} \Big)\: \xi_j \right) J_L\:. \]
In this way, the off-diagonal components of the Dirac current enter the
diagonal matrix entries of~${\mathscr{K}}_n$ and thus the EL equations~\eqref{l:Knccond}.
Since the gauge currents have the same behavior under gauge transformations
(see~\eqref{l:jphase}), their off-diagonal elements enter the EL equations
in the same way. For the mass terms, there is the complication that they
have a {\em{different}} behavior under gauge transformations
(for the logarithmic terms, this was studied in~\eqref{l:phaselog},
whereas for the contributions of the second order perturbation calculation, the dependence
on the gauge phases can be read off from the formulas given in~\cite[Definition~7.2.1]{PFP}).
This different behavior under gauge transformation does not cause problems for the
logarithmic poles, because we saw in~\S\ref{l:seclogAA} that
the logarithmic poles on the light cone can be arranged to vanish.
Thus the only effect of the different gauge behavior of the mass terms is that
it modifies the values of the bosonic mass corresponding to the
off-diagonal gauge potentials. The easiest method to describe this effect quantitatively
is to again work with the EL equations~\eqref{l:Kcond}, but to modify the off-diagonal matrix
elements of~${\mathscr{K}}_L$ and~${\mathscr{K}}_R$ by multiplying the
contributions~\eqref{l:mterm1}--\eqref{l:mterm7} with numerical
factors which take into account the linear behavior under off-diagonal left-handed
gauge transformations.
It is planned to work out the masses and coupling constants for a specific example of an
admissible regularization in a separate publication.

One might ask whether all coupling constants and masses in Theorem~\ref{l:thmfield}
should be the same, i.e.\ if also
\[ M_1=M_3 \qquad \text{and} \qquad c_1 = c_3 \:. \]
Indeed, the contribution by the bosonic current to~${\mathcal{Q}}_L$ in~\eqref{l:Jterm}
suggests that the derivative terms in the bosonic Lagrangian can be written as
\beq \label{l:traceform}
\Tr_{\C^2} \big( (\partial_j \hat{A}_L) (\partial^j \hat{A}_L) \big) \:.
\eeq
However, since the microlocal chiral transformation involves the masses of the
Dirac particles, which may be different in the two isospin components, there is no reason why
the more elegant form~\eqref{l:traceform} should be preserved when the microlocal chiral transformation
is taken into account. For the mass terms, on the other hand,
it is obvious from~\eqref{l:mterm1}--\eqref{l:mterm7} that the masses of the Dirac particles are
involved. Thus again, there is no reason why there should be a simple relation between
the masses~$M_1$ and~$M_3$.

\section{The Einstein Equations} \label{l:secgrav}
Our first task is to compute the symmetric tensor~${\mathcal{Q}}^{kl}$ as defined by~\eqref{l:Ikldef}.
If we used the form of~$\Delta Q$ in Corollary~\ref{l:cor42}, we would get zero, because
\[ \sum_{n,c} \Tr_{\C^2} \!\big( I_n \,{\mathcal{Q}}_c \big) \,\Tr_{\C^2}{I_n}
= \sum_{n,c} \Tr_{\C^2} \!\big( I_n \,{\mathcal{Q}}_c \big)
= \sum_{c} \Tr_{\C^2} \!\big({\mathcal{Q}}_c \big) = 0  \:, \]
where in the last step we used~\eqref{l:Rdef}. This means that in order to compute~\eqref{l:Qeffgrav},
we need to evaluate~\eqref{l:EL4} to higher order in~$(m \varepsilon)^{p_\reg}$
(note that these contributions were neglected in~\S\ref{l:sec40} according to~\eqref{l:higherneglect}).

Expanding~\eqref{l:EL4} to higher oder in powers of~$(m \varepsilon)^{p_\reg}$ is
a bit subtle because there might be contributions to~$\Delta |\lambda^{xy}_{ncs}|$
which are linear in~$(m \varepsilon)^{p_\reg}$ but do not involve curvature.
In this case, we would have to take into account the effect of curvature on the factors
\beq \label{l:lastfact}
\frac{\overline{\lambda^{xy}_{ncs}}}{|\lambda^{xy}_{ncs}|} \: F_{ncs}^{xy}\: P(x,y)
\eeq
in~\eqref{l:EL4}. The resulting contributions to~$Q(x,y)$ would not be proportional to~$\slashed{\xi}$,
giving rise to additional equations to be satisfied in the continuum limit.
Moreover, we would have to take into account the effect of the Dirac and bosonic currents
to~\eqref{l:lastfact}, giving rise to even more equations to be satisfied in the continuum limit.
For this reason, we must assume that our regularization is such that in the vacuum,
the quantities~$|\lambda^{xy}_{ncs}|$ coincide pointwise up to the order~$(m \varepsilon)^{2 p_\reg}$.
Such a regularization condition was already imposed in~\eqref{l:T0point}.
Now we need to complement it by a similar condition for the upper index minus one,
\beq \label{l:Tm1point}
\big| L^{(-1)}_{[0]} \big| = \big| T^{(-1)}_{[0]} \big| \left( 1 + \O \big( (m \varepsilon)^{2 p_\reg} \big) \right)
\qquad \text{pointwise}\:.
\eeq
We note that this is compatible with~\eqref{l:rcond} and poses
an additional condition on the regularization in the case~$n=-1$ and~$p=0$.
Similar as explained after~\eqref{l:T0point}, the pointwise condition~\eqref{l:Tm1point} could be
replaced by a number of conditions to be satisfied in a weak evaluation on the light cone,
but we do not enter this analysis here. Generally speaking,
the conditions~\eqref{l:T0point} and~\eqref{l:Tm1point} seem to indicate that the right-handed
neutrino states should affect the factors~$L^{(-1)}_{[0]}$ and~$L^{(0)}_{[0]}$
only by phase factors, up to errors of the order~$\O((m \varepsilon)^{2 p_\reg})$.

\begin{Lemma} \label{l:lemmaIT}
Assume that the regularization has the properties~\eqref{l:Tm1point} and~\eqref{l:T0point}.
Then the energy-momentum tensor gives the following contribution to~${\mathcal{Q}}^{kl}$,
\begin{align*}
{\mathcal{Q}}^{kl} \asymp\;& \frac{1}{2}\: K_8\: \bigg\{
\Big( (\hat{T}^{kl}_L)^1_1 - 3 \,(\hat{T}^{kl}_R)^1_1 + (\hat{T}^{kl}_L)^2_2 + (\hat{T}^{kl}_R)^2_2 \Big) \\
&+\frac{L^{(0)}_{[0]}}{T^{(0)}_{[0]}}\;
\Big(  - (\hat{T}^{kl}_L)^1_1 + 3\, (\hat{T}^{kl}_R)^1_1 -(\hat{T}^{kl}_L)^2_2 - (\hat{T}^{kl}_R)^2_2 \Big) \bigg\}  \\
&+ \O \big( (m \varepsilon)^{2 p_\reg} \big) \,(\deg = 4) + (\deg < 4)\:.
\end{align*}
\end{Lemma}
\Proof The matrices~${\mathscr{K}}_L$ and~${\mathscr{K}}_R$
were computed in Lemma~\ref{l:lemmaT}.
Substituting these formulas into the representation of Corollary~\ref{l:cor42}
and computing the trace in~\eqref{l:Qeffgrav} gives zero.
More generally, one sees from~\eqref{l:EL4} that the trace in~\eqref{l:Qeffgrav}
vanishes no matter what the perturbations of the eigenvalues~$\Delta \lambda^{xy}_{ncs}$
are, provided that we approximate the last term in~\eqref{l:EL4} by its leading asymptotics on
the light cone
\begin{align}
\frac{\overline{\lambda^{xy}_{ncs}}}{|\lambda^{xy}_{ncs}|}
\: F_{ncs}^{xy}\: P(x,y)
&= \delta_{s-} \:\frac{i}{2}\: \frac{T^{(0)}_{[0]} \overline{T^{(-1)}_{[0]}}}{\big|T^{(0)}_{[0]} T^{(-1)}_{[0]} \big|}
\: g\,T^{(-1)}_{[0]} \: I_n\:\chi_c\, \slashed{\xi} + (\deg < 4) \label{l:factlead} \\
&\qquad + \text{(higher orders in~$\varepsilon/\ell_\text{macro}$ and~$(m \varepsilon)^{p_\reg}$)}\:, \notag
\end{align}
where for clarity we wrote out the error terms in~\eqref{l:higherneglect}.
Therefore, it suffices to compute the correction to~\eqref{l:factlead} to next order
in~$(m \varepsilon)^{p_\reg}$. To this end, we must carefully distinguish between
factors~$T^{(n)}_{[p]}$ and~$L^{(n)}_{[p]}$, similar as done in~\eqref{l:cone}--\eqref{l:cthree}.
A straightforward calculation using~\eqref{l:Tm1point} and~\eqref{l:T0point} gives the result.
\QED

\begin{Lemma} \label{l:lemmaIR}
Curvature gives the following contribution to~${\mathcal{Q}}^{kl}$,
\begin{align*}
{\mathcal{Q}}^{kl} \asymp\;& \frac{\tau_\reg}{2 \delta^2}\: R^{kl}\: K_{16} \:
\bigg( 1 - \frac{L^{(0)}_{[0]}}{T^{(0)}_{[0]}} \bigg) \\
&+ \O \big( (m \varepsilon)^{2 p_\reg} \big) \,(\deg = 4) + o \big( |\vec{\xi}|^{-4} \big) + (\deg < 4)\:.
\end{align*}
\end{Lemma}
\Proof The matrices~${\mathscr{K}}_L$ and~${\mathscr{K}}_R$ were computed in Lemma~\ref{l:lemmaR}.
Again, substituting these formulas into the representation of Corollary~\ref{l:cor42}
and computing the trace in~\eqref{l:Qeffgrav} gives zero. Therefore, just as in the proof of
Lemma~\ref{l:lemmaIT}, we need to take into account the correction to~\eqref{l:factlead} to next order
in~$(m \varepsilon)^{p_\reg}$. Denoting the contributions to~${\mathscr{K}}_{L\!/\!R}$
in Lemma~\ref{l:lemmaR} leaving out the factors~$\xi^k \xi^l$ by~${\mathscr{K}}_{L\!/\!R}^{kl}$, we thus
obtain (cf.~\eqref{l:DelQ}, \eqref{l:Rdef} and~\eqref{l:Ikldef})
\begin{align*} 
{\mathcal{Q}}^{kl} \asymp\;& \frac{1}{2}\:\bigg\{
\Big( ({\mathscr{K}}^{kl}_L)^1_1 - 3 ({\mathscr{K}}^{kl}_R)^1_1 + ({\mathscr{K}}^{kl}_L)^2_2 + ({\mathscr{K}}^{kl}_R)^2_2 \Big) \\
&+\frac{L^{(0)}_{[0]}}{T^{(0)}_{[0]}}\;
\Big(  - ({\mathscr{K}}^{kl}_L)^1_1 + 3\, ({\mathscr{K}}^{kl}_R)^1_1 -({\mathscr{K}}^{kl}_L)^2_2 - ({\mathscr{K}}^{kl}_R)^2_2 \Big) \bigg\}  \\
&+ \O \big( (m \varepsilon)^{2 p_\reg} \big) \,(\deg = 4) + (\deg < 4)\:.
\end{align*}
The term~\eqref{l:curv0} drops out everywhere. Computing the contribution
by~\eqref{l:curv3} gives the result.
\QED

The next step is to satisfy~\eqref{l:varELgrav}. In fact, the results of the previous lemmas
are compatible with~\eqref{l:varELgrav} if we choose the parameter~$\tau$
in the Dirac Lagrangian~\eqref{l:LDirpar} as
\[ \tau = -4 \]
and the Lagrangian~$\LEH$ according to~\eqref{l:Lcurv} with
\beq \label{l:kappaval}
\kappa = \frac{\delta^2}{\tau_\reg}\: \frac{K_{17}}{K_{18}} \:,
\eeq
\nindex{dc0@$\kappa$ -- gravitational constant}%
\sindex{coupling constant!gravitational}%
where~$K_{17}$ and~$K_{18}$ are the composite expressions
\[ K_{17} = \frac{1}{2}\: K_{16} \;\bigg( 1 - \frac{L^{(0)}_{[0]}}{T^{(0)}_{[0]}} \bigg)  \qquad \text{and} \qquad
K_{18} = \frac{1}{2}\: K_8\: \bigg( 1 - \frac{L^{(0)}_{[0]}}{T^{(0)}_{[0]}} \bigg) \]
(which are both to be evaluated weakly on the light cone~\eqref{l:asy}).
These findings are summarized as follows.
\begin{Thm} \label{l:thmEinstein}
Assume that the parameters~$\delta$ and~$p_\reg$ satisfy the scaling~\eqref{l:deltascale2},
and that the regularization satisfies the conditions~\eqref{l:T0point} and~\eqref{l:Tm1point}.
Then the EL equations in the continuum limit~\eqref{l:Qform} can be expressed in terms of the effective
action~\eqref{l:Seff}.
The parameter~$\tau$ in the Dirac Lagrangian~\eqref{l:LDirpar}
is determined to have the value~$\tau=-4$.
The gravitational constant~$\kappa$ is given by~\eqref{l:kappaval}.
\end{Thm} \noindent
Combined with the equations for the chiral gauge fields in Theorem~\ref{l:thmfield},
this theorem shows that the structure of the interaction is described completely by the underlying EL
equations~\eqref{l:EL} corresponding to the causal action principle~\eqref{l:actprinciple}.

We point out that our results imply that the right-handed component of the neutrinos
must couple to the Einstein equations with a relative factor of~$-3$. In particular,
the right-handed component of the neutrinos has a negative energy density,
thus violating the usual energy conditions. This might give a possible explanation
for the anomalous acceleration of the universe.
\sindex{Einstein equations}%

One should keep in mind that the effective Lagrangian is determined only up to terms
which contribute to the EL equations~\eqref{l:Qform} to degree three or lower.
In particular, if the Ricci tensor is a multiple of the metric,
the term~$R_{jk} \xi^j \xi^k$ in Lemma~\ref{l:lemmaR} is of degree one on the light cone,
giving rise to a contribution which can be absorbed in the error term.
In other words, to the considered degree four on the light cone, the Ricci tensor
is determined only up to multiples of the metric. This gives precisely the freedom
to add the cosmological Lagrangian
\[ \int_\scrM \frac{2 \Lambda}{\kappa} \: \sqrt{-\det g} \: d^4x \]
\sindex{cosmological constant}%
\nindex{db8@$\Lambda$ -- cosmological constant}%
for an arbitrary value of the cosmological constant~$\Lambda$.
In principle, the cosmological constant could be determined in our approach
by evaluating the EL equations to degree three on the light cone. But this analysis goes
beyond the scope of the present work.

We point out that our results
exclude corrections to the Einstein-Hilbert action of higher order in the curvature tensor.
Note that the simple fractions~$K_{17}$ and~$K_{18}$ are both of degree four,
and thus their quotient is of the order one. Hence
\[ \kappa \sim \delta^2 \:. \]
This means that the Planck length is to be identified with the length scale~$\delta$
describing the shear and general surface states (see~\eqref{l:Texp} and~\eqref{l:ncontract}).

We next explain how this theorem could be extended to the case
\[ \delta \simeq \frac{1}{m}\: (m \varepsilon)^{\frac{p_\reg}{2}} \:. \]
In this case, the terms~$\sim m^2 R_{jk} \xi^j \xi^k$ in Lemma~\ref{l:lemmaR}
are of the same order as those~$\sim \tau_\reg/\delta^2 R_{jk} \xi^j \xi^k$ and
must be taken into account.
They can be obtained by a straightforward computation. The statement of 
Theorem~\ref{l:thmEinstein} will remain the same, except that the form of~$K_{16}$
will of course be modified.
The only structural difference is that~\eqref{l:K16} will then involve factors~$T^{(1)}_{[0]}$,
which have logarithmic poles on the light cone. It does not seem possible to compensate
these logarithmic poles by a microlocal transformation. Therefore, in order
for the logarithmic poles to drop out of the EL equations, one must impose that
\[ \sum_{\alpha=1}^3 m_\alpha^2 = \sum_{\alpha=1}^3 \tilde{m}_\alpha^2 \:. \]
This constraint for the neutrino masses can be understood similar as in Remark~\ref{l:remlargemass}.
Working out the detailed computations seems an interesting project for the future.

We next point out that our method of perturbing the regularized Minkowski
vacuum by a gravitational field implies in particular that the unperturbed regularization
is homogeneous, so that the parameters~$\varepsilon$ and~$\delta$ are
constant in space-time. Although this seems a good approximation locally,
it is conceivable that the regularization does change on the astrophysical
or cosmological scale. In this case, the gravitational constant would 
no longer be constant in space-time, but would become dynamical.
We refer the reader interested in this effect of ``dynamical gravitational
coupling'' to the paper~\cite{dgc}.
\sindex{dynamical gravitational coupling}%

We finally note how the energy-momentum tensor of the gauge field enters the
Einstein equations. The contribution of this energy-momentum tensor to the EL-equations
to degree four on the light cone was computed in Lemma~\ref{lemmaFFT}.
Since our gauge field is left-handed, only the term~\eqref{billog}
contributes. Due to the factor~$T^{(1)}_{[0]}$, this contribution has a logarithmic pole
on the light cone. As shown in Lemma~\ref{lemmaTcomp}, this logarithmic pole
on the light cone can be compensated by a suitable microlocal transformation.
As a result, the energy-momentum tensor enters the EL equations.
After this has been done, the energy-momentum tensor of the gauge field
enters the EL equations similar to the energy-momentum tensor of the Dirac field in~\eqref{l:JKL1}.
The only difference is that, since the simple fraction in~\eqref{billog} has a different form
than the simple fraction~$K_8$ in Lemma~\ref{l:lemmaT}, the energy-momentum tensor
of the gauge field comes with an additional regularization parameter.
A-priori, this regularization parameter is to be treated as a free empirical parameter
of the effective continuum theory. However, it can be fixed uniquely by the following mathematical consistency
condition: The Einstein equations imply that the total energy-momentum tensor must be
divergence-free. On the other hand, the energy-momentum tensor obtained by varying the
metric in the effective action~\eqref{l:Seff}
gives rise to a divergence-free vector field being a specific linear combination of the
energy-momentum tensor of the Dirac field and the energy-momentum tensor of the
gauge field. In general, these two energy-momentum tensors are not divergence-free separately,
only the specific linear combination is divergence-free as a consequence of the
Dirac equation and the field equations for the gauge fields.
Therefore, in order to allow for non-trivial solutions, we
are forced by mathematical consistency of the equations
to choose the regularization parameter in such a way that the energy-momentum tensors
enter the EL equations to degree four in the same linear combination as obtained
by varying the effective action~\eqref{l:Seff}.

These considerations complete the analysis of the EL equations
to degree four on the light cone up to errors of the order
\[ Q(x,y) = (\deg = 4) \cdot o\big( |\vec{\xi}|^{-2} \big) + (\deg < 4) \:. \]

\chapter[A System Involving Leptons and Quarks]
{The Continuum Limit of a Fermion System Involving Leptons and Quarks:
Strong, Electroweak and Gravitational Interactions} \label{quark}

\begin{abstract}
The causal action principle is analyzed for a system of relativistic fer\-mions
composed of massive Dirac particles and neutrinos.
In the continuum limit, we obtain an effective interaction described by
classical gravity as well as the strong and electroweak gauge fields of the standard model.
\end{abstract}

\section{Introduction}
In this chapter, we consider a system which is composed of seven massive sectors
and one neutrino sector, each containing three generations of fermions.
Analyzing the Euler-Lagrange equations of the causal action principle in the
continuum limit, we obtain a unification of gravity with the strong and
electroweak forces of the standard model.

More precisely, we obtain three main results.
The first main result is the so-called {\em{spontaneous block formation}} (see Theorem~\ref{q:thmsbf}),
stating that the eight sectors form pairs, so-called blocks.
\sindex{spontaneous block formation}%
The block involving the neutrinos can be regarded as the leptons, whereas the three other
blocks correspond to the quarks (in three colors). The index distinguishing the two sectors
within each block can be identified with the isospin. The interaction can be
described effectively by $\U(1) \times \SU(2) \times \SU(3)$-gauge potentials acting on the
blocks and on the isospin index.
In this way, one recovers precisely the gauge potentials of the standard model together with their
correct coupling to the fermions.

Our second main result is to derive the {\em{field equations}}
\sindex{field equations in the continuum limit}%
for the gauge fields. Theorem~\ref{q:thmgendyn} gives the general structure of the electroweak
theory of the standard model after spontaneous symmetry breaking, but the masses and coupling
constants involve more free parameters than in the standard model.
In Theorem~\ref{q:thmelectroweak} it is shown that one gets precise agreement with the electroweak
theory if one imposes three additional relations between the free parameters. 
Finally, in Proposition~\ref{q:prpasy} it is shown that these three additional relations 
hold in
the limit when the mass of the top quark is much larger than the lepton masses.
We thus obtain agreement with the strong and electroweak theory up to small corrections.
These corrections are discussed, and some of them are specified quantitatively.

Our third main result is to derive the gravitational interaction and the {\em{Einstein equations}}
(see Theorem~\ref{q:thmEinstein}).

We point out that the continuum limit gives the correspondence to the standard model
and to general relativity on the level of second-quantized fermionic fields coupled to
classical bosonic fields. For the connection to second-quantized bosonic fields 
we refer to~\cite{qft, qftlimit}. We also point out that we do not consider a Higgs field. This is why
we get the correspondence to the standard model after spontaneous symmetry breaking without the
Higgs field (i.e.\ for a constant Higgs potential). But in Section~\ref{q:sechiggs} it is explained
that the Higgs potential can possibly be identified with scalar potentials in the Dirac equation.

\section{Preliminaries}
In this section we repeat constructions used in Chapters~\ref{sector} and~\ref{lepton}
and adapt them to the system of Dirac seas to be considered here.

\subsectionn{The Fermionic Projector and its Perturbation Expansion} \label{q:secperturb}
We want to extend the analysis in Chapters~\ref{sector} and~\ref{lepton} to a system involving quarks.
Exactly as explained in~\cite[Section~5.1]{PFP}, the quarks are described
by additional sectors of the fermionic projector. More precisely, we describe the vacuum 
similar to~\eqref{l:Pvac} by the fermionic projector
\beq \label{q:Pvac}
P(x,y) = P^N(x,y) \oplus P^C(x,y) \:,
\eeq
\nindex{da0@$P^N(x,y)$ -- neutrino sector of the vacuum fermionic projector}%
\nindex{da2@$P^C(x,y)$ -- charged component of the vacuum fermionic projector}%
where the {\em{charged component}} $P^C$ is formed as the direct sum of seven
identical sectors, each consisting of a sum of three Dirac seas,
\beq \label{q:PC}
P^C(x,y) = \bigoplus_{a=1}^7 \sum_{\beta=1}^3 P^\text{vac}_{m_\beta}(x,y) \:,
\eeq
where~$m_\beta$ are the masses of the fermions and~$P^\text{vac}_m$ is the distribution
\[ P^\text{vac}_m(x,y) = \int \frac{d^4k}{(2 \pi)^4}\: (\slashed{k}+m)\: \delta(k^2-m^2)\: \Theta(-k^0)\: e^{-ik(x-y)}\:. \]
Thus every massive sector has the form as considered in~\eqref{s:A}.
\nindex{ca3@$m_\beta$ -- masses of charged fermions}%
\nindex{ba0@$P^\text{vac}_m(x,y)$ -- fermionic projector corresponding to a vacuum Dirac sea of mass~$m$}%
For the {\em{neutrino sector}} $P^N$ we choose the ansatz of massive neutrinos
(cf.~\eqref{l:massneutrino})
\beq \label{q:massneutrino}
P^N(x,y) = \sum_{\beta=1}^3 P^\text{vac}_{\tilde{m}_\beta}(x,y) \:.
\eeq
The neutrino masses~$\tilde{m}_\beta \geq 0$ will in general be different from
the masses~$m_\beta$ in the charged sector.
\nindex{da4@$\tilde{m}_\beta$ -- neutrino masses}%
For a discussion of this ansatz we also refer to~\S\ref{l:secmassive}, where
the alternative ansatz of chiral neutrinos is ruled out.

We introduce an {\em{ultraviolet regularization}} on the length scale~$\varepsilon$.
\nindex{aj8@$\varepsilon$ -- regularization length}%
\sindex{regularization!ultraviolet (UV)}%
The regularized vacuum fermionic projector is denoted by~$P^\varepsilon$.
\nindex{an4@$P^\varepsilon(x,y)$ -- regularized kernel of fermionic projector}%
We again use the formalism of the continuum limit as developed in~\cite[Chapter~4]{PFP}
and described in Section~\ref{s:sec5}. In the neutrino sector, we work exactly 
as in~\S\ref{l:sec23} with a non-trivial regularization by
right-handed high-energy states.

In order to describe an interacting system, we proceed exactly as described
in~\cite[Section~2.3]{PFP}, Section~\ref{s:sec4} and~\S\ref{l:sec24}.
We first introduce the {\em{auxiliary fermionic projector}} by
\[ P^\text{aux} = P^N_\text{aux} \oplus P^C_\text{aux}\:, \]
\sindex{fermionic projector!auxiliary}%
\nindex{ca8@$P^\text{aux}$ -- auxiliary fermionic projector}%
where
\beq \label{q:Paux}
P^N_\text{aux} = \Big( \bigoplus_{\beta=1}^3 P^\text{vac}_{\tilde{m}_\beta} \Big) \oplus 0
\qquad \text{and} \qquad
P^C_\text{aux} = \bigoplus_{a=1}^7 \bigoplus_{\beta=1}^3 P^\text{vac}_{m_\beta} \:.
\eeq
\nindex{dc31@$P^N_\text{aux}$ -- neutrino sector of auxiliary fermionic projector}%
\nindex{dc32@$P^C_\text{aux}$ -- charged component of auxiliary fermionic projector}%
Note that~$P^\text{aux}$ is composed of~$25$ direct summands, four in the neutrino
and~$21$ in the charged sector. The fourth direct summand of~$P^N_\text{aux}$ has the purpose
of describing the right-handed high-energy states. Moreover, we introduce the
{\em{chiral asymmetry matrix}}~$X$ and the {\em{mass matrix}}~$Y$ by
(cf.~\eqref{l:treg1} and~\eqref{l:Ydef1})
\begin{align*}
X &= \left( \1_{\C^3} \oplus \tau_\reg \,\chi_R \right) \oplus \bigoplus_{a=1}^7 \1_{\C^3} \\
m Y &= \text{diag} \big( \tilde{m}_1, \tilde{m}_2, \tilde{m}_3, 0 \big)
\oplus \bigoplus_{a=1}^7 \text{diag} \big( m_1, m_2, m_3 \big) \:,
\end{align*}
where~$m$ is an arbitrary mass parameter.
\sindex{mass matrix}%
\nindex{bh3@$m$ -- parameter used for mass expansion}%
\nindex{bh4@$Y$ -- mass matrix}%
\sindex{chiral asymmetry matrix}%
\nindex{dc4@$X$ -- chiral asymmetry matrix}%
\nindex{dc6@$\tau_\reg$ -- dimensionless parameter for high-energy states}%
Here~$\tau_\reg \in (0,1]$ is a dimensionless parameter
for which we always assume the scaling
\[ \tau_\reg = (m \varepsilon)^{p_\reg} \qquad \text{with} \qquad
0 < p_\reg < 2\:. \]
This allows us to rewrite the vacuum fermionic projector as
\beq \label{q:Pauxdef}
P^\text{aux} = X t = t X^* \qquad \text{with} \qquad t := \bigoplus_{\beta=1}^{25} P^\text{vac}_{m Y^\beta_\beta} \:.
\eeq
Now~$t$ is a solution of the Dirac equation
\[ (i \Pdd - m Y) \,t = 0 \:. \]
\nindex{dc8@$t$ -- distribution composed of vacuum Dirac seas}%
In order to introduce the interaction, we insert an operator~$\B$ into the Dirac equation,
\beq \label{q:Dinteract}
(i \Pdd + \B - m Y) \,\tilde{t} = 0 \:.
\eeq
\nindex{ar4@$\B$ -- external potential}%
\sindex{potential!bosonic}%
\sindex{potential!external}%
The causal perturbation theory (see~\cite[Section~2.2]{PFP}, \cite{norm} or Section~\ref{secfpext})
defines~$\tilde{t}$ in terms of a unique perturbation series. The {\em{light-cone expansion}}
(see~\cite[Section~2.5]{PFP} and the references therein or Section~\ref{seclight}) is a method for analyzing
the singularities of~$\tilde{t}$ near the light cone. This gives a representation of~$\tilde{t}$
of the form
\begin{align*}
\tilde{t}(x,y) = & \sum_{n=-1}^\infty
\sum_{k} m^{p_k} 
{\text{(nested bounded line integrals)}} \times  T^{(n)}(x,y) \\
&+ \tilde{P}^\lec(x,y) + \tilde{P}^\hec(x,y) \:,
\end{align*}
where~$\tilde{P}^\lec(x,y)$ and~$\tilde{P}^\hec(x,y)$ are smooth to every order in perturbation theory.
For the resulting
light-cone expansion to involve only {\em{bounded}} line integrals, we need to assume the
{\em{causality compatibility condition}}
\beq \label{q:ccc}
(i \Pdd + \B - m Y)\, X = X^* \,(i \Pdd + \B - m Y) \qquad \text{for all~$\tau_\reg \in (0,1]\:$.}
\eeq
\sindex{causality compatibility condition}%
Then the auxiliary fermionic projector of the sea states~$P^\sea$ is obtained similar to~\eqref{q:Pauxdef} by
multiplication with the chiral asymmetry matrix.

As in~\S\ref{l:sec24} we built the regularization into the formulas of the light-cone expansion
by the formal replacements
\begin{align*}
m^p \,T^{(n)} &\rightarrow m^p \,T^{(n)}_{[p]}\:, \\
\tau_\reg \,T^{(n)} &\rightarrow \tau_\reg \sum_{k=0}^\infty \frac{1}{k!}\: \frac{1}{\delta^{2k}}
T^{(k+n)}_{[R,2n]} \:.
\end{align*}
Moreover, we introduce particles and anti-particles by occupying additional states or by removing
states from the sea, i.e.\
\[ P^\text{aux}(x,y) = P^\sea(x,y)
-\frac{1}{2 \pi} \sum_{k=1}^{\np} \psi_k(x) \overline{\psi_k(y)}
+\frac{1}{2 \pi} \sum_{l=1}^{\na} \phi_l(x) \overline{\phi_l(y)} \]
(for the normalization of the particle and anti-particle states we refer to~\cite[Section~2.8]{PFP} or~\S\ref{s:sec43} and~\cite{norm}).
Finally, we introduce the regularized fermionic projector~$P$ by forming the {\em{sectorial projection}}
\sindex{sectorial projection}%
(see also~\cite[Section~2.3]{PFP}, \eqref{s:pt} or~\eqref{l:partrace0}),
\beq \label{q:partrace0}
(P)^i_j = \sum_{\alpha, \beta} (\tilde{P}^\text{aux})^{(i,\alpha)}_{(j, \beta)} \:,
\eeq
where~$i,j \in \{1, \ldots, 8\}$ is the sector index, and the indices~$\alpha$ and~$\beta$ run over
the corresponding generations (i.e., $\alpha \in \{1, \ldots 4\}$ if~$i=1$ and~$\alpha \in \{1, 2, 3 \}$
if~$i=2, \ldots, 8$). We again indicate the sectorial projection of the mass matrices by accents
\nindex{cb6@$\hat{\;},\: \acute{\;} \ldots \grave{\;}$ -- short notation for sectorial projection}%
(see~\cite[Section~7.1]{PFP}, \eqref{s:tildedef} or~\eqref{l:accents}),
\[ \hat{Y} = \sum_{\alpha} Y^\alpha_\alpha\:, \qquad
\acute{Y} Y \cdots \grave{Y} = \sum_{\alpha,\beta, \gamma_1, \ldots, \gamma_{p-1}}
Y^\alpha_{\gamma_1} \cdots Y^{\gamma_1}_{\gamma_2}
\cdots Y^{\gamma_{p-1}}_\beta . \]

As in Chapter~\ref{lepton}, we need assumptions on the regularization. Namely, again setting
\[ L^{(n)}_{[p]} = T^{(n)}_{[p]} + \frac{1}{3}\:\tau_\reg\, T^{(n)}_{[R,p]}\:, \]
\nindex{de0@$L^{(n)}_{[p]} = T^{(n)}_{[p]} + \tau_\reg T^{(n)}_{[R,p]}/3$}%
we impose the following regularization conditions (see~\eqref{l:RC1}, \eqref{l:T0point} and~\eqref{l:Tm1point})
\begin{gather}
T^{(0)}_{[0]} T^{(-1)}_{[0]} \overline{T^{(0)}_{[0]}} = 0 \qquad
\text{in a weak evaluation on the light cone} \label{q:RC1} \\
\big| L^{(n)}_{[0]} \big| = \big| T^{(n)}_{[0]} \big| \left( 1 + \O \big( (m \varepsilon)^{2 p_\reg} \big) \right)
\qquad \text{for~$n=0, -1$ pointwise} \:.  \label{q:Tpoint}
\end{gather}
Here by {\em{weak evaluation}}
\sindex{evaluation on the light cone!weak}%
we mean that we multiply by a test function~$\eta$ and
integrate, staying away from the origin. More precisely, we use the weak evaluation formula
(for details see~\S\ref{s:sec51})
\beq \label{q:asy}
\int_{|\vec{\xi}|-\varepsilon}^{|\vec{\xi}|+\varepsilon} dt \; \eta(t,\vec{\xi}) \:
\frac{ T^{(a_1)}_\circ \cdots T^{(a_\alpha)}_\circ \:
\overline{T^{(b_1)}_\circ \cdots T^{(b_\beta)}_\circ} }
{ T^{(c_1)}_\circ \cdots T^{(c_\gamma)}_\circ \:
\overline{T^{(d_1)}_\circ \cdots T^{(d_\delta)}_\circ} }
= \eta(|\vec{\xi}|,\vec{\xi}) \:\frac{c_{\reg}}{(i |\vec{\xi}|)^L}
\;\frac{\log^k (\varepsilon |\vec{\xi}|)}{\varepsilon^{L-1}} \:,
\eeq
which holds up to
\[ \text{(higher orders in~$\varepsilon/\ell_\text{macro}$ and~$\varepsilon/|\vec{\xi}|$)}\:. \]
Here~$L$ is the degree defined by~$\deg T^{(n)}_\circ = 1-n$,
and~$c_{\reg}$ is a so-called {\em{regularization parameter}}
(for details see again~\cite[Section~4.5]{PFP} or~\S\ref{s:sec51}).
In~\eqref{q:Tpoint} by {\em{pointwise}} we mean that if we multiply~$T^{(n)}_{[p]} - L^{(n)}_{[p]}$ by
any simple fraction and evaluate weakly on the light cone, we get zero
up to an error of the specified order.
We remark that~\eqref{q:Tpoint} could be replaced by a finite number of equations
to be satisfied in a weak evaluation on the light cone. But in order to keep our analysis
reasonably simple, we always work with the easier pointwise conditions~\eqref{q:Tpoint}.

\subsectionn{Chiral Gauge Potentials and Gauge Phases} $\;\;\;$
Similar as in~\S\ref{s:sec72} and~\S\ref{l:sec32}
we consider chiral gauge potentials. Thus the operator~$\B$ in the Dirac equation~\eqref{q:Dinteract}
is chosen as
\beq \label{q:chiral}
\B = \chi_L\: \slashed{A}_R + \chi_R\: \slashed{A}_L\:,
\eeq
where~$A^j_L$ and~$A^j_R$ are Hermitian $25 \times 25$-matrices acting on the sectors.
A-priori, the chiral gauge potentials can be chosen according to the gauge group
\[ \U(25)_L \times \U(25)_R\:. \]
\sindex{potential!chiral}%
\nindex{bh2@$A_L, A_R$ -- chiral potentials}%
This gauge group is too large for mathematical and physical reasons.
First, exactly as in~\S\ref{l:sec32}, the causality compatibility condition~\eqref{q:ccc} inhibits that
non-trivial high-energy contributions are mixed with the Dirac seas, giving rise to the smaller gauge group
\beq \label{q:U24}
\U(24)_L \times \U(24)_R \times \U(1)_R\:,
\eeq
(where the $\U(24)$ acts on the first three direct summands of~$P^N_\text{aux}$
and on the 21 direct summands in~$P^M_\text{aux}$ in~\eqref{q:Paux}).
Similar as described in~\S\ref{s:sec72} and~\S\ref{l:sec32},
to degree three on the light cone the gauge potentials describe generalized phase
transformations of the left- and right-handed components of the fermionic projector,
\beq
P^\text{aux}(x,y) \rightarrow \big( \chi_L \,U_L(x,y) + \chi_R \,U_R(x,y)
\big) P^\text{aux}(x,y) + (\deg < 2 )\:, \label{q:Pchiral}
\eeq
where the unitary operators~$U_c$ are ordered exponentials
(for details see~\cite[Section~2.5]{PFP}, \cite[Section~2.2]{light} or~\eqref{l:Pchiral}),
\[ U_c = \Pexp \Big( -i \int_x^y A^j_c \,\xi_j \Big) \:. \]

The fermionic projector is obtained from~\eqref{q:Pchiral} by forming the sectorial projection~\eqref{q:partrace0}.
Summing over the generation indices has the effect that wave functions are superimposed
which may involve different gauge phases. In other words, the sectorial projection in general
involves relative gauge phases.
In order to simplify the form of the gauge potentials,
we now argue that such relative gauge phases should be absent.
In fact, if such relative phases occurred, the different Dirac seas forming the fermionic projector would get
out of phase, implying that all relations for the fermionic projector would have to be satisfied
for each Dirac sea separately. This would give rise to many additional constraints for the regularization,
which seem impossible to satisfy. 
We remark that a similar argument is given in~\S\ref{l:sec32}.
Moreover, the physical picture is similar for the gravitational field, where
it was argued in Section~\ref{l:seccurv} that the metric tensor must be independent of the isospin
index.

The simplest method to avoid such relative phases would be to choose gauge potentials
which do not depend on the generation index, i.e.
\beq \label{q:nogeneration}
(A_L)^{(i,\alpha)}_{(j, \beta)} = (A^{\sec}_L)^i_j\: \delta^\alpha_\beta
\eeq
(where the superscript ``sec'' clarifies that the potential carries only sector indices).
\nindex{ea0@$A^{\sec}_L, A^{\sec}_R$ -- chiral potentials carrying only sector indices}%
In order to be compatible with the $U(1)_R$-subgroup in~\eqref{q:U24} acting on the
right-handed high-energy states in the neutrino sector, we need to choose the
potentials in~\eqref{q:nogeneration} corresponding to the gauge group
\beq \label{q:U8}
\U(8)_L \times \U(1)_R \times \U(7)_R \:,
\eeq
where the $\U(7)$ acts on the seven direct summands in~\eqref{q:PC}
but is trivial on the neutrinos~\eqref{q:massneutrino}.
The ansatz~\eqref{q:nogeneration} can be slightly generalized by allowing for unitary
transformations in each sector. This leads to the ansatz
\beq \label{q:Bansatz}
\B = \chi_R\: U^\mix_L \slashed{A}^{\sec}_L (U^\mix_L)^* + \chi_L\: U^\mix_R \slashed{A}^{\sec}_R (U^\mix_R)^* \:,
\eeq
\sindex{mixing matrix!general}%
\nindex{ea2@$U^\mix_L, U^\mix_R$ -- general mixing matrices}%
where the potentials~$A_c^{\sec}$ are again of the form~\eqref{q:nogeneration},
and the matrices~$U^\mix_c$ are constant unitary matrices which are diagonal in the sector index,
\beq \label{q:Umixform}
(U^\mix_c)^{(i,\alpha)}_{(j, \beta)} = \delta^i_j\: (U^i_c)^\alpha_\beta \qquad \text{with} \qquad
U^i_c \in \U(3)\:.
\eeq
Thus we allow for a different mixing matrix for every sector. Also, the mixing matrices may be
different for the left- and right-handed components of the spinors.
The fact that the mixing matrices are constant could be justified by using arguments similar
to those worked out for two sectors in Lemma~\ref{l:lemmablock}.
Here we do not enter such arguments again but simply take~\eqref{q:Bansatz} as our ansatz
for the chiral gauge potentials. It seems the most general ansatz which
avoids relative phases when forming the sectorial projection. Specializing the chiral gauge fields to
the ansatz~\eqref{q:Bansatz}, the matrices~$U_c$ in~\eqref{q:Pchiral} become
\beq
U_c = U^\mix_c\: \Pexp \Big( -i \int_x^y A^j_c \,\xi_j \Big) (U^\mix_c)^*\:.  \label{q:Lambda2}
\eeq

\subsectionn{The Microlocal Chiral Transformation}
Exactly as in~\S\ref{s:secgennonloc} and~\S\ref{s:sec44}, our method is to compensate
the logarithmic singularities of the current and mass terms by a microlocal chiral transformation.
\sindex{transformation of the fermionic projector!microlocal chiral}%
To this end, one considers a Dirac equation of the form
\beq \label{q:Dirnon}
(U^{-1})^* (i \Pdd +{\mathcal{B}} - m Y) \,U^{-1} \tilde{P}^\text{aux} = 0\:,
\eeq
where~$U$ is an integral operator with an integral kernel~$U(x,y)$,
which we write in the microlocal form
\nindex{cg8@$U(x,y)$ -- microlocal chiral transformation}%
\[ U(x,y) = \int \frac{d^4k}{(2 \pi)^4}\: U\Big( k,\frac{x+y}{2} \Big)\: e^{-i k(x-y)}\:, \]
where~$U(k, z)$ is a chiral transformation
\beq \label{q:Uans}
U(k,z) = \1 + \frac{i}{\sqrt{\Omega}} \:Z(k,z) \qquad \text{with} \qquad
Z(z) = \chi_L\, L^j(k,z) \gamma_j + \chi_R \,R^j(k,z) \gamma_j \:.
\eeq
\nindex{ea4@$Z(k,z)$ -- generator of microlocal chiral transformation}%
\nindex{ea6@$L(k,z), R(k,z)$ -- chiral components of~$Z(k,z)$}%
Writing the Dirac equation~\eqref{q:Dirnon} in the form~\eqref{q:Dinteract} with a
nonlocal operator~$\B$, the perturbative methods of \S\ref{q:secperturb} again apply.

More specifically, the matrices~$L$ and~$R$ in~\eqref{q:Uans} are chosen such that the 
matrices~$L[k,x] := \acute{L}_j(k,x)\, k^j$ and~$R[k,x] := \acute{L}_j(k,x)\, k^j$ satisfy the conditions
\begin{align}
L[k,x]\, L[k,x]^* &= R[k,x]\, R[k,x]^* = {\mathfrak{c}}_0(k,x)\, \1_{\C^2} \label{q:4con} \\
L[k,x]\, m^2 Y^2\, L[k,x]^* &= \frac{\Omega}{2}\: v_L(x) + {\mathfrak{c}}_2(k,x)\, \1_{\C^2} \\
R[k,x]\, m^2 Y^2\, R[k,x]^* &=  \frac{\Omega}{2}\: v_R(x) + {\mathfrak{c}}_2(k,x)\, \1_{\C^2} \:,
\end{align}
where~${\mathfrak{c}}_0$ and~${\mathfrak{c}}_2$ are real parameters, and~$\Omega=|k^0|$
denotes the frequency of the four-momentum~$k$.
\nindex{dh7@${\mathfrak{c}}_0(k), {\mathfrak{c}}_2(k)$ -- parameters in microlocal chiral transformation}%
\nindex{dh2@$\Omega$ -- absolute value of energy}%
The vector fields~$v_L$ and~$v_R$ are 
the currents or potentials which multiply the logarithmic singularities to be compensated.

Writing the Dirac equation~\eqref{q:Dirnon} raises the question how the potential~${\mathcal{B}}$
is to be chosen. The most obvious procedure would be to choose~${\mathcal{B}}$
equal to the chiral potentials in~\eqref{q:chiral}. However, as shown in~\S\ref{s:secshearmicro}
and~\S\ref{l:secshear}, this is not the correct choice, intuitively speaking because the
microlocal chiral transformation in~\eqref{q:Dirnon} has contributions which flip the chirality, making it
necessary to also modify the potentials in the Dirac operator.
To this end, we write down the Dirac equation for the auxiliary fermionic projector as
\[ \Dir_\text{flip} \, \tilde{P}^\text{aux} = 0 \:, \]
\nindex{di0@$\Dir_\text{flip}$ -- Dirac operator including microlocal chiral transformation}%
where~$\Dir_\text{flip}$ is obtained from the Dirac operator with chiral gauge fields by
\beq \label{q:Deflip}
\Dir_\text{flip} :=  (U_\text{flip}^{-1})^* \,\big(i \Pdd_x + \chi_L \slashed{A}_R + \chi_R \slashed{A}_L - m Y \big)\, U_\text{flip}^{-1} \:.
\eeq
Here~$U_\text{flip}$ is obtained from the operator~$U$ in~\eqref{q:Dirnon} by
\[ U_\text{flip} = \1 + (\U-\1)\, V \:, \]
and~$V$ is the unitary perturbation flow which changes the gauge potentials from~$A_{L\!/\!R}$
to~$A^\even_{L\!/\!R}$,
\[ V = U_\text{flow}[\chi_L \slashed{A}^\even_R + \chi_R \slashed{A}^\even_L]\;
U_\text{flow}[\chi_L \slashed{A}_R + \chi_R \slashed{A}_L]^{-1} \:. \]
For details we refer to~\eqref{s:scrDdef}, \eqref{s:Uflip} and~\eqref{a:Vflowdef}.

\subsectionn{The Causal Action Principle}
We again consider the causal action principle introduced in~\cite{PFP}.
The action is
\[ \Sact[P] = \iint_{\scrM \times \scrM} \L[A_{xy}] \:d^4 x\: d^4y \]
\nindex{aa4@$\Sact(\rho)$ -- causal action}%
\sindex{causal action}%
with the Lagrangian
\[ \L[A_{xy}] = |A_{xy}^2| - \frac{1}{32}\: |A_{xy}|^2 \:, \]
\nindex{aa2@$\L(x,y)$ -- Lagrangian}%
\sindex{Lagrangian!causal}%
where~$A_{xy} = P(x,y)\, P(y,x)$ denotes the closed chain
\sindex{closed chain}%
\nindex{ad0@$A_{xy}$ -- closed chain}%
and~$|A| = \sum_{i=1}^8 |\lambda_i|$ is the spectral weight.
As shown in~\S\ref{s:secELC}, the Euler-Lagrange equations in the continuum limit
can be written as
\beq \label{q:Qweak}
Q(x,y) = 0 \quad \text{if evaluated weakly on the light cone}\:,
\eeq
where~$Q(x,y)$ is defined as follows.
Similar as explained in~\S\ref{l:sec31}, we count the eigenvalues of the
the closed chain~$A_{xy}$ with algebraic multiplicities and denote them by~$\lambda^{xy}_{ncs}$,
where~$n \in \{1,\ldots, 8\}$, $c \in \{L,R\}$ and~$s \in \{+,-\}$.
The corresponding spectral projectors are denoted by~$F^{xy}_{ncs}$.
Then~$Q(x,y)$ is given by
\begin{align}
Q(x,y) &= \frac{1}{2} \sum_{ncs} \frac{\partial \L}{\partial \lambda^{xy}_{ncs}}
\:F_{ncs}^{xy}\: P(x,y) \nonumber \\
&= \sum_{n,c,s} \bigg[  |\lambda^{xy}_{ncs}| - \frac{1}{8} \sum_{n',c',s'}
|\lambda^{xy}_{n'c's'}| \bigg] \: \frac{\overline{\lambda^{xy}_{ncs}}}{|\lambda^{xy}_{ncs}|}
\: F_{ncs}^{xy}\: P(x,y) \:. \label{q:Qform}
\end{align}
\nindex{aq6@$Q(x,y)$ -- first variation of the Lagrangian}%

The equation~\eqref{q:Qweak} is satisfied in the vacuum (see~\S\ref{s:sec71}
and~\S\ref{l:sec31}).
When evaluating the EL equations in the interacting situation, in most cases it will be sufficient to
consider~\eqref{q:Qweak} for perturbations of the eigenvalues,
\beq \label{q:EL4}
0 = \Delta Q(x,y) := \sum_{n,c,s} \bigg[  \Delta |\lambda^{xy}_{ncs}| - \frac{1}{8} \sum_{n',c',s'}
\Delta |\lambda^{xy}_{n'c's'}| \bigg] \: \frac{\overline{\lambda^{xy}_{ncs}}}{|\lambda^{xy}_{ncs}|}
\: F_{ncs}^{xy}\: P(x,y) \:.
\eeq
\nindex{df8@$\Delta Q(x,y)$ -- first order perturbation of~$Q(x,y)$}%

\section{Spontaneous Block Formation} \label{q:secsbf}
The goal of this section is to derive constraints for the form of the admissible gauge fields.
The arguments are similar in style to those in~\cite[Chapter~7]{PFP}. However, as a main difference,
we here consider the effect of the sectorial projection and the mixing of the generations,
whereas in~\cite[Chapter~7]{PFP}
the contributions of higher order in a mass expansion (which are of lower degree on the light cone)
were analyzed. The analysis given here supersedes the arguments in~\cite[Chapter~7]{PFP},
which with the present knowledge must be regarded as being preliminary.

\subsectionn{The Statement of Spontaneous Block Formation}
Analyzing the EL equations to degree five and degree four on the light cone
gives rise to a number of equations which involve the chiral potentials without
derivatives. These equations do not describe a dynamics of the potentials
and fields, but merely pose constraints for the structure of the possible interactions.
We refer to these equations as the {\em{algebraic constraints}} for the gauge potentials.
\sindex{gauge potential!algebraic constraint for}%
The algebraic constraints trigger a mechanism where the eight sectors form pairs,
the so-called {\em{blocks}}.
\sindex{block}%
Describing the interaction within and among the four blocks
by chiral gauge fields gives rise to precisely the gauge groups and couplings in the
standard model.

In order to introduce a convenient notation, we denote those chiral potentials
of the form~\eqref{q:Bansatz} which satisfy all the algebraic constraints as {\em{admissible}}.
\sindex{potential!admissible}%
Since ordered exponentials of the chiral potentials appear (see for example~\eqref{q:Pchiral}
and~\eqref{q:Lambda2}), it seems necessary for mathematical consistency
to consider a set of admissible chiral gauge potentials which forms a Lie algebra,
the so-called {\em{dynamical gauge algebra}}~$\g$.
\sindex{gauge algebra!dynamical}%
\nindex{ea8@$\g$ -- dynamical gauge algebra}%
\nindex{eb0@$\A=(A_L, A_R)$ -- element of dynamical gauge algebra}%
More precisely, the commutator
of two elements~$\A=(A_L, A_R)$ and~$\tilde{A}=(\tilde{A}_L, \tilde{A}_R)$ in~$\g$ is
defined by
\[ [\A, \tilde{\A}] = \Big( [A_L, \tilde{A}_L], \: [A_R, \tilde{A}_R] \Big) \]
(where the brackets $[.,.]$ is the commutator of symmetric $8 \times 8$-matrices; note that
the mixing matrices in~\eqref{q:Bansatz} drop out of all commutators).
The assumption that~$\g$ is a Lie algebra is the implication~$\A, \tilde{\A} \in \g
\Rightarrow i [\A, \tilde{\A}] \in \g$.
The corresponding Lie group will be a Lie subgroup of the gauge group in~\eqref{q:U8}.
We denote this Lie group by~$\G \subset \U(8)_L \times \U(1)_R \times \U(7)_R$
and refer to it as the {\em{dynamical gauge group}}.
\sindex{gauge group!dynamical}%
\nindex{dk8@$\G$ -- dynamical gauge group}%

The potentials in the dynamical gauge algebra should be regarded as describing the physical interactions
of the system. In order to understand the algebraic constraints, we clearly want to find
{\em{all}} the potentials which satisfy the algebraic constraints. Therefore, we always
choose~$\G$ {\em{maximal}} in the sense that~$\G$ has no Lie group extension
extension~$\tilde{\G}$ with~$\G \subsetneqq \tilde{\G} \subset \U(8)_L \times \U(1)_R \times \U(7)_R$
which is also generated by admissible chiral potentials.

We begin with the following definition.
\sindex{gauge potential!free}%
\begin{Def} An admissible chiral potential~$\A = (A_L, A_R) \in \g$ is a {\bf{free gauge potential}}
if it has the following properties:
\begin{itemize}
\item[(a)] The potential is vectorial: $A_L = A_R =:A$.
\item[(b)] The potential does not depend on the generation index:
$A^{(i, \alpha)}_{(j, \beta)} = \delta^\alpha_\beta \,(A^{\sec})^i_j$.
\item[(c)] The potential commutes with the mass matrix: $[A, mY ]=0$.
\end{itemize}
The Lie group generated by all free gauge potentials is referred to as the
{\bf{free gauge group}}~$\G_\free \subset \G$.
\end{Def} \noindent
\sindex{gauge group!free}%
\nindex{eb2@$\G_\free \subset \G$ -- free gauge group}%
Since the conditions~(a)--(c) are linear and invariant under forming the Lie bracket,
$\G_\free$ is indeed a Lie subgroup of~$\G$.

A free gauge potential has the desirable property that it corresponds to a gauge
symmetry of the system (because it describes
isometries of the spin spaces; see~\S\ref{s:sec72}).
As a consequence, the mass terms vanish, implying that the
corresponding bosonic fields are necessarily massless (see~\S\ref{s:secnohiggs}).
Moreover, chiral potentials with the above properties~(a)--(c) 
satisfy all algebraic constraints (see~\S\ref{q:secpt}--\S\ref{q:secftt} below)
and are thus admissible.

Here is the main result of this section:
\begin{Thm} {\bf{(spontaneous block formation)}} \label{q:thmsbf}
\sindex{spontaneous block formation}%
Consider the setting introduced in~\S\ref{q:secperturb} and assume that the following conditions hold:
\begin{itemize}
\item[(i)] The admissible gauge potentials involve non-abelian left- or right-handed
gauge potentials.
\item[(ii)] The mixing matrices $U^\text{mix}_c$ in~\eqref{q:Bansatz}
are chosen such that the dimension of the free gauge group is maximal.
\end{itemize}
Then the effective gauge group is given by
\beq \label{q:U123}
\G = \U(1) \times \SU(2) \times \SU(3) \:.
\eeq
By relabeling the massive sectors and performing constant phase transformations of the
wave functions, we can arrange that the
corresponding gauge potentials~$A^\text{\tiny{\rm{em}}} \in \u(1)$,
$W \in \su(2)$ and~$G \in \su(3)$ enter the
operator~$\B$ in the Dirac equation~\eqref{q:Dinteract} as follows,
\begin{align}
\B[A^\text{\tiny{\rm{em}}}] &=
\slashed{A}^\text{\tiny{\rm{em}}} \:\text{\rm{diag}} \Big(0,-1, \frac{2}{3}, -\frac{1}{3}, \frac{2}{3}, -\frac{1}{3}, 
\frac{2}{3}, -\frac{1}{3} \Big) \label{q:em} \\
\B[W] &=
\chi_R \left( \slashed{W}_\text{\rm{\tiny{MNS}}} \oplus \slashed{W}_\text{\rm{\tiny{CKM}}} \oplus \slashed{W}_\text{\rm{\tiny{CKM}}} \oplus \slashed{W}_\text{\rm{\tiny{CKM}}} \right) \label{q:weak} \\
\B[G] &=  \left( \1 \oplus \slashed{G} \right) \otimes \1_{\C^2} ,
\label{q:strong}
\end{align}
where
\nindex{eb4@$A^\text{\tiny{\rm{em}}}$ -- free $\u(1)$-potential}%
\nindex{eb6@$W$ -- left-handed $\su(2)$-potential}%
\nindex{eb8@$G$ -- free~$\su(3)$-potential}%
\[ W_\text{\rm{\tiny{MNS}}} = \begin{pmatrix} (W)^{11} & 
(W)^{12}\, \UMNS^* \\[0.2em]
(W)^{21}\, \UMNS & (W)^{22} \end{pmatrix} \:,\qquad
W_\text{\rm{\tiny{CKM}}} = \begin{pmatrix} (W)^{11} & 
(W)^{12}\, \UCKM^* \\[0.2em]
(W)^{21}\, \UCKM & (W)^{22} \end{pmatrix} , \]
\nindex{ec0@$W_\text{\rm{\tiny{MNS}}}$ -- $W$-potential in neutrino block}%
\nindex{ec2@$W_\text{\rm{\tiny{CKM}}}$ -- $W$-potential in quark blocks}%
and~$\UMNS, \UCKM \in \U(3)$ are fixed unitary matrices.
If one of these matrices is non-trivial, the other is also non-trivial and
\beq \label{q:hU}
\hat{U}_\text{\rm{\tiny{MNS}}} = \hat{U}_\text{\rm{\tiny{CKM}}}\:.
\eeq
If the masses of the charged leptons and neutrinos~\eqref{q:PC} and~\eqref{q:massneutrino}
are different in the sense that
\beq \label{q:massdiff}
\sum_{\beta=1}^3 m_\beta^2 \neq \sum_{\beta=1}^3 \tilde{m}_\beta^2 \:,
\eeq
then the mixing matrices are necessarily non-trivial,
\beq \label{q:notrivmix}
\UMNS, \UCKM \neq \1_{\C^3} \:.
\eeq
\end{Thm} \noindent
\nindex{da6@$\UMNS$ -- Maki-Nakagawa-Sakata (MNS) matrix}%
\sindex{mixing matrix!Maki-Nakagawa-Sakata (MNS) matrix}%
\nindex{ec4@$\UCKM$ -- Cabibbo-Kobayashi-Maskawa (CKM) matrix}%
\sindex{mixing matrix!Cabibbo-Kobayashi-Maskawa (CKM) matrix}%
To clarify the notation, we first note that we always identify~$\u(n)$ with the Hermitian
$n \times n$-matrices, and~$\su(n)$ are the corresponding traceless matrices.
Next, the diagonal matrix in~\eqref{q:em} acts on the
eight sectors. The potential in~\eqref{q:weak} only couples to the left-handed component
of the spinors. Each of the four direct summands acts on two sectors
(i.e.\ $W_\text{\rm{\tiny{MNS}}}$ on the first and second sector, the next summand~$W_\text{\rm{\tiny{CKM}}}$ on the third and fourth sector, etc.). 
In~\eqref{q:strong} the direct sum $\1 + \slashed{G}$
is a $4 \times 4$ matrix acting on pairs of sectors as indicated by the factor~$\1_{\C^2}$
(i.e.\ the first column acts on the first and second
sector, the second column on the third and fourth sector, etc.).

The specific form of the potentials in the above theorem can be understood
as a mechanism where the sectors form pairs, referred to as {\em{blocks}}.
\sindex{block}%
\sindex{block!lepton}%
\sindex{block!quark}%
Thus the first
two sectors form the first block (referred to as the {\em{lepton block}}), the third and fourth sectors form
the second block (referred to as the first {\em{quark block}}), and so on. The potentials in~\eqref{q:weak}
are the same in each of the four blocks, except for the mixing matrices~$\UMNS$ and~$\UCKM$,
which may be different in the lepton and in the quark blocks.
The potentials in~\eqref{q:strong} describe an interaction of the three quark blocks.
Clearly, the potentials~$A^\text{\tiny{\rm{em}}}$ and~$G$ correspond to the electromagnetic and the strong
potentials in the standard model. The potential~$W$ corresponds to the gauge potentials
of the weak isospin. 
The reduction from the large gauge group~\eqref{q:U8} to
its subgroup~\eqref{q:U123} and to gauge potentials of the specific form~\eqref{q:em}--\eqref{q:strong}
can be regarded as a spontaneous breaking of the gauge symmetry.
We refer to this effect as {\em{spontaneous block formation}}.
\sindex{spontaneous block formation}%

We point out that without any additional assumptions (like (i) and (ii) above),
the dynamical gauge group will not be uniquely determined.
This is due to the fact that the algebraic constraints are nonlinear, and therefore
these constraints will in general be satisfied by different Lie algebras. Thus in general, there
will be a finite (typically small) number of possible
dynamical gauge groups, leaving the freedom to choose one of them as being the
``physical'' one. The above assumptions~(i) and~(ii) give a way to
single out a unique dynamical gauge group,
corresponding to the choice which we consider to be physically relevant.
Clearly, this procedure can be criticized
as not deriving the structure of the physical interactions purely from the causal action principle
and the form of the vacuum.
But at least, the choice of the dynamical gauge group is {\em{global}} in space-time,
i.e.\ it is to be made once and forever. Moreover, our procedure clarifies the following points:
\begin{itemize}[leftmargin=2em]
\itemD The gauge groups and couplings of the gauge fields to the fermion as used in the standard model
follow uniquely from general assumptions on the interaction, which do not involve any
specific characteristics of the groups or of the couplings.
\itemD The gauge groups of the standard model are maximal in the sense that no
additional chiral potentials are admissible. Thus we get an explanation why
there are {\em{not more}} physical gauge fields than those in the standard model.
\end{itemize}
As an example of a dynamical gauge group which we do not consider as being physically
relevant, one could choose~$\G_\free$ as the Lie group $\U(7)$ acting on
the~$7$ massive sectors. Forming~$\G$ as a maximal extension gives 
a dynamical gauge group where the corresponding left- and right-handed gauge potentials are all
abelian. This explains why an assumption like~(i) above is needed.

We remark that the specific form of assumption~(i) is a major simplification of our analysis,
because it makes it possible
to disregard the situation that there are non-abelian admissible potentials,
but that every such potential is a mixture of a left- and right-handed component.
We expect that assumption~(i) could be weakened by refining our methods,
but we leave this as a problem for future research.

The remainder of this section is devoted to the proof of Theorem~\ref{q:thmsbf}.
We first work out all the constraints for the gauge potentials
(\S\ref{q:secpt}--\S\ref{q:secftt}) and then combine our findings to infer the theorem
(\S\ref{q:secproof}).

\subsectionn{The Sectorial Projection of the Chiral Gauge Phases} \label{q:secpt}
Similar as explained in Section~\ref{l:sec3}, we shall now analyze the effect
of the gauge phases in the EL equations to degree five on the light cone.
Combining~\eqref{q:Pchiral}, \eqref{q:Lambda2} and~\eqref{q:partrace0},
the closed chain is computed by (see also~\S\ref{l:sec32})
\beq \label{q:LA}
\chi_L A_{xy} = \chi_L\: \hat{U}_L\: \hat{U}_R^* \: A^\text{vac}_{xy} + (\deg < 3) \:.
\eeq
Here~$A^\text{vac}_{xy}$ is the closed chain in the vacuum.
\nindex{ec6@$A^\text{vac}_{xy}$ -- closed chain of the vacuum}%
It is diagonal in the sector index and has the form (cf.~\S\ref{l:sec31})
\[ \chi_L A^\text{vac}_{xy} = \left\{ \begin{array}{cl}
\displaystyle
\frac{3}{4} \: \chi_L \left( 3 \slashed{\xi} T^{(-1)}_{[0]} \overline{\slashed{\xi} T^{(-1)}_{[0]}}
+ \tau_\reg \:\slashed{\xi} T^{(-1)}_{[0]} \overline{\slashed{\xi} T^{(-1)}_{[R, 0]}} \right) & \text{on the neutrino sector} \\
\frac{3}{4} \: \chi_L \,3 \slashed{\xi} T^{(-1)}_{[0]} \overline{\slashed{\xi} T^{(-1)}_{[0]}} 
 & \text{on the massive sectors}\:,
\end{array} \right. \]
up to contributions of the form~$\slashed{\xi}\, (\deg < 3) + (\deg < 2)$.
In~\S\ref{l:sec32} the size of~$\tau$ is discussed, leading to the two cases~{\bf{(i)}}
and~{\bf{(ii)}} (see~\eqref{l:casesiii}).
For brevity, we here only consider Case~{\bf{(i)}}, noting that Case~{\bf{(ii)}} can be
treated exactly as in~\S\ref{l:sec32}. Thus we assume that~$\tau$ is so small that the
factor~$T^{(-1)}_{[R, 0]}$ may be disregarded,
so that the closed chain of the vacuum simplifies to
\beq \label{q:Avacs}
\chi_L A^\text{vac}_{xy} = \frac{9}{4}\: \chi_L \,\slashed{\xi} T^{(-1)}_{[0]} \overline{\slashed{\xi} T^{(-1)}_{[0]}} \:.
\eeq
In order to satisfy the EL equations to degree five, the non-trivial eigenvalues of
the matrix~\eqref{q:LA} must all have the same absolute value.
Since the matrix~\eqref{q:Avacs} commutes with the matrices~$\hat{U}_L$ and~$\hat{U}_R^*$,
the eigenvalues of the closed chain are simply the products of the eigenvalues of~$\chi_L A^\text{vac}_{xy}$
and the eigenvalues of~$\hat{U}_L\: \hat{U}_R^*$. Since the nontrivial
eigenvalues of~$\chi_L A^\text{vac}_{xy}$ form a complex conjugate pair, the EL equations
to degree five are satisfied if and only if
\[ \text{the eigenvalues~$\hat{U}_L\: \hat{U}_R^*$ all have the same absolute value$\:$.} \]
This leads to constraints for the gauge potentials, which we now work out.

In preparation, we introduce a convenient notation. Our goal is to determine the
dynamical gauge group~$\G$. At the moment, we only know that it should be a Lie subgroup of the
group in~\eqref{q:U8}. The admissible chiral gauge potentials are vectors in the corresponding
Lie algebra~${\mathfrak{g}} = T_e \G$. More precisely, in view of~\eqref{q:Bansatz}, these
chiral potentials have the form
\[ {\mathfrak{g}} \ni \A = (A_L, A_R) \qquad \text{and} \qquad
A_c = U^\mix_c A_c^{\sec} (U^\mix_c)^* \:, \]
where~$A_c^{\sec}$ are Hermitian $8 \times 8$-matrices acting on the sectors.
Moreover, the matrix~$A_R$ does not mix the first with the other 7 sectors, i.e.
\beq \label{q:ARform}
A_R = \begin{pmatrix} (A_R)^1_1 & 0 & \cdots & 0 \\
0 & (A_R)^2_2 & \cdots & (A_R)^2_8 \\
\vdots & \vdots & \ddots & \vdots \\
0 & (A_R)^8_2 & \cdots & (A_R)^8_8 \end{pmatrix} \:.
\eeq

\begin{Lemma} \label{q:lemma31} Assume that for any~$(U_L, U_R) \in \G$, 
the eigenvalues of the matrix~$\hat{U}_L\: \hat{U}_R^*$ all have the same absolute value.
Then for any~$\A = (A_L, A_R) \in {\mathfrak{g}}$ there is a real number~$c(\A)$
such that the matrix
\beq \label{q:ALRcond}
\acute{A}_L \grave{A}_L + \acute{A}_R \grave{A}_R - \hat{A}_L^2 - \hat{A}_R^2 
- c(\A)\: \1_{\C^8}
\eeq
vanishes on all the eigenspaces of the matrix~$\hat{A}_L - \hat{A}_R$.
\end{Lemma}
\Proof For simplicity, we only consider the situation that the group element~$(U_L, U_R)$ is in a neighborhood
of the identity~$e \in \G$. Then, since~$\G$ is assumed to be a Lie group, we can represent the group element
with the exponential map,
\[ U_c = \exp(-i A_c) = \1 - i A_c - \frac{1}{2}\: A_c^2 + \O(\A^3) \:. \]
Forming the sectorial projection, we obtain
\[ \hat{U}_c = \exp(-i A_c) = \1 - i \hat{A}_c - \frac{1}{2}\: \acute{A}_c  \grave{A}_c + \O(\A^3) \:. \]
The effect of the sectorial projection becomes clearer when comparing with the unitary matrix obtained
by exponentiating the sectorial projection of~$A_c$,
\[ \exp(-i \hat{A}_c) = \1 - i \hat{A}_c - \frac{1}{2}\: \hat{A}_c^2 + \O(\A^3) \:. \]
This gives
\begin{align*}
\hat{U}_c &= \exp(-i \hat{A}_c) + \frac{1}{2} \left( \hat{A}_c^2 -  \acute{A}_c \grave{A}_c \right) + \O(\A^3) \\
&=\exp(-i \hat{A}_c) \left( \1 + \frac{1}{2} \big( \hat{A}_c^2 -  \acute{A}_c \grave{A}_c \big) \right) + \O(\A^3) \:,
\end{align*}
showing that~$\hat{U}_c$ is unitary up to a contribution to second order which is Hermitian.
As a consequence,
\beq \label{q:ULR}
\hat{U}_L\, \hat{U}_R^* = \exp(-i \hat{A}_L)
\left\{ \1 +\frac{1}{2} \big( \hat{A}_L^2 -  \acute{A}_L \grave{A}_L + 
\hat{A}_R^2 -  \acute{A}_R \grave{A}_R \big) \right\}  \exp(i \hat{A}_R) + \O(\A^3) \:.
\eeq
The curly brackets enclose a Hermitian matrix. Moreover, to the considered second order in~$\A$,
the curly brackets can be commuted to the left or right. This shows that the matrix~$\hat{U}_L\, \hat{U}_R^*$
is normal (i.e.\ it commutes with its adjoint). Therefore, the eigenvalues can be computed with
a standard perturbation calculation with degeneracies.
To first order in~$\A$, we need to diagonalize the matrix~$\hat{A}_L - \hat{A}_R$.
The exponentials in~\eqref{q:ULR} are unitary and thus
only change the eigenvalues by a phase. Therefore, the change of the absolute values
of the eigenvalues is described by a first order perturbation calculation
for the matrix in the curly brackets. This gives the result.
\QED
The condition~\eqref{q:ALRcond} arising from this lemma is difficult to analyze because the eigen\-spa\-ces of
the matrix~$\hat{A}_L - \hat{A}_R$ are unknown and depend on the potential in a complicated non-linear way.
A good strategy for satisfying the conditions for all~$\A \in {\mathfrak{g}}$
is to demand that the matrix in~\eqref{q:ALRcond} vanishes identically, i.e.
\beq \label{q:ALRcond2}
\acute{A}_L \grave{A}_L + \acute{A}_R \grave{A}_R - \hat{A}_L^2 - \hat{A}_R^2
= c(\A)\: \1_{\C^8} \:.
\eeq
Clearly, this is a stronger condition than~\eqref{q:ALRcond}.
But by perturbing the potentials in~$\g$,
one could also get information on the matrix elements of~\eqref{q:ALRcond2}
which mix different eigenspaces of~$\hat{A}_L - \hat{A}_R$, suggesting that
the assumptions of Lemma~\ref{q:lemma31} even imply that~\eqref{q:ALRcond2} holds.
Making this argument precise would make it necessary to study third order perturbations.
In order to keep our analysis reasonably simple, we shall not enter higher oder perturbation theory.
Instead, in what follows we take~\eqref{q:ALRcond2} as a necessary condition
which all admissible potentials~$\A = (A_L, A_R) \in \g$ must satisfy.

Let us reformulate~\eqref{q:ALRcond2} in a convenient notation. First, we let~$\sproj : \C^3 \rightarrow \C^3$
be the orthogonal projection onto the subspace spanned by the vector~$(1,1,1)$.
\nindex{ec8@$\sproj : \C^3 \rightarrow \C^3$ -- orthogonal projection to $\text{span}(1,1,1)$}%
We introduce the vector space
\[ T := \C^8 \times \C^3 \]
of vectors carrying a sector and a generation index.
We also consider~$\sproj$ as an operator on~$T$ which acts on the second factor (i.e.\ on the
generation index). Then the sectorial projections in~\eqref{q:ALRcond2}
can be written as
\beq \label{q:ALRcond3}
\sum_{c=L,R} \sproj A_c (\1-\sproj) A_c \,\sproj = c(\A)\: \1_T \:.
\eeq
We introduce the subspaces~$I:=\sproj(T)$ and~$J:=(\1-\sproj)(T)$; they are $8$-
respectively $16$-dimensional. Moreover, we introduce the operators
\beq \label{q:Bcdef}
B_c = (\1-\sproj) A_c \,\sproj \::\: I \rightarrow J \:.
\eeq
Combining the left- and right-handed matrices,
\beq \label{q:Bdef}
B := \begin{pmatrix} B_L \\ B_R \end{pmatrix} : I \rightarrow K := J \oplus J \:,
\eeq
we can write the condition~\eqref{q:ALRcond3} as
\beq \label{q:Bu}
\la B u | B u \ra = c(\A)\, \|u\|^2 \qquad \text{for all~$u \in I$}
\eeq
(where the scalar product and the norm refer to the canonical scalar products on~$K$ and~$I$,
respectively). In other words, the matrix~$B$ must be a multiple of an isometry.
We denote the possible values of~$B$ by~${\mathcal{B}}$,
\beq \label{q:calBdef}
{\mathcal{B}} := \left\{ \begin{pmatrix} (1-\sproj) A_L \sproj \\ (1-\sproj) A_R \sproj \end{pmatrix} : I \rightarrow K
\quad \text{with} \quad \A \in {\mathfrak{g}} \right\} .
\eeq
Then~${\mathcal{B}}$ is a real vector space of matrices. The condition~\eqref{q:Bu}
must hold on the whole vector space,
\beq \label{q:Bu2}
\la B u | B u \ra = c(B)\, \|u\|^2 \qquad \text{for all~$B \in {\mathcal{B}}$ and~$u \in I$}\:.
\eeq

The analysis of~\eqref{q:Bu2} bears some similarity to the ``uniform splitting lemma''
used in~\cite[Lemma~7.1.3]{PFP}. In fact, if~${\mathcal{B}}$ were a complex vector space, we
could polarize~\eqref{q:Bu2} to conclude that
\[ \la B u | B' u \ra = c(B,B')\, \|u\|^2 \qquad \text{for all~$B,B' \in {\mathcal{B}}$ and~$u \in I$}\:, \]
making it possible to apply~\cite[Lemma~7.1.3]{PFP}.
However, there is the subtle complication that~${\mathcal{B}}$ is only a {\em{real}} vector space,
implying that the above polarization is in general wrong.
This makes it necessary to modify the method such that we work purely with real vector spaces.
To this end, we consider~$I$ and~$K$ as real vector spaces, for clarity denoted by a subscript~$\R$.
These vector spaces have the real dimensions 16 respectively 64.
On~$I_\R$ and~$K_\R$ we introduce the scalar product
\[ \la .|. \ra_\R := \re \la .|. \ra \:. \]
We encode the complex structure in a real linear operator~$\I$ acting on~$I_\R$ and~$K_\R$ with the properties
\[ \I^* = -\I \qquad \text{and} \qquad \I^2 = -\1 \:. \]
Next, we let~$\re I$ be the subspace of~$I$ formed of all vectors with real components.
We also consider~$\re I$ as an $8$-dimensional subspace of~$I_\R$.
Moreover, we let~$\re : I_\R \rightarrow \re I$ be the orthogonal projection to the real part.
By restricting to~$\re I$, every operator~$B \in {\mathcal{B}}$ gives rise to a mapping
\[ B_\R := B|_{\re I} \::\: \re I \rightarrow K_\R\:. \]
Note that the operator~$B_\R$ is represented by a $64 \times 8$-matrix.
Knowing~$B_\R$, we can uniquely reconstruct the corresponding~$B$ by ``complexifying'' according to
\[ B u = B \re u - \I B \re(\I u)\:. \]

\begin{Lemma} \label{q:lemmaMrep} There is an isometry~$V : K_\R \rightarrow K_\R$ and  
a basis~$B_1, \ldots B_L$ of~${\mathcal{B}}$ (with $L \geq 0$) such that
the matrices~$(B_\ell)_\R$ have the representation
\[ (B_\ell)_\R = V \, M_\ell \]
with operators~$M_\ell : I_\R \rightarrow K_\R$ which in the canonical bases
have the block matrix representation
\[ M_1 = \begin{pmatrix} \1 \\ 0 \\ \vdots \\ 0 \\ 0 \end{pmatrix} \:, \quad
M_2 = \begin{pmatrix} 0 \\ \1 \\ \vdots \\ 0 \\ 0 \end{pmatrix} \:, \;\ldots\: , \quad
M_L = \begin{pmatrix} 0 \\ 0 \\ \vdots \\ \1 \\ 0 \end{pmatrix} \:. \]
Here the upper $L$ matrix entries are~$8 \times 8$-matrices, whereas the
lowest matrix entry is a $(64-8L) \times 8$-matrix.
\end{Lemma}
\Proof We rewrite~\eqref{q:Bu2} in real vector spaces as
\[ \la B_\R u | B_\R u \ra_\R = c(B)\, \|u\|_R^2 \qquad \text{for all~$B \in {\mathcal{B}}$ and~$u \in \re I$} \:. \]
Using the symmetry of the real scalar product, polarization gives
\beq \label{q:polar}
\la B_\R u | B'_\R u \ra_\R = c(B,B')\, \la u | v \ra_\R^2 \qquad \text{for all~$B,B' \in {\mathcal{B}}$
and~$u \in \re I$} \:.
\eeq
Now we can proceed as in the proof of~\cite[Lemma~7.1.3]{PFP}: 
Let $(e_1,\ldots,e_8)$ be the canonical basis of~$\re I$. 
We introduce the subspaces
\[ E_i = \text{span} \{ B_\R e_i {\mbox{ with }} B \in {\mathcal{B}} \}  \subset
K_\R \]
as well as the mappings
\[ \kappa_i \::\: {\mathcal{B}} \rightarrow E_i \:,\quad  B \mapsto B_\R e_i \:. \]
The property~\eqref{q:polar} implies that for all $B, B' \in {\mathcal{B}}$,
\beq \label{q:e:4y}
\langle B_\R e_i | B'_\R e_j \rangle_\R = c(B,B')\: \delta_{ij}\:.
\eeq
If $i \neq j$, this relation shows that the subspaces $(E_i)_{i=1,\ldots,p_1}$
are orthogonal. Moreover, in the case $i=j$, the relation~\eqref{q:e:4y} yields that the scalar products
$\langle \kappa_i(B') | \kappa_i(B') \rangle_\R$ are independent of $i$. Thus the
mappings $\kappa_i$ are isometrically equivalent, and so we can arrange by an isometry~$V$
that the~$\kappa_i$ have the matrix representations
\[ \kappa_1 = \begin{pmatrix} \kappa  \\ \vdots \\ 0 \\ 0 \end{pmatrix} \:, \;\ldots\: , \quad
\kappa_L = \begin{pmatrix} 0 \\ \vdots \\ \kappa \\ 0 \end{pmatrix} \:, \]
where~$\kappa : {\mathcal{B}} \rightarrow \R^8$.

Finally, we choose a basis~$B_1, \ldots, B_L$ of~${\mathcal{B}}$ such that~$\kappa(B_1) = (1,\ldots, 0)$,
$\kappa(B_2)=(0,1,\ldots, 0)$, etc. This gives the result.
\QED
Counting dimensions, this lemma shows in particular that the dimension
of~${\mathcal{B}}$ is at most~$8$.
In our applications we need the following refined counting of dimensions.
\begin{Corollary} \label{q:corcount}
Assume that the images of the matrices~$B_1, \ldots, B_L : I \rightarrow K$ span
an~$N$-dimensional subspace of~$K$. Then the dimension of~${\mathcal{B}}$ is bounded from above by
\beq \label{q:countin}
L \leq \frac{N}{4} \:.
\eeq
\end{Corollary}
\Proof Note that the real dimension of the image of~$(B_\ell)|_\R : I_\R \rightarrow K_\R$
is twice the complex dimension of~$B_\ell : I \rightarrow K$.
\QED

\subsectionn{The Bilinear Logarithmic Terms} \label{q:secblt}
In~\S\ref{l:seclogAA} the left-handed component of the bilinear logarithmic terms quadratic in the mass matrices 
were computed by (see~\eqref{l:BLform})
\beq \label{q:BLform}
\begin{split}
B_L :=\:& -\frac{m^2}{4} \Big\{ A_R^\even[\xi], \left( A_L[\xi]\, YY - 2 Y A_R[\xi]\, Y + YY A_L[\xi]
\right) \Big\} \:T^{(1)} \\
&+\frac{m^2}{8} \Big( A_L[\xi]^2  YY + 2 A_L[\xi] YY A_L[\xi]
+ YY A_L[\xi]^2 \xi_k) \Big)\, T^{(1)} \\
&-\frac{m^2}{2}\: Y A_R[\xi]^2\: Y \:T^{(1)} \:.
\end{split}
\eeq
\sindex{fermionic projector!bilinear logarithmic term}%
The right-handed component is obtained similarly by the replacements~$L \leftrightarrow R$.
Exactly as shown in Lemma~\ref{l:lemmamass2}, the EL equations in the continuum limit
are satisfied only if the matrices~$\hat{B}_L$ and~$\hat{B}_R$
coincide and are multiples of the matrix~$\acute{Y} \grave{Y}$.

Let us specify the potentials~$A^\even_c$ in~\eqref{q:Deflip}.
Exactly as shown in~\S\ref{l:secshear}, the shear contributions vanish only
if, in a suitable basis, the matrix~$A^\even_L$ coincides with~$A_R$ and~$A^\even_R$
coincides with~$A_L$, up to the choice of the mixing matrices. More precisely,
in order to introduce~$A^\even_L$, we
let~${\mathfrak{e}}_{i \alpha}(k,x)$ with~$i \in \{1, \ldots, 8\}$ and~$\alpha \in \{1,2,3\}$
be an orthonormal basis of~$\C^{8 \times 3}$ such that the vectors~${\mathfrak{e}}_{i 1}$
are multiples of the eight columns of the matrix~$L[k,x]^*$
(note that these column vectors are orthogonal according to~\eqref{q:4con}).
In this basis, the potential~$A^\even_L$ is defined by
\beq \label{q:Aeven}
A^\even_L = V_R A^{\sec}_R V_R^* \:,
\eeq
where~$A_R^{\sec}$ is the potential in~\eqref{q:Bansatz} (in the standard basis), and~$V_R$
are unitary matrices which are diagonal in the sector index,
\[ (V_R)^{(i,\alpha)}_{(j, \beta)}(x) = \delta^i_j\: (V^i_R)^\alpha_\beta(x) \qquad \text{with} \qquad
V^i_c(x) \in \U(3)\:. \]
This is analogous to~\eqref{q:Bansatz} and~\eqref{q:Umixform}, with the only difference
that different mixing matrices~$V^i_c$ appear, which may even depend on the space-time point~$x$.
In order to introduce~$A^\even_R$, one chooses similarly a basis~${\mathfrak{e}}_{i \alpha}(k,x)$
such that the vectors~${\mathfrak{e}}_{i 1}$ are multiples of the eight columns of the matrix~$R[k,x]^*$,
and in this basis one sets
\beq \label{q:Aeven2}
A^\even_R = V_L A^{\sec}_L V_L^*
\eeq
with a sector-diagonal unitary matrix~$V_L(x)$.
We point out that the construction of the potentials~$A^\even_{L\!/\!R}$ depends on the momentum~$k$
of the microlocal chiral transformation. As a consequence, these potentials are non-local
operators (for details see the discussion in~\S\ref{l:secshear}).

When using~\eqref{q:Aeven} and~\eqref{q:Aeven2}
in~\eqref{q:BLform}, the freedom in choosing the matrices~$V^i_c$
gives many free parameters to modify~$B_L$ and~$B_R$, making the situation rather complicated.
In order to derive necessary conditions, it suffices to consider particular choices for the potentials
for which the matrices~$V^i_c$ do not come into play.
One possibility is to assume that~$\g$ contains a right-handed potential~$\A = (0, A_R) \in \g$.
Then~$A_R^\even$ and~$A_L$ vanish, so that
\beq \label{q:BLcomp}
B_L =-\frac{m^2}{2}\: \acute{Y} A_R[\xi]^2\: \grave{Y} \:T^{(1)} \:.
\eeq
This must be a multiple of the matrix~$\acute{Y} \grave{Y}$. Proceeding similarly for left-handed potentials
gives the following result.

\begin{Lemma} \label{q:lemmaid}
Suppose that~$\A=(A_L, 0) \in \g$ (or~$\A=(0, A_R) \in \g$) is a left-handed
(respectively right-handed) admissible gauge potential. Then the
matrix~$A_L[\xi]^2$ (respectively~$A_R^2[\xi]$) is a
multiple of the identity matrix at every space-time point and for all directions~$\xi$.
\end{Lemma}

The next lemma gives additional information on left-handed or right-handed admissible gauge potentials.
For notational simplicity, we only state the result for the left-handed potentials.
\begin{Lemma} Suppose that~$\A=(A_L, 0) \in \g$ does not depend on the generation index, i.e.
\beq \label{q:genind}
(A_L)^{(i, \alpha)}_{(j, \beta)} = \delta^\alpha_\beta \,(A^{\sec})^i_j \:.
\eeq
Then
\[ \acute{Y} \,A_L[\xi]^2\, \grave{Y} = \acute{A}_L[\xi] \,Y^2\, \grave{A}_L[\xi]\:. \]
\end{Lemma}
\Proof According to~\eqref{q:genind}, we may compute~$B_L$ according to~\eqref{q:BLform}
with~$A_R^\even = A_L$. Then
\[ B_L =-\frac{m^2}{8} \left( A_L[\xi]^2\, Y^2 + 2 A_L[\xi] \,Y^2\, A_L[\xi] +  Y^2\, A_L[\xi]^2 \right) T^{(1)}. \]
This matrix must coincide with~$B_R$, which is computed similar to~\eqref{q:BLcomp} by
\[ B_R =-\frac{m^2}{2}\: Y A_L[\xi]^2\: Y \:T^{(1)} \:. \]
Applying Lemma~\ref{q:lemmaid}, the matrix~$A_L[\xi]^2$ is a multiple of the identity and thus
commutes with~$Y$. This gives the result.
\QED

\subsectionn{The Field Tensor Terms} \label{q:secftt}
The methods in~\S\ref{l:secfield} also apply to the present situation of
eight sectors. In particular, Proposition~\ref{l:prp58} can be restated as follows:
\sindex{fermionic projector!field tensor term}%

\begin{Prp} \label{q:prpfield}
Taking into account the contributions by the field tensor terms, the EL equations to degree
four can be satisfied only if the regularization satisfies the
conditions~\eqref{q:Tpoint} and~\eqref{q:RC1}.
If no further regularization conditions are imposed, then the chiral potentials must satisfy
at all space-time points the conditions
\beq \label{q:FTcond}
\Tr( {\mathfrak{I}}_1 A_R ) = 0 \qquad \text{and} \qquad \Tr( A_L + A_R ) = 0 \:,
\eeq
where~${\mathfrak{I}}_1$ is the projection on the neutrino sector.
\nindex{ed0@${\mathfrak{I}}_1$ -- projection on neutrino sector}%
If conversely the conditions~\eqref{q:Tpoint}, \eqref{q:RC1} and~\eqref{q:FTcond} are satisfied, then
the field tensor terms do not contribute to the EL equations of degree four.
\end{Prp}

\subsectionn{Proof of Spontaneous Block Formation} \label{q:secproof}
Instead of working with gauge groups, it will usually be more convenient to
consider the corresponding Lie algebras. This is no restriction because the
corresponding Lie groups can then be recovered by exponentiation.
When forming the Lie algebra of a product of groups, 
this gives rise to the direct sum of the algebras, like for example
\[ T_e \big( \U(8)_L \times \U(1)_R \times \U(7)_R \big) = 
\u(8)_L \oplus \u(1)_R \oplus \u(7)_R \:. \]
The proof of spontaneous block formation
will be given in several steps, which are organized in separate paragraphs.
\nsubsubsection{Left-handed $\su(2)$-potentials}
We now evaluate our assumption~(i) that~$\g$ should contain left- or right-handed
non-abelian potentials.

We first note that~$\g$ cannot contain right-handed potentials:
\begin{Lemma} \label{q:lemmanoright}
The dynamical gauge algebra~$\g$ does not contain potentials of the form
$(0, A_R)$ with~$A_R \neq 0$.
\end{Lemma}
\Proof Assume conversely that~$\A=(0,A_R) \in \g$ is a non-trivial
admissible right-handed potential. It follows from Lemma~\ref{q:lemmaid}
that~$A^2$ is a multiple of the identity. On the other hand, combining~\eqref{q:ARform}
with the fact that the right-handed potential vanishes on the neutrino sector
(see the first equation in~\eqref{q:FTcond}), we find that~$A_R$ must be of the form
\beq \label{q:ARform2}
A_R = \begin{pmatrix} 0 & 0 & \cdots & 0 \\
0 & (A_R)^2_2 & \cdots & (A_R)^2_8 \\
\vdots & \vdots & \ddots & \vdots \\
0 & (A_R)^8_2 & \cdots & (A_R)^8_8 \end{pmatrix} \:.
\eeq
As a consequence, $A_R^2$ cannot be a multiple of the identity, a contradiction.
\QED

Thus it remains to consider the case that~$\g$ contains non-abelian left-handed potentials.
The left-handed potentials form a Lie subalgebra of~$\g$,
\beq \label{q:gLdef}
\g_L := \big\{ \A = (A_L, 0) \in \g \big\} \subset \g \:.
\eeq
\nindex{ed2@$\g_L \subset \g$ -- left-handed dynamical gauge potentials}%
Again applying Lemma~\ref{q:lemmaid}, we know that every~$\A=(A_L, 0) \in \g$
has the property that~$A^2$ is a multiple of the identity.
The following general lemma gives an upper bound for the dimension of~$\g_L$.
\begin{Lemma} \label{q:lemmasu2}
Let~$\h \subset \su(N)$ be a Lie algebra with the additional property that
\beq \label{q:multiid}
A^2 \sim \1_{\C^N} \qquad \text{for all~$A \in \h$}\:.
\eeq
Then~$\h$ is isomorphic to a subalgebra of~$\su(2)$.
\end{Lemma}
\Proof Polarizing~\eqref{q:multiid}, we find that for all~$A, A' \in \h$,
\[ \left\{ A, A' \right\} = k(A,A') \, \1_{\C^N} \]
with a bilinear form~$k : \h \times \h \rightarrow \R$.
Since the square of a non-zero Hermitian matrix is positive semi-definite and non-zero,
we conclude that~$k$ is positive definite and thus defines a scalar product on~$\h$.
Hence~$\h$ generates a Clifford algebra~$\Cl(\h, \R)$.
Since~$\h$ is also a Lie algebra, the commutator of two elements in~$\h$ is again
an element of~$\h$. This means for the Clifford algebra that the bilinear covariants
$[u, v]$ with~$u,v \in \h$ are all multiples of the generators of the Clifford algebra.
This in turn implies that the dimension of the Clifford algebra is at most three
(for details see the classification of Clifford algebras
in~\cite{lawson+michelsohn}). Moreover, $\h$ is a Lie algebra isomorphic to
a subalgebra of~$\su(2)$.
\QED
Since every proper Lie subalgebra of~$\su(2)$ is abelian we immediately obtain the following result.
\begin{Corollary} The left-handed dynamical gauge group~$\g_L$, \eqref{q:gLdef},
is Lie algebra isomorphic to~$\su(2)$.
\end{Corollary}

We now write~$\g_L$ more explicitly as matrices.
\begin{Lemma} \label{q:lemmaALa}
There is a unitary matrix~$V \in \U(8)$ (acting on the generations) and
a basis~$(A_{L, \alpha})_{\alpha=1,2,3}$ of~$\g_L$ such that
\beq \label{q:Arep}
A_{L, \alpha} = U_L^\mix \:V \begin{pmatrix} \sigma_\alpha & 0 & 0 & 0 \\
0 & \sigma_\alpha & 0 & 0 \\ 0 & 0 & \sigma_\alpha & 0 \\ 0 & 0 & 0 & \sigma_\alpha
\end{pmatrix} V^* \:(U_L^\mix)^* \:,
\eeq
where~$\sigma^\alpha$ are the Pauli matrices, and~$U_L^\mix$ is the matrix in~\eqref{q:Umixform}.
\end{Lemma}
\Proof
Using~\eqref{q:Bansatz}, we
can pull out the mixing matrices and work with $8 \times 8$-matrices. Since~$\g_L$
is Lie algebra isomorphic to~$\su(2)$, it can be regarded as a representation of~$\su(2)$
on~$\C^8$. We decompose this representation into irreducible components.
The fact that the matrix $(A_{L, \alpha})^2$ is a multiple of the identity implies that
every irreducible component is the fundamental representation
(because all the other irreducible representations are not generators of a Clifford algebra).
This gives the result.
\QED

\nsubsubsection{Arranging the free gauge group of maximal dimension}
We denote the {\em{commutant}} of~$\g_L$ by~$\g_L'$,
\[ g_L' = \{ \A' \in \u(8)_L \oplus \u(1)_R \oplus \u(7)_R \text{ with }
[\A, \A'] = 0 \quad\forall \A \in \g_L \}\:. \]
\nindex{ed4@$\g_L'$ -- commutant of~$\g_L$}%

\begin{Lemma} The dynamical gauge algebra is contained in the direct sum
\[ \g \subset \g_L \oplus \g_L' \:. \]
\end{Lemma}
\Proof
For any~$\A = (A_L, 0) \in \g_L$ and~$\tilde{\A} = (\tilde{A}_L, \tilde{A}_R) \in \g$, the commutator is left-handed
\[ [\A, \tilde{\A}] = \big( [A_L, \tilde{A}_L], 0 \big) \:. \]
Therefore, this commutator must be an element of~$\g_L$. In this way, every~$\tilde{\A} \in \g$
gives rise to a linear endomorphism of~$\g_L$. In view of the structure
equations for~$\su(2)$ (which in the usual notation with Pauli matrices can be written
as~$[\sigma_\alpha, \sigma_\beta] = i \varepsilon_{\alpha \beta \gamma} \,\sigma_\gamma$ with~$\varepsilon$
the totally anti-symmetric Levi-Civita symbol), this endomorphism of~$\g_L$
can be realized uniquely as an inner endomorphism, i.e.\ there is a unique~$\hat{\A} =
(\hat{A}_L, 0) \in \g_L$ with
\[ [\A, \tilde{\A}] = [\A, \hat{\A}] \qquad \text{for all~$\A \in \g_L$}\:. \]
As a consequence, the potential~$\A^c := \tilde{\A}-\hat{\A}$ lies in the commutant~$\g_L'$.
We thus obtain a unique decomposition
\[ \tilde{\A} = \hat{\A} + \A^c \qquad \text{with} \qquad \hat{\A} \in \g_L \text{ and } \A^c \in \g_L' \:. \]
This concludes the proof.
\QED

In particular, this lemma gives information on the left-handed component of~$\g$, denoted by
\[ \pr_L \g = \left\{ A_L \text{ with } (A_L, A_R) \in \g \right\} \subset \u(8) \:. \]
\nindex{ed6@$\pr_L$ -- projection on left-handed component}%
Note that the projection~$\pr_L \g$ is a Lie algebra which contains~$\g_L$.
\begin{Lemma} \label{q:lemmadirsum}
The left-handed component of the dynamical gauge algebra satisfies the inclusion
\beq \label{q:dirsum}
\pr_L \g \subset \g_L \oplus \u(4) \:.
\eeq
Here the elements of~$\u(4)$ come with the matrix representation
\beq \label{q:Amat}
U_L^\mix \:V \:\big( A \otimes \1_{\C^2} \big)\: V^* \:(U_L^\mix)^* \:,\qquad A \in \u(4)\:,
\eeq
where the factor~$\1_{\C^2}$ acts on the block matrix entries in~\eqref{q:Arep}.
\end{Lemma}
\Proof The commutant of the matrices in~\eqref{q:Arep} is computed to be all matrices
whose block matrix entries are the identity. Taking a unitary transformation gives the result.
\QED
Let us consider what this lemma tells us about the possible form of
the free gauge algebra. Since~$\g$ does not contain right-handed gauge potentials
(see Lemma~\ref{q:lemmanoright}), the left-handed component of the free gauge potentials
is disjoint from~$\g_L$. Hence, using~\eqref{q:dirsum},
\beq \label{q:prLgfree}
\pr_L \g_\free \subset \u(4) \:,
\eeq
where the potentials in~$\u(4)$ are again of the form~\eqref{q:Amat}, plus possibly
vectors of~$\g_L$. The right-handed component of~$\g_\free$, on the other hand, must be of
the form~\eqref{q:ARform2}. Combining these findings gives the following result.

\begin{Lemma} \label{q:lemmamin}
Choosing the mixing matrices such that the free gauge group has the maximal
dimension, we obtain
\[ \g_\free = \u(1) \times \su(3) \:. \]
In a suitable global gauge, the gauge potentials in~$\g_L$ (see~\eqref{q:gLdef}
and Lemma~\ref{q:lemmaALa}) have a basis~$(A_{L, \alpha})_{\alpha=1,2,3}$ with
\beq \label{q:Arep3}
A_{L, \alpha} = U_L^\mix \:\begin{pmatrix} \sigma_\alpha & 0 & 0 & 0 \\
0 & \sigma_\alpha & 0 & 0 \\ 0 & 0 & \sigma_\alpha & 0 \\ 0 & 0 & 0 & \sigma_\alpha
\end{pmatrix} (U_L^\mix)^* \:,
\eeq
where the mixing matrix~$U_L$ is a diagonal matrix on the sectors of the form
\beq \label{q:ULform}
U_L^\mix = \text{\rm{diag}}(U_1, U_2, \1, \UCKM, \1, \UCKM, \1, \UCKM)
\eeq
with unitary matrices~$U_1, U_2, \UCKM \in \U(3)$.
The free $\u(1)$- and $\su(3)$-potentials, on the other hand, have the respective matrix representations
\begin{align}
A_L = A_R &= B \:\text{\rm{diag}} \Big(0,-1, \frac{2}{3}, -\frac{1}{3}, \frac{2}{3}, -\frac{1}{3}, 
\frac{2}{3}, -\frac{1}{3} \Big) && \text{with~$B \in \u(1) = \R$} \label{q:u1} \\
A_L = A_R &= \begin{pmatrix} 0 & 0 \\ 0 & C \end{pmatrix} \otimes \1_{\C^2} 
&& \text{with~$C \in \su(3)$} \:. \label{q:su3}
\end{align}
\end{Lemma}
\Proof The free gauge potentials are those vectorial potentials which 
are compatible with both~\eqref{q:prLgfree} and~\eqref{q:ARform2}.
The zero matrix entries in~\eqref{q:ARform2} imply that at least one of the rows and columns of~$\u(4)$
must be zero. But it is possible to realize the subgroup~$\u(3)$ by considering the
potentials of the form~\eqref{q:su3} (but with~$C \in \u(3)$). In order to get consistency,
we must make sure that all mixing matrices drop out of~\eqref{q:Amat}. This forces us to
choose the mixing matrix of the form
\[ U_L^\mix = \text{\rm{diag}}(U_1, U_2, U_3, U_4, U_3, U_4, U_3, U_4) \]
with four unitary matrices~$U_j \in \U(3)$. Since a joint unitary transformations of all
sectors has no influence on the potentials in~\eqref{q:Bansatz}, it is no loss of generality
to choose~$U_3=\1$. This gives~\eqref{q:ULform}.
In order to satisfy the second relation in~\eqref{q:FTcond}, the matrix~$C$ must be
trace-free. This gives~\eqref{q:su3}.

In order to find all the remaining vectorial potentials which are compatible with both~\eqref{q:prLgfree}
and~\eqref{q:ARform2}, one must keep in mind that the left-handed component may be formed
of linear combinations of~\eqref{q:Amat} and the potentials in~$\g_L$ of the form~\eqref{q:Arep3}.
In order for the first row and column to vanish, the only possibility is to form the linear combinations
of~$A_{L,3}$ and~$\1 \in \u(4)$
\[ B\, \text{\rm{diag}}(0,-1,0,-1,0,-1,0,-1)\:. \]
In view of the second relation in~\eqref{q:FTcond}, we must remove the trace by
adding a multiple of the potential~\eqref{q:su3} with~$C = \1_{\C^3}$. This gives~\eqref{q:u1}
and concludes the proof.
\QED

\nsubsubsection{Proof that~$\g$ is maximal}
So far, we constructed the dynamical gauge algebra
\[ \g_L \oplus \g_\free \simeq \su(2) \oplus \u(1) \oplus \su(3)\:. \]
Let us show that this dynamical gauge algebra is maximal.
To this end, assume that there is a chiral potential~$\A^\text{new} = (A_L^\text{new}, A_R^\text{new}) \in \u(8) \oplus \u(7)$ which is an element of~$\g$ but not of~$\g_L \oplus \g_\free$,
\[ \A^\text{new} \in \g \setminus \big( \g_L \oplus \g_\free \big) \:. \]
We want to deduce a contradiction.
Since right-handed potentials were excluded in Lemma~\ref{q:lemmanoright}, we can assume
that~$A_L^\text{new} \neq 0$. According to Lemma~\ref{q:lemmadirsum}, it suffices to consider
the potentials in~$\g_L \oplus \u(4)$ (where the $\u(4)$-potentials are represented similar to~\eqref{q:su3}
as~$C \times \1_{\C^2}$ with~$C \in \u(4)$). Moreover, by adding suitable potentials in~$\g_L \oplus \g_\free$
and using the freedom to conjugate with exponentials of potentials in~$\g_\free$, we can arrange
that~$A_L^\text{new}$ is of the form
\[ A_L^\text{new} = \alpha
\begin{pmatrix} \1_{\C^2} & 0 & 0 & 0 \\ 0 & 0 & 0 & 0 \\ 0 & 0 & 0 & 0 \\ 0 & 0 & 0 & 0
\end{pmatrix} + \beta\:
U_L^\mix \: \begin{pmatrix} 0 & \1_{\C^2} & 0 & 0 \\ \1_{\C^2} & 0 & 0 & 0 \\ 0 & 0 & 0 & 0 \\ 0 & 0 & 0 & 0
\end{pmatrix} \:(U_L^\mix)^* \]
with real parameters~$\alpha$ and~$\beta$.
By subtracting multiples of the potentials~$A_{L,3}$ and the $\u(1)$-potential, we can even arrange
that~$A_L^\text{new}$ is of the form
\beq \label{q:Anew}
A_L^\text{new} = \alpha
\begin{pmatrix} 0 & 0 & 0 & 0 \\ 0 & \1_{\C^2} & 0 & 0 \\ 0 & 0 & \1_{\C^2} & 0 \\ 0 & 0 & 0 & \1_{\C^2}
\end{pmatrix} + \beta\:
U_L^\mix \: \begin{pmatrix} 0 & \1_{\C^2} & 0 & 0 \\ \1_{\C^2} & 0 & 0 & 0 \\ 0 & 0 & 0 & 0 \\ 0 & 0 & 0 & 0
\end{pmatrix} \:(U_L^\mix)^* \:.
\eeq
The corresponding right-handed component~$A_R^\text{new}$ is unknown,
except that it must be of the form~\eqref{q:ARform2}. In particular, $A_R^\text{new}$ might involve
a non-trivial mixing matrix.

Our strategy is to evaluate off-diagonal matrix elements of~$\hat{B}_R$ 
(see~\eqref{q:BLform} with~$L$ and~$R$ exchanged)
for specific choices of the potential. The vanishing of these matrix elements
gives us constraints for~$\A^\text{new}$ which in turn imply that the
parameters~$\alpha$ and~$\beta$ in~\eqref{q:Anew} must vanish.
We begin with the parameter~$\beta$.
\begin{Lemma} \label{q:lemmaUtriv}
If~$\g$ contains a potential~$\A=(A_L, A_R)$ with~$A_L$ of the form~\eqref{q:Anew}
with~$\beta \neq 0$, then necessarily
\beq \label{q:Utriv}
U_1 = \1_{\C^3} \qquad \text{and} \qquad U_2 \UCKM^* = \1_{\C^3}\:.
\eeq
\end{Lemma}
\Proof By conjugating with $\SU(3)$-transformations in the free gauge group,
we can transform the potential~$A_L^\text{new}$ from~\eqref{q:Anew}
to the matrix
\beq \label{q:ALab}
A_L = \alpha
\begin{pmatrix} 0 & 0 & 0 & 0 \\ 0 & \1_{\C^2} & 0 & 0 \\ 0 & 0 & \1_{\C^2} & 0 \\ 0 & 0 & 0 & \1_{\C^2}
\end{pmatrix} + \beta\:
U_L^\mix M (U_L^\mix)^* \:,
\eeq
where~$M$ can be any of the six matrices
\begin{gather*}
\begin{pmatrix} 0 & \1 & 0 & 0 \\ \1 & 0 & 0 & 0 \\ 0 & 0 & 0 & 0 \\ 0 & 0 & 0 & 0 \end{pmatrix} ,\quad\quad\;\;
\begin{pmatrix} 0 & 0 & \1 & 0 \\ 0 & 0 & 0 & 0 \\ \1 & 0 & 0 & 0 \\ 0 & 0 & 0 & 0 \end{pmatrix} ,\quad\quad\;\;
\begin{pmatrix} 0 & 0 & 0 & \1 \\ 0 & 0 & 0 & 0 \\ 0 & 0 & 0 & 0 \\ \1 & 0 & 0 & 0 \end{pmatrix} , \\
\begin{pmatrix} 0 & -i \1 & 0 & 0 \\ i \1 & 0 & 0 & 0 \\ 0 & 0 & 0 & 0 \\ 0 & 0 & 0 & 0 \end{pmatrix} ,\quad
\begin{pmatrix} 0 & 0 & -i \1 & 0 \\ 0 & 0 & 0 & 0 \\ i \1 & 0 & 0 & 0 \\ 0 & 0 & 0 & 0 \end{pmatrix} ,\quad
\begin{pmatrix} 0 & 0 & 0 & -i \1 \\ 0 & 0 & 0 & 0 \\ 0 & 0 & 0 & 0 \\ i \1 & 0 & 0 & 0 \end{pmatrix} .
\end{gather*}
If~\eqref{q:Utriv} is violated, then the corresponding matrices~$(1-\sproj) A_L \sproj$ are
obviously non-trivial and linearly independent. As a consequence, the vector space~${\mathcal{B}}$
(see~\eqref{q:calBdef}) has dimension at least six.
If the mixing matrix~$\UCKM$ is non-trivial, the operators~$B$ corresponding
to the left-handed potentials~$A_{L,1}$ and~$A_{L,2}$ in~\eqref{q:Arep3} are also non-zero,
increasing the dimension of~${\mathcal{B}}$ to at least eight.
We now show that these dimensions contradict the upper bound of Corollary~\ref{q:corcount}.
We treat the cases separately when~$\UCKM$ is trivial and non-trivial.

In the case that the matrix~$\UCKM$ is non-trivial, the form of the right-handed component of the
potentials~\eqref{q:ARform2} implies that the first row cannot contribute to the operators~$B$.
As a consequence, the dimension~$N$ in Corollary~\ref{q:corcount} is at most~$2 \times (8+7) = 30$.
Hence~\eqref{q:countin} yields that the dimension of~${\mathcal{B}}$ is at most~$7$,
a contradiction.

In the case that the matrix~$\UCKM$ is trivial, the mixing of the generations comes about only
as a consequence of the matrices~$U_1$ and~$U_2$ in~\eqref{q:ULform}. Hence
the matrices~$B_L$ (see~\eqref{q:Bcdef}) are of the general form
\[ (1 - \sproj) \begin{pmatrix} 0 & \star\, U_1 U_2^* & \star\, U_1 & \star\, U_1 & \star\, U_1 & \star\, U_1 & \star\, U_1 & \star\, U_1 \\
\star\, U_2 U_1^* & 0 & \star\, U_2 & \star\, U_2 & \star\, U_2 & \star\, U_2 & \star\, U_2 & \star\, U_2 \\
\star\, U_1^* & \star\, U_2^* & 0 & 0 & 0 & 0 & 0 & 0 \\
\star\, U_1^* & \star\, U_2^* & 0 & 0 & 0 & 0 & 0 & 0 \\
\star\, U_1^* & \star\, U_2^* & 0 & 0 & 0 & 0 & 0 & 0 \\
\star\, U_1^* & \star\, U_2^* & 0 & 0 & 0 & 0 & 0 & 0 \\
\star\, U_1^* & \star\, U_2^* & 0 & 0 & 0 & 0 & 0 & 0 \\
\star\, U_1^* & \star\, U_2^* & 0 & 0 & 0 & 0 & 0 & 0 \end{pmatrix} \sproj \:, \]
where the stars~$\star$ denote complex factors. Evaluating more specifically
the six possible choices for the matrix~$M$ in~\eqref{q:ALab},
one immediately verifies that the span of the images of the corresponding matrices~$B_L$
has dimension~$8$ (note that the terms involving~$\alpha$ drop out, and that the 
first two rows have a six-dimensional image, whereas the other six rows have a two-dimensional image). Hence the dimension~$N$ in
Corollary~\ref{q:corcount} is at most~$8 + 2 \times 7 =22$.
The inequality~\eqref{q:countin} implies that the dimension of~${\mathcal{B}}$ is at most~$5$.
This is again a contradiction.
\QED

\begin{Lemma} Assume that the dynamical gauge algebra~$\g$ contains
a potential~$\A^\text{\rm{new}} = (A_L^\text{\rm{new}}, A_R^\text{\rm{new}}) \in \u(8) \oplus \u(7)$
with~$A_L^\text{\rm{new}}$ of the form~\eqref{q:Anew}. Then~$\beta$ vanishes.
\end{Lemma}
\Proof For a parameter~$\varepsilon>0$, we consider the family of potentials
\[ \A = \A^\text{old} + \varepsilon \, \A^\text{new} \qquad \text{with} \qquad \A^\text{old}
= (A^\text{old}_L=A_{L,3}, 0) \]
and~$A_{L,3}$ as in~\eqref{q:Arep3}.
We compute the terms linear in~$\varepsilon$. Moreover, we consider the
matrix entry of~$B_R$ in the third row and first column
(where we again consider~$B_R$ as an $8 \times 8$-matrix on the sectors).
Using that~$A_L^\text{old}$ is sector-diagonal and the matrix component~$(A_R)^1_3$ vanishes,
the matrix~$A_R$ drops out. Similarly, the matrix~$A_L^\text{even}$ drops out (see~\eqref{q:Aeven}),
which also has the desirable effect that the corresponding mixing matrix~$V_R$ does not
enter. We thus obtain
\begin{align*}
(B_R)^3_1 &= -\frac{m^2}{2}\:\varepsilon \left( Y 
\big( A^\text{old}_L A^\text{new}_L + A^\text{new}_L A^\text{old}_L \big)\: Y \right)^3_1 \:T^{(1)} 
+ \O(\varepsilon^2) \\
&= -\frac{m^2}{2}\:\varepsilon\: Y^3_3 
\Big( (A^\text{old}_L)^3_3 \,(A^\text{new}_L)^3_1 + (A^\text{new}_L)^3_1\, (A^\text{old}_L)^1_1 \Big)\:
Y^1_1 \:T^{(1)} + \O(\varepsilon^2) \\
&= m^2\,\varepsilon\: Y^3_3 (A^\text{new}_L)^3_1 \:Y^1_1 \:T^{(1)} + \O(\varepsilon^2) \\
&= m^2 \beta\, \varepsilon\: Y^3_3 \:U_1^* Y^1_1 \:T^{(1)} + \O(\varepsilon^2) \:,
\end{align*}
where in the last step we used~\eqref{q:Anew} together with the form of the mixing matrix~\eqref{q:ULform}. Taking the sectorial projection and applying Lemma~\ref{q:lemmaUtriv} gives the result.
\QED

The argument which shows that~$\alpha$ vanishes is somewhat different:
\begin{Lemma} Assume that the dynamical gauge algebra~$\g$ contains
a potential~$\A^\text{\rm{new}} = (A_L^\text{\rm{new}}, A_R^\text{\rm{new}}) \in \u(8) \oplus \u(7)$
with~$A_L^\text{\rm{new}}$ of the form~\eqref{q:Anew} and~$\beta=0$. Then~$\alpha$ vanishes.
\end{Lemma}
\Proof Assume conversely that there is an admissible potential~$\A = (A_L, A_R) \in \g$ with
\[ A_L = \begin{pmatrix} 0 & 0 \\ 0 & \1_{\C^3} 
\end{pmatrix} \otimes \1_{\C^2} \:. \]
This potential involves no mixing matrices. Thus we may compute~$B_L$ according to~\eqref{q:BLform}
with~$A_R^\even=A_L$. Since the left-handed component is sector-diagonal, we obtain
\[ B_L = -\frac{m^2}{2}\: Y\, (A_L-A_R)^2\, Y \:. \]
The sectorial projection of this matrix must be a multiple of~$\acute{Y} \grave{Y}$.
In view of~\eqref{q:ARform2}, this implies that~$(A_L-A_R)^2=0$, and thus~$A_L=A_R$.
But the resulting potential violates the second equation in~\eqref{q:FTcond} and is thus not admissible. This
is a contradiction.
\QED

The previous lemmas show that the parameters~$\alpha$ and~$\beta$ in~\eqref{q:Anew}
are both zero. But then~$A_L^\text{new}$ vanishes, a contradiction.
We conclude that~$\g = \g_L \oplus \g_\free$ is maximal.

\nsubsubsection{Non-triviality of the mixing matrices}
We now analyze the properties of the mixing matrices~$\UMNS$ and~$\UCKM$.
Suppose that one of these matrices is non-trivial. Then the matrix~$B$
(see~\eqref{q:Bdef})
corresponding to the left-handed potentials~$A_{L,1}$ and~$A_{L,2}$
in~\eqref{q:Arep3} is non-zero. The representation of Lemma~\ref{q:lemmaMrep}
yields in particular that the columns of the matrix~$B$ all have the same norm.
This implies that
\[ \big\| (\1-\sproj) \UMNS \sproj \big\| = \big\| (\1-\sproj) \UCKM \sproj \big\| \:. \]
This shows that if one of the matrices~$\UMNS$ and~$\UCKM$ is non-trivial,
the other is also non-trivial. Moreover, by a global phase transformation, we can arrange
that the relation~\eqref{q:hU} holds.

In the next lemma we show that~\eqref{q:massdiff} implies~\eqref{q:notrivmix}.
\begin{Lemma} \label{q:lemma5319}
Suppose that the masses of the charged leptons and neutrinos
are different in the sense~\eqref{q:massdiff}. Then the mixing matrices~$\UMNS$ and~$\UCKM$
are non-trivial.
\end{Lemma}
\sindex{mixing matrix!necessity of}%
\Proof Assume conversely that the potentials in~$\g_L$ do not involve mixing matrices.
Then for any~$\A \in \g_L$, we can compute~$B_L$ according to~\eqref{q:BLform}
with~$A_R^\even=A_L$. Considering the first sector and taking the sectorial projection, we obtain
\[ {\mathfrak{I}}_1\, \hat{B}_L \, {\mathfrak{I}}_1 = -\frac{m^2}{4}\: {\mathfrak{I}}_1 \Big( \acute{Y} \grave{Y}\, A_L[\xi]^2
+ A_L[\xi] \,\acute{Y} \grave{Y}\, A_L[\xi] \Big)\:{\mathfrak{I}}_1 T^{(1)} \:, \]
where~${\mathfrak{I}}_1$ denotes the projection on the neutrino sector.
Since the right-handed component of~$\A$ vanishes, we can also compute~$B_R$
with~$A_L^\even=A_R$. This gives
\[ {\mathfrak{I}}_1\, \hat{B}_R \, {\mathfrak{I}}_1 = -\frac{m^2}{2}\: {\mathfrak{I}}_1 \: \acute{Y} \grave{Y}\, A_L[\xi]^2 \:{\mathfrak{I}}_1 T^{(1)} \:. \]
We conclude that
\begin{align*}
{\mathfrak{I}}_1\, \big( \hat{B}_L - \hat{B}_R \big)\, {\mathfrak{I}}_1
&= \frac{m^2}{4}\: {\mathfrak{I}}_1 \Big( \acute{Y} \grave{Y}\, A_L[\xi]^2
- A_L[\xi] \,\acute{Y} \grave{Y}\, A_L[\xi] \Big)\:{\mathfrak{I}}_1\, T^{(1)} \\
&= \frac{m^2}{4}\: {\mathfrak{I}}_1A_L[\xi]^2 \:{\mathfrak{I}}_1 \,T^{(1)} \:\sum_{\alpha=1}^3 \big( \tilde{m}_\alpha^2 - m_\alpha^2 \big) \:.
\end{align*}
This is non-zero for the potentials~$A_{L,1}$ and~$A_{L,2}$ in~\eqref{q:Arep3}, a contradiction.
\QED

\nsubsubsection{Proof that~$\g$ is admissible}
We have shown that, under the assumptions of Theorem~\ref{q:thmsbf}, the
dynamical gauge potentials are necessarily of the form~\eqref{q:em}--\eqref{q:strong}.
It remains to show that these potentials are indeed admissible in the sense that they
satisfy all algebraic constraints.
This can be done explicitly as follows:
In order to study the sectorial projection of the gauge phases, one can use the fact
that the phases of the free gauge potentials~\eqref{q:em} and~\eqref{q:strong} drop out
of the closed chain and thus do not enter the EL equations. Therefore, it suffices
to consider the weak potentials~\eqref{q:weak}.
Since the weak potentials are block-diagonal, one may consider the lepton block
and the charged blocks separately. For the lepton block, the computations were
carried out in~\S\ref{l:sec32}. In the charged blocks, the computation is even easier
because there are no shear contribution; it reduces to applying Lemma~\ref{l:lemma32}.
For the bilinear logarithmic terms one can again use that the free gauge potentials
drop out. Therefore, one can analyze the neutrino block and the charged blocks
exactly as in~\S\ref{l:seclogAA}.
For the field tensor terms, the relevant contributions are linear in the field.
Therefore, one may choose a basis on the sectors where the field tensor is diagonal
and use the computations and results of~\S\ref{l:secfield}.

This completes the proof of Theorem~\ref{q:thmsbf}.

\section{The Effective Action}
In this section we rewrite the EL equations in the continuum limit~\eqref{q:EL4} as
the Euler-Lagrange equations of an effective action.
We adapt the methods introduced in Chapter~\ref{lepton}.
This adaptation is straightforward because the
dynamical gauge potentials as obtained in Theorem~\ref{q:thmsbf} 
either act separately within each block (the weak and electromagnetic gauge potentials),
or else they mix identical blocks (the strong gauge potentials). This makes it possible to analyze
the blocks separately. For each block, we can proceed just as in Sections~\ref{l:sec5}
and~\ref{l:sec6}.

\subsectionn{The General Strategy}
Our goal is to rewrite~\eqref{q:EL4} as the Euler-Lagrange equations of 
an effective action of the form
\beq \label{q:Seff}
\Sact_\text{eff} = \int_\scrM \left( \LDirac + \LYM + \LEH \right) \sqrt{-\deg g}\, d^4x \:,
\eeq
\sindex{effective action}%
\nindex{dl2@$\Sact_\text{eff}$ -- effective action}%
involving a Dirac Lagrangian (which describes the coupling of the Dirac wave functions
to the gauge fields and gravity), a Yang-Mills-type Lagrangian for the gauge fields
and the Einstein-Hilbert action. Moreover, the Dirac wave functions should satisfy the Dirac
equation (see~\eqref{q:Dinteract})
\beq \label{q:direx0}
(i \Pdd + \B - m Y) \psi = 0 \:.
\eeq
Exactly as explained in~\S\ref{l:sec52}, one must take into account that the
Dirac equation~\eqref{q:direx0} holds for the auxiliary fermionic projector (without
taking the sectorial projection), whereas the current and the energy-momentum tensor,
which are to be obtained by varying the Dirac Lagrangian in~\eqref{q:Seff},
involve a sectorial projection. This leads us to choose the effective Dirac Lagrangian as
\beq \label{q:LDirpar}
\LDirac = \re \Big( \overline{\psi} \:3 \sproj_\tau (i \Pdd +
\sproj \B \sproj - mY) \psi \Big) \:,
\eeq
\sindex{effective Lagrangian!Dirac}%
where the operator~$\sproj_\tau$ has the form
\beq \label{q:sprojt}
\sproj_\tau := (1+\tau \chi_L {\mathfrak{I}}_1)
\: \sproj \qquad \text{with~$\tau \in \R$} \:,
\eeq
and~${\mathfrak{I}}_1$ is again the projection on the neutrino sector.
The sectorial projections in~\eqref{q:LDirpar} are needed in order to get the correct
coupling of the Dirac wave functions to the bosonic fields. The parameter~$\tau$
in~\eqref{q:sprojt} gives us the freedom to modify the coupling of the right-handed
component of the neutrinos to the gravitational field.

Our goal is to choose the Lagrangians~$\LYM$ and~$\LEH$ such that their
first variation is compatible with~\eqref{q:EL4}. In order to treat the gauge fields,
one first rewrites~$\Delta Q$ as
\[ \Delta Q(x,y) = \frac{i}{2} \sum_{n,c} \Tr_{\C^2} \!\big( I_n \,{\mathcal{Q}}_c \big)
\,I_n\: \chi_c\, \slashed{\xi} \:, \]
and represents the factors~${\mathcal{Q}}_c$ by
\[ {\mathcal{Q}}_L := {\mathscr{K}}_L - \frac{1}{4}\: \Tr_{\C^2} \!\big(
{\mathscr{K}}_L+{\mathscr{K}}_R \big)\: \1_{\C^2} \]
(and similarly for~${\mathcal{Q}}_R$),
where the matrices~${\mathscr{K}}_c$ are defined by
\[ \Tr_{\C^2} \left( I_n \,{\mathscr{K}}_c \right) = \frac{\Delta |\lambda^{xy}_{nc-}|}{|\lambda_-|}\;3^3\:
T^{(0)}_{[0]} T^{(-1)}_{[0]} \:\overline{T^{(-1)}_{[0]}} + (\deg < 4)\:. \]
In Chapter~\ref{lepton}, the matrices~${\mathcal{K}}_c$ are computed in the neutrino block,
and these computations apply just as well to the quark blocks.
Next, in order to treat the tensor indices properly, one writes~${\mathscr{K}}_c$ as
\beq \label{q:Jcrep}
{\mathscr{K}}_c =  i \xi_k \:\mathfrak{J}_c^k \:+\: (\deg < 4) + o \big( |\vec{\xi}|^{-3} \big)
\eeq
and sets
\[ {\mathcal{Q}}_c^k := {\mathscr{K}}_c^k - \frac{1}{4}\: \Tr_{\C^2} \!\big(
{\mathscr{K}}^k_L+{\mathscr{K}}^k_R \big)\: \1_{\C^2} \:. \]
The Lagrangian~${\mathcal{L}}_\text{YM}$ must be chosen such as to satisfy the conditions
\beq \label{q:varEL}
K(\varepsilon, \xi)\, \frac{\delta}{\delta \A} \Big( {\mathcal{L}}_\text{Dirac} + {\mathcal{L}}_\text{YM} \Big)
= \Tr_{\C^8} \big( {\mathcal{Q}}^k_L[\hat{\mathscr{J}}, \A]\: (\delta \hat{A}_R)_k + {\mathcal{Q}}^k_R[
\hat{\mathscr{J}}, \A]\: (\delta \hat{A}_L)_k \big) \:,
\eeq
where~$K$ is a constant and~$\delta \A = (\delta A_L, \delta A_R) \in \g$
is a dynamical gauge potential. Here the square brackets~$[\hat{\mathscr{J}}, \A]$ clarify the
dependence on the chiral potentials and on the sectorial projection of the Dirac current.
\sindex{effective Lagrangian!Yang-Mills}%

In order to treat the gravitational field, we rewrite the trace component of~$\Delta Q$ as
\[ \Tr_{\C^{8 \times 4}} \big( \Delta Q\; \slashed{u} \big) = i \xi_j u^j \, {\mathcal{Q}}^{kl}[\hat{T}, g] \,\xi_k \xi_l \:. \]
Our task is to satisfy the relation
\beq \label{q:varELgrav}
i K(\varepsilon, \xi)\; \frac{\delta}{\delta g} \Big(
\left( {\mathcal{L}}_\text{Dirac} + {\mathcal{L}}_\text{YM} + {\mathcal{L}}_\text{curv} \right)
\sqrt{-\deg g} \Big) = {\mathcal{Q}}^{kl}[\hat{T}, g] \;\delta g_{kl} \:.
\eeq
\sindex{effective Lagrangian!Einstein-Hilbert}%

Once we have arranged~\eqref{q:varEL} and~\eqref{q:varELgrav}, we may consider~\eqref{q:Seff}
as the effective action in the continuum limit.
Varying the chiral potentials in~$\g$ gives the bosonic field equations,
whereas varying the metric gives the equations for gravity.
We point out that the variation of the effective action must always be performed under
the constraint that the Dirac equation~\eqref{q:direx0} holds. As explained in~\S\ref{l:secvary},
this gives rise to the so-called {\em{sectorial corrections}} to the field equations.
Since these corrections are computed exactly as in Chapter~\ref{lepton}, we do not enter the
calculations again. Instead, we restrict attention to deriving the effective action and to
discussing our findings.

\subsectionn{The Effective Lagrangian for Chiral Gauge Fields} \label{q:secefflag}
\nsubsubsection{General structure of the effective Lagrangian}
We begin with a general result on the structure of the effective Lagrangian for the gauge fields.
The connection to the Lagrangian of the standard model will be explained afterwards.
\begin{Thm} \label{q:thmgendyn}
Denoting the dynamical gauge potentials as in Theorem~\ref{q:thmsbf}
and decomposing the weak potentials as~$W = \sum_{\alpha=1}^3 W^\alpha \sigma^\alpha \in \su(2)$,
the EL equations in the continuum limit are of variational form~\eqref{q:varEL}, where the effective
Lagrangians is of the form~\eqref{q:LDirpar} and
\begin{align}
\LYM =\;& c_1 \left( \Tr_{\C^3} \big( (\partial_j G)(\partial^j G) \big) 
+ \frac{4}{3}\, (\partial_j A^\text{\rm{\tiny{em}}})(\partial^j A^\text{\rm{\tiny{em}}}) \right) \label{q:el1} \\
& +c_2 \,\Big( (\partial_j W^1)(\partial^j W^1) + (\partial_j W^2)(\partial^j W^2) \Big)
+ c_3 \:(\partial_j W^3)(\partial^j W^3) \label{q:el2} \\
&+ c_4 \: (\partial_j A^\text{\rm{\tiny{em}}})(\partial^j W^3)
+ M_1^2 \Big( W^1 W^1 + W^2 W^2 \Big) + M_3^2 \: W^3 W^3 \:. \label{q:el3}
\end{align}
\sindex{effective Lagrangian!Yang-Mills}%
Here~$c_1, c_2, c_3, c_4$ and~$M_1, M_3$ are parameters which depend on the regularization.
\end{Thm}
\Proof The matrix-valued vector fields~${\mathfrak{J}}_c$ in~\eqref{q:Jcrep}
were computed in~\S\ref{l:seccurrent}.
Combining~\eqref{q:RC1} with the integration-by-parts rule
\[ 0 = \nabla \left( T^{(0)}_{[0]}T^{(0)}_{[0]} \overline{T^{(0)}_{[0]}} \right)
= 2\, T^{(0)}_{[0]}T^{(-1)}_{[0]} \overline{T^{(0)}_{[0]}}
+ T^{(0)}_{[0]}T^{(0)}_{[0]} \overline{T^{(-1)}_{[0]}} \:, \]
one sees that the following simple fraction vanishes,
\[ K_2 := \frac{3}{4} \: \frac{1}{\overline{T^{(0)}_{[0]}}}
 \Big[ T^{(0)}_{[0]} T^{(0)}_{[0]} \:\overline{T^{(-1)}_{[0]} T^{(0)}_{[0]} }
- c.c. \Big] = 0 \:. \]
As a consequence, one obtains
\begin{align}
\!\!\!{\mathfrak{J}}^k_L =&\; \hat{J}_R^k\: K_1  + \hat{\jmath}_R^k \:K_3
\label{q:Jterm} \\ 
&-3 m^2  \left( \acute{A}^k_L Y \grave{Y} + \acute{Y} Y \grave{A}^k_L \right) 
K_4 \label{q:mterm1} \\ 
&+ m^2 \left( \hat{A}_L^k \: \acute{Y} \grave{Y} + \acute{Y} \grave{Y}\: \hat{A}_L^k \right)
K_4 \label{q:mterm2} \\ 
&-3 m^2 \left( \acute{A}_R^k Y \grave{Y} - 2 \acute{Y} A_L^k \,\grave{Y}
+ \acute{Y} Y \grave{A}_R^k \right) K_5 \label{q:mterm3} \\ 
&-6 m^2 \left( \acute{A}_L^k \grave{Y}\: \hat{Y} + \hat{Y}\: \acute{Y} \grave{A}_L^k \right)
K_6 \label{q:mterm4} \\ 
&+6 m^2 \left( \hat{Y} 
\acute{A}_L^k \grave{Y} + \acute{Y} \grave{A}_L^k \: \hat{Y} \right) K_7 \label{q:mterm5} \\ 
&+m^2 \left( \hat{A}_L^k \hat{Y} \hat{Y} + 2 \hat{Y} \hat{A}_R^k \hat{Y}
+ \hat{Y} \hat{Y} \hat{A}_L^k \right) K_6 \label{q:mterm6} \\ 
&-m^2 \left( \hat{A}_R^k \hat{Y} \hat{Y} + 2 \hat{Y} \hat{A}_L^k \hat{Y}
+ \hat{Y} \hat{Y} \hat{A}_R^k \right) K_7 \label{q:mterm7} 
\end{align}
with the simple fractions~$K_1, \ldots, K_7$ as given in~\eqref{l:Jterm}--\eqref{l:mterm7}
(and~${\mathfrak{J}}^k_R$ is obtained by the obvious replacements~$L \leftrightarrow R$).
For the Dirac current, we thus obtain
\[ \Tr_{\C^8} \big( {\mathcal{Q}}^k_L[\hat{\mathscr{J}}, \A]\: (\delta \hat{A}_R)_k + {\mathcal{Q}}^k_R[
\hat{\mathscr{J}}, \A]\: (\delta \hat{A}_L)_k \big)
\asymp K_1 \Tr_{\C^8} \big(J^k_R\: (\delta \hat{A}_R)_k + J^k_L\: (\delta \hat{A}_L)_k \big) \:. \]
This is compatible with~\eqref{q:varEL} and the variation of the Dirac Lagrangian~\eqref{q:LDirpar}
(for fixed wave functions) if we choose
\[ K(\varepsilon, \xi) = 3 \, K_1\:. \]
For the bosonic current and mass terms, one must compensate the logarithmic poles
on the light cone by a microlocal chiral transformation, just as described in~\S\ref{l:secmicroloc}.
For the free gauge potentials~$(A^\text{\tiny{\rm{em}}}, G) \in \u(1) \oplus \su(3)$,
the mass terms vanish. A direct computation gives
\begin{align*}
\Tr_{\C^8} & \big( {\mathcal{Q}}^k_L[\hat{\mathscr{J}}, \A]\: (\delta \hat{A}_R)_k + {\mathcal{Q}}^k_R[
\hat{\mathscr{J}}, \A]\: (\delta \hat{A}_L)_k \big) \\
\asymp\;& K_{19} \, \Tr_{\C^8} \left( j^k[A^\text{\rm{\tiny{em}}}]\, \delta A^\text{\rm{\tiny{em}}}
+ j^k[G]\, \delta G \right)
\end{align*}
for a suitable simple fraction~$K_{19}$, where~$j[A]^k = \partial^k_{\;j} A^j - \Box A^k$ is the bosonic current.
If only the potential~$W$ is considered, we can compute the right side of~\eqref{q:varEL}
exactly as in Section~\ref{l:sec6} to obtain
\begin{align*}
\Tr_{\C^8} & \big( {\mathcal{Q}}^k_L[\hat{\mathscr{J}}, \A]\: (\delta \hat{A}_R)_k + {\mathcal{Q}}^k_R[
\hat{\mathscr{J}}, \A]\: (\delta \hat{A}_L)_k \big) \\
\asymp\;& K_{20} \, \Big( j^k[W^1]\: (\delta W^1)_k + j^k[W^2]\: (\delta W^2)_k \Big)
+ K_{21} \, j^k[W^3]\: (\delta W^3)_k \\
&+m^2 K_{22} \, \Big( (W^1)^k\: (\delta W^1)_k + (W^2)^k\: (\delta W^2)_k \Big)
+ m^2 K_{23} \, (W^3)^k\: (\delta W^3)_k
\end{align*}
for suitable simple fractions~$K_\ell$.
Finally, we must take into account cross terms of~$A^\text{\tiny{\rm{em}}}$ and~$W^3$.
These have the form
\begin{align}
\Tr_{\C^8} & \big( {\mathcal{Q}}^k_L[\hat{\mathscr{J}}, \A]\: (\delta \hat{A}_R)_k + {\mathcal{Q}}^k_R[
\hat{\mathscr{J}}, \A]\: (\delta \hat{A}_L)_k \big) \notag \\
\asymp\;& K_{24} \: j^k[W^3]\: (\delta A^\text{\tiny{\rm{em}}})_k 
+ K_{25} \: j^k[A^\text{\tiny{\rm{em}}}]\: (\delta W^3)_k \label{q:t1} \\
&+m^2 K_{26}\: (W^3)^k\: (\delta A^\text{\tiny{\rm{em}}})_k
+ m^2 K_{27}\: (A^\text{\tiny{\rm{em}}})^k\: (W^3)_k \:. \label{q:t2}
\end{align}

Let us consider the terms~\eqref{q:t1}. In order to be compatible with the variational ansatz~\eqref{q:varEL},
we must impose that
\beq \label{q:mlcond}
K_{24}=K_{25} \:.
\eeq
This relation is automatically satisfied if we use the form of the current terms in~\eqref{q:Jterm}.
However, one should keep in mind that~$K_3$ has a logarithmic pole which must be compensated
by a microlocal chiral transformation. We thus obtain the condition for the microlocal chiral
transformation that it should preserve~\eqref{q:mlcond}.

Moreover, the factors~$K_{26}$ and~$K_{27}$ vanish, as we now explain.
First, the potential~$A^\text{\tiny{\rm{em}}}$ does not contribute to the mass terms,
implying that~$K_{27}$ is zero. Moreover, direct inspection of the
contributions~\eqref{q:mterm1}--\eqref{q:mterm7} shows that for a sector-diagonal potential
which does not involve a mixing matrix, the mass terms depends only on the combination~$A_L-A_R$.
This also implies that
\beq \label{q:mlcond2}
{\mathcal{Q}}_L^k = -{\mathcal{Q}}_R^k \:.
\eeq
On the other hand, for a variation by an electromagnetic potential,
$\delta A_L = \delta A_R$. Therefore, the right side of~\eqref{q:varEL} vanishes by symmetry,
implying that~$K_{26}$ is zero. Similar as explained above for~\eqref{q:mlcond},
the microlocal chiral transformation must be performed in such a way that~\eqref{q:mlcond2} is respected.

Combining all the terms gives the result.
\QED

\nsubsubsection{Correspondence to electroweak theory}
Let us discuss the form of the effective Lagrangian obtained in Theorem~\ref{q:thmgendyn}.
The first summand in~\eqref{q:el1} is precisely the Lagrangian of the strong interaction in the
standard model. The second summand~\eqref{q:el1} is the Lagrangian of the electromagnetic field.
One difference to the standard model is that the coupling constants of the strong and electromagnetic
fields are not independent, but are related to each other by an algebraic relation.
In order to understand this relation, one should keep in mind that the masses and coupling
constants appearing in Theorem~\ref{q:thmgendyn} should be regarded as the ``naked'' constants,
which coincide with the physical constants only at certain energy scale which can be thought
of as being very large (like for example the Planck energy). Thus a relation for the ``naked'' constants
does not mean that this relation should be valid also for the physical constants.
\sindex{coupling constant!naked}%
This situation is indeed very similar to that in grand unified theories (GUTs); we refer the reader
for example to the textbook~\cite{ross}.
\sindex{grand unified theory (GUT)}%
The terms in~\eqref{q:el2} and~\eqref{q:el3} have a similarity to the Lagrangian of the weak potential
after spontaneous symmetry breaking. Indeed, one obtains complete agreement for specific values
of the constants:
\begin{Thm} \label{q:thmelectroweak}
Assume that the parameters in the effective 
Lagrangian of Theorem~\ref{q:thmgendyn} satisfy the conditions
\beq \label{q:addrel}
c_2 = c_3 =c_4 \qquad \text{and} \qquad M_1 = M_3 \:.
\eeq
Then the effective Lagrangian coincides with the Lagrangian of the
standard model after spontaneous symmetry breaking excluding the Higgs field.
The coupling constants of the strong and weak gauge potentials as well of
weak hypercharge are given by
\[ g_\text{\rm{\tiny{strong}}} = \frac{1}{2 \sqrt{c_1}}\:,\qquad
g_\text{\rm{\tiny{weak}}} = \frac{2}{\sqrt{c_2}} \:,\qquad
g_\text{\rm{\tiny{hyp}}} = \frac{1}{2} \left(\frac{16}{3}\: c_1 - c_2 \right)^{-\frac{1}{2}} \:. \]
\end{Thm} \noindent
\sindex{coupling constant!strong}%
\sindex{coupling constant!electroweak}%
Under the assumptions of this theorem, one can introduce the $Z$ and~$W^\pm$-potentials
by forming the usual linear combinations of the weak potential and the potential of weak
hypercharge. The masses~$m_Z$ and~$m_W$ of the corresponding gauge bosons are related
to each other by
\[ m_Z = \frac{m_W}{\cos \Theta_W}\:, \]
\nindex{ed8@$m_W, m_Z$ -- masses of $W$ and $Z$ bosons}%
where the Weinberg angle~$\Theta_W$ is given as usual by
\[ \cos \Theta_W = \frac{g_\text{\rm{\tiny{weak}}}}{\sqrt{g_\text{\rm{\tiny{weak}}}^2
+ g_\text{\rm{\tiny{hyp}}}^2}}\:. \]
\nindex{ee0@$\Theta_W$ -- Weinberg angle}%
\Proof So far, we parametrized the isospin diagonal electroweak potentials
by the electromagnetic potential~$A^\text{\rm{\tiny{em}}}$ and the weak potential~$W^3$.
The standard model, however, is usually formulated instead in terms of the
potential of weak hypercharge~$A^\text{\rm{\tiny{hyp}}}$ and the weak potential.
Since the transformation from one parametrization to the other also changes the
weak potential, we denote the weak potential in the parametrization of the standard
model by an additional tilde. Then the potentials are related by
\[ A^\text{\rm{\tiny{em}}} = 2\, A^\text{\rm{\tiny{hyp}}}\:,\qquad
W^3 = \tilde{W}^3 - A^\text{\rm{\tiny{hyp}}} \:. \]
Using these relations in~\eqref{q:el1}--\eqref{q:el3}, the relevant part of the Lagrangian transforms
to 
\begin{align*}
\LYM \asymp\;& \frac{16}{3}\: c_1 (\partial_j A^\text{\rm{\tiny{hyp}}})(\partial^j A^\text{\rm{\tiny{hyp}}}) \\
& + c_3 \:(\partial_j A^\text{\rm{\tiny{hyp}}})(\partial^j A^\text{\rm{\tiny{hyp}}})
- 2 c_3 \:(\partial_j A^\text{\rm{\tiny{hyp}}})(\partial^j \tilde{W}^3)
+ c_3 \:(\partial_j \tilde{W}^3)(\partial^j \tilde{W}^3) \\
&+ 2 c_4 \: (\partial_j A^\text{\rm{\tiny{hyp}}})(\partial^j \tilde{W}^3) 
- 2 c_4 \: (\partial_j A^\text{\rm{\tiny{hyp}}})(\partial^j A^\text{\rm{\tiny{hyp}}}) \\
&+ M_3^2 \: (\tilde{W}^3 - A^\text{\rm{\tiny{hyp}}}) (\tilde{W}^3 - A^\text{\rm{\tiny{hyp}}})
\end{align*}
and thus
\beq \label{q:Lef}
\begin{split}
\LYM \asymp\;& \Big( \frac{16}{3}\: c_1 + c_3 - 2 c_4 \Big) (\partial_j A^\text{\rm{\tiny{hyp}}})(\partial^j A^\text{\rm{\tiny{hyp}}})
+ c_3 \:(\partial_j \tilde{W}^3)(\partial^j \tilde{W}^3) \\
& - 2 (c_3 - c_4)\:(\partial_j A^\text{\rm{\tiny{hyp}}})(\partial^j \tilde{W}^3) 
+ M_3^2 \: (\tilde{W}^3 - A^\text{\rm{\tiny{hyp}}}) (\tilde{W}^3 - A^\text{\rm{\tiny{hyp}}}) \:.
\end{split}
\eeq
This differs from the Lagrangian of the standard model in that
the kinetic term of the standard model does not involve the cross
terms~$\sim (\partial_j A^\text{\rm{\tiny{hyp}}}) (\partial^j \tilde{W}^3)$.
But using the equation~$c_3=c_4$ in~\eqref{q:addrel}, this cross term vanishes. We thus obtain for the full Lagrangian
\begin{align}
\LYM =\;& c_1 \Tr_{\C^3} \big( (\partial_j G)(\partial^j G) \big)
+\Big( \frac{16}{3}\: c_1 - c_3 \Big) (\partial_j A^\text{\rm{\tiny{hyp}}})(\partial^j A^\text{\rm{\tiny{hyp}}}) \\
&+c_2 \,\Big( (\partial_j W^1)(\partial^j W^1) + (\partial_j W^2)(\partial^j W^2) \Big)
+ c_3 \:(\partial_j \tilde{W}^3)(\partial^j \tilde{W}^3) \\
& + M_1^2 \Big( W^1 W^1 + W^2 W^2 \Big)
+ M_3^2 \: (\tilde{W}^3 - A^\text{\rm{\tiny{hyp}}}) (\tilde{W}^3 - A^\text{\rm{\tiny{hyp}}}) \:.
\label{q:massmatrix}
\end{align}
The constants in front of the quadratic derivative terms can be absorbed into the coupling
constants by rescaling the potentials. To this end, we introduce the coupling constants
\[ g_\text{\rm{\tiny{strong}}} = \frac{1}{2 \sqrt{c_1}}\:,\quad
g_\text{\rm{\tiny{hyp}}} = \frac{1}{2} \left(\frac{16}{3}\: c_1 - c_3 \right)^{-\frac{1}{2}} \:,\quad
g_2 = \frac{1}{2 \,\sqrt{c_2}} \:,\qquad g_3 = \frac{1}{2 \,\sqrt{c_3}} \:. \]
Rescaling the potentials according to
\beq \label{q:potres}
G \rightarrow g_\text{\rm{\tiny{strong}}} \, G\:,\quad
A^\text{\rm{\tiny{hyp}}} \rightarrow g_\text{\rm{\tiny{hyp}}} \, A^\text{\rm{\tiny{hyp}}}\:,\quad
W^{1\!/\!2} \rightarrow g_2\: W^{1\!/\!2} \:,\quad
W^3 \rightarrow g_3\: W^3 \:,
\eeq
the Lagrangian becomes
\begin{align*}
\LYM =\;& \frac{1}{4}\, \Tr_{\C^3} \big( (\partial_j G)(\partial^j G) \big)
+\frac{1}{4}\: (\partial_j A^\text{\rm{\tiny{hyp}}})(\partial^j A^\text{\rm{\tiny{hyp}}}) \\
&+\frac{1}{4}\: \Big( (\partial_j W^1)(\partial^j W^1) + (\partial_j W^2)(\partial^j W^2) 
+ (\partial_j \tilde{W}^3)(\partial^j \tilde{W}^3) \Big) \\
& + M_1^2 \,g_2^2 \Big( W^1 W^1 + W^2 W^2 \Big)
+ M_3^2 \: ( g_3 \tilde{W}^3 - g_\text{\rm{\tiny{hyp}}} A^\text{\rm{\tiny{hyp}}})
(g_3 \tilde{W}^3 - g_\text{\rm{\tiny{hyp}}} A^\text{\rm{\tiny{hyp}}}) \:.
\end{align*}
Now the kinetic term of the Lagrangian looks just as in the standard model.
Clearly, the rescaling of the potentials~\eqref{q:potres} must also be performed in the
Dirac Lagrangian~\ref{q:LDirpar}. This amounts to inserting coupling constants into the
gauge covariant derivative, which thus becomes
\[ D_j = \partial_j - i g_\text{\rm{\tiny{strong}}} \, G_j -i g_\text{\rm{\tiny{hyp}}} \, A^\text{\rm{\tiny{hyp}}}_j\,
{\mathscr{Y}}
-i g_2 \,\chi_L \big( W^1_j \sigma^1_\text{\rm{\tiny{iso}}} + W^2_j \sigma^2_\text{\rm{\tiny{iso}}} \big)
-i g_3 \,\chi_L W^3_j \sigma^3_\text{\rm{\tiny{iso}}} \:, \]
where~$\sigma^\alpha_\text{\rm{\tiny{iso}}}$ are the Pauli matrices acting on isospin,
and~${\mathscr{Y}}$ is the generator of the weak hypercharge,
\[ {\mathscr{Y}} = \chi_L\: \text{\rm{diag}} \Big(-1, -1, \frac{1}{3}, \frac{1}{3}, \frac{1}{3}, \frac{1}{3}, \frac{1}{3}, \frac{1}{3} \Big)
+ \chi_R\: \text{\rm{diag}} \Big(0, -2, \frac{4}{3}, -\frac{2}{3}, \frac{4}{3}, -\frac{2}{3}, \frac{4}{3}, -\frac{2}{3} \Big) \:. \]

In the standard model, there is only one coupling constant for the $\su(2)$-potentials.
This leads us to impose the equation~$c_2=c_3$ in~\eqref{q:addrel}.
Then~$g_2=g_3=: g_\text{\rm{\tiny{weak}}}$.
The last relation in~\eqref{q:addrel} is needed in order for the mass matrix
after spontaneous symmetry breaking to be compatible with~\eqref{q:massmatrix}
(see for example~\cite[Section~20.2]{peskin+schroeder}).
This concludes the proof.
\QED

\nsubsubsection{Additional relations between the regularization parameters}
The remaining important question is whether the relations~\eqref{q:addrel}
hold for suitable regularizations of the fermionic projector. Do they always hold?
Or are there in general deviations?

The general answer is that the relations~\eqref{q:addrel} do not need to hold in general.
But as will be specified in Proposition~\ref{q:prpasy} below,
the relations~\eqref{q:addrel} do hold in the limiting cases when the masses of the leptons are much larger
than the masses of the neutrinos, and the mass of the top quark is much larger than
the mass of the leptons. Therefore, using the hierarchy of the fermion masses of the standard
model, we obtain agreement with the standard model. In view of the experimental observations
\beq \label{q:error}
\frac{m_{\nu_\tau}^2}{m_\tau^2} \lesssim 8 \times 10^{-5} \qquad \text{and} \qquad
\frac{m_\tau^2}{m_\text{\tiny{top}}^2} \approx 10^{-4}  \:,
\eeq
it seems that our limiting case should be an excellent approximation.
But for general regularizations, we expect deviations for the masses and coupling constants
of electroweak theory of the order~\eqref{q:error}. Unfortunately, since at the moment we do not have
detailed information on how the microscopic structure of the physical regularization is,
we cannot make a prediction for the deviations.

\begin{Prp} \label{q:prpasy} Assume that all the mass parameters in~\eqref{q:PC}
and~\eqref{q:massneutrino} are dominated by the mass of the heaviest charged fermion, i.e.
\beq \label{q:m3dominate}
m_3 \gg m_1, m_2
\eeq
and
\beq \label{q:hierarchy1}
m_3 \gg \tilde{m}_1, \tilde{m}_2, \tilde{m}_3 \:.
\eeq
Moreover, assume that the physical (=renormalized) mass of the top quark is much larger than that
of the leptons,
\beq \label{q:hierarchy2}
m_\text{\rm{\tiny{top}}} \gg m_e, m_\nu, m_\tau \:.
\eeq
Then the parameters in the effective Lagrangian of Theorem~\ref{q:thmgendyn}
satisfy the relations in~\eqref{q:addrel} up to relative errors of the order
\[ \frac{m_1^2+m_2^2}{m_3^2} \:,\qquad \frac{\tilde{m}_1^2+\tilde{m}_2^2+\tilde{m}_3}{m_3^2}
\qquad \text{and} \qquad \frac{m_e^2+m_\nu^2+m_\tau^2}{m_\text{\rm{\tiny{top}}}^2}\:. \]
\end{Prp}
The remainder of this section is devoted to the derivation of this proposition.
Our derivation will not be a mathematical proof. Instead, we are content with explaining
the involved approximations in the non-rigorous style common in theoretical physics.
We begin by noting that the term involving the bosonic currents in~\eqref{q:Jterm}
contributes to the right side of~\eqref{q:varEL} by
\begin{align*}
\Tr_{\C^8} & \big( {\mathcal{Q}}^k_L[\hat{\mathscr{J}}, \A]\: (\delta \hat{A}_R)_k + {\mathcal{Q}}^k_R[
\hat{\mathscr{J}}, \A]\: (\delta \hat{A}_L)_k \big) 
\asymp K_3\: \Tr_{\C^8} \left( \hat{\jmath}_R^k \:(\delta \hat{A}_R)_k + 
\hat{\jmath}_L^k \:(\delta \hat{A}_L)_k \right) \\
&= \frac{16}{3}\: K_3 \:9\: j[A_{\text{\rm{\tiny{em}}}}]^k \:(\delta A_{\text{\rm{\tiny{em}}}})_k 
+4\,K_3 \:9 \left( j[A_{\text{\rm{\tiny{em}}}}]^k \:(\delta W^3)_k  + 
 j[W^3]^k \:(\delta A_{\text{\rm{\tiny{em}}}})_k  \right)
\\
&\quad+ 8\,K_3 \Big( 9\, j[W^3]^k \:(\delta W_3)_k +  j[\hat{W}^1]^k
\:(\delta \hat{W}_1)_k +  j[\hat{W}^2]^k \:(\delta \hat{W}_2)_k \Big)
\end{align*}
(the factors of~$9$ come up whenever we leave out the sectorial projection).
This is of variational form, leading us to choose
\beq \label{q:Lagkin}
\begin{split}
{\mathcal{L}}_\text{YM}
&\asymp \frac{8}{3}\: K_3 \:9\, (\partial_j A_{\text{\rm{\tiny{em}}}}) \:(\partial^j A_{\text{\rm{\tiny{em}}}})
+4\,K_3 \:9 \left(\partial_j A_{\text{\rm{\tiny{em}}}})(\partial^j W^3) \right) \\
&\quad+ 4\,K_3 \Big( 9\, (\partial_j W^3)(\partial^j W_3) + (\partial_j\hat{W}^1)(\partial_j\hat{W}^1)
+ (\partial_j\hat{W}^2)(\partial_j\hat{W}^2) \Big) \:.
\end{split}
\eeq
This is of the general form of Theorem~\ref{q:thmgendyn}, but with~$c_3=c_4$.
Thus one of the relations in~\eqref{q:addrel} is automatically satisfied. Moreover,
the coupling constants~$c_1$ and~$c_3$ are related by
\beq \label{q:c13rel}
c_1 = \frac{c_3}{2} \:.
\eeq
The relation~$c_2=c_3$ is violated because of the sectorial projection of the mixing matrix.
However, keeping in mind that the Dirac Lagrangian~\eqref{q:LDirpar} as well
as the mass terms also involve sectorial projections,
these sectorial projections indeed play no role. This will be explained at the
end of this section. If we disregard the sectorial projection for the moment, the relation~$c_2=c_3$
is also satisfied. We conclude that the structure of how the bosonic currents enter
the EL equations in the continuum limit is consistent with the relations on the
left of~\eqref{q:addrel}. Moreover, one has the additional relation~\eqref{q:c13rel}.

The subtle point is that~$K_3$ has a logarithmic pole which must be compensated by
a microlocal chiral transformation. Thus in order to decide if the
relations on the left of~\eqref{q:addrel} or the relation~\eqref{q:c13rel} remain valid, we need to
analyze whether the microlocal chiral transformation respects these relations.
This is not easy to tell because the analysis in~\S\ref{l:secmicroloc} depends in
a complicated way on the ratios of the fermion masses. Moreover, the parameters~${\mathfrak{c}}_0$
and~${\mathfrak{c}}_2$ were not determined explicitly.
But at least, we can analyze the behavior of the microlocal chiral transformation
if we make use of the mass hierarchies, as we now explain.

Before beginning, we need to adapt our method of compensating the logarithmic poles
to the construction of the effective Lagrangian in~\eqref{q:varEL}.
Recall that when introducing the microlocal chiral transformation in~\S\ref{s:secnonlocaxial},
\S\ref{s:secgennonloc}
and~\S\ref{l:secmicroloc}, we always compensated {\em{all}} the logarithmic poles.
However, in the construction of the effective Lagrangian as introduced in~\S\ref{l:secalgebra},
we argued that the EL equations in the continuum limit~\eqref{q:EL4} should be satisfied
only in the ``directions parallel to the bosonic degrees of freedom.'' This is implemented
mathematically by the fact that~\eqref{q:varEL} involves testing
with a dynamical gauge potential~$\delta \A \in \g$. As a consequence, it is no longer necessary
to compensate the logarithmic poles completely. It suffices to arrange that the logarithmic poles
drop out of~\eqref{q:varEL}. More precisely, the contributions~$(j_L, j_R)$ involving logarithmic poles
which remain after the microlocal chiral transformation must satisfy the condition
\beq \label{q:gperp}
\Tr_{\C^8} ( j_L \:\delta A_L + j_R \:\delta A_R)=0 \qquad \forall\, \delta \A \in \g \:.
\eeq
In order to express this condition in a convenient way, we
introduce the real vector space
\[ {\mathfrak{S}}_8 := \Symm(\C^8) \oplus \Symm(\C^8) \:, \]
where~$\Symm(\C^8)$ denotes the Hermitian $8 \times 8$-matrices.
Moreover, we introduce the bilinear form
\beq \label{q:Sbil}
\la .,. \ra_{\mathfrak{S}_8} \::\: {\mathfrak{S}}_8 \times \g \rightarrow \C \:,\qquad
\la (j_L, j_R), \A \ra_{\mathfrak{S}_8} = \Tr_{\C^8}(j_L A_L + j_R A_R) \:.
\eeq
Then~\eqref{q:gperp} can be expressed by saying that the logarithmic contribution
must be orthogonal to~$\g$ with respect to the bilinear form~\eqref{q:Sbil}.

We begin by considering {\em{sector-diagonal transformations}}. The
microlocal chiral transformation is worked out explicitly in Example~\ref{l:example45}.
The transformation involves the eigenvalues~$\mu_1, \ldots, \mu_4$ of the matrix~$S_0^{-1} S_2$.
In the lepton block, these eigenvalues are given by
(see also~\eqref{s:772})
\begin{align*}
\mu_{1\!/\!2} &= \frac{1}{3} \left( \tilde{m}_1^2 + \tilde{m}_2^2 + \tilde{m}_3^2 \mp \sqrt{
\tilde{m}_1^4 + \tilde{m}_2^4 + \tilde{m}_3^4 - \tilde{m}_1^2 \,\tilde{m}_2^2 - \tilde{m}_2^2 \,\tilde{m}_3^2
- \tilde{m}_1^2 \,\tilde{m}_3^2} \right) \\
\mu_{3\!/\!4} &= \frac{1}{3} \left( m_1^2 + m_2^2 + m_3^2 \mp \sqrt{
m_1^4 + m_2^4 + m_3^4 - m_1^2 \,m_2^2 - m_2^2 \,m_3^2
- m_1^2 \,m_3^2} \right) .
\end{align*}
In the quark blocks, one has similarly the eigenvalues~$\mu_{3\!/\!4}$, both with multiplicity two.
As explained in Example~\ref{l:example45}, the amplitude~$\kappa$ of the microlocal chiral transformation
in each sector can vary in the range (see~\eqref{l:lamineq})
\beq \label{q:mctrange}
{\mathfrak{c}}_0 \,\mu_1 \leq \kappa \leq {\mathfrak{c}}_0 \,\mu_2 \qquad \text{and} \qquad
{\mathfrak{c}}_0 \,\mu_3 \leq \kappa \leq {\mathfrak{c}}_0 \,\mu_4\:.
\eeq
The general strategy is to compensate the logarithmic poles choosing~${\mathfrak{c}}_0$
as small as possible. The eigenvalues~$\mu_1, \ldots, \mu_4$ scale like the masses squared.
Therefore, if the neutrino masses are much smaller than the masses of the lepton and
quarks~\eqref{q:hierarchy1},
then the microlocal chiral transformation has no effect in the neutrino sector.
Let us assume in addition that one of the masses of the charged leptons dominates~\eqref{q:m3dominate}.
Then
\beq \label{q:muapprox}
\mu_3 = \O \Big(\frac{m_1^2+m_2^2}{m_3^2} \Big) \:,\qquad
\mu_4 = \frac{2}{3}\: m_3^2 + \O \Big(\frac{m_1^2+m_2^2}{m_3^2} \Big) .
\eeq
As a consequence, the inequalities in~\eqref{q:mctrange} reduce to the interval
\beq \label{q:ainterval}
0 \leq \kappa \leq \frac{2}{3}\: {\mathfrak{c}}_0 \:m_3^2 \:.
\eeq

We conclude that for a sector-diagonal transformation,
our freedom in choosing the microlocal chiral transformation 
reduces to selecting for the left- and right-handed component
of every charged sector a parameter~$\kappa$ in the range~\eqref{q:ainterval}.
We denote these parameters by~$\kappa_{ac}$ with~$a \in \{2, \ldots, 8\}$ and~$c \in \{L, R\}$.
In order to minimize~${\mathfrak{c}}_0$, the best strategy is to choose
every parameter~$\kappa_{ac}$ at one of the boundary points of the interval, i.e.
\beq \label{q:kappabound}
\kappa_{ac} =0 \qquad \text{or} \qquad \kappa_{ac} =\frac{2}{3}\: {\mathfrak{c}}_0 \:m_3^2
\eeq
(with errors as specified in~\eqref{q:muapprox}).
Let us try this strategy for the current corresponding to~$A^\text{\rm{\tiny{hyp}}}$.
As this current is sector-diagonal, testing in~\eqref{q:varEL} gives zero if~$\delta \A$
is the potential~$\A[\delta W^1]$ or~$\A[\delta W^2]$. Moreover, this current is invariant under the action
of the strong~$\SU(3)$, implying that~\eqref{q:varEL} also vanishes if~$\delta \A$ is a strong
potential. Therefore, it suffices to consider the cases that~$\delta \A$ is~$\A[\delta A^\text{\rm{\tiny{hyp}}}]$
or~$\A[\delta \tilde{W}^3]$ (the tilde again clarifies that we parametrize the potentials
by~$(A^\text{\rm{\tiny{hyp}}}, \tilde{W}^3)$). The terms with logarithmic poles generated 
by the current of weak hypercharge are collinear to~$\A[\delta A^\text{\rm{\tiny{hyp}}}]$
and orthogonal to~$\A[\delta \tilde{W}^3]$ (with respect to the bilinear form~\eqref{q:Sbil}).
Thus we need to make sure that the logarithmic pole is compensated
when testing with~$\A[\delta A^\text{\rm{\tiny{hyp}}}]$, but that we get no contribution when
testing with~$\A[\delta \tilde{W}^3]$. This can be arranged by the two choices
\begin{align*}
(\kappa_{aL}) &= \frac{2}{3}\: {\mathfrak{c}}_0 \:m_3^2\; (0,0,0,0,0,0,0,0) \qquad\text{and} 
\qquad (\kappa_{aR}) = \frac{2}{3}\: {\mathfrak{c}}_0 \:m_3^2\;(0,0,1,0,1,0,1,0) \\
\intertext{or alternatively}
(\kappa_{aL}) &= \frac{2}{3}\: {\mathfrak{c}}_0 \:m_3^2\; (0,0,0,0,0,0,0,0) \qquad\text{and}
\qquad (\kappa_{aR}) = \frac{2}{3}\: {\mathfrak{c}}_0 \:m_3^2\; (0,0,0,1,0,1,0,1) \:.
\end{align*}
Indeed, since the contributions generated by the microlocal chiral transformation have 
a definite sign (see~\eqref{l:5con}), we need both cases, depending on whether
the bosonic current is future or past directed. By adjusting~${\mathfrak{c}}_0$, we can arrange that
the contributions involving logarithmic poles satisfy the condition~\eqref{q:gperp} and thus drop
out of~\eqref{q:varEL}.

The next step is to compute the corresponding
smooth contributions generated by the microlocal chiral transformation.
Again using that the largest mass dominates~\eqref{q:m3dominate}, the contribution
by the microlocal chiral transformation is simply given by the corresponding Dirac sea, i.e.
\beq \label{q:Psimple}
P(x,y) \sim \log |m^2 \xi^2| + c
+ i \pi \:\Theta(\xi^2) \:\epsilon(\xi^0)
\eeq
with a numerical constant~$c$ (see~\cite[Section~2.5]{PFP} or~\S\ref{s:sec44}).
Therefore, the smooth contribution is explicit. It is proportional to the original contribution involving
the logarithmic pole. This is very useful because we conclude that the current term
after compensating the logarithmic pole is again orthogonal to~$\A(\delta \tilde{W}^3)$
(with respect to the bilinear form~\eqref{q:Sbil}).
This means that in the kinetic term of the resulting Lagrangian, there is no cross
term of~$A^\text{\rm{\tiny{hyp}}}$ and~$\tilde{W}^3$. Comparing with~\eqref{q:Lef},
this gives precisely the relation~$c_1=c_3$. We conclude that the logarithmic
pole of weak of the bosonic current corresponding to weak hypercharge is compensated
such that the relation~$c_1=c_3$ is preserved (up to error terms as mentioned above).

We now proceed similarly for the current corresponding to~$\tilde{W}^3$.
Thus we want to choose parameters~$\kappa_{ac}$ of the form~\eqref{q:kappabound}
which respect the strong $\SU(3)$-symmetry, such that the logarithmic poles of the
current are removed, but the resulting contribution is orthogonal to~$\A[\delta A^\text{\rm{\tiny{hyp}}}]$.
A short computation shows that the only two solutions are
\beq \label{q:nogo}
\begin{split}
(\kappa_{aL}) &= \frac{2}{3}\: {\mathfrak{c}}_0 \:m_3^2\; (0,1,1,0,1,0,1,0) \:,\quad
(\kappa_{aR}) = \frac{2}{3}\: {\mathfrak{c}}_0 \:m_3^2\;(0,0,0,0,0,0,0,0) \\
(\kappa_{aL}) &= \frac{2}{3}\: {\mathfrak{c}}_0 \:m_3^2\;(0,1,0, 1,0,1,0,1) \:,\quad
(\kappa_{aR}) = \frac{2}{3}\: {\mathfrak{c}}_0 \:m_3^2\;(0,0,0,0,0,0,0,0) \:.
\end{split}
\eeq
Note that these ans\"atze have a contribution in the charged lepton sector.
As will be explained below, this leads to difficulties. The only method for avoiding these difficulties
is to give up~\eqref{q:kappabound} and to allow for the parameters~$\kappa_{ac}$ to
take values in the interior of the interval~\eqref{q:ainterval}. This makes it possible to
choose the parameters~$\kappa_{ac}$ such that they vanish in the lepton block.
Namely, a direct computation gives the solutions
\beq \label{q:go}
\begin{split}
(\kappa_{aL}) &= \frac{2}{3}\: {\mathfrak{c}}_0 \:m_3^2\; (0,0,a,0,a,0,a,0) \:,\quad
(\kappa_{aR}) = \frac{2}{3}\: {\mathfrak{c}}_0 \:m_3^2\;(0,0,b,c,b,c,b,c) \\
(\kappa_{aL}) &= \frac{2}{3}\: {\mathfrak{c}}_0 \:m_3^2\;(0,0,0,a,0,a,0,a) \:,\quad
(\kappa_{aR}) = \frac{2}{3}\: {\mathfrak{c}}_0 \:m_3^2\;(0,0,b,c,b,c,b,c) \:,
\end{split}
\eeq
where the parameters~$a,b,c$ are to be chosen such that
\[ a+4b-2c=0 \:,\qquad 0 \leq a,b,c \leq 1 \qquad \text{and} \qquad \max(a,b,c)=1 \:. \]
Let us explain the consequence of these different solutions.
In the case~\eqref{q:nogo}, the relation~\eqref{q:Psimple} again holds. This implies that
the relation~\eqref{q:c13rel} will hold after removing the logarithmic poles.
In the case~\eqref{q:go}, however, the relation~\eqref{q:Psimple} no longer holds,
because all three Dirac seas contribute substantially to the microlocal chiral transformation.
This makes the situation much more complicated, and we do want to not enter the details here.
For our purposes, it suffices to make the following remarks. First,
the parameters~$\kappa_{ac}$ must necessarily be chosen in accordance to the
relation~$c_1=c_3$, because otherwise~\eqref{q:varEL} could not be satisfied, and the
EL equations in the continuum limit would no longer be of variational form.
Moreover, since~\eqref{q:Psimple} is violated, 
the relation~\eqref{q:c13rel} will no longer hold after the logarithmic
poles have been removed. This makes it necessary to treat~$c_1$ and~$c_3$ as
independent effective parameters, giving rise to independent effective coupling
constants~$g_\text{\rm{\tiny{hyp}}}$ and~$g_\text{\rm{\tiny{weak}}}$.

We next consider {\em{non-sector\-dia\-go\-nal transformations}}.
Since the ansatz~\eqref{q:go} only affects the quark blocks, it can immediately be generalized
to non-sectordiagonal transformations.
Namely, since the microlocal transformation can be performed independently for the two chiral
components, it suffices to consider for example the left-handed component.
Then one can use an~$\U(2)$-transformation to diagonalize the logarithmic contribution.
Using the degeneracy of the masses in each block, this~$\U(2)$-transformation can also
be performed for the local chiral transformation by
\[ L[k] \rightarrow U\, L[k]\, U^* \qquad \text{with~$U \in \U(2)$} \:. \]
In this way, the constructions and results of Example~\ref{l:example45}
can also be used for the non-sectordiagonal transformations in the quark blocks.
This implies in particular that the constant~$c_2$ in the dynamical term of the gauge fields~$W^1$ and~$W^2$
in~\eqref{q:el2} coincides with the corresponding constant~$c_3$ for the gauge field~$W^3$.
We point out that this $\U(2)$-transformation cannot be used in the lepton block because the
masses of the neutrinos are different from those of the charged leptons.
In particular, it is not clear if and how the ansatz~\eqref{q:nogo} can be generalized to
non-sectordiagonal transformations.

Next, we need to analyze the {\em{mass terms}}. This is considerably more complicated
because we must analyze the contributions~\eqref{q:mterm1}--\eqref{q:mterm7}.
The only contribution with a logarithmic pole is the term~\eqref{q:mterm3}.
For the~$W^3$-potential, we can compensate the logarithm as explained above, choosing
for example
\begin{align*}
(\kappa_{aL}) &= \frac{2}{3}\: {\mathfrak{c}}_0 \:m_3^2\; \Big(0,0,\frac{1}{2},0,\frac{1}{2},0,\frac{1}{2},0 \Big)
\:,& (\kappa_{aR}) &= \frac{2}{3}\: {\mathfrak{c}}_0 \:m_3^2\;(0,0,0,1,0,1,0,1) \\
(\kappa_{aL}) &= \frac{2}{3}\: {\mathfrak{c}}_0 \:m_3^2\;(0,0,0,1,0,1,0,1) \:,&
(\kappa_{aR}) &= \frac{2}{3}\: {\mathfrak{c}}_0 \:m_3^2\;\Big(0,0,\frac{1}{2},0,\frac{1}{2},0,\frac{1}{2},0 \Big) \:.
\end{align*}
The resulting contribution is orthogonal to the electromagnetic component,
implying that the parameter~$K_{27}$ in~\eqref{q:t2} again vanishes.
Hence we only need to take into account the contributions where the mass terms are
tested by the left-handed weak potentials. In view of~\eqref{q:varEL}, it thus suffices
to consider~${\mathscr{J}}_R$. Moreover, as the mass terms vanish identically for free
gauge fields, it suffices to consider~\eqref{q:mterm1}--\eqref{q:mterm7} for a left-handed weak potential.
Hence the relevant contribution by the mass terms reduces to
\begin{align}
{\mathfrak{J}}_R^k \asymp&-3 m^2 \left( \acute{A}_L^k Y \grave{Y}
+ \acute{Y} Y \grave{A}_L^k \right) K_5 \label{q:mass1} \\ 
&+2 m^2\:\hat{Y} \hat{A}_L^k \hat{Y}\: K_6 -m^2 \left( \hat{A}_L^k \hat{Y} \hat{Y} 
+ \hat{Y} \hat{Y} \hat{A}_L^k \right) K_7 \:. \label{q:mass2} 
\end{align}

The following argument shows that the contribution~\eqref{q:mass1} drops out of the effective
EL equations:
In view of the hierarchy~\eqref{q:m3dominate}, the logarithmic pole of the mass term
is of the form~\eqref{q:Psimple}. Since the contribution by the microlocal chiral transformation
is of the same form, it cancels the contribution by~\eqref{q:mass1} including the smooth
contributions. As a result, \eqref{q:mass1} drops out of the effective EL equations.

The remaining contribution~\eqref{q:mass2} has the following structure.
In the three quark blocks, the factors~$\hat{Y}$ are constants, so that the mass term
can be written as~$c \hat{W}$. In the lepton block, however, the fact that the neutrino
masses are different from the masses of the charged leptons implies that
the mass terms for~$W^3$ have a different structure than those for~$W^1$ and~$W^2$.
This implies that the constants~$M_1$ and~$M_3$ in~\eqref{q:el3} will in general be different.

We now give an argument which shows that~$M_1$ and~$M_3$ coincide in the limiting case~\eqref{q:hierarchy2}
when the quark masses are much larger than the lepton masses.
This argument will also explain why the solution~\eqref{q:nogo} must be dismissed,
leaving us with the ansatz~\eqref{q:go} for the microlocal chiral transformation.
Our argument makes use of the concept that the masses~$m_\beta$ in~\eqref{q:Pvac}
are the ``naked'' masses, and that these masses are modified by the self-interaction
to the physical masses.
Having this concept in mind, it is a natural idea that the physical mass of the gauge bosons
should again be described by~\eqref{q:mass2} if only the masses of in the mass matrix~$mY$
are replaced by the physical fermion masses.
This idea is motivated by the renormalization program which states that for a renormalizable
theory the self-interaction describes a shift of the masses and coupling constants but
leaves the structure of the interaction unchanged.
However, it must be said that the renormalization of the fermionic projector
is work in progress.
If we take the results of the normalization program for granted and combine them
with the mass hierarchy~\eqref{q:hierarchy2},
then we conclude that all the contributions involving the fermion masses are much
smaller in the lepton block than in the quark blocks.
In particular, in the ansatz~\eqref{q:nogo} we must replace the
sequences~$0,1,\ldots$ by~$0, \delta, \ldots$ with~$\delta \ll 1$.
But then the resulting contribution is no longer orthogonal to~$\A[\delta A^\text{\rm{\tiny{hyp}}}]$.
Therefore, the ans\"atze~\eqref{q:nogo} must be dismissed.
For the mass terms in~\eqref{q:mass2}, we conclude that
the main contribution comes from the
quark sectors, giving rise to an effective mass Lagrangian of the form
\beq \label{q:Lagmass}
M^2 \Big( \hat{W}^1 \hat{W}^1 + \hat{W}^2 \hat{W}^2 + 9 \,W^3 W^3 \Big)
\eeq
which involves only one mass parameter.

It remains to analyze the effect of the sectorial projection of the potentials~$W^1$ and~$W^2$.
For notational simplicity, we only consider the potential~$W^1$. By inspecting~\eqref{q:LDirpar},
\eqref{q:Lagkin} and~\eqref{q:Lagmass}, one sees that only the sectorial projection of
the potential~$W^1$ enters.
Thus varying~$\hat{W}^1$, one sees that the rest mass of the bosonic field remained unchanged
if all the sectorial projections were left out. Moreover, varying the Dirac Lagrangian as
explained in~\S\ref{l:secvary}, one sees that the coupling to the Dirac particles
has the same form as without the sectorial projection, except for the sectorial corrections
mentioned after~\eqref{q:varELgrav}. This explains the last equation in~\eqref{q:addrel}
and thus establishes Proposition~\ref{q:prpasy}.

\subsectionn{The Effective Lagrangian for Gravity} \label{q:secgrav}
\begin{Thm} \label{q:thmEinstein}
Assume that the parameters~$\delta$ and~$p_\reg$ satisfy the scaling
\[ \varepsilon \ll \delta \ll \frac{1}{m}\: (m \varepsilon)^{\frac{p_\reg}{2}} \:, \]
and that the regularization satisfies the conditions~\eqref{q:Tpoint}.
Then the EL equations in the continuum limit~\eqref{q:Qform} can be expressed in terms of the effective
action~\eqref{q:Seff} with the Einstein-Hilbert action
\[ \LEH = \frac{1}{\kappa(\varepsilon, \delta)}\: (R+2 \Lambda) \]
\sindex{effective Lagrangian!Einstein-Hilbert}%
(where~$R$ denotes scalar curvature and~$\Lambda \in \R$ is the cosmological constant).
Here the gravitational constant~$\kappa$ is given by
\[ \kappa = \frac{\delta^2}{\tau_\reg}\: \frac{K_{17}}{K_{18}} \:, \]
\nindex{dc0@$\kappa$ -- gravitational constant}%
\sindex{coupling constant!gravitational}%
\sindex{cosmological constant}%
\nindex{db8@$\Lambda$ -- cosmological constant}%
where~$K_{17}$ and~$K_{18}$ are the simple fractions
\[ K_{17} = \frac{1}{2}\: K_{16} \;\bigg( 1 - \frac{L^{(0)}_{[0]}}{T^{(0)}_{[0]}} \bigg)  \qquad \text{and} \qquad
K_{18} = \frac{1}{2}\: K_8\: \bigg( 1 - \frac{L^{(0)}_{[0]}}{T^{(0)}_{[0]}} \bigg) \]
with~$K_{16}$ and~$K_8$ as in~\eqref{l:curv3} and~\eqref{l:JKL1}
(here all expressions are to be evaluated weakly on the light cone~\eqref{q:asy}).
The parameter~$\tau$ in the Dirac Lagrangian~\eqref{q:LDirpar} is determined to have the value~$\tau=-16$.
\end{Thm} \noindent
\Proof One proceeds exactly as in Section~\ref{l:secgrav}. The variation of the
matrices~${\mathcal{Q}}^{kl}$ is computed as in Lemma~\ref{l:lemmaIT} and~\ref{l:lemmaIR}.
In order to satisfy~\eqref{q:varELgrav} one must choose~$\tau=-16$.
Then the result follows immediately.
\QED

We finally explain how the energy-momentum tensor of the gauge fields comes into play.
The effect of the field tensor of the gauge fields was computed in Lemma~\ref{lemmaFFT}.
\sindex{energy-momentum tensor!of gauge fields}%
In Lemma~\ref{lemmaTcomp} we saw how the resulting logarithmic poles can be compensated
for an axial potential. We now need to generalize these results to the electroweak
and strong potentials as they appear in Theorem~\ref{q:thmsbf}.
If only a left-handed weak potential is present, we can apply Lemma~\ref{lemmaFFT} directly,
because after compensating the axial component of the weak potentials, the remaining contribution
is vectorial and is the same in all sectors. Hence it drops out of the EL equations.
If more general gauge potentials are present, the situation is more involved because
the contribution to be compensated is different in every sector.
In the following lemma we explain how to treat this situation in the
simple setting that only an electromagnetic potential is present.
\begin{Lemma} \label{lemmacomp2}
Assume that the chiral potentials have the form~$\B[A^\text{\tiny{\rm{em}}}]$ given in~\eqref{q:em}.
Then the logarithmic poles of the contribution to the fermionic projector~\eqref{FFlog}
can be compensated by the shear contributions corresponding to a microlocal chiral transformation
for a suitable choice of the potentials~$A^\even_{L\!/\!R}$ in~\eqref{l:Vflow}.
\end{Lemma}
\Proof As in the proof of Lemma~\ref{lemmaTcomp}, the first step is to specify the
microlocal transformation such that, similar to~\eqref{Pn1} and~\eqref{Pn2},
the contribution to the fermionic projector is of the form
\begin{align}
\tilde{P}(k) &= P(k) + \text{(vectorial)}\:\1_{\C^8}\, T^{(1)}_{[3, \mathfrak{c}]} \Big(1+ \O \big( \Omega^{-\frac{1}{2}} \big) \Big) \label{Pn3} \\
&\quad+ \text{(vectorial)}\:\1_{\C^8} \:\delta(k^2) \Big(1+ \O(\Omega^{-1}) \Big) \label{Pn4} \\
&\quad + \text{(pseudoscalar or bilinear)} \:\sqrt{\Omega} \:\delta'(k^2) \Big(1+ \O(\Omega^{-1}) \Big)
+ \text{(higher orders in~$\varepsilon/|\vec{\xi}|$)}\:. \notag
\end{align}
Since all the following transformations will be diagonal on the sectors, we may consider
the sectors separately. Thus, omitting the sector index, our transformations only act on the
generation index. We denote the masses in the considered sector by~$m_1$, $m_2$ and~$m_3$.

We again employ the ansatz~\eqref{l:Uans}, but now with a pure vector component, i.e.\
\[ U(k) = \1 + \frac{i}{\sqrt{\Omega}}\: L^j \gamma_j \]
with $3 \times 3$-matrices~$L^j$. Next, we choose the matrices~$L^j$ as diagonal matrices
involving one vector field~$v$,
\beq \label{Lvec}
L^j \gamma_j = \slashed{v}\: \text{diag} \big( \lambda_1, \lambda_2,  \lambda_3 \big) \:,
\eeq
for simplicity with real parameters~$\lambda_\alpha$.
Then the conditions~\eqref{l:1cond}--\eqref{l:3cond} can be arranged by imposing that
\[ \lambda_1 + \lambda_2 + \lambda_3 = 0 \:. \]
Moreover, the formulas~\eqref{l:4con} and~\eqref{l:5con} give rise to the conditions
\begin{align*}
|\lambda_1|^2 + |\lambda_2|^2 + |\lambda_3|^2 &= {\mathfrak{c}}_0 \\
m_1^2\, |\lambda_1|^2 + m_2^2\, |\lambda_2|^2 + m_3^2\, |\lambda_3|^2 &= {\mathfrak{c}}_2 \:,
\end{align*}
where the parameters~${\mathfrak{c}}_0$ and~${\mathfrak{c}}_2$ should be the same in every sector.
By choosing the parameters~$\lambda_\alpha$ according to the above
conditions, we can arrange a contribution to the fermionic projector of the form~\eqref{Pn3}
and~\eqref{Pn4}.

Next, we need to specify the gauge potentials~$A^\even_{L\!/\!R}$ in the considered sector.
We choose this gauge potential to be vectorial and make similar to~\eqref{Apsi} the ansatz
\[ A^\even_L = A^\even_R = |\psi \ra \la \psi| \:, \]
but now with a real vector~$\psi \in \R^3$.
It is convenient to choose the vectors in~$\C^3$
\[ f_1 = \begin{pmatrix} 1 \\ 1 \\ 1 \end{pmatrix}\:,\qquad
f_2 = \begin{pmatrix} \lambda_1 \\ \lambda_2 \\ \lambda_3 \end{pmatrix} \:. \]
Then we impose the conditions
\begin{align}
\la \psi , f_1 \ra_{\C^3} &= 0 \label{i1} \\
\la \psi, f_2 \ra_{\C^3} &= {\mathfrak{c}}_3 \label{i2}
\end{align}
(where the parameter~${\mathfrak{c}}_3$ is again the same in all sectors).
These two linear equations reduce the degrees of freedom in~$\psi$
to one free parameter. This free parameter may be used to prescribe the expectation value
\beq
\la f_2 , Y^2\, A^\even_{L\!/\!R} f_2 \ra_{\C^3} =
\la f_2 , Y^2\, \psi \ra_{\C^3}\: \la \psi,  f_2 \ra_{\C^3} \:. \label{i3}
\eeq
With~\eqref{i1} and~\eqref{i2}
(and keeping in mind that the parameters~$\lambda_\alpha$ and~$\psi_\alpha$ are all real)
we have arranged that the gauge phases drop out of~\eqref{l:1cond} and~\eqref{l:2cond}
(note that the relations similar to~\eqref{zero1}, \eqref{zero2} and~\eqref{zero3} are satisfied).
The contribution~\eqref{i3} can be used to compensate the logarithmic pole
of an energy-momentum tensor of the form~$T_{jk} \sim v_j A^\even_k + v_k A^\even_j$.

Exactly as explained in the proof of Lemma~\ref{lemmaTcomp},
the above construction generalizes to energy-momentum tensors
of the form~\eqref{Tcomp}.
\QED
If combinations of electroweak and strong gauge potentials are present, the resulting
contributions with logarithmic poles are no longer diagonal on the sectors.
As a consequence, the microlocal chiral transformation must involve off-diagonal
contributions in the sector index.
Since the resulting computations are rather tedious and not very
instructive, we do not give them here.

After the logarithmic poles have been compensated, we are in the same situation
as explained at the end of Section~\ref{l:secgrav}: The energy-momentum tensor of the
gauge fields enters the EL equations similar to the energy-momentum tensor of the
Dirac field, albeit potentially with a different coupling as determined by the
corresponding regularization parameters. In order to get mathematically consistent equations, these
regularization parameters must be adjusted such as to give the same couplings as obtained
by varying the effective action with respect to the metric. In this way, the effective field equations
in the continuum limit include the Einstein equations with the energy-momentum tensors
of both the Dirac field and the gauge fields.

\section{The Higgs Field} \label{q:sechiggs}
As explained in~\S\ref{s:secnohiggs},
the Higgs potential of the standard model can be
identified with suitable sca\-lar/pseu\-do\-sca\-lar potentials in the Dirac equation.
As shown in Lemma~\ref{s:lemmascal}, the contributions by the pseudoscalar potentials
to the fermionic projector drop out of the EL equations.
The scalar potentials, on the other hand, contribute to the EL equations to degree
three on the light cone. As the detailed computations are rather involved,
we postpone the analysis of these contributions to a future publication.
\sindex{Higgs mechanism}%
\sindex{Higgs field}%

\begin{appendix}
\chapter{Testing on Null Lines} \label{s:appnull}
\sindex{testing on null lines}%
\sindex{ultrarelativistic wave packet}%
In this appendix we justify the EL equations in the continuum limit~\eqref{s:ELcl}
by specifying the wave functions~$\psi_1$ and~$\psi_2$ used for testing the EL equations~\eqref{s:Q12}
in the setting with a general interaction and for systems involving several generations.
Our method is to adapt the causal perturbation expansion~\eqref{s:cpower} to
obtain corresponding expansions of~$\psi_1$ and~$\psi_2$. We then consider the
scaling of these terms to every order in perturbation theory. We rely on results from~\cite{PFP}
and~\cite{grotz}, also using the same notation.

We begin with the Dirac equation for the auxiliary fermionic projector of the general
form~\eqref{s:diracPaux}, where we assume that~$\B$ is a multiplication operator which is
smooth and decays so fast at infinity that
\beq \label{s:Bregular}
\int_\scrM \left| x^I \partial_x^J \B(x) \right| d^4x < \infty \qquad
\text{for all multi-indices~$I$ and~$J$ with~$|I| \leq 2$}\:.
\eeq
Under these assumptions,
every Feynman diagram of the causal perturbation expansion~\eqref{s:cpower}
is well-defined and finite (see~\cite[Lemma~2.2.2]{PFP} or Lemma~\ref{l:lemma0}). As in~\cite[Section~2.6]{PFP},
we denote the spectral projectors of the operator~$(i \Pdd + \B - mY - \mu \1)$
by~$\tilde{p}_{+\mu}$. In contrast to~\eqref{s:psi1}, where we cut out an $\omega$-strip
around the mass shell, it is here more convenient to remove a neighborhood in the
mass parameter by setting
\beq \label{s:Psi1}
\psi_1 = \eta - \int_{-\Delta m}^{\Delta m} \widehat{\tilde{p}}_{+\mu} \,\eta\: d\mu \:,
\eeq
where the tilde again denotes the sectorial projection.
When taking the product~$P \psi_1$, we get
cross terms involving different generations. However, as in the proof of~\cite[Theorem~2.6.1]{PFP}
one sees that these cross terms vanish in the infinite volume limit. Thus~$\psi_1$ indeed lies
in the kernel of the Dirac operator. Moreover,
by choosing~$\Delta m$ sufficiently small, we can make the difference~$\psi_1 - \eta$
as small as we like.

The construction of~$\psi_2$ is a bit more involved. In order to get into the framework
involving several generations, we first extend the wave packet in~\eqref{s:psi2} to an object with
$4g$ components,
\beq \label{s:psiwp}
\psi := (i \Pdd + m Y) \,\theta \quad \text{with} \quad
\theta = \left( e^{-i \Omega (t+x)} \:\phi(t+x-\ell,y,z) \right)
\oplus \underbrace{0 \oplus \cdots \oplus 0}_{g-1 \text{ summands}} .
\eeq
For~$\psi_2$ we make an ansatz involving a sectorial projection,
\beq \label{s:Psi2def}
\psi_2 = \sum_{\beta=1}^g \left( \psi_\beta + \Delta \psi_\beta^\text{D} + \Delta \psi_\beta^\text{E}
\right) ,
\eeq
where the corrections~$\psi_\beta^\text{D}$ and~$\Delta \psi_\beta^\text{E}$ should take into
account that the auxiliary Dirac equation must hold and that the generalized energy must
be negative, respectively.
In order to specify~$\Delta \psi_\beta^\text{D}$, we first apply the free Dirac operator to~$\psi$,
\begin{align*}
(i &\Pdd - m Y)\, \psi = (i \Pdd - mY)  (i \Pdd + mY)
\left( e^{-i \Omega (t+x)} \:\phi(t+x-\ell,y,z) \oplus 0 \oplus \cdots \oplus 0 \right) \\
&= -(\Box+m^2) \left( e^{-i \Omega (t+x)} \:\phi(t+x-\ell,y,z) \right)
\oplus 0 \oplus \cdots \oplus 0 = 
\left(\partial_y^2 + \partial_z^2 - m^2 \right) \theta\:,
\end{align*}
where~$m \equiv m_1$ is the mass of the first generation. Note that the $x$- and~$y$-parameters
dropped out. This implies that the obtained expression depends on the ``large'' parameter~$\Omega$
only via a phase; in this sense it is a small error term. In order for~$\psi + \Delta \psi_\beta^\text{D}$ to satisfy
the auxiliary Dirac equation~\eqref{s:diracPaux}, the wave function~$\Delta \psi_\beta^\text{Dirac}$
must be a solution of the inhomogeneous Dirac equation
\[ (i \Pdd + \B - m Y)\,  \Delta \psi^\text{D} = - \left(\partial_y^2 + \partial_z^2 -m^2
\right) \theta - \B \psi \:. \]
Solutions of this equation could be constructed rigorously with energy estimates
(see for example~\cite{john}). But here we are content with a perturbative treatment.
Denoting the Green's function of the zero mass free Dirac operator by~$s$, i.e.
\[ i \Pdd \,s = 1 \:, \]
we can solve for~$\Delta \psi^\text{D}$ in terms of the perturbation series
\beq \label{s:psiD}
\Delta \psi^\text{D} = - \sum_{k=0}^\infty \left(-s (\B-mY) \right)^k s \left[
\left(\partial_y^2 + \partial_z^2 -m^2 \right) f + \B \psi \right] .
\eeq
We point out that~$\Delta \psi^\text{D}$ is not uniquely determined, and this non-uniqueness
is reflected in the fact that there is the freedom in choosing different Green's functions,
like the advanced or retarded Green's functions or the Feynman propagator
(for details on the above operator expansions and the different Green's functions see~\cite{PFP}
or~\cite{grotz}). For our purpose, it is preferable to work with the retarded Green's
function~$s^\wedge$, whose kernel~$s^\wedge(x,y)$ is given explicitly by (see~\cite[Section~2.5]{PFP}
or~\eqref{8b} and~\eqref{l:10}, \eqref{l:12})
\beq \label{s:sretard}
s^\wedge(x,y) = -\frac{1}{2 \pi} \:i \Pdd_x \delta(\xi^2) \: \Theta(-\xi^0) \:.
\eeq
This has the advantage that the support of~$\Delta \psi^D$ lies in the future of~$\psi$,
and thus it is disjoint from the support of~$\psi_1$ (see Figure~\ref{s:fig1} on page~\pageref{s:fig1}).

The function~$\psi+\Delta \psi^\text{D}$ solves the auxiliary Dirac equation,
but in general it will have a component of generalized positive energy.
This positive-energy contribution must be subtracted in order to obtain a vector in the
image of~$P$. Formally, this can be achieved by setting
\beq \label{s:psiE}
\Delta \psi^\text{E} = -(1-P) \left( \psi+\Delta \psi^\text{D} \right) .
\eeq
In order to give this equation a meaning, one must keep in mind that the normalization
of the fermionic projector involves a $\delta$-distribution in the mass parameters,
i.e.\ $P_{+\mu} P_{+\mu'} = \delta(\mu-\mu') P_{+\mu}$.
Thus using the formalism introduced in~\cite[Section~2]{grotz}, we can make sense of~\eqref{s:psiE}
as an operator product simply by omitting all resulting $\delta$-distributions
(see also Section~\ref{secfpext}).
With~\eqref{s:Psi2def} as well as~\eqref{s:psiwp}, \eqref{s:psiD} and~\eqref{s:psiE}, we have
introduced~$\psi_2$ in terms of a well-defined perturbation series.

We now estimate~$\Delta \psi^\text{E}$ for large~$|\Omega|$, with a similar method as previously
used in~\cite[Theorem~3.4]{light} for the estimate of the non-causal high energy contribution.
\begin{Lemma} \label{s:lemmahighenergy}
To very order~$n$ in perturbation theory and for every~$\nu \in \N$,
there is a constant~$C(n, \nu)$ such that
the wave function~$\Delta \psi^\text{E}$ as defined by~\eqref{s:psiE} satisfies the inequality
\beq \label{s:DelpsiE}
\sup_{x \in \scrM} \left| (\Delta \psi^\text{E})^{(n)}(x) \right| \leq \frac{C(n, \nu)}{|\Omega|^\nu}\:.
\eeq
\end{Lemma}
\Proof The $n^\text{th}$ order contribution~$(\Delta \psi^\text{E})^{(n)}$
can be written as a finite number of terms of the form
\beq \label{s:opprod}
g := C_n \, \B \, C_{n-1} \,\B\, \cdots \, \B \,C_0 \,\phi\:,
\eeq
where every factor~$C_l$ is a linear combination of the operators~$p$, $k$, and~$s$.
Here~$\phi$ stands either for the wave function~$\psi$ in~\eqref{s:psiwp} or for
the square bracket in~\eqref{s:psiD}.
In either case, $\psi$ is given explicitly and involves the free parameter~$\Omega$.
It is preferable to proceed in momentum space. The regularity and decay
assumption~\eqref{s:Bregular} implies that
\beq \label{s:regmom}
\sup_{k \in \hscrM} \left| k^J \partial^I \hat{\B}(k) \right| < \infty \qquad
\text{for all multi-indices~$I$ and~$J$ with~$|I| \leq 2$}\:.
\eeq
Setting~$F_0=\hat{\phi}$ and
\beq \label{s:Fldef}
F_l(k) = \int \frac{d^4q}{(2 \pi)^4}\: \hat{\B}(k-q)\: C_l(q)\, F_{l-1}(q) \qquad
\text{(where $1 \leq l \leq n$)} \:,
\eeq
we can write the Fourier transform of~$g$ as
\beq \label{s:gfinal}
\hat{g}(k) = C_n(k)\: F_n(k)\:.
\eeq

It clearly suffices to prove the lemma for~$\nu$ an even number. Let us 
show inductively that the functions~$F_l$ satisfy the bounds
\beq \label{s:induct}
\sup_{(\omega, \vec{k}) \in \hscrM} (\omega-\Omega)^\nu 
\Big( |F_l(\omega, \vec{k})| + \sum_{i=0}^3 |\partial_i F_l(\omega, \vec{k})| \Big) <
C(l, \nu) \qquad \text{uniformly in~$\Omega$}\:.
\eeq
In the case~$l=0$, the claim follows immediately from the
explicit form of~$\phi$. To prove the induction step, we use the inequality
\[ (\omega-\Omega)^\nu \leq c(\nu) \left( (\omega-\omega')^\nu + (\omega'-\Omega)^\nu \right) \]
to obtain the estimate
\begin{align*}
\big| &(\omega-\Omega)^\nu\, F_l(\omega, \vec{k}) \big| \nonumber \\
&\leq c(\nu) \:\bigg| \int_{-\infty}^\infty \frac{d\omega'}{2 \pi} \int_{\R^3} \frac{d \vec{k}'}{(2 \pi)^3}\:
\left[ (\omega-\omega')^\nu \: \hat{\B}(\omega-\omega', \vec{k}-\vec{k}') \right]
\: C_l(\omega', \vec{k}')\, F_{l-1}(\omega', \vec{k}') \bigg| \\
&\;\;\;\; +c(\nu) \bigg| \int_{-\infty}^\infty \frac{d\omega'}{2 \pi} \int_{\R^3} \frac{d \vec{k}'}{(2 \pi)^3}\:
\hat{\B}(\omega-\omega', \vec{k}-\vec{k}')\: C_l(\omega', \vec{k}') \, 
\Big[ (\omega'-\Omega)^\nu F_{l-1}(\omega', \vec{k}') \Big] \bigg| \:.
\end{align*}
Furthermore, the factors~$C_l$ involve at most first derivatives; more precisely, they
are bounded in terms of Schwartz norms by (see for example~\cite[Proof of Lemma~2.2.2]{PFP})
\[ |C_l(f)| \leq \text{const}\, \|f\|_{4,1} \qquad \text{for all~$f \in {\mathcal{S}}$}\:. \]
Combining these inequalities, we can use the induction hypothesis together with~\eqref{s:regmom}
to bound the expression~$|(\omega-\Omega)^\nu\, F_l(\omega, \vec{k})|$ uniformly in~$\Omega$
and~$(\omega, \vec{k})$.
The expression~$|(\omega-\Omega)^\nu\, \partial_i F_l(\omega, \vec{k})|$ can be estimated in exactly
the same way if one keeps in mind that if we differentiate~\eqref{s:Fldef} with respect to~$k$,
the derivative acts only on the potential~$\hat{\B}$, but not on the factor~$F_{l-1}$.
This proves~\eqref{s:induct}.

We next consider the operators in~\eqref{s:opprod} in more detail. The factor~$(1-P)$ in~\eqref{s:psiE}
can be regarded as a projection operator onto the generalized positive-energy solutions of the Dirac equation.
The perturbation expansion of the fermionic projector can be arranged in such a way that
each operator product involves at least one factor~$p-k$ which projects onto the negative-energy
solutions (for details see~\cite{grotz}). Similarly, the factor~$(1-P)$ in~\eqref{s:psiE}
implies that we can arrange the operator products such that every contribution~\eqref{s:opprod}
involves at least one factor~$p+k$, being supported on the upper mass cone.
Thus in the corresponding induction step, we may replace an arbitrary even number
of factors~$(\omega - \Omega)$ by factors of~$|\Omega|$. In the following induction steps we
proceed as in~\eqref{s:induct}. At the end, we apply~\eqref{s:gfinal} to obtain the result.
\QED
We can now prove the main result of this appendix.
\begin{Prp} \label{s:prpnull}
Consider a fermion system in Minkowski space with an interaction~$\B$ which is a multiplication
operator satisfying the regularity and decay assumptions~\eqref{s:Bregular}.
Assume furthermore that the pole of~$Q$ is of order~$o(|\vec{\xi}|^{-4})$ at the origin
(see Definition~\ref{s:def82}).
Then the EL equations~\eqref{s:ELeqns} imply that the operator~$Q$ vanishes identically
in the continuum limit~\eqref{s:ELcl}.
\end{Prp}
\sindex{Euler-Lagrange equations!in the continuum limit}%
\Proof We introduce the wave functions~$\psi_1$ and~$\psi_2$ perturbatively via~\eqref{s:Psi1}
and \eqref{s:Psi2def} with~$\psi$, $\Delta \psi^\text{D}$ and~$\Delta \psi^\text{E}$ according
to~\eqref{s:psiwp}, \eqref{s:psiD} and~\eqref{s:psiE}. Evaluating the commutator~$[P,Q]$ as in~\eqref{s:PQeval}
gives the condition~\eqref{s:Q12}. Following the arguments in~\S\ref{s:secELC}, the leading
terms give~\eqref{s:ELcl}, and thus it remains to consider all correction terms.
The corrections of~$\psi_1$ can be made arbitrarily small by choosing the parameter~$\Delta m$
in~\eqref{s:Psi1} sufficiently small.
Working in~\eqref{s:psiD} with the retarded Green's function~\eqref{s:sretard}, the
support of~$\Delta \psi^\text{D}$ does not intersect the support of~$\eta$,
so that the corresponding contribution to~\eqref{s:Q12} is well-defined in the continuum limit.

The wave function~$\Delta \psi^\text{E}$ is more problematic, because it will in general not vanish
on the support of~$\eta$. But according to Lemma~\ref{s:lemmahighenergy}, we can make~$\Delta \psi^\text{E}$
arbitrarily small by choosing~$|\Omega|$ sufficiently large.
This is not quite good enough for two reasons: First, the integrand in~\eqref{s:Q12} becomes
more and more oscillatory as~$|\Omega|$ is increased, so that the leading contribution to~\eqref{s:Q12}
will also become small as~$|\Omega|$ gets large. And secondly, even if~$\Delta \psi^\text{E}$ is small,
it gives rise to a contribution at~$x=y$ where~$Q(x,y)$ is ill-defined.
The first problem can be treated by noting that the oscillations in the integrand of~\eqref{s:Q12}
will give rise to a polynomial decay in~$\Omega$ (typically a $1/\Omega$ behavior),
whereas according to~\eqref{s:DelpsiE}, the wave function $\Delta \psi^\text{E}$ decays in~$\Omega$
even rapidly. Thus we can indeed arrange that $\eqref{s:DelpsiE}$ is arbitrarily small compared to the leading
contribution in~\eqref{s:Q12}.
For the second problem we need to use that the pole of~$Q$ is of order~$o(|\vec{\xi}|^{-4})$ at the origin:
Due to this assumption, the integrand in~\eqref{s:Q12} will be at most logarithmically divergent at~$x=y$.
By modifying~$\psi_2$ by a suitable negative-energy solution of the Dirac equation (for example a
wave packet of negative energy, whose amplitude is fine-tuned),
one can arrange that this logarithmic divergence drops out. Then the integrals in~\eqref{s:Q12}
become finite, and by choosing~$|\Omega|$ sufficiently large, we can arrange that
the contribution of~$\Delta \psi^\text{E}$ to~\eqref{s:Q12} is much smaller than the
leading contribution which yields~\eqref{s:ELcl}.
\QED
Before discussing the result of this proposition, we estimate an
operator product which is similar to~\eqref{s:opprod} but involves a
nonlocal potential as considered in~\S\ref{s:sec113}.
\begin{Lemma} \label{s:lemmaopnonloc} We consider the expression
\[ g = C_n \, B_n \, C_{n-1} \,B_{n-1}\, \cdots \, B_1 \,C_0 \,\phi\:, \]
where every factor~$C_l$ stands for one of the operators~$p$, $k$, or~$s$.
As in the proof of Lemma~\ref{s:lemmahighenergy}, the function~$\phi$
is either the wave function~$\psi$ in~\eqref{s:psiwp} or the square bracket in~\eqref{s:psiD}.
Each factor~$B_l$ either stands for the multiplication operator~$\B$
satisfying~\eqref{s:Bregular}, or else it is a nonlocal operator~$\nf$
in the Schwartz class~\eqref{s:ndecay}. We assume that at least one factor~$B_l$
is a nonlocal operator. Then for every integer~$\nu$ there is a constant~$C(\nu)$ such that
\beq \label{s:grapid}
\sup_{x \in \scrM} \left| g(x) \right| \leq \frac{C(\nu)}{|\Omega|^\nu}\:.
\eeq
\end{Lemma}
\Proof As in the proof of Lemma~\ref{s:lemmahighenergy}, we proceed inductively in
momentum space. Suppose that~$p$ is the smallest index such that~$B_p=\nf$.
Then for all~$l < p$, only the potential~$\B$ is involved, and the functions~$F_l$
defined by~\eqref{s:Fldef} again satisfy the inequalities~\eqref{s:induct}.
In the $p^\text{th}$ induction step, we must replace~\eqref{s:Fldef} by
\[ F_p(k) = \int \frac{d^4q}{(2 \pi)^4} \: \hat{n}(k,q)\, C_l(q)\, F_{l-1}(q) \:, \]
where~$\hat{n} \in {\mathcal{S}}(\hscrM \times \hscrM)$
denotes the Fourier transform of~$n(x,y)$. Using the induction hypothesis~\eqref{s:induct}
together with the rapid decay of~$\hat{n}(p,q)$ in the variable~$q$,
we obtain a factor~$|\Omega|^{-\nu}$. In the remaining induction steps, we can use the
simpler method of Lemma~\ref{l:lemma0} to obtain the result.
\QED

Clearly, the setting of Proposition~\ref{s:prpnull} is too special for our applications.
But the methods and results can readily be extended in the following ways:
\begin{itemize}[leftmargin=2em]
\itemD Taking into account the {\em{wave functions}} of the particles and anti-particles:
We first note that, being solutions of the Dirac equation, the wave functions
of the particles and anti-particles in~\eqref{s:particles} are orthogonal to the wave function~$\psi_1$
(as is obvious from~\eqref{s:Psi1}). Furthermore, by choosing~$|\Omega|$ much larger than the energies
of all particle and anti-particle wave functions, we can arrange that these wave functions are
also orthogonal to~$\psi_2$. Then all the wave functions drop out of~\eqref{s:Q12}, so that 
we are back in the setting of Proposition~\ref{s:prpnull}.
\itemD Handling the {\em{microlocal chiral transformation}}:
\sindex{transformation of the fermionic projector!microlocal chiral}%
Following the constructions in~\S\ref{s:secgennonloc}, we must apply the
microlocal chiral transformation~\eqref{s:Umicro}
to the fermionic projector before forming the sectorial projection. Likewise, we here apply this
transformation to the wave functions~$\psi_1$ and~$\psi_2$ before forming the sectorial projection.
If this is done, all our arguments still go through.
\itemD Arranging the right order of the pole of~$Q$ at the origin:
As we saw in Section~\ref{s:sec7}, the operator~$Q$ vanishes identically to degree five on the
light cone. Thus the leading contribution to~$Q$ is of degree four on the light cone.
Since~$Q$ always involves a factor~$\slashed{\xi}$ (see~\eqref{s:Qxidef}), the pole of~$Q$
at the origin is indeed of the required order~$o(|\vec{\xi}|^{-4})$.

One might object that near the origin~$x=y$, where the continuum limit of~$Q(x,y)$
is not well-defined, the arguments of Section~\ref{s:sec7} do not apply, and thus there
might be a non-zero contribution to~$Q$ which scales like~\eqref{s:Qvac}, thus having
a pole~$\sim |\vec{\xi}|^{-4}$. However, as explained after~\eqref{s:Scrit}, we may assume that
in the vacuum, the operator~$Q$ vanishes identically, even at the origin where the
formalism of the continuum limit does not apply.
Since to degree five, an interaction only leads to phase transformations (see~\eqref{s:unperturb}),
the operator~$Q$ will then again vanish identically. As a consequence, the pole
of~$Q$ will indeed scale like~$|\vec{\xi}|^{-3}$, even without relying on the formalism
of the continuum limit.
\itemD Handling {\em{nonlocal potentials}}:
\sindex{potential!nonlocal}%
Proposition~\ref{s:prpnull} does not apply
to nonlocal potentials as introduced in~\S\ref{s:sec113}. 
Another difficulty is that the support argument used for~$\Delta \psi^\text{D}$
no longer applies.
But as shown in Lemma~\ref{s:lemmaopnonloc}, any contribution to~$\Delta \psi^\text{D}$
or~$\Delta \psi^\text{E}$ which involves a nonlocal potential satisfies the inequality~\eqref{s:grapid}
and can thus be made arbitrarily small by choosing~$|\Omega|$ sufficiently large.
Following the arguments in the proof of Proposition~\ref{s:prpnull}, this
gives us control of all error terms due to the nonlocal potentials, to every order
in perturbation theory.
\end{itemize}
We conclude that with the help of Proposition~\ref{s:prpnull}, we can justify the
EL equations of the continuum limit~\eqref{s:ELcl} for all fermion systems considered
in this book.

We finally analyze the scalings in a universe of finite life time.
\begin{Remark} (A universe of finite life-time) \label{s:remlife} {\em{
\sindex{life-time of universe}%
Suppose that instead of Minkowski space, we consider a more realistic universe of finite life time
$t_\text{max}$, like a cosmology with a ``big bang'' and a ``big crunch.''
In this case, the Fourier integral~\eqref{s:onesea} still gives a good local
description of a Dirac sea (this is made precise in the example of a closed FRW geometry
in~\cite[Theorem~5.1]{moritz}). However, one can no longer expect a continuum of states,
and therefore the condition~$P \psi_1 = 0$ can no longer be satisfied by removing an
arbitrarily thin strip around the mass shell. More precisely, the width~$\Delta \omega$ of
the strip in~\eqref{s:psi1} should be at least as large as the ``coarseness'' of the states in
momentum space. This gives rise to the scaling (for details in the example of the closed
FRW geometry see~\cite[Section~5]{moritz})
\[ \Delta \omega \sim \frac{1}{t_\text{max}} \:. \]
The corresponding contribution to the Fourier integral~\eqref{s:psi1} scales as follows,
\[ \Delta \psi_1 := \psi_1(x) - \eta(x) \sim \sup |\hat{\eta}|\: \frac{\Delta \omega}{\delta^3}
\sim \Delta \omega\:\delta\: |\eta(0)| \sim \frac{\delta}{t_\text{max}}\: |\psi_1(0)|\:. \]
As a consequence, the wave function~$\psi_1$ no longer vanishes on~$\mathfrak{L}$, but
\[ \psi_1|_{\mathfrak{L}} \sim \frac{\delta}{t_\text{max}}\: \sup |\psi_1| \:. \]
Furthermore, since~$\Delta \psi_1$ is supported near the lower mass shell in momentum space,
it decays in position space at infinity like the fundamental solution~$(p_m-k_m)(0,y)$,
smeared out on the scale~$\delta$.
Combining these statements, we find that the corresponding contribution to
the expectation value~\eqref{s:Q12} scales like
\[  \bra \Delta \psi_1 | Q \,\psi_2 \ket \sim \sup |\psi_1| \: \sup |\psi_2| \:
\delta^4\: \frac{\delta}{t_\text{max}}\: \frac{1}{\varepsilon^{L-1}}\: \varepsilon^{3-p} \:. \]
where~$p$ denotes the order of the pole at the origin, being defined as the smallest integer~$p$ such that
\[ \limsup_{x \rightarrow y} \:(|\xi^0| + |\vec{\xi}|)^{L-p} \:|\eta(x,y)| < \infty \:. \]
In comparison, the main contribution on the light cone around the origin scales like
\[  \bra \psi_1 | Q \,\psi_2 \ket \sim \sup |\psi_1| \: \sup |\psi_2| \:
\delta^4\: \frac{1}{\varepsilon^{L-1}}\: \ell^{-p}\: \frac{\delta^2}{|\Omega|} \:, \]
and thus
\beq \label{s:quotient}
\frac{\bra \Delta \psi_1 | Q \,\psi_2 \ket}{\bra \psi_1 | Q \,\psi_2 \ket} \sim
\frac{\varepsilon^{3-p}\: \ell^p\,|\Omega|}{t_\text{max}\: \delta}
= \left(\frac{\varepsilon}{\ell} \right)^{3-p} \varepsilon \,|\Omega|
\: \frac{\ell^3}{\varepsilon l_\text{max}\:\delta}\:.
\eeq
This equation involves the fundamental length scale $\sqrt{\varepsilon l_\text{max}}$.
The time since the big bang is estimated to about 13 billion years, which is the
same order of magnitude as the size of the visible universe, estimated to $28$ billion parsec.
Thus it seems reasonable to assume that
\[ t_\text{max} > 10^{10} \,\text{years} \sim 10^{26} \,\text{meters}\:. \]
Taking for~$\varepsilon$ the Planck length~$\varepsilon \sim 10^{-35} \,\text{meters}$,
we obtain
\[ \sqrt{\varepsilon\, l_\text{max}} \sim 10^{-4}\, \text{meters}\:. \]
It is remarkable that this is about the length scale of macroscopic physics.
Thus by choosing~$\varepsilon |\Omega|$ sufficiently small, we can make the quotient~\eqref{s:quotient}
arbitrarily small without violating the scalings~\eqref{s:scales}.
We conclude that even if the life time of our universe were finite, this would have no
effect on the statement of Proposition~\ref{s:prpnull}. \QEDrem }}
\end{Remark}

\chapter{Spectral Analysis of the Closed Chain} \label{s:appspec}
In this appendix we analyze how different contributions to the fermionic projector
influence the EL equations. In particular, we shall give the proofs of Lemmas~\ref{s:lemmalc1},
\ref{s:lemmalc2}, \ref{s:lemmalogterm1} and~\ref{s:lemmalogterm2}.
Furthermore, we will analyze a pseudoscalar differential potential (see~\eqref{s:sdlight}
and~\eqref{s:sdlight3}) and a scalar/pseudoscalar potential (Lemma~\ref{s:lemmascal}).
Finally, we prove Proposition~\ref{s:prpflip} on the shear contributions
caused by the microlocal chiral transformation.
\sindex{closed chain!spectral analysis of}%

\section{The General Procedure}
We first review the methods and the general procedure. The behavior of the fermionic projector
near the light cone is described by the light-cone expansion~\eqref{s:fprep}.
We concentrate on the singular behavior on the light cone as described by the series in~\eqref{s:fprep},
disregarding the smooth non-causal contributions~$\tilde{P}^\lec$ and~$\tilde{P}^\hec$
(for the smooth contributions see Appendix~\ref{s:appresum} and~\S\ref{s:sechighorder};
also cf.\ the end of~\S\ref{s:sec51}).
The terms of the light-cone expansion can be computed as described in~\cite[Section~2.5]{PFP} 
and Section~\ref{seclight} (for more details see~\cite{firstorder} and~\cite{light}).
The main task is to calculate the corresponding perturbation of the eigenvalues~$\lambda^{L\!/\!R}_\pm$,
because then the effect on the EL equations is given by Lemma~\ref{s:lemma81}.
In principle, the perturbation of the eigenvalues can be determined in a straightforward manner
by substituting the summands of the light-cone expansion into the closed chain~$A_{xy}$ \eqref{s:Adef},
and by performing a standard perturbation calculation for the eigenvalues of
the $(4 \times 4)$-matrix~$A_{xy}$. However, the combinatorics of the tensor contractions
inside the closed chain makes this direct approach so complicated that it is preferable to
use a more efficient method developed in~\cite[Appendix~G]{PFP}. We now outline this method,
giving at the same time a somewhat different viewpoint.

The first step is to perform the light-cone expansion of the
fermionic projector (as introduced in Section~\ref{seclight}).
For the spectral analysis of the closed chain we use the methods introduced
in Section~\ref{secspeccc}. More precisely, we first compute the matrix elements of the fermionic
projector in the double null spinor frame~$\f^c_s$ (see~\eqref{e:Ce2} in \S\ref{secpertspec}).
Transforming to the double null spinor frame
at such an early stage has the advantage that the contractions of the tensor indices 
(which arise by taking traces of products of Dirac matrices) are relatively easy to compute.
After forming the closed chain~$A_{xy}$ in our double-null spinor basis, we can compute
the eigenvalues of~$A_{xy}$ with a standard perturbation calculation
as explained in~\S\ref{secpertspec}.
As the unperturbed operator we choose the closed chain~\eqref{s:clc} which involves the axial
phases. This is particularly convenient because the unperturbed operator is diagonal in
our double null spinor basis, and moreover the unperturbed eigenvalues are non-degenerate
according to~\eqref{s:unperturb}. Thus it suffices to use simple perturbation theory without degeneracies.
Next, it is useful that the unperturbed eigenvalues~\eqref{s:unperturb} form two complex conjugate
pairs. This will remain true if perturbations of lower degree are taken into account,
so that~$\overline{\lambda^c_s} = \lambda^{\overline{c}}_{\overline{s}}$.
Therefore, it suffices to consider the eigenvalue~$\lambda^L_+$. The eigenvalue~$\lambda^R_+$
is then obtained by the replacement~$L \leftrightarrow R$, whereas the
eigenvalues~$\lambda^{L\!/\!R}_-$ are obtained by complex conjugation.
Expressing the perturbation calculation for~$\lambda^L_+$ in terms of the traces~\eqref{e:Ce2},
one finds that to the considered degree on the light cone, the vector~$v$ drops out.

In order to avoid computational errors, the light-cone expansion was carried out with the help
of the C++ program \textsf{class\_commute}, \label{s:classcommute}
which was originally developed for the calculations
in~\cite{firstorder} and~\cite{light}.
\sindex{computer algebra}%
The traces in~\eqref{e:Ce2}, which involve
the contractions of tensor indices, are also computed with the help of \textsf{class\_commute}.
The resulting matrix elements of the fermionic projector are exported to the
computer algebra program \textsf{Mathematica} (from this moment on, the tensor indices are
simply treated as fixed text strings). The perturbation calculation as well
as the expansions around the origin are then carried out by an algorithm implemented
in \textsf{Mathematica}. This also has the advantage that the standard simplification algorithms
of \textsf{Mathematica} and the comfortable front end are available\footnote{The
C++ program \textsf{class\_commute} and its computational output as well as
the \textsf{Mathematica} worksheets were included as ancillary files to the arXiv submission
arXiv:1211.3351 [math-ph].}.

\section{Vector and Axial Contributions} \label{appvecax}
We now list the results of these calculations, also giving some intermediate steps.
We note that some of these results were already obtained in~\cite[Appendix~G.3]{PFP},
however without using the algorithm implemented in \textsf{Mathematica}.
\sindex{potential!vector}%
\sindex{potential!axial}%

\Proof[Proof of Lemma~\ref{s:lemmalc1}] 
Using for line integrals as in~\cite{light} the short notation~\eqref{s:Lambda} and
(see also~\eqref{l:29x})
\beq \label{s:shortline}
\int_x^y [p,q \,|\, r]\, f := \int_0^1 \alpha^p\, (1-\alpha)^q\, (\alpha-\alpha^2)^r\:
f \big( \alpha y + (1-\alpha)x \big)\: d\alpha \:,
\eeq
\nindex{bh6@$\int_x^y [l,r \:\vert\: n] \cdots$ -- short notation for line integrals}%
the relevant contributions to the light-cone expansion can be written as
(cf.~\cite[Appendix~B]{PFP} and~\cite[Appendix~A]{light})
\sindex{light-cone expansion!explicit formulas}%
\begin{align}
\chi_L \,P(x,y) =\:&
\frac{i}{2}\: \chi_L \,e^{-i \Lambda^{xy}_L}\:\slashed{\xi}\: T^{(-1)} \label{s:Pgag} \\
&-\frac{1}{2} \:\chi_L\:\slashed{\xi}\, \xi_i  \int_x^y [0,0 \,|\, 1]\: j_L^i\: T^{(0)} \label{s:xij} \\
&+\frac{1}{4} \:\chi_L\:\slashed{\xi} \int_x^y F_L^{ij} \, \gamma_i \gamma_j \: T^{(0)} \label{s:FT1} \\
&- \chi_L \:\xi_i \int_x^y [0,1 \,|\, 0]\, F_L^{ij}\, \gamma_j \: T^{(0)} \label{s:FT2} \\
&-\chi_L\: \xi_i \int_x^y [0,1 \,|\, 1]\, \Pdd j_L^i\: T^{(1)} \label{s:dj }\\
&-\chi_L\: \int_x^y [0,2 \,|\, 0]\, j_L^i\,\gamma_i\: T^{(1)} \label{s:jLi} \\
&-i m\, \chi_L\:\xi_i \int_x^y Y A_R^i \: T^{(0)} \label{mT1} \\
&+\frac{im}{2}\:\chi_L\:\slashed{\xi} \int_x^y (Y \slashed{A}_R - \slashed{A}_L Y)\: T^{(0)} \label{mT2} \\
&+i m\,\chi_L\:\xi_i \int_x^y [0,0 \,|\, 1]\, Y \, j_R^i\: T^{(1)} \\
&-\frac{im}{2}\,\chi_L \int_x^y [1,0 \,|\, 0]\, Y  F_R^{ij} \gamma_i \gamma_j \: T^{(1)} \\
&-\frac{im}{2}\,\chi_L \int_x^y [0,1 \,|\, 0]\, F_L^{ij} \gamma_i \gamma_j \,Y\: T^{(1)} \\
&+im\,\chi_L \int_x^y [0,1 \,|\, 0] \left( Y (\partial_j A_R^j) - (\partial_j A_L^j) \,Y \right) T^{(1)} \\
&+\frac{m^2}{2}\:\chi_L\:\slashed{\xi} \,\xi_i \int_x^y [1,0 \,|\, 0]\, Y Y A_L^i \: T^{(0)} \label{mT3} \\
&+\frac{m^2}{2}\:\chi_L\:\slashed{\xi} \,\xi_i \int_x^y [0,1 \,|\, 0]\, A_L^i Y Y \: T^{(0)} \label{mT4} \\
&-m^2\,\chi_L\:\xi_i \int_x^y [0,0 \,|\, 1]\, Y Y F_L^{ij} \gamma_j \: T^{(1)} \\
&-m^2\,\chi_L\:\xi_i \int_x^y [0,2 \,|\, 0]\, F_L^{ij} \gamma_j \,Y Y \: T^{(1)} \\
&+m^2\,\chi_L \int_x^y [1,0 \,|\, 0]\, Y Y \slashed{A}_L\: T^{(1)} \label{s:Pmass1} \\
&-m^2\,\chi_L \int_x^y [0,0 \,|\, 0]\, Y \slashed{A}_R \,Y\: T^{(1)} \\
&+m^2\,\chi_L \int_x^y [0,1 \,|\, 0]\, \slashed{A}_L \,Y Y \: T^{(1)} \label{s:Pendgag} \\
&+ \slashed{\xi} \, (\deg < 1) + (\deg < 0) + \O(A_{L\!/\!R}^2) \:, \nonumber
\end{align}
where~$F_c^{jk} = \partial^j A_c^k - \partial^k A_c^j$ is the chiral field tensor
and~$j^k_c = \partial^k_{\;j} A^j_c - \Box A^k_c$ is the corresponding chiral current.
\nindex{di9@$F^{jk}_{\LR}$ -- chiral field tensor}%
\nindex{dg6@$j_{\LR}$ -- chiral bosonic current}%
The term~\eqref{s:Pgag} is our unperturbed fermionic projector~\eqref{s:Pchiral};
all other summands form our perturbation.
We form the sectorial projection, expand in powers of~$\xi$ and compute the matrix elements~\eqref{e:Ce2}.
Contracting the field tensor in~\eqref{s:FT1} and~\eqref{s:FT2} with~$\xi$,
we get the term~$F_{jk} \xi^j \xi^k$ which vanishes because we treat both factors~$\xi$
as outer factors (for a more careful analysis of these field tensor terms see~\S\ref{l:secfield}).
Moreover, the error terms of the form~$\slashed{\xi} \, (\deg < 1)$ are contracted in the computations
with another factor~$\slashed{\xi}$ or~$\overline{\slashed{\xi}}$, giving rise to a term of lower degree.
Likewise, the higher orders in~$A_{L\!/\!R}$ give rise to terms either of lower degree on the light cone
or of higher order in the expansion around the origin.
Computing~$\Delta \lambda^L_+$ by a first order perturbation calculation gives
\begin{align*}
\Delta \lambda^L_+ =& \frac{i g^2}{3}\:j_L^i \,\xi_i\:T^{(1)}_{[0]} \overline{T^{(-1)}_{[0]}}
-\frac{i g^2}{6}\: j_R^i \,\xi_i \:T^{(0)}_{[0]} \overline{T^{(0)}_{[0]}} \\
& - 2 i g m^2 \,A^i_\text{\rm{a}}\, \xi_i \:
\acute{Y} \grave{Y} \left(T^{(1)}_{[2]} \overline{T^{(-1)}_{[0]}}
+T^{(0)}_{[0]} \overline{T^{(0)}_{[2]}} \right) \\
& -2 i m^2 \,A^i_\text{\rm{a}}\, \xi_i\: \hat{Y}^2 \:\frac{T^{(0)}_{[1]} \overline{T^{(-1)}_{[0]}}
\Big( T^{(0)}_{[1]} \overline{T^{(0)}_{[0]}}+ c.c. \Big)  - c.c. }
{T^{(0)}_{[0]} \overline{T^{(-1)}_{[0]}} - T^{(-1)}_{[0]} \overline{T^{(0)}_{[0]}}}
+ (\deg < 2) + o \big( |\vec{\xi}|^{-1} \big)\: .
\end{align*}
The other eigenvalues are obtained by the replacement~$L \leftrightarrow R$ and by
complex conjugation. Substituting the resulting formulas into~\eqref{s:RRdef} gives the result
for the current terms.

For the computation of the mass term, we must consider another contribution where the mass
expansion of the fermionic projector of the vacuum
\[ P(x,y) \asymp m Y \,T^{(0)} + \frac{i}{2}\, m^2 Y^2\: \slashed{\xi} \,T^{(0)} \]
is taken into account in a perturbation calculation to second order,
\begin{align*}
\Delta \lambda^L_+ \asymp&\; m^2\: \hat{Y}^2\: T^{(0)}_{[1]} \overline{T^{(0)}_{[1]}}
+ g m^2\, \acute{Y} \grave{Y}\: T^{(0)}_{[0]} \overline{T^{(0)}_{[2]}} \\
&\; + \frac{g^2 m^2 \, \hat{Y}^2}{\lambda^L_+ - \lambda^R_-} \:\Big(
T^{(0)}_{[1]} \overline{T^{(-1)}_{[0]}} - T^{(-1)}_{[0]} \overline{T^{(0)}_{[1]}} \Big)
\Big( T^{(0)}_{[0]} \overline{T^{(0)}_{[1]}} - T^{(0)}_{[1]} \overline{T^{(0)}_{[0]}} \Big) .
\end{align*}
The analogous formulas for the other eigenvalues are obtained by the replacements~$L \leftrightarrow R$
and~$\nu \leftrightarrow \overline{\nu}$ as well as by complex conjugation.
In order to compute~$|\Delta \lambda^L_+|$, we use the identity
\[ \Delta |\lambda^L_+| \asymp \frac{1}{2\,|\lambda^L_+|}
\left( \lambda^R_-\: \Delta \lambda^L_+ + \lambda^L_+\: \Delta \lambda^R_- \right) \]
and express the unperturbed eigenvalues in terms of~\eqref{s:unperturb}.
Expanding the phase factors~$\nu$ and~$\overline{\nu}$ in powers of the axial
potential (see~\eqref{s:nudef}) and keeping the linear term in~$A_\text{\rm{a}}$, we obtain
\begin{align*}
\Delta |\lambda^L_+| \asymp&\; 
\frac{g^2 m^2 \hat{Y}^2}{|\lambda_+|}\: A^i_\text{\rm{a}}\, \xi_i\:
T^{(0)}_{[1]} \overline{T^{(0)}_{[1]}} \Big( T^{(-1)}_{[0]} \overline{T^{(0)}_{[0]}}
- T^{(0)}_{[0]} \overline{T^{(-1)}_{[0]}} \Big) \\
&\; + \frac{g^3 m^2 \acute{Y} \grave{Y}}{|\lambda_+|}\: A^i_\text{\rm{a}}\, \xi_i\:
T^{(0)}_{[0]} \overline{T^{(0)}_{[0]}}
\Big( T^{(-1)}_{[0]} \overline{T^{(0)}_{[2]}}
- T^{(0)}_{[2]} \overline{T^{(-1)}_{[0]}} \Big) ,
\end{align*}
and similarly for the other eigenvalues.
Substituting these formulas into~\eqref{s:RRdef} completes the proof.
\QED

\Proof[Proof of Lemma~\ref{s:lemmalc2}] According to~\eqref{s:Jadef} and~\eqref{s:particles},
the perturbation~$\Delta P(x,y)$ by the Dirac current satisfies the relations
\sindex{current!chiral Dirac}%
\[ \Tr \left( \gamma^j \, \Delta P(x,y) \right) = -\frac{1}{2 \pi}\: J_\text{\rm{v}}^j \:,\qquad
\Tr \left( \pseudo \gamma^j \, \Delta P(x,y) \right) = -\frac{1}{2 \pi}\: J_\text{\rm{a}}^j \]
and thus
\[ \Delta P(x,x) = -\frac{1}{8 \pi}\: \gamma_j J^j_\text{\rm{v}} + \frac{1}{8 \pi}\: \pseudo \gamma_j
J^j_\text{\rm{a}} \:. \]
The corresponding perturbation of the eigenvalue~$\Delta \lambda^L_+$ is computed to be
\beq \label{s:DelPJ}
\Delta \lambda^L_+ = \frac{ig}{8 \pi}\: J^i_L\, \xi_i \:\overline{T^{(-1)}_{[0]}}\:.
\eeq
The formula for~${\mathcal{R}}$ follows by a direct calculation.
\QED

\Proof[Proof of Lemma~\ref{s:lemmalammicro}]
The result follows exactly as in the proof of Lemma~\ref{s:lemmalc2}, noting that~\eqref{s:DelPmicro}
is obtained from~\eqref{s:DelPJ} by setting~$J_L=J_R=v$ and multiplying
by~$-8 \pi c/m^2$. Moreover, \eqref{s:Amic} tells us how to insert the gauge phases.
\QED

We now come to the pseudoscalar differential potential~\eqref{s:pdp}.
\sindex{potential!pseudoscalar differential}%
The contribution to the fermionic projector linear in~$v$ has the form
\beq \label{s:sdlight}
P(x,y) \asymp \frac{1}{2}\: \pseudo \slashed{\xi}\, \xi_i \int_x^y \Pdd v^i\: T^{(-1)} 
+\pseudo\: \xi_i\: v^i(x)\: T^{(-1)} + (\deg<2) \:.
\eeq
After applying the relation~$2 \slashed{\xi} \Pdd v^i = 2 \xi^j \partial_j v^i + [\slashed{\xi}, \Pdd v^i]$,
in the first term we can integrate by parts to obtain~\eqref{s:sdlead}.
The light-cone expansion to lower degree involves many terms, which we shall not
give here. To higher order in the mass, the contributions become less singular on the light cone.
In particular, the leading term cubic in the mass takes the form
\sindex{light-cone expansion!explicit formulas}%
\beq \begin{split} \label{s:sdlight3}
P(x,y) \asymp\:& -\frac{i m^3}{2}\: \pseudo \slashed{\xi}\, \xi_i \int_x^y [0,1 \,|\, 0]\,
\Big(Y v^i Y Y - v^i Y Y Y \Big)\: T^{(0)} \\
&-\frac{i m^3}{2}\: \pseudo \slashed{\xi}\, \xi_i \int_x^y [1,0 \,|\, 0]\,
\Big(Y Y Y v^i - Y Y v^i Y \Big)\: T^{(0)} \\
&+i m^3 \pseudo \int_x^y [0,1 \,|\, 0]\,
\Big(Y \slashed{v} Y Y - \slashed{v} Y Y Y \Big)\: T^{(1)} \\
&+i m^3 \pseudo \int_x^y [1,0 \,|\, 0]\,
\Big(Y Y Y \slashed{v} - Y Y \slashed{v} Y \Big)\: T^{(1)} \\
&+ \slashed{\xi} \, (\deg < 1) + (\deg < 0)\:. \\
\end{split}
\eeq
Expanding in powers of~$\xi$, we obtain~\eqref{s:v3} and~\eqref{s:v3def}.

We point out that for the scalar differential potential, the higher orders in perturbation theory
are difficult to handle because they are {\em{not}} of lower degree on the light cone.
Moreover, a resummation procedure similar to that for chiral potential does not seem to work.
For a constant potential, this problem corresponds to the effect of the ``deformation of the
light cone'' as discussed after~\eqref{s:Div}. In the more general setting here, this problem means
that the scalar differential potential cannot be treated perturbatively in a convincing way.
This serious difficulty was our original motivation for introducing the vector differential
potential~\eqref{s:vdp}, and to rewrite the combination of these potentials by the
local axial transformation~\eqref{s:Udef}.

For the local axial transformation~\eqref{s:Udef},
\sindex{transformation of the fermionic projector!local axial}%
the fermionic projector can easily be computed
non-perturbatively. To first order in~$v$, we obtain the contribution~\eqref{s:Pabasy}.
We now give a general symmetry argument which shows that this contribution drops out
of the EL equations.
\Proof[Proof of Lemma~\ref{s:lemmalogterm1}]
In order to avoid a case-by-case analysis of the different orders in a mass expansion,
as the unperturbed fermionic projector~$P_0(x,y)$ we choose the fermionic projector
of the vacuum~\eqref{s:6.0}. After regularization, we then obtain a fermionic projector
with a vector-scalar structure which we write in the form
\beq \label{s:Pformneu}
P(x,y) = \slashed{g} + h \:,\qquad P(y,x) = \overline{\slashed{g}} + \overline{h}
\eeq
with a complex vector field~$g(x,y)$ and a complex scalar~$h(x,y)$. The corresponding closed
chain~$A_{xy} = P(x,y)\, P(y,x)$ has two eigenvalues~$\lambda_\pm$,
both with multiplicity two, and the corresponding spectral projectors can be written as
\beq \label{s:Fpmform}
F_\pm = \frac{1}{\lambda_\pm - \lambda_\mp} \left( A_{xy} - \lambda_\mp \right)
\eeq
(for details see~\cite[Section~5.3]{PFP}). Consequently, a linear mapping on the eigenspace corresponding
to~$\lambda_s$ (with~$s \in \{+,-\}$) can be written as
\beq \label{s:Xdef}
X F_s \quad \text{with} \quad X = \alpha \1 + \beta_1 \pseudo \slashed{u}_1
+ \beta_2 \pseudo \slashed{u}_2 + i \beta_3\: \slashed{u}_1 \slashed{u}_2 \:,
\eeq
where~$\alpha, \beta_1, \beta_2, \beta_3$ are complex parameters and~$u_1, u_2$ are two
vector fields which are orthogonal to~$g$ and~$\overline{g}$. Our task is to show that the
linear perturbation of the fermionic projector by the functions~$c_\beta$ in~\eqref{s:Pabasy}
has no effect on the eigenvalues of the closed chain. To this end, we must show that~$\Delta A_{xy}$
vanishes on the degenerate subspaces, i.e.\
\[ \Tr \left( X F_s\, \Delta A_{xy} \right) = 0 \:. \]
Writing~$P(y,x)$ as~$P(x,y)^*$ and omitting the arguments~$(x,y)$, we have
\beq \label{s:Delpcom}
\Delta A_{xy} = (\Delta P) \,P^* + P \,(\Delta P)^* \qquad \text{and} \qquad
\Delta P = -i \sum_{\beta=1}^g \left[ c_\beta \pseudo \slashed{v}, P_\beta \right] .
\eeq
Combining these formulas, one sees that it suffices to show that
\[ \Tr \left( X F_s \: [ \pseudo \slashed{v}, P_\beta] \, P^* \right) = 0 \:, \]
because all other contributions are then obtained by conjugation and by taking linear combinations.

By cyclically commuting the operators and using that~$X$ commutes with~$P^*$, we obtain
\beq \label{s:trace}
\Tr \left( X F_s \: [ \pseudo \slashed{v}, P_\beta ]\, P^* \right) =  
\Tr \left( \pseudo \slashed{v}\: [P_\beta, \: X P^* F_s] \right) .
\eeq
It suffices to consider the vector component of~$P_\beta$ (otherwise the
commutator vanishes). Moreover, since the trace of an odd number of Dirac matrices vanishes,
we may restrict attention to the even contribution to~$X P^* F_s$. The bilinear contribution
to~$X P^* F_s$ gives rise to a vector contribution to the commutator, so that the trace in~\eqref{s:trace}
vanishes. Moreover, the scalar contribution to~$X P^* F_s$ drops out of the commutator. Hence
it remains to consider the {\em{pseudoscalar contribution}} to~$X P^* F_s$.

In view of~\eqref{s:Xdef}, \eqref{s:Fpmform} and~\eqref{s:Pformneu}, the only way to obtain a pseudoscalar
contribution is to combine four linearly independent Dirac matrices,
\[ X P^* F_s \asymp i \beta_3\: \slashed{u}_1 \slashed{u}_2 \:(P^* F_s)_\text{\tiny{bilinear}} \:, \]
where the subscript means that we only take into account the bilinear contribution.
Again using~\eqref{s:Fpmform} and~\eqref{s:Pformneu}, this bilinear contribution can be written as
\[ (P^* F_\pm)_\text{\tiny{bilinear}} = \frac{1}{\lambda_\pm - \lambda_\mp}\:
(P^* P P^*)_\text{\tiny{bilinear}}
= \frac{1}{\lambda_\pm - \lambda_\mp} \left(
\overline{h}\, \slashed{g} \, \overline{\slashed{g}} + \overline{\slashed{g}} \,h\, \overline{\slashed{g}}
+ \overline{\slashed{g}}\, \slashed{g}\, \overline{h} \right)_\text{\tiny{bilinear}} . \]
Using the anti-commutation relations, one finds that the first and third summands combine to a scalar,
and that the second summand is also a scalar. Thus the bilinear component of~$P^* F_s$ vanishes.
\QED

\Proof[Proof of Lemma~\ref{s:lemmalogterm2}]
Let us consider the different contributions in~\eqref{s:Ptres}.
The vector component clearly drops out of~\eqref{s:RRdef}.
The pseudoscalar and bilinear contributions are even. Thus in the perturbation
calculation for~$\lambda^c_s$, they only effect the odd powers in the mass expansion.
As a consequence, the leading contribution is of the form~$\Delta \lambda^c_s
= m \, (\deg < 2)$ and can thus be omitted.
It remains to consider the axial contribution,
\[ P(x,y) \asymp \pseudo \slashed{v}\: T^{(1)}_{[3]} + (\deg < 0)\:. \]
The corresponding perturbation of the eigenvalue~$\Delta \lambda^L_+$ is given by
\[ \Delta \lambda^L_+ \asymp
-i g\:v_j \xi^j \: T^{(1)}_{[3]} \overline{T^{(-1)}_{[0]}} + (\deg < 2) \:. \]
The other eigenvalues are again obtained by the replacement~$L \leftrightarrow R$ and by complex conjugation. Substituting the resulting formulas into~\eqref{s:RRdef} gives the result.
\QED

\section{Scalar and Pseudoscalar Contributions}
We next consider the perturbation of the fermionic projector by the scalar and pseudoscalar
Dirac current~\eqref{s:Jspdef}.
\sindex{potential!scalar}%
\sindex{potential!pseudoscalar}%
\sindex{current!scalar Dirac}%
\sindex{current!pseudoscalar Dirac}%
 According to~\eqref{s:particles}, the corresponding perturbation
of the fermionic projector is given by
\beq \label{s:DPscal}
\Delta P(x,y) = -\frac{1}{8 \pi} \left( J_\text{s} + i \pseudo \, J_\text{\rm{a}} \right) 
+ o \big( |\vec{\xi}|^0 \big)\:.
\eeq

\begin{Lemma} \label{s:lemmascal}
The first order contribution of the perturbation~\eqref{s:DPscal} to the operator~$Q(x,y)$
is of degree two on the light cone.
\end{Lemma}
\Proof A first order perturbation calculation yields
\beq \label{s:DelPps}
\Delta \lambda^L_+ =  \frac{J_\text{s}}{4 \pi}\: m \hat{Y}
\: \frac{ T^{(0)}_{[0]} \Big(  T^{(-1)}_{[0]} \overline{T^{(0)}_{[1]}} - c.c. \Big)
+ \overline{T^{(-1)}_{[0]}} \Big(  T^{(0)}_{[1]} \overline{T^{(0)}_{[0]}} - c.c. \Big)
}{T^{(0)}_{[0]} \overline{T^{(-1)}_{[0]}} - T^{(-1)}_{[0]} \overline{T^{(0)}_{[0]}}}
 + (\deg < 1)\:.
\eeq
Note that the pseudoscalar current dropped out.
This cancellation can also be understood from the following consideration.
The pseudoscalar contribution in~\eqref{s:DPscal} can be written symbolically as
\[ \Delta P(x,y) = i \,[\pseudo \slashed{v}, P(x,y)] \:, \]
where~$\slashed{v}(x,y)$ is a suitable function.
This perturbation has the same form as that in~\eqref{s:Delpcom}. Proceeding as in the
proof of Lemma~\ref{s:lemmalogterm1}, we conclude that the pseudoscalar
contribution drops out of the EL equations.

The resulting first order contribution to the EL equations is obtained by considering first
variations of the Lagrangian of the form~\eqref{s:Lcrit} and by substituting the
formulas for~$\Delta \lambda^c_s$. Counting degrees, this contribution is expected to
be of degree three on the light cone. However, since~\eqref{s:DelPps} has even parity,
whereas the first variation of the Lagrangian involves factors~$(|\lambda^L_s|-|\lambda^R_s|)$
of odd parity, this expected contribution of degree three vanishes.
\QED

\section{Shear Contributions}
We now come to the analysis of the Dirac equation~\eqref{s:Dflip}, where we interchanged
the chirality of the potentials in the even component of the Dirac operator.
\sindex{shear contribution!analysis of}%

\Proof[Proof of Proposition~\ref{s:prpflip}.] 
It suffices to consider the homogeneous transformation~\eqref{s:Zansatz} because then the result
immediately carries over to the microlocal transformation by considering just as in~\S\ref{s:secgennonloc}
the corresponding quasi-homogeneous an\-satz~\eqref{s:Umicro}.
Moreover, it suffices to show that the auxiliary fermionic projector defined
in analogy to~\eqref{s:Pmicroloc} by
\[ \tilde{P}^\text{aux} = \acute{U}_\text{flip}\: P^\text{aux}\: \grave{U}_\text{flip}^* \]
has the desired properties (where~$P^\text{aux}$ satisfies the Dirac equation~$(i \Pdd_x + \chi_L \slashed{A}_R
+ \chi_R \slashed{A}_L - m Y ) P^\text{aux}=0$). Namely, this fermionic operator obviously satisfies the
Dirac equation~\eqref{s:Dirflip}. This implies that it differs from the fermionic projector defined
from~\eqref{s:Dirflip} via the causal perturbation expansion only by smooth contributions, giving the result.

Expanding~\eqref{s:Zansatz} in powers of~$1/\sqrt{\Omega}$, we obtain
\begin{align*}
U_\text{flip} &= \1 + \left( \frac{i Z}{\sqrt{\Omega}} - \frac{Z^2}{2\Omega} \right) V + \O \big(
\Omega^{-\frac{3}{2}} \big) \\
\tilde{P}^\text{aux} &= P^\text{aux} + \frac{i}{\Omega} \Big( Z V P^\text{aux} - P^\text{aux} V^* Z \Big) \\
&\qquad + \frac{1}{\Omega} \:Z\, V P^\text{aux} V^*\, Z
- \frac{1}{2\Omega} \:\Big( Z^2 V \,P^\text{aux} + P^\text{aux} \,V Z^2 \Big)
+ \O \big( \Omega^{-\frac{3}{2}} \big) \:.
\end{align*}
Performing the light-cone expansion order by order in perturbation theory,
one finds that the factors~$V$ and~$V^*$ modify the fermionic projector by phases
and give rise to additional contributions of lower degree on the light cone.
In particular, the contributions of order~$\O (\Omega^{-\frac{3}{2}})$ are of order~$o(|\vec{\xi}|) \,(\deg <2)$
on the light cone. Moreover, the relations~\eqref{s:nonc1} imply that the linear terms in~$Z$ again drop out.
Moreover, the contributions involving~$Z^2$ can be compensated just as explained
for the contribution~\eqref{s:Uk3}. We conclude that
\[ \tilde{P} = P + \frac{1}{\Omega} \:\acute{Z}\, V P^\text{aux} V^*\, \grave{Z}
+ o(|\vec{\xi}|) \,(\deg <2) \:. \]
Hence, as desired, the transformation~$V$ only modifies the contribution generated by the
microlocal chiral transformation.

In order to specify~$V$, it is most convenient to work with the unitary perturbation flow,
which makes it possible to obtain the fermionic projector in the external potential~$P^\text{aux}$
from the vacuum fermionic projector~$P^\text{vac}$ by conjugation with an operator~$U_\text{flow}$,
\[ P^\text{aux} = U_\text{flow}\, P^\text{vac}\, U_\text{flow}^* \]
(see~\cite[Section~5]{grotz} or the more explicit and systematic treatment
with spatial or mass normalization in~\cite[Section~4]{norm}). For clarity, we denote the dependence
of~$U_\text{flow}$ on the external potential in square brackets, i.e.\ $U_\text{flow} =
U_\text{flow}[\chi_L \slashed{A}_R + \chi_R \slashed{A}_L]$. In order for~$V$ to flip the chirality of the
potentials, we simply choose
\beq \label{a:Vflowdef}
V = U_\text{flow}[\chi_L \slashed{A}_L + \chi_R \slashed{A}_R] \;U_\text{flow}[\chi_L \slashed{A}_R + \chi_R \slashed{A}_L]^{-1}
\eeq
(note that~$U_\text{flow}$ can be inverted with a Neumann series as a formal
power series in the external potential).
Then the operator~$V P^\text{aux} V^*$ coincides with the fermionic projector in the
presence of the chiral potential~$\chi_L \slashed{A}_L + \chi_R \slashed{A}_R$.
This concludes the proof.
\QED

\section{The Energy-Momentum Tensor of Chiral Gauge Fields} \label{appEMT}
We proceed by computing the energy-momentum tensor of chiral gauge fields.
The relevant contributions to the kernel of the fermionic projector are given by
\sindex{light-cone expansion!explicit formulas}%
\sindex{energy-momentum tensor!of gauge fields}%
\begin{align}
\chi_L\: P(x,y) &\asymp -i\: \chi_L\, \slashed{\xi} \int_x^y [0,1 \,|\, 1]\: dz_1 \int_{z_1}^y [0,1 \,|\, 0] \:dz_2\;
F^L_{ki}(z_1)\: F_L^{kj}(z_2)\, \xi^i\, \xi_j\: T^{(0)}_{[0]} + (\deg < 1) \notag \\
&= -\frac{i}{24}\: \chi_L\, \slashed{\xi} \: F^L_{ki}\: F_L^{kj}\, \xi^i\, \xi_j\: T^{(0)}_{[0]} + (\deg < 1) 
+ o \big(|\vec{\xi}|^2 \big)\:. \label{T10}
\end{align}
From the contributions involving~$T^{(1)}_{\circ}$, it suffices to compute the contraction with~$\slashed{\xi}$:
\begin{align}
\frac{1}{4}\, \Tr & \big( \slashed{\xi}\, \chi_{L\!/\!R}\: P(x,y) \big) \notag \\
\;\asymp &\pm \frac{1}{2} \int_x^y [0,2 \,|\, 0]\: dz_1 \int_{z_1}^y [0,0 \,|\, 0] \:dz_2\;
\epsilon_{ijkl}\: \xi^i\: F_{L\!/\!R}^{ja}(z_1) \,\xi_a\: F_{L\!/\!R}^{kl}(z_2)\: T^{(1)}_{[0]} \notag \\
&\pm \frac{1}{2} \int_x^y [0,2 \,|\, 0]\: dz_1 \int_{z_1}^y [0,1 \,|\, 0] \:dz_2\;
\epsilon_{ijkl}\: \xi^i\: F_{L\!/\!R}^{jk}(z_1) \: F_{L\!/\!R}^{la}(z_2)\,\xi_a\: T^{(1)}_{[0]} \notag \\
&-i \int_x^y [0,2 \,|\, 0]\: dz_1 \int_{z_1}^y [1,0 \,|\, 0] \:dz_2\; F^{L\!/\!R}_{ki}(z_1)\: F_{L\!/\!R}^{kj}(z_2)\, \xi^i\, \xi_j \: T^{(1)}_{[0]} \notag \\
&-4i \int_x^y [0,1 \,|\, 1]\: dz_1 \int_{z_1}^y [0,1 \,|\, 0] \:dz_2\; F^{L\!/\!R}_{ki}(z_1)\: F_{L\!/\!R}^{kj}(z_2)\, \xi^i\, \xi_j
\: T^{(1)}_{[0]} \notag \\
=&-\frac{i}{3} F^{L\!/\!R}_{ki}\: F_{L\!/\!R}^{kj}\, \xi^i\, \xi_j \: T^{(1)}_{[0]} + o \big(|\vec{\xi}|^2 \big) \:, \label{T11}
\end{align}
where~$\epsilon^{ijkl}$ is the totally anti-symmetric Levi-Civita symbol.
Note that the terms involving~$\epsilon^{ijkl}$ vanish to leading order at the origin.
This is due to the following lemma.
\begin{Lemma} For any anti-symmetric tensor~$F$ and any~$\xi$ on the light cone,
\beq \label{epsform}
\epsilon_{ijkl}\: \xi^i\: F^{ja} \,\xi_a\: F^{kl} = 0 \:.
\eeq
\end{Lemma}
\Proof We extend~$\xi$ to a basis~$(\xi=e_1, e_2, e_3, e_4)$ being a null frame in the sense that
\beq \label{nullframe}
\la e_1, e_2 \ra = \la e_2, e_1 \ra =1\:,\qquad \la e_3, e_3 \ra = \la e_4, e_4 \ra = -1 \:,
\eeq
and all other Minkowski inner products vanish. Then~$F$ can be represented as
\[ F = \sum_{i,j=1}^4 F^{ij} \, e_i \wedge e_j \]
(where for ease in notation we assume that~$F^{ij}=-F^{ji}$).
Since~$\xi$ is null, the index~$a$ in~\eqref{epsform} must not be equal to one.
Moreover, due to the total anti-symmetrization, none of the indices~$j,k,l$ must be equal to one.
Therefore, we may assume that all indices of~$F$ are not equal to one, i.e.
\[ F = \sum_{i,j=2}^4 F^{ij} \, e_i \wedge e_j \:. \]
Next, by performing a rotation of the basis vectors~$e_3$ and~$e_4$, we can arrange that~$F^{23}$ vanishes, i.e.
\beq \label{Fsimpform}
F = 2 \,F^{24} \,e_2 \wedge e_4 + 2 \,F^{34} \, e_3 \wedge e_4 \:.
\eeq
Now, in view of the left side of~\eqref{nullframe}, the index~$a$ in~\eqref{epsform} must be equal to two.
Consequently, the index~$j$ is equal to four.
On the other hand, we see from~\eqref{Fsimpform} that one of the indices~$k$ or~$l$ in~\eqref{epsform}
must also be equal to four. Hence anti-symmetrizing in the indices~$j$, $k$  and~$l$ gives zero.
\QED

In order to compute the perturbation of the eigenvalues of the closed chain, one
needs to take into account the contributions~\eqref{T10} and~\eqref{T11} to first order,
and the contributions~\eqref{s:FT1} and~\eqref{s:FT2} to second order in perturbation theory.
This gives
\begin{align}
\Delta \lambda^L_+ \;=& -\frac{2}{3}\:g\: 
\acute{F}^L_{ki}\: \grave{F}_L^{kj}\, \xi^i\, \xi_j\: T^{(1)}_{[0]} \overline{T^{(-1)}_{[0]}} \label{ord11} \\
&-\frac{g}{12}\: \acute{F}^R_{ki}\: \grave{F}_R^{kj}\, \xi^i\, \xi_j\: T^{(0)}_{[0]} \overline{T^{(0)}_{[0]}} \label{ord12} \\
&- \frac{1}{4}\: \left( \hat{F}^L_{ki}\: \hat{F}_R^{kj} \,\xi^i\, \xi_j
+ \epsilon_{ijkl}\: \xi^i\, \xi_a \: \hat{F}_L^{aj}\: \hat{F}_R^{kl} \right) T^{(0)}_{[0]} \overline{T^{(0)}_{[0]}}
\label{ord21} \\
&-\frac{1}{4}\: \Big( \hat{F}^L_{ki}\: \hat{F}_L^{kj} + \hat{F}^R_{ki}\: \hat{F}_L^{kj} \Big) \,\xi^i\, \xi_j\:
\frac{T^{(0)}_{[0]} T^{(0)}_{[0]} \overline{T^{(-1)}_{[0]} T^{(0)}_{[0]}}}{T^{(0)}_{[0]} \overline{T^{(-1)}_{[0]}} - T^{(-1)}_{[0]} \overline{T^{(0)}_{[0]}}} \label{ord22} \\
&-\frac{1}{2}\: \hat{F}^L_{ki}\: \hat{F}_R^{kj}\, \xi^i\, \xi_j\:
\frac{T^{(-1)}_{[0]} T^{(0)}_{[0]} \overline{T^{(0)}_{[0]} T^{(0)}_{[0]}}}{T^{(0)}_{[0]} \overline{T^{(-1)}_{[0]}} - T^{(-1)}_{[0]} \overline{T^{(0)}_{[0]}}} \\
&-\frac{1}{4}\: \epsilon_{ijkl}\: \xi^i\, \xi_a \: \hat{F}_R^{aj}\: \hat{F}_L^{kl}\,
\frac{T^{(0)}_{[0]} T^{(0)}_{[0]} \overline{T^{(-1)}_{[0]} T^{(0)}_{[0]}}}{T^{(0)}_{[0]} \overline{T^{(-1)}_{[0]}} - T^{(-1)}_{[0]} \overline{T^{(0)}_{[0]}}} \\
&-\frac{1}{4}\: \epsilon_{ijkl}\: \xi^i\, \xi_a \: \hat{F}_L^{aj}\: \hat{F}_R^{kl}\,
\frac{T^{(-1)}_{[0]} T^{(0)}_{[0]} \overline{T^{(0)}_{[0]} T^{(0)}_{[0]}}}{T^{(0)}_{[0]} \overline{T^{(-1)}_{[0]}} - T^{(-1)}_{[0]} \overline{T^{(0)}_{[0]}}} \:. \label{ord2fin}
\end{align}
More precisely, the contributions~\eqref{ord11} and~\eqref{ord12} are obtained by a
first order perturbation calculation for~$\Delta A = \Delta P(x,y)\, P(y,x) + P(x,y)\, \Delta P(y,x)$
with~$\Delta P$ being quadratic in the field strength.
The terms~\eqref{ord21} arises in a first order perturbation calculation for~$\Delta A
= \Delta P(x,y)\, \Delta P(y,x)$ with~$\Delta P$ being linear in the field strength.
The remaining contributions~\eqref{ord22}--\eqref{ord2fin} are the result of a
second order perturbation calculation for~$\Delta A = \Delta P(x,y)\, P(y,x) + P(x,y)\, \Delta P(y,x)$
with~$\Delta P$ being linear in the field strength.

Now Lemma~\ref{lemmaFFT} follows by a straightforward calculation using~\eqref{l:Kncdef}.
It is worth noting that the contribution~\eqref{ord22} can be written as
\[ i c \, \lambda_+ \qquad \text{with} \qquad
c = \frac{i}{36}\: \Big( \hat{F}^L_{ki}\: \hat{F}_L^{kj} + \hat{F}^R_{ki}\: \hat{F}_L^{kj} \Big) \,\xi^i\, \xi_j\:
\frac{T^{(0)}_{[0]} \overline{T^{(0)}_{[0]}}}{T^{(0)}_{[0]} \overline{T^{(-1)}_{[0]}} - T^{(-1)}_{[0]} \overline{T^{(0)}_{[0]}}} \in \R \]
and~$\lambda_+$ as in~\eqref{s:lpm}. Hence this contribution changes the eigenvalue only by a phase.
As a consequence, it drops out of~\eqref{l:Kncdef}. This explains why the formula
in Lemma~\ref{lemmaFFT} does not involve contributions~$\sim \hat{F}_L \hat{F}_L$
or~$\sim \hat{F}_R \hat{F}_R$ in which both field tensors have the same chirality.

\chapter{Ruling out the Local Axial Transformation} \label{s:applocaxial}
\sindex{transformation of the fermionic projector!local axial!ruling out this transformation}%
We saw in~\S\ref{s:secprobaxial} that the local axial transformation
yields a shear contribution to the closed chain which is of degree four on the light cone
(see~\eqref{s:Pcont} and~\eqref{s:Acont}). The resulting contribution~$\kappa$ to
the eigenvalues~$\lambda^{L\!/\!R}$ violates the EL equations
(see~\eqref{s:lambdak} and Figure~\ref{s:figabslambda}). We now analyze
the same mechanism for the general local transformation~\eqref{s:UPxy}.
We will find that only a very restrictive class of local transformations respects the
EL equations. In particular, this class involves does not include local axial transformations.

We first observe that the shear contributions appear only if the
two factors~$\xi$ and~$\overline{\xi}$ contained in~$P(x,y)$ and~$P(y,x)$
are not contracted to each other (because otherwise we get a
factor~$\la \xi, \overline{\xi} \ra$ which decreases
the degree on the light cone; see the calculation~\eqref{s:clcvac}--\eqref{s:lpm}).
Therefore, the factors~$\xi$ and~$\overline{\xi}$ are contracted to the local transformation.
Thus our task is to analyze how the operator~$\slashed{\xi}$ (and similarly~$\overline{\slashed{\xi}}$)
is affected by a local transformation.
Moreover, to the leading degree on the light cone, it suffices to fix a space-time point~$x$
and to consider the transformation~\eqref{s:UPxy} in the limiting case~$U(y)=U(x) \equiv U$.
Finally, as we are interested in the singularities on the light cone, we may clearly
disregard the smooth corrections in~\eqref{s:UPxy}.
This leads us to consider the transformation
\beq \label{s:xitrans}
\slashed{\xi} \rightarrow \acute{U} \slashed{\xi} \grave{U}^*\:.
\eeq
As~$U$ enters this formula only via~$\acute{U}$ and its adjoint, it is convenient to write
the generations in components,
\beq \label{s:Uadef}
\acute{U} = (U_1, \ldots, U_g) \:.
\eeq
Then the transformation~\eqref{s:xitrans} can be written as
\[ \slashed{\xi} \rightarrow \sum_{a=1}^g U_a \,\slashed{\xi}\, U_a^*\:. \]

In the formalism of the continuum limit, the vector~$\xi$ is null on the light cone, i.e.
\[ \Tr \big( \slashed{\xi}^2 \big) = 4\, \la \xi, \xi \ra = (\deg <0)\:. \]
Likewise, a general lightlike vector~$v$ satisfies the relation~$\Tr \slashed{v}^2=0$.
The next proposition specifies under which
assumptions on~$U$ this relation is preserved by the transformation~\eqref{s:xitrans}.

\begin{Prp} \label{s:prpC} 
Assume that for every light-like vector~$v$,
\beq \label{s:trzero}
\Tr \left( \big( \acute{U} \slashed{v} \grave{U}^* \big)^2 \right) = 0 \:.
\eeq
Then for all~$a, b \in \{1, \ldots, g\}$, there are real parameters~$\alpha$ and~$\beta$
such that
\beq \label{s:Uabrep}
U_a^* \,U_b = \alpha \1 + \beta \pseudo\:.
\eeq
Under the additional assumption that one of the operators~$U_a$ invertible,
there are complex parameters~$\alpha_1, \ldots, \alpha_g$ and~$\beta_1, \ldots, \beta_g$
as well as an invertible linear mapping~$A$ on the spinors at~$x$ such that
\beq \label{s:Uarep}
U_a = A \, (\alpha_a + \beta_a \, \pseudo)\:.
\eeq
\end{Prp}

For the proof we need a preparatory lemma.
We denote the space of four-component spinors at a given space-time point~$x$ by~$(V, \Sl .|. \Sr)$
(where~$\Sl .|. \Sr$ again denotes the spin scalar product~$\Sl \phi | \psi \Sr = \phi^\dagger \gamma^0 \psi$).
We say that a linear operator~$B$
on~$V$ is {\em{positive}} if
\[ \Sl \psi | B \psi \Sr \geq 0 \qquad \text{for all~$\psi \in V$}\:. \]

\begin{Lemma} \label{s:lemmaC1}
For every positive linear operator~$B$ on~$V$ the following implication holds:
\[ \Tr(B^2)=0 \quad \Longrightarrow \quad B^2 = 0\:. \]
\end{Lemma}
\Proof According to~\cite[Lemma~4.2]{discrete}, the zeros of the characteristic polynomial
of~$B$ are all real. Hence the assumption~$\Tr(B^2)=0$ implies that~$B$ is nilpotent.
Thus it remains to show that the Jordan chains of~$B$ have length at most two.

Let us assume conversely that~$B$ has a Jordan chain of length three. 
Then there is a spinor basis~$(\mathfrak{f}_\alpha)_{\alpha=1,\ldots, 4}$ in which~$B$
has the matrix representation
\beq \label{s:B3rep}
B = \begin{pmatrix} 0 & 1 & 0 & 0 \\ 0 & 0 & 1 & 0 \\ 0 & 0 & 0 & 0 \\ 0 & 0 & 0 & 0
\end{pmatrix} .
\eeq
The positivity of~$B$ clearly implies that~$B$ is symmetric with respect to
the spin scalar product. An elementary consideration shows that by
a suitable change of basis which respects~\eqref{s:B3rep}, one can arrange that
the signature matrix~$S$ defined by~$S_{\alpha \beta} = \Sl f_\alpha | f_\beta \Sr$
has the form
\[ S = \pm
\begin{pmatrix} 0 & 0 & 1 & 0 \\ 0 & 1 & 0 & 0 \\ 1 & 0 & 0 & 0 \\ 0 & 0 & 0 & -1
\end{pmatrix} \]
(for details see the proof of Lemma~4.4 in~\cite{continuum}).
A direct computation shows that the matrix~$SB$ has a negative eigenvalue,
in contradiction to the positivity of~$B$.

In the case that~$B$ has a Jordan chain of length four, we repeat the
last argument for the matrices
\[ B = \begin{pmatrix} 0 & 1 & 0 & 0 \\ 0 & 0 & 1 & 0 \\ 0 & 0 & 0 & 1 \\ 0 & 0 & 0 & 0
\end{pmatrix} \qquad \text{and} \qquad
S = \pm \begin{pmatrix} 0 & 0 & 0 & 1 \\ 0 & 0 & 1 & 0 \\ 0 & 1 & 0 & 0 \\ 1 & 0 & 0 & 0
\end{pmatrix} . \]
This concludes the proof.
\QED

\Proof[Proof of Proposition~\ref{s:prpC}] Choosing a future-directed lightlike vector~$v$, the bilinear
form~$\Sl .| \slashed{v}\, . \Sr$ is positive semi-definite. As a consequence, the
product~$\acute{U} \slashed{v} \grave{U}^*$ is a positive operator
on~$(V, \Sl .|. \Sr)$. In view of the assumption~\eqref{s:trzero}, we can apply Lemma~\ref{s:lemmaC1}
to conclude that
\beq \label{s:quzero}
(\acute{U} \slashed{v} \grave{U}^*)^2 = 0 \:.
\eeq

Assume that~$\Sl \psi | \slashed{v} \psi \Sr$ vanishes
for a given spinor~$\psi$. Then~$\psi$ lies in the null space of the inner product~$\Sl .| \slashed{v}\,. \Sr$.
Since the spin scalar product is non-degenerate,
it follows that~$\slashed{v} \psi=0$. We conclude that the following implication holds:
\beq \label{s:posimpl}
\Sl \psi | \slashed{v} \psi \Sr = 0 \quad \Longrightarrow \quad \slashed{v} \psi = 0 \:.
\eeq

We now apply this implication to~\eqref{s:quzero}. Choosing~$\psi =
\grave{U}^* \acute{U} \slashed{v} \grave{U}^* \chi$ with an arbitrary spinor~$\chi$,
the equation~\eqref{s:quzero} implies that~$\Sl \psi | \slashed{v} \psi \Sr = 0$.
Applying~\eqref{s:posimpl} and using that~$\chi$ is arbitrary, we obtain the equation
\[ \slashed{v} \,\grave{U}^* \acute{U} \,\slashed{v}\, \grave{U}^* = 0 \:. \]
Taking the adjoint of this equation and choosing~$\psi = \slashed{v} \grave{U}^* \acute{U} \slashed{v} \chi$,
we can again apply~\eqref{s:posimpl} to obtain
\[ \slashed{v} \,\grave{U}^* \acute{U}\, \slashed{v} = 0 \:. \]
Using the notation~\eqref{s:Uadef}, we obtain for all~$a,b \in \{1, \ldots, g\}$ the equations
\[ \slashed{v} \,U_a^* U_b\, \slashed{v} = 0 \:. \]
Multiplying from the left and right by chiral projectors, one sees that the even and odd components
of~$U_a U_b^*$ can be treated separately. A short calculation using the anti-commutation relations
shows that the operators~$U^*_a U_b$ can have only scalar
and pseudoscalar components. This proves~\eqref{s:Uabrep}.

To prove~\eqref{s:Uarep}, we assume that~$U_a$ is invertible. We set~$A = (U_a^*)^{-1}$.
Then for every~$b \in \{1, \ldots, g\}$, we can apply~\eqref{s:Uabrep} to obtain
\[ U_b = (A U_a^*) \,U_b = A \,(\alpha \1 + \beta \, \pseudo)\:. \]
Setting~$\alpha_b=\alpha$ and~$\beta_b=\beta$ gives the claim.
\QED

We now work out the consequences of Proposition~\ref{s:prpC}.
We first note that in the vacuum, $U(x)$ is the identity, and thus~$U_a = \1$ for all~$a$.
Therefore, we can assume that, at least for a weakly interacting system,
one of the matrices~$U_a$ is invertible. Thus Proposition~\ref{s:prpC} implies that
if the condition~\eqref{s:trzero} holds, then~$U(x)$ must have the representation~\eqref{s:Uadef}
and~\eqref{s:Uarep}.

The representation~\eqref{s:Uarep} means in words that the local transformation consists
of the transformation~$A$ which acts trivially on the generation index, and of additional
scalar and pseudoscalar transformations acting on the components of the generation index.
If we start with a local axial transformation~\eqref{s:Uexp} and add other potentials (like scalar, pseudoscalar, 
vector or bilinear components) plus possibly higher order contributions, we will never get
a transformation of the form~\eqref{s:Uarep}. Namely, it was essential for compensating
the logarithmic poles that the local axial transformation acts non-trivially on the generation index,
contrary to~\eqref{s:Uarep}. We conclude that local transformations of the form~\eqref{s:Uadef}
and~\eqref{s:Uarep} do not include local axial transformations.

This consideration also shows that if a local axial transformation is present, then the condition~\eqref{s:trzero}
is necessarily violated at some space-time point~$x$ for some lightlike vector~$v$.
The leading contribution to the closed chain~\eqref{s:clcvac} transforms to
\beq \label{s:Axyloc}
A_{xy} \asymp \frac{1}{4}\: \acute{U}(x) \slashed{\xi} \grave{U}(y)\: \acute{U}(y) \overline{\slashed{\xi}} \grave{U}(x)
\: \big| T^{(-1)}_{[0]} \big|^2 \:.
\eeq
Choosing a point~$y$ in a small neighborhood of~$x$ such that the difference vector~$\xi=y-x$ is a
multiple of~$v$, we obtain a shear contribution to the closed chain.
This shows that the formalism of the continuum limit is no longer valid. In particular, we get contributions
to the eigenvalues of the closed chain which are more singular on the light cone than in the vacuum.

The above argument does not necessarily imply that the EL equations are violated. Namely,
it leaves the possibility that the gauge phases enter the shear contributions
in agreement with~\eqref{s:Pchiral}, in which case the argument of Figure~\ref{s:figabslambda}
would not apply and the eigenvalues~$\lambda^{L\!/\!R}_\pm$ would still have the same absolute
value. In order to rule out this case, we now take into account the gauge phases and
restrict attention to those contributions to the closed chain for which these phases drop out.

We first point out that the gauge phases enter the factors~$\xi$ in~\eqref{s:Axyloc} only
via the transformation~\eqref{s:Pchiral}. The local transformation, however, cannot involve
any gauge phases, because the gauge phases are obtained as integrals along
a straight line joining two space-time points~$x$ and~$y$ (see~\eqref{s:Lambda})
and thus cannot be encoded in a function~$U(x)$ of one variable.
Therefore, to select the contributions to the closed chain which do not involve gauge phases,
we simply take those contributions to~\eqref{s:Axyloc} for which both factors~$\xi$ have the same chirality,
\[ A_{xy} \asymp \frac{1}{4}\: \sum_{c=L,R}
\acute{U}(x) \,\chi_c \,\slashed{\xi}\, \grave{U}(y)\: \acute{U}(y)\, \chi_c \,\overline{\slashed{\xi}} \,\grave{U}(x)
\: \big| T^{(-1)}_{[0]} \big|^2 \:. \]
Again considering the limiting case~$y=x$, we are led to the condition that the equation
\[ \sum_{c=L,R} \Tr \left( \big( \acute{U} \chi_c \slashed{v}\, \grave{U}^* \big)^2 \right) = 0 \]
should hold for every lightlike vector~$v$. As the operator~$\acute{U} \,\chi_c \slashed{v}\, \grave{U}^*$
is positive, the trace of its square is necessarily positive. Hence both summands must vanish separately,
\[ \Tr \left( \big( \acute{U} \chi_c \slashed{v}\, \grave{U}^* \big)^2 \right)
\qquad \text{for~$c=L$ or~$R$}\:. \]
We now combine the chiral projectors with the local transformation to obtain
(for example in the case~$c=L$),
\[ 0 = \Tr \left( \big( \acute{U} \,\chi_L \slashed{v}\, \grave{U}^* \big)^2 \right)
= \Tr \left( \big( (\acute{U} \chi_L) \,\slashed{v}\, (\acute{U} \chi_L)^* \big)^2 \right) . \]
For the new local transformation~$\acute{U} \chi_L$, the components~$U_a \chi_L$
are clearly not invertible, so that~\eqref{s:Uarep} no longer applies.
But we can still apply the first part of Proposition~\ref{s:prpC} to obtain
\[ \chi_R \:U_a^* \,U_b\: \chi_L = \alpha \1 + \beta \pseudo \:. \]
This equation is satisfied if and only if the product~$U_a^* U_b$ is even, i.e.
\[  [\pseudo,\,U_a^* \,U_b] = 0 \:. \]
This condition is considerably weaker than the representation~\eqref{s:Uarep}. In particular,
it is indeed fulfilled for special choices of~$U_a$ which do involve axial fields.
But we can nevertheless rule out local axial transformations in a perturbation expansion
around the vacuum. Namely, inserting the perturbation ansatz
\[ U_a = \1 + i E_a + \O(E_a^2) \:, \]
we obtain the condition
\[ [\pseudo,\,E_a^* - E_b] = 0 \:. \]
Thus the odd contribution to~$E_a$ must be of the form
\[ E_a = i (\slashed{v} + \pseudo\, \slashed{u}) + \text{(even)} \qquad \text{for all~$a = 1, \ldots, g$} \]
with two vector fields~$u$ and~$v$. In particular, the axial transformation acts trivially on
the generation index, making it impossible to compensate the logarithmic poles on the light cone.

\chapter{Resummation of the Current and Mass Terms at the Origin} \label{s:appresum}
\sindex{resummation!of smooth contributions}%
\sindex{resummation!of current and mass terms}%
As pointed out in~\S\ref{s:sec44}, the distribution~$T_a$ is {\em{not}} a power series in~$a$,
and thus it cannot be expanded in a Taylor series around~$a=0$ (see~\eqref{s:Taser}
and the explanation thereafter). The method of subtracting suitable counter terms~\eqref{s:Tacounter}
has the shortcoming that the subsequent calculations are valid only {\em{modulo smooth contributions}}
on the light cone. This method is suitable for analyzing the singularities on the light cone,
but it is not sufficient when smooth contributions to the fermionic projector become important
(cf.\ the discussion after~\eqref{s:multdist} and the beginning of~\S\ref{s:secfield1}).
We now present a convenient method for computing the smooth contributions to the fermionic
projector.
Our method is based on the resummation technique developed in~\cite[Section~4]{firstorder}
and is outlined as follows. We first perform the mass expansion not around zero mass,
but around a given mass parameter~$a>0$. Then, according to~\eqref{s:Taser}, the distribution~$T_a$
is smooth in~$a$, and we may set
\beq \label{s:Tandef}
T_a^{(n)} = \left( \frac{d}{da} \right)^n T_a \:.
\eeq
\nindex{fa0@$T_a^{(n)}$ -- term of mass expansion for~$a>0$}%
Adapting the method of the light-cone expansion, we can express any Feynman
tree diagram as a sum of terms of the form
\beq \label{s:Pla}
P^\sea(x,y) = \sum_{n=-1}^\infty
\sum_{k} m^{p_k} {\text{(phase-inserted nested line integrals)}} \times T_a^{(n)}(x,y) \:,
\eeq
where for each~$n$, the $k$-sum is finite, whereas the $n$-sum is to be understood
as a formal power series. Note that, in contrast to the series in~\eqref{s:fprep}, the infinite sum
in~\eqref{s:Pla} is {\em{not}} a light-cone expansion in the sense of Definition~\ref{s:l:def1},
because the distributions~$T^{(n)}_a$ all involve smooth contributions
and are thus only of the order~$\O((y-x)^0)$.
We proceed by partially carrying out the series in~\eqref{s:Pla} to obtain explicit
smooth contributions on the light cone. After this resummation has been performed,
we recover~\eqref{s:fprep}, but now with an explicit formula for~$\tilde{P}^\lec(x,y)$.

For simplicity, we develop the method only for the contributions to the fermionic projector
needed in this book: the vector and axial components of the fermionic projector
perturbed by chiral potentials to first order. But the method generalizes in a straightforward
way to arbitrary Feynman tree diagrams. Furthermore, we begin by considering a single
Dirac sea (the generalization to several generalizations will then be straightforward;
see the proof of Lemma~\ref{s:lemmasmooth} below). We thus consider the contribution to the
fermionic projector
\beq \label{s:Pfirst}
\Delta P = - s_m (\chi_L \slashed{A}_R + \chi_R \slashed{A}_L) t_m
- t_m (\chi_L \slashed{A}_R + \chi_R \slashed{A}_L) s_m
\eeq
with the spectral projector~$t_m$ and the Green's function~$s_m$ as in~\eqref{s:smdef1}.
In order to concentrate on the vector and axial components, we want to consider the
expression~$\Tr (\slashed{\xi} \chi_L\, \Delta P(x,y))$, being a well-defined distribution.
The singular part of this distribution on the light cone can be computed
by inserting the formulas of the light-cone expansion~\eqref{s:Pgag}--\eqref{s:Pendgag}
and using the contraction rule
\beq \label{s:simplecontract}
\xi^2\: T^{(n)}(x,y) = -4 n  T^{(n+1)}(x,y) + \text{(smooth contribution)}\:,\qquad n \in \{-1,0\}
\eeq
(which is immediately verified from the explicit formulas~\eqref{s:Taser}--\eqref{s:Tm1def}). We thus obtain
\begin{align}
\frac{1}{2}\, \Tr &\left( \slashed{\xi}\: \chi_L\: \Delta P(x,y) \right) = 
2 \int_x^y \xi_k\, A_L^k \: T^{(0)}_{[0]}(x,y) \label{s:P1gauge} \\
&-2 \int_x^y (\alpha-\alpha^2)\: \xi_k\, j_L^k \: T^{(1)}_{[0]}(x,y)  + m^2 \int_x^y \xi_k \left(A_L^k - A_R^k \right) \: T^{(1)}_{[2]}(x,y) \label{s:P1curmass} \\
&+\: \xi^k \,f_k(x,y) + (\deg < 0)\:, \label{s:fkdef}
\end{align}
\nindex{fa2@$f(x,y)$ -- smooth contribution to~$\chi_L P(x,y)$}%
where we added subscripts~$[.]$ in order to indicate how these factors are to
be regularized (although we do not need a regularization at this point),
and~$f_k(x,y)$ are yet undetermined smooth functions
(clearly, the summand in~\eqref{s:P1gauge} is the gauge term as
discussed in~\S\ref{s:sec72}, whereas the summands in~\eqref{s:P1curmass}
correspond to the current and mass terms considered in~\S\ref{s:sec81}). Our goal is to compute
the functions~$f_k(x,y)$ at the origin~$x=y$.

Our first step is to perform a mass expansion of the Feynman diagram~\eqref{s:Pfirst}
around a given~$a \neq 0$. To this end, we need suitable calculation rules which are
derived in the next lemma.
\begin{Lemma} The distributions~$T_a^{(n)}$, \eqref{s:Tandef}, satisfy for all~$n \in \N_0$ the calculation rules
\begin{align}
(-\Box_x - a) \,T^{(n)}_a(x,y) &= n\, T^{(n-1)}_a(x,y) \label{s:ra1} \\
\frac{\partial}{\partial x^k}\: T^{(n+1)}_a(x,y) &= \frac{1}{2}\: \xi_k\: T_a^{(n)}(x,y)
\label{s:ra2} \\
\xi^2\: T^{(n)}_a(x,y) &= -4 n  T^{(n+1)}_a(x,y) - 4 a  T^{(n+2)}_a(x,y) \:. \label{s:ra3}
\end{align}
In the case~$n\!=\!-1$, the rule~\eqref{s:ra2} can be used to define the
distribution~$\xi_k T_a^{(-1)}$.
Using this definition, the rule~\eqref{s:ra3} also holds in the case~$n\!=\!-1$.
\end{Lemma}
\Proof The relations~\eqref{s:ra1} and~\eqref{s:ra2} were already derived
in~\cite{light} (see~\cite[eqs~(3.5) and~(3.6)]{light}). For self-consistency we here repeat the proof.
Clearly, $T_a$ is a distributional solution of the Klein-Gordon equation,
\[ (-\Box_x-a)\, T_a(x,y) = 0 \:. \]
Differentiating~$n$ times with respect to~$a$ gives~\eqref{s:ra1}.
Next, we differentiate the identity in momentum space
\[ T_a(p) = \delta(p^2-a) \: \Theta(-p^0) \]
with respect to~$p_k$ to obtain
\[ \frac{\partial}{\partial p^k} T_a(p) = 2 p_k\: T_a^{(1)}(p)\:. \]
Using that differentiation in momentum space corresponds to multiplication
in position space and vice versa, we find
\[ \xi_k \: T_a(x,y) = 2 \frac{\partial}{\partial x^k} T_a^{(1)}(x,y)\:. \]
Differentiating~$n$ times with respect to~$a$ gives~\eqref{s:ra2}.

To derive~\eqref{s:ra3}, we first combine~\eqref{s:ra2}
with the product rule to obtain
\[ \Box_x T_a^{(1)} = \partial_x^k \left( \frac{1}{2}\, \xi_k T_a^{(0)} \right) 
= -2 T_a^{(0)} + \frac{1}{2}\, \xi_k\:\partial_x^k T_a^{(0)} 
= -2 T_a^{(0)} + \frac{1}{4}\, \xi^2\:T_a^{(-1)} \:. \]
On the other hand, we know from~\eqref{s:ra1} that
\[ \Box_x T_a^{(1)} = - T^{(0)}_a - a T^{(1)}_a \:. \]
Solving for~$\xi^2\:T_a^{(-1)}$, we obtain
\[ \xi^2\:T_a^{(-1)} = 4 T_a^{(0)} - 4 a T^{(1)}_a \:. \]
We finally differentiate this relation $n+1$ times with respect to~$a$, giving~\eqref{s:ra3}.
\QED
Alternatively, this lemma could be proved working with the series represen\-ta\-tion~\eqref{s:Taser}.
We also remark that in the case~$n\!=\!-1$, the rule~\eqref{s:ra2} is consistent with our earlier
definition~\eqref{s:Tm1def}.

Using the relations~\eqref{s:ra1} and~\eqref{s:ra2}, the 
mass expansion of the first order Feynman diagram~\eqref{s:Pfirst} was first
performed in~\cite{firstorder} (see~\cite[Theorem~3.3]{firstorder}, where the
mass expansion is referred to as the ``formal light-cone expansion'').
More generally, for the advanced and retarded Green's function, we have the
expansion
\beq \begin{split}
(S&^{(l)}_a \:V\: S^{(r)}_a)(x,y) \\
&= \sum_{n=0}^\infty 
        \frac{1}{n!} \int_0^1 \alpha^{l} \:(1-\alpha)^{r} \:
        (\alpha - \alpha^2)^n \: (\Box^n V)_{|\alpha y + (1-\alpha) x} \:d\alpha \;
        S^{(n+l+r+1)}_a(x,y) \:,
\end{split} \label{s:Slight}
\eeq
which is proved exactly as in the case~$a=0$ (see~\cite[Lemma~2.1]{light}, \cite[Lemma~2.5.2]{PFP}
or Lemma~\ref{l:lemma1}). The residual argument (cf.~\cite[Section~3.1]{light})
also generalizes immediately to the case~$a>0$, making it possible to
extend~\eqref{s:Slight} to the so-called {\em{residual
fermionic projector}} (the non-residual part of the fermionic projector is
precisely the non-causal high energy contribution, which can be analyzed
as indicated in~\S\ref{s:sechighorder}; for details see~\cite[Section~3.2]{light}).
Applied to our problem, we obtain the expansion
\beq \label{s:Tlight}
(S_a \:V\: T_a + T_a \:V\: S_a)(x,y)
= \sum_{n=0}^\infty \frac{1}{n!} \int_x^y
(\alpha - \alpha^2)^n \: (\Box^n V) \:d\alpha \; T^{(n+1)}_a(x,y) \:,
\eeq
where for the line integrals we again used the short notation~\eqref{s:shortline},
and~$S_a$ is the symmetric Green's function~\eqref{s:Samom}.
The mass expansion of~\eqref{s:Pfirst} is now readily obtained by
applying the differential operators~$(i \Pdd+m)$ and simplifying the Dirac
matrices using the rules~\eqref{s:ra2} and~\eqref{s:ra3}.
Multiplying by~$\slashed{\xi} \chi_L$ and taking the trace, a
straightforward calculation using again~\eqref{s:ra3} yields
\begin{align}
\frac{1}{2} \, \Tr \Big( &\slashed{\xi}\:\chi_L P(x,y) \Big) \;\asymp\; 2 \sum_{n=0}^\infty \frac{1}{n!}
\int_x^y (\alpha-\alpha^2)^n\: \xi_k\, \Big(\Box^n A_L^k \Big) \: T_{m^2}^{(n)}(x,y) \label{s:LP1} \\
&-2  \sum_{n=0}^\infty \frac{1}{n!} \int_x^y (2 \alpha-1)\: (\alpha-\alpha^2)^n\,
\Big( \Box^n \partial_i A_L^i \Big)\: T_{m^2}^{(n+1)}(x,y) \\
&- m^2 \sum_{n=0}^\infty \frac{1}{n!}
\int_x^y (\alpha-\alpha^2)^n\: \xi_k\, \Big(\Box^n A_L^k+ \Box^n A_R^k \Big) \: T_{m^2}^{(n+1)}(x,y) \\
&+2  m^2 \sum_{n=0}^\infty \frac{1}{n!} \int_x^y (2 \alpha-1)\: (\alpha-\alpha^2)^n\,
\Big( \Box^n \partial_i A_L^i \Big)\: T_{m^2}^{(n+2)}(x,y) \label{s:LP4}
\end{align}
(this result was again obtained with the help of  {\textsf{class\_commute}};
see page~\pageref{s:classcommute}).
Integrating the line integrals by parts,
\[ \int_x^y (2 \alpha-1)\: (\alpha-\alpha^2)^n\, \Big( \Box^n \partial_i A_L^i \Big) = 
\frac{1}{n+1} \int_x^y (\alpha-\alpha^2)^{n+1}\, \xi^k
\Big( \Box^n \partial_{ik} A_L^i \Big) , \]
the divergence terms can be rewritten to recover the chiral currents.
In particular, in the case~$A_L=A_R$ of a vector potential,
one immediately verifies that~\eqref{s:LP1}--\eqref{s:LP4} has the correct behavior
under gauge transformations. Furthermore, one readily sees that
the expansion~\eqref{s:LP1}--\eqref{s:LP4} is compatible with~\eqref{s:P1gauge}--\eqref{s:fkdef}
in the sense that the singularities on the light cone coincide.
We now subtract~\eqref{s:LP1}--\eqref{s:LP4}
from~\eqref{s:P1gauge}--\eqref{s:fkdef} and solve for~$\xi^k f_k(x,y)$.

In order to compute~$f_k(x,x)$, it suffices to take into account the
constant counter term in~\eqref{s:Tacounter}, as can be done by the replacement 
\beq \label{s:Ndef}
T_a(x,y)  \longrightarrow T^\reg_a + N(a) 
\qquad \text{with} \qquad N(a):= \frac{1}{32 \pi^3}\: a \log |a| \:,
\eeq
\nindex{fa4@$N(a)$ -- smooth contribution to~$T_a$}%
and similarly for the $a$-derivatives. Moreover, in the line integrals we may set~$x=y$.
In order to keep the formulas simple, we also specialize to the situation
where {\em{only the axial potential}} in~\eqref{s:axialdef} is present.
We thus obtain
\begin{align*}
f^k&(x,x) =
2 \sum_{n=0}^\infty \frac{1}{n!}
\int_0^1 (\alpha-\alpha^2)^n\: \Big(\Box^n A_\text{\rm{a}}^k(x) \Big) \: N^{(n)}(m^2)\: d\alpha \\
&-2 \sum_{n=0}^\infty \frac{1}{(n+1)!} \int_0^1 (\alpha-\alpha^2)^{n+1}
\Big( \Box^n \partial^k_{\;\,i} A_\text{\rm{a}}^i(x) \Big)\: 
\left( N^{(n+1)}(m^2) - m^2 N^{(n+2)}(m^2) \right) d\alpha \:.
\end{align*}
By linearity, it suffices to consider the case that~$A_\text{\rm{a}}$ is a plane wave of momentum~$q$,
\[ A^k_\text{\rm{a}}(z) = \hat{A}^k_\text{\rm{a}}\: e^{-i q (z-x)} \:. \]
Then the above sums are recognized as Taylor series,
\beq \label{s:taylor}
\begin{split}
\sum_{n=0}^\infty \frac{\lambda^n}{n!} \:N^{(n)}(m^2)
&= N^{(0)}(m^2+\lambda) \\
\sum_{n=0}^\infty \frac{\lambda^n}{(n+1)!}\: N^{(n+\ell)}(m^2)
&=  \frac{N^{(\ell-1)} (m^2 +\lambda)- N^{(\ell-1)}(m^2)}{\lambda} \:,
\end{split}
\eeq
where we introduced the abbreviation~$\lambda = -(\alpha - \alpha^2)\, q^2$
(the second equation in~\eqref{s:taylor} can be derived from the first by integration
over~$\lambda$). We thus obtain
\nindex{fa4@$N(a)$ -- smooth contribution to~$T_a$}%
\nindex{fa2@$f(x,y)$ -- smooth contribution to~$\chi_L P(x,y)$}%
\begin{align*}
f^k(x,x) =\;& 2 
\int_0^1 A_\text{\rm{a}}^k(x)\: N \Big(m^2 + \lambda \Big) d\alpha \\
&- \frac{2}{\lambda} \int_0^1 (\alpha-\alpha^2)\: \partial^k_{\;\,i} A_\text{\rm{a}}^i(x)\:
\left. \Big( N^{(0)} - m^2 N^{(1)} \Big) (m^2+\nu)
\right|_{\nu=0}^{\nu=\lambda}\: d\alpha \\
=\;& \frac{1}{16 \pi^3}
\int_0^1 A_\text{\rm{a}}^k(x)\; \big(m^2+\lambda \big) \:\log\big|m^2+\lambda \big| \:d\alpha \\
&-\frac{1}{16 \pi^3} \int_0^1 (\alpha-\alpha^2)\: \partial^k_{\;\,i} A_\text{\rm{a}}^i(x)\,
\log \big| m^2+\lambda \big| \: d\alpha \:,
\end{align*}
where in the last step we substituted the explicit formula for~$N(a)$ in~\eqref{s:Ndef}.
In order to rewrite the last result in terms of the axial current, we use
the identity~$\lambda A^k_\text{\rm{a}} = (\alpha-\alpha^2) \Box A^k_\text{\rm{a}}$ to conclude
\begin{align*}
f^k(x,x) =\;& \frac{m^2}{16 \pi^3}\: A_\text{\rm{a}}^k(x)
\int_0^1 \log\Big| m^2 - (\alpha-\alpha^2) q^2 \Big| \:d\alpha \\
&-\frac{1}{16 \pi^3}\:  j_\text{\rm{a}}^k(x) \int_0^1 (\alpha-\alpha^2)\:
\log \Big| m^2 - (\alpha-\alpha^2) q^2 \Big| \: d\alpha \:.
\end{align*}
Substituting this result into the light-cone expansion~\eqref{s:P1gauge}--\eqref{s:fkdef}
evaluated at the origin and using that~$\int_0^1 (\alpha-\alpha^2)=1/6$,
one sees that the term~$\xi_k f^k(x,x)$ can be incorporated into the formulas of the
light-cone expansion by the replacements
\beq \label{s:T1rep}
\left. \begin{split}
T^{(1)}_{[0]} &\rightarrow T^{(1)}_{[0]}
+ \frac{\log (m^2)}{32 \pi^3}
+\frac{6}{32 \pi^3} \int_0^1 (\alpha-\alpha^2)\:
\log \Big| 1 - (\alpha-\alpha^2) \,\frac{q^2}{m^2} \Big| \: d\alpha \\
T^{(1)}_{[2]} &\rightarrow T^{(1)}_{[2]}
+ \frac{\log (m^2)}{32 \pi^3}
+\frac{1}{32 \pi^3}
\int_0^1 \log \Big| 1 - (\alpha-\alpha^2) \,\frac{q^2}{m^2} \Big| \:d\alpha\:.
\end{split} \qquad \right\}
\eeq

To clarify the above construction, we point out that the radius of convergence of the
Taylor series in~\eqref{s:taylor} is~$|\lambda|=m^2$. Thus in the case~$|\lambda|>m^2$,
these series do {\em{not}} converge absolutely, so that~\eqref{s:taylor} can be understood
only on the level of formal Taylor series. For the reader who feels uncomfortable with
formal power series, we remark that all formal expansions could be avoided by
regularizing the distribution~$T_a$ according to~\eqref{s:Tacounter} {\em{before}}
performing the light-cone expansion, making a later resummation unnecessary.
However, this method seems technically complicated and has not yet been carried
out (see also the discussions in~\cite[Section~3.3]{light} and after~\eqref{s:DelT}).
In this book, we will be content with the formal character of~\eqref{s:taylor}.

We are now ready to prove the main result of this appendix.

\Proof[Proof of Lemma~\ref{s:lemmasmooth}]
We return to the situation with three generations and a general axial potential~$A_\text{\rm{a}}(z)$. 
As the axial potential is diagonal on the generation index, the auxiliary fermionic projector
splits into the direct sum of three fermionic projectors, corresponding to the Dirac seas
of masses~$m_1$, $m_2$, and~$m_3$. Thus the sectorial projection~\eqref{s:pt} reduces to a sum
over the generation index.
Decomposing~$A_\text{\rm{a}}$ into Fourier modes,
\[ A_\text{\rm{a}}(z) = \int_\scrM \frac{d^4z}{(2 \pi)^4}\: \hat{A}_\text{\rm{a}}(q)\: e^{-i q (z-x)}\:, \]
for every~$\hat{A}_\text{\rm{a}}(q)$ and for every generation we may apply the replacement
rules~\eqref{s:T1rep}. Rewriting the multiplication in momentum space by a convolution in
position space gives the formulas~\eqref{s:s0def}--\eqref{s:fb2def}.

In order to check the prefactors, it is convenient to verify whether the
arguments of the logarithms can be combined to give dimensionless quantities.
This is indeed the case with the expressions
\begin{align*}
\log |\xi^2| + \frac{1}{3} \sum_{\beta=1}^3 \log(m_\beta^2)
&= \frac{1}{3} \sum_{\beta=1}^3 \log \left| m_\beta^2 \xi^2 \right| \\
\log |\xi^2| + \frac{1}{m^2 \acute{Y} \grave{Y}} \sum_{\beta=1}^3 m_\beta^2 \log(m_\beta^2)
&= \frac{1}{m^2 \acute{Y} \grave{Y}} \sum_{\beta=1}^3 m_\beta^2 \log |m_\beta^2 \xi^2|\:,
\end{align*}
explaining the prefactors in~\eqref{s:s0def} and~\eqref{s:s2def} relative to those in~\eqref{s:logpole2}.

We finally need to verify that the smooth contributions which were disregarded in the
formalism of~\S\ref{s:sec51} really enter the EL equations according to the simple replacement
rules~\eqref{s:T1rep}. The subtle point is that the contraction rule in the continuum limit~\eqref{s:eq54}
is not the same as the corresponding distributional identity~\eqref{s:simplecontract}, and this
might give rise to additional terms which are not captured by~\eqref{s:T1rep}.
Fortunately, such additional terms do not appear, as the following consideration shows:
To degree four on the light cone, the smooth contributions to~$P(x,y)$ enter the EL equations
only if the smooth term is contracted with a factor~$\slashed{\xi}$ without generating a factor~$\xi^2$
(the contributions involving~$\xi^2$ are of degree three on the light cone). Thus for the smooth
contributions, the contraction rule~\eqref{s:eq54} is not applied, and therefore it could here
be replaced by the simpler distributional identities~\eqref{s:simplecontract} and~\eqref{s:ra3}.
\QED

We finally carry out the $\alpha$-integrals in~\eqref{s:T1rep} in closed form
and discuss the result. This result will not be used in this book. But it is nevertheless worth stating,
because it gives more explicit information on the structure of the non-causal correction terms.
\begin{Lemma} \label{s:lemmaclosed}
The functions~$\hat{f}^\beta_{[p]}$ defined by~\eqref{s:fb0def} and~\eqref{s:fb2def}
can be written as
\beq \label{s:gpdef}
\hat{f}^\beta_{[p]}(q) = \lim_{\varepsilon \searrow 0} g_{[p]} \Big( \frac{q^2 + i \varepsilon}{4 m_\beta^2} \Big) ,
\eeq
\nindex{ch6@$\hat{f}^\beta_{[p]}$ -- Fourier transform of~$f^\beta_{[p]}$}%
where the functions~$g_{[p]}(z)$ are defined in the upper half plane by
\begin{align*}
g_{[0]}(z) &= -\frac{3+5z}{3z} + \frac{1+z-2z^2}{2 z \,\sqrt{z (z-1)}}
\left[ \log \Big( 1-2z+\sqrt{z (z-1)} \Big) - i \pi \,\Theta(z-1) \right] \\
g_{[2]}(z) &= -2 - \frac{\sqrt{z(z-1)}}{z}
\left[ \log \Big( 1-2z+\sqrt{z (z-1)} \Big) - i \pi \,\Theta(z-1) \right] ,
\end{align*}
where the logarithm in the complex plane is cut along the ray~$-i \R^+$ (and~$\Theta$
is the Heaviside function, extended continuously to the upper half plane).
\end{Lemma}
\Proof Writing the logarithm of the absolute value for any~$x \in \R$ as
\[ \log | 1 - x | = \lim_{\delta \searrow 0} \Big(
\log \big(1 -  (x+i \delta) \big) + i \pi \,\Theta \big( (x+i \delta)-1 \big) \Big) \:, \]
we obtain the representation~\eqref{s:gpdef} with
\begin{align*}
g_{[0]}(z) &= 6 \int_0^1 (\alpha-\alpha^2) \Big(
\log \big( 1 - 4 (\alpha-\alpha^2) z \big) + i \pi \,\Theta \big( 4 (\alpha-\alpha^2) z-1 \big) \Big) \,d\alpha \\
g_{[2]}(z) &= \int_0^1 \Big(
\log \big(1 - 4 (\alpha-\alpha^2) z \big) + i \pi \,\Theta \big(4 (\alpha-\alpha^2) z-1 \big) \Big) \,d\alpha\;.
\end{align*}
It remains to calculate these integrals for~$z$ in the upper half plane, thus avoiding the
singularities on the real line. The term involving the Heaviside function is readily computed in closed form.
Thus it remains to consider for~$\ell=0,1$ the integrals
\[ \int_0^1 (\alpha-\alpha^2)^\ell \,\log \Big( 1 - 4 (\alpha-\alpha^2)\, z \Big) \,d\alpha 
= \frac{1}{2} \int_0^1 \log \Big( 1 - x z \Big)
\left\{ \left(\frac{x}{4} \right)^\ell  \frac{1}{\sqrt{1-x}} \right\} dx \:, \]
where in the last step we transformed to the integration variable~$x:=4(\alpha-\alpha^2)$.
After computing the indefinite integral of the expression inside the curly brackets,
we can integrate by parts.
Then the logarithm in the integrand disappears, and the calculation
of the integral becomes elementary.
\QED

\begin{figure}
\begin{center}
\includegraphics[width=8cm]{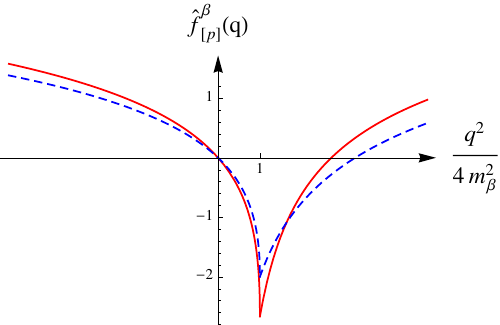}
\caption{The functions~$\hat{f}^\beta_{[0]}$ (red, solid) and~$\hat{f}^\beta_{[2]}$ (blue, dashed)}
\label{s:fig3}
\end{center}
\end{figure}

In Figure~\ref{s:fig3} the functions~$\hat{f}^\beta_{[0]}$ and~$\hat{f}^\beta_{[2]}$ are
plotted.
One sees that these functions attain their minimum if~$q^2 = 4 m_\beta^2$, and for this
value of~$q^2$ the function has a cusp. The asymptotics for large~$|q^2|$ is obtained
by dropping the summand one in the argument of the logarithm in~\eqref{s:fb0def}
and~\eqref{s:fb2def},
\begin{align*}
\hat{f}^\beta_{[0]}(q) &\sim 6\, \int_0^1 (\alpha-\alpha^2)\: \log \bigg| (\alpha-\alpha^2)
\,\frac{q^2}{m_\beta^2} \bigg| \, d\alpha = -\frac{5}{3} + \log \Big( \frac{q^2}{m_\beta^2} \Big) \\
\hat{f}^\beta_{[2]}(q) &\sim \int_0^1 \log \bigg| (\alpha-\alpha^2) \,\frac{q^2}{m_\beta^2} \bigg|
\, d\alpha  = -2 + \log \Big( \frac{q^2}{m_\beta^2} \Big) \,,
\end{align*}
revealing a logarithmic divergence as~$q^2 \rightarrow \pm \infty$.
For small momenta, the functions have the asymptotics
\[ \hat{f}^\beta_{[0]}(q) = -\frac{q^2}{5 m_\beta^2} + \O(q^4)\:,\qquad
\hat{f}^\beta_{[2]}(q) = -\frac{q^2}{6 m_\beta^2} + \O(q^4) \:, \]
describing a non-trivial low energy effect.

\chapter{The Weight Factors~$\rho_\beta$} \label{s:apprho}
In~\cite{reg} the ansatz for the vacuum~\eqref{s:A} was generalized by introducing
so-called {\em{weight factors}}~$\rho_\beta$ for the Dirac seas,
\[ P(x,y) = \sum_{\beta=1}^g \rho_\beta \int \frac{d^4k}{(2 \pi)^4}\: (\slashed{k} +m_\beta)\:
\delta(k^2-m_\beta^2)\: \Theta(-k^0)\: e^{-ik(x-y)}\:. \]
\sindex{weight factor}%
\nindex{fb0@$\rho_\beta$ -- weight factor}%
This generalization turns out to be useful when considering an action principle for the
masses of Dirac particles~\cite{vacstab}; for a physical discussion see~\cite[Appendix~A]{reg}.
All the constructions in this book could immediately be extended to the setting with
weight factors, as we now explain.

The weight factors are introduced into the auxiliary fermionic projector of the vacuum~\eqref{s:Pauxvac}
by the replacement
\[ \bigoplus_{\beta=1}^g  \;\rightarrow\; \bigoplus_{\beta=1}^g \rho_\beta\:. \]
For our systems, the causality compatibility condition (see~\cite[eq.~(A.1)]{reg} or~\eqref{l:ccc}) does not
cause problems, because all our potentials are either diagonal in the generation index,
or else they can be described by a microlocal chiral transformation (see~\S\ref{s:secgennonloc}),
in which case the causality compatibility condition is irrelevant. We conclude that the causal
perturbation series as well as the light-cone expansion remain well-defined.
The weight factors are taken into account simply by inserting them into the resulting
formulas. More precisely, the number of generations is to be replaced by the sum of the weights,
\[ g \;\rightarrow\; \sum_{\beta=1}^g \rho_\beta \:. \]
Moreover, the weights must be introduced into the sectorial projections by the replacements
\[ \hat{Y} \rightarrow \sum_{\alpha, \beta=1}^g \rho_\alpha Y^\alpha_\beta\:,\qquad
 \underbrace{\acute{Y} Y \cdots \grave{Y}} _{\text{$p$ factors~$Y$}} \;\rightarrow 
\sum_{\alpha,\beta, \gamma_1, \ldots, \gamma_{p-1}=1}^g \!\!\!\!\!\!\!\! \rho_\alpha\:
Y^\alpha_{\gamma_1} \cdots Y^{\gamma_1}_{\gamma_2}
\cdots Y^{\gamma_{p-1}}_\beta \:, \]
or more generally using the rule
\[ \acute{B} \rightarrow \sum_{\beta=1}^g \rho_\beta \,B^\beta_. \]
(to avoid confusion, we note that, due to the causality compatibility condition,
the weight factors could just as well be inserted at the last instead of the first summation index).
When performing the microlocal chiral transformation, the weight factors must
be inserted in the obvious way into~\eqref{s:Pktrans}.
After these straightforward modifications, all our formulas and results remain valid.
It seems a promising strategy for the construction of realistic physical models to choose the fermion
masses and the weight factors according to state stable vacuum configurations
as exemplified in~\cite{vacstab}.

\chapter{The Regularized Causal Perturbation Theory with Neutrinos} \label{l:appA}
\sindex{causal perturbation expansion!regularized with neutrinos}%
\section{The General Setting}
For clarity, we begin with a single Dirac sea (i.e.\ with one direct summand of~\eqref{l:Paux}
or~\eqref{l:dirsum2}). Thus without regularization, the vacuum is described
as the product of the Fourier integral~\eqref{l:Pmdef} with a chiral asymmetry matrix,
\beq \label{l:Xdefn}
P = X t \qquad \text{with~$t=P_m$ and~$X = \1, \chi_L$ or~$\chi_R$}\:,
\eeq
under the constraint that~$X=\1$ if~$m>0$. We again denote the regularization by an index~$\varepsilon$.
We always assume that the regularization is {\em{homogeneous}}, so that~$P^\varepsilon$ is a
multiplication operator in momentum space, which sometimes we denote for clarity
by~$\hat{P}^\varepsilon$. If~$m>0$, we assume that the regularization satisfies all the conditions
in~\cite[Chapter~4]{PFP}; see also the compilation in Section~\ref{s:sec3}. In the case~$m=0$, we relax
the conditions on the shear and allow for general surface states, as explained in \S\ref{l:sec22}.
In the low-energy regime, $P^\varepsilon$ should still be of the form~\eqref{l:Xdefn}, i.e.
\beq \label{l:dirlow}
\hat{P}^\varepsilon(k) = 
\left\{ \begin{aligned} & (\slashed{k} + m)
\: \delta \!\left( k^2 - m^2 \right) && \text{if $m>0$} \\
& X \,\slashed{k}\, \delta(k^2) && \text{if $m=0$}
\end{aligned} \right. 
\qquad\quad \text{($|k^0|+|\vec{k}| \ll \varepsilon^{-1}$)}\:.
\eeq
However, in the high-energy regime, $P^\varepsilon$ will no longer satisfy the Dirac equation.
But in preparation of the perturbation expansion, we need to associate the states~$P^\varepsilon$
to eigenstates of the Dirac operator (not necessarily to the eigenvalue~$m$).
To this end, we introduce two operators~$V_\text{shift}$ and~$V_\text{shear}$
with the following properties. The operator~$V_\text{shift}$ has the purpose of changing the
momentum of states such that general surface states (as in Figure~\ref{l:fig1}~(B)) are mapped
onto the mass cone, i.e.
\beq \label{l:Vshiftdef}
\big( V_\text{shift} \psi \big)(k) := \psi \big( v_\text{shift}(k) \big) \:,
\eeq
where~$ v_\text{shift} : \hscrM \rightarrow \hscrM$ is a diffeomorphism.
The operator~$V_\text{shear}$, on the other hand, is a unitary multiplication operator in momentum
space, which has the purpose of introducing the shear of the surface states (i.e.\ it should map the
states in Figure~\ref{l:fig1}~(B) to those in Figure~\ref{l:fig1}~(C)),
\[ \big( V_\text{shear} \psi \big)(k) = \hat{V}_\text{shear}(k) \:\psi(k) \qquad \text{with}
\qquad \text{$\hat{V}_\text{shear}(k)$ unitary} \:. \]
\nindex{fb2@$V_\text{shift}, V_\text{shear}$ -- operators in the regularized causal perturbation theory}%
These operators are to be chosen such that the operator~$\check{P}^\varepsilon$ defined by
\beq \label{l:Pedef}
P^\varepsilon = V_\text{shear} \,V_\text{shift}\: \check{P}^\varepsilon \:V_\text{shift}^{-1}\,
V_\text{shear}^{-1}
\eeq
is of the following form,
\beq \label{l:Ptan}
\check{P}^\varepsilon(k) = \left\{ \begin{aligned} & d(k) \Big( \slashed{k} + m(k)\,\1 \Big)
\delta \!\left( k^2 - m(k)^2 \right) &\qquad& \text{if $m>0$} \\
& d(k) \,X \,\slashed{k}\, \delta(k^2) &\qquad& \text{if $m=0$}
\end{aligned} \right. 
\eeq
with~$X$ as in~\eqref{l:Xdefn}.
Thus in the massive case, $\check{P}^\varepsilon$ should be composed of Dirac eigenstates
corresponding to an energy-dependent mass~$m(k)>0$, and it should have vector-scalar
structure. In the massless case, we demand that~$k^2=0$, so that the states of~$\check{P}^\varepsilon$
are all neutral. The ansatz~\eqref{l:Ptan} is partly a matter of convenience, and partly a requirement
needed for the perturbation expansion (see Proposition~\ref{l:prpbelow} below).
Moreover, we assume for convenience that~$\check{P}^\varepsilon$ is composed only of states
of negative energy,
\beq \label{l:ebed}
\check{P}^\varepsilon(k) = 0 \qquad \text{if~$k^2<0$ or~$k^0 >0$}\:.
\eeq
In view of~\eqref{l:dirlow}, it is easiest to assume that~$V_\text{shift}$ and~$V_\text{shear}$
are the identity in the low-energy regime, i.e.
\beq \label{l:Vlow}
\hat{V}_\text{shear}(k) = \1 \quad \text{and} \quad
v_\text{shift}(k) = k \qquad \text{if~$|k^0|+|\vec{k}| \ll \varepsilon^{-1}$}\:.
\eeq
Then, by comparing~\eqref{l:dirlow} with~\eqref{l:Ptan}, one finds that
\beq \label{l:alphalow}
d(k) = \1 \quad \text{and} \quad
m(k) = m \qquad \text{if~$|k^0|+|\vec{k}| \ll \varepsilon^{-1}$}\:.
\eeq
The required regularization of~$P^\varepsilon(x,y)$ on the scale~$\varepsilon$
is implemented by demanding that
\beq \label{l:ddecay}
\text{$d(k)$ decays on the scale~$|k^0|+|\vec{k}| \sim \varepsilon^{-1}$}\:.
\eeq
In view of their behavior in the low-energy regime, it is natural to assume that
the functions in~\eqref{l:Vlow} and~\eqref{l:alphalow} should be smooth in momentum space
and that their derivatives scale in powers of the regularization length, i.e.\
\beq \label{l:kscale}
\begin{aligned}
\big|\nabla_k^\gamma d(k) \big| &\sim \varepsilon^{|\gamma|} |d(k)| \:, &\qquad
\big|\nabla_k^\gamma m(k) \big| &\sim \varepsilon^{|\gamma|} |m(k)| \\
\big|\nabla_k^\gamma \hat{V}_\text{shear}(k) \big| &\sim \varepsilon^{|\gamma|} |\hat{V}_\text{shear}(k)|
\:, & \quad \big|\nabla_k^\gamma v_\text{shift}(k) \big| &\sim \varepsilon^{|\gamma|}
|v_\text{shift}(k)|\:.
\end{aligned}
\eeq
Clearly, the above conditions do not uniquely determine the function~$d$ and the
operators~$V_\text{shift}$ and~$V_\text{shear}$. But we shall see that the results of our
analysis will be independent of the choice of these operators.
We remark that the transformation~$V_\text{shear}$ is analogous to
the transformations~$U_l$ considered in~\cite[Appendix~D]{PFP} (see~\cite[eq.~(D.22)]{PFP}),
except that here we consider only one unitary transformation.

The last construction immediately generalizes to a system of Dirac seas. Namely, suppose
that without regularization, the auxiliary fermionic projector of the vacuum
is a direct sum of Dirac seas (see for example~\eqref{l:Paux} or~\eqref{l:dirsum2}),
\[ P^\text{aux} = \bigoplus_{\ell=1}^{\ell_\text{max}} X_\ell \,t_\ell \:. \]
Then we introduce~$P^\varepsilon$ simply by taking the direct sum of the corresponding
regularized seas
\begin{align*}
P^\text{aux} := \bigoplus_{\ell=1}^{\ell_\text{max}} P^\varepsilon_\ell \:,\qquad
V_\text{shift} := \bigoplus_{\ell=1}^{\ell_\text{max}} V^\ell_\text{shift}\:, \qquad
V_\text{shear} := \bigoplus_{\ell=1}^{\ell_\text{max}} V^\ell_\text{shear} \:.
\end{align*}
Setting
\[ P^\text{aux} = V_\text{shear} \,V_\text{shift}\: \check{P}^\varepsilon \:V_\text{shift}^{-1}\,
V_\text{shear}^{-1} \:, \]
the operator~$\check{P}^\varepsilon$ satisfies the Dirac equation in momentum space
\beq \label{l:Dirk}
(\slashed{k} - m Y(k))\, \widehat{\check{P}^\varepsilon}(k) = 0 \:,
\eeq
where the mass matrix is given by (cf.~\eqref{l:Ydef1} or~\eqref{l:Ydef2})
\[ m Y(k) = \bigoplus_{\ell=1}^{\ell_\text{max}} m_\ell \:. \]
In the low-energy regime, we know furthermore that
\[ \left. \begin{array}{c} \hat{V}_\text{shear}(k) = \1 \quad \text{and} \quad
v_\text{shift}(k) = k \\[.5em]
\hat{P}^\text{aux}(k) = X \,t \end{array} \right\}
\qquad \text{if~$|k^0|+|\vec{k}| \ll \varepsilon^{-1}$} \:, \]
where~$X$ and~$t$ are given as in~\eqref{l:Pauxdef}. Clearly, the regularity
assumptions~\eqref{l:kscale} are imposed similarly for~$\hat{P}^\text{aux}$.
Finally, we need to specify what we mean by saying that two Dirac seas are regularized
in the same way. The difficulty is that, as mentioned above, different choices of~$d$,
$\hat{V}_\text{shear}$ and~$\hat{V}_\text{shift}$ may give rise to the same regularization
effects. In order to keep the situation reasonably simple, we use the convention that
if we want two Dirac seas to show the same regularization effects, we choose the
corresponding functions~$d$ as well as~$\hat{V}_\text{shear}$ and~$\hat{V}_\text{shift}$ to be
exactly the same. If conversely two Dirac seas should show different regularization effects,
we already choose the corresponding functions~$d$ to be different.
Then we can say that two Dirac seas labeled by~$a$ and~$b$ are {\em{regularized
in the same way}} if~$d_a \equiv d_b$. \label{l:samereg}
In this case, our convention is that also~$(\hat{V}_\text{shear})_a =
(\hat{V}_\text{shear})_b$ and~$(\hat{V}_\text{shift})_a =
(\hat{V}_\text{shift})_b$. This notion gives rise to an equivalence relation on the Dirac seas.
In the formalism of \S\ref{l:sec24}, the equivalence classes will be labeled by
the parameters~$\tau^\reg_i$ (see~\eqref{l:treg1} and~\eqref{l:treg2}).

\section{Formal Introduction of the Interaction}
Now the interaction can be introduced most conveniently by using the {\em{unitary perturbation
flow}}~\cite[Section~5]{grotz} (see also~\cite[Section~4]{norm}). In order not to get confused with the
mass matrix, we introduce an additional spectral parameter~$\mu$ into the free Dirac equation, which
in momentum space reads
\[ (\slashed{k} - mY(k) - \mu \1) \hat{\psi}(k) = 0 \:. \] 
For this Dirac equation, we can introduce the spectral projectors~$p$, the causal fundamental
solutions~$k$ and the symmetric Green's functions~$s$ can be introduced just as
in~\cite[Section~2.2]{PFP}, if only in the formulas in momentum space we replace~$m$
by~$m Y(k)$. For clarity, we denote the dependence on~$\mu$ by an
subscript~$+\mu$ (this notation was used similarly in~\cite[Section~2.6]{PFP};
see also~\cite[Appendix~C.3]{PFP} for an additional ``modified mass scaling'', which we will for
simplicity not consider here).
We describe the interaction by inserting
an operator~$\B$ into the Dirac operator,
\[ {\mathcal{D}} = i \Pdd + \B - m Y(k)\:. \]
After adding the subscript~$+\mu$ to all factors~$p$, $k$ or~$s$ in the operator
products in~\cite[Section~5]{grotz}, we obtain an operator~$U$ which
associates to every solution~$\psi$ of the free Dirac equation~$(i \Pdd - mY - \mu\, \1) \psi=0$
a corresponding solution~$\tilde{\psi}$ of the interacting Dirac equation~$(i \Pdd + \B- mY - \mu\, \1)
\psi=0$,
\[ U(\B) \::\: \psi \mapsto \tilde{\psi} \:. \]
The operator~$U$ is uniquely defined in terms of a formal power series in~$\B$.
Taking~$\mu$ as a free parameter, in~\cite[Section~5]{grotz} the operator~$U$ is 
shown to be unitary with respect to the indefinite inner product~\eqref{l:iprod}.
We now use~$U$ to unitarily transform all the Dirac states contained in the
operator~$\check{P}^\varepsilon$ and set
\beq \label{l:Pformal}
\boxed{ \quad \tilde{P}^\text{aux} = V_\text{shear} \,V_\text{shift}\: 
U(\B)\:\check{P}^\varepsilon\: U(\B)^{-1} \:V_\text{shift}^{-1}\,
V_\text{shear}^{-1} \:. \quad }
\eeq
This construction uniquely defines the regularized auxiliary fermionic projector with
interaction~$\tilde{P}^\text{aux}$ in terms of a formal power expansion in~$\B$. The fermionic
projector is then obtained by forming the sectorial projection (see~\eqref{l:partrace0} or~\eqref{l:partrace}).

\section{Compatibility Conditions for the Interaction} \label{l:appcompatible}
In order to derive the structure of the admissible~$\B$, we first consider a perturbation calculation
to first order and assume that~$\B$ is a multiplication operator in position space having the form
of a plane wave of momentum~$q$,
\beq \label{l:Bplane}
\B(x) = \B_q \: e^{-i q x} \:.
\eeq
In this case  (cf.~\cite[eq.~(D.14)]{PFP}),
\[ \Delta \check{P}^\text{aux} = -\int_{-\infty}^\infty d\mu \left(
s_{+\mu} \:{\B}\:p_{+\mu}\:\check{P}^\varepsilon \:+\: \check{P}^\varepsilon \:p_{+\mu}
\:{\B}\:s_{+\mu} \right) \:. \]
Using a matrix notation in the direct sums with indices~$a,b \in \{1, \ldots, \ell_\text{max}\}$,
we obtain in momentum space (for the notation see~\cite[Chapter~2]{PFP} or~\cite{grotz})
\beq \label{l:DelP}
\begin{split}
(\Delta \check{P}^\varepsilon)^a_b(k+q, k) &= -\int_{-\infty}^\infty d\mu \left\{
s_{m_a+\mu}(k+q) \:({\B}_q)^a_b\; p_{m_b+\mu}(k)\: (\check{P}^\varepsilon)^b_b(k) \right. \\
& \qquad\qquad\quad \left. + (\check{P}^\varepsilon)^a_a(k+q)\: p_{m_a+\mu}(k+q)
\:({\B}_q)^a_b \;s_{m_b+\mu}(k) \Big) \right\} .
\end{split}
\eeq
This equation was already considered in~\cite[Section~3]{firstorder} and~\cite[Appendix~D]{PFP}.
However, here we analyze the situation more systematically and in a more general context,
pointing out the partial results which were obtained previously.

For clarity, we analyze~\eqref{l:DelP} step by step, beginning with the diagonal elements.
For ease in notation, we assume that~$\B_q$ has only one non-trivial component, which is
on the diagonal,
\beq \label{l:Bdiagonal}
(\B_q)^a_b = \delta^{a \ell} \:\delta_{b \ell}\: {\mathcal{B}}
\eeq
with~$\ell \in \{1, \ldots, \ell_\text{max}\}$ and~$ {\mathcal{B}}$ a matrix acting on Dirac spinors.
Shifting the integration variable according to~$m_\ell + \mu \rightarrow \mu$, we obtain
(cf.~\cite[eq.~(D.15)]{PFP})
\begin{align*}
\lefteqn{ \hspace*{-0.2cm} (\Delta \check{P}^\varepsilon)^\ell_\ell(k+q, k) } \\
&= -\int_{-\infty}^\infty d\mu \left\{
s_{\mu}(k+q) \:{\mathcal{B}}\; p_{\mu}(k)\: (\check{P}^\varepsilon)^\ell_\ell(k) 
 + (\check{P}^\varepsilon)^\ell_\ell(k+q)\: p_{\mu}(k+q)\:{\mathcal{B}} \:s_{\mu}(k) \right\} \\
&=-\int_{-\infty}^\infty d\mu\: \epsilon(\mu) \left\{ \frac{\mbox{PP}}{(k+q)^2 - \mu^2} \:(\slashed{k} +\slashed{q} + \mu) \:{\mathcal{B}}\,(\slashed{k}+\mu)\:\delta(k^2-\mu^2)
\: (\check{P}^\varepsilon)^\ell_\ell(k) \right. \\
&\hspace*{3cm} \left. +\: (\check{P}^\varepsilon)^\ell_\ell(k+q) \:\delta((k+q)^2-\mu^2) \:(\slashed{k} + \slashed{q}+\mu) \:{\mathcal{B}}\, (\slashed{k} + \mu)
\:\frac{\mbox{PP}}{k^2 - \mu^2} \right\} \\
&=-\int_{-\infty}^\infty d\mu\: \epsilon(\mu) \; \frac{\text{PP}}{2 k q + q^2}\:
\Big\{ (\slashed{k} +\slashed{q} + \mu) \:{\mathcal{B}}\:(\slashed{k}+\mu)\:\delta(k^2-\mu^2)
\: (\check{P}^\varepsilon)^\ell_\ell(k) \\
&\hspace*{3cm} -\: (\check{P}^\varepsilon)^\ell_\ell(k+q) \:\delta((k+q)^2-\mu^2) \:(\slashed{k} + \slashed{q}+\mu) \:{\mathcal{B}}\: (\slashed{k} + \mu) \Big\} , \quad
\end{align*}
where in the last step we used that the argument of the $\delta$-distribution vanishes.
Carrying out the $\mu$-integration gives (cf.~\cite[eq.~(D.15)]{PFP})
\beq \label{l:B7}
\begin{split}
(\Delta \check{P}^\varepsilon)^\ell_\ell(k+q, k) &=-\frac{\text{PP}}{4 k q + 2 q^2}\:
\left\{ \Big( (\slashed{k} +\slashed{q}) \:{\mathcal{B}} +
{\mathcal{B}}\: \slashed{k} \Big) \: (\check{P}^\varepsilon)^\ell_\ell(k)  \right. \\
& \qquad\qquad\qquad\quad\;\; \left. - (\check{P}^\varepsilon)^\ell_\ell(k+q) \Big(
(\slashed{k} + \slashed{q}) \:{\mathcal{B}}
+ {\mathcal{B}}\: \slashed{k} \Big) \right\} .
\end{split}
\eeq
Here the principal part has poles if~$2 k q + q^2=0$, leading to a potential divergence
of~$\Delta \check{P}^\varepsilon$. In order to explain the nature of this divergence, we
first point out that if~$\B$ had been chosen to be a smooth function with rapid decay,
then~$\Delta \check{P}$ would have been finite (see the proof of~\cite[Lemma~2.2.2]{PFP}).
Thus the potential divergence is related to the fact that the plane wave in~\eqref{l:Bplane}
does {\em{not}} decay at infinity. A more detailed picture is obtained
by performing the light-cone expansion (see~\cite{firstorder} and~\cite[Appendix~F]{PFP}).
Then one can introduce the notion that~\eqref{l:B7} is causal if its light-cone expansion
only involves integrals along a line segment~$\overline{xy}$. Since such integrals
are uniformly bounded, it follows immediately that all contributions to the light-cone expansion
are finite for all~$q$. If conversely~\eqref{l:B7} diverges, then the analysis in~\cite[Appendix~F]{PFP}
reveals that individual contributions to the light-cone expansion do diverge, so that unbounded
line integrals must appear (see also the explicit light-cone expansions in~\cite{lightint}).
In this way, one gets a connection between the boundedness of~\eqref{l:B7}
and the {\em{causality of the light-cone expansion}}.

Unbounded line integrals lead to contributions to the EL equations whose scaling behavior
in the radius is different from all other contributions. Therefore, the EL equations are satisfied
only if all unbounded line integrals drop out. The easiest way to arrange this is to demand that
the fermionic projector itself should not involve any unbounded line integrals. This is our motivation
for imposing that
\beq \label{l:bcond}
\text{ $(\Delta \hat{P})^\varepsilon(k+q, k)$ should be bounded locally uniformly in~$q$. }
\eeq

Let us analyze this boundedness condition for~\eqref{l:B7}. Since the denominator in~\eqref{l:B7}
vanishes as~$q \rightarrow 0$, we clearly get the necessary condition that the curly brackets
must vanish at~$q=0$,
\beq \label{l:C2}
\big[ \{\slashed{k}, {\mathcal{B}} \}, \check{P}^\varepsilon(k) \big] = 0 \:.
\eeq
Using~\eqref{l:Ptan} together with the identity
\[ \big[ \{\slashed{k}, {\mathcal{B}} \}, \slashed{k} \big] = \left[ k^2, {\mathcal{B}} \right]
+ \slashed{k} {\mathcal{B}} \slashed{k} - \slashed{k} {\mathcal{B}} \slashed{k} = 0 \:, \]
we find that~\eqref{l:C2} is automatically satisfied in the case~$X = \1$. The situation is more
interesting if a chiral asymmetry is present. If for example~$X=\chi_L$, we get the condition
\[ \big[ \{\slashed{k}, {\mathcal{B}} \}, \chi_L \slashed{k} \big] = 0 \:. \]
This condition is again trivial if~${\mathcal{B}}$ is odd (meaning that~$\{{\mathcal{B}}, \pseudo\}=0$).
However, if~${\mathcal{B}}$ is even, we conclude that
\[ 0 = \frac{\pseudo}{2} \big\{ \{\slashed{k}, {\mathcal{B}} \}, \slashed{k} \big\}
= \pseudo\: \slashed{k} \,{\mathcal{B}}\, \slashed{k} \]
(where in the last step we used that~$k^2=0$ in view of~\eqref{l:Ptan}).
As~$k$ is any state on the lower mass shell, this rules out that~${\mathcal{B}}$ is a bilinear potential,
leaving us with a scalar or a pseudoscalar potential. In order to rule out these potentials, we
next choose a vector~$\hat{q}$ with~${\hat{q}} k =0$, set~$q=\varepsilon \hat{q}$ and
consider~\eqref{l:B7} in the limit~$\varepsilon \rightarrow 0$. Then the denominator in~\eqref{l:B7}
diverges like~$\varepsilon^{-2}$, so that the curly brackets must tend to zero even~$\sim \varepsilon^2$.
\beq \label{l:C3}
\Big( (\slashed{k} + \varepsilon \hat{\slashed{q}}) {\mathcal{B}} + {\mathcal{B}} \slashed{k} \Big) \check{P}^\varepsilon(k)
- \check{P}^\varepsilon(k+ \varepsilon \hat{q})
\Big( (\slashed{k} + \varepsilon \hat{\slashed{q}}) {\mathcal{B}} + {\mathcal{B}} \slashed{k} \Big) = \O(\varepsilon^2) \:.
\eeq
Using that~$\check{P}^\varepsilon(k)$ is left-handed and that~${\mathcal{B}}$ is even,
we find that the first summand in~\eqref{l:C3} is right-handed, whereas the second summand
is left-handed. Hence both summand must vanish separately, and thus
\[ 0 = \Big( (\slashed{k} + \hat{\slashed{q}}) {\mathcal{B}} + {\mathcal{B}} \slashed{k} \Big) \check{P}^\varepsilon(k) = 
\hat{\slashed{q}} \,{\mathcal{B}} \, \slashed{k}\: d(k)\: \delta(k^2) \:. \]
This condition implies that~${\mathcal{B}}$ must vanish. We conclude that if~$X=\chi_L$, only
odd potentials may occur. We can write this result more generally as
\beq \label{l:cccpot}
\boxed{ \quad \B \,X = X^* \,\B\:. \quad }
\eeq
\sindex{causality compatibility condition}%
We have thus derived the {\em{causality compatibility condition}}~\eqref{l:ccc}
from our boundedness condition~\eqref{l:bcond}. This derivation is an alternative to the
method in~\cite[Section~2.3]{PFP},
where the same condition was introduced by the requirement that it should be possible to
commute the chiral asymmetry matrix through the perturbation expansion.

So far, we considered~\eqref{l:bcond} in the limit~$q \rightarrow 0$. We now analyze
this condition for general~$q$. Using~\eqref{l:Ptan} and~\eqref{l:cccpot}, a short calculation gives
\[ (\Delta \check{P}^\varepsilon)^\ell_\ell(k+q, k) = 
-X \,\frac{(\slashed{k} +\slashed{q}) \:{\mathcal{B}}\: \slashed{k}}{4 k q + 2 q^2} \: \Big( 
d(k)\: \delta(k^2-m^2) - d(k+q)\: \delta((k+q)^2-m^2) \Big) , \]
where we set~$m=m_\ell$. If~$d(k)=d(k+q)$, the transformations
\begin{align*}
\int_0^1 \delta' \big( k^2-m^2 + \tau (2 k q + q^2) \big)\: d\tau
&= \frac{1}{2 k q + q^2} \int_0^1 \frac{d}{d\tau} \delta \big( k^2-m^2 + \tau (2 k q + q^2) \big)\: d\tau \\
&= \frac{1}{2 k q + q^2 }\Big(\delta((k+q)^2-m^2) - \delta(k^2-m^2) \Big)
\end{align*}
show that~$\Delta \check{P}^\varepsilon$ is indeed a bounded distribution for any~$q$.
Thus it remains to be concerned about the contribution if~$\delta(k) \neq \delta(k+q)$,
\beq \label{l:singular}
X\, \frac{(\slashed{k} +\slashed{q}) \:{\mathcal{B}}\: \slashed{k}}{4 k q + 2 q^2} \:
\delta(k^2-m^2) \: \Big( d(k+q)-d(k) \Big)\:.
\eeq
Unless in the trivial case~${\mathcal{B}}=0$, this contribution is infinite at
the poles of the denominator.
We conclude that in order to comply with the condition~\eqref{l:bcond}, we must impose
that the weight function~$d(k)$ in~\eqref{l:Ptan} is constant on the mass shell~$k^2=m(k)^2$.
This is indeed the case in the low-energy regime~\eqref{l:alphalow}. However, in the high-energy region,
the function~$d(k)$ is in general not a constant (and indeed, assuming that~$d(k)$ is constant
would be in contradiction to~\eqref{l:ddecay}).
Our way out of this problem is to observe that~\eqref{l:singular}
implies that the light-cone expansion of~\eqref{l:B7} is in general not causal, in the sense that
it involves unbounded line integrals. However, using that~$d(k+q)-d(k) \sim q |\nabla d|$,
the scalings $q \sim \ell_\text{macro}^{-1}$ and~\eqref{l:kscale} show that these
non-causal contributions to the light-cone expansion are of
\beq \label{l:error}
\text{higher order in~$\varepsilon/\ell_\text{macro}$}\:.
\eeq
This consideration shows that the perturbation expansion will give rise to error terms
of higher order in~$\varepsilon/\ell_\text{macro}$. In what follows, we will always neglect
such error terms. If this is done, the above assumptions are consistent and in agreement
with~\eqref{l:bcond}, provided that the causality compatibility condition~\eqref{l:cccpot} holds.

Before moving on to potentials which mix different Dirac seas, we remark that
the above arguments can also be used to derive constraints for the possible
form of~$\hat{P}^\varepsilon(k)$, thus partly justifying our ansatz~\ref{l:Ptan}.
\begin{Prp}[Possible form of~$\check{P}^\varepsilon$]
\label{l:prpbelow}
Suppose that $\Delta \check{P}^\varepsilon$ as given by~\eqref{l:B7} satisfies the
condition~\eqref{l:bcond}. We renounce the assumptions on~$\check{P}^\varepsilon$
(see~\eqref{l:Ptan}, \eqref{l:ebed}, \eqref{l:alphalow} and~\eqref{l:kscale}), but we
assume instead that the admissible interaction includes chiral or axial potentials. 
Then for every~$k$, there are complex coefficients~$a,b,c,d \in \C$ such that
\beq \label{l:pc1}
\check{P}^\varepsilon(k) = a \1 + i b \pseudo + c \slashed{k} + d \pseudo \slashed{k} \:.
\eeq
\end{Prp}
\Proof We again analyze the necessary condition~\eqref{l:C2}. This condition is clearly
satisfied if~${\mathcal{B}}$ is a vector potential. Thus by linearity, we may assume that~${\mathcal{B}}
= \pseudo \slashed{A}$ is a vector potential. It follows that
\[ 0 = \big[ \{\slashed{k}, \pseudo \slashed{A} \}, \check{P}^\varepsilon(k) \big] 
= \big[ \pseudo  [\slashed{A}, \slashed{k} ], \check{P}^\varepsilon(k) \big] . \]
Decomposing~$\check{P}^\varepsilon(k)$ into its even and odd components,
by linearity we can again consider these components after each other. If~$\check{P}^\varepsilon(k)$
is odd, i.e.\
\[ \check{P}^\varepsilon(k) = \slashed{u} + \pseudo\, \slashed{v}\:, \]
we obtain
\[ 0 =  \pseudo \,\big\{ [\slashed{A}, \slashed{k}], \check{P}^\varepsilon(k) \big\} = 
4 \pseudo \big((ku) \,\slashed{A} - (A u)\, \slashed{k} \big) + 4 \big( (kv) \,\slashed{A} - (A v)\, \slashed{k} \big) \:. \]
Since~$A$ is arbitrary, it follows that~$u$ and~$v$ must be multiples of~$k$.

If~$\check{P}^\varepsilon(k)$ is even, we obtain the condition
\[ \big[ [\slashed{A}, \slashed{k}], \check{P}^\varepsilon(k) \big] = 0\:. \]
This condition is obviously satisfied if~$\check{P}^\varepsilon(k)$ is a scalar or a pseudoscalar.
Writing the bilinear component of~$\check{P}^\varepsilon(k)$ in the form
$F_{ij} \gamma^i \gamma^j$ with an anti-symmetric tensor field~$F$, we get the condition
\[ 0 = \big[ [\slashed{A}, \slashed{k}], F_{ij} \gamma^i \gamma^j \big]  = 
2 F_{ij} k^i \,\big[ \slashed{A}, \gamma^j \big] - 2 F_{ij} A^i \left[ \slashed{k}, \gamma^j \right] . \]
Since~$A$ is arbitrary, it follows that~$F=0$, concluding the proof.
\QED
The step from~\eqref{l:pc1} to our the stronger assumption~\eqref{l:dirlow} could be justified
by the assumption that the image of~$P^\varepsilon$ should be negative definite
or neutral, and furthermore by assuming that without chiral asymmetry the matrix~$\pseudo$
is absent, whereas the chiral asymmetry is then introduced simply by multiplying
with~$\chi_L$ or~$\chi_R$. We finally remark that in~\cite[Appendix~D]{PFP},
the condition~\eqref{l:C2} is analyzed for~${\mathcal{B}}$ a scalar potential to conclude
that~$\check{P}^\varepsilon(k)$ should commute with the Dirac operator
(see~\cite[eq.~(D.16) and~eq.~(D.17)]{PFP}). This is consistent with our ansatz~\eqref{l:Ptan},
which is even a solution of the Dirac equation~\eqref{l:Dirk}.
However, we here preferred to avoid working with scalar potentials, which do not seem
crucial for physically realistic models.

Next, instead of~\eqref{l:Bdiagonal} we consider a general potential~$\B_q$ which may have
off-diagonal terms in the direct summands. Then~\eqref{l:DelP} can be evaluated similar
as in the computation after~\eqref{l:Bdiagonal}, but the calculation is a bit more complicated.
Therefore, we first compute the integral of the first summand in~\eqref{l:DelP},
\begin{align}
\int_{-\infty}^\infty &d\mu \:s_{m_a+\mu}(k+q) \:({\B}_q)^a_b\; p_{m_b+\mu}(k)\:
(\check{P}^\varepsilon)^b_b(k) \nonumber \\
&=\int_{-\infty}^\infty d\mu\: \epsilon(m_b + \mu) \;\frac{\text{PP}}{(k+q)^2 - (m_a +\mu)^2}\:
\delta(k^2-(m_b + \mu)^2) \nonumber \\
&\qquad\qquad\qquad\qquad\qquad \times
(\slashed{k} +\slashed{q} + m_a + \mu)\:({\B}_q)^a_b \: (\slashed{k}+m_b + \mu)\: (\check{P}^\varepsilon)^b_b(k)
\nonumber \\
&= \sum_{\mu = \pm |k| - m_b}  \frac{1}{m_b + \mu}\: 
 {\mathfrak{B}}^a_b\, (\check{P}^\varepsilon)^b_b(k)\:, \label{l:B21}
\end{align}
where we set
\beq \label{l:sBdef}
{\mathfrak{B}}^a_b = \frac{1}{2}\, \text{PP} \bigg(
\frac{(\slashed{k} +\slashed{q} + m_a + \mu)\:({\B}_q)^a_b \: (\slashed{k}+m_b + \mu)}
{2 k q + q^2 - (m_a^2-m_b^2) -2 \mu (m_a-m_b)} \bigg)
\eeq
and~$|k| = \sqrt{k^2}$ (note that, in view of our assumption~\eqref{l:ebed}, the
factor~$(\check{P}^\varepsilon)^b_b(k)$ guarantees that the above expression vanishes
if~$k^2<0$). Treating the second summand in~\eqref{l:DelP} similarly, we obtain
\beq
(\Delta \check{P}^\varepsilon)^a_b(k+q, k) =
\sum_{\mu = \pm |k+q| - m_a} \frac{(\check{P}^\varepsilon)^a_a(k+q)}{m_a + \mu}\:
 \:\mathfrak{B}^a_b
\;- \sum_{\mu = \pm |k| - m_b} 
\mathfrak{B}^a_b \,\frac{(\check{P}^\varepsilon)^b_b(k)}{m_b + \mu}\:. \label{l:B23}
\eeq
This formula is rather involved, but fortunately we do not need to enter a detailed analysis.
It suffices to observe that~\eqref{l:sBdef} has poles in~$q$, which lead to singularities
of~\eqref{l:B21}. Thus the only way to satisfy the condition~\eqref{l:bcond} is
to arrange that contributions of the first and second expression
on the right of~\eqref{l:B23} cancel each other. In view of~\eqref{l:Ptan}, the first
expression involves~$d_a(k+q)$, whereas in the second expression the term~$d_b(k)$ appears.
This shows that in order to get the required cancellations, the functions~$d_a$ and~$d_b$ must
coincide. Using the notion introduced on page~\pageref{l:samereg},
we conclude that~$\B$ may describe an interaction of Dirac seas only if they
are regularized in the same way. An interaction of Dirac seas with different
regularization, however, is prohibited by the causality condition for the light-cone expansion.
For brevity, we also say that the interaction must be {\em{regularity compatible}}.

\section{The Causal Perturbation Expansion with Regularization}
We are now ready to perform the causal perturbation expansion. In~\cite[Section~5]{grotz},
the unitary perturbation flow is introduced in terms of an operator product expansion.
Replacing the Green's functions and fundamental solutions in this expansion
by the corresponding operators of the free Dirac equation~\eqref{l:Dirk}, we can write
the operator~$U(\B)\,\check{P}^\varepsilon\, U(\B)^{-1}$ as a series of
operator products of the form
\[  {\mathscr{Z}} := C_1\:\B\: \cdots \: \B\: C_p \:\B\; \check{P}^\varepsilon\;
\B\: C_{p+1} \:\B\: \cdots \:\B\: C_k \:, \]
where the factors~$C_l$ are the Green's functions or fundamental solutions of the
free Dirac equation~\eqref{l:Dirk}.
The operators~$C_l$ are diagonal in momentum space, whereas the potential~$\B$
varies on the macroscopic scale and thus changes the momentum only on the
scale~$\ell_\text{macro}^{-1}$. Thus all the factors~$C_l$ will be evaluated at the
same momentum~$p$, up to errors of the order~$\ell_\text{macro}^{-1}$.
We refer to this momentum~$p$, determined only up to summands of the
order~$\ell_\text{macro}^{-1}$, as the considered {\em{momentum scale}}.
In view of the regularity assumptions on the functions~$d$ and~$m$ in~\eqref{l:kscale},
we may replace them by the constants~$d(p)$ and~$m(p)$, making an error of the order~\eqref{l:error}.
This evaluation of the regularization functions is referred to as the {\em{fixing of the momentum scale}},
and we indicate it symbolically by~$|_{\text{scale $p$}}$.
Since~$\B$ is regularity compatible, we may then
commute the constant matrix~$d$ to the left. Moreover, we can apply the causality compatibility
condition~\eqref{l:cccpot} together with the form of~$X$ in~\eqref{l:Ptan} and~\eqref{l:dirlow} to
also commute the chiral asymmetry matrix~$X$ to the left. We thus obtain the expansion
\begin{align}
 U(\B)\,\check{P}^\varepsilon\, U(\B)^{-1} \big|_{\text{scale $p$}}
= \sum_{k=0}^\infty \sum_{\alpha=0}^{\alpha_{\max}(k)} c_\alpha \,&
X d  \:C_{1,\alpha} \, \B\,C_{2,\alpha} \,\B\, \cdots \,\B\, C_{k+1, \alpha}
\big|_{\text{scale $p$}} \label{l:opex} \\
& + \text{(higher orders in~$\varepsilon/\ell_\text{macro}$)}\:, \nonumber
\end{align}
where we set
\[ X = \bigoplus_{\ell=1}^{\ell_\text{max}} X_\ell \qquad \text{and} \qquad
d = \bigoplus_{\ell=1}^{\ell_\text{max}} d_\ell(p) \:, \]
and~$c_\alpha$ are combinatorial factors.
Here the combinatorics of the operator products coincides precisely with that
of the causal perturbation expansion for the fermionic projector
as worked out in detail in~\cite{grotz, norm}.

\section{The Behavior under Gauge Transformations}
\sindex{gauge transformation!behavior of regularized fermionic projector}%
In order to analyze the behavior of the above expansion under $U(1)$-gauge transformations,
we consider the case of a pure gauge potential, i.e.\ $\B = \Pdd \Lambda$ with a real-valued
function~$\Lambda$. Then the gauge invariance of the causal perturbation expansion yields
\beq \label{l:UPUm}
U(\B)\,\check{P}^\varepsilon\, U(\B)^{-1} \big|_{\text{scale $p$}}
= (e^{i \Lambda} \check{P}^\varepsilon e^{-i \Lambda})\big|_{\text{scale $p$}}
+ \text{(higher orders in~$\varepsilon/\ell_\text{macro}$)}\:.
\eeq
According to~\eqref{l:Pformal}, we obtain~$\tilde{P}^\text{aux}$ by applying
with~$V_\text{shift}$ and~$V_\text{shear}$.
The transformation~$V_\text{shift}$ is a subtle point which requires a detailed explanation.
We first consider its action on a multiplication operator in momentum space~$M(k)$.
Then, according to the definition~\eqref{l:Vshiftdef},
\begin{align}
(V_\text{shift} \,M\, V_\text{shift}^{-1} \psi)(k)
&= (M V_\text{shift}^{-1} \psi) \big(v_\text{shift}(k) \big) \nonumber \\
&= M\big(v_\text{shift}(k) \big) \: (V_\text{shift}^{-1} \psi) \big(v_\text{shift}(k) \big)
= M\big(v_\text{shift}(k) \big) \: \psi(k) \:, \label{l:momtrans}
\end{align}
showing that the transformation again yields a multiplication operator, but with a transformed argument.
In order to derive the transformation law for multiplication operators in position space, we first
let~$f = e^{-i q x}$ be the operator of multiplication by a plane wave. Then
\begin{align*}
(V_\text{shift} \,f\, V_\text{shift}^{-1} \psi)(k)
&= (f V_\text{shift}^{-1} \psi)\big(v_\text{shift}(k) \big) \\
&= (V_\text{shift}^{-1} \psi)\big(v_\text{shift}(k) - q\big)
= \psi\Big(v_\text{shift}^{-1} \big( v_\text{shift}(k) - q \big) \Big) .
\end{align*}
This can be simplified further if we assume that the momentum~$q \sim \ell_\text{macro}^{-1}$
is macroscopic. Namely, the scaling of the function~$v_\text{shift}$ in~\eqref{l:kscale} allows us
to expand in a Taylor series in~$q$,
\[ (V_\text{shift} \,f\, V_\text{shift}^{-1} \psi)(k)
= \psi\Big( k - Dv_\text{shift}^{-1} \big|_{v_\text{shift}(k)} \,q \Big)
+ \text{(higher orders in~$\varepsilon/\ell_\text{macro}$)} \:. \]
Thus~$V_\text{shift} \,f\, V_\text{shift}^{-1}$ is again a multiplication operator in
position space, but now corresponding to the new momentum
\[ L(k) \, q \qquad \text{with} \qquad L(k) := Dv_\text{shift}^{-1}
\big|_{v_\text{shift}(k)} \:. \]
Again in view of the regularity assumptions~\eqref{l:kscale}, when fixing the momentum scale
we may replace the argument~$k$ by~$p$, i.e.
\[ V_\text{shift} \,e^{-i q x}\, V_\text{shift}^{-1} \big|_\text{scale~$p$} = 
e^{-i (L(p) q)\,x} + \text{(higher orders in~$\varepsilon/\ell_\text{macro}$)} \:. \]
Using the relation~$(L(p) q)\,x = q\,L(p)^* x$, we can rewrite this transformation law
simply as a linear transformation of the space-time coordinates. Then the transformation law
generalizes by linearity to a general multiplication operator by a function~$f$ which varies on
the macroscopic scale, i.e.
\beq \label{l:postrans}
V_\text{shift} \,f(x)\, V_\text{shift}^{-1} \big|_\text{scale~$p$} = 
f \big( L(p)^* x \big) + \text{(higher orders in~$\varepsilon/\ell_\text{macro}$)} \:.
\eeq

With~\eqref{l:momtrans} and~\eqref{l:postrans}, we can transform~\eqref{l:UPUm} to the
required form~\eqref{l:Pformal}. Using that~$V_\text{shear}$ commutes with scalar
and macroscopic multiplication operators (again up to higher orders in~$\varepsilon/\ell_\text{macro}$),
we obtain in view of~\eqref{l:Pedef}
\beq P^\varepsilon(x,y) \big|_\text{scale p}
= e^{i \Lambda (L(p)^*x)} \:P^\varepsilon(x,y)\: e^{-i \Lambda (L(p)^*y )}
+ \text{(higher orders in~$\varepsilon/\ell_\text{macro}$)} \:. \label{l:Petrans}
\eeq
Except for the factors~$L(p)^*$, this formula describes the usual behavior of the fermionic
projector under gauge transformations. In particular, if we do not consider general
surface states and~$V_\text{shift}=\1$, then our perturbation expansion is gauge invariant.
However, if we consider general surface states described by a non-trivial operator~$V_\text{shift}$,
then the matrix~$L$ will in in general not be the identity, and the transformation law~\eqref{l:Petrans}
violates gauge invariance.

Our method to recover gauge invariance is to replace the gauge potential~$A$ by a
more general operator~$\mathscr{A}$, which in momentum space has the form
\beq \label{l:Anonlocal}
\mathscr{A} \Big( p+\frac{q}{2}, p - \frac{q}{2} \Big) := \hat{A} \big( L(p)^{-1}q \big) \:,
\eeq
where~$\hat{A}$ is the Fourier transform of the classical potential to be used when no
regularization is present. In the case~$A = \Pdd \Lambda$ of a pure gauge field
and fixing the momentum scale, we then find that~$\mathscr{A}$ coincides
with the multiplication operator~$A(\big( (L(p)^*)^{-1} x \big))$,
just compensating the factors~$L(p)^*$ in~\eqref{l:Petrans}.
In view of the regularity assumptions~\eqref{l:kscale}, the
matrix~$L$ scales in powers of the regularization length. Thus~$A$
is a {\em{nonlocal}} operator, but only {\em{on the microscopic scale}}~$\varepsilon$.
On the macroscopic scale, however, it coincides with the classical local potential.
We also point out that the compatibility conditions worked out in Section~\ref{l:appcompatible}
under the assumption that~$\B$ is a multiplication operator are valid just as well
for the nonlocal potential~\eqref{l:Anonlocal}, because after fixing the momentum scale,
$\mathscr{A}$ reduces to a multiplication operator, so that our previous considerations again apply.

\section{The Regularized Light-Cone Expansion}
\sindex{light-cone expansion!regularization}%
We are now in the position to perform the light-cone expansion. Our starting point is the
operator product expansion~\eqref{l:opex}. We choose~$\hat{\B}$ to be the Fourier transform
of a general multiplication or differential operator and introduce~$\B$ in analogy to~\eqref{l:Anonlocal}
as a non-local operator. After fixing the momentum scale, this operator
reduces again to a multiplication or differential operator. Then the light-cone expansion 
can be performed exactly as described
in~\cite{firstorder} and~\cite[Section~2.5]{PFP} (see also Section~\ref{seclight}).
Finally, one can transform the obtained formulas
with~$V_\text{shift}$ and~$V_\text{shear}$ (again using the rules~\eqref{l:momtrans}
and~\eqref{l:postrans}). Since the resulting line integrals do not depend on the momentum scale,
the regularization only affects the factors~$T^{(n)}$. The condition that~$\B$ should be
regularity compatible can be described by the parameters~$\tau_i^\reg$.
We thus obtain the formalism described in \S\ref{l:sec23} and~\S\ref{l:sec24}.

We finally compare our constructions with those in~\cite[Appendix~D]{PFP}. Clearly,
the constructions here are much more general because they apply to any order in perturbation
theory and may involve a chiral asymmetry.
Moreover, the momentum shift operator~$V_\text{shift}$ makes it possible to
describe general surface states, and also we allow for a large shear of the surface states
(whereas in~\cite[Appendix~D]{PFP} we always assumed the shear to be small).
Nevertheless, the basic idea that in order to preserve the gauge
invariance in the presence of a regularization,
one should replace the classical potentials by operators which are nonlocal on the
microscopic scale already appeared in~\cite[Appendix~D]{PFP} (see the explanation
after~\cite[eq.~(D.26)]{PFP}). Thus~\cite[Appendix~D]{PFP}
can be regarded as a technical and conceptual preparation which is superseded by the
constructions given here.
\end{appendix}

%


\backmatter


\begin{thebibliography}{10}

\bibitem[BF]{baer+fredenhagen}
C.~B\"ar and K.~Fredenhagen~(eds), \emph{Quantum {F}ield {T}heory on {C}urved
  {S}pacetimes}, Lecture Notes in Physics, vol. 786, Springer Verlag, Berlin,
  2009.

\bibitem[Ba]{baumgaertel}
H.~Baumg{\"a}rtel, \emph{Analytic {P}erturbation {T}heory for {M}atrices and
  {O}perators}, Operator Theory: Advances and Applications, vol.~15,
  Birkh\"auser Verlag, Basel, 1985.

\bibitem[BF]{lagrange}
Y.~Bernard and F.~Finster, \emph{On the structure of minimizers of causal
  variational principles in the non-compact and equivariant settings},
  arXiv:1205.0403 [math-ph], Adv. Calc. Var. \textbf{7} (2014), no.~1, 27--57.

\bibitem[BD]{bjorken}
J.D. Bjorken and S.D. Drell, \emph{Relativistic {Q}uantum {M}echanics},
  McGraw-Hill Book Co., New York, 1964.

\bibitem[B1]{bogachev}
V.I. Bogachev, \emph{Measure theory. {V}ol. {I}}, Springer-Verlag, Berlin,
  2007.

\bibitem[B2]{bognar}
J.~Bogn{\'a}r, \emph{Indefinite {I}nner {P}roduct {S}paces}, Springer-Verlag,
  New York, 1974, Ergebnisse der Ma\-the\-matik und ihrer Grenzgebiete, Band
  78.

\bibitem[BLMS]{sorkin}
L.~Bombelli, J.~Lee, D.~Meyer, and R.D. Sorkin, \emph{Space-time as a causal
  set}, Phys. Rev. Lett. \textbf{59} (1987), no.~5, 521--524.

\bibitem[BDF]{brunetti+dutsch+fredenhagen}
R.~Brunetti, M.~D{\"u}tsch, and K.~Fredenhagen, \emph{Perturbative algebraic
  quantum field theory and the renormalization groups}, arXiv:0901.2038
  [math-ph], Adv. Theor. Math. Phys. \textbf{13} (2009), no.~5, 1541--1599.

\bibitem[Ch]{christensen}
S.M. Christensen, \emph{Vacuum expectation value of the stress tensor in an
  arbitrary curved background: the covariant point-separation method}, Phys.
  Rev. D (3) \textbf{14} (1976), no.~10, 2490--2501.

\bibitem[CL]{coddington}
E.A. Coddington and N.~Levinson, \emph{Theory of {O}rdinary {D}ifferential
  {E}quations}, McGraw-Hill Book Company, Inc., New York-Toronto-London, 1955.

\bibitem[C1]{collins}
J.C. Collins, \emph{Renormalization}, Cambridge Monographs on Mathematical
  Physics, Cambridge University Press, Cambridge, 1984.

\bibitem[C2]{connes}
A.~Connes, \emph{Noncommutative {G}eometry}, Academic Press Inc., San Diego,
  CA, 1994.

\bibitem[DDMS]{merkl}
D.-A. Deckert, D.~D{\"u}rr, F.~Merkl, and M.~Schottenloher,
  \emph{Time-evolution of the external field problem in quantum
  electrodynamics}, arXiv:0906.0046v2 [math-ph], J. Math. Phys. \textbf{51}
  (2010), no.~12, 122301, 28.

\bibitem[DFS]{small}
A.~Diethert, F.~Finster, and D.~Schiefeneder, \emph{Fermion systems in discrete
  space-time exemplifying the spontaneous generation of a causal structure},
  arXiv:0710.4420 [math-ph], Int.\ J.\ Mod.\ Phys. A \textbf{23} (2008),
  no.~27/28, 4579--4620.

\bibitem[D1]{dieudonne1}
J.~Dieudonn{\'e}, \emph{Foundations of {M}odern {A}nalysis}, Academic Press,
  New York-London, 1969, Enlarged and corrected printing, Pure and Applied
  Mathematics, Vol. 10-I.

\bibitem[D2]{dirac3}
P.A.M. Dirac, \emph{Discussion of the infinite distribution of electrons in the
  theory of the positron}, Proc. Camb. Philos. Soc. \textbf{30} (1934),
  150--163.

\bibitem[EG]{epstein+glaser}
H.~Epstein and V.~Glaser, \emph{The role of locality in perturbation theory},
  Ann. Inst. H. Poincar\'e Sect. A (N.S.) \textbf{19} (1973), 211--295.

\bibitem[FS]{fierz+scharf}
H.~Fierz and G.~Scharf, \emph{Particle interpretation for external field
  problems in {QED}}, Helv. Phys. Acta \textbf{52} (1979), no.~4, 437--453.

\bibitem[F1]{endlich}
F.~Finster, \emph{Derivation of field equations from the principle of the
  fermionic projector}, arXiv:gr-qc/9606040 (unpublished preprint in German)
  (1996).

\bibitem[F2]{lightint}
\bysame, \emph{Light-cone expansion of the {D}irac sea with light cone
  integrals}, arXiv:funct-an/9707003, unpublished preprint (1997).

\bibitem[F3]{sea}
\bysame, \emph{Definition of the {D}irac sea in the presence of external
  fields}, arXiv:hep-th/9705006, Adv. Theor. Math. Phys. \textbf{2} (1998),
  no.~5, 963--985.

\bibitem[F4]{U22}
\bysame, \emph{Local {$\rm U(2,2)$} symmetry in relativistic quantum
  mechanics}, arXiv:hep-th/9703083, J. Math. Phys. \textbf{39} (1998), no.~12,
  6276--6290.

\bibitem[F5]{firstorder}
\bysame, \emph{Light-cone expansion of the {D}irac sea to first order in the
  external potential}, arXiv:hep-th/9707128, Michigan Math. J. \textbf{46}
  (1999), no.~2, 377--408.

\bibitem[F6]{light}
\bysame, \emph{Light-cone expansion of the {D}irac sea in the presence of
  chiral and scalar potentials}, arXiv:hep-th/9809019, J. Math. Phys.
  \textbf{41} (2000), no.~10, 6689--6746.

\bibitem[F7]{PFP}
\bysame, \emph{The {P}rinciple of the {F}ermionic {P}rojector}, hep-th/0001048,
  hep-th/0202059, hep-th/0210121, AMS/IP Studies in Advanced Mathematics,
  vol.~35, American Mathematical Society, Providence, RI, 2006.

\bibitem[F8]{rev}
\bysame, \emph{The principle of the fermionic projector: An approach for
  quantum gravity?}, arXiv:gr-qc/0601128, {Q}uantum {G}ravity (B.~Fauser,
  J.~Tolksdorf, and E.~Zeidler, eds.), Birkh\"auser Verlag, Basel, 2006,
  pp.~263--281.

\bibitem[F9]{osymm}
\bysame, \emph{Fermion systems in discrete space-time---outer symmetries and
  spontaneous symmetry breaking}, arXiv:math-ph/0601039, Adv. Theor. Math.
  Phys. \textbf{11} (2007), no.~1, 91--146.

\bibitem[F10]{discrete}
\bysame, \emph{A variational principle in discrete space-time: Existence of
  minimizers}, arXiv:math-ph/0503069, Calc. Var. Partial Differential Equations
  \textbf{29} (2007), no.~4, 431--453.

\bibitem[F11]{reg}
\bysame, \emph{On the regularized fermionic projector of the vacuum},
  arXiv:math-ph/0612003, J. Math. Phys. \textbf{49} (2008), no.~3, 032304, 60.

\bibitem[F12]{lrev}
\bysame, \emph{From discrete space-time to {M}inkowski space: Basic mechanisms,
  methods and perspectives}, arXiv:0712.0685 [math-ph], Quantum {F}ield
  {T}heory (B.~Fauser, J.~Tolksdorf, and E.~Zeidler, eds.), Birkh\"auser
  Verlag, Basel, 2009, pp.~235--259.

\bibitem[F13]{continuum}
\bysame, \emph{Causal variational principles on measure spaces},
  arXiv:0811.2666 [math-ph], J. Reine Angew. Math. \textbf{646} (2010),
  141--194.

\bibitem[F14]{entangle}
\bysame, \emph{Entanglement and second quantization in the framework of the
  fermionic projector}, arXiv:0911.0076 [math-ph], J. Phys. A: Math. Theor.
  \textbf{43} (2010), 395302.

\bibitem[F15]{dice2010}
\bysame, \emph{The fermionic projector, entanglement, and the collapse of the
  wave function}, arXiv:1011.2162 [quant-ph], J. Phys.: Conf. Ser. \textbf{306}
  (2011), 012024.

\bibitem[F16]{srev}
\bysame, \emph{A formulation of quantum field theory realizing a sea of
  interacting {D}irac particles}, arXiv:0911.2102 [hep-th], Lett. Math. Phys.
  \textbf{97} (2011), no.~2, 165--183.

\bibitem[F17]{qft}
\bysame, \emph{Perturbative quantum field theory in the framework of the
  fermionic projector}, arXiv:1310.4121 [math-ph], J. Math. Phys. \textbf{55}
  (2014), no.~4, 042301.

\bibitem[F18]{cfsrev}
\bysame, \emph{Causal fermion systems -- an overview}, arXiv:1505.05075
  [math-ph], {Q}uantum {M}athematical {P}hysics: A {B}ridge
  between {M}athematics and {P}hysics (F.~Finster, J.~Kleiner, C.~Röken, and
  J.~Tolksdorf, eds.), Birkh\"auser Verlag, Basel, 2016, pp.~313--380.

\bibitem[F19]{index}
\bysame, \emph{The chiral index of the fermionic signature operator},
  arXiv:1404.6625 [math-ph], to appear in Math. Res. Lett. (2016).

\bibitem[F20]{qftlimit}
F.~Finster et al, \emph{The quantum field theory limit of causal
  fermion systems}, in preparation.

\bibitem[FG1]{grotz}
F.~Finster and A.~Grotz, \emph{The causal perturbation expansion revisited:
  Rescaling the interacting {D}irac sea}, arXiv:0901.0334 [math-ph], J. Math.
  Phys. \textbf{51} (2010), 072301.

\bibitem[FG2]{lqg}
\bysame, \emph{A {L}orentzian quantum geometry}, arXiv:1107.2026 [math-ph],
  Adv. Theor. Math. Phys. \textbf{16} (2012), no.~4, 1197--1290.

\bibitem[FG3]{cauchy}
\bysame, \emph{On the initial value problem for causal variational principles},
  arXiv:1303.2964 [math-ph], to appear in J. Reine Angew. Math. (2016).

\bibitem[FGS]{rrev}
F.~Finster, A.~Grotz, and D.~Schiefeneder, \emph{Causal fermion systems: A
  quantum space-time emerging from an action principle}, arXiv:1102.2585
  [math-ph], Quantum {F}ield {T}heory and {G}ravity (F.~Finster, O.~M\"uller,
  M.~Nardmann, J.~Tolksdorf, and E.~Zeidler, eds.), Birkh\"auser Verlag, Basel,
  2012, pp.~157--182.

\bibitem[FH]{vacstab}
F.~Finster and S.~Hoch, \emph{An action principle for the masses of {D}irac
  particles}, arXiv:0712.0678 [math-ph], Adv. Theor. Math. Phys. \textbf{13}
  (2009), no.~6, 1653--1711.

\bibitem[FK]{topology}
F.~Finster and N.~Kamran, \emph{Spinors on singular spaces and the topology of
  causal fermion systems}, arXiv:1403.7885 [math-ph] (2014).

\bibitem[FK1]{dice2014}
F.~Finster and J.~Kleiner, \emph{Causal fermion systems as a candidate for a unified physical
  theory}, arXiv:1502.03587 [math-ph], J. Phys.: Conf. Ser. \textbf{626}
  (2015), 012020.

\bibitem[FK2]{noether}
\bysame, \emph{Noether-like theorems for causal variational principles},
  arXiv:1506.09076 [math-ph], Calc. Var. Partial Differential Equations
  \textbf{55:35} (2016), no.~2, 41.

\bibitem[FK3]{jet}
\bysame, \emph{A Hamiltonian formulation of causal variational principles}, in preparation.

\bibitem[FKT]{intro}
F.~Finster, J.~Kleiner, and J.-H. Treude, \emph{An {I}ntroduction to the
  {F}ermionic {P}rojector and {C}ausal {F}ermion {S}ystems}, in preparation.

\bibitem[FM]{drum}
F.~Finster and O.~M\"uller, \emph{Lorentzian spectral geometry for globally
  hyperbolic surfaces}, arXiv:1411.3578 [math-ph], to appear in Adv. Theor.
  Math. Phys. (2017).

\bibitem[FMR]{hadamard}
F.~Finster, S.~Murro, and C.~R\"oken, \emph{The fermionic projector in a
  time-dependent external potential: Mass oscillation property and {H}adamard
  states}, arXiv:1501.05522 [math-ph], J. Math. Phys. \textbf{57} (2016), 072303.
  
\bibitem[FP]{ssymm}
F.~Finster and W.~Plaum, \emph{A lattice model for the fermionic projector in a
  static and isotropic space-time}, arXiv:0712.067 [math-ph], Math. Nachr.
  \textbf{281} (2008), no.~6, 803--816.

\bibitem[FR1]{moritz}
F.~Finster and M.~Reintjes, \emph{The {D}irac equation and the normalization of
  its solutions in a closed {F}riedmann-{R}obertson-{W}alker universe},
  arXiv:0901.0602 [math-ph], Classical Quantum Gravity \textbf{26} (2009),
  no.~10, 105021.

\bibitem[FR2]{finite}
\bysame, \emph{A non-perturbative construction of the fermionic projector on
  globally hyperbolic manifolds {I} -- {S}pace-times of finite lifetime},
  arXiv:1301.5420 [math-ph], Adv. Theor. Math. Phys. \textbf{19} (2015), no.~4,
  761--803.

\bibitem[FR3]{infinite}
\bysame, \emph{A non-perturbative construction of the fermionic projector on
  globally hyperbolic manifolds {II} -- {S}pace-times of infinite lifetime},
  arXiv:1312.7209 [math-ph], to appear in Adv. Theor. Math. Phys. (2016).

\bibitem[FR]{dgc}
F.~Finster and C.~R\"oken, \emph{Dynamical gravitational coupling as a modified
  theory of general relativity}, arXiv:1604.03872 [gr-qc] (2016).

\bibitem[FS]{support}
F.~Finster and D.~Schiefeneder, \emph{On the support of minimizers of causal
  variational principles}, arXiv:1012.1589 [math-ph], Arch. Ration. Mech. Anal.
  \textbf{210} (2013), no.~2, 321--364.

\bibitem[FT1]{loop}
F.~Finster and J.~Tolksdorf, \emph{Bosonic loop diagrams as perturbative solutions of the classical
  field equations in $\phi^4$-theory}, arXiv:1201.5497 [math-ph], J. Math.
  Phys. \textbf{53} (2012), no.~5, 052305.

\bibitem[FT2]{norm}
\bysame, \emph{Perturbative description of the fermionic projector:
  Normalization, causality and {F}urry's theorem}, arXiv:1401.4353 [math-ph],
  J. Math. Phys. \textbf{55} (2014), no.~5, 052301.

\bibitem[Fr]{friedlander2}
F.G. Friedlander, \emph{Introduction to the {T}heory of {D}istributions},
  second ed., Cambridge University Press, Cambridge, 1998, With additional
  material by M. Joshi.

\bibitem[FSW]{fulling+sweeny+wald}
S.A. Fulling, M.~Sweeny, and R.M. Wald, \emph{Singularity structure of the
  two-point function quantum field theory in curved spacetime}, Comm. Math.
  Phys. \textbf{63} (1978), no.~3, 257--264.

\bibitem[GLR]{GLR}
I.~Gohberg, P.~Lancaster, and L.~Rodman, \emph{Indefinite {L}inear {A}lgebra
  and {A}pplications}, Birkh\"auser Verlag, Basel, 2005.

\bibitem[GR]{gradstein}
I.S. Gradshteyn and I.M. Ryzhik, \emph{Table of {I}ntegrals, {S}eries, and
  {P}roducts}, Fourth edition prepared by Ju. V. Geronimus and M. Ju. Ceitlin,
  Academic Press, New York, 1965.

\bibitem[G]{drgrotz}
A.~Grotz, \emph{A {L}orentzian quantum geometry}, Dissertation Universit\"at
  Regensburg, urn:nbn:de:bvb:355-epub-231289, 2011.

\bibitem[HLS1]{hainzl+sere2}
C.~Hainzl, M.~Lewin, and E.~S{\'e}r{\'e}, \emph{Existence of a stable polarized
  vacuum in the {B}ogoliubov-{D}irac-{F}ock approximation},
  arXiv:math-ph/0403005, Comm. Math. Phys. \textbf{257} (2005), no.~3,
  515--562.

\bibitem[HLS2]{hainzl+sere1}
\bysame, \emph{Self-consistent solution for the polarized vacuum in a no-photon
  {QED} model}, arXiv:physics/0404047, J. Phys. A: Math. Theor. \textbf{38}
  (2005), no.~20, 4483--4499.

\bibitem[Ha]{halmosmt}
P.R. Halmos, \emph{Measure {T}heory}, Springer, New York, 1974.

\bibitem[He]{heisenberg2}
W.~Heisenberg, \emph{Bemerkungen zur {D}iracschen {T}heorie des {P}ositrons},
  Z. Phys. \textbf{90} (1934), 209--231.

\bibitem[J]{john}
F.~John, \emph{Partial {D}ifferential {E}quations}, fourth ed., Applied
  Mathematical Sciences, vol.~1, Springer-Verlag, New York, 1991.

\bibitem[JZKGKS]{joos}
E.~Joos, H.D. Zeh, C.~Kiefer, D.~Giulini, J.~Kupsch, and I.-O. Stamatescu,
  \emph{Decoherence and the {A}ppearance of a {C}lassical {W}orld in {Q}uantum
  {T}heory}, second ed., Springer-Verlag, Berlin, 2003.

\bibitem[Ka]{kato}
T.~Kato, \emph{Perturbation {T}heory for {L}inear {O}perators}, Classics in
  Mathematics, Springer-Verlag, Berlin, 1995.

\bibitem[K1]{klaus}
M.~Klaus, \emph{Nonregularity of the {C}oulomb potential in quantum
  electrodynamics}, Helv. Phys. Acta \textbf{53} (1980), no.~1, 36--39.

\bibitem[KS1]{klaus+scharf1}
M.~Klaus and G.~Scharf, \emph{The regular external field problem in quantum
  electrodynamics}, Helv. Phys. Acta \textbf{50} (1977), no.~6, 779--802.

\bibitem[KS2]{klaus+scharf2}
\bysame, \emph{Vacuum polarization in {F}ock space}, Helv. Phys. Acta
  \textbf{50} (1977), no.~6, 803--814.

\bibitem[K2]{kleinert}
H.~Kleinert, \emph{Path {I}ntegrals in {Q}uantum {M}echanics, {S}tatistics,
  {P}olymer {P}hysics, and {F}inancial {M}arkets}, fourth ed., World Scientific
  Publishing Co. Pte. Ltd., Hackensack, NJ, 2006.

\bibitem[LL]{landau2}
L.D. Landau and E.M. Lifshitz, \emph{The {C}lassical {T}heory of {F}ields},
  Revised second edition. Course of Theoretical Physics, Vol. 2. Translated
  from the Russian by Morton Hamermesh, Pergamon Press, Oxford, 1962.

\bibitem[L]{langer}
H.~Langer, \emph{Spectral functions of definitizable operators in {K}rein
  spaces}, Functional {A}nalysis ({D}ubrovnik, 1981), Lecture Notes in Math.,
  vol. 948, Springer, Berlin, 1982, pp.~1--46.

\bibitem[LM]{lawson+michelsohn}
H.B. Lawson, Jr. and M.-L. Michelsohn, \emph{Spin {G}eometry}, Princeton
  Mathematical Series, vol.~38, Princeton University Press, Princeton, NJ,
  1989.

\bibitem[Lax]{lax}
P.D. Lax, \emph{Functional {A}nalysis}, Pure and Applied Mathematics (New
  York), Wiley-Interscience [John Wiley \& Sons], New York, 2002.

\bibitem[NS]{nenciu+scharf}
G.~Nenciu and G.~Scharf, \emph{On regular external fields in quantum
  electrodynamics}, Helv. Phys. Acta \textbf{51} (1978), no.~3, 412--424.

\bibitem[OLBC]{DLMF}
F.W.J. Olver, D.W. Lozier, R.F. Boisvert, and C.W. Clark (eds.), \emph{Digital
  {L}ibrary of {M}athematical {F}unctions}, National Institute of Standards and
  Technology from http://dlmf.nist.gov/ (release date 2011-07-01), Washington,
  DC, 2010.

\bibitem[PS]{peskin+schroeder}
M.E. Peskin and D.V. Schroeder, \emph{An {I}ntroduction to {Q}uantum {F}ield
  {T}heory}, Addison-Wesley Publishing Company Advanced Book Program, Reading,
  MA, 1995.

\bibitem[P]{pokorski}
S.~Pokorski, \emph{Gauge {F}ield {T}heories}, second ed., Cambridge Monographs
  on Mathematical Physics, Cambridge University Press, Cambridge, 2000.

\bibitem[Ra]{radzikowski}
M.J. Radzikowski, \emph{Micro-local approach to the {H}adamard condition in
  quantum field theory on curved space-time}, Comm. Math. Phys. \textbf{179}
  (1996), no.~3, 529--553.

\bibitem[RS1]{reed+simon}
M.~Reed and B.~Simon, \emph{Methods of {M}odern {M}athematical {P}hysics. {I}, {F}unctional
  analysis}, second ed., Academic Press Inc., New York, 1980.

\bibitem[RS2]{reed+simon2}
\bysame, \emph{Methods of modern mathematical physics. {II}.
  {F}ourier analysis, self-adjointness}, Academic Press, New York-London, 1975.

\bibitem[Ro]{ross}
G.G. Ross, \emph{Grand {U}nified {T}heories}, Frontiers in Physics, vol.~60,
  Benjamin/Cummings Publishing Co., Inc., Advanced Book Program, Reading, MA,
  1984.

\bibitem[Ru]{rudinprinciples}
W.~Rudin, \emph{Principles of {M}athematical {A}nalysis}, third ed.,
  McGraw-Hill Book Co., New York-Auckland-D\"usseldorf, 1976, International
  Series in Pure and Applied Mathematics.

\bibitem[S]{sakuraiadv}
J.J. Sakurai, \emph{Advanced {Q}uantum {M}echanics}, Addison-Wesley Publishing
  Company, 1967.

\bibitem[S1]{scharf}
G.~Scharf, \emph{Finite {Q}uantum {E}lectrodynamics}, Texts and Monographs in
  Physics, Springer-Verlag, Berlin, 1989.

\bibitem[S2]{schwabl1}
F.~Schwabl, \emph{Quantum {M}echanics}, third ed., Springer-Verlag, Berlin,
  2002.

\bibitem[Se]{serber}
R.~Serber, \emph{Linear modifications of the {M}axwell field equations}, Phys.
  Rev. \textbf{48} (1935), 49--54.

\bibitem[T]{taylor3}
M.E. Taylor, \emph{Partial {D}ifferential {E}quations. {III}}, Applied
  Mathematical Sciences, vol. 117, Springer-Verlag, New York, 1997.

\bibitem[U]{uehling}
E.A. Uehling, \emph{Polarization effects in the positron theory}, Phys. Rev.
  \textbf{48} (1935), 55--63.

\end{thebibliography}
\providecommand{\bysame}{\leavevmode\hbox to3em{\hrulefill}\thinspace}
\providecommand{\MR}{\relax\ifhmode\unskip\space\fi MR }
\providecommand{\MRhref}[2]{%
  \href{http://www.ams.org/mathscinet-getitem?mr=#1}{#2}
}
\providecommand{\href}[2]{#2}

\ifx\nocomments\undefined
	\Printindex{notation}{Notation Index}
	\Printindex{subject}{Subject Index}
\else
	\Printindex{notation}{Notation Index}
	\Printindex{subject}{Subject Index}
\fi
\newpage

\chapter*{Back Cover}
%
%

\centerline{\large{\bf{About this book}}}

\vspace*{.5em}
This monograph introduces the basic concepts of the theory of causal fermion systems, a recent approach to the description of fundamental physics. The theory yields quantum mechanics, general relativity and quantum field theory as limiting cases and is therefore a candidate for a unified physical theory. From the mathematical perspective, causal fermion systems provide a general framework for describing and analyzing non-smooth geometries and ``quantum geometries.'' The dynamics is described by a novel variational principle, called the causal action principle.

In addition to the basics, the book provides all the necessary mathematical background and explains how the causal action principle gives rise to the interactions of the standard model plus gravity on the level of second-quantized fermionic fields coupled to classical bosonic fields. The focus is on getting a mathematically sound connection between causal fermion systems and physical systems in Minkowski space.

The book is intended for graduate students entering the field, and is furthermore a valuable reference work for researchers in quantum field theory and quantum gravity.

\vspace*{3em}
\centerline{\bf{\large{About the author}}}
\vspace*{.5em}

Felix Finster studied physics and mathematics at the University of Heidelberg, where he graduated in 1992 with Claus Gerhardt and Franz Wegner. In 1992-1995 he wrote his PhD thesis at ETH Z\"urich with Konrad Osterwalder. In 1996-1998 he was a post-doc with Shing-Tung Yau at Harvard University. From 1998-2002 he was member of the Max Planck Institute for Mathematics in the Sciences in Leipzig in the group of Eberhard Zeidler. Since 2002 he has been full professor of mathematics at the University of Regensburg. He works on problems in general relativity and quantum field theory. 

\end{document}